# The Belle II Physics Book

E. Kou[75,¶,†], P. Urquijo[145,§,†], W. Altmannshofer[135,¶], F. Beaujean[79,¶], G. Bell[122,¶], M. Beneke[114,¶], I. I. Bigi[148,¶], F. Bishara[150,16,¶], M. Blanke[49,51,¶], C. Bobeth[113,114,¶], M. Bona[152,¶], N. Brambilla[114,¶], V. M. Braun[50,¶], J. Brod[112,135,¶], A. J. Buras[115,¶], H. Y. Cheng[43,¶], C. W. Chiang[92,¶], M. Ciuchini[59,¶], G. Colangelo[128,¶], A. Crivellin[102,¶], H. Czyz[156,29,¶], A. Datta[146,¶], F. De Fazio[53,¶], T. Deppisch[51,¶], M. J. Dolan[145,¶], J. Evans[135,¶], S. Fajfer[109,141,¶], T. Feldmann[122,¶], S. Godfrey[7,¶], M. Gronau[62,¶], Y. Grossman[15,¶], F. K. Guo[45,134,¶], U. Haisch[150,11,¶], C. Hanhart[21,¶], S. Hashimoto[30,26,¶], S. Hirose[89,¶], J. Hisano[89,90,¶], L. Hofer[127,¶], M. Hoferichter[168,¶], W. S. Hou[92,¶], T. Huber[122,¶], T. Hurth [63,¶], S. Jaeger[159,¶], S. Jahn[83,¶], M. Jamin[126,¶], J. Jones[104,¶], M. Jung[113,¶], A. L. Kagan[135,¶], F. Kahlhoefer[1,¶], J. F. Kamenik[109,141,¶], T. Kaneko[30,26,¶], Y. Kiyo[64,¶], A. Kokulu[114,140,¶], N. Kosnik[109,141,¶], A. S. Kronfeld[20,¶], Z. Ligeti[19,¶], H. Logan[7,¶], C. D. Lu[41,¶], V. Lubicz[153,¶], F. Mahmoudi[142,¶], K. Maltman[173,¶], S. Mishima[30,¶], M. Misiak[166,¶], K. Moats[7,¶], B. Moussallam[74,¶], A. Nefediev[39,88,77,¶], U. Nierste[51,¶], D. Nomura[30,¶], N. Offen[50,¶], S. L. Olsen[133,¶], E. Passemar[37,118,¶], A. Paul[16,31,¶], G. Paz[170,¶], A. A. Petrov[170,¶], A. Pich[164,¶], A. D. Polosa[58,¶], J. Pradler[40,¶], S. Prelovsek[109,141,50,¶], M. Procura[123,¶], G. Ricciardi[54,¶], D. J. Robinson[132,19,¶], P. Roig[9,¶], J. Rosiek[166,¶], S. Schacht[60,15,¶], K. Schmidt-Hoberg[16,¶], J. Schwichtenberg[51,¶], S. R. Sharpe[167,¶], J. Shigemitsu[117,¶], D. Shih[105,¶], N. Shimizu[162,¶], Y. Shimizu[69,¶], L. Silvestrini[58,¶], S. Simula[59,¶], C. Smith[76,¶], P. Stoffer[131,¶], D. Straub[113,¶], F. J. Tackmann[16,¶], M. Tanaka[98,¶], A. Tayduganov[112,¶], G. Tetlalmatzi-Xolocotzi[95,¶], T. Teubner[140,¶], A. Vairo[114,¶], D. van Dyk[114,¶], J. Virto[82,114,¶], Z. Was[93,¶], R. Watanabe[147,¶], I. Watson [155,¶], S. Westhoff[139,¶], J. Zupan[135,¶], R. Zwicky[136,¶], F. Abudinén[83,§], I. Adachi[30,26,§], K. Adamczyk[93,§], P. Ahlburg[129,§], H. Aihara[162,§], A. Aloisio[54,§], L. Andricek[84,§], N. Anh Ky[44,§], M. Arndt[129,§], D. M. Asner[5,§], H. Atmacan[158,§], T. Aushev[87,§], V. Aushev[110,§], R. Ayad[161,§], T. Aziz[111,§], S. Baehr[48,§], S. Bahinipati[33,§], P. Bambade[75,§], Y. Ban[103,§], M. Barrett[170,§], J. Baudot[47,§], P. Behera[36,§], K. Belous[38,§], M. Bender[78,§], J. Bennett[8,§], M. Berger[40,§], E. Bernieri[59,§], F. U. Bernlochner[48,§], M. Bessner[138,§], D. Besson[88,§], S. Bettarini[57,§], V. Bhardwaj[32,§], B. Bhuyan[34,§], T. Bilka[10,§], S. Bilmis[86,§], S. Bilokin[47,§], G. Bonvicini[170,§], A. Bozek[93,§], M. Bračko[144,109,§], P. Branchini[59,§], N. Braun[48,§], R. A. Briere[8,§], T. E. Browder[138,§], L. Burmistrov[75,§], S. Bussino[59,§], L. Cao[48,§], G. Caria[145,§], G. Casarosa[57,§], C. Cecchi[56,§], D. Červenkov[10,§], M.-C. Chang[22,§], P. Chang[92,§], R. Cheaib[146,§], V. Chekelian[83,§], Y. Chen[154,§], B. G. Cheon[28,§], K. Chilikin[77,§], K. Cho[70,§], J. Choi[14,§], S.-K. Choi[27,§], S. Choudhury[35,§], D. Cinabro[170,§], L. M. Cremaldi[146,§], D. Cuesta[47,§], S. Cunliffe[16,§], N. Dash[33,§],

E. de la Cruz Burelo[9,§], E. de Lucia[52,§], G. De Nardo[54,§], M. De Nuccio[16,§], G. De Pietro[59,§], A. De Yta Hernandez[9,§], B. Deschamps[129,§], M. Destefanis[60,§], S. Dey[116,§], F. Di Capua[54,§], S. Di Carlo[75,§], J. Dingfelder[129,§], Z. Doležal[10,§], I. Domínguez Jiménez[125,§], T. V. Dong[30,26,§], D. Dossett[145,§], S. Duell[129,§], S. Eidelman[6,96,77,§], D. Epifanov[6,96,§], J. E. Fast[100,§], T. Ferber[16,§], S. Fiore[18,§], A. Fodor[85,§], F. Forti[57,§], A. Frey[24,§], O. Frost[16,§], B. G. Fulsom[100,§], M. Gabriel[83,§], N. Gabyshev[6,96,§], E. Ganiev[61,§], X. Gao[3,§], B. Gao[23,§], R. Garg[101,§], A. Garmash[6,96,§], V. Gaur[169,§], A. Gaz[90,§], T. Geßler[65,§], U. Gebauer[24,§], M. Gelb[48,§], A. Gellrich[16,§], D. Getzkow[65,§], R. Giordano[54,§], A. Giri[35,§], A. Glazov[16,§], B. Gobbo[61,§], R. Godang[157,§], O. Gogota[110,§], P. Goldenzweig[48,§], B. Golob[141,109,§], W. Gradl[63,§], E. Graziani[59,§], M. Greco[60,§], D. Greenwald[114,§], S. Gribanov[6,96,§], Y. Guan[17,§], E. Guido[60,§], A. Guo[41,§], S. Halder[111,§], K. Hara[30,26,§], O. Hartbrich[138,§], T. Hauth[48,§], K. Hayasaka[94,§], H. Hayashii[91,§], C. Hearty[130,§], I. Heredia De La Cruz[9,§], M. Hernandez Villanueva[9,§], A. Hershenhorn[130,§], T. Higuchi[66,§], M. Hoek[63,§], S. Hollitt[124,§], N. T. Hong Van[44,§], C.-L. Hsu[160,§], Y. Hu[41,§], K. Huang[92,§], T. Iijima[89,90,§], K. Inami[90,§], G. Inguglia[40,§], A. Ishikawa[119,§], R. Itoh[30,26,§], Y. Iwasaki[30,§], M. Iwasaki[97,§], P. Jackson[124,§], W. W. Jacobs[37,§], I. Jaegle[137,§], H. B. Jeon[73,§], X. Ji[41,§], S. Jia[3,§], Y. Jin[162,§], C. Joo[66,§], M. Künzel[16,§], I. Kadenko[110,§], J. Kahn[78,§], H. Kakuno[121,§], A. B. Kaliyar[36,§], J. Kandra[10,§], K. H. Kang[73,§], Y. Kato[90,§], T. Kawasaki[68,§], C. Ketter[138,§], M. Khasmidatul[143,§], H. Kichimi[30,§], J. B. Kim[71,§], K. T. Kim[71,§], H. J. Kim[73,§], D. Y. Kim[108,§], K. Kim[172,§], Y. Kim[172,§], T. D. Kimmel[169,§], H. Kindo[30,26,§], K. Kinoshita[135,§], T. Konno[68,§], A. Korobov[6,96,§], S. Korpar[144,109,§], D. Kotchetkov[138,§], R. Kowalewski[165,§], P. Križan[141,109,§], R. Kroeger[146,§], J.-F. Krohn[145,§], P. Krokovny[6,96,§], W. Kuehn[65,§], T. Kuhr[78,§], R. Kulasiri[67,§], M. Kumar[81,§], R. Kumar[101,§], T. Kumita[121,§], A. Kuzmin[6,96,§], Y.-J. Kwon[172,§], S. Lacaprara[55,§], Y.-T. Lai[30,§], K. Lalwani[81,§], J. S. Lange[65,§], S. C. Lee[73,§], J. Y. Lee[106,§], P. Leitl[83,§], D. Levit[114,§], S. Levonian[16,§], S. Li[3,§], L. K. Li[41,§], Y. Li[41,§], Y. B. Li[103,§], Q. Li[103,§], L. Li Gioi[83,§], J. Libby[36,§], Z. Liptak[138,§], D. Liventsev[169,§], S. Longo[165,§], A. Loos[158,§], G. Lopez Castro[9,§], M. Lubej[109,§], T. Lueck[57,§], F. Luetticke[129,§], T. Luo[23,§], F. Müller[16,§], Th. Müller[48,§], C. MacQueen[145,§], Y. Maeda[90,§], M. Maggiora[60,§], S. Maity[33,§], E. Manoni[56,§], S. Marcello[60,§], C. Marinas[129,§], M. Martinez Hernandez[4,§], A. Martini[59,§], D. Matvienko[6,96,77,§], J. A. McKenna[130,§], F. Meier[160,§], M. Merola[54,§], F. Metzner[48,§], C. Miller[165,§], K. Miyabayashi[91,§], H. Miyake[30,26,§], H. Miyata[94,§], R. Mizuk[77,88,87,§], G. B. Mohanty[111,163,§], H. K. Moon[71,§], T. Moon[106,§], A. Morda[55,§], T. Morii[66,§], M. Mrvar[109,§], G. Muroyama[90,§], R. Mussa[60,§], I. Nakamura[30,26,§], T. Nakano[99,§], M. Nakao[30,26,§], H. Nakayama[30,26,§], H. Nakazawa[92,§], T. Nanut[109,§], M. Naruki[72,§], K. J. Nath[34,§], M. Nayak[116,§], N. Nellikunnummel[151,§], D. Neverov[90,§], C. Niebuhr[16,§], J. Ninkovic[84,§], S. Nishida[30,26,§], K. Nishimura[138,§], M. Nouxman[143,§], G. Nowak[93,§], K. Ogawa[94,§], Y. Onishchuk[110,§], H. Ono[94,§], Y. Onuki[162,§], P. Pakhlov[77,88,§], G. Pakhlova[87,§], B. Pal[5,§], E. Paoloni[57,§], H. Park[73,§], C.-S. Park[172,§], B. Paschen[129,§], A. Passeri[59,§], S. Paul[114,§], T. K. Pedlar[80,§], M. Perelló[46,§], I. M. Peruzzi[52,§], R. Pestotnik[109,§], L. E. Piilonen[169,§], L. Podesta Lerma[125,§], V. Popov[87,§], K. Prasanth[111,§], E. Prencipe[21,§], M. Prim[48,§], M. V. Purohit[158,§], A. Rabusov[114,§], R. Rasheed[47,§], S. Reiter[65,§], M. Remnev[6,96,§], P. K. Resmi[36,§], I. Ripp-Baudot[47,§], M. Ritter[78,§],

M. Ritzert[139,§], G. Rizzo[57,§], L. Rizzuto[141,109,§], S. H. Robertson[85,§], D. Rodriguez Perez[125,§], J. M. Roney[165,§], C. Rosenfeld[158,§], A. Rostomyan[16,§], N. Rout[36,§], S. Rummel[78,§], G. Russo[54,§], D. Sahoo[111,§], Y. Sakai[30,26,§], M. Salehi[143,78,§], D. A. Sanders[146,§], S. Sandilya[135,§], A. Sangal[135,§], L. Santelj[47,§], J. Sasaki[162,§], Y. Sato[30,§], V. Savinov[151,§], B. Scavino[63,§], M. Schram[100,§], H. Schreeck[24,§], J. Schueler[138,§], C. Schwanda[40,§], A. J. Schwartz[135,§], R. M. Seddon[85,§], Y. Seino[94,§], K. Senyo[171,§], O. Seon[90,§], I. S. Seong[138,§], M. E. Sevior[145,§], C. Sfienti[63,§], M. Shapkin[38,§], C. P. Shen[3,§], M. Shimomura[91,§], J.-G. Shiu[92,§], B. Shwartz[6,96,§], A. Sibidanov[165,§], F. Simon[83,113,§], J. B. Singh[101,§], R. Sinha[42,§], S. Skambraks[83,§], K. Smith[145,§], R. J. Sobie[165,§], A. Soffer[116,§], A. Sokolov[38,§], E. Solovieva[77,87,§], B. Spruck[63,§], S. Stanič[149,§], M. Starič[109,§], N. Starinsky[147,§], U. Stolzenberg[24,§], Z. Stottler[169,§], R. Stroili[55,§], J. F. Strube[100,§], J. Stypula[93,§], M. Sumihama[25,§], K. Sumisawa[30,26,§], T. Sumiyoshi[121,§], D. Summers[146,§], W. Sutcliffe[48,§], S. Y. Suzuki[30,26,§], M. Tabata[13,§], M. Takahashi[16,§], M. Takizawa[107,§], U. Tamponi[60,§], J. Tan[145,§], S. Tanaka[30,26,§], K. Tanida[2,§], N. Taniguchi[30,§], Y. Tao[137,§], P. Taras[147,§], G. Tejeda Munoz[4,§], F. Tenchini[16,§], U. Tippawan[12,§], E. Torassa[55,§], K. Trabelsi[30,26,§], T. Tsuboyama[30,26,§], M. Uchida[120,§], S. Uehara[30,26,§], T. Uglov[77,87,§], Y. Unno[28,§], S. Uno[30,26,§], Y. Ushiroda[30,26,162,§], Y. Usov[6,96,§], S. E. Vahsen[138,§], R. van Tonder[48,§], G. Varner[138,§], K. E. Varvell[160,§], A. Vinokurova[6,96,§], L. Vitale[61,§], M. Vos[46,§], A. Vossen[17,§], E. Waheed[145,§], H. Wakeling[85,§], K. Wan[162,§], M.-Z. Wang[92,§], X. L. Wang[23,§], B. Wang[135,§], A. Warburton[85,§], J. Webb[145,§], S. Wehle[16,§], C. Wessel[129,§], J. Wiechczynski[93,§], P. Wieduwilt[24,§], E. Won[71,§], Q. Xu[41,§], X. Xu[41,§], B. D. Yabsley[160,§], S. Yamada[30,§], H. Yamamoto[119,§], W. Yan[3,§], W. Yan[154,§], S. B. Yang[71,§], H. Ye[16,§], I. Yeo[70,§], J. H. Yin[41,§], M. Yonenaga[121,§], T. Yoshinobu[94,§], W. Yuan[55,§], C. Z. Yuan[41,§], Y. Yusa[94,§], S. Zakharov[77,87,§], L. Zani[57,§], M. Zeyrek[86,§], J. Zhang[41,§], Y. Zhang[23,§], Y. Zhang[154,§], X. Zhou[3,§], V. Zhukova[77,§], V. Zhulanov[6,96,§], and A. Zupanc[141,109,§]

[1] *RWTH, Aachen University, D-52056 Aachen, Germany*
[2] *Advanced Science Research Center, Japan Atomic Energy Agency, Naka 319-1195, Japan*
[3] *Beihang University, Beijing 100191, China*
[4] *Benemerita Universidad Autonoma de Puebla, Puebla 72570, Mexico*
[5] *Brookhaven National Laboratory, Upton, New York 11973, USA*
[6] *Budker Institute of Nuclear Physics SB RAS, Novosibirsk 630090, Russian Federation*
[7] *Ottawa-Carleton Institute of Physics, Department of Physics, Carleton University, Ontario K1S 5B6, Canada*
[8] *Carnegie Mellon University, Pittsburgh, Pennsylvania 15213, USA*
[9] *Centro de Investigacion y de Estudios Avanzados del Instituto Politecnico Nacional, Mexico City 07360, Mexico*
[10] *Faculty of Mathematics and Physics, Charles University, 121 16 Prague, Czech Republic*
[11] *CERN, Theoretical Physics Department, CH-1211 Geneva 23, Switzerland*
[12] *Chiang Mai University, Chiang Mai 50202, Thailand*
[13] *Chiba University, Chiba 263-8522, Japan*
[14] *Chonnam National University, Kwangju 660-701, South Korea*

[15]Department of Physics, LEPP, Cornell University, Ithaca, NY 14853, USA
[16]Deutsches Elektronen–Synchrotron, 22607 Hamburg, Germany
[17]Duke University, Durham, North Carolina 27708, USA
[18]ENEA Casaccia, I-00123 Roma, Italy
[19]Ernest Orlando Lawrence Berkeley National Laboratory, University of California, Berkeley, CA 94720, USA
[20]Theoretical Physics Department, Fermi National Accelerator Laboratory, Batavia, IL 60510, USA
[21]Forschungszentrum Jülich, 52425 Jülich, Germany
[22]Department of Physics, Fu Jen Catholic University, Taipei 24205, Taiwan
[23]Key Laboratory of Nuclear Physics and Ion-beam Application (MOE) and Institute of Modern Physics, Fudan University, Shanghai 200443, China
[24]II. Physikalisches Institut, Georg-August-Universität Göttingen, 37073 Göttingen, Germany
[25]Gifu University, Gifu 501-1193, Japan
[26]The Graduate University for Advanced Studies (SOKENDAI), Hayama 240-0193, Japan
[27]Gyeongsang National University, Chinju 660-701, South Korea
[28]Hanyang University, Seoul 133-791, South Korea
[29]Helmholtz-Institute 55128 Mainz, Germany
[30]High Energy Accelerator Research Organization (KEK), Tsukuba 305-0801, Japan
[31]Institut für Physik, Humboldt-Universität zu Berlin, D-12489 Berlin, Germany
[32]Indian Institute of Science Education and Research Mohali, SAS Nagar, 140306, India
[33]Indian Institute of Technology Bhubaneswar, Satya Nagar 751007, India
[34]Indian Institute of Technology Guwahati, Assam 781039, India
[35]Indian Institute of Technology Hyderabad, Telangana 502285, India
[36]Indian Institute of Technology Madras, Chennai 600036, India
[37]Indiana University, Bloomington, Indiana 47408, USA
[38]Institute for High Energy Physics, Protvino 142281, Russian Federation
[39]Institute for Theoretical and Experimental Physics, B. Cheremushkinskaya 25, 117218 Moscow, Russia
[40]Institute of High Energy Physics, Vienna 1050, Austria
[41]Institute of High Energy Physics, Chinese Academy of Sciences, Beijing 100049, China
[42]Institute of Mathematical Sciences, Chennai 600113, India
[43]Institute of Physics, Academia Sinica Taipei, Taiwan 115, Republic of China
[44]Institute of Physics, VAST, Hanoi, Vietnam
[45] Institute of Theoretical Physics, Chinese Academy of Sciences, Beijing 100190, China
[46]Instituto de Fisica Corpuscular, Paterna 46980, Spain
[47]Institut Pluridisciplinaire Hubert Curien, Strasbourg 67037, France
[48]Institut für Experimentelle Teilchenphysik, Karlsruhe Institute of Technology, 76131 Karlsruhe, Germany
[49]Institut für Kernphysik, Karlsruhe Institute of Technology, Hermann-von-Helmholtz-Platz 1, D-76344 Eggenstein-Leopoldshafen, Germany

[50]*Institut für Theoretische Physik, Universität Regensburg, D-93040 Regensburg, Germany*
[51]*Institut für Theoretische Teilchenphysik, Karlsruhe Institute of Technology, Wolfgang-Gaede-Str. 1, D-76131 Karlsruhe, Germany*
[52]*INFN Laboratori Nazionali di Frascati, I-00044 Frascati, Italy*
[53]*INFN Sezione di Bari, Via Orabona 4, I-70126 Bari, Italy*
[54]*INFN Sezione di Napoli and Dipartimento di Scienze Fisiche, Università di Napoli Federico II, I-80126 Napoli, Italy*
[55]*INFN Sezione di Padova and Dipartimento di Fisica, Università di Padova, I-35131 Padova, Italy*
[56]*INFN Sezione di Perugia and Dipartimento di Fisica, Università di Perugia, I-06123 Perugia, Italy*
[57]*INFN Sezione di Pisa and Dipartimento di Fisica, Università di Pisa, I-56127 Pisa, Italy*
[58]*INFN Sezione di Roma and Università di Roma "La Sapienza", I-00185 Roma, Italy*
[59]*INFN Sezione di Roma Tre and Università di Roma Tre, I-00146 Roma, Italy*
[60]*INFN Sezione di Torino and Dipartimento di Fisica, Università di Torino, I-10125 Torino, Italy*
[61]*INFN Sezione di Trieste and Dipartimento di Fisica, Università di Trieste, I-34127 Trieste, Italy*
[62]*Physics Department, Technion, Israel Institute of Technology, 32000 Haifa, Israel*
[63]*Johannes Gutenberg-Universität Mainz, Institut für Kernphysik, D-55099 Mainz, Germany*
[64]*Department of Physics, Juntendo University, Inzai, Chiba 270-1695, Japan*
[65]*Justus-Liebig-Universität Gießen, 35392 Gießen, Germany*
[66]*Kavli Institute for the Physics and Mathematics of the Universe (WPI), University of Tokyo, Kashiwa 277-8583, Japan*
[67]*Kennesaw State University, Kennesaw, Georgia 30144, USA*
[68]*Kitasato University, Tokyo 108-0072, Japan*
[69]*Kogakuin University, 2665-1 Nakano, Hachioji, Tokyo 192-0015, Japan*
[70]*Korea Institute of Science and Technology Information, Daejeon 305-806, South Korea*
[71]*Korea University, Seoul 136-713, South Korea*
[72]*Kyoto University, Kyoto 606-8502, Japan*
[73]*Kyungpook National University, Daegu 702-701, South Korea*
[74]*L'institut de physique nucléaire d'orsay, IN2P3/CNRS et Université Paris-Sud 11, Centre Scientifique d'Orsay, F-91898 Orsay Cedex, France*
[75]*Laboratoire de l'Accélérateur Linéaire, IN2P3/CNRS et Université Paris-Sud 11, Centre Scientifique d'Orsay, F-91898 Orsay Cedex, France*
[76]*Laboratoire de Physique Subatomique et Cosmologie, IN2P3/CNRS et Université Grenoble Alpes, 38000 Grenoble, France*
[77]*P.N. Lebedev Physical Institute of the Russian Academy of Sciences, Moscow 119991, Russian Federation*
[78]*Ludwig Maximilians University, 80539 Munich, Germany*
[79]*Excellence cluster universe, Ludwig Maximilians University, 80539 Munich, Germany*

[80]Luther College, Decorah, Iowa 52101, USA
[81]Malaviya National Institute of Technology Jaipur, Jaipur 302017, India
[82]Center for Theoretical Physics, Massachusetts Institute of Technology, Cambridge, MA 02139, USA
[83]Max-Planck-Institut für Physik, 80805 München, Germany
[84]Semiconductor Laboratory of the Max Planck Society, Max-Planck-Institut für Physik, 81739 München, Germany
[85]McGill University, Montréal, Québec, H3A 2T8, Canada
[86]Middle East Technical University, 06531 Ankara, Turkey
[87]Moscow Institute of Physics and Technology, Moscow Region 141700, Russian Federation
[88]Moscow Physical Engineering Institute, Moscow 115409, Russian Federation
[89]Kobayashi-Maskawa Institute, Nagoya University, Nagoya 464-8602, Japan
[90]Graduate School of Science, Nagoya University, Nagoya 464-8602, Japan
[91]Nara Women's University, Nara 630-8506, Japan
[92]Department of Physics, National Taiwan University, Taipei 10617, Taiwan
[93]H. Niewodniczanski Institute of Nuclear Physics, Krakow 31-342, Poland
[94]Niigata University, Niigata 950-2181, Japan
[95]Nikhef, Science Park 105, NL-1098 XG Amsterdam, Netherlands
[96]Novosibirsk State University, Novosibirsk 630090, Russian Federation
[97]Osaka City University, Osaka 558-8585, Japan
[98]Department of Physics, Graduate School of Science, Osaka University, Toyonaka, Osaka 560-0043, Japan
[99]Research Center for Nuclear Physics, Osaka University, Osaka 567-0047, Japan
[100]Pacific Northwest National Laboratory, Richland, Washington 99352, USA
[101]Panjab University, Chandigarh 160014, India
[102]Paul Scherrer Institut, CH-5232 Villigen PSI, Switzerland
[103]Peking University, Beijing 100871, China
[104]Seccion Fisica, Departamento de Ciencias, Pontificia Universidad Catolica del Peru, Apartado 1761, Lima, Peru
[105]NHETC, Rutgers University, Piscataway, NJ 08854 USA
[106]Seoul National University, Seoul 151-742, South Korea
[107]Showa Pharmaceutical University, Tokyo 194-8543, Japan
[108]Soongsil University, Seoul 156-743, South Korea
[109]J. Stefan Institute, 1000 Ljubljana, Slovenia
[110]Taras Shevchenko National University of Kyiv, Kiev, Ukraine
[111]Tata Institute of Fundamental Research, Mumbai 400005, India
[112]Technische Universität Dortmund, Fakultät Physik, Otto-Hahn-Str. 4, D-44227 Dortmund, Germany
[113]Excellence Cluster Universe, Technische Universität München, 85748 Garching, Germany
[114]Department of Physics, Technische Universität München, 85748 Garching, Germany
[115]Institute for Advanced Study, Technische Universität München, Lichtenbergstr. 2a,D-85748 Garching
[116]Tel Aviv University, School of Physics and Astronomy, Tel Aviv, 69978, Israel
[117]The Ohio State University, Department of Physics, Columbus, OH 43210, USA

[118] *Theory Center, Thomas Jefferson National Accelerator Facility, Newport News, VA 23606, USA*
[119] *Department of Physics, Tohoku University, Sendai 980-8578, Japan*
[120] *Tokyo Institute of Technology, Tokyo 152-8550, Japan*
[121] *Tokyo Metropolitan University, Tokyo 192-0397, Japan*
[122] *Theoretische Physik 1, Naturwissenschaftlich-Technische Fakultät, Universität Siegen, D-57068 Siegen, Germany*
[123] *Fakultät für Physik, Universität Wien, Boltzmanngasse 5, Vienna 1090, Austria*
[124] *Department of Physics, University of Adelaide, Adelaide, South Australia 5005, Australia*
[125] *Universidad Autonoma de Sinaloa, Sinaloa 80000, Mexico*
[126] *IFAE, Autonomous University of Barcelona, Spain*
[127] *Departament de Fisica Quántica i Astrofisica (FQA), Institut de Ciéncies del Cosmos (ICCUB), Universitat de Barcelona (UB), Spain*
[128] *Albert Einstein Center for Fundamental Physics, Institute for Theoretical Physics, University of Bern, Sidlerstrasse 5, 3012 Bern, Switzerland*
[129] *University of Bonn, 53115 Bonn, Germany*
[130] *University of British Columbia, Vancouver, British Columbia, V6T 1Z1, Canada*
[131] *Department of Physics, University of California at San Diego, 9500 Gilman Drive, La Jolla, CA 92093-0319, USA*
[132] *Santa Cruz Institute for Particle Physics and Department of Physics, University of Calfornia at Santa Cruz, Santa Cruz CA 95064, USA*
[133] *University of the Chinese Academy of Science, Beijing, 100049, People's Republic of China*
[134] *School of Physical Sciences, University of Chinese Academy of Sciences, Beijing 100049, China*
[135] *University of Cincinnati, Cincinnati, Ohio 45221, USA*
[136] *Higgs Centre For Theoretical Physics, School of Physics and Astronomy, University of Edinburgh, Edinburgh EH9 3JZ, Scotland*
[137] *University of Florida, Gainesville, Florida 32611, USA*
[138] *University of Hawaii, Honolulu, Hawaii 96822, USA*
[139] *University of Heidelberg, 68131 Mannheim, Germany*
[140] *Department of Mathematical Sciences, University of Liverpool, Liverpool L69 3BX, UK*
[141] *Faculty of Mathematics and Physics, University of Ljubljana, 1000 Ljubljana, Slovenia*
[142] *Université Lyon 1, ENS de Lyon, CNRS, Centre de Recherche Astrophysique de Lyon UMR5574, F-69230 Saint-Genis-Laval, France*
[143] *University of Malaya, 50603 Kuala Lumpur, Malaysia*
[144] *University of Maribor, 2000 Maribor, Slovenia*
[145] *School of Physics, The University of Melbourne, Victoria 3010, Australia*
[146] *University of Mississippi, University, Mississippi 38677, USA*
[147] *Université de Montréal, Physique des Particules, Montréal, Québec, H3C 3J7, Canada*
[148] *Department of Physics, University of Notre Dame du Lac Notre Dame, IN 46556, USA*
[149] *University of Nova Gorica, 5000 Nova Gorica, Slovenia*

[150] *Rudolf Peierls Centre for Theoretical Physics, University of Oxford, OX1 3NP Oxford, UK*
[151] *University of Pittsburgh, Pittsburgh, Pennsylvania 15260, USA*
[152] *Queen Mary University of London, School of Physics and Astronomy, Mile End Road, London E1 4NS, UK*
[153] *Dipartimento di Matematica e Fisica, Università di Roma Tre, Via della Vasca Navale 84, I-00146 Roma, Italy*
[154] *University of Science and Technology of China, Hefei 230026, China*
[155] *University of Seoul, 130-743 Seoul, South Korea*
[156] *Institute of Physics, University of Silesia, PL-41500 Chorzow, Poland*
[157] *University of South Alabama, Mobile, Alabama 36688, USA*
[158] *University of South Carolina, Columbia, South Carolina 29208, USA*
[159] *Department of Physics and Astronomy, University of Sussex, Falmer, Brighton BN1 9QH, UK*
[160] *School of Physics, University of Sydney, New South Wales 2006, Australia*
[161] *Department of Physics, Faculty of Science, University of Tabuk, Tabuk 71451, Saudi Arabia*
[162] *Department of Physics, University of Tokyo, Tokyo 113-0033, Japan*
[163] *Utkal University, Bhubaneswar 751004, India*
[164] *IFIC, University of Valencia - CSIC, Spain*
[165] *University of Victoria, Victoria, British Columbia, V8W 3P6, Canada*
[166] *Institute of Theoretical Physics, Faculty of Physics, University of Warsaw, 02-093 Warsaw, Poland*
[167] *Physics Department, University of Washington, Seattle WA 98195, USA*
[168] *Institute for Nuclear Theory, University of Washington, Seattle, WA 98195-1550, USA*
[169] *Virginia Polytechnic Institute and State University, Blacksburg, Virginia 24061, USA*
[170] *Wayne State University, Detroit, Michigan 48202, USA*
[171] *Yamagata University, Yamagata 990-8560, Japan*
[172] *Yonsei University, Seoul 120-749, South Korea*
[173] *Department of Mathematics and Statistics, York University, Toronto, Ontario M3J 1P3, Canada*
[†] *Editor.*
[§] *Belle II Collaborator.*
[¶] *Theory or external contributing author.*

# 1. Preface

The Belle II-Theory interface Platform (B2TiP) was created as a physics prospects working group of the Belle II collaboration in June 2014. It offered a platform where theorists and experimentalists work together to elucidate the potential impacts of the Belle II program, which includes a wide scope of physics topics: $B$ physics, charm, $\tau$, quarkonium physics, electroweak precision measurements and dark sector searches. It is composed of nine working groups (WGs), which are coordinated by teams of theory and experiment conveners: WG1) Semileptonic and leptonic $B$ decays, WG2) Radiative and Electroweak penguins, WG3) $\phi_1$ and $\phi_2$ (time-dependent $CP$ violation) measurements, WG4) $\phi_3$ measurement, WG5) Charmless hadronic $B$ decay, WG6) Charm, WG7) Quarkonium(like), WG8) $\tau$ and low-multiplicity processes, WG9) New Physics. We organised workshops twice a year from 2014 until 2016, which moved from KEK in Japan to Europe and the Americas, gathering experts in the respective fields to discuss with Belle II members.

One of the goals for B2TiP was to propose so-called “golden- and silver-channels”: we asked each working group to choose among numerous possible measurements, those that would have the highest potential impact and to focus on them for the writeup. Theorists scrutinised the role of those measurements in terms of understanding the theory behind them, and estimated the theoretical uncertainties, now achievable as well as prospects for the future. For flavour physics, having tight control of hadronic uncertainties is one of the most crucial aspects in the field, and this is considered as an important criteria to determine the golden or silver-channels. Experimentalists, on the other hand, investigated the expected improvements with data from Belle II. For the channels where the errors are dominated by statistical uncertainties, or where systematic errors are reducible, the errors can decrease rapidly as more data becomes available. The impact of the upgraded performance from Belle II is a crucial element for reducing the uncertainties: we therefore include the latest available studies of the detector efficiency using Monte Carlo simulated events. We list the golden(silver)-channel table in this first chapter, as a guide for the chapters that follow.

The book is not a collection of reports based on talks given at the workshops. The working group conveners endeavoured to construct a coherent document that can be used by Belle II collaborators, and others in the field of flavour physics, as a reference. There were two books of a similar type written in the past, “The BaBar book” [1] and “The Physics of the B factories” [2]. In order to avoid too much repetition with respect to those references, we refer to them as possible for introductory material.

We would like to thank the section editors and contributing authors for the many stimulating discussions and their tremendous effort to bring the book together.

## *1.1. Working Groups*

The Belle II Theory Interface Platform working groups and convenors were assigned as follows.

**Leptonic and Semileptonic $B$ decays**

Experiment: G. De Nardo (Naples), A. Zupanc (IJS)

Theory: F. Tackmann (DESY), A. Kronfeld (FNAL, *LQCD*), R. Watanabe (Montreal)

**Radiative and Electroweak Penguin $B$ decays**
Experiment: A. Ishikawa (Tohoku), J. Yamaoka (PNNL)
Theory: U. Haisch (Oxford), T. Feldmann (Siegen)

**Time Dependent $CP$ Violation of $B$ mesons**
Experiment: A. Gaz (Nagoya), L. Li Gioi (MPI Munich)
Theory: S. Mishima (Rome/KEK), J. Zupan (Cincinnati)

**Determination of the Unitarity Triangle angle $\phi_3$**
Experiment: J. Libby (IIT Madras)
Theory: Y. Grossman (Cornell), M. Blanke (CERN)

**Hadronic $B$ decays and direct $CP$ Violation**
Experiment: P. Goldenzweig (KIT)
Theory: M. Beneke (TUM), C-W. Chiang (NCU)

**Charm flavour and spectroscopy**
Experiment: G. Casarosa (Pisa), A. Schwartz (Cincinnati)
Theory: A. Petrov (Wayne), A. Kagan (Cincinnati)

**Quarkonium(like) physics**
Experiment: B. Fulsom (PNNL), R. Mizuk (ITEP), R. Mussa (Torino), C-P. Shen (Beihang)
Theory: N. Brambilla (TUM), C. Hanhart (Juelich), Y. Kiyo (Juntendo), A. Polosa (Rome), S. Prelovsek (Ljubljana, *LQCD*)

**Tau decays and low-multiplicity physics**
Experiment: K. Hayasaka (Nagoya), T. Ferber (DESY)
Theory: E. Passemar (Indiana), J. Hisano (Nagoya)

**New physics and global analyses**
Experiment: F. Bernlochner (Bonn), R. Itoh (KEK)
Theory: J. Kamenik (Ljubljana), U. Nierste (KIT), L. Silvestrini (Rome)

Further direct contributors to the chapters are given in the chapter headers.

## *1.2. Committees*

The B2TiP workshop and book organising committee is comprised of

- Emi Kou (LAL)
- Phillip Urquijo (Melbourne)

An international advisory committee assisted in steering the coordination of the workshops and report.

- Marco Ciuchini (Rome)
- Tim Gershon (Warwick)

- Bostjan Golob (IJS)
- Shoji Hashimoto (KEK)
- Francois Le Diberder (LAL)
- Zoltan Ligeti (LBNL)
- Thomas Mannel (Siegen)
- Hitoshi Murayama (IPMU)
- Matthias Neubert (Mainz)
- Junko Shigemitsu (Ohio)

The Belle II experiment ex-officio comprised of

- Francesco Forti (Pisa)
- Thomas Browder (Hawaii)
- Yoshihide Sakai (KEK)

## *1.3. Workshops*

This report is the culmination of a two-year workshop series held to develop the physics program for Belle II. The schedule for the workshops was as follows.

| | |
|---|---|
| KEK, Kickoff meeting | 16th–17th June 2014 |
| KEK, Joint KEK-FF / 1st B2TiP workshop | 30th–31st October 2014 |
| Krakow, 2nd B2TiP workshop | 27th–28th April 2015 |
| KEK, Joint KEK-FF / 3rd B2TiP workshop | 28th–29th October 2015 |
| Pittsburgh, 4th B2TiP workshop | 23rd–25th May 2016 |
| Munich, 5th B2TiP workshop and editorial meeting | 15th–17th November 2016 |

*Contents* PAGE

# 2. Introduction

*Section author(s): P. Urquijo, E. Kou*

## 2.1. Introduction

The primary physics goals of Belle II, as a next generation flavour factory, are to search for new physics (NP) in the flavour sector at the intensity frontier, and to improve the precision of measurements of Standard Model (SM) parameters. The SuperKEKB facility is designed to collide electrons and positrons at centre-of-mass energies in the regions of the $\Upsilon$ resonances. Most of the data will be collected at the $\Upsilon(4S)$ resonance, which is just above threshold for $B$-meson pair production where no fragmentation particles are produced. The accelerator is designed with asymmetric beam energies to provide a boost to the centre-of-mass system and thereby allow for time-dependent charge-parity ($CP$) symmetry violation measurements. The boost is slightly less than that at KEKB, which is advantageous for analyses with neutrinos in the final state that require good detector hermeticity, although it requires better vertex reconstruction resolution. SuperKEKB has a design luminosity of $8 \times 10^{35}\text{cm}^{-2}\text{s}^{-1}$, about 40 times larger that of KEKB's recorded peak and 80 times KEKB's design luminosity. This luminosity will produce a total of $5 \times 10^{10}$ $b$, $c$ and $\tau$ pairs over a period of 8 years. The first collision data taking run was undertaken in 2018, used primarily for accelerator and detector commissioning.

The SM is, at the current level of experimental precision and at the energies reached so far, the best tested theory of nature at a fundamental level. Despite its tremendous success in describing the fundamental particles and their interactions, excluding gravity, it does not provide answers to many fundamental questions. The SM does not explain why there should be only three generations of elementary fermions and why there is an observed hierarchy in the fermion masses. The origin of mass of fundamental particles is explained within the SM by spontaneous electroweak symmetry breaking, resulting in the Higgs boson. However, it is not clear whether the Higgs boson can account for neutrino masses. It is also not yet clear whether there is a only single SM Higgs boson or whether there may be a more elaborate Higgs sector with other Higgs-like particle as in supersymmetry or other NP models. At the cosmological scale, there is the unresolved problem with the matter-antimatter asymmetry in the universe. While the violation of $CP$ symmetry is a necessary condition for the evolution of a matter-dominated universe, the observed $CP$ violation within the quark sector that originates from the complex phase of the Cabibbo-Kobayashi-Maskawa (CKM) matrix is many orders of magnitude too small to explain the dominance of matter in the universe. Hence, there must exist undiscovered sources of the $CP$ asymmetry. Furthermore, the elements of the CKM matrix exhibit a roughly diagonal hierarchy, even though the SM does not require this. This may indicate the presence of a new mechanism, such as a flavour symmetry, that exists unbroken at a higher energy scale. Considering the open questions that in the SM remain unanswered, it is fair to conclude that the present theory is an extremely successful but phenomenological description of subatomic processes at the energy scales up to $\mathcal{O}(1\text{ TeV})$. Many New Physics (NP) scenarios have been proposed to explain these shortcoming of the SM, where new particles and new processes arise.

Experiments in high energy physics are designed to address the above questions through searches of NP using complementary approaches. At the energy frontier, the LHC experiments are able to discover new particles produced in proton-proton collisions at a centre-of-mass energy of up to 14 TeV. Sensitivity to the direct production of a specific new particle depends on the cross section and on the size of the data sample. At the intensity frontier, signatures of new particles or processes can be observed through measurements of suppressed flavour physics reactions or from deviations from SM predictions. An observed discrepancy can be interpreted in terms of NP models. This is the approach of Belle II.

The sensitivity of Belle II to NP depends on the strength of the flavour violating couplings of the NP. The mass reach for new particle/process effects can be as high as $\mathcal{O}(100 \text{ TeV})$ if the couplings are not as suppressed as in the SM [3]. In the past, measurements of processes quantum corrections have given access to high mass scale physics before accelerators were available to directly probe these scales. Belle II and SuperKEKB will exploit our strengths at the intensity frontier by moving beyond a simple observation of a NP effect to its detailed characterisation through over-constraining measurements in several related flavour physics reactions.

## *2.2. New physics search strategy after the B-factories and LHC run I and run II first data*

The LHC experiments, ATLAS, CMS, LHCb, have been operating extremely well since its commencement in 2009 and are rapidly changing the scene of particle physics. Needless to say, the discovery of the Higgs boson in 2012 was the most significant event in particle physics in recent years. Its mass, 125 GeV, and its production and decay patterns being SM-like not only provides a confirmation of the SM but also puts very strong constraints on the Higgs sector of various new physics models (especially those that contain more than one neutral Higgs). The 125 GeV Higgs has excluded a large parameter space of minimal SUSY models, from which we expected some signals in Belle II observables. The mass constraints on the direct searches of new particles are also advancing as well. For example, the lower mass bounds of the new gauge bosons , $m_{Z',W'}$ within the sequential model (*i.e.* SM-like) has been pushed up to ~3 TeV and the vector like fermion masses now exceed ~ 800 GeV. Since new physics effects in Belle II observables are roughly proportional to the inverse of the mass of these particles (with powers of 2, 3, 4 etc. depending on the observable), the chance to observe a signal from such generic models is diminishing. However, it is important to note that, such minimal or generic models are often quite unnatural from the theoretical point of view since we need to impose a very high degree of symmetry (*i.e.* to mimic SM) to realise them. New physics models that we search for in Belle II are those that include more specific flavour couplings, for which indirect searches can push the new physics scale much higher than the direct search programs. Hints of new physics in previous and on-going experiments may provide us with some indication of the kind of new flavour phenomena that we should look for.

An important flavour coupling structure to examine for new physics are $b \to s$ transitions, which been a focus of both theory and experiment in recent years. Since the outset of the $B$-factory experiments precise $CP$ violation measurements in the $B_d$ system have been done using tree level $b \to c\bar{c}s$ transitions (such as the golden mode $B \to J/\psi K^0_S$ final state), and as time went by, the B factories started observing $CP$ violation through the loop induced $b \to s$ transitions, such as $B \to \phi K_S^0$ or $B \to \eta' K_S^0$ processes (the first observation in 2003 had shown a small tension as well). The $b \to s\bar{q}q$ transition is induced by gluon penguin diagrams. In the SM, $CP$ violation in $b \to s$ transitions is expected to be very small. Thus, any significant observation of $CP$ violation can be interpreted a signal beyond the SM. This is a new area of research in B physics as the precision is still very far from the measurements using the tree level processes and there is a lot room for new physics contributions. It is worth mentioning that a small tension has also been observed in another type of $CP$ violation, direct $CP$ violation, in $B \to K\pi$ decays, which also occurs in part due to penguin $b \to s\bar{q}q$ transitions. The LHCb experiment is in the right position to tackle this question from a different direction, by measuring the parameters of $B_s - \overline{B}_s$ mixing, which occurs due to another type of loop diagram, the $b \to s$ box process. So far, the LHCb results for this observable are consistent with the SM. However, since 2013, LHCb has started observing a few very interesting deviations from the SM in the other $b \to s$ transitions such as in the $B \to K^*\mu^+\mu^-$ angular distribution and the ratio of rates of $B \to K^{(*)}e^+e^-$ to $B \to K^{(*)}\mu^+\mu^-$ (so-called $R(K^{(*)})$). Those excesses are said to be reaching the 4-5 $\sigma$ level. The very specific appearances of these anomalies were not predicted and they are opening a new trend in particle physics: the new particles with very distinct flavour couplings in $b \to s$ transition as well as a possible lepton universality violation.

Another important hint of new physics was in the measurement of the branching ratio of $B \to \tau\nu$, which in 2006 had shown a deviation from SM expectation (in particular $|V_{ub}|$ measured from other channels). There is now some tension between measurements by Belle and BaBar of this rate. This is a tree level annihilation $b \to u$ transition and the final state includes at least two neutrinos so it is experimentally quite challenging: so far the B factories reconstructed only a few hundred events. As $B \to \tau\nu$ is particularly sensitive to the charged Higgs that in general couples more strongly to a heavier particles, this result is somehow natural from the new physics point of view. Even more intriguingly, other anomalies were reported in similar channels, $B \to D{*}\tau\nu$ and $B \to D\tau\nu$ by the BaBar, Belle and LHCb collaborations. The tension with the SM is now reaching to the $\sim 4\sigma$ level. These results may be indicating that tau leptons have a very unique sensitivity to new physics. As mentioned above, the identification of the decay modes involving tau leptons is challenging, but they will become readily accessible at Belle II. Thus, the flavour structure with distinguished tau lepton coupling will be tested at the Belle II at a higher precision.

### *2.3. Flavour physics questions to be addressed by Belle II*

Further study of the quark sector is necessary to reveal NP at high mass scales, even beyond the direct reach of the LHC, that may manifest in flavour observables. There are several important questions that can only be addressed by further studies of flavour physics, as described below. Belle II will access a large number of new observables to test for NP in flavour transitions in the quark and lepton sectors.

- *Are there new $CP$ violating phases in the quark sector?* The amount of $CP$ violation in the SM quark sector is orders of magnitude too small to explain the baryon-antibaryon asymmetry. New insights will come from examining the difference between $B^0$ and $\bar{B}^0$ decay rates, namely via measurements of time-dependent $CP$ violation in penguin transitions of $b \to s$ and $b \to d$ quarks, such as $B \to \phi K^0$ and $B \to \eta' K^0$. $CP$ violation in charm mixing, which is negligible in the SM, will also provide information on new phenomena in the up-type quark sector. Another key area will be to understand the mechanisms that produced large amounts of $CP$ violation in the time integrated rates of charmless hadronic $B$ decays, such as $B \to K\pi$ and $B \to K\pi\pi$, observed by the $B$-factories and LHCb.

- *Does nature have multiple Higgs bosons?* Many extensions to the SM, such as two-Higgs-doublet models, predict charged Higgs bosons in addition to a neutral SM-like Higgs. The charged Higgs will be searched for in flavour transitions to $\tau$ leptons, including $B \to \tau\nu$ and $B \to D^{(*)}\tau\nu$. Deviations from the SM have been observed in the latter with significance greater than $3\sigma$. Extended Higgs mechanisms can also introduce extra sources of $CP$ violation.
- *Does nature have a left-right symmetry, and are there flavour-changing neutral currents beyond the SM?* Approaches include measurements of time-dependent $CP$ violation in $B \to K^{*0}(\to K^0_S\pi^0)\gamma$, triple-product $CP$ violation asymmetries in $B \to VV$ decays, and semileptonic decays $B \to V\ell\nu$, $V = D^*,\ \rho$. It is of great interest to measure $b \to s\nu\bar{\nu}$ transitions such as $B \to K^{(*)}\nu\bar{\nu}$, part of a class of decays with large missing energy. It is also important to improve FCNCs measurements of $b \to d$, $b \to s$ and $c \to u$ transitions. It is crucial to measure forward-backward asymmetries as a function of the $q^2$ of the dilepton, $A_{\text{FB}}(q^2)$, in inclusive $b \to s\ell^+\ell^-$ decays and in charged weak interactions.
- *Are there sources of lepton flavour violation (LFV) beyond the SM?* Neutrino experiments have found large mixing involving $\tau$ neutrinos, raising the question: are there flavour changing processes such as $\tau \to \mu\gamma$ visible at the $10^{-8}$ level? LFV in charged lepton decay at such rates are key predictions in many neutrino mass generation mechanisms and other models of physics beyond the SM. The expected sensitivities to $\tau$ decays will be unrivalled due to correlated production with minimal collision background. Belle II will analyse $\tau$ leptons in for LF, $CP$ violation, measurements of the electric dipole moment, and $(g-2)$ of the $\tau$.

It is also worth noting that Belle II will measure the current array of CKM observables, the matrix elements and their phases, with unprecedented precision.

## *2.4. Non-flavour program physics case*

Belle II will be able to address fundamental questions not directly related to flavour physics, leveraging from the clean environment of $e^+e^-$ collisions, and the large data set. Two of the driving questions are as follows.

- *Is there a dark sector of particle physics at the same mass scale as ordinary matter?* Belle II has unique sensitivity to dark matter via missing energy decays. While most searches for new physics at Belle II are indirect, there are models that predict new particles at the MeV to GeV scale - including weakly and non-weakly interacting massive particles that couple to the SM via new gauge symmetries. These models often predict a rich sector of hidden particles that include dark-matter candidates and gauge bosons. Belle II is implementing new trigger strategies, such as a single photon trigger, to capture these elusive events.

- *What is the nature of the strong force in binding hadrons?* With $B$ factories and hadron colliders having discovered a large number of states that were not predicted by the conventional meson interpretation, changing our understanding of QCD in the low-energy regime, study of quarkonia is high on the agenda at Belle II. New particles can be produced near resonance, achievable by adjusting the machine energy, or by initial state radiation, which effectively provides a continuum of centre of mass energies. Belle II has near hermetic coverage and good detection capabilities for all neutral and charged particles, and can play a central role in these analyses.

### *2.5. Advantages of SuperKEKB and Belle II*

There are many experimental reasons that make SuperKEKB and Belle II perfectly suited to address these puzzles in particles physics.

- Running on the $\Upsilon(4S)$ resonance produces a very clean sample of $B^0\bar{B}^0$ pairs in a quantum correlated $1^{--}$ state. The low background environment allows for reconstruction of final states containing photons from decays of $\pi^0$, $\rho^{\pm}$, $\eta$, $\eta'$ etc.. Neutral $K_L^0$ mesons are also efficiently reconstructed.
- Detection of the decay products of one $B$ allows the flavour of the other $B$ to be tagged.
- Flavour production asymmetry is zero, while the detector hermeticity and azimuthal asymmetry make charged asymmetries in reconstruction very small.
- Due to low track multiplicities and detector occupancy, the $B$, $D$ and $\tau$ reconstruction efficiency is high and the trigger bias is very low. This reduces correction and systematic uncertainties in many types of measurements, *e.g.* Dalitz plot analyses.
- With asymmetric beam energies the Lorentz boost of the $e^+e^-$ system is large enough so that $B$ or $D$ mesons travel an appreciable distance before decaying, allowing precision measurements of lifetimes, mixing parameters, and $CP$ violation.
- Since the absolute delivered luminosity is measured with Bhabha scattering, the experiment is able to measure absolute branching fractions.
- Since the initial state is known, "missing mass" analyses can be performed to infer the existence of new particles via energy/momentum conservation rather than reconstructing their final states. By fully reconstructing a $B$ or $D$ decay in a hadronic or semileptonic final state, rare decays with neutrinos can be observed or measured with minimal model dependence.
- In addition to producing large samples of $B$ and $D$ decays, an $e^+e^-$ machine produces large samples of $\tau$ leptons allowing for measurements of rare $\tau$ decays and searches for lepton flavour and lepton number violation $\tau$ decays in a very low background environment.
- The high output rate and relatively low background environment allows for highly efficient triggers of low multiplicity and dark sector signatures.
- The precisely known interaction centre-of-mass energy and excellent detector hermeticity are key for searches for bottomonium transitions using recoil techniques.
- Production of resonances through initial state radiation processes allows for clean and complete probes of the charmonium sector through a continuum of production energies.

The legacy of the $B$-factories laid the groundwork for many areas that will be further exploited at SuperKEKB. Their results provided a theoretically clean measurement the unitarity triangle (UT) angle $\phi_1$. After the accumulation of ~1 ab$^{-1}$ of data, it proved to be a precise calibration for NP. To check the consistency of the SM, Belle measured the other two angles of the UT, $\phi_2$ and $\phi_3$. The results for the sides and angles of the UT are consistent. However, NP contributions of order 10% the size of the SM amplitude are still allowed. In parallel to fixing the weak interaction parameters of the UT, Belle also completed a decade of studies and publications on rare decays and QCD. Belle II builds on this experience, shifting focus to NP exploration beyond the SM.

## *2.6. Overview of SuperKEKB*

The target luminosity of SuperKEKB is a factor 40 greater than the recorded peak of KEKB, requiring a substantial upgrade to the accelerator complex [4]. The essential elements in the increase of the luminosity are a reduction in the the beam size at the collision point by a factor of 20, from about 1 $\mu$m to 50 nm, and an increase in the currents by a factor of 2 compared to the KEKB values. (Table 1). This is known as a 'nano-beam' scheme, and was invented by P. Raimondi for the Italian super $B$ factory [5]. Compared to KEKB, the two beams collide at an even larger angle of 83 mrad (22 mrad in KEKB). A somewhat lower beam energy asymmetry of 7 GeV (electrons) and 4 GeV (positrons), instead of 8 GeV and 3.5 GeV, was chosen to reduce the beam losses due to Touschek scattering in the lower energy beam. This is expected to reduce the spatial separation between $B$-mesons, studied in time dependent $CP$ violation measurements, but leads to slight improvements in solid angle acceptance for missing energy decays.

Table 1: SuperKEKB: design parameters of the low energy (LER) and high energy (HER) accelerator rings [4].

| | LER ($e^+$) | HER ($e^-$) | |
|---|---|---|---|
| Energy | 4.000 | 7.007 | GeV |
| Half crossing angle | 41.5 | | mrad |
| Horizontal emittance | 3.2 | 4.6 | nm |
| Emittance ratio | 0.27 | 0.25 | % |
| Beta functions at IP (x/y) | 32 / 0.27 | 25 / 0.30 | mm |
| Beam currents | 3.6 | 2.6 | A |
| Beam-beam parameter | 0.0881 | 0.0807 | |
| Luminosity | $8 \times 10^{35}$ | | cm$^{-2}$s$^{-1}$ |

The modifications to the accelerator complex include: a new electron injection gun, a new target for positron production, and a new additional damping ring for the positron beam. The upgrade of the accelerator also includes a redesign of the lattices of the low energy and high energy rings, replacing short dipoles with longer ones (in the low energy ring), installing TiN-coated beam pipes with ante-chambers, modifications to the RF system, and a completely redesigned interaction region.

Figure 1 shows the flexibility in the allowed beam energies of the LER and HER respectively. The range of beam energies covers the $\Upsilon(1S)$ and $\Upsilon(6S)$ resonance states for physics operation. The maximum achievable centre of mass energy is 11.24 GeV at SuperKEKB due to the limited power of the injector linac, but even if this limit could be circumvented the beam transport limit (due magnet strength) would only allow an increase to about 12 GeV. With beam energies much lower than $\Upsilon(1S)$, for example near the $\tau$ production threshold, the current lattice design is not sufficient.

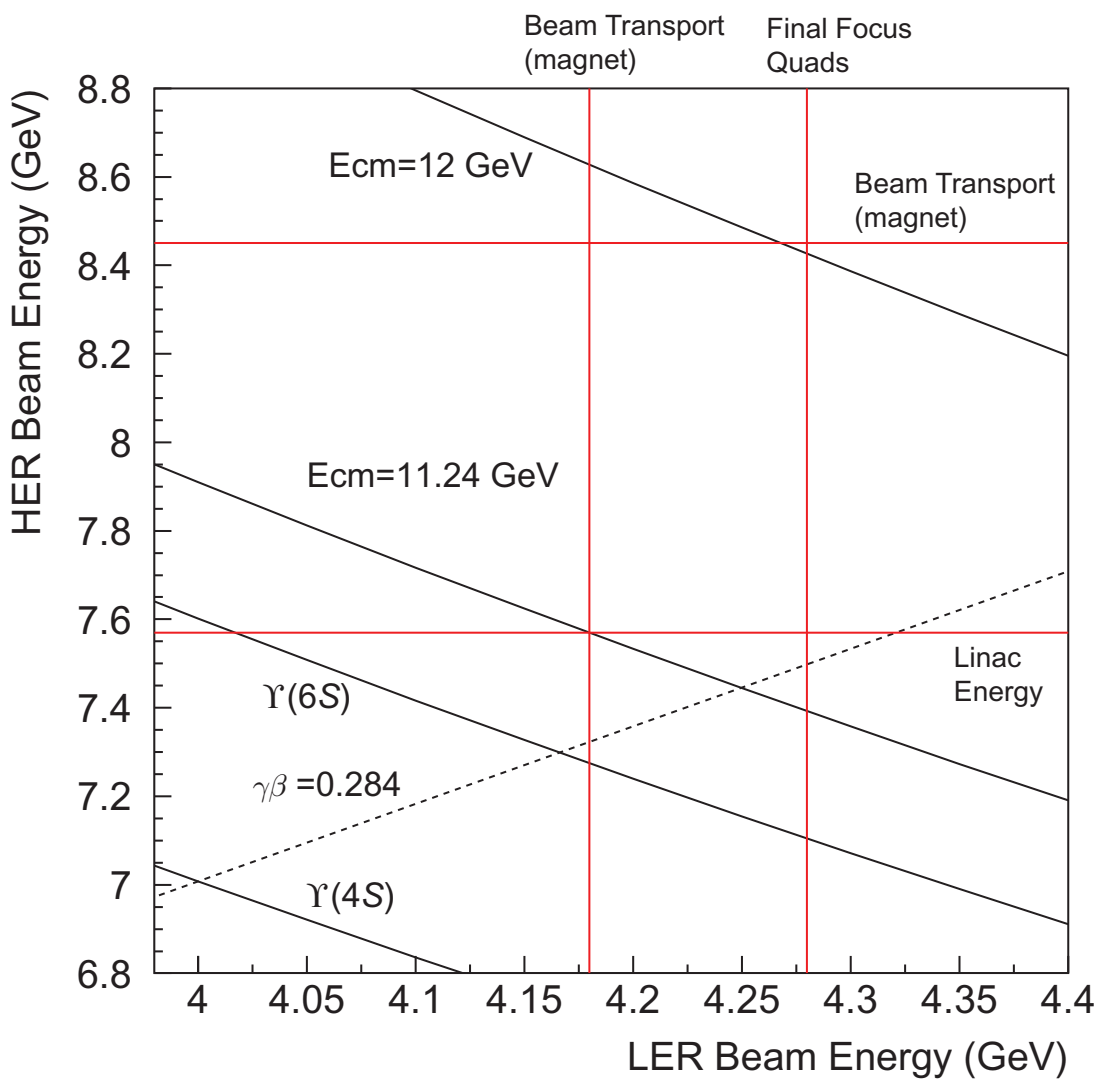

Fig. 1: Beam energies required to achieve centre of mass energies for $\Upsilon(4S)$, $\Upsilon(6S)$, 11.24 GeV, and 12 GeV. The horizontal axis is the LER beam energy and the vertical axis is the HER beam energy.

## 2.7. *Data taking overview*

The SuperKEKB accelerator will have the capacity to deliver $e^+e^-$ collisions in the centre of mass energy range from just below the $\Upsilon(1S)$ (9.46 GeV) to just above the $\Upsilon(6S)$ (11.24 GeV). While the vast majority of the data will be taken at $\Upsilon(4S)$, a program of data taking at other centre-of-mass energies will be undertaken as was done at Belle. The existing $B$-factory data sets are given in Table 2.

Table 2: Existing $e^+e^-$ datasets collected near $\Upsilon$ resonances.

| Exp. | Scans / Off-res. | $\Upsilon(5S)$ 10876 MeV | | $\Upsilon(4S)$ 10580 MeV | | $\Upsilon(3S)$ 10355 MeV | | $\Upsilon(2S)$ 10023 MeV | | $\Upsilon(1S)$ 9460 MeV | |
|---|---|---|---|---|---|---|---|---|---|---|---|
| | $fb^{-1}$ | $fb^{-1}$ | $10^6$ | $fb^{-1}$ | $10^6$ | $fb^{-1}$ | $10^6$ | $fb^{-1}$ | $10^6$ | $fb^{-1}$ | $10^6$ |
| CLEO | 17.1 | 0.4 | 0.1 | 16 | 17.1 | 1.2 | 5 | 1.2 | 10 | 1.2 | 21 |
| BaBar | 54 | $R_b$ scan | | 433 | 471 | 30 | 122 | 14 | 99 | – | |
| Belle | 100 | 121 | 36 | 711 | 772 | 3 | 12 | 25 | 158 | 6 | 102 |

There are a multitude of physics topics unique to the physics program of Belle II: with rare decays and $CP$ asymmetries in $B$ decays at the forefront. The program provides simultaneous studies of a wide range of areas in $b$-quark, $c$-quark, $\tau$-lepton, two-photon, quarkonium and exotic physics. The latter two topics have come to the fore in recent times, particularly concerning puzzles in our understanding of QCD in describing 4 (and 5)-quark states, and the searches for a dark sector. Open questions will be addressed with extended run periods at $\Upsilon(1S)$, $\Upsilon(2S)$, $\Upsilon(3S)$, $\Upsilon(5S)$, near the $\Upsilon(6S)$, and fine energy scans in intermediate regions. Measurements at $\Upsilon(5S)$ also offer useful insights into $B_s$ decays.

Data taking at SuperKEKB will be performed in two main phases.

- In the first collision data taking phase (called "phase 2" as "phase 1" denoted the accelerator commissioning phase in 2016 without the final focus and Belle II detector), commencing February 2018 and running until July 2018, SuperKEKB and the interaction region was commissioned before the installation of the sensitive silicon inner detectors. The peak luminosity delivered by SuperKEKB reached $0.5 \times 10^{34}/\text{cm}^2/\text{s}$, and a data set of order 0.5 $\text{fb}^{-1}$ was collected at the $\Upsilon(4S)$ resonance. This small data set may be used for searches of dark sectors that were previously limited by a lack of efficient triggers.
- The second collision phase will see the full detector and will allow for the full flavour program to commence, expected to start in early 2019. The expected projected peak instantaneous, and integrated luminosities at SuperKEKB through to 2025 are shown in Fig. 2. The full data taking program for samples at the different centre of mass energies is under development, and a subject of many working group chapters. It is clear that well motivated studies with non $\Upsilon(4S)$ data taking could have substantial statistical gains even in early data taking. The physics program at $\Upsilon(4S)$ is covered in most chapters. The program at $\Upsilon(5S)$ is covered over several chapters: $B_s$ decays are covered in the semileptonic $B$ and hadronic $B$ chapters, while bottomonium is covered in the quarkonium chapter.

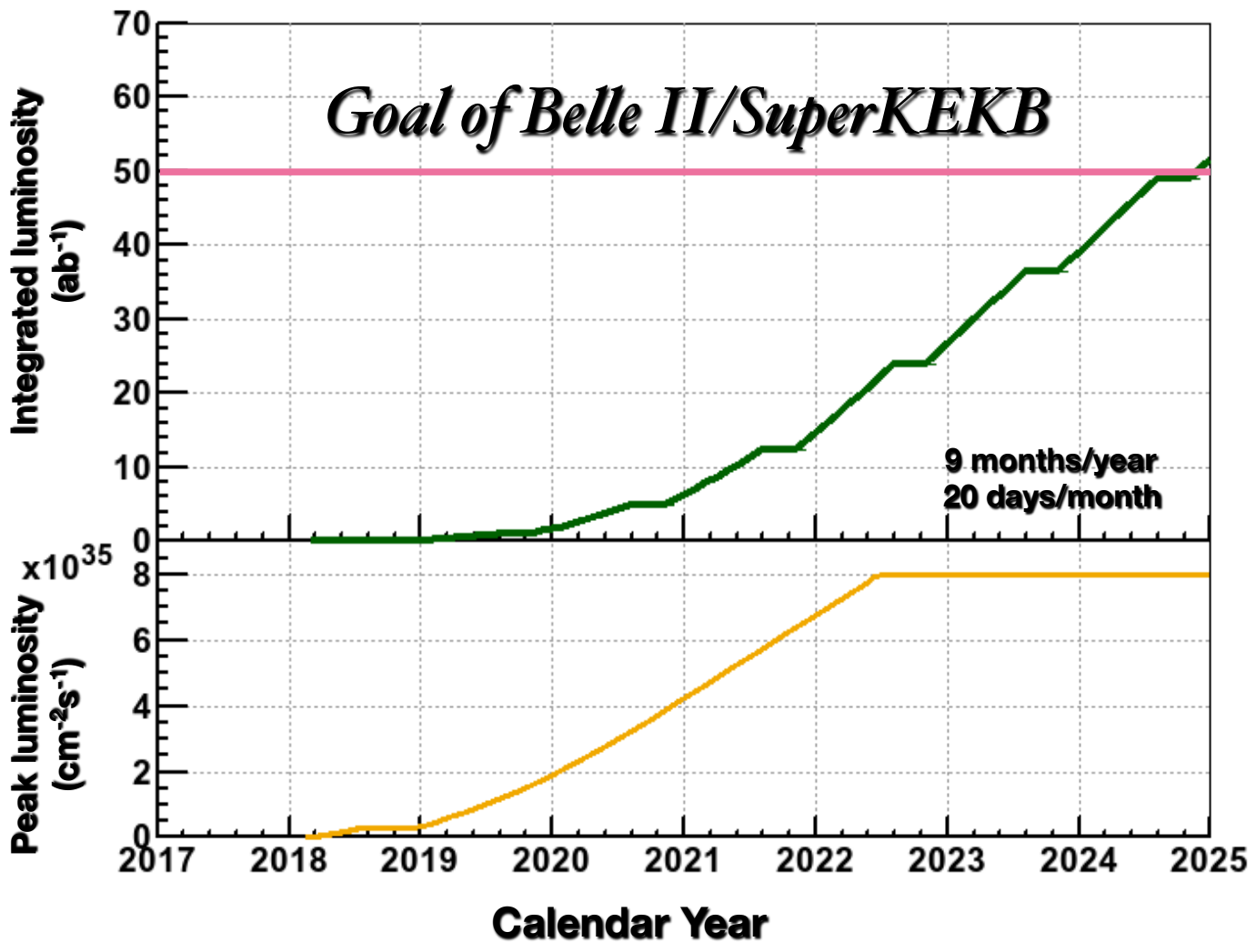

Fig. 2: The projected peak instantaneous, and integrated luminosities at SuperKEKB through to 2025 assuming nine months operation per year.

*Belle II Detector, Simulation, Reconstruction, Algorithms.* In the first few chapters, we cover the detector design, detector simulation, beam induced background, particle reconstruction and analysis algorithms of Belle II. The performance of Belle II for particle reconstruction and robustness against higher beam background is shown, which are critical in assessing the reach of the experiment. New algorithms for flavour tagging, $B$ full reconstruction, and vertex reconstruction are also presented.

*Theory.* Fundamentals of flavour interactions and strong interaction dynamics are presented. A recap of the CKM picture and effective Hamiltonians for flavour interactions is provided, followed by a detailed assessment of the prospects of lattice QCD calculations over the coming decade. Finally we provide a primer on resonances, relevant for many hadronic decay analyses at Belle II.

*Semileptonic and leptonic B decays.* This chapter presents prospects for leptonic and semileptonic B decays to electron, muon and tau leptons, summarised in Tables 3 and 4. There is significant interest in the sensitivity to lepton flavour universality violating (LFUV) new phenomena, such as a charged Higgs-like coupling to tau leptons, where Belle II can make substantial advances. The chapter also details the experimental and theoretical advances for precision measurements of the CKM matrix elements, $|V_{ub}|$ and $|V_{cb}|$. Full simulation studies of Belle II in $B \to \pi \ell \nu$ and $B \to \tau \nu$ are presented. It is expected that 5 $\sigma$ discovery level measurements of $B \to \tau \nu$ and $B \to \mu \nu$ are possible with less than 5 ab$^{-1}$ at SM branching fractions.

Table 3: Expected errors on several selected observables in leptonic and semileptonic $B$ decays.

| Observables | Belle (2017) | Belle II 5 ab$^{-1}$ | Belle II 50 ab$^{-1}$ |
|---|---|---|---|
| $\|V_{cb}\|$ incl. | $42.2 \cdot 10^{-3} \cdot (1 \pm 1.8\%)$ | 1.2% | – |
| $\|V_{cb}\|$ excl. | $39.0 \cdot 10^{-3} \cdot (1 \pm 3.0\%_{\text{ex.}} \pm 1.4\%_{\text{th.}})$ | 1.8% | 1.4% |
| $\|V_{ub}\|$ incl. | $4.47 \cdot 10^{-3} \cdot (1 \pm 6.0\%_{\text{ex.}} \pm 2.5\%_{\text{th.}})$ | 3.4% | 3.0% |
| $\|V_{ub}\|$ excl. (WA) | $3.65 \cdot 10^{-3} \cdot (1 \pm 2.5\%_{\text{ex.}} \pm 3.0\%_{\text{th.}})$ | 2.4% | 1.2% |
| $\mathcal{B}(B \to \tau\nu)$ $[10^{-6}]$ | $91 \cdot (1 \pm 24\%)$ | 9% | 4% |
| $\mathcal{B}(B \to \mu\nu)$ $[10^{-6}]$ | $< 1.7$ | 20% | 7% |
| $R(B \to D\tau\nu)$ (Had. tag) | $0.374 \cdot (1 \pm 16.5\%)$ | 6% | 3% |
| $R(B \to D^*\tau\nu)$ (Had. tag) | $0.296 \cdot (1 \pm 7.4\%)$ | 3% | 2% |

*Radiative and Electroweak Penguin B decays.* The prospects for flavour changing neutral current B decays to radiative and rare dilepton final states are presented, summarised in Tables 5 and 6. There are several clear strengths of the Belle II program: the use of full $B$ reconstruction allows for precise studies of missing energy decays such as $B \to K^{(*)}\nu\bar{\nu}$ which should be accessible with the Belle II data set, improved particle identification detectors will be used for precision studies of $b \to d\gamma$ transitions, inclusive transitions will be studied through various techniques, and lepton flavour universality violation will be studied thanks to the low radiation length in the tracking volume allowing for precise reconstruction of electrons, muons and tau leptons.

Table 4: Belle II Golden/Silver observables for the pure-leptonic and the semi-leptonic $B$ decays. The theory column indicates the robustness against the theory uncertainties. The discovery column shows the integrated luminosity at which the discovery of new physics is possible. The Sys. limit column indicates at which integrated luminosity the experimental systematic uncertainty dominates. The vs LHCb/BESIII, Belle columns show the originality and the competitiveness against those experiments. The anomaly column indicates whether there is an existing hint of new physics and the NP columnshows whether the observable is sensitive to new physics models.

| Process | Observable | Theory | Sys. dom. (Discovery) [$ab^{-1}$] | vs LHCb | vs Belle | Anomaly | NP |
|---|---|---|---|---|---|---|---|
| $B \to \pi \ell \nu_\ell$ | $\lvert V_{ub}\rvert$ | ★★★ | 10-20 | ★★★ | ★★★ | ★★ | ★ |
| $B \to X_u \ell \nu_\ell$ | $\lvert V_{ub}\rvert$ | ★★ | 2-10 | ★★★ | ★★ | ★★★ | ★ |
| $B \to \tau \nu$ | $Br.$ | ★★★ | >50 (2) | ★★★ | ★★★ | ★ | ★★★ |
| $B \to \mu \nu$ | $Br.$ | ★★★ | >50 (5) | ★★★ | ★★★ | ★ | ★★★ |
| $B \to D^{(*)} \ell \nu_\ell$ | $\lvert V_{cb}\rvert$ | ★★★ | 1-10 | ★★★ | ★★ | ★★ | ★ |
| $B \to X_c \ell \nu_\ell$ | $\lvert V_{cb}\rvert$ | ★★★ | 1-5 | ★★★ | ★★ | ★★ | ★★ |
| $B \to D^{(*)} \tau \nu_\tau$ | $R(D^{(*)})$ | ★★★ | 5-10 | ★★ | ★★★ | ★★★ | ★★★ |
| $B \to D^{(*)} \tau \nu_\tau$ | $P_\tau$ | ★★★ | 15-20 | ★★★ | ★★★ | ★★ | ★★★ |
| $B \to D^{**} \ell \nu_\ell$ | $Br.$ | ★ | - | ★★ | ★★★ | ★★ | - |

*Time dependent $CP$ violation in $B$ decays.* The prospects for time-dependent CP violation of $B$ mesons and the determination of the CKM angles $\phi_1$ and $\phi_2$ are presented in this chapter, summarised in Tables 7 and 8. Sensitivity studies based on Belle II simulation for $\phi_1$ measurement with the penguin dominated modes, $B \to \phi K^0_S, \eta' K^0_S, \pi^0 K^0_S$, are performed. The theoretical progress on *the penguin pollution* for high precision measurement of $\phi_1$ with the tree level processes is discussed. A Belle II sensitivity study on the challenging $B \to \pi^0\pi^0$ time-dependent CP asymmetry measurement for $\phi_2$ determination is performed. The subsequent $\phi_2$ measurement will rely on isospin relations: theoretical estimates of the isospin breaking effects on the $\phi_2$ determination are reviewed.

*Measurement of the UT angle $\phi_3$.* The prospects for measuring the CKM UT angle $\phi_3$ with tree-level measurements of $B \to D^{(*)} K^{(*)}$ decays are presented in this chapter, summarised in Tables 9 and 10. It is expected that Belle II will ultimately reach a precision of 1 to 2 degrees on this angle through use of a variety of channels and extraction techniques.

Table 5: Expected errors on several selected observables in radiative and electroweak penguin $B$ decays. Note that 50 $ab^{-1}$ projections for $B_s$ decays are not provided as we do not expect to collect such a large $\Upsilon(5S)$ data set.

| Observables | Belle | Belle II | |
|---|---|---|---|
| | (2017) | 5 $ab^{-1}$ | 50 $ab^{-1}$ |
| $\mathcal{B}(B \to K^{*+}\nu\overline{\nu})$ | $< 40 \times 10^{-6}$ | 25% | 9% |
| $\mathcal{B}(B \to K^{+}\nu\overline{\nu})$ | $< 19 \times 10^{-6}$ | 30% | 11% |
| $A_{CP}(B \to X_{s+d}\gamma)$ $[10^{-2}]$ | $2.2 \pm 4.0 \pm 0.8$ | 1.5 | 0.5 |
| $S(B \to K_S^0\pi^0\gamma)$ | $-0.10 \pm 0.31 \pm 0.07$ | 0.11 | 0.035 |
| $S(B \to \rho\gamma)$ | $-0.83 \pm 0.65 \pm 0.18$ | 0.23 | 0.07 |
| $A_{FB}(B \to X_s\ell^+\ell^-)$ $(1 < q^2 < 3.5\ \mathrm{GeV}^2/c^4)$ | 26% | 10% | 3% |
| $Br(B \to K^+\mu^+\mu^-)/Br(B \to K^+e^+e^-)$ $(1 < q^2 < 6\ \mathrm{GeV}^2/c^4)$ | 28% | 11% | 4% |
| $Br(B \to K^{*+}(892)\mu^+\mu^-)/Br(B \to K^{*+}(892)e^+e^-)$ $(1 < q^2 < 6\ \mathrm{GeV}^2/c^4)$ | 24% | 9% | 3% |
| $\mathcal{B}(B_s \to \gamma\gamma)$ | $< 8.7 \times 10^{-6}$ | 23% | – |
| $\mathcal{B}(B_s \to \tau\tau)$ $[10^{-3}]$ | – | $< 0.8$ | – |

*Hadronic $B$ decays.* This chapter presents at the prospects for charmless hadronic $B$ decays and direct $CP$ violation, summarised in Tables 9 and 11. The theoretical computation of the branching ratio and $CP$ asymmetry of the $B \to PP$, $PV$, $VV$ ($P$ and $V$ denote pseudoscalar and vector mesons, respectively) processes using QCD and SU(3) symmetry is reviewed. The theoretical prediction is partially data-driven and each decay mode plays different role to reduce the theoretical uncertainties. The phenomenology of the angular analysis of three body final state for new physics search is also reviewed. Experimental measurement for these channels will be reduced significantly at Belle II, since those are currently dominated by statistical or reducible systematical errors

*Charm physics.* This chapter presents the prospects for charm meson physics, summarised in Tables 12 and 13. Charm is a large area of opportunity for Belle II, covering $CP$ violation, FCNC, tree level and missing energy decay transition measurements. Novel techniques for tagging in $CP$ violation measurements are shown.

*Quarkonium.* This chapter presents the prospects for quarkonium(like) physics, providing a detailed theoretical overview of perturbative QCD computation, lattice QCD as well as models for unconventional states (Tetraquark, Hybrid mesons and Hadronic molecule) is presented. At Belle II, charmonium(-like) states can be produced from $B$ decays, initial state radiation, two photon collisions, and double charmonium production, which allow for detailed studies of the nature of any observed states. The motivations for dedicated non-$\Upsilon(4S)$ runs are detailed: to provide us with a deeper understanding of bottomonium(-like) states. Light Higgs and lepton universality violation searches using decays of $\Upsilon(1S, 2S, 3S)$ are also reviewed.

Table 6: Belle II Golden/Silver observables for the radiative and the electroweak penguin $B$ decays. See the caption in Table 4 for more details. The precision limit of the $B \to X_s\gamma$ measurement estimated simply by estimating the point where the statistic uncertainties dominate. However, the systematic uncertainties may be further reduced by adding more data.

| Process | Observable | Theory | Sys. dom. (Discovery) [$ab^{-1}$] | vs LHCb | vs Belle | Anomaly | NP |
|---|---|---|---|---|---|---|---|
| $B \to K^{(*)}\nu\nu$ | $Br., F_L$ | ⋆⋆⋆ | >50 | ⋆⋆⋆ | ⋆⋆⋆ | ⋆ | ⋆⋆ |
| $B \to X_{s+d}\gamma$ | $A_{\mathrm{CP}}$ | ⋆⋆⋆ | >50 | ⋆⋆⋆ | ⋆⋆⋆ | ⋆ | ⋆⋆ |
| $B \to X_d\gamma$ | $A_{\mathrm{CP}}$ | ⋆⋆ | >50 | ⋆⋆⋆ | ⋆⋆⋆ | - | ⋆⋆ |
| $B \to K^0_S\pi^0\gamma$ | $S_{K^0_S\pi^0\gamma}$ | ⋆⋆ | >50 | ⋆⋆ | ⋆⋆⋆ | ⋆ | ⋆⋆⋆ |
| $B \to \rho\gamma$ | $S_{\rho\gamma}$ | ⋆⋆ | >50 | ⋆⋆⋆ | ⋆⋆⋆ | - | ⋆⋆⋆ |
| $B \to X_s\ell^+\ell^-$ | $Br.$ | ⋆⋆⋆ | >50 | ⋆⋆⋆ | ⋆⋆ | ⋆⋆ | ⋆⋆⋆ |
| $B \to X_s\ell^+\ell^-$ | $R_{X_s}$ | ⋆⋆⋆ | >50 | ⋆⋆⋆ | ⋆⋆⋆ | ⋆⋆ | ⋆⋆⋆ |
| $B \to K^{(*)}e^+e^-$ | $R(K^{(*)})$ | ⋆⋆⋆ | >50 | ⋆⋆ | ⋆⋆⋆ | ⋆⋆⋆ | ⋆⋆⋆ |
| $B \to X_s\gamma$ | $Br.$ | ⋆⋆ | 1-5 | ⋆⋆⋆ | ⋆ | ⋆ | ⋆⋆ |
| $B_{d,(s)} \to \gamma\gamma$ | $Br., A_{\mathrm{CP}}$ | ⋆⋆ | >50 | ⋆⋆ | ⋆⋆ | - | ⋆⋆ |
| $B \to K^* e^+e^-$ | $P_5'$ | ⋆⋆ | >50 | ⋆⋆⋆ | ⋆⋆ | ⋆⋆⋆ | ⋆⋆⋆ |
| $B \to K\tau\ell$ | $Br.$ | ⋆⋆⋆ | >50 | ⋆⋆ | ⋆⋆⋆ | ⋆⋆ | ⋆⋆⋆ |

Table 7: Expected errors on several selected observables related to the measurement of time dependent CP violation in $B$ decays and the measurement of the UT angles $\phi_1$ and $\phi_2$.

| Observables | Belle | Belle II | |
|---|---|---|---|
| | (2017) | 5 $ab^{-1}$ | 50 $ab^{-1}$ |
| $\sin 2\phi_1(B \to J/\psi K^0)$ | $0.667 \pm 0.023 \pm 0.012$ | 0.012 | 0.005 |
| $S(B \to \phi K^0)$ | $0.90^{+0.09}_{-0.19}$ | 0.048 | 0.020 |
| $S(B \to \eta' K^0)$ | $0.68 \pm 0.07 \pm 0.03$ | 0.032 | 0.015 |
| $S(B \to J/\psi\pi^0)$ | $-0.65 \pm 0.21 \pm 0.05$ | 0.079 | 0.025 |
| $\phi_2$ [°] | $85 \pm 4$ (Belle+BaBar) | 2 | 0.6 |
| $S(B \to \pi^+\pi^-)$ | $-0.64 \pm 0.08 \pm 0.03$ | 0.04 | 0.01 |
| $Br.(B \to \pi^0\pi^0)$ | $(5.04 \pm 0.21 \pm 0.18) \times 10^{-6}$ | 0.13 | 0.04 |
| $S(B \to K^0\pi^0)$ | $-0.11 \pm 0.17$ | 0.09 | 0.03 |

*Tau and low multiplicity physics.* The prospects for tau and low multiplicity physics are presented in this chaper, summarised in Tables 14 and 15. The measurement of the lepton flavour violating $\tau$ decays will be improved by orders of magnitude by Belle II experiment. The sensitivity of different decay channels to theoretical models are discussed by using the effective couplings. The $CP$ violation in $\tau$ decay is possible both in the measurement of the cross section difference in $\tau^\pm$ as well as of various angular observables at Belle II. A improved measurement of the $e^+e^-$ hadronic cross section as well as hadron productions from two photon collisions at Belle II and its impact on the theoretical prediction of the muon anomalous magnetic moment $g-2$ are also discussed.

Table 8: Belle II Golden/Silver observables on the measurement of time dependent CP violation in $B$ decays and the measurement of the UT angles $\phi_1$ and $\phi_2$. See the caption in Table 4 for more details.

| Process | Observable | Theory | Sys. dom. (Discovery) [ab$^{-1}$] | vs LHCb | vs Belle | Anomaly | NP |
|---|---|---|---|---|---|---|---|
| $B \to J/\psi K_S^0$ | $\phi_1$ | ⋆⋆⋆ | 5-10 | ⋆⋆ | ⋆⋆ | ⋆ | ⋆ |
| $B \to \phi K_S^0$ | $\phi_1$ | ⋆⋆ | >50 | ⋆⋆ | ⋆⋆⋆ | ⋆ | ⋆⋆⋆ |
| $B \to \eta' K_S^0$ | $\phi_1$ | ⋆⋆ | >50 | ⋆⋆ | ⋆⋆⋆ | ⋆ | ⋆⋆⋆ |
| $B \to \rho^\pm \rho^0$ | $\phi_2$ | ⋆⋆⋆ | >50 | ⋆ | ⋆⋆⋆ | ⋆ | ⋆ |
| $B \to J/\psi \pi^0$ | $\phi_1$ | ⋆⋆⋆ | >50 | ⋆ | ⋆⋆⋆ | - | - |
| $B \to \pi^0 \pi^0$ | $\phi_2$ | ⋆⋆ | >50 | ⋆⋆⋆ | ⋆⋆⋆ | ⋆⋆ | ⋆⋆ |
| $B \to \pi^0 K_S^0$ | $S_{\rm CP}$ | ⋆⋆ | >50 | ⋆⋆⋆ | ⋆⋆⋆ | ⋆⋆ | ⋆⋆ |

Table 9: Expected errors on several selected hadronic $B$ decay observables, including direct $CP$ violation.

| Observables | Belle (2017) | Belle II 5 ab$^{-1}$ | Belle II 50 ab$^{-1}$ |
|---|---|---|---|
| $\phi_3$ GGSZ | $68 \pm 13$ | 4.7 | 1.5 |
| $\mathcal{A}_{\mathcal{CP}}(B \to K^0\pi^0)[\%]$ | $-0.05 \pm 0.14 \pm 0.05$ | 0.07 | 0.04 |
| $\mathcal{I}(B \to K\pi)[\%]$ | $0.27 \pm 0.14$ | 0.07 | 0.03 |
| $\mathcal{I}(B \to K\rho)[\%]$ | $-0.44 \pm 0.49$ | 0.25 | 0.06 |

*Beyond the standard model and global fit analyses.* The beyond standard model chapter describes new physics models that can be observed in flavour transitions, specifically those testable at Belle II. A variety of theoretical models are discussed, and the best decay modes to observe effects from those models.

In the global fit chapter, we provide prospects for Belle II in global fit analyses of the CKM unitarity triangle, based on studies by CKMFitter and UTFit groups. Global analyses of tree and FCNC $B$ decays are performed in effective operator approaches using projected constraints from Belle II on inclusive and exclusive decays.

Table 10: Belle II Golden/Silver observables for $\phi_3$ measurements. The GLW method utilises the $CP$-eigenstate final states and the ADS method the final states, $K^+X^-$ ($X^- = \pi^-, \pi^-\pi^0, \pi^-\pi^-, \pi^+$). The GGSZ method utilises the self-conjugate multi-body final states, $K_S^0 h^+h^-$ and GLS method, $K_S^0 K^+\pi^-$ final state.

| | Process | Observable | Theory | Sys. dom. (Discovery) [ab$^{-1}$] | vs LHCb | vs Belle | Anomaly | NP |
|---|---|---|---|---|---|---|---|---|
| | GGSZ | $\phi_3$ | ★★★ | >50 | ★★ | ★★★ | ★ | ★★ |
| | GLW | $\phi_3$ | ★★★ | >50 | ★★ | ★★★ | ★ | ★★ |
| | ADS | $\phi_3$ | ★★ | >50 | ★★ | ★★★ | ★ | ★★★ |
| | Time-dependent | $\phi_3 - \phi_2$ | ★★ | - | ★★ | ★★ | ★ | ★ |

Table 11: Belle II Golden/Silver observables for hadronic $B$ decay measurements.

| | Process | Observable | Theory | Sys. dom. (Discovery) [ab$^{-1}$] | vs LHCb | vs Belle | Anomaly | NP |
|---|---|---|---|---|---|---|---|---|
| | $B \to \pi^0 K^0$ | $A_{\rm CP}$, $I_{K\pi}$ | ★★ | >50 | ★★★ | ★★★ | ★★★ | ★★ |
| | $B \to \rho K$ | $A_{\rm CP}$, $I_{K\rho}$ | ★ | >50 | ★★ | ★★★ | - | ★★ |
| | $B \to \ell\nu\gamma$ | $\lambda_B$ | ★★ | >50(10) | ★★★ | ★★★ | ★ | ★★ |
| | $B \to \rho K^*$ | $f_L$ | ★★ | >50 | ★★ | ★★ | - | ★★★ |
| | $B \to K^+K^-/\pi^+\pi^-$ | $Br.$, $A_{CP}$ | ★★ | >50 | ★ | ★★★ | ★★ | ★★ |
| | $B \to K\pi\pi, KKK$ | $A_{CP}$ | ★★ | >50 | ★★ | ★ | ★★★ | ★ |
| | $B_s \to K^0\overline{K}^0$ | Lifetime | ★ | >5 | ★★ | ★★★ | - | ★★ |

### *2.9. The Belle II Golden Flavour Channels*

A summary of the expected sensitivities for key flavour observables at selected integrated luminosities is given in Table 16. In this table we indicate modes where LHCb will be in close competition with Belle II.

LHCb will have high statistics samples all $b$ and $c$ hadrons and are particularly sensitive to modes to all charged particle final states. Belle II will be particularly sensitive to $B$ and $D_{(s)}$ measurements where final states contain neutrinos, multiple photons, $\pi^0$ mesons, or neutral kaons. The $e^+e^-$ program of Belle II also includes an extensive scope for studies of $\tau$-leptons and a number of other non-flavour physics topics (not shown in this table).

Table 12: Expected errors on several selected charm physics observables.

| Observables | Belle (2017) | Belle II 5 $\text{ab}^{-1}$ | Belle II 50 $\text{ab}^{-1}$ |
|---|---|---|---|
| $x(D^0 \to K_S^0\pi^+\pi^-)$ $[10^{-2}]$ | $0.56 \pm 0.19 \pm {}^{0.07}_{0.13}$ | 0.16 | 0.11 |
| $y(D^0 \to K_S^0\pi^+\pi^-)$ $[10^{-2}]$ | $0.30 \pm 0.15 \pm {}^{0.05}_{0.08}$ | 0.10 | 0.05 |
| $\lvert q/p\rvert(D^0 \to K_S^0\pi^+\pi^-)$ | $0.90 \pm {}^{0.16}_{0.15} \pm {}^{0.08}_{0.06}$ | 0.12 | 0.07 |
| $A_{CP}(D^+ \to \pi^+\pi^0)$ $[10^{-2}]$ | $2.3 \pm 1.2 \pm 0.2$ | 0.54 | 0.17 |
| $A_{CP}(D^0 \to \pi^0\pi^0)$ $[10^{-2}]$ | $-0.03 \pm 0.64 \pm 0.10$ | 0.28 | 0.09 |
| $A_{CP}(D^0 \to K_S^0\pi^0)$ $[10^{-2}]$ | $-0.21 \pm 0.16 \pm 0.09$ | 0.08 | 0.02 |
| $A_{CP}(D^0 \to K_S^0K_S^0)$ $[10^{-2}]$ | $0.02 \pm 1.53 \pm 0.17$ | 0.66 | 0.23 |
| $A_{CP}(D^0 \to \phi\gamma)$ $[10^{-2}]$ | $-9.4 \pm 6.6 \pm 0.1$ | $\pm 3.0$ | $\pm 1.0$ |
| $f_{D_s}$ | 2.5% | 1.1% | 0.3% |

Table 13: Belle II Golden/Silver observables for charm physics.

| | Process | Observable | Theory | Sys. dom. (Discovery) $[\text{ab}^{-1}]$ | vs LHCb/BESIII | vs Belle | Anomaly | NP |
|---|---|---|---|---|---|---|---|---|
| ● | $D^0 \to K_s\pi^+\pi^-$ | $x$, $y$,$\lvert q/p\rvert$ | ⋆⋆⋆ | 20 | ⋆⋆ | ⋆⋆⋆ | - | ⋆⋆ |
| ● | $D^0 \to K_S^0K_S^0$ | $A_{\text{CP}}$ | ⋆⋆ | >50 | ⋆⋆⋆ | ⋆⋆⋆ | ⋆ | ⋆ |
| ● | $D^0 \to \pi^0\pi^0$ | $A_{\text{CP}}$ | ⋆⋆⋆ | >50 | ⋆⋆⋆ | ⋆⋆ | ⋆ | ⋆ |
| ● | $D^+ \to \pi^+\pi^0$ | $A_{\text{CP}}$ | ⋆⋆ | >50 | ⋆⋆⋆ | ⋆⋆ | ⋆ | ⋆⋆ |
| ● | $D_s \to \ell^+\nu$ | $f_{D_s}$ | ⋆⋆⋆ | 1-5 | ⋆⋆⋆ | ⋆ | - | ⋆⋆ |
| ● | $D^0 \to V\gamma$ | $A_{\text{CP}}$ | ⋆ | >50 | ⋆⋆ | ⋆⋆ | ⋆⋆ | ⋆⋆ |
| ● | $D^0 \to \gamma\gamma$ | $Br.$ | ⋆ | >50 | ⋆⋆ | ⋆⋆ | ⋆⋆ | ⋆⋆ |
| ● | $D^0 \to \nu\bar{\nu}$ | $Br.$ | ⋆⋆⋆ | >50 | ⋆⋆⋆ | ⋆⋆ | ⋆⋆⋆ | ⋆⋆⋆ |
| ● | $D \to \ell^+\nu$ | $f_D$ | ⋆⋆⋆ | 1-5 | ⋆ | ⋆ | - | ⋆⋆ |

Table 14: Expected limits on several selected $\tau$ LFV searches.

| Observables | Belle (2014) | Belle II 5 $\text{ab}^{-1}$ | Belle II 50 $\text{ab}^{-1}$ |
|---|---|---|---|
| $Br.(\tau \to \mu\gamma)$ $[10^{-9}]$ | $< 45$ | $< 15$ | $< 5$ |
| $Br.(\tau \to e\gamma)$ $[10^{-9}]$ | $< 120$ | $< 39$ | $< 12$ |
| $Br.(\tau \to \mu\mu\mu)$ $[10^{-9}]$ | $< 21$ | $< 3$ | $< 0.3$ |
| $Br.(\tau \to eee)$ $[10^{-9}]$ | $< 27$ | $< 4$ | $< 0.4$ |
| $Br.(\tau \to eKK)$ $[10^{-9}]$ | $< 33$ | $< 6$ | $< 0.6$ |
| $Br.(\tau \to \mu\pi^0)$ $[10^{-9}]$ | $< 120$ | $< 34$ | $< 11$ |
| $\lvert\Im(\eta_s)\rvert(\tau \to K_S^0\pi\nu)$ | 0.026 | 0.010 | 0.003 |

Table 15: Belle II Golden/Silver observables for $\tau$ physics and low multiplicity.

| Process | Observable | Theory | Sys. dom. (Discovery) [$\text{ab}^{-1}$] | vs LHCb/BESIII | vs Belle | Anomaly | NP |
|---|---|---|---|---|---|---|---|
| $\tau \to \mu\gamma$ | $Br.$ | ⋆⋆⋆ | >50 | ⋆⋆⋆ | ⋆⋆⋆ | ⋆ | ⋆⋆⋆ |
| $\tau \to \ell\ell\ell$ | $Br.$ | ⋆⋆⋆ | >50 | ⋆⋆⋆ | ⋆⋆⋆ | ⋆ | ⋆⋆⋆ |
| $\tau \to K^0_\mathrm{S}\pi\nu$ | $\lvert\Im(\eta_s)\rvert$ | ⋆⋆⋆ | >50 | ⋆⋆⋆ | ⋆⋆⋆ | ⋆⋆ | ⋆⋆ |
| $e^+e^- \to \gamma A'(\to$invisible$)$ | $\sigma$ | ⋆⋆⋆ | >50 | ⋆⋆⋆ | ⋆⋆⋆ | ⋆ | ⋆⋆⋆ |
| $e^+e^- \to \gamma A'(\to \ell^+\ell^-)$ | $\sigma$ | ⋆⋆⋆ | >50 | ⋆⋆⋆ | ⋆⋆⋆ | ⋆ | ⋆⋆⋆ |
| $e^+e^- \to \gamma a'(\to \gamma^+\gamma^-)$ | $\sigma$ | ⋆⋆⋆ | >50 | ⋆⋆⋆ | ⋆⋆⋆ | ⋆ | ⋆⋆⋆ |
| $\Upsilon(1S) \to$ invisible | ⋆⋆⋆ | $Br.$ | >50 | ⋆⋆⋆ | ⋆⋆⋆ | ⋆ | ⋆⋆⋆ |
| $\chi_{b0}(1P) \to \tau\tau$ | ⋆⋆⋆ | $Br.$ | >50 | ⋆⋆⋆ | ⋆⋆⋆ | ⋆ | ⋆⋆⋆ |
| $\pi$ form factor | $g-2$ | ⋆⋆ | - | ⋆⋆⋆ | ⋆⋆ | ⋆⋆ | ⋆⋆⋆ |
| ISR $e^+e^- \to \pi\pi$ g-2 | $g-2$ | ⋆⋆ | - | ⋆⋆⋆ | ⋆⋆⋆ | ⋆⋆ | ⋆⋆⋆ |

Table 16: Expected errors on several selected flavour observables with an integrated luminosity of 50 $ab^{-1}$ of Belle II data. Errors given in % represent relative errors. In the final column we denote where LHCb is expected to reach a highly competitive level of precision: if one experiment is expected to be slightly more accurate we list it first.

| Observables | Expected the. accuracy | Expected exp. uncertainty | Facility (2025) |
|---|---|---|---|
| UT angles & sides | | | |
| $\phi_1$ [°] | *** | 0.4 | Belle II |
| $\phi_2$ [°] | ** | 1.0 | Belle II |
| $\phi_3$ [°] | *** | 1.0 | LHCb/Belle II |
| $\lvert V_{cb}\rvert$ incl. | *** | 1% | Belle II |
| $\lvert V_{cb}\rvert$ excl. | *** | 1.5% | Belle II |
| $\lvert V_{ub}\rvert$ incl. | ** | 3% | Belle II |
| $\lvert V_{ub}\rvert$ excl. | ** | 2% | Belle II/LHCb |
| $CP$ Violation | | | |
| $S(B \to \phi K^0)$ | *** | 0.02 | Belle II |
| $S(B \to \eta' K^0)$ | *** | 0.01 | Belle II |
| $\mathcal{A}(B \to K^0\pi^0)[10^{-2}]$ | *** | 4 | Belle II |
| $\mathcal{A}(B \to K^+\pi^-)$ $[10^{-2}]$ | *** | 0.20 | LHCb/Belle II |
| (Semi-)leptonic | | | |
| $\mathcal{B}(B \to \tau\nu)$ $[10^{-6}]$ | ** | 3% | Belle II |
| $\mathcal{B}(B \to \mu\nu)$ $[10^{-6}]$ | ** | 7% | Belle II |
| $R(B \to D\tau\nu)$ | *** | 3% | Belle II |
| $R(B \to D^*\tau\nu)$ | *** | 2% | Belle II/LHCb |
| Radiative & EW Penguins | | | |
| $\mathcal{B}(B \to X_s\gamma)$ | ** | 4% | Belle II |
| $A_{CP}(B \to X_{s,d}\gamma)$ $[10^{-2}]$ | *** | 0.005 | Belle II |
| $S(B \to K^0_S\pi^0\gamma)$ | *** | 0.03 | Belle II |
| $S(B \to \rho\gamma)$ | ** | 0.07 | Belle II |
| $\mathcal{B}(B_s \to \gamma\gamma)$ $[10^{-6}]$ | ** | 0.3 | Belle II |
| $\mathcal{B}(B \to K^*\nu\overline{\nu})$ $[10^{-6}]$ | *** | 15% | Belle II |
| $R(B \to K^*\ell\ell)$ | *** | 0.03 | Belle II/LHCb |
| Charm | | | |
| $\mathcal{B}(D_s \to \mu\nu)$ | *** | 0.9% | Belle II |
| $\mathcal{B}(D_s \to \tau\nu)$ | *** | 2% | Belle II |
| $A_{CP}(D^0 \to K^0_S\pi^0)$ $[10^{-2}]$ | ** | 0.03 | Belle II |
| $\lvert q/p\rvert(D^0 \to K^0_S\pi^+\pi^-)$ | *** | 0.03 | Belle II |
| $A_{CP}(D^+ \to \pi^+\pi^0)$ $[10^{-2}]$ | ** | 0.17 | Belle II |
| Tau | | | |
| $\tau \to \mu\gamma$ $[10^{-10}]$ | *** | $< 50$ | Belle II |
| $\tau \to e\gamma$ $[10^{-10}]$ | *** | $< 100$ | Belle II |
| $\tau \to \mu\mu\mu$ $[10^{-10}]$ | *** | $< 3$ | Belle II/LHCb |

# 3. Belle II Detector

*Section author(s): B. Fulsom, P. Križan, P. Urquijo, C. H. Li*

## 3.1. Introduction

The tool for discoveries at the new generation (super) B-factory will be the Belle II detector (Fig. 3). While the new detector clearly fits the same shell as its predecessor, the super-conducting soleniod magnet with the iron return yoke, all components are either new or considerably upgraded [6].

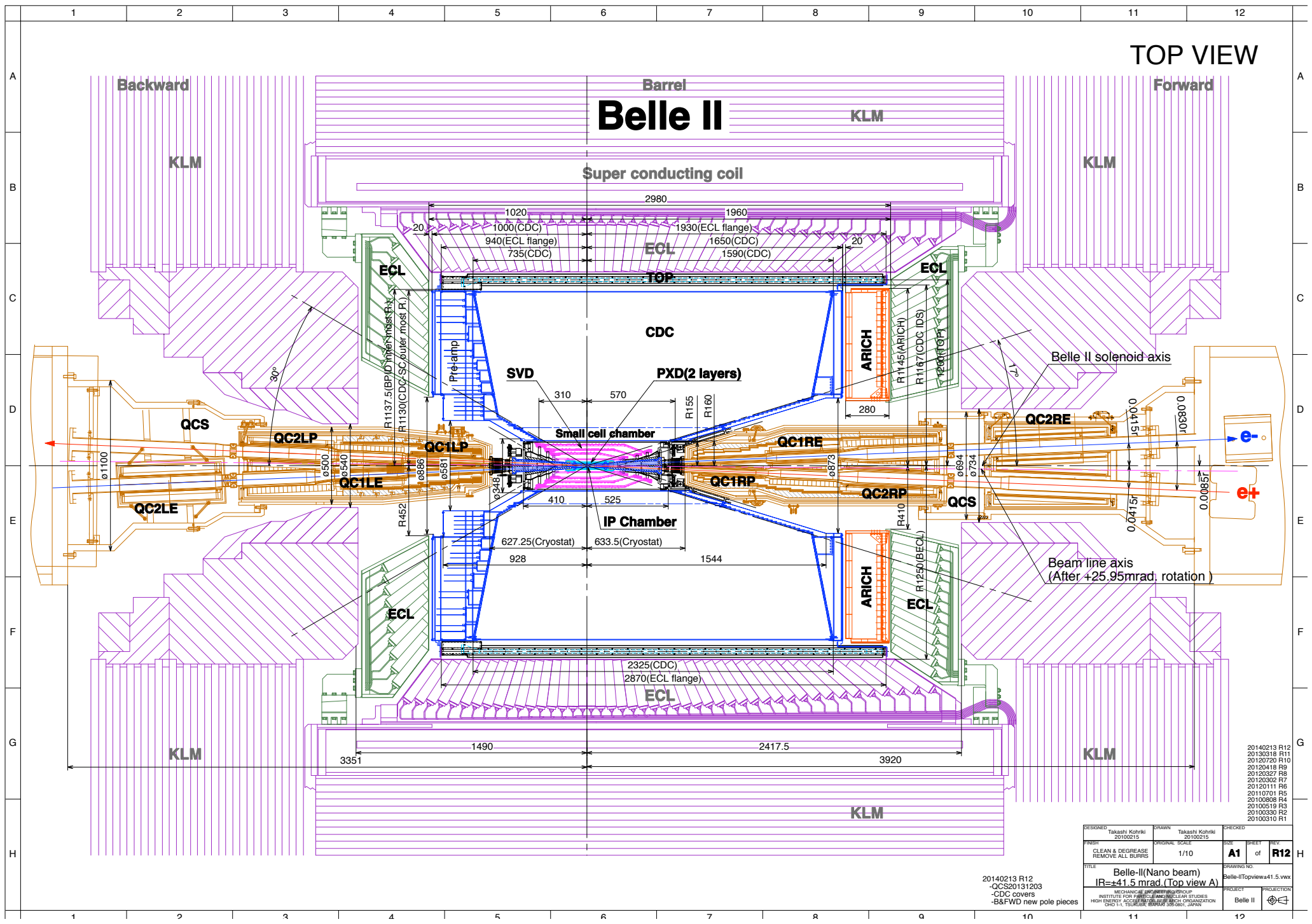

Fig. 3: Belle II top view.

Compared to Belle, the Belle II detector will be taking data at an accelerator with a 40 times higher luminosity, and thus has to be able to operate at 40 times higher event rates, as well as with backgrounds rates higher by a factor of 10 to 20 [6]. To maintain the excellent performance of the spectrometer, the critical issue will be to mitigate the effects of higher background levels, which lead to an increase in occupancy and radiation damage, as well as to fake hits and pile-up noise in the electromagnetic calorimeter, and to neutron induced hits in the muon detection system. Higher event rates also require modifications to the trigger scheme, data acquisition system and computing with respect to the precursor experiment. The trigger and DAQ have also been adapted to support a broader low-multiplicity (dark sector) physics analysis program. In addition, improved hadron identification is needed, and a hermeticity at least as good as in the original Belle detector is required.

The requirements for a $B$ factory detector can be summarised as follows. The apparatus should meet the following criteria:

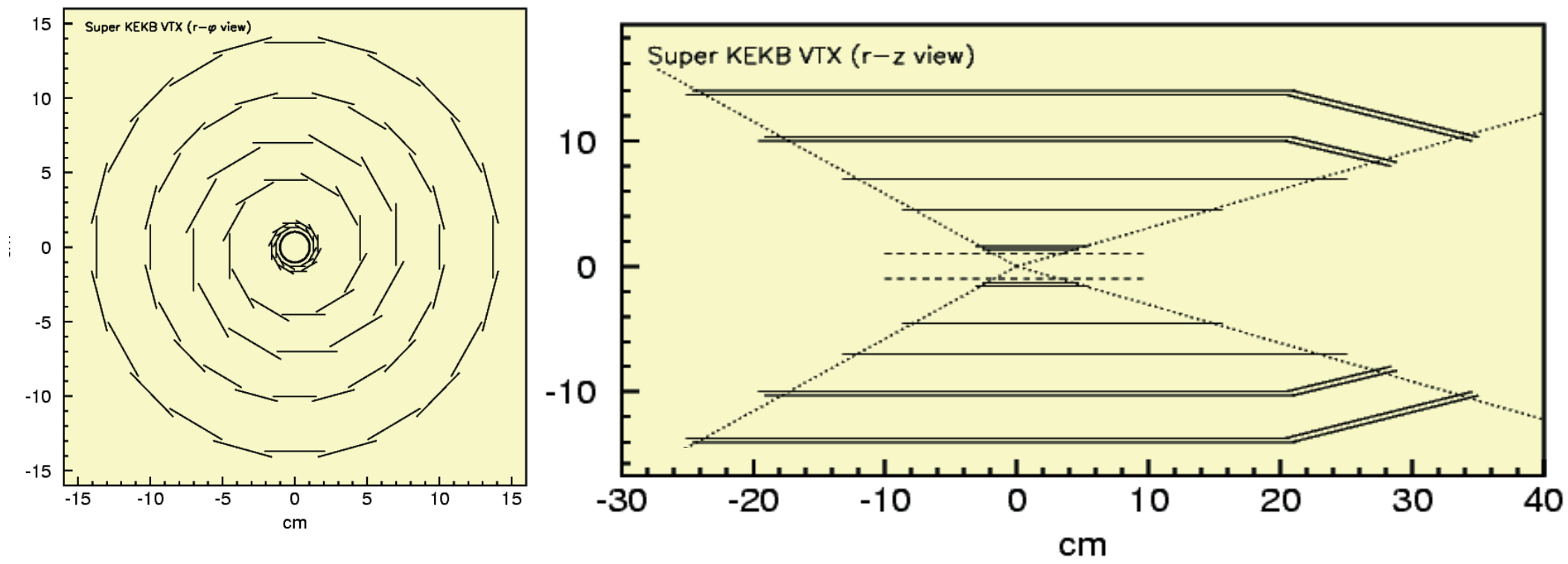

Fig. 4: A schematic view of the Belle II vertex detector with a Be beam pipe, two pixelated layers and four layers with silicon strip sensors.

- Excellent vertex resolution ($\approx 50\mu$m);
- Very high reconstruction efficiencies for charged particles with momenta down to a few hundred MeV/$c$, and improved efficiency for charged particles with momenta down to 50 MeV/$c$;
- Very good momentum resolution over the whole kinematic range of the experiment, *i.e.* up to $\approx$ 8 GeV/$c$;
- Precise measurements of photon energy and direction from a few tens of MeV to $\approx$ 8 GeV, and efficient detection from 30 MeV onwards;
- Highly efficient particle identification system to separate pions, kaons, protons, electrons and muons over the full kinematic range of the experiment;
- Cover the (almost) full solid angle;
- Fast and efficient trigger system, as well as a data acquisition system capable of storing large quantities of data.

The design choices of the Belle II experiment are summarised in Table 17, and are discussed in some detail below. A full discussion can be found in the Technical Design Report (TDR) [6].

The expected Belle II detector performance of some of the critical components, including the track reconstruction efficiency and particle identification capabilities, are discussed in Sec. 5.

### *3.2. Vertex detector (VXD)*

The new vertex detector is comprised of two devices, the silicon Pixel Detector (PXD) and Silicon Vertex Detector (SVD), with altogether six layers (Fig. 4) around a 10 mm radius Be beam pipe. The first layers at $r = 14$ mm and $r = 22$ mm will use pixelated sensors of the DEPFET type [7, 8].

The remaining four layers at radii of 38 mm, 80 mm, 115 mm, and 140 mm will be equipped with double-sided silicon strip sensors. In comparison, in Belle the outermost vertex detector

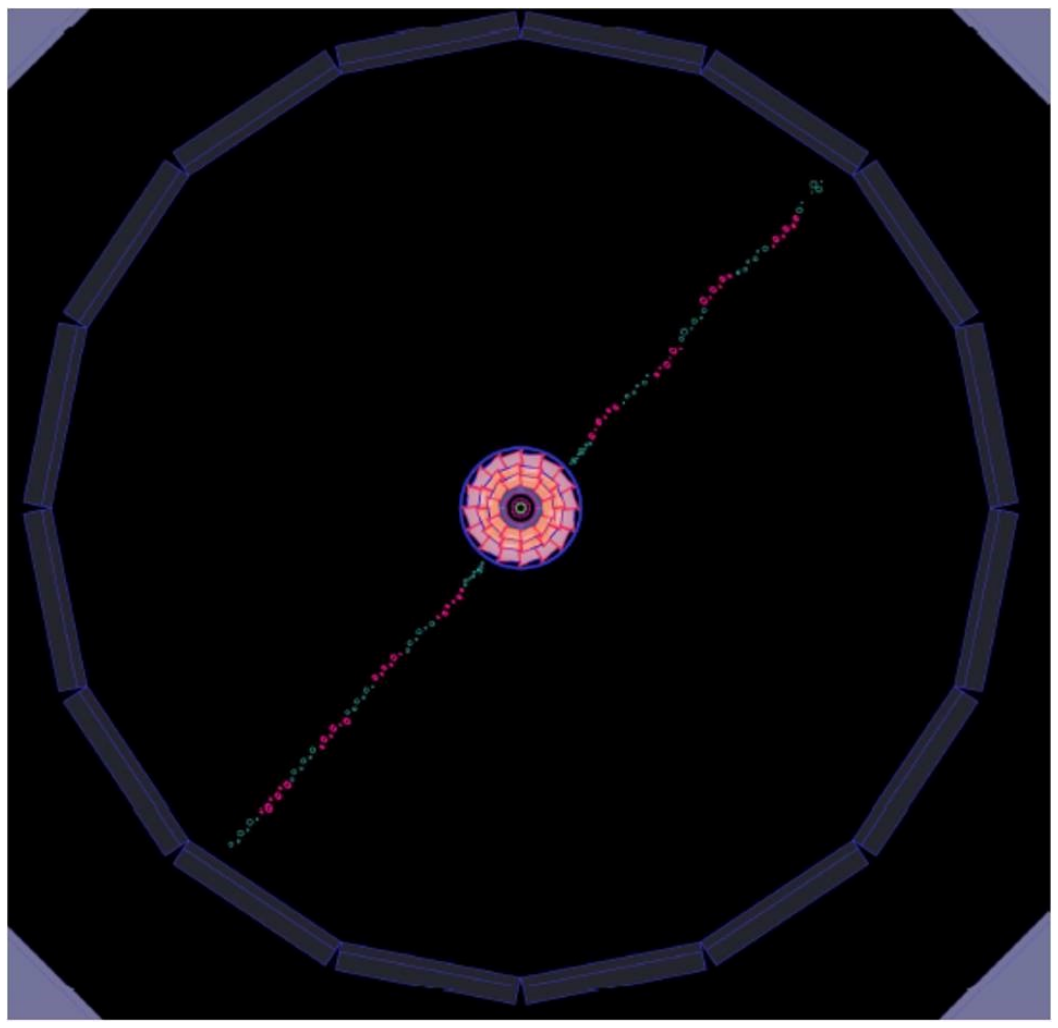

Fig. 5: A cosmic muon as recorded by the Belle II Central Drift Chamber (CDC).

layer was at a radius of 88 mm. The summary table (Table 17) lists the sensor strip pitch sizes.

Compared to the Belle vertex detector, the beam pipe and the first two detector layers are closer to the interaction point, and the outermost layer is at a considerably larger radius. As a result, significant improvement is expected with respect to Belle in the vertex resolution, as well as in the reconstruction efficiency for $K_S^0 \to \pi^+\pi^-$ decays with hits in the vertex detector [6].

### *3.3. Central Drift Chamber (CDC)*

One of the core instruments of the Belle II spectrometer is the central tracking device, a large volume drift chamber with small drift cells. Compared to Belle, it extends to a larger radius (1130 mm compared to 880 mm) due to the upgrade to a much thinner PID device in the barrel region. To be able to operate at high event rates with increased background levels, the chamber has smaller drift cells than the one used in Belle. In total, the CDC contains 14 336 sense wires arranged in 56 layers, either in "axial" orientation (aligned with the solenoidal magnetic field) or "stereo" (skewed with respect to the axial wires). By combining information from axial and stereo layers it is possible to reconstruct a full 3D helix track. The chamber gas is comprised of a He-$C_2H_6$ 50:50 mixture with an average drift velocity of 3.3 cm/$\mu$s and a maximum drift time of about 350 ns for 17 mm cell size.

The drift chamber is by now fully constructed and installed in the Belle II detector and has been commissioned with cosmic rays (Fig. 5).

### *3.4. Particle identification system (TOP and ARICH)*

For particle identification in the barrel region, a time-of-propagation (TOP) counter is used [9, 10]. This is a special kind of Cherenkov detector where the two dimensional information of a Cherenkov ring image is given by the time of arrival and impact position of Cherenkov photons at the photo-detector at one end of a 2.6 m long quartz bar (Fig. 6). Each detector module (16 in total) consists of a 45 cm wide and 2 cm thick quartz bar with a small expansion volume (about 10 cm long) at the sensor end of the bar. The expansion wedge introduces some additional pinhole imaging, relaxes slightly the precision timing requirements and reduces the hit occupancy at the photo-detector [10]. At the exit window of the wedge, two rows of sixteen fast multi-anode photon detectors are mounted. The TOP counter requires photo-sensors with a single photon time resolution of about 100 ps, which can be achieved with a 16-channel MCP PMT [10, 11] specially developed for this purpose. For precision timing required in this type of counter, a custom-made waveform sampling read-out electronics is used [12]. Note that for this identification method the starting (particle production) time has to be known with a precision of about 50 ps; this is indeed challenging, but was already achieved for the time-of-flight (TOF) counter of Belle [13].

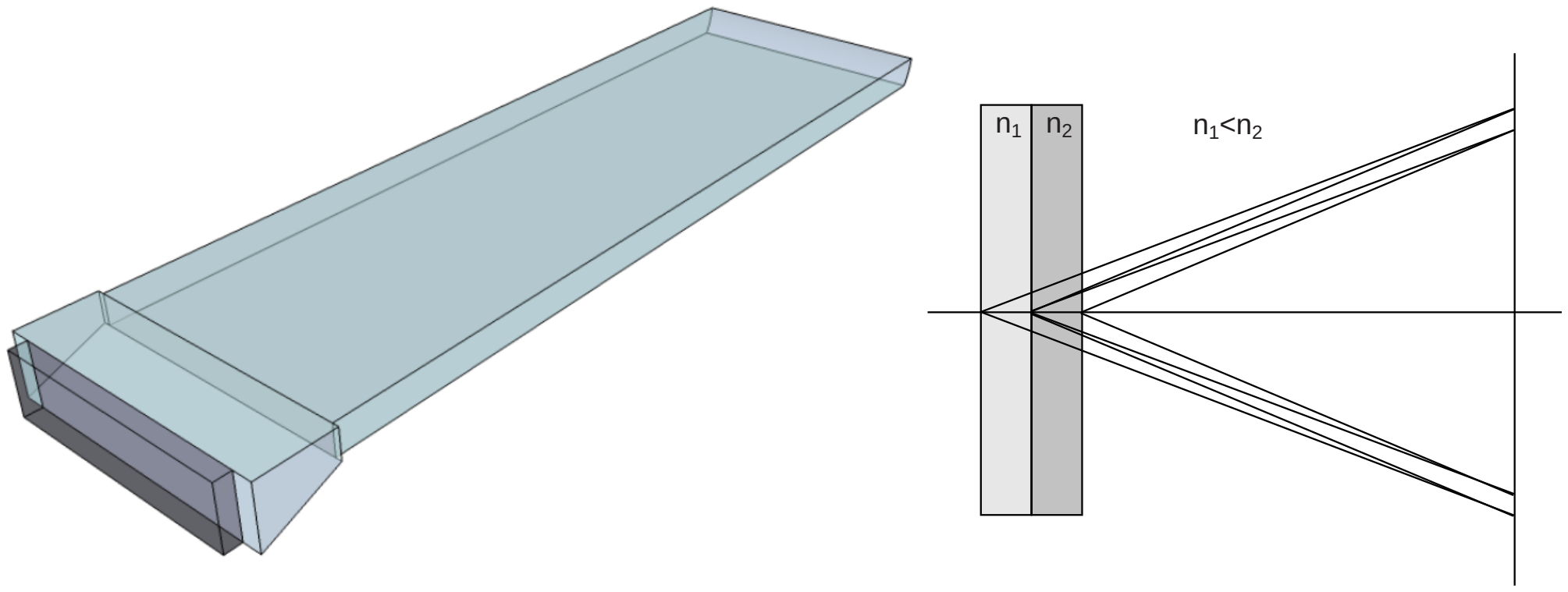

Fig. 6: Belle-II PID systems: one of the modules of the TOP counter (left), principle of operation of the proximity focusing RICH with a non-homogeneous aerogel radiator in the focusing configuration (right).

In the forward end-cap region, ARICH, a proximity focusing Cherenkov ring imaging detector with aerogel as Cherenkov radiator will be employed to identify charged particles. The design requirements include a low momentum threshold for pions and good separation of pions and kaons from 0.4 GeV/$c$ up to about 4 GeV/$c$.

A key parameter of the RICH, the number of detected Cherenkov photons, is increased by a novel method (Fig. 6). Two 2 cm thick layers of aerogel with different refractive indices ($n = 1.045$ upstream, $n = 1.055$ downstream) are used to increase the yield without degrading the Cherenkov angle resolution [14, 15]. As the single photon sensitive high granularity sensor, the hybrid avalanche photon detector (HAPD) is used, developed in a joined effort with Hamamatsu [16, 17]. In this 73 × 73 mm$^2$ sensor with 144 channels, photo-electrons are accelerated over a potential difference of 8 kV, and are detected in avalanche photodiodes (APD). Sensor production was optimised (thicknesses of p and p+ layers, additional intermediate electrode) following radiation tolerance tests [17] with neutrons and gamma rays. All 16 modules of the TOP counter have been installed, and are being commissioned. The ARICH detector is fully installed; all photo-sensor modules (HAPD light sensors and read-out electronics boards) have by now been installed and are being commissioned. With a partially equipped detector, the first Cherenkov rings observed are shown in Fig. 7.

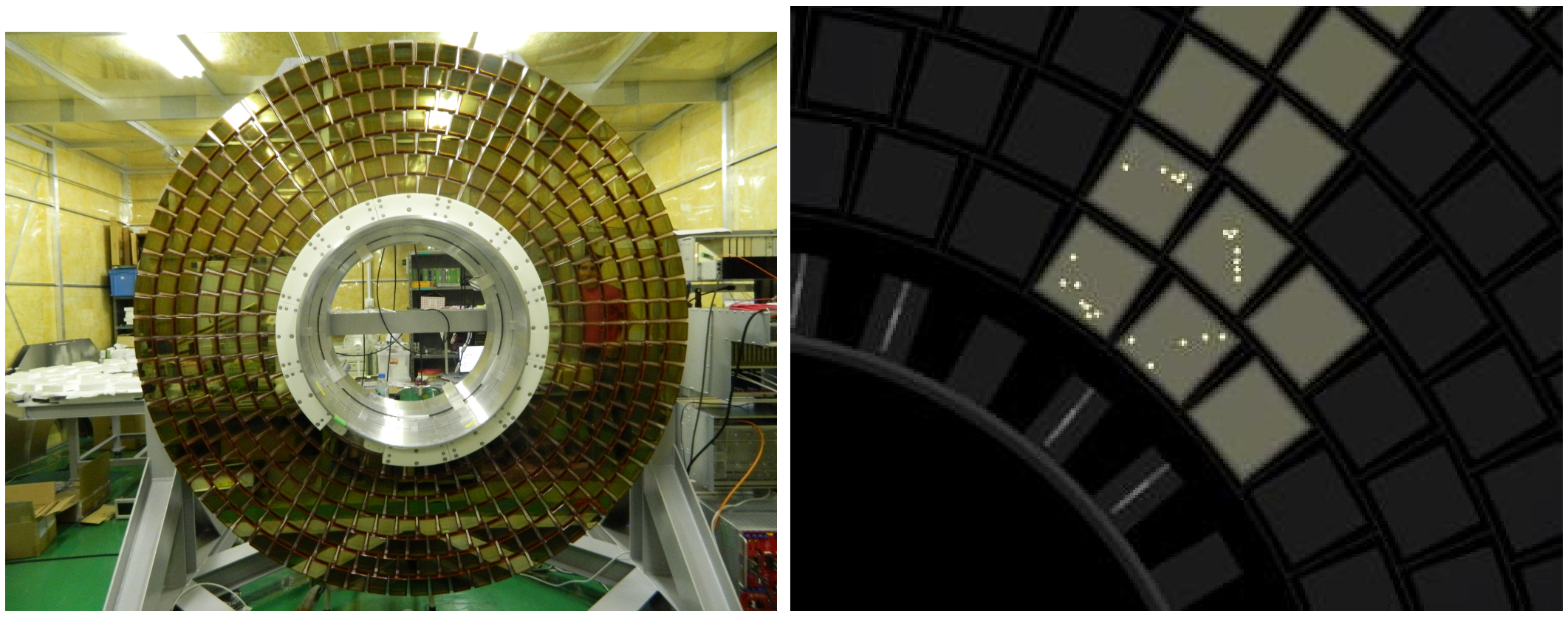

Fig. 7: ARICH detector: photon detector plane with HAPD sensors (left); a ring produced by a cosmic muon (right).

### *3.5. Electromagnetic Calorimeter (ECL)*

The electromagnetic calorimeter is used to detect gamma rays as well as to identify electrons, *i.e.* separate electrons from hadrons, in particular pions. It is a highly-segmented array of thallium-doped caesium iodide CsI(Tl) crystals assembled in a projective geometry (Fig. 3). All three detector regions, barrel as well as the forward and backward end-caps, are instrumented with a total of 8736 crystals, covering about 90% of the solid angle in the centre-of-mass system. The CsI(Tl) crystals, preamplifiers and support structures have been reused from Belle, whereas the readout electronics and reconstruction software have been upgraded. In the Belle experiment, the energy resolution observed with the same calorimeter was $\sigma_E/E = 4\%$ at 100 MeV, 1.6% at 8 GeV, and the angular resolution was 13 mrad (3 mrad) at low (high) energies; $\pi^0$ mass resolution was 4.5 MeV/$c^2$ [2]; in absence of backgrounds a very similar performance would also be expected in Belle II.

In the presence of considerably elevated background levels as compared to the operation in Belle, the relatively long decay time of scintillations in CsI(Tl) crystals will considerably increase the overlapping of pulses from neighbouring (background) events. To mitigate the resulting large pile-up noise, scintillator photo-sensors were equipped with wave-form-sampling read-out electronics. In the forward region of the detector, close to the beam pipe, much higher background rates are expected, such that even with the new wave-form-sampling electronics the pile-up noise will degrade the performance. Some further degradation could come from a reduction of the light yield due to radiation damage, although this effect seems to be less significant than originally anticipated [18]. As a possible solution for this region of the spectrometer, a replacement of CsI(Tl) with considerably faster and radiation tolerant pure CsI is under study [19].

### *3.6. $K_{\mathrm{L}}$- Muon Detector (KLM)*

The $K_L^0$ and muon detector (KLM) consists of an alternating sandwich of 4.7 cm thick iron plates and active detector elements located outside the superconducting solenoid. The iron plates serve as the magnetic flux return for the solenoid. They also provide 3.9 interaction lengths or more of material, beyond the 0.8 interaction lengths of the calorimeter, in which $K_L^0$ mesons can shower hadronically.

The Belle KLM, based on glass-electrode resistive plate chambers (RPC), has demonstrated good performance during the entire data taking period of the Belle experiment. Contrary to Belle, in Belle II in some KLM detector areas (both endcaps and the innermost layers in the barrel region) large background rates are expected due to neutrons that are mainly produced in electromagnetic showers from background reactions (*e.g.*, radiative Bhabha scattering). The long dead time of the RPCs during the recovery of the electric field after a discharge significantly reduces the detection efficiency under such background fluxes. The resulting fake muon identification probability would become so high in the endcap region of the spectrometer and in the two inner layers of the barrel, that such a counter would be useless [6]. To mitigate this problem, RPCs have been replaced by layers of scintillator strips with wavelength shifting fibers, read out by silicon photomultiplier (SiPMs, Geiger mode operated APDs) as light sensors [20]. Note that the high neutron background will also cause damage to the SiPMs, and will therefore considerably increase the dark count rate in the light sensor; irradiation tests have shown, however, that such a detector system can be reliably operated by appropriately setting the discrimination threshold.

### *3.7. Trigger System*

The trigger system of Belle II has a non trivial role to identify events of interest during data taking. The scope of physics analysis topics that require dedicated triggers will be broad at Belle II. These triggers must work efficiently in the presence of the much higher background rates expected from SuperKEKB, and satisfy the limitations of the data acquisition system (DAQ). A well-designed trigger system unlocks a broad variety of topics not probed in the previous generation B-factories. Excellent examples of triggers for new phenomena include the single photon trigger for dark sector searches, and the two- and three- photon triggers for axion-like particle searches.

The dominant beam background sources are discussed in detail in the next chapter, and are namely from the Touschek effect, Beam-gas Scattering, Synchrotron radiation, Radiative Bhabha process, two-photon process, and beam-beam effects. The rates of these background processes are correlated with multiple factors *e.g.* beam size, beam current, luminosity, accelerator status, vacuum conditions, and so on.

Most of these processes are characterised by the presence of fewer than two charged particle tracks in CDC, accompanied with one or two clusters in the ECL. These topologies are similar to those of primary collision events to low-multiplicity production modes, and are therefore a large problem for such physics studies.

The flagship measurements for Belle II in $B$- and $D$- flavour physics are expected to be highly robust to trigger implementation, where events will be easily identified from the presence of at least 3 tracks in the CDC trigger and a large deposition of energy in the ECL. Similarly to Belle, the trigger for most $B$-decays will be close to 100% efficient, for events that are reconstructed by offline algorithms.

The long list of new low multiplicity and dark sector triggers under development at Belle II will increase the physics scope but present a large challenge to the DAQ system. In addition to $B$ physics, Belle II is also an excellent ground for the study of many other important topics *e.g.* $\tau$ physics, dark sector searches, two-photon physics, and precision measurements of low-multiplicity and ISR processes. Precision measurements of luminosity from low-multiplicity events are also important input to precision flavour physics measurements. The low-multiplicity topology of these processes is however similar to the background processes mentioned above, leading to low purity and must be tackled using online algorithms.

The scheme of Belle II trigger system is composed of two levels: hardware based low level trigger (L1) and software based high level trigger (HLT). Key design features of each level are described in turn below.

The nominal L1 trigger has a latency of 5 $\mu$s, and maximum trigger output rate of 30 kHz, limited by the read-in rate of the DAQ. To cope with the high event rate and background level at Belle II, a series of upgrades are implemented at L1. The key areas of improvement involve the implementation of firmware based reconstruction algorithms and trigger logic.

- *Tracking.* Novel 3D tracking algorithms (based on 3D-fitting, and Multi Layer Perceptron (MLP) respectively) have been developed to provide the vertex position in the direction of the beam-line ($z-$axis). This is used to suppress beam background that does not originate from the interaction point. At Belle only 2D information was derived in the L1 trigger. The 3D track information allows for matching the CDC track with associated ECL clusters, and therefore improve particle identification at the trigger level.
- *Calorimeter.* High rate background from radiative Bhabha scattering, which has a cross section of of 74 nb in the CDC acceptance, will be reduced with improved online reconstruction techniques. Bhabha vetoes in the $B$-factories tended to remove a substantial rate of interesting low-multiplicity processes. To better suppress Bhabha events, a 3D Bhabha logic has been developed in the ECL trigger which uses 3D ECL clustering information.
- *Global reconstruction.* The trigger information from each sub-detector trigger is combined using an FPGA based Global Reconstruction Logic (GRL) to perform low level particle and event reconstruction *e.g.* matching between tracks found in the CDC and clusters found by the ECL trigger. The GRL is one of the key new components of the Belle II L1, and will be critical for controlling rates at high luminosity.
- *Trigger menu.* Belle II will have a new trigger menu, or set of trigger lines, to satisfy a variety physics analysis targets. For hadronic processes *e.g.* B decays and continuum, they will be triggered with high efficiency by requiring that there are at least three tracks in CDC. Low multiplicity processes are easily mimicked by radiative Bhabha or beam background events, and are therefore difficult to efficiently trigger on.
- *Trigger conditions.* The trigger menu will be designed for specific periods of data taking at varying collision centre of mass energies and instantaneous luminosity. It will be tuned to take into account varying background flux as a function of the polar angle of the detector. The regions of the detector close to beam pipe suffers high beam-induced background. Background from beam gas is more prevalent at the beginning of data taking due to the beam vacuum conditions at startup. The background processes with scattering rates proportional to the luminosity *e.g.* Bhabha, will be more prevalent as luminosity rises.
- *Dark sector trigger.* Dark matter searches are a big challenge for the trigger, which can be characterised by the presence of only one energetic photon in the final state. Bhabha and $e^+e^- \to \gamma\gamma$ are the dominant background in the endcaps and at high luminosity.

Consequently loose triggers are applied for the photon in the barrel of ECL, and tight conditions are applied in the endcaps. Some trigger lines may need to eventually be pre-scaled but this will be decided later. These triggers are detailed further in the dark sector physics section.

As a key component of the DAQ, the HLT must reduce online event rates to 10 kHz for offline storage, and it must identify track regions of interest for PXD readout to reduce data flux. The HLT reconstructs the event with offline reconstruction algorithms, allowing access to full granularity event reconstruction using all detectors except for the PXD.

- *Architecture.* The HLT will suppress the event rate to 15 kHz firstly with the information from the CDC track finding and ECL reconstruction, which have been optimised for fast online operation. Only the events passing this first step are considered for full event reconstruction. This step typically rejects residual beam background not found by the L1. The event rate is further reduced to 10 kHz by using full reconstruction information.
- *Trigger menu.* A robust trigger menu for the HLT is in development. As with L1, Bhabha scattering is a dominant background.
- *CPU farm.* To process at nominal 30 kHz a total of 6000 CPU cores are employed. This is the required rate for nominal instantaneous luminosity.

### *3.8. Detector commissioning phases*

The Belle II experiment is scheduled to begin its first “physics” run in 2019. As a prelude to this, two commissioning periods known as “Phase 1” (February 2016-June 2016) and “Phase 2” (Februray 2018 - July 2018) were scheduled where a collection of detectors, known as BEAST 2 (Beam Exorcism for A Stable Belle II Experiment) were deployed for measuring background rates and operating conditions. During Phase 1, the solenoid was not active, and no collisions took place. However, for Phase 2 all subsystems except for the full vertex detectors were employed, for colliding beams to produce useful physics and calibration events.

Given the expected luminosity profile, it will likely take until at least mid-2019 for Belle II to collect an $\Upsilon(4S)$ dataset large enough to equal that of the $B$-factory experiments. Data collected at different centre-of-mass energies is a consideration to ensure Belle II accesses unique data sets from early in its program.

During phase 2, Belle II will contain only one octant of the pixel detector (PXD) and silicon vertex detector (SVD), consisting of 2 and 4 ladders, respectively. They will be placed in the +X direction, which is expected to have the highest beam background radiation. The final focusing magnets, QCSL and QCSR, will be installed such that combined with the Belle II solenoid, the final magnetic field configuration will be present for charged particle track reconstruction. An exact copy of the final physics run beam pipe with final geometry and composition will be installed (the exception is the gold foil thickness, which will be 6.6 $\mu$m instead of the nominal 10 $\mu$m in order to measure synchrotron radiation). Most of the BEAST2 commissioning detectors will not be included in the Belle II DAQ, and are used solely for beam background characterisation. All of the outer detector elements will be present and operational in Phase 2: the CDC, TOP, ARICH, ECL, and KLM.

The main aim of Phase 2 is to commission the SuperKEKB accelerator to a point where integrating the full VXD is deemed safe. The majority of time will be spent towards achieving this aim. The nominal operating energy is 7 GeV on 4 GeV (centre-of-mass energy at $\Upsilon(4S)$), but the machine should be capable of operating anywhere in the range from $\Upsilon(2S)$ (at 10.02 GeV) up to 11.25 GeV. The beam energy spread is expected to be fairly close to the nominal value of approximately 5 MeV, even during this stage. The goal is to reach an instantaneous luminosity of approximately $1 \times 10^{34}$ cm$^{-2}$s$^{-1}$, and to measure the luminosity dependence of the leading background processes. The first few months of Phase 2 will be devoted towards machine commissioning goals, BEAST background studies, and ramp up of the instantaneous luminosity to reach its target. If these tasks are accomplished in a timely manner, the remainder can be used for physics data collection.

Table 17: Summary of the detector components.

| Purpose | Name | Component | Configuration | Readout channels | $\theta$ coverage |
|---|---|---|---|---|---|
| Beam pipe | Beryllium | | Cylindrical, inner radius 10 mm, 10 $\mu$m Au, 0.6 mm Be, 1 mm paraffin, 0.4 mm Be | | |
| Tracking | PXD | Silicon Pixel (DEPFET) | Sensor size: 15×(L1 136, L2 170) mm$^2$, Pixel size: 50×(L1a 50, L1b 60, L2a 75, L2b 85) $\mu$m$^2$; two layers at radii: 14, 22 mm | 10M | [17°;150°] |
| | SVD | Silicon Strip | Rectangular and trapezoidal, strip pitch: 50(p)/160(n) - 75(p)/240(n) $\mu$m, with one floating intermediate strip; four layers at radii: 38, 80, 115, 140 mm | 245k | [17°;150°] |
| | CDC | Drift Chamber with He-$C_2H_6$ gas | 14336 wires in 56 layers, inner radius of 160mm outer radius of 1130 mm | 14k | [17°;150°] |
| Particle ID | TOP | RICH with quartz radiator | 16 segments in $\phi$ at $r \sim 120$ cm, 275 cm long, 2cm thick quartz bars with 4×4 channel MCP PMTs | 8k | [31°;128°] |
| | ARICH | RICH with aerogel radiator | 2×2 cm thick focusing radiators with different $n$, HAPD photodetectors | 78k | [14°;30°] |
| Calorimetry | ECL | CsI(Tl) | Barrel: $r = 125 - 162$cm, end-cap: $z = -102 - +196$cm | 6624 (Barrel), 1152 (FWD), 960 (BWD) | [12.4°;31.4°], [32.2°;128.7°], [130.7°;155.1°] |
| Muon ID | KLM | barrel:RPCs and scintillator strips | 2 layers with scintillator strips and 12 layers with 2 RPCs | $\theta$ 16k, $\phi$ 16k | [40°;129°] |
| | KLM | end-cap: scintillator strips | 12 layers of (7-10)×40 mm$^2$ strips | 17k | [25°;40°], [129°;155°] |

The lack of the VXD elements is expected to have the largest impact on physics during Phase 2. Due to the missing VXD, track reconstruction in Phase 2 is entirely dependent on the CDC, therefore tracks must be able to reach this detector and produce sufficient hits in order to be reconstructed. This leads to efficiency losses at low $p_{\mathrm{T}}$ due to acceptance. These effects are seen with momenta below 1 GeV/$c$ and become most pronounced below 150 MeV/$c$, with almost no sensitivity to tracks with $p_T < 75$ MeV in Phase 2. Losses are approximately uniform in the azimuthal angle $\phi$, except in the $\phi = 0$ direction where PXD and SVD elements will be partially installed. Efficiency in the polar angle is roughly constant except at the CDC edge and SVD wedge regions. Therefore analyses requiring these tracks (*e.g.* soft pions in $\Upsilon(3S) \to \pi^+\pi^-\Upsilon(2S)$ decays, detecting all tracks in an event) will be affected. Studies of photon efficiency indicate that no appreciable difference is expected in performance between Phases 2 and 3. The full ECL will be present and operational during both phases. Even though the VXD will be absent, the BEAST2 components contribute a nearly equivalent amount of material. This has been reduced as much as possible to avoid affecting performance for physics and commissioning. As a result, analyses relying on photon detection are expected to be as effective in Phase 2 as in Phase 3.

The physics potential for these phases will be discussed further in the WG7-Quarkonium(like) physics and WG8-Tau decay and low-multiplicity physics chapters.

# 4. Belle II Simulation

*Section author(s): T. Ferber, D. Kim, H. Nakayama, M. Ritter, M. Staric*

## 4.1. Introduction

This chapter describes the simulation tools used in the studies presented in this report. This includes a brief review of the main event generators, the detector simulation and an overview of the expected beam backgrounds. Some analyses require very specific event generators whose description is given in the respective subsections. The reference cross sections for various physics processes are provided as well.

All simulations start with at least one event generator that simulates the primary physics process, followed by a detailed detector simulation. Some studies include the effects of beam background, which is simulated in specific background simulations and added to the physics event simulation.

The studies presented, and the performance reported, throughout this report use different versions of the Belle II software basf2. This software is under active development and the performance (*e.g.* resolution, efficiency, background tolerance) typically improves with each software revision. Most of the studies make use of centrally produced MC campaigns: MC5 is based on release-00-05-03, MC6 and MC7 are based on release-00-07-02, and MC8 is based on release-00-08-00. The latest basf2 version used in the publication is release-00-09-01, which is referenced in some performance outlooks.

## 4.2. Cross Sections

Cross sections for the most important physics processes are given in Table 18 at the default beam energy. In addition to the normalisation values, a rough estimate for observable cross sections within acceptance and some typical generator–level selection criteria are given by the indented values. The selection criteria (if any) for the non-indented cross section values correspond to typical event generator selections.

## 4.3. Generators

Most studies in this report are based on three main event generators: `EvtGen 1.3` [21] is used to model the decays of B and D mesons into exclusive final states. `PYTHIA 8.2` [22] is used for inclusive decay final states and for the continuum production of light quark pairs. $\tau$ pair production is generated using `KKMC 4.15` [23, 24] with the decays handled by `TAUOLA` [25]. In addition, the large cross section QED background processes $e^+e^- \to e^+e^-(\gamma)$ and $e^+e^- \to \gamma\gamma(\gamma)$ are simulated using `BABAYAGA.NLO` [26–30], and $e^+e^- \to e^+e^-e^+e^-$ and $e^+e^-\mu^+\mu^-$ are simulated using `AAFH`[1] [31–33].

All event generators use the same beam parameters, such as the mean beam energies and the vertex position, which are provided by a central data base. The default beam energies are $E_{HER} = 7.004\,\mathrm{GeV}$ and $E_{LER} = 4.002\,\mathrm{GeV}$. The effect of beam energy smearing is included in `EvtGen` and `BABAYAGA.NLO` only. The smearing is modelled as single Gaussian for the HER and LER beams individually, with a width of $\sigma_{HER} = 5.13\,\mathrm{MeV}$ and $\sigma_{LER} = 2.375\,\mathrm{MeV}$, respectively. The default vertex position is the detector centre $(0, 0, 0)$. The vertex smearing covariance matrix is calculated from the horizontal ($x$) and vertical ($y$) beam size at the IP, with the bunch lengths ($z$) of the LER ($\sigma_x$=10.2 $\mu$m, $\sigma_y$=0.059 $\mu$m, $\sigma_z$=5 mm) and HER ($\sigma_x$=7.75 $\mu$m, $\sigma_y$=0.059 $\mu$m, $\sigma_z$=6 mm). The beam angles with respect to the $z$-axis are $\theta_{HER} = 0.0415$ and $\theta_{LER} = -0.0415$. Normally–distributed bunch densities are assumed for the calculation, and the probability density functions for the two bunches are multiplied to get a resulting beam spot. Vertex position smearing is included for all generators.

[1] This generator is sometimes also called `BDK` or `DIAG36`.

Table 18: Total production cross section from various physics processes from collisions at $\sqrt{s} = 10.58\,\mathrm{GeV}$. $W_{\ell\ell}$ is the minimum invariant secondary fermion pair mass.

| Physics process | Cross section [nb] | Selection Criteria | Reference |
|---|---|---|---|
| $\Upsilon(4S)$ | $1.110 \pm 0.008$ | - | [2] |
| $u\bar{u}(\gamma)$ | 1.61 | - | `KKMC` |
| $d\bar{d}(\gamma)$ | 0.40 | - | `KKMC` |
| $s\bar{s}(\gamma)$ | 0.38 | - | `KKMC` |
| $c\bar{c}(\gamma)$ | 1.30 | - | `KKMC` |
| $e^+e^-(\gamma)$ | $300 \pm 3$ (MC stat.) | $10° < \theta_e^* < 170°$, $E_e^* > 0.15\,\mathrm{GeV}$ | `BABAYAGA.NLO` |
| $e^+e^-(\gamma)$ | 74.4 | $p_e > 0.5\,\mathrm{GeV}/c$ and e in ECL | - |
| $\gamma\gamma(\gamma)$ | $4.99 \pm 0.05$ (MC stat.) | $10° < \theta_\gamma^* < 170°$, $E_\gamma^* > 0.15\,\mathrm{GeV}$ | `BABAYAGA.NLO` |
| $\gamma\gamma(\gamma)$ | 3.30 | $E_\gamma > 0.5\,\mathrm{GeV}$ in ECL | - |
| $\mu^+\mu^-(\gamma)$ | 1.148 | - | `KKMC` |
| $\mu^+\mu^-(\gamma)$ | 0.831 | $p_\mu > 0.5\,\mathrm{GeV}/c$ in CDC | - |
| $\mu^+\mu^-\gamma(\gamma)$ | 0.242 | $p_\mu > 0.5\,\mathrm{GeV}$ in CDC, $\geq 1\ \gamma$ ($E_\gamma > 0.5\,\mathrm{GeV}$) in ECL | - |
| $\tau^+\tau^-(\gamma)$ | 0.919 | - | `KKMC` |
| $\nu\bar{\nu}(\gamma)$ | $0.25 \times 10^{-3}$ | - | `KKMC` |
| $e^+e^-e^+e^-$ | $39.7 \pm 0.1$ (MC stat.) | $W_{\ell\ell} > 0.5\,\mathrm{GeV}/c^2$ | `AAFH` |
| $e^+e^-\mu^+\mu^-$ | $18.9 \pm 0.1$ (MC stat.) | $W_{\ell\ell} > 0.5\,\mathrm{GeV}/c^2$ | `AAFH` |

`EvtGen` is an event generator originally developed for BaBar and CLEO. `EvtGen` accounts for cascade decays involving multiple vertices and spin configurations. Input data for each decay process is passed to the code as a complex amplitude. In cases where a number of complex amplitudes are invoked for the same process, these are added before the decay probabilities are calculated and consequently the interference terms, which are of significant importance in many $B$-physics studies, are included. `EvtGen` is controlled by means of a fairly complete decay table (`DECAY.DEC`), which lists all possible decay processes, their branching ratios, and the model (amplitude) which is to be used to decay them. Belle II currently uses the simplified default EvtGen decay file for generic B events, which lacks some improvements that were included in the Belle or BaBar decay files. Since EvtGen only handles exclusive final states, PYTHIA 8.2 is used to produce final states not included in the decay file. Double counting is avoided by rejecting decays produced by PYTHIA 8.2 that are already included in the decay file. PHOTOS is used to simulate final state radiation correction in decays [34]. Up to MC8, EvtGen is also used to simulate $u\bar{u}$, $d\bar{d}$, $s\bar{s}$, and $c\bar{c}$ continuum events that are fragmented into final states using PYTHIA 8.2. Unlike at Belle, the continuum light quark production in EvtGen does not include initial state radiation. Starting with release-00-09-01, continuum events are produced using KKMC and PYTHIA 8.2 and include ISR.

In general, it is not straightforward to translate the Belle fragmentation settings to Belle II since the PYTHIA version has been changed from PYTHIA 6 to PYTHIA 8.2 and not all PYTHIA 6 parameters have PYTHIA 8.2 equivalents and vice versa. All currently used non–default PYTHIA 8.2 parameters are listed in Table 19 and were chosen to approximate the settings used in Belle. It is planned to tune the PYTHIA 8.2 parameters that control the fragmentation process of light (uds) and charm quarks for Belle II based on Belle data before the start of Belle II phase 3 data taking. The parameters will be tuned separately with and without the ones responsible for excited meson production.

Table 19: PYTHIA 8.2 parameters with changed values in Belle II.

| Parameter name | Default | Belle II |
|---|---|---|
| StringFlav:etaSup | 0.60 | 0.27 |
| StringFragmentation:stopMass | 1.0 | 0.3 |
| StringZ:aLund | 0.68 | 0.32 |
| StringZ:bLund | 0.98 | 0.62 |
| StringZ:rFactC | 1.32 | 1.0 |

KKMC is the default generator to simulate the two fermion final states $e^+e^- \to \mu^+\mu^-(\gamma)$ and $e^+e^- \to \tau^+\tau^-(\gamma)$. The currently implemented version is based on the Belle implementation of KKMC4.19 including a modified interface for tau decays. KKMC generates multi–photon initial state radiation (ISR), final state radiation (FSR), and the interference of initial and final state radiation (IFI). These QED corrections are complete NLO for ISR, IFI, and FSR, and almost complete NNLO for ISR and FSR within the framework of exclusive coherent exponentiation based on Yennie–Frautschi–Suura exclusive exponentiation. $\tau$ decays are handled by TAUOLA-exp-11-10-2005, taking into account spin polarisation effects and transverse spin correlations in $\tau$ decays. The hadronic currents for $\tau \to 4\pi$ are taken from CMD–2, all others from CLEO. Electroweak corrections within KKMC are implemented using the DIZET6.21 library of the ZFITTER project [35, 36]. The DIZET6.21 routine REPI for the calculation of the time–like real part of the electromagnetic coupling $\alpha_{QED}(s)$ has been replaced as described in [37]. The electroweak corrections are complete one–loop with some higher–order extensions. The theoretical precision of the generator for lepton pairs is stated to be better than 0.5 % for both cross section and inclusive differential distributions within the detector acceptance for beam energies at and above the $\Upsilon(4S)$, including uncertainties due to vacuum polarisation [37].

`BABAYAGA.NLO` is the default generator to simulate large angle (above about 5° in the CM frame) $e^+e^- \to e^+e^-(\gamma)$ (Bhabha) and $e^+e^- \to \gamma\gamma(\gamma)$ final states. `BABAYAGA.NLO` generates multi–photon ISR, FSR, and IFI based on the matching of exact NLO corrections with a parton shower algorithm. $Z$ exchange and $\gamma - Z$ interference are included at the Born level. Narrow resonances and vacuum polarisation corrections are included but no other electroweak corrections. The theoretical precision of the generator is stated to be about 0.1 % for both cross section and inclusive differential distributions within the detector acceptance.

The non–radiative four fermion final states $e^+e^- \to e^+e^-e^+e^-$ and $e^+e^- \to e^+e^-\mu^+\mu^-$ are simulated using `AAFH`. `AAFH` includes all LO QED diagrams and their interference, but no higher–order QED corrections, no weak corrections, and no $Z$–exchange. The leading order calculation is exact and includes final state mass kinematics. The leading order divergency of the process is controlled using a selection criterion on the minimum invariant secondary fermion pair mass, with typically $W_{\ell\ell} > 0.5\,\text{GeV}/c^2$.

### *4.4. Beam–induced background*

We begin by giving an overview of the five main beam background sources at SuperKEKB. We include luminosity-dependent backgrounds such as radiative Bhabha scattering and production of two-photon events.

*4.4.1. Touschek scattering.* The first background source is the Touschek effect, which is enhanced at SuperKEKB due to the Nano-beam scheme. The Touschek effect is an intra-bunch scattering process, where Coulomb scattering of two particles in the same beam bunch changes the particles' energies to deviate from the nominal energy of the bunch. One particle ends up with an energy higher than nominal, the other with lower energy than nominal.

The Touschek scattering probability is calculated using Bruck's formula, as described in [4]. The total scattering rate, integrated around the ring, is proportional to the number of filled bunches and the second power of the bunch current, and inversely proportional to the beam size and the third power of the beam energy. Simple extrapolation based on beam size predicts that the Touschek background at SuperKEKB will be a factor of ~20 higher than at KEKB.

Touschek-scattered particles are subsequently lost at the beam pipe inner wall after they propagate further around the ring. If the loss position is close to the detector, the resulting shower might reach the detector. To mitigate Touschek background, we utilise horizontal and vertical movable collimators and metal shields. The collimators, located at different positions around the ring, stop particles that deviate from nominal trajectories and prevent them from reaching Belle II. While we had horizontal collimation only from the inner side of the beams at KEKB, Touschek background can be reduced effectively by collimating the beam horizontally from both the inner and outer side.The horizontal collimators located just before to the interaction region play an important role in minimising the beam loss rate inside the detector. The nearest LER collimator is only 18 m upstream of the interaction point. In phase 3, there will also be heavy-metal shields in the vertex detector (VXD) volume and on the superconducting final focus cryostat, to prevent shower particles from entering the Belle II acceptance.

*4.4.2. Beam-gas scattering.* The second beam background source is the so-called beam-gas scattering, *i.e.* scattering of beam particles by residual gas molecules in the beam pipe. This can occur via two processes, Coulomb scattering, which changes the direction of the beam particle, and Bremsstrahlung scattering, which decreases the energy of the beam particles. The rate of beam-gas scattering is proportional to the beam current and to the vacuum pressure in the beam pipe. At SuperKEKB, the beam currents will be approximately two times higher than at KEKB, while the vacuum level, except for the interaction region, will be similar to that at KEKB.

The rate of Beam-gas Bremsstrahlung losses in the detector is well suppressed by horizontal collimators and is negligible compared to the Touschek loss rate in the detector. However, the beam-gas Coulomb scattering rate is expected to be a factor of $\sim 100$ higher than at KEKB, because the SuperKEKB beam pipe radius inside the detector is smaller, and the maximum vertical beta function is larger. Beam-gas scattered particles are lost by hitting the beam pipe inner wall while they propagate around the ring, just like Touschek-scattered particles.

The countermeasures used for Touschek background, movable collimators and heavy-metal shields, are also effective at reducing beam-gas background. In particular, vertical collimators are essential for reducing Coulomb scattering backgrounds. However, potential Transverse Mode Coupling (TMC) instabilities caused by vertical collimators should be carefully examined, since the vertical beta function is larger than horizontal beta function. Therefore, the collimator width must satisfy two conditions at the same time:

- narrow enough to avoid beam loss in the detector
- wide enough to avoid TMC instability

The only way to achieve this is to use vertical collimators with $\sim 2$ mm width in locations where the vertical beta function is relatively small. This is different from horizontal collimators, which are installed where the horizontal beta function is large.

*4.4.3. Synchrotron radiation.* The third background source is synchrotron radiation (SR) emitted from the beam. Since the SR power is proportional to the beam energy squared and magnetic field strength squared, the HER beam is the main source of this type of background. The energy spectrum of SR photons ranges from a few keV to tens of keV.

During early running of KEKB, the inner layer of the Belle Silicon Vertex Detector (SVD) was severely damaged by x-rays with E $\sim 2$ keV from the HER. To absorb SR photons before they reach the Belle II inner detectors (PXD/SVD), the inner surface of the beryllium beam pipe is coated with a gold layer. The shape of IR beam pipe is designed to avoid direct SR hits at the detector. Ridge structures on the inner surface of incoming pipes prevent scattered photons from reaching the interaction point.

*4.4.4. Radiative Bhabha process.* The fourth background source is Radiative Bhabha scattering. Photons produce by the radiative Bhabha process propagate along the beam axis direction and interact with the iron of magnets. In these interactions, there is a very large production rate of neutrons via the photo-nuclear resonance mechanism.

Such neutrons are the main background source for the outermost Belle II detector, the $K_L$ and muon detector (KLM), situated in the return yoke of the experiment's solenoid magnet. The rate of neutron production by radiative Bhabha events is proportional to the luminosity, which is 40 times higher at SuperKEKB than at KEKB. Additional neutron shielding in the accelerator tunnel is required to stop these neutrons.

Both the electron and positron energy decrease after radiative Bhabha scattering. KEKB employed shared QCS magnets for the incoming and outgoing beams, and as a result the scattered particles were over-bent by the QCS magnets. The particles then hit the wall of magnets and electromagnetic showers were generated.

In SuperKEKB we use two separate quadrupole magnets and both orbits for incoming and outgoing beams are centred in the Q-magnets. We therefore expect the radiative Bhabha background due to over-bent electrons and positrons to be small, and only the small fraction with very large energy loss ($\Delta E$) is lost inside the detector. However, since the design luminosity of SuperKEKB is 40 times higher than that of KEKB, the rate of those large $\Delta E$ particles is still not negligible and will be comparable to Touschek and Beam-gas background after installation of collimators. The transverse kick from the solenoid field due to a finite crossing angle is the crucial and inevitable cause of these beam losses. The intrinsic angular beam divergence at the IP, angular diffusion by the radiative Bhabha process, and leak fields from the other ring's Q-magnets also play a role, but are less crucial than the solenoid kick.

In addition, radiative Bhabha losses within $|s| < 65$ cm of the IP are particularly dangerous because we cannot put enough shielding material in that region to prevent showers from entering the acceptance region. The cryostat is located at $|s| > 65$ cm.

*4.4.5. Two photon process.* The fifth beam background results from very low momentum electron-positron pairs produced via the two-photon process $ee \to eeee$. Such pairs can spiral around the solenoid field lines and leave multiple hits in the inner Belle II detectors.

In addition to the emitted pairs, primary particles which lose large amount of energy or scatter with large angle can be lost inside the detector, in the same way as explained in the radiative Bhabha section. Losses within $|s| < 65$ cm from the IP are also dangerous.

*4.4.6. Simulated samples.* According to beam background simulation provided by the accelerator group, the most important sources are radiative Bhabha scattering, Touschek scattering, and beam–gas interactions. These backgrounds are simulated with a dedicated accelerator group software called SAD [38] which is not part of basf2.

SAD simulates the transportation of particles through the accelerator. If a particle leaves the nominal beam trajectory and collides with the beam pipe or collimator in the Belle II experimental region, its position and momentum vector are saved to a file. The files normally correspond to one $\mu$s of running the accelerator at the nominal SuperKEKB luminosity. The data from SAD simulation are then passed to the Geant4 simulation software [39, 40] within basf2 to produce background samples of a given type.[2] The samples are saved in the standard basf2 ROOT format [41]. The events in these files correspond to the interaction of a single beam particle in the material of the interaction region and consist of simulated hits (SimHits) of all detector components. The equivalent accelerator running time and the background type are also saved within the files.

[2] The set of physics models used for the Geant4 simulation of the background events is different from the one used for physics events. This is to reproduce the behaviour of neutrons more precisely.

The two-photon QED background has been studied for the inner tracking detectors but is not yet included in the default background mixing. It is generated within basf2 using the generator AAFH (see Section 4.3) followed by Geant4 simulation, and the output is saved in the same file format. The earlier versions of the simulation library did not have adequate description of the magnetic field, so only PXD and SVD SimHits were included in the output files. Later with the improvement of the magnetic field description, SimHits for outer detectors will be also included in the output. Other backgrounds like synchrotron radiation and gammas from radiative Bhabha scattering events are less intense and are currently not included in background mixing.

The background types are listed in Table 20. The rate of events is calculated from the number of events in the sample and the equivalent accelerator running time.

Table 20: Beam background types (12th background campaign).

| Type | Source | Rate [MHz] |
|---|---|---|
| radiative Bhabha | HER | 1320 |
| radiative Bhabha | LER | 1294 |
| radiative Bhabha (wide angle) | HER | 40 |
| radiative Bhabha (wide angle) | LER | 85 |
| Touschek scattering | HER | 31 |
| Touschek scattering | LER | 83 |
| beam-gas interactions | HER | 1 |
| beam-gas interactions | LER | 156 |
| two-photon QED | - | 206 |

*Background mixing* The simulated background samples are used to add background to the simulated events. Adding background to simulated events is done by adding SimHits; digitisation is done after that. Possible pile–up of hits is therefore inherently included. The average number of background events of a given type to be added to a single simulated event is determined from the rate $R$ of a particular background sample and the time window $\Delta t$ in which the background is mixed

$$\bar{N} = sR\Delta t, \tag{1}$$

where $s$ is an optional scaling factor. The number of background events added to a particular simulated event is then generated according to a Poisson distribution with the mean $\bar{N}$. To simulate contributions from a different bunch, the background events are shifted in time randomly within the time window. This means that all SimHits of a given background event are shifted by the same time and therefore the correlations between detector components are preserved. The discrete bunch nature is however neglected because of sufficiently small bunch spacing.

The size of the time window depends on the detector component. It ranges from 100 ns (TOP) to 26 $\mu$s (ECL). To reduce CPU time we chose the time window of $[-1.0,\ 0.8]$ $\mu$s, which fits the most detector components, except PXD and ECL; these two have time windows of $[-17.6,\ 8.5]$ $\mu$s and $[-10.0,\ 10.0]$ $\mu$s, respectively. Additional background samples are used for mixing the background outside the default time window in these two cases.

Table 21 shows a comparison of the number of digitised hits (clusters for PXD and SVD) per event from beam–induced background with those from generic $B\overline{B}$ events.

Table 21: Number of digitised hits per event for beam-induced background (12th background campaign) and for generic $B\overline{B}$ events without background. For PXD and SVD the clusters are counted instead of digits. Numbers in parenthesis are without two–photon QED background.

| component | background | generic $B\overline{B}$ |
|---|---|---|
| PXD | 10000 (580) | 23 |
| SVD | 284 (134) | 108 |
| CDC | 654 | 810 |
| TOP | 150 | 205 |
| ARICH | 191 | 188 |
| ECL | 3470 | 510 |
| BKLM | 484 | 33 |
| EKLM | 142 | 34 |

*Background Overlay* When experimental data become available we will use a different method. Instead of using simulated beam background, the background overlay method will add background measured by random trigger. The background overlay is therefore done by adding the measured background event to the simulated one using digitised hits. Possible pile-up of hits must be taken into account with dedicated methods. These methods can model the pile-up only approximately since the measured background includes only the hits above the detection threshold.

A framework for background overlay has been designed to unify the method for all detector components. It consists of two basf2 modules and a base class for digitised hits (or clusters of hits). The first module, which must run in a single process mode, reads the data from a standard basf2 ROOT background file, and the second module, which can run in a multi-process mode, performs the overlay. Each class for digitised hits must implement two base class methods: the one that returns the unique channel identifier of the hit and the one that implements the pile-up method, which is usually detector specific. The first method is used to identify channels where background hits are added to the existing simulated hit. If this happens, the second method is called. The return value then signals whether the pile-up criterion was fulfilled. If not, the background hit is added to the collection of simulated hits.

### 4.5. *Detector Simulation*

The simulation package of basf2 is based on the Geant4 software [39, 40], with the version number 10.1.2.[3] There are two methods to supply the primary event to Geant4: one can use the particle gun class, which is part of the Geant4 package, or one can employ a specific generator software. For the latter case, the particles created by the generator package are sent to Geant4 for simulation via the interface implemented in the basf2 simulation package. Most of the decay processes of particles are described by the generator software. Short lived particles such as $K_S^0$ are usually decayed by Geant4. Exchange bosons and initial particles such as $e^-$ and $e^+$ are not passed to Geant4. During the simulation, Geant4 transports each primary particle step–by–step inside the detector and creates secondary particles. Digitisation of hit information in the sensitive volume of the detectors is handled by separate basf2 modules, rather than using software objects incorporated into Geant4 [6]. The result from the Geant4 simulation is sent to a persistent data storage (DataStore) to be used by other basf2 modules.

To simulate propagation of particles in the detector, physics processes of the interactions between the particles and the detector materials must be specified. These physics models can be either supplied by users or selected from the physics lists provided by the Geant4 group. We use the recommended physics list by the Geant4 group for the high energy physics experiments, `FTFP_BERT` [42].[4] The `FTFP` and `BERT` acronyms stand for hadronic shower models at different energies: the Fritiof quark–gluon string model at high energy, and the Bertini intra–nuclear cascade model at low energy. The transition area between the two models depends on each particle type, typically from 4 to 5 GeV [42, Section 3],[43–46]. `FTFP_BERT` contains all the standard electromagnetic processes provided by the Geant4 group [47].

For the production threshold for secondary particles inside the detector material, we use the default level set by the Geant4 software [48].

### 4.6. *Magnetic field in basf2*

Uncertainties in magnetic fields will affect Belle II analyses in several ways. The magnetic field is an input to reconstruction of charged tracks. To obtain the optimal resolution of the charged track momentum, the magnetic field must be understood precisely. The reconstruction efficiencies of particles depend on the accuracy of magnetic field information. Differences between the magnetic field used for the detector simulation and the one used for collision data sets may result in systematic bias. Differences between the magnetic field used for the offline reconstruction and the true magnetic field may create a systematic bias as well.

Inside the Belle II detector, there are two sources of magnetic fields: the detector solenoid, and the final focus system (QCS). The detector solenoid, which is comprised of an iron yoke and a superconducting solenoid, creates a uniform magnetic field of 1.5 T at the centre of the detector [6]. The iron yoke is interlaced with the KLM detector. The QCS is an extension of the SuperKEKB collider, whose purpose is to focus the incoming $e^+$ and $e^-$ beams at the collision point [49]. The main components of the QCS are eight superconducting quadrupole magnets. In addition, there are secondary superconducting magnets used for correction and compensation. On the surface, the magnetic fields generated by all the components of the QCS can be added linearly and used for simulation. However, due to the ferromagnetic yokes and shields around the main quadrupole magnets, non–linear characteristics are introduced in the magnetic field [49].

[3] Geant4 version 10.1.2 was included in basf2 release 00–06–00 on December 2015. Before, version 9.6.2 was used.

[4] Included in basf2 release 00–04–00 since May 2014.

The Opera3D/TOSCA software [50] was used to produce precision models of the magnetic field. The resulting 3D magnetic field map has been incorporated into basf2 on April 2016,[5] which replaced the constant field of 1.5 T as the default map for simulation and reconstruction (see Fig. 8). Note that earlier analysis results are based on the constant field map. Detailed studies are being conducted to improve the precision of the 3D magnetic field map. In–situ measurements of the Belle II magnetic field have been carried out on September 2015 to provide references for the model. More in–situ measurements and further analysis are planned to improve the precision of the field map to 0.1%.

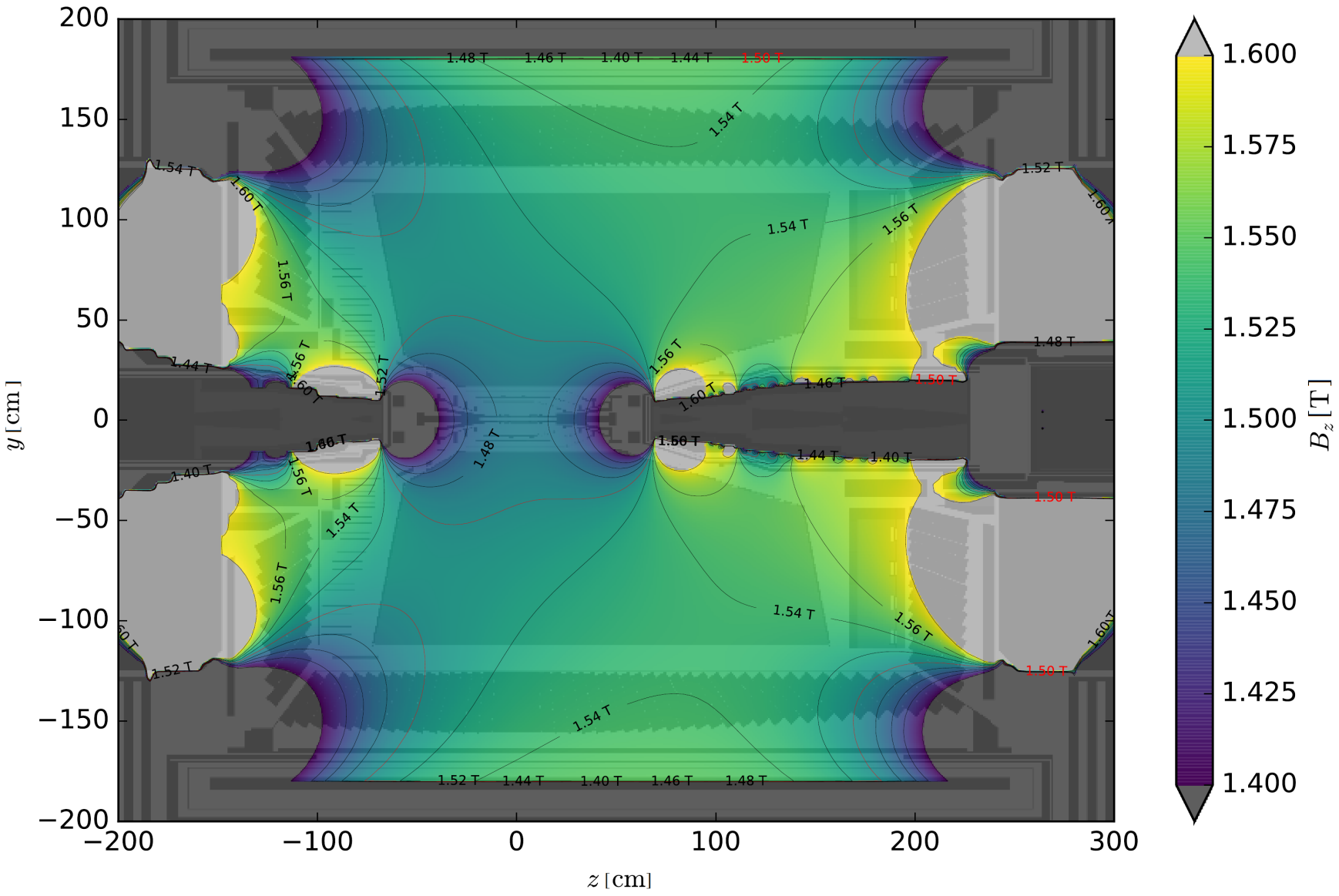

Fig. 8: The z component of the 3D magnetic field map as used in basf2 release version 00–07–00.

[5] basf2 release 00–07–00

# 5. Reconstruction Software

*Section author(s): F.Abudinen, J. Bennett, T. Bilka, G. Casarosa, T. Ferber, J-F. Krohn, C. MacQueen, L. Piilonen, L. Santelj, M. Staric*

## 5.1. Introduction

The Belle II detector will build upon the success of the first generation B-factories to establish a strong physics program. While many components of the Belle II detector are based on the design of the Belle detector, many improvements have been made in order to maintain similar performance in a much higher background environment. Significant efforts have also been made in improving reconstruction software to this end. The reconstruction algorithms and their performance characteristics are summarised in this chapter.

## 5.2. Software overview

Online and offline data handling is performed by the Belle II analysis software framework (basf2). The framework is designed to allow independent processing blocks called modules to perform relatively small tasks, which are executed linearly within a defined path. The configuration of modules for a specific purpose is defined using steering files. Modules communicate by passing information to and from a common object store, which also keeps track of relationships between objects in each event.

Given the enormous data output rate at Belle II, a robust and efficient framework for data analysis is vital. Data sets will be processed in several phases, with a reduction and enhancement occurring in each phase. The raw data is reconstructed to provide physical quantities from detector information like track hits and calorimeter clusters. This information can then be used to construct high-level objects such as charged tracks. The hit- and cell-level information is then discarded and the event size is reduced by a factor of approximately 40. The reduced information, including the high-level objects, is then used to determine particle level information such as four-momentum and event shape variables.

As this book contains projections and preliminary studies based on samples produced in several different MC campaigns, the performance of reconstruction algorithms is sometimes given for multiple software releases. Unless otherwise noted, the performance plots and reconstruction algorithms described in this chapter are based on basf2 release-00-05-03, which was used in the fifth MC campaign (MC5). Performance characterisation is also given for more recent basf2 software libraries, including release-00-07-00, release-00-07-02, and release-00-08-00, which were used in the sixth (MC6), seventh (MC7), and eighth (MC8) MC campaigns, respectively.

## 5.3. Tracking

The main task of the tracking is the reconstruction of charged particles originating from the primary and secondary decay vertices. Simply speaking, it consists of firstly identifying the VXD and CDC hits due to ionisation of a given charged particle in a sea of background hits from other particles, machine background or detector noise, and secondly obtaining the trajectory from a fit to the hit positions. Most of the tracks originate from inside the beam pipe, except for the charged decay products of the long-lived $V^0$-like particles ($K_S^0$, $\Lambda$, and converted photons) that are created outside the beampipe. The tracking algorithms must identify the two oppositely charged decay products of $K^0_S$, $\Lambda$, and photons decaying inside the tracking volume and pair them. In subsection 5.3.1 we describe the steps of charged particle reconstruction, while in subsection 5.3.2, more information specific to $V^0$-like particle reconstruction is provided.

Reconstructed particle trajectories are also used for the alignment of the detector. An optimally aligned detector is crucial to perform high precision unbiased measurements of flavour quantities with time dependence. The details of alignment are explained in section 5.3.3.

Finally, run-dependent knowledge of the spatial distribution of primary interactions (beamspot) can be used as a powerful constraint when fitting decay chains. The beamspot can be inferred from the reconstruction of $e^+e^- \to \mu^+\mu^-$ events. This is foreseen but is not yet implemented.

*5.3.1. Charged particle reconstruction.* The tracking software provides the analyst with lists of charged particle trajectories that have been fitted with an associated mass hypothesis. At the analysis level, a track is represented by $\{\vec{p},\vec{x}\}$, where $\vec{x}$ is the point of closest approach to the origin of the coordinate system, and $\vec{p}$ is the particle momentum in $\vec{x}$. The detector hits associated to the track are not propagated after tracking, in order to reduce the size of the analysis files (mDST). Additional information is also preserved for the analyst, *e.g.* the number of hits in each detector layer of the VXD and CDC that has been used to fit the track. This is important for selecting high quality tracks in analysis.

Charged particle reconstruction can be divided in two main parts:

- *track finding*, where detector hits belonging to a single track are collected together into a track candidate, and
- *track fitting*, where the track trajectory is determined by fitting the track candidate.

In the following we report the details these reconstruction steps.

*Track finding.* Track finding consists of applying pattern recognition algorithms to determine track candidates. The features of the detector hits in the CDC and the VXD are different, therefore dedicated pattern recognition algorithms for each detector have been developed.

The VXD track finder algorithm is based on the cellular automaton (CA) model [51, 52]. The large number of combinatorial track candidates in this approach is reduced by applying filters of increasing sophistication. Firstly, track segments are built, connecting pairs of hits in adjacent layers. They are the core units of the CA, known as cells. In this approach only compatible hits are combined into cells by consulting a look-up table, called the sector map, which is created by simulating a large number of tracks in the VXD. The second stage consists of determining whether cells that share a hit are neighbours passing a set of geometrical requirements. As in the first stage, the selection criteria are obtained from the sector map. This process is iterated and the track candidates are then identified as threads of neighbouring cells. In order to obtain a set of non-overlapping track candidates, a Hopfield network using a quality indicator is employed. The sector maps may vary according to the momentum of the particle, therefore it is possible to run the track finder multiple times by using sector maps for different momentum regions.

Two complementary algorithms for CDC track finding are employed: a global and a local track finder. The global track finder uses all hits at the same time by applying a Hough transformation [53] to the hit positions and looking for intersections in the Legendre space using a quad tree search. It is fast and highly efficient for high $p_t$ tracks originating from the origin, and can treat cases with missing hits. The local track finder searches for segments and tracks using a cellular automaton and the neighbourhood relations between hits. The track finder is robust against energy losses and tracks that do not originate from the IP. The combination of these two track finders results in excellent reconstruction efficiency.

The track candidates from the VXD and the CDC are then merged together according to the distance between the VXD and CDC track candidates extrapolated to the CDC outer wall. In the future we foresee cross-detector searches, for example the extrapolation of the CDC track candidates toward the VXD detector planes in order to add VXD hits to CDC track candidates, and vice-versa from the VXD to the CDC.

*Track fitting.* A track propagating in a vacuum in a constant magnetic field moves along a helix described by five parameters, defined at a point $\vec{P}$ of the trajectory. In Belle II the point $\vec{P}$ is identified with the perigee, the point of closest approach to the origin in the $r/\phi$ plane. The five parameters employed in the Belle II tracking software are the following:

- $d_0$: the signed distance of the perigee from the origin in the transverse plane. The sign depends on the direction of the angular momentum of the track at the perigee with respect to the magnetic field.
- $z_0$: the longitudinal signed distance of the perigee from the origin.
- $\phi_0$: the angle between the transverse momentum at the perigee and the $x$ axis.
- $\tan\lambda$: the tangent of the angle between the momentum at the perigee and the transverse plane.
- $\omega$: the curvature, where the sign corresponds to the charge of the track.

The trajectories of tracks in Belle II are not ideal helices, as the charged particles interact with both passive and active detector material inside the tracking volume, losing a fraction of their energy and undergoing multiple scattering. In addition, the magnetic field provided by the superconducting solenoid is not constant in space. These effects are all taken into account in the tracking algorithms, in particular in the track fitting and track extrapolation stages. In order to correctly treat the interaction of particles with matter, a hypothesis on the mass of the particle must be made. The version of the software used for sensitivity studies shown in this book only supports the pion mass hypothesis. However, we have implemented the functionality to account for different mass hypotheses (electron, muon, pion, kaon or proton) depending on the momentum of the track. For example, high energy pions and kaons have very similar interactions with matter, therefore a single mass hypothesis is sufficient at momenta above 1 GeV/$c$, whereas different mass hypotheses yield better resolution below that threshold.

The main track fitting algorithm used in our reconstruction is the deterministic annealing filter (DAF)[54]. The DAF is based on a standard track fitting algorithm, the Kalman filter (KF)[55]. The latter is equivalent to a least squares method, where it takes into account the interactions with the material but has no means of dealing with false hit assignments or incorrect assumptions about wire passage. Both DAF and KF are implemented in the GENFIT package[56, 57]. To deal with these shortcomings, Belle II uses a DAF in which the points are weighted according to their residual to the smoothed track and hits with large residuals are suppressed with an annealing procedure.

*Combined performance.* The tracking efficiency for charged particle reconstruction is reported in Fig. 9 as a function of the transverse momentum and the polar angle, with and without beam induced background. The efficiency is defined as the ratio between the number of fitted tracks and the number of generated charged primary particles. The efficiency at low transverse momentum in the presence of beam induce background is lower than ideal. There are ongoing improvements to the algorithms that are being incorporated into the tracking software as the experiment evolves, to mitigate the impact of such background. The efficiency of a more recent version of tracking algorithms is reported below.

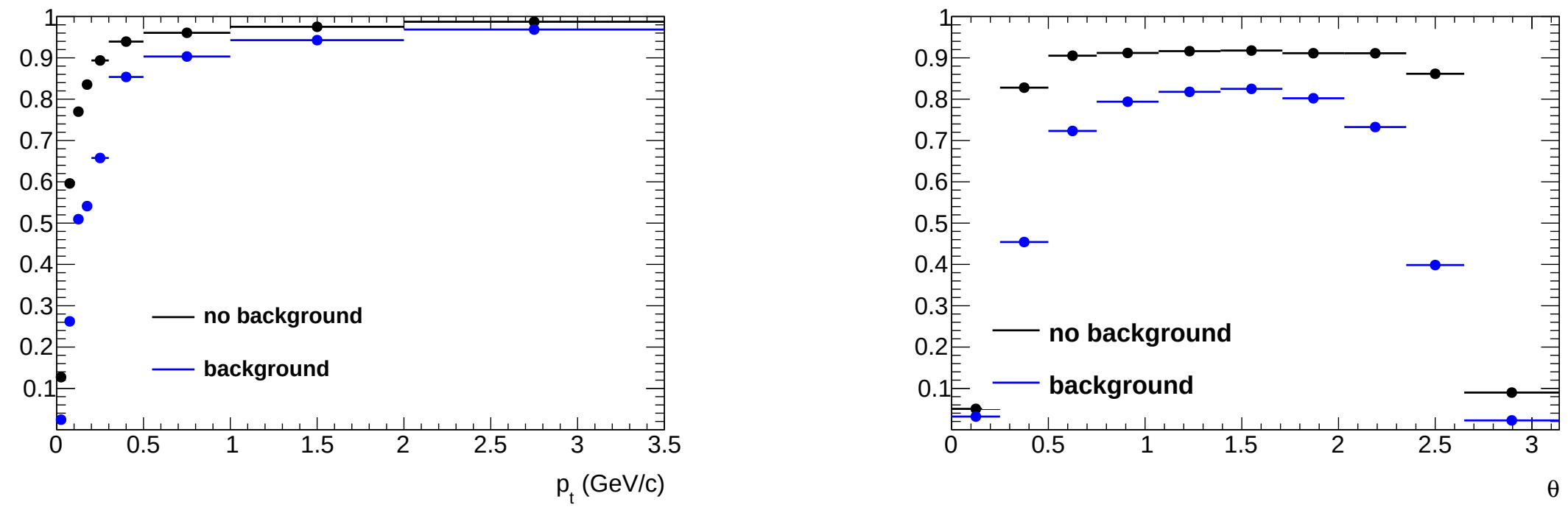

Fig. 9: Tracking Efficiency as a function of transverse momentum (left) and polar angle (right) evaluated on $\Upsilon(4S)$ events.

In Fig. 10 we show the fitted impact parameter pull distributions. The core gaussian shows very little bias and within the nominal width for both parameters. A small fraction of events, below 10%, show a positively biased pull distribution and a width a factor two larger.

The average particle boost $\langle\beta\gamma\rangle$ for $B$-mesons produced at the predecessor collider KEKB was about 0.425. Here $\beta = v/c$ is the ratio between the velocity of the particle $v$ and the speed of light $c$, and $\gamma$ is the Lorentz factor. Due to the lower boost at SuperKEKB ($\langle\beta\gamma\rangle \approx 0.284$), we need more precise track reconstruction than achieved by the predecessor experiment Belle to reach a comparable resolution in the measurement of the decay time of primary particles (Sec. 6.2.3). In Fig. 11 we show the resolutions of the transverse $d_0$ and the longitudinal $z_0$ impact parameters as functions of the pseudo-momenta $p\beta\sin(\theta)^{3/2}$ and $p\beta\sin(\theta)^{5/2}$. The pseudo-momenta are chosen to take into account the effect of multiple scattering of charged particles [58]. A precision of about $10\,\mu$m on both impact parameters is expected for high momentum tracks matching the expectations in the Technical Design Report [6]. Figure 11 shows that on both track impact parameters we improve the resolution by almost a factor two with respect to Belle.

*Improvements on Tracking Efficiency.* In this section we report the efficiency of the VXD and the CDC standalone pattern recognition. The VXD pattern recognition algorithm has been re-designed and re-implemented since the majority of the physics MC studies were performed for this book. As an example, we report in the left plot of Fig. 12 the track finding efficiency using only SVD hits. The overall efficiency is higher, and the degradation of the performance due to background is much less pronounced than earlier versions of the tracking algorithms.

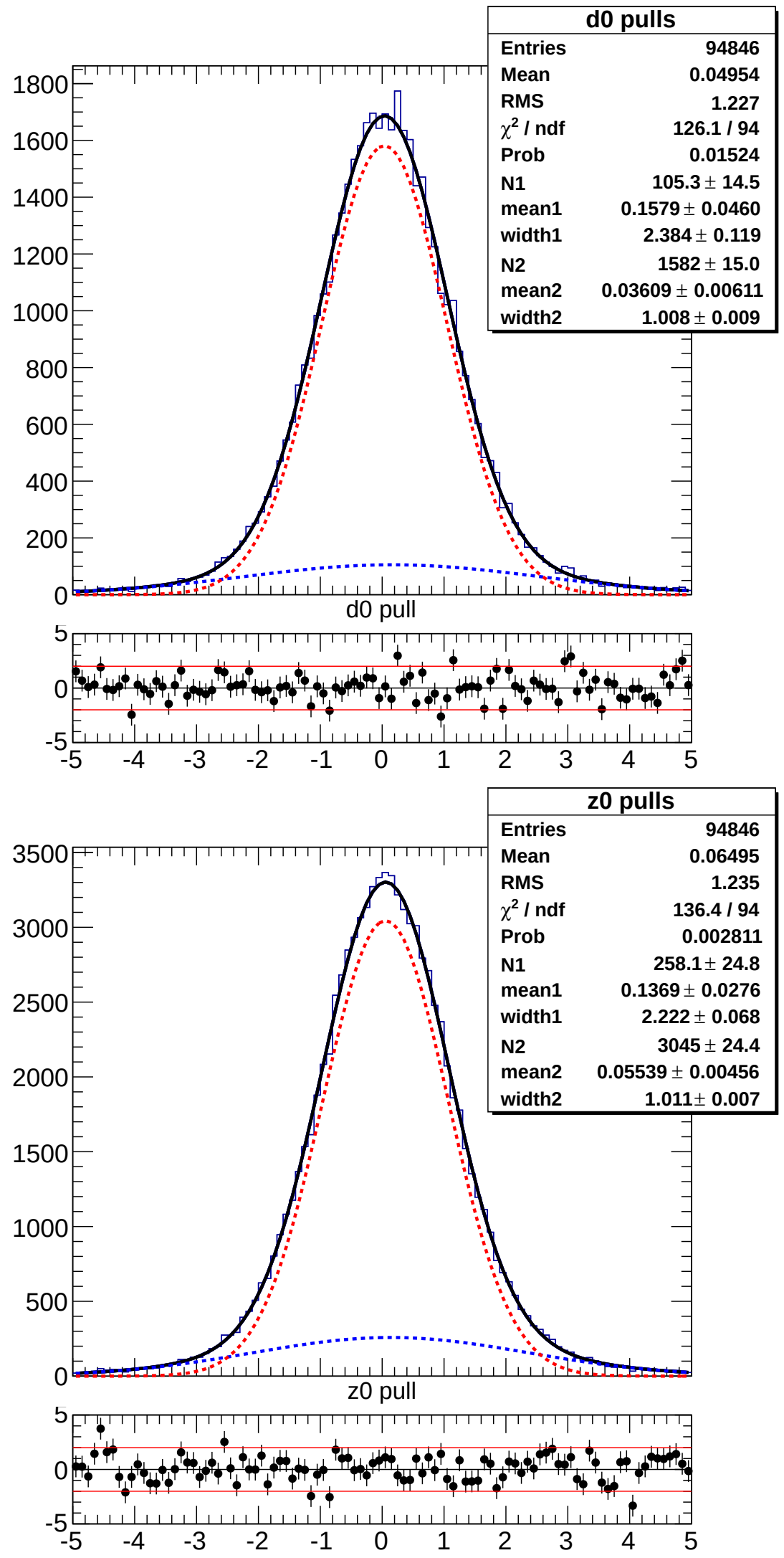

Fig. 10: Distribution of the transverse (left) and longitudinal (right) impact parameters pulls fitted with the sum of two Gaussians, evaluated on $e + e^- \to \Upsilon(4S)$ events.

*5.3.2. $V^0$-like particle reconstruction.* Long-lived neutral particles that decay into two charged particles at some distance away from the interaction point are reconstructed using a dedicated algorithm. This $V^0$ reconstruction takes place after the reconstruction of charged particles and is intended to avoid extrapolation through material on the analysis level, where the actual $V^0$ selection takes place. This is in accordance with the design goal of removing dependence of analysis level information on knowledge of the detector material.

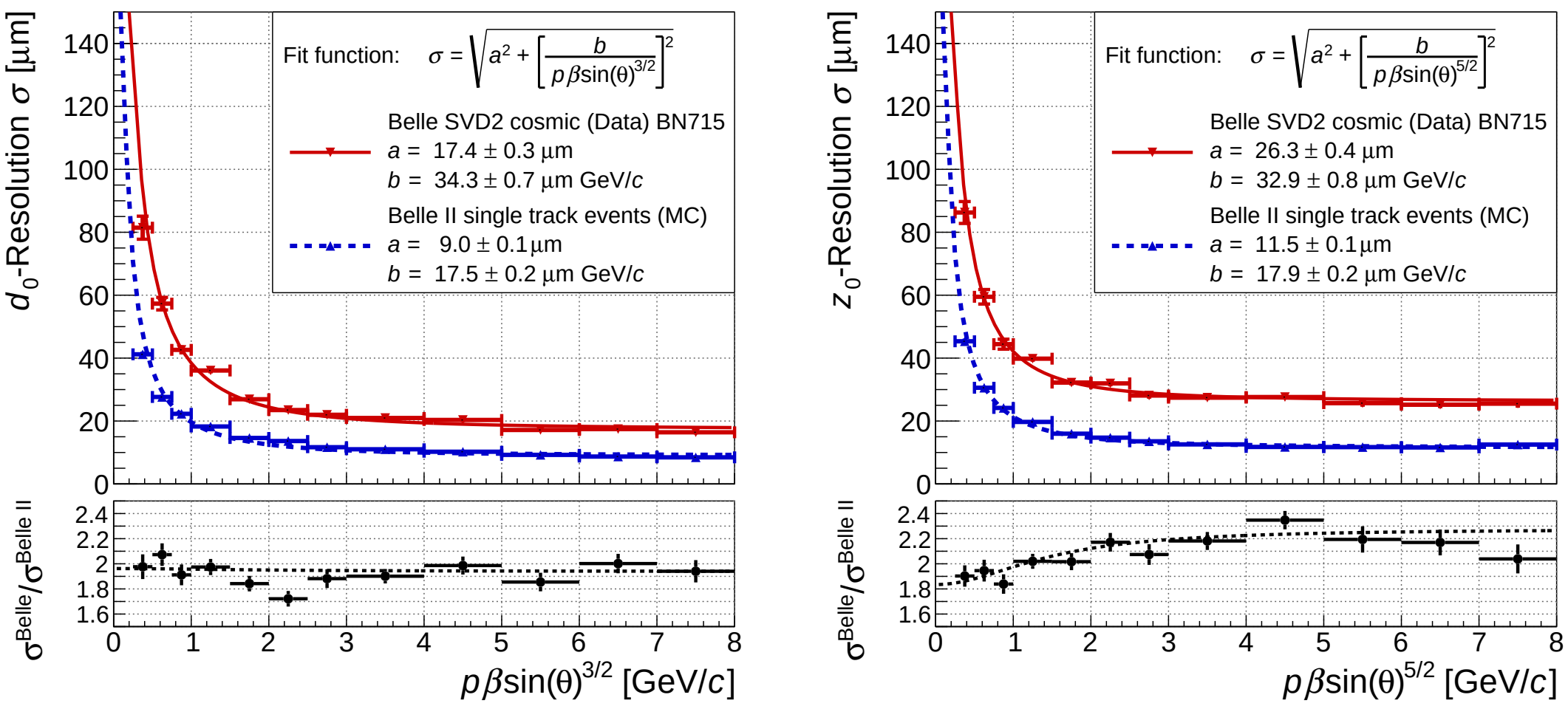

Fig. 11: Resolution of the transverse $d_0$ (left) and longitudinal $z_0$ (right) impact parameters. The results for MC events with a single muon track using the Belle II tracking algorithm are compared with the results for Belle cosmic events [58]. The resolution in each bin is estimated using the $\sigma$ value of a single Gaussian function fitted in a region containing 90% of the data around the mean value of the distributions.

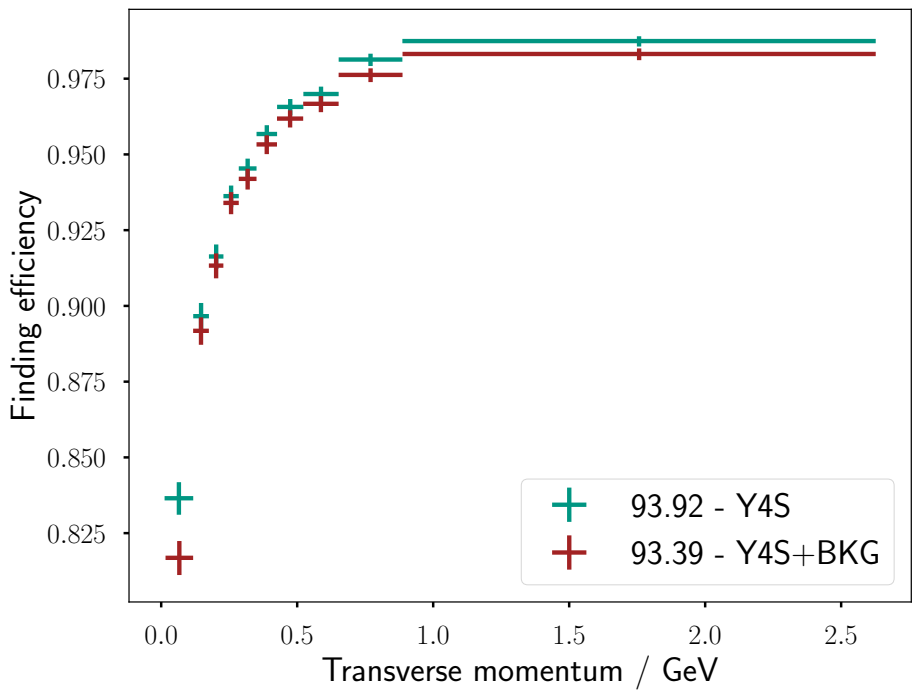

Fig. 12: SVD-only pattern recognition efficiency versus the transverse momentum, for $e^+e^- \to \Upsilon(4S)$ events with and without beam induced background.

The goal of $V^0$ reconstruction is to keep good $V^0$ candidates originating from outside the beam pipe, as well as those from inside the beam pipe whose reconstructed mass is close to the mass of the mother particle. Unlikely track combinations may be suppressed by restricting the $\chi^2$ from the vertex fit or the radius of the $V^0$ vertex.

The $V^0$ reconstruction algorithm pairs all oppositely charged tracks and extrapolates each track to the innermost hit of either track. If the extrapolation fails, the combination is rejected. Studies show that this restriction has no effect on efficiency. Each accepted combination is processed by the vertex reconstruction package RAVE [59]. If the vertex fit fails, the combination is rejected. Each surviving combination is then subject to selection criteria that depend on the vertex fit $\chi^2$ (less than 50) and, for vertices inside the beam pipe (a vertex radius less than 1 cm), the mass window (within 30 MeV/c$^2$ of the nominal $V^0$ mass).

The $V^0$ algorithm will be used to reconstruct $K^0_S \to \pi\pi$, $\Lambda$, and photon conversions. Figure 13 shows the efficiency of $K^0_S$ reconstruction efficiency as a function of transverse momentum, with and without beam background effects.

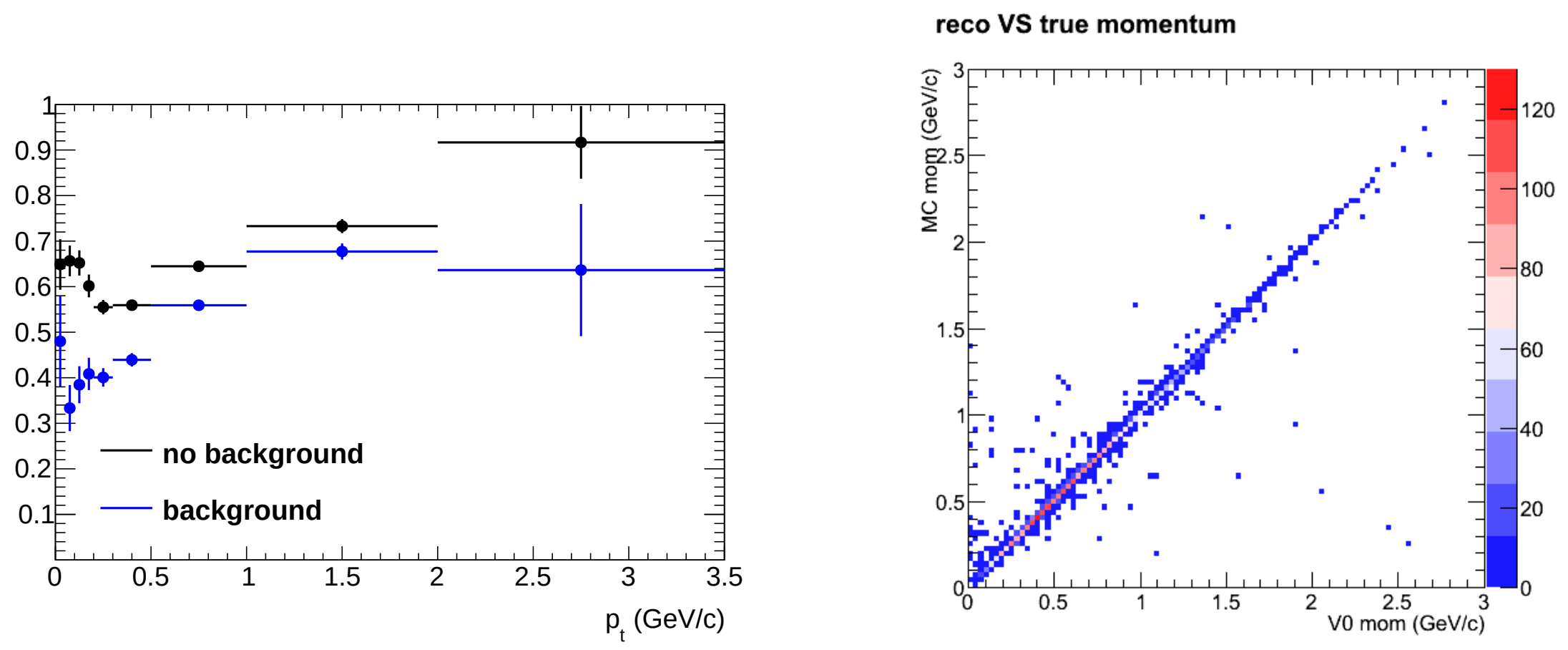

Fig. 13: The $K^0_s \to \pi^+\ \pi^-$ reconstruction efficiency as a function of transverse momentum (left) and a 2D histogram of the generated versus reconstructed $K^0_s$ momentum (right).

*5.3.3. Alignment.* To reach the design performance of the detector, various calibration constants must be determined. For the VXD, many of these constants describe the position and orientation of the silicon sensors. This calibration is commonly referred to as alignment. To determine the alignment constants, a so-called global approach using the Millepede II tool [60, 61] has been chosen for use at Belle II. The alignment is computed through minimisation of track-to-hit residuals by means of a linear least squares method. Because Millepede fits all track and alignment parameters simultaneously, all correlations are kept in the solution. Therefore it is desirable to simultaneously determine as many constants as possible. For this reason, the CDC is also integrated into the procedure and its alignment and some calibration constants can be determined together with the VXD alignment. Investigation of the possibility to integrate other sub-detectors into the procedure is ongoing, *e.g.* the alignment of the muon system.

*Track Parameterisation.* Reconstructed tracks and decays, as well as cosmic muons, can be used as input to the alignment procedure. All such tracks are first re-fitted by the General Broken Lines algorithm (GBL) [62], which is integrated into the GENFIT toolkit [63] and basf2. GENFIT is a generic track fitting toolkit for HEP experiments. It allows for the generic treatment of measurements of various type and dimension (planar 1D or 2D measurements, wire hits etc.) and offers all needed capabilities for track extrapolation in detector material and inhomogeneous magnetic fields. The GBL parameterisation carefully treats multiple scattering effects, adding additional fit parameters (kink angles) to an initial reference trajectory derived from the result of the standard reconstruction output. The additional degrees of freedom are removed by constraining the variance of multiple scattering angles from the moments of the radiation length distribution along the reference particle's trajectory. Tracks are locally parameterised using five or four parameters at each measurement plane depending on the presence or absence of a magnetic field, respectively. For the drift chamber, a virtual measurement plane is constructed by means of the GENFIT formalism. Combined particle candidates, composed of multiple particle tracks constrained to originate from a common vertex can be an input of the alignment as well. Optionally, a beam constraint can be added for decays originating from the primary interaction point, such as di-muon events. In a similar manner, *e.g.* two body decays with an invariant mass constraint can be introduced in the procedure.

*Alignment Parameterisation.* In the VXD, the sensors are parameterised as planes with six rigid body alignment parameters: for the 212 sensors, we have 1276 parameters. Sensor deformations or additional calibrations of the Lorentz angle in the magnetic field can be included in the procedure as well. For the treatment of correlated movements of sensors, a hierarchy of alignment objects can be defined. This allows for the treatment of time dependence of the larger structures, which is different to internal sensor alignments during simultaneous minimisation.

For the CDC, the alignment of the layers and larger structures, *e.g.* end plates, is considered. The $x - t$ relation as well as channel timing offsets or time walk corrections can be calibrated. For the muon system, the modules are treated as rigid planar bodies in the initial stage.

Beam and vertex constraints rely on the estimation of the primary vertex position (based on beam spot parameters), and corrections to it can be determined during the simultaneous minimisation. Only the position of the primary vertex is calibrated. The covariance matrix of the primary beam spot is an input to the procedure and is calibrated via other means.

*Alignment Workflow.* Millepede II is integrated into the common calibration framework, which makes use of dedicated basf2 modules to collect samples and run calibration algorithms. During the collection step for alignment, reconstructed tracks are re-fitted using the nominal detector positions corrected with previously determined alignment constants. Each detector interfaces this procedure via a special class representing the local-to-global transformation. This class also provides the derivatives of local residuals with respect to its assigned calibration parameters. Various track samples (primary decays, background, cosmic rays, etc.) from different operating conditions (cosmics without magnetic field) are under investigation.

Ultimately some constants may be determined in a time-dependent way, especially those affecting many measurements, such as the positions of large structures, while keeping the procedure computationally manageable. The procedure as a whole can also be applied locally, for example only for the PXD alignment or to determine the relative alignment of the PXD and SVD. If such corrections are determined online, they will serve as initial values for the global procedure when enough data and track samples are accumulated.

### *5.4. Calorimeter reconstruction*

The electromagnetic calorimeter is used to reconstruct the energy and position of depositions from neutral and charged particles with the best possible resolution and efficiency. While the energy and position reconstruction is primarily needed for photons and neutral hadrons, it may also aid the electron and charged hadron reconstruction in regions without, or with only limited, tracking coverage. The sum of all reconstructed showers is used to constrain the missing energy in decays involving neutrinos. A special case is the reconstruction of highly energetic $\pi^0 \to \gamma\gamma$ decays where the two photon showers overlap or merge.

The second task of the calorimeter is particle identification for electrons, muons, charged hadrons, neutral hadrons and photons based on shower shape variables and tracks matched to clusters.

A critical aspect of calorimeter cluster reconstruction, and electron reconstruction is the material budget in front of the calorimeter. In Belle II the number of radiation lengths (or thickness) $X/X_0$ is approximately 0.3 in the barrel and higher in the endcaps and in regions with service material. The material budget is depicted in Fig. 14.

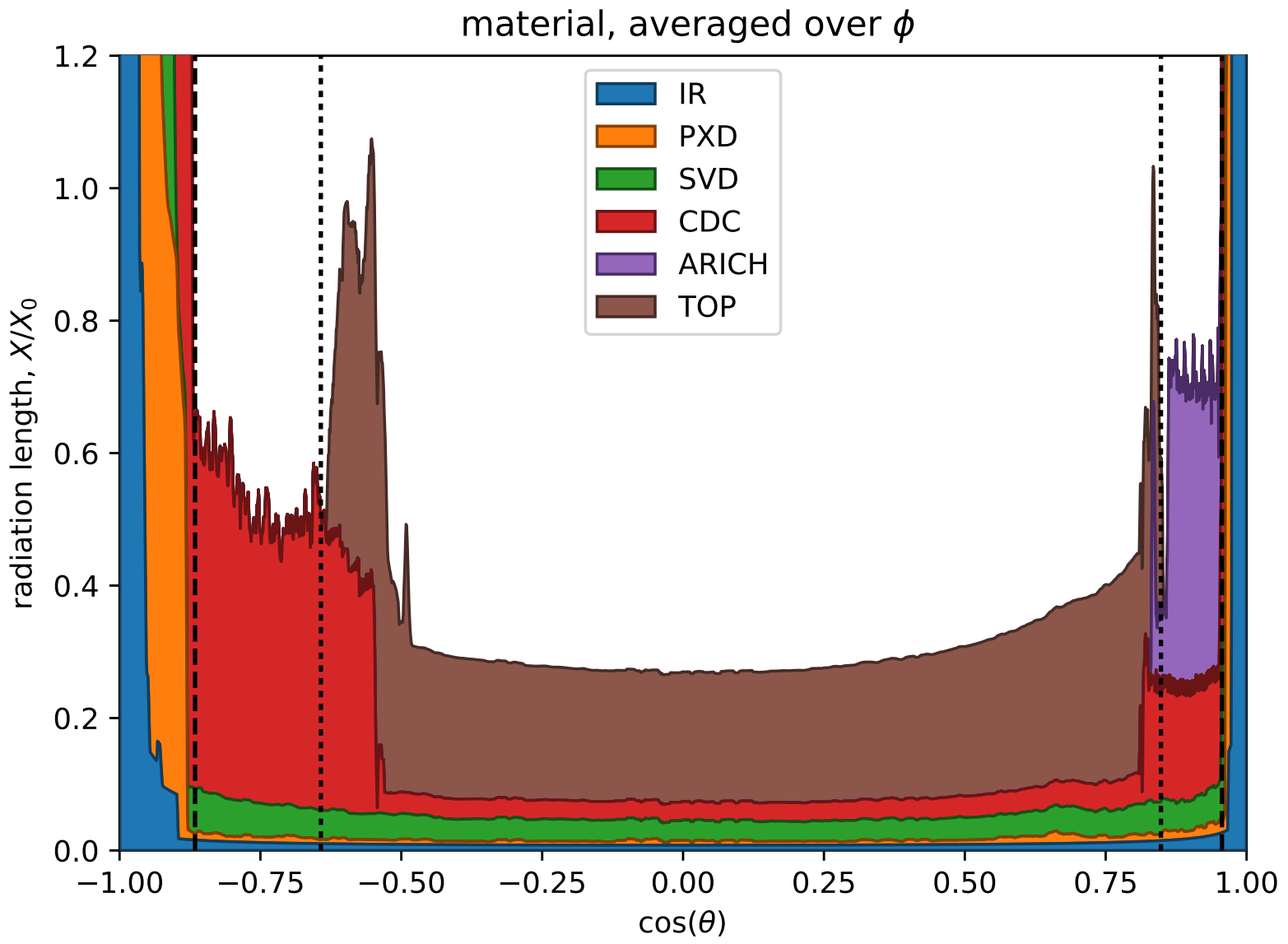

Fig. 14: The number of EM radiation lengths (or thickness) $X/X_0$ in front of the calorimeter as a function of $\cos\theta$, averaged over $\phi$.

The clustering used up to release-00-07-02 is an incomplete adaptation of the Belle clustering code which was developed for a low background environment. It starts from an initial list of crystals with energy deposits above some threshold, nominally 0.5 MeV, which is about twice the expected level from electronics noise. To obtain some robustness against high beam backgrounds, the energy threshold was raised as a function of crystal polar angle to between 1.28 MeV (barrel) and 2.5 MeV (outer endcap rings). A cluster starts with a seed crystal with at least 10 MeV that is a local energy maximum amongst its nearest neighbour crystals. A nearest neighbour touches either the side or the corner of the crystal and a local maximum is a crystal whose energy exceeds that of its next neighbours. All crystals from the initial list that are nearest or next-to-nearest neighbours of the seed crystal are added to the cluster. In the barrel, the size of a cluster is thus limited to a square arrangement of $5 \times 5$ crystals. If clusters share crystals after this step, their energies are split according to the ratio of energy of each cluster to the sum of energies of all crystals in the overlapping clusters. This energy splitting does not provide the correct position nor the correct weighted list of crystals for subsequent shower shape calculations. The centroid $\vec{x}$ of each cluster is calculated by using linear weights of all crystals in a cluster,

$$\vec{x} = \frac{\sum_i E_i \vec{x}_i}{\sum_i E_i}, \tag{2}$$

where $E_i$ is the energy of the $i$–th crystal and $x_i$ is the geometric centre of the $i$–th crystal. It should be noted that this position reconstruction is known to be biased towards the crystal centre of the highest energy crystal in the shower. The cluster energy is reconstructed as the linear sum over all included crystals. The peak position of the reconstructed photon energy is corrected to the true value in a subsequent step as a function of reconstructed polar angle and energy. The cluster time $t_{\text{cluster}}$ is the time of the highest energetic crystal in a cluster with respect to the collision time. In order to reduce out-of-time beam backgrounds, clusters with $|t_{\text{cluster}}| < 125\,\text{ns}$ are rejected. Clusters are matched with tracks using a GEANT based extrapolation routine. A cluster that contains a crystal hit that is consistent with an extrapolated track is matched to that track.

The described calorimeter reconstruction does not perform optimally in a high background environment and has various shortcomings, *e.g.* biased position reconstruction, simplistic track matching, and oversimplified cluster splitting. The average dose caused by various background sources as function of polar angle $\theta$ in the ECL is shown in Fig. 15. Several improvements have been introduced to the ECL reconstruction with release-00-08-00. The new cluster algorithm reconstructs connected regions (CR) starting with single crystals with an energy of at least 10.0 MeV as seeds, as before. Surrounding crystals are added if their energy is above 0.5 MeV. This procedure is continued if the added crystal energy is at least 1.5 MeV. If two CRs share a crystal, they are merged. The optimal CR contains all deposited energy for a particle and merges CRs from different particles only if different particles deposit energy in the shared crystals. Each CR is then split into one or more clusters as follows:

Each crystal in a CR that is a local energy maximum amongst its nearest neighbouring crystals serves as seed for one cluster. Iteratively, all crystals of the CR are assigned to each local maximum using shared weights that are normalised to unity per crystal. The weight per crystal is given by

$$w_i = \frac{E_i e^{(-Cd_i)}}{\sum_k E_k e^{(-Cd_k)}}, \tag{3}$$

where $C = 0.7$ is a constant determined from MC, $d_i$ is the distance between the $i$–th crystal and the cluster centroid, and the denominator sums over all crystals, $k$, in the cluster. This weight is motivated by the lateral distribution of an electromagnetic shower which decreases exponentially from the cluster centre. The centroid $\vec{x}$ of each cluster is then calculated by using logarithmic weights of all crystals,

$$\vec{x} = \frac{\sum_i w'_i \vec{x}_i}{\sum_i w'_i}, \tag{4}$$

where $w'_i = 4.0 + \log(w_i \times E_i/E_{all})$, and $E_{all}$ is the sum of all weighted crystals with weights $w_i > 0$. Only crystals with $w'_i > 0$ are included in the calculation of $\vec{x}$. This procedure is iterated until the average centroid positions of all clusters in the CR are stable within 1 mm.

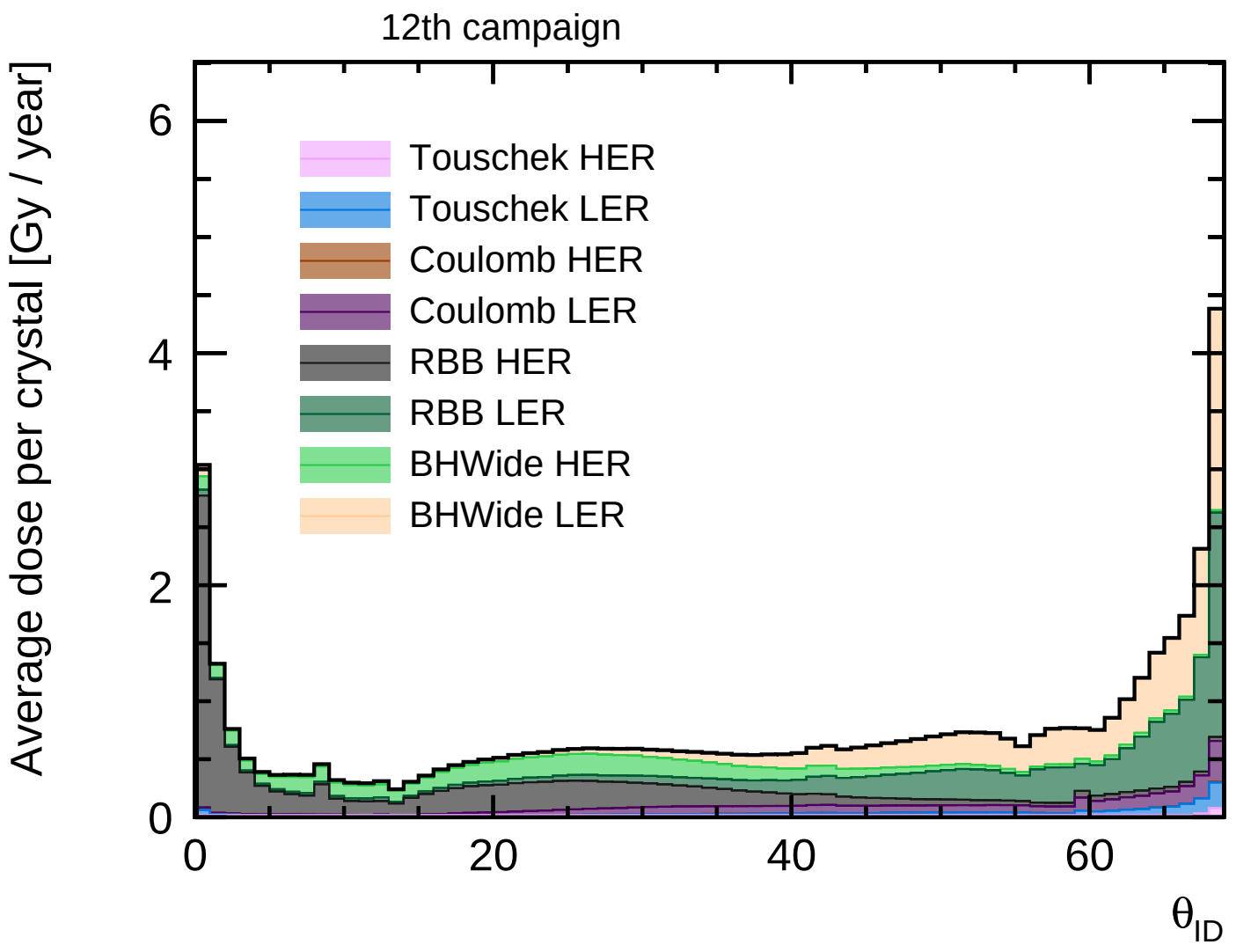

Fig. 15: Average dose per crystal as function of ECL $\theta$–ring. $\theta_{ID}$=0 corresponds to the most forward crystals, and $\theta_{ID}$=68 corresponds to the most backward crystals.

The cluster energy $E$ is reconstructed by summing the $n$ crystals with highest weighted crystal energies $E = \sum_n w_i \times E_i$ of up to 21 nearest neighbours. The number of crystals, $n$, depends on the background levels per event and an energy estimation based on the sum of the 8 nearest neighbours. The reconstructed peak photon energy is corrected to the true value as a function of reconstructed polar angle, azimuthal angle, energy, and expected background level. The cluster time resolution is determined to contain 99 % of all signal clusters based on MC. Showers below 50 MeV with a reconstructed time above this value are not stored.

A comparison of the photon energy resolution obtained using the clustering code of release-00-05-03 (MC5) and release-00-08-00 for different background levels is shown in Fig. 16. The new reconstruction offers a significantly improved energy resolution at low energies. The photon reconstruction efficiency is shown in Fig. 17. The new ECL reconstruction can be extended to reconstruct multiple hypotheses based on the particle type that created the shower and additional shower shape variables are available. In addition the dedicated reconstruction of merged $\pi^0$ mesons, where the two photons cannot be separated into two different clusters, will be performed. Track matching will be based on a likelihood of the nearest track to a cluster using the covariance matrix of the track fit.

### *5.5. Charged particle identification*

Effective and efficient charged particle identification (PID) is vital to the physics goals of the Belle II experiment. Good PID information is necessary to isolate hadronic final states, reduce backgrounds and enable flavour-tagging techniques. The Belle II detector, described in Chapter 3, contains an upgraded PID system, including a Time-Of-Propagation (TOP) counter in the barrel region of the detector and a proximity-focusing Aerogel Ring-Imaging Cherenkov (ARICH) detector in the forward endcap region, to provide information on charged particles over the full kinematic range. The information from these detector systems is combined with that from specific ionisation ($dE/dx$) measurements from the SVD and CDC to act as the primary sources of information for charged hadron PID. In a similar way, the ECL provides the primary information for use in electron identification and the KLM provides that for muon identification. Charged hadron and lepton PID are described in more detail in the following sections.

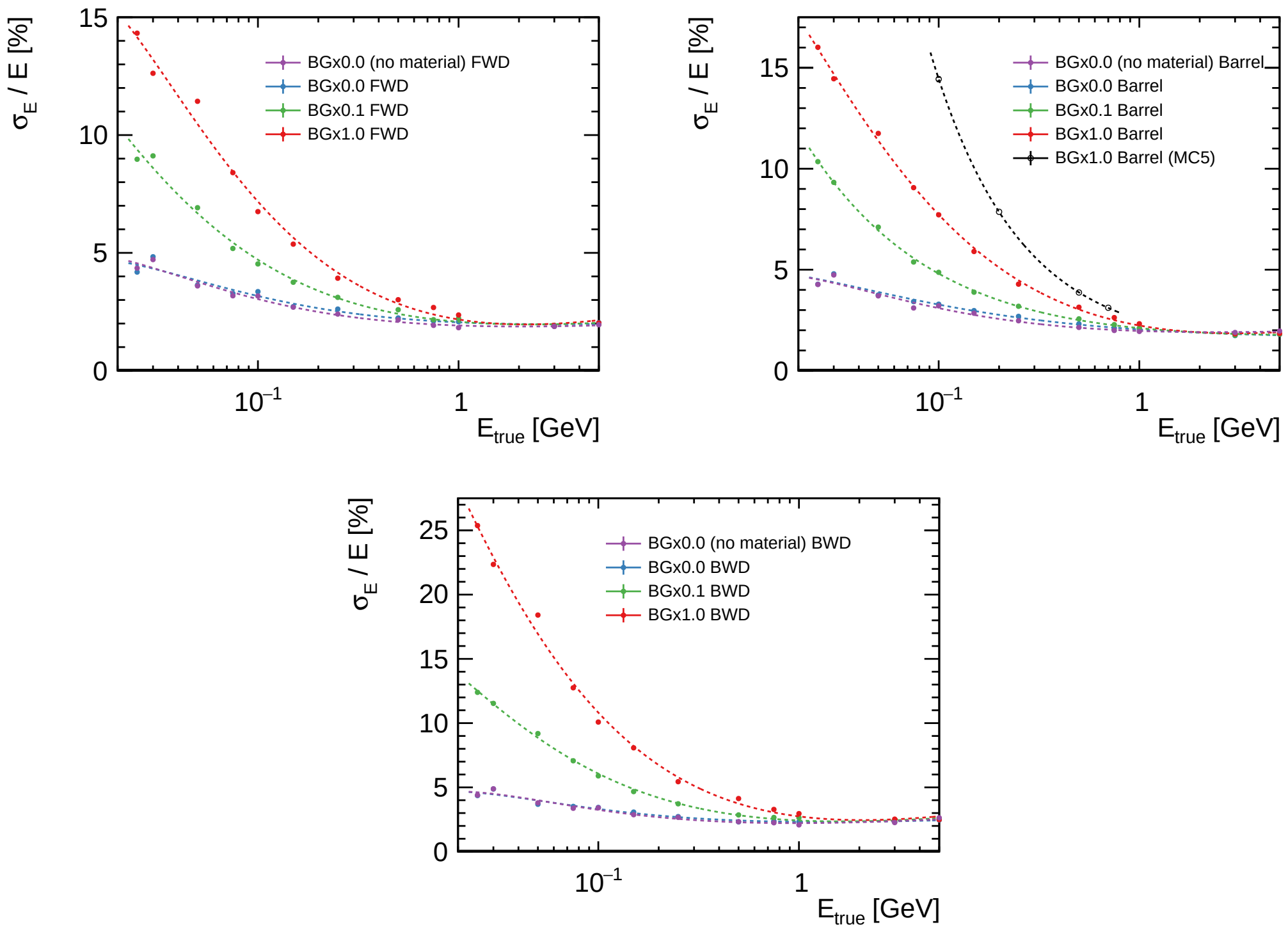

Fig. 16: Photon energy resolution as a function of true photon energy for the FWD endcap (a), barrel (b), and BWD endcap (c) regions. Note the different $y$–axis ranges of the plots. A smooth curve has been fit to the points to guide the eye. An older implementation of the ECL reconstruction software (used in MC5) is also plotted in (b).

Charged particle identification at Belle II relies on likelihood based selectors. Information from each PID system is analysed independently to determine a likelihood for each charged particle hypothesis. These likelihoods may then be used to construct a combined likelihood ratio. Analysis specific criteria may be used to construct prior probabilities. When combined with the likelihoods, the priors allow for the construction of the probability for a charged track to have a particular identity. This provides the optimal PID performance, but comes at the cost of requiring analysis specific optimisation. The uncertainty on the selection efficiency cannot be pre-determined using this method.

The likelihood selectors rely on likelihood ratios constructed in the following way. First, the PID log likelihoods from each detector are summed to create a combined PID likelihood for each of six long-lived charged particle hypotheses: electron, muon, pion, kaon, proton and deuteron. Next, the difference in log likelihood between two particle hypotheses is used to construct a PID value $L(\alpha : \beta)$ according to

$$\mathcal{L}(\alpha : \beta) = \frac{1}{1 + e^{\ln \mathcal{L}_\alpha - \ln \mathcal{L}_\beta}} = \frac{\prod_{\text{det}} \mathcal{L}(\alpha)}{\prod_{\text{det}} \mathcal{L}_\alpha + \prod_{\text{det}} \mathcal{L}_\beta}, \tag{5}$$

where $\alpha$ and $\beta$ represent two different particle types and the product is over the active detectors for the PID type of interest. The value $\mathcal{L}(\alpha : \beta)$ is greater than 0.5 for a charged track that more closely resembles a particle of type $\alpha$ than one of type $\beta$ and is less than 0.5 otherwise. More details on the PID types are given in the following sections.

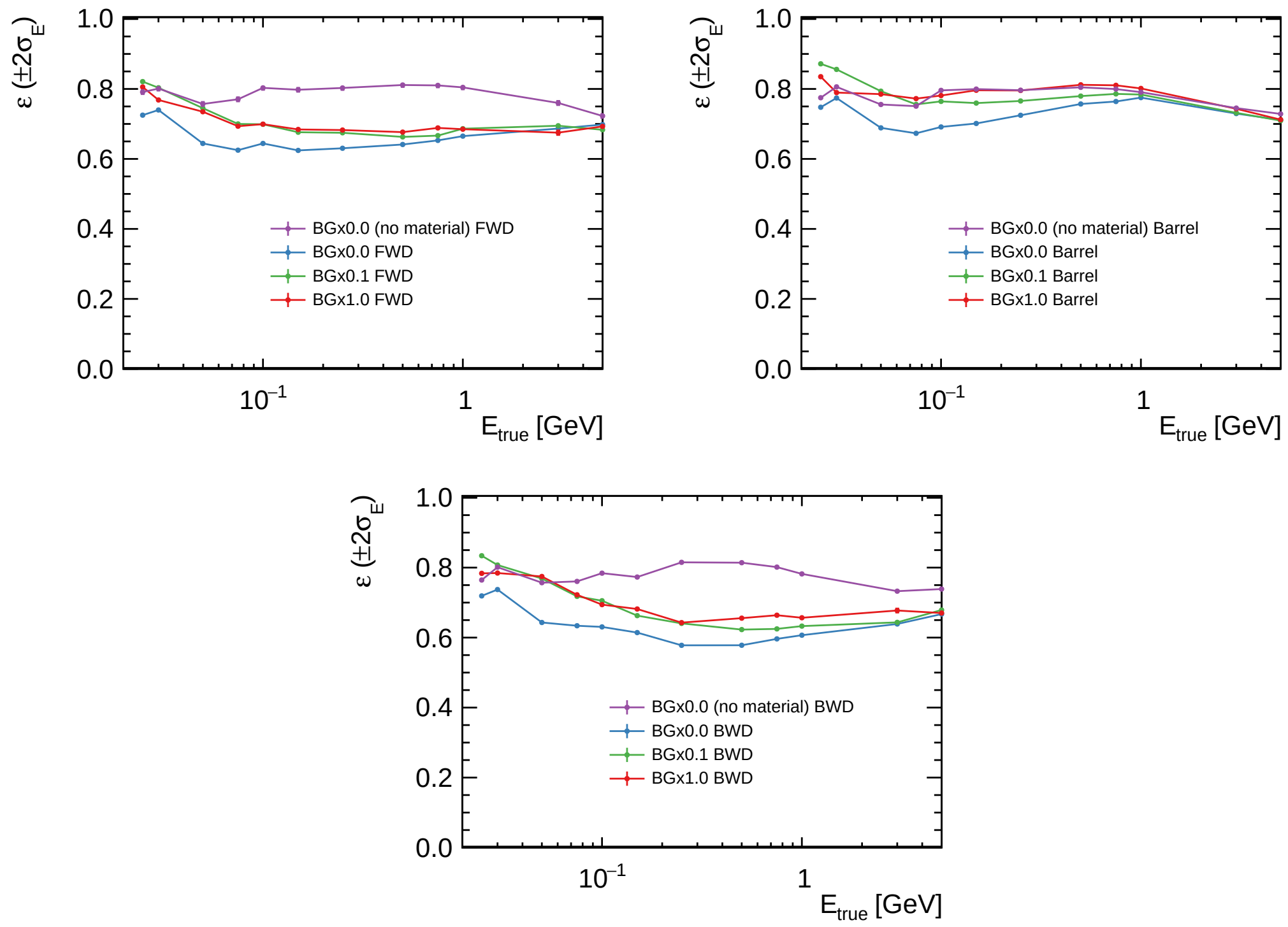

Fig. 17: ECL reconstruction efficiency for single photons for different background levels, with and without material in front of the ECL for the FWD endcap (a), barrel (b), and BWD endcap (c) regions. The points are connected by straight lines to guide the eye.

The performance plots included in this section were generated from inclusive samples of $10^6$ $c\bar{c}$ events generated during the fifth and sixth MC campaigns. These samples were reconstructed with release-00-05-03 and release-00-07-00 of the Belle II software, respectively.

*5.5.1. $dE/dx$ measurements.* The ionisation energy loss, $dE/dx$, of a particle travelling through the Belle II detector is determined from measurements in the VXD and CDC. It is expected that the $dE/dx$ measurement should depend only on the particle velocity, $\beta$, or equivalently $\beta\gamma = p/m$. Thorough calibration is required to avoid systematic effects that break this dependence. The $\beta\gamma$ universality of the $dE/dx$ response for pions and kaons at Belle II is displayed in Fig. 18. In general, the $dE/dx$ information is provides better discrimination power for particle momenta below about 1 GeV/$c$ (Fig. 19).

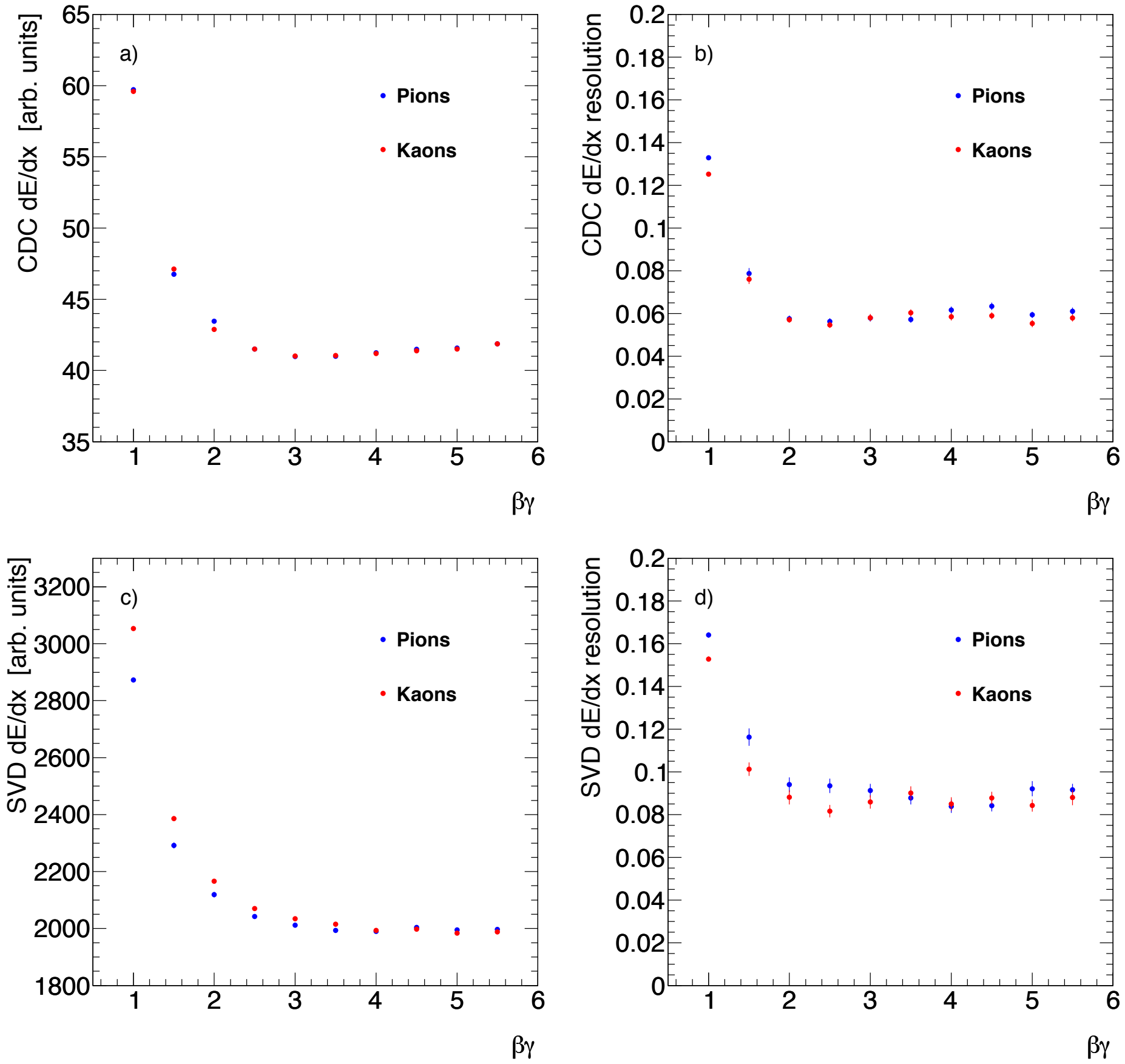

Fig. 18: Truncated $dE/dx$ means (a,c) and resolutions (b,d) for pion and kaon samples generated at specific values of $\beta\gamma$. The residual non-universality in the SVD is due to the fact that the measured momentum at the IP is different than the momentum in each SVD layer.

*Determination of Likelihoods.* The VXD and CDC detectors make independent $dE/dx$ measurements and require different calibration procedures. At the time this document is written, the $dE/dx$ reconstruction algorithms in both subsystems construct likelihood values using information from individual hits. A likelihood value is determined for each particle hypotheses, including pion, kaon, proton, muon, electron, and deuteron, using a lookup table constructed from large MC samples. To reduce the effect of non-Gaussian tails, the lowest 5% and highest 25% $dE/dx$ measurements of each track are not used in the likelihood determination.

Future versions of the software will use a parameterisation of the truncated mean and resolution to determine $dE/dx$ PID discriminators. A $\chi$ variable is determined by comparing the measured $dE/dx$ truncated mean to a predicted value and resolution. The predicted values are calculated from a parameterisation of $dE/dx$ as a function of $\beta\gamma$. The predicted resolutions depend on the $dE/dx$ measurement, the number of hits on the track, and the polar angle of the track. After determining the parameterisation for the predicted means and resolutions, a $\chi$ value is determined according to

$$\chi_h = \frac{I_{\text{meas}} - I_{\text{pred},h}}{\sigma_{\text{pred},h}} \tag{6}$$

where $h$ is the particle type, $I$ is the $dE/dx$ truncated mean, and $\sigma$ is the resolution for the given particle type. As the distributions of this $\chi$ variable are approximately Gaussian, it may be converted to a likelihood and combined with the output of other PID systems. The performance of such an algorithm is generally similar to the current method, but will enable a better charaterisation of the resolution.

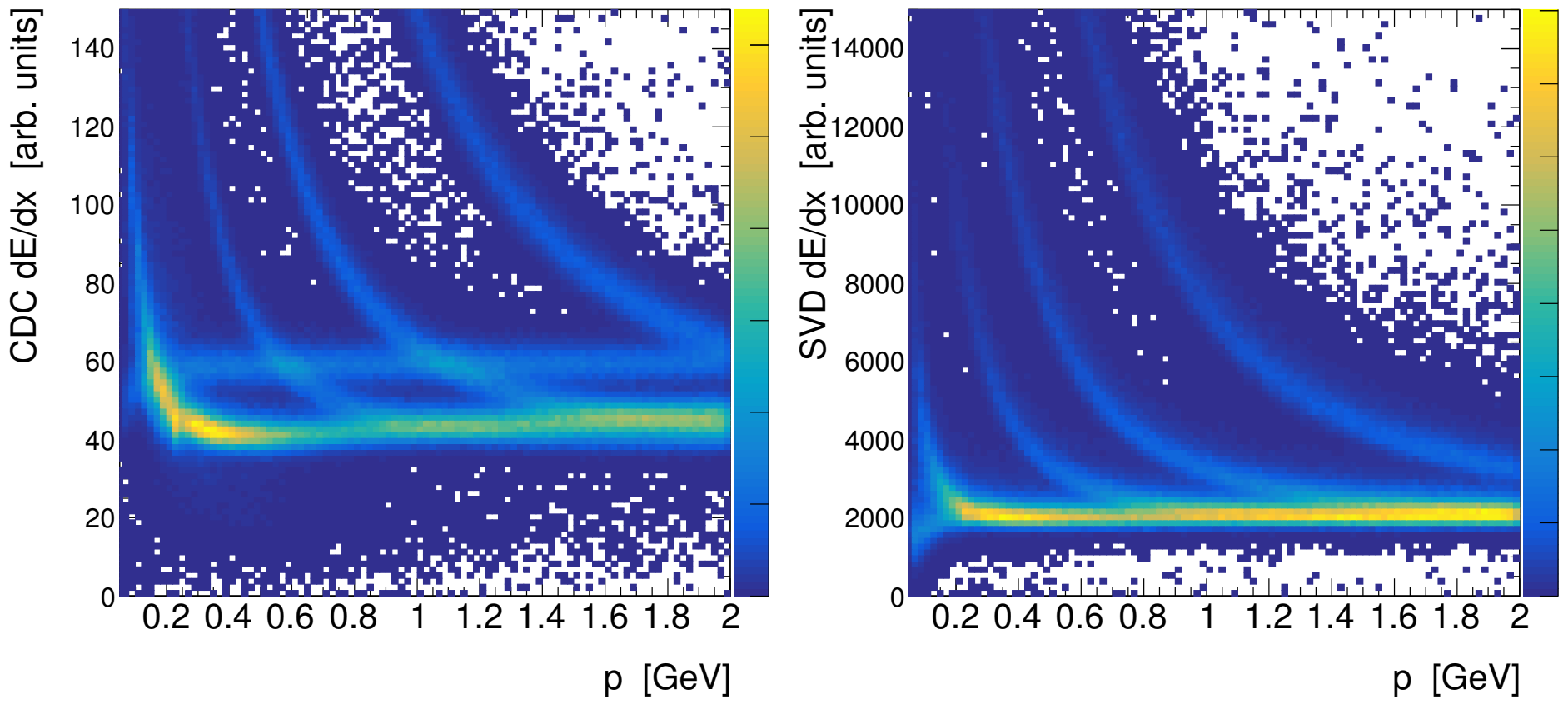

Fig. 19: Truncated $dE/dx$ means as a function of momentum for the CDC (left) and SVD (right). Distinct bands are evident for the various particle species below about 1 GeV/c.

*Performance.* Defining the signal efficiency as the fraction of events relative to the generated quantity that have a likelihood of being identified as the true particle type greater than that of being identified as another particle type (*e.g.*, $L_K > L_\pi$), the average kaon efficiency from dE/dx in the SVD below 700 MeV/c is about 96%. The comparable value in the CDC is nearly identical. The fraction of pions misidentified as kaons under the same criteria is about 3.1% in the SVD and about 1.1% in the CDC. Combining the information from these two detectors yields an average kaon efficiency of about 99.5% below 700 MeV/c, with a fake rate of about 0.2%.

*5.5.2. Charged hadron identification.* Particle identification for charged hadrons, which in this context include pions, kaons, protons and deuterons, $\{\pi, K, p, d\}$, depends primarily on likelihood information from the CDC, TOP, and ARICH detectors. These detectors also contribute to the particle identification of charged leptons, $\{e, \mu\}$. The methods to construct the likelihoods for each of these detector systems are briefly described here.

*TOP Likelihoods.* The TOP counter is a novel type of PID device that combines time-of-flight measurements with the Cherenkov ring imaging technique [9]. The dominant contribution to the resolution of this detector is the dispersion of light while propagating in the quartz bar. This effect is mitigated by focusing the Cherenkov light onto the photon detector with a spherical mirror and measuring the coordinates of the photon impact position. To further improve the resolution, an expansion prism is added at the bar exit window.

The TOP counter consists of sixteen 2.7 m long modules positioned in the space between the CDC and the ECL and covers the polar angles from 32° to 120°. The gaps between the modules account for about 5% of the uncovered area.

An extended likelihood method is used to determine log likelihoods for the six long-lived charged particle types. The extended log likelihood probability for a given charged particle hypothesis $h$ is defined as

$$\ln \mathcal{L}_h = \sum_{i=1}^{N} \ln\Big(\frac{S_h(x_i, y_i, t_i) + B(x_i, y_i, t_i)}{N_e}\Big) + \ln P_N(N_e), \tag{7}$$

where $S_h(x, y, t)$ is the signal distribution for the hypothesis, $h$, $B(x, y, t)$ is the distribution of background and $N_e = N_h + N_B$ is the expected number of detected photons, being a sum of the expected number of signal photons $N_h$ for hypothesis $h$ and the expected number of background photons, $N_B$. The channel coordinates are given by $x$ and $y$, and the integration is performed over the full range $t$ of the time-of-arrival measurement. The second term in Eq. 7 is the Poisson probability to obtain $N$ photons if the mean is $N_e$.

The normalisations of $S_h(x, y, t)$ and $B(x, y, t)$ are:

$$\sum_{j=1}^{n_{ch}} \int_0^{t_m} S_h(x_j, y_j, t)dt \quad = \quad N_h, \tag{8}$$

$$\sum_{j=1}^{n_{ch}} \int_0^{t_m} B(x_j, y_j, t)dt \quad = \quad N_B, \tag{9}$$

where the sum runs over all channels $n_{ch}$ of the photon detector array.

The ring image of the TOP counter is a complicated pattern, which, besides the Cherenkov angle, also depends on the particle impact position and the angles with respect to the quartz bar. The distribution for a particular detection channel $j$ can be parametrised as a sum of Gaussian distributions

$$S_h(x_j, y_j, t) = \sum_{k=1}^{m_j} n_{kj} g(t - t_{kj}; \sigma_{kj}), \tag{10}$$

where $n_{kj}$ is the number of photons in the $k$-th peak of channel $j$; $t_{kj}$ is the position and $\sigma_{kj}$ the width of the peak, and $g(t - t_{kj}; \sigma_{kj})$ is the normalised Gaussian distribution; and $m_j$ counts the number of peaks in the channel $j$ for $t < t_m$.

The quantities $n_{kj}$, $t_{kj}$ and $\sigma_{kj}$ are functions of the Cherenkov angle $\theta_c$, the photon emission point $(x_0, y_0, z_0)$ given by the particle impact position, the particle impact angles $(\theta, \phi)$, and the unfolded channel coordinate $x_D^{kj} = ka \pm x_j$, where $k$ represents the number of internal reflections at the side walls and $a$ the width of the quartz bar. Using the above input data it is possible to solve for the unknown Cherenkov azimuthal angle $\phi_c^{kj}$ and thus determine the photon directional vector [10, 64].

Once the photon direction is known, $t_{kj}$ is obtained by ray-tracing. The number of photons in the peak is calculated with

$$n_{kj} = N_0 \ell \sin^2 \theta_c \frac{\Delta\phi_c^{kj}}{2\pi}, \tag{11}$$

where $N_0$ is the figure of merit of the Cherenkov counter, $\ell$ is the length of the particle trajectory in the quartz bar and $\Delta\phi_c^{kj}$ is the range of the Cherenkov azimuthal angle covered by the measuring channel $j$. The peak width $\sigma_{kj}$ is obtained by summing in quadrature the following contributions:

- photon emission point spread (parallax error), $\sigma_\ell = dt_{kj}/d\lambda \cdot \ell/\sqrt{12}$, where $\lambda$ is the running parameter of the particle trajectory inside the quartz bar ($0 \leq \lambda \leq \ell$),
- multiple scattering of the particle in the quartz, $\sigma_{\text{scat}} = dt_{kj}/d\theta_c \cdot \theta_0(\ell/2)$, where $\theta_0(\ell/2)$ is calculated with the well known multiple scattering approximation [65],
- dispersion (chromatic error), $\sigma_{\text{disp}} = dt_{kj}/de\sigma_e$, where $\sigma_e$ is the r.m.s of the energy distribution of detected photons in the channel,
- channel size, $\sigma_{\text{ch}} = dt_{kj}/dx_D \cdot \Delta x_j/\sqrt{12}$, where $\Delta x_j$ is the channel width,
- and transit time spread of the photon detector, $\sigma_{\text{TTS}}$.

The derivatives $dt_{kj}/d\lambda$, $dt_{kj}/d\theta_c$, $dt_{kj}/de$ and $dt_{kj}/dx_D$ are calculated numerically according to the method described in detail in Refs. [10, 64].

Identification and mis-identification efficiencies have been studied with MC simulations. Using $e^+e^- \to c\bar{c}$ samples, we obtain the performance shown in Fig. 20. The efficiency is defined as the proportion of tracks that are properly identified according to the generated information for all tracks that fall within the TOP acceptance. In the momentum region below 2 GeV/$c$, the efficiency of identifying a kaon is about 94% with a probability of approximately 4% of being mis-identified as a pion. Above 2 GeV/$c$ the performance slowly decreases and gives about 85% efficiency with a 15% fake rate at 3 GeV/$c$. Figure 20 also shows that when the nominal beam background is included, the performance of the counter is not appreciably degraded. Other interesting studies are discussed in Ref. [66]. Discrimination of multiple charged particle hypotheses is possible with the TOP.

*ARICH Likelihoods.* In the Belle II detector, PID in the forward endcap is achieved with the aerogel ring imaging Cherenkov counter (ARICH). The ARICH covers the polar angle range from 17° to 35°. Reconstructed tracks from the CDC are extrapolated to the ARICH detector volume and a likelihood function is constructed for each of the six different particle type hypotheses for tracks that pass through the aerogel layer. The likelihood function is based on a comparison of the observed spatial distribution of Cherenkov photons on the photodetector plane with the expected distribution for the given track parameters (position and momentum vector on the aerogel plane) for a given particle type.

The ARICH likelihood functions are constructed based on the method described in Ref. [67, 68]. For each of the particle type hypotheses, $h$, a likelihood function is calculated as $\mathcal{L}_h = \prod_i p_i^h$, where $i$ runs over all pixels of the detector and $p_i^h$ is the probability for pixel $i$ to record the observed number of hits (1 or 0) assuming particle type $h$. As the $p_i^h$ is a Poisson distribution, one can show that $\mathcal{L}_h$ can be rewritten as $\ln \mathcal{L} = -N + \sum_i (n_i + \ln(1 - e^{-n_i}))$, where $N$ is the expected total number of hits, $n_i$ is the expected (calculated) average number of hits on pixel $i$, and the sum runs only over the pixels that were hit in an event[6].

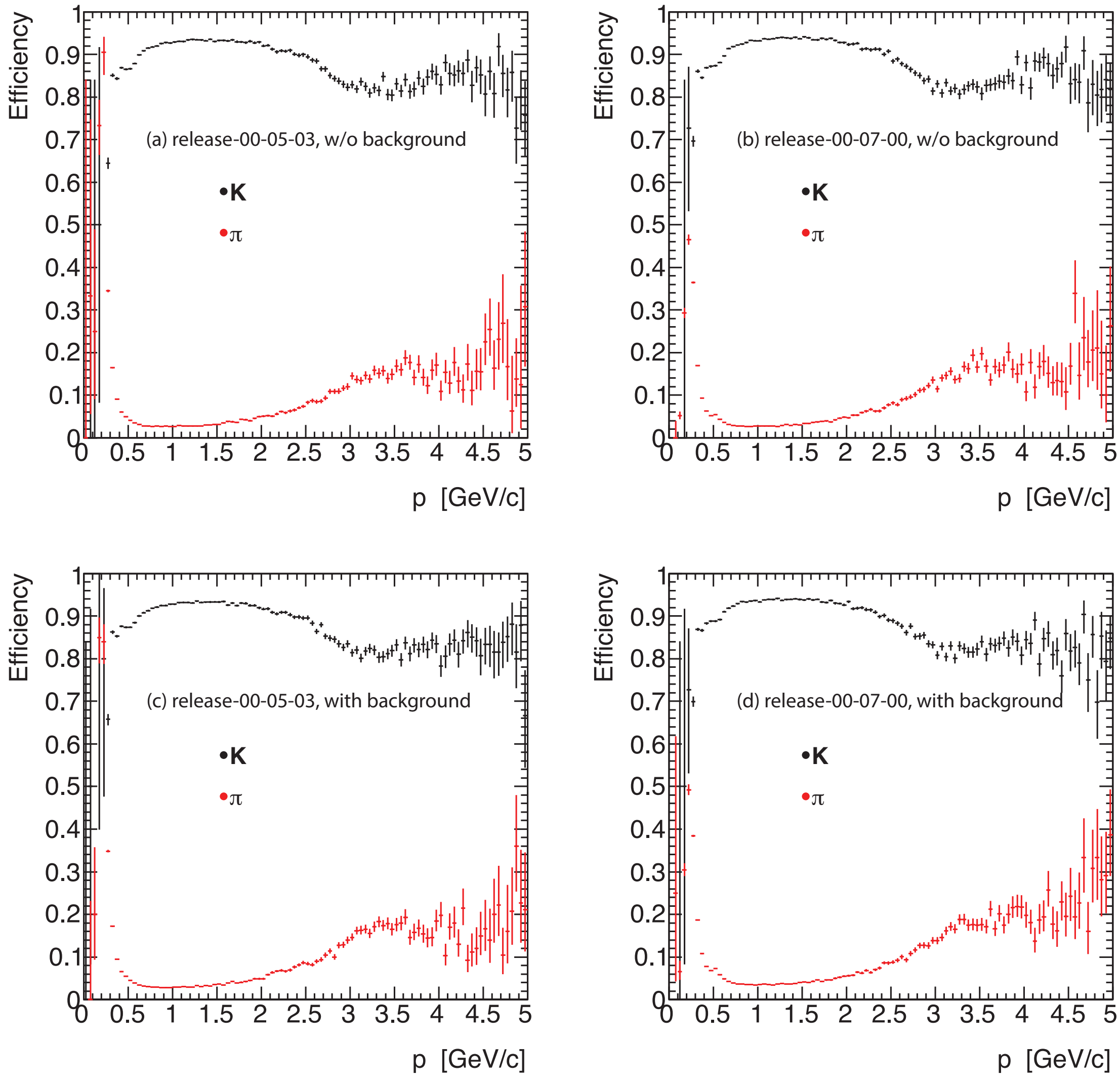

Fig. 20: TOP counter kaon identification efficiency (black markers) and pion fake rate (red markers) as a function of momentum for $\ln \mathcal{L}_{\mathcal{K}} > \ln \mathcal{L}_{\pi}$, obtained with a MC simulation without beam background (a,b) and with beam background (c,d). The performance using release-00-05-03 is given in (a) and (c) while that for release-00-07-00 is given in (b) and (d). Only tracks that fall within the TOP acceptance are considered.

The expected average number of hits on pixel $i$, $n_i$, is obtained as a sum of contributions from signal and background hits, $n_i = n_i^s + n_i^b$, where the signal contribution is divided into contributions from the first and second aerogel layers, $n_i^s = n_i^{s,1} + n_i^{s,2}$. The contribution of

[6] For brevity, index $h$ is omitted, but note that $N$ and $n_i$ depend on $h$.

each aerogel layer ($r$) is calculated as

$$n_i^{s,r} = \epsilon_{det} N^{s,r} \int_{\Omega_i} \frac{1}{2\pi} G(\theta, \theta_h^r, \sigma_h^r) d\theta d\phi \tag{12}$$

where $\epsilon_{\text{det}}$ is the photon detection efficiency and $N^{s,r}$ is the number of photons emitted from aerogel layer, $r$ (theoretically calculated). The integral gives the probability for a Cherenkov photon being emitted by particle type $h$ from aerogel layer $r$ into the solid angle covered by pixel $i$ ($\theta$ and $\phi$ are the polar and azimuthal angles with respect to the track direction). A Gaussian function $G$ with a mean at the expected Cherenkov angle ($\theta_h^r$) and width $\sigma_h^r$ (due to uncertainty in photon emission position) for a track of particle type $h$ is used to describe Cherenkov angle distribution (*i.e.* $\theta$). To obtain the number of photons emitted from the aerogel layer ($N^{s,r}$) a general expression is used for the Cherenkov photon yield, where the track path length in the aerogel, Rayleigh scattering, and the photon loss on the edges of aerogel tiles are taken into account. A constant (pixel-independent) value is assumed for the background contribution $n_i^b$, set to correctly describe the observed distribution.

The expected total number of hits, $N$, is obtained as $\epsilon_{det}\epsilon_{acc}(N^{s,1} + N^{s,2})$, where $\epsilon_{acc}$ is the geometrical acceptance correction factor (*i.e.* what fraction of the Cherenkov ring falls on the photo-sensitive surface). The acceptance correction factor is calculated using a simple ray tracing simulation in which 200 rays, uniformly distributed in $\phi$ and at the expected Cherenkov angle $\theta_h^r$, are propagated from the mean emission point in the aerogel to the detector plane. The number of track lines that hit the photo-sensitive surface is used to determine the correction factor.

The above procedure is carried out for all six particle hypotheses. The log-likelihood difference, $\ln \mathcal{L}_{h_1} - \ln \mathcal{L}_{h_2}$, is used to distinguish $h_1$ and $h_2$ particle types. On average about 12 signal photons are detected per saturated track (*i.e.* with $\theta_{\text{Cherenkov}} = \theta_{\text{Cherenkov}}^{\text{max}}$). The PID performance is mainly degraded by tracks with poor track information (position and direction on the aerogel plane), which result either from poor reconstruction or rescattering in the CDC aluminium endplate, and by tracks that produce a very low number of photons. The latter are mainly tracks that pass through the gap between two adjacent aerogel tiles, or tracks producing a Cherenkov ring that largely misses the photosensitive area.

The PID performance of the ARICH detector for $K/\pi$ separation is depicted in Fig. 21.

*5.5.3. Muon identification.* Muon identification (MuID) in the KLM uses the differences in longitudinal penetration depth and transverse scattering of the extrapolated track. The MuID reconstruction module in the tracking package of basf2 proceeds in two steps: (1) track extrapolation using the muon hypothesis and (2) likelihood extraction for each of six particle hypotheses: muon, pion, kaon, proton, deuteron, and electron.

The six likelihoods that are assigned to a given track are stored as unnormalised log likelihood values and normalised likelihood values in the MuID data-object. In the post-reconstruction analysis, the log likelihood differences may be used to select or reject the muon hypothesis for a given track.

The KLM geometry exhibits several features: the barrel has 15 detector layers with no iron before the innermost layer; the forward (backward) endcap has 14 (12) detector layers with iron before the innermost layer. The iron plates are about 4.7 cm thick and are separated by detector-filled 4.4-cm gaps. The KLM has less iron and detector coverage in the forward and backward overlap regions since the endcaps' outer radius is about 310 cm: there may be as few as 8 detector layers for some polar angles. Thus, the separation power between muons and non-muons is weaker here.

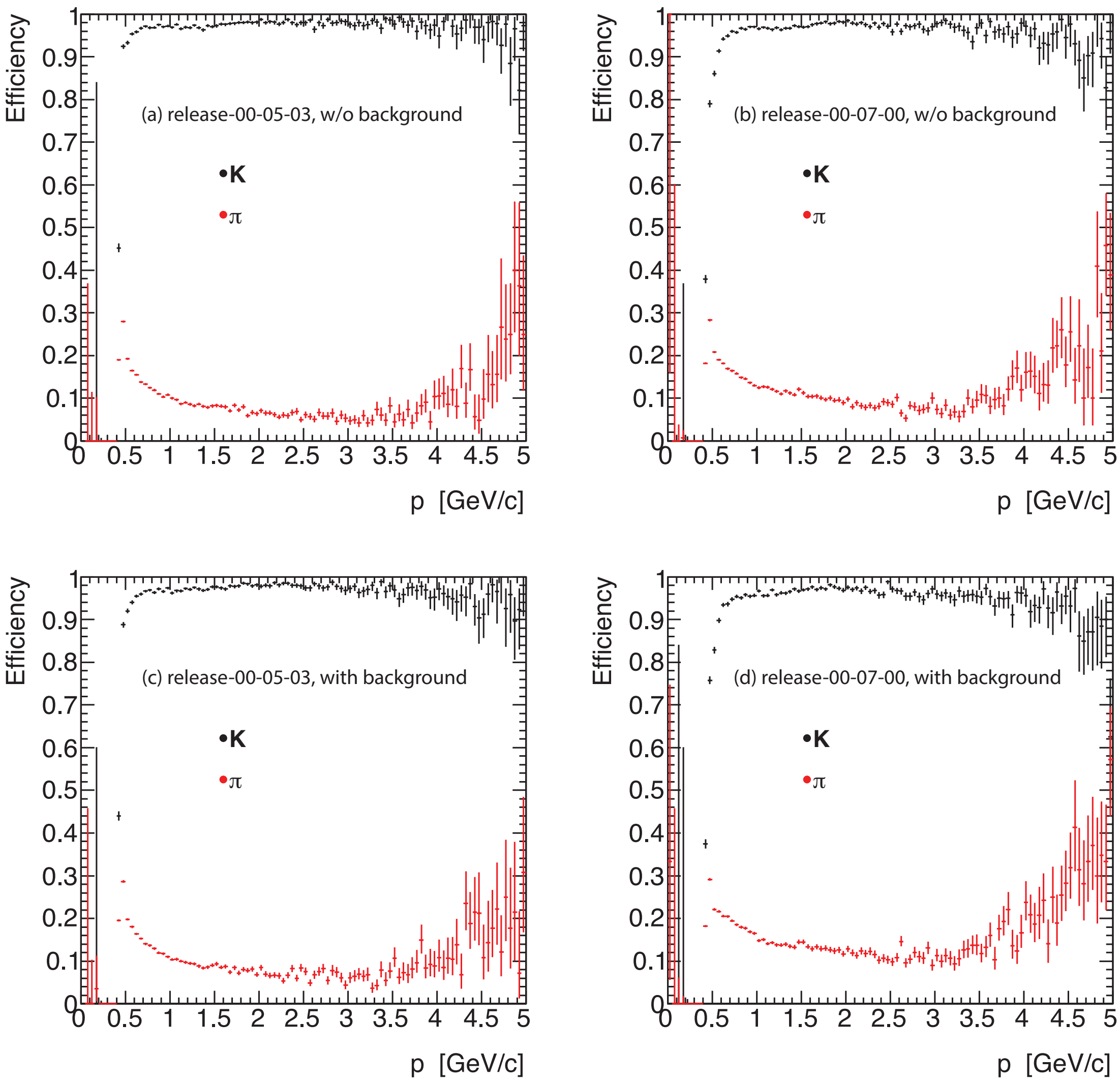

Fig. 21: ARICH kaon identification efficiency (black markers) and pion fake rate (red markers) as a function of momentum for $\ln \mathcal{L}_{\mathcal{K}} > \ln \mathcal{L}_{\pi}$, obtained with a MC simulation without beam background (a,b) and with beam background (c,d). The performance using release-00-05-03 is given in (a) and (c) while that for release-00-07-00 is given in (b) and (d). Only tracks that fall within the ARICH acceptance are considered.

*Track Extrapolation.* Each charged track that is reconstructed in the tracking detectors (CDC, SVD and PXD) is extrapolated outward using GEANT4E [39], starting at the last hit on the reconstructed track in the CDC, assuming the muon hypothesis. During this extrapolation, GEANT4E reduces the track's momentum by the mean integrated specific-ionisation ($dE/dx$) energy loss in the intervening material and inflates the elements of the phase-space covariance matrix due to (elastic) multiple scattering and fluctuations in $dE/dx$. Particles are assumed to not decay during this extrapolation.

Extrapolation through the non-KLM sections by GEANT4E does not consider actual hits in any of the sensitive elements. In contrast, extrapolation through the KLM uses each matching hit in a Kalman-filtering adjustment of the track parameters and covariance matrix [69].

The extrapolation proceeds step by step through the detector geometry, starting at the outermost point on the reconstructed track's trajectory (usually in the CDC) and with the reconstructed track's phase-space coordinates and covariance matrix. Upon crossing a KLM detector layer, the nearest two-dimensional hit—if any—in that layer is considered for association with the track. If the hit is within about $3\sigma$ in either of the two local coordinate directions (where $\sigma$ is the sum in quadrature of the extrapolation's position uncertainty and the hit-measurement uncertainty) then it is declared a matching hit and the Kalman filter uses it to adjust the track properties before the next step in the extrapolation. At the same time, the Kalman filter's fit quality ($\chi^2$) is accumulated for the track. A given 2D hit may be associated with more than one track. The hit-matching algorithm begins with the inner layers and moves outward such that a given KLM detector layer is examined at most once for a given track. The extrapolation ends when the kinetic energy falls below a user-defined threshold (nominally 2 MeV) or the track curls inward to a cylindrical radius below 40 cm or escapes from the KLM.

*Determination of Likelihoods.* If the track reached the KLM, it is classified according to how and where the extrapolation ended (stopped in or exited from the barrel or the endcap). The likelihood of having the matched-hit range and transverse-scattering $\chi^2$ distribution (based on the distance between the measured and extrapolated tracks) is obtained from pre-calculated probability density functions (PDFs). There are separate PDFs for each charged-particle hypothesis and charge and for each extrapolation outcome.

The longitudinal-profile PDF value $\mathcal{P}_L(\vec{x};\ O,\ \ell,\ H)$ for extrapolation-ending outcome $O$ and outermost layer $\ell$ and for particle hypothesis $H \in \{\mu^\pm,\ \pi^\pm,\ K^\pm,\ e^\pm,\ p,\ \bar{p},\ d,\ \bar{d}\}$ is sampled according to the measurement vector $\vec{x}$ given by (a) the pattern of all KLM layers touched during the extrapolation—not just the outermost such layer—and (b) the pattern of matched hits in the touched layers. Note that the momentum and direction are not included in the measurement vector; the outermost extrapolation layer $\ell$ is a proxy for these. The extrapolation outcome $O$ accounts for the KLM geometry by classifying a track as stopping in or exiting (1) the barrel only, (2) the forward endcap only, (3) the backward endcap only, (4) the overlap region between the forward barrel and endcap, and (5) the overlap region between the backward barrel and endcap.

The transverse-scattering probability density function $\mathcal{P}_T(\chi^2,\ n;\ D,\ H)$ for KLM region $D$ (barrel-only, endcap-only, or overlap) and particle hypothesis $H$ is sampled according to the measurements of $\chi^2$ from the Kalman filter and the number $n \in \{2,\ 4,\ ...,\ 36\}$ of degrees of freedom—twice the count of matching-hit layers since there are two independent measurements per layer. The muon hypothesis PDF is very close to the ideal $\chi^2$ distribution for the given number of degrees of freedom while the non-muon hypothesis PDFs are considerably broader for low degrees of freedom—the most likely scenario for a true non-muon.

For each track, the likelihood for a given particle hypothesis is the product of the corresponding longitudinal-profile and transverse-scattering PDF values:

$$\mathcal{L}(H;\ O,\ \ell,\ D,\ \vec{x},\ \chi^2,\ n) = \mathcal{P}_L(\vec{x};\ O,\ \ell,\ H) \cdot \mathcal{P}_T(\chi^2,\ n;\ D,\ H)\,. \tag{13}$$

The natural logarithm of this value is stored in the MuID data-object. Then, the six likelihood values are normalised by dividing by their sum and stored in the MuID data-object.

Presently, significance levels, $\mathcal{S}$, are not available. Such values might be used, for example, to remove tracks that are not consistent with *any* hypothesis by requiring $\mathcal{S} > \mathcal{S}_{\rm min}$. This feature will be added in a future release.

*Muon Efficiency and Pion Fake Rate.* The log likelihood difference

$$\Delta \equiv \ln(\mathcal{L}(\mu^+; O,\ \ell,\ D,\ \vec{x},\ \chi^2,\ n)) - \ln(\mathcal{L}(\pi^+; O,\ \ell,\ D,\ \vec{x},\ \chi^2,\ n)) \tag{14}$$

is the most powerful discriminator between the competing hypotheses. The requirement $\Delta > \Delta_{\rm min}$ for a user-selected $\Delta_{\rm min}$ provides the best signal efficiency for the selected background rejection.

Log likelihood differences for true muons and pions are shown in Fig. 22 as a function of the track momentum. Clearly, choosing a momentum-independent threshold on $\Delta_{\rm min}$ that is non-zero (and positive) will reject low-momentum muons. Similar behaviour is seen when choosing a criterion that is independent of the polar or azimuthal angles since the log likelihood differences are softer in the uninstrumented azimuthal cracks between sectors and in the barrel-endcap overlap regions where the KLM is thinner (with only around 8 detector and iron layers).

Muon efficiency and pion fake rate are shown in Fig. 23 as functions of momentum, polar angle, and azimuthal angle for three values of the log likelihood-difference threshold. The black curves exhibit the behaviour for the nominal cut of $\Delta > 0$: the muon efficiency is 90–98% for momenta above 1.0 GeV/$c$ while the pion fake rate is 2.5–6%; the muon efficiency is flat at 96% in $\theta$ while the pion fake rate is 2–6%; the muon efficiency is 92–98% in $\phi$ (with dips at each octant boundary and at the solenoid cryostat chimney) while the pion fake rate is roughly flat at 3.5% (or 4% at the chimney). The red curves exhibit more pronounced differences as a function of $p$, $\theta$ and $\phi$ for the much tighter cut of $\Delta > 20$, where muon efficiency is sacrificed somewhat—and unevenly in each of these variables—for much better purity.

*5.5.4. Electron identification.* Global electron identification (EID) combines individual likelihoods from the ECL, $dE/dx$ measurements taken from the SVD and CDC, and Cherenkov PID information from the TOP, and ARICH. The $E/p$ value, however, is the primary feature for separating electrons from other charged particles (namely, muons and pions), where $E$ is the energy measured in the ECL and $p$ is the absolute track momentum. The $E/p$ distributions for electrons, muons, and pions are shown in Fig. 24 for a variety of momentum ranges. For $p \geq 1$ GeV/$c$, there is sufficient distinction between electrons and other charged particles in this distribution, making it a useful parameter for a fit-based likelihood profile for EID.

The ECL Electron ID Module is responsible for using momentum and polar angle dependent probability distribution function fit parameters to find the best fit to the $E/p$ distribution. It then derives a fit-based electron log likelihood. This log likelihood can then be combined with EID log likelihoods from other sub-detectors to create the combined global EID log likelihood used in analyses. The $E/p$ distribution is fit using a Gaussian convoluted with a Crystal Ball (CB) function in bins of track momentum and polar angle of the ECL shower associated with the electron.

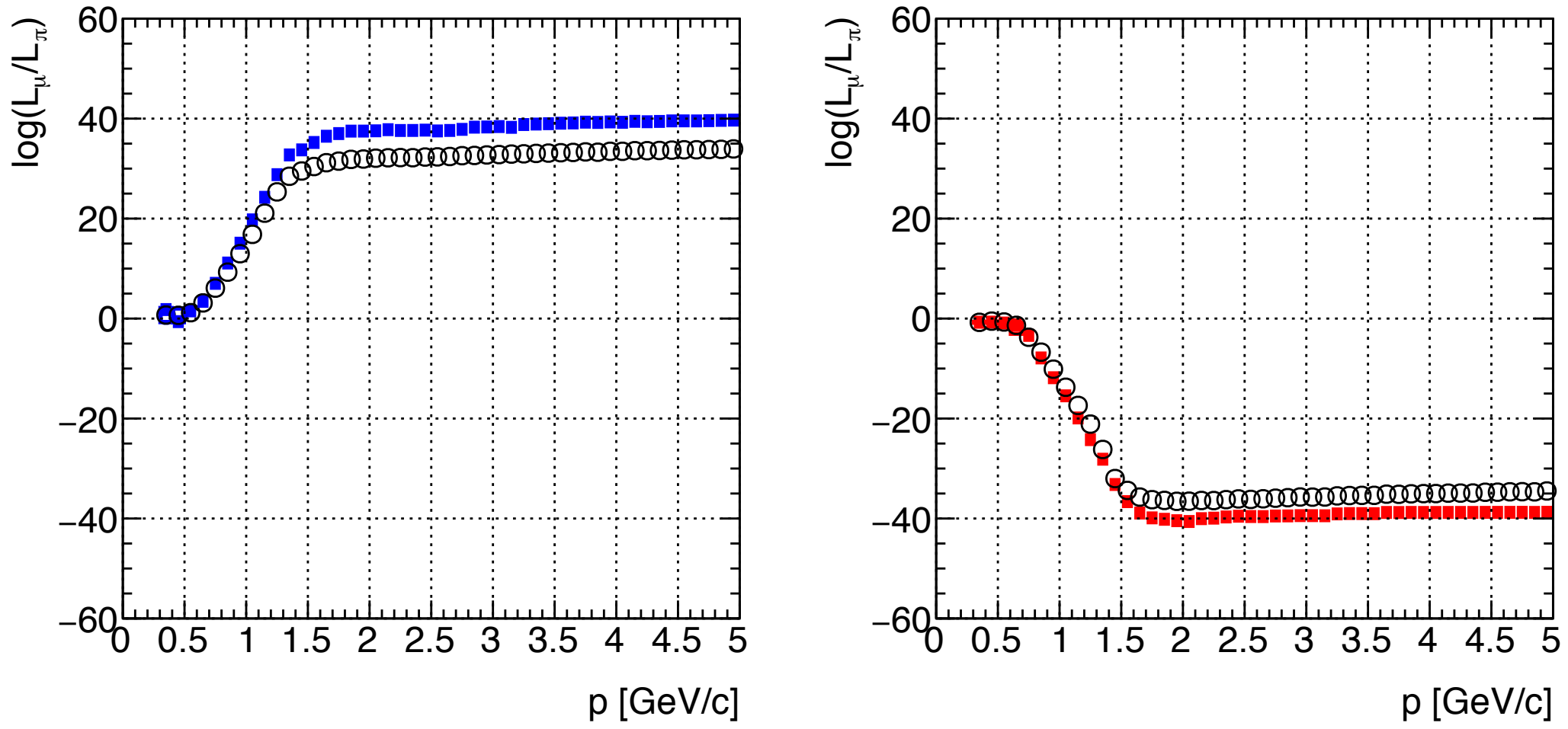

Fig. 22: Log likelihood difference between muon and pion hypotheses for true muons (left) and true pions (right) as a function of the track momentum in GeV/$c$. In each bin, the open circle (box) represents the mean (median) of the distribution at a given momentum. The mean [or median] deviates increasingly from zero with rising momentum (*i.e.*, with increasing number of crossed KLM layers) and then saturates at about $\pm 40$ for exiting tracks.

Separation between electrons and muons is quite good for sufficiently energetic particles (*i.e.* muons with $p > 0.3$ GeV/$c$ which are able to reach the KLM). Separation between electrons and pions, however, is much more difficult. This is particularly true for low-momentum particles where, as is seen in Fig. 24, the $E/p$ distributions for differing particle types are very similar. The difficulty in distinguishing electrons from pions is further exemplified in Fig. 25, which shows the electron efficiency for true electrons and true pions as function of momentum. We see a high electron efficiency and low pion misidentification for momenta $1 \leq p \leq 3$ GeV/$c$. At low momentum, the electron efficiency drastically drops off as the radius of curvature of a low momentum electron in the presence of the magnetic field is very small. Therefore, for low-momentum electrons, the particle often fails to reach the ECL.

A useful quantity for charged PID is the delta log likelihood value, defined as $\Delta = \ln(\mathcal{L}_e) - \ln(\mathcal{L}_\pi)$, where $\mathcal{L}_e$ is the global electron likelihood and $\mathcal{L}_\pi$ is the global pion likelihood. For true electrons, this quantity takes on positive values, while for true pions, it takes on negative values. Moreover, we can represent the separation between electrons and pions by considering receiver operating characteristic (ROC) curves of electron efficiency against pion fake rate for various momentum ranges, calculated using the delta log likelihood distributions for true electrons and true pions. Figure 26, shows that the separation between electrons and pions becomes more ambiguous at lower momenta. More sophisticated electron ID algorithms will be considered for Belle II analyses, based on Zernike moments (image moments using Zernike polynomials as basis functions) of lateral shower shapes, longitudinal shower information, track–cluster matching, and corrections for bremsstrahlung.

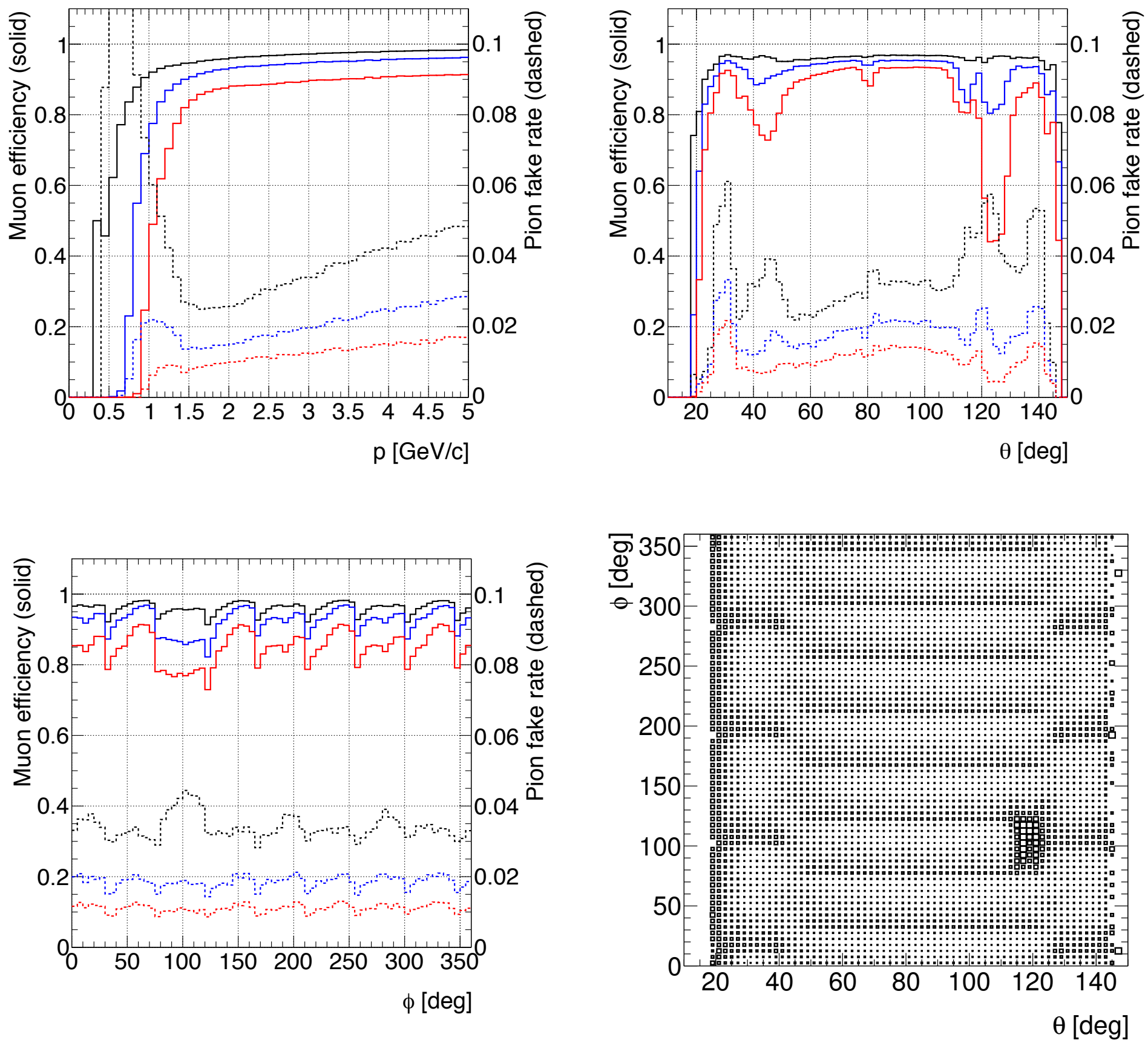

Fig. 23: Muon efficiency (solid, left-axis scale) and pion fake rate (dashed, right-axis scale) for three values of the log likelihood-difference cut: $\Delta_{\rm min} = 0$ (black), 10 (blue), and 20 (red) as a function of momentum (top left), polar angle (top right), and azimuthal angle (bottom left). Muon inefficiency as a function of $\phi$ vs $\theta$ (bottom right), illustrating the geometric inefficiencies at the sector boundaries (8 horizontal enhancements in the barrel; 4 horizontal enhancements in each endcap) and in the vicinity of the solenoid chimney.

*5.5.5. Combined PID performance.* The performance of Belle II PID is estimated using inclusive $e^+e^- \to c\bar{c}$ MC samples. Minimal track quality restrictions are applied. Using the generated information, a sample of each particle type is constructed. The PID efficiency for a sample of particles of type $\alpha$ ($\pi$, $K$, $p$, $e$, $\mu$) is determined by taking the ratio of events that have $L(\alpha : \beta) > 0.5$ to the total sample size, for a given $\beta$ ($\pi$, $K$, $p$, $e$, $\mu$). For example, the $K/\pi$ selection efficiency is given by the fraction of a sample of true kaon tracks that have $L(K : \pi) > 0.5$. In a similar fashion, the pion fake rate is the fraction of a sample of true pion tracks that have $L(K : \pi) > 0.5$. The selection efficiency for various pairs of particle types are given in Figs. 27 and 28. The slight difference in PID performance between the two releases is primarily due to errors in the $dE/dx$ PDFs in release-00-05-03. The performance for release-00-07-00 is a more accurate estimate.

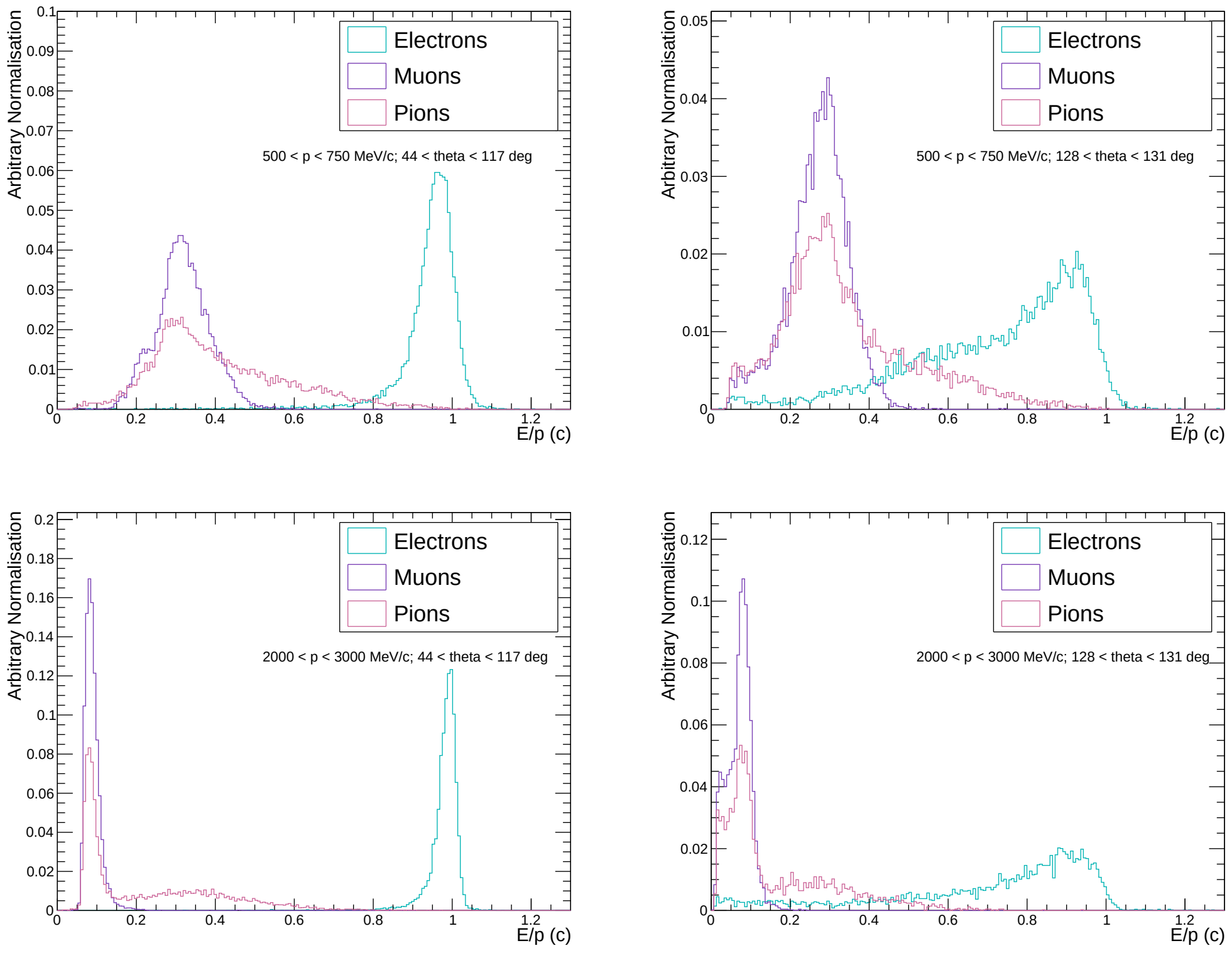

Fig. 24: The $E/p$ distributions for two momentum ranges ($500$ MeV$/c < p < 750$ MeV$/c$ and $2000$ MeV$/c < p < 3000$ MeV$/c$) and polar angle ranges ($44^\circ < \theta < 117^\circ$ and $128^\circ < \theta < 131^\circ$). This is an excellent discriminator for EID when $p > 1$ GeV$/c$ and in the barrel region. However, for low-momentum particles and particles near the crack region, the separation is less distinct.

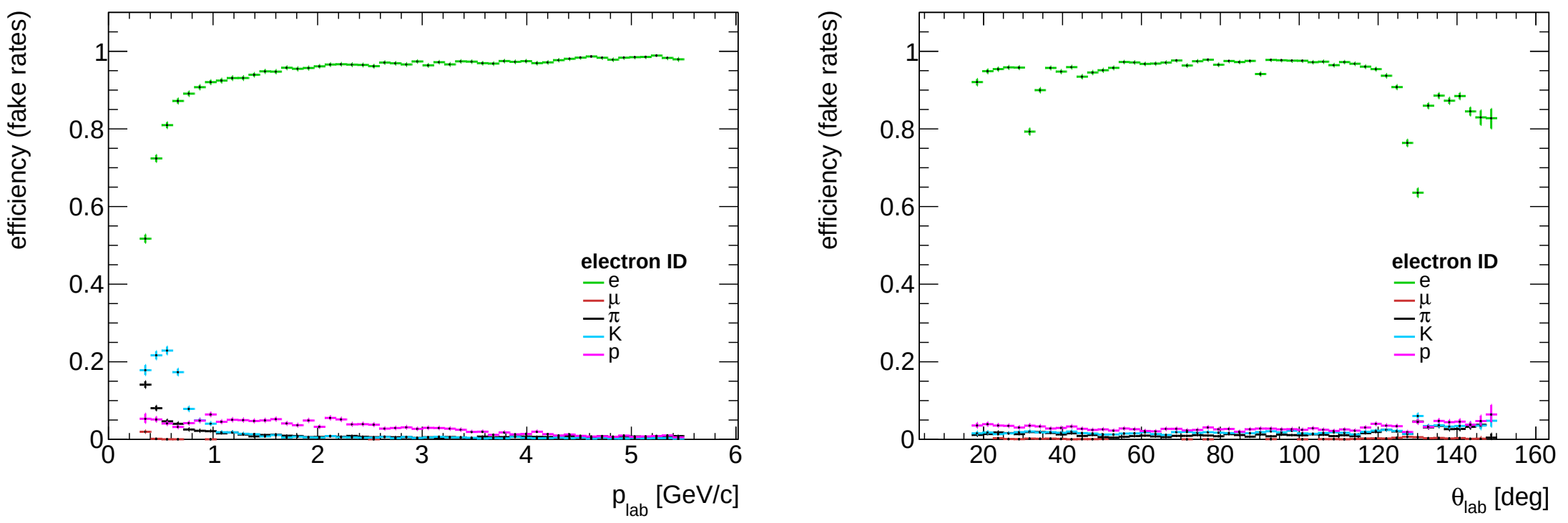

Fig. 25: ECL-only EID efficiency as a function of momentum (left) and as a function of polar angle (right).

In addition to the efficiency plots, Fig. 29 and Fig. 30 show ROC plots for $K/\pi$ and $\pi/K$ separation in release-00-05-03 and release-00-07-00, respectively. These plots use kaon and pion samples from $D^*$ decays to $D^0\pi$ in $\Upsilon(4S)$ MC, where the $D^0$ decays to $K\pi$. In this way, a relatively clean sample of each particle type can be obtained with minimal selection criteria and without truth information.

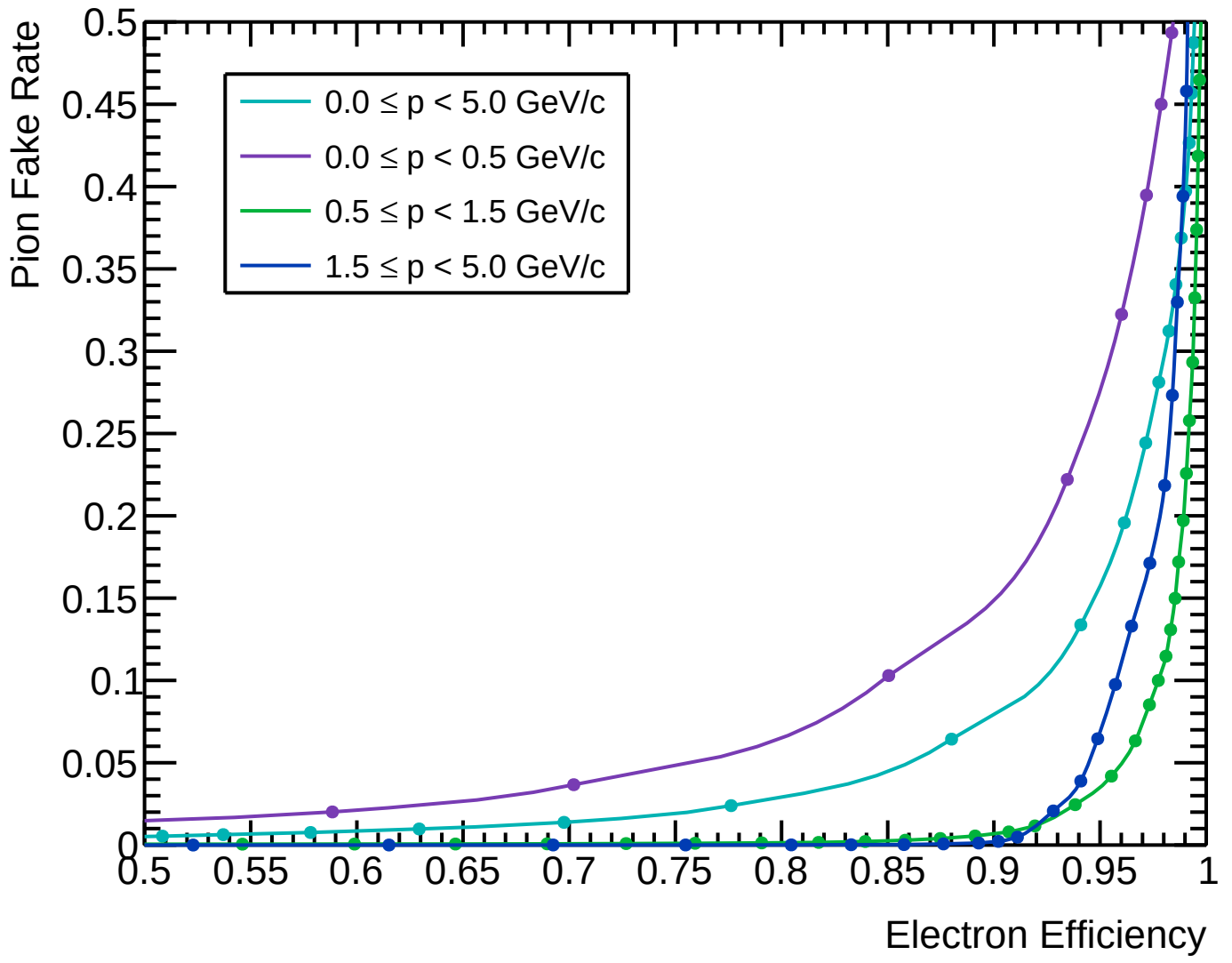

Fig. 26: Electron efficiency against pion fake rate as calculated using the delta log likelihood. This is shown for all particles, low-momentum particles, mid-momentum particles, and high-momentum particles.

The kaon efficiency and pion fake rates using only $dE/dx$, TOP and ARICH information respectively is given in Fig. 31. For each subdetector we show the performance in their respective solid angle acceptance regions.

### *5.6. Neutral particle identification*

*5.6.1. Photon and $\pi^0$ identification.* The identification of photons in the ECL is based on parameters that describe the shower shape of ECL clusters not matched to a reconstructed track. The identification relies on the fact that electromagnetic showers caused by an incident photon is cylindrically symmetric in the lateral direction and the energy deposition decreases exponentially with the distance from the incident axis. The ECL reconstruction up to release-00-07-02 provides only the energy ratio of the nearest $3 \times 3$ to the nearest $5 \times 5$ crystals around a local maximum which is close to unity for true photons. The main background for photon cluster reconstruction comes from neutral or charged hadron interactions. These interactions create asymmetric shower shapes and often result in more than one ECL cluster that is not matched to a charged track, hadronic splitoffs, which yield a large number of fake photon candidates if not identified. The minimal energy of an ECL cluster for physics studies in the presence of nominal backgrounds is 100 MeV in the forward endcap, 90 MeV in the barrel and 160 MeV in the backward endcap for the original ECL reconstruction. Starting with release release-00-08-00, the ECL reconstruction provides additional shower shape variables and the improved clustering algorithm allows to lower the energy threshold to about 25 MeV. Photon likelihoods based on kinematics, shower shapes and timing information can be used in the future to provide particle lists of different efficiency and purity.

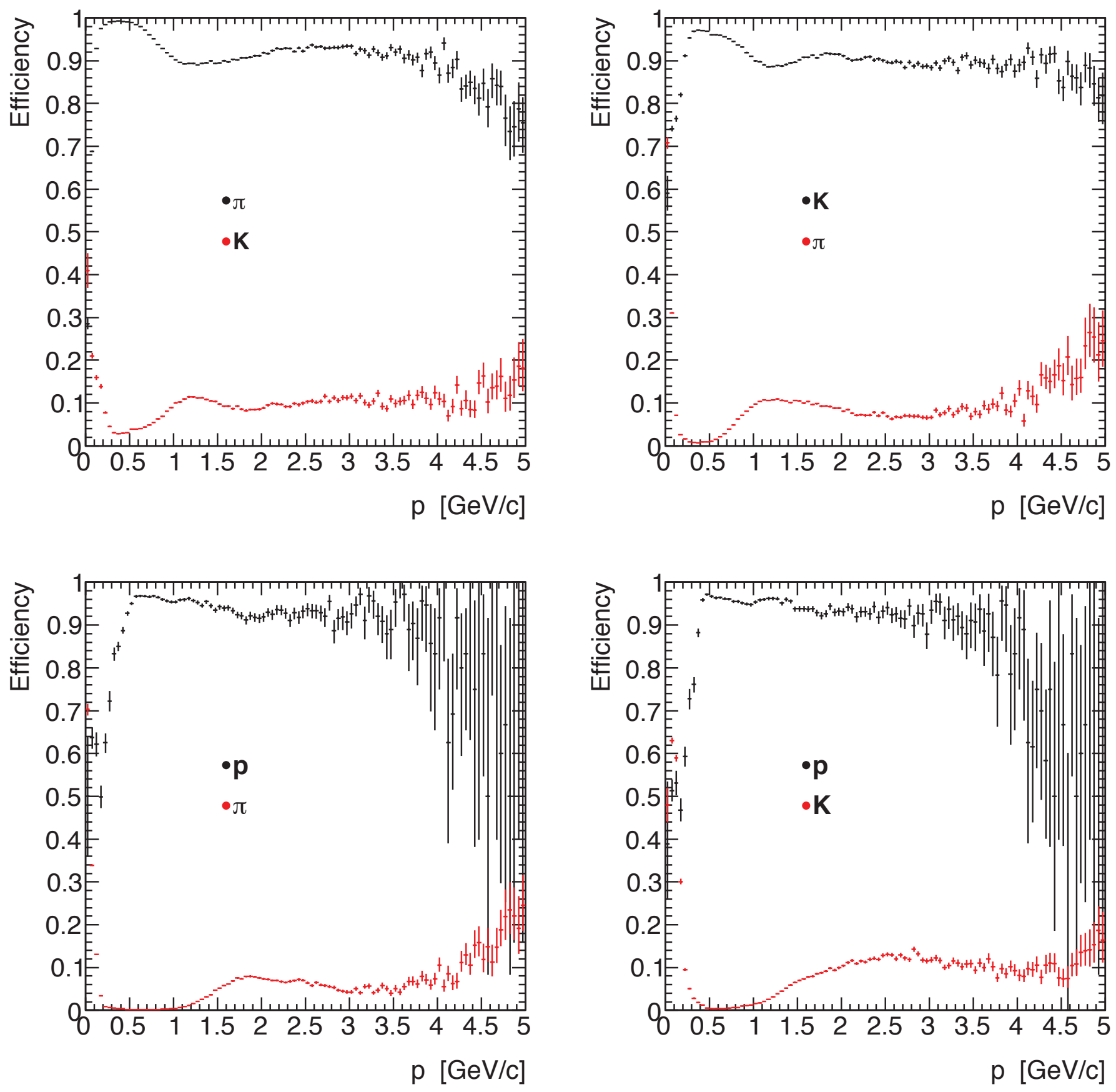

Fig. 27: Charged particle identification selection efficiency for various pairs of particle types as a function of momentum in release-00-05-03. The black markers show the selection efficiency as determined from an inclusive MC sample without beam backgrounds, while the red markers show the fake rate. Only tracks that fall within the acceptance of at least one of the PID detectors or the CDC are considered.

The reconstruction of $\pi^0$ mesons in the decay mode $\pi^0 \to \gamma\gamma$ is based on the combination of two photon candidates. For $\pi^0$ energies below about 1 GeV, the angular separation between the two photons is usually large enough to produce two non–overlapping ECL clusters. For $\pi^0$ energies above about 1 GeV but below about 2.5 GeV, the ECL clusters from the two photons overlap but can still be reconstructed as two separate photon candidates in the ECL. The $\pi^0$ energy can be directly reconstructed from the photon 4-momenta. The $\pi^0$ energy resolution is improved by performing a mass constrained fit of the two photon candidates to the nominal $\pi^0$ mass. Multivariate classifiers can be used to provide higher purity $\pi^0$ particle lists. A low photon energy threshold is mandatory to obtain a high $\pi^0$ efficiency for generic $B$ decays: A 50 MeV threshold for both photons results in a $\pi^0$ efficiency of 76 %, 30 MeV in 93 % and 20 MeV in 98 %.

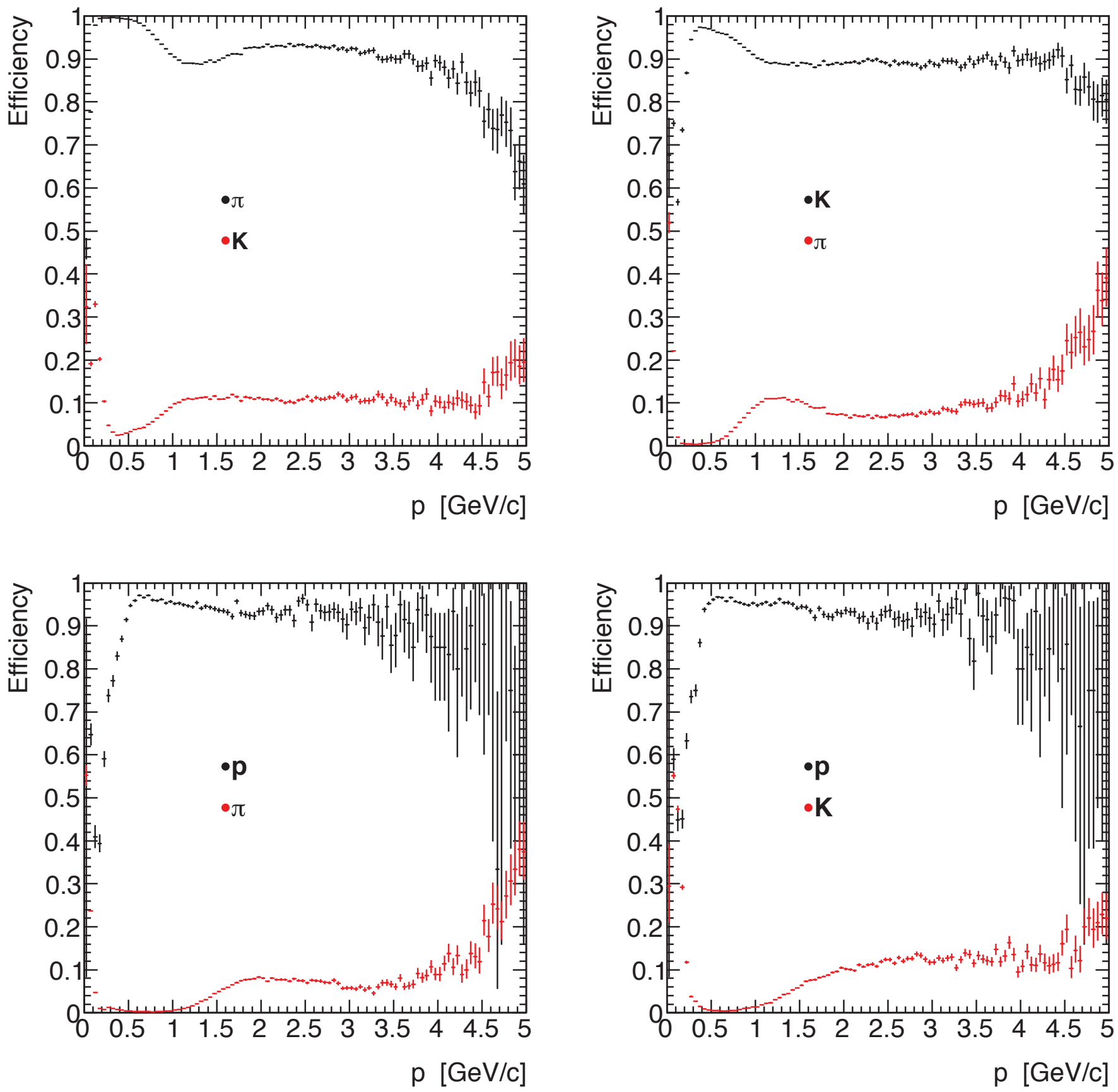

Fig. 28: Charged particle identification selection efficiency for various pairs of particle types as a function of momentum in release-00-07-00. The black markers show the selection efficiency as determined from an inclusive MC sample without beam backgrounds, while the red markers show the fake rate. Only tracks that fall within the acceptance of at least one of the PID detectors or the CDC are considered.

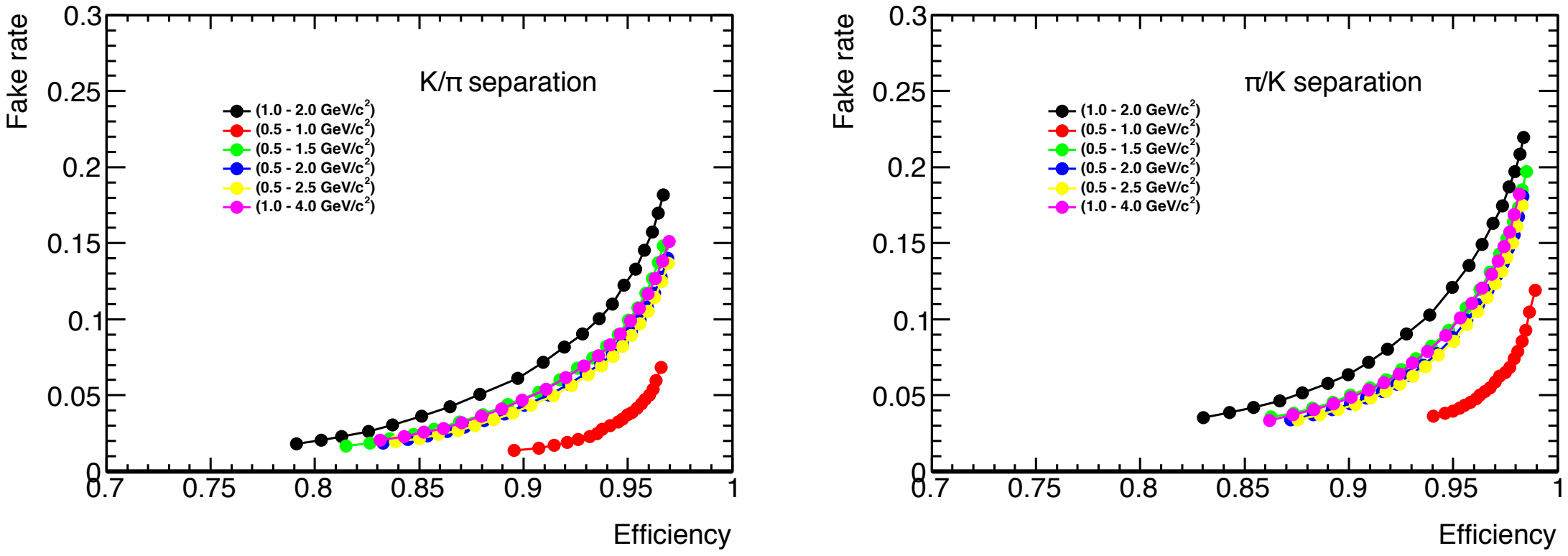

Fig. 29: Fake rates versus efficiencies for $K/\pi$ (left) and $\pi/K$ (right) separation in release-00-05-03. The coloured lines show the ROC curves for different momentum regions. The markers represent different cuts on the likelihood ratio.

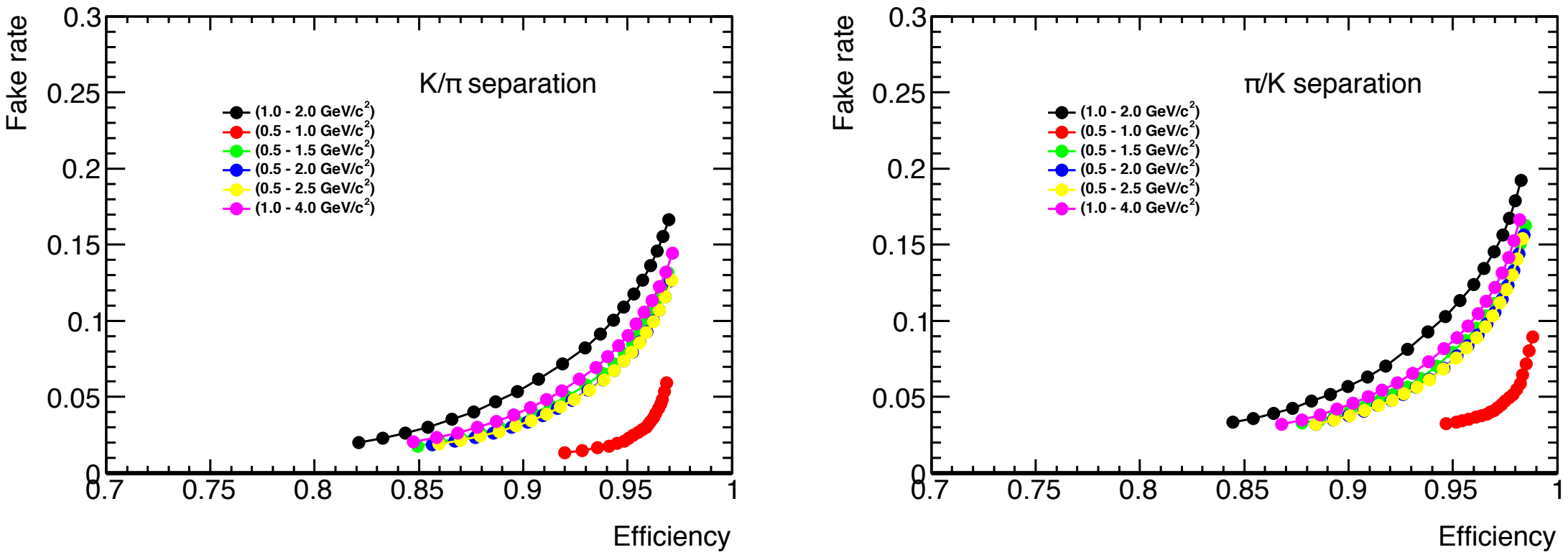

Fig. 30: Fake rates versus efficiencies for $K/\pi$ (left) and $\pi/K$ (right) separation in release-00-07-00. The coloured lines show the ROC curves for different momentum regions. The markers represent different cuts on the likelihood ratio.

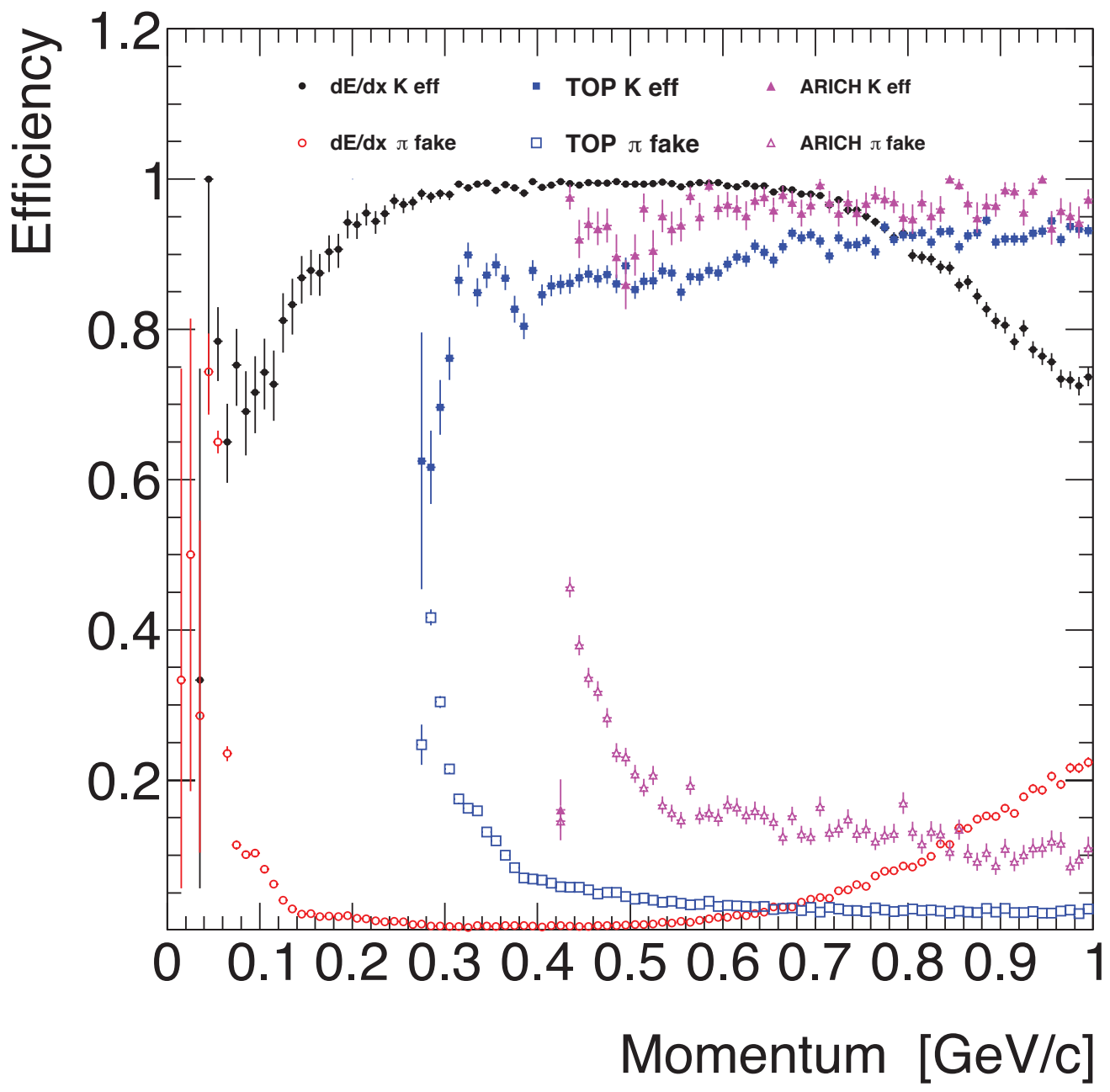

Fig. 31: Kaon detection efficiency and pion fake rate for low momentum tracks from release-00-05-03. The performance is determined using only those tracks that are within the acceptance of the detector of interest. That is, the denominator for the efficiency is different for each detector.

For $\pi^0$ energies above about 2.5 GeV, *e.g.* from $B \to \pi^0\pi^0$, the two photon induced showers often do not have separate local maxima anymore and are reconstructed as one photon candidate. The $\pi^0$ energy can be deferred from the shower's second moment shape variable.

*5.6.2. $K_L^0$ identification.* The identification of $K_L^0$ mesons is based on information collected by the KLM and ECL detectors. The detector material of the KLM provides $> 3.9$ hadronic interaction lengths $\lambda_0$ and the ECL provides $\approx 0.8\ \lambda_0$.
Multivariate methods are used to classify ECL clusters and KLM clusters according to their probability to originate from a $K_L^0$.
Classifiers are constructed from Stochastically Gradient Boosted Decision Trees (BDT), implemented as described in Ref. [70]. The classification is performed separately for the KLM and ECL in the reconstruction package of basf2. The classifiers are trained on a $K_L^0$ target in $\Upsilon(4S)$ events and their output is normalised to $x \in (0, 1)$. In general $K_L^0$ mesons are not easy to classify as their signal in the detector is not mutually different from other more common neutral particles such as neutrons, $\pi^0$ and photons (in the ECL). The largest contributions to the background are from neutrons and photons originating from beam interactions with detector or beam-pipe material, followed by neutral particles from the primary interaction. However, the classifiers still outperform the $K_L^0$ identification algorithms of Belle by a factor of $\approx 2$, assuming generic events without preselection.

*Variables used in the classifications.* The classifiers are fed with all information that is available, including cluster shapes, kinematic variables, and information gained from other detectors and algorithms. There is no single variable that provides significant separation power alone. The three most important variables that have the best proven separation in the KLM are as follows:

- Distance to the next track: neutral clusters are not likely to have a nearby track.
- Cluster timing: fake clusters from beam background are likely to be not in time with the primary collision.
- Number of innermost layers hit: hadronic clusters are likely to have a wider radius than electromagnetic clusters.

In total, 19 variables are used for this classifier.

In the ECL the most significant variables are as follows:

- Distance to the next track: neutral clusters should only rarely have a track close by.
- Energy in central nine crystals divided by the energy in the outer 21: the shape depends on whether it is a hadronic or electromagnetic cluster.
- Total energy deposition in the cluster: each $K_L^0$ deposits very little energy in the clusters, typically in the $< 50$ MeV range.

For this classifier, a total of 38 variables including shower shapes, ECL cluster Zernike-polynomials and kinematic variables were investigated.

*Performance of the classifiers.* The classifier performances are evaluated on $\Upsilon(4S) \to \bar{B}B$ events. The background rejection versus efficiency behaviour of the classifiers is depicted in Fig. 32. Fake rates and efficiencies of the classifications depend on the chosen working point (threshold). The optimal threshold value depends on the desired performance and type or size of background. The fake rates and efficiencies for the arbitrary threshold are depicted in Fig. 33. In general, the $K_L^0$ classification performance depends on the background level and composition, the magnetic field map, and the tracking performance.

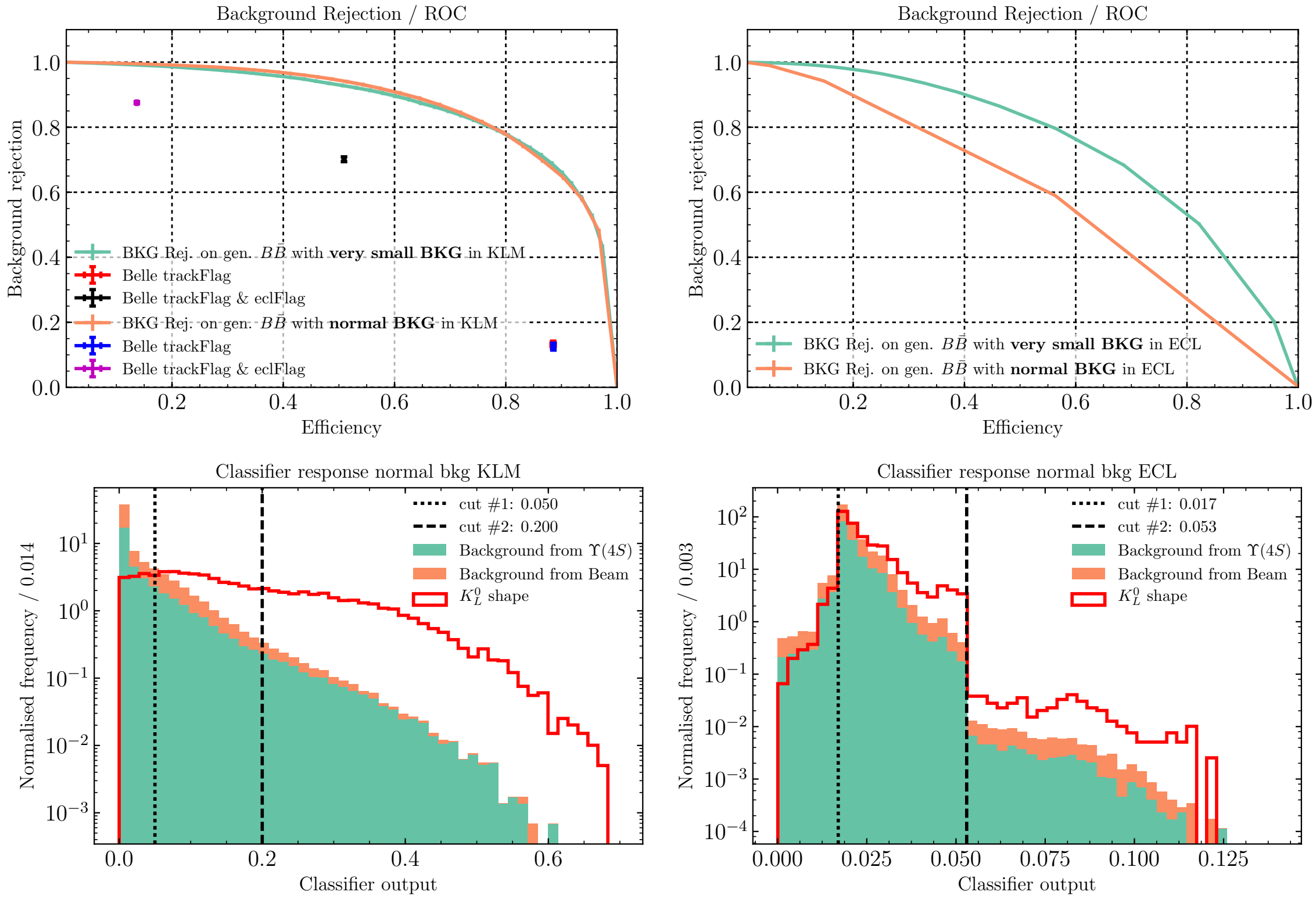

Fig. 32: a) Efficiency-background rejection behaviour of the $K_L^0$ KLM classifiers for nominal (orange) and small (green) background levels. For each background level, a classifier was trained and tested. The expected performance using the deprecated Belle algorithms for small and nominal background levels are depicted as data points. b) Efficiency-background rejection behaviour of the ECL classifiers for different levels of beam background. c) Output distribution of the KLM classifier. d) Output distribution of the ECL classifier. The vertical lines (cuts) displayed in the classifier outputs will be used for performance evaluation in Fig 33.

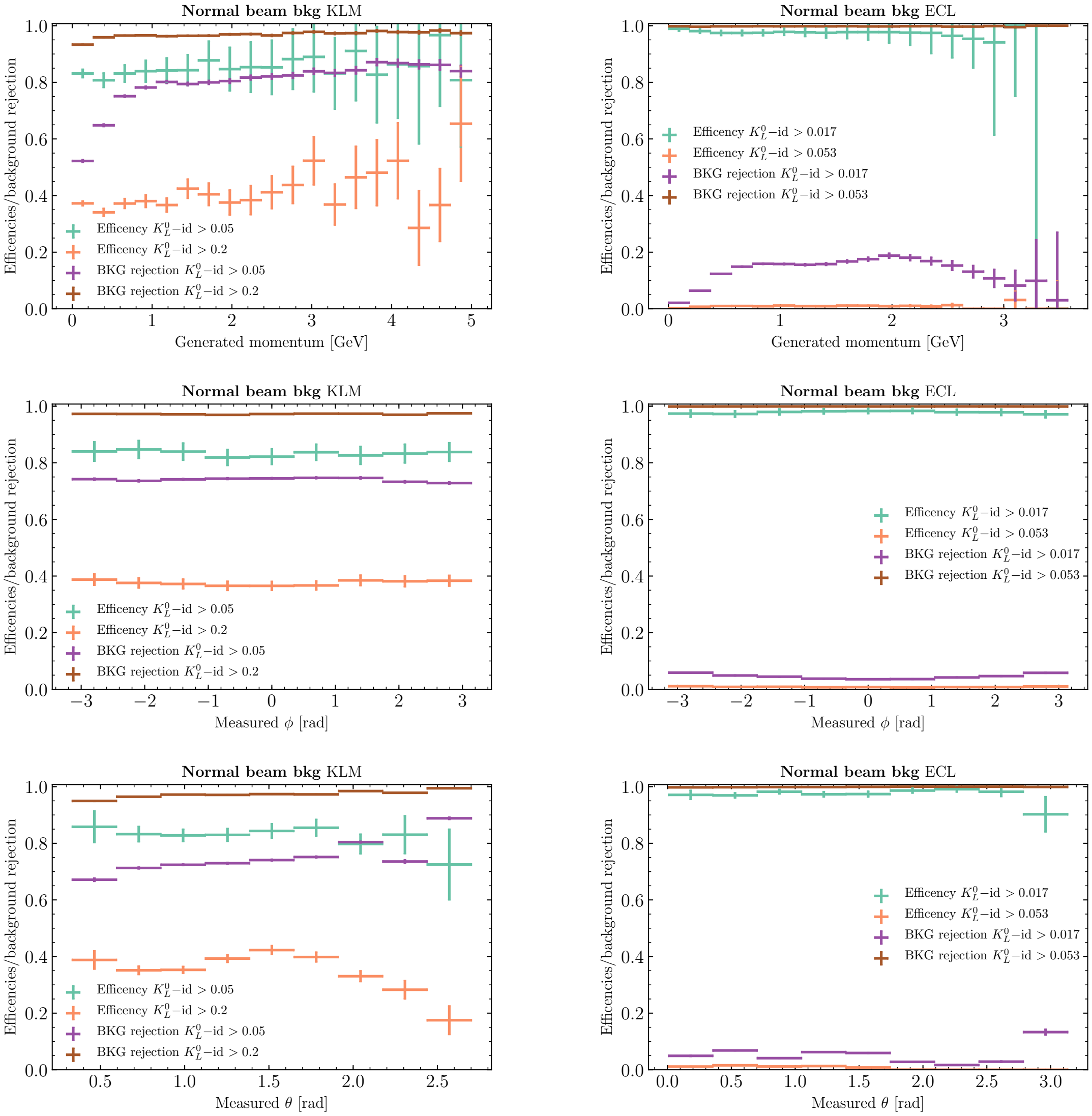

Fig. 33: $K_{\mathrm{L}}^0$ classification performance in KLM (left column) and ECL (right column): background rejection and efficiencies are depicted for arbitrary thresholds on the classifier output, measured in bins of generated momentum (first row), $\phi$ (second row) and $\theta$ (last row). The classifiers used were trained and evaluated on nominal beam background levels. The choice of the thresholds is depicted in Fig. 32.

# 6. Physics Analysis Software

*Section author(s): F. Abudinen, L. Li Gioi, P. Goldenzweig, T. Keck, F. Tenchini, D. Weyland, A. Zupanc.*

## 6.1. Introduction

The physics analysis software of Belle II makes use of the available data to perform efficient, reproducible measurements of physical quantities of interest; minimising both statistical and systematic uncertainties. The analysis software contains commonly used analysis tools allowing an easy, efficient and accurate data analysis. Here we focus on novel innovations on vertex fitting, $q\bar{q}$ continuum suppression, flavour tagging, and full $B$ reconstruction algorithms for missing energy analyses. For each of these algorithms we provide the respective expected performance at Belle II, and in some cases make comparisons to Belle.

## 6.2. Vertex reconstruction

Vertex reconstruction is the procedure by which the parameters of a decay vertex or interaction vertex are determined from the reconstructed parameters of the outgoing particles. It deals both with finding (pattern recognition) and with fitting the interaction vertices. It extracts the vertex position and recalculates the momentum and the invariant mass of the decaying particle, using the modified daughters' momenta after the vertex reconstruction. The decay length of an unstable particle inside a decay chain or the decay time difference, $\Delta t$, between the two $B$ mesons from an $\Upsilon(4S)$ decay can also be computed using a vertex fitter.

*6.2.1. Vertex finding algorithms.* The Belle II experiment has deployed three implementations of a vertex fit: KFitter, developed for the Belle experiment, RAVE [59], a standalone package originating from the CMS vertex fitting libraries, and TreeFitter [71], initially conceived by the BaBar collaboration. We use both KFitter and RAVE for kinematic fits and RAVE for geometric fits. TreeFitter is used for the fitting entire decay chains.

*Kinematic fits.* Kinematic fitting uses the known properties of a specific decay chain to improve the measurements of the process. Lagrangian multipliers are used in order to impose the kinematic constraints to the fit. Given the measurements, $\mathbf{q} = (q_1, ..., q_n)$, with a covariance matrix, $V$, and kinematic constraints, $\mathbf{h}(\mathbf{q})$, the function to be minimised in terms of the most suitable vertex is:

$$\chi^2 = (\mathbf{q} - \bar{\mathbf{q}})^T V^{-1}(\mathbf{q} - \bar{\mathbf{q}}) + +2\lambda^T(\boldsymbol{D}\delta\boldsymbol{y} + \boldsymbol{h}(boldsymbolq)) \tag{15}$$

where $\lambda$ is the Lagrange multiplier, $h(\bar{\mathbf{q}}) = 0$ and $\mathbf{D} = \partial\mathbf{h}/\partial\mathbf{y}$. Here $\bar{\mathbf{q}}$ represents the improved measurements, $\mathbf{d}$ is the kinematic constraint at the starting value and $\delta\mathbf{y}$is the difference between the improved measurement and the starting value.

*Adaptive Vertex Fit.* The RAVE libraries [59, 72], introduce the concept of soft assignment: a track is associated to a specific vertex with an assignment probability, or weight $w_i$ [72]:

$$w_i(\chi_i^2) = \frac{e^{-\chi_i^2/2T}}{e^{-\chi_i^2/2T} + e^{-\sigma_{cut}^2/2T}} \tag{16}$$

where $\chi_i^2$ is the square of the standardised residual, $\sigma_{cut}$ is defined as the standardised residual for which $w_i = 0.5$ and the temperature parameter $T$ defines the softness of the weight function.

The fitter is then implemented as an iterated, re-weighted Kalman filter [69]: in every iteration new track weights are computed and the vertex is estimated using these weights. This weight can be interpreted as a track-to-vertex assignment probability. Instead of minimising the least squares sum, as is expected from a Kalman fitting method, the algorithm minimises the weighted least squares sum. In order to avoid falling prematurely into local minima, a deterministic annealing schedule is introduced; in each iteration step the temperature parameter is lowered [72]:

$$T_i = 1 + r \cdot (T_{i-1} - 1), \tag{17}$$

where $T_i$ is the temperature parameter $T$ at iteration $i$ and $r$ denotes the annealing ratio. For convergence, $0 < r < 1$ is needed.

*Decay Chain Fitting.* The typical approach when reconstructing a decay chain is to start fitting final state particles, building the tree from the bottom up until the head of the decay is reached. This approach is generally valid, but may not always be optimal.

The TreeFitter module implements an alternate approach [71] where the decay tree is parameterised and fitted globally. This allows for the sharing of information across the tree, improving vertices that would otherwise be badly resolved or even impossible to fit without additional conditions such as mass constraints. This approach is especially useful to succesfully fit decay channels that are rich in neutral or missing particles and also provides the analyst with the full decay covariance matrix, which is beneficial for error treatment in time-dependent Dalitz analyses.

Since the whole tree is parameterised, a naive $\chi^2$ minimisation would naturally involve the inversion of large matrices, making the fit computationally very expensive. This problem is solved by applying the constraints to the fit individually using a Kalman gain formalism, which mitigates execution times down to a manageable level. The computational speed of TreeFitter then becomes comparable to KFitter when fitting individual vertices, and scales roughly quadratically with the complexity of the decay. When fitting a typical-sized decay tree, execution times are comparable to RAVE.

*6.2.2. Decay vertex.* Decay vertex positions are usually determined using kinematic vertex fits. As a benchmark for testing the Belle II vertex reconstruction performance, we use the decay vertex of the $J/\psi$ in the $B^0 \to J/\psi\, K_S^0$ decay mode. Figure 34 shows the fit residuals of the $z$ component of the fitted $J/\psi$ vertex position. A resolution of $26\,\mu$m is obtained. The same vertex fit performed using Belle MC returns a resolution of $46\,\mu$m, which is almost twice as large. The improvement seen is consistent with the expected improvement in the impact parameter resolution (Sec. 5.3.1) due to the Belle II Pixel Vertex Detector (PXD).

*6.2.3. Tag-side B vertex fitting.* To be sensitive to time-dependent $CP$ violating effects, the vertex resolution must be sufficient to resolve the oscillations of neutral $B$ mesons. The measurement of the distance between their decay vertices uses the relation

$$\Delta t \equiv t_{rec} - t_{tag} = \Delta l / \beta\gamma c \tag{18}$$

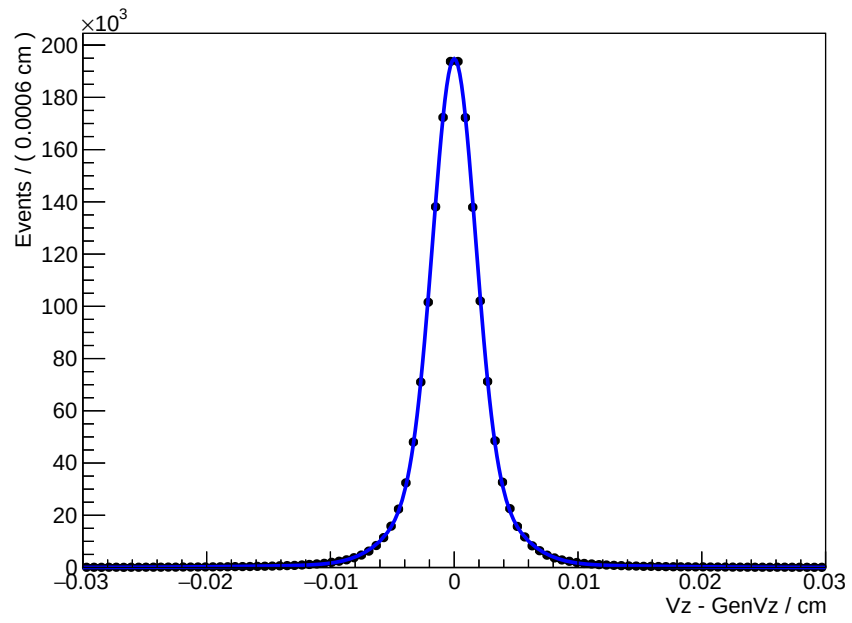

Fig. 34: $J/\psi \to \mu^+\mu^-$ vertex fit residuals: resolution $= 26\,\mu$m. The fit is performed using the sum of three Gaussian functions. The values of the shift and resolution are defined as the weighted averages of the mean values and the standard deviations of the three functions.

where $t_{rec}$ ($t_{tag}$) is the decay time of the fully reconstructed (tag) $B$ meson and $\Delta l$ is the distance between the two $B$ decay vertices in the boost direction. The largest contribution to the $\Delta t$ resolution comes from the tag $B$ vertex fit. In the decay tree of a $B$ meson we can divide the tracks in three groups: tracks originating from the $B$ decay, including the ones coming from decay vertices indistinguishable from the $B$ (*e.g.* $\mu^+$ and $\mu^-$ in $B^0 \to [J/\psi \to \mu^+\mu^-]K^0_S$); tracks originating from $D$ mesons; and tracks originating from $K^0_S$ decays. We fit the tag side $B$ using the RAVE Adaptive Vertex Fit algorithm. This takes as input all tracks with at least one hit in the PXD that are not used for the signal side $B$ candidate, except from the tracks attributed to $K^0_S$ candidates. In the case of a non-converging fit, tracks that do not have associated PXD hits are also used.

To reduce the weight of the tracks originating from $D$ mesons, we constrain the fit to a region defined by an ellipsoid around the boost direction (Fig. 35), where the $B$ is more likely to decay rather than $D$ mesons that carry more transverse momentum. This constraint acts as a weight in the final $\chi^2$.In the case of non-converging fit, the constraint is redefined, enlarging its size in the boost direction, becoming virtually equivalent to a cylinder.

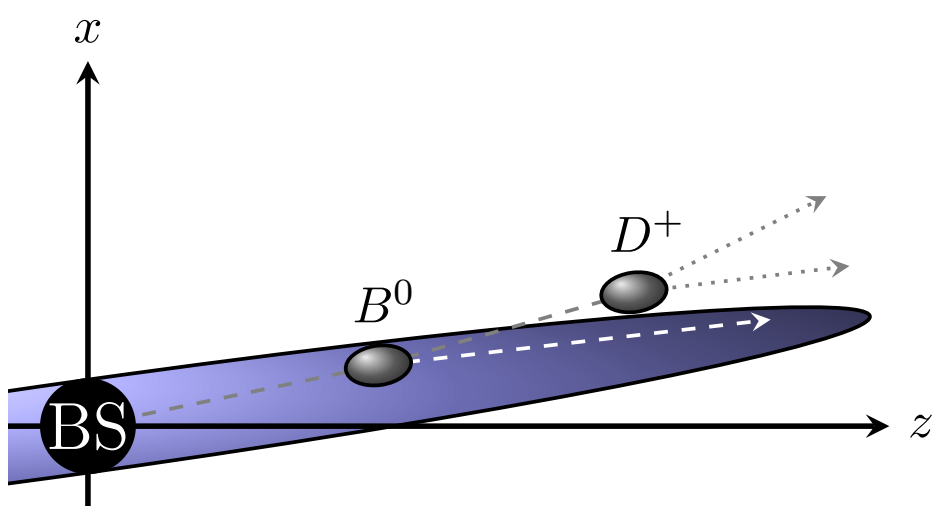

Fig. 35: Schematic representation of the tag $B$ vertex fit. The $B$ meson will have a higher vertex fit probability than the $D$ meson originating from the $B$. BS denotes the beam spot.

We obtain, for the tag side vertex fit of correctly reconstructed $B$ mesons, a bias of $6\,\mu$m and a resolution of $53\,\mu$m, independent of the signal $B$ decay mode. The total efficiency is 96%, constant as a function of $\Delta t$. Figure 36 (left) shows the residuals of the tag $B$ vertex fit of fully reconstructed $B^0 \to [J/\psi \to \mu^+\mu^-][K^0_S \to \pi^+\pi^-]$.

The sensitivity to the time-dependent $CP$ violating parameters of Eq. 301 in Sec. 10.1 strongly depends on the resolution of the $\Delta t$ measurement (Eq. 18). The lower boost of SuperKEKB, with respect to KEKB, produces an average distance between the two $B$ mesons of about 130 $\mu$m, 35% compared to 200 $\mu$m at KEKB. This makes it more difficult to resolve the decay vertices of the two $B$ mesons and it is one of the main motivations for the development of the Belle II PXD. The new hardware, together with the new vertex reconstruction algorithms, provides an improvement in the vertex resolution of both $B$ mesons. This translates to a $\Delta t$ resolution of 0.77 ps and a bias of $-0.03$ ps, which provides superior separation capability compared to Belle (resolution = 0.92 ps, bias = 0.2 ps), exceeding then the design requirements. Figure 36 (right) shows the $\Delta t$ residuals of $B^0 \to [J/\psi \to \mu^+\mu^-][K^0_S \to \pi^+\pi^-]$.

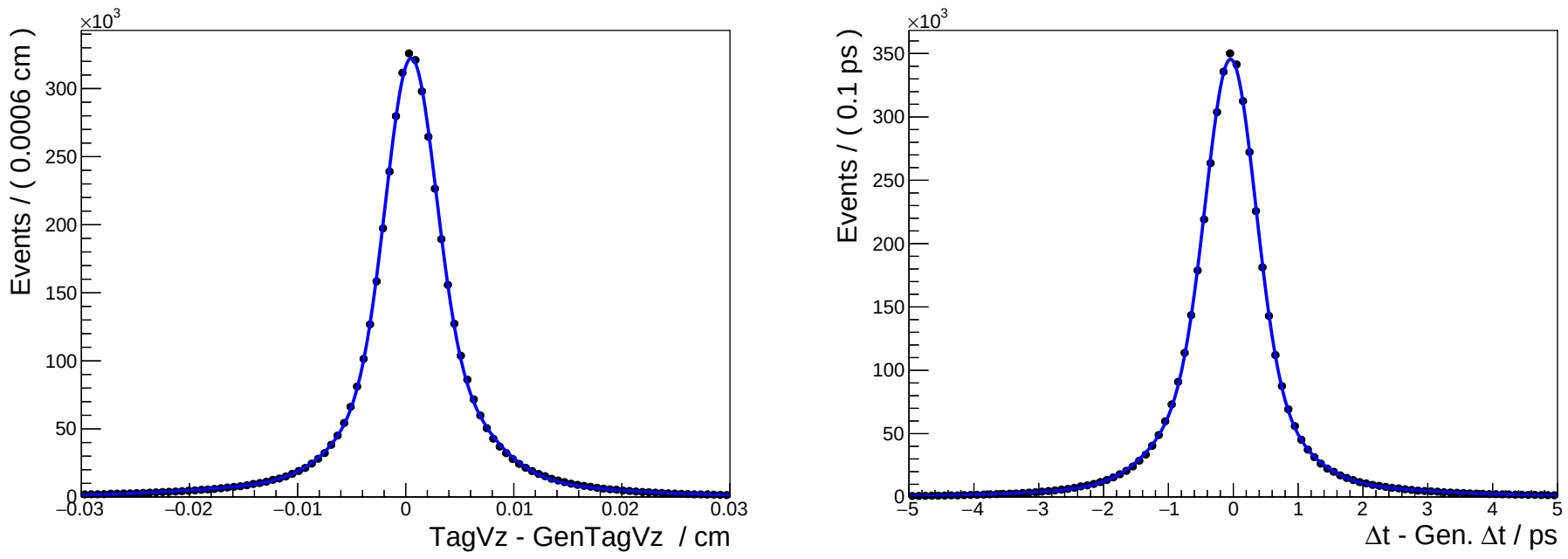

Fig. 36: The tag-side vertex fit residuals (left), with bias = 6 $\mu$m and resolution = 53 $\mu$m, and the $\Delta t$ residuals (right), with bias = $-0.03$ ps, resolution = 0.77 ps, for the fully reconstructed $B^0 \to [J/\psi \to \mu^+\mu^-][K^0_S \to \pi^+\pi^-]$. Both fits are performed using the sum of three Gaussian functions.

### *6.3. Composite Particle Reconstruction*

In the Belle II experiment, short-lived particles decaying at or near the interaction point (such as $B$ mesons or charm hadrons) cannot be measured directly by the sub-detectors, but instead must be reconstructed from the four-momenta of their long-lived decay products. Discriminating variables sensitive to composite particle properties can be subsequently built from final state information in order to perform background separation. A few such quantities are discussed in this section.

*6.3.1. Invariant Mass Resolution.* One of the simplest ways to reduce background, in particular that of the combinatorial kind, is to introduce selection criteria based on the invariant mass of intermediate particles in the decay chain.

The relative mass resolution achievable by Belle II can be estimated by performing a vertex fit of multiple resonances ($J/\psi$, $\psi(2S)$, $\Upsilon(1S)$ and $\Upsilon(2S)$) decaying into a di-muon state, as shown in Fig. 37. This takes advantage of the common kinematics shared by the final states. True muons from MC are selected from well-reconstructed tracks originating from the interaction region.

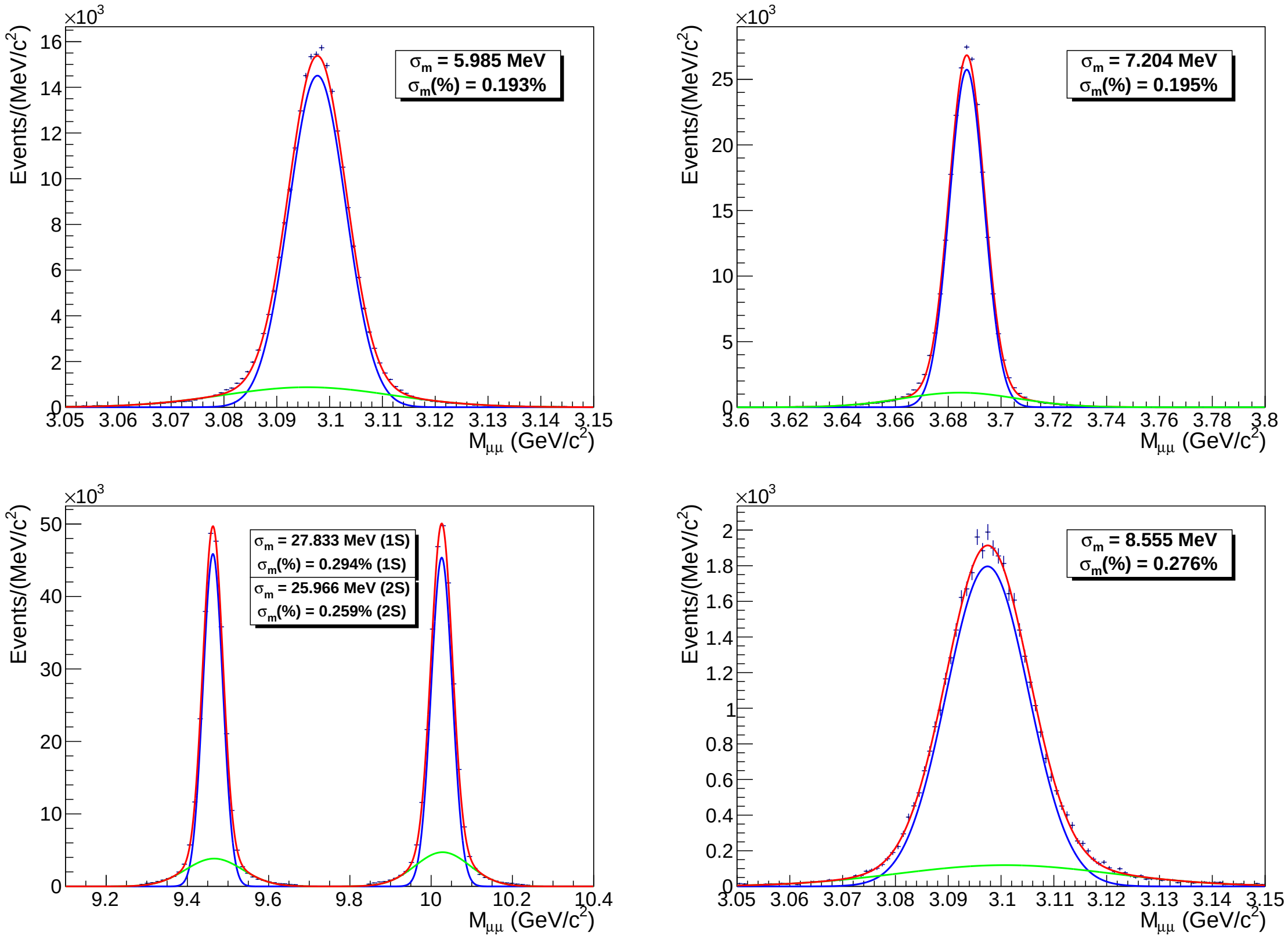

Fig. 37: Mass vertex fit residuals for di-muon resonances in Belle II: $J/\psi$ (*top left*), $\psi(2S)$ (*top right*), $\Upsilon(1S/2S)$ (*bottom left*) and Belle: $J/\psi$ (*bottom right*). The fit is performed using the sum of two Gaussian functions.

The resulting mass resolution is summarised as a function of mass, in Fig. 38. The projected resolution is $\sim 0.2\%$ for charmonium and $\lesssim 0.3\%$ for bottomonium resonances, with a 30% improvement compared to Belle. This largely originates from the improved transverse momentum, $p_{\mathrm{T}}$, resolution provided by the Belle II CDC.

*6.3.2. Beam-Constrained Fits.* When performing vertex reconstruction, knowledge of the production vertex can be used as an additional constraint to improve the fit resolution if the vertex is not well constrained. One example is the $D^{*+} \to D^0\pi^+$ decay, where the low momentum pion track is very sensitive to multiple scattering, while the $D^0$ vertex fit only provides a directional constraint. Requiring the $D^{*+}$ to originate from the beamspot substantially improves the vertex resolution.

We test the impact of this constraint by using the $D^{*+} \to [D^0 \to K^-\pi^+]\pi^+$ decay. The mass difference resolution is pictured in Fig. 39 with and without the beam constraint applied; the fit resolution is significantly improved. A comparison to a beam-constrained Belle fit of the same decay is provided, where the achievable mass resolution shows a 27% improvement.

In time-dependent analyses, good vertex resolution in the boost direction is key for the measurement of $CP$ parameters. Performing the vertex fit with a constraint parallel to the boost direction can improve the vertex resolution in all cases where the direction of the momentum cannot be precisely extracted from the track fit results. This type of fit is also needed when only one (pseudo-)track is available for the determination of the vertex. The RAVE fitter defines the IPTube constraint as an ellipsoid with a very long axis in the direction of the boost with the size of the beamspot in the orthogonal directions. Figure 40 shows the results of the $B^0$ vertex resolution in the boost direction of $B^0 \to \pi^0\pi^0$ with one of the two $\pi^0$ mesons decaying via the Dalitz channel $\pi^0 \to e^+e^-\gamma$. An improvement of 85 % in the vertex resolution is observed when using the IPTube constraint.

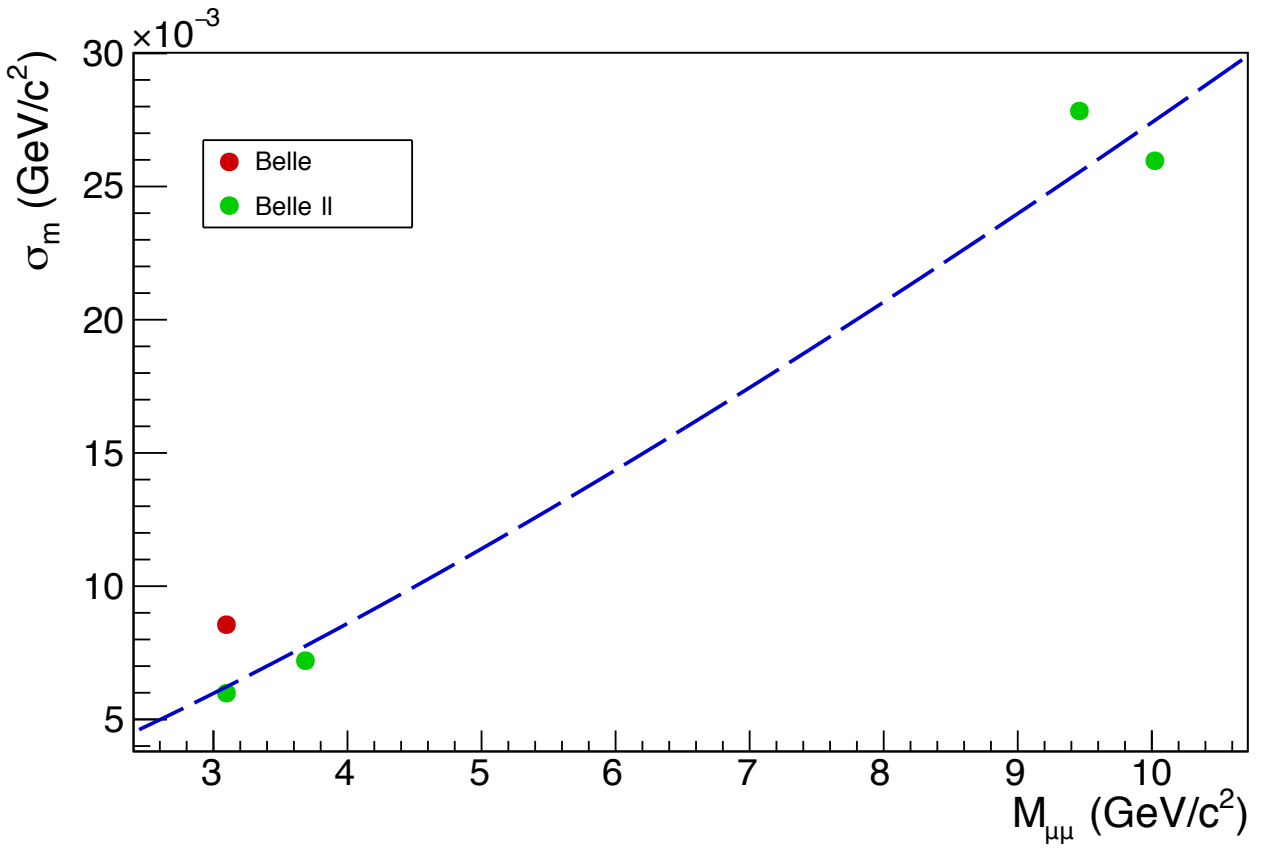

Fig. 38: Mass resolution as a function of resonant mass for Belle II (green) and Belle (red) MC. An empirical power law curve is fit through the points.

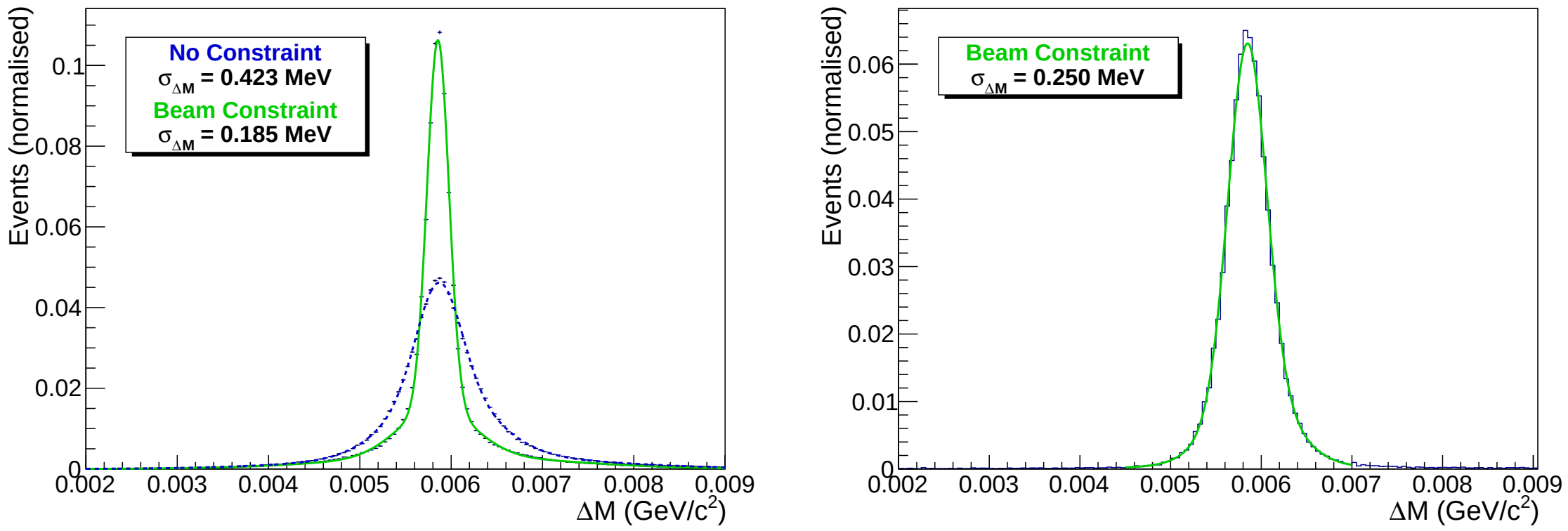

Fig. 39: Mass difference resolution for $D^{*+} \to [D^0 \to K^+\pi^-]\pi^+$ simulated events in Belle II (*left*) and Belle (*right*). The distribution for unconstrained vertices is fit are shown in blue while those with beamspot constraints are in shown in green.

*6.3.3. Beam-Constrained Observables.* Rather than using the $B$ invariant mass, $e^+e^-$ $B$ factories provide an additional energy- and mass-like constraint that can be used for background separation. $B$ mesons produced at the $\Upsilon(4S)$ resonance have well defined kinematics which are constrained by the mass of the $\Upsilon(4S)$ and by the properties of the beam.

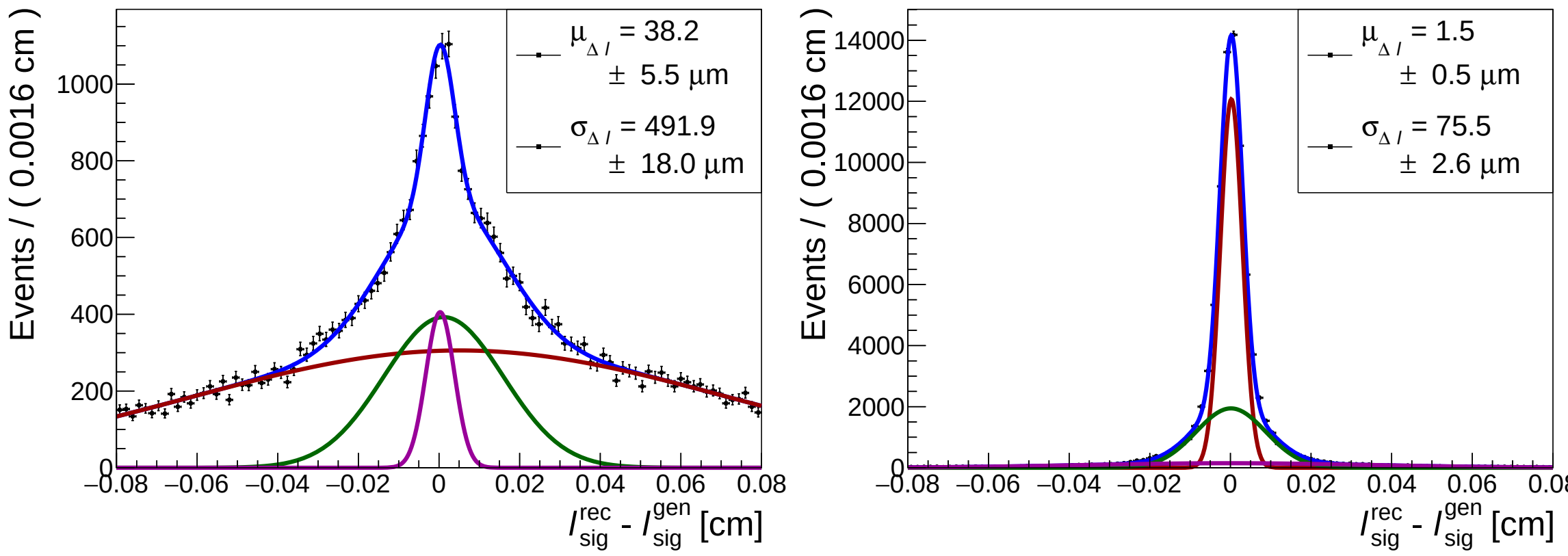

Fig. 40: $B^0$ vertex resolution in the boost direction of $B^0 \to \pi^0\pi^0$ with one of the two $\pi^0$ mesons decaying via the Dalitz channel $\pi^0 \to e^+e^-\gamma$ for the unconstrained fit (*left*), and IPTube constrained fit (*right*).

The reconstructed invariant mass of the single $B$ system must be equal to the nominal $B$ meson mass $m_B$, while the total energy should be equal to the beam energy when reconstructed in the rest frame, $E^*_{beam} = \sqrt{s}/2$. Two new variables can be defined using these constraints: the energy difference $\Delta E$ and the beam-energy constrained mass $M_{\rm bc}$. $\Delta E$ is defined as

$$\Delta E = E^*_B - E^*_{beam} = (2p^\mu_B p^\mu_{boost} - s)/2\sqrt{s}, \tag{19}$$

where $p^\mu_B$ and $p^\mu_{boost}$ are the four momenta of the $B$ meson and the $e^+e^-$ system, respectively. The beam-energy constrained mass is constructed by substituting the $B$ energy with the beam energy in the invariant mass calculation:

$$M_{\rm bc} = \sqrt{E^{*2}_{beam} - \mathbf{p}^{*\mathbf{2}}_{\mathbf{B}}}. \tag{20}$$

For a correctly reconstructed $B$ meson decay, the true values would be $\Delta E = 0$ and $M_{\rm bc} = m_B$. Figure 41 shows the distribution for these observables for two sample decay channels: $B^+ \to [D^0 \to K^-\pi^+]\pi^+$ and $B^+ \to [D^0 \to K^-\pi^+\pi^0]\pi^+$ for both Belle and Belle II MC. As we can see, the performance in channels such as the $D^0 \to K^-\pi^+$ is comparable between the two experiments. On the other hand, modes with neutral pions such as $D^0 \to K^-\pi^+\pi^0$ show a significant improvement in the latest versions of the Belle II software, owing to lower bias on the photon position determination (see Sec. 5.4). The improvement on the distribution core is $\sim 20\%$ on $M_{\rm bc}$ and $\sim 50\%$ on $\Delta E$ for this particular channel. Improvements in neutral particle reconstruction directly impacts reconstruction in modes where they are present.

## 6.4. *Continuum Suppression*

The dominant source of combinatorial background in $B$ physics analyses at Belle II comes from $e^+e^- \to q\bar{q}$ ($q = u, d, s, c$) events, where random combinations of particles in the final state may mimic the kinematic signatures of the signal. These events are referred to as "continuum background". In charmless $B$ decay ($b \to u, s$) channels, combinatorial background from continuum events is often the dominant source of background. Continuum suppression is also important in controlling potential systematic uncertainties in precision measurements of charm $b \to c$ decay modes.

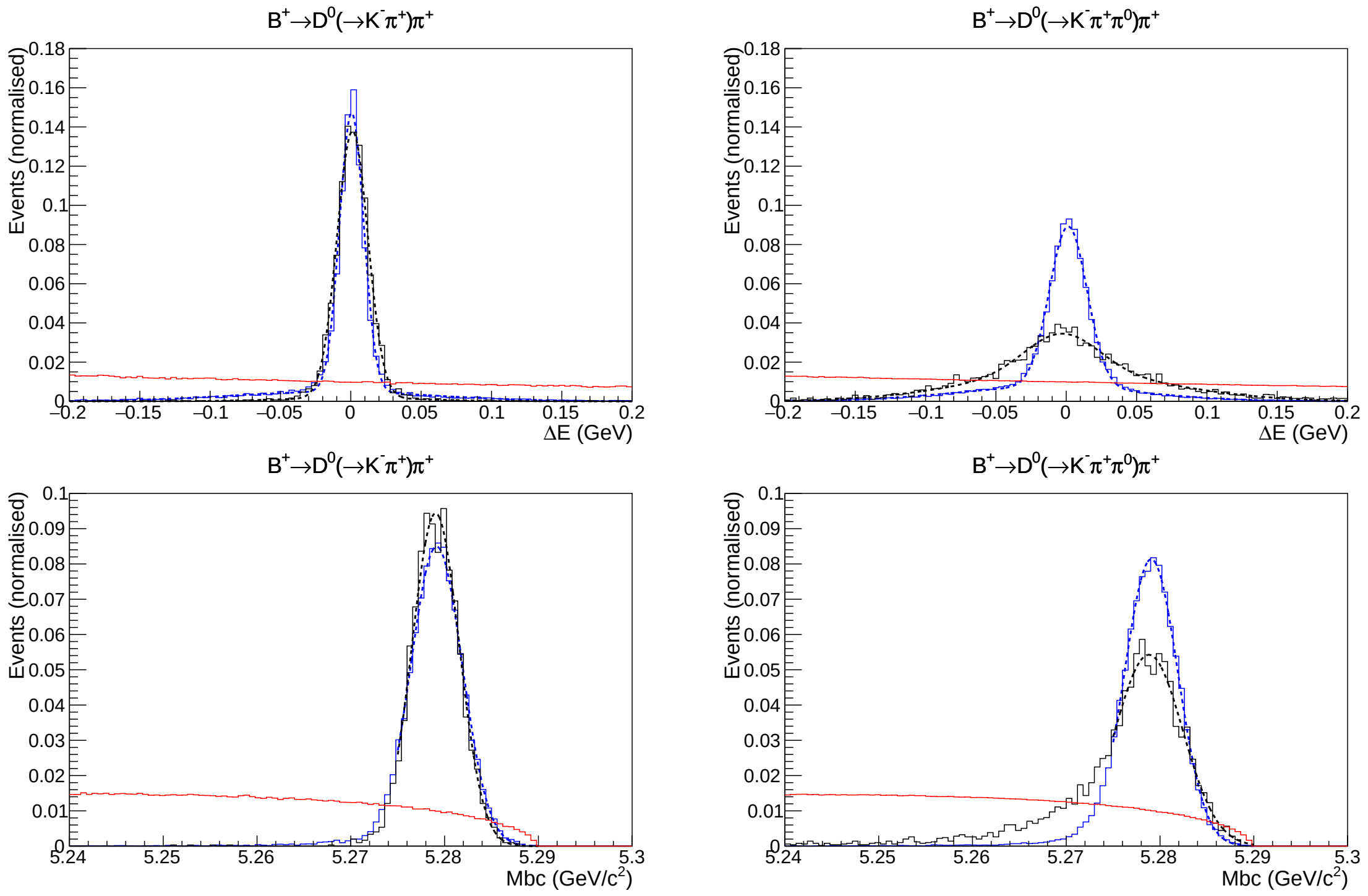

Fig. 41: $\Delta E$ and $M_{\rm bc}$ distributions for $B^+ \to [D^0 \to K^-\pi^+(\pi^0)]\pi^+$ simulated events in Belle II (*blue*) in comparison with Belle (*black*). The red curve shows the distribution for continuum background. Beam background is included in the simulation.

The Physics of the $B$ Factories book provides a comprehensive review of the variables and methods employed by the Belle and BaBar collaborations to suppress continuum background [2]. While the definition of many of the variables used in Belle II are identical to those in Belle, the implementation often differs. For example, in Belle II we introduce deep learning techniques that use detector-level (DL) information as input for classification. In the following subsections, we first briefly review the traditional, engineered variables (E) used in continuum suppression in both Belle and Belle II, as described in [2]. We then introduce the DL variables, and compare the performance of various configurations of input variables and classifier algorithms (hereby referred to as classifiers), namely Boosted Decision Trees (BDTs) and Deep Neural Networks (DNNs) [7]. All results are obtained using MC simulated $B^0 \to K^0_S\pi^0$ events for signal and MC simulated continuum events for background.

### *6.4.1. Engineered variables.*

- ***B* meson direction:** Using the angle $\theta_B$ between the reconstructed momentum of the $B$ candidate (computed in the $\Upsilon(4S)$ reference frame) and the beam axis, the variable $|\cos\theta_B|$ allows one to discriminate between signal $B$ decays and the $B$ candidates from continuum background. The spin-1 $\Upsilon(4S)$ decaying into two spin-0 $B$ mesons results in a $\sin^2\theta_B$ angular distribution with respect to the beam axis; in contrast for $e^+e^- \to f\bar{f}$ events, the spin-1/2 fermions $f\bar{f}$ and the two resulting jet-like structures, are distributed according to a $1 + \cos^2\theta_B$ distribution.

[7] For a comprehensive study, see [73].

- **Thrust:** For a collection of $N$ momenta $\vec{p}_i$ ($i = 1, \cdots N$), the thrust axis $\vec{T}$ is defined as the unit vector along which their total projection is maximal. The thrust scalar $T$ (or thrust) is defined as

$$T = \frac{\sum_{i=1}^{N} \left| \vec{T} \cdot \vec{p}_i \right|}{\sum_{i=1}^{N} |\vec{p}_i|}. \tag{21}$$

The thrust of both the $B$ ($T_B$) and the rest of the event (ROE) ($T_{\mathrm{ROE}}$) provide discrimination between signal and continuum background.

- **Thrust angles:** A useful related variable is $|\cos\theta_{\mathrm{T}}|$, where $\theta_{\mathrm{T}}$ is the angle between the thrust axis of the momenta of the $B$ candidate decay's final state particles (all evaluated in the $\Upsilon(4S)$ rest frame), and the thrust axis of the ROE. For a $B\bar{B}$ event, both $B$ mesons are produced almost at rest in the $\Upsilon(4S)$ rest frame, so their decay particles are isotropically distributed and their thrust axes are randomly distributed. Therefore, $|\cos\theta_{\mathrm{T}}|$ follows a uniform distribution in the range $[0, 1]$. In contrast, for $q\bar{q}$ events, the momenta of particles follow the direction of the jet-like structures in the event, and as a consequence the thrusts of both the $B$ candidate and the ROE are strongly directional and collimated, yielding a $|\cos\theta_{\mathrm{T}}|$ distribution strongly peaked at large values.

  Another thrust-related variable is $\theta_{T,B}$, the angle between the thrust axis of the $B$ decay particles and the beam axis; for $B\bar{B}$ events, $|\cos\theta_{\mathrm{T,B}}|$ is uniformly distributed, while for continuum events, the thrust of particle momenta from the $B$ candidate tends to follow a $1 + \cos^2\theta_{\mathrm{T,B}}$ distribution followed by the jet-like structures.

- **CLEO Cones:** The CLEO collaboration introduced variables based on the sum of the absolute values of the momenta of all particles within angular sectors around the thrust axis in intervals of 10°, resulting in 9 concentric cones [74]. The cone in the direction of the thrust axis is merged with the respective cone in the opposite direction. There are two options for constructing the CLEO cones in Belle II: they can be calculated from all final state particles in the event, or from only ROE particles.

- **Fox-Wolfram moments:** For a collection of $N$ particles with momenta $p_i$, the $l^{\mathrm{th}}$ order Fox-Wolfram moment $H_l$ [75] is defined as

$$H_l = \sum_{i,j}^{N} |p_i|\,|p_j|\, P_l\left(\cos\theta_{i,j}\right), \tag{22}$$

where $\theta_{i,j}$ is the angle between $p_i$ and $p_j$, and $P_l$ is the $l^{\mathrm{th}}$ order Legendre polynomial. In the limit of vanishing particle masses, $H_0 = 1$, which is why the normalised ratio $R_l = H_l/H_0$ is often used, so that for events with two strongly collimated jet-like structures, $R_l$ takes values close to zero (one) for odd (even) values of $l$. These sharp signatures provide a convenient discrimination between events with different topologies. The variable $R_2$ is a strongly discriminating variable, commonly used in $B$ physics analyses.

- **Modified Fox-Wolfram moments:** The Belle collaboration developed the modified Fox-Wolfram moments $H^{so}_{xl}$ and $H^{oo}_{l}$ ($l \in [0, 4]$), where all reconstructed particles in an event are divided into two categories: $B$ candidate daughters (denoted as $s$) and particles from the ROE (denoted as $o$). The $H^{so}_{xl}$ moments are decomposed into an additional three categories (denoted as $x$) depending on whether the particle is charged ($x = c$), neutral ($x = n$), or missing ($x = m$). Additionally, for $H^{so}_{xl}$, the missing momentum of an event is treated as an additional particle. For even $l$,

$$H^{so}_{xl} \;\; = \;\; \sum_{i}\sum_{jx} |p_{jx}| P_l(\cos\theta_{i,jx}), \tag{23}$$

where $i$ runs over the $B$ daughters; $jx$ runs over the ROE in the category $x$; $p_{jx}$ is the momentum of particle $jx$; and $P_l(\cos\theta_{i,jx})$ is the $l^{\rm th}$ order Legendre polynomial of the cosine of the angle between particles $i$ and $jx$. For odd $l$, we have $H^{so}_{nl} = H^{so}_{ml} = 0$ and

$$H^{so}_{\rm cl} = \sum_{i}\sum_{jx} Q_i Q_{jx} |p_{jx}| P_l(\cos\theta_{i,jx}), \tag{24}$$

where $Q_i$ and $Q_{jx}$ are the charges of particle $i$ and $jx$, respectively. There are a total of eleven $H^{so}_{xl}$ moments: two for $l = 1, 3$ and nine $(3 \times 3)$ for $l = 0, 2, 4$.

The definition of the five $H^{oo}_{l}$ moments are as follows:

$$H^{oo}_{l} = \begin{cases} \sum_j \sum_k |p_j||p_k| P_l(\cos\theta_{j,k}) & (l = \text{even}) \\ \sum_j \sum_k Q_j Q_k |p_j||p_k| P_l(\cos\theta_{j,k}) & (l = \text{odd}), \end{cases}$$

where $j$ and $k$ run over the ROE and the other variables are the same as in Eq. (24). The $H^{so}_{xl}$ and $H^{oo}_{l}$ moments are normalized to $H^{max}_{0}$ and $(H^{max}_{0})^2$, respectively, where $H^{max}_{0} = 2\,(E^*_{\rm beam} - \Delta E)$, to not depend on $\Delta E$.

There are two options for constructing the modified Fox-Wolfram moments: they can be calculated from the $B$ primary daughters, or from the final state particles from the $B$ decay. However, the latter is rarely employed due to potential biases introduced with the method.

- **Missing mass and transverse energy:** The missing mass squared is defined as

$$M^2_{miss} = \left( E_{\Upsilon(4S)} - \sum_{n=1}^{N_t} E_n \right)^2 - \sum_{n=1}^{N_t} |p_n|^2, \tag{25}$$

where $E_{\Upsilon(4S)}$ is the energy of $\Upsilon(4S)$, $E_n$ and $p_n$ are the energy and momentum of the reconstructed particle $n$ respectively, and $N_t$ is the total number of reconstructed final state particles. The transverse energy is the scalar sum of the transverse momentum of each particle $\sum_{n=1}^{N_t} |(P_{\rm t})_n|$.

- **Vertex separation of $B$ and $\overline{B}$:** Due to the relatively long lifetime of $B$ mesons with respect to strongly decaying lighter mesons, they have a longer average flight distance due to the boost between the $\Upsilon(4S)$ and laboratory systems. The quantity $\Delta z = z_{B_{\rm sig}} - z_{B_{\rm tag}}$ (the distance in the beam direction between the $B$ vertex, $z_{B_{\rm sig}}$, and the vertex from the ROE, $z_{B_{\rm tag}}$) has a broader distribution for $B\bar{B}$ events than for continuum events, and can be used to suppress continuum in analyses.

- **Flavour tagging:** The flavour tagging algorithm (Section 6.5) returns the flavour of the tagged meson, $q(=\pm 1)$, along with a flavour-tagging quality factor, $r$, which ranges from zero for no flavour discrimination to one for unambiguous flavour assignment. For signal events, $q$ is usually consistent with the flavour opposite to that of the signal $B$, while it is random for continuum events.

*6.4.2. Detector-level variables.* In contrast to the engineered variables, which represent the whole event, the detector-level variables are built on a track and calorimeter cluster basis. There are twelve variables used for tracks and ten variables used for calorimeter clusters. In addition to these variables, the classifier also contains information relating to the charge and to whether or not the cluster or track belong to the ROE.

- **Momentum (clusters and tracks)**: The momentum variables include the magnitude $p$, the azimuth angle $\phi$ and the cosine of the polar angle $\cos\theta$, as well as their uncertainties. Instead of using the center-of-mass system, the $z$-axis is also rotated to the thrust axis of the $B$ candidate. This rotated coordinate system is inspired by the CLEO Cones and is called the thrust frame.
- **Calorimeter cluster specific variables**: There are four variables: number of hits, timing, E9E21, and region. The number of hits of the cluster is employed as well as its timing, which is used to tell if the cluster occurred at the same time as the event. E9E21 is a ratio of the energy between the inner 9 and outer 21 cells of a calorimeter cluster. The region here refers to the ECL region in which the cluster was detected (forward, barrel, or backward).
- **Track specific variables**: The track specific variables include particle probabilities for kaons, electrons, muons and protons. The $\chi^2$ probability of the track fit is used, as well as the number of CDC hits.
- **Vertex variables (V)**: The use of vertex variables (available for tracks) adds discriminating power, but can create unwanted correlations between the classifier output and $\Delta z$. Due to these correlations, the vertex variables cannot be used in every analysis and are therefore treated separately. When used, they are calculated in the same thrust frame as the momentum variables.

*6.4.3. Variable sets.* Three different input variable sets are used for comparison. The engineered (E) variables are the variables used in the traditional approach and therefore serve as the baseline set containing 30 variables. In the second set, the variables are complemented by the detector-level variables without vertex information (E+DL) resulting in 470 variables. The vertex variables are added in the last set (E+DL+V) for a total of 530 variables.

*6.4.4. Hyper parameters.* Selecting the optimal hyper parameters is crucial to a classifier. To evaluate different hyper parameter sets, typically a statistical uncertainty based figure of merit is needed.

*6.4.5. Comparison between traditional and deep learning methods.* The comparison is done with two classifiers, each trained on three different variable sets. The first approach BDT (E), uses the FastBDT algorithm trained on the engineered continuum suppression variables often used at Belle II [70]. Therefore, it serves as the baseline. For the deep neural networks we use TensorFlow. [8].

[8] Comparisons with relational networks and adversarial networks can be found in [73].

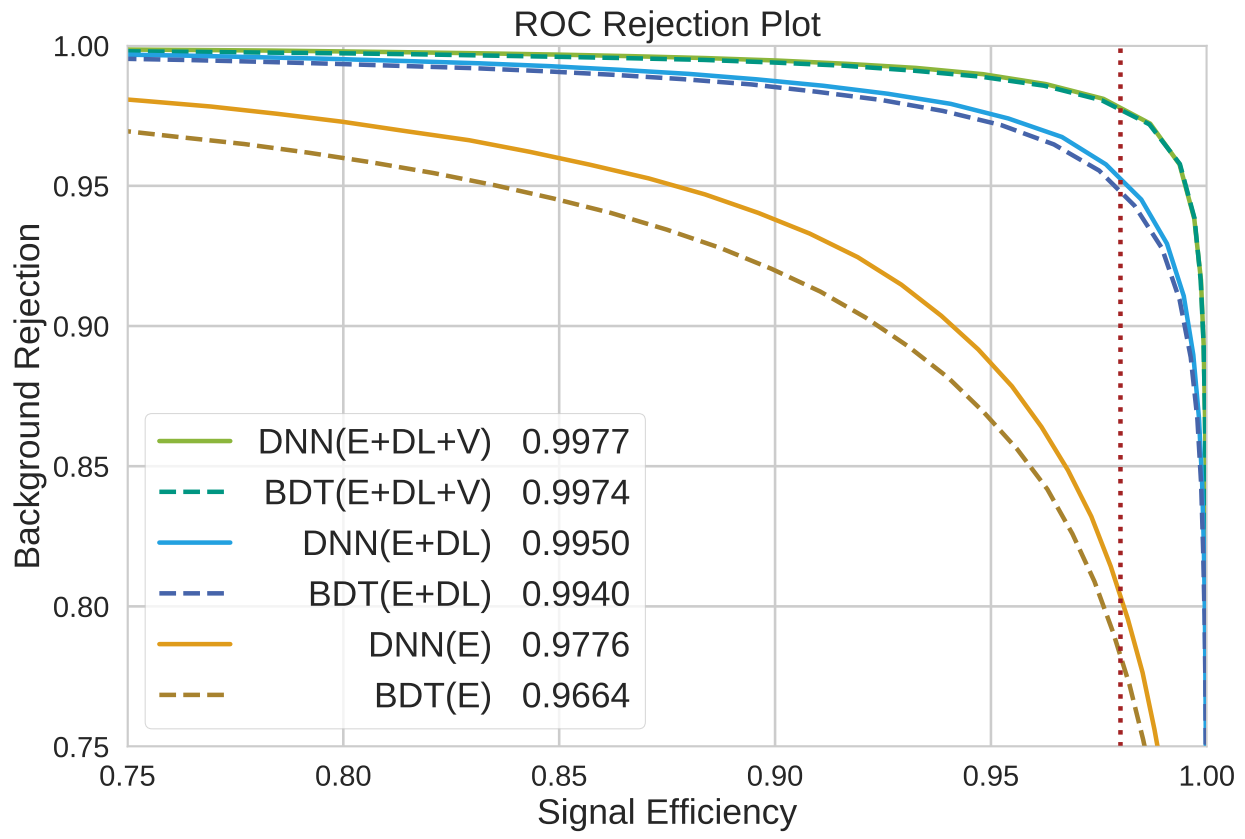

Fig. 42: ROC curve of the BDTs and DNNs for each variable set. Each training is performed five times and the best result is used for this plot. The corresponding AUC score is listed in the legend. The 98% signal efficiency cut used for Table 22 is shown in red.

We introduce a metric to compare classifiers relative to the baseline BDT (E), RB(98), the relative amount of background remaining after a 98% signal efficiency criterion. The amount of background remaining after this criterion is calculated relative to the amount of background of the baseline procedure. As an example, an RB(98) of 60% means that the user can expect only 60% of the background compared to the baseline classifier, losing 2% of signal in both cases.

The training time is chosen as the last metric for comparison. This should only serve as an approximation, as training time is hardware dependent and BDTs and DNNs are trained on different hardware. (For the training of the DNNs, GPUs were used to speed up the training, while for BDTs, only CPU-based training is possible.)

*6.4.6. Performance.* The ROC curves and their area-under-the-curve (AUC) scores are shown in Fig. 42. With each additional variable set, the classification result significantly improves, though aside from the first variable set E, there are only small differences between BDTs and DNNs.

The RB(98) scores and the training times are shown in Table 22. The RB(98) scores further confirm the large increase in classification capability using the new variable sets. With the variable set E+DL, the amount of background is only around 20%, relative to the amount from the BDT (E) approach. When including the vertex variables, the background is reduced by an additional factor of two, and is only 10% of the amount remaining in the BDT (E) approach. The largest difference between DNNs and BDTs are in the training times.

The discriminating variables described above are combined to a continuum suppression discriminant $C_{CS}$ using any of the methods described above. In the majority of analyses, a loose threshold is initially applied to the $C_{CS}$ to reject continuum events with minimal loss of signal, and the remaining events are transformed to a new observable

$$C'_{CS} = \log\left(\frac{C_{CS} - \text{cut}}{1 - C_{CS}}\right), \tag{26}$$

Table 22: The relative amount of background on a 98% signal efficiency cut (RB(98)) and the time needed for training for each classifier shown in Fig. 42. As a baseline, the traditional approach BDT based on the first variable set is chosen.

| Classifier | RB(98) in % | Training Time in h:min |
|---|---|---|
| DNN (E+DL+V) | 9.81 | 1:01 |
| BDT (E+DL+V) | 10.12 | 26:26 |
| DNN (E+DL) | 21.65 | 0:33 |
| BDT (E+DL) | 23.24 | 25:42 |
| DNN (E) | 90.35 | 0:54 |
| BDT (E) | 100 | 1:39 |

which has a Gaussian-like shape that can be described analytically by a single or double Gaussian function.

It should be noted that given the large number of input parameters in the most advanced configurations of the classifiers, correlations can both be very high between input parameters, and between the classifier itself and signal observables. The depth and complexity of the continuum suppression classifiers employed in analyses will need to be chosen judiciously and their application will be analysis dependent. Adversarial network approaches to penalise approaches for high correlations to signal observables may be considered in the future.

### *6.5. Flavour Tagger*

Flavour tagging is required for measurements of $B$-meson mixing and for measurements of $CP$ violation, where the decay of a neutral $B$ meson ($B^0_{\rm sig}$) is fully reconstructed and the flavour of the accompanying $B^0$ meson ($B^0_{\rm tag}$) has to be determined. The task of the flavour tagger is to determine the flavour of $B^0_{\rm tag}$ at the time of its decay. At the $\Upsilon(4S)$ resonance, $B$ mesons pairs are produced in isolation, *i.e.* without additional particles. Therefore, reconstructed tracks and neutral ECL and KLM clusters remaining after the full reconstruction of $B^0_{\rm sig}$ can be assumed to a good approximation to belong to the decay of $B_{\rm tag}$.

$B$ mesons exhibit a large number of possible decay channels. Many of them provide unambiguous flavour signatures through flavour-specific final states. Flavour signatures correspond to signed characteristics of the decay products that are correlated with the charge sign of the $b$-quark in the $B$ meson. Because of the wide range of possible decay channels it is not feasible to fully reconstruct a large fraction of flavour-specific $B_{\rm tag}$ decays. Instead of a full reconstruction, the flavour tagger applies inclusive techniques based on multivariate methods to maximally exploit the information provided by the different flavour-specific signatures in flavour-specific decays.

The Belle II flavour tagger has been developed by adopting several useful concepts of previous algorithms used by the Belle and the BaBar collaborations [2].

*6.5.1. Definitions.* Given a total number of events $N$, the efficiency $\varepsilon$ is defined as the fraction of events to which the flavour tagging algorithm can assign a flavour tag, *i.e.*

$$\varepsilon = \frac{N^{\text{tag}}}{N}, \tag{27}$$

where $N^{\text{tag}}$ is the number of tagged events. The fraction of wrong identifications over the number of tagged events is denoted by $w$. Thus, the number of tagged $B^0$ and $\overline{B}^0$ events is given by

$$\begin{aligned} N_{B^0}^{\text{tag}} &= \varepsilon(1-w)N_{B^0} + \varepsilon w N_{\overline{B}^0} \\ N_{\overline{B}^0}^{\text{tag}} &= \varepsilon(1-w)N_{\overline{B}^0} + \varepsilon w N_{B^0}, \end{aligned} \tag{28}$$

where $N_{B^0}$ and $N_{\overline{B}^0}$ are the true number of $B^0$ and $\overline{B}^0$ mesons on the tag side, respectively. The observed $CP$ violation asymmetry is then

$$a_{CP}^{\text{obs}} = \frac{N_{B^0}^{\text{tag}} - N_{\overline{B}^0}^{\text{tag}}}{N_{B^0}^{\text{tag}} + N_{\overline{B}^0}^{\text{tag}}} = (1-2w) \cdot \frac{N_{B^0} - N_{\overline{B}^0}}{N_{B^0} + N_{\overline{B}^0}} = (1-2w) \cdot a_{CP}, \tag{29}$$

where $a_{CP}$ corresponds to the $CP$ violation asymmetry measured in $CP$ violation analyses (Eq. 301 in Sec. 10.1). To reduce systematic uncertainties, the value of $w$ should be precisely measured. The size of the observed $CP$ asymmetry is proportional to $|1-2w|$, *i.e.* the $CP$ asymmetry becomes "diluted" due to the wrong tag fraction. The so-called dilution factor is defined as

$$r \equiv |1-2w|, \tag{30}$$

where $r = 0$ means no flavour information ($w = 0.5$) and $r = 1$ corresponds to an unambiguous tag ($w = 0, 1$). The statistical uncertainty of $a_{CP}$ is

$$\delta a_{CP} = \frac{\delta a_{CP}^{\text{obs}}}{1-2w}. \tag{31}$$

Assuming that $a_{CP}^{\text{obs}}$ is small, *i.e.* $N_{B^0}^{\text{tag}} \approx N_{\overline{B}^0}^{\text{tag}}$, one obtains for the statistical uncertainty of $a_{CP}^{\text{obs}}$

$$\delta a_{CP}^{\text{obs}} \overset{N_{B^0}^{\text{tag}} \approx N_{\overline{B}^0}^{\text{tag}}}{=} \frac{1}{\sqrt{N^{\text{tag}}}}. \tag{32}$$

Thus, one finds that

$$\delta a_{CP} = \frac{1}{\sqrt{N^{\text{tag}}}(1-2w)}. \tag{33}$$

The effective tagging efficiency $\varepsilon_{\text{eff}}$ of a flavour tagging algorithm is defined such that the statistical uncertainty on the measured asymmetry $a_{CP}$ is related to the effective number of tagged events$N^{\text{eff}}$ by $1/\sqrt{N^{\text{eff}}} = 1/\sqrt{\varepsilon_{\text{eff}} \cdot N}$. Comparing this with Eq. 33, one obtains

$$\varepsilon_{\text{eff}} = \frac{N^{\text{tag}}}{N} \cdot (1-2w)^2 = \varepsilon \cdot r^2. \tag{34}$$

The maximisation of the effective efficiency results in a minimisation of the statistical uncertainty. In general, the scaling of $\delta a_{CP}$ with $\varepsilon_{\text{eff}}$ is only approximate. For a likelihood-based analysis, the expected statistical uncertainty of an estimated $CP$ or mixing asymmetry can be obtained with a maximum-likelihood estimator (Sec. 10.1).

Up to now, $w$ and $\varepsilon$ have been considered to be equal for $q = +1(-1)$. However, a slight difference can arise as a result of a charge-asymmetric detector performance. To take this effect into account, one redefines

$$\varepsilon = \frac{\varepsilon_{B^0} + \varepsilon_{\overline{B}^0}}{2}, \qquad w = \frac{w_{B^0} + w_{\overline{B}^0}}{2}, \tag{35}$$

and introduces the differences

$$\Delta\varepsilon = \varepsilon_{B^0} - \varepsilon_{\overline{B}^0}, \qquad \Delta w = w_{B^0} - w_{\overline{B}^0}, \tag{36}$$

where the index corresponds to the true flavour, *e.g.* $w_{B^0}$ is the fraction of true $B^0$ mesons that were wrongly classified as $\overline{B}^0$.

*6.5.2. Tagging Categories.* The flavour tagger is based on flavour-specific decay modes with relatively high branching fractions ( $\gtrsim 2$ %). Each decay mode exhibits a particular decay topology and provides a flavour-specific signature. Some additional signatures are obtained by combining similar or complementary decay modes. Within a so-called category, a particular flavour signature is considered separately. The current flavour tagger is based on 13 categories which are presented in Table 23.

Table 23: Tagging categories and their targets (left) with some characteristic examples of the considered decay modes (right). "P*" stands for momentum in the centre-of-mass frame and $l^{\pm}$ for charged leptons ($\mu$ or $e$).

| Categories | Targets for $\overline{B}^0$ |
|---|---|
| Electron | $e^-$ |
| Intermediate Electron | $e^+$ |
| Muon | $\mu^-$ |
| Intermediate Muon | $\mu^+$ |
| Kinetic Lepton | $l^-$ |
| Intermediate Kinetic Lepton | $l^+$ |
| Kaon | $K^-$ |
| Kaon-Pion | $K^-$, $\pi^+$ |
| Slow Pion | $\pi^+$ |
| Maximum P* | $l^-$, $\pi^-$ |
| Fast-Slow-Correlated (FSC) | $l^-$, $\pi^+$ |
| Fast Hadron | $\pi^-$, $K^-$ |
| Lambda | $\Lambda$ |

Underlying decay modes

$\overline{B}^0 \to D^{*+}\ \overline{\nu}_\ell\ \ell^-$
$\quad\hookrightarrow D^0\ \pi^+$
$\qquad\hookrightarrow X\ K^-$

$\overline{B}^0 \to D^+\ \pi^-\ (K^-)$
$\quad\hookrightarrow K^0\ \nu_\ell\ \ell^+$

$\overline{B}^0 \to \Lambda_c^+\ X^-$
$\quad\hookrightarrow \Lambda\ \pi^+$
$\qquad\hookrightarrow p\ \pi^-$

The decay modes are characterised by flavour-specific final state particles. These particles are treated as targets since their charges are correlated with the flavour of $B_{\text{tag}}$. In order to extract these flavour-specific signatures, the targets have to be identified among all available particle candidates. To accomplish this task, discriminating variables are calculated for each particle candidate. An overview of the discriminating variables for each category is presented in Table 24.

*Leptons.* Primary and secondary leptons from $B$ decays are used as target particles for different categories. In the first case, the leptons stem from $\overline{B}^0 \to X\ l^- \bar{\nu}_l$ decays via $b \to\ c(u)\ l^- \bar{\nu}_l$. A negatively (positively) charged primary lepton unambiguously tags a $\overline{B}^0$ ($B^0$) meson. Primary electrons and muons are the targets of the *Electron* and the *Muon* categories, respectively. Both are considered as targets in the *Kinetic Lepton* category.

Secondary leptons that are produced through the cascade decay of charmed mesons and baryons of the type $\overline{B}^0 \to\ X_c \left[\to\ l^+ \nu_l\ X_{s(d)}\right]\ X$ via transitions $b \to\ c \to\ s\ (d)\ l^+ \nu_l$ can be used to tag the flavour of the $B$ meson. In this case the charge-flavour correspondence is reversed, *i.e.* a positively (negatively) charged secondary lepton tags a $\overline{B}^0$ ($B^0$) meson. Since their momentum spectrum is much softer (lower) in comparison with the primary leptons, secondary leptons are referred to as intermediate leptons. Intermediate electrons and intermediate muons are the targets of the *Intermediate Electron* and the *Intermediate Muon* categories, respectively. Both are considered as targets in the *Intermediate Kinetic Lepton* category.

In order to distinguish primary and secondary leptons from all other candidates, kinematic and particle identification variables (PID), such as the electron and muon likelihoods $\mathcal{L}_e$ and $\mathcal{L}_\mu$ (Sec. 5.5), are used as discriminating variables. Within the kinematic variables, the momentum variables, such as the absolute momentum $p^*$ and the transverse momentum $p^*_{\mathrm{T}}$ in the $\Upsilon(4S)$ centre-of-mass frame as well as the absolute momentum $p$ and the transverse momentum $p_{\mathrm{t}}$ in the laboratory frame, have the highest discrimination power, especially for primary leptons. Intermediate leptons are more difficult to distinguish from other candidates because of their softer momentum spectrum. Additionally, the cosine of the polar angle of the momentum in the laboratory frame, $\cos\theta$ is considered.

Direct leptons are produced at the $B^0_{\mathrm{tag}}$ decay vertex and therefore have small impact parameters.

Further separation power is obtained from additional variables calculated in the $\Upsilon(4S)$ centre-of-mass frame, assuming that $B_{\mathrm{tag}}$ is produced at rest:

- $M^2_{\mathrm{rec}}$, the squared invariant mass of the recoiling system $X$ where the four-momentum is defined by:

$$p^\mu_X = \sum_{i \neq l} p^\mu_i, \tag{37}$$

  where the index $i$ runs over all charged and neutral candidates and $l$ corresponds to the index of the lepton candidate. Therefore

$$M^2_{\mathrm{rec}} = m^2_X = g_{\mu,\nu} p^\mu_X p^\nu_X\ . \tag{38}$$

- $p^*_{\mathrm{miss}}$, the absolute value of the missing momentum $\boldsymbol{p}^*_{\mathrm{miss}}$ which is defined by

$$\boldsymbol{p}^*_{\mathrm{miss}} = \boldsymbol{p}^*_{B^0} - \boldsymbol{p}^*_X - \boldsymbol{p}^*_l. \tag{39}$$

  Taking into account that the $B^0$ meson is produced at rest in the $\Upsilon(4S)$ frame, *i.e.* $\boldsymbol{p}^*_{B^0} \approx \mathbf{0}$, one obtains

$$\boldsymbol{p}^*_{\mathrm{miss}} \approx -\left(\boldsymbol{p}^*_X + \boldsymbol{p}^*_l\right). \tag{40}$$

- $\cos\theta^*_{\mathrm{miss}}$, the cosine of the angle between the momentum $\boldsymbol{p}^*_l$ of the lepton candidate and the missing momentum $\boldsymbol{p}^*_{\mathrm{miss}}$.

- $E_{90}^{W}$, the energy in the hemisphere defined by the direction of the momentum transfer (virtual $W^{\pm}$) in the $B$ meson decay. The momentum of the virtual $W^{\pm}$ is given by

$$\boldsymbol{p}_W^* \;=\; \boldsymbol{p}_l^* + \boldsymbol{p}_\nu^* \approx \; \boldsymbol{p}_l^* + \boldsymbol{p}_{\text{miss}}^* = -\boldsymbol{p}_X^*, \tag{41}$$

where the momentum $\boldsymbol{p}_\nu$ of the neutrino is estimated using the missing momentum $\boldsymbol{p}_{\text{miss}}^*$. The sum of energies for $E_{90}^{W}$ extends over all charged and neutral candidates in the recoiling system $X$ that are in the same hemisphere with respect to the $W^{\pm}$:

$$E_{90}^{W} = \sum_{i \in X,\; \boldsymbol{p}_i^* \cdot \boldsymbol{p}_W^* > 0} E_i. \tag{42}$$

- $\cos\theta_{\text{T}}^*$, the cosine of the angle between the lepton candidate's momentum $\boldsymbol{p}_l^*$ and the thrust axis of the $B_{\text{tag}}^0$ in the $\Upsilon(4S)$ centre-of-mass frame (Eq. 21).

Figure 43 shows example distributions of discriminating input variables for the Kinetic Lepton category (S. all distributions and correlation plots in [76]).

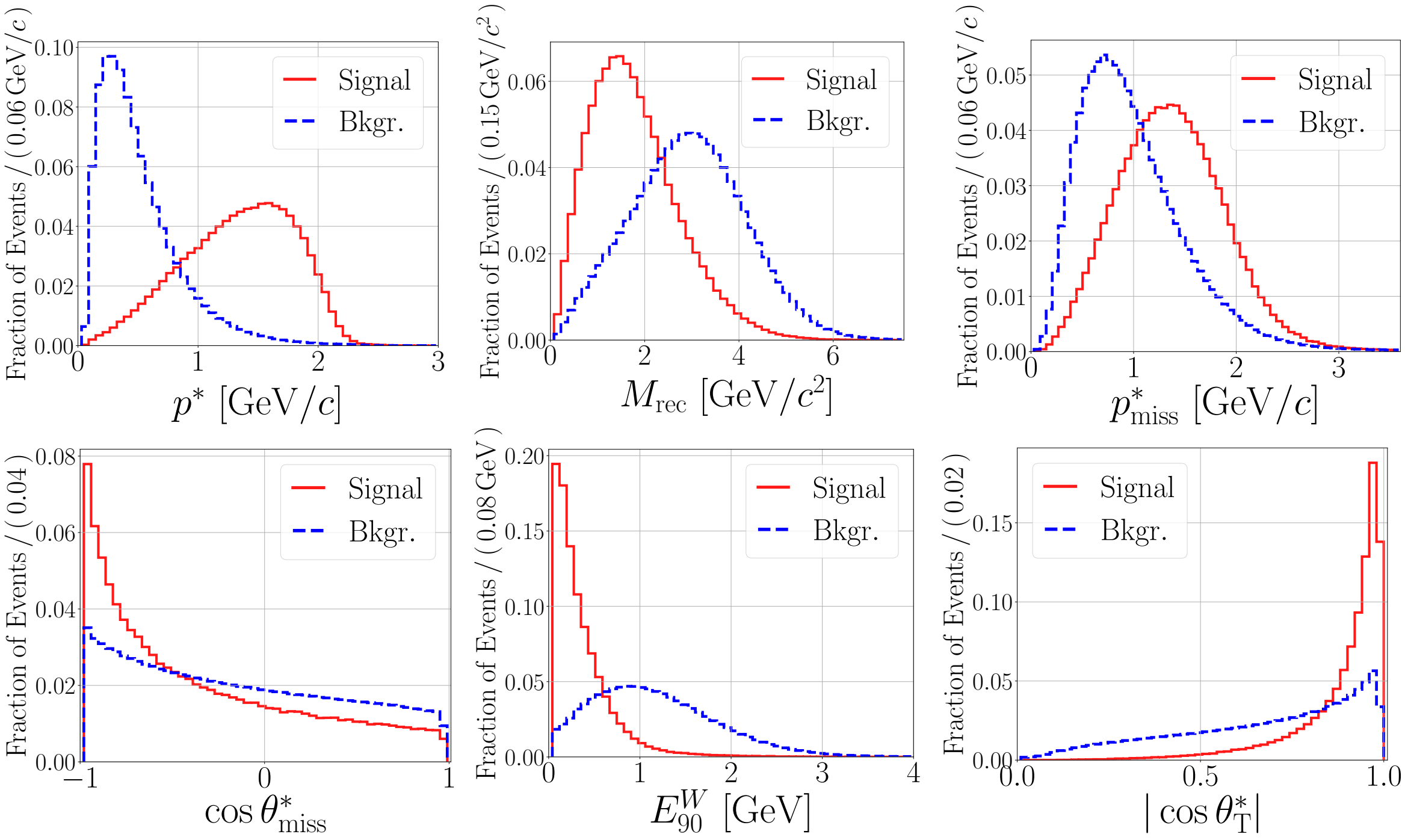

Fig. 43: Distributions of $p^*$, $M_{\text{rec}}$, $p_{\text{miss}}^*$, $\cos\theta_{\text{miss}}^*$, $E_{90}^{W}$ and $\cos\theta_{\text{T}}^*$ for the Kinetic Lepton category. Target muons and electrons are called signal, while all other particles are called background.

*Kaons.* Kaons are produced predominantly through decays of charmed mesons via $b \to c \to s$ transitions, $\overline{B}^0 \to D[\to K^- X]X$. Kaons stemming from such decays, and from decays of charmed baryons via $b \to c \to s$ transitions, tag a $\overline{B}^0$ ($B^0$) if they are negatively (positively) charged, and are referred to as "right sign" kaons.

The Kaon category has the highest flavour discrimination power due to the large inclusive branching fraction $\mathcal{B}\left(B^{\pm}/B^0 \to K^{\pm}\right) = (78.9 \pm 2.5)\%$ [77] and the fact that the fraction of right sign kaons $\mathcal{B}\left(B^{\pm}/B^0 \to K^{+}\right) = (66 \pm 5)\%$ is higher than the fraction of wrong sign kaons $\mathcal{B}\left(B^{\pm}/B^{0} \to K^{-}\right) = (13 \pm 4)\%$ [77]. For example kaons produced through $\bar{b} \to W^{+}\left[\to c\bar{s}/c\bar{d}\right] X$ with $c \to s \to K^{-}$.

In addition to the momentum variables ($p^*$, $p^*_{\rm t}$, $p$, $p_{\rm t}$, and $\cos\theta$) and the impact parameters ($d_0$ and $|\boldsymbol{x}|$), the following discriminating variables are used to identify target kaons:

- $\mathcal{L}_K$, the PID kaon likelihood (Sec. 5.5).
- $n_{K^0_S}$, the number of reconstructed $K^0_S$ on the tag side. A charged kaon produced through $b \to c\bar{c}s/c\bar{c}d$ transitions or through hadronisation of $s\bar{s}$ out of the vacuum is usually accompanied by one or more $K^0_S$.
- $\sum p_{\rm t}^2$, the sum of the squared transverse momentum of all tracks on the tag side in the laboratory frame. A high value of this quantity indicates a higher probability that the kaon candidate was produced through $b \to W^{-} c\,[\to s \to K^{-}]$. Lower values indicate a production through $b \to XW^{-}\,[\to \bar{c}s/\bar{c}d]$, $\bar{c} \to \bar{s} \to K^{+}$ which corresponds to a wrong sign kaon.
- $M^2_{\rm rec}$, $E^W_{90}$, $p^*_{\rm miss}$, $\cos\theta^*_{\rm miss}$ and $\cos\theta^*_{\rm T}$, the variables in the $\Upsilon(4S)$ frame which distinguish the lepton background.

Figure 44 shows example distributions of discriminating input variables for the Kaon category.

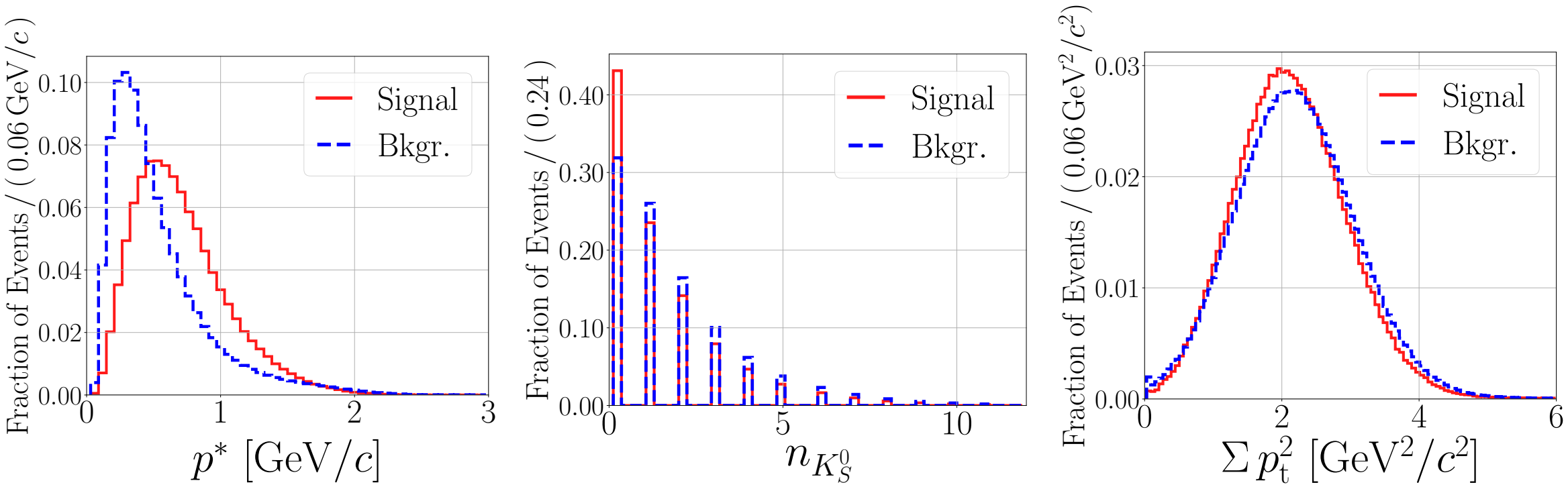

Fig. 44: Distributions of $p^*$, $n_{K^0_S}$ and $\sum p_{\rm t}^2$ for the Kaon category. Target kaons are called signal, while all other particles are called background.

*Slow Pions.* Primary and secondary pions are considered as target particles for several categories. The charge of secondary pions from $\bar{B}^0 \to XD^{*+}\left[\to D^0\pi^+\right]$ decays provide tagging information. Due to their soft (lower mean) momenta they are referred to as slow pions and considered as targets in the Slow Pion category. The Slow Pion category uses all the variables applied within the Kaon category, in order to distinguish the background from kaons and leptons. Additionally, the pion and the electron PID likelihoods $\mathcal{L}_\pi$ and $\mathcal{L}_e$ of each particle candidate are considered. The latter helps to distinguish the background from electrons created either through photon conversions or through $\pi^0 \to e^+e^-\gamma$ Dalitz decays.

The variable $\cos\theta^*_{\rm T}$ has considerable separation power. Slow pions are produced together with the $D^0$ nearly at rest in the $D^{*+}$ frame. Therefore, the flight direction of the target slow pions is close to the direction of the $D^0$ decay products and opposite to the other $B^0_{\rm tag}$ decay products. Low momentum background tracks can be distinguished by correlating the direction of the candidate and the direction of the remaining tracks from the $B^0_{\mathrm{tag}}$ decay, which corresponds to a good approximation to the $B^0_{\mathrm{tag}}$ thrust axis.

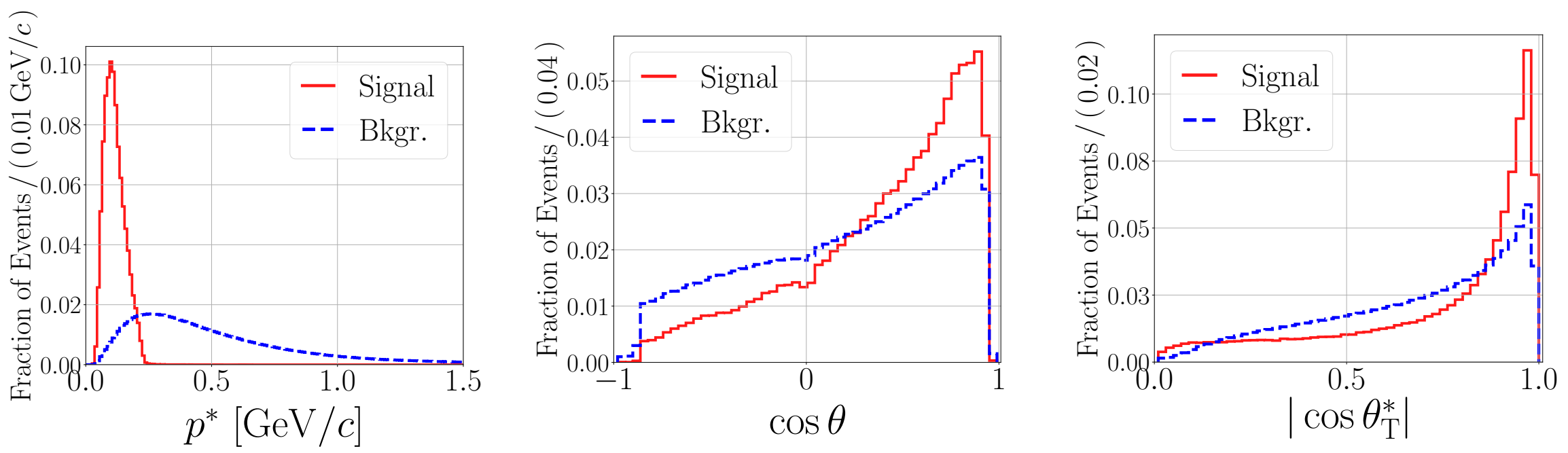

Fig. 45: Distributions of $p^*$, $\cos\theta$ and $\cos\theta^*_{\mathrm{T}}$ for the Slow Pion category. Target slow pions are called signal, while all other particles are called background.

Figure 45 shows example distributions of discriminating input variables for the Slow Pion category.

*Fast hadrons.* The targets of the Fast Hadron category are kaons and pions from the $W$ boson in $b \to c(u)\, W^-$ decays and from "1-prong" decays of primary tauons, *i.e.* $\overline{B}{}^0 \to \tau^- [\to \pi^-(K^-)\nu_\tau]\, \bar{\nu}_\tau\; X$. The category considers as targets also those kaons and pions that are produced through intermediate resonances, which decay via strong processes conserving the flavour information, *e.g.* $\overline{B}{}^0 \to K^{*-} \left[\to K^-\pi^0\right] X$ or $\overline{B}{}^0 \to \tau^- \left[\to \rho^- \left[\to \pi^-\pi^0\right] \nu_\tau\right] \bar{\nu}_\tau\; X$. The target kaons and pions are referred to as fast hadrons because of their hard (higher mean) momentum spectrum. A negatively (positively) charged fast hadron indicates a $\overline{B}{}^0$ ($B^0$) meson. The Fast Hadron category uses the same set of variables applied within the Slow Pion category.

*Correlation between kaons and slow pions (Kaon-Pion).* If an event contains both a target kaon and a target slow pion, *e.g.* a $\overline{B}{}^0 \to X D^{*+} \left[\to D^0 [\to K^-\; X]\, \pi^+\right]$ decay, the flavour tagging information from the individual categories is improved by exploiting the correlations between both targets.

The following variables are considered:

- $\mathcal{L}_K$, the PID kaon likelihood.
- $y_{\mathrm{Kaon}}$, the probability of being a target kaon obtained from the individual Kaon category.
- $y_{\mathrm{SlowPion}}$, the probability of being a target slow pion obtained from the individual Slow Pion category.
- $\cos\theta^*_{K\pi}$, the cosine of the angle between the kaon and the slow pion momentum in the $\Upsilon(4S)$ frame. If both targets are present, they are emitted in approximately the same direction in the $\Upsilon(4S)$ frame.
- $q_K \cdot q_\pi$, the charge product of the kaon and the slow pion candidates. A right sign kaon and the corresponding slow pion are produced with opposite charges in agreement with their individual flavour-charge correspondence.

*High momentum particles (Maximum $P^*$).* Hadrons and leptons from the $W$ boson in $b \to c(u)\, W^-$ are characterised by a very hard momentum spectrum. An inclusive tag can be performed by selecting the track with the highest momentum in the $\Upsilon(4S)$ frame and using its charge as a flavour tag. A negatively (positively) charged fast particle indicates a $\overline{B}^0$ ($B^0$) meson. The purpose is to recover flavour tagging information from primary particles that may not have been selected either as a primary lepton or as a fast hadron. The discriminating variables are the momentum variables ($p^*$, $p^*_\mathrm{t}$, $p$, $p_\mathrm{t}$, and $\cos\theta$), the impact parameters ($d_0$ and $|\boldsymbol{x}|$) and $\cos\theta^*_\mathrm{T}$.

*Correlation between fast and slow particles (FSC).* Events of the kind $\overline{B}^0 \to D^{*+}W^-$ contain both a target slow pion and a high-momentum primary particle originating from the $W$ boson. In that case, additional flavour tagging information can be gained by using the correlations between the slow pion and the high momentum particle.

The $W^\pm$ and the $D^{*+}$ are produced back-to-back in the $B^0_\mathrm{tag}$ centre-of-mass frame. Therefore, the angle between the track of the target fast particle and the target slow pion is expected to be very large. Useful discriminating variables are:

- $\mathcal{L}_{K\mathrm{Slow}}$, the PID kaon likelihood of the slow pion candidate.
- $p^*_\mathrm{Slow}$, the momentum of the slow pion candidate in the $\Upsilon(4S)$ frame.
- $p^*_\mathrm{Fast}$, the momentum of the high-momentum candidate in the $\Upsilon(4S)$ frame.
- $\cos\theta^*_\mathrm{T,\,Slow}$, the cosine of the angle between the thrust axis and the slow pion candidate in the $\Upsilon(4S)$ frame.
- $\cos\theta^*_\mathrm{T,\,Fast}$, the cosine of the angle between the thrust axis and the high-momentum candidate in the $\Upsilon(4S)$ frame.
- $\cos\theta^*_\mathrm{SlowFast}$, the cosine of the angle between the slow and the high-momentum candidates in the $\Upsilon(4S)$ frame.
- $q_\mathrm{Slow}\cdot q_\mathrm{Fast}$, the charge product of the slow pion and the high-momentum candidates. In agreement with their individual flavour-charge correspondence, the targets have to be produced with opposite charges.

*Lambda baryons.* Additional flavour tagging information can be obtained by considering the flavour of $\Lambda$ baryons, since they are likely to contain an $s$ quark from the cascade transition $b \to c \to s$. The presence of a $\Lambda$ ($\bar{\Lambda}$) baryon indicates a $\overline{B}^0$ ($B^0$). Although the fraction of events containing a target $\Lambda$ is rather small, they provide relatively clean flavour tagging information. The $\Lambda$ candidates are obtained by reconstructing $\Lambda \to p\pi^-$ ($\bar{\Lambda} \to \bar{p}\pi^+$) decays through combinations of proton and pion candidates on the tag side. In addition to the momentum variables of the reconstructed $\Lambda$, proton and pion used for the reconstruction, the following discriminating variables are used.

- $\mathcal{L}_p$, $\mathcal{L}_\pi$ the PID likelihoods of the proton and the pion.
- $q_\Lambda$, the flavour of the $\Lambda$ baryon.
- $M_\Lambda$, the reconstructed mass of the $\Lambda$.
- $n_{K^0_S}$, the number of reconstructed $K^0_S$ on the tag side.
- $\cos\theta_{\boldsymbol{x}_\Lambda,\boldsymbol{p}_\Lambda}$, the cosine of the angle between the $\Lambda$ momentum $\boldsymbol{p}_\Lambda$ and the direction from the IP to the reconstructed $\Lambda$ vertex $\boldsymbol{x}_\Lambda$ in the laboratory frame.
- $|\boldsymbol{x}_\Lambda|$, the distance between the $\Lambda$ vertex and the IP.

- $\sigma_{\Lambda}^{zz}$, the error of the $\Lambda$ vertex fit in $z$-direction.

A summary of the discriminating variables for each category is presented in Table 24.

Table 24: Discriminating input variables for each category. For some of the categories the $p$-value of the track fit is taken into account. For the Lambda category, the $p$-value of the reconstructed $\Lambda$ decay vertex is used. All variables are calculated for each considered particle candidate.

| Categories | Discriminating input variables |
|---|---|
| Electron<br>Int. Electron | $\mathcal{L}_e,\ p^*,\ p_{\rm t}^*,\ p,\ p_{\rm t},\ \cos\theta,\ d_0,\ \lvert\boldsymbol{x}\rvert,\ M_{\rm rec}^2,\ E_{90}^W,\ p_{\rm miss}^*,\ \cos\theta_{\rm miss}^*,\ \cos\theta_{\rm T}^*,\ p\text{-val.}$ |
| Muon<br>Int. Muon | $\mathcal{L}_\mu,\ p^*,\ p_{\rm t}^*,\ p,\ p_{\rm t},\ \cos\theta,\ d_0,\ \lvert\boldsymbol{x}\rvert,\ M_{\rm rec}^2,\ E_{90}^W,\ p_{\rm miss}^*,\ \cos\theta_{\rm miss}^*,\ \cos\theta_{\rm T}^*,\ p\text{-val.}$ |
| Kin. Lepton<br>Int. Kin. Lep. | $\mathcal{L}_e,\ \mathcal{L}_\mu,\ p^*,\ p_{\rm t}^*,\ p,\ p_{\rm t},\ \cos\theta,\ d_0,\ \lvert\boldsymbol{x}\rvert,\ M_{\rm rec}^2,\ E_{90}^W,\ p_{\rm miss}^*,\ \cos\theta_{\rm miss}^*,\ \cos\theta_{\rm T}^*,\ p\text{-val.}$ |
| Kaon | $\mathcal{L}_K,\ p^*,\ p_{\rm t}^*,\ p,\ p_{\rm t},\ \cos\theta,\ d_0,\ \lvert\boldsymbol{x}\rvert,\ n_{K_S^0},\ \sum p_{\rm t}^2,$<br>$M_{\rm rec}^2,\ E_{90}^W,\ p_{\rm miss}^*,\ \cos\theta_{\rm miss}^*,\ \cos\theta_{\rm T}^*,\ \chi^2$ |
| Slow Pion<br>Fast Hadron | $\mathcal{L}_\pi,\ \mathcal{L}_e,\ \mathcal{L}_K,\ p^*,\ p_{\rm t}^*,\ p,\ p_{\rm t},\ \cos\theta,\ d_0,\ \lvert\boldsymbol{x}\rvert,\ n_{K_S^0},\ \sum p_{\rm t}^2,$<br>$M_{\rm rec}^2,\ E_{90}^W,\ p_{\rm miss}^*,\ \cos\theta_{\rm miss}^*,\ \cos\theta_{\rm T}^*,\ p\text{-val.}$ |
| Kaon-Pion | $\mathcal{L}_K,\ y_{\rm Kaon},\ y_{\rm SlowPion},\ \cos\theta_{K\pi}^*,\ q_K\cdot q_\pi$ |
| Maximum P* | $p^*,\ p_{\rm t}^*,\ p,\ p_{\rm t},\ d_0,\ \lvert\boldsymbol{x}\rvert,\ \cos\theta_{\rm T}^*$ |
| FSC | $\mathcal{L}_{K{\rm Slow}},\ p_{\rm Slow}^*,\ p_{\rm Fast}^*,\ \cos\theta_{\rm T,\,Slow}^*,\ \cos\theta_{\rm T,\,Fast}^*,\ \cos\theta_{\rm SlowFast}^*,\ q_{\rm Slow}\cdot q_{\rm Fast}$ |
| Lambda | $\mathcal{L}_p,\ \mathcal{L}_\pi,\ p_\Lambda^*,\ p_\Lambda,\ p_p^*,\ p_p,\ p_\pi^*,\ p_\pi,\ q_\Lambda,\ M_\Lambda,\ n_{K_S^0},\ \cos\theta_{\boldsymbol{x}_\Lambda,\boldsymbol{p}_\Lambda},\ \lvert\boldsymbol{x}_\Lambda\rvert,\ \sigma_\Lambda^{zz},\ p\text{-val.}$ |

*6.5.3. Algorithm.* The Belle II flavour tagger is a modular algorithm based on multivariate methods which provide a flavour tag $q$ for the $B^0_{\rm tag}$ meson together with the corresponding flavour dilution factor $r$. It does so by analysing the tracks and the neutral clusters that remain after $B^0_{\rm sig}$ reconstruction. The output $y \in [-1, 1]$ of the flavour tagger is equivalent to $y = q \cdot r$, where $y = -1$ (1) corresponds to a perfectly tagged $\overline{B}^0$ ($B^0$).

The flavour of $B^0_{\rm tag}$ results from a combination of the thirteen flavour signatures discussed in the previous subsections. Each of these signatures corresponds to the output of a single category which can be understood as an individual "sub-tagger". A schematic overview of the information flow in the algorithm is presented in Fig. 46. The algorithm of the flavour tagger is a 2-level process comprised of event- and combiner- levels. The event-level process is performed within each individual category. On this level, a multivariate method assigns to each particle candidate a probability $y_{\rm cat}$, which is the probability of being the target of the corresponding category providing the correct flavour tag. The particle candidates correspond to the tracks that remain from the reconstruction of the signal $B_{\rm sig}$ meson. Since each track is fitted with 5 different mass hypotheses ($e$, $\mu$, $K$, $\pi$ and $p$), each category considers the mass hypotheses belonging to its own targets. To determine $y_{\rm cat}$, the event-level multivariate methods are given the discriminating variables of the corresponding category. In some calculations all reconstructed tracks and all neutral ECL and KLM clusters that remain after the full $B^0_{\rm sig}$ reconstruction are taken into account.Within each category, the particle candidates are ranked according to the values of $y_{\text{cat}}$. The candidate with the highest $y_{\text{cat}}$ is selected as the target. The exception is the Maximum P* category, where the target is the candidate with the largest momentum in the $\Upsilon(4S)$ frame.

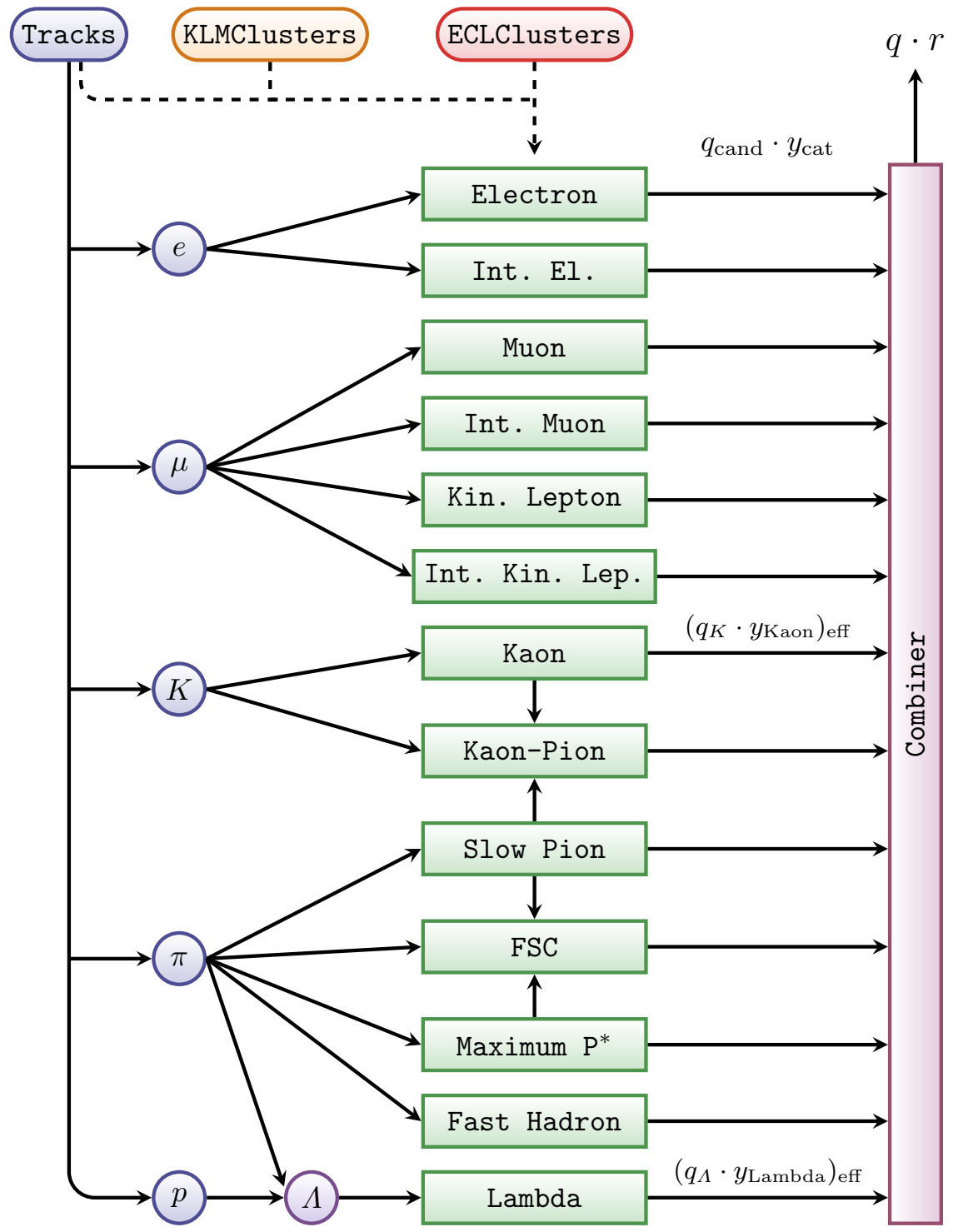

Fig. 46: Schematic overview of the Flavour Tagger: Reconstructed tracks are available for five different mass hypotheses. Each green box corresponds to a category. The charge $q_{\text{cand}}$ and the probability $y_{\text{cat}}$ are explained in the text. The values $(q_{\text{cand}} \cdot y_{\text{cat}})_{\text{eff}}$ are defined in Eq. 43.

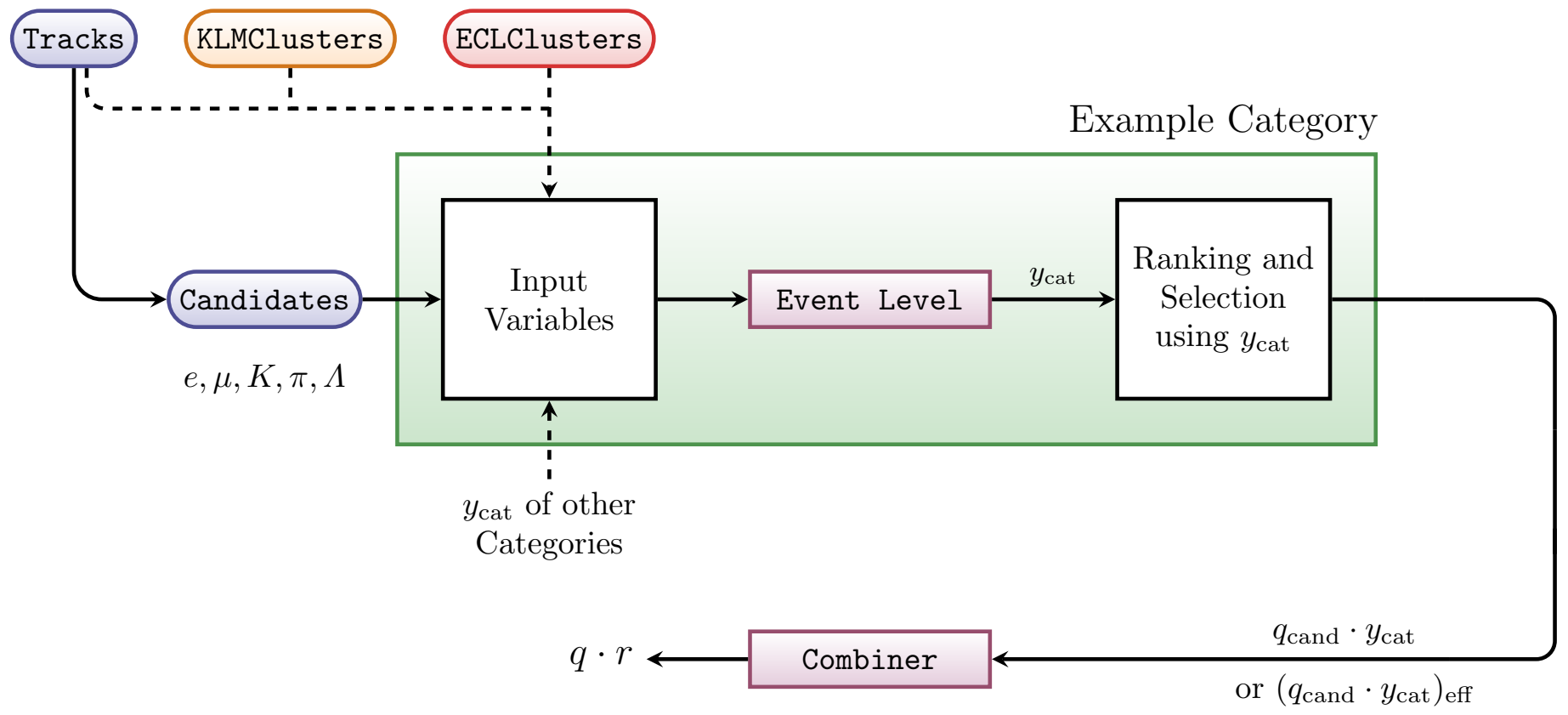

Fig. 47: Procedure for each single category (green box): the candidates correspond to the reconstructed tracks for a specific mass hypothesis. The input variables are presented in Table 24; some of them consider all reconstructed tracks and all neutral ECL and KLM clusters on the tag side. The magenta boxes represent multivariate methods: $y_{\text{cat}}$ is the output of the event-level. The output of the combiner is equivalent to the product $q \cdot r$.

Table 25: Performance of individual categories for the flavour tagging algorithm. All values are given as a percentage considering only statistical uncertainties.

| | Belle II MC | | Belle MC | |
|---|---|---|---|---|
| Categories | $\varepsilon_{\text{eff}} \pm \delta\varepsilon_{\text{eff}}$ | $\Delta\varepsilon_{\text{eff}} \pm \delta\Delta\varepsilon_{\text{eff}}$ | $\varepsilon_{\text{eff}} \pm \delta\varepsilon_{\text{eff}}$ | $\Delta\varepsilon_{\text{eff}} \pm \delta\Delta\varepsilon_{\text{eff}}$ |
| Electron | $5.53 \pm 0.01$ | $0.20 \pm 0.02$ | $5.80 \pm 0.01$ | $-0.04 \pm 0.03$ |
| Int. Electron | $1.51 \pm 0.01$ | $-0.18 \pm 0.01$ | $0.74 \pm 0.01$ | $0.00 \pm 0.01$ |
| Muon | $5.51 \pm 0.01$ | $0.22 \pm 0.02$ | $5.74 \pm 0.01$ | $0.08 \pm 0.03$ |
| Int. Muon | $0.40 \pm 0.01$ | $-0.03 \pm 0.01$ | $0.33 \pm 0.01$ | $0.00 \pm 0.01$ |
| KinLepton | $11.36 \pm 0.02$ | $0.31 \pm 0.03$ | $11.70 \pm 0.02$ | $0.08 \pm 0.04$ |
| Int. Kin. Lep. | $1.41 \pm 0.01$ | $-0.10 \pm 0.01$ | $0.56 \pm 0.01$ | $0.00 \pm 0.01$ |
| Kaon | $22.41 \pm 0.02$ | $-0.84 \pm 0.04$ | $19.28 \pm 0.02$ | $-0.29 \pm 0.04$ |
| Kaon-Pion | $14.85 \pm 0.01$ | $-0.25 \pm 0.03$ | $15.15 \pm 0.02$ | $-0.26 \pm 0.04$ |
| Slow Pion | $10.16 \pm 0.01$ | $0.07 \pm 0.02$ | $9.27 \pm 0.01$ | $-0.05 \pm 0.03$ |
| FSC | $14.16 \pm 0.02$ | $-0.13 \pm 0.03$ | $11.54 \pm 0.01$ | $-0.11 \pm 0.03$ |
| Maximum P* | $13.05 \pm 0.01$ | $1.36 \pm 0.03$ | $11.96 \pm 0.02$ | $0.05 \pm 0.03$ |
| Fast Hadron | $4.61 \pm 0.01$ | $1.14 \pm 0.01$ | $1.54 \pm 0.01$ | $-0.04 \pm 0.01$ |
| Lambda | $2.77 \pm 0.01$ | $0.79 \pm 0.01$ | $1.53 \pm 0.01$ | $0.24 \pm 0.01$ |

The procedure within each single category is illustrated in Fig. 47. The combiner level is the last step in the process. It corresponds to a multivariate method that takes thirteen input values, and gives $y = q \cdot r$ as the output. Each input value is the product $q_{\text{cand}} \cdot y_{\text{cat}}$ of each category, where the charge $q_{\text{cand}}$ and the probability $y_{\text{cat}}$ correspond to the particle candidate selected as target. For two special cases, the Kaon and the Lambda categories, the input value is the effective product

$$(q_{\text{cand}} \cdot y_{\text{cat}})_{\text{eff}} = \frac{\prod_i \left(1 + (q_{\text{cand}} \cdot y_{\text{cat}})_i\right) - \prod_i \left(1 - (q_{\text{cand}} \cdot y_{\text{cat}})_i\right)}{\prod_i \left(1 + (q_{\text{cand}} \cdot y_{\text{cat}})_i\right) + \prod_i \left(1 - (q_{\text{cand}} \cdot y_{\text{cat}})_i\right)} \tag{43}$$

where the products extend over the three particles with the highest $y_{\text{cat}}$ probability. For the Lambda category, $q_{\text{cand}}$ corresponds to the $B^0$ flavour tagged by the $\Lambda$ candidate, *i.e.* $q_\Lambda = -1(+1)$ for $\Lambda(\bar{\Lambda})$.

The multivariate method chosen for the event- and the combiner-level is a fast boosted decision tree (FBDT) [70]. For the combiner-level, an independent multivariate method, a multi-layer perceptron (MLP) [78, 79], is employed to cross-check the result of the FBDT.

The flavour tagger is trained using two statistically independent MC samples: one for the event-level and one for the combiner-level to avoid bias from a possible statistical correlation. At each training step, one half of the sample is used as the training sample and the other half as a test and validation sample. The event-level is trained first and each category is trained independently. The FBDT and the MLP combiners are trained afterwards.

*6.5.4. Performance.* The performance of the Belle II flavour tagger has been evaluated using Belle II MC, as well as using Belle MC and Belle collision data. The MC events used for training and testing correspond to $B^0\overline{B}^0$ pairs in which one meson ($B^0_{\rm sig}$) decays to $J/\psi\, K^0_S$ while $B^0_{\rm tag}$ decays to any possible final state according to the known branching fractions [77]. Only events where the decay channel $B^0_{\rm sig} \to J/\psi\,[\to \mu^+\mu^-]\,K^0_S\,[\to \pi^+\pi^-]$ could be fully reconstructed and correctly matched with the MC decay chain are selected for training and testing. After the selection, the size of the Belle II and the Belle training samples is approximately $2 \times 1.3$ and $2 \times 1$ million MC events, respectively, and the size of the Belle II and the Belle testing samples is approximately 2.6 and 2 million MC events, respectively.

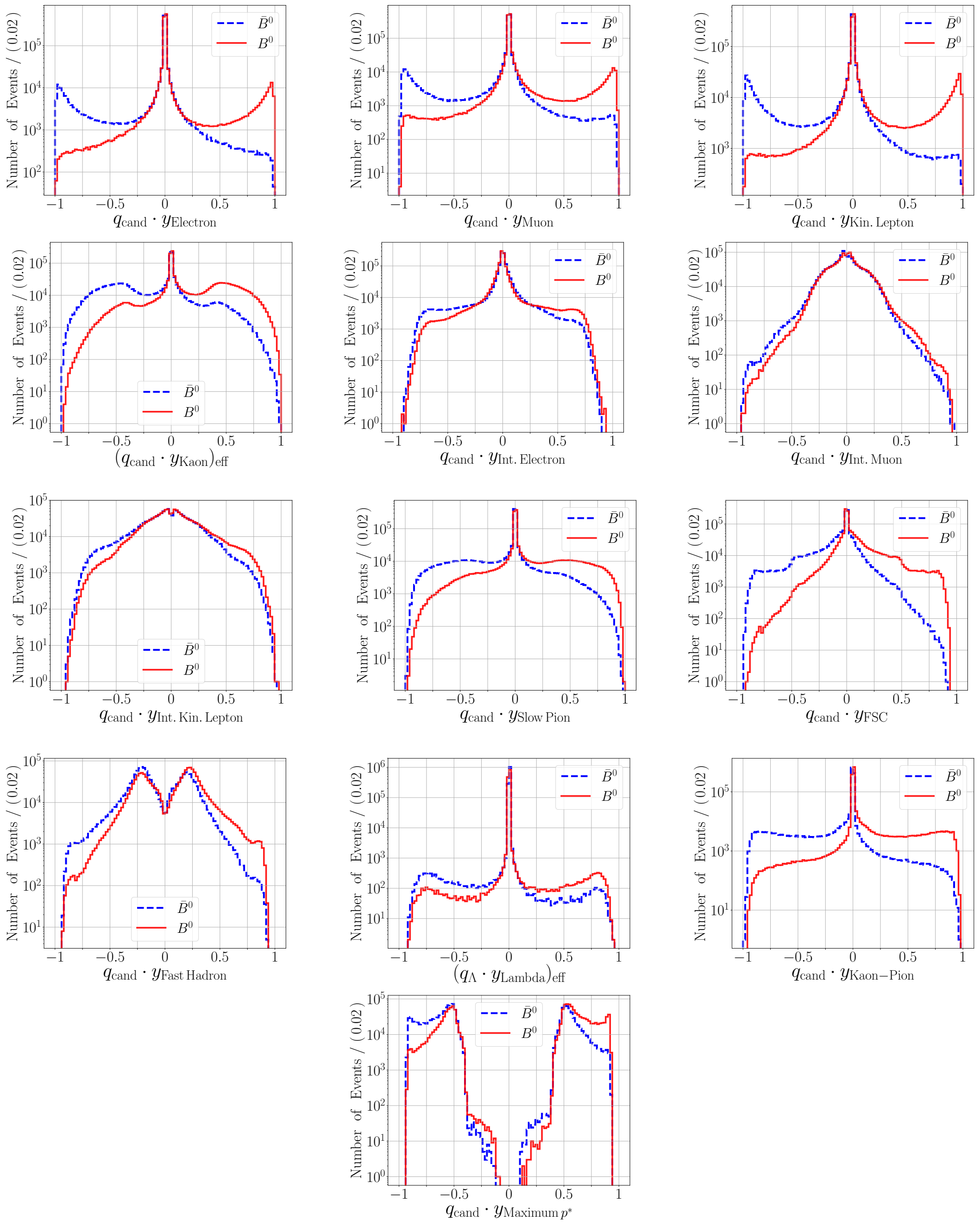

Fig. 48: Combiner input distributions for all categories.

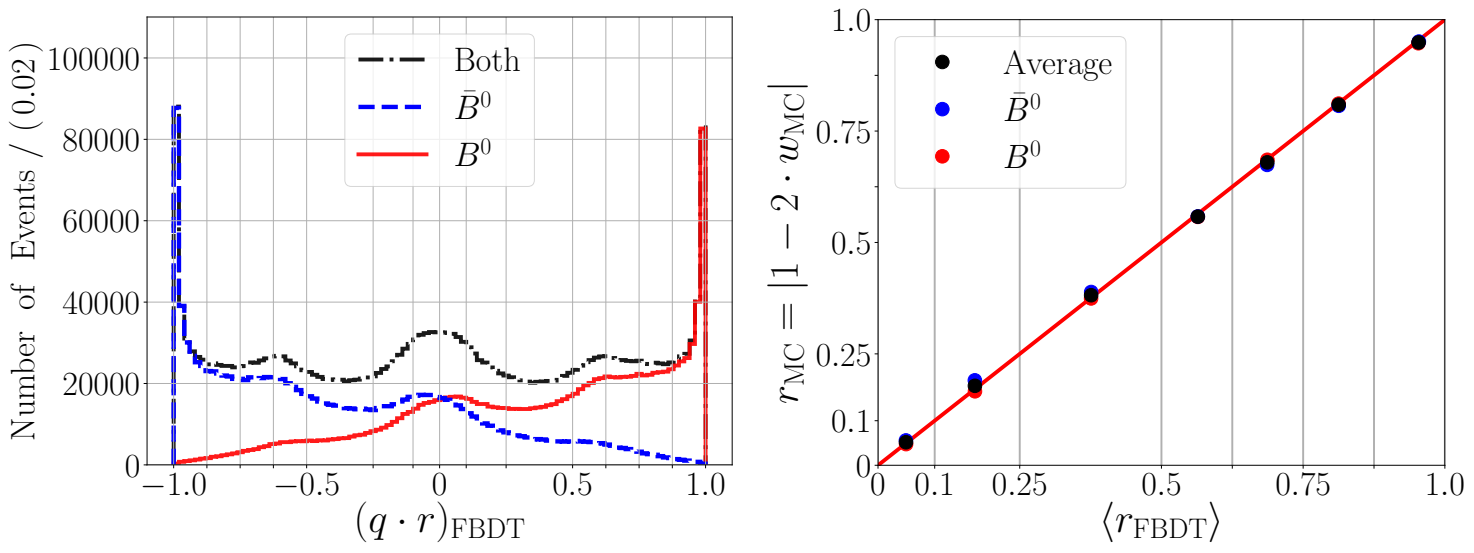

a) Belle II MC

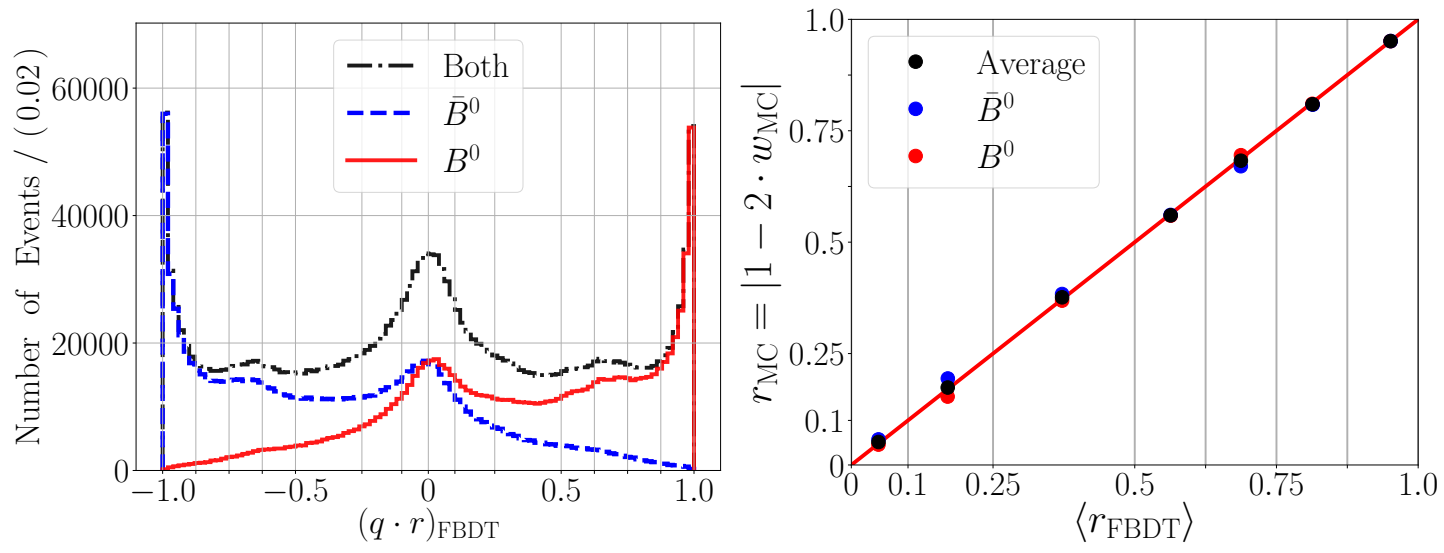

b) Belle MC

Fig. 49: Results of the FBDT combiner using a) Belle II MC and b) Belle MC. Left: Distributions of the output $q \cdot r$. Right: Correlations between the dilution $r_{\mathrm{MC}} = 1 - 2w_{\mathrm{MC}}$ taken from MC truth and the mean absolute value of the combiner output $r = |q \cdot r|$ in each $r$-bin. The errors on both axes are not visible due to their small size. The red diagonal line is a guideline and the vertical grey lines correspond to the limits of the $r$-bins.

The test with Belle collision data is performed on a set of $B^0 \overline{B}^0$ pairs, where the same decay channel $B^0_{\mathrm{sig}} \to J/\psi\, [\to \mu^+\mu^-]\, K^0_S\, [\to \pi^+\pi^-]$ is reconstructed on the signal side. The signal selection is performed following previous Belle analyses [80] using the full Belle data sample which corresponds to $711\,\mathrm{fb}^{-1}$. The obtained signal yield is 8508 events.

The distributions of the thirteen combiner input values, which are derived from the outputs, $y_{\mathrm{cat}}$, of the individual categories, are presented in Fig. 48. The large peaks at zero are due to cases where the target provides no flavour information. In general, a value close to zero indicates that the probability of finding a certain flavour-specific signature within the $B^0_{\mathrm{tag}}$ final state is very low. A value closer to $\pm 1$ indicates a more reliable flavour tag.

The output $y = q \cdot r$, which corresponds to the product of tagged flavour $q$ and the dilution factor $r$, can be found in Fig. 49 (left) for the FBDT combiner on MC. Figure 49 (right) also shows a linearity check between the true dilution $r_{\mathrm{MC}}$ determined using MC information and the mean $\langle r \rangle$ of the dilution provided by the combiners. The dilution determined using MC information corresponds to $r_{\mathrm{MC}} = 1 - 2w_{\mathrm{MC}}$, where the wrong tag fraction $w_{\mathrm{MC}}$ is determined by comparing the MC truth with the combiner output, *i.e.* an event is wrongly tagged if $q_{\mathrm{MC}} \neq q = \mathrm{sgn}(q \cdot r)$. The mean dilution $\langle r \rangle$ of the combiner output is simply the mean of $|q \cdot r|$ for each $r$-bin. Figure 49 (right) shows the results of the linearity check for events where $B_{\mathrm{tag}}$ is a $B^0$, for events where $B_{\mathrm{tag}}$ is a $\overline{B}^0$, and in average.

The results using Belle data and Belle MC are shown together in Figure 50 by superimposing the normalised $q \cdot r$ output distributions. Within the uncertainties, the shapes of the normalised $q \cdot r$ distributions for Belle data and Belle MC show good agreement.

Table 26: Performance of FBDT combiner on MC. All values are given as a percentage considering only statistical uncertainties.

| FBDT Combiner on Belle II MC | | | | | | |
|---|---|---|---|---|---|---|
| $r$- Interval | $\varepsilon_i$ | $\Delta\varepsilon_i$ | $w_i \pm \delta w_i$ | $\Delta w_i \pm \delta\Delta w_i$ | $\varepsilon_{\text{eff},i} \pm \delta\varepsilon_{\text{eff},i}$ | $\Delta\varepsilon_{\text{eff},i} \pm \delta\Delta\varepsilon_{\text{eff},i}$ |
| $0.000 - 0.100$ | 12.0 | $-0.28$ | $47.42 \pm 0.09$ | $0.41 \pm 0.18$ | $0.0320 \pm 0.0022$ | $-0.0110 \pm 0.0044$ |
| $0.100 - 0.250$ | 14.3 | 0.07 | $41.09 \pm 0.08$ | $1.23 \pm 0.16$ | $0.4537 \pm 0.0082$ | $-0.1234 \pm 0.0163$ |
| $0.250 - 0.500$ | 19.9 | 0.12 | $30.92 \pm 0.06$ | $0.72 \pm 0.13$ | $2.8961 \pm 0.0196$ | $-0.2012 \pm 0.0392$ |
| $0.500 - 0.625$ | 11.9 | 0.06 | $22.10 \pm 0.07$ | $0.05 \pm 0.15$ | $3.7022 \pm 0.0205$ | $0.0040 \pm 0.0410$ |
| $0.625 - 0.750$ | 12.0 | 0.14 | $16.01 \pm 0.06$ | $-0.57 \pm 0.13$ | $5.5576 \pm 0.0231$ | $0.2529 \pm 0.0462$ |
| $0.750 - 0.875$ | 11.6 | 0.21 | $9.54 \pm 0.05$ | $-0.25 \pm 0.11$ | $7.5934 \pm 0.0236$ | $0.2267 \pm 0.0473$ |
| $0.875 - 1.000$ | 18.3 | $-0.32$ | $2.55 \pm 0.02$ | $0.19 \pm 0.05$ | $16.4389 \pm 0.0265$ | $-0.4199 \pm 0.0529$ |
| Total | | | $\varepsilon_{\text{eff}} = \sum_i \varepsilon_i \cdot \langle 1 - 2w_i \rangle^2 = 36.67 \pm 0.05$ | | $\Delta\varepsilon_{\text{eff}} = -0.27 \pm 0.10$ | |
| FBDT Combiner on Belle MC | | | | | | |
| $0.000 - 0.100$ | 15.4 | 0.06 | $47.61 \pm 0.09$ | $0.59 \pm 0.18$ | $0.0354 \pm 0.0001$ | $0.0002 \pm 0.0002$ |
| $0.100 - 0.250$ | 16.1 | 0.01 | $41.50 \pm 0.09$ | $2.03 \pm 0.17$ | $0.4667 \pm 0.0009$ | $-0.0011 \pm 0.0017$ |
| $0.250 - 0.500$ | 20.0 | $-0.16$ | $31.41 \pm 0.07$ | $0.74 \pm 0.15$ | $2.7591 \pm 0.0042$ | $-0.0410 \pm 0.0085$ |
| $0.500 - 0.625$ | 9.9 | 0.01 | $21.83 \pm 0.09$ | $0.08 \pm 0.18$ | $3.1384 \pm 0.0067$ | $0.0101 \pm 0.0134$ |
| $0.625 - 0.750$ | 10.4 | 0.14 | $15.64 \pm 0.08$ | $-1.24 \pm 0.16$ | $4.9015 \pm 0.0102$ | $0.1380 \pm 0.0203$ |
| $0.750 - 0.875$ | 10.3 | 0.04 | $9.32 \pm 0.06$ | $-0.10 \pm 0.13$ | $6.7843 \pm 0.0141$ | $0.0418 \pm 0.0283$ |
| $0.875 - 1.000$ | 17.9 | $-0.13$ | $2.43 \pm 0.03$ | $0.01 \pm 0.05$ | $16.1464 \pm 0.0244$ | $-0.2362 \pm 0.0487$ |
| Total | | | $\varepsilon_{\text{eff}} = \sum_i \varepsilon_i \cdot \langle 1 - 2w_i \rangle^2 = 34.26 \pm 0.03$ | | $\Delta\varepsilon_{\text{eff}} = -0.09 \pm 0.06$ | |

At Belle II, differences between the performance for $B^0$ and for $\overline{B}^0$ can occur due to $CP$ asymmetries on the tag side. Due to the initial entanglement of the neutral $B$-meson pair at the $\Upsilon(4S)$, $CP$ asymmetries on the signal-side $B$-meson decay cause $CP$ asymmetries on the tag side. $CP$ asymmetries on the tag side result into different decay-vertex distributions for $B^0$ and for $\overline{B}^0$ causing different impact-parameter distributions for positively and for negatively charged target particles. Thus, if the MC events used to train the flavor tagger are generated with $CP$ violation on the signal side, the flavor tagger learns the corresponding $CP$ asymmetry on the tag side to artificially distinguish between $B^0$ and $\overline{B}^0$. Belle II is sensitive to $CP$ asymmetries on the tag-side due to the novel small size of the interaction region (about 20 times smaller than at Belle in beam direction). To avoid that the flavor tagger learns these asymmetries, it must be trained with MC events generated without built-in CP violation [76]. Another possible cause for differences between $B^0$ and $\overline{B}^0$ are asymmetries in the detector performance for positively and for negatively charged particles.

Considering now the effective efficiency $\varepsilon_{\text{eff}}$, it has been defined by sorting the tagged events into bins of the dilution factor, $r$, adopting the same $r$-binning applied by the Belle experiment [81]. The expression for the effective efficiency in Eq. 34 becomes

$$\varepsilon_{\text{eff}} = \varepsilon \sum_i \frac{n_i}{N_{\text{tag}}} \langle r \rangle_i^2 = \sum_i \frac{n_i}{N} \langle r \rangle_i^2 = \sum_i \varepsilon_i \langle r \rangle_i^2, \quad \text{using} \quad \varepsilon_i = \frac{n_i}{N}, \tag{44}$$

where the sum extends over all $r$- bins, and the tagging efficiency corresponds to $\varepsilon = \frac{N_{\text{tag}}}{N}$, where $N_{\text{tag}}$ is the number of tagged events. The measured value of $\varepsilon$ on Belle data is 99.8 %, which is equal to the previous value measured by Belle using the Belle flavour tagger [81] and is consistent with the value of 99.9% obtained using the Belle II flavour tagger on Belle MC and on Belle II MC.

For each individual category, an effective efficiency can be calculated if the corresponding combiner input value $q_{\text{cand}} \cdot y_{\text{cat}}$ is taken as a single flavour tag, *i.e.* if each category is considered as a "sub-tagger". These effective efficiencies are presented in Table 25. The Lepton, the Kaon and the Pion categories provide relatively clean flavour signatures and relatively high branching fractions.

In general, the fraction $w$ can be extracted from the combiner output $y = q \cdot r$ through

$$w = \frac{1 - |y|}{2}, \tag{45}$$

if the dilution $r = |y|$ provided by the combiner is linear with respect to the true dilution $r_{\text{MC}} = 1 - 2w_{\text{MC}}$, determined using MC information. This linearity has been verified for Belle II and for Belle MC in Fig. 49 (right).

Table 26 lists the global performance quantities of the flavour tagger on MC for both combiner-level multivariate methods. The FBDT combiner achieves a total effective efficiency $\varepsilon_{\text{eff}}$ of $(36.67 \pm 0.05)\%$ on Belle II MC and $(34.26 \pm 0.03)\%$ on Belle MC. The performance of the flavour tagger on Belle data is presented in Table 27. The FBDT combiner achieves a total effective efficiency of $(33.6 \pm 0.5)\%$ on Belle data.

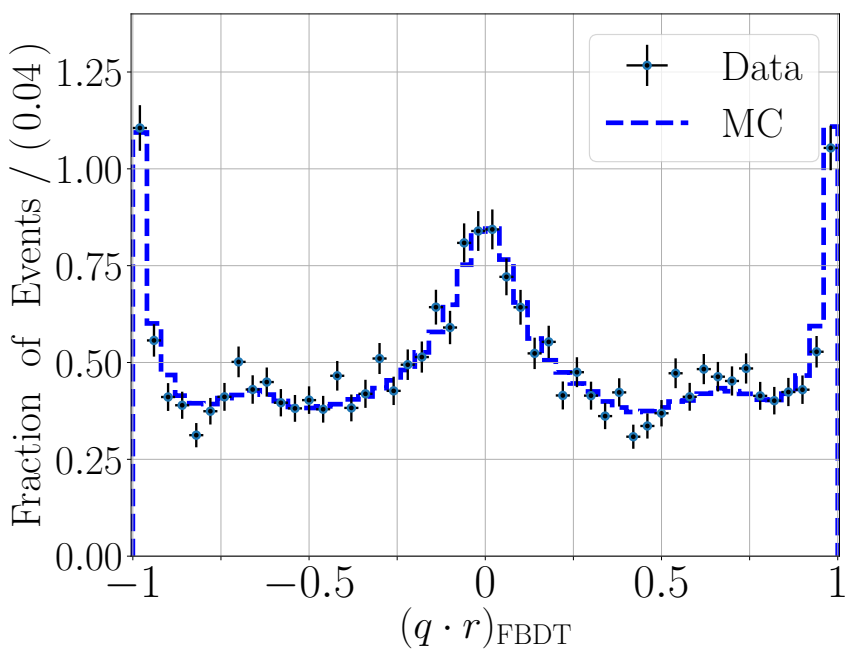

Fig. 50: Normalized $q \cdot r$ distributions on Belle data and on Belle MC.

Table 27: Performance of the Belle II flavour tagger on Belle data. All values are given in percent considering only statistical uncertainties.

| FBDT Combiner | | | |
|---|---|---|---|
| $r$- Interval | $\varepsilon_i$ | $w_i \pm \delta w_i$ | $\varepsilon_{\text{eff},i} \pm \delta\varepsilon_{\text{eff},i}$ |
| 0.000 − 0.100 | 15.19 | $47.64 \pm 0.04$ | $0.034 \pm 0.001$ |
| 0.100 − 0.250 | 16.53 | $41.50 \pm 0.06$ | $0.477 \pm 0.013$ |
| 0.250 − 0.500 | 20.28 | $31.39 \pm 0.09$ | $2.803 \pm 0.066$ |
| 0.500 − 0.625 | 10.04 | $21.74 \pm 0.06$ | $3.204 \pm 0.105$ |
| 0.625 − 0.750 | 11.07 | $15.63 \pm 0.06$ | $5.222 \pm 0.162$ |
| 0.750 − 0.875 | 10.34 | $9.40 \pm 0.06$ | $6.807 \pm 0.218$ |
| 0.875 − 1.000 | 16.38 | $2.33 \pm 0.05$ | $14.863 \pm 0.366$ |
| Total | $\varepsilon_{\text{eff}} = \sum_i \varepsilon_i \cdot \langle 1 - 2w_i \rangle^2 = 33.6 \pm 0.5$ | | |

*6.5.5. Novel aspects of the Belle II flavour tagger.* The major improvements in the Belle II flavour tagger with respect to the Belle flavour tagger consist of: the inclusion of three complementary flavour signatures corresponding to the Kaon-Pion, the FSC and the Maximum P* categories; the consideration of fast kaons as targets in the Fast Hadron category (Belle used only fast pions); the use of more tagging variables within each category; and the employment of robust FBDT and MLP multivariate methods. The Belle flavour tagger is based on multi-dimensional lookup tables and considers 10 flavour signatures, which correspond to the same used by the Belle II flavour tagger apart from the three complementary signatures mentioned above. The flavour signatures used by the Belle flavour tagger are sorted into four categories (Lepton, Kaon, Slow Pion and Lambda). In comparison, the Belle II flavour tagger considers 13 flavour signatures. Furthermore, in the Fast Hadron category, fast pions and fast kaons are considered together. In the Belle approach, each particle candidate could be used only once as a candidate within a certain category according to a specific classification criterion, while in the Belle II flavour tagger each particle candidate is used as a candidate within all categories (disregarding the Lambda category).

In comparison with the previous Belle flavour tagger, which reached an effective efficiency of $(30.1 \pm 0.4)\%$ on Belle data [2], the Belle II flavour tagger reaches an effective efficiency of $(33.6 \pm 0.5)\%$ on Belle data. An additional increase of about 3% in effective efficiency is observed with the Belle II flavour tagger on Belle II MC which is due to the improved track reconstruction and the improved PID performance at Belle II.

## *6.6. Full Event Interpretation*

*6.6.1. Introduction.* Measurements of decays including neutrinos, in particular rare decays, suffer from missing kinematic information. The Full Event Interpretation [82] (FEI) algorithm partially recovers this information by reconstructing one of the $B$ mesons from the decay of the $\Upsilon(4S)$ meson. Information from the reconstructed $B$ meson together with the precisely known initial state infers kinematic and flavour constraints on the remaining $B$ meson, which is considered for a subsequent signal analysis. The two mesons are denoted the tag-side $B_{\text{tag}}$ and the signal-side $B_{\text{sig}}$, respectively as illustrated in Figure 51.

FEI is an essential component in a wide range of analyses, including: the measurement of the CKM matrix element $|V_{ub}|$ through the semileptonic decay $b \to u\ell\nu$; the search for a charged-Higgs effect in $B \to D\tau\nu$; and the precise measurement of the branching fraction of $B \to \tau\nu$.

Belle employed a similar algorithm for tag-side $B$ meson reconstruction called Full Reconstruction (FR) [83]. As further developments FEI includes more decay modes, best candidate selections and provides more automation.

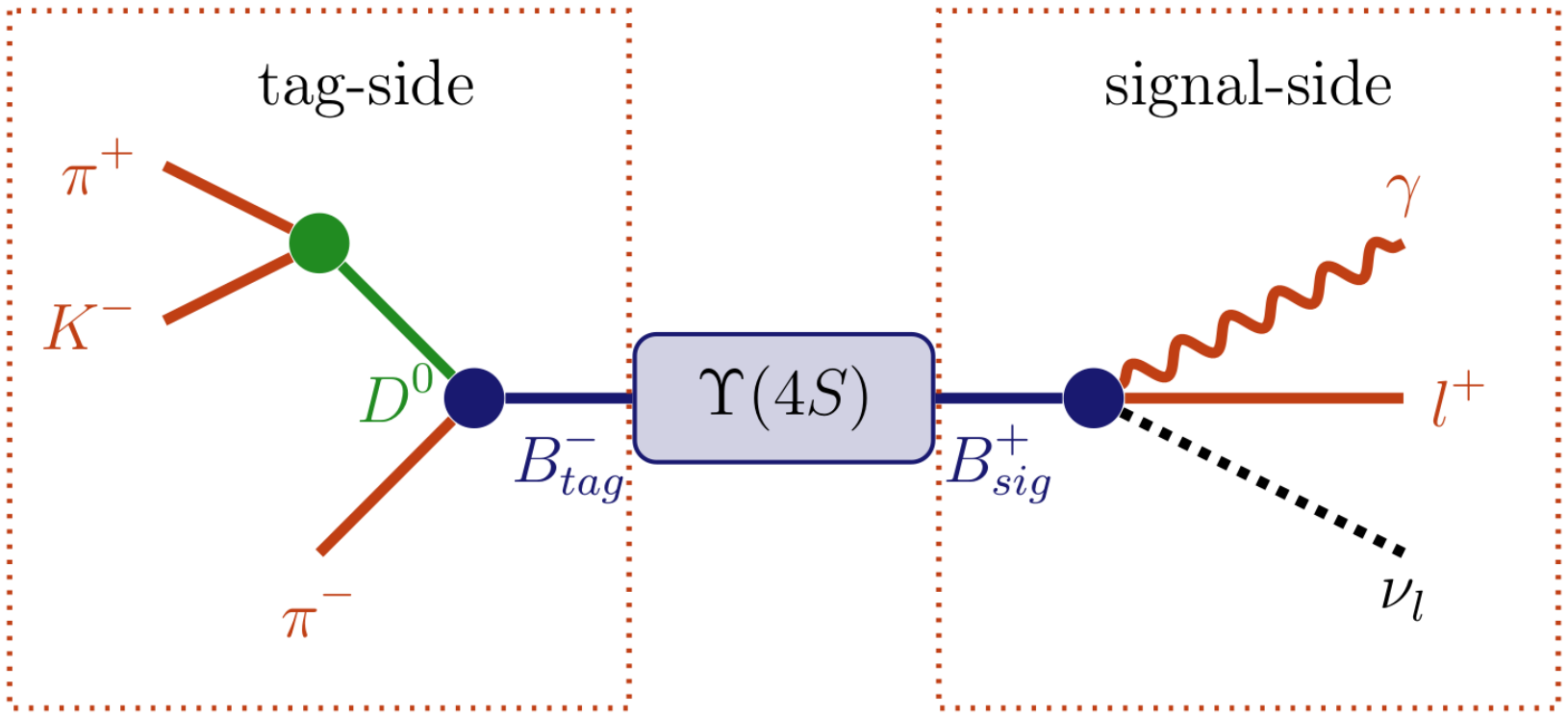

Fig. 51: Diagram showing the use of the FEI in a search for the decay $B^+ \to l^+\nu_l\gamma$. The FEI is used to reconstruct the tag-side $B$ meson, $B^-_{\text{tag}}$, which in this case decays as $B^-_{\text{tag}} \to (D^0 \to \pi^+K^-)\pi^-$. Meanwhile, the signal-side $B$ meson, $B^-_{\text{signal}}$, decays to $l^+\nu_l\gamma$.

*6.6.2. Algorithm.* The basic idea of the FEI is to reconstruct individual particle decay channels, which occur in the decay chain of the $B$ meson, in a hierarchical manner. For each unique decay channel of a particle a Multivariate Classifier (MVC) is trained using simulated events.

The algorithm starts by selecting final-state particle candidates, which include electrons, muons, pions, $K^0_L$ mesons and photons, and training a MVC for each of these using detector information. Building on this, intermediate particle candidates, which include $J/\psi$, $\pi^0$, $D$ and $D^*$ mesons, are reconstructed and a MVC is trained for each reconstructed decay channel. Finally, the $B$ candidates are reconstructed and the corresponding classifiers are trained.

The hierarchical structure of the FEI is shown in Figure 52 and a full list of decay modes reconstructed by the algorithm is given in Table 28.

An important aspect of the FEI is that the MVC of a given particle utilises as an input the MVC outputs of any daughters in addition to kinematic information (e.g daughter 4-momenta, vertex fit information). The MVC for a candidate, in effect, combines all information about a candidate, into a single value, the signal probability. Consequently, candidates from different decay channels can be treated equally in the following reconstruction steps. For instance, FEI reconstructs 15 decay channels of the $D^0$. Afterwards, the generated $D^0$ candidates are used to reconstruct $D^{*0}$ in 2 decay channels. All information about the specific $D^0$ decay channel of the candidate is encoded in its signal probability, which is available to the $D^{*0}$ classifiers. Thereby, the hierarchical approach reconstructs $2 \times 15 = 30$ exclusive decay channels and provides a signal probability for each candidate, which makes use of all available information.

It is computationally unfeasible to handle all possible $B$ meson candidates from all possible particle candidates. The FEI handles this issue of combinatorics by applying pre and post selections on each particle candidate. The pre-selection is performed before computationally demanding tasks such as vertex fits or application of MVCs and consists of a basic selection to remove wrongly reconstructed candidates and a best candidate selection. For the best candidate selection, the candidates are ranked according to a discriminating variable and only the $n$ highest ranked candidates are accepted, where $n$ depends on the particle species and is usually is between 10 and 20. The post-candidate selection makes a tighter selection on the signal probability of the candidate itself and hence uses all available information about the candidate. In addition, it includes a best candidate selection accepting the $m$ (usually in the range 10-20) highest ranked candidates according to signal probability.

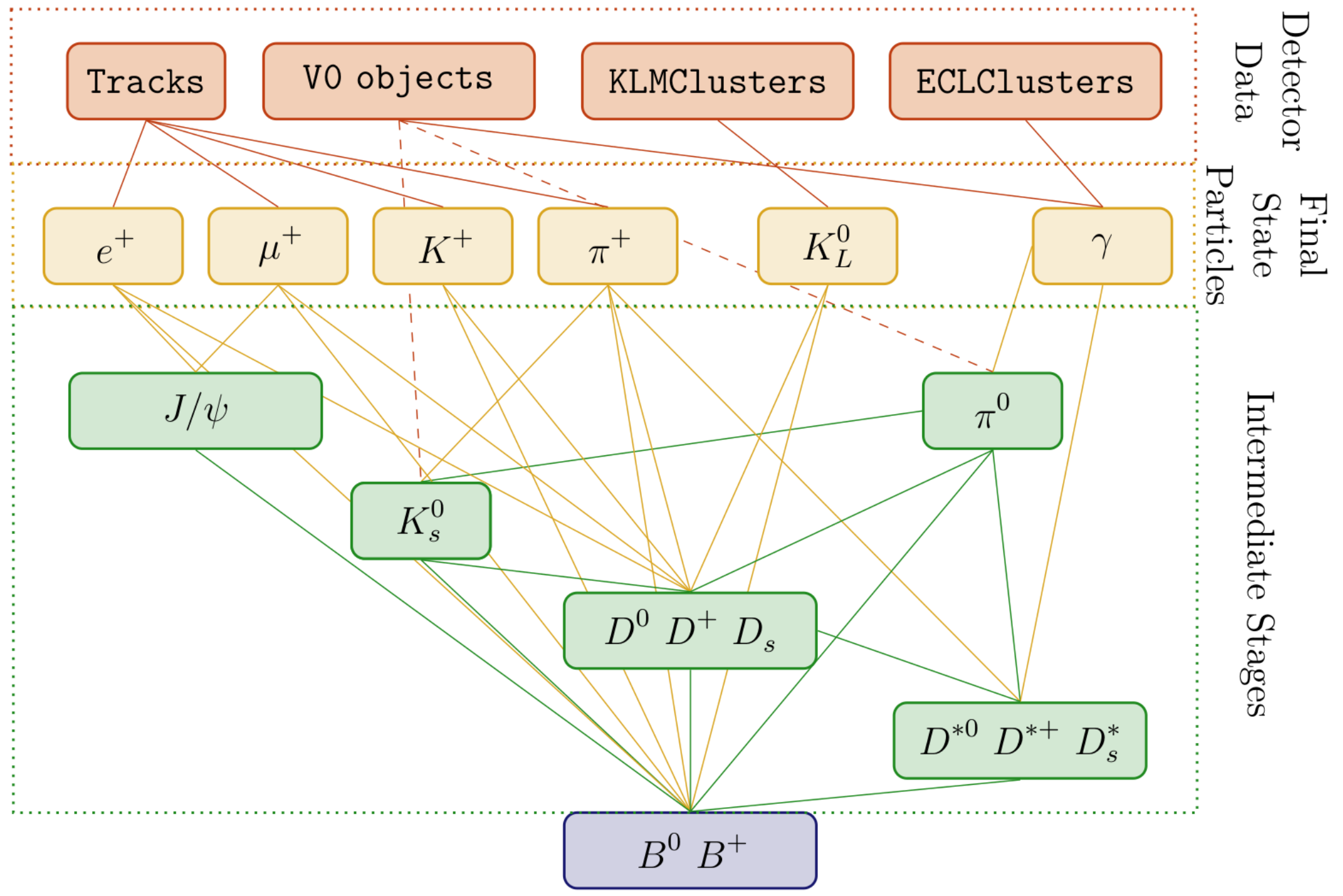

Fig. 52: Hierarchy of the Full Event Interpretation algorithm.

Table 28: $B^+$, $B^0$ and $D$ decay modes included in FEI. The modes listed in the lower parts of the tables were not considered in the Belle FR.

| $B^+$ modes | $B^0$ modes |
|---|---|
| $B^+ \to \overline{D}^0\pi^+$ | $B^0 \to D^-\pi^+$ |
| $B^+ \to \overline{D}^0\pi^+\pi^0$ | $B^0 \to D^-\pi^+\pi^0$ |
| $B^+ \to \overline{D}^0\pi^+\pi^0\pi^0$ | $B^0 \to D^-\pi^+\pi^+\pi^-$ |
| $B^+ \to \overline{D}^0\pi^+\pi^+\pi^-$ | $B^0 \to D_s^+ D^-$ |
| $B^+ \to D_s^+\overline{D}^0$ | $B^0 \to D^{*-}\pi^+$ |
| $B^+ \to \overline{D}^{*0}\pi^+$ | $B^0 \to D^{*-}\pi^+\pi^0$ |
| $B^+ \to \overline{D}^{*0}\pi^+\pi^0$ | $B^0 \to D^{*-}\pi^+\pi^+\pi^-$ |
| $B^+ \to \overline{D}^{*0}\pi^+\pi^+\pi^-$ | $B^0 \to D^{*-}\pi^+\pi^+\pi^-\pi^0$ |
| $B^+ \to \overline{D}^{*0}\pi^+\pi^+\pi^-\pi^0$ | $B^0 \to D_s^{*+}D^-$ |
| $B^+ \to D_s^{*+}\overline{D}^0$ | $B^0 \to D_s^+ D^{*-}$ |
| $B^+ \to D_s^+\overline{D}^{*0}$ | $B^0 \to D_s^{*+}D^{*-}$ |
| $B^+ \to \overline{D}^0 K^+$ | $B^0 \to J/\psi K_S^0$ |
| $B^+ \to D^-\pi^+\pi^+$ | $B^0 \to J/\psi K^+\pi^+$ |
| $B^+ \to J/\psi K^+$ | $B^0 \to J/\psi K_S^0\pi^+\pi^-$ |
| $B^+ \to J/\psi K^+\pi^+\pi^-$ | |
| $B^+ \to J/\psi K^+\pi^0$ | |
| $B^+ \to D^-\pi^+\pi^+\pi^0$ | $B^0 \to D^-\pi^+\pi^0\pi^0$ |
| $B^+ \to \overline{D}^0\pi^+\pi^+\pi^-\pi^0$ | $B^0 \to D^-\pi^+\pi^+\pi^-\pi^0$ |
| $B^+ \to \overline{D}^0 D^+$ | $B^0 \to \overline{D}^0\pi^+\pi^-$ |
| $B^+ \to \overline{D}^0 D^+ K_S^0$ | $B^0 \to D^- D^0 K^+$ |
| $B^+ \to \overline{D}^{*0} D^+ K_S^0$ | $B^0 \to D^- D^{*0} K^+$ |
| $B^+ \to \overline{D}^0 D^{*+} K_S^0$ | $B^0 \to D^{*-} D^0 K^+$ |
| $B^+ \to \overline{D}^{*0} D^{*+} K_S^0$ | $B^0 \to D^{*-} D^{*0} K^+$ |
| $B^+ \to \overline{D}^0 D^0 K^+$ | $B^0 \to D^- D^+ K_S^0$ |
| $B^+ \to \overline{D}^{*0} D^0 K^+$ | $B^0 \to D^{*-} D^+ K_S^0$ |
| $B^+ \to \overline{D}^0 D^{*0} K^+$ | $B^0 \to D^- D^{*+} K_S^0$ |
| $B^+ \to \overline{D}^{*0} D^{*0} K^+$ | $B^0 \to D^{*-} D^{*+} K_S^0$ |
| $B^+ \to \overline{D}^{*0}\pi^+\pi^0\pi^0$ | $B^0 \to D^{*-}\pi^+\pi^0\pi^0$ |

| $D^+$, $D^{*+}$, $D_s^+$ modes | $D^0$,$D^{*0}$ modes |
|---|---|
| $D^+ \to K^-\pi^+\pi^+$ | $D^0 \to K^-\pi^+$ |
| $D^+ \to K^-\pi^+\pi^+\pi^0$ | $D^0 \to K^-\pi^+\pi^0$ |
| $D^+ \to K^-K^+\pi^+$ | $D^0 \to K^-\pi^+\pi^+\pi^-$ |
| $D^+ \to K^-K^+\pi^+\pi^0$ | $D^0 \to \pi^-\pi^+$ |
| $D^+ \to K_S^0\pi^+$ | $D^0 \to \pi^-\pi^+\pi^0$ |
| $D^+ \to K_S^0\pi^+\pi^0$ | $D^0 \to K_S^0\pi^0$ |
| $D^+ \to K_S^0\pi^+\pi^+\pi^-$ | $D^0 \to K_S^0\pi^+\pi^-$ |
| $D^{*+} \to D^0\pi^+$ | $D^0 \to K_S^0\pi^+\pi^-\pi^0$ |
| $D^{*+} \to D^+\pi^0$ | $D^0 \to K^-K^+$ |
| $D_s^+ \to K^+K_S^0$ | $D^0 \to K^-K^+K_S^0$ |
| $D_s^+ \to K^+\pi^+\pi^-$ | $D^{*0} \to D^0\pi^0$ |
| $D_s^+ \to K^+K^-\pi^+$ | $D^{*0} \to D^0\gamma$ |
| $D_s^+ \to K^+K^-\pi^+\pi^0$ | |
| $D_s^+ \to K^+K_S^0\pi^+\pi^-$ | |
| $D_s^+ \to K^-K_S^0\pi^+\pi^+$ | |
| $D_s^+ \to K^+K^-\pi^+\pi^+\pi^-$ | |
| $D_s^+ \to \pi^+\pi^+\pi^-$ | |
| $D_s^{*+} \to D_s^+\pi^0$ | |
| $D^+ \to \pi^+\pi^0$ | $D^0 \to K^-\pi^+\pi^0\pi^0$ |
| $D^+ \to \pi^+\pi^+\pi^-$ | $D^0 \to K^-\pi^+\pi^+\pi^-\pi^0$ |
| $D^+ \to \pi^+\pi^+\pi^-\pi^0$ | $D^0 \to \pi^-\pi^+\pi^+\pi^-$ |
| $D^+ \to K^+K_S^0K_S^0$ | $D^0 \to \pi^-\pi^+\pi^0\pi^0$ |
| $D^{*+} \to D^+\gamma$ | $D^0 \to K^-K^+\pi^0$ |
| $D_s^+ \to K_S^0\pi^+$ | |
| $D_s^+ \to K_S^0\pi^+\pi^0$ | |
| $D_s^{*+} \to D_s^+\pi^0$ | |

*6.6.3. Hadronic, Semileptonic and Inclusive Tagging.* There are three distinct tag-side reconstruction methods in common use: hadronic, semileptonic and inclusive tag-side reconstruction.

- Hadronic tagging solely uses hadronic decay channels for $B$ reconstruction. Hence, the kinematics of the reconstructed candidates are well known and the tagged sample is pure. The method is typically low in efficiency, at the order of 0.1%.
- Semileptonic tagging uses semileptonic $B$ decays. Due to the high branching fraction of semileptonic decays this approach usually has a higher tagging efficiency. This method suffers from missing kinematic information due to the neutrino in the final state of the decay. Hence, the sample is not as pure as in the hadronic case.
- Inclusive tagging combines the four-momenta of all particles in the rest of the event of the signal-side $B$ candidate. The achieved tagging efficiency is usually one order of magnitude above the hadronic and semileptonic tagging approaches. However, the decay topology is not explicitly reconstructed, and the method suffers from high background.

FEI combines the first two methods, hadronic and semileptonic tag-side reconstruction, into a single algorithm.

*6.6.4. Training modes.* FEI is trained on MC and subsequently applied to collision data. There are three different types of events one has to consider in the training and application of FEI.

- $B\bar{B}$ events: $\Upsilon(4S) \to B\bar{B}$ for charged and neutral $B\bar{B}$ pairs.
- Continuum events; $e^+e^- \to c\bar{c}, s\bar{s}, d\bar{d}, u\bar{u}$.
- Signal events; $e^+e^- \to \Upsilon(4S) \to B\bar{B}$, where one $B$ decays generically and the other decays in a signal channel such as $B^+ \to \tau^+\nu$.

The final classifier output for the $B_{\text{tag}}$ mesons is used to separate signal from background. The dominant background is typically from $B\bar{B}$ events. In the study presented here, continuum suppression criteria are not applied and the FEI is trained solely using simulated $B\bar{B}$ events.

*6.6.5. Performance estimations.* The performance of FEI can quantified by the tag-side efficiency, which is defined as the number of correctly reconstructed tag-side $B$ mesons divided by the total number of $\Upsilon(4S)$ events. While in MC events correctly reconstructed events can be determined using MC matching in data a maximum likelihood fit in a discriminating variable must be used to distinguish correctly reconstructed tag-side $B$ mesons. Another important performance estimate of the FEI is the purity, which is defined as the number of correctly reconstructed tag-side $B$ mesons divided by the number of events in which a tag-side $B$ meson is reconstructed. Tighter selections on the signal probability of the tag-side $B$ meson increase purity with the trade off of a lower tag-side efficiency. The maximal achievable tag-side efficiency is particularly important as it is directly related to maximal achievable signal efficiency for a given measurement.

The maximum tag-side efficiency of the FEI as estimated on Belle and Belle II MC is compared with that of the predecessor algorithm FR in Table 29. We observe much higher efficiency with the Belle II algorithm in Belle MC, for both hadronic and semileptonic channels, particularly for loose selection. This is largely attributable to the newly added modes in FEI, which are less clean than those used commonly with FR. Figure 53 shows a comparison between the beam-energy constrained mass distributions of the FEI and FR with Belle data for tight and loose selections on the signal probability. We observe a much higher efficiency with the Belle II algorithm for purities below 50% (75%) for charged (neutral) $B_{\text{tag}}$ reconstruction. This is mostly likely due to the newly added modes in FEI, which are less clean than those used commonly with FR. This, additionally, results in the algorithm having a much higher maximum tag-side efficiency as shown in Table 29.

Figure 54 shows the tag-side efficiency against purity for $B^+$ and $B^0$ tag-side candidates in Belle II MC. The points correspond to the scan of the tag-side $B$ meson signal probability starting from 0.01 with steps of 0.04. We find that beam-induced background has a large effect and reduces the achievable tag-side efficiency for a given purity.

*6.6.6. Calibration.* There can be substantial differences between the tag-side reconstruction efficiency in data and simulation given the large number of decay modes considered by FEI, the large number of MVCs and the reliance on simulation to train FEI. An important systematic error in analyses using full $B$ reconstruction methods is the FEI efficiency calibration, which directly affects absolute branching fraction measurements. In the case of semileptonic $B$ decay measurements, it has become the dominant systematic uncertainty and hence is a limiting factor in precision measurements at Belle II. Several techniques for calibration have been used in Belle, and are described in turn.

Table 29: Tag-side efficiency defined as the number of correctly reconstructed tag-side $B$ mesons divided by the total number of $\Upsilon(4S)$ events. Belle II has much higher beam background, which can affect the total efficiency.

| Tag | FR, Belle | FEI, Belle MC | FEI, Belle II MC |
|---|---|---|---|
| Hadronic $B^+$ | 0.28 % | 0.49 % | 0.61 % |
| Semileptonic $B^+$ | 0.67 % | 1.42 % | 1.45 % |
| Hadronic $B^0$ | 0.18 % | 0.33% | 0.34 % |
| Semileptonic $B^0$ | 0.63 % | 1.33% | 1.25 % |

- $B \to D^{(*)}\ell\nu$ calibration. Events are double tagged, where the signal side is reconstructed in a known semileptonic decay mode, in bins of the tag quality variables. This has been used in $B \to X_u\ell\nu$ analyses. The systematic errors were approximately 4.5%, shared between statistical (1.5%), reconstruction (2.7%), and branching fraction uncertainties (3%) [84]. The detection uncertainties are mostly based on data driven techniques, while the branching fractions are more difficult to improve in the future.
- $B \to X\ell\nu$ calibration. Events are also double tagged, however the signal side selected only via the presence of a charged lepton originating from a semileptonic $B$ decay. This has been used in precision exclusive $B \to D^{(*)}\ell\nu$ decay analyses [85]. The technique is systematics limited but higher precision than the $B \to D^{(*)}\ell\nu$ calibration approach.
- Control mode calibration. An analysis sideband region is chosen that is enhanced in a well known decay mode, and calibrated accordingly. This technique has been used by rare decay analyses where it is useful to calibrate tag efficiencies with topologies similar to the signal process.

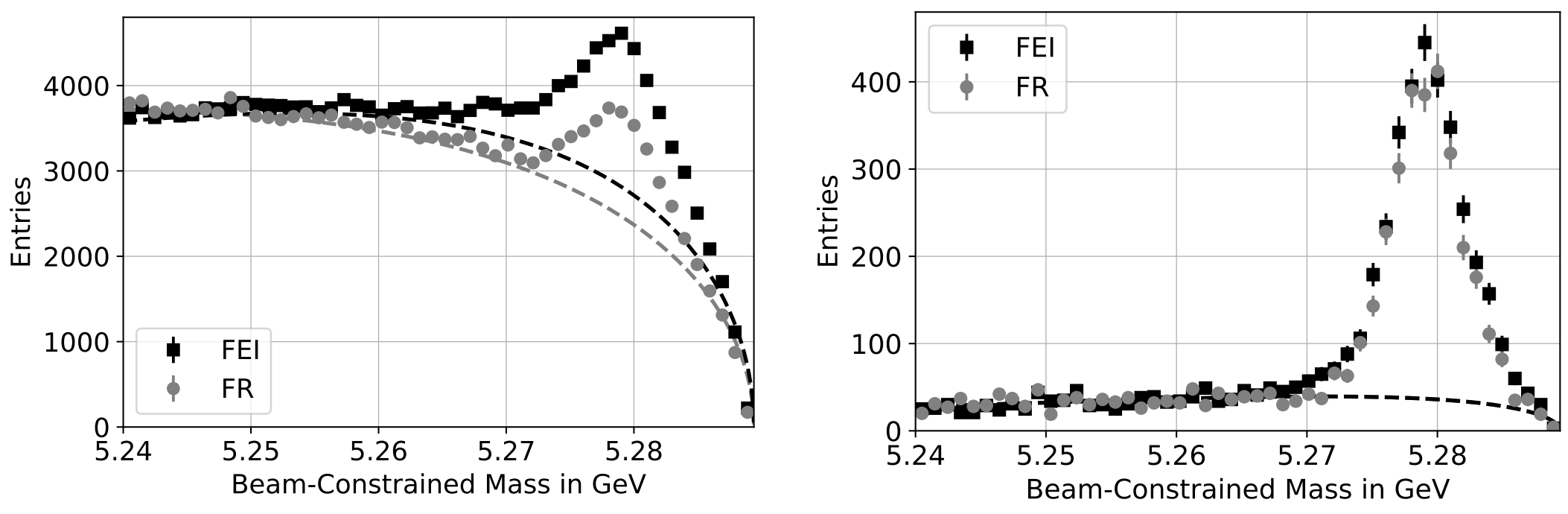

a) Beam-energy constrained mass of charged hadronic tag candidates with a loose threshold on the classifier output.

b) Beam-energy constrained mass of charged hadronic tag candidates with a tight threshold on the classifier output.

Fig. 53: Performance of $B$ candidate reconstruction in hadronic decay modes with the FEI and FR algorithms on Belle data.

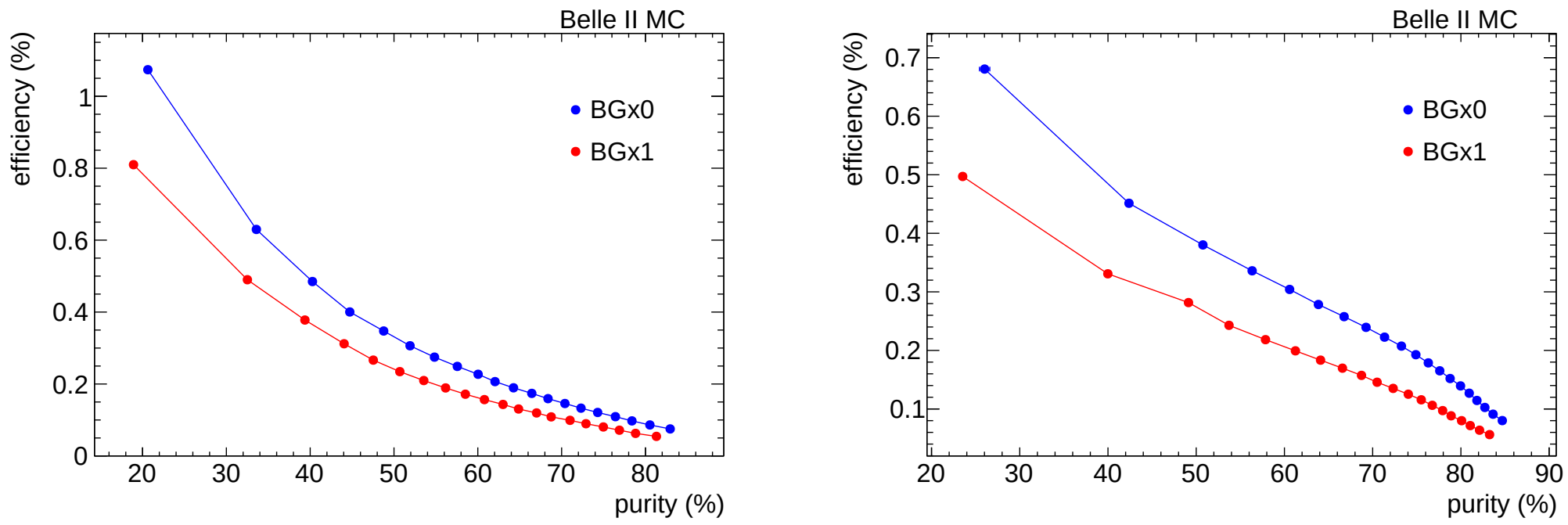

Fig. 54: Efficiency versus purity of charged $B$ (left) and neutral $B$ (right) candidates reconstructed with FEI in hadronic modes. Two beam-induced background scenarios are compared, zero background (BGx0) and nominal background (BGx1).

## 7. Theory overview

*Section author(s): C. Hanhart, S. Hashimoto, T. Kaneko, E. Kou, A.S. Kronfeld, U. Nierste, S. Prelovsek, S.R. Sharpe, J. Shigemitsu, S. Simula*

### 7.1. Introduction

The source of flavour violation in the Standard Model (SM) is the Yukawa interaction between fermions and the Higgs doublet

$$\Phi = \begin{pmatrix} \phi^+ \\ v + \frac{H+i\chi}{\sqrt{2}} \end{pmatrix} . \tag{46}$$

Here $H$ is the field of the physical Higgs fields, $v = 174\,\mathrm{GeV}$ is the vacuum expectation value (vev), and $\phi^+$, $\chi$ are the pseudo-Goldstone fields related to longitudinally polarised $W^+$ and $Z$ bosons. The quark Yukawa Lagrangian reads

$$\mathcal{L}^q_{\text{Yuk}} = -\,\overline{Q}_j\, Y^d_{jk}\, \Phi\, d'_{Rk} \;-\; \overline{Q}_j\, Y^u_{jk}\, \epsilon\, \Phi^* u'_{Rk} + \text{h.c.}\,, \tag{47}$$

where $j, k = 1, 2, 3$ labels the generation (repeated indices are summed over) and

$$\epsilon = \begin{pmatrix} 0 & 1 \\ -1 & 0 \end{pmatrix} .$$

The right-handed quark fields $u'_{Rk}$, $d'_{Rk}$ are singlets of the electroweak gauge group SU(2), while the left-handed quarks form SU(2) doublets:

$$Q_j = \begin{pmatrix} u'_{Lj} \\ d'_{Lj} \end{pmatrix} .$$

The arbitrary complex 3×3 Yukawa matrices $Y^{u,d}$ give rise to the two quark mass matrices $M^{u,d} = Y^{u,d} v$. To diagonalise these matrices we perform unitary rotations of the fields $u'_{L,Rk}$, $d'_{L,Rk}$ (called "weak eigenstates") to a new basis of "mass eigenstates": $u'_{L,Rj} = S^u_{L,Rjk}\, u_{L,Rk}$, $d'_{L,Rj} = S^d_{L,Rjk}\, d_{L,Rk}$.

The unprimed fields correspond to the physical particles, and where convenient we write $u_R = u_{R1}$, $c_R = u_{R2}$, and $t_R = u_{R3}$ with an analogous notation for the left-handed and down-type quark fields. The piece of $\mathcal{L}^q_{\text{Yuk}}$ containing the —now diagonal— mass matrices reads:

$$\begin{aligned} \mathcal{L}_m &= -\, m_{u_j} \left[\bar{u}_{Lj} u_{Rj} + \bar{u}_{Rj} u_{Lj}\right] - m_{d_j} \left[\bar{d}_{Lj} d_{Rj} + \bar{d}_{Rj} d_{Lj}\right] \\ &\equiv -\sum_{q=u,d,s,c,b,t} m_q\, \overline{q} q\, . \end{aligned}$$

Here $m_{u,\ldots t}$ are the quark masses and we have introduced the usual four-component Dirac field $q \equiv q_L + q_R$ (recalling $\overline{q}_R q_R = \overline{q}_L q_L = 0$). The four unitary matrices $S^{u,d}_{L,R}$ drop out everywhere with one important exception: The Cabibbo-Kobayashi-Maskawa (CKM) matrix

$$V = S^{u\dagger}_L S^d_L \tag{48}$$

appears in the couplings of the $W$ boson to quarks:

$$\mathcal{L}^q_W = \frac{g}{\sqrt{2}} \left[ V_{jk}\, \bar{u}_{Lj}\, \gamma^\mu d_{Lk}\, W^+_\mu + V^*_{jk}\, \bar{d}_{Lk}\, \gamma^\mu u_{Lj}\, W^-_\mu \right] . \tag{49}$$

CKM elements are commonly labeled with the quark flavours, so that *e.g.* $V_{cb} \equiv V_{u_2 d_3} \equiv V_{23}$. $\mathcal{L}^q_W$ violates the discrete symmetries parity ($P$), time reversal ($T$), and charge conjugation ($C$). The parity transformation $\vec{x} \to -\vec{x}$ exchanges the left-handed quark fields in the Lagrangian (49) with their right-handed counterparts. Since the $W$ boson does not couple at all to right-handed quarks, $P$ violation in the SM is maximal. The same is true for $C$ violation, because $C$ maps left-handed fermion fields onto right-handed anti-fermion fields. However, the combination of the two transformations, $CP$, does not change the chirality of the fermion fields in Eq. (49):

$$\bar{u}_{Lj}\,\gamma^\mu d_{Lk}\,W^+_\mu \quad \stackrel{CP}{\longleftrightarrow} \quad \bar{d}_{Lk}\,\gamma^\mu u_{Lj}\,W^-_\mu\,. \tag{50}$$

Apparently $\mathcal{L}^q_W$ conserves $CP$ if $V_{jk}$ is real. However, $\mathrm{Im}V_{jk}\neq 0$ does not imply that $CP$ is violated: If we can make $V_{jk}$ real by multiplying the quark fields with unphysical phase factors, $d_k \to d_k \exp(i\phi_{d_k})$ and $u_j \to u_j \exp(i\phi_{u_j})$, $CP$ is conserved as well. You may easily check that this rephasing of the quark fields changes $V_{jk}$ in Eq. (49) to $V_{jk}\exp(i\phi_{d_k} - i\phi_{u_j})$. In a world with just two fermion generations it is always possible to render $V_{jk}$ real. Kobayashi and Maskawa realised that this is no more true once you add a third generation and thereby correctly identified the dominant mechanism of $CP$ violation in flavour-changing transitions [86]. A unitary $3\times 3$ matrix involves 6 complex phases, five of which can be removed by the re-phasing transformation described above. The remaining phase is a physical, $CP$ violating parameter, the Kobayashi-Maskawa (KM) phase $\delta_{\mathrm{KM}}$.

Flavour-changing transitions among fermions with the same electric charge are called flavour-changing neutral current (FCNC) processes. The unitarity of the matrices $S^{u,d}_{L,Rjk}$ and the CKM matrix $V$ in Eq. (48) leads to a dramatic suppression of FCNC transitions, which is referred to as the Glashow-Iliopoulos-Maiani (GIM) mechanism. The tree-level GIM mechanism renders the couplings of the neutral gauge bosons ($Z$, photon $\gamma$, gluon $g$) flavour-diagonal. We exemplify this for $Z$ coupling to right-handed down-type quarks here:

$$\begin{aligned} Z^\mu\,\bar{d}'_{Rj}\gamma_\mu d'_{Rj} &= Z^\mu\,\bar{d}_{Rk}S^{d\dagger}_{Rkj}\gamma_\mu S^d_{Rjl}d_{Rl} \\ &= Z^\mu\,\bar{d}'_{Rk}\gamma_\mu d_{Rk}. \end{aligned}$$

In the last step the unitarity relation $S^{d\dagger}_{Rkj}S^d_{Rjl}=\delta_{kl}$ has been used. Historically, the aim to understand the suppression of the FCNC process $s\to d\mu^+\mu^-$ led Glashow, Iliopoulos, and Maiani to postulate the existence of a fourth quark, charm, to build an SU(2) doublet $Q_2=(c_L,s_L)^T$ in analogy to $Q_1=(u_L,d_L)^T$: The GIM mechanism only works if the gauge interactions treat all fermion generations on the same footing, so that the described unitary rotations are meaningful. While FCNC processes are forbidden at tree level, they nevertheless occur through loop diagrams. Figs. 55 and 56 show two prominent examples, the $B^0_d - \overline{B}^0_d$ mixing box and the gluon penguin diagrams. The GIM mechanism also affects such loop-induced FCNC transitions: The diagrams of Figs. 55 and 56 involve contributions from different quarks on the internal lines, namely $u$, $c$, and $t$. These contributions differ from each other only by the CKM elements accompanying the $W$ couplings and by the masses of the virtual up-type quarks, *e.g.* for the penguin amplitude of Fig. 56 we may write

$$\mathcal{A} = \sum_{q=u,c,t} V^*_{qb}V_{qd}\,f\left(\frac{m_q^2}{M_W^2}\right), \tag{51}$$

where $M_W$ is the mass of the $W$ boson. Now CKM unitarity implies $V^*_{cb}V_{cd} = -V^*_{tb}V_{td} - V^*_{ub}V_{ud}$ and we may eliminate $V^*_{cb}V_{cd}$ from Eq. (51):

$$\mathcal{A} = \; V^*_{tb}V_{td}\left[f\left(\frac{m_t^2}{M_W^2}\right) - f\left(\frac{m_c^2}{M_W^2}\right)\right] + V^*_{ub}V_{ud}\left[f\left(\frac{m_u^2}{M_W^2}\right) - f\left(\frac{m_c^2}{M_W^2}\right)\right] \tag{52}$$

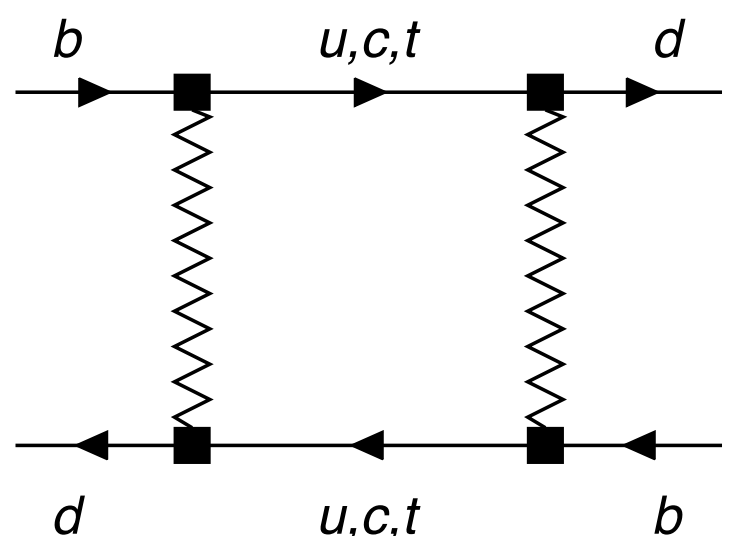

Fig. 55: Box diagram for $B_d^0$–$\overline{B}_d^0$ mixing.

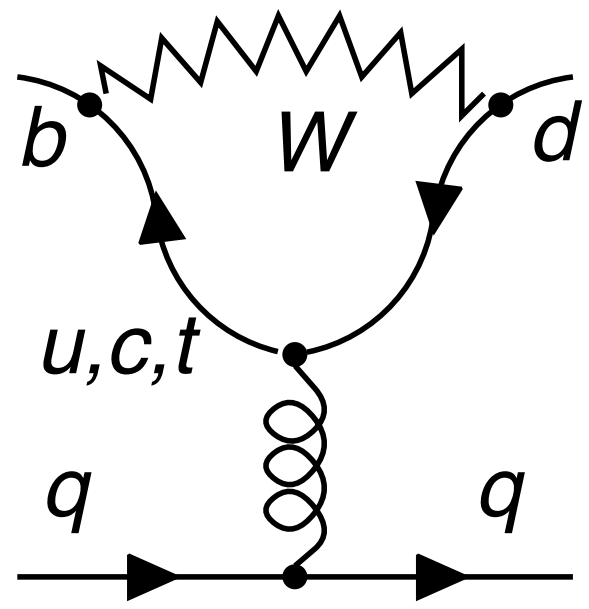

Fig. 56: Penguin diagram for the decay $b \to dq\overline{q}$ with the curly line representing a gluon. For $q = u$ or $q = c$ there is also a tree diagram.

We observe the terms in square brackets vanish if the two masses involved are equal. Nowadays this feature is usually meant when people refer to the GIM mechanism. Since $m_c - m_u \ll M_W$, the second term in Eq. (52) is GIM suppressed. We realise that the large value of $m_t$ makes the first term unsuppressed. Historically, the unexpectedly large $B_d^0$–$\overline{B}_d^0$ mixing observed in 1987 at the ARGUS detector at DESY was the first hint of a heavy top quark. The situation is different in charm physics: Here the quarks on the internal line are $d, s, b$ and moreover the diagrams with virtual $t$ come with the tiny CKM factor $V_{cb}^* V_{ub}$. Thus the SM predictions for FCNC transitions of charm quarks are tiny.

In summary, flavour physics probes the Yukawa sector of the Standard Model. Theories going beyond the SM (BSM models) may contain a larger Higgs sector with new Yukawa couplings or may involve flavour-violating parameters which are unrelated to Higgs-fermion couplings. FCNC transitions are suppressed by a loop factor and small CKM elements. In a large class of FCNC observables (including all FCNC charm transitions and FCNC decays of charged leptons) there is an additional GIM suppression. These features make FCNC transitions very sensitive to new physics, with the power to probe virtual effects of particles with masses above 100 TeV (and the actual sensitivity depending on the considered model). To date flavour physics is the only field in which $CP$ violation has been observed. The Standard Model accommodates $CP$ violation in flavour-changing transitions through a single parameter, the Kobayashi-Maskawa (KM) phase in the CKM matrix.

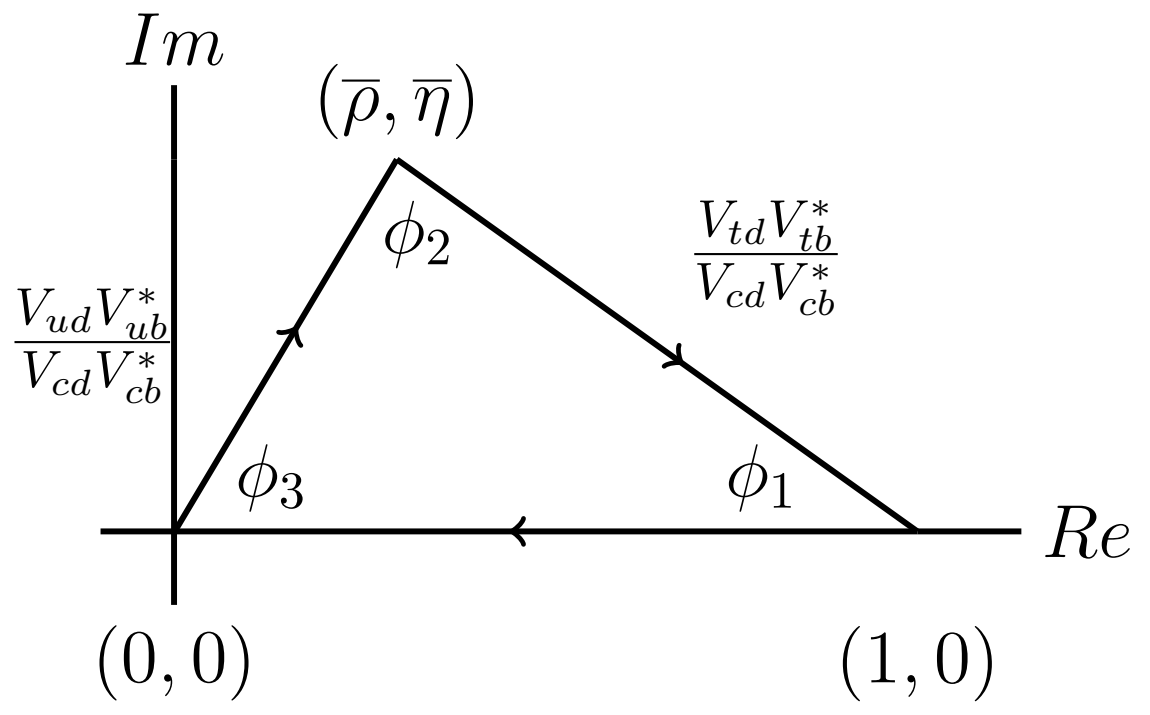

Fig. 57: The unitarity triangle.

### *7.2. CKM matrix and unitarity triangle*

The CKM matrix is the $3 \times 3$ unitary matrix

$$\mathbf{V} = \begin{pmatrix} V_{ud} & V_{us} & V_{ub} \\ V_{cd} & V_{cs} & V_{cb} \\ V_{td} & V_{ts} & V_{tb} \end{pmatrix}, \quad \mathbf{V}\mathbf{V}^\dagger = 1, \tag{53}$$

which can be parameterised by four free parameters. The flavour physics program at Belle II, just like at its predecessors, will have an ability to over-constrain these parameters and a potential to discover significant deviations from SM expectations.

The standard choice of the CKM matrix is obtained as a product of three rotation matrices ordered as [87, 88]:

$$\mathbf{V} = \begin{pmatrix} c_{12}c_{13} & s_{12}c_{13} & s_{13}e^{-i\delta} \\ -s_{12}c_{23} - c_{12}s_{23}s_{13}e^{i\delta} & c_{12}c_{23} - s_{12}s_{23}s_{13}e^{i\delta} & s_{23}c_{13} \\ s_{12}s_{23} - c_{12}c_{23}s_{13}e^{i\delta} & -c_{12}s_{23} - s_{12}c_{23}s_{13}e^{i\delta} & c_{23}c_{13} \end{pmatrix} \tag{54}$$

where $c_{ij} = \cos\theta_{ij}, s_{ij} = \sin\theta_{ij}$ and $\delta$ is the $CP$ violating phase. With experimental knowledge of the hierarchy $|V_{ub}|^2 \ll |V_{cb}|^2 \ll |V_{us}|^2$, an expansion was introduced [89]. By defining [90]

$$s_{12} \equiv \lambda, \quad s_{23} \equiv A\lambda^2, \quad s_{13}e^{-i\delta} = A\lambda^3(\rho - i\eta), \tag{55}$$

where $\lambda \simeq 0.22$, we can re-write the CKM matrix in terms of the four new parameters, $A, \lambda, \rho, \eta$

$$\mathbf{V} = \begin{pmatrix} 1 - \frac{1}{2}\lambda^2 & \lambda & A\sqrt{\rho^2+\eta^2}e^{-i\delta}\lambda^3 \\ -\lambda & 1 - \frac{1}{2}\lambda^2 & A\lambda^2 \\ A(1-\sqrt{\rho^2+\eta^2}e^{i\delta})\lambda^3 & -A\lambda^2 & 1 \end{pmatrix} + \mathcal{O}(\lambda^4) \tag{56}$$

which is, up to $\mathcal{O}(\lambda^4)$, equivalent to the Wolfenstein parameterisation [89]. Notice that the definition in Eq. (55) implies that the unitarity condition can be written in terms of $A, \lambda, \rho, \eta$ at all order in $\lambda$.

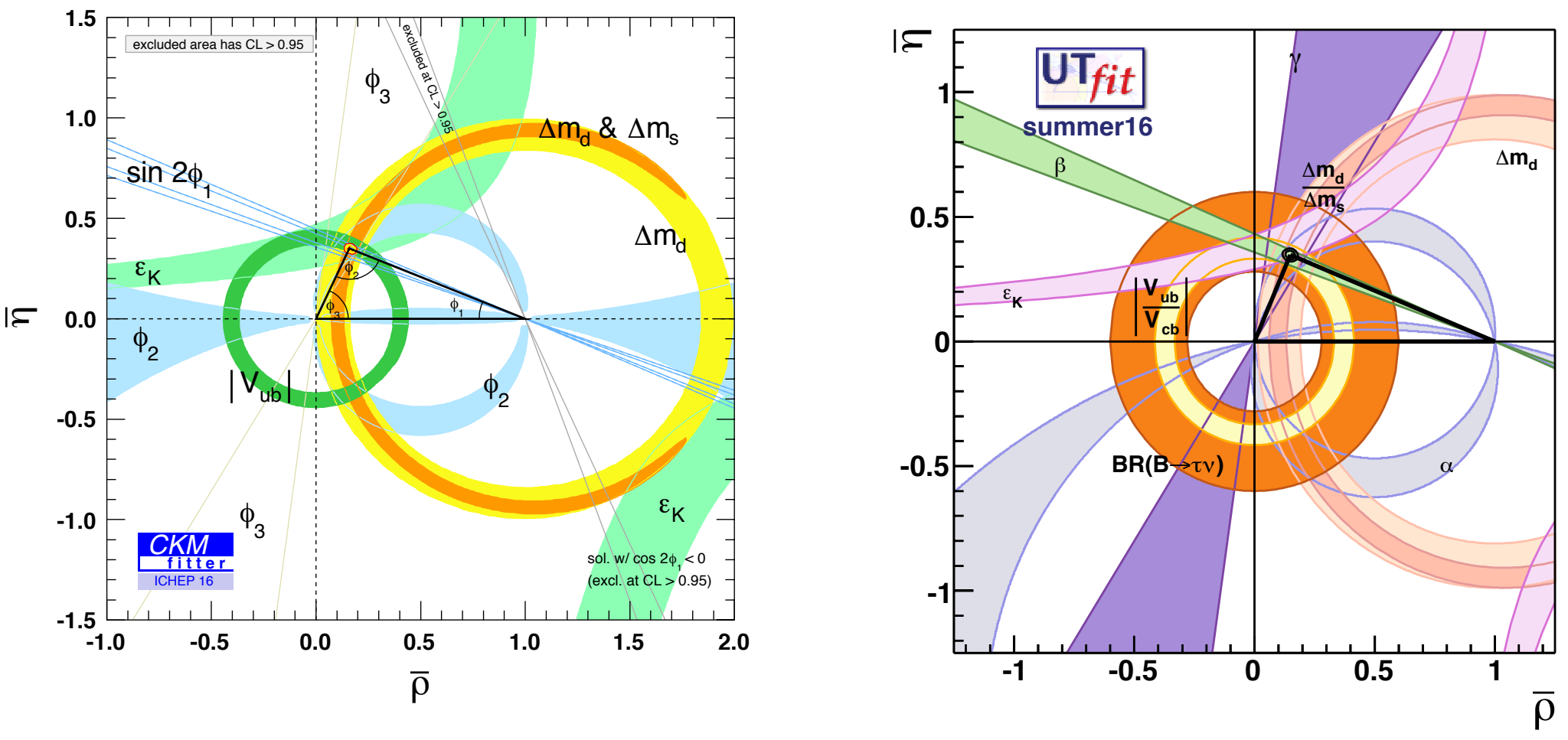

Fig. 58: The current situation of the unitarity triangle constraints by CKMfitter (left) and UTfit (right).

The unitarity condition of the CKM matrix leads to nine independent equations . The one most relevant to $B$ physics is:

$$V_{ud}^* V_{ub} + V_{cd}^* V_{cb} + V_{td}^* V_{tb} = 0 \tag{57}$$

In order to form the *unitarity triangle* (UT), we divide this equation by $V_{cd}V_{cb}^*$[9]

$$1 + \frac{V_{ud}^* V_{ub}}{V_{cd} V_{cb}^*} + \frac{V_{td}^* V_{tb}}{V_{cd} V_{cb}^*} = 0 \tag{58}$$

and then, introduce new parameters [91]:

$$\bar{\rho} + i\bar{\eta} \equiv -\frac{V_{ud}V_{ub}^*}{V_{cd}V_{cb}^*}, \quad 1 - (\bar{\rho} + i\bar{\eta}) \equiv -\frac{V_{td}V_{tb}^*}{V_{cd}V_{cb}^*} \tag{59}$$

which is related to $\rho, \eta$ in Eq. (55) as[10]

$$\rho + i\eta = \frac{\sqrt{1 - A^2\lambda^4}(\bar{\rho} + i\bar{\eta})}{\sqrt{1-\lambda^2}[1 - A^2\lambda^4(\bar{\rho} + i\bar{\eta})]} \tag{60}$$

The *unitarity triangle* is then obtained by drawing Eq. (58) on the $\bar{\rho} - \bar{\eta}$ plane (see Fig. 57). The three angles are defined as:

$$\phi_1 \equiv \arg\left[-\frac{V_{cd}V_{cb}^*}{V_{td}V_{tb}^*}\right], \quad \phi_2 \equiv \arg\left[-\frac{V_{td}V_{tb}^*}{V_{ud}V_{ub}^*}\right], \quad \phi_3 \equiv \arg\left[-\frac{V_{ud}V_{ub}^*}{V_{cd}V_{cb}^*}\right] \tag{61}$$

These angles are also known as $\phi_1 = \beta, \phi_2 = \alpha$ and $\phi_3 = \gamma$.

The latest results of the global fit to UT parameters is shown in Fig. 58. Two sides of the triangle are determined from measurements of decay rates $|V_{ub}|/|V_{cb}|$ and mixing $\Delta M_d/\Delta M_s$.

[9] Note that $V_{cd}V_{cb}^* = -A\lambda^3 + \mathcal{O}(\lambda^7)$, but in practice we often assume that $V_{cd}V_{cb}^*$ is real and $|V_{cd}V_{cb}^*| = A\lambda^3$.

[10] Note that the definition of $\bar{\rho}, \bar{\eta}$ in Eq. (59) and the relation in Eq.(60) are to all orders in $\lambda$. The $\bar{\rho}, \bar{\eta}$ here are equivalent to those in [90] up to $\mathcal{O}(\lambda^4)$.

These constraints, as well those coming from indirect $CP$ violation measurements in Kaons, $\epsilon_K$, depend strongly on the hadronic parameter inputs. The theoretical computations of these parameters will be reviewed in the next subsection. We emphasise that improvements in the lattice QCD computations become crucial for a more precise determination of the sides of the triangle.

The bound from the $|V_{ub}|/|V_{cb}|$ constraint shown in these figures is obtained through combinations of various measurements. However, as reviewed in chapter 8 there are tensions between $V_{ub}$ determinations with exclusive and inclusive semi-leptonic $b \to ul\nu$ decays in addition to a hint of a deviation from the SM in the tauonic pure-leptonic $B \to \tau\nu$ decay. New physics contributions can be different for these three types of processes (see chapter 17) and improved Belle II measurements will provide us with a more detailed picture. We note in passing that the $V_{us}$ element, which determines the $\lambda$ parameter, also shows some disagreement between its determination by the exclusive $K$ decays and the inclusive $\tau$ decay. The latter can be studied at Belle II, which can help to improve the $V_{us}$ determination (see chapter 15).

The angles $\phi_1, \phi_2$ are determined from measurements of time-dependent $CP$ asymmetries, as detailed in chapter 10. The angle $\phi_1$ has been measured from the interference between $B_d$ oscillation with $b \to c\bar{c}s$ decays, and was an outstanding success of the BaBar and Belle collaborations. Contrary to the other oscillation parameter, $\Delta M_d$, most of the hadronic uncertainties cancel out in this $CP$ violating observable and it therefore provides a very clean and precise determination of $\phi_1$. $B_d$ oscillation arises from a FCNC $b\bar{d} - \bar{b}d$ coupling, which is induced by the $W$-boson box diagram in the SM as shown in the previous section. Various new physics models predict extra contributions to the $b\bar{d} - \bar{b}d$ coupling, via either tree or loop diagrams. For either process, we should keep in mind that there is correlation between new physics contributions to $\phi_1$ measured in $b \to c\bar{c}s$ decays, and those to $\Delta M_d$ since both come from the $b\bar{d} - \bar{b}d$ coupling. Currently there is reasonable agreement between the average value of $|V_{ub}|$, and $\phi_1$. If we take for example the $|V_{ub}|$ value derived from inclusive semi-leptonic decays or $B \to \tau\nu$, there is a tension. Belle II will clarify this situation.

The angle $\phi_2$ is measured from interference between the $b \to ud\bar{u}$ tree and the $b \to dq\bar{q}$ penguin ($q = u, d$) process, with the decays such as $B \to \pi\pi, \pi\rho, \rho\rho$. In minimal models, new physics contributions to the $b \to dq\bar{q}$ penguin loop diagram and $b\bar{d} - \bar{b}d$ diagram can be strongly correlated while there are many new physics models which contribute to only one of them. The experimental error on $\phi_2$ is still very large and more precise measurements by Belle II have the potential to reveal a deviation from the other UT fit inputs.

The third angle $\phi_3$ is measured via the $CP$ asymmetry which occurs due to the interference between different tree level diagrams (see chapter 11). Decay modes of the type $B \to D^{(*)}K^{(*)}$ and $B \to D^{(*)}\pi$ , where the $D$ meson decays to a variety of final states, can be used to obtain a very precise determination of $\phi_3$. The theoretical uncertainty, which comes from the loop diagrams, can be very well under control. The measurement of $\phi_3$ is highly statistics limited, and will be greatly improved in the era of Belle II. If $\phi_3$ turns out to be inconsistent with the other UT constraints, there is a possibility of new physics contributing to tree level $B \to D^{(*)}K^{(*)}(\pi)$ processes. On the other hand new physics contributions could be in the other measurements, especially those in loop induced observables (see chapter 11 for more detail).

In summary, there is excellent potential at Belle II to discover new physics through precision tests of the unitarity triangle. In order to clarify the significance of the agreement or deviation, global fits may be necessary. A more detailed discussion on this aspect can be found in chapter 18

## *7.3. Effective Hamiltonian*

Flavour-changing amplitudes involve widely separated mass scales, ranging from $\Lambda_{\rm QCD} \sim 350\,{\rm MeV}$ over $\overline{m}_c \sim 1.25\,{\rm GeV}$ and $\overline{m}_b \sim 4.3\,{\rm GeV}$ to $M_W = 80.4\,{\rm GeV}$ and $\overline{m}_t \sim 165\,{\rm GeV}$. The QCD coupling $\alpha_s = g_s^2/(4\pi)$ changes dramatically over this range of energies: While we can use perturbation theory (*i.e.* calculate Feynman diagrams with quarks and gluons) for QCD effects associated with scales of $\overline{m}_b$ and above, this is not possible for the dynamics associated with the energy scale $\Lambda_{\rm QCD}$ related to genuine non-perturbative effects like the confinement of quarks and gluons into colourless hadrons. In a given calculation, we must separate the physics of the different scales to apply different calculational methods to the different energy regimes. To this end an important theoretical tool is the effective weak Hamiltonian. For the description of the decay of $b$-flavoured hadrons we need the $|\Delta B| = 1$ Hamiltonian $H^{|\Delta B|=1}$. Here $B$ denotes the beauty quantum number which changes by one unit if the $b$ or $\bar{b}$ decays into lighter quarks. $H^{|\Delta B|=1}$ is constructed in a way that it reproduces the decay amplitudes amplitudes of the full Standard Model up to corrections of order $m_b^2/M_W^2$. An important feature of the effective theory described by $H^{|\Delta B|=1}$ is the absence of $W$ and top-quark fields. To find the interaction vertices of $H^{|\Delta B|=1}$ one contracts the lines with heavy $W$ and $t$ lines in the SM Feynman diagrams to a point. For instance, to lowest order in QCD the $W$-mediated decay $b \to c\overline{u}d$ is described by the effective operator $Q_2^{c\bar{u}d} = \bar{d}_L^\alpha \gamma_\mu u_L^\alpha \, \bar{c}_L^\beta \gamma^\mu b_L^\beta$, where $\alpha$ and $\beta$ are colour indices. Beyond leading order in $\alpha_s$ we can exchange a gluon between the $b-c$ and $u-d$ quark lines. To accomodate this in $H^{|\Delta B|=1}$ we need another operator, $Q_1^{c\bar{u}d} = \bar{d}_L^\alpha \gamma_\mu u_L^\beta \, \bar{c}_L^\beta \gamma^\mu b_L^\alpha$. The piece of $H^{|\Delta B|=1}$ responsible for $b \to c\overline{u}d$ decays is

$$H^{b \to c\bar{u}d} = \frac{4G_{\rm F}}{\sqrt{2}} V_{cb} V_{ud}^* \sum_{j=1,2} C_j Q_j^{c\bar{u}d}. \qquad (62)$$

Here the Fermi constant $G_F$ and the CKM elements are factored out by convention. The Wilson coefficients $C_j$ are the coupling constants of the effective operators $Q_j$. These coefficient contain the full short-distance information of the theory, *i.e.* the full dependence on the heavy masses $M_W$ and $m_t$. The $C_j$ can be calculated in perturbation theory; the order of $\alpha_s$ is referred to as “LO” (leading order), “NLO” (next-to-leading order), and so on. The calculation involves two steps: Firstly, a given decay amplitude is calculated in the SM and compared to the same amplitude calculated with the effective Hamiltonian (matching calculation). Requiring both results to be the same up to terms of $m_{bq}^2/M_W^2$ then fixes $C_j$ at a chosen renormalisation scale, the matching scale $\mu_W$. This scale must be chosen of the order of the heavy masses $M_W$ and $m_t$ to ensure that the perturbative calculation makes sense (*i.e.* that corrections decrease with the order of $\alpha_s$). A physical process does not depend on the numerical value of $\mu_W$ and the $\mu_W$ dependence of a given amplitude decreases order-by-order in $\alpha_s$. Secondly, one calculates the $C_j$ at a low energy scale $\mu_b$, where $\mu_b$ is of the order of $m_b$ which sets the energy scale for decays of $b$-flavoured hadrons. This second step is called renormalisation-group evolution. The result can be written as

$$\vec{C}(\mu_b) = U(\mu_b, \mu_W)\vec{C}(\mu_w) \tag{63}$$

where

$$\vec{C}(\mu) \equiv \begin{pmatrix} C_1(\mu) \\ C_2(\mu) \end{pmatrix}. \tag{64}$$

As for the matching calculation we can use established perturbative methods to determine the evolution matrix $U(\mu_b, \mu_W)$. Let's now apply this framework to a given physical process, taking $B^- \to D^0\pi^-$ as example. The decay amplitude reads

$$\langle D^0\pi^- | H^{b\to c\bar{u}d} | B^- \rangle = \frac{G_{\rm F}}{\sqrt{2}} V_{cb}V_{ud}^* \times \sum_{j=1,2} C_j(\mu_b) \langle D^0\pi^- | Q_j^{c\bar{u}d}(\mu_b) | B^- \rangle.$$

An important feature of the effective Hamiltonian is the independence of the Wilson coefficients from the actual physical process. If we study other $b \to c\bar{u}d$ modes such as $\overline{B}^0 \to D^+\pi^-$ or $\Lambda_b \to \Lambda_c\pi-$, we will encounter the same coefficients $C_j$, with all process-dependence residing in the hadronic matrix elements of the operators $Q_{1,2}$. The calculation of the hadronic matrix elements from first principles is difficult. In our example we can express $\langle D^0\pi^- | Q_{1,2}^{c\bar{u}d}(\mu_b) | B^- \rangle$ in terms of the $B \to D$ form factor and the pion decay constant in certain limits of QCD (considering either an infinite number $N_c$ of colours or an infinitely heavy $b$ quark). The corrections to these limts are not calculable with present techniques. It is often possible to relate different hadronic matrix elements to each other by using symmetries of QCD like flavour-SU(3). This approximate symmetry connects matrix elements which are related by unitary rotations of the three light quark fields $u$,$d$,$s$. Flavour-SU(3) would be an exact symmetry, if these quarks had the same mass. The SU(2) subgroup related to unitary rotations of $(u, d)^T$ corresponds to isospin symmetry and holds with an accuracy of 2% or better. Most importantly, QCD respects the $CP$ symmetry. In our example this entails $\langle D^0\pi^- | Q_j^{c\bar{u}d}(\mu_b) | B^- \rangle = \langle \overline{D}^0\pi^+ | Q_j^{c\bar{u}d\,\dagger}(\mu_b) | B^+ \rangle$. The $CP$ symmetry of QCD is a key feature allowing us to eliminate all hadronic matrix elements from the $CP$ asymmetries in several "golden modes". The full $|\Delta B| = 1$ Hamiltonian needed to describe SM physics reads

$$H^{|\Delta B|=1} = \quad H^{b\to c\bar{u}d} + H^{b\to u\bar{c}d} + H^{b\to c\bar{u}s} + H^{b\to u\bar{c}s} + H^{b\to s} + H^{b\to d}. \tag{65}$$

Here terms describing the tree-level semileptonic decays $b \to q\ell\overline{\nu}$, $q = u, c$, $\ell = e, \mu, \tau$ are omitted, as the effective-Hamiltonian picture is not really needed to describe these decays. (The relevant Wilson coefficients are equal to 1 at all scales.) The last two terms in Eq. (65) are the most interesting pieces of $H^{|\Delta B|=1}$. Most of the physics described in this report involves $H^{b\to s}$ or $H^{b\to d}$.

### *7.4. Remarks about Resonances*

*Section author(s): C. Hanhart*

*7.4.1. Introduction.* A detailed understanding of the concept of resonances and the non-perturbative interactions of QCD at low and intermediate energies will be crucial for a theoretically controlled analysis of various Belle II data. To see this observe, *e.g.*, that they not only shape the Dalitz plots of heavy meson decays — and therefore need to be controlled quantitatively, *e.g.*, for an effective hunt for $CP$ violation within and beyond the Standard Model in these observables (for a recent discussion see Ref. [92, 93]) — but also are interesting for their own sake: As of today we do not even understand what kinds of hadrons (= bound systems of quarks and gluons) do exist in nature. While Belle played a crucial role in establishing the existence of hadrons beyond the most simple quark-antiquark structure with the discovery of the charged charmonium-like states $Z_b(10610)$ and $Z_b(10650)$ in 2013, as of today we understand neither the structure of those states nor under which conditions they are produced — for details we refer to the chapter on quarkonia.

Therefore, to lift the last mysteries of the SM and beyond in the years to come, high precision data analysed with sophisticated theoretical tools are needed. In particular, the simple Breit-Wigner description that parameterises the invariant matrix-element $\mathcal{M}$ for some reaction in a given partial wave as

$$\mathcal{M}_{ab} = -\sum_r \frac{g_a^r g_b^r}{s - s_r} \ , \tag{66}$$

with $s_r = (M_r - i\Gamma_r/2)^2$ appears to be justified only under very special conditions, as explained below. In this section the concept of resonances (as well as other singularities of the scattering matrix) is introduced and possible parameterisations thereof are explained.

*7.4.2. What is a resonance?.* In a particle physics experiment in general transition rates are measured between defined in and out states. Theoretically, *e.g.*, transitions from the states $A, B$ to some multi-body final state are described by the so-called $S$-matrix (see, *e.g.*, Ref. [94], Chapter 4)

$$_{\text{out}}\langle \mathbf{p_1 p_2} ... | \mathbf{k_A k_B} \rangle_{\text{in}} \equiv \langle \mathbf{p_1 p_2} ... | S | \mathbf{k_A k_B} \rangle \ , \tag{67}$$

where the particles in both the initial and final state are characterized by their three momenta — all other possibly relevant quantum numbers like spin, charge etc. are not shown explicitly to keep the notation simple. While the 'in' and 'out' states that appear on the left are defined at some large negative and positive time, respectively, the states on the right may be defined at any common reference time. As a consequence of the conservation of probability the $S$-matrix is a unitary operator — $S^\dagger S = \mathbb{1}$. It describes the full scattering process including the piece where the two initial particles pass by without any interaction. It is useful to separate the interesting, interacting part from the full $S$-matrix via

$$\begin{aligned} \langle \mathbf{p_1 p_2} ... | S - \mathbb{1} | \mathbf{k_A k_B} \rangle \ &= \\ (2\pi)^4 \delta^{(4)}(k_A + k_B - \textstyle\sum p_f) i\mathcal{M}(k_A, k_B \to p_f) \ , \end{aligned} \tag{68}$$

where $\mathcal{M}$ denotes the invariant matrix element. Particles manifest their existence as poles of the $S$–matrix or, equivalently, as poles of $\mathcal{M}$. Thus one needs to map out the singularities of the scattering matrix in order get access to the particle content of a given reaction. In general it is assumed that the $S$–matrix is analytic up to

- *branch points*: On the one hand they occur at each threshold for a kinematically allowed process (*e.g.* at the $\bar{K}K$ threshold in the $\pi\pi$ scattering amplitude) — these are called right–hand cuts. On the other hand there might also be left–hand cuts, which occur when reactions in the crossed channel become possible. Those are often located in the unphysical regime for the reaction studied but can still influence significantly, *e.g.*, the energy dependence of a reaction. When a reaction goes via an intermediate state formed by one or more unstable states, branch points can also be located inside the complex plane of the unphysical sheet [95];

- *bound states*: They appear as poles on the physical sheet and are only allowed to occur on the real $s$–axis below the lowest threshold. Narrow unstable states which correspond to poles on the physical sheet for not the lowest threshold behave very similarly in many aspects. Classic examples in this context are the $f_0(980)$ located on the physical sheet for the $\bar{K}K$–channel which couples also to the much lighter $\pi\pi$ channel[11] and $D_{s0}(2317)$ and $D^*_{s1}(2460)$ located on the physical sheet for the $KD$ and $KD^*$ channels, respectively, but decaying via isospin violation into $D_s\pi$ and $D^*_s\pi$, respectively;
- *virtual states*: As bound states they appear on the real $s$–axis below the lowest threshold, however, on the unphysical sheet. Probably the most famous example of this kind of $S$–matrix singularity is the pole in $S$–wave proton-proton or neutron-neutron scattering (as well as the isovector part of proton-neutron scattering). The corresponding pole is located within about 1 MeV of the threshold giving rise to a scattering length of about 20 fm. However, in contrast to the isoscalar channel, where the deuteron appears as bound state, in the isovector channel the interaction is too weak to form a bound state. There is also evidence that the $X(3872)$ is a virtual state [98];
- and last but not least *resonances* which appear as poles on an unphysical sheet close to the physical one.

For a discussion of the analytic structure of the $S$–matrix with focus on scattering experiments we refer to Ref. [95] and references therein. In what follows the focus will be on the physics of resonances and how to parametrise them. For a detailed discussion on the subject we refer to the resonance review in the Review of Particle Physics by the Particle Data Group [88].

*7.4.3. A comment on Breit-Wigner functions.* A pole the $S$-matrix and thus any resonance is uniquely characterised by its pole position and its residues. Thus a parameterisation of the kind given in Eq. (66) appears natural and one may identify the couplings $g_a$ with the residues $\mathrm{res}^{\mathrm{r}}_{\mathrm{a}}$ [12]. This expression is nothing but a sum over Breit-Wigner functions, which is not only commonly used in very many experimental analyses but also in recent theoretical works — see, *e.g.* Ref. [99]. This kind of parameterisation in general allows for a high quality description of data (as long as enough terms are included in the sum). However, it should be used with care for it may introduce various uncontrollable systematic uncertainties into the analysis as detailed below.

First of all Breit-Wigner functions with a constant width violate analyticity, since the analyticity of the $S$–matrix leads to the Schwarz reflection principle, $S(s^*) = S^*(s)$. Therefore, a pole at $s = s_0$ is necessarily accompanied by a pole at $s = s_0^*$. As illustrated in Figure 59, for narrow, isolated resonances it is only the pole in the lower half plane of the unphysical sheet that is relevant near the resonance peak and it is this pole that it is accounted for by the Breit-Wigner function in the vicinity of the pole. However, at the threshold clearly both poles are equally distant and thus equally relevant. Thus, as soon as amplitudes are to be described over a larger energy range the relevant cuts need to be included properly, *e.g.* by the well known Flatte parametrisation [100] or variants thereof. However, there are resonances where even this modification is not sufficient. An example is the $f_0(500)$ or $\sigma$-meson which has a line shape that deviates significantly even from that of a Breit-Wigner with an energy dependent width [101]. In these cases more sophisticated forms need to be used. We come back to this point below.

[11] For a detailed discussion on this aspect of the $f_0(980)$, see Refs. [96, 97].

[12] For simplicity we do not discuss possible angular distributions of the decay particles here which may be included in a straightforward way. See, *e.g.*, Ref. [88].

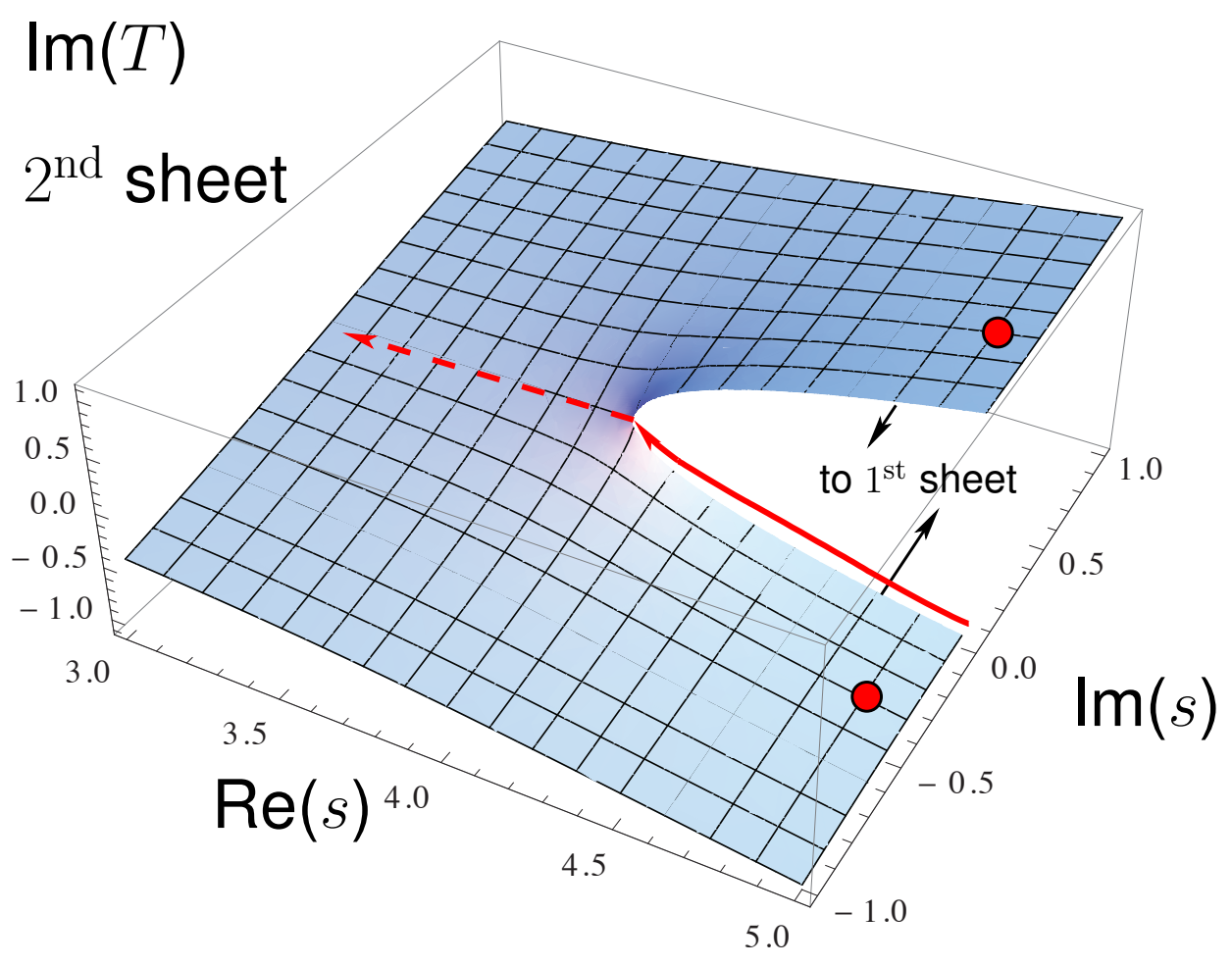

Fig. 59: Sketch of the imaginary part of the scattering amplitude on the unphysical sheet of the complex $s$–plane close to the opening of a threshold. The red solid line shows the physical axis, located on the physical sheet very close to the lower part of sheet. The red dots show the possible location of the resonance poles.

Second, a sum of Breit-Wigners necessarily violates unitarity. To see this we focus on elastic two-body scattering. Then it is straightforward to derive from the unitarity of the $S$-matrix

$$\mathcal{M} - \mathcal{M}^* = 2i\sigma\mathcal{M}^*\mathcal{M} \ , \tag{69}$$

where $\sigma$ denotes the two-body phase space. Furthermore, if we assume that

$$\mathcal{M} = -\frac{(\mathrm{res}^{(1)})^2}{s - M_1^2 + iM_1\Gamma_1} - \frac{(\mathrm{res}^{(2)})^2}{s - M_2^2 + iM_2\Gamma_2} \ . \tag{70}$$

we get

$$\begin{aligned} \mathrm{Im}(\mathcal{M}) - \sigma|\mathcal{M}|^2 \ &= \ \frac{(\mathrm{res}^{(1)})^2(\Gamma_1\mathrm{M}_1 - \sigma(\mathrm{res}^{(1)})^2)}{(s-M_1^2)^2+M_1^2\Gamma_1^2} + \frac{(\mathrm{res}^{(2)})^2(\Gamma_2\mathrm{M}_2 - \sigma(\mathrm{res}^{(2)})^2)}{(s-M_2^2)^2+M_2^2\Gamma_2^2} \\ &+ \ \mathrm{Re}\left(\frac{2\sigma(\mathrm{res}^{(1)}\mathrm{res}^{(2)})^2}{(s-M_1^2+iM_1\Gamma_1)(s-M_2^2-iM_1\Gamma_2)}\right) . \end{aligned}$$

Unitarity requires this expression to vanish. While the first two terms might be removed by choosing $\Gamma_i M_i = \sigma \mathrm{res}^{(\mathrm{i})2}$, which is the unitarity condition for a single resonance [13], it appears not possible to remove the interference term shown in the last line of Equation (71) with constant residues. Thus using Eq. (66) for a single partial wave amplitude with two (or more) resonances is justified only if $M_1 - M_2 \gg (M_1\Gamma_1 + M_2\Gamma_2)/(M_1 + M_2)$. Since the production rate of the individual resonances depends on the source term the resonance parameters extracted using Eq. (66) necessarily get reaction dependent.

[13] This implies that the residue is real — a condition already used to write Equation (71)

Most experiments in particle physics are not scattering but production experiments. For these the unitarity relation reads

$$[\mathcal{A}_a - \mathcal{A}_a^*] = 2i \sum_c \mathcal{M}_{ca}^* \sigma_c \mathcal{A}_c \ . \tag{71}$$

Since $\mathcal{A}$ and $\mathcal{M}$ have identical poles, also this relation can not be fulfilled by a simple Breit-Wigner ansatz. Moreover, a channel and energy dependent production mechanism might distort the line shape of a particular resonance significantly, such that any fit with a symmetric function (as a Breit-Wigner) will deliver channel dependent parameters. For example, if one fits the two–pion invariant mass distribution of $\eta \to \pi\pi\gamma$ (most recently measured at KLOE [102]) with a Breit-Wigner amplitude, one can get a decent fit, however, with a quite low mass parameter for the $\rho$-meson. What is often done in analyses to cure this is to add to the $\rho$-Breit-Wigner distribution a contact term, which is then interpreted as a non-resonant contribution. However, also this violates unitarity for then the phase of the scattering amplitude, in the example above assumed to be given by the $\rho$–amplitude, deviates from the phase of the production amplitude — in conflict with the Watson theorem [103] [14]. Note that the logic presented is not in conflict with the presence of a particle production at tree level: As soon as the final state interaction (*e.g.* in from for a resonant rescattering) is taken into account for this term the tree level term gets canceled. This is discussed within a resonance model in Ref. [104] and in more general terms in Ref. [105].

The only sensible way to account for non-constant production operators is via multiplying the $\rho$–distribution with, *e.g.*, a polynomial — for the case of $\eta \to \pi\pi\gamma$ this is discussed in detail in Ref. [106]. This may be improved further by inclusion of the leading left–hand singularity induced by the $a_2$ meson in the crossed channel [107]. An even more striking energy dependence of the production mechanism can be induced by triangle singularities. This is demonstrated on the example of $\eta(1405)$ and $\eta(1475)$ in Ref. [108] where both signals are explained by a single pole accompanied by a triangle singularity (for a recent discussion of triangle singularities see Ref. [109])[15].

*7.4.4. How to do better.* One way to improve is to construct coupled channel models consistent with the fundamental principles — especially multi-channel unitarity. This approach is developed best for meson–baryon scattering as discussed in Ref. [112]. For the particular case of the very near threshold states $Z_b(10610)$ and $Z_b(10650)$ the coupled channel equations are solved analytically in Refs. [113, 114].

Alternatively one may use the unitarity relation presented in Eq. (71) as the basis for a dispersion theoretical approach. In the single channel case there is a straightforward analytic solution, the Omnès function, for the production amplitude in terms of the scattering phase shift $\delta(s)$ in the corresponding channel [115]

$$\mathcal{A}(s) = P(s)\Omega(s) \ , \tag{72}$$

[14] The Watson theorem may be read off from Eq. (71) immediately: In the single channel case the left-hand side denotes $2i$ times the imaginary part of $\mathcal{A}$, which is purely imaginary. Accordingly the phase of $\mathcal{A}$ needs to match the phase of $\mathcal{M}$.

[15] Triangle singularities can also enhance transition amplitudes in certain kinematic regimes as discussed in Ref. [110].

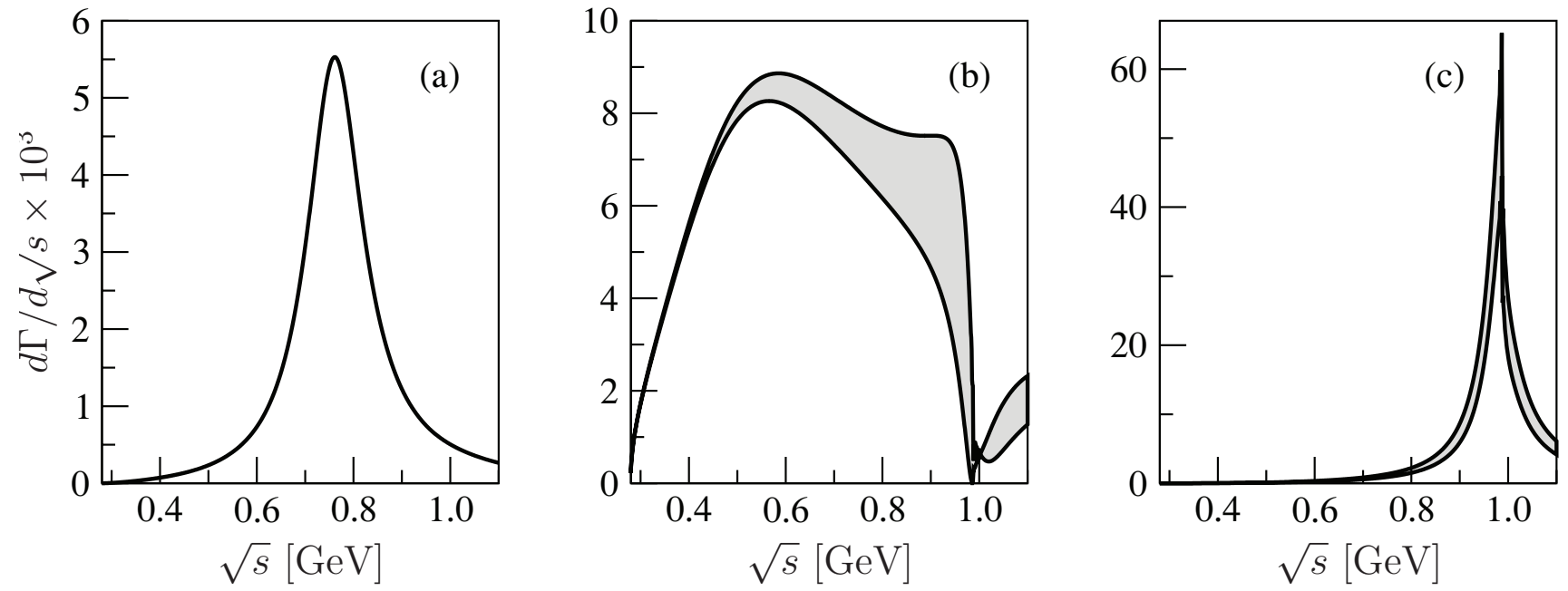

Fig. 60: The predicted signals individually for the currents (a) $\bar{q}\gamma^\alpha q$, (b) $(\bar{u}u + \bar{d}d)/2$, and (c) $\bar{s}s$ calculated for the kinematics relevant for the transition $\tau \to \mu\pi\pi$. In all cases the effective coupling constant is set to 1 GeV$^{-2}$. For the uncertainty bands reflect the uncertainty in the form factor normalisation. The figure is adapted from Ref. [111].

with

$$\Omega(s) = \exp\left(\frac{s}{\pi}\int \frac{ds'\ \delta(s')}{s'(s'-s-i\epsilon)}\right) , \tag{73}$$

where the presence of the polynomial $P(s)$ acknowledges the fact that the unitarity relation of Eq. (71) only fixes the amplitude up to a function that does not have a right hand discontinuity. For the $\pi\pi$ P–waves, where the phase shifts show a prominent resonant structure driven by the $\rho$–meson, the resulting Omnès function resembles a pronounced $\rho$–peak — see left panel of Fig. 60. Note that the two pion phase shifts are known very well due to sophisticated analyses based on Roy equations and variants thereof [116, 117]. An illustrative example that using the Omnès solutions is not only theoretically more sound than using Breit-Wigner functions but also beneficial in data analyses is presented Ref. [118] where recent data on $\overline{B}^0_{d/s} \to J/\psi\pi\pi$ by the LHCb collaboration [119, 120] are studied. For example for pion invariant energies up to 1 GeV the $\bar{B}^0$ decays can be described with only three free parameters equally well compared to the LHCb Breit-Wigner fit that required 14 parameters to analyse the same energy region.

As soon as the first relevant inelasticity enters the above solution no longer applies. Then possible strategies are to match the low energy Omnès solution to a resonance description of the N/D type at higher energies [105] or to solve the corresponding coupled channel problem [121]. In the isovector–vector channel ($\pi\pi$ P wave) the first inelasticity formally enters at the four pion threshold — however, in reality this channel provides a visible inelasticity only well above 1 GeV [122]. The situation is different in the scalar–isoscalar channel, since the $\pi\pi$–system couples strongly to $\bar{K}K$. Chiral perturbation theory allows one to fix the value of the light quark part of the pion scalar form factor at $s = 0$ to sufficient accuracy, however, the normalisation of the strangeness pion scalar form factor is not that well known. Figure 60 shows the results obtained for the modulus of the pion vector form factor, the non-strange and the strange scalar from factor in panel $(a)$, $(b)$ and $(c)$, respectively, here shown as predicted for the BSM process $\tau \to \mu\pi\pi$ in Ref. [111]. The sensitivity due to the uncertainty in the strange form factor normalisation is illustrated by the uncertainty bands. The strange form factor exhibits a peak around 1 GeV, which is produced by the $f_0(980)$ resonance. On the contrary in the pion scalar non-strange form factor the $\sigma$ or $f_0(500)$ meson appears as a broad bump (notice the non-Breit–Wigner shape) around 500 MeV and the $f_0(980)$ appears as a dip rather than a peak. The very different line shapes of the different form factors shown in Fig. 60 can be exploited to disentangle different BSM source terms. The ideas of Ref. [111] were generalised in Refs. [123, 124].

So far we discussed two hadron interactions only and largely ignored left–hand cut contributions. The formalism can be extended by means of the Khuri–Treiman equations [125] to include also crossed–channel singularities as well as three–body dynamics [126–132], but discussing this goes beyond the scope of this presentation.

### *7.5. Lattice QCD*

*7.5.1. Introduction.* The intensity frontier probes new physics through quantum loop effects by a strict comparison between precise theoretical predictions and experimental measurements. For many quantities, the accuracy of the comparison is currently limited by the theoretical uncertainties from the hadronic matrix elements describing non-perturbative QCD effects in the underlying processes. Moreover, as the heavy-flavour factories accumulate high statistics data, many new quarkonium and exotic states have been observed. Non-perturbative dynamics of QCD is also essentially important to understand their nature including the spectra, quantum numbers and decay properties. Lattice QCD is a powerful method to study non-perturbative aspects of QCD with controlled and systematically-improvable accuracy. It is expected to play a key role to the success of the SuperKEKB / Belle II experiment by timely providing theoretical inputs with commensurate uncertainties.

Lattice QCD is a regularisation of QCD on a discrete Euclidean space-time lattice. On a finite-volume lattice, the path integral is reduced into a finite-dimensional integral and can be numerically evaluated by a Monte Carlo sampling of gauge field configurations on a computer. This numerical simulation does not rely on the perturbative expansion, and enables us to non-perturbatively study QCD.

In principle, uncertainties due to the lattice formulation and numerical simulation can be systematically reduced by a large-scale simulation: namely, by generating many configurations on a fine and large lattice. While such a realistic simulation is computationally intensive, continuous development of powerful computers and simulation techniques has led to increasingly precise and wide applications of lattice QCD. These include physics of the QCD vacuum, hadron spectrum and structure, QCD at finite temperature and density, *ab-initio* nuclear physics, and simulations of theories beyond QCD.

For instance, the energy of a hadron stable under QCD can be calculated from the asymptotic behavior of a two-point function

$$\langle \mathcal{O}_H(t)\mathcal{O}_H^\dagger(0)\rangle \quad \to \quad \frac{|Z_H|^2}{2E_H} e^{-E_H t} \quad (t \to \infty) \tag{74}$$

towards the large temporal separation $t$. Here $\mathcal{O}_H$ is an interpolating field of the hadron, and $Z_H = \langle 0|\mathcal{O}_H|H\rangle$ represents the overlap of $\mathcal{O}_H$ with the physical state $|H\rangle$. The low-lying hadron spectrum calculated in this way is in impressive agreement with experiment [133]. The permille-level neutron − proton mass splitting has been also reproduced by taking account of the mass difference of up and down quarks as well as the electro-magnetic (EM) corrections [134, 135].

Precise study of flavour physics is one of the most important applications of lattice QCD from its early stage. When a process has at most one hadron stable under QCD both in the initial and final states, the relevant hadronic matrix element can be straightforwardly extracted from correlation function. For the leptonic decays, for instance, the overlap factor in (74) gives $\langle 0|A_\mu|H\rangle$ by using the axial current as $\mathcal{O}_H$. The matrix element $\langle H'|O_{\rm int}|H\rangle$ for the semileptonic decays and neutral meson mixings can be obtained from three-point function

$$\langle \mathcal{O}_{H'}(t')\mathcal{O}_{\rm int}(t)\mathcal{O}_H^\dagger(0)\rangle \quad \to \quad \frac{Z_{H'}Z_H^*}{4E_{H'}E_H}\langle H'|O_{\rm int}|H\rangle e^{-E_{H'}(t'-t)-E_H t} \quad (t, t'-t\to\infty), \tag{75}$$

where $\mathcal{O}_{\rm int}$ represents the interaction operator. Recent realistic simulations can accurately calculate these two- and three-point functions, and we refer to the relevant matrix elements as "gold-plated". A main thrust of recent lattice efforts is improving the accuracy of the gold-plated quantities. We summarize the current status in Sec. 7.5.2, and make forecasts for the future precision in Sec. 7.5.5.

We have to take account of the final state interaction to study hadronic decays. In this case, however, the amplitudes of the correlation functions are not directly related to the hadronic matrix elements as we discuss in Sec. 7.5.3. Theoretical framework to study the hadronic decays is under active development, and is being applied to the quarkonium and exotic states, which generally lie above thresholds (Sec. 7.5.4).

*7.5.2. (Semi)leptonic decays and mixing.* The hadronic matrix element for the leptonic decay is parametrised by using the decay constant

$$\langle 0|A_\mu|B_{(s)}(p)\rangle \quad = \quad p_\mu f_{B_{(s)}}, \tag{76}$$

and vector and scalar form factors for the $B\to\pi\ell\nu$ and $D\ell\nu$ semileptonic decays

$$\langle H(p')|V_\mu|B_{(s)}(p)\rangle \quad = \quad \left(p+p'-\frac{M_B^2-M_H^2}{q^2}q\right)_\mu f_+(q^2)+\frac{M_B^2-M_H^2}{q^2}q_\mu f_0(q^2), \tag{77}$$

where $q^2=(p'-p)^2$ is the momentum transfer. The $B_{(s)}$ meson mixing matrix element is written by using the bag parameter as

$$\langle \bar{B}^0_{(s)}|\mathcal{O}_1|B^0_{(s)}\rangle \quad = \quad \frac{8}{3}f^2_{B_{(s)}}M^2_{B_{(s)}}B_{B_{(s)}}, \tag{78}$$

where $\mathcal{O}_1=\left[\bar{b}\gamma_\mu(1-\gamma_5)q\right]\left[\bar{b}\gamma_\mu(1-\gamma_5)q\right]$ and $q=d$ $(s)$ for $B$ $(B_s)$. Precise knowledge of these gold-plated quantities is essential in the search for new physics at Belle II. Their accuracy can be straightforwardly improved by a large-scale simulation accumulating high statistics on a fine and large lattice at the physical point, where quark masses are set to their physical values.

Such a realistic simulation is computationally so demanding, because the simulation cost quite rapidly increases as we approach the continuum limit and decrease the up and down quark masses to the physical point. Previous lattice simulations have often employed unphysically heavy up and down quarks, and extrapolated their results to the physical point. This procedure is referred to as the chiral extrapolation. However, thanks to recent advances in computer power and improvements in simulation algorithms, gauge field ensembles including effects of dynamical up, down, strange and even charm quarks are becoming available near and at the physical point.

Typical lattice spacings are larger than or comparable to the Compton wave lengths of the bottom quarks $m_b^{-1}$. The control of discretisation errors arising from bottom valence quarks is therefore an essential issue in the current and future precision study of $B$ physics. We note that the lattice action is not unique: it can be improved to have reduced discretisation errors by, for instance, adding irrelevant operators. Heavy quark actions on the lattice have been developed based on the heavy quark effective theory, non-relativistic QCD, and the so-called Fermilab formalism [136–138] to directly simulate $m_b$ at the currently available lattice spacings and to describe the discretisation errors of simulation results. Another good strategy is to compute suitable ratios of physical observables by using a relativistic lattice action, and interpolate them between available heavy quark masses and their known static limit [139].

As discussed in Chapter 8 in detail, the gold-plated quantities are now being calculated with fully controlled errors. The $B_{(s)}$ meson decay constants have been calculated with an accuracy of a few percent, and confirmed by several independent calculations with different actions [140]. The accuracies of the state-of-the-art studies of the $B\to\pi\ell\nu$ [141–143], $B\to D^{(*)}\ell\nu$ [144–147] decays, and the $B_{(s)}$ mixing [148–151] are approaching to the same level, although the number of such precision computations is rather limited.

Over the next decade, we expect more independent calculations with even better accuracies by simulating the physical point on finer lattices. In Sec. 7.5.5, we make forecasts for the future lattice precision, which is used in this report to discuss interplay between the precise lattice calculations and Belle II measurements in the search for new physics.

So far, the $B$ meson matrix elements have been usually calculated in the isospin limit without the EM corrections. As the precision approaches the percent level, control of the isospin corrections becomes increasingly important and is actively being pursued [152]. Recently, a method has been proposed to compute the EM effects in hadronic processes where infrared divergences are present [153]. The isospin corrections to the leptonic decay rates $\Gamma(\pi, K\to\ell\nu)$ have been successfully calculated [154]. We note that this method is applicable to heavy meson (semi)leptonic decays.

The scope of the precision lattice calculation is expanding to other gold-plated processes. For instance, $B_s\to K\ell\nu$ provides an independent determination of $|V_{ub}|$, and $B\to K(\pi)ll$ mediated by FCNC is sensitive to new physics. Simulation techniques for $B\to\pi\ell\nu$ can be straightforwardly applied to these decays, and results with similar accuracies are becoming available [141, 155–159].

The baryon decays also provide independent determinations of the CKM matrix elements and constraints on new physics, but with systematics different from the meson decays. The first lattice calculations for the $\Lambda_b\to p\ell\nu$, $\Lambda_c\ell\nu$, and $\Lambda\ell\ell$ decays have been reported in Refs. [160–162]. However, baryons are known to be more challenging in controlling the chiral extrapolation and finite-volume effects. These issues can be addressed in the relatively short term by more realistic and/or independent calculations.

The gold-plated quantities are important inputs to determine relevant CKM matrix elements from given exclusive decays. As is well known, however, there is a long-standing tension between the exclusive and inclusive decays in $|V_{ub}|$ and $|V_{cb}|$ [88]. Although the analysis of the inclusive decay rate employs the heavy quark expansion (HQE) and hence has very different theory systematics from those for the exclusive decays, lattice QCD can contribute to the inclusive determinations as well. In the HQE, the expansion coefficients encode non-perturbative hadronic dynamics. Lattice calculation of the coefficients has been pursued for more than twenty years [163–169]. Another interesting future direction is to extract the inclusive decay rate on the lattice [170, 171]. In this approach, the relevant structure functions are accessible through the scattering matrix element of two weak currents between single-$B_{(s)}$-meson states $\langle B_{(s)}|T\{J_\mu^\dagger J_\nu\}|B_{(s)}\rangle$, which is gold-plated. While the numerical calculation of the relevant four-point functions is challenging, encouraging results for the $B_s \to X_c \ell\nu$ decay have been obtained in Ref. [170].

*7.5.3. Hadronic decays.* There are, however, many interesting observables that are not gold-plated. These involve final states with more than one strongly-interacting particle, *e.g.* $K \to \pi\pi$, $D \to \pi\pi$ and $B \to DK$ decay amplitudes, or require the calculation of long-distance contributions, *e.g.* $D^0 - \bar{D}^0$ mixing. To calculate these using lattice QCD requires new methods beyond those needed for gold-plated observables, and also requires, in general, significantly more computational resources. Such quantities lie at the frontier of present lattice efforts: some have been calculated with controlled errors, others are close to being controlled, while for others the required theoretical formalism does not yet exist.

We first discuss the issues that arise when calculating decay amplitudes. The key theoretical issue is that lattice calculations perforce are done in finite spatial volume $V$, so that the multiparticle states, *e.g.* $|DK\rangle_V$, differ from the infinite volume *out*-states, $|DK\rangle_{out}$, that are needed to define decay amplitudes. Thus while a lattice calculation can, in principle, calculate matrix elements such as $\langle B|H_W|DK\rangle_V$ (with $H_W$ the effective weak Hamiltonian), these differ in an essential way from the desired amplitudes, *e.g.* $\langle B|H_W|DK\rangle_{out}$. One difference is that the desired amplitude is complex (due to final state interactions) while the finite-volume amplitude is real.[16] A more significant difference is that multiparticle states such as $|DK\rangle_V$ contain a mixture of all the particle combinations that are accessible via strong interactions at the energy of the initial particle. For the $DK$ state with energy $M_B$, these combinations include $DK\pi\pi$, $D^*K^*$, and many other possibilities. These "contaminations" are not small, but rather are $\mathcal{O}(1)$ effects.

An additional, more practical, issue is that one must use a finite-volume $DK$ state that has the same energy as the initial $B$. This is therefore a highly excited state compared to the ground state in which the $D$ and $K$ are at rest (assuming that the total momentum vanishes). The signal for the ground state will dominate over that for the excited state by a factor of $e^{(M_B - M_D - M_K)\tau}$, where $\tau$ is the Euclidean time. This problem can be overcome in principle by using appropriate operators to couple to the $DK$ system, tuned to avoid couplings to lighter states. In this regard, it is encouraging that there have in recent years been tremendous advances in the methodology for extracting excited state energies, for example [173, 174].

While these issues are challenging, substantial progress has been made, particularly in the case of $K \to \pi\pi$ decays. This is based on seminal work by Lüscher relating the spectrum of two-particle finite-volume states below the inelastic threshold to the elastic phase shift [175, 176], and subsequent work by Lellouch and Lüscher showing how to relate the finite-volume matrix elements described above to the physical amplitudes [177].

This formalism has been successfully implemented in recent work by the RBC/UKQCD collaboration. They finesse the issue of excited states using tuned boundary conditions so that the lightest state is the desired one. They have a fully controlled result for the $\Delta I = 3/2$ $K \to \pi\pi$ amplitude [178], and a result at a single lattice spacing for the $\Delta I = 1/2$ amplitude [179]. Fully controlled results for the latter are expected soon. They also have determined the imaginary parts of these amplitudes, albeit with larger errors, and thus can provide the SM prediction for the direct $CP$ violation parameter $\mathrm{Re}\,[\epsilon'/\epsilon]$ [179]. A slight tension with the experimental value [88] is of great phenomenological interest [179–181].

[16] This is related to the Maiani-Testa no-go theorem [172].

Subsequent to the work of Lüscher and Lellouch, the theoretical framework for studying two particle systems in lattice QCD has been generalised to a moving frame, to non-identical particles with arbitrary spin, and to multiple two-particle channels [182–187]. These extensions have been applied successfully in lattice studies of resonance physics (see Sec. 7.5.4). They are, however, not yet sufficient to allow lattice simulations to study $D$ or $B$ decays, because of the prevalence of states containing three or more particles. Producing the required generalisation is an active area of research, with significant progress made for three particles [188–191], but further developments are needed to have a general theory. It is not unreasonable to hope that such a theory will be available in 3-5 years.

We close this subsection by commenting briefly on prospects for lattice calculations of $D^0 - \bar{D}^0$ mixing amplitudes. The short-distance contributions require gold-plated calculations and are under good control [192, 193]. However, the mixing is dominated by long distance contributions from many intermediate states, and for these new methodology is needed. Significant progress has been made on the analogous, although simpler, case of $K^0 - \bar{K}^0$ mixing [194]. Here a new technique has been developed involving the insertion of two factors of $H_W$ integrated over their relative time separation, and first results indicate that the method works. The extension to $D^0$ mesons faces two major challenges: the need to control many exponentially growing intermediate states with sufficient accuracy, and the need to make a finite-volume correction. The latter will require the completion of the multiparticle formalism discussed above.

*7.5.4. Quarkonium and exotic states.* High statistics data of the $e^+e^-$ collision at $B$ factories brought about rich outcome for the spectroscopy of hadrons containing heavy quarks. One of the most interesting news is the discovery of the exotic hadrons. Chapter 14 considers lattice studies of interesting quarkonium-like states, while a brief summary of the status is given here.

Lattice QCD is a powerful method to study heavy hadron spectroscopy from first principles: it can study properties of experimentally observed states, and can also provide valuable reference spectra for yet-unobserved states.

Quarkonium spectra below open flavour thresholds are gold-plated, and recent precise lattice calculations show good agreement with experiment. The main remaining uncertainty for these comes from the omission of $\bar{c}c$ or $\bar{b}b$ disconnected diagrams; these remain a great challenge as they lead to intermediate states with multiple light hadrons.

Until recently, all quarkonium(-like) states above thresholds were treated as stable under the strong interaction; the most extensive excited and hybrid charmonium spectrum with this approach has been obtained in [195]. This unphysical assumption is now being removed by developments described below.

Many interesting hadrons and in particular all candidates for the exotic hadrons lie near or above thresholds. Properties of such unstable particles are not gold-plated and are encoded in their scattering and transition amplitudes. Among those, lattice can most easily treat hadrons that lie above only one two-particle threshold $M_{H_1} + M_{H_2}$, or lie slightly below it; such cases are (unfortunately) rare in Nature. The most rigorous way to extract the scattering matrix $S(E)$ for elastic $H_1H_2$ scattering is based on the Lüscher's formalism discussed in Sec. 7.5.3. One determines energies of $H_1H_2$ eigenstates $E$ from a lattice simulation in a finite volume. This gives the infinite-volume scattering matrix $S(E)$ at that energy via Lüscher's relation [196]. This leads to $S(E)$ only for specific values of $E$ since the spectrum in a finite volume is discrete. A hadronic resonance $R \to H_1H_2$ is inferred from the pole of $S(E)$ on the unphysical Riemann sheet. Likewise, the bound state is inferred from a pole on the real axis below threshold as discussed in Sec. 7.4.1.

This approach has been extensively verified on elastic resonances like $\rho$ and $K^*$, where it leads to masses and widths close to the experimental values. In the quarkonium sector, for instance, the mass and width of the vector charmonium $\psi(3770)$ were extracted by considering $D\bar{D}$ scattering [197]. An experimentally-established charmonium-like state $X(3872)$ seems not to fit into the simple quark model but its existence is theoretically confirmed from lattice QCD [198, 199]. It is found as a pole in $D\bar{D}^*$ scattering just below threshold. For X(5568), reported by the D0 collaboration [200], a lattice simulation of the relevant $B_s\pi^+$ scattering [201] does not find any evidence in accordance with the recent LHCb measurement [202].

The radiative and weak transitions $\langle H_2|J^\mu|H_1\rangle$ for $H_{1,2}$ that are strongly-stable are gold-plated. Considerably more challenging are transitions where initial or final hadrons are strongly decaying resonances. The general strategy to treat those was proposed, for example, in Ref. [203]. This has been employed only for $\langle\rho|J^\mu_{em}|\pi\rangle$ transition [204, 205], which in practice implied the determination of the $\pi\pi \to \pi\gamma$ amplitude and its evaluation at the $\rho$-meson pole.

Most of the interesting and exotic hadrons actually lie above two or more thresholds, *i.e.* they can decay to several two-hadron final states. The rigorous way to address this problem is via generalised Lüscher formalism [189]. Each energy of the lattice eigenstate $E$ leads to one equation with several unknown $S^{ij}(E)$. The direct extraction of $S^{ij}(E)$ becomes practically impossible. The Hadron Spectrum Collaboration managed to extract $2\times 2$ [206] and $3\times 3$ [174] scattering matrices by parametrising $S^{ij}(E)$ as a function of $E$ using certain number of parameters. The $S$ matrix was continued to the complex plane: its poles on the unphysical Riemann sheet indicate masses and widths of the resonances, while poles on the real axis indicate bound states. This challenging strategy was applied only for "non-exotic" channels when the scattering particles did not carry spin. Most of the exotic hadrons have $J = 1$ and involve scattering of particles with spin, which brings additional complications.

One can expect rigorous results in the next 5 years for hadrons that can decay via few (two or three) two-hadron final states. That applies for example to $Z_c^+(3900)$, while $Z_c^+(4430)$ and $Z_b^+$ lie above many more two-hadron thresholds and it is difficult to envisage rigorous progress along these lines there. Many interesting hadrons can strongly decay also to three-hadron final states, which presents an even greater challenge. Theoretical framework to address those is being constructed [189–191], but no QCD simulations employed it so far.

Another possibility to extract $S(E)$ is the HALQCD approach [207], which starts by determining the two-hadron Bethe-Salpeter wave function and two-hadron potential from lattice QCD. The scattering matrix $S(E)$ is then obtained using the Schrödinger equation for given two-hadron potential. This approach has not been verified on conventional resonances yet. Recently the HALQCD collaboration employed the coupled-channel version of this approach to determine the $3 \times 3$ matrix $S(E)$ relevant for the $Z_c(3900)$ channel [208].

The Born-Oppenheimer approach may be applied for the systems with heavy quarks $Q$, where the static heavy-quark sources are surrounded by the light degrees of freedom. The potential $V(r)$ is calculated as a function of distance $r$ between a static pair $Q(0)Q(r)$ or $Q(0)\bar{Q}(r)$ in the presence of the light degrees of freedom. The potential $V(r)$ is used in the Schrödinger equation to search for bound states and resonances. This has been considered for low-lying bottomonia, quenched hybrids [209], $BB^{(*)}$ and recently also for closed-bottom $B\bar{B}^{(*)}$ [210]. Many interesting Born-Oppenheimer potentials [211] remain to be explored.

*7.5.5. Current lattice inputs and forecasts for future precision.* In this subsection, we summarise the current lattice inputs and make forecasts for the future lattice precision, which are used in this report to discuss the Belle II sensitivity to new physics.

*Assumptions for forecasts.* We provide the following five types of the lattice inputs:

- "current": As the current lattice input, we quote the world average by the Flavour Lattice Averaging Group (FLAG) in Ref. [140], where available. Note that Table 30 lists the updated average for the decay constants and mixing parameters by including recent precise results in Refs. [151, 212]. For details, we refer the readers to the web update of the FLAG review in `http://itpwiki.unibe.ch/flag` [17].
- "5 yr w/o EM": We assume a factor of 2 reduction of the lattice QCD uncertainty in the next five years and that the uncertainty of the EM correction is negligible (*e.g.* for processes insensitive to the EM correction).
- "5 yr w/ EM": The lattice QCD uncertainty is reduced by a factor of 2, but we add in quadrature 1% uncertainty from the EM correction[18].
- "10 yr w/o EM": We assume a factor of 5 reduction of the lattice QCD uncertainty in the next ten years. It is also assumed that the EM correction will be under control and its uncertainty is negligible.
- "10 yr w/ EM": We assume lattice QCD uncertainties reduced by a factor of 5, but add in quadrature 1% uncertainty from the EM correction.

Note that recent precision lattice calculations start to provide their estimate of the QED uncertainty. The entries "5 yr w/ EM" and "10 yr w/ EM" suggest that the control of this uncertainty will become increasingly important in the future.

*Leptonic decays and $B_{(s)}$ meson mixing.* The hadronic matrix elements for the $B_{(s)}$ meson leptonic decays and mixing are parameterised by using the decays constants $f_{B_{(s)}}$ and bag parameters $B_{B_{(s)}}$ as Eqs. (76) and (78). These gold-plated quantities have been calculated in $N_f = 2+1$ QCD, which includes degenerate up and down sea quarks as well as strange sea quarks. Results with dynamical charm quarks are also available for $f_{B_{(s)}}$.

[17] The latest review quote $B_B = 1.30(0.10)$ and $B_{B_s}/B_B = 1.032(38)$, which have slightly larger uncertainty than those in Table 30 due to a change in estimating the correlation among different calculations.

[18] For the $B \to D^*$ form factor in Table 38, we assume 0.5 % uncertainty estimated in Ref. [144].

Table 30: Lattice inputs for decay constants $f_{B_{(s)}}$ and bag parameters $B_{B_{(s)}}$ in the SM. The current average of $f_{B_{(s)}}$ for $N_f = 2+1$ and 2+1+1 are obtained from Refs. [150, 213–216] and Refs. [212, 217], respectively. The average of $B_{B_{(s)}}$ is obtained from Refs. [148, 150, 151]. $f_{B_{(s)}}\sqrt{B_{B_{(s)}}}$ is in units of MeV.

| $N_f$ | input | $f_B$[MeV] | $f_{B_s}$[MeV] | $f_{B_s}/f_B$ |
|---|---|---|---|---|
| 2+1+1 | current | 188(3) | 227(4) | 1.203(0.007) |
| | 5 yr w/o EM | 188(1.5) | 227(2.0) | 1.203(0.0035) |
| | 5 yr w/ EM | 188(2.4) | 227(3.0) | 1.203(0.013) |
| | 10 yr w/o EM | 188(0.60) | 227(0.80) | 1.203(0.0014) |
| | 10 yr w/ EM | 188(2.0) | 227(2.4) | 1.203(0.012) |
| 2+1 | current | 192.0(4.3) | 228.4(3.7) | 1.201(0.016) |
| | 5 yr w/o EM | 192.0(2.2) | 228.4(1.9) | 1.201(0.0080) |
| | 5 yr w/ EM | 192.0(2.9) | 228.4(2.9) | 1.201(0.014) |
| | 10 yr w/o EM | 192.0(0.86) | 228.4(0.74) | 1.201(0.0032) |
| | 10 yr w/ EM | 192.0(2.1) | 228.4(2.4) | 1.201(0.012) |
| $N_f$ | input | $f_B\sqrt{B_B}$ | $f_{B_s}\sqrt{B_{B_s}}$ | $\xi$ |
| 2+1 | current | 225(9) | 274(8) | 1.206(0.017) |
| | 5 yr w/o EM | 225(4.5) | 274(4.0) | 1.206(0.0085) |
| | 5 yr w/ EM | 225(5.0) | 274(4.8) | 1.206(0.015) |
| | 10 yr w/o EM | 225(1.8) | 274(1.6) | 1.206(0.0034) |
| | 10 yr w/ EM | 225(2.9) | 274(3.2) | 1.206(0.013) |
| $N_f$ | input | $B_B$ | $B_{B_s}$ | $B_{B_s}/B_B$ |
| 2+1 | current | 1.30(0.09) | 1.35(0.06) | 1.032(0.036) |
| | 5 yr w/o EM | 1.30(0.045) | 1.35(0.030) | 1.032(0.018) |
| | 5 yr w/ EM | 1.30(0.047) | 1.35(0.033) | 1.032(0.021) |
| | 10 yr w/o EM | 1.30(0.018) | 1.35(0.012) | 1.032(0.0072) |
| | 10 yr w/ EM | 1.30(0.022) | 1.35(0.018) | 1.032(0.013) |

Table 30 summarises the current lattice inputs and the forecasts. We note that there are several definitions for the bag parameters in the literature. Here, as in the lattice papers, the same definitions as for kaons are used (see, *e.g.*, Ref. [151] and references therein).

In the SM, only the matrix element (78) contributes to the mass difference $\Delta M_{(s)}$ between the $B_{(s)}$ meson mass eigenstates. Beyond the SM, however, $\Delta M_{(s)}$ receives contributions from additional four operators

$$\mathcal{O}_2 = \bar{b}^a(1-\gamma_5)q^a\bar{b}^b(1-\gamma_5)q^b, \qquad \mathcal{O}_3 = \bar{b}^a(1-\gamma_5)q^b\bar{b}^b(1-\gamma_5)q^a, \tag{79}$$

$$\mathcal{O}_4 = \bar{b}^a(1-\gamma_5)q^a\bar{b}^b(1+\gamma_5)q^b, \qquad \mathcal{O}_5 = \bar{b}^a(1-\gamma_5)q^b\bar{b}^b(1+\gamma_5)q^a, \tag{80}$$

where $a$ and $b$ denote color indices, and $q = d$ $(s)$ for $B$ $(B_s)$. Their matrix elements can be parametrised as

$$\langle \bar{B}^0_{(s)}|\mathcal{O}_i|B^0_{(s)}\rangle \;=\; c_i\left\{\left(\frac{M_{B_{(s)}}}{m_b+m_q}\right)^2 + d_i\right\} f^2_{B_{(s)}} M^2_{B_{(s)}} B_{B_{(s)}} \tag{81}$$

with

$$(c_2, d_2) = (-5/3, 0), \quad (c_3, d_3) = (1/3, 0), \quad (c_4, d_4) = (2, 1/6), \quad (c_5, d_5) = (2/3, 3/2). \quad (82)$$

These bag parameters have been calculated in Ref. [149] ($N_f = 2$, namely only with dynamical up and down quarks) and Ref. [151] ($N_f = 2+1$). As seen in Table 31, however, there is a tension in $B^{(5)}_{B_{(s)}}$ between $N_f = 2$ and 2+1, the cause of which has to be understood in the future. We use results for $N_f = 2+1$ to make the forecast.

Table 31: Lattice inputs for bag parameters beyond the SM from $N_f = 2$ [149] and 2+1 [151] QCD. $f_{B_{(s)}}\sqrt{B^{\{(2),\ldots,(5)\}}_{B_{(s)}}}$ is in units of MeV.

| $N_f$ | input | $f_B\sqrt{B^{(2)}_B}$ | $f_B\sqrt{B^{(3)}_B}$ | $f_B\sqrt{B^{(4)}_B}$ | $f_B\sqrt{B^{(5)}_B}$ |
|---|---|---|---|---|---|
| 2+1 | current | 169(8) | 200(19) | 197(7) | 190(9) |
| | 5 yr w/o EM | 169(4.0) | 200(9.5) | 197(3.5) | 190(4.5) |
| | 5 yr w/ EM | 169(4.3) | 200(9.7) | 197(4.0) | 190(4.9) |
| | 10 yr w/o EM | 169(1.6) | 200(3.8) | 197(1.4) | 190(1.8) |
| | 10 yr w/ EM | 169(2.3) | 200(4.3) | 197(2.4) | 190(2.6) |
| 2 | current | 160(8) | 177(17) | 185(9) | 229(14) |

| $N_f$ | input | $f_{B_s}\sqrt{B^{(2)}_{B_s}}$ | $f_{B_s}\sqrt{B^{(3)}_{B_s}}$ | $f_{B_s}\sqrt{B^{(4)}_{B_s}}$ | $f_{B_s}\sqrt{B^{(5)}_{B_s}}$ |
|---|---|---|---|---|---|
| 2+1 | current | 205(7) | 240(16) | 231(7) | 222(8) |
| | 5 yr w/o EM | 205(3.5) | 240(8.0) | 231(3.5) | 222(4.0) |
| | 5 yr w/ EM | 205(4.1) | 240(8.4) | 231(4.2) | 222(4.6) |
| | 10 yr w/o EM | 205(1.4) | 240(3.2) | 231(1.4) | 222(1.6) |
| | 10 yr w/ EM | 205(2.5) | 240(4.0) | 231(2.7) | 222(2.7) |
| 2 | current | 195(7) | 215(17) | 220(9) | 285(14) |

| $N_f$ | input | $B^{(2)}_B$ | $B^{(3)}_B$ | $B^{(4)}_B$ | $B^{(5)}_B$ |
|---|---|---|---|---|---|
| 2+1 | current | 0.76(0.08) | 1.07(0.22) | 1.04(0.09) | 0.96(0.10) |
| | 5 yr w/o EM | 0.76(0.040) | 1.07(0.11) | 1.04(0.045) | 0.96(0.050) |
| | 5 yr w/ EM | 0.76(0.041) | 1.07(0.11) | 1.04(0.046) | 0.96(0.051) |
| | 10 yr w/o EM | 0.76(0.016) | 1.07(0.044) | 1.04(0.018) | 0.96(0.020) |
| | 10 yr w/ EM | 0.76(0.018) | 1.07(0.045) | 1.04(0.021) | 0.96(0.022) |
| 2 | current | 0.72(0.03) | 0.88(0.13) | 0.95(0.05) | 1.47(0.12) |

| $N_f$ | input | $B^{(2)}_{B_s}$ | $B^{(3)}_{B_s}$ | $B^{(4)}_{B_s}$ | $B^{(5)}_{B_s}$ |
|---|---|---|---|---|---|
| 2+1 | current | 0.81(0.06) | 1.10(0.16) | 1.02(0.07) | 0.94(0.07) |
| | 5 yr w/o EM | 0.81(0.030) | 1.10(0.080) | 1.02(0.035) | 0.94(0.035) |
| | 5 yr w/ EM | 0.81(0.031) | 1.10(0.081) | 1.02(0.036) | 0.94(0.036) |
| | 10 yr w/o EM | 0.81(0.012) | 1.10(0.032) | 1.02(0.014) | 0.94(0.014) |
| | 10 yr w/ EM | 0.81(0.014) | 1.10(0.034) | 1.02(0.017) | 0.94(0.017) |
| 2 | current | 0.73(0.03) | 0.89(0.12) | 0.93(0.04) | 1.57(0.11) |

*Semileptonic decays.* The $B_{(s)} \to H\ell\nu$ semileptonic decays have been studied in $N_f = 2+1$ QCD. If the daughter meson $H$ is pseudoscalar, only the weak vector current contributes due to parity symmetry, and the matrix element (77) is parametrised by the vector and scalar form factors, which depend on the momentum transfer $q^2$. In Ref. [140], FLAG fits available lattice data into a model independent parametrisation of the $q^2$ dependence proposed by Bourrely, Caprini and Lellouch [218]

$$f_+(q^2) = \frac{1}{B_+(q^2)} \sum_{n=0}^{N_+-1} a_n^+ \left\{ z^n - (-1)^{n-N_+} \frac{n}{N_+} z^{N_+} \right\}, \tag{83}$$

$$f_0(q^2) = \frac{1}{B_0(q^2)} \sum_{n=0}^{N_0-1} a_n^0 \, z^n. \tag{84}$$

The Blaschke factor $B_{+(0)}$ is chosen as $B_{+(0)}(q^2) = 1 - q^2/M_{\mathrm{pole},+(0)}$, if there exists the lowest resonance in the vector (scalar) channel with its mass $M_{\mathrm{pole},+(0)}$ below the threshold $\sqrt{t_+} = M_{B_{(s)}} + M_H$. For $B \to \pi\ell\nu$, this factor is set to $B_0(q^2) = 1$. The expansion parameter $z$ is defined as

$$z(q^2, t_0) = \frac{\sqrt{t_+ - q^2} - \sqrt{t_+ - t_0}}{\sqrt{t_+ - q^2} + \sqrt{t_+ - t_0}}, \tag{85}$$

where $t_0 = (M_{B_{(s)}} + M_H)(\sqrt{M_{B_{(s)}}} - \sqrt{M_H})^2$. The FLAG analysis employs $N_+ = N_0 = 3$. The expansion coefficients $a^+_{\{0,1,2\}}$ and $a^0_{\{0,1\}}$ are fit parameters, whereas $a^0_2$ is expressed in term of all remaining coefficients to impose the kinematical constraint $f_+(0) = f_0(0)$.

In this report, we quote the current inputs for the coefficients $a_n^{\{+,0\}}$ and their correlation matrix, and make forecasts for $a_n^{\{+,0\}}$. Those for the $B \to \pi\ell\nu$ and $B_s \to K\ell\nu$ decays are summarised in Tables 32 – 35.

Table 32: Current input for $B \to \pi\ell\nu$ obtained from Refs. [141, 143].

| $a_n^i$ | "current" | correlation matrix | | | | |
|---|---|---|---|---|---|---|
| $a_0^+$ | 0.404(13) | 1 | 0.404 | 0.118 | 0.327 | 0.344 |
| $a_1^+$ | −0.68(13) | 0.404 | 1 | 0.741 | 0.310 | 0.900 |
| $a_2^+$ | −0.86(61) | 0.118 | 0.741 | 1 | 0.363 | 0.886 |
| $a_0^0$ | 0.490(21) | 0.327 | 0.310 | 0.363 | 1 | 0.233 |
| $a_1^0$ | −1.61(16) | 0.344 | 0.900 | 0.886 | 0.233 | 1 |

Table 33: Forecasts for $B \to \pi\ell\nu$.

| forecast | $a_0^+$ | $a_1^+$ | $a_2^+$ | $a_0^0$ | $a_1^0$ |
|---|---|---|---|---|---|
| 5 yr w/o EM | 0.404(0.0065) | −0.68(0.065) | −0.86(0.31) | 0.490(0.011) | −1.61(0.080) |
| 5 yr w/ EM | 0.404(0.0077) | −0.68(0.065) | −0.86(0.31) | 0.490(0.012) | −1.61(0.082) |
| 10 yr w/o EM | 0.404(0.0026) | −0.68(0.026) | −0.86(0.12) | 0.490(0.0042) | −1.61(0.032) |
| 10 yr w/ EM | 0.404(0.0048) | −0.68(0.027) | −0.86(0.12) | 0.490(0.0065) | −1.61(0.036) |

Due to imperfect knowledge of the resonance spectrum, the Blaschke factors are set to $B_{\{+,0\}} = 1$ for the $B \to D\ell\nu$ decay. We list the lattice inputs in Tables 36 and 37. Results for the ratio $R_D = \mathrm{Br}(B \to D\tau\nu)/\mathrm{Br}(B \to D\ell\nu)$ are available in Refs. [145, 146]. Their average is

$$R(D) = 0.300(8). \tag{86}$$

Table 34: Current input for $B_s \to K\ell\nu$ obtained from Refs. [141, 155].

| $a_n^i$ | "current" | correlation matrix | | | | |
|---|---|---|---|---|---|---|
| $a_0^+$ | 0.360(14) | 1 | 0.098 | -0.216 | 0.730 | 0.345 |
| $a_1^+$ | −0.828(83) | 0.098 | 1 | 0.459 | 0.365 | 0.839 |
| $a_2^+$ | 1.11(55) | -0.216 | 0.459 | 1 | 0.263 | 0.6526 |
| $a_0^0$ | 0.233(10) | 0.730 | 0.365 | 0.263 | 1 | 0.506 |
| $a_1^0$ | 0.197(81) | 0.345 | 0.839 | 0.652 | 0.506 | 1 |

Table 35: Forecasts for $B_s \to K\ell\nu$.

| forecast | $a_0^+$ | $a_1^+$ | $a_2^+$ | $a_0^0$ | $a_1^0$ |
|---|---|---|---|---|---|
| 5 yr w/o EM | 0.360(0.0070) | −0.828(0.042) | 1.11(0.28) | 0.233(0.0050) | 0.197(0.041) |
| 5 yr w/ EM | 0.360(0.0079) | −0.828(0.042) | 1.11(0.28) | 0.233(0.0055) | 0.197(0.041) |
| 10 yr w/o EM | 0.360(0.0028) | −0.828(0.017) | 1.11(0.11) | 0.233(0.0020) | 0.197(0.016) |
| 10 yr w/ EM | 0.360(0.0046) | −0.828(0.019) | 1.11(0.11) | 0.233(0.0031) | 0.197(0.016) |

Table 36: Current input for $B \to D\ell\nu$ obtained from Refs. [145, 146].

| $a_n^i$ | "current" | correlation matrix | | | | |
|---|---|---|---|---|---|---|
| $a_0^+$ | 0.909 (14) | 1 | 0.737 | 0.594 | 0.976 | 0.777 |
| $a_1^+$ | −7.11(65) | 0.737 | 1 | 0.940 | 0.797 | 0.992 |
| $a_2^+$ | 66 (11) | 0.594 | 0.940 | 1 | 0.666 | 0.938 |
| $a_0^0$ | 0.794 (12) | 0.976 | 0.797 | 0.666 | 1 | 0.818 |
| $a_1^0$ | −2.45(65) | 0.777 | 0.992 | 0.938 | 0.818 | 1 |

Table 37: Forecasts for $B \to D\ell\nu$.

| forecast | $a_0^+$ | $a_1^+$ | $a_2^+$ | $a_0^0$ | $a_1^0$ |
|---|---|---|---|---|---|
| 5 yr w/o EM | 0.909(0.0070) | −7.11(0.33) | 66(5.5) | 0.794(0.0060) | −2.45(0.33) |
| 5 yr w/ EM | 0.909(0.011) | −7.11(0.33) | 66(5.5) | 0.794(0.010) | −2.45(0.33) |
| 10 yr w/o EM | 0.909(0.0028) | −7.11(0.13) | 66(2.2) | 0.794(0.0024) | −2.45(0.13) |
| 10 yr w/ EM | 0.909(0.0095) | −7.11(0.15) | 66(2.3) | 0.794(0.0083) | −2.45(0.13) |

The $B \to D^*\ell\nu$ decay rate receives contributions both from the weak vector and axial-vector currents. However, modern lattice calculation [144] focuses on the zero recoil point, where the matrix element reduces to a single form factor $\mathcal{F}$ from the axial current

$$\langle D^*|A_\mu|B\rangle \;=\; i\sqrt{4M_B M_{D^*}}\epsilon_\mu\mathcal{F}. \tag{87}$$

Here $\epsilon$ represents the polarisation of $D^*$. The current input and forecasts are summarised in Table 38, where we assume 0.5% uncertainty of the EM correction estimated in Ref. [144].

Table 38: Lattice inputs for $B\to D^*\ell\nu$ from Ref. [144].

| input | current | 5 yr w/o EM | 5 yr w/ EM | 10 yr w/o EM | 10 yr w/ EM |
|---|---|---|---|---|---|
| $\mathcal{F}$ | 0.906(0.013) | 0.906(0.0065) | 0.906(0.0079) | 0.906(0.0026) | 0.906(0.0052) |

The $\Lambda_b \to p\ell\nu$ and $\Lambda_b \to \Lambda_c\ell\nu$ decays provide an independent determination on $|V_{ub}|/|V_{cb}|$. So far only one modern, unquenched calculation exists for the relevant form factors [161] leading to $|V_{ub}|/|V_{cb}|=0.083(0.004)_{\rm ex}(0.004)_{\rm lat}$. As mentioned in Sec. 7.5.2, baryons are more challenging in controlling the chiral extrapolation and finite-volume effects. While these systematics have to be checked by independent calculations, these issues can be addressed in the relatively short term.

# 8. Leptonic and Semileptonic $B$ Decays

*Editors: A. S. Kronfeld, G. De Nardo, F. J. Tackmann, R. Watanabe, A. Zupanc*
*Additional section writers: F. Bernlochner, M. Gelb, P. Goldenzweig, S. Hirose, Z. Ligeti, M. Merola, F. Metzner, G. Ricciardi, Y. Sato, C. Schwanda, M. Tanaka, P. Urquijo*

## *8.1. Introduction*

In this chapter, we consider leptonic and semileptonic $B$ meson decays that proceed in the SM via a first-order weak interactions and are mediated by the $W$ boson. $B$ meson decays involving electrons and muons are expected to be dominated by the tree-level $W$ boson decays and any new physics contributions are expected to be highly suppressed with respect to the SM. Semileptonic decays involving light leptons therefore provide an excellent laboratory for measurement of the magnitudes of the CKM-matrix elements $V_{cb}$ and $V_{ub}$. They are fundamental parameters of the SM and have to be determined experimentally. The magnitude of $V_{cb}$ normalises the Unitarity Triangle, and the ratio of magnitudes of $V_{ub}$ and $V_{cb}$ determines the side opposite to the angle $\phi_1$ ($\beta$). Thus, they play a central role in tests of the CKM sector of the SM, and complement the measurements of $CP$ asymmetries in $B$ decays. Leptonic and semileptonic decays involving the heavier $\tau$ lepton provide additional information on SM processes and can also be sensitive to non-SM contributions such as charged Higgs bosons.

The rest of the chapter is structured as follows. In this introduction, we briefly present an overview of the experimental techniques used in studies of $B$ decay modes involving neutrinos or missing energy. Then, in Sec. 8.2, we establish notation for matrix elements appearing in leptonic and semileptonic $B$ decays. In the remainder of the chapter, we present the Belle II prospects for measuring various observables in purely leptonic $B$ meson decays (Sec. 8.3), leptonic decays radiating a hard (high energy) photon (Sec. 8.4), favoured and suppressed semitauonic decays (Sec. 8.5), exclusive favoured and suppressed semileptonic decays (Sec. 8.6), and inclusive favoured and suppressed semileptonic decays (Sec. 8.7) and discuss their potential to uncover new physics. The Belle II prospects are partly based on studies performed on simulated Belle II and Belle MC samples and partly on estimates obtained by projecting existing results from Belle and Babar, taking into account improvements in the detector and software algorithms. The beam-induced background (see Sec. 4) is expected to be much higher in Belle II compared to Belle or BaBar, which represents a more challenging environment for studies of decays with missing energy. One of the major goals of all studies performed on Belle II MC samples that are presented in the rest of the chapter is to show that we can successfully and effectively suppress the much higher beam-induced background with the improved capabilities of the upgraded Belle II detector (see Chapter 3).

*Experimental techniques.* Semileptonic and leptonic decays have at least one neutrino in the final state, which escapes the detector undetected, and limits the use of kinematic constraints to reject background, which are constructed from the measured momenta and energies of the visible decay products (*e.g.* the beam-energy-constrained mass, $M_{\rm bc}$ (Eq. 190), and the energy difference, $\Delta E$ (Eq. 191)). Semileptonic and leptonic decays or other $B$ meson decays with missing energy can be measured at the $B$ factories, due to the unique experimental conditions: known production process of $B\overline{B}$ pairs and the fact that the detector encloses the interaction region almost hermetically. These two properties allow us to infer the 4-momentum of undetected particles, such as neutrinos, from measured momenta and energies of all other particles produced in the $e^+e^-$ collision (except for the neutrino) and imposing energy-momentum conservation. Such a measurement technique is commonly referred to as an untagged measurement. In the case where both the signal $B$ meson (denoted as $B_{\rm sig}$) and the other $B$ meson (denoted tag $B_{\rm tag}$ meson) are reconstructed in the event, two powerful constraints can be constructed and exploited to suppress the background or to identify signal decays.

- The missing mass squared is defined as

$$M^2_{\rm miss} = (p_{e^+e^-} - p_{B_{\rm sig}} - p_{B_{\rm tag}})^2, \tag{88}$$

  where $p_{e^+e^-}$ is the known 4-momentum of the colliding beams, and the $p_{B_{\rm sig}}$ and $p_{B_{\rm tag}}$ are the measured 4-momenta of the reconstructed signal and tag $B$ mesons, respectively. In the case of semileptonic decay of the signal $B$ meson, such as $B \to \pi\ell\nu$ or $B \to D\ell\nu$, only one neutrino is missing and hence the $M^2_{\rm miss} = m^2_\nu$ peaks at zero. Note that in this section $\ell$ typically denotes $e$ or $\mu$.
- Extra energy in the calorimeter, $E_{\rm extra}$, is defined as the sum of the energy deposits in the calorimeter that cannot be directly associated with the reconstructed daughters of the $B_{\rm tag}$ or the $B_{\rm sig}$. For signal events, $E_{\rm extra}$ (or $E_{\rm ECL}$) must be either zero or a small value arising from beam background hits and detector noise, since neutrinos do not interact in the calorimeter. On the other hand, most background events (whether $B$ decays or $q\bar{q}$ continuum) are distributed toward higher $E_{\rm extra}$ due to the contribution from additional clusters, produced by unassigned tracks and neutrals from the mis-reconstructed tag and/or signal $B$ mesons.

Measurements of leptonic and semileptonic decays have in the past been performed using three different experimental techniques, differing only in the way that the tag $B$ meson in the event is reconstructed. In untagged analyses, the missing energy and momentum of the whole event are used to determine the 4-momentum of the missing neutrino from the signal (semi)leptonic decay as described above. Measurements where the tag $B$ meson is reconstructed in well-defined decays are commonly denoted as tagged measurements. Semileptonic tagging involves partial reconstruction of a $B_{\rm tag} \to D^{(*)}\ell\nu_\ell$ decay as the tagging mode. In this case, two neutrinos are present in the event and the 4-momentum of the $B_{\rm sig}$ cannot be fully constrained. In full reconstruction tagging, a hadronically decaying $B_{\rm tag}$ meson is reconstructed, against which the signal decay recoils. The improvements to the detector acceptance, efficiency of particle detection, and the tag $B$ meson reconstruction efficiency expected in Belle II have a large impact on physics potential. The slightly reduced beam-energy asymmetry at Super KEKB compared to KEKB leads to a small increase in solid angle coverage. Improved particle identification, and $K^0_S$ reconstruction efficiency improves separation between $b \to u$ and $b \to c \to s$ transitions. Dedicated low-momentum tracking algorithms will improve tagging efficiencies and identification of events that have slow pions from $D^*$ decays. The latter is also very important for $b \to c$ background rejection in inclusive $b \to u\ell\nu$ analyses. See Sec. 6.6 for more details on tag $B$ meson reconstruction and expected performance at Belle II.

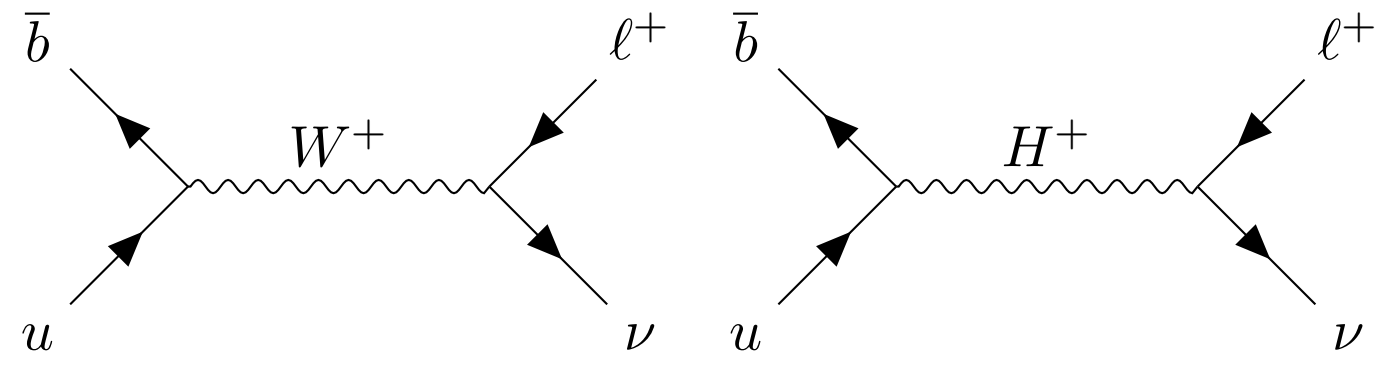

Fig. 61: Feynman diagrams of purely leptonic $B^+$ decays, mediated by a charged weak boson (left) or a charged Higgs as predicted in new physics models (right).

## *8.2. Matrix Elements of Electroweak Currents*

Author: A. S. Kronfeld (th.)

As hadronic matrix elements in exclusive leptonic and semileptonic decays are used in Chapter 9, as well as here, it is convenient to standardise notation by collecting the necessary formulae in one place. To keep the notation general, we write the definitions of decay constants and form factors using $B$ mesons in the initial state decaying to either pseudoscalar mesons ($P = D$, $\pi$, $K$) or vector mesons ($V = D^*$, $\rho$, $K^*$) in the final state. The CKM elements for the tree-level decays will be abbreviated $V_{qb}$, where $q = c$, $u$.

*8.2.1. Leptonic Decays $B^+ \to \ell^+\nu$ and $B^0 \to \ell^+\ell^-$.* At leading order in the electroweak interaction, the amplitude for the leptonic decay contains a hadronic factor,

$$\langle 0|A^\mu|B(p)\rangle = ip^\mu f_B, \tag{89}$$

where $A^\mu$ is an axial-vector current (for the charged current, $A^\mu = \bar{b}\gamma^\mu\gamma^5 u$), and the decay constant $f_B$ is a useful parameterisation, because the only Lorentz structure available is the $B$-meson 4-momentum $p^\mu$. By conservation of angular momentum, the only other non-vanishing matrix element for $B \to$ no-hadrons is

$$\langle 0|P|B(p)\rangle = -i\frac{M_B^2}{m_b + m_u} f_B, \tag{90}$$

where $P$ is the pseudoscalar density (here $P = \bar{b}\gamma^5 u$), $M_B$ is the $B$-meson mass, and $m_b$ and $m_u$ are renormalised quark masses.[19] The decay constant $f_B$ is the same in Eqs. (89) and (90) owing to the partial conservation of the axial-vector current (PCAC), $\partial \cdot A = i(m_b + m_u)P$, which holds when $A^\mu$, $P$, and the masses are renormalised consistently. These considerations apply amplitudes both to the charged-current decay $B^+ \to \ell^+\nu_\ell$ and to the flavour-changing neutral-current (FCNC) decay $B^0_{(s)} \to \ell^+\ell^-$. In each formula in this section, $M_B$ and $f_B$ are the mass and decay constant of the $B^\pm$, $B^0$, or $B_s$ meson, as the case may be. Feynman diagrams of SM and beyond SM leptonic $B^+$ decays are shown in Fig. 61.

The partial width for either decay is (assuming axial contributions only)

$$\Gamma(B \to \ell_1\ell_2) = \frac{M_B}{4\pi}|G|^2 f_B^2 \zeta_{12} \frac{\lambda_{12}^{1/2}}{M_B^2}, \tag{91}$$

[19] We use lower case $m$ for masses of elementary particles (quarks and leptons) and upper case $M$ for hadron masses.

where $G$ contains couplings and (for FCNCs) loop factors, $m_1$ and $m_2$ are the lepton masses, and

$$\lambda_{12} = (M_B^2 - m_1^2 - m_2^2)^2 - 4m_1^2 m_2^2, \tag{92}$$

$$\zeta_{12} = m_1^2 + m_2^2 - \frac{(m_1^2 - m_2^2)^2}{M_B^2}. \tag{93}$$

These formulas do not hold when the final-state leptons' masses differ unless the interaction boils down to $V \pm A$. In a general setting, $|G|^2\zeta_{12}$ must be replaced with a more complicated expression. Processes such as $B^0 \to \mu^\pm\tau^\mp$ have unmeasurably small rates in the SM, so the general formula is not important.

In the SM, one finds

$$G = \frac{G_F}{\sqrt{2}} V_{ub}, \quad (m_{\nu_\ell} \to 0), \qquad \text{charged-current decay } B^+ \to \ell^+\nu_\ell, \tag{94}$$

$$G = \frac{G_F^2 m_W^2}{\pi^2} V_{tb}^* V_{tq} C_A, \qquad \text{FCNC decay } B^0_{(s)} \to \ell^+\ell^-,\ q \in \{d, s\}, \tag{95}$$

where $G_F$ is the Fermi constant, $V$ is the CKM matrix, $m_W$ is the $W$-boson mass, and $C_A$ is the Wilson coefficient obtained from integrating out the massive $W$, $Z$, and top quark. Reference [219] contains results for $C_A$ including QED corrections.

The factor of the lepton mass in the leptonic-decay amplitude arises because the lepton has to flip its spin to conserve angular momentum. This helicity suppression (for $\ell = e, \mu$) does not apply to the radiative leptonic decay $B^+ \to \ell^+\nu_\ell\gamma$. This feature is relevant for $D^+_{(s)} \to \mu^+\nu_\mu(\gamma)$ and important for $B^+ \to \mu^+\nu_\mu(\gamma)$ [220]. (For the $D_{(s)}$ decay, Ref. [221] estimates a 1% effect for photon cuts used in existing measurements.) Once measurements of the $B^+ \to \mu^+\nu_\mu$ branching fraction are made with a precision of a few percent, theorists should revisit the radiative corrections; for light mesons these issues are under control [222]. As discussed in Sec. 8.4.1, when the photon is hard, $E_\gamma \sim \frac{1}{2}M_B$, these decays can be used to extract information about $B$-meson structure and non-leptonic decays [223].

*8.2.2. Semileptonic Decay to a Pseudoscalar Meson.* The amplitudes for the semileptonic decays $B^0 \to P^-\ell^+\nu_\ell$ and $B^+ \to P^0\ell^+\nu_\ell$, at leading order in the electroweak interaction, contain the hadronic factor

$$\langle P(k)|V^\mu|B(p)\rangle = \left(p^\mu + k^\mu - \frac{M_B^2 - M_P^2}{q^2} q^\mu\right) f_+(q^2) + \frac{M_B^2 - M_P^2}{q^2}\, q^\mu\, f_0(q^2), \tag{96}$$

where $V^\mu$ is the vector part of the weak current ($V^\mu = \bar{b}\gamma^\mu u$ for $B \to \pi$ and $B_s \to K$, and $V^\mu = \bar{b}\gamma^\mu c$ for $B \to D$ and $B_s \to D_s$). Two 4-vectors appear in this process, and, hence, two form factors, which are functions of $q^2$ (where $q = p - k$). The vector (scalar) form factor $f_+$ ($f_0$) arises when the $\ell\nu_\ell$ system has $J^P = 1^-$ ($0^+$). At $q^2 = 0$, $f_0(0) = f_+(0)$.

Beyond the SM, scalar and tensor currents can mediate these decays. Such contributions to the decay amplitude entail the scalar and tensor form factors

$$\langle P(k)|S|B(p)\rangle = \frac{M_B^2 - M_P^2}{m_b - m_q} f_0(q^2), \tag{97}$$

$$\langle P(k)|T^{\mu\nu}|B(p)\rangle = \frac{2}{M_B + M_P}\,(p^\mu k^\nu - p^\nu k^\mu)\, f_T(q^2), \tag{98}$$

where $S$ and $T^{\mu\nu}$ are scalar and tensor currents (here $S = \bar{b}q$, $T^{\mu\nu} = \bar{b}i\sigma^{\mu\nu}q$, $q = c, u$). The scalar form factor in Eq. (97) is the same as that Eq. (96), owing to the partial conservation of the vector current (PCVC), $i\partial \cdot V = (m_b - m_q)S$. Feynman diagrams of SM and beyond SM semileptonic $B$ decays are shown in Fig. 62.

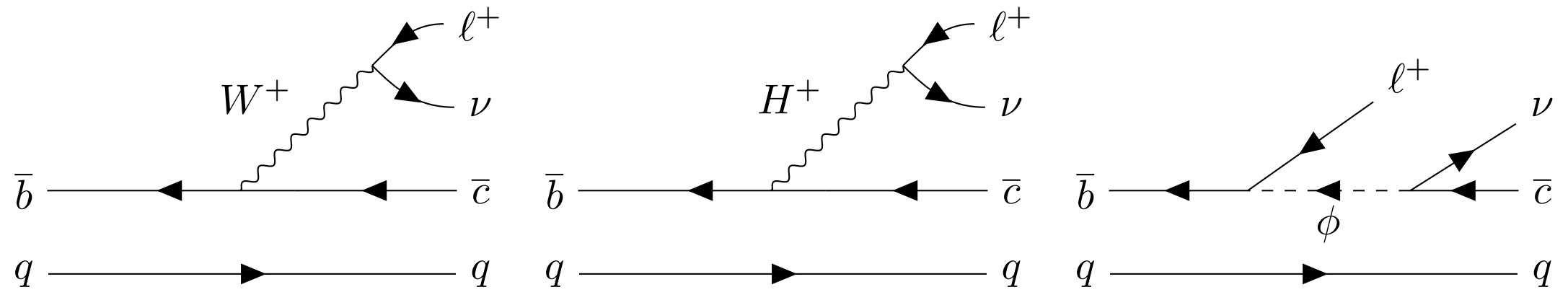

Fig. 62: Feynman diagrams of semileptonic $B$ decays, mediated by a charged weak boson (left) as well as mediators predicted in new physics models: a charged Higgs (middle), and a leptoquark (right).

The doubly differential partial width for $B \to P\ell^\pm\nu_\ell$ (assuming no scalar or tensor current) is [224]

$$\frac{d^2\Gamma}{dq^2\, d\cos\theta_\ell} = C_q|\eta_{\rm EW}|^2 \frac{G_F^2|V_{qb}|^2}{(2\pi)^3} \frac{\lambda^{1/2}}{8M_B^3} \frac{\lambda_{12}^{1/2}}{q^2} \left[\left(q^2 - m_1^2 - m_2^2 - \frac{\lambda_{12}}{q^2}\cos^2\theta\right)\frac{\lambda}{q^2}|f_+|^2 + \right. \tag{99}$$

$$\left. + \zeta_{12}\frac{(M_B^2 - M_P^2)^2}{q^2}|f_0|^2 \mp 2(m_1^2 - m_2^2)(M_B^2 - M_P^2)\frac{\lambda^{1/2}}{q^2}\frac{\lambda_{12}^{1/2}}{q^2}\cos\theta\, \Re\left(f_+ f_0^*\right)\right],$$

where $C_q = 1/2$ for $\pi^0$ and 1 otherwise[20], $\eta_{\rm EW}$ is an electroweak correction discussed below, $\lambda_{12}$ and $\zeta_{12}$ are obtained from Eqs. (92) and (93) by substituting $M_B^2 \to q^2$, and

$$\lambda = (M_B^2 + M_P^2 - q^2)^2 - 4M_B^2 M_P^2, \tag{100}$$

$$\cos\theta = 4\lambda^{-1/2}\left(1 - \frac{m_\ell^2}{q^2}\right)^{-1}\left(\frac{p_B \cdot q\, p_\ell \cdot q}{q^2} - p_B \cdot p_\ell\right), \tag{101}$$

the last being the angle in the centre-of-mass of the $\ell\ell$ system between the $B$ meson and lepton 1 with charge $\pm 1$. Quantities such as $\lambda$, $\lambda_{12}$ are sometimes known as the Källén functions.

Integrating over $\cos\theta$,

$$\frac{d\Gamma}{dq^2} = C_q|\eta_{\rm EW}|^2\frac{G_F^2|V_{qb}|^2}{(2\pi)^3}\frac{\lambda^{1/2}}{4M_B^3}\frac{\lambda_{12}^{1/2}}{q^2}\left\{\lambda\beta_{12}|f_+|^2 + \zeta_{12}\frac{(M_B^2 - M_P^2)^2}{q^2}|f_0|^2\right\}, \tag{102}$$

where

$$\beta_{12} = 1 - \frac{m_1^2 + m_2^2}{q^2} - \frac{\lambda_{12}}{(q^2)^2}. \tag{103}$$

For a massless neutrino,

$$\beta_{\ell 0} = \left(1 - \frac{m_\ell^2}{q^2}\right)\left(\frac{2}{3} + \frac{m_\ell^2}{3q^2}\right), \tag{104}$$

$$\zeta_{\ell 0} = m_\ell^2\left(1 - \frac{m_\ell^2}{q^2}\right), \tag{105}$$

$$\frac{\lambda_{\ell 0}^{1/2}}{q^2} = \left(1 - \frac{m_\ell^2}{q^2}\right). \tag{106}$$

[20] This factor stems from the fact that a $b \to u$ current produces only the $\bar{u}u$ component of the $\pi^0$.

The kinematic factors for $\ell^+\ell^-$ are obtained by setting $m_2 = m_1 = m_\ell$:

$$\beta_{\ell\ell} = \frac{2}{3}\left(1 - \frac{m_\ell^2}{q^2}\right), \tag{107}$$

$$\zeta_{\ell\ell} = 2m_\ell^2, \tag{108}$$

$$\frac{\lambda_{\ell\ell}^{1/2}}{q^2} = \left(1 - \frac{4m_\ell^2}{q^2}\right)^{1/2}. \tag{109}$$

These formulas hold for a $V - A$ lepton current; in general the pattern of lepton masses and couplings is more complicated. Of course, any measurable signal with lepton flavour violation, *i.e.*, $m_1 \neq m_2$, is a major discovery whatever the $V$ and $A$ couplings are.

When the scalar and tensor currents contribute (beyond the SM), the expression for $d\Gamma/dq^2$ becomes very complicated. See Ref. [225] or the arXiv version of Ref. [226] for the full formula (written for $D \to K\ell\nu$ decay). See also Ref. [227].

The decay amplitude has two interesting corrections. In Eq. 99, $\eta_{\text{EW}}$ denotes the leading logarithmic contribution of two-loop electroweak diagrams,

$$\eta_{\text{EW}} = 1 + \frac{\alpha}{\pi}\left[\ln\frac{M_W}{\mu} + \tan^2\theta_W \frac{M_W^2}{M_Z^2 - M_W^2}\ln\frac{M_Z}{M_W}\right], \tag{110}$$

where $\mu$ is a scale separating this contribution and contributions that depend on hadron structure. In the leading-logarithmic approximation, the same hadronic matrix elements arise, so the correction is multiplicative; in this context, it is reasonable to set $\mu = M_B$. In Eqs. 99, 102, and 122 (below), $G_F$ is defined via the muon lifetime. Second, final states with two charged particles have a Coulomb attraction that increases the rate. For a discussion, see Ref. [228, 229]. More theoretical work may be needed, but it is clear that experimental results must be reported separately for $B^0 \to P^-\ell^+\nu$ and $B^+ \to P^0\ell^+\nu$.

*8.2.3. Semileptonic Decay to a Vector Meson.* Last, let us consider the amplitude for the semileptonic decay $B^+ \to V^0\ell^+\nu_\ell$ at leading order in the electroweak interaction. Now there are three 4-vectors in the process, so the decomposition of the amplitude into form factors reads ($\varepsilon^{0123} = +1$)

$$\langle V(k,\epsilon^{(V)})|V^\mu|B(p)\rangle = i\bar\epsilon_\rho^{(V)}\frac{2\varepsilon^{\mu\rho}{}_{\sigma\tau}p^\sigma k^\tau}{M_B + M_V}V(q^2), \tag{111}$$

$$\begin{aligned}\langle V(k,\epsilon^{(V)})|A^\mu|B(p)\rangle = \bar\epsilon_\rho^{(V)}\bigg[&2M_V\frac{q^\rho q^\mu}{q^2}A_0(q^2) + (M_B + M_V)\left(g^{\rho\mu} - \frac{q^\rho(p+k)^\mu}{M_B^2 - M_V^2}\right)A_1(q^2) \\ &+ 2M_V q^\rho\left(\frac{(p+k)^\mu}{M_B^2 - M_V^2} - \frac{q^\mu}{q^2}\right)A_3(q^2)\bigg],\end{aligned} \tag{112}$$

with the same vector current, $V^\mu$, as above and the axial current $A^\mu = \bar b\gamma^\mu\gamma^5 u$ or $\bar b\gamma^\mu\gamma^5 c$. Here, $\bar\epsilon^{(V)}$ denotes the polarisation vector of the final state. Note that $M_B^2 - M_V^2 = q\cdot(p + k)$. Sometimes $A_3$ is eliminated in favour of a form factor $A_2$ via $2M_V A_3 = (M_B + M_V)A_1 - (M_B - M_V)A_2$. At $q^2 = 0$, $A_0(0) = A_3(0)$.

It is more useful to decompose the amplitude according the helicity of the virtual $W$ [224]. There are several notations of form factors in the literature. Whatever one chooses on the right hand side of Eqs. (111) and (112), it is straightforward to relate the matrix elements to the helicity amplitudes.

Being off shell, the $W$ has four polarisations: scalar (spin 0), longitudinal, and two transverse (the last three spin 1). In the frame with the $B$ at rest and the $V$ flying out along the $+z$ axis, the polarisation vectors, respectively, are ($q^0 = M_B - E_V$, $E_V = p \cdot k/M_B$)

$$\epsilon_s^{(W)} = \frac{1}{\sqrt{q^2}}\left(q^0, 0, 0, -|\boldsymbol{k}|\right) = \frac{q}{\sqrt{q^2}}, \tag{113}$$

$$\epsilon_0^{(W)} = \frac{1}{\sqrt{q^2}}\left(|\boldsymbol{k}|, 0, 0, -q^0\right), \tag{114}$$

$$\epsilon_\pm^{(W)} = \frac{1}{\sqrt{2}}\left(0, \pm 1, -i, 0\right), \tag{115}$$

where $\boldsymbol{k}$ is the three-momentum of the final-state vector meson in the rest frame of the $B$. The subscript $s$ denotes the $J = 0$ partial wave (for historical reasons), and 0 and $\pm$ denote the $J_z$ component of the $J = 1$ partial wave. Similarly

$$\epsilon_0^{(V)} = \frac{1}{M_V}\left(|\boldsymbol{k}|, 0, 0, E_V\right), \tag{116}$$

$$\epsilon_\pm^{(V)} = \frac{1}{\sqrt{2}}\left(0, \mp 1, -i, 0\right) \tag{117}$$

provide the polarisation vectors for the final-state vector meson. In Eqs. (111)–(112), a bar on a polarisation vector denotes complex conjugation in Minkowski space, and complex conjugation of only the spatial components in Euclidean space (useful in lattice QCD).

The helicity amplitudes $H_a = \langle V(k, \epsilon^{(V)})|\bar{\epsilon}_a^{(W)} \cdot (V - A)|B(p)\rangle$ are then

$$H_s(q^2) = -\frac{\lambda^{1/2}}{\sqrt{q^2}} A_0(q^2), \tag{118}$$

$$H_0(q^2) = -\frac{\sqrt{q^2}(M_B^2 + 3M_V^2 - q^2)}{2M_V(M_B - M_V)} A_1(q^2) - \frac{\lambda}{(M_B^2 - M_V^2)\sqrt{q^2}} A_3(q^2), \tag{119}$$

$$H_\pm(q^2) = -(M_B + M_V)A_1(q^2) \pm \frac{\lambda^{1/2}}{M_B + M_V} V(q^2), \tag{120}$$

where the Källén function $\lambda$ is the same as before, except with $M_V$ instead of $M_P$. In $H_s$ and $H_0$, the final-state vector meson has $J_z = 0$; in $H_\pm$, it has $J_z = \pm 1$. Note that in lattice QCD, it is most straightforward to compute $A_1$, $V$, and two more linear combinations of $A_0$, $A_1$, and $A_3$. The full amplitude is then proportional to

$$\sum_{ab} g^{ab} L_a H_b = L_s H_s - L_0 H_0 - L_+ H_+ - L_- H_-, \quad a \in \{s, 0, +, -\}, \tag{121}$$

with lepton helicity amplitudes $L_a = \bar{u}(\nu)\gamma \cdot \epsilon_a^{(W)}(1 - \gamma_5)v(\ell)$.

The triple differential rate (in $q^2$, $\cos\theta$, and $\phi$, which is the angle between the decay planes of $B$ and $V$) for the semileptonic decay $B^+ \to V^0 \ell^+ \nu_\ell$ can be found in Refs. [224]. Integrating over all angles,

$$\frac{d\Gamma}{dq^2} = C_q |\eta_{\mathrm{EW}}|^2 \frac{G_F^2 |V_{qb}|^2}{(2\pi)^3} \frac{\lambda^{1/2}}{4M_B^3} \frac{\lambda_{12}^{1/2}}{q^2} \left\{ q^2 \beta_{12} \left[ |H_+|^2 + |H_-|^2 + |H_0|^2 \right] + \zeta_{12} |H_s|^2 \right\}, \tag{122}$$

where $C_q = 1/2$ for $\rho^0$ and 1 otherwise, $\lambda_{12}$ and $\zeta_{12}$ are obtained from Eqs. (92) and (93) by substituting $M_B^2 \to q^2$, and

$$\beta_{12} = 1 - \frac{m_1^2 + m_2^2}{q^2} - \frac{\lambda_{12}}{(q^2)^2}. \tag{123}$$

Note that the differential rate for the semileptonic decay $B^+ \to P^0\ell^+\nu_\ell$ is the same after dropping the $H_\pm$ terms.[21] These formulas again hold for a $V-A$ lepton current; in general the pattern of lepton masses and couplings is more complicated.

Beyond the SM, the pseudoscalar and tensor currents can mediate these decays, in addition to the SM vector and axial-vector currents. The matrix element for the pseudoscalar follows in analogy to Eq. (90):

$$\langle V(k,\epsilon^{(V)})|P|B(p)\rangle = \frac{2M_V}{m_b+m_q}\bar{\epsilon}^{(V)}\cdot qA_0(q^2) = \frac{\lambda^{1/2}}{m_b+m_q}A_0(q^2), \tag{124}$$

with the last equality holding only in the polarisation, namely $\epsilon^{(V)} = \epsilon_0^{(V)}$, with a nonzero amplitude. The tensor current has the matrix element

$$\begin{aligned}\langle V(k,\epsilon^{(V)})|T^{\mu\nu}|B(p)\rangle = i\varepsilon^{\mu\nu}{}_{\sigma\tau}\bar{\epsilon}_\rho^{(V)}\Bigg\{ & g^{\rho\sigma}(p+k)^\tau T_1(q^2) - g^{\rho\sigma}q^\tau\frac{M_B^2-M_V^2}{q^2}\left[T_1(q^2)-T_2(q^2)\right] \\ & + q^\rho\frac{(p+k)^\sigma q^\tau}{q^2}\left[T_1(q^2)-T_2(q^2)-\frac{q^2}{M_B^2-M_V^2}T_3(q^2)\right]\Bigg\}.\end{aligned} \tag{125}$$

In penguin amplitudes, the combinations $q_\nu T^{\mu\nu}$ and $\varepsilon^{\mu\nu\alpha\beta}q_\nu T_{\alpha\beta}$ appear, leading straightforwardly to additional terms in the helicity amplitudes. See also Ref. [227].

The discussion of electroweak and Coulomb correction in the paragraph with Eq. 110 applies here too.

### *8.3. Leptonic B decays*

Authors: G. De Nardo (exp.), M. Merola (exp.), R. Watanabe (th.)

The branching fraction of $B^+ \to \ell^+\nu$, $\mathcal{B}_\ell$, is proportional to the mass squared of the charged lepton, cf. Eqs. 91 and 93. Hence, $B_\tau$, $B_\mu$, and $B_e$ are hierarchical in the respective lepton mass in the absence of new physics. We take $|V_{ub}| = (3.55\pm0.12)\times10^{-3}$, determined from exclusive semileptonic $B$ decays by HFLAV [230], and $f_B = (186\pm4)\,\text{MeV}$ from Ref. [217], which is the only entry in the 2016 FLAG [140] average with four active flavours.[22] The predicted values for the SM are then found to be

$$B_\tau = (7.7\pm0.6)\times10^{-5},\quad B_\mu = (3.5\pm0.3)\times10^{-7},\quad B_e = (8.1\pm0.6)\times10^{-12}. \tag{126}$$

This class of decays is of interest not only to test the SM but also search for new physics at Belle II.

Past measurements of $\mathcal{B}(B\to\tau\nu_\tau)$ by Belle and BaBar were performed with two independent approaches to reconstruct $B_{\text{tag}}$: using semileptonic and hadronic decays [232–236]. At present, no single experiment finds a significance greater than $5\sigma$. Combining the measurements by Belle and BaBar, the world average is given as $(1.06\pm0.19)\times10^{-4}$ [230], which has over $5\sigma$ significance. This is consistent with the prediction ($B_\tau = (7.7\pm0.6)\times10^{-5}$) at $2\sigma$.

---

[21] In $B\to P\ell\nu$, $H_0(q^2) = (\lambda/q^2)^{1/2}f_+(q^2)$ and $H_s = [(M_B^2-M_P^2)/\sqrt{q^2}]f_0(q^2)$.

[22] FLAG will update its averages in 2018. For decay constants, the most significant new result is $f_B = 189.4\pm1.4$ MeV from Ref. [231].

The light-leptonic modes $B \to \ell\nu_\ell$ for $\ell = e, \mu$ are two-body decays, which implies that the charged lepton momentum in the rest frame of $B_{\rm sig}$ is $m_B/2$. Thus, this unique 2-body decay topology can be exploited in search analyses. The light-leptonic modes have not yet been observed [237, 238]. The upper limit on $B_\mu$ is then summarised as $< 1 \times 10^{-6}$ at 90% CL, whereas that on $B_e$ is also given as $< 0.98 \times 10^{-6}$ [77].

The above summary shows that the present branching fraction measurement of $B^+ \to \tau^+\nu_\tau$ and upper limit of $B^+ \to \mu^+\nu_\mu$ are already close to their SM predictions. We expect that these processes will eventually be observed with more than $5\sigma$ significance at SuperKEKB/Belle II. The decay $B^+ \to e^+\nu_e$ can be observed only if new physics greatly enhances its decay rate.

In the absence of new physics, purely leptonic decays can provide direct determinations of $|V_{ub}|$ with relatively small theoretical uncertainty. Since discrepancies amongst the $|V_{ub}|$ determinations from exclusive and inclusive processes are long standing, leptonic decays can provide important orthogonal information as is done in the determination of $|V_{cd}|$ and $|V_{cs}|$.

The presence of new physics with different chiral structure would primarily be observed through modifications to $B^+ \to \ell^+\nu_\ell$ rates. Namely, we can describe the branching fraction as.

$$\mathcal{B}(B^+ \to \ell^+\nu_\ell)_{\rm NP} = \mathcal{B}(B^+ \to \ell^+\nu_\ell)_{\rm SM} \times \left|1 + r^\ell_{\rm NP}\right|^2, \tag{127}$$

for the new physics model. Comparing the current data and the SM reference values shown above, we can see the present constraints as

$$\left|1 + r^\tau_{\rm NP}\right| = 1.17 \pm 0.12\,, \quad \left|1 + r^\mu_{\rm NP}\right| < 1.7 \ (90\% \ {\rm CL})\,, \quad \left|1 + r^e_{\rm NP}\right| < 348 \ (90\% \ {\rm CL})\,. \tag{128}$$

Theoretical uncertainties are not taken into account in the latter two results as they are considered negligible.

### *8.3.1. $B \to \tau\nu_\tau$.*

*Belle II Full Simulation Study.* The study presented here aims at estimating the precision of Belle II on the measurement of the branching fraction of $B \to \tau\nu_\tau$ with 1, 5 and 50 ab$^{-1}$ of data respectively. The analysis is performed on the MC5 Belle II production (see Sec. 4) corresponding to 1 ab$^{-1}$ of generic $B^+B^-/B^0\overline{B}^0$, $u\overline{u}$, $d\overline{d}$, $s\overline{s}$, $c\overline{c}$ background processes and $100 \times 10^6$ signal events. In these samples the expected machine background (see Sec. 4) is superimposed with the simulated primary collision events.

The analysis strategy is to use a hadronic tag method through the Full Event Interpretation (FEI) algorithm (Sec. 6.6). It makes use of thousands of $B$ meson decay modes and builds up a multivariate discriminant to assign to each $B$ candidate (tag) a probability of correct reconstruction. In order to reject mis-reconstructed $B_{\rm tag}$ candidates, a criterion is placed on the FEI discriminant corresponding to purities of 49% and 93% for correctly reconstructed $B$ mesons in the background and signal samples, respectively. In the case that multiple candidates are reconstructed in the event, the one with the highest FEI discriminant value is chosen. The purity of the samples is evaluated after continuum background rejection by means of a fit to the $M_{bc}$ distribution. The $M_{bc}$ distribution is modelled with an Argus function for the combinatorial background plus a Crystal Ball function for the correctly reconstructed $B$ candidates. The number of events under the Crystal Ball is then counted above an $M_{bc}$ threshold of 5.275 GeV/$c^2$.

After the reconstruction of the $B_{\text{tag}}$ side, the presence of only one additional track in the event is required, consistent with a 1-prong $\tau$ decay. Particle identification criteria (PID) are applied to select four $\tau$ decay modes: $\mu\nu\overline{\nu}$, $e\nu\overline{\nu}$, $\pi\nu$, and $\rho\nu$. The PID criteria are based on likelihood ratios, described in detail in Sec. 5.5. Candidate charged $\rho$ mesons are required to originate from $\pi\pi^0$ pairs in the mass window $0.625 < m_{\pi\pi^0} < 0.925$ GeV/$c^2$; in turn the $\pi^0$ candidates are reconstructed by pairing two neutral clusters and applying an invariant mass window on the $\gamma\gamma$ pair of $0.12 < m_{\gamma\gamma} < 0.16\,\text{GeV}/c^2$. Mis-reconstructed $B_{\text{tag}}$ candidates are suppressed at this stage applying the following selection criteria: $5.275 < M_{bc} < 5.290\,\text{GeV}/c^2$ and $-0.20 < \Delta E < 0.04$ GeV.

Due to the high level of machine background in Belle II (due to the smaller beam size of SuperKEKB) a dedicated study has been performed on MC simulated events to optimally select the photon candidates from $e^+e^-$ collisions (from now on called "physics" photons) and reject beam induced background photon candidates (from now on called "background" photons). Several cluster-related discriminating variables have been exploited for this purpose, among which the most important are the cluster energy, the cluster timing and the ratio between the energy deposited in a 3×3 and in 5×5 square of crystals around the centre of the cluster, E9/E25. (Note that E9/E21 will be used in future Belle II physics analyses). Physics photon candidates are required to satisfy a minimum energy threshold since they have a harder energy spectrum than background photons. Beam-induced photon production is not correlated with bunch crossings, and so the cluster time distribution shows a uniform distribution for background photons and a peak near the bunch crossing time for physics photons. A tight time window is selected corresponding to a 90–95% efficiency for physics photons. Physics photon candidates are expected to have a relatively narrow E9/E25 distribution consistent with a single photon, while beam induced photon showers exhibit a larger spread of energy deposits. As background photons are expected to have large impact on the forward region of the detector, different selection criteria are imposed for the forward, barrel and backward detector regions. These photon candidates are used in $\pi^0$ reconstruction and for determining the remaining energy deposition in the calorimeter from physics photons, denoted $E_{\text{ECL}}$.

In order to reduce contamination from continuum background events (mainly $e^+e^- \to q\overline{q}$), several topological variables (Sec. 6.4) have been considered: normalised second Fox-Wolfram moments, $\cos\theta_{\text{thrust}}$, CLEO Cones and KSFW moments, exploiting the different topology of events with spherical symmetry as $B^+B^-$ over events with back-to-back symmetry, as $\ell^+\ell^-$ and, to a lesser extent, $q\overline{q}$. Keeping only the variables that are weakly correlated with $E_{\text{ECL}}$, two multivariate discriminants using Boosted Decision Trees (BDTs) have been trained on continuum background and signal $B \to \tau\nu_\tau$ events, in the signal window $5.27 < M_{bc} < 5.29\,\text{GeV}/c^2$ and $E_{\text{ECL}} < 0.3\,\text{GeV}$, using the TMVA toolkit [239]. Leptonic and hadronic $\tau$ decay modes are trained separately, since the latter are most affected by continuum background. Continuum events are rejected by placing a threshold on the BDT discriminant at the maximum point in the figure of merit (FOM) $S/\sqrt{S+B}$, where $S$ and $B$ are the number of signal and background events, respectively. The thresholds are found to be $\text{BDT}_{\text{had}} > 0.2$ corresponding to 99% continuum rejection and 47% signal efficiency, and $\text{BDT}_{\text{lep}} > 0.04$, corresponding to 93% continuum rejection and 65% signal efficiency. Figure 63 shows the BDT discriminant output for signal and background events, separated by hadronic and leptonic modes.

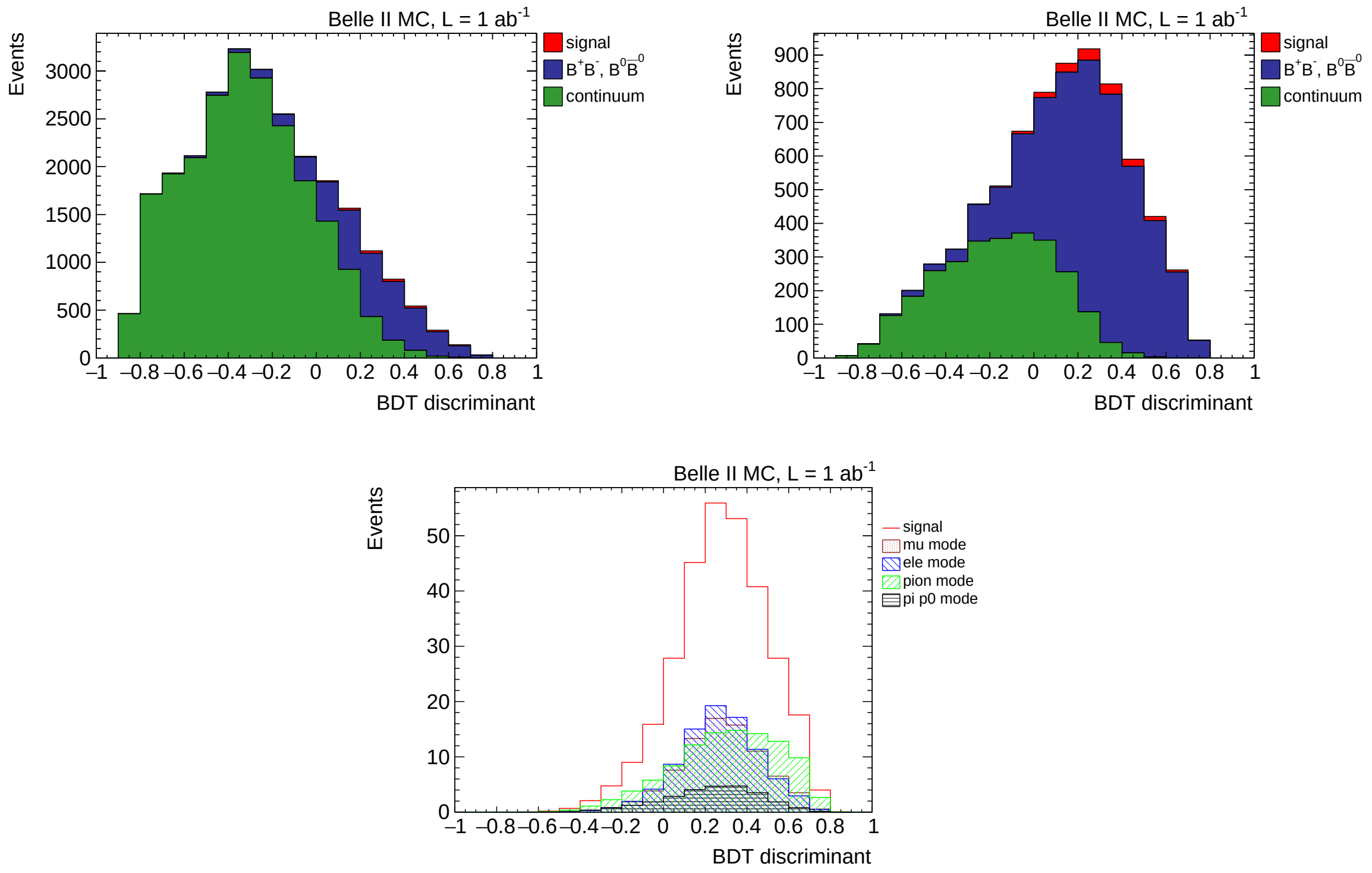

Fig. 63: Top left and right: BDT discriminant distributions for the $B \to \tau\nu_\tau$ analysis, depicting the signal (red), $B\bar{B}$ background (blue) and continuum events (green), in the hadronic (left) and leptonic (right) $\tau$ decay channels. Bottom: BDT discriminant distribution for signal events, showing separately the contribution of the four decay modes. The events are normalised to an integrated luminosity of 1 $\text{ab}^{-1}$.

A characteristic feature of $B \to \tau\nu_\tau$ decays is the presence of two or three neutrinos in the final state. This property can be used in the analysis by requiring that a significant amount of missing energy and momentum is present for leptonic channels, $M^2_{\text{miss}} > 12(\,\text{GeV}/c^2)^2$, and less for the hadronic channels, $M^2_{\text{miss}} < 12(\,\text{GeV}/c^2)^2$. In addition the reconstructed momentum of the $\pi$ and $\rho$ on the signal side in the CMS frame is required to satisfy $p^*_{\text{sig}} > 1.6$ GeV/$c$. The thresholds listed for $M_{\text{miss}}$ and $p^*_{\text{sig}}$ have been chosen based on FOM optimisation. It should be noted that Belle used $M_{\text{miss}}$ in the signal yield fit, however we have taken a simplified 1-D fit approach for this sensitivity study.

In Fig. 64(a), the $E_{\text{ECL}}$ distribution is shown after applying all selection criteria. The comparison of signal $E_{\text{ECL}}$ distribution of this analysis with the one obtained by the Belle Collaboration measurement with hadronic tag [234] is also shown in figure 64(b). The extra energy resolution at Belle II is slightly better than Belle despite the increased beam background levels. The one-sided 68 percentile of the $E_{\text{ECL}}$ distribution is found to be 0.22 GeV for Belle II and 0.28 GeV for Belle.

The signal efficiencies and expected signal and background yields in two $E_{\text{ECL}}$ windows are reported in Table 39 for this analysis and compared to the Belle MC from hadronic tag measurement [234].

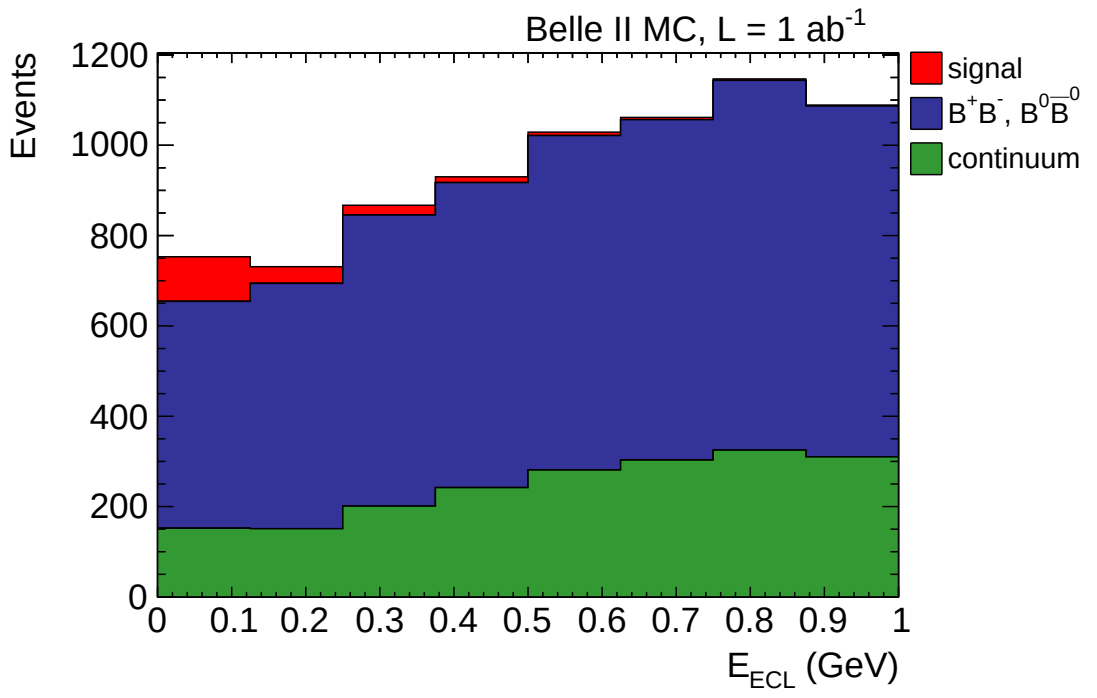

(a) $E_{\mathrm{ECL}}$ distribution for signal (red), $B\overline{B}$ background (blue) and continuum (green). The events are normalised to an integrated luminosity of 1 ab$^{-1}$

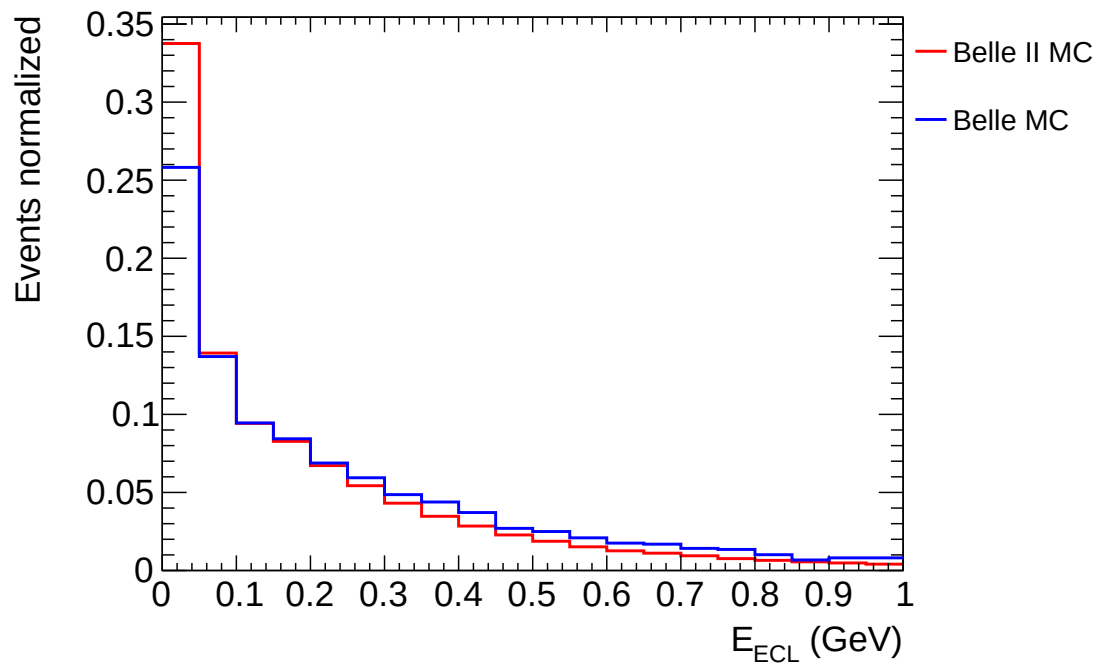

(b) Comparison of signal $E_{\mathrm{ECL}}$ distribution for this analysis (red) and the Belle measurement with hadronic tag (blue).

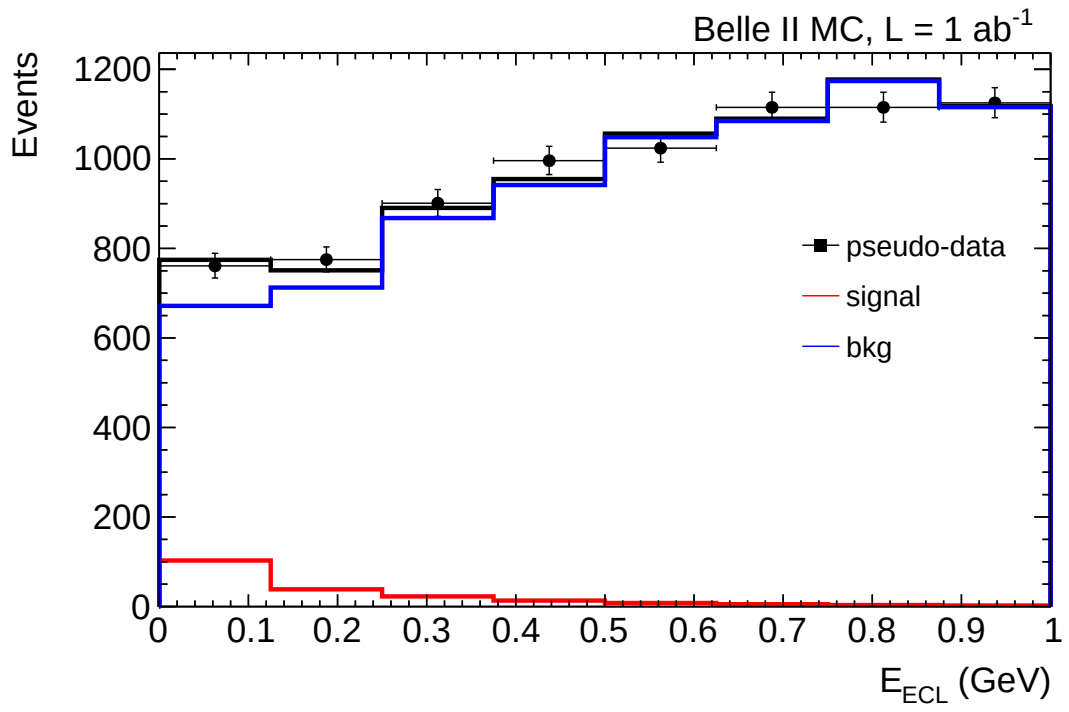

(c) Maximum likelihood fit to pseudo-data $E_{\mathrm{ECL}}$ distribution sampled from simulation. The red and blue histograms represent the signal and background fit functions (templates from simulation). The events correspond to an integrated luminosity of 1 ab$^{-1}$.

Fig. 64: $E_{\mathrm{ECL}}$ distributions for signal and background in the analysis of $B \to \tau\nu$ .

In order to estimate the expected precision of the $B \to \tau\nu_\tau$ branching fraction measurement, a toy MC study has been performed generating a high-statistics sample of pseudo-experiments. For each experiment, a pseudo-dataset has been generated according to the signal and background MC expectations, and a binned maximum likelihood fit is performed using a two-component parameterised function where the $E_{\mathrm{ECL}}$ distributions for signal and background events are taken from simulation. In Fig. 64(c) an illustrative plot of the fit to one pseudo-dataset is shown.

Assuming a branching fraction of $0.82\times10^{-4}$ (December 2016 result from the CKMfitter group [91]) the mean uncertainty is found to be $\sim$29%, with 1 ab$^{-1}$ of equivalent integrated luminosity. A large number of pseudo-experiments has been generated to estimate the expected significance of the branching fraction measurement, according to the following procedure: a likelihood ratio test statistic $Q$ has been defined and evaluated on pseudo-datasets sampled from signal plus background (S+B) and background only $E_{\mathrm{ECL}}$ distributions. Then the p-value of the background null hypothesis is evaluated as the ratio between the number of pseudo-experiments which give a value of $Q$ lower than the expected test statistics (for a S+B hypothesis), and the total number of pseudo-experiments. The calculation led to a p-value in the background-only hypothesis of $3.8 \times 10^{-4}$ corresponding to a statistical significance of $3.4\sigma$.

Table 39: Expected Belle II yields of signal and background events in $1\,\mathrm{ab}^{-1}$ for two different $E_{\mathrm{ECL}}$ windows, compared to Belle MC.

| $E_{\mathrm{ECL}}$ | | $< 1\,\mathrm{GeV}$ | $< 0.25\,\mathrm{GeV}$ |
|---|---|---|---|
| Belle II | Background yield | 7420 | 1348 |
| | Signal yield | 188 | 136 |
| | Signal efficiency ($10^{-3}$) | 2.2 | 1.6 |
| Belle | Background yield | 2160 | 365 |
| | Signal yield | 97 | 60 |
| | Signal efficiency ($10^{-3}$) | 1.2 | 0.7 |

*Systematic uncertainties.* Based on Belle measurements [234], the main sources of systematic uncertainties are the signal and background $E_{\mathrm{ECL}}$ PDFs, the uncertainty on the relative contributions from $B$ decays that peak near zero $E_{\mathrm{ECL}}$ (i.e. peaking background), the tagging efficiency, and the $K_L^0$ veto efficiency, followed by the minor uncertainties due to the number of $B\bar{B}$ pairs, the signal efficiency (PID efficiency, $\tau$ branching fractions, $\pi^0$ efficiency and tracking efficiency), and MC sample sizes used for background PDFs.

The uncertainties on PDFs and tagging efficiency are limited by statistical precision in the $B \to D^{*0}\ell\nu$ control sample on data, and so are expected to scale with luminosity similarly to the statistical uncertainty. The uncertainty due to modelling of the peaking background components, typically due to an undetected $K_{\mathrm{L}}^0$, is estimated by varying the branching fractions of the contributing decays within the experimental uncertainties. This will be a major concern in Belle II analyses. We expect to reach a systematic uncertainty of better than 3% from this contribution. The uncertainty on the $K_L^0$ veto efficiency is obtained from control samples in data, comparing yields of $\phi \to K_L^0 K_S^0$ to $\phi \to K^- K^+$ in a $D^0 \to \phi K_S^0$ sample. Such calibrations were found to be very large in Belle, where data and MC efficiencies differed by approximately 40%. The discrepancy is attributable to the inaccuracy in modelling hadronic interactions in the KLM. GEANT4, which Belle II will use instead of GEANT3 as in Belle, may provide more accurate simulation, but given the calibration is large it may be difficult to improve the systematic uncertainty to better than 2% on the $B \to \tau\nu_\tau$ branching fraction.

The uncertainty on the signal efficiency is expected to scale with luminosity as in the case of the statistical uncertainty. The uncertainties due to the $\tau$ branching fractions ($> 0.6\%$) are not expected to improve substantially. Finally, the uncertainty on the number of $B\bar{B}$ pairs is expected to be limited to about 1%.

The expected systematic uncertainty on the $B \to \tau\nu_\tau$ branching fraction with an integrated luminosity of $1\,\mathrm{ab}^{-1}$ is calculated to be 13%, based on a scaling of the uncertainties of the Belle measurement with hadronic tag [234].

Anticipating the results detailed in Table 41, the luminosity needed to reach a $5\sigma$ discovery of $B \to \tau\nu_\tau$ including statistic and systematic uncertainties is about $2.6\,\mathrm{ab}^{-1}$.

Table 40: Yields of expected signal and background events in the $B \to \tau\nu_\tau$ study for two different $E_{\rm ECL}$ windows, with and without beam background, with a data set of $L = 1\,\mathrm{ab}^{-1}$.

| $E_{\rm ECL}$ | | $< 1\,\mathrm{GeV}$ | $< 0.25\,\mathrm{GeV}$ |
|---|---|---|---|
| | Background yield | 12835 | 2062 |
| without background | Signal yield | 332 | 238 |
| . | Signal efficiency (‰) | 3.8 | 2.7 |
| | Background yield | 7420 | 1348 |
| with background | Signal yield | 188 | 136 |
| | Signal efficiency (‰) | 2.2 | 1.6 |

Table 41: Expected uncertainties on the $B \to \tau\nu_\tau$ branching fraction for different luminosity scenarios with hadronic and semileptonic tag methods.

| | Integrated Luminosity ( $\mathrm{ab}^{-1}$) | 1 | 5 | 50 |
|---|---|---|---|---|
| | Statistical uncertainty (%) | 29 | 13 | 4 |
| Hadronic tag | Systematic uncertainty (%) | 13 | 7 | 5 |
| | Total uncertainty (%) | 32 | 15 | 6 |
| | Statistical uncertainty (%) | 19 | 8 | 3 |
| Semileptonic tag | Systematic uncertainty (%) | 18 | 9 | 5 |
| | Total uncertainty (%) | 26 | 12 | 5 |

*Beam background.* In order to estimate the impact of machine background on the branching fraction measurement, the analysis is repeated on a MC5 Belle II production where no machine background is superimposed on physics events. Continuum background suppression and the signal side selection have been re-optimised for this configuration and the statistical evaluation with toy MC is performed as above. The results are shown in Table 40, compared to the case including the expected machine background, and in Figure 66. The higher selection efficiency in absence of beam background is due to higher $B$-tag reconstruction efficiency (see Figure 65) and that, in order to maximise the FOM, a looser selection is applied on the signal side. It may also be due to a greater abundance of fake tracks in the presence of beam background, which must be further studied at Belle II. For a more general discussion of the FEI tagging performance we refer to Sec. 6.6. The mean uncertainty on the $B \to \tau\nu_\tau$ branching fraction is found to be $\sim$20% with 1 $\mathrm{ab}^{-1}$ of equivalent integrated luminosity, corresponding to a statistical significance of approximately $5\sigma$.

*Summary.* Table 41 summarises the results and projections of the uncertainties on the branching fraction measurement with 1, 5 and 50 $\mathrm{ab}^{-1}$ data sets, using hadronic and semileptonic tagging respectively. These approaches are statistically independent. The projections of measurements using semileptonic tags are based entirely on Belle measurements [240], since no dedicated studies have been performed with Belle II simulation.

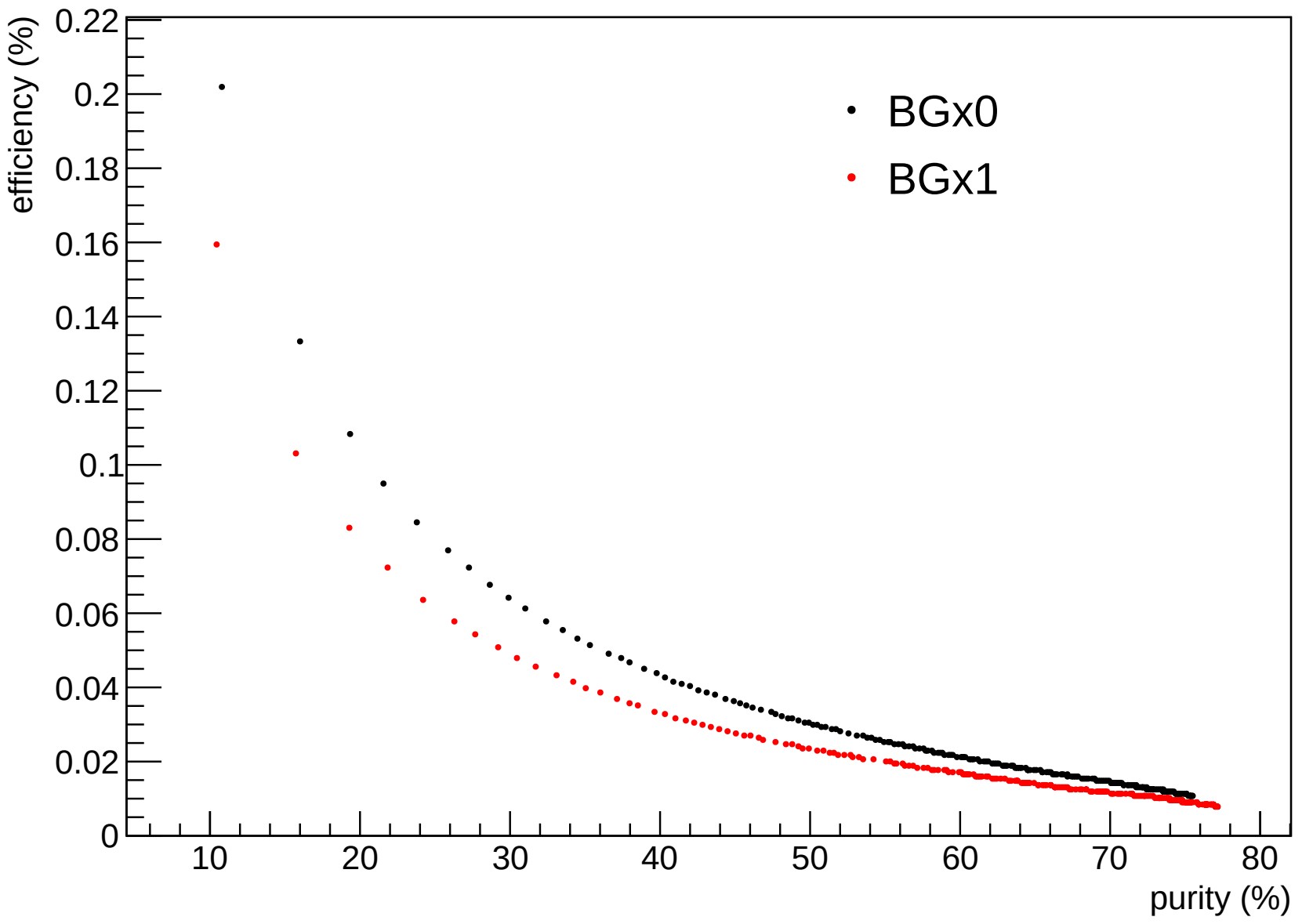

Fig. 65: $B$-tag ROC curves with (BGx1) and without (BGx0) nominal beam background in the $B \to \tau\nu_\tau$ study. The points correspond to a scan of the FEI discriminant output. The efficiency is evaluated as the ratio between the $B$-tag reconstructed candidates (i.e., passing the FEI discriminant cut) and the total generated candidates, and the purity as the ratio between the correctly reconstructed $B$-tags and the total reconstructed candidates. The curves are evaluated on $B^+B^-$ events requiring the presence of only 1 track and PID quality criteria on the signal side.

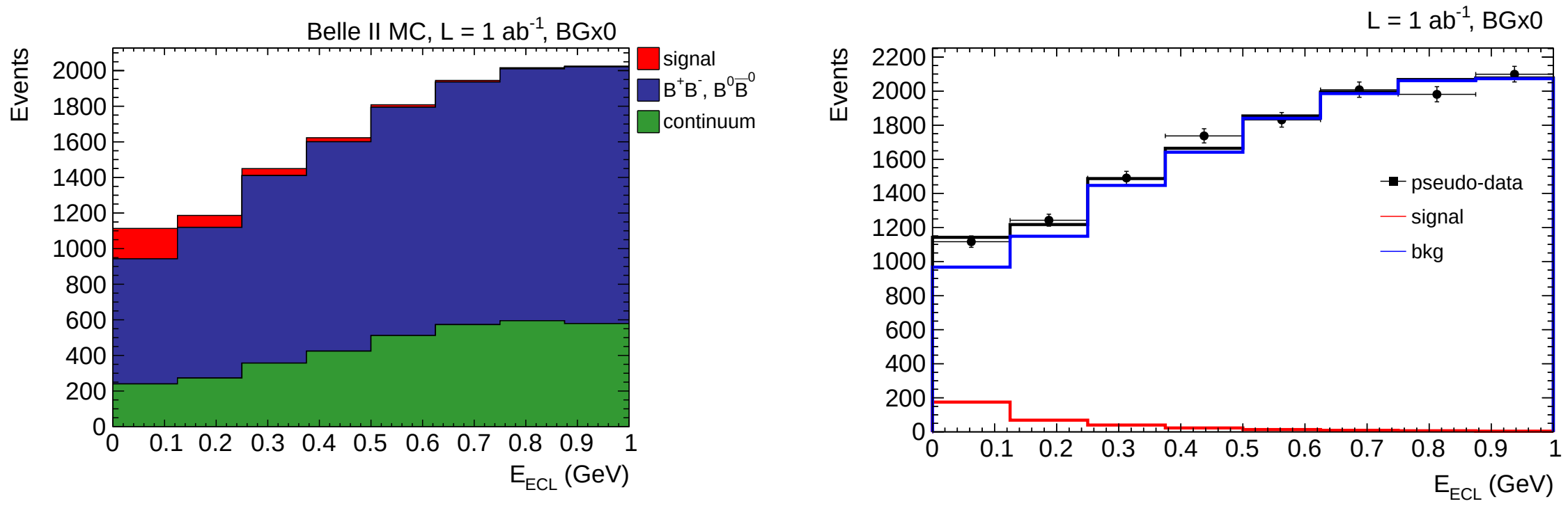

(a) $E_{\mathrm{ECL}}$ distribution for signal (red), $B\overline{B}$ background (blue) and continuum (green) events.

(b) Maximum likelihood fit to pseudo-data $E_{\mathrm{ECL}}$ distribution sampled from simulation. The red and blue histograms represent the signal and background fit functions.

Fig. 66: $E_{\mathrm{ECL}}$ distributions for the $B \to \tau\nu_\tau$ study without machine background. The events correspond to an integrated luminosity of 1 $\mathrm{ab}^{-1}$.

*8.3.2. $B \to \mu\nu_\mu$.* There have been several searches for the $B \to \mu\nu_\mu$ decay to date and the most recent ones [238, 241, 242] are summarised in Table 42.The most stringent limits to date are set by untagged searches.

Table 42: The results of searches for the decay $B \to \mu\nu_\mu$.

| Experiment | Upper limit at 90% C.L. | Comment |
|---|---|---|
| Belle [241] | $2.7 \times 10^{-6}$ | Fully reconstructed hadronic tag, 711 $\text{fb}^{-1}$ |
| Belle [242] | $1.1 \times 10^{-6}$ | Untagged analysis, 711 $\text{fb}^{-1}$ |
| BaBar [238] | $1.0 \times 10^{-6}$ | Untagged analysis, $468 \times 10^6$ $B\bar{B}$ pairs |

Table 43: The branching fractions for leptonic $B$ decays in the SM calculations, and the respective event yields with the full Belle data sample and the expected Belle II data sets.

| $\ell$ | $\mathcal{B}_{\text{SM}}$ | 711 $\text{fb}^{-1}$ | 5 $\text{ab}^{-1}$ | 50 $\text{ab}^{-1}$ |
|---|---|---|---|---|
| $\tau$ | $(7.71 \pm 0.62) \times 10^{-5}$ | $61200 \pm 5000$ | $430000 \pm 35000$ | $4300000 \pm 350000$ |
| $\mu$ | $(3.46 \pm 0.28) \times 10^{-7}$ | $275 \pm 23$ | $1930 \pm 160$ | $19300 \pm 1600$ |
| $e$ | $(0.811 \pm 0.065) \times 10^{-11}$ | $0.0064 \pm 0.0005$ | $0.0453 \pm 0.0037$ | $0.453 \pm 0.037$ |

The expected branching fractions and event yields in the full Belle data set as well as expected Belle II milestones using the value of $|V_{ub}| \times 10^3 = 3.55 \pm 0.12$ from the 2016 HFLAV report [230] and $f_B = 186 \pm 4$ MeV from Ref. [217], which is the only entry in the 2016 FLAG average [140],[22] are shown in Table 43. The process $B^\pm \to \mu^\pm\nu_\mu$ may be observed with $3\sigma$ significance with around 2 $\text{ab}^{-1}$, whereas the $B^\pm \to e^\pm\nu_e$ process is not measurable even with the Belle II data set, and only an upper limit is expected for SM-like scenarios.

The clean environment of an $e^+e^-$ machine, where only one $B\bar{B}$ pair is expected in an event, allows for two main search approaches: *untagged* and *full reconstruction*. The latter leads to very good purity at the cost of very low efficiency. In the untagged analysis the products of the signal decay are selected first and the rest of the event (RoE) is used to build various shape parameters that discriminate $B$-meson decays from other hadronic modes. The efficiency of the untagged method can be rather high.

A 2014 Belle study [241] searched for the $B \to \mu\nu_\mu$ process using one fully reconstructed $B$ meson as a tag. In the signal $B$-meson rest frame the momentum of the $\mu$ is monochromatic due to two-body decay kinematics, with good momentum resolution of $\sim 14$ MeV/$c$ that separates the signal from other $B$ decays. This analysis demonstrated the drawback of the method – extremely low signal selection efficiency of $\sim 10^{-3}$ which leads to the result shown in Table 42 and only $\sim 21$ signal events with the full Belle II integrated luminosity are expected.

The most recent untagged analysis of $B \to \mu\nu_\mu$ with Belle data has much higher signal selection efficiency of 39% but suffers from much higher background. It can be used to anticipate results with the Belle II data set. To separate signal from background events a simple neural network has been developed and trained using various event kinematic parameters. The projections of the muon momentum $p^*_\mu$ in the centre-of-mass frame and the neural net output variable for the full Belle data set in the signal enhanced region is shown in Fig. 67. For $2.6\ \text{GeV}/c < p^*_\mu < 2.85\ \text{GeV}/c$ and $NN_{\text{out}} > 0.84$ the figure-of-merit is $\text{FOM}_{\text{Belle}} = N_{\text{sig}}/\sqrt{N_{\text{sig}} + N_{\text{bkg}}} = 31.5/\sqrt{31.5 + 300} \approx 1.73$ and can be scaled to the full Belle II expected data set as $\text{FOM}_{\text{BelleII}} = \text{FOM}_{\text{Belle}} \times \sqrt{50\ \text{ab}^{-1}/0.711\ \text{ab}^{-1}} \approx 14.5$. We therefore expect approximately 7% statistical precision on the branching fraction. Naively, to reach $5\sigma$ significance Belle II should collect approximately 6 ab$^{-1}$. With a much larger data set at Belle II, the systematic uncertainties will be as good or better than the statistical uncertainty in this channel.

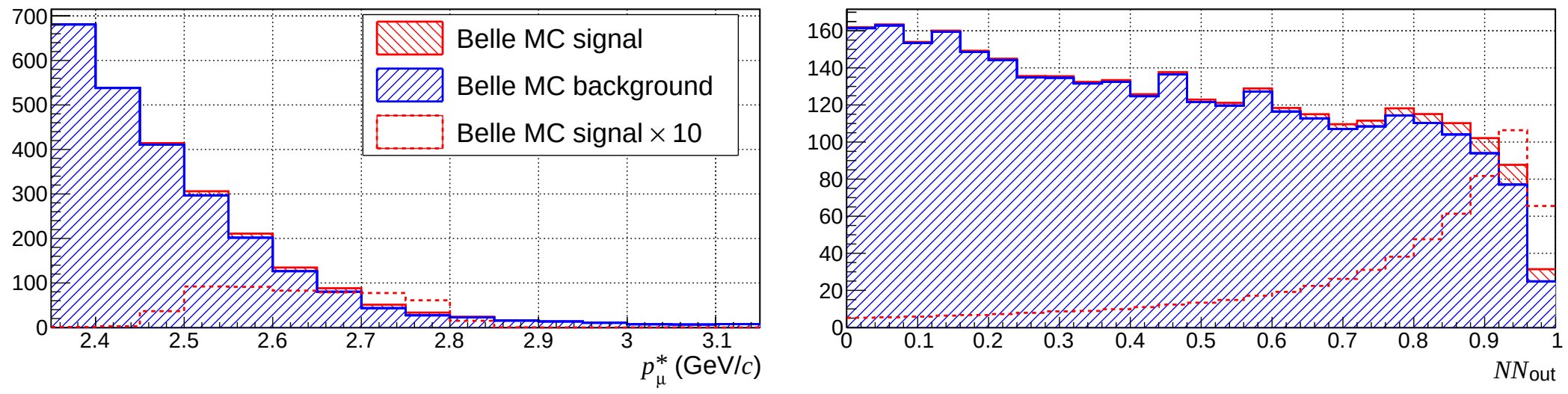

Fig. 67: The distributions of the muon momentum $p^*_\mu$ in the centre-of-mass and the neural net output variable $NN_{\rm out}$ in the signal enhanced region 2.6 GeV/$c < p^*_\mu <$ 2.85 GeV/$c$, and $NN_{\rm out} > 0.84$, respectively, based on Belle MC and equivalent to the full Belle data set of 711 fb$^{-1}$.

*8.3.3. Sensitivity to new physics.* In the following, we will consider the scenario that new physics only measurably affects the tau mode, that is, $r^\mu_{\rm NP} = r^e_{\rm NP} = 0$. The dominant sources of theoretical uncertainty in $B^+ \to \ell^+\nu_\ell$ are $f_B$ and $|V_{ub}|$, therefore to mitigate them, we can form ratios to light leptonic modes defined as.

$$R_{\rm ps} = \frac{\tau_{B^0}}{\tau_{B^+}} \frac{\mathcal{B}(B^+ \to \tau^+\nu_\tau)}{\mathcal{B}(B^0 \to \pi^-\ell^+\nu_\ell)}, \quad R_{\rm pl} = \frac{\mathcal{B}(B^+ \to \tau^+\nu_\tau)}{\mathcal{B}(B^+ \to \mu^+\nu_\mu)}. \tag{129}$$

Experimentally, $R_{\rm ps}$ has the advantage that $B^0 \to \pi^-\ell^+\nu_\ell$ is experimentally well known, whereas $R_{\rm pl}$ is not yet measured. On the theoretical side, $R_{\rm ps}$ contains theoretical uncertainties from $f_B/f_+$, while $R_{\rm pl}$ has a very precise theoretical prediction in the SM.

Predictions for these ratios are calculated in Ref. [243] and are as follows:

$$R^{\rm NP}_{\rm ps} = (0.539 \pm 0.043)\left|1 + r^\tau_{\rm NP}\right|^2, \tag{130}$$

$$R^{\rm NP}_{\rm pl} = \frac{m^2_\tau}{m^2_\mu} \frac{(1 - m^2_\tau/m^2_B)^2}{(1 - m^2_\mu/m^2_B)^2} \left|1 + r^\tau_{\rm NP}\right|^2 \simeq 222.37 \left|1 + r^\tau_{\rm NP}\right|^2. \tag{131}$$

The current experimental constraints on $B^+ \to \tau^+\nu_\tau$ [77] and $B^0 \to \pi^-\ell^+\nu_\ell$ [230] result in $R^{\rm exp}_{\rm ps} = 0.73 \pm 0.14$. This is compared with Eq. 130 to find the following constraint on $r^\tau_{\rm NP}$:

$$\left|1 + r^\tau_{\rm NP}\right| = 1.16 \pm 0.11\,. \tag{132}$$

We find that $R_{\rm ps}$ provides a slightly tighter bound than the direct branching fraction measurement. The present experimental uncertainty in $R^{\rm exp}_{\rm ps}$ of 0.14 is expected to improve substantially, as discussed in Sec.8.3.1. The purely muonic mode has only upper limits on $\mathcal{B}(B^+ \to \mu^+\nu_\mu)$.The upper limit is approaching the SM prediction, and we expect that the muonic mode will be precisely measured at Belle II. Therefore, $R_{\rm pl}$ may also play an important role for new physics searches in $B^+ \to \tau^+\nu_\tau$. The following study discusses the future sensitivities of $R_{\rm ps}$ and $R_{\rm pl}$ to new physics contributions, $r^\tau_{\rm NP}$, with 5 ab$^{-1}$ and 50 ab$^{-1}$ of Belle II data.

Table 44: Expected 95% CL limits on $r^{\tau}_{\rm NP}$ from $R_{\rm ps}$ and $R_{\rm pl}$ at Belle II with $5\,{\rm ab}^{-1}$ and $50\,{\rm ab}^{-1}$ of accumulated data. The new physics contribution is assumed to be real and no larger than the SM contribution ($|r^{\tau}_{\rm NP}| < 1$).

| Luminosity | $R_{\rm ps}$ | $R_{\rm pl}$ |
|---|---|---|
| $5\,{\rm ab}^{-1}$ | $[-0.22, 0.20]$ | $[-0.42, 0.29]$ |
| $50\,{\rm ab}^{-1}$ | $[-0.11, 0.12]$ | $[-0.12, 0.11]$ |

To determine the sensitivity to new physics through $r^{\tau}_{\rm NP}$, we assume that experimental central values of the ratios are at the SM expectation and that new physics contributions are no greater than the SM contributions ($|r^{\tau}_{\rm NP}| < 1$) unless otherwise stated. The expected experimental errors on $R_{\rm ps}$ and $R_{\rm pl}$ are then determined by taking the Belle II estimates of $B \to \tau\nu_\tau$, $B \to \mu\nu_\mu$, and $B \to \pi\ell\nu_\ell$ with luminosities of $5\,{\rm ab}^{-1}$ and $50\,{\rm ab}^{-1}$:

$$R_{\rm ps}^{5\,{\rm ab}^{-1}} = 0.54 \pm 0.11\,, \quad R_{\rm ps}^{50\,{\rm ab}^{-1}} = 0.54 \pm 0.04\,, \tag{133}$$

$$R_{\rm pl}^{5\,{\rm ab}^{-1}} = 222 \pm 76\,, \quad R_{\rm pl}^{50\,{\rm ab}^{-1}} = 222 \pm 26\,. \tag{134}$$

With the use of the above expected constraints, the 95% CL expected limits on $r^{\tau}_{\rm NP}$ are given in Table 44. A new physics contribution to $B \to \tau\nu_\tau$ with $r^{\tau}_{\rm NP} \gtrsim O(0.1)$ can be tested at 95% CL. The observable $R_{\rm pl}$ has low sensitivity at $5\,{\rm ab}^{-1}$, but with $50\,{\rm ab}^{-1}$ it will be comparable with $R_{\rm ps}$. Further improvements in the sensitivity of $R_{\rm pl}$ may be achieved through more direct measurements.

### 8.4. *Radiative Leptonic*

#### 8.4.1. $B^+ \to \ell^+\nu_\ell\gamma$. Authors: F. Metzner, M. Gelb, P. Goldenzweig (Exp.)

The radiative leptonic decay $B^+ \to \ell^+\nu_\ell\gamma$ yields important information for the theoretical predictions of non-leptonic $B$ meson decays into light-meson pairs. The emission of the photon probes the first inverse moment $\lambda_B$ of the light-cone distribution amplitude (LCDA) of the $B$ meson. This parameter is a vital input to QCD factorisation schemes for the non-perturbative calculation of non-leptonic $B$ meson decays [244, 245] (see Section 12.3.2). The importance of $B^+ \to \ell^+\nu_\ell\gamma$ decays in the empirical determination of the parameter $\lambda_B$ can be found in Refs. [223, 246], where detailed theoretical calculations of the decay are presented.

The partial branching ratio $\Delta\mathcal{B}$ is given by the integral of the differential decay width over the photon energies relevant for the considered selection, multiplied by the lifetime of the $B$ meson $\tau_B$:

$$\Delta\mathcal{B}(B^+ \to \ell^+\nu_\ell\gamma) = \tau_B \int_{\rm Selection} dE_\gamma\, \frac{d\Gamma}{dE_\gamma}, \tag{135}$$

where photon energies below 1 GeV are considered as unsafe since the factorisation approach for the calculation of the form factors requires the condition $2E_\gamma \sim m_b$. Thus, only photons with energies above this threshold were considered in the most recent Belle analysis [247]. The theoretical relation between the value of $\lambda_B$ and the respective partial branching ratio for two selection criteria for the signal photon energy is shown in Fig. 68.

Belle obtained limits of $\Delta\mathcal{B}(B^+ \to \ell^+\nu_\ell\gamma) < 3.5 \times 10^{-6}$ and $\lambda_B > 238\,{\rm MeV}$ (90% C.L.) for photons above $1\,{\rm GeV}$ with the full $\Upsilon(4S)$ dataset of $711\,{\rm fb}^{-1}$ [247] (see [248] for an updated Belle result). In preparation for Belle II, a new analysis has been prepared with Belle MC in the basf2 framework, where the signal-specific FEI is employed (Sec. 6.6). To enable a comparison of the two analysis methods, the expected yield (determined from MC with a partial branching fraction of $\Delta\mathcal{B}(B^+ \to \ell^+\nu_\ell\gamma) = 5.0 \times 10^{-6}$) for both methods is displayed in Table 45. The new tagging algorithm results in three times the expected signal yield with the same dataset. The yield is extracted from a simultaneous fit to the squared missing mass distributions of the electron and muon channels. The results for the improved analysis are shown in Fig. 69.

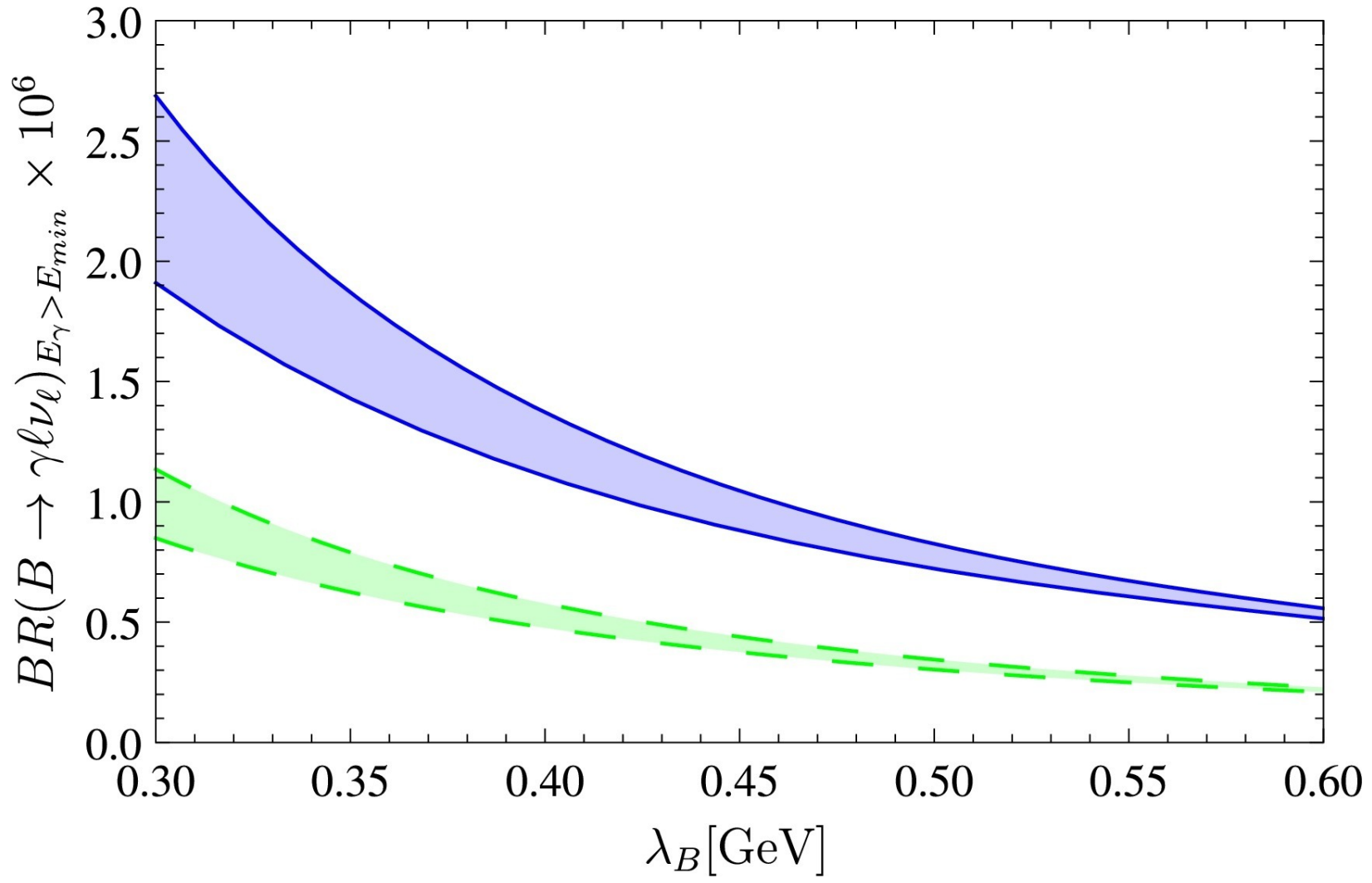

Fig. 68: Dependence of the theoretical prediction for the partial branching fraction $\Delta\mathcal{B}(B^+ \to \ell^+\nu_\ell\gamma)$ on the value of the QCD factorisation parameter $\lambda_B$ for two signal photon selection criteria: the threshold with lower theoretical uncertainties $E_\gamma > 1.7\,\mathrm{GeV}$ (lower, dashed); and the threshold $E_\gamma > 1.0\,\mathrm{GeV}$ which yields a significantly higher efficiency [246].

Table 45: Expected signal yields determined with Belle MC for the new analysis using the signal-specific FEI in basf2 ($N_{\mathrm{New}}$). The values are compared to the expected yields in the published Belle analysis ($N_{\mathrm{Published}}$) [247]. Both MC studies assume a partial branching fraction of $\Delta\mathcal{B}(B^+ \to \ell^+\nu_\ell\gamma) = 5.0 \times 10^{-6}$, to enable a comparison of the expected yields with the different analysis frameworks.

| | $B^+ \to e^+\nu_e\gamma$ | $B^+ \to \mu^+\nu_\mu\gamma$ | Combined |
|---|---|---|---|
| $N_{\mathrm{New}}$ | 24.8 | 25.7 | 50.5 |
| $N_{\mathrm{Published}}$ | 8.0 | 8.7 | 16.5 |

Further constraints on the energy of the neutrino would enable the experimental examination of the difference between the axial and vector form factor, and thus the impact of the power-suppressed contributions to the decay width [223]. However, the selection required for this study—the neutrino has to receive the majority of the $B$ meson's energy—reduces the statistics significantly, rendering it unfeasible with the Belle dataset. However, with the large Belle II dataset, this aspect of the decay will also be addressed.

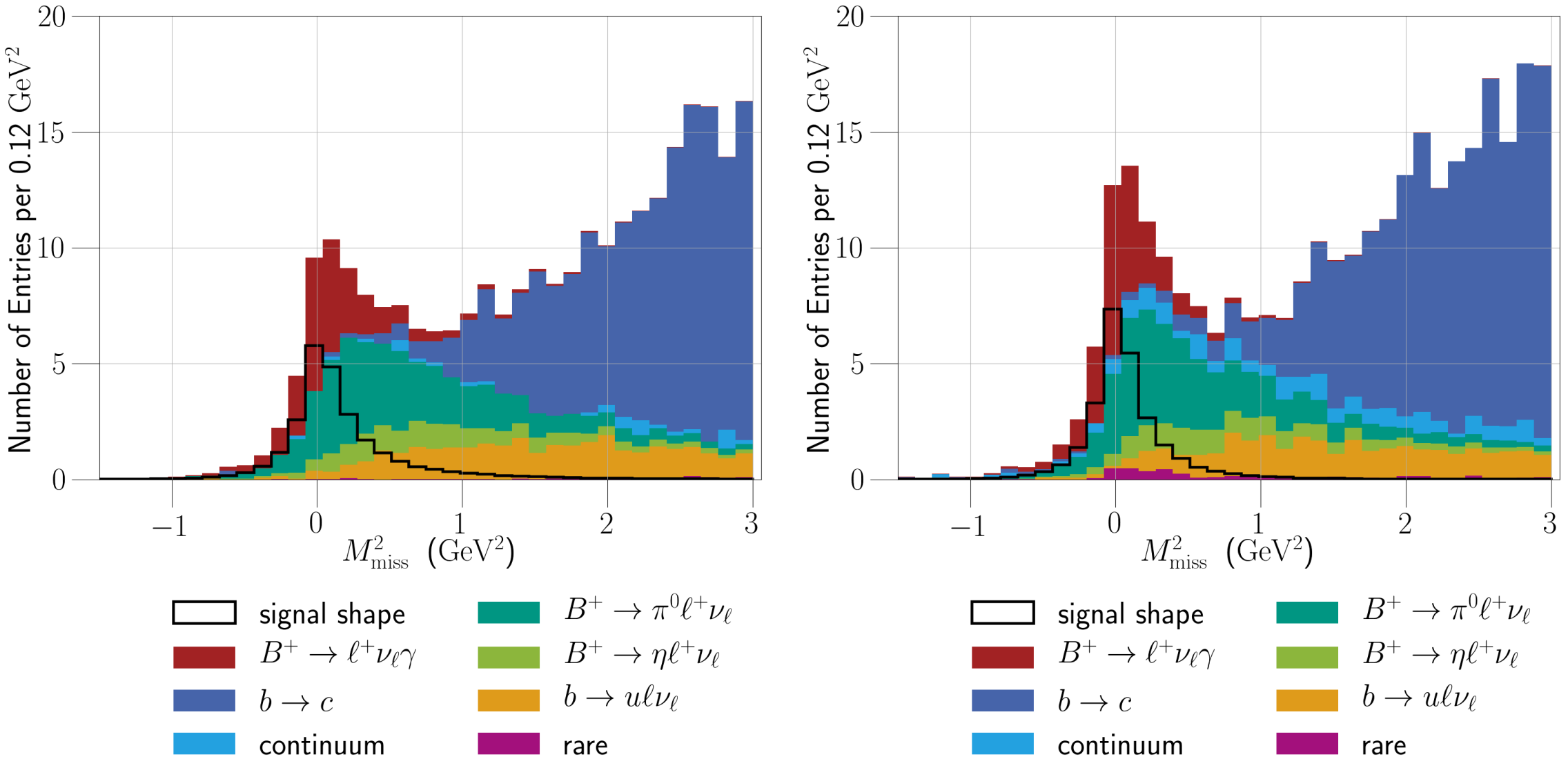

Fig. 69: The Belle MC squared missing mass distribution for $B^+ \to e^+\nu_e\gamma$ (left) and $B^+ \to \mu^+\nu_\mu\gamma$ (right) of the new analysis using the signal-specific FEI in basf2. The signal yield ($N_{\rm New}$) is reported in Table 45.

Table 46: Expected statistical error in $10^{-6}$ for Belle and Belle II for a simulated partial branching fraction of $\Delta\mathcal{B}(B^+ \to \ell^+\nu_\ell\gamma) = 5.0 \times 10^{-6}$.

| Belle<br>New analysis | Belle II<br>$5\,\mathrm{ab}^{-1}$ | Belle II<br>$50\,\mathrm{ab}^{-1}$ |
|---|---|---|
| +1.48 | +0.56 | +0.18 |
| -1.39 | -0.53 | -0.17 |

In addition, a toy MC study is used to estimate the expected statistical error with $5\,\mathrm{ab}^{-1}$ and $50\,\mathrm{ab}^{-1}$ of Belle II data. The statistical errors are expected to be significantly reduced with the increased dataset. The results can be found in Table 46.

## *8.5. Semitauonic decays*

*8.5.1. $B \to D^{(*)}\tau\nu$.* Authors: S. Hirose, Y. Sato (exp.), M. Tanaka (th.), R. Watanabe (th.)

The decays $B \to D^{(*)}\tau\nu$ are described at the quark level as $b \to c\tau\nu$ tree-level transitions that proceed in the SM through the exchange of a virtual $W$ boson. The ratios, defined as

$$R_{D^{(*)}} = \frac{\mathrm{Br}(B \to D^{(*)}\tau\nu_\tau)}{\mathrm{Br}(B \to D^{(*)}\ell\nu_\ell)}, \tag{136}$$

are useful observables to test for new physics (NP) as theoretical uncertainties in form factors and $|V_{cb}|$ largely cancel out. First measured by Belle [249], these ratios have since been more precisely measured by BaBar [250], Belle [251], and LHCb [252]. The combination of these measurements shows a tendency towards larger values than SM expectation with a deviation of nearly $4\sigma$. A better understanding of these anomalous results is of high priority at Belle II, because the discrepancy could be due to NP contributions. In addition to $R(D^{(*)})$, measurements of the polarisations of the $\tau$ lepton and the $D^*$ meson respectively are also good probes for the NP. They are defined by.

$$P_\tau(D^{(*)}) = \frac{\Gamma^+ - \Gamma^-}{\Gamma^+ + \Gamma^-}, \tag{137}$$

$$P_{D^*} = \frac{\Gamma_\mathrm{L}}{\Gamma_\mathrm{L} + \Gamma_\mathrm{T}}, \tag{138}$$

where $\Gamma^{+(-)}$ and $\Gamma_{\mathrm{L(T)}}$ are the decay rate with the $\tau$ helicity $+1/2$ $(-1/2)$ and that with the longitudinally (transversely) polarised $D^*$, respectively. In the future, Belle II should also be able to perform precise differential measurements in $q^2$, and will measure the decay angles, $\theta_\ell$, $\theta_V$, $\chi$, as was done for $B \to D^{(*)}\ell\nu$ decays at Belle and BaBar (see Section ). It will be challenging to reconstruct $\theta_\ell$ and $\chi$ in the $\tau$ channels due to the presence of at least two neutrinos, but there will be some experimental sensitivity.

*SM predictions.* Processes of the type $B \to D^{(*)}\tau\nu$ have a rich phenomenology that can be used to probe the nature of any NP contributions. For example, differential distributions, such as the momentum transfer to the lepton pair, $q^2 = (p_\tau + p_\nu)^2 = (p_B - p_{D^{(*)}})^2$ can be modified in the presence of a charged Higgs. Form factors of $B \to D^{(*)}$ transitions and the CKM matrix element magnitude $|V_{cb}|$ are extracted through measurements of $B \to D^{(*)}\ell\nu$ for $\ell = e, \mu$. The differential decay rates are described in Sections 8.2.2 and 8.2.3 and the form factors $f_+(q^2)$ for $B \to D$ and $V(q^2)$ and $A_i(q^2)$ for $B \to D^*$ can be determined from experimental data combined with results from lattice-QCD. (For notation, see Eqs. 96, 111, and 112.) The differential decay rates of $B \to D^{(*)}\tau\nu$ decays contain additional form factors, $f_0(q^2)$ and $A_0(q^2)$, from terms proportional to $m_\tau^2$. These additional form factors can also be computed with lattice-QCD. At present, there are lattice-QCD results for $f_+(q^2)$ and $f_0(q^2)$ [145, 146], while the calculations of the $q^2$ dependence of the $B \to D^*$ form factors is underway [253].

Various SM predictions for $R_D^{(*)}$ have been calculated [145, 146, 250, 254–261]. In addition to the lattice QCD study, the heavy quark effective theory (HQET) and QCD sum rules have been applied to evaluate the form factors. For example, in Ref. [259, 261], theoretical inputs from lattice and QCD sum rule are combined to be fitted to the experimental data of $B \to D^{(*)}\ell\nu$, up to $\mathcal{O}(\Lambda_{\mathrm{QCD}}/M_Q)$ and $\mathcal{O}(\alpha_s)$ corrections to the form factors in HQET (A higher order contribution of $\mathcal{O}(\Lambda_{\mathrm{QCD}}^2/M_Q^2)$ could be important [262]). Another approach, based on the Boyd-Grinstein-Lebed parameterisation [258, 260], has also been considered. Taking into account all recent SM predictions, HFLAV presently (2018 value) quotes:

$$R_D^{\mathrm{SM}} = 0.299 \pm 0.003, \tag{139}$$

$$R_{D^*}^{\mathrm{SM}} = 0.258 \pm 0.005. \tag{140}$$

At present, the $\tau$ and $D^*$ polarisations have been estimated in the SM as follows: $P_\tau(D) = 0.34 \pm 0.03$ [263] $(0.325 \pm 0.009$ [255]$)$, $P_\tau(D^*) = -0.47 \pm 0.04$ [260] $(-0.497 \pm 0.013$ [264]$)$, and $P_{D^*} = 0.46 \pm 0.04$ [265], where the values in parentheses are older estimations with only the leading order contributions of the HQET expansion.)

Table 47: Summary of experimental measurements of semitauonic $B$ decays.

| Exp. | Tag method | $\tau^-$ decays | Observables | Fit variables |
|---|---|---|---|---|
| Belle [249] | Untagged | $e^-\nu_\tau\bar{\nu}_e, \pi\nu_\tau$ | $\mathcal{B}(\bar{B}^0 \to D^{*+}\tau^-\bar{\nu}_\tau)$ | $M_{\rm bc}^{\rm tag}$ |
| Belle [266] | Untagged | $\ell^-\nu_\tau\bar{\nu}_\ell, \pi\nu_\tau$ | $\mathcal{B}(B^- \to D^{(*)0}\tau^-\bar{\nu}_\tau)$ | $M_{\rm bc}^{\rm tag}$ and $p_{D^0}$ |
| Belle [251] | Hadronic | $\ell^-\nu_\tau\bar{\nu}_\ell$ | $R_D$, $R_{D^*}$, $q^2$, $\lvert p_\ell^*\rvert$ | $M_{\rm miss}^2$ and $\mathcal{O}_{NB}$ |
| Belle [267] | Semileptonic | $\ell^-\nu_\tau\bar{\nu}_\ell$ | $R_{D^*}$, $\lvert p_\ell^*\rvert$ $\lvert p_{D^*}^*\rvert$ | $E_{\rm ECL}$ and $\mathcal{O}_{NB}$ |
| Belle [268] | Hadronic | $h^-\nu_\tau$ | $R_{D^*}$, $P_\tau(D^*)$ | $E_{\rm ECL}$ and $\cos\theta_{\rm hel}$ |
| Belle [269] | Semileptonic | $\ell^-\nu_\tau\bar{\nu}_\ell$ | $R_D$, $R_{D^*}$ | $E_{\rm ECL}$ and $\mathcal{O}_{BDT}$ |
| BaBar [250, 270] | Hadronic | $\ell^-\nu_\tau\bar{\nu}_\ell$ | $R_D$, $R_{D^*}$, $q^2$ | $M_{\rm miss}^2$ and $p_\ell$ |
| LHCb [252] | – | $\ell^-\nu_\tau\bar{\nu}_\ell$ | $E_\mu^*$, $m_{\rm miss}^2$, $q^2$ | |
| LHCb [271] | – | $h^-h^+h^-\nu_\tau$ | $q^2$, $t_\tau$, $\mathcal{O}_BDT$ | |

*Experimental data.* The strategy of $R_{D^{(*)}}$ measurements is to evaluate the ratio of the efficiency corrected yields of $B \to D^{(*)}\tau\nu_\tau$ and $B \to D^{(*)}\ell\nu_\ell$ events. The experimental approach for measuring $B \to D^{(*)}\tau\nu_\tau$ decays is similar to that used in $B \to \tau\nu_\tau$, owing to the presence of two or more neutrinos in the final state. Three different methods are used: hadronic tag, semileptonic tag and untagged (or inclusive tag). The experimental methods at the $B$-factory experiments are summarised in Table 47.

In the untagged method, the most important variable to extract signal events is the beam-energy constrained mass of the tag $B$ meson, $M_{\rm bc}^{\rm tag}$. In the Belle analysis of $B \to D^{(*)}\tau\nu_\tau$ with a hadronic tag, the most important variables are related to the missing momentum such as $M_{\rm miss}^2$ and the extra energy in the calorimeter, $E_{\rm ECL}$. The $M_{\rm miss}^2$ distribution is used to separate $B \to D^{(*)}\tau\nu_\tau$ signals from $B \to D^{(*)}\ell\nu_\ell$. In the higher $M_{\rm miss}^2$ region, where most the $B \to D^{(*)}\tau\nu_\tau$ signal populates, a fit to a multivariate classifier is performed, based on several input variables such as $E_{\rm ECL}$, the lepton momentum, $p_\ell$. In the BaBar analysis, the lepton momentum, $p_\ell$, is used for the fit as well as $M_{\rm miss}^2$; $E_{\rm ECL}$ is used only for background suppression prior to the final fit.

In the analyses with a semileptonic tag, $E_{\rm ECL}$ is used to separate $B^0 \to D^*\tau\nu_\tau$ and $B^0 \to D^*\ell\nu_\ell$ from other background, and a multivariate classifier is formed using the signal-side $\cos\theta_{B-D^*\ell}$, $m_{\rm miss}^2$ and $E_{\rm visible}$. Ultimately two-dimensional binned fit approaches are used to determine the signal yields of the signal and normalisation modes.

In the above analyses, leptonic $\tau$ decay modes were chosen to achieve better background rejection. The most important background in these studies originate from $B \to D^{**}\ell\nu_\ell$, where $D^{**}$ mesons are excited charmed mesons higher than the $D^*(2010)$. If we fail to reconstruct particles (mainly $\pi^0$'s) from $D^{**}$ in the $B \to D^{**}\ell\nu_\ell$ decay, such events can mimic the signal, and so the identification of $B \to D^{**}\ell\nu_\ell$ background is critical. We have limited knowledge of the branching fractions to $B \to D^{**}\ell\nu_\ell$ and the $D^{**}$ decay itself, which needs to be addressed at Belle II.

In the most recently published Belle analysis of $B \to D^*\tau\nu_\tau$ [272], hadronic $\tau$ decay modes $\tau^- \to h^-\nu_\tau$ $h^- = \pi^-$, $\rho^-$) have been used, which are statistically independent from the other measurements and can be determined with useful precision. The main background originates from hadronic $B$ decays, $B \to D^*\pi^+X$, where $X$ consists of one or more unreconstructed $\pi^0$, $\eta$, $\gamma$ or pairs of charged particles. Large uncertainties in branching fractions of the exclusive hadronic $B$ decay modes may introduce a sizeable systematic uncertainty. Nevertheless, it is advantageous that $B \to D^* \tau \nu_\tau$ is measured with a different composition of systematic uncertainties with respect to all other measurements using $\tau^- \to \ell^- \bar{\nu}_\ell \nu_\tau$. To extract the signal yields, a similar approach to the previous hadronic-tag analysis with $\tau^- \to \ell^- \bar{\nu}_\ell \nu_\tau$ has been employed; $M^2_{\text{miss}}$ and $E_{\text{ECL}}$ are used for determination of the $B \to D^* \ell \nu_\ell$ and the $B \to D^* \tau \nu_\tau$ yields, respectively. Another advantage of this analysis is the capability to measure $P_\tau(D^*)$ using the kinematics of two-body $\tau$ decays. $P_\tau(D^*)$ is measured using the distribution of $\cos\theta_{\text{hel}}$, the cosine of the angle between the momentum of the $\tau$-daughter meson and the direction opposite the momentum of the $\tau\nu_\tau$ system.

Table 48: Results of $R_{D^{(*)}}$ measurements by BaBar, Belle and LHCb. The correlation column lists the linear $R_D - R_{D^*}$ statistical, systematic and total correlations respectively. The averages are taken from the 2018 HFLAV combination [230]. The analysis method and the $\tau$ decay are indicated: Had for the hadronic tag, SL for the semileptonic tag, $\ell^-$ for $\tau^- \to \ell^- \bar{\nu}_\ell \nu_\tau$ and $h^-$ for $\tau^- \to h^- \nu_\tau$.

| | $R_D$ | $R_{D^*}$ | Correlation |
|---|---|---|---|
| BaBar (Had, $\ell^-$) | $0.440 \pm 0.058 \pm 0.042$ | $0.332 \pm 0.024 \pm 0.018$ | $-0.45/-0.07/-0.27$ |
| Belle (Had, $\ell^-$) | $0.375 \pm 0.064 \pm 0.026$ | $0.293 \pm 0.038 \pm 0.015$ | $-0.56/-0.11/-0.49$ |
| Belle (SL, $\ell^-$) | $0.307 \pm 0.037 \pm 0.016$ | $0.283 \pm 0.018 \pm 0.014$ | $-0.53/-0.51/-0.51$ |
| LHCb ($\ell^-$) | NA | $0.336 \pm 0.027 \pm 0.030$ | NA |
| LHCb ($h^-$) | NA | $0.291 \pm 0.019 \pm 0.029$ | NA |
| Belle (Had, $h^-$) | NA | $0.270 \pm 0.035^{+0.028}_{-0.025}$ | NA |
| Average | $0.340 \pm 0.027 \pm 0.013$ | $0.295 \pm 0.011 \pm 0.008$ | $-0.38$ |
| | | $P_\tau(D^*)$ | |
| Belle (Had, $h^-$) | | $-0.38 \pm 0.51^{+0.21}_{-0.16}$ | |

The current experimental results for $R_D$, $R_{D^*}$ and $P_\tau(D^*)$ are summarised in Table 48. Typical figures for the $B \to D \tau \nu_\tau$ mode in the hadronic tag analysis with $\tau^- \to \ell^- \bar{\nu}_\ell \nu_\tau$ are shown in Fig. 70. In addition to Belle and BaBar, LHCb has also demonstrated its capability to measure $R_{D^*}$ at the Large Hadron Collider. The world-average $R_{D^{(*)}}$ shows about $4\sigma$ deviation from the SM predictions. The result of $P_\tau(D^*)$ is consistent with the SM prediction and excludes $P_\tau(D^*) > +0.5$ at 90% confidence level (CL). NP effects can be thoroughly probed in kinematic distributions as well as the total branching fraction. So far, only the measured $q^2$ spectrum and the momenta of the $D^*$ and the charged lepton have been compared to NP scenarios. The spectra are generally consistent with SM predictions although they are highly statistics limited at this stage.

Table 49 shows the composition of the systematic uncertainties in each recent Belle measurement of $R_{D^{(*)}}$. Currently, the dominant source of systematic uncertainty is the limited MC sample size, which affects the estimations of reconstruction efficiency, the understanding of the cross-feed components, and the description of the PDFs used in the fit. These uncertainties are clearly reducible using larger MC sample sizes.

Apart from MC sample size, a significant systematic uncertainty arises from branching fractions of the decays of $B \to D^{**} \ell \nu_\ell$ and $D^{**}$ for the analyses with $\tau^- \to \ell^- \bar{\nu}_\ell \nu_\tau$, and hadronic $B$ decays for the analysis with $\tau^- \to h^- \nu_\tau$. Belle and BaBar take different approaches to determine the yields of $B \to D^{**}\ell\nu_\ell$. In the BaBar analysis, the yield of $B \to D^{**}\ell\nu_\ell$ background is constrained with control samples in which an additional neutral pion is required with respect to the nominal event selection. This approach assumed that the $D^{**}$ branching ratio is saturated by $D^{**} \to D^{(*)}\pi$ modes (*i.e.* single pion transitions), which is not the case and may have caused some bias, although a corresponding systematic uncertainty was applied. On the other hand, in the Belle analyses, the yield of $B \to D^{**}\ell\nu_\ell$ background, where $D^{**}$ decays to a variety of allowed modes, is floated in the fit for the signal sample. For precision measurements at Belle II, dedicated measurements of $B \to D^{**}\ell\nu_\ell$ and hadronic $B$ decays with a large data sample are essential. Other non-negligible systematic uncertainties arise from the form factors of $B \to D^{(*)}\ell/\tau\nu$ decays, background from $B \to X_c D^{(*)}$, and large cross-feed from $B \to D^*\ell/\tau\nu$ to $B \to D\ell/\tau\nu$. Ultimately Belle II must also constrain $B \to D^{**}\tau\nu_\tau$ through dedicated measurements.

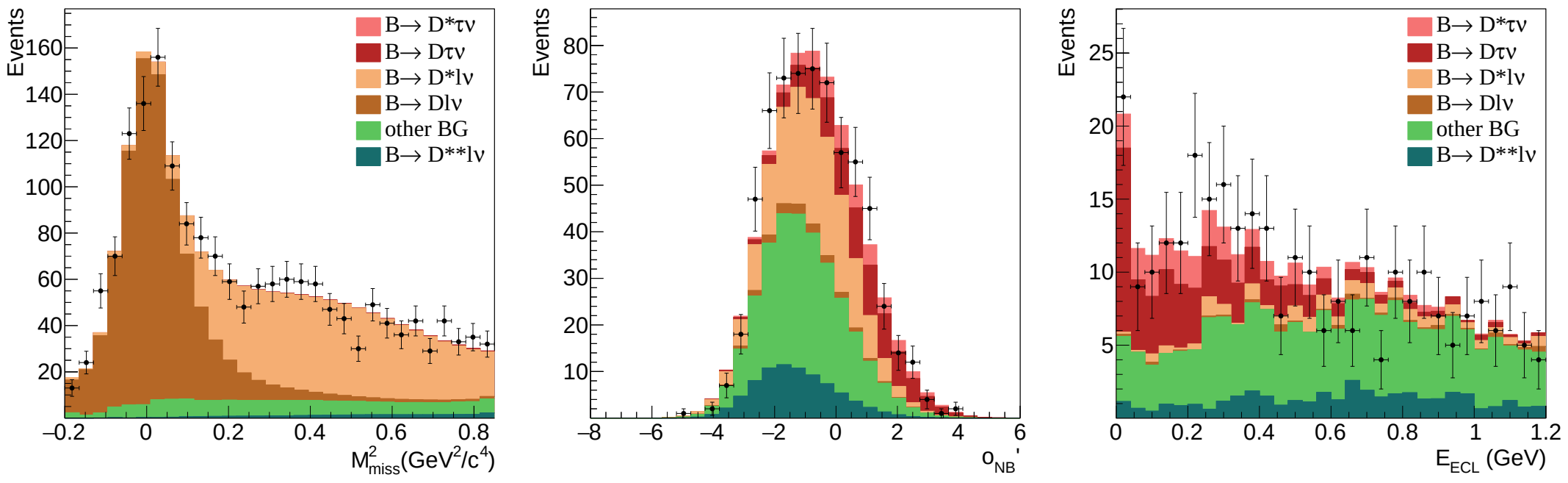

Fig. 70: Fit projections and data points with statistical uncertainties for the $B \to D\tau\nu_\tau$ mode in the Belle hadronic tag analysis with $\tau^- \to \ell^-\bar{\nu}_\ell\nu_\tau$ [251]. The left, the centre and the right panels show the distributions of $M^2_{\rm miss}$, $\mathcal{O}'_{NB}$ ($M^2_{\rm miss} > 0.85$ GeV$^2/c^4$) and $E_{\rm ECL}$ ($M^2_{\rm miss} > 2.0$ GeV$^2/c^4$), respectively.

Table 49: Composition of the systematic uncertainty in each Belle analysis. Relative uncertainties in percent are shown. The analysis method and the $\tau$ decay mode are indicated in the parentheses; their meaning is explained in the caption of Table 48.

| Source | Belle (Had, $\ell^-$) $R_D$ | Belle (Had, $\ell^-$) $R_{D^*}$ | Belle (SL, $\ell^-$) $R_{D^*}$ | Belle (Had, $h^-$) $R_{D^*}$ |
|---|---|---|---|---|
| MC statistics | 4.4% | 3.6% | 2.5% | $^{+4.0}_{-2.9}$% |
| $B \to D^{**}\ell\nu_\ell$ | 4.4% | 3.4% | $^{+1.0}_{-1.7}$% | 2.3% |
| Hadronic $B$ | 0.1% | 0.1% | 1.1% | $^{+7.3}_{-6.5}$% |
| Other sources | 3.4% | 1.6% | $^{+1.8}_{-1.4}$% | 5.0% |
| Total | 7.1% | 5.2% | $^{+3.4}_{-3.5}$% | $^{+10.0}_{-9.0}$% |

*Theoretical interpretation: model independent.* In the presence of NP, semitauonic decays, $B \to D^{(*)}\tau\nu_\tau$ decays can be described by the most general effective Lagrangian of $b \to c\tau\bar{\nu}$:

$$-\mathcal{L}_{\rm eff} = 2\sqrt{2}G_F V_{cb}\big[(1 + C_{V_1})\mathcal{O}_{V_1} + C_{V_2}\mathcal{O}_{V_2} + C_{S_1}\mathcal{O}_{S_1} + C_{S_2}\mathcal{O}_{S_2} + C_T\mathcal{O}_T\big], \quad (141)$$

where the four-fermion operators are defined as

$$\mathcal{O}_{V_1} = \bar{c}_L \gamma^\mu b_L \, \bar{\tau}_L \gamma_\mu \nu_L, \tag{142}$$

$$\mathcal{O}_{V_2} = \bar{c}_R \gamma^\mu b_R \, \bar{\tau}_L \gamma_\mu \nu_L, \tag{143}$$

$$\mathcal{O}_{S_1} = \bar{c}_L b_R \, \bar{\tau}_R \nu_L, \tag{144}$$

$$\mathcal{O}_{S_2} = \bar{c}_R b_L \, \bar{\tau}_R \nu_L, \tag{145}$$

$$\mathcal{O}_T = \bar{c}_R \sigma^{\mu\nu} b_L \, \bar{\tau}_R \sigma_{\mu\nu} \nu_L, \tag{146}$$

and the $C_X$ terms ($X = V_{1,2}, S_{1,2}, T$) denote the Wilson coefficients of the operators, $\mathcal{O}_X$, which represent possible NP contributions. The SM condition requires that $C_X = 0$ for all $X$ types.

Here we will consider NP scenarios where only one $C_X$ at a time is non-zero. In addition, two scenarios of non-vanishing $C_{S_2} = \pm 7.8 C_T$, predicted by the $S_1^{\mathrm{LQ}}$ or $R_2^{\mathrm{LQ}}$ leptoquark model [225, 273], are considered. The BaBar study in Ref. [250] showed that the anomalous values of the ratios are unlikely to be explained by a type II 2HDM charged Higgs, corresponding to the $S_1$ scenario ($C_{S_1} \neq 0$) above (it of course also disfavours the SM). Their study showed that the acceptance and the measured shapes of the kinematic distributions are affected by the existence of NP effects: most notably the presence of a charged Higgs influences the $q^2$ spectrum in $B \to D\tau\nu$.

Constraints on the other scenarios, based on a rough estimate, comparing the existing $R_D$ and $R_{D^*}$ measurements in Table 48, are presented in Fig. 71. In each scenario, shaded regions in (light) red are allowed at 68(95)% CL. The allowed regions are well away from the SM points in all cases ($C_X = 0$), implying that current experimental data favours a large contribution from NP.

*Theoretical interpretation: model dependent.* Semitauonic $B$ decays are mediated by tree-level processes in the SM and the current experimental values differ quite significantly from the SM predictions (at the level of 20% for the $V_1$ scenario as can be seen in Fig. 71). BSM physics close to the weak scale is required in order to explain the deviations from the SM if the NP is perturbative. There are two classes of models which can give the desired effects at tree level: ($i$) models with an extended Higgs sector providing a charged scalar and ($ii$) models with leptoquarks. Dedicated analyses of constraints on these models are described in Section 18.3.1, with projections for future constraints expected at Belle II. Here, phenomenological aspects of models relevant for $R_D$ and $R_{D^*}$ are briefly reviewed.

Due to the heavy $\tau$ lepton in the final state, tauonic $B$ decays are sensitive to charged Higgs bosons, which contribute to the scalar effective operators ($\mathcal{O}_{S_{1,2}}$) [274–278]. A simultaneous explanation of $R_D$ and $R_{D^*}$ is only possible for the $S_2$ scenario with a sizeable contribution ($C_{S_2} \sim -1.5$). The 2HDM of type II generates a dominant contribution to $\mathcal{O}_{S_1}$ (for large $\tan\beta$). It can neither explain $R_D$ and $R_{D^*}$ simultaneously [250] nor $R_{D^*}$ alone without violating bounds from other flavour observables [279]. Other 2HDMs such as I, X (leptospecific), Y (flipped), and aligned type also cannot accommodate the $R_{D^{(*)}}$ anomaly within other flavour constraints. A comprehensive study of flavour constraints for the 2HDMs with natural flavour conservation is given in Refs. [280–282]. The 2HDM of type III is still capable of explaining the data, because a charged Higgs contribution to $\mathcal{O}_{S_2}$ can be sizeable if the coupling of the third-generation quark doublet to a right-handed $c$ quark is large [283–285].

Fig. 71: Current constraints on NP scenarios based on measurements from Belle, Babar and LHCb. The (light) red regions are allowed at 68(95)% CL. The "$V_i$ scenario" means that $C_{V_i} \neq 0$ and all other $C_X = 0$.

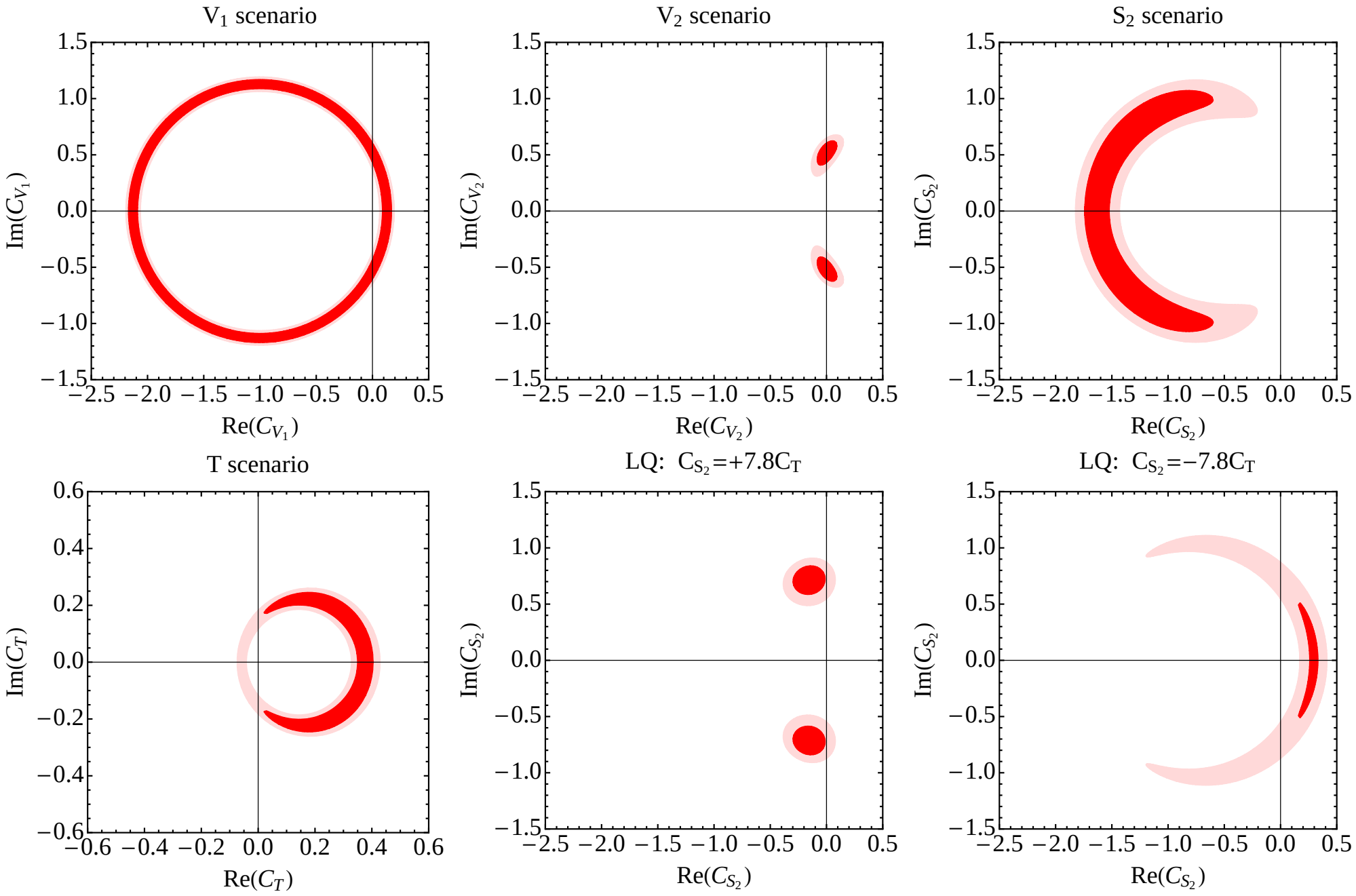

The current results for $R_{D^{(*)}}$ are $R_D^{\rm exp}/R_D^{\rm SM} \simeq R_{D^*}^{\rm exp}/R_{D^*}^{\rm SM}$ within uncertainty. Such a relation is naturally given in scenarios that contain a non-zero contribution to $\mathcal{O}_{V_1}$, *i.e.* a left-handed current. A straightforward realisation of the left-handed current is given by a $W'$ boson implemented in a new $SU(2)_L$ gauge group. This class of model can also address the $R_K$ anomaly (lepton flavour non-universality in $B \to K\ell^+\ell^-$), as well as $R_{D^{(*)}}$, see Refs. [286–290]. Some types of leptoquark models can also induce $\mathcal{O}_{V_1}$ [225, 288, 291–295] and explain $R_K$ and $R_{D^{(*)}}$ at the same time [288, 289].

*Future prospects.* Based on the existing results from Belle and the expected statistical and experimental improvements at Belle II, we provide estimates of the precision on $R_{D^{(*)}}$ and $P_\tau(D^*)$ in Table 50 for two integrated luminosities. In Fig. 72, the expected precisions at Belle II are compared to the current results and SM expectations. They will be comparable to the current theoretical uncertainty. Furthermore, precise polarisation measurements, $P_\tau(D^*)$, and decay differentials will provide further discrimination of NP scenarios (see e.g. Refs. [227, 263] for a detailed discussion). In the estimates for $P_\tau(D^*)$, we take the pessimistic scenario that no improvement to the systematic uncertainty arising from hadronic $B$ decays with three or more $\pi^0$, $\eta$ and $\gamma$ can be achieved. However, although challenging, our understanding of these modes should be improved by future measurements at Belle II and hence the systematic uncertainty will be further reduced. As shown in Fig. 70, the Belle analyses of $B \to D^{(*)}\tau\nu_\tau$ largely rely on the $E_{\rm ECL}$ shape to discriminate between signal and background events. One possible challenge at Belle II is therefore to understand the effects from the large beam-induced background on $E_{\rm ECL}$. From studies of $B \to \tau\nu$, shown earlier in this Section, $E_{\rm ECL}$ should be a robust observable.

Table 50: Expected precision for $R_{D^{(*)}}$ and $P_\tau(D^*)$ at Belle II, given as the relative uncertainty for $R_{D^{(*)}}$ and absolute for $P_\tau(D^*)$. The values given are the statistical and systematic errors respectively.

| | 5 ab$^{-1}$ | 50 ab$^{-1}$ |
|---|---|---|
| $R_D$ | $(\pm 6.0 \pm 3.9)\%$ | $(\pm 2.0 \pm 2.5)\%$ |
| $R_{D^*}$ | $(\pm 3.0 \pm 2.5)\%$ | $(\pm 1.0 \pm 2.0)\%$ |
| $P_\tau(D^*)$ | $\pm 0.18 \pm 0.08$ | $\pm 0.06 \pm 0.04$ |

With the Belle II data set, NP scenarios can be precisely tested with $q^2$ and other distributions of kinematic observables. Figure 73 demonstrates the statistical precision of the $q^2$ measurement with 50 ab$^{-1}$ data based on a toy-MC study with the hadron tag based analysis. A quantitative estimation of the future sensitivity to a search for NP in $\bar{B} \to D^{(*)}\tau\bar{\nu}$ is shown in Fig. 74 [296]: it shows the regions of $C_X$ that are probed by the ratios (red) and the $q^2$ distributions (blue) at Belle II with 5 ab$^{-1}$ (dashed lines) and 50 ab$^{-1}$ (solid lines) respectively, at 95% CL.One finds that the distributions are very sensitive to all NP scenarios, including those with new scalar or tensors mediators. NP contributions that enter in $C_X$ can be described as

$$C_X \approx \frac{1}{2\sqrt{2}G_F V_{cb}} \frac{gg'}{M_{\rm NP}^2}, \tag{147}$$

where $g$ and $g'$ denote the couplings of new heavy particles to quarks and leptons respectively (at the NP mass scale $M_{\rm NP}$). Assuming couplings of $g, g' \sim 1$, one finds that the Belle II NP mass scale reach, $M_{\rm NP} \sim (2\sqrt{2}G_F V_{cb} C_X)^{-1/2}$, is about 5–10 TeV/c$^2$.

*8.5.2.* $B \to \pi\tau\nu$. Authors: R. Watanabe (th.), F. Bernlochner (exp.)

It is natural to expect that any NP contributions in $b \to c\tau\nu$ may also show up in $b \to u\tau\nu$ processes. A limit on the branching fraction of $B \to \pi\tau\nu$ has been determined by the Belle collaboration, Ref. [297]. They observed no significant signal and obtained the 90% CL upper limit as $\mathcal{B}(B \to \pi\tau\nu) < 2.5 \times 10^{-4}$. The measured branching fraction obtained was $\mathcal{B}(B \to \pi\tau\nu) = (1.52 \pm 0.72 \pm 0.13) \times 10^{-4}$, where the first error (along with the central value) is statistical and the second is systematic.

Evaluation of the form factors for the $B \to \pi$ transition have been performed using QCD predictions and experimental data. In the recent lattice studies of Refs. [143, 159], the authors have computed the vector and tensor amplitudes for $B \to \pi$. In their studies, the form factors are parameterised in the model independent Bourelly-Caprini-Lellouch (BCL) expansion approach [143, 159], defined as

$$f_0(q^2) = \sum_{n=0}^{N_z-1} b_n^0 z^n \,, \tag{148}$$

$$f_j(q^2) = \frac{1}{1-q^2/M_{B^*}^2} \sum_{n=0}^{N_z-1} b_n^j \left[ z^n - (-1)^{n-N_z} \frac{n}{N_z} z^{N_z} \right] \,, \tag{149}$$

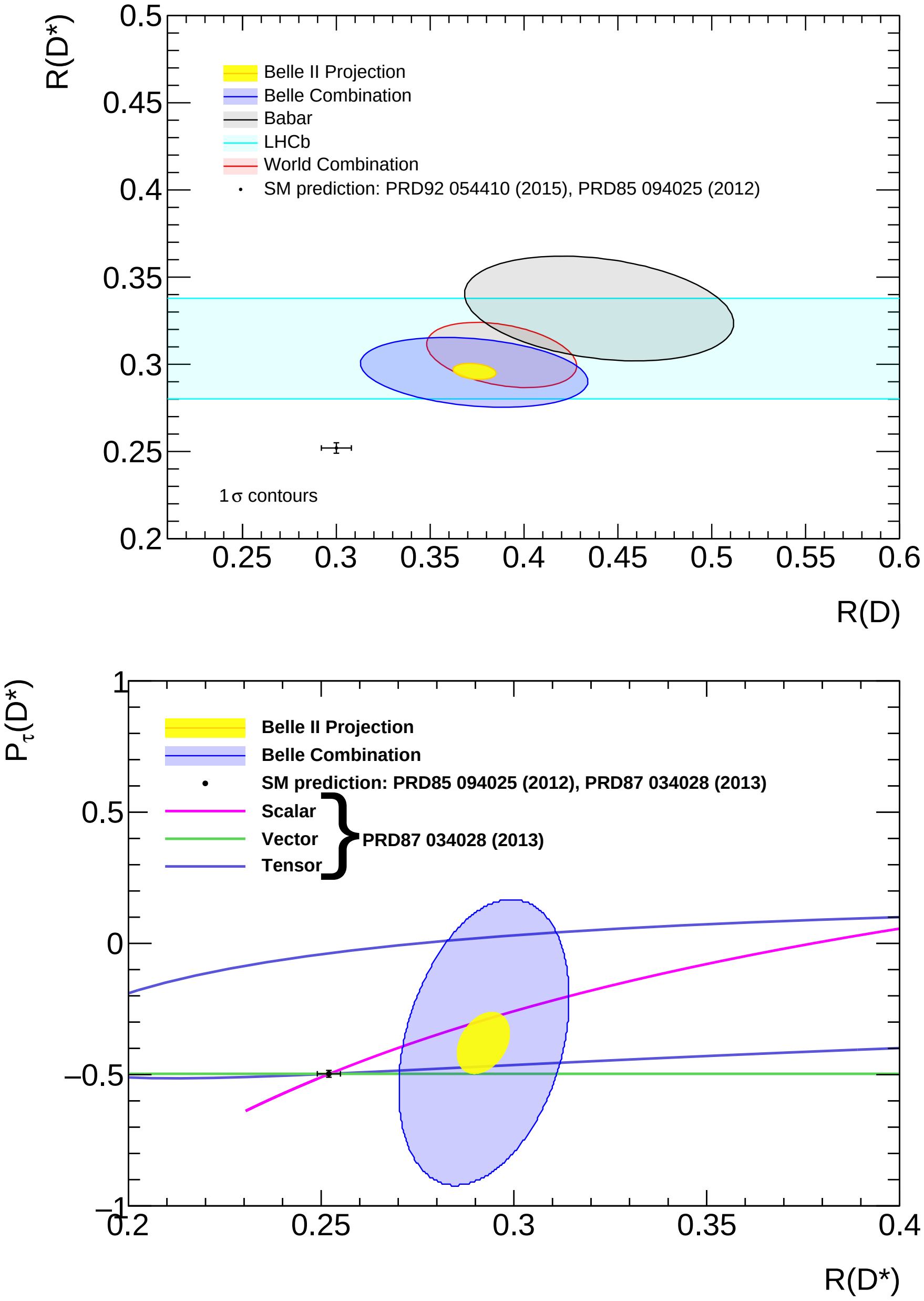

Fig. 72: Expected Belle II constraints on the $R_D$ vs $R_{D^*}$ plane (left) and the $R_{D^*}$ vs $P_\tau(D^*)$ plane (right) compared to existing experimental constraints from Belle. The SM predictions are indicated by the black points with theoretical error bars. In the right panel, the NP scenarios "Scalar", "Vector" and "Tensor" assume contributions from the operators $\mathcal{O}_{S_1}$, $\mathcal{O}_{V_1}$ and $O_T$, respectively.

for polarisations $j = +$ and $T$, where $M_{B^*} = 5.325\,\mathrm{GeV}$, $b_n^{0,+,T}$ are expansion parameters, and $N_z(= 4)$ is the expansion order. The analytical variable $z$ is defined as

$$z \equiv z(q^2) = \frac{\sqrt{t_+ - q^2} - \sqrt{t_+ - t_0}}{\sqrt{t_+ - q^2} + \sqrt{t_+ - t_0}}\,, \tag{150}$$

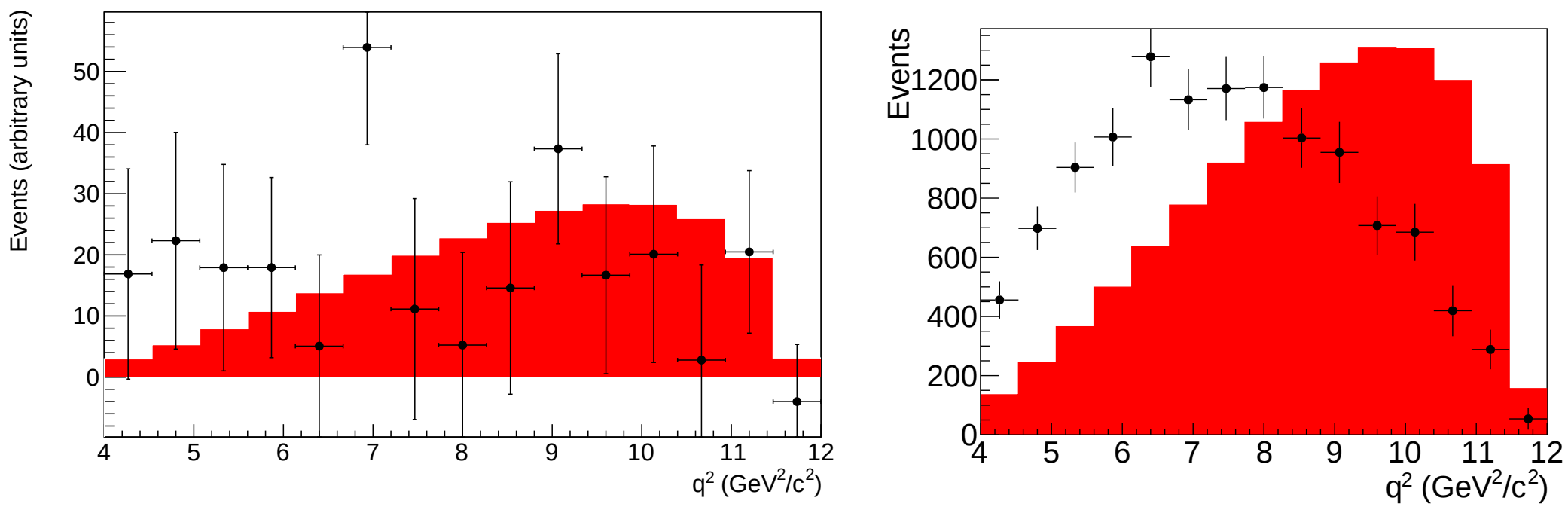

Fig. 73: On the left is the $B \to D\tau\nu$ $q^2$ distribution in the hadronic tag analysis and $\tau^- \to \ell^-\bar{\nu}_\ell\nu_\tau$ with the full Belle data sample [251]. On the right is the projection to the 50 ab$^{-1}$ Belle II data seta. In both panels, the solid histograms show the predicted distribution shape with the 2HDM of type II at $\tan\beta/m_{H^\pm} = 0.5$ (GeV$/c^2)^{-1}$. In the right panel, pseudo-data are shown based on the SM hypothesis.

where $t_0 = (M_B + M_\pi)(\sqrt{M_B} - \sqrt{M_\pi})^2$ and $t_+ = (M_B + M_\pi)^2$. The expansion parameters have been determined in fits to lattice simulations and experimental data on light leptonic modes $B \to \pi\ell\nu_\ell$ [84, 298–300]. The scalar form factor, present in $\tau$ modes, has been obtained in lattice QCD via the vector matrix element; cf. Eqs. 96 and 97.

We consider the ratio of branching fractions to test for NP contributions:

$$R_\pi \equiv \frac{\mathcal{B}(B \to \pi\tau\nu_\tau)}{\mathcal{B}(B \to \pi\ell\nu_\ell)} \equiv \frac{\mathcal{B}_\tau}{\mathcal{B}_\ell}\,, \tag{151}$$

where $|V_{ub}|$ cancels out. Possible NP scenarios can be described by

$$-\mathcal{L}_{\rm eff} = 2\sqrt{2}G_F V_{ub}\Big[(1 + C_{V_1})\mathcal{O}_{V_1} + C_{V_2}\mathcal{O}_{V_2} + C_{S_1}\mathcal{O}_{S_1} + C_{S_2}\mathcal{O}_{S_2} + C_T\mathcal{O}_T\Big], \tag{152}$$

similarly to the $b \to c$ case above, where $C_X$ (for $X = V_{1,2}, S_{1,2}$, and $T$) indicates a NP contribution in terms of the Wilson coefficient of $\mathcal{O}_X$ normalised by $2\sqrt{2}G_F V_{ub}$. The differential branching fractions for each tau helicity, $\lambda_\tau = \mp 1/2$, are then written as [243]

$$\frac{d\mathcal{B}_\tau^-}{dq^2} = N_B\Big|(1 + C_{V_1} + C_{V_2})\sqrt{q^2}H_{V,+} + 4C_T m_\tau H_T\Big|^2, \tag{153}$$

$$\begin{aligned}\frac{d\mathcal{B}_\tau^+}{dq^2} = \frac{N_B}{2}\Bigg[&\Big|(1 + C_{V_1} + C_{V_2})m_\tau H_{V,+} + 4C_T\sqrt{q^2}H_T\Big|^2 \\ &+ 3\Big|(1 + C_{V_1} + C_{V_2})m_\tau H_{V,0} + (C_{S_1} + C_{S_2})\sqrt{q^2}H_S\Big|^2\Bigg],\end{aligned} \tag{154}$$

with

$$N_B = \frac{\tau_B G_F^2 V_{ub}^2}{192\pi^3 M_B^3}\sqrt{Q_+Q_-}\left(1 - \frac{m_\tau^2}{q^2}\right)^2, \tag{155}$$

where $Q_\pm = (M_B \pm M_\pi)^2 - q^2$ and the quantities $H$ contain the hadron transition form factors. The differential branching fractions for $B \to \pi\ell\nu_\ell$ are given by

$$\frac{d\mathcal{B}_\ell^-}{dq^2} = \frac{d\mathcal{B}_\tau^-}{dq^2}\bigg|_{m_\tau \to 0,\, C_X = 0}, \qquad \frac{d\mathcal{B}_\ell^+}{dq^2} = 0\,. \tag{156}$$

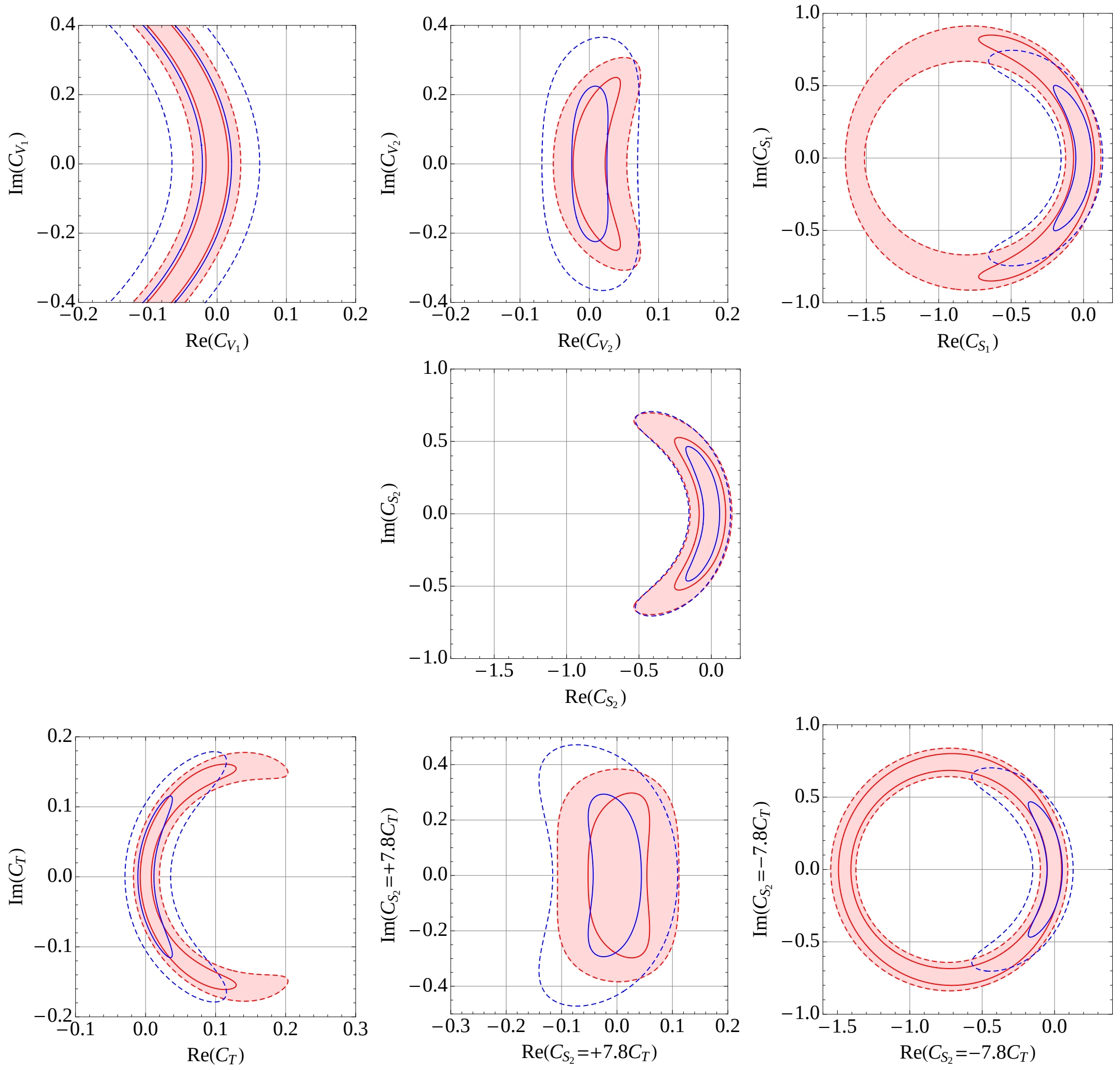

Fig. 74: Expected sensitivity of Belle II to constrain NP coefficients, $C_X$ at 95% CL. Regions inside the boundaries in red and blue can be probed at Belle II by measurement of the ratios and the distributions, respectively, with 5 ab$^{-1}$ (dashed lines) and 50 ab$^{-1}$ (solid lines).

Finally, $R_\pi$ is given as

$$R_\pi = \frac{\int_{m_\tau^2}^{q^2_{\max}} (d\mathcal{B}_\tau^+ + d\mathcal{B}_\tau^-)/dq^2}{\int_0^{q^2_{\max}} d\mathcal{B}_\ell^-/dq^2}\,, \tag{157}$$

where $q^2_{\max} = (M_B - M_\pi)^2$.

Given the above formula and input for $b_n^{0,+}$, the SM predicts $R_\pi^{\rm SM} = 0.641 \pm 0.016$, whereas the experimental data suggests $R_\pi^{\rm exp.} \simeq 1.05 \pm 0.51$ by using $\mathcal{B}(B \to \pi \ell \bar{\nu}_\ell) = (1.45 \pm 0.05) \times 10^{-4}$ [77]. Thus, at present the experimental result is consistent with the SM prediction given the large uncertainty. We can place loose bounds on NP scenarios from $R_\pi$. In Fig. 75, the constraints on $C_{V_1}$, $C_{V_2}$, $C_{S_1}$, $C_{S_2}$, and $C_T$ are shown, where it is assumed that the NP contribution comes from only one effective operator $\mathcal{O}_X$ for $X = V_1$, $V_2$, $S_1$, $S_2$, or $T$. As can be seen, the current data has already constrained NP contributions to be roughly $|C_X| \lesssim O(1)$, which implies that a contribution larger than that of the SM ($2\sqrt{2}G_F V_{ub}$) is disfavoured.

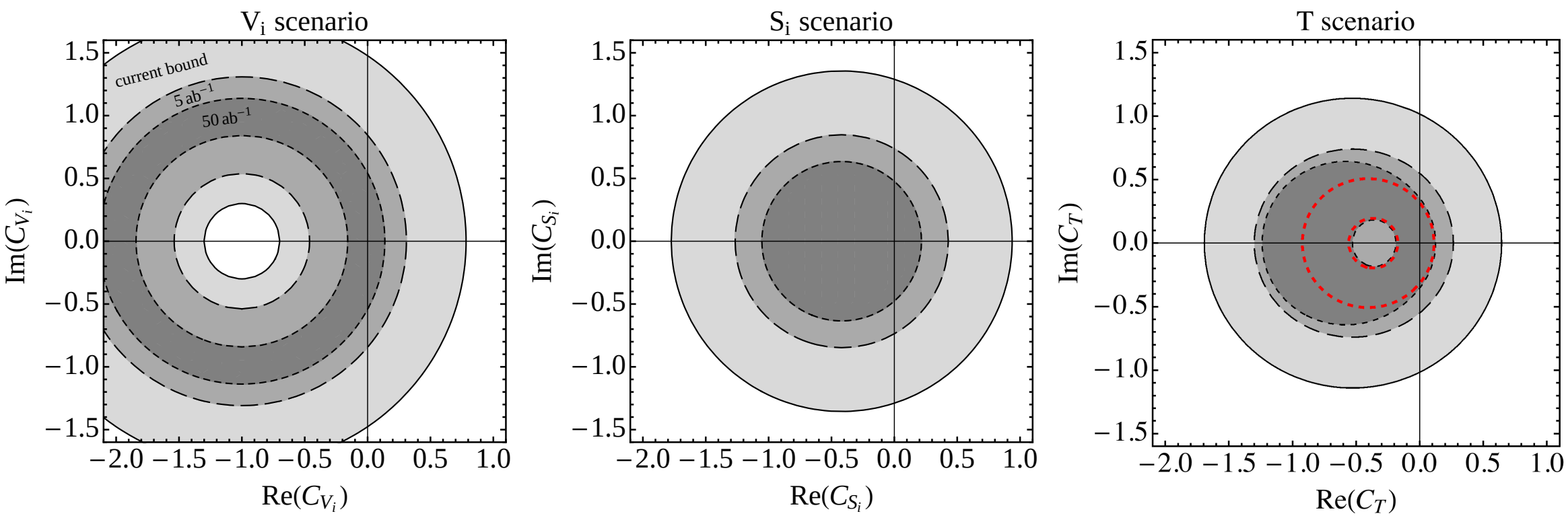

Fig. 75: Allowed regions for $V_1$, $V_2$, $S_1$, $S_2$, and $T$ scenarios based on the measurement of $R_\pi$. The light grey region is allowed from the measurement of $R_\pi$ by the Belle experiment at 95% CL. The $V_1$ and $V_2$ ($S_1$ and $S_2$) scenarios have the same region since their contributions are identical. The dark (darker) gray regions with black dashed curves denote a possible reach of 95% CL constraint expected at the Belle II, when $\mathcal{L} = 5\,\mathrm{ab}^{-1}$ ($50\,\mathrm{ab}^{-1}$) data is accumulated. For these results, the theoretical uncertainties given in Refs. [143, 159] are taken into account. The thick dashed red lines for the tensor case show the exclusion limit when the theoretical uncertainty is reduced by a factor of two.

A key reason for measuring $B \to \pi\tau\nu$ is that the tensor type interaction of NP that affects $b \to u\tau\nu$ cannot be constrained from $B \to \tau\nu$. The current results for $b_n^T$ for the tensor form factor still have large uncertainties [159]. Nevertheless, the constraint on $C_T$ is comparable to the other NP scenarios. Improvements in the evaluation of the tensor form factor will be significant for the future measurement of this process at Belle II.

The following study determined the future sensitivity of $R_\pi$ to NP scenarios with $5\,\mathrm{ab}^{-1}$ and $50\,\mathrm{ab}^{-1}$ of Belle II data, based on Ref. [243]. To estimate exclusion limits on the Wilson coefficient $C_X$, it is assumed that the experimental central value is identical to the SM prediction and the expected experimental errors at $5\,\mathrm{ab}^{-1}$ and $50\,\mathrm{ab}^{-1}$ are extrapolated from the Belle measurement [297]. The expected constraints from Belle II are therefore

$$R_\pi^{5\,\mathrm{ab}^{-1}} = 0.64 \pm 0.23\,, \tag{158}$$

$$R_\pi^{50\,\mathrm{ab}^{-1}} = 0.64 \pm 0.09\,. \tag{159}$$

The above values are compared with each NP scenario to determine constraints on $C_X$, as shown in Fig. 75. Focusing on the vicinity of the origin of $C_X$, we see that $|C_X| \gtrsim O(0.1)$ can be tested by the $R_\pi$ measurement for vector and tensor scenarios. A large negative contribution to $C_{V_i} \sim -2$ for example, will always be allowed within the uncertainty. For the tensor case, we expect to constrain $|C_T| \lesssim 1$, which can be improved if the theoretical uncertainties are addressed. In the figure, a scenario where the theoretical uncertainty is reduced by half is also presented, indicating improved sensitivity to tensor interactions. As for the scalar scenarios, $B \to \tau\nu$ has better sensitivity than $B \to \pi\tau\nu$ due to the chiral enhancement of the pseudoscalar contribution in the purely leptonic decay.

*8.5.3. $B \to X_c\tau\nu$.* Authors: F. Bernlochner (exp.), J. Hasenbusch (exp.), Z. Ligeti (th.)

*Introduction.* The anomalously large rates of $B \to D^{(*)}\tau\nu$ measured by BaBar, Belle and LHCb demand additional, independent measurements of $b \to c\tau\nu$ transitions. The measurement of inclusive $B \to X_c\tau\bar{\nu}$ decays could provide such additional information. This rate has not been directly measured, except for the related LEP measurements usually quoted as the average rate of an admixture of $b$ flavoured hadrons to decay semileptonically to $\tau + X$ [88]

$$\mathcal{B}(b \to X\tau^+\nu) = (2.41 \pm 0.23)\%. \tag{160}$$

The LEP analyses selected large missing energy events in the hemisphere opposite to a $b$-tagged jet, so the measurements constrain a linear combination of $b \to X\tau\bar{\nu}$, $\tau\bar{\nu}$, $X\nu\bar{\nu}$, and $X\tau\bar{\tau}$ [301, 302], of which, in the SM, the $X_c\tau\bar{\nu}$ rate dominates. The approaches for modelling semileptonic $B$ decays were inconsistent in the the LEP measurements. ALEPH for example, claims the most precise constraint by far on $R(X_c)$ yet does not explicitly quote any uncertainty for the $B \to D^{**}\ell\nu$ background. The LEP result (Eq. 160) is nevertheless in good agreement with the SM, as a more recent update of the SM prediction for $R(X_c)$ yields [293].

$$R(X_c) = 0.223 \pm 0.004, \tag{161}$$

which, combined with the world average, $\mathcal{B}(B^- \to X_c e\bar{\nu}) = (10.92 \pm 0.16)\%$ [230, 303], yields [293]

$$\mathcal{B}(B^- \to X_c\tau\bar{\nu}) = (2.42 \pm 0.05)\%. \tag{162}$$

The above prediction does not include $O(1/m_b^3)$ and $O(\alpha_s^2)$ effects, which turn out to be quite sizeable. More recent publications have evaluated these effects: see for example Ref. [304] which gives $R(X_c) = 0.212 \pm 0.003$ and Ref. [305] which gives $R(X_c) = 0.214 \pm 0.004$. This prediction is nevertheless precise, thus the inclusive measurement can provide information complementary to the exclusive channels.

There is a tension between the exclusive and inclusive measurements [293], as the isospin-constrained fit for the sum of branching ratios [250]

$$\mathcal{B}(\bar{B} \to D^*\tau\bar{\nu}) + \mathcal{B}(\bar{B} \to D\tau\bar{\nu}) = (2.78 \pm 0.25)\ \%, \tag{163}$$

is above the inclusive measured rate. This tension is further exacerbated by the 0.15% SM prediction for the branching fractions to the four $D^{**}\tau\bar{\nu}$ modes [306]. Measuring the inclusive rate should be feasible. Uncertainties of the individual $\bar{B} \to D^{(*)}\tau\bar{\nu}$ branching ratios are expected to be reduced to about 3% with Belle II data.

*Theoretical results.* Here, we briefly discuss the SM predictions (following Ref. [293]) for differential distributions in inclusive $B \to X_c\tau\bar{\nu}$ decays, including $1/m_b^2$ and $\alpha_s$ corrections. These results can improve the sensitivity of $b \to c\tau\bar{\nu}$ related observables to BSM physics. The inclusive $B \to X_c\tau\bar{\nu}$ decay rates can be computed model-independently in an operator product expansion (OPE) just like for $B \to X_c\ell\bar{\nu}$, see Sec. 8.7. The perturbative and non-perturbative corrections can be systematically incorporated, and are modest if one measures the total inclusive rate without substantial phase space cuts. We outline here how these corrections become large near endpoint regions of these spectra. The triple differential $B \to X\tau\bar{\nu}$ distribution has long been known, including the leading nonperturbative corrections of order $1/m_b^2$ [307–309]. Until recently [310], the theoretical predictions were not available using a well-defined short-distance quark mass scheme.

One often uses the dimensionless kinematic variables

$$\hat{q}^2 = \frac{q^2}{m_b^2}, \quad v \cdot \hat{q} = \frac{v \cdot q}{m_b}, \quad y = \frac{2E_\tau}{m_b}, \quad x = \frac{2E_\nu}{m_b}, \tag{164}$$

where $q = p_\tau + p_\nu$ is the dilepton momentum, $v = p/M_B$ is the four-velocity of the $B$ meson, and $E_{\tau,\nu}$ are the $\tau, \nu$ energies in the $B$ rest frame. The mass ratios

$$\rho_\tau = m_\tau^2/m_b^2, \qquad \rho = m_c^2/m_b^2 \tag{165}$$

are also needed.

The triple differential decay rate in the $B$ rest frame is

$$\begin{aligned} &\frac{1}{\Gamma_0} \frac{d\Gamma}{d\hat{q}^2\, dy\, dv \cdot \hat{q}} \\ &\quad = 24\, \theta\big[(2v \cdot \hat{q} - y_+)y_+ - \hat{q}^2\big]\, \theta\big[\hat{q}^2 - (2v \cdot \hat{q} - y_-)y_-\big] \\ &\quad \times \Big\{ 2(\hat{q}^2 - \rho_\tau)\hat{W}_1 + \big[y(2v \cdot \hat{q} - y) - \hat{q}^2 + \rho_\tau\big]\, \hat{W}_2 \\ &\qquad + 2\big[\hat{q}^2(y - v \cdot \hat{q}) - \rho_\tau v \cdot \hat{q}\big]\, \hat{W}_3 + \rho_\tau(\hat{q}^2 - \rho_\tau)\, \hat{W}_4 + 2\rho_\tau(2v \cdot \hat{q} - y)\, \hat{W}_5 \Big\}, \end{aligned} \tag{166}$$

where

$$\Gamma_0 = |V_{cb}|^2 G_F^2 \frac{m_b^5}{192\pi^3} \tag{167}$$

is the tree-level free-quark decay rate. The $\hat{W}_i$ are the structure functions of the hadronic tensor [307, 311], which in the local OPE of $\Lambda_{\rm QCD}^2/m_b^2$ contain Dirac $\delta$, $\delta'$, and $\delta''$ functions of $(1 + \hat{q}^2 - 2v \cdot \hat{q} - \rho)$. The key kinematical difference between zero and non-zero lepton mass arises from the fact that

$$y_\pm = \frac{1}{2}\Big(y \pm \sqrt{y^2 - 4\rho_\tau}\Big), \tag{168}$$

which enters the phase space boundaries in Eq. (166), simplifies in the $m_\tau \to 0$ limit: $y_+ \to y$ and $y_- \to 0$ (in general, $y_+ y_- = \rho_\tau$). In the massive lepton case, the appearance of the second nontrivial Heaviside $\theta$ function in Eq. (166) sets a lower limit on $\hat{q}^2$. For fixed $\hat{q}^2$ and $v \cdot \hat{q}$,

$$\hat{q}_- + x_\tau \hat{q}_+ \le y \le \hat{q}_+ + x_\tau \hat{q}_-, \tag{169}$$

where $\hat{q}_\pm = v \cdot \hat{q} \pm \sqrt{(v \cdot \hat{q})^2 - \hat{q}^2}$. At the parton level, $v \cdot \hat{q} = (1 + \hat{q}^2 - \rho)/2$ gives the partonic phase space in the $\hat{q}^2 - y$ plane at tree level. The limits on $\hat{q}^2$ for fixed $y$ are

$$y_-\Big(1 - \frac{\rho}{1 - y_-}\Big) \le \hat{q}^2 \le y_+\Big(1 - \frac{\rho}{1 - y_+}\Big), \tag{170}$$

This is illustrated in Fig. 76, where $\rho = (1.3/4.7)^2$ and $\rho_\tau = (1.777/4.7)^2$ were used. Beyond tree level, the lower limit of the $\hat{q}^2$ integration and the lower limit of $y$ integration for $\hat{q}^2 < \hat{q}_0^2 = \sqrt{\rho_\tau}\big[1 - \rho/(1 - \sqrt{\rho_\tau})\big]$ is replaced by $\hat{q}^2 > \rho_\tau$ and $y > 2\sqrt{\rho_\tau}$.

Integrating over $\hat{q}^2$, the limits on $y$ are

$$2\sqrt{\rho_\tau} < y < 1 + \rho_\tau - \rho. \tag{171}$$

Integrating over $y$, the limits of $\hat{q}^2$ are

$$\rho_\tau < \hat{q}^2 < (1 - \sqrt{\rho})^2. \tag{172}$$

These are the partonic phase space limits which enter the OPE calculation, while the physical phase space limits are determined by the hadron masses.

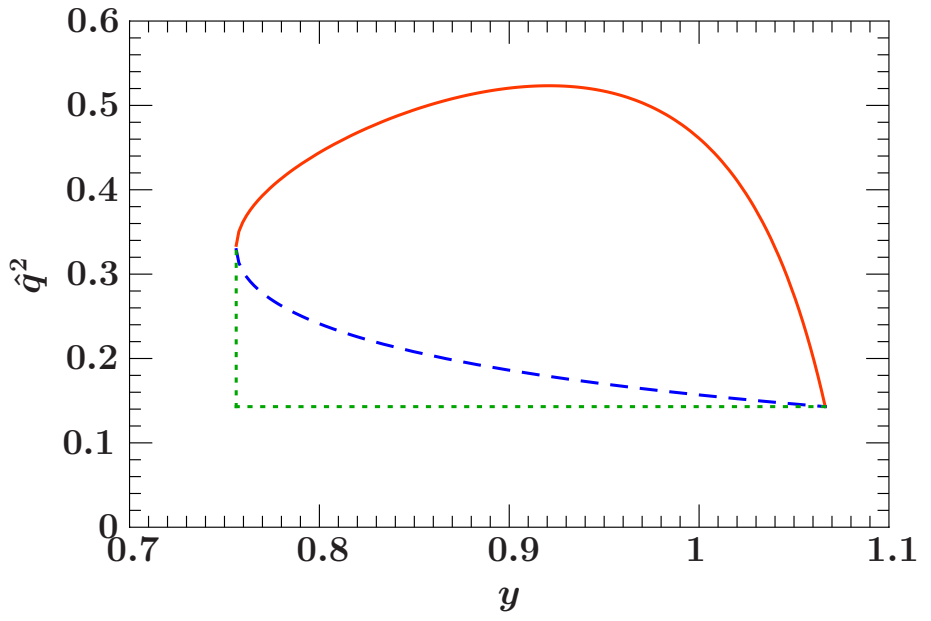

Fig. 76: The $b \to c\tau\bar{\nu}$ Dalitz plot for free quark decay. The solid orange boundary comes from the first $\theta$ function in Eq. (166), the dashed blue boundary from the second one [310].

Besides the total rate, the $q^2$ and $E_\tau$ spectra have been studied in detail [310]. The OPE breaks down for large values of $E_\tau$. Similar to $B \to X_s\gamma$ and $B \to X_u l\bar{\nu}$ (see Sec. 8.7), the terms in the OPE that are enhanced near the endpoint can be re-summed into a nonperturbative shape function. The shape function expansion can be rendered valid away from the endpoint region as well, such that it smoothly recovers the local OPE result [312–314]. One obtains at leading order [310]

$$\begin{aligned}\frac{1}{\Gamma_0}\frac{d\Gamma}{dy} = 2\sqrt{y^2-4\rho_\tau}\int d\hat{\omega} m_b F(m_b\hat{\omega}+m_B-m_b) \\ \times\,\theta(y-2\rho_\tau)\theta(1-R_\omega)(1-R_\omega)^2\left\{y\rho\frac{1-R_\omega}{R_\omega}\right. \\ \left.+\,(1+2R_\omega)\big[y-\hat{\omega}y_- - 2\rho_\tau\big](2-y-\hat{\omega})\right\}.\end{aligned} \tag{173}$$

where $\Gamma_0 = |V_{cb}|^2 G_F^2 m_b^5/(192\pi^3)$ is the tree-level free-quark decay rate, $R_\omega = \rho/[(1-y_+-\hat{\omega})(1-y_-)]$, and $F(k)$ is the leading order universal shape function. The endpoint region of the $E_\tau$ spectrum is given by $1-y_+ \sim \Lambda_{\rm QCD}/m_b$. For smaller values of $E_\tau$, the usual local OPE is reliable.

The order $1/m_b^2$ corrections reduce the $B \to X_c\tau\bar{\nu}$ rate by about 7 − 8%, concentrated mainly at higher values of $\hat{q}^2$, dominated by the terms proportional to $\lambda_2$. As for large values of $E_\tau$, the OPE also breaks down for large values of $\hat{q}^2$. Near maximal $\hat{q}^2$, the $\lambda_2$ terms behave as $(\hat{q}^2_{\rm max}-\hat{q}^2)^{-1/2}$, and the differential rate becomes negative. This breakdown of the OPE occurs because the hadronic final state gets constrained to the lightest few hadronic resonances, which are not calculable in the OPE. Thus, integration over some range of $\Delta\hat{q}^2$ is necessary near maximal $\hat{q}^2$ to obtain a reliable result.

Restoring the dimensions of the variables, the phase space limits are

$$m_\tau < E_\tau < \frac{m_b^2-m_c^2+m_\tau^2}{2m_b}, m_\tau^2 < q^2 < (m_b-m_c)^2. \tag{174}$$

One can see using $m_{b,c} = m_{B,D} - \bar{\Lambda} + \mathcal{O}(\Lambda^2_{\rm QCD}/m^2_{b,c})$ that the difference of the upper limit of $q^2$ at the parton level, $(m_b-m_c)^2$, and at the hadronic level, $(m_B-m_D)^2$, is suppressed by $\Lambda^2_{\rm QCD}$. However, the lepton energy endpoint receives an $\mathcal{O}(\Lambda_{\rm QCD})$ correction, although numerically only about 100 MeV (it is $\sim$ 300 MeV for $B \to X_u e\bar{\nu}$).

Writing $m_c = m_b^{1S} - \delta m_{bc}$, and treating $\delta m_{bc} = m_b - m_c$ as an independent parameter is practical, as it is well constrained by measured $B \to X_c\ell\bar{\nu}$ spectra, and is the dominant source of formally $\mathcal{O}(\lambda_1/m_c^2)$ corrections [315]. We use a conservative $\pm 20$ MeV/c$^2$ uncertainty on $\delta m_{bc}$. We also use $\lambda_1 = -0.3$ GeV$^2$ and $\lambda_2 = 0.12$ GeV$^2$, and vary both by 25% to account for uncertainties and higher order effects.

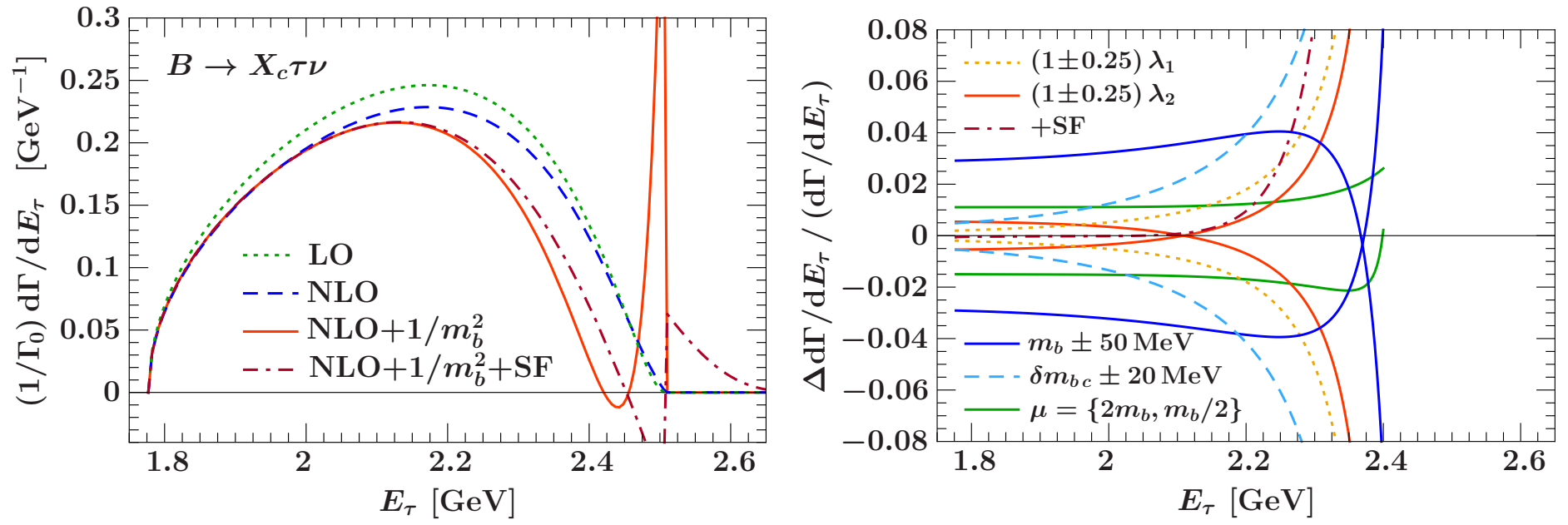

Fig. 77: The OPE predictions for $d\Gamma/dE_\tau$ in $B \to X_c\tau\bar{\nu}$ [310]. Left: the dotted green curve shows the free-quark decay result, the dashed blue curve includes $\mathcal{O}(\alpha_s)$ corrections, and the solid orange curve includes both $\alpha_s$ and $1/m_b^2$ corrections. The dot-dashed dark red curve combines $\mathcal{O}(\alpha_s, 1/m_b^2)$ with the leading shape function. Right: the solid blue curve shows the fractional uncertainty due to the variation of $m_b^{1S}$ by $\pm 50$ MeV/c$^2$, the dashed light blue curves show that of $\delta m_{bc}$ by $\pm 20$ MeV/c$^2$, the solid green curves show the $\mu$ variation between $m_b/2$ and $2m_b$, and the solid red (dotted light orange) curves show the variation of $\lambda_2$ ($\lambda_1$) by $\pm 25\%$. The dot-dashed dark red curve shows the correction from the leading shape function.

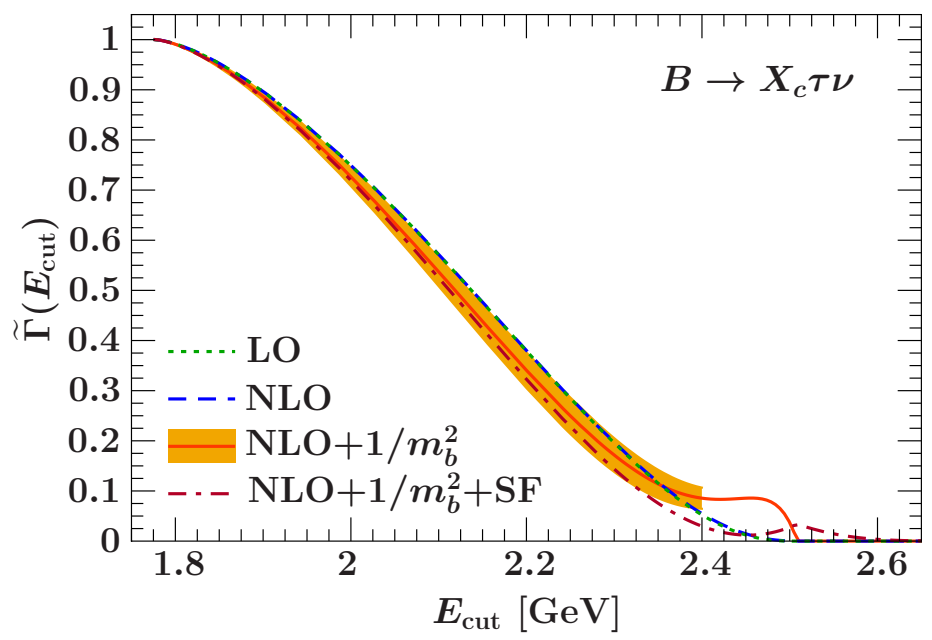

Fig. 78: Fraction of events above a given $\tau$ energy threshold $d\Gamma/dE_\tau$. The shaded band shows the total uncertainty.

Figure 77 shows the predictions for $d\Gamma/dE_\tau$ in the $1S$ mass scheme for the $b$ quark. The $1/m_b^2$ corrections are negligible at low values of $E_\tau$ (and $q^2$), while their effects become important for larger values. The peculiar shape of $d\Gamma/dE_\tau$ including the $\mathcal{O}(1/m_b^2)$ terms is due to the fact that near the endpoint both the $\lambda_1$ and $\lambda_2$ terms are large, and the $\lambda_1$ term changes sign. The dot-dashed (dark red) curve combines the $\mathcal{O}(\alpha_s)$ and $\mathcal{O}(1/m_b^2)$ corrections with the tree-level leading shape function result in Eq. (173). For this, the fit result from Ref. [316] was used (and for consistency also $m_b^{1S}$, with a conservative $\pm 50$ MeV/c$^2$ uncertainty).

Figure 77 (right) shows the fractional uncertainties from varying the parameters. The variations from $m_b = (4.71 \pm 0.05)$ GeV/c$^2$ keeping $\delta m_{bc} = 3.4$ GeV/c$^2$ fixed and $\delta m_{bc} =$ $\pm 20$ MeV/c$^2$ dominate at low and high values, respectively. Variations on the renormalisation scale, $\mu$, between $m_b/2$ and $2m_b$, the coefficients of $\lambda_2$ and $\lambda_1$ and relative corrections due to shape function effects are shown.

Since the spectrum cannot be calculated reliably near maximal $E_\tau$, Fig. 78 shows the rate above a cut, normalised to the total rate, $\widetilde{\Gamma}(E_{\text{cut}}) = (1/\Gamma)\int_{E_{\text{cut}}} d\Gamma/dE_\tau$, at different orders in the OPE. The $\mathcal{O}(\alpha_s)$ corrections have a negligible effect on these distributions since they affect the shapes very mildly. The yellow band shows the total uncertainty obtained by adding all uncertainties in quadrature. The dot-dashed (dark red) curve shows the effect of including the leading shape function.

*Experimental challenges.* The analysis of $B \to X\tau\nu$ at the $B$-factories is a tremendous challenge, even at Belle II. Here we discuss the main experimental considerations.

To study $B \to X\tau\nu$, tagged samples are a big advantage, particularly tags that are fully reconstructed in hadronic decay modes. Furthermore, it is useful to restrict the analysis to $\tau \to \ell\nu\nu$ modes, as the background level is lower than the hadronic modes. An important discriminant is the lepton momentum in the rest frame of the decaying $B$ meson, which can be determined from the kinematics of the hadronic tag. The lepton momentum is lower for $B \to X\tau\nu$ signal events compared to $B \to X\ell\nu$ events as the latter produces higher energy prompt (decay daughters of the $B$) leptons and thus separates this background very well. However, background from hadronic $B$ decays, producing secondary leptons ($B \to D \to \ell$) or hadrons faking leptons, will contribute with similar momenta to the signal.

Other important properties of the signal are the three undetectable neutrinos in the decay, which carry away momentum and energy. From the known initial $e^+e^-$ state kinematics and under the assumption that the tag side decayed in a hadronic decay mode, the missing energy and momentum can be derived. Most of the relevant background contributions are expected to have less missing momentum and energy than the signal decay, which is typically probed using the missing mass squared $m^2_{\text{miss}}$ observable. All major background contributions peak at zero missing mass as they decay with a single neutrino, while the signal resides in the tail regions. Combined, the lepton momentum and missing mass squared can provide a powerful two-dimensional discriminant. This is a high statistics analysis with an overwhelmingly high background level. Thus, even small deviations between data and MC cannot be treated as statistical fluctuations and need to be well understood. This leads directly to the challenges and needs of this analysis.

The analysis relies heavily on the modelling of signal and background semileptonic $B$ decay processes, particularly at low lepton momentum. Mis-modelling of the inclusive semileptonic $B$ decay spectrum can significantly bias the result, demanding accurate form-factors for all semileptonic decay contributions and accurate composition of the $X_c$ components. The biggest challenge is to describe the poorly measured high mass excited charm state modes, which can behave similarly to the signal. However, even the well known $B \to D^*\ell\nu$ decay is a source of uncertainty since it is the largest individual contribution to the lepton momentum spectrum.

The modelling of secondary leptons must also be accurate. Secondary leptons that arise through upper vertex cascade transitions ($B \to D \to \ell$) will have similar momenta to the signal and therefore pose a challenge. Furthermore, hadronic $B$ decays where hadrons fake leptons may contribute through a diverse set of decay chains of hadronic $B$ decays. Such hadronic decays are typically not well constrained. Further work must be done to reduce the hadron fakes by improving lepton identification separation power at low momentum regions.

The analysis is sensitive to the accuracy of detector modelling. This leads to slight efficiency differences that become significant in tails of missing mass distributions.

Belle II should consider strategies to isolate $B \to D^{**}\tau\nu$ decay modes by first reconstructing a $D$ or $D^*$ and looking for an additional $M_X$ component. It will be challenging due to the lower rates of these modes, and the lower efficiency of explicitly reconstructing the charm mesons.

## *8.6. Exclusive semileptonic*

### *8.6.1. $B \to D^{(*)}\ell\nu$.* Authors: A. S. Kronfeld (th.), C. Schwanda (exp.)

*Experimental status.* The decays $B \to D^*\ell\nu$ and $B \to D\ell\nu$ are currently the preferred modes for determining the CKM element magnitude $|V_{cb}|$ using exclusive decays. The experiments measure the differential decay rate of $B \to D^{(*)}\ell\nu$ as a function of the recoil variable $q^2$ or, equivalently, $w = (M_B^2 + M_{D^{(*)}}^2 - q^2)/2M_BM_{D^{(*)}}$. The formulas for differential decays rates in $q^2$ (and $\cos\theta$, which is an angle between the hadrons and the charged lepton) are given in Sec. 8.2.

The experimental analyses have to date used a simplifying parameterisation of the form factors, from Caprini, Lellouch and Neubert (CLN) [317]. For the form factor and $f_+(q^2)$ and $A_1(q^2)$, this parameterisation reads[23]

$$A_1(q^2) = A_1(q^2_{\rm max})\big[1 - 8\rho^2_{D^*}z + (53\rho^2_{D^*} - 15)z^2 - (231\rho^2_{D^*} - 91)z^3\big], \tag{175}$$

$$f_+(q^2) = f_+(q^2_{\rm max})\big[1 - 8\rho^2_{D}z + (51\rho^2_{D} - 10)z^2 - (252\rho^2_{D} - 84)z^3\big], \tag{176}$$

where

$$z = \frac{\sqrt{w+1} - \sqrt{2}}{\sqrt{w+1} + \sqrt{2}}, \tag{177}$$

and $q^2_{\rm max} = (M_B - M_{D^{(*)}})^2$ corresponds to $w = 1$ and $z = 0$. The CLN parameterisation for the other form factors (or, equivalently, helicity amplitudes) for $B \to D^*$ are given by

$$R_1(w) = R_1(1) - 0.12(w-1) + 0.05(w-1)^2, \tag{178}$$

$$R_2(w) = R_2(1) + 0.11(w-1) - 0.06(w-1)^2, \tag{179}$$

for certain form-factor ratios. It is important to bear in mind that the numerical coefficients in Eqs. 175, 176, 178, and 179 are estimates with (omitted) uncertainties. Before using CLN in future work, the coefficients would have to be reevaluated with modern inputs and their uncertainty propagated. It is advisable, however, to move to a model-independent parameterisation; see below.

Tables 51 and 52 summarise the most significant measurements of $B \to D^*\ell\nu$ and $B \to D\ell\nu$. They report $\eta_{\rm EW}\mathcal{F}(1)|V_{cb}|$, $\rho^2_{D^*}$, $\eta_{\rm EW}\mathcal{G}(1)|V_{cb}|$, and $\rho^2_D$, where $\mathcal{F}(1) \propto A_1(q^2_{\rm max})$, $\mathcal{G}(1) \propto f_+(q^2_{\rm max})$. Due to the cleanliness of the $D^{*+} \to D^0\pi^+$, $D^0 \to K^-\pi^+$ signal, untagged analyses of $B \to D^*\ell\nu$ yield the most precise results on $\Upsilon(4S)$ datasets of order 1 ab$^{-1}$. The systematic uncertainty in $\eta_{\mathrm{EW}}\mathcal{F}(1)|V_{cb}|$ is five times larger than the statistical one, with leading systematics arising from tracking, lepton and hadron identification efficiencies. Background uncertainties are not a leading source of uncertainty. For $B \to D\ell\nu$ however, background is a major concern. On $B$ factory data sets the most precise analyses used hadronic tagging and a large number of reconstructed $D$ modes. In the most precise analysis [85], statistical and systematic uncertainty are of similar size, with the leading source of systematic uncertainty being the hadronic tag calibration. This can be controlled further at Belle II if high purity tag side decay modes are used.

[23] The literature on heavy-to-heavy transitions often introduces notation for quantities proportional to $A_1$ and $f_+$.

Table 51: Measurements of $\eta_{\mathrm{EW}}\mathcal{F}(1)|V_{cb}|$ and of $\rho^2_{D^*}$ in the CLN parameterisation of the form factor [317].

| Experiment | $\eta_{\mathrm{EW}}\mathcal{F}(1)\lvert V_{cb}\rvert[10^{-3}]$ | $\rho^2_{D^*}$ |
|---|---|---|
| BaBar [318] | $34.4 \pm 0.3_{\mathrm{stat}} \pm 1.1_{\mathrm{syst}}$ | $1.191 \pm 0.048_{\mathrm{stat}} \pm 0.028_{\mathrm{syst}}$ |
| BaBar [319] | $35.9 \pm 0.2_{\mathrm{stat}} \pm 1.2_{\mathrm{syst}}$ | $1.22 \pm 0.02_{\mathrm{stat}} \pm 0.07_{\mathrm{syst}}$ |
| Belle [320] | $35.1 \pm 0.2_{\mathrm{stat}} \pm 0.6_{\mathrm{syst}}$ | $1.106 \pm 0.031_{\mathrm{stat}} \pm 0.007_{\mathrm{syst}}$ |

Table 52: Measurements of $\eta_{\mathrm{EW}}\mathcal{G}(1)|V_{cb}|$ and of $\rho^2_D$ in the CLN parameterisation of the form factor [317].

| Experiment | $\eta_{\mathrm{EW}}\mathcal{G}(1)\lvert V_{cb}\rvert$ $[10^{-3}]$ | $\rho^2_D$ |
|---|---|---|
| BaBar [319] | $43.1 \pm 0.8_{\mathrm{stat}} \pm 2.3_{\mathrm{syst}}$ | $1.20 \pm 0.04_{\mathrm{stat}} \pm 0.07_{\mathrm{syst}}$ |
| BaBar [321] | $43.0 \pm 1.9_{\mathrm{stat}} \pm 1.4_{\mathrm{syst}}$ | $1.20 \pm 0.09_{\mathrm{stat}} \pm 0.04_{\mathrm{syst}}$ |
| Belle [85] | $42.29 \pm 1.37$ | $1.09 \pm 0.05$ |

The Heavy Flavour Averaging Group (HFLAV) has performed a fit to these measurements [230] and obtains for $D^*\ell\nu$,

$$\eta_{\mathrm{EW}}\mathcal{F}(1)|V_{cb}| = (35.61 \pm 0.43) \times 10^{-3}, \tag{180}$$

$$\rho^2_{D^*} = 1.205 \pm 0.026, \tag{181}$$

and for $D\ell\nu$,

$$\eta_{\mathrm{EW}}\mathcal{G}(1)|V_{cb}| = (41.57 \pm 1.00) \times 10^{-3}, \tag{182}$$

$$\rho^2_D = 1.128 \pm 0.033. \tag{183}$$

To convert these results into measurements of $|V_{cb}|$, theory input for the form factor normalisation at zero recoil ($w = 1$) is needed. Using the most recent lattice-QCD calculations from the Fermilab Lattice and MILC collaborations [144, 145] for $\mathcal{F}(1)$ and $\mathcal{G}(1)$, HFLAV obtains

$$|V_{cb}|_{D^*\ell\nu} = (39.05 \pm 0.47_{\mathrm{exp}} \pm 0.58_{\mathrm{th}}) \times 10^{-3}, \tag{184}$$

$$|V_{cb}|_{D\ell\nu} = (39.18 \pm 0.94_{\mathrm{exp}} \pm 0.36_{\mathrm{th}}) \times 10^{-3}. \tag{185}$$

There is good consistency between $|V_{cb}|$ determined from $B \to D^*\ell\nu$ and $B \to D\ell\nu$ decays, but the exclusive measurement is at odds with the inclusive determination of $|V_{cb}|$ (Sect. 8.7) by approximately $3\sigma$ ($3.2\sigma$ for $|V_{cb}|$ from $B \to D^*\ell\nu$ and $2.4\sigma$ for $|V_{cb}|$ from $B \to D\ell\nu$), which clearly calls for further studies at Belle II.

As discussed in Sec. 8.5, lattice QCD already provides the full kinematic dependence of the $B \to D\ell\nu$ form factors [145, 146, 257], and corresponding work for $B \to D^*\ell\nu$ is underway [253]. Instead of CLN, these studies use the parameterisation of Boyd, Grinstein and Lebed (BGL) [322], which uses the same variable $z$ but with no assumptions on the coefficients, apart from mild constraints stemming from unitarity in quantum mechanics.

There are indications that a change from CLN to BGL might shed light on the inclusive/exclusive discrepancy. Note that reporting only the CLN parameters, instead of the form factors bin-by-bin, impedes a simultaneous fit with lattice QCD data at $w > 1$. This is especially problematic for $B \to D\ell\nu$ where the experimental rate approaches zero at zero recoil with $\lambda^{3/2}$ instead of $\lambda^{1/2}$; cf. Eqs. 102 and 122. In Ref. [85], it was shown that the change from CLN to BGL together with the inclusion of lattice-QCD results at $w > 1$ shifts $\eta_{\mathrm{EW}}|V_{cb}|$ from $(40.12 \pm 1.34) \times 10^{-3}$ to $(41.10 \pm 1.14) \times 10^{-3}$, *i.e.* towards the inclusive result. See also Refs. [145, 146, 257] for similar results. Furthermore, Ref. [323] presents a reanalysis of the preliminary Belle data of Ref. [324] and found that a change from CLN to BGL changes the fit result for $|V_{cb}|$ from $(38.2 \pm 1.5) \times 10^{-3}$ to $(41.7 \pm 2.0) \times 10^{-3}$, again compatible with $|V_{cb}|$ inclusive measurements. An investigation of unitarity bounds on the form factor parameters was performed in Ref. [260]. The recent untagged analysis of Belle [320] finds compatible results for the CLN and BGL parameterisations: $|V_{cb}|(\mathrm{CLN}) = (38.4 \pm 0.9) \times 10^{-3}$, and $|V_{cb}|(\mathrm{BGL}) = (38.3 \pm 1.0) \times 10^{-3}$.

*Opportunities at Belle II.* The goal of Belle II for exclusive $|V_{cb}|$ is to see whether fits to lattice-QCD and experimental results from the full kinematics agree in shape and, if so, obtain a robust determination. With current $\Upsilon(4S)$ data sets of order 1 ab$^{-1}$, and so the limitation has been systematic uncertainties. Unless the detector performance is better understood at Belle II, the experimental uncertainties cannot be reduced. This is feasible, but requires careful examination of tracking efficiencies and particle identification. Recently, a tagged analysis of $B \to D^*\ell\nu$ using the full Belle data set has become available, although its results are still preliminary [324]. Here, the experimental uncertainty in $|V_{cb}|$ is 3.5% compared to 2.9% in Ref. [325]. It should be noted that the main systematic uncertainty in the latter paper was on tracking efficiencies, which has since been improved threefold at Belle. In any case, the tagged analysis of $B \to D^*\ell\nu$ is approaching the precision of untagged measurements while still being limited systematically, due to the calibration of the hadronic tag. In summary, a reduction of systematic uncertainties at Belle II, namely of the hadronic tag calibration for tagged measurements, is required to improve current measurements of $|V_{cb}|$ exclusive. Belle II's analyses of $B \to D^{(*)}\ell\nu$ can emphatically address the discrepancy between $|V_{cb}|$ inclusive and exclusive.

These modes can be precisely probed for new physics currents that may modify angular distributions or introduce phenomena such as lepton flavour universality violation [262]. Dedicated studies by experiment are yet to be performed, despite the rich offering of experimental information in these high rate, high purity decay modes. Crucially, Belle II must report results of measurements in parameterisation independent ways, *i.e.* by providing unfolded spectra or similar.

*8.6.2. $B \to D^{**}\ell\nu$.* Authors: G. Ricciardi (th.) The study of semileptonic decays to excited charm modes was an ongoing challenge in the $B$-factories, yet knowledge of their contribution to the total decay width is a limiting uncertainty in $|V_{cb}|$ and semitauonic $B$ decays.

The orbitally excited charm states $D_1(2420)$ and $D_2^*(2460)$ have relatively narrow widths, about 20-30 MeV/c$^2$, and have been observed and studied by a number of experiments (see Ref. [326] for the most recent study). The other two states, $D_0^*(2400)$, $D_1'(2430)$, are more difficult to detect due to their large widths, about 200-400 MeV/c$^2$ [327–331]. The theoretical expectation is that the states with large width should correspond to $j^l = 1/2^+$ states, which decay as $D_{0,1}^* \to D^{(*)}\pi$ through $S$ waves by conservation of parity and angular momentum. Similarly, the states with small width should correspond to $j^l = 3/2^+$ states, since $D_2^* \to D^{(*)}\pi$ and $D_1 \to D^*\pi$ decay through $D$ waves. Decays such as $D_1 \to D^*\pi$ may occur through both $D$ and $S$ waves, but the latter are disfavoured by heavy-quark symmetry.

The spectroscopic identification of heavier states is less clear. In 2010 Babar observed candidates for the radial excitations of the $D^0$, $D^{*0}$ and $D^{*+}$, as well as the $L = 2$ excited states of the $D^0$ and $D^+$ [332]. Resonances in the 2.4–2.8 GeV/$c^2$ region of hadronic masses have also been identified at LHCb [333–336].

Most calculations, using sum rules [337, 338], quark models [339–342], OPE [343, 344] (but not constituent quark models [345]), indicate that the narrow width states should dominate over the broad $D^{**}$ states. This is in contrast to experimental results: a tension known as the "1/2 vs 3/2" puzzle. One possible weakness common to these theoretical approaches is that they are derived in the heavy-quark limit and corrections might be large. For instance, it is expected that $1/m_c$ corrections induce a significant mixing between $D_1$ amd $D_1'$, which could soften the 1/2 vs 3/2 puzzle at least for the $1^+$ states [346]. However, no real conclusion can be drawn until more data on the masses, widths, and absolute branching ratios of the orbitally-excited $D$ meson states become available. The other puzzle is that the sum of the measured semileptonic exclusive rates with a $D^{(*)}$ in the decay chain is less than the inclusive one (the "gap" problem) [331, 347]. Decays into $D^{(*)}$ make up $\sim 70\%$ of the total inclusive $B \to X_c l\bar{\nu}$ rate and decays into $D^{(*)}\pi$ make up another $\sim 15\%$, leaving a gap of about 15%. In 2014 the full Babar data set was used to improve the precision on decays involving $D^{(*)}\pi\,\ell\,\nu$ and to search for decays of the type $D^{(*)}\pi\,\pi\ell\,\nu$. Preliminary results assign about 0.7% to $D^{(*)}\pi\,\pi\ell\,\nu$, reducing the significance of the gap from $7\sigma$ to $3\sigma$ [348].

The theoretical description of $B \to D^{**}\ell\nu$ channels have been investigated in Ref. [306]. Lattice-QCD studies are in progress with realistic charm mass, while preliminary results on $\bar{B} \to D^{**}\ell\nu$ form factors are available [349–351].

Belle II must precisely isolate all four orbitally excited modes and characterise their sub-decay modes as accurately as possible, to ultimately constrain and measure the branching ratios. Form factors must be determined in all modes through precise differential measurements. Complementary information on the decay rates of orbitally excited modes should be extracted from hadronic $B$ decays and include multi-pion and other light quark meson transitions.

*8.6.3. $B \to \pi\ell\nu$.* Authors: A. S. Kronfeld (th.), M. Lubej (exp.), A. Zupanc (exp.) The aim for Belle II is to reach one percent-level determinations of $|V_{ub}|$ through a variety of experimental and theoretical approaches. In the case of $B \to \pi\ell\nu$, the challenge is that the experimental branching fraction measurements are most precise at low $q^2$, whereas the lattice-QCD form factors are best determined at high $q^2$. Interpolating the results will rely on constraining form factor parameterisations. To obtain the best possible $|V_{ub}|$, the numerical lattice-QCD form-factor data must be extended to the full kinematic range, while the experimental measurement is expected to greatly improve with improved statistical power. An order of magnitude more data will allow precise tests of these lattice-QCD predictions for the $q^2$ dependence. If the $q^2$ shapes of experiment and lattice QCD agree, it is straightforward to fit the relative normalisation to obtain $|V_{ub}|$.

Measurements of decay rates of exclusive $B \to X_u \ell \nu_\ell$ decays, where $X_u$ denotes a light meson containing a $u$ quark, such as $\pi$, $\rho$, $\omega$, $\eta^{(\prime)}$, etc., and $\ell$ an electron or muon, have in the past been performed using three different experimental techniques that differ only in the way the tag $B$ meson in the event is reconstructed: tagged hadronic, tagged semileptonic, or untagged. In the rest of this subsection, we present results of sensitivity studies on the determination of $|V_{ub}|$ through exclusive $B \to \pi \ell^+ \nu_\ell$ decays using the untagged and hadronic tagged reconstruction techniques of the $B_{\rm tag}$ performed with the Belle II MC5 sample.

*Untagged measurement.* To reconstruct signal $B$ candidates, good pion and lepton candidates are selected based on the responses of particle identification sub-detectors and by requiring that their momenta in the laboratory system exceeds 1 GeV/$c$. Improved $K/\pi$ separation in Belle II allows for better $b \to c \to s$ rejection than in Belle or BaBar. The two charged daughter particles are required to originate from the same vertex.Under the assumption that the neutrino is the only missing particle, the cosine of the angle between the inferred direction of the reconstructed $B$ and that of the $Y = \pi\ell$ system is

$$\cos\theta_{BY} = \frac{2E_B^* E_Y^* - M_B^2 - M_Y^2}{2p_B^* p_Y^*}, \tag{186}$$

where $E^*$ and $p^*$ are energy and 3-momentum magnitude in the CMS system of the $B$ and $Y$, respectively. The energy and momentum magnitude of the $B$ meson are given by energy-momentum conservation and can be calculated as $E_B^* = E_{CMS}/2$ and $p_B^* = \sqrt{E_B^{*2} - M_B^2}$. Correctly reconstructed candidates should strictly populate the interval $|\cos\theta_{B,Y}| \leq 1$, although due to the detector resolution, a small fraction of signal is reconstructed outside this interval. Background processes usually have more than one missing particle and therefore the angle is not constrained between $-1$ and 1. Due to detector resolution effects, we require $-1.2 < \cos\theta_{BY} < 1.1$ to ensure high efficiency.

The Belle II detector geometry hermetically covers a large portion of the full solid angle (approximately 90%), so we assume all remaining tracks and clusters in the rest of event (ROE) originate from the tag $B_{\rm tag}$ meson. This $B_{\rm tag}$ candidate is not reconstructed in the same way as in tagged analyses, but rather just by adding the 4-momenta of the remaining tracks and clusters as

$$\mathrm{p}_{ROE} = \sum_i (E_i, \boldsymbol{p}_i) + \sum_j (\sqrt{m_j^2 + p_j^2}, \boldsymbol{p}_j), \tag{187}$$

where $i$ and $j$ indices run over all clusters and tracks not used in the reconstruction of the signal side, respectively. The mass hypothesis of track $j$, $m_j$, is determined based on the response of the particle ID sub-detectors. We take it to be the one with the highest posterior probability.

The missing 4-momentum of the event is given as

$$\mathrm{p}_{\mathrm{miss}} = (E_{\mathrm{miss}}, \boldsymbol{p}_{\mathrm{miss}}) = \mathrm{p}_{e^+e^-} - \mathrm{p}_Y - \mathrm{p}_{ROE}, \tag{188}$$

where the $\mathrm{p}_{e^+e^-} = (E_{e^+e^-}, \boldsymbol{p}_{e^+e^-})$ denotes the 4-momentum of the colliding beam particles. For correctly reconstructed candidates and only one missing neutrino, $\mathrm{p}_{\mathrm{miss}}$ should be equal to $\mathrm{p}_\nu$, with $\mathrm{p}^2_{\mathrm{miss}} = m^2_\nu = 0$. Due to resolution effects, the signal distribution peaks at zero with a non-zero spread.

With the neutrino momentum determined, we can attempt to correct the $Y$ momentum to obtain the momentum of the signal $B$ meson as

$$\mathrm{p}_B = \mathrm{p}_Y + (p_{\mathrm{miss}}, \boldsymbol{p}_{\mathrm{miss}}) = (E_B, \boldsymbol{p}_B), \tag{189}$$

where we have substituted the missing energy $E_{\mathrm{miss}}$ with the missing momentum magnitude $p_{\mathrm{miss}}$ due to better energy resolution.

With the $B$ meson 4-momentum we can calculate the $B$ meson specific variables: $M_{\mathrm{bc}}$, the beam-energy constrained mass, and $\Delta E$, the beam energy difference, defined in the laboratory frame as

$$M_{\mathrm{bc}} = \sqrt{\frac{(s/2 + \boldsymbol{p}_B \cdot \boldsymbol{p}_{e^+e^-})^2}{E^2_{e^+e^-}} - p^2_B}, \tag{190}$$

$$\Delta E = \frac{\mathrm{p}_B \cdot \mathrm{p}_{e^+e^-} - s/2}{\sqrt{s}}, \tag{191}$$

where $\sqrt{s}$ is the CMS energy.

The momentum transferred from the $B$ meson to the leptonic part is calculated as $q^2 = (\mathrm{p}_B - \mathrm{p}_\pi)^2$. The $q^2$ resolution function is shown in Fig. 81. The resolution can be even further improved (reducing the root-mean-square of the resolution by around 20%) by taking into account the fact that the $B$ momentum is kinematically constrained to lie on a cone around the $Y$ pseudo-particle's momentum and taking the weighted average over four possible configurations of the direction of the $B$ meson.

It is not optimal to sum over all remaining calorimeter clusters and charged tracks in the event as indicated in Eq. 187 due to extra tracks and extra clusters from back-splashes, beam background induced interactions, and secondary interactions of primary particles produced in $e^+e^-$ collisions. To select good tracks and reject those produced in secondary interactions of primary particles with the detector material, we train a boosted decision tree (BDT) using the following input: impact parameters in radial ($d_0$) and $z$ directions ($z_0$) and their uncertainties, the track momentum $p$, cosine of the polar angle of the track ($\cos\theta$), number of hits in the vertex detector, track fit $p$-value and the distance to the nearest cluster at the calorimeter radius. To reject beam background induced energy deposits in the calorimeter and back-splashes we train another BDT with the following input: the energy deposited in a $3 \times 3$ block over that in a $5 \times 5$ block of calorimeter crystals, $E_9/E_{25}$, cosine of the polar angle of the cluster, $\cos\theta$, cluster timing, cluster energy, number of hits in the calorimeter, probability of the cluster coming from a $\pi^0$ particle, and distance to the nearest track hit at the calorimeter radius. As mentioned before, improper summation of tracks and clusters leads to degraded $B$ candidate distributions as shown in Fig. 79. To optimise the ROE selection based on an MVA output discriminant we use the criteria that maximises the signal purity in the $\Delta E$ signal region.

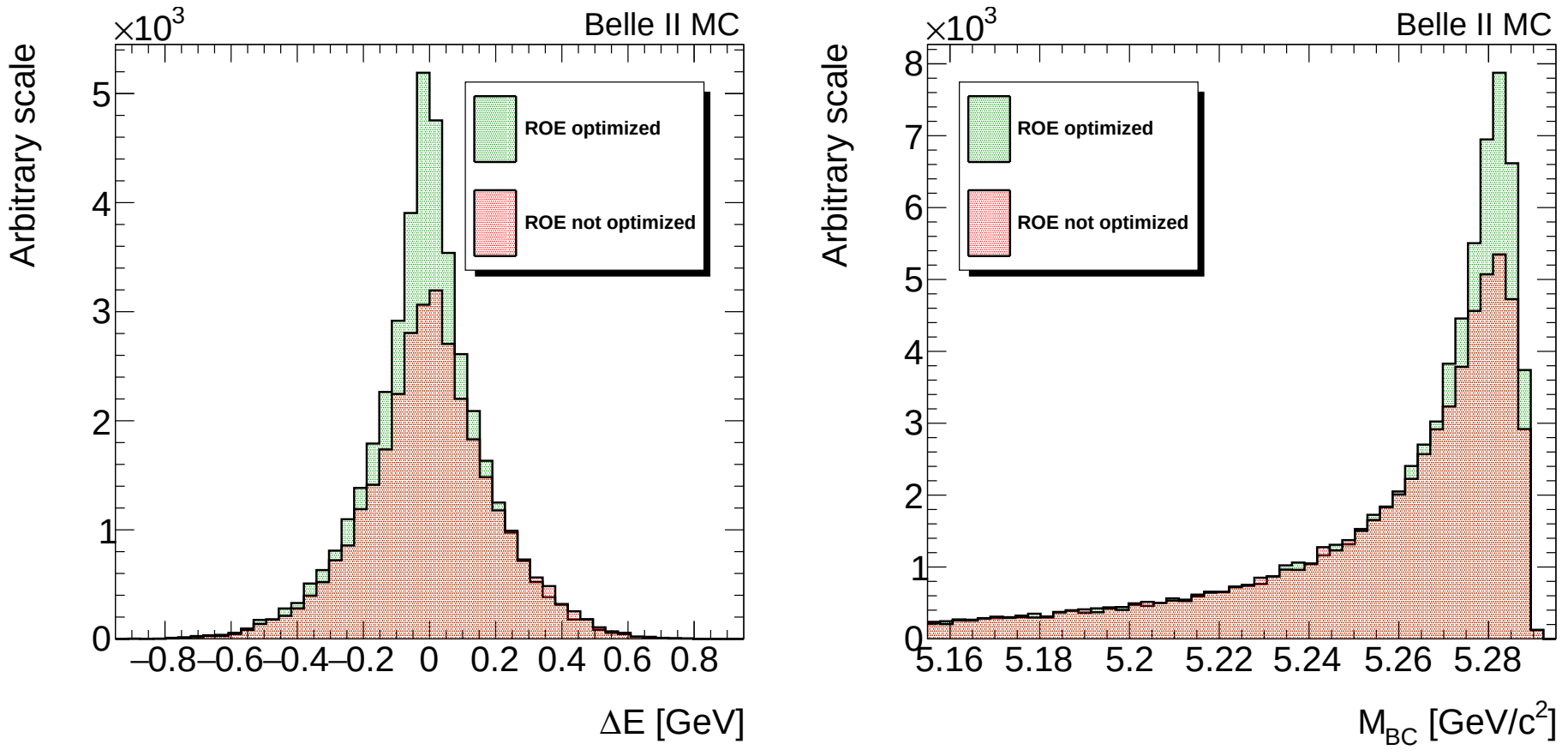

Fig. 79: The distributions of $\Delta E$ and $M_{\rm bc}$ for the untagged $B \to \pi\ell\nu$ analysis, with and without rest of event (ROE) optimisation.

There are three major sources of background: quark continuum ($e^+e^- \to q\overline{q},\ q = u, d, s, c$), Cabibbo favoured processes ($B \to X_c\ell\nu$) and other Cabibbo suppressed processes other than $B \to \pi\ell\nu$ ($B \to X_u\ell\nu$). Suppressing each type of background requires separate treatment. Continuum events represent the easiest background category to suppress, since the event shape of continuum events is more jet-like, whereas $B\overline{B}$ events have a more isotropic event shape. For this background type we train four different BDT classifiers (defined in Sec. 6.4): (i) CLEO cones (event topology), (ii) Kakuno-Super-Fox-Wolfram moments (event topology), (iii) output from (1) and (2) with additional thrust-axis variables, (iv) and output from (3) with additional $B$ meson selection variables.

Additional $B$ meson selection variables include the pion identification probability, the lepton helicity angle $\cos\theta_\ell$, the missing momentum polar angle $\theta_{\rm miss}$, the difference between flight distances of the $B$ mesons $dz$, the angle between the $Y$ pseudo-particle and the $z$-axis $\cos\theta_{BY}$, and an improved version of the $m^2_{\rm miss}$ variable, $m^2_{\rm miss}/2E_{\rm miss}$, where its resolution does not decrease with $E_{\rm miss}$. Each input variable was checked for its correlation with $q^2$ and all variables with a significant correlation were discarded. The optimal BDT output selection requirement is determined by maximising a statistical power figure of merit. To suppress $b \to u\ell\nu_\ell$ background, we train another BDT. The final sample composition, after all selection criteria are applied, is shown in Fig. 80. The $q^2$-averaged signal efficiency is found to be around 20%. We identify signal candidates by performing a 2-dimensional fit to $M_{bc}$ and $\Delta E$. The sample is then split into 13 bins of $q^2$ from 0 to $q^2_{\rm max} = 26.4$ GeV$^2/c^2$. We define the fit region as $M_{\rm bc} > 5.095$ GeV$/c^2$ and $|\Delta E| < 0.95$ GeV and perform fits to extract the raw signal yield in each $q^2$ bin.

*Tagged measurement.* In the tagged measurement we first require that the tag $B$ meson is fully reconstructed in one of many potential hadronic decay modes. After finding a good $B_{\rm tag}$ candidate, we require that the rest of the event is consistent with the signature of the signal decay; it contains only two additional oppositely charged tracks, one being consistent with the pion and one with the lepton hypothesis based on particle identification sub-detectors. The lepton charge must be consistent with the flavour of the decaying $B$. As in the case of the $B^+ \to \tau^+\nu_\tau$ study, we use $B_{\rm tag}$ candidates provided by the Full Event Interpretation algorithm (see Sec. 6.6) with a signal probability exceeding 0.1%. In the case of multiple $B_{\rm tag}B_{\rm sig}$ candidates we keep the combination with the $B_{\rm tag}$ candidate that has the highest signal probability. Since we measure the 4-momentum of the tag $B$ meson, we can infer the signal $B$ meson 4-momentum, the missing 4-momentum of the neutrino produced in the signal decay, and the momentum transfer to the lepton system squared, $q^2$, as

$$\mathrm{p}_{B_{\rm sig}} = \mathrm{p}_{\Upsilon(4S)} - \mathrm{p}_{B_{\rm tag}}, \tag{192}$$

$$\mathrm{p}_{\rm miss} = \mathrm{p}_\nu = \mathrm{p}_{\Upsilon(4S)} - \mathrm{p}_{B_{\rm tag}} - \mathrm{p}_\pi - \mathrm{p}_\ell, \tag{193}$$

$$q^2 = (\mathrm{p}_\ell + \mathrm{p}_\nu)^2 = (\mathrm{p}_{B_{\rm sig}} - \mathrm{p}_\pi)^2 = (\mathrm{p}_{\Upsilon(4S)} - \mathrm{p}_{B_{\rm tag}} - \mathrm{p}_\pi)^2, \tag{194}$$

where we take the tag $B$ meson 4-momentum in the $\Upsilon(4S)$ frame to be

$$\mathrm{p}_{B_{\rm tag}} = (E_{\rm CMS}/2, \boldsymbol{p}_{B_{\rm tag}}). \tag{195}$$

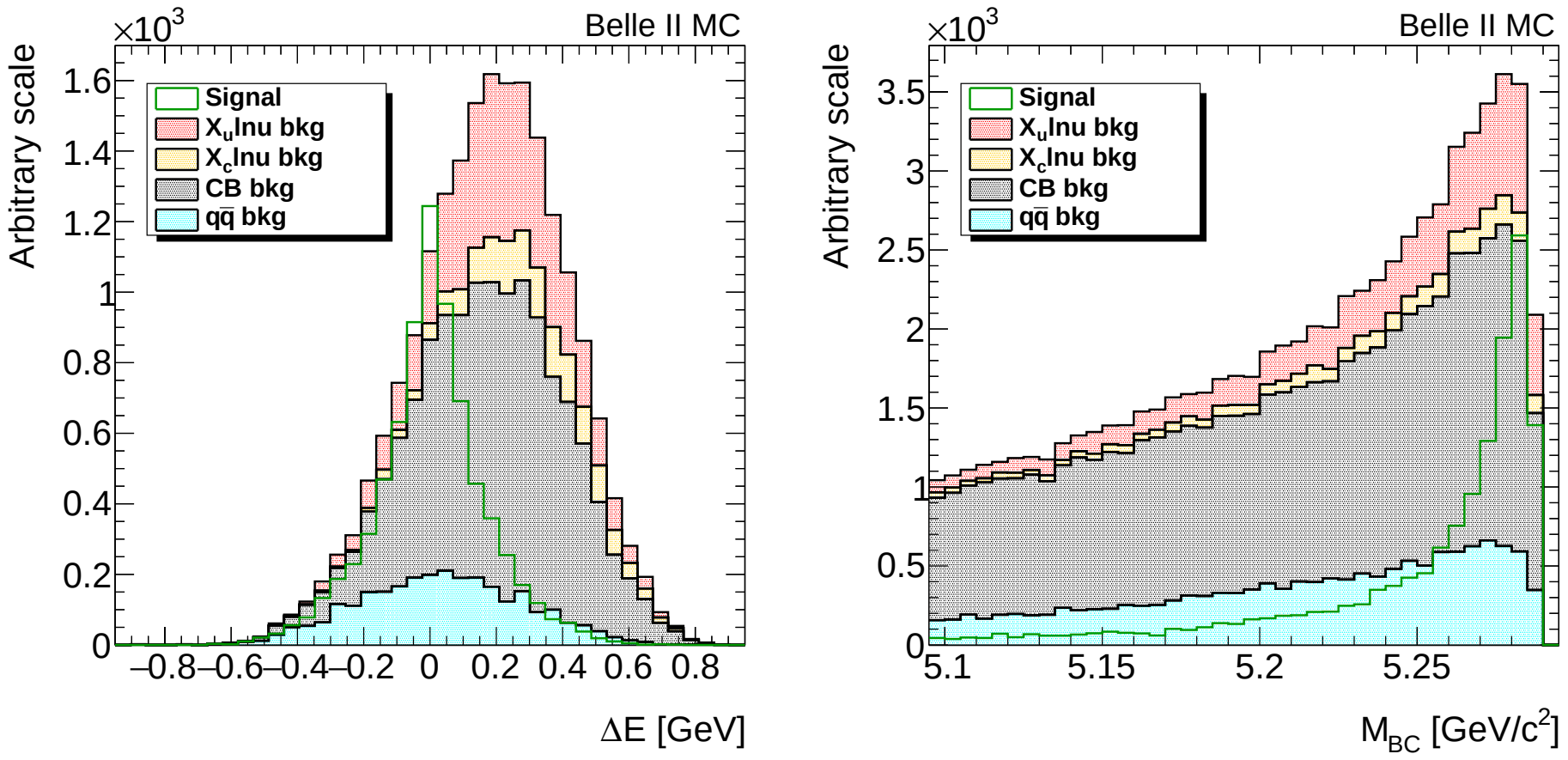

Fig. 80: The distributions of $M_{\rm bc}$ (with $|\Delta E| < 0.15$ GeV) and $\Delta E$ (with $M_{\rm bc} > 5.27$ GeV/$c^2$) with signal and background components for the full $q^2$ range. The signal is shown separately and set to the expected yield.

The precise measurement of the momentum of the tag $B$ meson results in an improved measurement of $q^2$ compared to the untagged measurement, as shown in Fig. 81. The overall reconstruction efficiency is found to be 0.55% in the MC sample, which is considerably above the reconstruction efficiency (0.3%) of the tagged measurement reported by Belle [84]

The signal is extracted from the missing mass squared distribution ($M^2_{\rm miss} = \mathrm{p}^2_{\rm miss}$), where the signal is expected to be located in a narrow peak near zero, while background from other $b \to u\ell\nu$ transitions populates a wider region towards higher missing mass, due to extra missing particles in the decay, as shown in Fig. 82. These processes can be further suppressed by requiring that there is little energy deposited in the calorimeter that can not be associated to the decay products of the signal nor to the tag $B$. Alternatively, the signal can be extracted by performing a 2-dimensional fit to $M^2_{\rm miss}$ and $E_{\rm extra}$ (see Fig. 83).

*Systematic uncertainties.* A full breakdown of the systematic uncertainties of the tagged and untagged measurements are given in Table 53. In the tagged method, most systematic

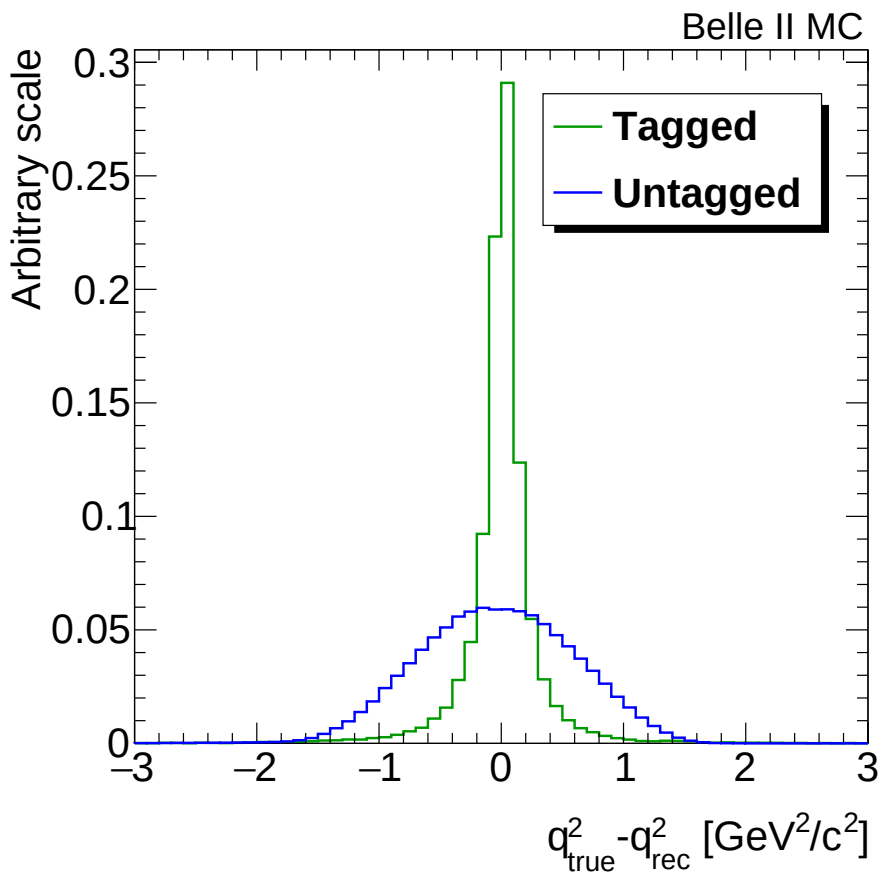

Fig. 81: Resolution of $q^2$ from untagged and tagged measurement of $B^0 \to \pi^- \ell^+ \nu_\ell$ decays.

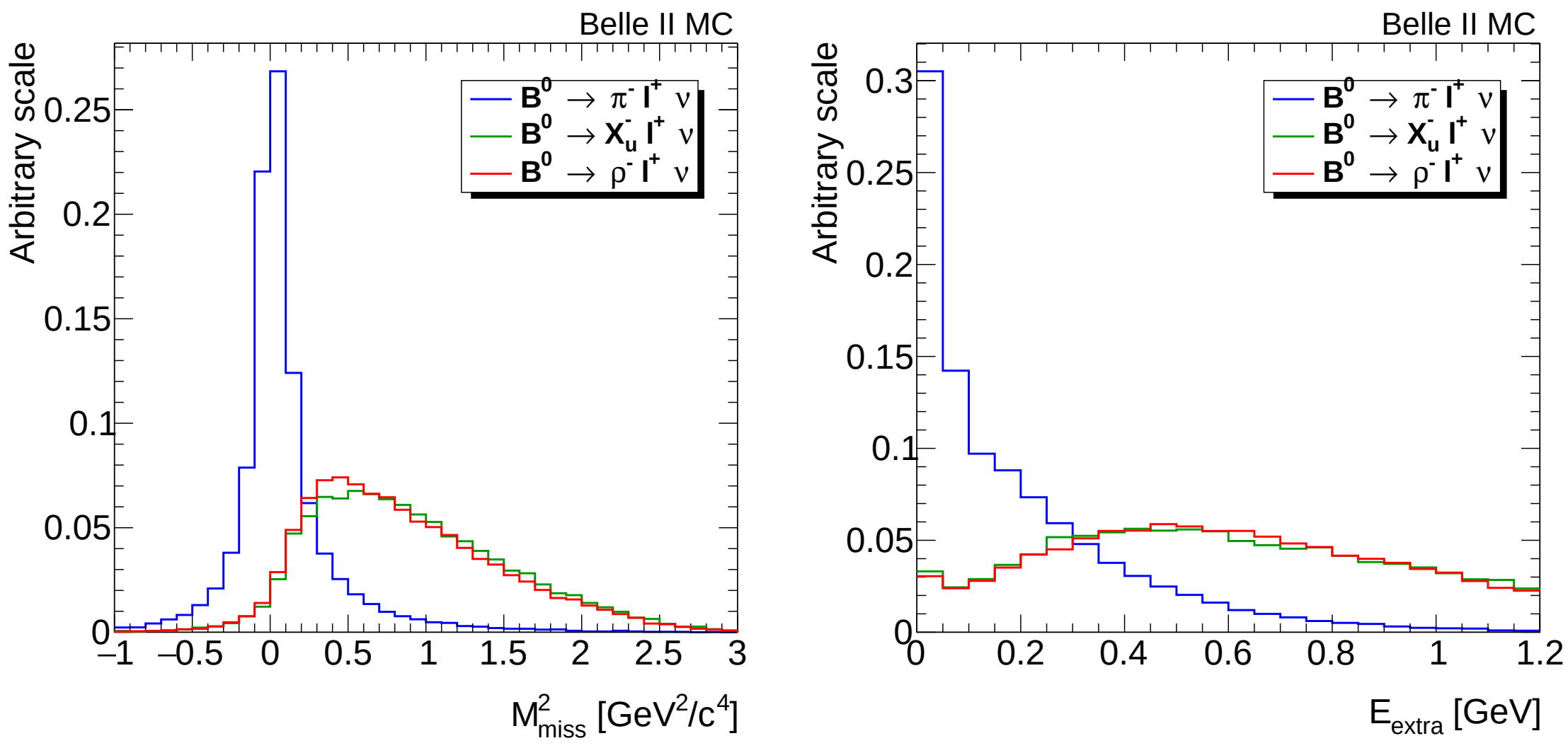

Fig. 82: The $M^2_{\text{miss}} = \text{p}^2_{\text{miss}}$ (left) and $E_{\text{extra}}$ (right) distributions of tagged $B^0 \to \pi^- \ell^+ \nu_\ell$ candidates as obtained from events containing signal $B^0 \to \pi^- \ell^+ \nu_\ell$ decays, and background $B^0 \to \rho^- \ell^+ \nu_\ell$ and $B^0 \to X_u^- \ell^+ \nu_\ell$ decays.

j

uncertainties are determined from purely data driven techniques. Systematics due to background modelling from $B \to X_c \ell \nu$ and $B \to X_u \ell \nu$ (cross-feed) are reasonably small due to the high purity of the method. The untagged method suffers from low purity, which makes it more difficult to isolate signal from poorly understood cross-feed background. Although the quoted model uncertainties are already small, totalling less than 2% on the branching fraction, it would require far more detailed studies of the full $B \to X_u \ell \nu$ rate across $q^2$ to reliably reduce them further. The remaining irreducible uncertainty is derived from the normalisation to the number of $B$ mesons produced, shared between the Belle $N(B\overline{B})$ measurement and the production fraction $f_{00}/f_{+-}$. Although these are systematics-limited quantities, they can be improved with better experimental detection systematics and more orthogonal measurements. A conservative limit of 1% is assigned.

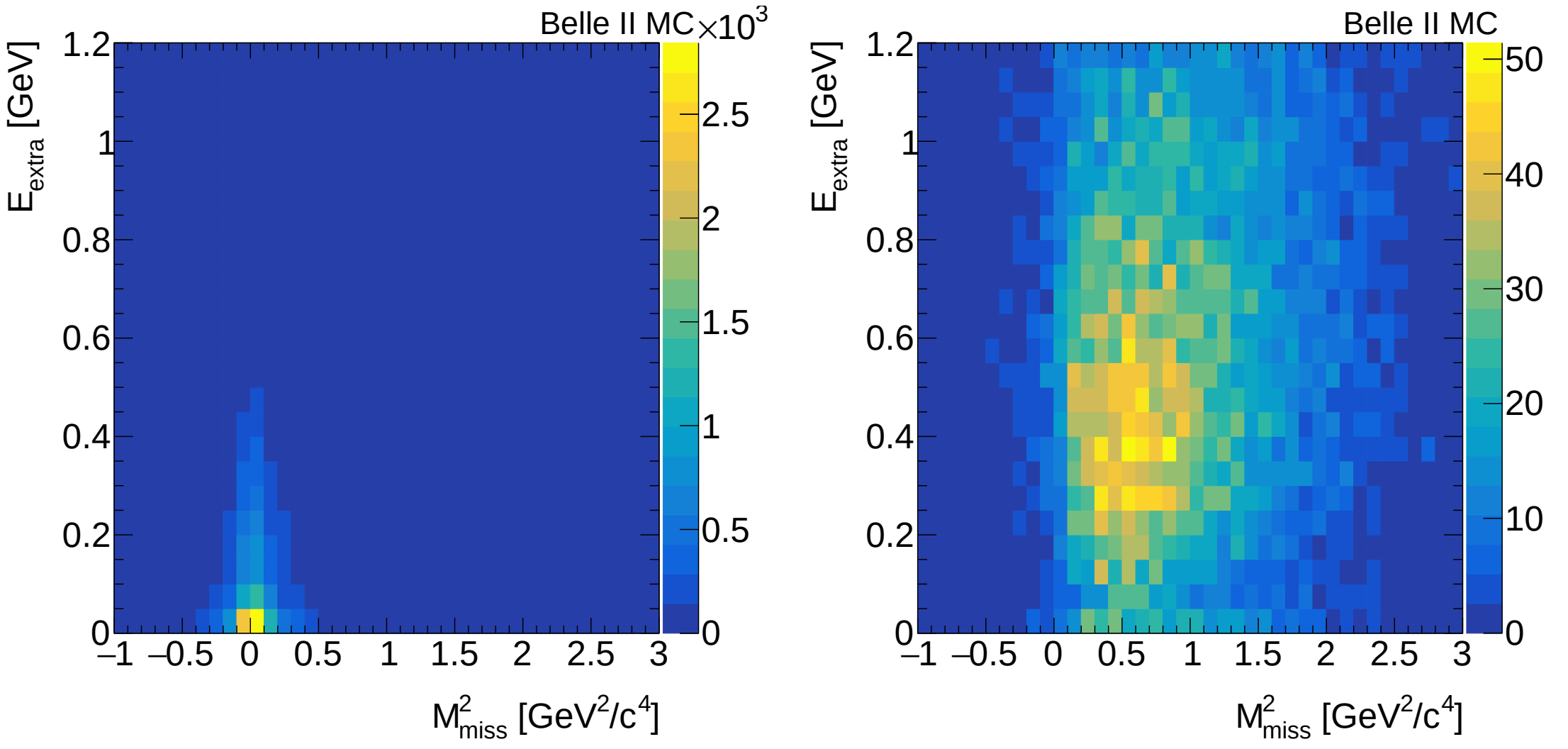

Fig. 83: The 2D $M^2_{\text{miss}}$ and $E_{\text{extra}}$ distribution of tagged $B^0 \to \pi^- \ell^+ \nu_\ell$ candidates as obtained from events containing signal (left) and background $B^0 \to \rho^- \ell^+ \nu_\ell$ decays (right).

Table 53: Summary of systematic uncertainties on the branching fractions of $B^0 \to \pi^- \ell^+ \nu_\ell$ decays in hadronic tagged and untagged Belle analyses with 711 fb$^{-1}$ [84] and 605 fb$^{-1}$ [299] data samples, respectively. The estimated precision limit for some sources of systematic uncertainties is given in parentheses.

| Source | Error (Limit) [%] | |
|---|---|---|
| | Tagged [%] | Untagged |
| Tracking efficiency | 0.4 | 2.0 |
| Pion identification | – | 1.3 |
| Lepton identification | 1.0 | 2.4 |
| Kaon veto | 0.9 | – |
| Continuum description | 1.0 | 1.8 |
| Tag calibration and $N_{B\overline{B}}$ | 4.5 (2.0) | 2.0 (1.0) |
| $X_u \ell \nu$ cross-feed | 0.9 | 0.5 (0.5) |
| $X_c \ell \nu$ background | – | 0.2 (0.2) |
| Form factor shapes | 1.1 | 1.0 (1.0) |
| Form factor background | – | 0.4 (0.4) |
| Total | 5.0 | 4.5 |
| (reducible, irreducible) | (4.6, 2.0) | (4.2, 1.6) |

*8.6.4. $B_s \to K \ell \nu$.* The decay $B_s^0 \to K^- \ell^+ \nu_\ell$ proceeds at the tree-level in the SM via the flavour-changing charged-current $b \to u$ transition. The only difference between this decay and the $B \to \pi \ell \nu$ decays is in the spectator quark: a strange quark in $B_s^0 \to K^- \ell^+ \nu_\ell$ and a down (up) quark in $B_{d(u)} \to \pi^{-(0)} \ell \nu_\ell$ decays. Recently, several groups have performed lattice-QCD calculations of the form factors in $B_s^0 \to K^- \ell^+ \nu_\ell$ decays [141, 155]. Thus, precise measurements of the rate and $q^2$ dependence will provide an independent way to determine $|V_{ub}|$.

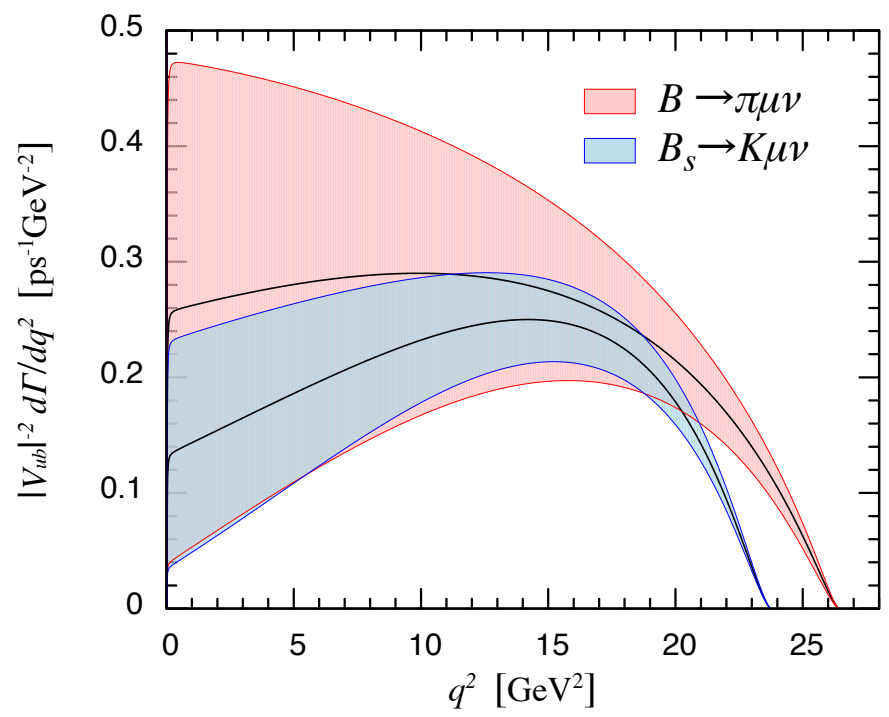

Fig. 84: SM predictions for the differential decay rate divided by $|V_{ub}|^2$ for $B_s^0 \to K^-\mu^+\nu_\mu$ and $B^0 \to \pi^-\mu^+\nu_\mu$ decays from Ref. [141].

As can be seen from the SM predicted differential decay rate for $B_s^0 \to K^-\ell^+\nu_\ell$ and $B \to \pi\ell\nu$ decays by the RBC and UKQCD collaborations [141] in Fig. 84, the predictions for $B_s^0 \to K^-\ell^+\nu_\ell$ are more precise than those made for $B \to \pi\ell\nu$ decays. The decay $B_s^0 \to K^-\ell^+\nu_\ell$ has not yet been measured, but will be possible to measure using data from an $\Upsilon(5S)$ run at Belle II.

The number of produced $B_s^{(*)}\overline{B}_s^{(*)}$ pairs in $e^+e^-$ collisions at CMS energies near the $\Upsilon(5S)$ resonance is more than an order of magnitude lower than the number of $B\overline{B}$ pairs produced near the $\Upsilon(4S)$ centre-of-mass energy ($E_{\rm CM}$) per ab$^{-1}$. The reason is due to the lower cross-section for $b\bar{b}$ production at $E_{\rm CM}(\Upsilon(5S))$ (approximately 0.3 nb) and the low probability for $b\bar{b}$ to hadronise to $B_s^{(*)}\overline{B}_s^{(*)}$ pairs (about 20%). A data sample collected at $E_{\rm CM}(\Upsilon(5S))$ corresponding to 1 ab$^{-1}$ would contain only around 60 million $B_s^{(*)}\overline{B}_s^{(*)}$ pairs, which makes the measurement of $B_s^0 \to K^-\ell^+\nu_\ell$ much more challenging, due to a degraded signal to noise ratio and the high rate of $B_{u/d} \to X_c\ell\nu$ and $B_{u/d} \to X_u\ell\nu$ background. The untagged measurement approach is best suited for the study of $B_s^0 \to K^-\ell^+\nu_\ell$ decays.

The untagged measurement strategy described here follows the strategy described earlier for $B^0 \to \pi^-\ell^+\nu_\ell$ decays. The major difference with respect to the $B^0 \to \pi^-\ell^+\nu_\ell$ study is in the simulated sample. Here, we used Belle's simulated sample of $e^+e^- \to \Upsilon(5S) \to B_s^{(*)}\overline{B}_s^{(*)}$, $B^{(*)}\overline{B}^{(*)}$, $B^{(*)}\overline{B}^{(*)}\pi$, $B\overline{B}\pi\pi$, and $e^+e^- \to q\overline{q}$ as such samples were not yet available for Belle II at the time of writing. The Belle experiment's simulated samples, corresponding to a data sample of around 720 fb$^{-1}$, were converted to Belle II's data format and analysed with Belle II analysis software.

The reconstruction efficiency for signal $B_s^0 \to K^-\ell^+\nu_\ell$ decays is found to be 9.2%, while the background suppression rate for other processes is similar to the one reported by previous $B^0 \to \pi^-\ell^+\nu_\ell$ untagged studies. The efficiencies for background $B_s^{(*)}\overline{B}_s^{(*)}$, $B\overline{B}X$, and $q\overline{q}$ events are found to be $1.9 \times 10^{-4}$, $3.2 \times 10^{-4}$, and $2.5 \times 10^{-6}$, respectively. The $M_{\rm bc}$ and $\Delta E$ distributions for accepted events are shown in Fig. 85. Fits to the $M_{\rm bc}$ and $\Delta E$ distributions in 6 bins of $q^2$ yields in total $2196 \pm 165$ signal events (setting $\mathcal{B}(B_s^0 \to K^-\ell^+\nu_\ell) = 1.5 \times 10^{-4}$ in the simulation) indicating that measurement of the decay rate can reach 5–10% precision at Belle II with a 1 ab$^{-1}$ data sample collected at $E_{\rm CM}(\Upsilon(5S))$.

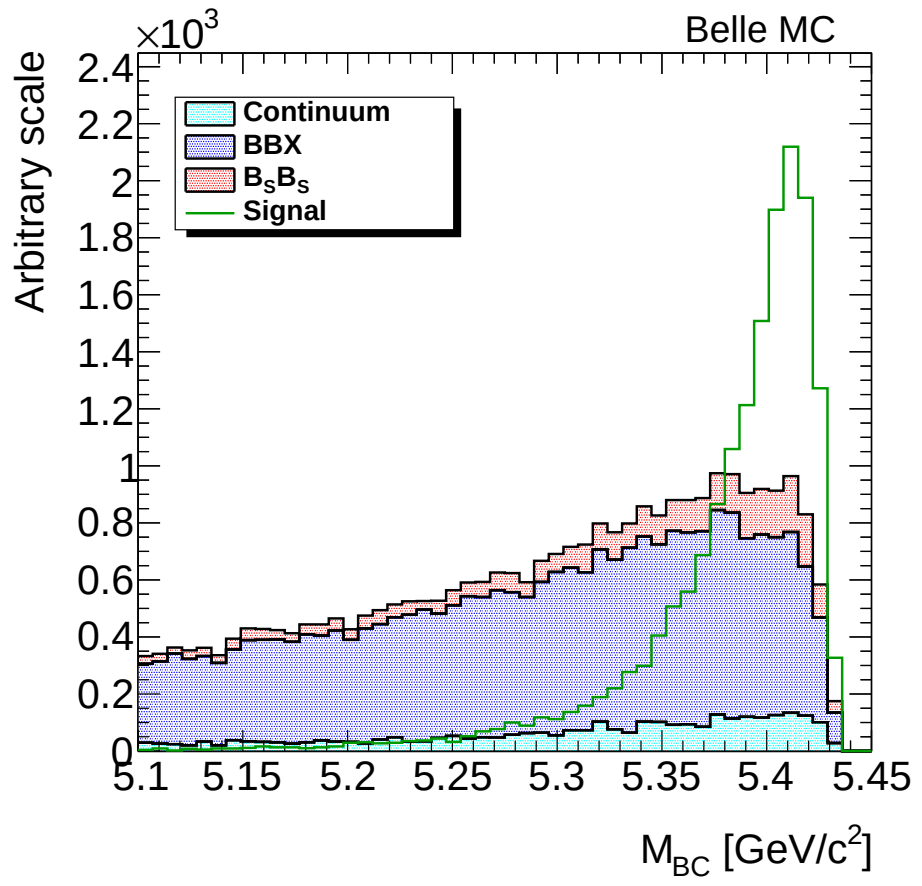

Fig. 85: $M_{\rm bc}$ and $\Delta E$ distributions of the $B_s \to K\ell\nu$ analysis over the full $q^2$ range, with signal and background components depicted separately and with arbitrary normalisation. The signal component is not to scale with the background.

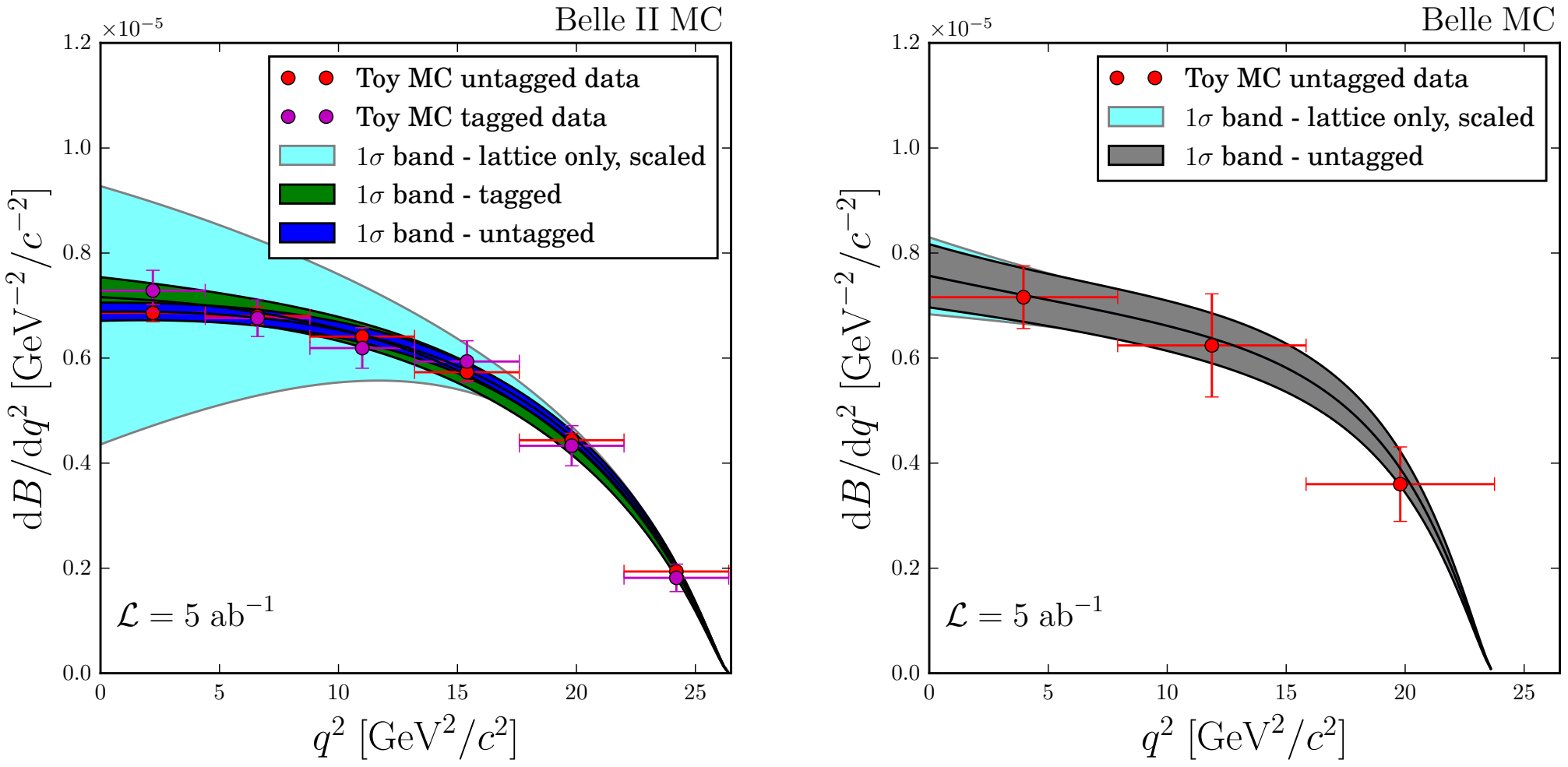

Fig. 86: Model independent BCL fits ($N_{\rm par} = 3 + 1$) for $B \to \pi\ell\nu$ tagged and untagged (left) and $B_s \to K\ell\nu$ untagged (right) with 5 ab$^{-1}$ data samples, and lattice-QCD error forecasts in 5 years (w/ EM).

*8.6.5. $|V_{ub}|$ extraction.* The value of $|V_{ub}|$ and its expected precision are extracted via a simultaneous fit to simulated data and lattice-QCD predictions. Both inputs were used to construct a $\chi^2 = \chi^2_{\rm data} + \chi^2_{\rm QCD}$ function that was minimised. The fits of all three modes for $\mathcal{L} = 5$ ab$^{-1}$ of simulated data are shown in Fig. 86. The values of $\sigma_{V_{ub}}$ for all three modes and projections to various values of integrated luminosity are shown in Tables 54 and 55. Lattice-QCD uncertainties also have a large impact on the precision of $|V_{ub}|$, so efforts to reduce the lattice-QCD uncertainties are expected in the future (see Sec. 7.5). Projections of $\sigma_{V_{ub}}$ for various cases of lattice-QCD forecasts can be seen in Fig. 87.

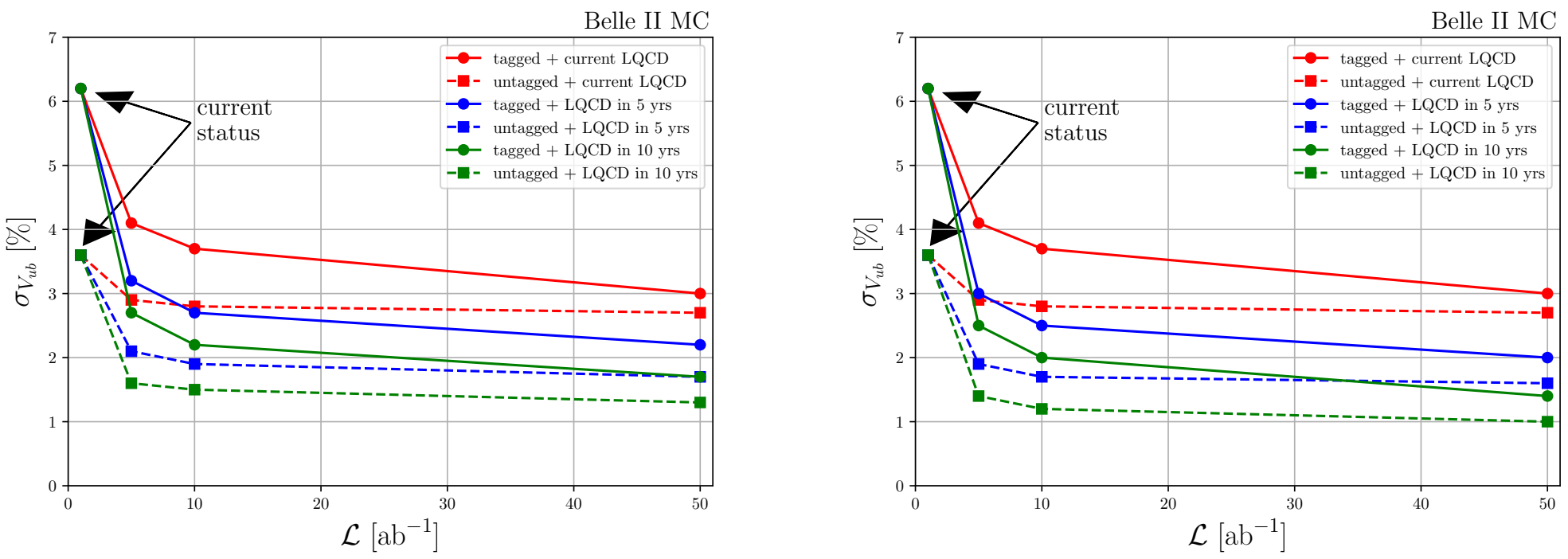

Fig. 87: Projections of the $|V_{ub}|$ uncertainty for various luminosity values and lattice-QCD error forecasts for $B \to \pi \ell \nu$ tagged and untagged modes. The figure on the left is obtained by using lattice forecasts with EM corrections and the figure on the right by forecasts without these corrections.

Table 54: Projections of $|V_{ub}|$ uncertainties at various Belle II luminosities for $B \to \pi \ell \nu$ tagged (T) and untagged (UT) modes. All uncertainties are in %. Lattice-QCD error forecasts were taken into account according to Sec. 7.5. The error in the second right-most column corresponds to forecasts with uncertainties due to EM corrections (Sec. 7), and the final column corresponds to forecasts without uncertainties due to EM corrections.

| $\mathcal{L}$ [ab$^{-1}$] | Tag | $\sigma_{\mathcal{B}}$ (stat, sys) | $\sigma_{\text{QCD}}^{\text{forecast}}$ | $\sigma_{V_{ub}}$(EM) | $\sigma_{V_{ub}}$(no EM) |
|---|---|---|---|---|---|
| 1 | T | 3.6, 4.4 | current | 6.2 | - |
| | UT | 1.3, 3.6 | | 3.6 | 3.6 |
| 5 | T | 1.6, 2.7 | in 5 yrs | 3.2 | 3.0 |
| | UT | 0.6, 2.2 | | 2.1 | 1.9 |
| 10 | T | 1.2, 2.4 | in 5 yrs | 2.7 | 2.6 |
| | UT | 0.4, 1.9 | | 1.9 | 1.7 |
| 50 | T | 0.5, 2.1 | in 10 yrs | 1.7 | 1.4 |
| | UT | 0.2, 1.7 | | 1.3 | 1.0 |

Table 55: Projections of $|V_{ub}|$ uncertainties at various Belle II luminosities for the $B_s \to K \ell \nu$ untagged mode. All uncertainties are in %. Lattice-QCD error forecasts were taken into account according to to Sec. 7.5.

| $\mathcal{L}$ [ab$^{-1}$] | $\sigma_{\mathcal{B}}$ (stat, sys) | $\sigma_{\text{QCD}}^{\text{forecast}}$ | $\sigma_{V_{ub}}$(EM) | $\sigma_{V_{ub}}$(no EM) |
|---|---|---|---|---|
| 1 | 6.5, 3.6 | current | 6.5 | |
| 5 | 2.9, 2.2 | in 5 yrs | 4.7 | 4.5 |

### *8.7. Inclusive semileptonic*

Authors: G. Ricciardi (th.), F. J. Tackmann (th.), P. Urquijo (exp.)

*8.7.1. Overview.* In inclusive semileptonic $B \to X \ell \nu$ decays one considers the sum over all possible kinematically allowed hadronic final states $X$. In the theoretical description the optical theorem then allows one to replace the sum over hadronic final states with a sum over partonic final states, which eliminates any long-distance sensitivity to the final state. The short-distance QCD corrections, which appear at the typical scale $\mu \sim m_b$ of the decay, can be computed in perturbation theory.

The remaining long-distance corrections are related to the initial $B$ meson. They can be expanded in the heavy-quark expansion (HQE) in powers of $\Lambda_{\rm QCD}/m_b \sim 0.1$, where $\Lambda_{\rm QCD}$ is a typical hadronic scale of order $M_B - m_b \sim 0.5\,\text{GeV}$. This expansion systematically expresses the decay rate in terms of non-perturbative parameters that describe the universal properties of the $B$ meson.

The non-perturbative parameters affect the differential decay rates from which $|V_{cb}|$ and $|V_{ub}|$ are extracted. Their dominant effect is on the shapes of the distributions while $|V_{cb}|$ and $|V_{ub}|$ only enter through the overall normalisation. Hence, the strategy for a precise determination of $|V_{cb}|$ and $|V_{ub}|$ is to fit them together with the relevant non-perturbative parameters, as well as the $b$-quark mass, from the experimental measurements.

The present inclusive $|V_{cb}|$ and $|V_{ub}|$ determinations are theoretically limited by the imprecise knowledge of the required non-perturbative parameters. Hence, a key goal for Belle II will be to reduce this systematic limitation, in conjunction with theoretical improvements, by exploiting the large data set to obtain precise and detailed measurements of differential distributions, ultimately mapping out the complete triple-differential decay rate: in $p_\ell$, $m_X^2$, and $q^2$. In the case of $|V_{cb}|$, this effort will be focused on extending the scope of existing moments measurements. For $|V_{ub}|$, spectral information will be compared to theory for the first time in global analyses.

*8.7.2. Inclusive $|V_{cb}|$ from $B \to X_c\ell\nu$.* The perturbative calculations of the $B \to X_c\ell\nu$ differential decay rates are well established. The current global fits for $|V_{cb}|$ are performed to the measured moments of the lepton energy, $E_\ell$, and hadronic mass, $m_X^2$ (with various minimum energy thresholds on the lepton) [230]. The most recent HFLAV global fit (in the kinetic scheme) extracts $|V_{cb}|$ together with the local OPE parameters appearing at $1/m_b^2$ and $1/m_b^3$ as well as the quark masses, yielding $|V_{cb}| = (42.19 \pm 0.78) \times 10^{-3}$.

The total uncertainty of about 2% is limited by the theoretical uncertainties, dominated by perturbative and power correction uncertainties. The fit is also very sensitive to the precise treatment of the theory uncertainty correlations in the predictions for the different moments [352]. The HFLAV fit uses theory predictions up to NNLO, while the $1/m_b^2$ and higher corrections are included at tree level. The complete power corrections up to $O(\alpha_s \times 1/m_b^2)$ are known, and including them in the global fit [353] leads to $|V_{cb}| = (42.19 \pm 0.78) \times 10^{-3}$. A similar result in Ref. [354] using external information on $m_b$ finds $|V_{cb}| = (42.00 \pm 0.65) \times 10^{-3}$. The effect of the $1/m_b^4$ and $1/m_b^5$ corrections in the global fit have also been estimated [354] by constraining the large number of new parameters with the so-called Lowest-Lying State Approximation (LLSA) [354–356]. They are found to have a minor effect, giving $|V_{cb}| = (42.11 \pm 0.74) \times 10^{-3}$. Further theoretical improvements are feasible through the calculation of the $O(\alpha_s \times 1/m_b^3)$ corrections and eventually the $O(\alpha_s^3)$ corrections.

Although the current global $|V_{cb}|$ fit is theoretically limited, more precise measurements of inclusive $B \to X_c\ell\nu$ at Belle II will be very valuable to scrutinise the inclusive $|V_{cb}|$ determinations, and help to resolve the tension between the inclusive and exclusive determinations. In particular, precise measurements of hadronic mass moments directly in bins of $E_\ell$ instead of a lower cut on $E_\ell$ would be useful to avoid unnecessary large statistical correlations in the measurements. Precise measurements of the $E_\ell$ spectrum, including the kinematic end-point, should be performed, which will provide valuable insight into the eventual breakdown of the local OPE description. It may also be possible to obtain nontrivial constraints on the shape functions that are of primary relevance for inclusive $|V_{ub}|$ determinations [312]. In addition, measurements of other single-differential spectra, such as the hadronic energy, $E_X$, neutrino-energy, $E_\nu$, and $q^2$ spectra will be useful to provide complementary kinematic information.

*$B_s \to X_c \ell\nu$ from $\Upsilon(5S)$ data.* Both inclusive and semi-inclusive measurements of $B_s \to X_c \ell\nu$ have been performed by experiments to date. This class of measurements is not typically used for the extraction of $|V_{cb}|$ but rather the determination of $B_s$ production rates at $B$-factories and hadron colliders. A precise measurement of the $B_s$ production fraction at the $\Upsilon(5S)$ will allow measurements of the absolute branching fractions of channels that are used as normalisation modes at LHCb. It is also an important background to any future measurements of charmless semileptonic $B_s$ decays at Belle II. Measurements of inclusive and semi-inclusive rates at Belle II would be based on data taken near the $\Upsilon(5S)$ resonance. Due to the relatively small values of the $B_s$ production fraction, $f_s \approx 0.2$, and $\sigma(\Upsilon(5S)) \approx 0.3$ nb, these analyses suffer from a larger background than those of $B$ mesons at the $\Upsilon(4S)$ resonance.

The inclusive semileptonic branching fraction of $B_s \to X\ell\nu$ decays was measured by BaBar and Belle [357, 358] and found to be in agreement with the expectations from SU(3) flavour symmetry [359, 360], which is also interesting to test more precisely. Such tests are crucial for understanding branching fraction predictions for $B_s$ decays. Semi-inclusive analyses of $B_s \to D_s^- X\ell^+\nu$ and $B_s \to D_s^{*-} X\ell^+\nu$ decays and measurements of their branching fractions have been performed by the D0 [361], LHCb [362], and Belle experiments [363]. Belle also reported the first measurement of the semi-inclusive branching fractions $\mathcal{B}(B_s \to D_s X\ell\nu)$ and $\mathcal{B}(B_s \to D_s^* X\ell\nu)$ using its entire 121 fb$^{-1}$ $\Upsilon(5S)$ dataset.

These measurements were limited by production rate uncertainties. In Belle II, $B_s$ tagging methods on data samples in excess of 1 ab$^{-1}$ will circumvent $B_s$ normalisation limitations and mitigate background from $B$ mesons. With a sample of about 5 ab$^{-1}$ we should expect to reach a statistical precision of 2% and systematic precision of about 4% using a hadronic tag. Other methods may be of greater use with smaller data sets, such as a $D_s$ and/or lepton tag, but ultimately the hadronic tag is most effective in accurate absolute branching ratio measurements.

*8.7.3. Inclusive $|V_{ub}|$.* The current $|V_{ub}|$ measurements from BaBar, Belle and CLEO generally exhibit tension with exclusive and CKM global fit determinations. The inclusive values vary depending on the kinematic fiducial region, which may be due to differences in theory treatment in these regions, or to experimental signal and background modelling imperfections. Above all, the goal for the measurement of $|V_{ub}|$ from inclusive $B \to X_u \ell\nu$ decays is to understand the persistent tension between exclusive and inclusive determinations. The large data set at Belle II must be exploited to constrain the dominant sources of uncertainties, namely non-perturbative parameters in decay modelling, and final state hadronisation effects.

*Theoretical Overview.* The theoretical description of inclusive $B \to X_u \ell\nu$ decays is based on the same underlying principles that are used for inclusive $B \to X_c \ell\nu$ decays. The total $B \to X_u \ell\nu$ rate can in principle be calculated in an OPE in terms of local operators, which has a similar structure as for the total $B \to X_c \ell\nu$ rate, with non-perturbative corrections first appearing at $O(1/m_b^2)$.

However, the primary challenge for the inclusive $|V_{ub}|$ determination is the overwhelming background from $B \to X_c \ell\nu$. As a result, the dominant experimental sensitivity to $B \to X_u \ell\nu$ and $|V_{ub}|$ is in the region of phase space where the $B \to X_c \ell\nu$ background is kinematically forbidden, namely the region where the hadronic $X_u$ system has invariant mass $m_X \leq M_D$. Due to the much larger $B \to X_c \ell\nu$ rate, the residual background from mis-reconstructed $B \to X_c \ell\nu$ decays is still important in this region.

In this phase-space region, non-perturbative corrections are kinematically enhanced, and as a result the non-perturbative dynamics of the decaying $b$ quark inside the $B$ meson becomes an $O(1)$ effect.

In addition to the lepton energy, $E_\ell$, the decay kinematics can be described with the hadronic variables

$$p_X^+ = E_X - |\vec{p}_X| \,, \qquad p_X^- = E_X + |\vec{p}_X| \,, \tag{196}$$

where $E_X$ and $\vec{p}_X$ are the energy and momentum of the hadronic system in the $B$ meson rest frame. In terms of these, the total hadronic and leptonic invariant masses are given by

$$m_X^2 = p_X^+ p_X^- \,, \quad q^2 = (M_B - p_X^+)(M_B - p_X^-) \,. \tag{197}$$

The differential decay rate is given by

$$\frac{d^3\Gamma}{dp_X^+ \, dp_X^- \, dE_\ell} = \frac{G_F^2 |V_{ub}|^2}{192\pi^3} \int dk \, C(E_\ell, p_X^-, p_X^+, k) \, F(k) + O\Big(\frac{\Lambda_{\rm QCD}}{m_b}\Big) \,. \tag{198}$$

The photon energy spectrum in the inclusive rare decay $B \to X_s \gamma$ plays an important role in determinations of $|V_{ub}|$, as it is given in terms of the same leading shape function appearing in Eq. (198),

$$\frac{d\Gamma}{dE_\gamma} = |C_7^{\rm incl}|^2 |V_{tb} V_{ts}^*|^2 m_b^2 \int dk \, C(E_\gamma, k) \, F(k) + O\Big(\frac{\Lambda_{\rm QCD}}{m_b}\Big) \,. \tag{199}$$

The "shape-function" $F(k)$ is a non-perturbative function that describes the momentum distribution of the $b$ quark in the $B$ meson [364, 365]. For $p_X^+ \sim k \sim \Lambda_{\rm QCD}$, which includes the endpoint region of the lepton energy spectrum as well as a large portion of the small-$m_X$ region, the full shape of the non-perturbative component of $F(k)$ is necessary to obtain an accurate description of the differential decay rate. An essential property is that $F(k)$ is normalised to unity, such that it only affects the shape of the decay rate but not the normalisation.

In addition to $F(k)$, several additional sub-leading shape functions appear at $O(\Lambda_{\rm QCD}/m_b)$ [366], and an even larger number of unknown shape functions appear at $O(\alpha_s \Lambda_{\rm QCD}/m_b)$ [367]. The differential decay rate contains three underlying hadronic structure functions, so there are effectively three independent combinations entering the description of $B \to X_u \ell\nu$.

For $p_X^+ \gg k \sim \Lambda_{\rm QCD}$, only the first few moments of $F(k)$ are needed, which recovers the expansion in terms of local OPE parameters. In practice, the experimental measurements can lie anywhere between these two kinematic regimes, which makes it important to have a consistent description across phase space.

The coefficient $C(E_\ell, p_X^-, p_X^+, k)$ in Eq. 198 describes the partonic quark decay $b \to u\ell\nu$ and can be computed in QCD perturbation theory. It is known up to NNLO. In the $b \to u$ sensitive region, $p_X^+ \ll p_X^-$, it also contains Sudakov double logarithms, $\ln^2(p_X^+/p_X^-)$, which can be re-summed up to NNLL.

The unknown form of the shape function is a dominant systematic limitation in the inclusive $|V_{ub}|$ determination. An important parametric uncertainty is due to $m_b$. While the total decay rate scales as $m_b^5$, in the shape-function region the dependence can be much stronger. A substantial part of the $m_b$ dependence is entangled with $F(k)$ and enters indirectly via its first moment, which makes a consistent treatment of $F(k)$ important.

For highly inclusive analyses the local OPE applies and the shape function becomes irrelevant, but the experimental analyses depend crucially on the signal modelling. References [368] and [369–371] have attempted to estimate the uncertainty stemming from the functional form of the shape function(s). They consistently found relatively small errors, and more advanced analyses [372] end up with only slightly larger uncertainties, at the level of a few percent. While we certainly need to understand the shape function(s), we also need to validate the existing theory frameworks on experimental data and the best way to do this is to measure kinematic distributions.

*Measurements.* Existing inclusive $|V_{ub}|$ determinations are typically based on measurements of partial branching fractions in various fiducial kinematic regions. These regions have been chosen to balance between experimental statistical uncertainties, and theoretical uncertainties, and to probe for inconsistencies in predictions of non-perturbative effects. Several theoretical approaches have been used to translate the measurements into $|V_{ub}|$, which differ in their treatment of perturbative corrections and the parametrisation of non-perturbative effects, in particular in the shape-function region. These are BLNP (Bosch, Lange, Neubert, Paz) [369–371], GGOU (Gambino, Giordano, Ossola, Uraltsev) [368], DGE (dressed gluon exponentiation by Andersen and Gardi) [373], and ADFR (Aglietti, Di Lodovico, Ferrera, Ricciardi) [374–376]. The former two use non-perturbative model functions to parameterise the shape functions, where the model parameters are adjusted to obtain the correct first non-perturbative moments such that the local OPE result is reproduced outside the shape-function region. The latter two use perturbative models for the shape function. A detailed review can be found in Ref. [2].

Selected results are summarised in Table 56 with the HFLAV average [230]. Currently the most precise $|V_{ub}|$ determinations by both BaBar and Belle appear to come from the most inclusive measurements, which use 467 and 657 million $B\bar{B}$ pairs, respectively. These analyses rely on hadronic tagging, which provides flavour and kinematic information for inclusive reconstruction of the signal side. The signal is reconstructed by identifying a charged lepton then summing all tracks and neutral clusters in the event to form a hadron candidate. Selection criteria includes charged and neutral kaon vetoes (events with $K^\pm$, $K_S$ but not $K_L$ on the signal side are rejected), vetoes for events that contain slow-pions likely to have originated from $D^{*+}$ decays, and requirements for small missing mass. The signal yields are determined from simultaneous fits of the $b \to u$ signal and the dominant $b \to c$ background in the two-dimensional hadron mass $m_X$-$q^2$ distribution. The only explicit phase-space restriction on the extracted $B \to X_u \ell \nu$ branching ratio is the lower threshold on the lepton momentum, $E_\ell > E_{\rm min}$ with $E_{\rm min}$ as low as 0.8 GeV. However these analyses do have many selection criteria that induce non-trivial dependence of the efficiency on decay dynamics. Therefore the fit and the detection efficiency both require knowledge of the $b \to u$ signal model, and since the sensitivity to $b \to u$ comes from the shape-function region, this leads to direct dependence on the theoretical decay model. Direct sensitivity to the underlying theory model used in MC was studied in a recent BaBar analysis of the lepton energy spectrum [381]. A breakdown of the systematic uncertainties in the most recent Belle analysis is shown in Table 57, broken into reducible and irreducible components.

Table 56: Status of inclusive $|V_{ub}|$ determinations from HFLAV [230].

| | $(|V_{ub}| \times 10^3)$ | | | |
|---|---|---|---|---|
| Measurement | BLNP [369–371] | GGOU [368] | DGE [373] | ADFR [374–376] |
| HFLAV 2016 | $4.44 \pm 0.15^{+0.21}_{-0.22}$ | $4.52 \pm 0.16^{+0.15}_{-0.16}$ | $4.52 \pm 0.15^{+0.11}_{-0.14}$ | $4.08 \pm 0.13^{+0.18}_{-0.12}$ |
| Belle $p^*_\ell > 1$ GeV/$c$ [377] | $4.50 \pm 0.27^{+0.20}_{-0.22}$ | $4.62 \pm 0.28^{+0.13}_{-0.13}$ | $4.62 \pm 0.28^{+0.09}_{-0.10}$ | $4.50 \pm 0.30^{+0.20}_{-0.20}$ |
| BaBar $p^*_\ell > 1$ GeV/$c$ [378] | $4.33 \pm 0.24^{+0.19}_{-0.21}$ | $4.45 \pm 0.24^{+0.12}_{-0.13}$ | $4.44 \pm 0.24^{+0.09}_{-0.10}$ | $4.33 \pm 0.24^{+0.19}_{-0.19}$ |
| CLEO $2.1 < E_e < 2.6$ GeV [379] | $4.22 \pm 0.49^{+0.29}_{-0.34}$ | $3.86 \pm 0.45^{+0.25}_{-0.27}$ | $4.23 \pm 0.49^{+0.22}_{-0.31}$ | $3.42 \pm 0.40^{+0.17}_{-0.17}$ |
| Belle $1.9 < E_e < 2.6$ GeV [380] | $4.93 \pm 0.46^{+0.26}_{-0.29}$ | $4.82 \pm 0.45^{+0.23}_{-0.23}$ | $4.95 \pm 0.46^{+0.16}_{-0.21}$ | $4.48 \pm 0.42^{+0.20}_{-0.20}$ |
| BaBar $2.0 < E_e < 2.6$ GeV [381] | $4.51 \pm 0.12^{+0.41}_{-0.34}$ | $3.92 \pm 0.10^{+0.23}_{-0.29}$ | $3.81 \pm 0.10^{+0.18}_{-0.16}$ | – |

Normalisation for $|V_{ub}|$ may reach a precision limitation due to calibration of the tagging method, although it can be measured as a ratio with $B \to X_c \ell \nu$ which will cancel some uncertainties. Systematics related to reconstruction efficiencies, fake leptons, and continuum are data driven and expected to improve with a larger data set. Belle II's hadron tag is expected to perform better than that used in the previously published Belle inclusive analysis with about 3-4 times better efficiency.

A large fraction of the residual background is due to $B \to X_c \ell \nu$ events where the charm meson decays to a $K^0_L$. It is difficult to reconstruct $K^0_L$ mesons and model their hadronic interactions with the KLM and ECL. If precise measurements and reliable calibration of $K^0_L$ identification can be performed in Belle II via use of high-rate control modes, it would greatly aid in purifying this analysis in the high $M_X$ region. Very few analyses to date have attempted to veto on the presence of $K^0_L$ in the signal due to the large differences between data and MC simulation in hadronic interactions. Belle II should also look into vetoes of slow $\pi^0$ mesons from $D^{*0}$ decays, and exploit improved low-momentum tracking for more efficient slow $\pi^+$ reconstruction.

*Decay modelling and fragmentation.* Systematic uncertainties and biases introduced through model dependence are a very important consideration for Belle II measurements of this channel. Measurements must improve modelling, and improve robustness to fluctuations in modelling choices. $B \to X_u \ell \nu$ modelling is performed via an admixture of exclusive and inclusive contributions depicted in Fig. 88. Typically the exclusive component is comprised of well measured contributions, such as $X_u = \pi,\ \rho,\ \eta,\ \omega$ but this is only around 20% of the total rate. The remainder is left to be modelled by an inclusive generator. Further measurements of the specific hadronic contributions to the semileptonic decay width are crucial.

Table 57: Systematic errors (in percent) on the branching fractions for $B \to X_u \ell\nu$ in the hadron tagged sample, with 605 fb$^{-1}$ of Belle data. The precision limit for some systematics is given in brackets.

| Source | Error on $\mathcal{B}$ (irreducible limit) |
|---|---|
| $\mathcal{B}(D^{(*)}\ell\nu)$ | 1.2 (0.6) |
| Form factors $(D^{(*)}\ell\nu)$ | 1.2 (0.6) |
| Form factors & $\mathcal{B}(D^{(**)}\ell\nu)$ | 0.2 |
| $B \to X_u\ell\nu$(SF) | 3.6 (1.8) |
| $B \to X_u\ell\nu(g \to s\bar{s})$ | 1.5 |
| $\mathcal{B}(B \to \pi/\rho/\omega\ell\nu)$ | 2.3 |
| $\mathcal{B}(B \to \eta^{(\prime)}\ell\nu)$ | 3.2 |
| $\mathcal{B}(B \to X_u\ell\nu)$ unmeasured/fragmentation | 2.9 (1.5) |
| Continuum & Combinatorial | 1.8 |
| Secondaries, Fakes & Fit | 1.0 |
| PID& Reconstruction | 3.1 |
| BDT/Normalisation | 3.1 (2.0) |
| Total | 8.1 |
| (Total reducible) | 7.4 |
| (Total irreducible) | 3.2 |

Another effect not yet effectively addressed in previous $B \to X_u\ell\nu$ analyses is the fragmentation of the $X_u$ system. Studies in recent Belle exclusive $B \to X_u\ell\nu$, and a semi-exclusive $B \to X_s\gamma$ analyses demonstrate that the nominal light quark fragmentation is different to that found in data. Both found that the probability for low multiplicity final states to be produced is overestimated by PYTHIA (JETSET), as shown in Table 58. This can impact substantially on reconstruction efficiencies and PDF shapes for branching fraction fits. To further constrain this effect, inclusive analyses will need to allow for a degree of freedom in hadron multiplicity, similar to the semi-exclusive approach pioneered in $B \to X_s\gamma$. Strange-anti-strange production, *i.e.* $B \to K\bar{K}\ell\nu$, is not constrained by experiment and yet kaon vetoes are commonly used in inclusive analyses. Such channels must be measured to reduce bias, as listed in Table 57. The large data set at Belle II will allow for differential measurements in kinematic observables, such as $M_X^2$, $q^2$, and $p_\ell$, separately for both charged and neutral $B$ decays. This provides important information to constrain uncertainties on shape functions, weak annihilation and signal modelling. The inclusive analyses performed to date provide insufficient information to rule out any of the theoretical frameworks used in the extraction of $|V_{ub}|$ and hence new and better shape information is critical.

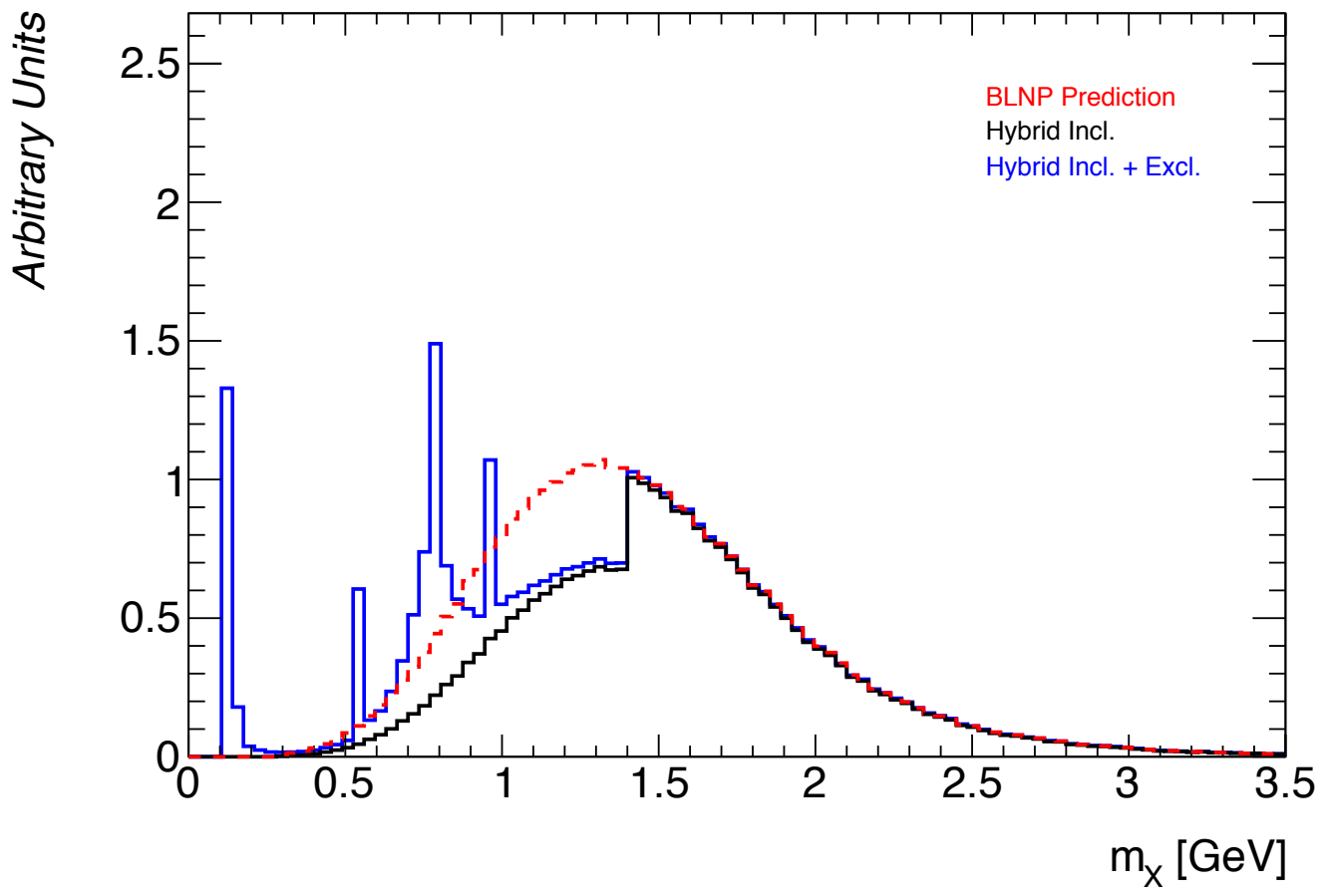

Fig. 88: Modeling of the $B \to X_u \ell \nu$ decay hadronic invariant mass, based on the BLNP [369–371] inclusive prediction (dashed red line) as well as the inclusive (solid black line) and inclusive plus exclusive cocktail (solid blue line) used commonly in MC simulations.

.

Table 58: The relative proportion of each $B \to X_s \gamma$ mode in the range 1.15 GeV/$c^2 < M_{X_s} <$ 2.8 GeV/$c^2$ in the data and default MC. The striking difference between default PYTHIA MC and data multiplicity must be addressed directly in inclusive $b \to u \ell \nu$ measurements.

| Mode | Data | Default MC |
|---|---|---|
| $K\pi$ without $\pi^0$ | $4.2 \pm 0.4$ | 10.3 |
| $K\pi$ with $\pi^0$ | $2.1 \pm 0.2$ | 5.4 |
| $K2\pi$ without $\pi^0$ | $14.5 \pm 0.5$ | 12.9 |
| $K2\pi$ with $\pi^0$ | $24.0 \pm 0.7$ | 15.2 |
| $K3\pi$ without $\pi^0$ | $8.3 \pm 0.8$ | 5.9 |
| $K3\pi$ with $\pi^0$ | $16.1 \pm 1.8$ | 15.7 |
| $K4\pi$ | $11.1 \pm 2.8$ | 12.3 |
| $K2\pi^0$ | $14.4 \pm 3.5$ | 14.4 |
| $K\eta$ | $3.2 \pm 0.8$ | 4.9 |
| $3K$ | $2.0 \pm 0.3$ | 3.0 |

*Model independent measurements.* To take advantage of future theory improvements, measurements at Belle II should be performed and reported as independent of theoretical assumptions as possible. This will require measurements of differential spectra that fully characterise the transitions as in exclusive decays, *e.g.* $q^2$, $\theta_\ell$, $m_X^2$, $p_\ell$. Such measurements have not been performed accurately by the $B$-factories to date.

One of the quoted HFLAV averages is $|V_{ub}| = 4.62 \pm 0.20 \pm 0.29$ [230], which is obtained using the alternative BLL (Bauer, Ligeti and Luke) [382] approach based on performing a local OPE calculation at large $q^2$, and is hence limited to measurements that use a cut in the $m_X$-$q^2$ plane. Weak annihilation contributions, which are concentrated at maximal $q^2$, seem to be strongly constrained by semileptonic charm decays [383–385]. Nevertheless, they remain a source of theoretical uncertainty that is hard to quantify here. Hence, precise separate measurements of charged and neutral $B$ meson decays to constrain these contributions are well motivated at Belle II, as are direct searches for weak annihilation effects at high $q^2$.

Key input to these extractions are the values of the HQE parameters: the $b$-quark mass, $m_b$, and the Fermi motion quantity, $\mu_\pi^2$. These quantities are typically obtained from fits to moments in $B \to X_c \ell \nu$ inclusive decays, with additional constraints from either QCD calculations for $m_c$ in the kinetic scheme, or from $B \to X_s \gamma$ inclusive decays. They can also be extracted from the heavy-quark-mass dependence of meson masses containing heavy and light quark pairs, computed in lattice QCD [167, 169, 386, 387]. Measurement of the HQE parameters is limited by experimental precision and can be improved with dedicated analyses at Belle II with a larger data set and smaller experimental systematic uncertainties.

*$|V_{ub}|$ global fit.* Due to the intrinsic trade-off between experimental and theoretical cleanliness, there is no simple prescription for an optimal region of phase-space in which to measure the partial branching fraction. Instead, the most precise and reliable inclusive $|V_{ub}|$ determination should exploit all available experimental and theoretical information. This is accomplished with a global fit to the full spectrum information to simultaneously extract the overall normalisations ($|V_{ub}|$ for $B \to X_u \ell \nu$ and $|C_7^{\rm incl}|$ for $B \to X_s \gamma$) together with the required parameters such as $m_b$ and the leading (and eventually sub-leading) non-perturbative shape functions $F(k)$. In this way one minimises the uncertainties and makes maximal use of all available data, and the fit automatically "chooses" the most sensitive region given the experimental and theoretical uncertainties.

Compared to the global $|V_{cb}|$ fit, a global $|V_{ub}|$ fit is more involved, since the non-perturbative quantities to be fitted are now continuous functions rather than a few numbers. For this reason it will be important to combine both $B \to X_u \ell \nu$ and $B \to X_s \gamma$ data as well as constraints on the shape function moments from the non-perturbative parameters extracted from $B \to X_c \ell \nu$.

Experimentally, this requires the precise measurement of as many independent differential spectra as possible to maximise the available shape information, which will be key to constraining sub-leading corrections. Interesting possibilities would be double-differential measurements in $E_\ell$ and $m_X$, but also in other variables such as $p_X^+$, $q^2$, $E_X$, or $E_\nu$. Ultimately one should aim to measure the full spectra of each of these quantities as precisely as possible. The separation of charged and neutral $B$ mesons will also be important to understand effects such as weak annihilation.

Theoretically, the central ingredient for a global $|V_{ub}|$ fit is a model-independent treatment of the shape function, as was first proposed in Ref. [314]. More recently, artificial neural networks have been used to provide a very flexible and essentially model-independent parameterisation of the shape function [372]. The important requirement is that it must be possible within the global fit to let the form of $F(k)$ as well as its uncertainties be characterised solely by the uncertainties in the included experimental measurements, such that any intrinsic limitations from model-dependent assumptions are avoided.

Using this approach, a global fit to all available $B \to X_s \gamma$ measurements extracting $|C_7^{\rm incl}|$ along with $F(k)$ has been performed in Ref. [316], demonstrating the feasibility of this approach.

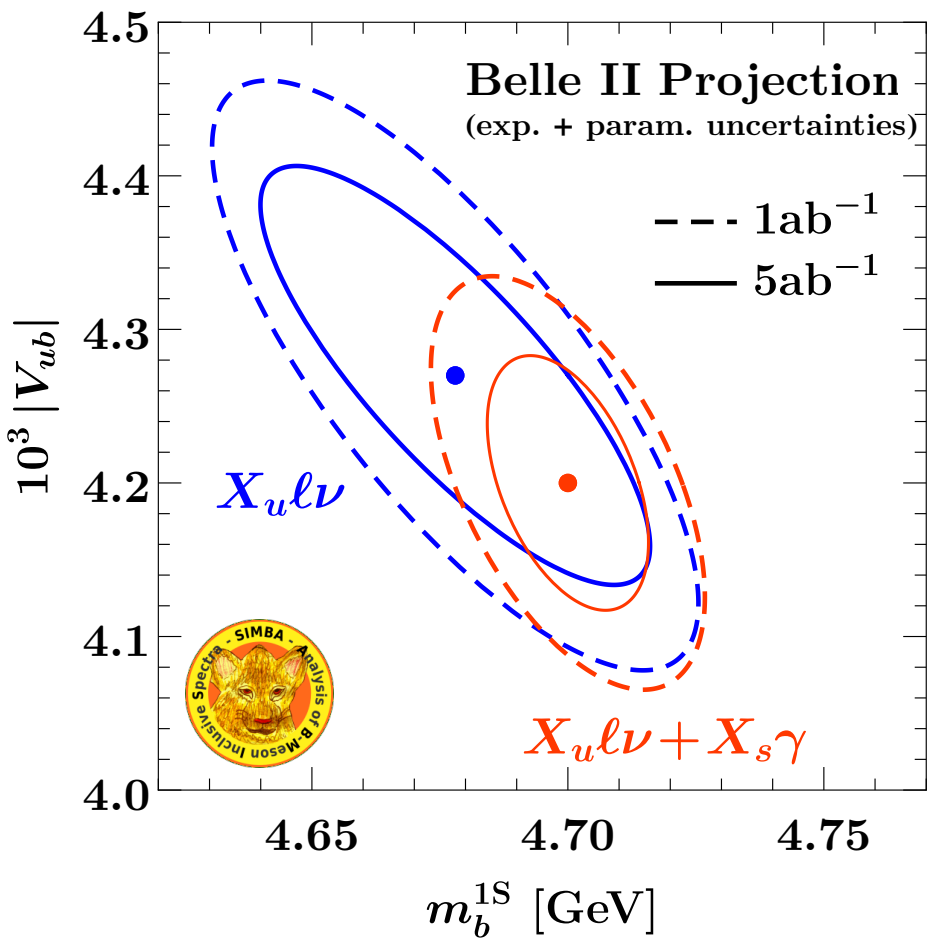

Fig. 89: Projections for a global $|V_{ub}|$ fit at Belle II with 1 ab$^{-1}$ and 5 ab$^{-1}$. No theory uncertainties are included in the fit, which can be expected to be of similar size to the experimental ones.

Projections for a global fit using two projected single-differential spectra in $m_X$ and $E_\ell$ for $B \to X_u \ell\nu$ and a $E_\gamma$ spectrum in $B \to X_s\gamma$ from Belle II at 1 ab$^{-1}$ and 5 ab$^{-1}$ are shown in Fig. 89. Projections with even higher integrated luminosity are hard to obtain, because they will require improvements on the experimental systematics.

The achievable precision will strongly depend on the precision and number of available spectra. The projected fit uncertainties at 1 ab$^{-1}$ (5 ab$^{-1}$) are about 4.5% (3%) for the fit to $B \to X_u \ell\nu$ only and 3% (2%) for the combined fit to $B \to X_s\gamma$ and $B \to X_u\ell\nu$. These fit uncertainties already include the dominant parametric uncertainties from $m_b$ and $F(k)$, and are constrained in the fit by the data. These projections do not include sub-leading shape function effects, which are expected to become relevant at this level of precision, but can also be constrained by measurements. Such effects were discussed and evaluated in Ref. [372]. In general, one can expect that Belle II data can and should be exploited to reduce the current theoretical uncertainties.

A summary of projections for inclusive $|V_{ub}|$ is given in Table 59, in addition to the exclusive and leptonic decay based determinations described earlier. Large ranges on the theoretical uncertainty of the inclusive method are due to variations in results from the contributing theory groups.

## *8.8. Conclusions*

Belle II will have a lot to say on leptonic and semileptonic $B$ meson decays. Precise measurements of the CKM matrix element magnitudes are crucial for pinning down the allowed level of $CP$ violation in the SM, but much work must be done to resolve inconsistencies in approaches for both $|V_{ub}|$ and $|V_{cb}|$. Prospects are particularly good for improvements to $|V_{ub}|$, on inclusive and exclusive approaches, owing to more data and better particle reconstruction performance at Belle II. Highly significant anomalies in semi-tauonic modes should be confirmed or refuted after only 5 ab$^{-1}$ of data. This will only be achievable if substantial effort is made to measure and carefully characterise the $B \to D^{**}\ell\nu$ background. Differential spectra will be measured with great precision, to probe possible new physics models. Measurements of leptonic $B$ decays are yet to be seen with $5\,\sigma$ significance in either the tau or muon modes by a single experiment. The former is achievable with approximately 2 ab$^{-1}$ at Belle II, and the latter is achievable with about 5 ab$^{-1}$ (assuming the SM branching ratio). Many new opportunities for new physics searches will be opened up with more data.

Table 59: Expected uncertainties in $|V_{ub}|$ measurements, given in percent, with the Belle full data sample, 5 ab$^{-1}$ and 50 ab$^{-1}$ Belle II data. Note that the statistical uncertainty quoted for exclusive $|V_{ub}|$ is that of the branching fraction, however a fit to the spectrum information is typically used to determine $|V_{ub}|$. While we expect to eventually perform a $|V_{ub}|$ measurement with $B \to \mu\nu$, we do not list an estimate for total precision: there is no clearly established signal with the $B$ factory data, and the searches suffered significant systematic uncertainties making it difficult to perform a reliable projection. We use the lattice-QCD projected precision for the future data sets.

| | Statistical | Systematic (reducible, irreducible) | Total Exp | Theory | Total |
|---|---|---|---|---|---|
| $\lvert V_{ub}\rvert$ exclusive (had. tagged) | | | | | |
| 711 fb$^{-1}$ | 3.0 | (2.3, 1.0) | 3.8 | 7.0 | 8.0 |
| 5 ab$^{-1}$ | 1.1 | (0.9, 1.0) | 1.8 | 1.7 | 3.2 |
| 50 ab$^{-1}$ | 0.4 | (0.3, 1.0) | 1.2 | 0.9 | 1.7 |
| $\lvert V_{ub}\rvert$ exclusive (untagged) | | | | | |
| 605 fb$^{-1}$ | 1.4 | (2.1, 0.8) | 2.7 | 7.0 | 7.5 |
| 5 ab$^{-1}$ | 1.0 | (0.8, 0.8) | 1.2 | 1.7 | 2.1 |
| 50 ab$^{-1}$ | 0.3 | (0.3, 0.8) | 0.9 | 0.9 | 1.3 |
| $\lvert V_{ub}\rvert$ inclusive | | | | | |
| 605 fb$^{-1}$ (old $B$ tag) | 4.5 | (3.7, 1.6) | 6.0 | 2.5−4.5 | 6.5−7.5 |
| 5 ab$^{-1}$ | 1.1 | (1.3, 1.6) | 2.3 | 2.5−4.5 | 3.4−5.1 |
| 50 ab$^{-1}$ | 0.4 | (0.4, 1.6) | 1.7 | 2.5−4.5 | 3.0−4.8 |
| $\lvert V_{ub}\rvert$ $B \to \tau\nu$ (had. tagged) | | | | | |
| 711 fb$^{-1}$ | 18.0 | (7.1, 2.2) | 19.5 | 2.5 | 19.6 |
| 5 ab$^{-1}$ | 6.5 | (2.7, 2.2) | 7.3 | 1.5 | 7.5 |
| 50 ab$^{-1}$ | 2.1 | (0.8, 2.2) | 3.1 | 1.0 | 3.2 |
| $\lvert V_{ub}\rvert$ $B \to \tau\nu$ (SL tagged) | | | | | |
| 711 fb$^{-1}$ | 11.3 | (10.4, 1.9) | 15.4 | 2.5 | 15.6 |
| 5 ab$^{-1}$ | 4.2 | (4.4, 1.9) | 6.1 | 1.5 | 6.3 |
| 50 ab$^{-1}$ | 1.3 | (2.3, 1.9) | 2.6 | 1.0 | 2.8 |

## 9. Radiative and Electroweak Penguin $B$ Decays

*Editors: T. Feldmann, U. Haisch, A. Ishikawa and J. Yamaoka*
*Additional section writers: W. Altmannshofer, G. Bell, C. Bobeth, S. Cunliffe, T. Huber, J. Kamenik, A. Kokulu, E. Kou, E. Manoni, M. Misiak, G. Paz, C. Smith, D. Straub, J. Virto, S. Wehle and R. Zwicky*

### *9.1. Introduction*

Flavour-changing neutral current (FCNC) $b \to s$ and $b \to d$ processes continue to be of great importance to precision flavour physics. The FCNC processes proceed to lowest order via one-loop diagrams (called penguin or box diagrams) in the Standard Model (SM). Since new-physics particles may enter the loop diagrams or even mediate FCNCs at tree level, the $b \to s$ and $b \to d$ transitions are sensitive to physics beyond the SM. Since final states involving photons or lepton pairs are both theoretically and experimentally clean, radiative and electroweak (EW) penguin $B$ decays are ideal place to search for new physics. The Belle II physics program in this area will focus on processes such as the inclusive $B \to X_{s,d}\gamma$ and $B \to X_{s,d}\ell^+\ell^-$ channels, as well as rare decays involving photons or neutrinos like $B_{d,s} \to \gamma\gamma$, $B \to K^{(*)}\nu\bar{\nu}$, $B_{d,s} \to \tau^+\tau^-$ and $B \to K^{(*)}\tau^+\tau^-$. Fully-inclusive measurements of the $b \to s, d\,\gamma$ and $b \to s, d\,\ell^+\ell^-$ transitions are very difficult at LHCb and so is the detection of $B$-meson decays into final states containing photon pairs, neutrinos or taus. As a result, Belle II is the only experiment that can provide detailed information on the latter FCNC processes in the near future.

A second important physics goal of Belle II in the area of radiative and EW penguin $B$ decays will be to provide independent tests of the anomalies recently uncovered by the LHCb and Belle experiments in the angular analysis of $B \to K^*\ell^+\ell^-$ [388–390] as well as in the determination of $R_K = \mathrm{Br}(B^+ \to K^+\mu^+\mu^-)/\mathrm{Br}(B^+ \to K^+e^+e^-)$ [391] and $R_{K^*} = \mathrm{Br}(B^0 \to K^{*0}\mu^+\mu^-)/\mathrm{Br}(B^0 \to K^{*0}e^+e^-)$ [392]. Some of these measurements have also been performed by ATLAS and CMS, although with less sensitivity [393–395]. In order to shed further light on the possible origin of the existing flavour anomalies, additional independent measurements are needed. Given that the reconstruction efficiency for electrons is comparable to that for muons thanks to the excellent electromagnetic calorimeter, the Belle II experiment is the natural place to perform such measurements.

In this section, we discuss the theoretical basics and the Belle II sensitivity to the aforementioned decay modes. The chapter is organised as follows. In section 9.1.1, the theoretical framework is provided, namely the effective Hamiltonian as well as a brief overview of the hadronic effects relevant to the radiative and the EW penguin decays. In section 9.2, the inclusive and exclusive radiative decays, $b \to s\gamma$ and $b \to d\gamma$, are discussed. It becomes apparent in this section that at Belle II, a separation of $B \to \rho\gamma$ from $B \to K^*\gamma$ becomes more accurate due to the improved particle identification. In section 9.3, double-radiative decays are examined. A first observation of $B \to \gamma\gamma$ decay may be possible during the early data taking of Belle II. In section 9.4, the inclusive and exclusive EW penguin decays, $b \to s\ell^+\ell^-$ decays, are reviewed. The Belle II experiment can play an important role to test the anomalies observed by LHCb in the angular observable of $B \to K^*\mu^+\mu^-$. Furthermore, Belle II will have access to the $B \to K^*e^+e^-$ channel with nearly the same sensitivity as $B \to K^*\mu^+\mu^-$, which will provide crucial additional information. The interplay of the inclusive and exclusive $B \to K^* \ell^+ \ell^-$ and $B \to X_s \ell^+ \ell^-$ decays is also stressed. In section 9.5, decay channels which involve missing energy such as $B \to K^{(*)} \nu \bar{\nu}$ and $B_{d,s} \to \nu \bar{\nu}$ are discussed. An early discovery of $B \to K^{(*)} \nu \bar{\nu}$ is possible at Belle II. Possible dark matter interpretations of the missing energy signatures are also briefly analysed.

*9.1.1. Theoretical basics.* *(Contributing authors: T. Feldmann and U. Haisch)*

*Effective Hamiltonian.* After decoupling the top quark, the Higgs boson and the EW gauge bosons, flavour-changing weak interactions relevant for the $b \to q\gamma$ transitions with $q = d, s$ can be described in the SM by the following effective Hamiltonian (see e.g. [396, 397])

$$\mathcal{H}_{\text{eff}}^{\text{SM}} = -\frac{4G_F}{\sqrt{2}}\, \lambda_t^{(q)} \left[ \sum_{i=1}^{8} C_i\, Q_i + \kappa_q \sum_{i=1}^{2} C_i\, (Q_i - Q_i^u) \right] . \tag{200}$$

Here $G_F$ is the Fermi constant and we have defined $\kappa_q = \lambda_u^{(q)}/\lambda_t^{(q)} = (V_{uq}^* V_{ub})/(V_{tq}^* V_{tb})$. The crucial difference between the transitions with $d$-quarks and $s$-quarks in the final state stems from the distinct Cabibbo-Kobayashi-Maskawa (CKM) hierarchy

$$\begin{aligned} \lambda_u^{(s)} : \lambda_c^{(s)} : \lambda_t^{(s)} &= \mathcal{O}(\lambda^4 : \lambda^2 : \lambda^2) \, , \\ \lambda_u^{(d)} : \lambda_c^{(d)} : \lambda_t^{(d)} &= \mathcal{O}(\lambda^3 : \lambda^3 : \lambda^3) \, , \end{aligned} \tag{201}$$

with the Wolfenstein parameter $\lambda \simeq 0.225$ governing the size of branching ratios and the respective hierarchies of different decay topologies.

Expressions for the current-current ($Q_{1,2}$), four-quark ($Q_{3-6}$), photonic dipole ($Q_7$) and gluonic dipole ($Q_8$) operators can be found for instance in [397]. Let us quote here the most important ones:

$$\begin{aligned} Q_1 &= (\bar{q}_L \gamma_\mu T^a c_L)(\bar{c}_L \gamma^\mu T^a b_L) \, , \\ Q_2 &= (\bar{q}_L \gamma_\mu c_L)(\bar{c}_L \gamma^\mu b_L) \, , \\ Q_7 &= \frac{e}{16\pi^2}\, m_b (\bar{q}_L \sigma^{\mu\nu} b_R) F_{\mu\nu} \, , \\ Q_8 &= \frac{g_s}{16\pi^2}\, m_b (\bar{q}_L \sigma^{\mu\nu} T^a b_R) G^a_{\mu\nu} \, , \end{aligned} \tag{202}$$

where $e$ and $g_s$ are the electromagnetic and strong coupling, $F_{\mu\nu}$ and $G^a_{\mu\nu}$ the $U(1)_{\text{em}}$ and $SU(3)_c$ field-strength tensors, $T^a$ are colour generators, and the indices $L, R$ denote the chirality of the quark fields. The operators $Q_{1,2}^u$ appearing in (200) are obtained from $Q_{1,2}$ by replacing $c$-quark by $u$-quark fields.

The Wilson coefficients $C_i$ in (200) contain the short-distance (SD) dynamics, *i.e.* physics from high energies, and can thus be calculated in perturbation theory. In the SM, they are first evaluated at the scale $\mu_w = \mathcal{O}(m_W)$ and then evolved down to $\mu_b = \mathcal{O}(m_b)$ using the renormalisation group equations (RGEs) in the effective theory. At present, all the low-energy Wilson coefficients $C_i(\mu_b)$ relevant for $b \to q\gamma$ are known to next-to-next-to-leading order (NNLO) in QCD, and include a resummation of logarithmically-enhanced effects of $\mathcal{O}(\alpha_s^2)$ contributions [398].

In the case of the rare decays into two charged leptons $b \to q\ell^+\ell^-$ with $\ell = e, \mu, \tau$, the SM operator basis in (200) has to be extended by two additional operators

$$
\begin{aligned}
Q_9 &= \frac{e}{16\pi^2}(\bar{q}_L\gamma_\mu b_L)(\bar{\ell}\gamma^\mu\ell)\,, \\
Q_{10} &= \frac{e}{16\pi^2}(\bar{q}_L\gamma_\mu b_L)(\bar{\ell}\gamma^\mu\gamma_5\ell)\,,
\end{aligned} \tag{203}
$$

while for the $b \to q\nu\bar{\nu}$ transitions only the single operator

$$
Q_L^\ell = (\bar{q}_L\gamma_\mu b_L)(\bar{\nu}_{\ell L}\gamma^\mu\nu_{\ell L})\,, \tag{204}
$$

is relevant. Also in the case of the $b \to q\ell^+\ell^-$ modes the relevant low-energy Wilson coefficients $C_i(\mu_b)$ are known to NNLO accuracy within the SM [399–401], while in the case of $b \to q\nu\bar{\nu}$ only the next-to-leading order (NLO) corrections are fully known [402, 403].[24]

The effect of physics beyond the SM (BSM) to radiative and rare $b \to q$ transitions can enter (200) in essentially two ways: ($i$) through modified values for the high-scale Wilson coefficients $C_i$ not necessarily aligned with the flavour coefficients $\lambda_t^{(q)}$ and/or ($ii$) through additional operators with different chirality and/or flavour structures compared to the SM.

*Hadronic Effects.* As it stands, the effective Hamiltonian (200) only describes the weak decays at the parton level. The physics associated to long-distance (LD) dynamics requires to evaluate hadronic matrix elements

$$
\langle X_{d,s}\gamma\,(\ell^+\ell^-)|Q_i|B\rangle \tag{205}
$$

of the operators $Q_i$, which contain non-perturbative QCD effects. A particular subtlety arises from the fact that in case of purely hadronic operators, the final state can also be generated by (real or virtual) photon radiation from internal lines during the hadronic transition. The theoretical description of hadronic corrections to the partonic decay crucially depends on the way these transitions are probed in terms of one or the other hadronic observable. In all cases one exploits the fact that the mass $m_b$ of the decaying $b$-quark is significantly larger than the typical hadronic scale set by (multiples of) the fundamental QCD scale $\Lambda_{\rm QCD} = \mathcal{O}(200\,\text{MeV})$.

For fully-inclusive observables, the heavy-quark expansion (HQE) is equivalent to a local operator product expansion (OPE) [311, 407] by which total decay rates can be expressed in terms of forward $B$-meson matrix elements of local operators. Here, the partonic decay represents the leading term in a simultaneous expansion in powers of $\Lambda_{\rm QCD}/m_b$ and $\alpha_s(m_b)$. The OPE breaks down when one tries to calculate differential inclusive decay distributions near phase-space boundaries. A twist expansion involving forward matrix elements of non-local light-cone operators (so-called shape functions) is then required to properly account for non-perturbative effects [364, 365, 408]. It was generally believed that all non-local operators reduce to local ones when the differential decay distributions are integrated over the entire phase-space, but then shown in [409, 410] for $B \to X_s\gamma$ that this is not always the case. These non-local power corrections can be expressed in terms of soft functions or subleading shape functions. At present our knowledge of these functions is limited to their asymptotic behaviour as well as constraints on their moments. In consequence, the precise impact of non-local power corrections is difficult to estimate in practice.

[24] The smallness of NNLO effects in $B_s \to \mu^+\mu^-$ [404] suggests that also in the case of $b \to q\nu\bar{\nu}$ such contributions should have a very limited phenomenological impact. NLO EW effects similar to those studied in [405, 406] are instead more relevant.

In case of exclusive decay observables, $B$-meson decays involving no energetic light hadrons can be described in Heavy Quark Effective Theory (HQET). At first approximation, the relevant hadronic quantities are given by $B \to X$ transition form factors which can be obtained with reasonable accuracy from lattice-QCD simulations, see [411] and references therein. In recent years, various lattice results became available e.g. $B \to \pi$ form factors [143, 159], $B \to K$ form factors [158, 412], $B \to K^*$ and $B_s \to \phi$ form factors [413, 414]. The lattice simulations are performed for high-momentum transfer, $q^2 \geq 14\,\text{GeV}^2$, *i.e.* small hadronic recoil. Predictions for smaller values of the invariant mass $q^2$ of the lepton pair are then obtained by employing well-motivated extrapolations.

In many cases (notably for $B \to V\gamma$ decays), however, we are interested in situations where the energy transfer $E_{\text{recoil}}$ to light hadrons in the final state is large of the order of $m_b/2$. In these cases, the systematic heavy-mass expansion leads to the concept of QCD-(improved) factorisation (QCDF) (cf. [415, 416]). The predictive power of QCDF is limited by hadronic uncertainties related to the transition form factors and the light-cone distribution amplitudes for the leading Fock states in the involved hadrons, as well as by power corrections in $\Lambda_{\text{QCD}}/m_b$. Form factors at large hadronic recoil can, for instance, be calculated with QCD light-cone sum rules (LCSRs), for a review see e.g. [417, 418]. Recent LCSR estimates include twist-three radiative and twist-four tree-level contributions, but have an accuracy of not better than 10%, which implies an uncertainty of at least 20% on the level of branching ratios (see for instance [419] for a recent discussion). More troublesome is the issue of power corrections. A naive dimensional estimate indicates that such contributions should be of the order of $\Lambda_{\text{QCD}}/E_{\text{recoil}}$, but the exact number is hard to quantify.

## *9.2. Inclusive and Exclusive Radiative Penguin Decays*

### *9.2.1. Inclusive $B \to X_q\gamma$ decays.* *(Contributing authors: M. Misiak and G. Paz)*

*Experimental Status.* The inclusive $B \to X_q\gamma$ decays provide important constraints on masses and interactions of many possible BSM scenarios such as models with extended Higgs sectors or supersymmetric (SUSY) theories. Measurements of their $CP$-averaged and isospin-averaged branching ratios by BaBar [420–423] and Belle [424, 425] lead to the following combined results

$$\text{Br}_{s\gamma}^{\text{exp}} = (3.27 \pm 0.14) \cdot 10^{-4}\,, \tag{206}$$

$$\text{Br}_{d\gamma}^{\text{exp}} = (1.41 \pm 0.57) \cdot 10^{-5}\,. \tag{207}$$

They are in perfect agreement with the corresponding SM predictions [426, 427]

$$\text{Br}_{s\gamma}^{\text{SM}} = (3.36 \pm 0.23) \cdot 10^{-4}\,, \tag{208}$$

$$\text{Br}_{d\gamma}^{\text{SM}} = \left(1.73^{+0.12}_{-0.22}\right) \cdot 10^{-5}\,. \tag{209}$$

The results in (206) to (209) correspond to the photon energy cut $E_\gamma > E_0 = 1.6\,\text{GeV}$ in the decaying meson rest frame. The measurements have been performed at $E_0 \in [1.7, 2.0]\,\text{GeV}$ for $\text{Br}_{s\gamma}$, and at $E_0 \simeq 2.24\,\text{GeV}$ for $\text{Br}_{d\gamma}$. Next, extrapolations down to $E_\gamma > E_0 = 1.6\,\text{GeV}$ were applied. Such extrapolations are unavoidable because the experimental background subtraction errors rapidly grow with decreasing $E_0$, while the theoretical non-perturbative uncertainties grow with increasing $E_0$.

In the average for $\mathrm{Br}_{s\gamma}^{\mathrm{exp}}$ given in (206), only the measurements at $E_0 = 1.9\,\mathrm{GeV}$ have been chosen as an input, and the extrapolation factors from [428] have been used. The question whether uncertainties in these factors have been properly estimated awaits a devoted study [316, 429, 430], especially in view of the upcoming more precise measurements at Belle II. The necessary extrapolation for $\mathrm{Br}_{d\gamma}^{\mathrm{exp}}$ (207) was performed in [431], following the method of [428]. In this case, the precision is much less of an issue given the large uncertainties in the original experimental result [421].

*Basic Formulas.* Theoretical calculations of $\mathrm{Br}_{q\gamma}$ within and beyond the SM are based on the equality

$$\Gamma(\bar{B} \to X_q\gamma) = \Gamma(b \to X_q^p\gamma) + \delta\Gamma_{\text{non-per}}\,, \tag{210}$$

where $\Gamma(b \to X_q^p\gamma)$ stands for the perturbative $b$-quark decay rate with only charmless partons in the final state $X_s^p$ (strangeness $= -1$) or $X_d^p$ (strangeness $= 0$). As long as the photon energy cut $E_0$ is large (i.e. $E_0 \gtrsim 1\,\mathrm{GeV}$) but not too close to the endpoint $E_{\max} \simeq 2.56\,\mathrm{GeV}$ (i.e. $E_{\max} - E_0 \gtrsim \Lambda_{\mathrm{QCD}}$), the non-perturbative effects accounted for by $\delta\Gamma_{\text{non-per}}$ remain under control, and constitute a correction at the few percent level [410, 432]. However, to discuss their size in a meaningful manner, one needs to get rid of $m_{b,\mathrm{pole}}^5$ from the leading perturbative contribution $\Gamma(b \to X_q^p\gamma)$, as on-shell masses of quarks are ill-defined. For this purpose, a normalisation to the semi-leptonic decay rate can be used. The SM results quoted in (208) to (209) have been derived from the formula [433]

$$\mathrm{Br}_{q\gamma} = \mathrm{Br}_{c\ell\nu}\,\xi_q\,\frac{6\alpha}{\pi C}\left[P_q(E_0) + N_q(E_0)\right], \tag{211}$$

where $\xi_q = |V_{tq}^* V_{tb}/V_{cb}|^2$ is the relevant CKM factor, $\alpha = \alpha(0)$ is the electromagnetic coupling constant renormalised at $q^2 = 0$, $\mathrm{Br}_{c\ell\nu}$ stands for the $CP$-averaged and isospin-averaged branching ratio of the semi-leptonic $\bar{B} \to X_c\ell\bar{\nu}$ decay, and $C$ represents the so-called semi-leptonic phase-space factor

$$C = \left|\frac{V_{ub}}{V_{cb}}\right|^2 \frac{\Gamma(\bar{B} \to X_c\ell\bar{\nu})}{\Gamma(\bar{B} \to X_u\ell\bar{\nu})}\,. \tag{212}$$

The function $P_q(E_0)$ is defined by the ratio

$$\frac{\Gamma(b \to X_q^p\gamma) + \Gamma(\bar{b} \to X_{\bar{q}}^p\gamma)}{|V_{cb}/V_{ub}|^2\,\Gamma(b \to X_u^p e\bar{\nu})} = \xi_q\frac{6\alpha}{\pi}P_q(E_0)\,. \tag{213}$$

In the $q = s$ case, the non-perturbative effects accounted for by $N_s(1.6\,\mathrm{GeV})$ in (211) enhance the central value of $\mathrm{Br}_{s\gamma}^{\mathrm{SM}}$ by around 3% [434], while the corresponding uncertainty amounts to about $\pm 5\%$ [410]. In the $q = d$ case, one encounters additional sources of uncertain hadronic effects that originate from the CKM-unsuppressed $b \to du\bar{u}\gamma$ transitions [432]. We shall come back to the issue of non-perturbative corrections after discussing the dominant perturbative term $P_q(E_0)$.

*Theoretical Calculations of $P_s(E_0)$.* For $b \to s\gamma$, the CKM element ratio $\kappa_s$ in (200) is small, changing $\mathrm{Br}_{s\gamma}^{\mathrm{SM}}$ by less than 0.3%. Barring this effect and the higher-order EW ones,

$P_s(E_0)$ is given within the SM by

$$P_s(E_0) = \sum_{i,j=1}^{8} C_i^{\rm eff}(\mu_b) C_j^{\rm eff}(\mu_b)\, K_{ij}\,, \tag{214}$$

where $C_i^{\rm eff}$ are certain linear combinations of the Wilson coefficients $C_i$ (cf. [426]). They differ from $C_i$ only for $i = 7, 8$, and are fixed by the requirement that the leading-order (LO) $b \to s\gamma$ and $b \to sg$ amplitudes are proportional to $C_7^{\rm eff}$ and $C_8^{\rm eff}$.

To match the experimental precision, the symmetric matrix $K_{ij}$ needs to be determined up to $\mathcal{O}(\alpha_s^2)$ in its perturbative expansion

$$K_{ij} = \sum_{n=0}^{\infty} \left(\frac{\alpha_s(\mu_b)}{4\pi}\right)^n K_{ij}^{(n)}\,. \tag{215}$$

The quantities $K_{ij}^{(0)}$ and $K_{ij}^{(1)}$ are already known in a practically complete manner, with the latest contributions coming from [435, 436]. As far as $K_{ij}^{(2)}$ are concerned, it is sufficient to restrict to the operators listed in (202) because the remaining ones are negligible at the NNLO level due to their small Wilson coefficients and other suppression factors. Currently complete NNLO expressions are available for $K_{77}^{(2)}$ [437–439] and $K_{78}^{(2)}$ [440, 441] only. For $K_{ij}^{(2)}$ with $i, j \in \{1, 2, 8\}$, the two-body final-state contributions are known in a complete manner, while the three-body and four-body contributions have been evaluated [442–444] in the Brodsky-Lepage-Mackenzie (BLM) [445] approximation.

It remains to discuss $K_{17}^{(2)}$ and $K_{27}^{(2)}$. The BLM approximations for these quantities have been established for some time [442, 446]. The same is true for effects due to non-vanishing quark masses in loops on the gluon lines [447]. However, the generic non-BLM parts of $K_{17}^{(2)}$ and $K_{27}^{(2)}$ have been found so far only in two limiting cases for the $c$-quark mass, namely $m_c \gg m_b/2$ [448, 449] and $m_c = 0$ [426]. An interpolation between these two limits was performed in [426], leading to the conclusion that the considered non-BLM corrections are sizeable, enhancing ${\rm Br}_{s\gamma}^{\rm SM}$ by about 5%.

An uncertainty of ±3% due to the interpolation in $m_c$ was included in the error budget of (208). It was added in quadrature to the other three uncertainties of ${\rm Br}_{s\gamma}^{\rm SM}$: non-perturbative (±5%), higher-order (±3%) and parametric (±2%). Future improvements in the accuracy of the perturbative calculations of $P_s(E_0)$ will require determining $K_{17}^{(2)}$ and $K_{27}^{(2)}$ for the physical value of $m_c$ without any interpolation. A first step in this direction has been made in [450]. One should investigate as well whether the quoted size of the non-perturbative uncertainties can be reduced by combining lattice inputs with measurements of observables like the $CP$ and/or the isospin asymmetry (IA) in $B \to X_s\gamma$ (cf. the discussion after (220)). Improving both the uncertainty due to the interpolation in $m_c$ and the non-perturbative errors by a factor of 2, the total uncertainty of ${\rm Br}_{s\gamma}^{\rm SM}$ would be reduced from around 7% to about 4%. Whether such theoretical improvements are possible in the near future remains to be seen.

*The Case of* ${\rm Br}_{d\gamma}^{\rm SM}$. Extending the NNLO calculation to the case of ${\rm Br}_{d\gamma}^{\rm SM}$, one needs to take into account that, contrary to $\kappa_s$, the ratio $\kappa_d$ is not numerically small. The global

CKM fit in [451] implies that

$$\kappa_d \;=\; \left(0.007^{+0.015}_{-0.011}\right) + i\left(-0.404^{+0.012}_{-0.014}\right). \tag{216}$$

Due to the small value of $\mathrm{Re}\,\kappa_d$, terms proportional to $|\kappa_d|^2$ turn out to give the dominant $\kappa_d$ effects in the $CP$-averaged $\mathrm{Br}^{\mathrm{SM}}_{d\gamma}$. In such terms, perturbative two-body and three-body final-state contributions arise only at $\mathcal{O}(\alpha_s^2)$ and $\mathcal{O}(\alpha_s)$, respectively. They vanish for $m_c = m_u$, which implies that they are suppressed by $m_c^2/m_b^2 \simeq 0.1$. As a result, the main $\kappa_d$ effect comes from four-body final states, namely from the $b \to d u\bar{u}\gamma$ mode that appears already at tree level.

One way to calculate these contributions consist in evaluating the $b \to d u\bar{u}\gamma$ diagrams including a common light-quark mass $m_q$ inside the collinear logarithms [444], and then to vary $m_b/m_q$ between $10 \sim m_B/m_K$ and $50 \sim m_B/m_\pi$ to estimate the uncertainty. Such an approach leads to an effect of 2% to 11% on $\mathrm{Br}_{d\gamma}$. A more involved analysis with the help of fragmentation functions gives an almost identical range [432]. As a result, the SM prediction for $\mathrm{Br}_{d\gamma}$ in (209) is essentially insensitive to which of the two methods is used. The central value in that equation corresponds to the first method with $m_b/m_q = 50$.

*Non-Perturbative Effects in $B \to X_q\gamma$.* In discussing the non-perturbative effects in $B \to X_q\gamma$, one has to distinguish contributions from the interference of $Q_7$ with itself, and contributions from other operators. It is convenient to express the quantity $N_s(E_0)$ that was defined in (211) in terms of the Wilson coefficients, by analogy to (214)

$$N_s(E_0) = \sum_{i,j=1}^{8} C_i^{\mathrm{eff}}(\mu_b) C_j^{\mathrm{eff}}(\mu_b)\, S_{ij}\,. \tag{217}$$

For $E_0$ far from the endpoint region, $S_{77}$ is parameterised by matrix elements of higher-dimensional local operators. These matrix elements are universal in the sense that they contribute also to semi-leptonic $B$ decays. In consequence, one finds

$$S_{77} = \sum_{n=2}^{\infty} \frac{1}{m_b^n} \sum_k c_{k,n} \langle O_{k,n}\rangle\,. \tag{218}$$

The $\langle O_{k,n}\rangle$ matrix elements scale as $\Lambda^n_{\mathrm{QCD}}$, which implies that the power corrections start at power $\Lambda^2_{\mathrm{QCD}}/m_b^2$. The coefficients $c_{k,2}$ were calculated up to $\mathcal{O}(\alpha_s)$ in [452, 453]. Their $\mathcal{O}(\alpha_s^0)$ parts [454, 455] turn out to vanish due to accidental cancellation of corrections of this order to the radiative and semi-leptonic $B \to X_u \ell\bar{\nu}$ decays. The quantity $S_{77}$ affects the SM prediction for $\mathrm{Br}_{s\gamma}$ (208) by around $-0.3\%$ only, which includes the effect of the $\mathcal{O}(\alpha_s^0)$ coefficients $c_{k,3}$ [456]. The coefficients $c_{k,4}$ and $c_{k,5}$ have also been calculated at $\mathcal{O}(\alpha_s^0)$ [355], but the corresponding matrix elements are poorly constrained, and the resulting small correction has been neglected in (208).

In the endpoint region, the $(Q_7, Q_7)$ interference part of the photon energy spectrum is described by the following symbolic factorisation formula:

$$\frac{d\Gamma_{77}}{dE_\gamma} \sim H \cdot J \otimes S + \frac{1}{m_b}\sum_i H \cdot J \otimes s_i + \frac{1}{m_b}\sum_i H \cdot j_i \otimes S + \mathcal{O}\left(\frac{\Lambda^2_{\mathrm{QCD}}}{m_b^2}\right). \tag{219}$$

The hard functions $H$ and jet functions $J, j_i$ are calculable in perturbation theory. The shape functions $S$ and $s_i$ are non-perturbative and given in terms of non-local matrix elements.

At the leading power, there is only a single shape function $S$. It is universal in the sense that it also appears for the endpoint region of semi-leptonic $B$ decays [364, 365, 408]. The subleading shape functions $s_i$ contribute also to the endpoint region of semi-leptonic $B$ decays, but in a different linear combination. For the first term in (219), $H$ [438] and $J$ [457] are known up to $\mathcal{O}(\alpha_s^2)$. For the second term, $H$ and $J$ are known explicitly at $\mathcal{O}(\alpha_s^0)$ only [367, 371, 458] (see also [366]). For the third term, $H$ is known at $\mathcal{O}(\alpha_s^0)$ and $j_i$ at $\mathcal{O}(\alpha_s)$ [459]. As one integrates over the photon energy in (219), the shape functions reduce to local operators, and one obtains (218). Measurements of the $B \to X_s\gamma$ photon spectrum are being used in calculations that are necessary to extract $|V_{ub}|$ from $B \to X_u \ell\bar{\nu}$ [364, 369, 430, 460, 461]. These computations currently do not include uncertainties stemming from the resolved photon contributions (see below).

Non-perturbative effects from other pairs of operators are more complicated. Apart from "direct" photon contribution arising from diagrams in which the photon couples directly to the weak vertex, there are also "resolved" photon contribution in which the photon couples to light partons. For example, $Q_8$ gives rise to the process $b \to sg \to s\bar{q}q\gamma$, and $Q_2$ leads to the process $b \to s\bar{c}c \to sg\gamma$. Such effects were discussed in the literature [409, 434, 462–467] but were only studied systematically in [410]. Taking them into account, the photon spectrum in the endpoint region can be factorised symbolically as [410]

$$\frac{d\Gamma}{dE_\gamma} \sim H \cdot J \otimes S + H \cdot J \otimes s \otimes \bar{J} + H \cdot J \otimes s \otimes \bar{J} \otimes \bar{J}\,. \tag{220}$$

The first term in (220) is the direct photon contribution, similar to (219), while the terms in the second line correspond to the resolved photon contributions that start at order $\Lambda_{\rm QCD}/m_b$. The jet functions $\bar{J}$ are perturbative. The soft functions $s$ are non-perturbative and, unlike the shape functions, they contain non-localities in two light-cone directions.

In the integrated rate, the resolved photon contributions leads to $\Gamma \sim \bar{J} \otimes h$, where $h$ are non-local matrix elements. At power $\Lambda_{\rm QCD}/m_b$, the only non-vanishing contributions to the integrated rate arise from $S_{27}$, $S_{78}$, and $S_{88}$. Conservative modelling gives a total of around 5% non-perturbative uncertainty in $\mathrm{Br}_{s\gamma}^{\rm SM}$ from the resolved photon contributions at $E_0 = 1.6\,\mathrm{GeV}$. Direct photon contributions to $S_{ij}$ are smaller, and can be included in the 5% uncertainty estimate.

The resolved photon contributions are more important in the case of the $CP$ asymmetry

$$A_{CP} = \frac{\Gamma(\bar{B} \to X_s\gamma) - \Gamma(B \to X_{\bar{s}}\gamma)}{\Gamma(\bar{B} \to X_s\gamma) + \Gamma(B \to X_{\bar{s}}\gamma)}\,. \tag{221}$$

As shown in [468], they dominate over perturbative effects [469–472]. Within the SM, one obtains the prediction

$$A_{CP}^{\rm SM} \in [-0.6, 2.8]\%\,, \tag{222}$$

while including perturbative effects alone would lead to an asymmetry of around 0.5%. Resolved photon contributions also imply that the difference between the $CP$ asymmetries for charged and neutral $B$ mesons are sensitive to new-physics effects [468]. To linear order in the new-physics contributions to the Wilson coefficients $C_7$ and $C_8$ and assuming that the dominant current-current contribution $C_2$ remains SM-like, one finds for the additive

Table 60: Observables accessible in $B \to X_q\gamma$ and the corresponding reconstruction methods. The table uses abbreviations for reconstruction (reco.), hadronic (had.), semi-leptonic and leptonic (SL and L), efficiency (effi.), signal to background ratio ($S/B$), if the spectator quark may be specified ($q$), and if the momentum of the signal $B$ meson is measured ($p_B$).

| reco. method | tagging | effi. | $S/B$ | $q$ | $p_B$ | $A_{CP}$ | $\Delta_{0+}$ | $\Delta A_{CP}$ |
|---|---|---|---|---|---|---|---|---|
| sum-of-exclusive | none | high | moderate | $s$ or $d$ | yes | yes | yes | yes |
| fully-inclusive | had. $B$ | very low | very good | $s$ and $d$ | yes | yes | yes | yes |
| | SL $B$ | very low | very good | $s$ and $d$ | no | yes | yes | yes |
| | L | moderate | good | $s$ and $d$ | no | yes | no | no |
| | none | very high | very bad | $s$ and $d$ | no | no | no | no |

new-physics contribution to $A_{CP}$ the following approximate expression

$$A_{CP}^{\rm NP} \simeq 0.05\,{\rm Re}\left(C_7^{\rm NP}\right) - 0.47\,{\rm Im}\left(C_7^{\rm NP}\right) + 0.24\,{\rm Im}\left(C_8^{\rm NP}\right) \,. \tag{223}$$

This result implies that $A_{CP}$ is a sensitive probe of new physics that leads to CP-violating contributions to the dipole operators $Q_7$ and $Q_8$. Such effects are only weakly constrained by the $B \to X_s\gamma$ branching ratio.

Currently, the main source of uncertainty in ${\rm Br}_{s\gamma}^{\rm SM}$ are the resolved photon contributions. The extraction of HQET parameters from $B \to X_c\ell\bar{\nu}$, as done in [354], can help to better control the $S_{27}$ contribution. By better measuring the IA in $B \to X_s\gamma$ defined as

$$\Delta_{0+} = \frac{\Gamma(B^0 \to X_s\gamma) - \Gamma(B^+ \to X_s\gamma)}{\Gamma(B^0 \to X_s\gamma) + \Gamma(B^+ \to X_s\gamma)}\,, \tag{224}$$

one can furthermore hope to pin down the $S_{78}$ contributions since these quantities are directly related [410, 473]. Employing the so-called vacuum-insertion approximation (VIA) to estimate the relevant hadronic matrix element leads to the following SM prediction

$$\Delta_{0+}^{\rm SM} \in [0.1, 7.4]\%\,, \tag{225}$$

where one should keep in mind that the VIA is a very rough approximation. Rather than comparing the SM prediction (225) to future precise measurements of $\Delta_{0+}$ to look for new physics it thus seems more advantages to exploit the relation

$$\frac{N_s^{78}(E_0)}{P_s(E_0)} = -\frac{\Delta_{0+}}{3}\,, \tag{226}$$

to experimentally constrain the size of the non-perturbative contribution $N_s^{78}(E_0)$ (or equivalent $S_{78}$) introduced in (211). New Belle II measurements of $\Delta_{0+}$ can therefore help to reduce the non-perturbative uncertainties in the SM prediction for $B \to X_s\gamma$, in particular if these measurements remain consistent with zero.

*9.2.2. Measurements of $B \to X_s\gamma$.* *(Contributing author: A. Ishikawa)*

There are two methods to reconstruct $B \to X_q\gamma$ decays. They will be referred to as the sum-of-exclusive method and the fully-inclusive method. In the sum-of-exclusive method, the hadronic system is reconstructed from many exclusive decays containing a kaon, such as $K n\pi$, $K\eta m\pi$ or $3Km\pi$, where $n \geq 1$ and $m \geq 0$. Hadronic candidates are then combined with a hard photon to reconstruct $B$-meson candidates. In the fully-inclusive method, the other $B$ meson is usually tagged to improve the $S/B$ ratio. One can require a fully reconstructed hadronic final state (hadronic tag), a fully reconstructed semi-leptonic decay (semi-leptonic tag), or only an energetic lepton (leptonic tag) from the $B$-meson decay.

The prompt photons are selected as isolated clusters in the ECL that are not matched to any charged tracks. The polar angle of the photon direction must be within the barrel ECL region. The cluster shape is required to be consistent with an electromagnetic shower. In order to reduce contaminations from asymmetric $\pi^0/\eta \to \gamma\gamma$ decays, the photon candidate is paired with all other photons in the event. When the pair is consistent with $\pi^0$ or $\eta$, the prompt photon candidate is discarded. In case of Belle, the signal yields for $E_\gamma > 1.9$ GeV with 711 fb$^{-1}$ are $13359 \pm 169$ for sum-of-exclusive and $8945 \pm 240$ for full inclusive with lepton tagging. The errors are statistical only.

The two reconstruction methods have their own pros and cons, and provide access to different observables, as summarised in Table 60. Only the sum-of-exclusive method can specify that the transition was $b \to s$ (or $b \to d$), whereas the fully-inclusive method can only ever measure the sum of $b \to s$ and $b \to d$ transitions. Reconstructing the other $B$-meson decay determines the charges of the $b$ quark and/or the spectator quark ($d$ or $u$) in the signal $B$ meson, which is required to measure direct $CP$ and/or isospin violation.

The branching ratio of $B \to X_s\gamma$ was measured by BaBar [420, 422, 423, 474], Belle [424, 475] and CLEO [476]. The uncertainties of the measured branching ratios are systematically dominated. Given the expected large Belle II data sample, a reduction of systematic uncertainties is of utmost importance. For instance, at Belle, the dominant source of systematic uncertainties in the inclusive analysis with lepton tagging arises from neutral hadrons faking photons. Dedicated studies of the cluster shape in the calorimeter, which were not performed at Belle, allow to constrain the contribtuion of the fake photons or even reduce the contribtuion. At Belle II, it should be possible to reduce this uncertainty from 3.7% to 1.9% by the studies. A conservative estimate gives that the total systematic uncertainty with a photon energy threshold of 1.9 GeV can be reduced from 5.3% to 3.2%.

So far, all measurements required a photon energy threshold in the range of $[1.7, 2.0]$ GeV, extrapolating to the photon energy threshold of 1.6 GeV assuming a theoretical model. At Belle II, the branching ratio with the photon energy threshold of 1.6 GeV is directly measurable, removing the need to perform the extrapolation and in turn the corresponding source of systematic uncertainties. Lowering the photon energy threshold leads, however, to an increase of the size of the systematic uncertainty due to hadronic backgrounds. Thus, several energy thresholds will need to be considered in the future experimental analyses to better control this systematics.

The photon spectrum in the $B$-meson rest frame can be directly measured with a fully-inclusive analysis with hadronic tagging, since the momentum of the $B$ meson is known. Note that unfolding of the Doppler effect due to a finite $B$-meson momentum in the $\Upsilon(4S)$ rest frame is needed in case a fully-inclusive analysis with lepton tagging is performed. The hadronic tagging provides a straightforward approach to measure the moments of the photon energy spectrum. The uncertainty on the branching ratio measured with hadronic tagging is expected to be dominated by statistics at Belle due to the limited number of tagged $B$ mesons. In view of the large data set at Belle II, systematic uncertainties will instead dominate. In fact, like in the case of lepton tagging, the dominant source of systematic uncertainty arises from mis-reconstruction of neutral hadrons as photons. As a result the uncertainties of the branching ratio measurements with hadronic tagging will be comparable and strongly correlated with the uncertainty in the lepton tagging analysis.

The branching ratio measurement with the sum-of-exclusive method has different systematics, compared to the fully-inclusive analysis. The dominant sources of systematic uncertainties will be due to fragmentation and missing decay modes. Given the large data set it should however be possible to reduce the latter source of uncertainty by including additional decay modes, but even then the accuracy of the branching ratio measurement via the sum-of-exclusive method is expected to be slightly lower than the uncertainty provided by fully-inclusive analyses.

As already mentioned around (224), measurements of the isospin asymmetry $\Delta_{0+}$ could be useful to reduce the theoretical uncertainties of the branching ratio of $B \to X_q \gamma$. It has been found in [409, 473] that if a more precise measurement of $\Delta_{0+}$ turns out to stay near zero, that would help to significantly reduce the theoretical uncertainty. The BaBar collaboration measured $\Delta_{0+}(B \to X_s \gamma) = (-0.6 \pm 5.8 \pm 0.9 \pm 2.4)\%$ with the sum-of-exclusive method [477] and $\Delta_{0+}(B \to X_{s+d} \gamma) = (-6 \pm 15 \pm 7)\%$ with the fully inclusive method [420] with partial data sets of $81.9\,\mathrm{fb}^{-1}$ and $210\,\mathrm{fb}^{-1}$, respectively. Recently, Belle also measured $\Delta_{0+}(B \to X_s \gamma) = (+1.70 \pm 1.39 \pm 0.87 \pm 1.15)\%$ with sum-of-exclusive method using a full data sample of $711\,\mathrm{fb}^{-1}$ [478]. In the measurements, the first error is statistical, the second is systematic and the third is due to the production ratio of $B^+B^-$ and $B^0\bar{B}^0$ from $\Upsilon(4S)$ decay ($f_{+-}/f_{00}$). At Belle II, both the sum-of-exclusive method and the fully-inclusive method with hadronic tagging can be performed. As an example, the sum-of-exclusive method can reduce the experimental uncertainty in $\Delta_{0+}$ down to 0.6% with $50\,\mathrm{ab}^{-1}$ of data (see Table 61).

The dominant uncertainty of $\Delta_{0+}(B \to X_s \gamma)$ at Belle II will be of systematic origin and related to the ratio $f_{+-}/f_{00}$. The most promising method to measure $f_{+-}/f_{00}$ without assuming isospin invariance in hadronic $B$ decays is the use of double semi-leptonic decays, $\bar{B} \to D^* \ell^- \bar{\nu}$, as has been done by BaBar [479]. Belle II measurements of $\Delta_{0+}(B \to X_{s+d} \gamma)$ will instead be statistically limited.

Direct $CP$ violation in $B \to X_{s+d} \gamma$ has also been measured in an inclusive analysis with lepton tagging. Belle has measured this quantity with the full data set and the result is dominated by statistics, $A_{CP}(B \to X_{s+d} \gamma) = (1.6 \pm 3.9 \pm 0.9)\%$ for $E_\gamma > 2.1\,\mathrm{GeV}$ [480]. At Belle II with $50\,\mathrm{ab}^{-1}$ the statistical uncertainty will amount to 0.5%. The dominant source of systematic uncertainty from the asymmetry of the background can be assessed using increased data in background regions (so-called sidebands). A conservative estimate shows that a systematic uncertainty of 0.4% is reachable.

Both the sum-of-exclusive reconstruction and the fully-inclusive reconstruction with hadronic tagging can determine the flavour and isospin of the parent in $B \to X_q \gamma$ decays. Such a separation is needed in order to study the direct $CP$ violation and the difference of direct $CP$ violation between the charged and neutral $B$ mesons $\Delta A_{CP}(B \to X_q \gamma) = A_{CP}(B^+ \to X_q^+ \gamma) - A_{CP}(B^0 \to X_q^0 \gamma)$. Given that $\Delta A_{CP}(B \to X_s \gamma) \propto \mathrm{Im}\,(C_8/C_7)$ [468], measurements of $\Delta A_{CP}(B \to X_s \gamma)$ provide sensitive probes of new physics.

As stated earlier the theoretical uncertainty of the $CP$ asymmetry (223) is dominated by the contribution from resolved photons [468]. Precise measurement of $A_{CP}$ hence allows to constrain the size of non-local power corrections. The existing measurements of $A_{CP}$ by BaBar and Belle with $429\,\mathrm{fb}^{-1}$ and $711\,\mathrm{fb}^{-1}$ use the sum-of-exclusive method and find $(+1.7 \pm 1.9 \pm 1.0)\%$ [481] and $(+1.71 \pm 1.26 \pm 0.21)\%$ [478], respectively. BaBar and Belle also measured $\Delta A_{CP} = (+5.0 \pm 3.9 \pm 1.5)\%$ for $E_\gamma > 2.1\,\mathrm{GeV}$ [481] and $\Delta A_{CP} = (+1.26 \pm 2.40 \pm 0.67)\%$ for $E_\gamma > 1.9\,\mathrm{GeV}$ [478], respectively.

Belle II can measure both $A_{CP}$ and $\Delta A_{CP}$ yet with a much larger data set. A reduction of the systematic uncertainties is therefore crucial at Belle II. The systematic uncertainty due to the detector asymmetry can be reduced, in part due to the statistics of the larger data sample, since it is in practice determined from control samples or sideband events. The bias from the asymmetry due to peaking background can be expressed as a product of the number of peaking background events and the difference of $A_{CP}$ between signal and peaking background. BaBar conservatively took all of the $B\bar{B}$ background events as contributing to the latter uncertainty. At Belle II it should be possible to obtain a more realistic estimate, since the $CP$ asymmetries of both charged and neutral $B \to X_s\gamma$ decays and the dominant peaking backgrounds can be measured precisely. As a result the achievable accuracy of the measurement of $\Delta A_{CP}$ is determined by the statistical uncertainty for which a precision of 0.3% is expected. BaBar and Belle usually assumed that the direct $CP$ violation does not depend on the specific $X_s$ decay mode while Belle II can also test this assumption with its large data set.

Belle II will also perform a measurement of $\Delta A_{CP}(B \to X_{s+d}\gamma)$ using the fully-inclusive reconstruction with hadronic tagging. With $711\,\mathrm{fb}^{-1}$ about $300 \pm 27$ signal events are expected at Belle with the neutral $B$ fraction of 52% which corresponds to a 16% precision on $\Delta A_{CP}$. At Belle II, the statistical uncertainty is still dominant even after including a factor of two improvement in the hadronic tagging efficiency.

### *9.2.3. Measurement of $B \to X_d\gamma$.* *(Contributing author: A. Ishikawa)*

In constrast to $B \to X_s\gamma$ the inclusive $B \to X_d\gamma$ decay is experimentally largely unexplored. In consequence, Belle II is in the near-term future the only place to study the various $B \to X_d\gamma$ observables.

Since a fully-inclusive analysis is impossible in the presence of the large $B \to X_s\gamma$ background, a measurement of $B \to X_d\gamma$ has to rely on the sum-of-exclusive method. BaBar in [421] has managed to reconstructed $7 X_d$ decay modes, $2\pi$, $3\pi$ and $4\pi$ modes with at most one neutral pion and $\pi^\pm\eta\,(\to \gamma\gamma)$ mode and applied a hadronic mass cut of $2.0\,\mathrm{GeV}$. At Belle II the statistical uncertainties will at some point be smaller than the systematic ones, and the increase in luminosity can be exploited to achieve a better understanding of the hadronic spectrum as well as the fragmentation of the $X_d$ system, including missing modes to reduce the systematic uncertainties as done by the $B$ factories in the sum-of-exclusive measurement of $B \to X_s\gamma$. In fact, the dominant systematic uncertainty from missing modes can be reduced to 10% by adding reconstructed decay modes, such as final states having five pions, two $\pi^0$, two kaons and an $\eta$ plus multiple pions or an $\eta'$ plus multiple pions, as well as by applying a looser hadronic mass cut. The second and third largest uncertainties are of statistical origin (6%) and the systematic uncertainty due to fragmentation (5%). The total uncertainty on $\mathrm{Br}_{d\gamma}$ is expected to be around 14% with $50\,\mathrm{ab}^{-1}$ of integrated luminosity.

Table 61: Sensitivities of observables for the radiative inclusive $B$ decay. A photon energy threshold of $E_\gamma > 1.9\,\mathrm{GeV}$ ($E_\gamma > 2.0\,\mathrm{GeV}$) is assumed for the $B \to X_s\gamma$ ($B \to X_d\gamma$) analysis. Some sensitivities at Belle are extrapolated to $0.71\,\mathrm{ab}^{-1}$. In the case of the branching ratios the quoted uncertainties are relative ones, while for what concerns $\Delta_{0+}$, $A_{CP}$ and $\Delta A_{CP}$ they are absolute numbers.

| Observables | Belle $0.71\,\mathrm{ab}^{-1}$ | Belle II $5\,\mathrm{ab}^{-1}$ | Belle II $50\,\mathrm{ab}^{-1}$ |
|---|---|---|---|
| $\mathrm{Br}(B \to X_s\gamma)^{\text{lep-tag}}_{\text{inc}}$ | 5.3% | 3.9% | 3.2% |
| $\mathrm{Br}(B \to X_s\gamma)^{\text{had-tag}}_{\text{inc}}$ | 13% | 7.0% | 4.2% |
| $\mathrm{Br}(B \to X_s\gamma)_{\text{sum-of-ex}}$ | 10.5% | 7.3% | 5.7% |
| $\Delta_{0+}(B \to X_s\gamma)_{\text{sum-of-ex}}$ | 2.1% | 0.81% | 0.63% |
| $\Delta_{0+}(B \to X_{s+d}\gamma)^{\text{had-tag}}_{\text{inc}}$ | 9.0% | 2.6% | 0.85% |
| $A_{CP}(B \to X_s\gamma)_{\text{sum-of-ex}}$ | 1.3% | 0.52% | 0.19% |
| $A_{CP}(B^0 \to X_s^0\gamma)_{\text{sum-of-ex}}$ | 1.8% | 0.72% | 0.26% |
| $A_{CP}(B^+ \to X_s^+\gamma)_{\text{sum-of-ex}}$ | 1.8% | 0.69% | 0.25% |
| $A_{CP}(B \to X_{s+d}\gamma)^{\text{lep-tag}}_{\text{inc}}$ | 4.0% | 1.5% | 0.48% |
| $A_{CP}(B \to X_{s+d}\gamma)^{\text{had-tag}}_{\text{inc}}$ | 8.0% | 2.2% | 0.70% |
| $\Delta A_{CP}(B \to X_s\gamma)_{\text{sum-of-ex}}$ | 2.5% | 0.98% | 0.30% |
| $\Delta A_{CP}(B \to X_{s+d}\gamma)^{\text{had-tag}}_{\text{inc}}$ | 16% | 4.3% | 1.3% |
| $\mathrm{Br}(B \to X_d\gamma)_{\text{sum-of-ex}}$ | 30% | 20% | 14% |
| $\Delta_{0+}(B \to X_d\gamma)_{\text{sum-of-ex}}$ | 30% | 11% | 3.6% |
| $A_{CP}(B^+ \to X^+_{u\bar{d}}\gamma)_{\text{sum-of-ex}}$ | 42% | 16% | 5.1% |
| $A_{CP}(B^0 \to X^0_{d\bar{d}}\gamma)_{\text{sum-of-ex}}$ | 84% | 32% | 10% |
| $A_{CP}(B \to X_d\gamma)_{\text{sum-of-ex}}$ | 38% | 14% | 4.6% |
| $\Delta A_{CP}(B \to X_d\gamma)_{\text{sum-of-ex}}$ | 93% | 36% | 11% |

The observables $\Delta_{0+}(B \to X_d\gamma)$, $A_{CP}(B \to X_d\gamma)$ and $\Delta A_{CP}(B \to X_d\gamma)$ have up to now not been measured. In the asymmetries, large parts of the systematic uncertainties cancel out and the corresponding measurements will therefore be statistically limited at Belle II. With $50\,\mathrm{ab}^{-1}$ of data, the precision on $\Delta_{0+}(B \to X_d\gamma)$ can be estimated to be about 14%. The accuracy of $A_{CP}$ is expected to be slightly worse than that of $\Delta_{0+}$ since flavour tagging of the other $B^0$ meson is needed for flavour non-eigenstate $B^0 \to X^0_{d\bar{d}}\gamma$ decays. By taking into account an effective flavour tagging efficiency of 30% and using the product of the mixing probability in the $B^0\bar{B}^0$ system, $\chi_d = 0.1875$, the anticipated precision of $A_{CP}(B \to X_d\gamma)$ is 5%. The quoted uncertainty is dominated by the statistical uncertainty on $A_{CP}(B^+ \to X^+_{u\bar{d}}\gamma)$, while the accuracy of a future $\Delta A_{CP}$ measurement is dominated by the statistical uncertainty on $A_{CP}(B^0 \to X^0_{d\bar{d}}\gamma)$ and amounts to roughly 11%.

The summary of the Belle II sensitivities for the various $B \to X_d\gamma$ channels is shown in Table 61.

### *9.2.4. Exclusive $b \to q\gamma$ decays.* *(Contributing authors: E. Kou and R. Zwicky)*

*Preliminaries.* Radiative decays into light vector mesons $B_{(q,s)} \to V\gamma$ with $V = K^*, \rho, \omega, \phi$, represent prototypes of FCNC transitions. Promising candidates are $B_{(q,s)} \to (K^*, \phi)\gamma$ for the $b \to s$ and $B_{(q,s)} \to (\rho/\omega, \bar{K}^*)\gamma$ for the $b \to d$ transitions.

To first approximation only the matrix elements of the photonic dipole operator $Q_7$ in (202) enter, which are described by hadronic transition form factors for the $b \to q$ tensor currents. The remaining operators describe LD physics contributions, from internal emission of the photon during the hadronic transition, and thus generically involve strong-interaction phases. There are three basic LD topologies. One originating from the gluonic dipole operator $Q_8$ and two from four-quark operators $Q_{1-6}$, referred to as weak annihilation (WA) and quark loop (QL) topologies in the following (see e.g. Section 2 of [482] for the relevant Feynman diagrams). The WA topology is only relevant if the valence quarks in the initial $B$ and light vector meson matches the flavour structure of the respective four-quark operator in (200). In the QL topology two quarks from the four-quark operators with the same flavour are contracted to a closed loop from which the external photon and/or additional gluons can be radiated.

In QCDF the LD processes have been shown to factorise at LO in $\Lambda_{\rm QCD}/m_b$ and $\mathcal{O}(\alpha_s)$ [482–484]. A LCSR computation for the contribution of $Q_8$ at leading twist has been performed in [485], where also a discussion of the relation to QCDF can be found. WA has been computed in the LCSR approach in [486–488]. The computation of QL in LCSR is involved, and a hybrid treatment of QCDF and LCSR has been presented in [488]. LD $c$-quark loop contributions are a topic in their own right and will be discussed in more detail later on.

Unlike their semi-leptonic counterparts, $B_{(q,s)} \to V\ell^+\ell^-$ to be discussed in Section 9.4.3, $B_{(q,s)} \to V\gamma$ decays do not lend themselves to a rich angular analysis. Instead, they are described by two helicity amplitudes corresponding to the two possible photon polarisations. Schematically, one has

$$H_\mp \propto \lambda_t^{(d,s)} \left\{ \begin{matrix} m_b \\ m_{(d,s)} \end{matrix} \right\} C_7 \,(1 + \delta_{\rm fac})\, T_1(0) + \sum_{U=u,c} \lambda_U^{(d,s)}\, L_\mp^U(0)\,, \tag{227}$$

where $T_1(0)$ is the relevant $B \to V$ transition form factor, $\delta_{\rm fac}$ denotes factorisable QCD corrections and $L_\mp^U$ stands for the previously discussed LD contributions (including the Wilson coefficients of the hadronic operators).

While in the SM the polarisation of the photon is predominantly left-handed, leading to the hierarchy $H_- \gg H_+$, in BSM models with right-handed currents this does not necessarily have to be the case. In fact, LHCb reported recently the first direct observation (with $5.2\sigma$ significance) that the photon is not unpolarised in $b \to s\gamma$ through a measurement of angular correlations in $B^\pm \to K^\pm\pi^\mp\pi^\pm\gamma$ [489]. This raises the question by how much Belle II can improve on this and future LHCb measurements. Concerning the sensitivity of the photon polarisation to new physics, one should compare the prospects that exclusive $b \to s\gamma$ measurements have to those that arise from $B \to K^*\ell^+\ell^-$. Relevant articles in this context are for instance [490–493].

The branching ratios for $B \to V\gamma$ decays are proportional to $|H_+|^2 + |H_-|^2$, where the form factor $T_1(0)$ in (227) provide a major part of the theoretical uncertainties. Numerically, they are estimated to be of $\mathcal{O}(4 \cdot 10^{-5})$ for the $b \to s$ transitions, while those for the $b \to d$ transitions are further suppressed by a factor of $\lambda^2 \simeq 0.05$. In contrast, WA turns out to be sizeable for the $b \to d\gamma$ modes [494] as a result of the CKM hierarchies (201).

*Observables.* Because of the rather large hadronic uncertainties of more than 20%, the branching ratios $B \to V\gamma$ are not considered to be the most promising candidates for discovering BSM physics. On the other hand, since the uncertainties of individual modes are strongly correlated, considering ratios of branching ratios such as $R_{K^*\gamma/\phi\gamma} = \mathrm{Br}(B \to K^*\gamma)/\mathrm{Br}(B_s \to \phi\gamma)$ is advantageous both from a theoretical and experimental point of view. The SM prediction for this ratio reads [488][25]

$$R^{\rm SM}_{K^*\gamma/\phi\gamma} = 0.78 \pm 0.18\,, \tag{228}$$

while the LHCb collaboration measured $R^{\rm exp}_{K^*\gamma/\phi\gamma} = 1.23 \pm 0.12$ [495, 496]. The observed deviation of $2\sigma$ cannot be regarded as statistically significant, but it would be interesting to understand if there can be a correlation to the discrepancies observed by LHCb in $B \to K^*\mu^+\mu^-$ and $B_s \to \phi\mu^+\mu^-$ (see e.g. [388, 389, 497–499]). Another ratio of interest is $R_{\rho\gamma/K^*\gamma}$, which has been used for the first determinations of $|V_{td}/V_{ts}|$ [482, 484, 500]. After the precision measurements of $B_s$–$\bar{B}_s$ mixing, the extractions of $|V_{td}/V_{ts}|$ via $R_{\rho\gamma/K^*\gamma}$ are however no longer competitive.

Other observables which are sensitive to BSM contributions to (200) are the IAs, and the direct and indirect $CP$-asymmetries. The IAs can be defined as

$$a_I^{\bar{0}-} = \frac{c_V^2\,\Gamma(\bar{B}^0 \to \bar{V}^0\gamma) - \Gamma(B^- \to V^-\gamma)}{c_V^2\,\Gamma(\bar{B}^0 \to \bar{V}^0\gamma) + \Gamma(B^- \to V^-\gamma)}\,, \tag{229}$$

where $c_{\rho^0} = \sqrt{2}$ and $c_{K^{*0}} = 1$ are isospin-symmetry factors. The IAs are essentially driven by two effects, both of them involving LD physics: $(i)$ photon emission from the spectator quark which probes the different charge factors for $u$-quarks and $d$-quarks and $(ii)$ matrix elements of isotriplet combinations of hadronic operators in the effective Hamiltonian (200). In order to accumulate more statistics one can define $CP$-averaged IAs through $\bar{a}_I = (a_I^{\bar{0}-} + a_I^{0+})/2$. Subtleties concerning the $CP$-averaging of the IAs are discussed in [488].

Early analyses of the IAs in the framework of QCDF can be found in [482, 484, 501]. It turns out that the dominant SM contribution to (229) for $B \to K^*\gamma$ arises as a subleading effect in the HQE and involves the Wilson coefficients of $Q_{1-6}$. Compared to this, the effect of $Q_8$ is numerically suppressed, but in QCDF suffers from endpoint divergences of convolution integrals, which leads to rather large uncertainties. The problem of endpoint divergences can be avoided by determining the relevant matrix elements directly in the LCSR approach which has been performed for the contributions of $Q_8$ in [485] and for the QL topologies in [488].

For exclusive $b \to d\gamma$ transitions, the situation is somewhat different because the current-current operators $Q^u_{1,2}$ enter with unsuppressed CKM factors $\lambda_u^{(d)}$. Their relatively large annihilation contribution thus interferes with the naively factorising contribution from the electromagnetic operator $Q_7$ proportional to $\lambda_t^{(d)}$. The resulting strong dependence of the IA of $B \to \rho\gamma$ on $\cos\phi_2$ was noted in [482, 484] where approximate formulas can be found.

[25] The quoted theory uncertainty is improvable as correlations have only partially been taken into account in [488].

The most up-to-date theoretical predictions for the IAs are [488]

$$\bar{a}_I^{\rm SM}(K^*\gamma) = (4.9 \pm 2.6)\,\% \,, \qquad \bar{a}_I^{\rm SM}(\rho\gamma) = (5.2 \pm 2.8)\,\% \,. \tag{230}$$

Notice that the former prediction is consistent with the HFLAV average $\bar{a}_I^{\rm exp}(K^*\gamma) = (6.3 \pm 1.7)\%$ [230], whereas the latter is in slight tension $\bar{a}_I^{\rm exp}(\rho\gamma) = (30^{-13}_{+16})\%$, albeit with considerable uncertainty. Notice that HFLAV uses the definition $\Delta_{\rho\gamma} = -2\bar{a}_I(\rho\gamma)/\left(1+\bar{a}_I(\rho\gamma)\right)$ instead of $\bar{a}_I(\rho\gamma)$. The closeness of the two values in (230) is a consequence of the CKM angle $\phi_2$ being roughly $90^\circ$ which suppresses the above-mentioned interference term. This can be exploited to define the observable [488]

$$1 - \delta_{a_I} = \frac{\bar{a}_I(\rho\gamma)}{\bar{a}_I(K^*\gamma)} \sqrt{\frac{\bar{\Gamma}(B \to \rho\gamma)}{\bar{\Gamma}(B \to K^*\gamma)}} \left| \frac{V_{ts}}{V_{td}} \right| \,, \tag{231}$$

where $\delta_{a_I}$ is close to zero, and the quantity $(1-\delta_{a_I})^{\rm SM} = 0.90 \pm 0.11$ shows a reduced uncertainty with respect to the individual $CP$-averaged IAs introduced in (230). The experimental average $\delta_{a_I}^{\rm exp} = -4.0 \pm 3.5$ [488] can be improved at Belle II through more statistics as well as taking into account experimental correlations. The sensitivity of (231) to BSM physics has been studied in [488] in a model-independent fashion.

At Belle II, one can study the time-dependent $CP$ asymmetries [502]

$$A_{\rm CP}(t) = \frac{\Gamma(\bar{B} \to f\gamma) - \Gamma(B \to f\gamma)}{\Gamma(\bar{B} \to f\gamma) + \Gamma(B \to f\gamma)} = \frac{S_{f\gamma}\sin(\Delta m_q t) - C_{f\gamma}\cos(\Delta m_q t)}{\cosh\left(\frac{\Delta\Gamma_q t}{2}\right) - H_{f\gamma}\sinh\left(\frac{\Delta\Gamma_q t}{2}\right)} \,, \tag{232}$$

where $f$ ought to be a $CP$ eigenstate as otherwise the effect washes out. Note that the width difference $\Delta\Gamma_q$ can be safely neglected for $B_d$ but that is not the case for $B_s$. This feature leads to the new observables $H_{f\gamma}$ [503]. The mixing-induced asymmetries $S_{f\gamma}$ arise from the interference between $B(\bar{B}) \to f\gamma$ and $B(\bar{B}) \to \bar{B}(B) \to f\gamma$ amplitudes and read

$$S_{V\gamma} = \frac{2\,\xi_V {\rm Im}\left[\frac{q}{p}\left(\bar{H}_+ H_+^* + \bar{H}_- H_-^*\right)\right]}{|H_+|^2 + |H_-|^2 + |\bar{H}_+|^2 + |\bar{H}_-|^2} \,, \tag{233}$$

where $\xi_V$ is the $CP$ eigenvalue of $V$, $p$, $q$ relate the physical and flavour eigenstates, $H_\pm$ have been defined in (227), and $\bar{H}_\pm$ are the corresponding amplitudes of the conjugate decay. At Belle II, one can expect a significant improvement in the determination of $A_{\rm CP}(t)$ in the channels such as $f = K_S^0\pi^0, \pi^+\pi^-$ mediated by $K^*$ and $\rho$ resonances, which will be discussed in some more detail.

Before embarking on the discussion of LD contributions, we first give predictions for (233) including SD effects only. Using $q/p \simeq e^{-2i\phi_1}$, one obtains

$$S^{\rm SM,SD}_{K^*(K_S^0\pi^0)\gamma} = -2\,\frac{m_s}{m_b}\sin 2\phi_1 \,, \qquad S^{\rm SM,SD}_{\rho^0(\pi^+\pi^-)\gamma} = 0 \,. \tag{234}$$

Numerically, $S^{\rm SM,SD}_{K^*(K_S^0\pi^0)\gamma} = \mathcal{O}(-3\%)$ while the quantity $S^{\rm SM,SD}_{\rho^0(\pi^+\pi^-)\gamma}$ vanishes because the $CP$-odd oscillation phase $\phi_1$ cancels exactly against the phase from the helicity amplitude. Examples of BSM models which can induce sizeable right-handed currents consistent with the constraint from ${\rm Br}(B \to X_s\gamma)$ include left-right symmetric models [502, 504, 505] and a supersymmetric (SUSY) $SU(5)$ grand unified theory with right-handed neutrinos [505].

A model-independent study can be found in [490]. In the presence of a right-handed magnetic penguin operator $Q_7'$, one obtains for instance

$$S^{\rm SD}_{K^*(K^0_S\pi^0)\gamma} \simeq \frac{{\rm Im}\left[e^{-2i\phi_1}\left(C_7^* C_7' + C_7 C_7'^*\right)\right]}{|C_7|^2 + |C_7'|^2}\,. \tag{235}$$

LD QCD contributions denoted by $L_U$ in (227) modify the predictions (234) and arise first at $\mathcal{O}(\Lambda_{\rm QCD}/m_b)$. The dominant corrections are expected to stem from $c$-quark loops [506], because such effects are due to the current-current operators $Q_{1,2}$ in (200) that have large Wilson coefficients. By using the corresponding contribution of the inclusive decays it has been concluded in the latter work that the LD contamination in (234) could be as large as 10%. By performing a kinematic decomposition it can however be shown that $H_- \gg H_+$ holds at leading twist for any local transition operator [485, 507]. The hierarchy of helicity amplitudes can therefore only be broken by higher-twist effects, and one such contributions comes from gluon exchange between the $c$-quark loop and the vector meson. An explicit evaluation of the LD corrections due to $c$-quark loops [494, 503, 508] yields a correction of $\mathcal{O}(1\%)$, which is considerably smaller than the inclusive calculation would suggest (see also [509]).[26] Further evidence for the smallness of LD $c$-quark effects arises from the fact that the corrections to the helicity hierarchy are of $\mathcal{O}(m_V^2/m_b^2)$. This indicates that the hierarchy is more badly broken by excited (i.e. heavier) vector meson states. Vertex corrections are treated in QCDF [482, 483] and automatically obey $H_- \gg H_+$. The evaluation of the vertex corrections beyond factorisation is challenging and remains a future task. Including both SD and LD contributions, the quantities in (234) turn into [494, 511]

$$S^{\rm SM}_{K^*(K^0_S\pi^0)\gamma} = (-2.3 \pm 1.6)\%\ , \qquad S^{\rm SM}_{\rho^0(\pi^+\pi^-)\gamma} = (0.2 \pm 1.6)\%\,. \tag{236}$$

The photon polarisation is one of the most challenging measurements in $B$ physics today and various modes have been proposed to further improve the precision — see [490] for more details. LHCb has already applied many of the proposed methods and Belle II should be able to further extend these studies. For instance at Belle II it should be possible to expand the recent LHCb analysis [489] of angular correlations in $B^\pm \to K^\pm\pi^\mp\pi^\pm\gamma$ [512, 513] by including the neutral modes as well as performing a Dalitz analysis [514]. The angular analysis of $B \to K^* e^+e^-$ has been performed by LHCb [515] at very low $q^2$ where the photonic dipole operator $Q_7$ and its chirality-flipped partner $Q_7'$ dominate. A similar analysis should be possible at Belle II and furthermore, the use of the angular distribution of the converted photon from $B \to K^*\gamma$ is under discussion [516].

The direct $CP$ asymmetries $C_{f\gamma}$ require weak $CP$-odd and strong $CP$-even phase differences of two amplitudes and are therefore by default sensitive to $CP$-odd phases beyond the SM. $CP$-even phases instead originate from LD QCD effects. In the SM the direct $CP$ asymmetry for $b \to s\gamma$ is small, since there is no $CP$-odd phase at $\mathcal{O}(\lambda^3)$. These observables can thus serve as null-tests. As an example we quote $C^{\rm SM}_{\phi(\to KK)\gamma} = (0.5 \pm 0.5)\,\%$ from [503]. For the $b \to d\gamma$ modes on the other hand the $t$-quark loop diagram induces a sizeable $CP$-odd phase. For example, in [517] a direct $CP$ asymmetry of 15% is predicted for $B_d \to \pi^+\pi^-\gamma$ within the SM.

[26] It was recently proposed in [510] that the LD contributions entering $H_+$ can be controlled by considering both the $B \to V(1^-)\gamma$ and the $B \to A(1^+)\gamma$ decay with $V$ and $A$ nearly degenerate states, such as the $\rho$ and the $a_1$ meson. By taking the sum $S_{B\to\rho\gamma} + S_{B\to a_1\gamma}$, one measures the LD contributions entering $H_+$, whereas the difference can measure the new-physics contribution with considerably improved precision depending on calculable ratios of LD effects. These methods also extend to the low-$q^2$ region of $B \to V\ell^+\ell^-$, with particular promise for the electron channels.

*9.2.5. Measurement of $B \to V\gamma$ decays.* *(Contributing author: A. Ishikawa)*

The $b \to s\gamma$ transition was first observed by CLEO via $B \to K^*\gamma$ in 1993 [518]. Two decades later this decay is still important in the search for new physics. The three most important observables in this channel are the photon polarisation, the isospin and the $CP$ asymmetries.

The $K^*$ mesons are reconstructed from either of the $K^-\pi^0$, $K^0_S\pi^-$, $K^-\pi^+$ and $K^0_S\pi^0$ decays. The $B$-meson candidate is reconstructed by combining the $K^*$ candidate and a hard photon reconstructed from an electromagnetic cluster in the electromagnetic calorimeter (ECL) which is not associated with any charged tracks in the tracking system. Exclusive modes are much cleaner than the fully-inclusive mode thanks to requirements imposed on the difference in energy, $\Delta E$, and the beam-constrained mass, $M_{bc}$. The $K^-\pi^0$, $K^0_S\pi^-$ and $K^-\pi^+$ modes are flavour eigenstates which can be used for measurements of $A_{\rm CP}$ while $K^0_S\pi^0$ with flavour tagging of the other $B$ meson can be used to measure the time-dependent $CP$ asymmetry (232) which is sensitive to the polarisation of the final-state photon.[27]

At Belle II with $5\,\mathrm{ab}^{-1}$ of data the measurement of $\bar{a}_I(K^*\gamma)$ may already be systematically limited. The dominant uncertainty is due to $f_{+-}/f_{00}$ and amounts to 0.5%. Notice that this uncertainty is smaller by a factor of five than that of the most up-to-date SM prediction (230). Measurements of the direct $CP$ asymmetries will instead still be statistically limited. The corresponding uncertainties are estimated to be 0.2% and 0.3% for $A_{\rm CP}(B^0 \to K^{*0}\gamma)$ and $A_{\rm CP}(B^+ \to K^{*+}\gamma)$, respectively, which constitutes a factor of eight improvement compared to the Belle result [523]. Notice that the theoretical uncertainty of the corresponding SM prediction $A_{\rm CP}^{\rm SM}(B^0 \to K^{*0}\gamma) = (0.3 \pm 0.1)\%$ [493] is smaller than the statistical uncertainty reachable at Belle II. A precision measurement of $A_{\rm CP}(B^0 \to K^{*0}\gamma)$ is nevertheless an important goal since it will allow to set stringent constraints on the imaginary part of the Wilson coefficient of $Q_7$ [493, 524], which otherwise is difficult to bound. Like $A_{\rm CP}$ also the measurement of $\Delta A_{\rm CP}$ will be statistically limited at Belle II and the projected uncertainty amounts to 0.4% with $50\,\mathrm{ab}^{-1}$ of luminosity.

The $b \to d\gamma$ process was first observed in 2006 [525] by Belle through the exclusive $B \to \rho\gamma$ and $B^0 \to \omega^0\gamma$ decays. All the branching ratios, isospin asymmetries, direct and time-dependent $CP$ asymmetries have been measured subsequently [526–528], but the achieved precision is not high enough to set stringent limits on new physics. This lack in precision is unfortunate since the measured value of $\bar{a}_I(\rho\gamma)$ shows a slight tension with the SM prediction, a fact that has already been mentioned in the context of (230). Thanks to the good particle identification (PID) system and the large integrated luminosity to be recorded at Belle II, precise measurement of $B \to (\rho, \omega)\gamma$ will be possible for the first time, which is crucial in view of the aforementioned tension.

[27] At Belle, the time-dependent $CP$ asymmetries were measured with $B \to K^*(K^0_S\pi^0)\gamma$ [519], $B \to K^0_S\eta\gamma$ [520], $B \to K^0_S\pi^+\pi^-\gamma$ [521] and $B \to K^0_S\phi\gamma$ [522].

Table 62: Sensitivities of observables for radiative exclusive $B$ decays. We assume that $5\,\mathrm{ab}^{-1}$ of data will be taken on the $\Upsilon(5S)$ resonance by Belle II. Some numbers at Belle are extrapolated to $0.71\,\mathrm{ab}^{-1}$ ($0.12\,\mathrm{ab}^{-1}$) for the $B_{u,d}$ ($B_s$) decay. As in Table 61 the quoted uncertainties are depending on the observable either relative or absolute ones.

| Observables | Belle $0.71\,\mathrm{ab}^{-1}$ ($0.12\,\mathrm{ab}^{-1}$) | Belle II $5\,\mathrm{ab}^{-1}$ | Belle II $50\,\mathrm{ab}^{-1}$ |
|---|---|---|---|
| $\Delta_{0+}(B \to K^*\gamma)$ | 2.0% | 0.70% | 0.53% |
| $A_{CP}(B^0 \to K^{*0}\gamma)$ | 1.7% | 0.58% | 0.21% |
| $A_{CP}(B^+ \to K^{*+}\gamma)$ | 2.4% | 0.81% | 0.29% |
| $\Delta A_{CP}(B \to K^*\gamma)$ | 2.9% | 0.98% | 0.36% |
| $S_{K^{*0}\gamma}$ | 0.29 | 0.090 | 0.030 |
| $\mathrm{Br}(B^0 \to \rho^0\gamma)$ | 24% | 7.6% | 4.5% |
| $\mathrm{Br}(B^+ \to \rho^+\gamma)$ | 30% | 9.6% | 5.0% |
| $\mathrm{Br}(B^0 \to \omega\gamma)$ | 50% | 14% | 5.8% |
| $\Delta_{0+}(B \to \rho\gamma)$ | 18% | 5.4% | 1.9% |
| $A_{CP}(B^0 \to \rho^0\gamma)$ | 44% | 12% | 3.8% |
| $A_{CP}(B^+ \to \rho^+\gamma)$ | 30% | 9.6% | 3.0% |
| $A_{CP}(B^0 \to \omega\gamma)$ | 91% | 23% | 7.7% |
| $\Delta A_{CP}(B \to \rho\gamma)$ | 53% | 16% | 4.8% |
| $S_{\rho^0\gamma}$ | 0.63 | 0.19 | 0.064 |
| $\lvert V_{td}/V_{ts}\rvert_{\rho/K^*}$ | 12% | 8.2% | 7.6% |
| $\mathrm{Br}(B_s^0 \to \phi\gamma)$ | 23% | 6.5% | – |
| $\mathrm{Br}(B^0 \to K^{*0}\gamma)/\mathrm{Br}(B_s^0 \to \phi\gamma)$ | 23% | 6.7% | – |
| $\mathrm{Br}(B_s^0 \to K^{*0}\gamma)$ | – | 15% | – |
| $A_{CP}(B_s^0 \to K^{*0}\gamma)$ | – | 15% | – |
| $\mathrm{Br}(B_s^0 \to K^{*0}\gamma)/\mathrm{Br}(B_s^0 \to \phi\gamma)$ | – | 15% | – |
| $\mathrm{Br}(B^0 \to K^{*0}\gamma)/\mathrm{Br}(B_s^0 \to K^{*0}\gamma)$ | – | 15% | – |

The $\rho$ and $\omega$ mesons are reconstructed from two-pion and three-pion final states. Hard photon candidates are combined with the light mesons to form $B$-meson candidates. A dominant continuum background can be suppressed by a multivariate analysis with event shape variables. The large $b \to s\gamma$ background which peaks in $\Delta E$ and $M_{bc}$ can be significantly suppressed by the new PID system, using the iTOP for the barrel region and the ARICH for the forward endcap region.

Assuming that that the current central experimental value of $\bar{a}_I(\rho\gamma)$ is confirmed, Belle II can observe a $5\sigma$ deviation from the SM prediction already with $6\,\mathrm{ab}^{-1}$. With $50\,\mathrm{ab}^{-1}$ of data the statistical uncertainty (1.7%) will dominate the measurement with the largest systematic uncertainties arising from $f_{+-}/f_{00}$ (0.5%) and background modelling (0.5%). In total a precision of 1.9% on $\bar{a}_I(\rho\gamma)$ will be achievable at Belle II, which compares favourably with the current theoretical SM uncertainty of 2.8% as quoted in (230).

The $CP$ asymmetries in the case of charged and neutral $B$ mesons are measured in different ways. The mode $B^+ \to \rho^+\gamma$ is self-flavour tagging thus allowing for a straightforward measurement of the direct $CP$ asymmetry. In contrast, $B^0 \to \rho^0\gamma$ is not a flavour eigenstate, yet a time-dependent measurement with flavour tagging will allow to extract both $A_{\mathrm{CP}}$ and the $S$ parameter. With $50\,\mathrm{ab}^{-1}$ of data one can expect to reach a precision of 3.0%, 3.8% and 6.4% for $A_{\mathrm{CP}}(B^+ \to \rho^+\gamma)$, $A_{\mathrm{CP}}(B^0 \to \rho^0\gamma)$ and $S_{\rho^0\gamma}$, respectively.

The magnitude of the ratio $V_{td}/V_{ts}$ of CKM matrix elements can be extracted by measuring appropriate ratios of branching ratios such as $\mathrm{Br}(B \to (\rho,\omega)\gamma)/\mathrm{Br}(B^0 \to K^*\gamma)$ [494]. However, already with $1\,\mathrm{ab}^{-1}$ of integrated luminosity the resulting uncertainty will be dominated by the theoretical uncertainties.

Radiative $B_s^0$ decays can be also studied at the $\Upsilon(5S)$ resonance. The Belle measurement of the branching ratio of $B_s^0 \to \phi\gamma$ [529] is limited by the uncertainty on the $B_s^0$ production ($f_s\sigma_{b\bar{b}}$) at the $\Upsilon(5S)$ resonance, which amounts to about 17%. The current precision of the world average of $f_s$ is dominated by the Belle measurement of the inclusive $B_s^0 \to D_{(s)}X$ decay [530] that uses $1.9\,\mathrm{fb}^{-1}$ of data at the $\Upsilon(5S)$ resonance. This measurement can be improved at Belle II with a few different approaches, namely the dilepton method, exclusive decays in $B_s^0$ tagged and untagged events as well as inclusive $B_s$ decays. Assuming that 4% precision on $f_s$ is achieved at Belle II, the sensitivity of $\mathrm{Br}(B_s^0 \to \phi\gamma)$ will be 6.5%, which is still dominated by the uncertainty on $f_s\sigma_{b\bar{b}}$.

The $B_s^0 \to K^{*0}\gamma$ decay mode was not searched for yet. The reconstruction of this decay is almost the same as for $B^0 \to K^{*0}\gamma$ and thus straightforward to perform. The $b \to s$ counterpart, i.e. $B_s^0 \to \phi\gamma$, serves as a peaking background, which however can be eliminated by studying the invariant mass of the hadronic system under a kaon-mass assumption as well as using the good PID information of Belle II. Other possible peaking backgrounds from $B_s^0 \to K^{*0}\pi^0/\eta$ with asymmetric decays of $\pi^0/\eta$ are also not measured yet. These can be suppressed by a $\pi^0/\eta$ veto and by examining the helicity angle distribution of the $K^{*0}$. The $B_s^0 \to K^{*0}\gamma$ decay can be observed at Belle II with an integrated luminosity of $3.5\,\mathrm{ab}^{-1}$, and the achievable precision on the branching ratio can be expected to be 15% with $5\,\mathrm{ab}^{-1}$. The ratios of the branching ratios and the direct CP asymmetries can also be measured with the same precision.

A summary of the Belle II sensitivities for the various exclusive $B \to V\gamma$ channels is provided in Table 62.

### *9.2.6. Importance of PID for $b \to d\gamma$.* *(Contributing author: S. Cunliffe)*

In both the inclusive and exclusive transition analyses, PID information plays an important role. Specifically PID is necessary to reduce the problematic background originating from misidentified kaons from $B \to X_s\gamma$ processes. To give a relevant example consider the case of $B^0 \to K^{*0}\gamma$ followed by $K^{*0} \to K^+\pi^-$. The latter decay rate is roughly by a factor of 30 larger than the dominant $b \to d\gamma$ process, i.e. $B^0 \to \rho^0\gamma$ with $\rho^0 \to \pi^+\pi^-$, meaning that a good PID is necessary to be able to separate signal from background.

A study based on the full Belle II simulation is performed to quantify the performance of the PID system. Samples of 1 million events of both $B^0 \to \rho^0\gamma$ and $B^0 \to K^{*0}\gamma$ are generated. After performing a full detector reconstruction a simple pre-selection criteria is applied to both samples. An optimisation for a cut on the pion probability (defined in Section 5.5) is performed to maximise the figure of merit, $S/\sqrt{S+B}$. Here $S$ is the number of correctly identified $B^0 \to \rho^0\gamma$ events, and $B$ is the number of $B^0 \to K^{*0}\gamma$ where the kaon track was mis-reconstructed as a pion. Both $S$ and $B$ are scaled to the expected number of events in $5\,\text{ab}^{-1}$ of data. The value of the optimal selection cut is found to give a figure of merit well above 10.

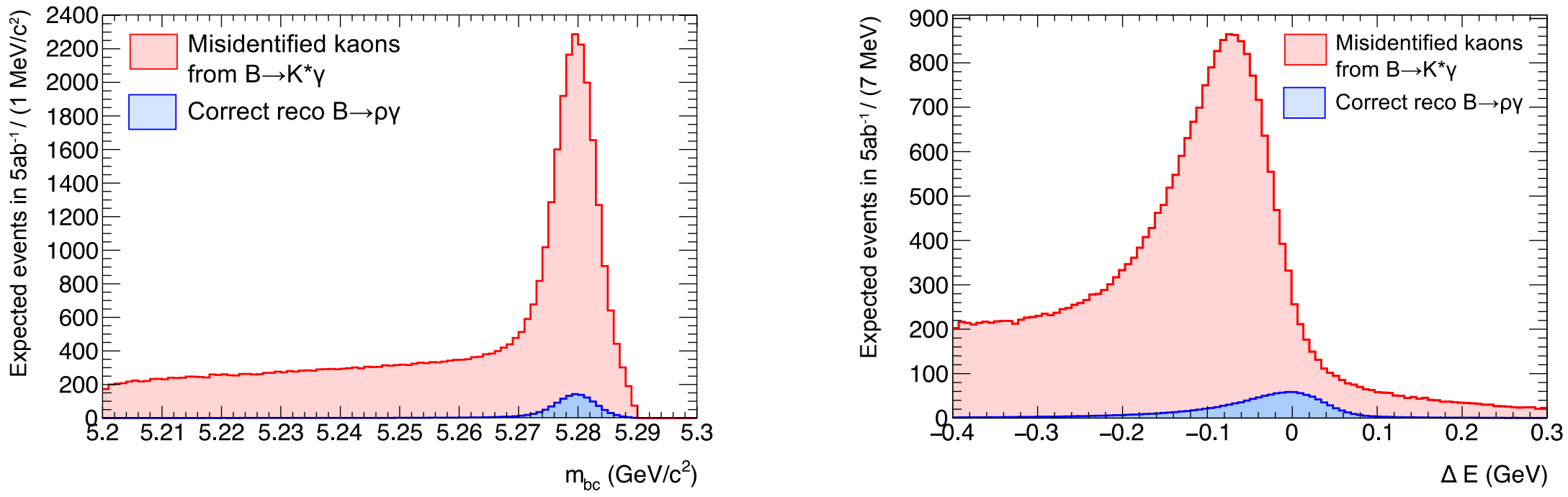

Fig. 90: Distributions of $M_{bc}$ and $\Delta E$ for correctly identified $B^0 \to \rho^0\gamma$ signal events (blue) overlaid with misidentified $B \to K^*\gamma$ where the kaon from the $K^{*0}$ decay is mis-reconstructed as a pion (red). With no PID selection cut the background swamps the signal.

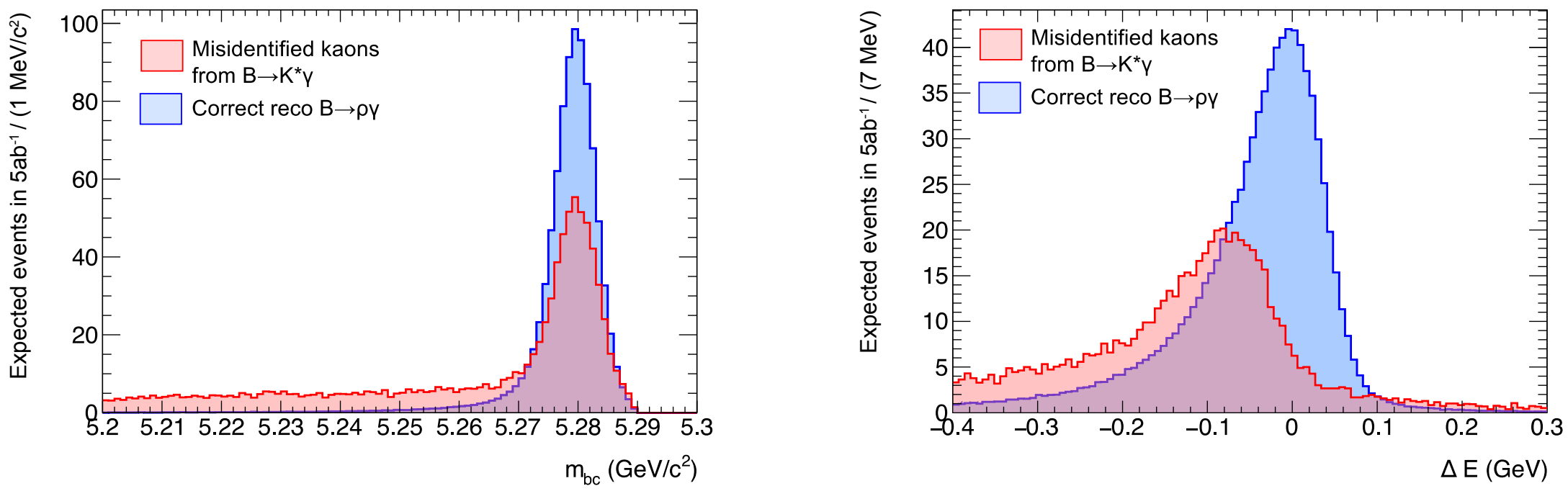

Fig. 91: Same as Figure 90 but employing PID information. Distributions of $M_{bc}$ and $\Delta E$ for correctly identified $B^0 \to \rho^0\gamma$ signal events (blue) overlaid with misidentified $B \to K^*\gamma$ where the kaon from the $K^{*0}$ decay is mis-reconstructed as a pion (red). After a simple optimisation of PID selection, the background is reduced significantly.

Figures 90 and 91 show overlaid distributions of the beam constrained-mass, $M_{bc}$, and energy difference, $\Delta E$, for both samples before, and after the selection cut at the optimal point. The importance of PID is evident from the two figures.

The above study is repeated using a simulation of the Belle detector, in order to compare to the associated Belle PID performance. The Belle optimisation is performed for the analogous PID likelihood variables described in Section 5.2.1 of [2]. The Belle II PID system is found to provide an improvement in the figure of merit by approximately 30%.

### *9.3. Double-Radiative Decays*

*(Contributing authors: C. Bobeth and A. Kokulu)*

*9.3.1. $B_q \to \gamma\gamma$ Decays.* In the SM, the branching ratios of the $B_q \to \gamma\gamma$ decays scale as the involved CKM elements $|V_{td}|^2$ and $|V_{ts}|^2$, predicting an enhancement of the $B_s \to \gamma\gamma$ decay over the $B_d \to \gamma\gamma$ decay by a factor of $|V_{ts}/V_{td}|^2 \simeq 20$. Using the full data set at $\Upsilon(5S)$ [529], Belle obtained the following 90% CL upper limit

$$\mathrm{Br}(B_s \to \gamma\gamma)_{\mathrm{exp}} < 3.1 \cdot 10^{-6}\,, \tag{237}$$

on the branching ratio of $B_s \to \gamma\gamma$. The searches for $B_d \to \gamma\gamma$ at $\Upsilon(4S)$ resulted instead in the 90% CL upper limits

$$\mathrm{Br}(B_d \to \gamma\gamma)_{\mathrm{exp}} < \begin{cases} 3.2 \cdot 10^{-7}\,, \\ 6.2 \cdot 10^{-7}\,, \end{cases} \tag{238}$$

from the full data set of BaBar [531], and a partial data set of $104\,\mathrm{fb}^{-1}$ of Belle [532] out of the available $711\,\mathrm{fb}^{-1}$. The corresponding SM predictions are given by [533]

$$\begin{aligned} \mathrm{Br}(B_s \to \gamma\gamma)_{\mathrm{SM}} &\in [0.5, 3.7] \cdot 10^{-6}\,, \\ \mathrm{Br}(B_d \to \gamma\gamma)_{\mathrm{SM}} &\in [1.0, 9.8] \cdot 10^{-8}\,, \end{aligned} \tag{239}$$

and are either close to or only by an order of magnitude below the bounds (237) and (238). The above comparison shows that Belle II will be able to discover $B_d \to \gamma\gamma$ with the anticipated 50 times larger data set at $\Upsilon(4S)$. Furthermore, an appropriately large $\Upsilon(5S)$ data set could provide an observation of $B_s \to \gamma\gamma$.

From a theoretical point of view, double radiative $B_q \to \gamma\gamma$ decays are complementary to the corresponding radiative inclusive $B \to X_q\gamma$ decay. They depend on the same Wilson coefficient $C_7$ of the photonic dipole operator (202), but the contribution of four-quark operators in $B_q \to \gamma\gamma$ is different compared to $B \to X_q\gamma$. This feature provides a complementary test of the Wilson coefficient $C_7$ which plays an important role in many BSM models.

As will be explained in more detail below, the main source of theoretical uncertainty in the QCDF approach arises due to the first negative moment, $\lambda_B$, of the $B$-meson distribution amplitude. This hadronic parameter can be determined from the radiative leptonic decay $B \to \ell\bar{\nu}\gamma$ [223, 246]. For the definition and a detailed discussion of the phenomenological impact on two-body hadronic decays, see Section 8.

The amplitude of the $\bar{B} \to \gamma(k_1, \epsilon_1)\,\gamma(k_2, \epsilon_2)$ decays — hereafter $B$ stands for both $B_q$ — has the general structure

$$\mathcal{A} = A_+ \left[2\,(k_1 \cdot \epsilon_2)\,(k_2 \cdot \epsilon_1) - m_B^2(\epsilon_1 \cdot \epsilon_2)\right] - A_-\, 2i\varepsilon_{\mu\nu\alpha\beta}\, k_1^\mu k_2^\mu \epsilon_1^\alpha \epsilon_2^\beta\,. \tag{240}$$

The $CP$ properties of the corresponding two-photon final states are indicated by the subscripts $\pm$ on the amplitudes $A_\pm$. The parallel spin polarisation of the photons is described by $A_+$, whereas the perpendicular one is encoded in $A_-$.

The decay rate is obtained after summation over photon polarisations

$$\Gamma(\bar{B} \to \gamma\gamma) = \frac{m_B^3}{16\pi}\left(|A_+|^2 + |A_-|^2\right). \tag{241}$$

In the absence of methods to tag the flavour of the initial $B$ meson, the $CP$-averaged branching ratio must be considered instead

$$\overline{\mathrm{Br}}_{\gamma\gamma} = \frac{\tau_B}{2}\left[\Gamma(\bar{B} \to \gamma\gamma) + \Gamma(B \to \gamma\gamma)\right], \tag{242}$$

where $\Gamma(B \to \gamma\gamma)$ is determined from (241) using the amplitudes $\bar{A}_\pm$ of the $CP$-conjugated decay $B \to \gamma\gamma$. Further, for the case of untagged $B_s$ decays the sizeable decay width leads to rapid mixing and requires to perform a time-integration [534] in order to obtain the experimentally measured $CP$-averaged and time-integrated branching ratio

$$\langle \overline{\text{Br}}_{\gamma\gamma} \rangle = \frac{1 + y_s A_{\Delta\Gamma}}{1 - y_s^2} \, \overline{\text{Br}}_{\gamma\gamma} \, . \tag{243}$$

It depends on $y_s = \Delta\Gamma_s/(2\Gamma_s) = 0.075 \pm 0.012$, where $\Gamma_s = 1/\tau_{B_s}$ the inverse of the lifetime, the $CP$-averaged branching ratio (242) at time $t = 0$ and the mass-eigenstate rate asymmetry $A_{\Delta\Gamma}$ in $B \to \gamma\gamma$. The latter quantity can in principle be determined in an untagged but time-dependent analysis via a measurement of the effective lifetime [534]. Notice that the prefactor appearing on the right-hand side of (243) amounts to around $[0.9, 1.1]$ if $A_{\Delta\Gamma}$ is varied in its physically allowed range of $[-1, 1]$. Given the sizeable uncertainties quoted in (239), one can hence simply use $\langle \overline{\text{Br}}_{\gamma\gamma} \rangle = \overline{\text{Br}}_{\gamma\gamma}$ when comparing to future Belle II data.

Direct $CP$ violation can be tested by a tagged measurement of

$$\begin{aligned} r_{\text{CP}} &= \frac{|\mathcal{A}|^2 - |\bar{\mathcal{A}}|^2}{|\mathcal{A}|^2 + |\bar{\mathcal{A}}|^2} \, , \\ r_{\text{CP}}^{\pm} &= \frac{|A_\pm|^2 - |\bar{A}_\pm|^2}{|A_\pm|^2 + |\bar{A}_\pm|^2} \, , \end{aligned} \tag{244}$$

where extractions of $r_{\text{CP}}^{\pm}$ also require the determination of the photon polarisations.

A systematic analysis of these decays in the heavy quark limit $m_b \gg \Lambda_{\text{QCD}}$ has been first given in [533]. In this limit, the hadronic matrix elements of operators $Q_i$ of the effective Hamiltonian (200) factorise

$$\langle \gamma\gamma | Q_i | \bar{B} \rangle = f_B \int_0^1 d\omega \, T_i^{\mu\nu}(\omega) \, \phi_B^+(\omega) \, \epsilon_{1\mu} \epsilon_{2\nu} \, . \tag{245}$$

The $T_i^{\mu\nu}$ are perturbatively calculable SD functions, whereas the non-perturbative effects are contained in the $B$-meson decay constant $f_B$ and the leading light-cone distribution amplitude (LCDA) of the $B$ meson in HQET, denoted as $\phi_B^+$. The latter depend on the light-cone momentum $\omega$ of the spectator quark inside the $B$ meson. Within the QCD factorisation setup [533], only the first negative moment,

$$\frac{1}{\lambda_B} = \int_0^1 d\omega \, \frac{\phi_B^+(\omega)}{\omega} \, , \tag{246}$$

of the LCDA of the $B$ meson appears.

The leading-power contribution arises from the emission of a hard photon from the $B$-meson spectator quark for the matrix element of the photonic dipole operator $Q_7$,

$$\begin{aligned} A_\pm &= \frac{G_F}{\sqrt{2}} \frac{\alpha}{3\pi} f_B \sum_{p=u,c} \lambda_p^{(q)} A_\pm^p \, , \\ A_\pm^p &= -C_7^{\text{eff}} \, \frac{m_B}{\lambda_B} \, , \end{aligned} \tag{247}$$

where $C_7^{\text{eff}}$ is the effective coupling of this operator at the low-energy scale $\mu_b$. At this order in the power expansion, one has $(r_{\text{CP}}^{\pm})_{\text{SM}} = 0$. Furthermore, since $\overline{\text{Br}}_{\gamma\gamma} \propto (f_B/\lambda_B)^2$ and given the accuracy of lattice predictions for $f_B$, in the case of the branching ratios the main theoretical uncertainty comes from $\lambda_B$.

At the subleading order in the power expansion, there are two types of contributions to the matrix element of $Q_7$: $(i)$ higher-twist contributions and $(ii)$ the one-particle reducible (1PR) diagram where the photon is emitted from the $b$-quark line. Both corrections naively represent a correction of $\mathcal{O}(\Lambda_{\rm QCD}/m_b) = \mathcal{O}(10\%)$ and have so far been neglected in the theoretical predictions. One-particle-irreducible (1PI) contributions of the four-quark operators in the effective Hamiltonian (200) also arise at $\mathcal{O}(\Lambda_{\rm QCD}/m_b)$. The corresponding matrix elements were shown to factorise in the heavy-quark limit to NLO in QCD, leading to $\langle\gamma\gamma|Q_i|\bar{B}\rangle = f_B T_i^{\mu\nu}\epsilon_{1\mu}\epsilon_{2\nu}$, independent of $\omega$. Numerically the largest contributions stem from the current-current operators $Q_{1,2}^p$. They give an additional contribution to the coefficient $A_-^p$ appearing in (247). One obtains

$$A_-^p = -C_7^{\rm eff}\,\frac{m_B}{\lambda_B} - \frac{2}{3}\,(C_1^p + N_c C_2^p)\,g(z_p), \tag{248}$$

where $C_{1,2}^p$ are the Wilson coefficients of $Q_{1,2}^p$ at the scale $\mu_b$ and $N_c = 3$. The function $g(z_p)$ with $z_p = m_p^2/m_b^2$ developes an imaginary part only for $p = c$ when setting $m_u$ to zero, which provides the leading contribution to $r_{\rm CP}^-$. The quantity $r_{\rm CP}^+$ on the other hand still remains zero. The QCD penguin operators $Q_{3-6}$ contribute equally to the $u$-quark and $c$-quark sectors and their overall effect is very small [535]. Feynman graphs that depict all the relevant contributions can be found in Sections 3 and 4 of [533].

Including all relevant effects, the $CP$ asymmetries have been estimated as [533, 536, 537]

$$\begin{aligned} &(r_{\rm CP})_{\rm SM}^s \simeq 0.5\%\,, \qquad &&(r_{\rm CP}^-)_{\rm SM}^s \simeq 0.4\%\,,\\ &(r_{\rm CP})_{\rm SM}^d \simeq -5\%\,, \qquad &&(r_{\rm CP}^-)_{\rm SM}^d \simeq -10\%\,, \end{aligned} \tag{249}$$

within the SM while $(r_{\rm CP}^+)_{\rm SM}^{s,d} \simeq 0\%$. Notice that the predictions for $B_d$ are larger than those for $B_s$ as a result of the CKM hierarchies (201).

The dependence of the branching ratios on $\lambda_B$ cancels almost completely in their ratio, leading to

$$\frac{{\rm Br}(B_s\to\gamma\gamma)_{\rm SM}}{{\rm Br}(B_d\to\gamma\gamma)_{\rm SM}} \simeq \left|\frac{V_{ts}}{V_{td}}\right|^2 \frac{\tau_{B_s} f_{B_s}^2 m_{B_s}^3}{\tau_{B_d} f_{B_d}^2 m_{B_d}^3}\,. \tag{250}$$

Compared to $\lambda_B$, other parametric uncertainties due to the CKM elements and $f_B$ are currently subdominant. Higher-order radiative QCD effects are estimated via factorisation-scale variation to be of $\mathcal{O}(30\%)$, and subleading power corrections are expected to be of $\mathcal{O}(10\%)$ [533].

In BSM models, the $B_q \to \gamma\gamma$ decays can receive two types of non-standard contributions:

$(i)$ Modifications of the Wilson coefficient $C_7$, which will also leave an imprint in $B \to X_q\gamma$.
$(ii)$ Modifications of the 1PI contributions due to four-fermion operators $b \to q f\bar{f}$, where $f$ stands for the five possible light quarks or the three charged leptons.

The first type has been studied in various models such as the two-Higgs-doublet-model of type II (2HDM-II) [538, 539], the minimal supersymmetric SM (MSSM) [540] and universal extra dimensions [541]. However, due to strong constraints on $C_7$ from $B \to X_q\gamma$, large modifications of ${\rm Br}(B_q \to \gamma\gamma)$ are by now already excluded.

The complementarity of $B_q \to \gamma\gamma$ comes therefore mainly from the second type of modifications due to non-standard four-fermion operators $b \to q f\bar{f}$ with vectorial and scalar Dirac structures, which contribute differently to $B_q \to \gamma\gamma$ and $B \to X_q\gamma$ [537], turning it into an interesting probe of such effects. Experimentally least constrained are the $b \to s\tau^+\tau^-$ operators, which have been studied model-independently in [537]. Currently large deviations from $(r_{\rm CP})^s_{\rm SM}$ are still allowed. Concerning the rate it might be enhanced up to a factor of order two, depending also on the exact value of $\lambda_B$, which determines the relative size of four-fermion operators versus the contribution of $Q_7$. Such effects arise for example in SUSY with broken $R$-parity [542] or leptoquark scenarios [537].

### 9.3.2. *Searches for $B_q \to \gamma\gamma$.* *(Contributing author: A. Ishikawa)*

Since the final states do not have charged particles, the $B_s \to \gamma\gamma$ and $B_d \to \gamma\gamma$ decays have so far only been searched for at $e^+e^-$ colliders [529, 531, 532]. The obtained upper limits (237) and (238) are several times larger than the corresponding SM predictions (239). Given its large data set, Belle II will be able to observe the $B_q \to \gamma\gamma$ decays and perform new-physics searches through precise measurements of these unique transitions.

The reconstruction of $B_q \to \gamma\gamma$ decays is straightforward. Two isolated clusters in the calorimeter, whose shower shapes are consistent with an electromagnetic shower, are combined to reconstruct the $B$-meson candidates. The $B$ meson is identified through the $\Delta E$ and $M_{bc}$ distributions. Since the calorimeter is about 16 radiation lengths, the $\Delta E$ distribution has a longer tail to lower $\Delta E$ values due to shower leakage. The dominant backgrounds are off-timing Bhabha events on top of hadronic events and continuum events with initial state radiation. The former can be reduced by requiring tight timing constraints on ECL and trigger hits (which is the default in the Belle II reconstruction), while the latter can be suppressed by the use of event shape variables.

Assuming that $\mathrm{Br}(B_d \to \gamma\gamma) = 3.1 \cdot 10^{-8}$, the decay should be observed with an integrated luminosity of $12\,\mathrm{ab}^{-1}$ and the relative uncertainty on the branching ratio is expected to be about 10% with $50\,\mathrm{ab}^{-1}$ of data. The given uncertainty is statistically dominated. After an observation, the direct $CP$ violation can be measured using flavour tagging. With $50\,\mathrm{ab}^{-1}$ it should be possible to measure $A_{\rm CP}(B_d \to \gamma\gamma)$ with a precision of about 25%.

The data taking strategy at $\Upsilon(5S)$ is not determined yet. If we make the standard assumption of this document that $5\,\mathrm{ab}^{-1}$ data will be accumulated, the data sample will contain about $2.9 \cdot 10^8$ $B_s^{(*)0}\bar{B}_s^{(*)0}$ pairs. The precision of $\mathrm{Br}(B_s^0 \to \gamma\gamma)$ with $5\,\mathrm{ab}^{-1}$ will be 23% which is a bit larger to claim an observation. To observe the $B_s \to \gamma\gamma$ decay, $7\,\mathrm{ab}^{-1}$ of integrated luminosity are needed. There is no reason not to add another few $\mathrm{ab}^{-1}$ of data for observation, which takes about a few months. Since flavour tagging of $B_s$ mesons is difficult due to fast $B_s$–$\bar{B}_s$ oscillations and the worse proper-time resolution compared to the $B_d$ case, a measurement of the direct $CP$ asymmetry of $B_s \to \gamma\gamma$ seems very difficult. An exception could be provided by $CP$ tagging of the other $B_s$ meson in the $\Upsilon(5S) \to B_s^0\bar{B}_s^0$ or $\Upsilon(5S) \to B_s^{*0}\bar{B}_s^{*0}$ processes. Further studies of the $CP$ tagging efficiency using full event interpretation are needed to clarify this issue.

The Belle II sensitivities for the $B_{d,s} \to \gamma\gamma$ modes are summarised in Table 63.

### 9.3.3. *$B \to X_s\gamma\gamma$ decay.* *(Contributing authors: C. Bobeth and A. Kokulu)*

Table 63: Belle II sensitivities for the $B_{d.s} \to \gamma\gamma$ modes. We assume that $5\,\mathrm{ab}^{-1}$ of data will be taken on the $\Upsilon(5S)$ resonance at Belle II. Some numbers at Belle are extrapolated to $0.71\,\mathrm{ab}^{-1}$ ($0.12\,\mathrm{ab}^{-1}$) for the $B_d$ ($B_s$) decay. The given branching ratio and asymmetry uncertainties are relative and absolute uncertainties, respectively.

| Observables | Belle $0.71\,\mathrm{ab}^{-1}$ ($0.12\,\mathrm{ab}^{-1}$) | Belle II $5\,\mathrm{ab}^{-1}$ | Belle II $50\,\mathrm{ab}^{-1}$ |
|---|---|---|---|
| $\mathrm{Br}(B_d \to \gamma\gamma)$ | $< 740\%$ | 30% | 9.6% |
| $A_{CP}(B_d \to \gamma\gamma)$ | – | 78% | 25% |
| $\mathrm{Br}(B_s \to \gamma\gamma)$ | $< 250\%$ | 23% | – |

Compared to $B \to X_s\gamma$, the double-radiative process $B \to X_s\gamma\gamma$ is suppressed by an additional factor of $\alpha/(4\pi)$, which leads to the naive expectation $\mathrm{Br}(B \to X_s\gamma\gamma)_{\mathrm{SM}} = \mathcal{O}(10^{-7})$. Given its small branching ratio it is unsurprising that the mode $B \to X_s\gamma\gamma$ has not been observed so far.

Even though it is very rare compared to the single radiative $B \to X_s\gamma$ decay, the double-radiative process has some features that make it worthwhile to study it at Belle II. These features are:

(*i*) In contrast to $B \to X_s\gamma$, the current-current operators $Q_{1,2}$ contribute to $B \to X_s\gamma\gamma$ via 1PI diagrams already at LO. As a result, measurements of the double-radiative decay mode would allow to put bounds on these 1PI corrections.

(*ii*) For $B \to X_s\gamma\gamma$ one can study more complicated distributions such as $d^2\Gamma/(dE_1 dE_2)$, where $E_{1,2}$ are the final state photon energies, or a forward-backward asymmetry ($A_{\mathrm{FB}}$) that can provide additional sensitivity to BSM physics.

In order to exploit these features in a clean way, SM predictions beyond the LO are needed. A first step towards achieving NLO accuracy has been made in [543, 544] by the calculation of the $(Q_7, Q_7)$ interference contribution to the differential distributions at $\mathcal{O}(\alpha_s)$. In the latter works it has been shown that the NLO corrections associated to $(Q_7, Q_7)$ are large and can amount to a relative change of around $\pm 50\%$ compared to the corresponding LO predictions [545–548]. Further progress towards $B \to X_s\gamma\gamma$ at NLO was made recently in [549] by providing the $(Q_8, Q_8)$ self-interference contribution. Although these corrections should be suppressed relative to those from $(Q_7, Q_7)$ by $\left|C_8^{\mathrm{eff}} Q_d / C_7^{\mathrm{eff}}\right|^2 \simeq 3\%$ the appearance of collinear logarithms $\ln(m_s/m_b)$ could upset this naive expectation. One important outcome of the work [549] is that the logarithmically-enhanced contributions stay small in the full phase-space, and as a result the $(Q_8, Q_8)$ interference represents only a subleading NLO correction. The NLO calculation of the numerically important $(Q_7, Q_7)$ interference contribution has very recently been extended to the case of a non-zero $s$-quark mass [550].

Including all known perturbative corrections the state-of-the-art SM prediction reads [550]

$$\mathrm{Br}(B \to X_s\gamma\gamma)_{\mathrm{SM}}^{c=0.02} = (0.9 \pm 0.3) \cdot 10^{-7}\,, \tag{251}$$

where $c$ represents a cut on the phase-space (for details see [550]) which guarantees that the two photons are not soft and also not parallel to each other. The quoted uncertainty is dominated by the error due to scale variations $\mu_b \in [m_b/2, 2m_b]$. Since scale ambiguities represent the largest theoretical uncertainty at present, a more reliable SM prediction can only be achieved by calculating further NLO corrections such as for instance the $(Q_{1,2}, Q_7)$ interference term. We add that LD resonant [547] and spectator quark [551] effects are small and have therefore not been included in (251).

The inclusive $B \to X_s\gamma\gamma$ decay has also been examined in extensions of the SM. Predictions for $\mathrm{Br}(B \to X_s\gamma\gamma)$ and $A_{\mathrm{FB}}$ have been obtained in 2HDMs [546, 548] and in the framework of $R$-parity violating SUSY [542]. In all cases it has been found that $\mathcal{O}(1)$ effects in $B \to X_s\gamma\gamma$ can arise in the models under consideration.

Experimentally, the inclusive $B \to X_s\gamma\gamma$ can be studied by the sum of the exclusive method, i.e. measuring $B \to (K, K^*, \cdots)\gamma\gamma$ decays. The challenge of this measurement is to eliminate the background of $\gamma\gamma$ originating from various resonances. The $B^\pm \to K^\pm h \to K^\pm\gamma\gamma$ decay is studied by the Belle Collaboration [552] in 2006, where the resonance contributions from $h = (\eta, \eta', \eta_c, \eta_c(2S), \chi_{c0}, \chi_{c2}, J/\psi, X(3872))$ are measured. The largest contributions come from $Br(B^\pm \to K^\pm\eta \to K^\pm\gamma\gamma) = (0.87^{+0.16+0.10}_{-0.15-0.07}) \times 10^{-6}$ and $Br(B^\pm \to K^\pm\eta' \to K^\pm\gamma\gamma) = (1.40^{+0.16+0.15}_{-0.15-0.12}) \times 10^{-6}$ while the branching ratios for the others channels are typically less than $\mathcal{O}(10^{-7})$. As the branching ratio measurements of $B^\pm \to K^\pm h$ as well as $h \to \gamma\gamma$ have been significantly improved since 2006 and it will be further improved in the future, we must be able to estimate these background contributions at a high accuracy. Further investigation is needed in order to quantify the sensitivity to the new physics in this channel while we see a great potential at Belle II.

## *9.4. Inclusive and Exclusive $b \to s\ell^+\ell^-$ Decays*

### *9.4.1. Inclusive $B \to X_q\ell^+\ell^-$ decay.* *(Contributing authors: G. Bell and T. Huber)*

Inclusive $B \to X_q\ell^+\ell^-$ decays provide information on the quark flavour sector that is complementary to inclusive $b \to q\gamma$ and exclusive $b \to q\ell^+\ell^-$ transitions. In contrast to $B \to X_q\gamma$, an angular analysis of the decay products entails a richer dependence on the SD Wilson coefficients. Compared to exclusive $b \to q\ell^+\ell^-$ decays, on the other hand, hadronic uncertainties are under better theoretical control for the inclusive modes. Measurements of the $B \to X_q\ell^+\ell^-$ decay distributions at Belle II will thus complement the LHCb studies of the exclusive $b \to q\ell^+\ell^-$ transitions, thereby providing important cross-checks of the deviations found by LHCb and Belle in $B \to K^*\mu^+\mu^-$ and related modes [389–391, 553].

The two main kinematic variables in inclusive $B \to X_s\ell^+\ell^-$ decays are the di-lepton invariant mass squared $m^2_{\ell\ell} = q^2$ and $z = \cos\theta$, where $\theta$ is the angle between the three-momenta of the positively charged lepton $\ell^+$ and the initial $B$ meson in the di-lepton centre-of-mass frame. In terms of these variables, the double differential decay width takes the form of a second-order polynomial in $z$ [554],

$$\frac{d^2\Gamma}{dq^2\,dz} = \frac{3}{8}\left[(1+z^2)H_T(q^2) + 2zH_A(q^2) + 2(1-z^2)H_L(q^2)\right]. \tag{252}$$

The functions $H_T$, $H_A$, and $H_L$ represent three independent observables. $H_A$ is up to a rational factor equivalent to the forward-backward asymmetry [555], while the $q^2$ spectrum is given by the sum of $H_T$ and $H_L$:

$$\begin{aligned} \frac{dA_{\mathrm{FB}}}{dq^2} &= \int_{-1}^{+1} dz\, \frac{d^2\Gamma}{dq^2 dz}\,\mathrm{sgn}(z) = \frac{3}{4}H_A(q^2)\,, \\ \frac{d\Gamma}{dq^2} &= \int_{-1}^{+1} dz\, \frac{d^2\Gamma}{dq^2 dz} = H_T(q^2) + H_L(q^2)\,. \end{aligned} \tag{253}$$

The observables mainly depend on the Wilson coefficients $C_7$, $C_9$ and $C_{10}$. Taking only these three coefficients into account and suppressing a common prefactor $G_F^2 m_b^5 \left|V_{ts}^* V_{tb}\right|^2 / (48\pi^3)$, one has (with $\hat{s} = q^2/m_b^2$)

$$\begin{aligned} H_T(q^2) &= 2\hat{s}\,(1-\hat{s})^2 \left[ \left|C_9 + \frac{2}{\hat{s}}\,C_7\right|^2 + \left|C_{10}\right|^2 \right], \\ H_L(q^2) &= (1-\hat{s})^2 \left[ \left|C_9 + 2\,C_7\right|^2 + \left|C_{10}\right|^2 \right], \\ H_A(q^2) &= -4\hat{s}\,(1-\hat{s})^2 \;\mathrm{Re}\left[ C_{10}\big(C_9 + \frac{2}{\hat{s}}\,C_7\big) \right]. \end{aligned} \tag{254}$$

The di-lepton invariant mass spectrum is dominated by charmonium resonances ($J/\psi$, $\psi(2S)$, etc.), which are usually removed by kinematic cuts. This leads to the so-called perturbative di-lepton invariant mass regions: the low-$q^2$ region for $q^2 \in [1,6]\,\mathrm{GeV}^2$ and the high-$q^2$ region for $q^2 > 14.4\,\mathrm{GeV}^2$. Within these regions, one expects that the theoretical uncertainties can be controlled to around 10%.

In the low-$q^2$ region, the observables can be computed within a local OPE in the heavy-quark limit. The perturbative calculation is well advanced and higher-order QCD [399, 556–564] and EW [564–567] corrections are available to NNLO and NLO, respectively. The leading power corrections of order $\Lambda^2_{\mathrm{QCD}}/m_b^2$ [455, 568–570], $\Lambda^3_{\mathrm{QCD}}/m_b^3$ [571, 572] and $\Lambda^2_{\mathrm{QCD}}/m_c^2$ [434] are also known. The latter can be considered as parts of the resolved photon contributions [573].

In the high-$q^2$ region, on the other hand, the heavy-mass expansion breaks down at the endpoint of the $q^2$ spectrum. For the integrated high-$q^2$ spectrum, however, there exists an effective expansion in inverse powers of $m_b^{\mathrm{eff}} = m_b\left(1 - \sqrt{\hat{s}_{\mathrm{min}}}\right)$ instead of $m_b$. This expansion converges less rapidly, and the convergence behaviour depends on the value of the $q^2$ cut, $\hat{s}_{\mathrm{min}} = q^2_{\mathrm{min}}/m_b^2$ [562].

The differential decay width is furthermore affected by QED corrections, which lead to two major modifications. First, the electron and muon channels get different contributions of the form $\ln(m_b^2/m_\ell^2)$, which stem from collinear photon emissions. Second, the simple $z$-dependence of the double differential decay distribution in (252) gets modified and becomes a complicated function of $z$ [567]. In the presence of QED radiation, the observables (254) are therefore defined by taking appropriate projections of the double differential rate [567]. In order to compare theoretical predictions with experimental data, it is important that the experimental analyses use the same prescriptions.

The theoretical uncertainties can be further reduced by normalising the observables to the inclusive semi-leptonic $B \to X_c \ell \bar{\nu}$ decay rate. The SM predictions for the $B \to X_s \mu^+\mu^-$ observables then become

$$\begin{aligned} H_T[1,6]_{\mu\mu} &= (4.03 \pm 0.28) \cdot 10^{-7}, \\ H_L[1,6]_{\mu\mu} &= (1.21 \pm 0.07) \cdot 10^{-6}, \\ H_A[1,6]_{\mu\mu} &= (-0.42 \pm 0.16) \cdot 10^{-7}, \\ \mathrm{Br}[1,6]_{\mu\mu} &= (1.62 \pm 0.09) \cdot 10^{-6}, \\ \mathrm{Br}[>14.4]_{\mu\mu} &= (2.53 \pm 0.70) \cdot 10^{-7}. \end{aligned} \tag{255}$$

Here the notation $O[q_0^2, q_1^2]$ with $O = H_T, H_L, H_A, \mathrm{Br}$ means that the relevant observable has been integrated over $q^2 \in [q_0^2, q_1^2]$. The complete list of theory predictions can be found in [567]. To tame the large uncertainty in the high-$q^2$ branching ratio, which mainly stems from poorly known parameters in the power corrections, a normalisation to the semi-leptonic $B \to X_u \ell \bar{\nu}$ rate with the same cut in $q^2$ was proposed [572],

$$\mathcal{R}(s_0) = \frac{\displaystyle\int_{\hat{s}_0}^{1} d\hat{s}\, \frac{d\Gamma_{B\to X_s \ell^+\ell^-}}{d\hat{s}}}{\displaystyle\int_{\hat{s}_0}^{1} d\hat{s}\, \frac{d\Gamma_{B\to X_u \ell\bar{\nu}}}{d\hat{s}}} \,. \tag{256}$$

Employing this normalisation results in

$$\mathcal{R}(14.4)_{\mu\mu} = (2.62 \pm 0.30) \cdot 10^{-3} \,. \tag{257}$$

Unfortunately, the achieved precision cannot yet be exploited, because the BaBar [574, 575] and Belle [576–578] measurements suffer from sizeable experimental uncertainties in the ballpark of 30% (cf. Eq. (258)). Furthermore, all measurements performed at the $B$-factories are based on a sum over exclusive final states, which makes a direct comparison to the theoretical predictions non-trivial.

Belle II can significantly improve upon this situation and with its two orders of magnitude larger data sample, it might for the first time be possible to perform a complete angular analysis of $B \to X_s \ell^+ \ell^-$ decays. In the beginning, Belle II will still have to rely on the sum-over-exclusive method, but a fully-inclusive analysis based on the recoil technique may be feasible in the long term.

The prospects for future improvements on the experimental side calls for refinements of the SM predictions. Some of the important questions to be addressed are:

(*i*) In the absence of a fully-inclusive analysis, one has to revisit the theoretical issues that arise from semi-inclusive analyses. In particular, a cut on the hadronic invariant mass $M_{X_s} \lesssim 1.8\,\mathrm{GeV}$ affects the low-$q^2$ region and induces additional theoretical uncertainties. The theoretical description in this "shape-function region" is similar to $B \to X_u \ell \nu$ and $B \to X_s \gamma$ decays [579, 580]. An analysis of the effects from sub-leading shape functions was presented in [581], and a prediction for the position of the zero of the forward-backward asymmetry in the presence of the $M_{X_s}$ cut was given in [582]. Similar studies for other observables as well as a detailed analysis of the impact of the $M_{X_s}$ cut on the extraction of the Wilson coefficients are yet to be performed.

(*ii*) Similar to inclusive $B \to X_s \gamma$ decays (see Section 9.2.1), a systematic analysis of hadronic non-local power corrections includes resolved contributions in which the virtual photon couples to light partons instead of connecting directly to the effective weak interaction vertex. These contributions stay non-local even when the hadronic mass cut is released and therefore represent an irreducible uncertainty independent of the cut. A first analysis that quantifies this uncertainty can be found in [573].

(*iii*) To estimate the impact of the charmonium resonances on the low-$q^2$ and high-$q^2$ regions, one may attempt to model the resonance structure explicitly. The most commonly used implementation via the Krüger-Sehgal approach [583] uses dispersion relations for the electromagnetic vacuum polarisation. The model is based on the assumption that the $c\bar{c}$ loop and the $b \to s$ transition factorise, which is not justified on theoretical grounds.

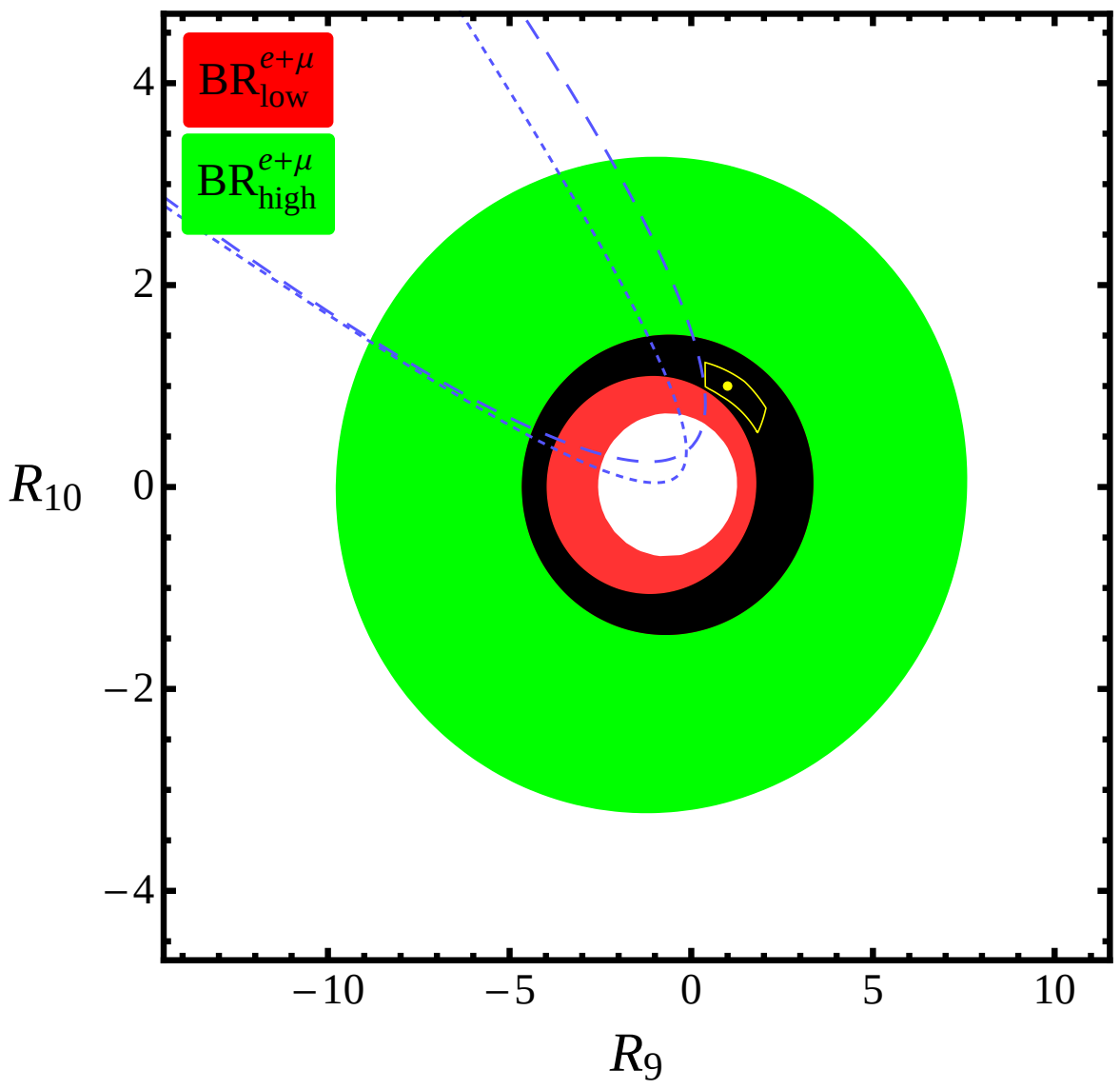

Fig. 92: 95% confidence level (CL) constraints on the Wilson coefficient ratios $R_{9,10} = C_{9,10}/C_{9,10}^{\rm SM}$. Shown are the branching ratio constraints at low-$q^2$ (red) and high-$q^2$ (green), together with their overlap (black). The region outside the dashed parabola shaped regions is allowed by the Belle measurement of the forward-backward asymmetry. The yellow dot is the SM point and the yellow contour is the future Belle II reach, assuming the central values of $R_{9,10}$ are unity as in the SM. See [567] for further details.

Since LHCb measurements of $B^+ \to K^+\mu^+\mu^-$ indeed suggest that non-factorisable corrections substantially modify the interference, theoretical investigations that go beyond the Krüger-Sehgal approach seem to be required.

(*iv*) The ratio $R_{X_s} = {\rm Br}_{\mu\mu}/{\rm Br}_{ee}$, in analogy to the quantity $R_{K^{(*)}}$ in the exclusive modes, is among the "golden modes" proposed for the early Belle II run. A measurement of $R_{X_s}$ will shed light on possible hints for lepton flavour non-universality recently observed by LHCb [391, 392]. Given the expected Belle II precision, a careful reanalysis of photon radiation will become important since collinear QED corrections represent the leading source of lepton flavour universality breaking in the SM. As the size of these contributions is sensitive to the imposed experimental cuts, a close interaction between experiment and theory is needed.

(*v*) The latest analyses of $B \to X_d\ell^+\ell^-$ decays date back more than ten years [584, 585]. An update with a decomposition into angular observables, including higher-order QCD and QED bremsstrahlung corrections, appears to be timely. Due to the different hierarchy of CKM elements, one expects larger $CP$-violating effects in $b \to d\ell^+\ell^-$ than in $b \to s\ell^+\ell^-$ transitions.

The experimental data can be used to constrain new-physics effects in a model-independent fashion, i.e. by constraining the Wilson coefficients (see Section 9.4.5 for further details). For the case of $C_9$ and $C_{10}$ the current situation as well as the potential impact of future Belle II measurements is illustrated in Figure 92 [554, 567]. From the figure it is evident that the new-physics potential of $B \to X_s\ell^+\ell^-$ decays has not yet been fully exploited. Furthermore, right-handed currents — which have been extensively studied in exclusive transitions — were not included in the latest theory studies, and the synergy and complementarity of inclusive and exclusive $b \to s\ell^+\ell^-$ analyses is yet to be explored. To this purpose, detailed Monte Carlo (MC) studies could be used in conjunction with realistic theory predictions to estimate how much statistics is needed at Belle II to reach or exceed the sensitivity of the LHCb measurements on the exclusive modes. Such analyses could build on the studies [554, 567].

*9.4.2. Measurement of $B \to X_s\ell^+\ell^-$.* *(Contributing author: A. Ishikawa)*

All existing measurements of the inclusive $B \to X_s\ell^+\ell^-$ mode have employed the sum-of-exclusive method [574–578]. In this method the hadronic system $X_s$ is reconstructed from $Kn\pi$ final states with $n \leq 4$, allowing for at most one neutral pion. The $X_s$ system is combined with the di-electron or di-muon pair to reconstruct the $B$ meson. The $B$ meson is identified by its $\Delta E$ and $M_{bc}$ distributions. Since the decay does not contain hard photons, the $\Delta E$ resolution is much better than that in $B \to X_s\gamma$. This allows one to adopt a tight $\Delta E$ selection which compared to the $B \to X_s\gamma$ analysis suppresses the likelihood of multiple candidates in a single event and the self-cross-feed. A hadronic mass selection is applied to reduce combinatorial backgrounds, i.e. $M_{X_s} < 1.8\,\mathrm{GeV}$ at BaBar [575] and $M_{X_s} < 2.1\,\mathrm{GeV}$ [576] or $M_{X_s} < 2.0\,\mathrm{GeV}$ [577, 578] at Belle. For the low-$q^2$ and high-$q^2$ region, the weighted averages of the experimental results read [567]

$$\begin{aligned} \mathrm{Br}[1,6]_{\ell\ell}^{\mathrm{exp}} &= (1.58 \pm 0.37) \cdot 10^{-6}\,, \\ \mathrm{Br}[> 14.4]_{\ell\ell}^{\mathrm{exp}} &= (4.8 \pm 1.0) \cdot 10^{-7}\,, \end{aligned} \tag{258}$$

respectively. Notice that compared to the theory prediction (255) the measurement of $\mathrm{Br}[1,6]_{\ell\ell}$ has an uncertainty that is larger by around a factor of 4. In addition, Belle presented a measurement of the forward-backward asymmetry [578] and BaBar a measurement of the $CP$ asymmetry [575].

Our study of the prospects of the $B \to X_s\ell^+\ell^-$ measurements at Belle II are based on a cut of $M_{X_s} < 2.0\,\mathrm{GeV}$, but we emphasise that this selection can be loosened in order to better understand the $X_s$ spectrum and to reduce theoretical uncertainties. There are three dominant backgrounds. The first one is associated to $c\bar{c}$ continuum events in which both charm quarks decay semi-leptonically, the second one arises from $B\bar{B}$ events with two leptons either from semi-leptonic $B$ or $D$ decays, and the third one is due to $B \to J/\psi\ \big(\psi(2S)\big)X_s$ backgrounds. The semi-leptonic backgrounds can be suppressed by missing energy information and vertex quality requirement, while the $B \to J/\psi\ \big(\psi(2S)\big)X_s$ backgrounds can be eliminated by applying appropriate cuts on the invariant mass of the di-lepton system.

The partial branching ratios in the low-$q^2$ and high-$q^2$ regions are under good theoretical control $\big($see (255) and (257)$\big)$ and thus precise measurements of the di-lepton spectra will allow to put stringent constraints on the Wilson coefficients $C_9$ and $C_{10}$. We define the following $q^2$ regions $[1.0, 3.5]\,\mathrm{GeV}^2$ (low1), $[3.5, 6.0]\,\mathrm{GeV}^2$ (low2) and $> 14.4\,\mathrm{GeV}^2$ (high). Given the large data sample expected at Belle II the reduction of systematic uncertainties is crucial. Thanks to the large branching fractions of the $B \to K^{(*)}\ell^+\ell^-$ modes and the good $\Delta E$ resolution compared to $B \to X_s\gamma$, missing-mode and fragmentation uncertainties can be reduced by adding additional reconstructed decays, such as three-kaon modes, that were not included in earlier studies. In the high (low) $q^2$ region, these uncertainties are expected to be as small as 1% (as large as 4%) due to the lower (higher) multiplicity of $X_s$ decays while $K^*$-$X_s$ transition uncertainty could be as large as 2% (as small as 1%) due to the larger (smaller) fraction of $K^*$. With $50\,\text{ab}^{-1}$ of data we expect total uncertainties of 6.6%, 6.4% and 4.7% for the partial branching ratios in the low1, low2 and high region as defined above.

Table 64: The Belle II sensitivities for the inclusive $B \to X_s \ell^+\ell^-$ observables corresponding to an invariant mass cut of $M_{X_s} < 2.0\,\text{GeV}$. The given sensitivities are relative or absolute uncertainties depending on the quantity under consideration.

| Observables | Belle $0.71\,\text{ab}^{-1}$ | Belle II $5\,\text{ab}^{-1}$ | Belle II $50\,\text{ab}^{-1}$ |
|---|---|---|---|
| $\text{Br}(B \to X_s \ell^+\ell^-)$ ($[1.0, 3.5]\,\text{GeV}^2$) | 29% | 13% | 6.6% |
| $\text{Br}(B \to X_s \ell^+\ell^-)$ ($[3.5, 6.0]\,\text{GeV}^2$) | 24% | 11% | 6.4% |
| $\text{Br}(B \to X_s \ell^+\ell^-)$ ($> 14.4$ GeV$^2$) | 23% | 10% | 4.7% |
| $A_{\text{CP}}(B \to X_s \ell^+\ell^-)$ ($[1.0, 3.5]\,\text{GeV}^2$) | 26% | 9.7 % | 3.1 % |
| $A_{\text{CP}}(B \to X_s \ell^+\ell^-)$ ($[3.5, 6.0]\,\text{GeV}^2$) | 21% | 7.9 % | 2.6 % |
| $A_{\text{CP}}(B \to X_s \ell^+\ell^-)$ ($> 14.4$ GeV$^2$) | 21% | 8.1 % | 2.6 % |
| $A_{\text{FB}}(B \to X_s \ell^+\ell^-)$ ($[1.0, 3.5]\,\text{GeV}^2$) | 26% | 9.7% | 3.1% |
| $A_{\text{FB}}(B \to X_s \ell^+\ell^-)$ ($[3.5, 6.0]\,\text{GeV}^2$) | 21% | 7.9% | 2.6% |
| $A_{\text{FB}}(B \to X_s \ell^+\ell^-)$ ($> 14.4$ GeV$^2$) | 19% | 7.3% | 2.4% |
| $\Delta_{\text{CP}}(A_{\text{FB}})$ ($[1.0, 3.5]\,\text{GeV}^2$) | 52% | 19% | 6.1% |
| $\Delta_{\text{CP}}(A_{\text{FB}})$ ($[3.5, 6.0]\,\text{GeV}^2$) | 42% | 16% | 5.2% |
| $\Delta_{\text{CP}}(A_{\text{FB}})$ ($> 14.4$ GeV$^2$) | 38% | 15% | 4.8% |

Belle II measurements of the forward-backward asymmetry $A_{\text{FB}}$ in $B \to X_s \ell^+\ell^-$ are expected to provide the most stringent limits on the Wilson coefficients $C_9$ and $C_{10}$. Since large parts of the theoretical and experimental systematic uncertainties cancel out in $A_{\text{FB}}$ the corresponding measurements will be statistically limited. The expected uncertainties on $A_{\text{FB}}$ in the low1, low2 and high region are 3.1%, 2.6% and 2.4%, respectively, assuming the SM.

A helicity decomposition of $B \to X_s \ell^+\ell^-$ provides the three observables $H_i$ defined in (252). While $H_A$ and the combination $H_T + H_L$ have been measured (cf. (253)) independent measurements of $H_T$ and $H_L$ have not been performed by BaBar and Belle, but will be possible at Belle II. As for the measurements of the branching ratios, the experimental determinations of the coefficients $H_i$ will not be systematically limited until $10\,\text{ab}^{-1}$ have been collected. Considering normalised observables might help to reduce the systematic uncertainties.

Measurement of the $CP$ asymmetries in $B \to X_s \ell^+\ell^-$ can be used to search for new source of $CP$ violation. Not only the rate asymmetry, but also the $CP$ asymmetry of angular distributions, such as forward-backward $CP$ asymmetry ($A_{\text{FB}}^{\text{CP}}$) are useful [586]. Since the denominator of the $A_{\text{FB}}^{\text{CP}}$ can be zero if $A_{\text{FB}}$ for $\bar{B}$ and $B$ are zero or have opposite sign, we consider the difference of the $A_{\text{FB}}$ between $\bar{B}$ and $B$ mesons defined as $\Delta_{\text{CP}}(A_{\text{FB}}) = A_{\text{FB}}^{\bar{B}} - A_{\text{FB}}^{B}$. Since most of systematic uncertainties calcel out by taking the ratio, dominant uncertainty is statistical.

Tests of lepton flavour universality can also be performed by measuring $R_{X_s}$. The Belle II detector has certainly a good resolution to the $e^+e^-$ mode and the $R_{X_s}$ measurement is promising. We can expect a performance similar to those of the exclusive channel (i.e. the $R_{K^{(*)}}$ measurement), which will be discussed in Section 9.4.4.

A summary of the Belle II sensitivities for the various $B \to X_s \ell^+ \ell^-$ observables is provided in Table 64.

*9.4.3. Exclusive $B \to K^{(*)}\ell^+\ell^-$ decays.* *(Contributing authors: W. Altmannshofer, U. Haisch and D. Straub)*

The $\bar{B} \to \bar{K}^* (\to \bar{K}\pi)\, \ell^+\ell^-$ transition

$$\frac{d^4\Gamma}{dq^2\, d\cos\theta_\ell\, d\cos\theta_K\, d\phi} = \frac{9}{32\pi}\, I(q^2, \theta_\ell, \theta_K, \phi)\,, \tag{259}$$

is completely described in terms of twelve angular coefficient functions $I_j$ [587–589], namely

$$\begin{aligned} I(q^2, \theta_\ell, \theta_K, \phi) = {} & I_1^s \sin^2\theta_K + I_1^c \cos^2\theta_K + (I_2^s \sin^2\theta_K + I_2^c \cos^2\theta_K)\cos 2\theta_\ell \\ & + I_3 \sin^2\theta_K \sin^2\theta_\ell \cos 2\phi + I_4 \sin 2\theta_K \sin 2\theta_\ell \cos\phi \\ & + I_5 \sin 2\theta_K \sin\theta_\ell \cos\phi + (I_6^s \sin^2\theta_K + I_6^c \cos^2\theta_K)\cos\theta_\ell \\ & + I_7 \sin 2\theta_K \sin\theta_\ell \sin\phi + I_8 \sin 2\theta_K \sin 2\theta_\ell \sin\phi \\ & + I_9 \sin^2\theta_K \sin^2\theta_\ell \sin 2\phi\,. \end{aligned} \tag{260}$$

The adopted angular conventions are illustrated in Figure 93 and follow [388] (see also [589]). The angle $\theta_\ell$ is the angle between the direction of the $\ell^-$ in the dilepton rest frame and the direction of the dilepton in the $\bar{B}$ rest frame. The angle $\theta_K$ is the angle between the direction of the kaon in the $\bar{K}^*$ rest frame and the direction of the $\bar{K}^*$ in the $\bar{B}$ rest frame. The angle $\phi$ is the angle between the plane containing the dilepton pair and the plane containing the kaon and pion from the $\bar{K}^*$.

The decay distribution for the $CP$-conjugate mode $B \to K^*(\to K\pi)\ell^+\ell^-$ is given by a formula analog to (259) with different angular functions, which we call $\bar{I}_j$. Note that for this decay, $\theta_\ell$ is the angle between the direction of the $\ell^+$ in the dilepton rest frame and the direction of the dilepton in the $B$ rest frame, while $\theta_K$ is the angle between the direction of the kaon in the $K^*$ rest frame and the direction of the $K^*$ in the $B$ rest frame. As a result, the functions $\bar{I}_j$ can be obtained by the replacements

$$I^{(a)}_{1,2,3,4,5,6} \to \bar{I}^{(a)}_{1,2,3,4,5,6}\,, \qquad I^{(a)}_{7,8,9} \to -\bar{I}^{(a)}_{7,8,9}\,. \tag{261}$$

with $a = s, c$. These quantities which encode the angular distribution of the exclusive decay can be expressed in terms of helicity (or transversity) amplitudes that depend on the dilepton invariant mass squared, the Wilson coefficients $C_7, C_9, C_{10}, C_S, C_P$ and their chirality-flipped counterparts as well as the $B \to K^*$ form factors that arise from the matrix elements $\langle K^*|Q_i|B\rangle$. The situation is much simpler for the $B \to K\ell^+\ell^-$ decay which gives rise to only three observables, namely the branching ratio, the forward-backward asymmetry $A_{\rm FB}$ and the flat term $F_H$ [590].

The self-tagging nature of the $\bar{B} \to \bar{K}^* (\to \bar{K}\pi)\, \ell^+\ell^-$ decay means that it is possible to determine both $CP$-averaged and $CP$-asymmetric quantities that depend on the

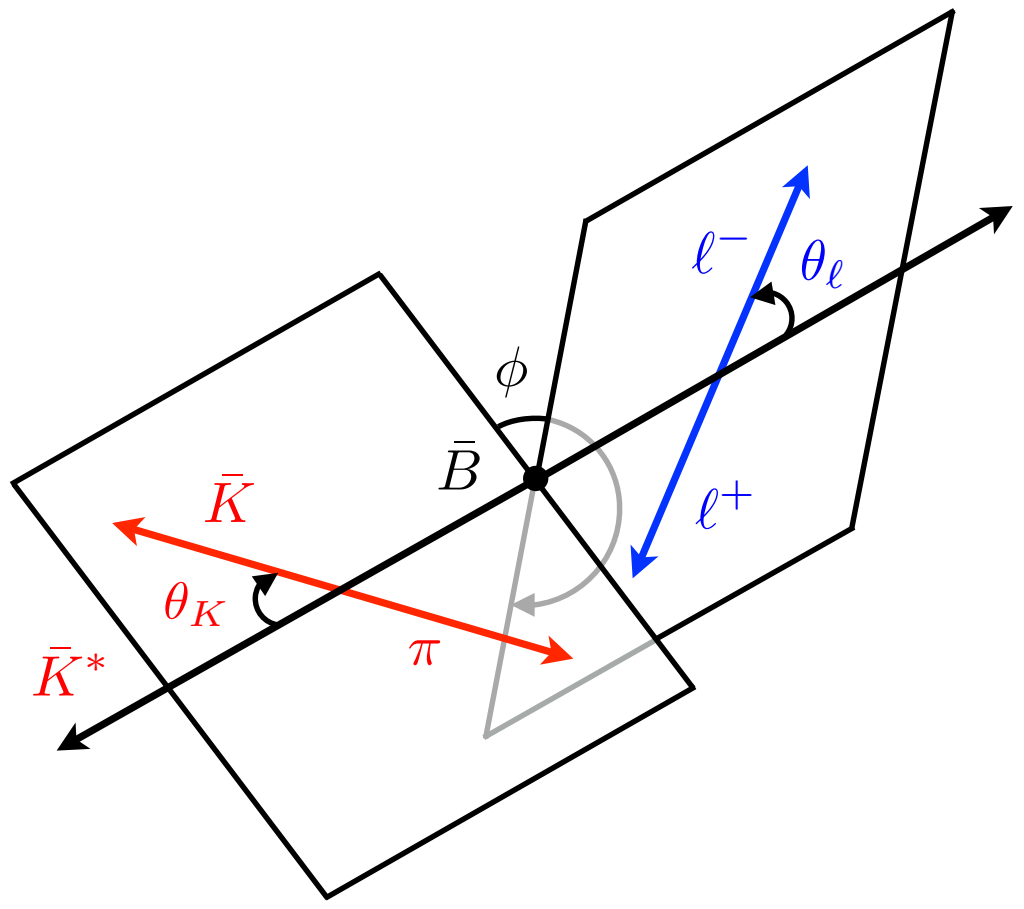

Fig. 93: Angular conventions used in the description of the $\bar{B} \to \bar{K}^* (\to \bar{K}\pi)\,\ell^+\ell^-$ decay.

coefficients [588]

$$S_j = \left(I_j + \bar{I}_j\right) \Big/ \frac{d\Gamma}{dq^2}\,, \qquad A_j = \left(I_j - \bar{I}_j\right) \Big/ \frac{d\Gamma}{dq^2}\,, \tag{262}$$

respectively. The two most measured angular observables are the forward-backward asymmetry and the $K^*$ longitudinal polarisation fraction:

$$A_{\rm FB} = \frac{3}{4} S_{6s} + \frac{3}{8} S_{6c}\,, \qquad F_L = -S_{2c}\,. \tag{263}$$

By exploiting symmetry relations it is also possible to construct $CP$-averaged observables that are largely insensitive to form-factor uncertainties [591–593]. These are

$$P_1 = \frac{S_3}{2S_{2s}}\,, \qquad P_2 = \frac{S_{6s}}{8S_{2s}}\,, \qquad P_3 = -\frac{S_9}{4S_{2s}}\,, \tag{264}$$

as well as

$$\begin{aligned} P_4' &= \frac{S_4}{2\sqrt{-S_{2s}S_{2c}}}\,, \qquad & P_5' &= \frac{S_5}{2\sqrt{-S_{2s}S_{2c}}}\,, \\ P_6' &= \frac{S_7}{2\sqrt{-S_{2s}S_{2c}}}\,, \qquad & P_8' &= \frac{S_8}{2\sqrt{-S_{2s}S_{2c}}}\,. \end{aligned} \tag{265}$$

The above definitions of the coefficients $S_j$ and the observables $P_i$ and $P_i'$ correspond to those used by LHCb [389]. Analog $CP$-violating observables $P_i^{\rm CP}$ and $P_i'^{\rm CP}$ can be defined by simply replacing the coefficient $S_j$ in the numerator of $P_i$ and $P_i'$ by the corresponding coefficient $A_j$. Notice that the observables $P_1$ and $P_2$ are commonly also called $A_T^{(2)} = P_1$ [594], $A_T^{(\rm re)} = 2P_2$ and $A_T^{(\rm im)} = -2P_3$ [595].

In order to illustrate the importance of Belle II measurements of the observables defined in (263) to (265), we consider the two cases $P_1$ and $P_5'$. At small di-lepton masses the angular variable $P_1$ is sensitive to the photon polarisation. In fact, in the heavy-quark and large-energy limit and ignoring $\alpha_s$ and $m_s/m_b$ suppressed effects, one finds

$$A_T^{(2)} \simeq \frac{2{\rm Re}\,(C_7 C_7')}{|C_7|^2 + |C_7'|^2}\,, \qquad A_T^{(\rm im)} \simeq \frac{2{\rm Im}\,(C_7 C_7')}{|C_7|^2 + |C_7'|^2}\,. \tag{266}$$

To maximise the sensitivity to the virtual photon, it is necessary to go to very small $q^2$ which is only possible in the case of the decay $B \to K^* e^+ e^-$. Precision measurement of $P_1$ as well as of $P_3$ in the di-electron channel are thus essential for probing possible BSM effects related to the right-handed magnetic penguin operator $Q_7'$ [491, 507, 596]. Consequently, decays like $B \to K^* e^+ e^-$ emerge as highly relevant for the Belle II programme.

The angular observable $P_5'$ is instead a sensitive probe of the semi-leptonic operators $Q_9$ and $Q_{10}$ and their interference with $Q_7$. In the same approximation that led to (266), one obtains the expression

$$P_5' \simeq \frac{\mathrm{Re}\left(C_{10}^* C_{9,\perp} + C_{9,||}^* C_{10}\right)}{\sqrt{\left(|C_{9,\perp}|^2 + |C_{10}|^2\right)\left(|C_{9,||}|^2 + |C_{10}|^2\right)}}\,, \tag{267}$$

if only contributions from SM operators are included. Here

$$C_{9,\perp} = C_9^{\rm eff}(q^2) + \frac{2m_b m_B}{q^2}\, C_7^{\rm eff}\,, \qquad C_{9,||} = C_9^{\rm eff}(q^2) + \frac{2m_b}{m_B}\, C_7^{\rm eff}\,. \tag{268}$$

Importantly the above results for $P_1$ and $P_5'$ are correct only in the infinite heavy-quark limit. While in the case of (266) the leading power-corrections are formally of $\mathcal{O}(\Lambda_{\rm QCD}^2/m_b^2)$, in the case of (267) a rather complex structure of $\Lambda_{\rm QCD}/m_b$ terms arises (see [491] for details). Since at present the relevant power corrections can only be modelled, assumption-free extractions of $C_9$ and $C_{10}$ as well as their chirality-flipped partners from measurements of $P_5'$ and other angular observables are not possible.

Additional information on $C_9$, $C_{10}$, $C_9'$ and $C_{10}'$ can fortunately be gleaned from the lepton flavour universality ratios

$$R_H[q_0^2, q_1^2] = \frac{\displaystyle\int_{q_0^2}^{q_1^2} dq^2\, \frac{d\Gamma(B \to H\mu^+\mu^-)}{dq^2}}{\displaystyle\int_{q_0^2}^{q_1^2} dq^2\, \frac{d\Gamma(B \to He^+e^-)}{dq^2}}\,, \tag{269}$$

with $H = K, K^*$. The SM predictions for these ratios are 1 with high precision. Phase space effects are small and can be taken into account. Theoretical uncertainties from CKM factors as well as from form factors and other hadronic effects cancel in the ratio. Corrections due to collinear photon emissions have been studied recently and appear to be well described by existing Monte Carlo tools [597]. Any deviation in $R_H$ from the SM prediction exceeding the few percent level would thus be a sign of new physics.

Including only the dominant linear BSM contributions from interference with the SM, the ratios $R_K$ and $R_{K^*}$ can be approximated by [598]

$$R_K[1,6] \simeq 1 + \Delta_+\,, \qquad R_{K^*}[1,6] \simeq 1 + \Delta_+ - p\,(\Delta_+ - \Delta_-)\,, \tag{270}$$

with

$$\Delta_\pm = \frac{2}{\left|C_9^{\rm SM}\right|^2 + \left|C_{10}^{\rm SM}\right|^2} \left\{ \sum_{i=9,10} \mathrm{Re}\left[C_i^{\rm SM}\left(C_i^{{\rm NP}\,\mu} \pm C_i'^{\,\mu}\right)\right] - (\mu \to e) \right\}, \tag{271}$$

and $p \simeq 0.86$ is the so-called polarisation fraction of the $K^*$ meson [590, 598]. The labels "SM" and "NP" denote the SM and new-physics contributions, respectively, and the index $\mu$ or $e$ indicates the flavour content of the corresponding operator. Under the assumption that new physics modifies the di-muon channels only and that the relevant corrections are real, one obtains numerically

$$\begin{aligned} R_K[1,6] &\simeq 1 + 0.24\left(C_{LL}^{\mathrm{NP}\,\mu} + C_{RL}^{\mu}\right)\,, \\ R_{K^*}[1,6] &\simeq 1 + 0.24\left(C_{LL}^{\mathrm{NP}\,\mu} - C_{RL}^{\mu}\right) + 0.07\,C_{RL}^{\mu}\,, \end{aligned} \tag{272}$$

where we have introduced the chiral Wilson coefficients

$$C_{LL}^{\mathrm{NP}\,\ell} = C_9^{\mathrm{NP}\,\ell} - C_{10}^{\mathrm{NP}\,\ell}\,, \qquad C_{RL}^{\ell} = C_9^{\prime\ell} - C_{10}^{\prime\ell}\,. \tag{273}$$

From (272) one observes that $R_K$ only probes the combination $C_{LL}^{\mathrm{NP}\,\ell} + C_{RL}^{\ell}$ of Wilson coefficients, while $R_{K^*}$ is mostly sensitive to $C_{LL}^{\mathrm{NP}\,\ell} - C_{RL}^{\ell}$. The observables $R_K$ and $R_{K^*}$ thus provide complementary information as they constrain different chirality structures of possible lepton flavour universality violating new physics in rare $B$ decays. Notice furthermore that measurements of lepton flavour universality double ratios such as $R_K/R_{K^*} \simeq 1 + 0.41\,C_{RL}^{\mu}$ directly probe right-handed currents in a theoretically clean way [598].

Belle II will also be able to perform lepton flavour universality tests using angular observables. Suitable variables include differences of angular observables in $B \to K^*\mu^+\mu^-$ and $B \to K^*e^+e^-$ [599, 600], for instance $\Delta_{A_{\mathrm{FB}}} = A_{\mathrm{FB}}(B \to K^*\mu^+\mu^-) - A_{\mathrm{FB}}(B \to K^*e^+e^-)$ or $Q_i = P_i^{\mu} - P_i^{e}$. The differences in angular observables are predicted to be zero in the SM with high accuracy. Non-zero values would therefore again be an indication of new physics.

The recent LHCb measurements of $R_K[1,6] = 0.745^{+0.090}_{-0.074} \pm 0.036$ [391] and $R_{K^*}[1.1,6] = 0.69^{+0.11}_{-0.07} \pm 0.05$ [392] deviate by $2.6\sigma$ and $2.4\sigma$ from their SM values. Previous measurements from BaBar [601] and Belle [602] have considerably larger uncertainties and are compatible with both the SM prediction and the LHCb results. New physics that only modifies the $b \to s\mu^+\mu^-$ transition but leaves $b \to se^+e^-$ unaffected can explain the deviations seen in the lepton flavour universality ratios $R_K$ and $R_{K^*}$ and simultaneously address other $B$-physics anomalies, like the discrepancy in $P_5'$ [389] and the too low $B_s \to \phi\mu^+\mu^-$ branching ratio [499]. Independent validations of the deviations observed in $P_5'$, $R_K$ and $R_{K^*}$ are needed to build a solid case for new physics. In the near future, Belle II is the only experiment that can perform such cross-checks.

#### *9.4.4. Measurements of $B \to K^{(*)}\ell^+\ell^-$.* *(Contributing authors: A. Ishikawa and S. Wehle)*

The $b \to s\ell^+\ell^-$ transition has first been observed in 2001 by Belle in the $B \to K\ell^+\ell^-$ channel [603]. Two years later in 2003, Belle then observed the $B \to K^*\ell^+\ell^-$ mode [604]. These observations opened the door for new-physics searches via EW penguin $B$ decays. The branching ratio and forward-backward asymmetry as a function of $q^2$ in $B \to K^{(*)}\ell^+\ell^-$ are important observables. A first measurement of the forward-backward asymmetry was also done by Belle in 2006 [605]. By now, several experiments have measured them [393, 394, 602, 606–609]. Due to the spin structure of the $K^*$ meson, a full angular analysis of $B \to K^*\ell^+\ell^-$ with optimised observables is a very powerful way to search for new physics. These optimised angular observables are less sensitive to form factor uncertainties that plague the theory calculations.

*$B \to K^*\mu^+\mu^-$ channel.* In 2013, the LHCb collaboration announced the observation of a tension in the optimised observable $P_5'$ with $1\,\mathrm{fb}^{-1}$ of data [610]. This tension has been confirmed two years later when LHCb presented their $B \to K^*\mu^+\mu^-$ angular analysis based on the full LHC Run I data set of $3\,\mathrm{fb}^{-1}$ [389]. Belle has recently also reported results of a angular analysis with its full data set using both charged and neutral $B$ mesons decaying to $K^*e^+e^-$ and $K^*\mu^+\mu^-$ [390]. The Belle results are consistent with the angular analyses by LHCb, which considered alone show a $3.3\sigma$ discrepancy from the SM [611].

The observed deviations make further independent measurements of the angular distributions in $B \to K^*\mu^+\mu^-$ mandatory. Our extrapolations for Belle II are based on the systematic uncertainties obtained at Belle. For example, the difference between simulation and data was estimated directly from $B \to J/\psi K^*$ decays as measured by Belle. Since at Belle II the mismodelling in the simulation will be improved, such an approach should lead to conservative projections. The uncertainty due to peaking backgrounds can be reduced by including the individual components in the fitted model. The individual components, which may be small, are more reliably modelled in a larger data set. The uncertainty that is associated to the efficiency modelling can be reduced by adding correlation between $q^2$ and the helicity angle $\cos\theta_\ell$ in the efficiency function. We find that with $2.8\,\mathrm{ab}^{-1}$ of Belle II data, the uncertainty on $P_5'$ in the $q^2 \in [4, 6]\,\mathrm{GeV}^2$ bin using both muon and electron modes will be comparable to the $3.0\,\mathrm{fb}^{-1}$ LHCb result [389] that uses the muon mode only. A naive extrapolation then leads to the conclusion that the Belle II with 50 $\mathrm{ab}^{-1}$ of data can reach to the accuracies of the optimised observables slightly higher than LHCb on a comparable time scale, i.e. with an integrated luminosity of approximately $22\mathrm{fb}^{-1}$. We add that at Belle II the iTOP and ARICH might be able to identify low momentum muons, which may increase the available data in the low-$q^2$ region. Our projections do not include such possible improvements.

*$B \to K^*e^+e^-$ channel.* As mentioned before, an angular analysis of $B \to K^*e^+e^-$ at very low $q^2$ is a sensitive probe of the photon polarisation [491, 507, 595, 596]. In fact, angular observables such as $P_1$ and $P_3^{\mathrm{CP}}$ or $A_T^{(2)}$ and $A_T^{\mathrm{Im}}$ are functions of different combinations of real and imaginary parts of $C_7$ and $C_7'$ and hence together with $S_{K^*\gamma}$ and $\mathrm{Br}(B \to X_s\gamma)$ form a basis of clean observables that allow to completely determine the contributions to $Q_7$ and $Q_7'$ from experiment.

LHCb has measured the angular observables using $3\,\mathrm{fb}^{-1}$ of data [515]. They reconstructed 124 signal events for the $q^2$ range $[0.002, 1.12]\,\mathrm{GeV}^2$ where the lower bound is limited by angular resolution on $\tilde{\phi}$, $\tilde{\phi}$, where $\tilde{\phi} = \phi + \phi$ if $\phi < 0$. At Belle II, the resolution in $\tilde{\phi}$ is better than at LHCb, and the reconstruction efficiency of the electron mode is higher than that of the muon mode at low $q^2$. These features will allow for precise Belle II measurement of $B \to K^*e^+e^-$ in the low $q^2$-region. With $50\,\mathrm{fb}^{-1}$ of data, the sensitivities of $A_{\mathrm{T}}^{(2)}$ and $A_{\mathrm{T}}^{(\mathrm{im})}$ will be 0.066 and 0.064, respectively. The quoted uncertainties are dominated by statistical errors.

Belle II will not only be able to perform measurements of $B \to K^*e^+e^-$ in the low-$q^2$ region, but has also a unique sensitivity to the high-$q^2$ region. As mentioned earlier, Belle has already measured the angular function of $B \to K^*e^+e^-$ in the high-$q^2$ region [390], which provides important additional information to understand the LHCb anomaly in $B \to K^*\mu^+\mu^-$ channel. From this study, it is expected that approximately the same sensitivity to the $B \to K^* \mu^+ \mu^-$ and the $B \to K^* e^+ e^-$ channels can be achieved at Belle II. This is possible since, in contrast to LHCb where the radiative photon recovery is difficult, the reconstruction efficiency for electrons is comparable to that for muons at Belle II thanks to the better electromagnetic calorimeter.

Table 65: The Belle II sensitivities for the observables in the low-$q^2$ region of the $B \to K^* \ell^+ \ell^-$ decay. Some numbers at Belle are extrapolated to $0.71\,\mathrm{ab}^{-1}$.

| Observables | Belle $0.71\,\mathrm{ab}^{-1}$ | Belle II $5\,\mathrm{ab}^{-1}$ | Belle II $50\,\mathrm{ab}^{-1}$ |
|---|---|---|---|
| $A_{\mathrm{T}}^{(2)}$ ($[0.002, 1.12]\,\mathrm{GeV}^2$) | – | 0.21 | 0.066 |
| $A_{\mathrm{T}}^{\mathrm{Im}}$ ($[0.002, 1.12]\,\mathrm{GeV}^2$) | – | 0.20 | 0.064 |

*Test of lepton flavour universality.* As mentioned above, we can expect a very high sensitivity to both muon and electron modes at Belle II. By taking the ratio between these two modes, almost all systematic uncertainties cancel out. In consequence, all the ratios $R_K$, $R_{K^*}$ and $R_{X_s}$ can be measured precisely. At present, LHCb has measured $R_K$ and $R_{K^*}$ only in the low-$q^2$ region, while Belle II will have access to both the low-$q^2$ and high-$q^2$ regions. The dominant source of uncertainty is due to the imperfect lepton identification which is expected to lead to a relative error of 0.4%. Given the smallness of this uncertainty, the Belle II measurements of $R_K$, $R_{K^*}$ and $R_{X_s}$ will all be statistically limited. It thus follows that with $20\,\mathrm{ab}^{-1}$ of data, Belle II should be able to confirm the $R_K$ anomaly observed by LHCb with a significance of $5\sigma$, if it is indeed due to new physics. We add that measurements of the observables $Q_{4,5} = P'^{\mu}_{4,5} - P'^{e}_{4,5}$ [600], which have been recently performed by Belle for the first time [390], are also statistically limited at Belle II.

The Belle II sensitivities for the $B \to K^{(*)} \ell^+ \ell^-$ channels are summarised in Table 65 (observables in the low-$q^2$ region), Table 66 (angular observables for different bins) and Table 67 (observables to test lepton flavour universality).

*9.4.5. Interplay of future inclusive and exclusive $b \to s \ell^+ \ell^-$ measurements. (Contributing authors: T. Huber, A. Ishikawa and J. Virto)*

In the following, we will study the phenomenological impact that future Belle II measurements of the branching ratio and forward-backward asymmetry in $B \to X_s \ell^+ \ell^-$ with $50\,\mathrm{ab}^{-1}$ of integrated luminosity may have. We consider three $q^2$ bins, namely $[1, 3.5]\,\mathrm{GeV}^2$, $[3.5, 6]\,\mathrm{GeV}^2$ and $> 14.4\ \mathrm{GeV}^2$, and derive model-independent constraints on the Wilson coefficients of the operators $Q_9$ and $Q_{10}$ introduced in (203). In particular, we will ask the following question: if the true values of the new-physics contributions are $C_9^{\mathrm{NP}}$ and $C_{10}^{\mathrm{NP}}$, respectively, with which significance will Belle II be able to exclude the SM?

This question is answered by the contours shown in Figure 94, which have been obtained from a $\chi^2$ fit based on the theory predictions of [567], but including an extra 5% uncertainty to account for non-perturbative effects [573]. Consider for example a point in the $C_9^{\mathrm{NP}}$–$C_{10}^{\mathrm{NP}}$ plane which resides on the contour labeled “5”. If this point represents the true values of the new-physics contributions then a fit including only the measurements $\mathrm{Br}(B \to X_s \ell^+ \ell^-)$ and $A_{\mathrm{FB}}(B \to X_s \ell^+ \ell^-)$ will result in a pull of the SM with respect to the best fit point by $5\sigma$.

Table 66: The Belle II sensitivities of the angular observables in $B \to K^* \ell^+ \ell^-$. Some numbers at Belle are extrapolated to $0.71\,\mathrm{ab}^{-1}$.

| Observables | Belle $0.71\,\mathrm{ab}^{-1}$ | Belle II $5\,\mathrm{ab}^{-1}$ | Belle II $\mathrm{ab}^{-1}$ |
|---|---|---|---|
| $F_L$ ([1.0, 2.5] GeV$^2$) | 0.19 | 0.063 | 0.025 |
| $F_L$ ([2.5, 4.0] GeV$^2$) | 0.17 | 0.057 | 0.022 |
| $F_L$ ([4.0, 6.0] GeV$^2$) | 0.14 | 0.046 | 0.018 |
| $F_L$ (> 14.2 GeV$^2$) | 0.088 | 0.027 | 0.009 |
| $P_1$ ([1.0, 2.5] GeV$^2$) | 0.59 | 0.24 | 0.078 |
| $P_1$ ([2.5, 4.0] GeV$^2$) | 0.53 | 0.21 | 0.071 |
| $P_1$ ([4.0, 6.0] GeV$^2$) | 0.43 | 0.17 | 0.057 |
| $P_1$ (> 14.2 GeV$^2$) | 0.33 | 0.12 | 0.040 |
| $P_2$ ([1.0, 2.5] GeV$^2$) | 0.32 | 0.12 | 0.040 |
| $P_2$ ([2.5, 4.0] GeV$^2$) | 0.30 | 0.11 | 0.036 |
| $P_2$ ([4.0, 6.0] GeV$^2$) | 0.24 | 0.090 | 0.029 |
| $P_2$ (> 14.2 GeV$^2$) | 0.086 | 0.034 | 0.011 |
| $P_3$ ([1.0, 2.5] GeV$^2$) | 0.32 | 0.12 | 0.040 |
| $P_3$ ([2.5, 4.0] GeV$^2$) | 0.30 | 0.11 | 0.036 |
| $P_3$ ([4.0, 6.0] GeV$^2$) | 0.24 | 0.090 | 0.029 |
| $P_3$ (> 14.2 GeV$^2$) | 0.18 | 0.068 | 0.022 |
| $P_4'$ ([1.0, 2.5] GeV$^2$) | 0.50 | 0.18 | 0.056 |
| $P_4'$ ([2.5, 4.0] GeV$^2$) | 0.45 | 0.15 | 0.049 |
| $P_4'$ ([4.0, 6.0] GeV$^2$) | 0.34 | 0.12 | 0.040 |
| $P_4'$ (> 14.2 GeV$^2$) | 0.26 | 0.099 | 0.032 |
| $P_5'$ ([1.0, 2.5] GeV$^2$) | 0.47 | 0.17 | 0.054 |
| $P_5'$ ([2.5, 4.0] GeV$^2$) | 0.42 | 0.15 | 0.049 |
| $P_5'$ ([4.0, 6.0] GeV$^2$) | 0.34 | 0.12 | 0.040 |
| $P_5'$ (> 14.2 GeV$^2$) | 0.23 | 0.088 | 0.027 |
| $P_6'$ ([1.0, 2.5] GeV$^2$) | 0.50 | 0.17 | 0.054 |
| $P_6'$ ([2.5, 4.0] GeV$^2$) | 0.45 | 0.15 | 0.049 |
| $P_6'$ ([4.0, 6.0] GeV$^2$) | 0.36 | 0.12 | 0.040 |
| $P_6'$ (> 14.2 GeV$^2$) | 0.27 | 0.10 | 0.032 |
| $P_8'$ ([1.0, 2.5] GeV$^2$) | 0.51 | 0.19 | 0.061 |
| $P_8'$ ([2.5, 4.0] GeV$^2$) | 0.47 | 0.17 | 0.056 |
| $P_8'$ ([4.0, 6.0] GeV$^2$) | 0.38 | 0.14 | 0.045 |
| $P_8'$ (> 14.2 GeV$^2$) | 0.27 | 0.10 | 0.032 |

The figure thus allows to determine the significance with which future Belle II measurements of $B \to X_s \ell^+ \ell^-$ can exclude the SM, depending on which are the true values of the Wilson coefficients $C_9$ and $C_{10}$.

For comparison, the $1\sigma$, $2\sigma$ and $3\sigma$ regions in the $C_9^{\mathrm{NP}}$–$C_{10}^{\mathrm{NP}}$ plane that are obtained from the global analysis [612] are also shown in Figure 94 as red contours. One can see that Belle II would exclude the SM by more than $5\sigma$ if the central value $C_9^{\mathrm{NP}} = -1$ preferred by the global fit turns out to be correct. Notice that since the underlying hadronic uncertainties in the inclusive mode are independent of those that enter exclusive transitions, precision measurements of the $B \to X_s\ell^+\ell^-$ channel provide important complementary information in the context of global fits. This shows that Belle II can play a decisive role in the search for new physics via $b \to s\ell^+\ell^-$ transitions.

Table 67: The Belle II sensitivities to $B \to K^{(*)}\ell^+\ell^-$ observables that allow to test lepton flavour universality. Some numbers at Belle are extrapolated to $0.71\,\mathrm{ab}^{-1}$.

| Observables | Belle $0.71\,\mathrm{ab}^{-1}$ | Belle II $5\,\mathrm{ab}^{-1}$ | Belle II $50\,\mathrm{ab}^{-1}$ |
|---|---|---|---|
| $R_K$ ($[1.0, 6.0]\,\mathrm{GeV}^2$) | 28% | 11% | 3.6% |
| $R_K$ ($> 14.4\,\mathrm{GeV}^2$) | 30% | 12% | 3.6% |
| $R_{K^*}$ ($[1.0, 6.0]\,\mathrm{GeV}^2$) | 26% | 10% | 3.2% |
| $R_{K^*}$ ($> 14.4\,\mathrm{GeV}^2$) | 24% | 9.2% | 2.8% |
| $R_{X_s}$ ($[1.0, 6.0]\,\mathrm{GeV}^2$) | 32% | 12% | 4.0% |
| $R_{X_s}$ ($> 14.4\,\mathrm{GeV}^2$) | 28% | 11% | 3.4% |
| $Q_{F_L}$ ($[1.0, 2.5]\,\mathrm{GeV}^2$) | 0.19 | 0.063 | 0.025 |
| $Q_{F_L}$ ($[2.5, 4.0]\,\mathrm{GeV}^2$) | 0.17 | 0.057 | 0.022 |
| $Q_{F_L}$ ($[4.0, 6.0]\,\mathrm{GeV}^2$) | 0.14 | 0.046 | 0.018 |
| $Q_{F_L}$ ($> 14.2\,\mathrm{GeV}^2$) | 0.088 | 0.027 | 0.009 |
| $Q_1$ ($[1.0, 2.5]\,\mathrm{GeV}^2$) | 0.59 | 0.24 | 0.078 |
| $Q_1$ ($[2.5, 4.0]\,\mathrm{GeV}^2$) | 0.53 | 0.21 | 0.071 |
| $Q_1$ ($[4.0, 6.0]\,\mathrm{GeV}^2$) | 0.43 | 0.17 | 0.057 |
| $Q_1$ ($> 14.2\,\mathrm{GeV}^2$) | 0.33 | 0.12 | 0.040 |
| $Q_2$ ($[1.0, 2.5]\,\mathrm{GeV}^2$) | 0.32 | 0.12 | 0.040 |
| $Q_2$ ($[2.5, 4.0]\,\mathrm{GeV}^2$) | 0.30 | 0.11 | 0.036 |
| $Q_2$ ($[4.0, 6.0]\,\mathrm{GeV}^2$) | 0.24 | 0.090 | 0.029 |
| $Q_2$ ($> 14.2\,\mathrm{GeV}^2$) | 0.086 | 0.034 | 0.011 |
| $Q_3$ ($[1.0, 2.5]\,\mathrm{GeV}^2$) | 0.32 | 0.12 | 0.040 |
| $Q_3$ ($[2.5, 4.0]\,\mathrm{GeV}^2$) | 0.30 | 0.11 | 0.036 |
| $Q_3$ ($[4.0, 6.0]\,\mathrm{GeV}^2$) | 0.24 | 0.090 | 0.029 |
| $Q_3$ ($> 14.2\,\mathrm{GeV}^2$) | 0.18 | 0.068 | 0.022 |
| $Q_4$ ($[1.0, 2.5]\,\mathrm{GeV}^2$) | 0.50 | 0.18 | 0.056 |
| $Q_4$ ($[2.5, 4.0]\,\mathrm{GeV}^2$) | 0.45 | 0.15 | 0.049 |
| $Q_4$ ($[4.0, 6.0]\,\mathrm{GeV}^2$) | 0.34 | 0.12 | 0.040 |
| $Q_4$ ($> 14.2\,\mathrm{GeV}^2$) | 0.26 | 0.099 | 0.032 |
| $Q_5$ ($[1.0, 2.5]\,\mathrm{GeV}^2$) | 0.47 | 0.17 | 0.054 |
| $Q_5$ ($[2.5, 4.0]\,\mathrm{GeV}^2$) | 0.42 | 0.15 | 0.049 |
| $Q_5$ ($[4.0, 6.0]\,\mathrm{GeV}^2$) | 0.34 | 0.12 | 0.040 |
| $Q_5$ ($> 14.2\,\mathrm{GeV}^2$) | 0.23 | 0.088 | 0.027 |
| $Q_6$ ($[1.0, 2.5]\,\mathrm{GeV}^2$) | 0.50 | 0.17 | 0.054 |
| $Q_6$ ($[2.5, 4.0]\,\mathrm{GeV}^2$) | 0.45 | 0.15 | 0.049 |
| $Q_6$ ($[4.0, 6.0]\,\mathrm{GeV}^2$) | 0.36 | 0.12 | 0.040 |
| $Q_6$ ($> 14.2\,\mathrm{GeV}^2$) | 0.27 | 0.10 | 0.032 |
| $Q_8$ ($[1.0, 2.5]\,\mathrm{GeV}^2$) | 0.51 | 0.19 | 0.061 |
| $Q_8$ ($[2.5, 4.0]\,\mathrm{GeV}^2$) | 0.47 | 0.17 | 0.056 |
| $Q_8$ ($[4.0, 6.0]\,\mathrm{GeV}^2$) | 0.38 | 0.14 | 0.045 |
| $Q_8$ ($> 14.2\,\mathrm{GeV}^2$) | 0.27 | 0.10 | 0.032 |

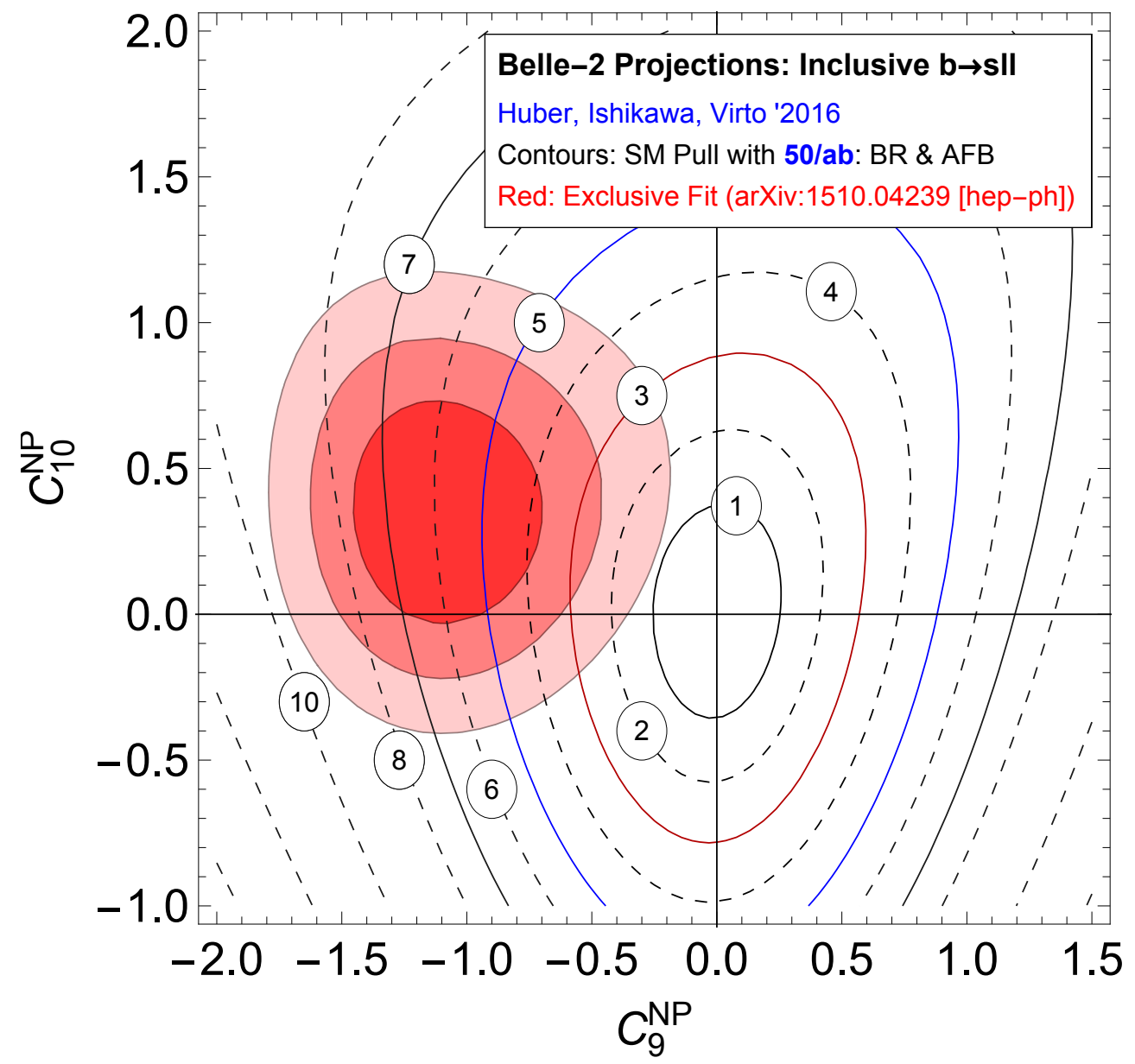

Fig. 94: Exclusion contours in the $C_9^{\rm NP}$–$C_{10}^{\rm NP}$ plane resulting from future inclusive $b \to s\ell^+\ell^-$ measurements at Belle II. For comparison the constraints on $C_9^{\rm NP}$ and $C_{10}^{\rm NP}$ following from the global fit presented in [612] is also shown.

## 9.5. *Missing Energy Channels: $B \to K^{(*)}\nu\bar{\nu}$ and $B_q \to \nu\bar{\nu}$*

### 9.5.1. *$B \to K^{(*)}\nu\bar{\nu}$ transitions.* *(Contributing author: D. Straub)*

The $B \to K^{(*)}\nu\bar{\nu}$ decays provide clean testing grounds for new dynamics modifying the $b \to s$ transition [613–615]. Unlike in other $B$-meson decays, factorisation of hadronic and leptonic currents is exact in the case of $B \to K^{(*)}\nu\bar{\nu}$ because the neutrinos are electrically neutral. Given the small perturbative and parametric uncertainties, measurements of the $B \to K^{(*)}\nu\bar{\nu}$ decay rates would hence in principle allow to extract the $B \to K^{(*)}$ form factors to high accuracy.

Closely related to the $B \to K^{(*)}\nu\bar{\nu}$ modes are the $B$ decays that lead to an exotic final state $X$, since the missing energy signature is the same. Studies of such signals are very interesting in the dark matter context and may allow to illuminate the structure of the couplings between the dark and SM sectors [616].

*$B \to K^{(*)}\nu\bar{\nu}$ in the SM.* Due to the exact factorisation, the precision of the SM prediction for the branching ratios of $B \to K^{(*)}\nu\bar{\nu}$ is mainly limited by the $B \to K^{(*)}$ form factors and by the knowledge of the relevant CKM elements. The relevant Wilson coefficient is known in the SM, including NLO QCD and NLO EW correction to a precision of better than 2% [402, 403, 405]. Concerning the form factors, combined fits using results from LCSRs at low $q^2$ and lattice QCD at high $q^2$ can improve the theoretical predictions.

Using $|\lambda_t^{(s)}| = (4.06 \pm 0.16) \cdot 10^{-2}$ for the relevant CKM elements, obtained using unitarity and an average of inclusive and exclusive tree-level determinations of $|V_{cb}|$, as well as a combined fit to LCSR [419] and lattice QCD [617] results for the $B \to K^*$ form factors, one obtains the following SM prediction for the $B \to K^* \nu\bar{\nu}$ branching ratio [618]

$$\mathrm{Br}(B \to K^* \nu\bar{\nu})_{\mathrm{SM}} = (9.6 \pm 0.9) \cdot 10^{-6} . \tag{274}$$

An angular analysis of the angle spanned by the $B$ meson and the $K^+$ meson resulting from the $K^* \to K^+\pi^-$ decay gives access to an additional observable, the $K^*$ longitudinal polarisation fraction $F_L$, which is sensitive to right-handed currents [614]. The corresponding SM prediction is $F_L^{\mathrm{SM}} = 0.47 \pm 0.03$ [613]. Even with the low number of events expected, it can been shown that such an analysis is quite possible at Belle II (details can be found below in the corresponding experimental section).

The $B \to K$ form factors are known to an even better precision from lattice QCD. Extrapolating the lattice result to the full $q^2$ range, one arrives at [618]

$$\mathrm{Br}(B^+ \to K^+ \nu\bar{\nu})_{\mathrm{SM}} = (4.6 \pm 0.5) \cdot 10^{-6} . \tag{275}$$

Since the isospin asymmetry vanishes for both decays (except for a presumably negligible difference in the charged and neutral form factors), the $B^0$ vs. $B^+$ branching ratios can be trivially obtained by rescaling with the appropriate lifetimes — once the tree-level $B^+ \to \tau^+\nu\,(\tau^+ \to K^+\bar{\nu})$ contribution is properly taken into account [615].

*BSM physics in $B \to K^{(*)}\nu\bar{\nu}$.* Within the SM, the $B \to K^{(*)}\nu\bar{\nu}$ decays are mediated by the effective operator (204) which involves a sum over the three neutrino flavours $\ell = e, \mu, \tau$. In BSM scenarios, there can be a left-handed operator for each neutrino flavour as well as right-handed one of the form

$$Q_R^\ell = (\bar{s}_R \gamma_\mu b_R)(\bar{\nu}_{\ell L} \gamma^\mu \nu_{\ell L}) . \tag{276}$$

In total there can hence be six different operators.

The two branching ratios give access to two combinations of the six Wilson coefficients, namely

$$\begin{aligned} \frac{\mathrm{Br}(B \to K\nu\bar{\nu})}{\mathrm{Br}(B \to K\nu\bar{\nu})_{\mathrm{SM}}} &= \frac{1}{3}\sum_\ell (1 - 2\,\eta_\ell)\,\epsilon_\ell^2 , \\ \frac{\mathrm{Br}(B \to K^*\nu\bar{\nu})}{\mathrm{Br}(B \to K^*\nu\bar{\nu})_{\mathrm{SM}}} &= \frac{1}{3}\sum_\ell (1 + \kappa_\eta \eta_\ell)\,\epsilon_\ell^2 , \end{aligned} \tag{277}$$

where $\kappa_\eta$ is a ratio of binned form factors [613] and

$$\epsilon_\ell = \frac{\sqrt{|C_L^\ell|^2 + |C_R^\ell|^2}}{|C_L^{\mathrm{SM}}|} , \qquad \eta_\ell = \frac{-\mathrm{Re}\left(C_L^\ell C_R^{\ell *}\right)}{|C_L^\ell|^2 + |C_R^\ell|^2} . \tag{278}$$

While in principle, no general constraint on the size of BSM effects in $B \to K^{(*)}\nu\bar{\nu}$ decays can be derived from other processes, in practice in many models there is a relation between semi-leptonic decays with neutrinos and the ones with charged leptons in the final state. This is because $SU(2)_L$ gauge symmetry relates left-handed neutrinos and charged leptons. This relation can be most conveniently studied in the SM effective field theory (SMEFT) [619, 620], based on an OPE in powers of the inverse new-physics scale. The relevant dimension-six operators read

$$
\begin{aligned}
Q_{Hq}^{(1)} &= (\bar{q}_L\gamma_\mu q_L)H^\dagger i\overleftrightarrow{D}^\mu H\,,\\
Q_{Hq}^{(3)} &= (\bar{q}_L\gamma_\mu\tau^a q_L)H^\dagger i\overleftrightarrow{D}^\mu \tau^a H\,,\\
Q_{Hd} &= (\bar{d}_R\gamma_\mu d_R)H^\dagger i\overleftrightarrow{D}^\mu H\,,\\
Q_{ql}^{(1)} &= (\bar{q}_L\gamma_\mu q_L)(\bar{l}_L\gamma^\mu l_L)\,,\\
Q_{ql}^{(3)} &= (\bar{q}_L\gamma_\mu\tau^a q_L)(\bar{l}_L\gamma^\mu\tau^a l_L)\,,\\
Q_{dl} &= (\bar{d}_R\gamma_\mu d_R)(\bar{l}_L\gamma^\mu l_L),
\end{aligned}
\tag{279}
$$

where $H$ denotes the Higgs doublet field, while $q_L$ and $l_L$ are the quark and lepton doublets, respectively, and we have suppressed flavour indices. The generators of $SU(2)_L$ are denoted by $\tau^a$. The SMEFT Wilson coefficients can be matched onto the low-energy Wilson coefficients $C^\ell_{L,R}$ and the ones relevant for $b\to s\ell^+\ell^-$ transitions as follows [613, 621, 622]

$$
\begin{aligned}
C_L &\propto C_{ql}^{(1)} - C_{ql}^{(3)} + C_Z\,,\\
C_R &\propto C_{dl} + C_Z'\,,\\
C_9 &\propto C_{qe} + C_{ql}^{(1)} + C_{ql}^{(3)} - \zeta C_Z\,,\\
C_9' &\propto C_{de} + C_{dl} - \zeta C_Z'\,,\\
C_{10} &\propto C_{qe} - C_{ql}^{(1)} - C_{ql}^{(3)} + C_Z\,,\\
C_{10}' &\propto C_{de} - C_{dl} + C_Z'\,,
\end{aligned}
\tag{280}
$$

where

$$
C_Z = \frac{1}{2}\left(C_{Hq}^{(1)} + C_{Hq}^{(3)}\right), \qquad C_Z' = \frac{1}{2}C_{Hd}\,, \tag{281}
$$

and $\zeta = 1-4s_w^2 \simeq 0.08$ is the accidentally small vector coupling of the $Z$ boson to charged leptons with $s_w$ the sine of the weak mixing angle. While in full generality, these relations are not very useful due to the larger number of operators in the SMEFT, they become useful in models where only a subset of the SMEFT operators are generated. For instance, in models with an additional $SU(2)_L$-singlet neutral heavy gauge boson ($Z'$), one has $C_{ql}^{(3)}=0$. If in addition $Z$–$Z'$ mixing is small, one obtains the prediction

$$
C_L = \frac{C_9 - C_{10}}{2}\,, \qquad C_R = \frac{C_9' - C_{10}'}{2}\,. \tag{282}
$$

In the opposite limit of a new-physics model where only the coefficients $C_Z$ and $C_Z'$ are generated, one obtains

$$
C_L = C_{10}\,, \qquad C_9 = -\zeta C_{10}\,, \tag{283}
$$

and

$$
C_R = C_{10}'\,, \qquad C_9' = -\zeta C_{10}'\,. \tag{284}
$$

In both cases, the existing data on $b \to s\ell^+\ell^-$ transitions limit the size of possible BSM effects in $B \to K^{(*)}\nu\bar{\nu}$. However, in models where new physics enters in the pattern $C_{ql}^{(1)} = -C_{ql}^{(3)}$, larger modifications are possible without any constraint from $b \to s\ell^+\ell^-$ processes. Indeed such a pattern is realised in a particular leptoquark model [613] up to loop effects [294]. Finally, we stress that the constraints from $b \to s\ell^+\ell^-$ processes can be weakened by the contributions of additional operators not relevant in $b \to s\nu\bar{\nu}$, like dipole operators or operators involving right-handed leptons.

In the discussion after (279), we have neglected lepton flavour so far. In fact, in the $B \to K^{(*)}\nu\bar{\nu}$ decays all three neutrino flavours contribute and cannot be distinguished experimentally. In $b \to s\ell^+\ell^-$ transitions, on the other hand, the most precise measurements have been done with muons, and the modes with electrons in the final state are less strongly constrained. Finally, $b \to s\tau^+\tau^-$ decays have not been observed at all to date due to the difficulty posed by the identification of tau leptons. This highlights another important feature of the $B \to K^{(*)}\nu\bar{\nu}$ decays: if new physics couples mostly to the third generation of leptons (and lepton neutrinos), it could cause large enhancements of the $B \to K^{(*)}\nu\bar{\nu}$ branching ratios without strongly affecting $b \to se^+e^-$ or $b \to s\mu^+\mu^-$ decays. Such a dominant coupling to third-generation lepton flavour has been put forward to explain various anomalies in $B$ physics recently [287, 623], cf. the related discussion in Section 9.6.

*Related $b \to q\nu\bar{\nu}$ decays.* The processes $B_s \to \phi\nu\bar{\nu}$ or $B_s \to \eta^{(\prime)}\nu\bar{\nu}$ are based on the same quark-level transition as $B \to K^{(*)}\nu\bar{\nu}$ and only differ in their form factors. In addition, there are also exclusive decays based on the $b \to d\nu\bar{\nu}$ transition, e.g. $B \to \rho\nu\bar{\nu}$, $B \to \omega\nu\bar{\nu}$ or $B \to \pi\nu\bar{\nu}$. In the SM, the SD contribution to these decay rates are parametrically suppressed by $|V_{td}/V_{ts}|^2 \simeq 0.05$ with respect to the $b \to s\nu\bar{\nu}$ modes and thus challenging to detect. Further, charged modes are polluted by the large Cabibbo-allowed tree-level contribution $B^+ \to \tau^+\nu\,(\tau^+ \to (\pi,\rho)^+\bar{\nu})$. Still, order-of-magnitude enhancements of these modes relative to the SM expectations are not excluded in a model-independent fashion.

*9.5.2. Measurements of $B \to K^{(*)}\nu\bar{\nu}$ . (Contributing authors: A. Ishikawa, E. Manoni and D. Straub)*

Searches for the $B \to K^{(*)}\nu\bar{\nu}$ charged and neutral channels have been performed by both BaBar and Belle using hadronic tagging [624, 625] and semi-leptonic tagging [626, 627]. The resulting upper limits at 90% CL are a factor of two to five above the SM predictions [613] for the $K^+$, $K^{*+}$ and $K^{*0}$ channels. Even if new physics does not contribute to the $b \to s\nu\bar{\nu}$ transitions, Belle II will be able to observe the $B \to K^{(*)}\nu\bar{\nu}$ decays.

We have estimated the sensitivities (i.e. reachable precisions) of $B^+ \to K^+\nu\bar{\nu}$, $B^0 \to K^{*0}\nu\bar{\nu}$ and $B^+ \to K^{*+}\nu\bar{\nu}$ by combining the hadronic tagging and semi-leptonic tagging analyses. The three decay modes will be observed with about $10\,\mathrm{ab}^{-1}$ of data and with $50\,\mathrm{ab}^{-1}$ the sensitivities on the branching ratio will be about 10%. Once the $K^*$ modes are observed, measurements of the differential branching ratio and $K^*$ polarisation are important subjects. We performed toy studies and found that it should be possible to measure $F_L$ with an uncertainty of 0.11 when the input $F_L$ value is 0.47 as predicted in the SM [613].

In order to evaluate the impact of machine background on the $B \to K^{(*)}\nu\bar{\nu}$ searches, we have studied signal and generic MC samples (from the MC5 central campaign, described in the "Belle II Simulation" Section 4), in two configuration: physics events superimposed to the nominal machine background ("BGx1" configuration), physics events without machine background ("BGx0" configuration). We considered the $B^\pm \to K^{*\pm}\nu\bar{\nu}$ channel with $K^{*\pm}$ reconstructed in the $K^\pm\pi^0$ final state.

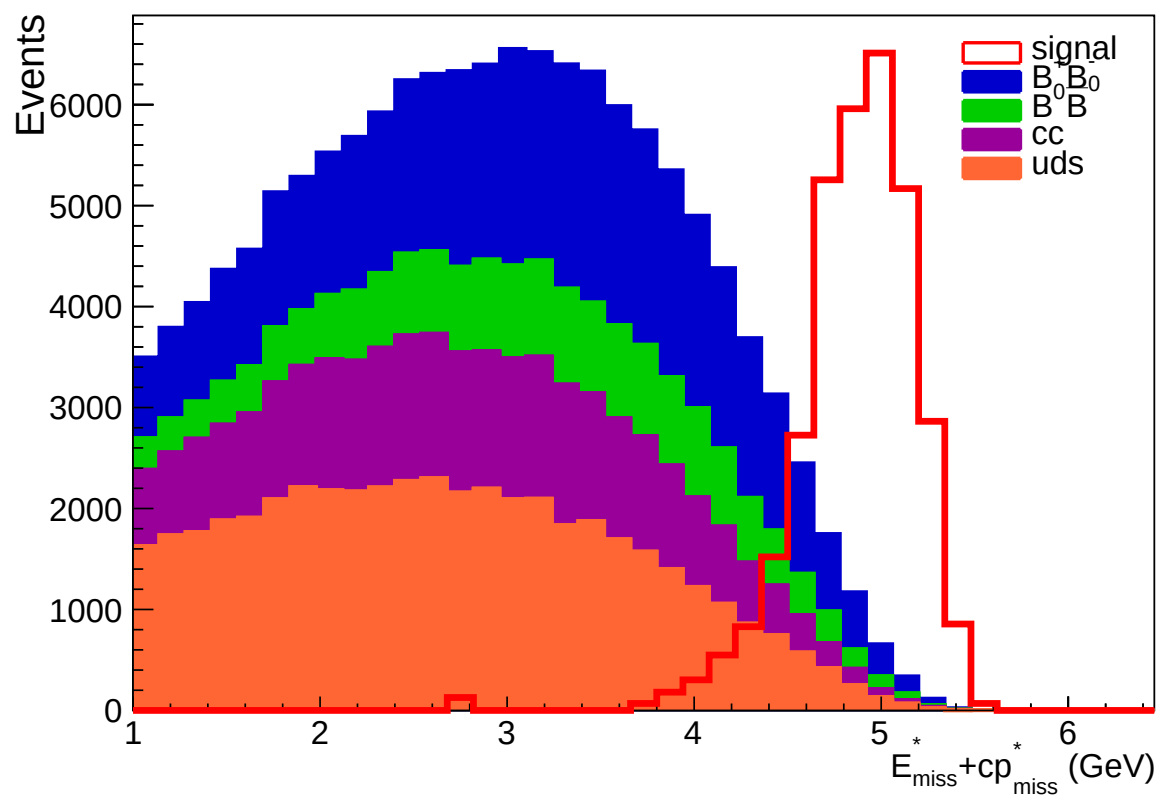

Fig. 95: Distribution of $E^*_{\text{miss}} + cp^*_{\text{miss}}$ for the signal (red) and for the generic MC samples (see legend) in the case of the $K^{*+} \to K^+\pi^0$ channel. The shown results correspond to the "BGx1" configuration after all the selection criteria have been applied but the ones on $E^*_{\text{miss}} + cp^*_{\text{miss}}$ and $E_{\text{ECL}}$. The number of generic MC events corresponds to an integrated luminosity of $1\,\text{ab}^{-1}$, while the signal normalisation is arbitrary.

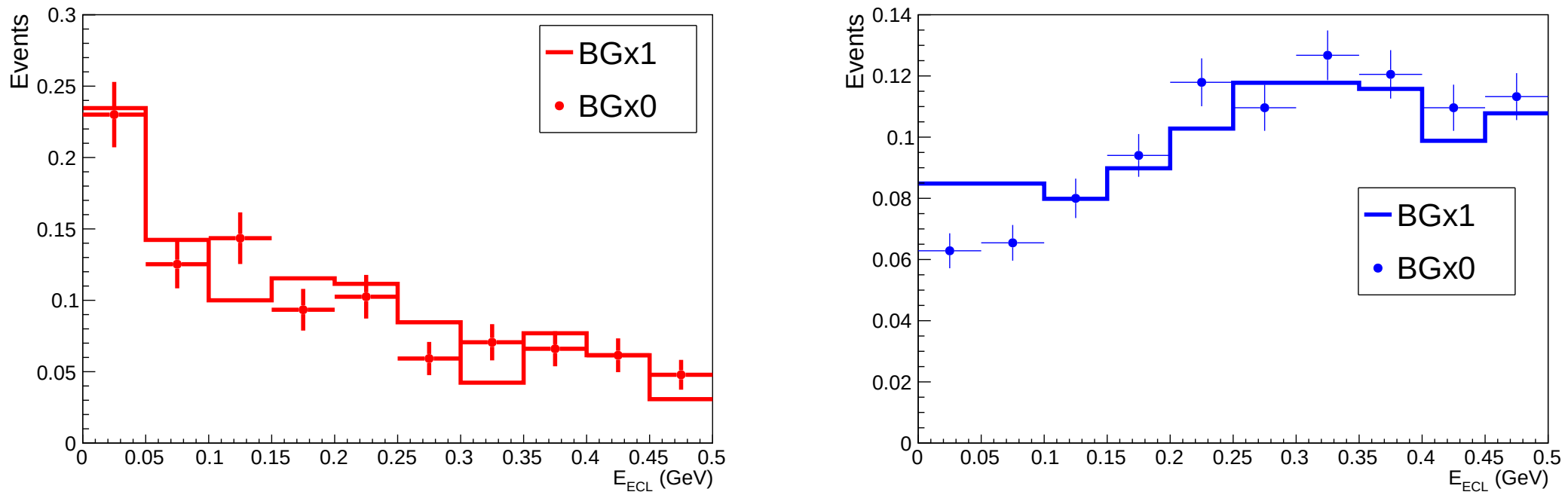

Fig. 96: Distribution of $E_{\text{ECL}}$ normalised to unitary area for the "BGx0" (dots) and the "BGx1" (line) configurations in the case of the $K^{*+} \to K^+\pi^0$ channel after applying all selection criteria. Left: signal MC sample. Right: charged $B^+B^-$ sample.

The used generic MC samples consist of a mixture of $B^+B^-$, $B^0\bar{B}^0$, $u\bar{u}$, $d\bar{d}$, $c\bar{c}$, and $s\bar{s}$ corresponding to $1\,\text{ab}^{-1}$ of data. About 1 million signal MC events, with $K^{*\pm}$ decaying to both $K^\pm\pi^0$ and $K^0_S\pi^\pm$ have also been generated. The signal signature in the recoil of a $B$ reconstructed in hadronic final states are searched for. To do that the official FEI algorithm (see Section 6.6) with ad-hoc refinements on particle identification and cluster cleaning, as done for the $B^+ \to \tau^+\nu$ analysis documented in Section 8.3, are used.

We select $\Upsilon(4S)$ candidates in which the $B_{\text{tag}}$ probability given by the FEI is higher than 0.5%. Moreover, no extra-tracks (tracks not associated to the signal $B$ meson nor to the tag-side $B$ meson) should be reconstructed. We select the best $\Upsilon(4S)$ candidate in the event according to the highest $B_{\rm tag}$ signal probablity and the smallest difference between the reconstructed $K^*$ mass and the PDG value.

Table 68: Number of generic events ($N_{\rm bkg}$), signal selection efficiency ($\varepsilon$), signal significance ($N_{\rm sig}/\sqrt{N_{\rm bkg}}$ with arbitrary normalisation of the signal), and expected upper limit (UL) at 90% CL extracted with a bayesian approach, for zero and nominal background configurations. The MC5 campaign with 1 ab$^{-1}$ of data is useds. The uncertainties reported and the ones used in the UL estimation are statistical only.

| | Background ×0 | Background ×1 |
|---|---|---|
| $N_{\rm bkg}$ | $6415 \pm 80$ | $3678 \pm 61$ |
| $\varepsilon$ ($10^{-4}$) | $10.3 \pm 0.3$ | $5.38 \pm 0.23$ |
| $N_{\rm sig}/\sqrt{N_{\rm bkg}}$ | 0.16 | 0.15 |
| UL ($10^{-4}$) | 2.6 | 3.8 |

Once the best $B\bar{B}$ pair is selected, we exploit variables related to the $B_{\rm tag}$ kinematics ($M_{bc}$ and $\Delta E$ variables) in order to remove mis-reconstructed candidates. Both requirements suppress events in which $B_{\rm tag}$'s originate from a wrong combination of charged and neutral particles, both in $B\bar{B}$ and $q\bar{q}$ events.

The continuum events can be further reduced by considering event shape variables such as $R_2$, i.e. the normalised second Fox-Wolfram moment. The goodness of the strange mesons reconstructed in the signal side is checked through a selection requirement on the difference between the reconstructed mass and the PDG value. Properties of the missing energy in the signal side are also exploited. We define the missing four-momentum in the centre-of-mass (CM) frame as the difference of the $\Upsilon(4S)$ four-momentum and the sum of the $B_{\rm tag}$ and $K^*$ four-momenta. Since no extra tracks are allowed, the missing momentum is related to actual neutrinos, extra-neutrals and particles escaping the detector acceptance. One of the most powerful selection variables of the analysis is the sum of the missing energy and of the modulus of the missing three-momentum in the CM frame ($E^*_{\rm miss} + cp^*_{\rm miss}$) which is required to be greater than 4.5 GeV.

Figure 95 shows the $E^*_{\rm miss} + cp^*_{\rm miss}$ distributions for the $K^{*+} \to K^+\pi^0$ channel, for signal and generic MC samples in the “BGx1” configuration. The quantity $E^*_{\rm miss} + cp^*_{\rm miss}$ is much less correlated to the $\nu\bar{\nu}$ invariant mass than $E^*_{\rm miss}$ or $p^*_{\rm miss}$ alone, making it suitable for a model-independent analysis. A signal region in the extra-neutral energy deposited in the calorimeter ($E_{\rm ECL}$) is also defined, requiring $E_{\rm ECL} < 0.5$ GeV. The distributions for signal MC and the dominant source of background surviving the selection, namely charged $B\bar{B}$ decays, are shown in Figure 96 in both “BGx1” and “BGx0” configurations. In Table 68 a comparison of the selection performances considering the two machine background configurations are reported. As can be noticed, both efficiency and background contamination is higher for the “BGx0” case. This is due to the fact that the optimisation of the selection at reconstruction level has been optimised using the “BGx1” sample and also for the “BGx0” configuration we have used a FEI training performed on the sample with machine background superimposed. The overall signal significance is higher in the background-free sample as expected. From this study we can conclude that, with the machine background campaign used in the MC5 production cycle, the detector performances and the reconstruction algorithms are robust against machine background. This has been tested on a $K^{*+}$ final state with both a neutral particle and charged tracks. In this respect, the analysis of finals states with $K^{*+} \to K_S^0(\pi^+\pi^-)\pi^+$ and $K^{*0} \to K^+\pi^-$, reconstructed with track only, should give similar or better results.

The above study (Figures 95 and 96 and Table 68) is performed only for testing the robustness of the analysis against beam background, which will become one of the major obstacles for physics as the luminosity increases. In order to obtain the ultimate sensitivity achievable at Belle II, we will need to impose more optimised selection criteria and to consider more realistic efficiency and systematic effects, which is beyond the scope of this book. For now, we can extrapolate the Belle sensitivity to Belle II by reducing the statistical errors by the increase in integrated luminosity (see Table 69). We can see that Belle II should be able to observe $B \to K^{(*)}\nu\bar{\nu}$ decays with approximately 5 ab$^{-1}$ of data, by combining the charged and neutral $B$ decay modes. The expected sensitivity of the branching ratios for $B \to K^{(*)}\nu\bar{\nu}$ with 50 ab$^{-1}$ are of order 10%, and thus comparable to the theoretical uncertainties of the SM predictions. A toy MC simulation of the extraction of the longitudinal polarisation fraction $F_L$ of the $K^*$ has been performed and the sensitivity reaches 0.08 for both charged and neutral $B$ decays. The corresponding uncertainty on the SM prediction is 0.03.

Figure 97 shows the constraints on the new-physics contributions to the Wilson coefficients $C_L^{\rm NP}$ and $C_R$ normalised to the SM value of $C_L$, assuming them to be real and independent of the neutrino flavour. The gray areas indicate the 90% CL excluded regions from the first generation $B$ factories, which rule out large enhancements of the Wilson coefficients with respect to the SM expectations. They also rule out a band where $C_L^{\rm NP} + C_R \simeq -C_L^{\rm SM}$. In this region the branching ratio of $B \to K\nu\bar{\nu}$, which is only sensitive to the sum $C_L + C_R$, is close to zero and the combination of the BaBar and Belle searches already rules out a vanishing branching ratio at 90% CL. The coloured bands show the regions allowed at 68% CL by the Belle II measurements with full statistics, assuming the sensitivities quoted in Table 69 and the SM central values for both $F_L$ and the branching ratios. The green band refers to the $B^+ \to K^+\nu\bar{\nu}$ measurement. For $B \to K^*\nu\bar{\nu}$, two bands are shown. The purple one accounts for constraints from the branching ratio only, while the orange one shows the constraint obtained by combining both the branching ratio and $F_L$. As can be seen, a large portion of the currently allowed parameter space will be excluded with the full Belle II statistics.

*9.5.3. Experimental search for $B_q \to \nu\bar{\nu}$ or invisible final states. (Contributing author: A. Ishikawa)*

The $B_d \to \nu\bar{\nu}$ decay and $B_d$-meson decays to invisible final states were searched for by BaBar with semi-leptonic tagging [628] and by Belle using hadronic tagging [629]. The resulting 90% CL upper limits on the branching ratios are $1.7 \cdot 10^{-5}$ and $1.3 \cdot 10^{-4}$, respectively. The $B_s \to \nu\bar{\nu}$ decay has instead not been searched for yet. These decays are helicity suppressed by the neutrino mass, so that the SM expectation is exactly zero (see [630] for predictions taking into account the neutrino masses).

Since there are no charged tracks nor photons in the final states, only the tag-side $B$ mesons can be used for the searches. The Belle analysis used an old hadronic tagging without hierarchical reconstruction method [631], which can increase the tagging efficiency by a factor of two. Another factor of two improvement can be obtained by introducing the FEI. Requirements on event shape variables using multivariate techniques to suppress continuum and $\tau^+\tau^-$ backgrounds are promising to improve the sensitivity further. In combination, an improvement by a factor of five on the efficiency of the hadronic tagging analysis is expected at Belle II. Such an improvement is still not sufficient to beat the semi-leptonic tagging analysis, which is expected to provide upper limits on the branching ratios that are three times better than those following from hadronic tagging. By combining hadronic and semi-leptonic tagging, Belle II is expected to set an upper limit on $\text{Br}(B_d \to \nu\bar{\nu})$ of $1.5 \cdot 10^{-6}$ with $50\,\text{ab}^{-1}$ of integrated luminosity.

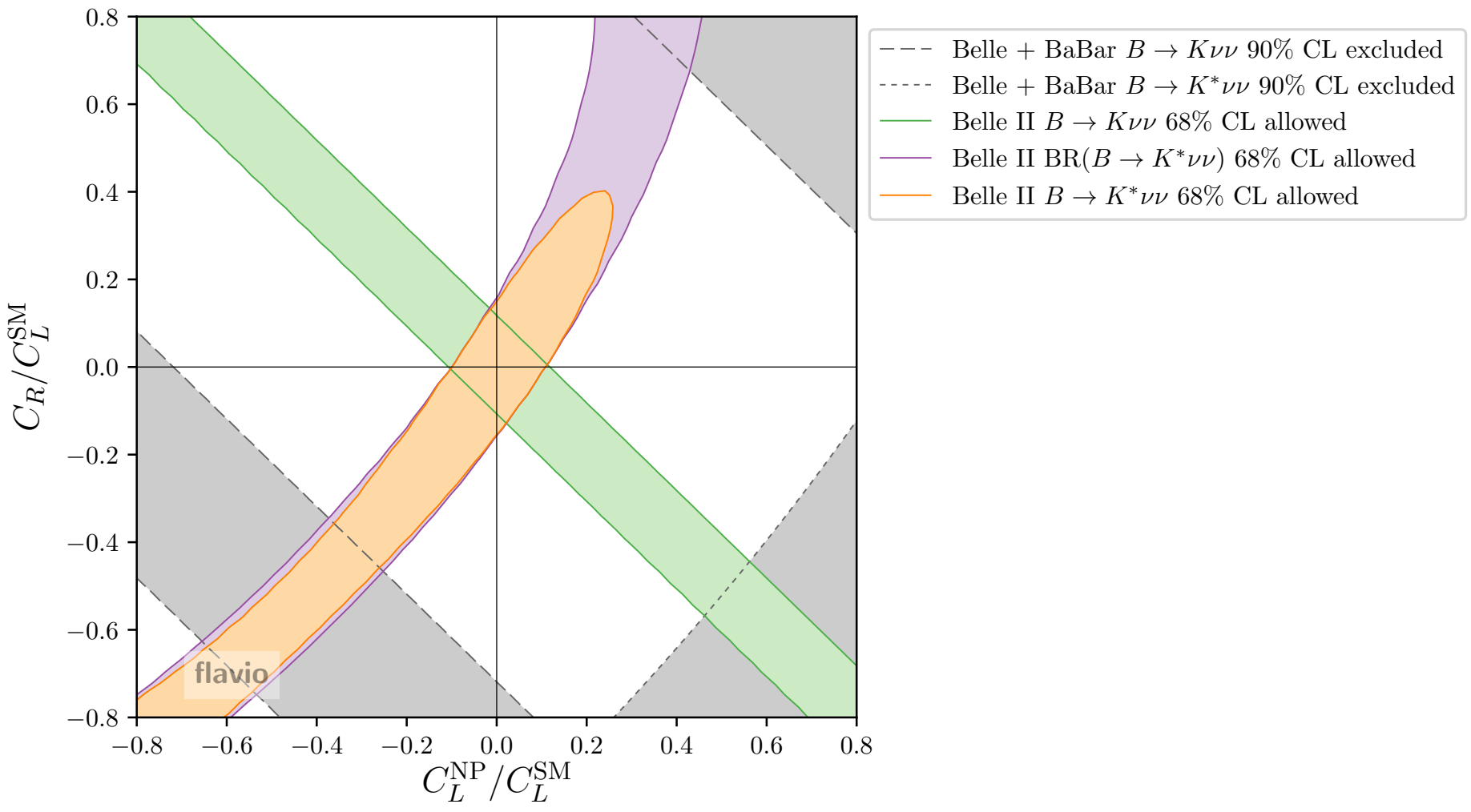

Fig. 97: Constraint on new-physics contributions to the Wilson coefficients $C_L^{\rm NP}$ and $C_R$ normalised to the SM value of $C_L$, assuming them to be real and independent of the neutrino flavour. Dashed (dotted) lines show 90% C.L. excluded regions from upper limits on $Br(B \to K^{(*)}\nu\bar{\nu})$ at Belle and Babar, green (purple) band represents 68% CL allowed region from expected measurements of $Br(B \to K^{(*)}\nu\bar{\nu})$ at Belle II, and orange band gives 68% CL allowed region from expected measurements of $F_L(K^*)$ and the branching ratio in $B \to K^*\nu\bar{\nu}$ at Belle II.

The hadronic $B_s$ tagging efficiency using a hierarchical reconstruction method gives an efficiency that is two times better than that for $B_d$. The semi-leptonic tagging is not tried yet, however it is expected that the tagging efficiency is smaller than that for $B_d$, since the dominant semi-leptonic decay $B_d^0 \to D^{*-}\ell^+\nu$ is clean due to the small mass splitting of $D^{*-}$ and $\bar{D}^0\pi^-$. We conservatively assume that the semileptonic $B_s$ tagging is three times worse than that for $B_d$. By combining the hadronic and semi-leptonic tagging, it is expected that an upper limit on $\text{Br}(B_s \to \nu\bar{\nu})$ of $1.1 \cdot 10^{-5}$ can be set with the full data set of $5\,\text{ab}^{-1}$ collected at $\Upsilon(5S)$.

A summary of the Belle II sensitivities for the modes with neutrinos in the final sates is presented in Table 69.

Table 69: Sensitivities to the modes involving neutrinos in the final states. We assume that $5\,\mathrm{ab}^{-1}$ of data will be taken on the $\Upsilon(5S)$ resonance at Belle II. Some numbers at Belle are extrapolated to $0.71\,\mathrm{ab}^{-1}$ $(0.12\,\mathrm{ab}^{-1})$ for the $B_{u,d}$ $(B_s)$ decay.

| Observables | Belle $0.71\,\mathrm{ab}^{-1}$ $(0.12\,\mathrm{ab}^{-1})$ | Belle II $5\,\mathrm{ab}^{-1}$ | Belle II $50\,\mathrm{ab}^{-1}$ |
|---|---|---|---|
| $\mathrm{Br}(B^+ \to K^+ \nu\bar{\nu})$ | $< 450\%$ | 30% | 11% |
| $\mathrm{Br}(B^0 \to K^{*0} \nu\bar{\nu})$ | $< 180\%$ | 26% | 9.6% |
| $\mathrm{Br}(B^+ \to K^{*+} \nu\bar{\nu})$ | $< 420\%$ | 25% | 9.3% |
| $F_L(B^0 \to K^{*0} \nu\bar{\nu})$ | – | – | 0.079 |
| $F_L(B^+ \to K^{*+} \nu\bar{\nu})$ | – | – | 0.077 |
| $\mathrm{Br}(B^0 \to \nu\bar{\nu}) \times 10^6$ | $< 14$ | $< 5.0$ | $< 1.5$ |
| $\mathrm{Br}(B_s \to \nu\bar{\nu}) \times 10^5$ | $< 9.7$ | $< 1.1$ | – |

*9.5.4. Interpreting missing energy signals as non-standard invisible states. (Contributing author: C. Smith)*

The successes of the SM do not rule out the presence of new light particles. Indeed, if they are sufficiently weakly interacting with SM particles, they could have evaded direct detection until now. One could think for example of the extreme situation in which a unique new particle, fully neutral under the whole SM gauge group, is added to the SM. Our only window to discover such a particle would be its gravitational interactions, and there would be no hope of an earth-based discovery in the foreseeable future. In a more realistic setting though, new neutral light particles would be accompanied by new dynamics at some scale. Presumably, this new dynamics would also affect the SM, and would thus indirectly couple the visible and hidden sectors.

There are many examples of such BSM models. The most well-known example is the axion [632–635], introduced to cure the strong $CP$ problem of the SM. More crucially, there are now very strong indications that the universe is filled with dark matter, so there should be at least one new electrically neutral colourless particle, possibly lighter than the EW scale. Once opening that door, it is not such a drastic step to imagine a whole dark sector, i.e. a set of darkly interacting dark particles only loosely connected to our own visible sector. For a recent review, including further physical motivations from string theory or extra dimensional settings, see for instance [636].

*Experimental Searches.* New light states could show up as missing energy in some process $A \to BX_{\mathrm{dark}}$, with $A$ and $B$ some SM particle states and $X_{\mathrm{dark}}$ representing one or more dark particles. Because of their very weak couplings, high luminosity is crucial to have any hope of discovery, and except in some special circumstances colliders cannot compete with low-energy experiments yet.

Several $B$ decay modes offer unique windows for the search of new dark states with masses up to a few GeV. Specifically, the most promising processes are

$$\begin{aligned} &B \to X_{\rm dark}\,,\\ &B \to (\pi,\rho) X_{\rm dark}\,,\\ &B \to (K,K^*) X_{\rm dark}\,, \end{aligned} \tag{285}$$

with $X_{\rm dark}$ made of at least two dark particles for the first mode, but possibly only one for the others. This also includes situations in which the dark particle is not stable but has cascade decays in the hidden sector, e.g. $X_{\rm dark} \to Y_{\rm dark} Y_{\rm dark}$.

In this context, the SM decays with $X_{\rm SM} = \nu\bar{\nu}$ act as an irreducible background. The relevant branching ratios are smaller than about $10^{-9}$ for the fully invisible mode, and $10^{-5}$ for those with $\pi$, $\rho$, $K$, or $K^*$. It is important to stress though that the kinematics may be different. The differential rate $d\Gamma/dq_X^2$ with $q_X^2$ the missing invariant mass, depends on the nature of $X_{\rm dark}$ and may strongly deviate from that with $X_{\rm SM}$. This is obvious if $X_{\rm dark}$ is a single particle, in which case $d\Gamma/dq_X^2$ would show a peak at $q_X^2 = m_X^2$, or when $X_{\rm dark}$ is made of two states $Y$ with $m_Y^2 \gg 0$ since $d\Gamma/dq_{YY}^2$ would vanish below $q_{YY}^2 = 4m_Y^2$. More generally, $d\Gamma/dq^2$ strongly depend on the Dirac structure(s) involved in the effective couplings of the dark states to the SM quark current $b \to q$, and thereby on whether these states are scalar, fermion or vector particles.

This caveat concerning the differential rate must be kept in mind when reinterpreting the bounds on the branching ratios for $B \to (\pi,\rho,K,K^*)\nu\bar{\nu}$ as bounds on the production of new light states. Not only are those limits obtained from measurements over a fraction of the phase-space, but the SM differential rate is explicitly assumed in the extrapolation. To be consistent, it is thus compulsory to use the same cuts on the produced meson momentum as in the experimental analysis. In this respect, it should be remarked that some recent experimental results [624] do perform differential analyses over the whole $q^2$ range. Those are the data most suitable to look for new light states.

Finally, it should be mentioned that these modes also constrain indirectly other observables. For example, since the branching ratios ${\rm Br}(B^+ \to K^{*+} J/\psi) = (0.143 \pm 0.008)\%$ or ${\rm Br}(B^+ \to \rho^+ \bar{D}) = (1.34 \pm 0.18)\%$ [77] being significantly larger than those for the decays with missing energy (285), the latter modes indirectly bound $J/\psi \to X_{\rm dark}$ or $\bar{D} \to X_{\rm dark}$ whenever $m_{J/\psi}^2$ or $m_D^2$ falls within the missing invariant mass window of the experimental search. This method has been used, and is the best available for charmonium, but remains to be applied for charmed mesons. It is not so promising for $K_{L,S} \to X_{\rm dark}$ because the $B$ decay branching ratios involving kaons are not much enhanced compared to those with a neutrino pair, and because the reach on $B(K \to X_{\rm dark})$ would in any case be very far from the $10^{-10}$ achievable for the golden mode $B(K \to \pi X_{\rm dark})$.

*Theoretical Classification and Expectations.* To organise the search for new light states as model-independently as possible, the strategy is to construct the equivalent of the SMEFT operator basis [619, 620, 637] once the SM particle content is extended, and then constrain all the operators involving the new state(s). This program is more involved than it seems. Clearly, the leading operators one has to consider, the so-called portals, strongly depend on generic assumptions on the nature of the new state. For example, its spin has to be specified, as well as whether it carries a dark charge and needs to be produced in pairs.

Importantly, the dimension of the leading effective operators depend on these assumptions, and we refer to [616, 638] for a complete list of leading interactions of the SM fields with a dark scalar, spin 1/2 or 3/2 fermion or vector boson. For each type of new particle, we separate the case in which it is neutral or charged under some dark symmetry, and then further distinguish the overall leading operators to those involving the quark currents. Indeed, from the point of view of flavour physics, whether the dark states couple dominantly to Higgs or gauge bosons, hence are flavour-blind, or when they couple to quarks and leptons, whether they are able to directly induce the flavour transition is crucial. Even if it is not favourable from a dimensionality point of view to couple $X_{\rm dark}$ directly to quarks, failure to do so means that the flavour transition must still proceed through the SM weak interaction, and ends up suppressed by $G_F$ and CKM factors. From these considerations, three classes of scenarios for a generic effective coupling of $X_{\rm dark}$ to quarks can be identified. We refer to [616] for the full classification of the effective operators, and here only illustrate these three classes for the case of the production of a dark fermion pair.

First, consider the SM contributions which constitutes the irreducible background for BSM production of dark states. It can be embodied into the generic dimension-six effective operators

$$\mathcal{H}_{\rm eff} = \sum_{q=s,d} \frac{c^{bq}}{\Lambda^2}\, \bar{b}\Gamma q\, \bar{\nu}\Gamma\nu \,, \tag{286}$$

where $\Gamma$ represents all possible Dirac structures and $c^{bq}$ denote the Wilson coefficients. We recall that in the SM one has $\Lambda \simeq m_W$, $c^{bs}_{\rm SM} \simeq \alpha_w/(4\pi)\,\lambda$ and $c^{bd}_{\rm SM} \simeq \alpha_w/(4\pi)\,\lambda^3$ with $\alpha_w = g^2/(4\pi)$ the $SU(2)_L$ coupling constant.

For the first scenario, imagine that the production of a dark fermion pair proceeds through the flavour-changing operator $\bar{Q}^I\gamma_\mu Q^J\bar{\psi}\gamma^\mu\psi$, where $Q$ is a left-handed quark doublet and $I, J$ denote flavour indices. The BSM rate will be of the order of the SM $b \to q\nu\bar{\nu}$ rate when

$$\frac{c^{bq}_{\rm dark}}{\Lambda^2} \simeq G_F\, \frac{\alpha_w}{4\pi}\, \lambda_t^{(q)} \,. \tag{287}$$

Provided the Wilson coefficient $c^{bq}_{\rm dark}$ is $\mathcal{O}(1)$, the reach in $\Lambda$ is rather high, i.e. about 40 TeV (20 TeV) for $b \to d$ ($b \to s$) transitions.

On the contrary, for the second scenario, imagine that the leading coupling is flavour-blind, say $H^\dagger \overset{\leftrightarrow}{D}_\mu H\bar{\psi}\gamma^\mu\psi \supset v^2\bar{\psi}\gamma_\mu\psi Z^\mu$ with $v \simeq 246\,\text{GeV}$ the Higgs vacuum expectation value. Then, the production of new states is driven by the SM $Z$ penguin. As a result, the relation (287) takes the form

$$c^{HH}_{\rm dark}\, \frac{v^2}{\Lambda^2}\, G_F\, \frac{\alpha_w}{4\pi}\lambda_t^{(q)} \simeq G_F\, \frac{\alpha_w}{4\pi}\, \lambda_t^{(q)} \,. \tag{288}$$

In this case, the reach in $\Lambda$ is around the EW scale at best, i.e. when $c^{HH}_{\rm dark} = \mathcal{O}(1)$, and is in general not competitive with other searches using EW precision observables, invisible Higgs boson decay or other flavour-blind searches. Note that even very low-energy probes are sensitive to $v^2\bar{\psi}\gamma_\mu\psi Z^\mu$ since the $Z$ boson couples to all SM fermions. A similar conclusion is valid for all the flavour blind operators, even when those arise at a much lower order and appear superficially less suppressed by the new-physics scale $\Lambda$ than those involving quark fields.

In between these two extreme situations, there is a third scenario. If the dark state couples dominantly to top-quark pairs, then all the flavour-blind low-energy searches would be inefficient, while high-energy collider searches relying for example on the associated productions of a top quark and a dark state would not be competitive yet. In this case, the FCNC processes still represent our best window, even if the reach in the BSM scale $\Lambda$ would not be much higher than the EW scale.

## *9.6. Tauonic EW Penguin Modes*

*(Contributing authors: W. Altmannshofer and J. Kamenik)*

$B$-meson decays to $\tau^+\tau^-$ final states are experimentally largely uncharted territory. While a few bounds like $\mathrm{Br}\,(B_d \to \tau^+\tau^-) < 1.3 \cdot 10^{-3}$ [639] and $\mathrm{Br}\,(B^+ \to K^+\tau^+\tau^-) < 2.25 \cdot 10^{-3}$ [640] do exist, they are all orders of magnitudes away from the corresponding SM predictions. In view of the fact that measurements of $\tau^+\tau^-$ final states remain a big challenge at LHCb, and that it is unclear whether a sensitivity beyond $\mathcal{O}(10^{-3})$ can be reached [641], Belle II might be the only next-generation machine that allows to explore these modes in some depth.

### *9.6.1. $b \to q\tau^+\tau^-$ and lepton flavour violating modes with taus.*

*Purely tauonic modes.* The most recent SM predictions for the branching ratios of the purely leptonic $B_s \to \tau^+\tau^-$ and $B_d \to \tau^+\tau^-$ decays include NNLO QCD corrections and NLO EW corrections [219, 404, 406]. They are given by

$$\begin{aligned} \mathrm{Br}(B_s \to \tau^+\tau^-)_{\mathrm{SM}} &= (7.73 \pm 0.49) \cdot 10^{-7}\,, \\ \mathrm{Br}(B_d \to \tau^+\tau^-)_{\mathrm{SM}} &= (2.22 \pm 0.19) \cdot 10^{-8}\,. \end{aligned} \tag{289}$$

These SM predictions refer to the average time-integrated branching ratios. The uncertainties are dominated by CKM elements and the $B$-meson decay constants $f_{B_q}$. The used input parameters are collected in [219].

*Semi-tauonic Modes.* Predictions for exclusive semi-leptonic decays depend on form factors. In the semi-tauonic decays the di-lepton invariant mass, $q^2$, is restricted to the range from $4m_\tau^2 \simeq 12.6\,\mathrm{GeV}^2$ to $(m_B - m_H)^2$, where $H = \pi, K, K^*, \ldots$ . To avoid contributions from the resonant decay through the narrow $\psi(2S)$ charmonium resonance, $B \to H\psi(2S)$ with $\psi(2S) \to \tau^+\tau^-$, SM predictions are typically restricted to a di-tau invariant mass $q^2 > 15\,\mathrm{GeV}^2$. In this kinematic regime, lattice computations are expected to provide reliable results for the form factors (see the discussion in Section 9.1.1).

Combining the uncertainties from the relevant CKM elements and form factors leads to SM predictions for the branching ratios of the semi-tauonic decays with an accuracy of around 10% to 15%. The presence of broad charmonium resonances above the open charm threshold is a source of additional uncertainty. Possible effects of the broad resonances are typically taken into account by assigning an additional error of a few percent following [642] (or possibly more [643] when the $B \to K\ell^+\ell^-$ LHCb data [644] is considered), which is subdominant compared to the CKM and form factor uncertainties.

SM predictions for the decay $B \to \pi\tau^+\tau^-$ have been presented in [159, 645] using form factors from the Fermilab/MILC collaboration [143, 159]. Results are given for the branching ratios and the "flat term" in the angular distributions (cf. [590, 645] for the definition of the latter observable)

$$\begin{aligned}\mathrm{Br}(B^+ \to \pi^+\tau^+\tau^-)_{\mathrm{SM}} &= (4.29 \pm 0.39) \cdot 10^{-9}\,,\\ \mathrm{Br}(B^0 \to \pi^0\tau^+\tau^-)_{\mathrm{SM}} &= (1.99 \pm 0.18) \cdot 10^{-9}\,,\\ F_H(B \to \pi\tau^+\tau^-)_{\mathrm{SM}} &= 0.80 \pm 0.02\,,\end{aligned} \tag{290}$$

where the prediction for $F_H(B \to \pi\tau^+\tau^-)_{\mathrm{SM}}$ holds for both $B^+$ and $B^0$ and all errors quoted in [159, 645] have been added in quadrature to obtain the final uncertainties. The above predictions correspond to a di-tau invariant mass squared $q^2 \in [15, 22]\,\mathrm{GeV}^2$. Predictions for additional $q^2$ bins are available in [159, 645]. The dominant uncertainties in the branching ratios come from the $B \to \pi$ form factors and the CKM input. Those uncertainties cancel to a large extent in the flat term.

Also for the $B \to K\tau^+\tau^-$ decays, SM predictions have been given in [645], using recent lattice determination of the $B \to K$ form factors from the Fermilab/MILC collaboration [158]. The SM predictions for the branching ratios and the flat terms read

$$\begin{aligned}\mathrm{Br}(B^+ \to K^+\tau^+\tau^-)_{\mathrm{SM}} &= (1.22 \pm 0.10) \cdot 10^{-7}\,,\\ \mathrm{Br}(B^0 \to K^0\tau^+\tau^-)_{\mathrm{SM}} &= (1.13 \pm 0.09) \cdot 10^{-7}\,,\\ F_H(B \to K\tau^+\tau^-)_{\mathrm{SM}} &= 0.87 \pm 0.02\,,\end{aligned} \tag{291}$$

where we added all uncertainties quoted in [645] in quadrature. As in the case of the $B \to \pi\tau^+\tau^-$, the value of $F_H(B \to K\tau^+\tau^-)_{\mathrm{SM}}$ applies to the charged and neutral channel and the above predictions refer to the $q^2$ range $[15, 22]\,\mathrm{GeV}^2$. Predictions for additional $q^2$ bins can be found in [645]. Again, the dominant source of uncertainty in the branching ratio arises from the $B \to K$ form factors and from the CKM elements, while in the flat terms these errors largely cancel.

The SM predictions for the $B \to K^*\tau^+\tau^-$ branching ratios read [618]

$$\begin{aligned}\mathrm{Br}(B^+ \to K^{*+}\tau^+\tau^-)_{\mathrm{SM}} &= (0.99 \pm 0.12) \cdot 10^{-7}\,,\\ \mathrm{Br}(B^0 \to K^{*0}\tau^+\tau^-)_{\mathrm{SM}} &= (0.91 \pm 0.11) \cdot 10^{-7}\,,\end{aligned} \tag{292}$$

where the di-tau $q^2$ ranges from $15\,\mathrm{GeV}^2$ to the kinematic endpoint around $19.2\,\mathrm{GeV}^2$. The used $B \to K^*$ form factors are based on a combined fit of lattice and LCSR results [419].

The SM prediction for the $B_s \to \phi\tau^+\tau^-$ branching ratio is given by [618]

$$\mathrm{Br}(B_s \to \phi\tau^+\tau^-)_{\mathrm{SM}} = (0.73 \pm 0.09) \cdot 10^{-7}\,, \tag{293}$$

where the di-tau invariant mass ranges from $15\,\mathrm{GeV}^2$ to the kinematic endpoint at roughly $18.9\,\mathrm{GeV}^2$. The used $B_s \to \phi$ form factors are based on a combined fit of lattice and LCSR results [419].

*Lepton flavour universality ratios with taus.* We define the lepton flavour universality ratios, in analogy to (269), $R_H^{\ell\ell'}[q_0^2, q_1^2] = \frac{\int_{q_0^2}^{q_1^2} dq^2 d\mathrm{Br}(B\to H\ell^+\ell^-)/dq^2}{\int_{q_0^2}^{q_1^2} dq^2 d\mathrm{Br}(B\to H\ell'^+\ell'^-)/dq^2}$. In these ratios uncertainties from CKM elements drop out. Also form factor uncertainties cancel almost exactly in ratios involving electrons and muons, while in ratios with taus, these uncertainties get reduced.

The SM predictions from [645] read

$$(R_\pi^{\mu\tau})_{\rm SM} = 1.18 \pm 0.06\,, \qquad (R_K^{\mu\tau})_{\rm SM} = 0.87 \pm 0.02\,, \tag{294}$$

for the $q^2 \in [15, 22]\,\text{GeV}^2$ bin. For the $B \to K^*$ decays we find [618]

$$(R_{K^*}^{\mu\tau})_{\rm SM} = 2.44 \pm 0.09\,, \tag{295}$$

where $q^2 \in [15, 19.2]\,\text{GeV}^2$. Within the quoted uncertainties, the results (294) and (295) apply to both charged and neutral decays.

*Probing BSM Physics.* Since the $b \to q\tau^+\tau^-$ decays involve third-generation fermions in the final state, one can envisage new-physics scenarios — such as models with extended Higgs or gauge sectors or scenarios with leptoquarks — that give rise to effects in the $\tau^+\tau^-$ modes, while leaving the $e^+e^-$ and/or $\mu^+\mu^-$ channels unaltered. In a model-independent approach, tau-specific new physics in rare $B$-meson decays can be described by an effective Hamiltonian that contains besides the operators $Q_7$, $Q_9$, $Q_{10}$ introduced in (202) and (203) their chirality-flipped partners $Q_7'$, $Q_9'$, $Q_{10}'$ as well as

$$Q_S = (\bar{q}_L b_R)(\bar{\tau}_R \tau_L)\,, \qquad Q_S' = (\bar{q}_R b_L)(\bar{\tau}_L \tau_R)\,. \tag{296}$$

To constrain all possible $\tau^+\tau^-$ operators, one should try to measure/bound both purely leptonic and semi-leptonic modes, since they have different blind directions in parameter space [537, 646]. In this respect it is also interesting to note that $b \to s\nu\bar{\nu}$ decays can constrain the operator combinations containing a left-handed tau current $Q_9 - Q_{10}$ and $Q_9' - Q_{10}'$, due to $SU(2)_L$ invariance. On the other hand, $b \to s\nu\bar{\nu}$ is blind to the orthogonal directions $Q_9 + Q_{10}$ and $Q_9' + Q_{10}'$, that contain right-handed tau currents.

Many BSM models can lead to modifications in the $b \to q\tau^+\tau^-$ channels. Interestingly, several models that address the LHCb anomalies in the $b \to s\mu^+\mu^-$ sector [389, 391, 497–499, 610, 647] or the evidence of lepton flavour universality violation in $B \to D^{(*)}\ell\nu$ decays [250–252, 267, 268, 270, 272], predict characteristic deviations in $b \to s\tau^+\tau^-$ transitions from the SM predictions.

The model proposed in [648] contains a $Z'$ boson, associated to the gauge symmetry of muon-number minus tau-number, $L_\mu - L_\tau$. Given the current anomalies in the $b \to s\mu^+\mu^-$ sector, the model predicts a suppression of all semi-leptonic $b \to s\mu^+\mu^-$ decays by about 20% [599]. The $L_\mu - L_\tau$ symmetry implies that all semi-leptonic $b \to s\tau^+\tau^-$ decays are instead enhanced. Translating the predictions for $b \to s\mu^+\mu^-$ transitions found in the minimal flavour violation (MFV) scenario of [599] to the tau sector using [618], we find

$$\begin{aligned}
\text{Br}(B^+ \to K^+\tau^+\tau^-)_{L_\mu - L_\tau} &= (1.46 \pm 0.13) \cdot 10^{-7}\,, \\
\text{Br}(B^0 \to K^0\tau^+\tau^-)_{L_\mu - L_\tau} &= (1.35 \pm 0.12) \cdot 10^{-7}\,, \\
\text{Br}(B^+ \to K^{*+}\tau^+\tau^-)_{L_\mu - L_\tau} &= (1.53 \pm 0.23) \cdot 10^{-7}\,, \\
\text{Br}(B^0 \to K^{*0}\tau^+\tau^-)_{L_\mu - L_\tau} &= (1.40 \pm 0.21) \cdot 10^{-7}\,,
\end{aligned} \tag{297}$$

where the $K^{+,0}$ branching ratios refer to the $q^2$ region $q^2 \in [15, 22]\,\mathrm{GeV}^2$, while the $K^*$ rates correspond to $q^2 \in [15, 19.2]\,\mathrm{GeV}^2$. The $B_s \to \tau^+\tau^-$ decay remains SM-like in the $L_\mu - L_\tau$ framework.

In the scenarios discussed in [286, 287, 623], the current $B$ physics anomalies are addressed by BSM physics in the form of left-handed currents involving mainly the third generation. In these scenarios the $b \to s\tau^+\tau^-$ decays can in principle be enhanced by an order of magnitude compared to the SM predictions. Left-handed (LH) currents imply a strong correlation between $b \to s\tau^+\tau^-$ and $b \to s\nu\bar{\nu}$ decays, see also the discussion in Section 9.5.1 above. Using the current upper bound on $\mathrm{Br}(B^+ \to K^+\nu\bar{\nu}) < 1.6 \cdot 10^{-7}$ [624], one finds the following maximal values for the tauonic branching ratios [618]

$$\begin{aligned}
\mathrm{Br}(B^+ \to K^+\tau^+\tau^-)_{\mathrm{LH}} &< 24.5 \cdot 10^{-7}\,, \\
\mathrm{Br}(B^0 \to K^0\tau^+\tau^-)_{\mathrm{LH}} &< 22.5 \cdot 10^{-7}\,, \\
\mathrm{Br}(B^+ \to K^{*+}\tau^+\tau^-)_{\mathrm{LH}} &< 22.8 \cdot 10^{-7}\,, \\
\mathrm{Br}(B^0 \to K^{*0}\tau^+\tau^-)_{\mathrm{LH}} &< 20.1 \cdot 10^{-7}\,, \\
\mathrm{Br}(B_s \to \tau^+\tau^-)_{\mathrm{LH}} &< 1.5 \cdot 10^{-5}\,.
\end{aligned} \tag{298}$$

The $q^2$ regions are chosen as in (297). Enhancements beyond the above bounds are possible in the presence of right-handed currents [613]. Measurement of $b \to s, d\tau^+\tau^-$ modes are thus likely to play an important role in the search for lepton non-universality and indirectly may also provide useful information on lepton flavour violation (see for instance the discussion in Ref. [649]).

*9.6.2. Experimental prospects for tauonic EW penguin decays. (Contributing author: A. Ishikawa)*

Studies of the $B^+ \to K^+\tau^+\tau^-$ and $B_{d,s} \to \tau^+\tau^-$ decay modes are interesting because they allow to search for new physics which affects EW penguin $B$ decays involving third-generation leptons. Since the final states contain multiple neutrinos, a tagging of the other $B$ meson is needed to search for these decays. Recently, Belle demonstrated that hadronic $B_s$ tagging for rare decays is possible, despite the dominant mode of production proceeding through intermediate excited states which degrades resolution. After tagging the other $B$ meson, tau leptons can be reconstructed in single prong decays. Even with the improved reconstruction techniques, observations of the SM branching ratios of $B^+ \to K^+\tau^+\tau^-$ and $B_{d,s} \to \tau^+\tau^-$ are unlikely. The expected upper limits on the branching ratios that Belle II should be able to place are of order $10^{-5}$ and $10^{-4}$ for $B_d$ and $B_s$ decays, respectively.

Searches for lepton flavour violating $B^+ \to K^+\tau^\pm\ell^\mp$ and $B_{s,d} \to \tau^\pm\ell^\mp$ ($\ell = e, \mu$) decays are relatively easy compared to the di-tau modes, since the $\tau$ can be reconstructed in hadronic $B$ tagged events. In the three-body $B^+ \to K^+\tau^\pm\ell^\mp$ decays, the $\tau$ four-momentum can be determined from the momentum of the $B$, $K$ and $\ell^\mp$, and in the two-body $B_{s,d} \to \tau^\pm\ell^\mp$ decays the $\ell$ is monochromatic . These clear signatures allow to set better upper limits of the order of $10^{-6}$ for $\mathcal{B}(B^+ \to K^+\tau\ell) = \mathcal{B}(B^+ \to K^+\tau^+\ell^-) + \mathcal{B}(B^+ \to K^+\tau^-\ell^+)$ and $\mathcal{B}(B_d \to \tau^\pm\ell^\mp)$, and of the order of $10^{-5}$ for $\mathcal{B}(B_s \to \tau^\pm\ell^\mp)$ decays.

Table 70: The Belle II sensitivities for the EW penguin $B$ decays involving taus in the final states. We assume that $5\,\mathrm{ab}^{-1}$ of data will be taken on the $\Upsilon(5S)$ resonance. Some numbers at Belle are extrapolated to $0.71\,\mathrm{ab}^{-1}$ $(0.12\,\mathrm{ab}^{-1})$ for the $B_{u,d}$ $(B_s)$ decay.

| Observables | Belle $0.71\,\mathrm{ab}^{-1}$ $(0.12\,\mathrm{ab}^{-1})$ | Belle II $5\,\mathrm{ab}^{-1}$ | Belle II $50\,\mathrm{ab}^{-1}$ |
|---|---|---|---|
| $\mathrm{Br}(B^+ \to K^+\tau^+\tau^-) \cdot 10^5$ | $< 32$ | $< 6.5$ | $< 2.0$ |
| $\mathrm{Br}(B^0 \to \tau^+\tau^-) \cdot 10^5$ | $< 140$ | $< 30$ | $< 9.6$ |
| $\mathrm{Br}(B^0_s \to \tau^+\tau^-) \cdot 10^4$ | $< 70$ | $< 8.1$ | – |
| $\mathrm{Br}(B^+ \to K^+\tau^\pm e^\mp) \cdot 10^6$ | – | – | $< 2.1$ |
| $\mathrm{Br}(B^+ \to K^+\tau^\pm \mu^\mp) \cdot 10^6$ | – | – | $< 3.3$ |
| $\mathrm{Br}(B^0 \to \tau^\pm e^\mp) \cdot 10^5$ | – | – | $< 1.6$ |
| $\mathrm{Br}(B^0 \to \tau^\pm \mu^\mp) \cdot 10^5$ | – | – | $< 1.3$ |

The Belle II sensitivities for the various observables which contain taus in the final sate are summarised in Table 70.

### 9.7. *Conclusions*

The study of radiative and EW penguin $B$ decays remains an important area of precision physics with its overarching goal to discover new physics indirectly by finding deviations between measurements and the corresponding SM predictions. This research direction has been established by a joined effort within the theory community and experimental results from BaBar, Belle and LHCb.

Belle II will contribute to the flavour precision program in two ways. Firstly, by improving the measurements of various FCNC key observables. The inclusive and exclusive $b \to s, d\gamma$ channels as well as the inclusive $b \to s\ell^+\ell^-$ transition (see Tables 61, 62 and 64 for the expected Belle II sensitivities) are well-known examples, but the studies performed in this chapter also show that Belle II can perform measurements that are competitive to those at LHCb for exclusive $b \to s\ell^+\ell^-$ channels, including obervables that test lepton flavour universality (see Tables 65, 66 and 67 for the Belle II prospects). As exemplified by Figure 94 for the case of the $b \to s\ell^+\ell^-$ transitions this complementarity and synergy can play a crucial role in indirectly discovering (or constraining) BSM physics.

Belle II will furthermore push the frontier in the field of radiative and EW penguin $B$ decays by measuring modes at the level predicted by the SM that have so far not been observed by any other experiment. The prime examples for such *discovery channels* are $B_d \to \gamma\gamma$ and $B \to K^{(*)}\nu\bar{\nu}$ (see Tables 63 and 69 for the expected Belle II sensitivities). In other cases, such as $B_s \to \nu\bar{\nu}$ or $B$ decays to final states containing $\tau^+\tau^-$, $\tau^\pm e^\mp$ or $\tau^\pm\mu^\mp$ pairs, Belle II will not be able to observe them at the SM level. However with $50\,\mathrm{ab}^{-1}$ of data the existing limits will be improved by orders of magnitude (see Table 70 for the future Belle II constraints) which will further constrain possible new-physics couplings to neutrinos and taus as well as flavour violation in the lepton sector.

## 10. Time Dependent $CP$ Asymmetries of $B$ mesons and the Determination of $\phi_1$, $\phi_2$

*Editors: A. Gaz, L. Li Gioi, S. Mishima, J. Zupan*

*Additional section writers: F. Abudinén, F. Bishara, M. Gronau, Y. Grossman, S. Jaeger, M. Jung, S. Lacaprara, A. Martini, A. Mordà, D. Robinson, A. Tayduganov*

### *10.1. Introduction*

The measurements of the CKM unitarity triangle angles, $\phi_1, \phi_2, \phi_3$, amount, within the SM, to different ways of measuring the single $CP$ violating phase in the CKM matrix. In the presence of NewPhysics (NP), additional phases might lead to an overall inconsistency of the constraints on the CKM Unitarity Triangle. This would be a clear indication of NP.

In this section we describe the methods for determining the angles

$$\phi_1 \equiv \arg[-V_{cb}^* V_{cd}/(V_{tb}^* V_{td})], \tag{299}$$

and

$$\phi_2 \equiv \arg[-V_{tb}^* V_{td}/V_{ub}^* V_{ud}]. \tag{300}$$

All of these methods are sensitive to the $B$–$\overline{B}$ mixing phase. In the SM this is induced at one loop level and can be modified by many NP models. The angle $\phi_3$ is determined from tree level processes, and is less sensitive to NP (see Section 10).

Precision measurements of angles $\phi_1$ and $\phi_2$ are crucial inputs into the CKM unitarity triangle fits. For the time-dependent $CP$ asymmetry parameter $S_{J/\psi K^0}$, the most precise measurement determining the $\phi_1$ angle, Belle II is projected to reduce the present world average error of 0.022 [230] down to 0.0052 (see Table 97). For penguin dominated modes, $\phi K^0$, $\eta' K^0$, $\omega K_S^0$, $K_S^0 \pi^0 \gamma$, $K_S^0 \pi^0$, that are particularly sensitive to NP from penguin contributions, Belle II is expected to improve the world average precision by a factor of two with 5 $\mathrm{ab}^{-1}$ (see Table 97). The experimental errors on the measurements that enter the determination of the angle $\phi_2$ will have errors reduced by factors between 2 and 10 depending on the sources of systematic uncertainties (see Tables 91 and 92). Additionally, the novel measurement of the time-dependent $CP$ asymmetry parameter $S_{\pi^0\pi^0}$ will help to reduce the ambiguities in the determination of $\phi_2$ within the $B \to \pi\pi$ decays. Considering the decays $B \to \pi\pi$ and $B \to \rho\rho$ together, the total uncertainty on $\phi_2$ is projected to be about $0.6^\circ$ (the current world average error is about $4.2^\circ$ [650]). An important requirement is that the theoretical uncertainties within the SM predictions – to be discussed below – are controlled sufficiently well.

The general strategies for extracting $\phi_{1,2}$ use time-dependent $CP$ asymmetries due to the interference between $B$–$\overline{B}$ mixing and $B$ decay amplitudes. Interference between the two neutral $B$ meson evolution eigenstates $|B_\pm\rangle = p|B\rangle \pm q|\overline{B}\rangle$ generates the time-dependent $CP$ asymmetry

$$a_{CP}(t) \equiv \frac{\Gamma(\bar{\mathcal{D}};t) - \Gamma(\mathcal{D};t)}{\Gamma(\bar{\mathcal{D}};t) + \Gamma(\mathcal{D};t)} = \frac{S_f \sin(\Delta m\, t) + A_f \cos(\Delta m\, t)}{\cosh(\Delta\Gamma\, t/2) + A_{\Delta\Gamma}^f \sinh(\Delta\Gamma\, t/2)}\,, \tag{301}$$

where

$$S_f = \frac{2\,\mathrm{Im}[\lambda_f]}{1+|\lambda_f|^2}\,, \quad -A_f \equiv C_f = \frac{1-|\lambda_f|^2}{1+|\lambda_f|^2}\,, \quad A_f^{\Delta\Gamma} = -\frac{2\,\mathrm{Re}[\lambda_f]}{1+|\lambda_f|^2}\,. \tag{302}$$

Here $\mathcal{D} : B(t) \to f$ and $\bar{\mathcal{D}} : \overline{B}(t) \to f$, with $f$ a common $CP$ eigenstate with eigenvalue $\eta_f = \pm 1$; $\Delta m \equiv m_H - m_L > 0$ and $\Delta\Gamma \equiv \Gamma_L - \Gamma_H$ are respectively the mass and decay rate splittings of the heavy (H) and light (L) eigenstates. The initial, $t = 0$, states are flavour tagged, *i.e.*, $B(0) = B$ and $\overline{B}(0) = \overline{B}$. In the $B_d$ system the decay width splitting $\Delta\Gamma$ can be safely set to zero up to sub percent precisions [230]. However, it is non-negligible in the $B_s$ system, discussed below.

The interference between mixing and decay is described by the parameter

$$\lambda_f \equiv (q/p)(\bar{\mathcal{A}}_f/\mathcal{A}_f), \tag{303}$$

where the decay amplitudes are $\mathcal{A}_f \equiv \langle f|H_{\rm ew}|B\rangle$ and $\bar{\mathcal{A}}_f \equiv \langle f|H_{\rm ew}|\overline{B}\rangle$ ($\mathcal{A}_f$ should not be confused with $A_f \equiv -C_f$). In the $B$–$\overline{B}$ system, $CP$ violation in mixing ($|q/p| \neq 1$) is measured separately and is negligible [230]. We can thus safely assume that $q/p = e^{-i\phi_d}$, where the $B$–$\overline{B}$ mixing phase is strictly

$$\phi_d = \arg[V_{tb}V_{td}^*/(V_{tb}^*V_{td})] \simeq 2\phi_1, \tag{304}$$

up to negligible corrections in the SM, but possibly large corrections if there are NP contributions.

In this section we present sensitivity studies based on Belle II simulation for the following four final states: $B^0 \to \phi K^0$, $\eta' K^0$, $K_S^0\pi^0\gamma$, and $\pi^0\pi^0$ decays. The complete analysis, from the reconstruction of intermediate resonances to the final maximum likelihood fit is performed. In estimating the final sensitivity we take into account the expected improvements, most notably those affecting the reconstruction efficiencies.

Based on these studies and on the reconstruction efficiencies obtained by the BaBar and Belle experiments we also extrapolate the present sensitivities to Belle II for the channels $B^0 \to J/\psi\pi^0$, $B^0 \to \omega K_S^0$, $B^0 \to K_S^0\pi^0$, which are related to the measurement of the angle $\phi_1$, and for the channels $B^0 \to \pi^+\pi^-$, $B^+ \to \pi^+\pi^0$, $B^0 \to \rho^0\rho^0$, $B^0 \to \rho^+\rho^-$ and $B^+ \to \rho^+\rho^0$, which are related to the measurement of $\phi_2$. We discuss in detail the systematic uncertainties that will affect the cleanest and highest statistics channel $B^0 \to J/\psi K^0$ for the measuerement of $\phi_1$. Based on this discussion we also estimate the systematic uncertainties which will affect the channels $B \to \pi\pi$ and $B \to \rho\rho$ for the measurement of $\phi_2$.

Figure 98 shows the time-dependent $CP$ asymmetry distributions that can be measured at Belle II in the $B^0 \to J/\psi\, K_S^0$ and $\eta' K_S^0$ channels with an integrated luminosity of 50 ab$^{-1}$. As inputs to the simulations we set $S_{J/\psi\, K_S^0} = 0.70$ and $S_{\eta' K_S^0} = 0.55$, see eq. (302). Such a difference between $S_{J/\psi\, K_S^0}$ and $S_{\eta' K_S^0}$ would be an unambiguous sign of New Physics and would be easily detectable by the Belle II experiment.

## *10.2. Determination of $\phi_1$*

*Contributing authors: M. Jung, L. Li Gioi, D. Robinson*

*10.2.1. Theory:* $\sin 2\phi_1$ *from* $b \to c\bar{c}s$. The angle $\phi_1$ is the most precisely measured $CP$ violating quantity to date. As such it is one of the most important inputs in the global CKM fits and a cornerstone input to the tests of the SM.

The sensitivity to $\phi_1$ comes from the $CP$ asymmetry parameter $S_f$ in eq. (302) which measures the sum of the mixing phase $-\phi_d$ and the relative phase $\arg(\bar{\mathcal{A}}_f/\mathcal{A}_f)$, see (303).

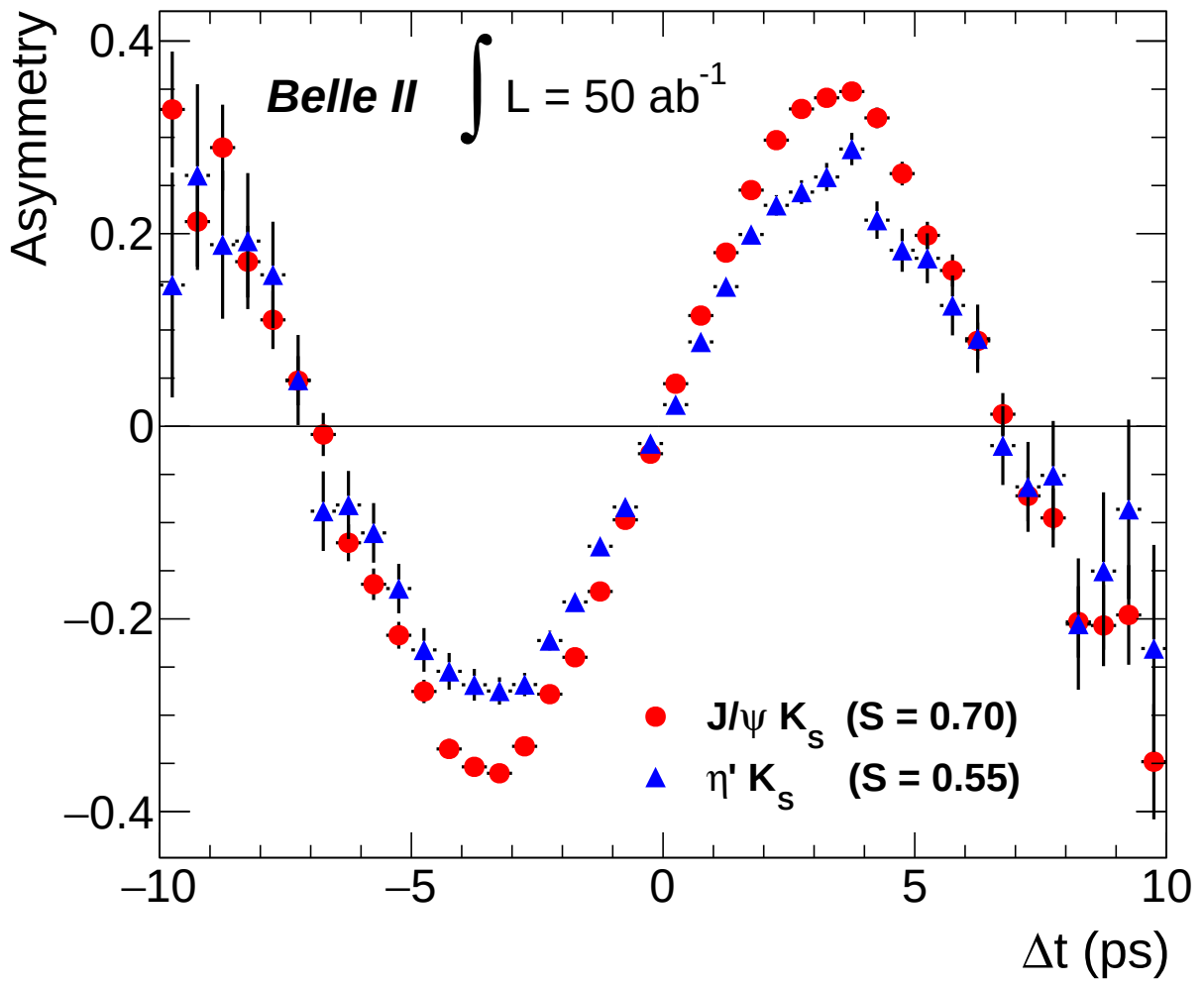

Fig. 98: Time-dependent $CP$ asymmetries for the final states $J/\psi K_S^0$ (red dots) and $\eta' K_S^0$ (blue triangles), using $S_{J/\psi K_S^0} = 0.70$ and $S_{\eta' K_S^0} = 0.55$ as inputs to the Monte Carlo. With the full integrated luminosity of 50 ab$^{-1}$ the two values would be unambiguously distinguishable, signifying the existence of New Physics.

For $b \to c\bar{c}s$ transitions, CKM unitarity permits the decay amplitudes to be written as [28]

$$\mathcal{A}_f = \lambda_c^s \, T_f + \lambda_u^s \, P_f \,, \qquad \lambda_i^q \equiv V_{ib}^* V_{iq} \,. \tag{305}$$

While $P_f$ and $T_f$ correspond at leading order to penguin and tree $b \to c\bar{c}s$ contributions, respectively (see also Fig. 99), for the sub-percent precision measurements of $S_f$ anticipated by Belle II subleading corrections become important, and such a diagrammatic interpretation of these contributions is no longer possible.

Since $\lambda_u^s$ is doubly CKM-suppressed compared to $\lambda_c^s$ one has $\bar{\mathcal{A}}_f/\mathcal{A}_f \simeq \eta_f \lambda_c^{s*}/\lambda_c^s$, and therefore

$$S_f \simeq -\eta_f \sin(\phi_d) + \mathcal{O}\big(\lambda_u^s/\lambda_c^s\big) \,, \tag{306}$$

while the direct $CP$ asymmetry $A_f \simeq 0$. The time-dependent $CP$ asymmetry in $b \to c\bar{c}s$ decays thus allows a theoretically clean extraction of $\phi_1$, up to doubly CKM-suppressed corrections. The control of the latter constitutes the main challenge with available and future precision data.

Despite this challenge, as we will show below, the determination of the $B$ mixing phase $\phi_1$ via $b \to c\bar{c}s$ transitions remains an excellent way to search for NP that gives additional contributions to meson mixing. The SM uncertainties need to be brought under control at the present level of experimental precision, and even more so with the precision aimed at with Belle II.

[28] Reparametrisation invariance permits the decay amplitude to always be expressed in terms of $\lambda_{u,c}^s$ and matrix elements, $A_{u,c}$, *i.e.* as $\mathcal{A}_f = \lambda_c^s \, A_u + \lambda_u^s \, A_c$, even in the presence of an additional NP contribution with an arbitrary weak phase [651–653]. However, in this case the interpretation of $A_{u,c}$ as matrix elements of SM currents does not hold anymore, and symmetry relations are potentially affected.

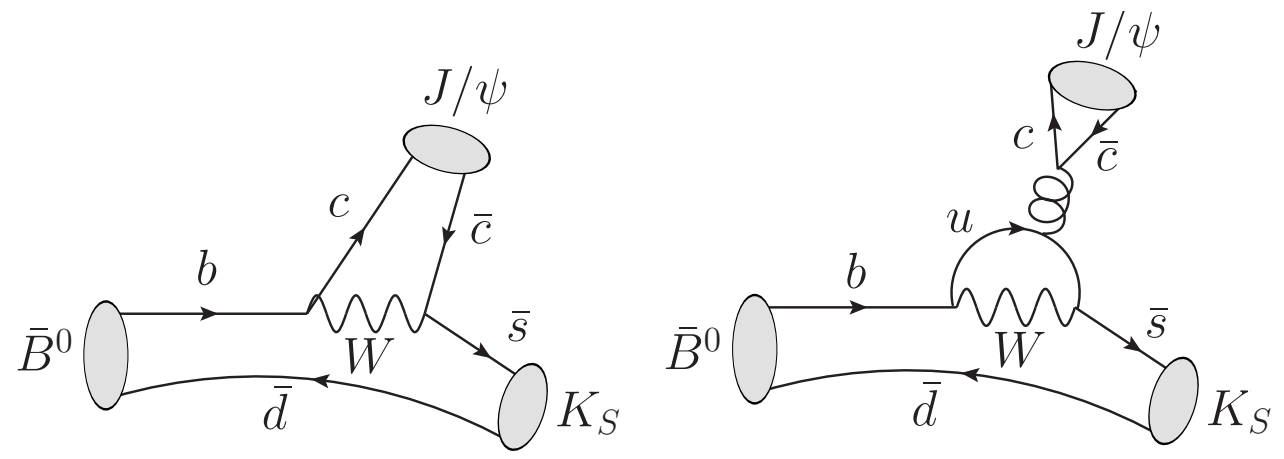

Fig. 99: Examples of diagrams for the $T_f$ (left) and $P_f$ (right) amplitudes in eq. (305) for the $B \to J/\psi K_S^0$ decay. Only part of the contributions to $T_f$ and $P_f$ are shown.

$\sin\phi_1$ *from* $B \to J/\psi K_S^0$. The "golden mode" for measuring $\sin\phi_d$ is $B \to J/\psi K_S^0$ because of the expected small theoretical uncertainty and clean experimental signature.[29] Let us denote the corresponding $\mathcal{O}(\lambda_u^s/\lambda_c^s)$ terms in eq. (306) by $\Delta S_{J/\psi K_S^0}$, so that

$$S_{J/\psi K_S^0} \equiv \sin\phi_d + \Delta S_{J/\psi K_S^0} \equiv \sin(\phi_d + \delta\phi_{J/\psi K_S^0})\,, \tag{307}$$

where $\delta\phi_{J/\psi K_S^0}$ is also $\mathcal{O}(\lambda_u^s/\lambda_c^s)$. The small parameters $\Delta S_{J/\psi K_S^0}$ or $\delta\phi_{J/\psi K_S^0}$ are often referred to as the "penguin pollution" in the extraction of $\phi_d$ from $S_{J/\psi K_S^0}$. The only potentially sizeable contribution to $\Delta S_{J/\psi K_S^0}$ is expected to arise from insertions of tree-level operators $\mathcal{O}_{1,2}^u$ in the SM effective Hamiltonian for non-leptonic $b \to s$ transitions, and then closing the up quark loop. This loop is expected to generate another suppression factor in addition to the CKM suppression, but the resulting net effect is hard to quantify.

A naïve estimate of $\Delta S_{J/\psi K_S^0}$ from the CKM-suppression alone yields $\delta\phi_{J/\psi K_S^0} \lesssim 2°$, comparable to the experimental uncertainty in the current world-average $S_{J/\psi K_S^0} = 0.682 \pm 0.021$ ($\phi_1 = (22.5 \pm 0.9)°$) [80, 655, 656]. Belle II is projected to improve the experimental precision to the sub-degree level with $50\,\mathrm{ab}^{-1}$ integrated luminosity, well inside the estimated theory uncertainty. In the remainder of this subsection we therefore discuss various strategies to either compute, bound or control $\Delta S_{J/\psi K_S^0}$.

To this end, it is helpful to write a parametrisation of $\Delta S_{J/\psi K_S^0}$ to leading order in CKM suppression. Using the definitions of eqs. (302) and (305), together with the definitions $\bar{\lambda}^2 \equiv -\lambda_u^s\lambda_c^d/\lambda_c^s\lambda_u^d$ and $\phi_3 \equiv \arg(-\lambda_u^d/\lambda_c^d)$,

$$\Delta S_{J/\psi K_S^0} = 2\bar{\lambda}^2 \mathrm{Re}\frac{P_f}{T_f}\sin\phi_3\cos\phi_d + \mathcal{O}(\bar{\lambda}^4)\,, \tag{308}$$

in which we have used the fact that $\bar{\lambda}^2$ ($\simeq 0.05$) [91, 657] is real up to higher-order CKM corrections, so that $\mathrm{Im}(\lambda_u^s/\lambda_c^s) \simeq \bar{\lambda}^2\sin\phi_3$. All terms are well known experimentally, apart from the reduced matrix element ratio $P_f/T_f$.

Theoretical calculations or estimations of $P_f/T_f$ provide one path to controlling $\Delta S_{J/\psi K_S^0}$. Formally, this term is not only CKM-suppressed, but is also expected to receive a loop suppression $\lesssim 5\%$ [658]. Attempts to calculate $\Delta S_{J/\psi K_S^0}$ via QCD factorisation (QCDF) techniques [659], or by combining QCDF with perturbative QCD [660], yield estimates in the range $\Delta S_{J/\psi K^0_S} \sim 10^{-3}$. The $\mathcal{O}(1/m_{c,b})$ correction may, however, be large. Specifically, the leading corrections that scale as $\Lambda_{\rm QCD}/(\alpha_s m_c)$, which is of order unity for realistic charm quark masses [244]. Long-distance rescattering effects from intermediate charmless states may enhance $P_f/T_f$, although perhaps only moderately [661].

[29] Note that $f = J/\psi K_S^0$ is a $CP$ eigenstate up to $\epsilon_K$ corrections that arise at the sub-percent level [654]. Note furthermore that $\eta_{J/\psi K_S^0} = -1$. The discussion here applies trivially also to the determination of $\sin\phi_d$ from $B \to J/\psi K_L^0$, except that $\eta_{J/\psi K_L^0} = 1$. This mode is harder to measure, however.

The ratio $P_f/T_f$ may also be estimated via an OPE-type approach, by integrating out the $u$-quark loop, on the basis that the typical momentum flow is large $\sim m_{J/\psi}$ [662]. This approach produces a factorisation formula for the penguin contributions, relying on the observation that soft and collinear divergences formally cancel or factorise up to $\Lambda_{\rm QCD}/m_{J/\psi}$ corrections. The remaining matrix elements are estimated in the large-$N_c$ limit ($N_c = 3$ is the number of colours in QCD). This leads to the estimate $|\delta\phi_{J/\psi K^0_S}| < 0.68°$ [662].

Light-quark flavour symmetries, either flavour $SU(3)$ or its subgroup $U$-spin, provide an alternate avenue to control $P_f/T_f$ and thereby $\Delta S_{J/\psi K^0_S}$. The simplest manifestations of this approach use a single mode related to $B \to J/\psi K^0_S$ by such a flavour symmetry. In the $U$-spin limit, the $B \to J/\psi K^0_S$ amplitude in eq. (305) is related to the $B_s \to J/\psi K^0_S$ amplitude, denoted $\mathcal{A}'$, via [663–665]

$$\mathcal{A}'_f = \lambda^d_c T_f + \lambda^d_u P_f\,. \tag{309}$$

Since $B_s \to J/\psi K^0_S$ constitutes a $b \to c\bar{c}d$ process, the $\lambda^d_u P_f$ term is no longer CKM-suppressed. Therefore both $\Delta S'_{J/\psi K^0_S}$ and $C'_{J/\psi K^0_S}$ can be sizeable and hence be more easily determined. Using external knowledge of the $B_s$ mixing phase, $\phi_s$, permits the extraction of an estimate for $\mathrm{Re}(P_f/T_f)$ and thus $\Delta S_{J/\psi K^0_S}$. Similarly, $B \to J/\psi \pi^0$ can be used as a partner mode [666–668]. This mode has the advantage that it is more easily measurable than $B_s \to J/\psi K^0_S$ at Belle II, at which the $B_s$ dataset will be limited. However, in this case a dynamical assumption regarding small annihilation contributions in $P_f$ is necessary in addition to $U$-spin to obtain the analogue of eq. (309). A drawback of using $b \to c\bar{c}d$ transitions in this procedure is that the rates of these modes are suppressed by a factor of $\lambda^2 \sim 1/20$. This problem is to be overcome with high-luminosity experiments like Belle II as discussed in section 10.2.2.

The $U$-spin limit approach is limited by the size of $U$-spin-breaking corrections parametrised by the parameter $\varepsilon \sim m_s/\Lambda_{\rm QCD} \sim 0.2$. The flavour breaking corrections cannot be controlled with a single partner mode, and hence such $U$-spin analyses require additional assumptions. More sophisticated flavour symmetry analyses therefore include fits from multiple decay modes, using a combination of $CP$ asymmetries and $CP$-averaged rates [665, 669]. Differences arise in the treatment of $SU(3)$ breaking in these fits. One approach uses a model-independent expansion to first order in the breaking [669]; another considers only factorisable breaking as a starting point, while an additional non-factorisable part is assumed to be smaller [665]. Differing dynamical assumptions, *e.g.*, about smallness of annihilation contributions, are also necessary to obtain a well-constrained fit. Importantly, these methods are data-driven, such that their precision improves with additional data. In Table 71, we summarise the theory expectations from analyses performed over the last decade.

The importance of measuring $CP$-averaged rates is emphasised by the $SU(3)$ relation [670],

$$(1+\bar{\lambda}^2)\sin\phi_d = S_{J/\psi K^0_S} - \bar{\lambda}^2 S_{J/\psi\pi^0} - 2(\Delta_K + \bar{\lambda}^2\Delta_\pi)\cos\phi_d\,\tan\phi_3\,, \tag{310}$$

in which penguin pollution effects are cancelled to $\mathcal{O}(\varepsilon)$ and leading corrections arise from isospin-breaking terms. Here $\Delta_{K,\pi}$ are splittings of the charged and neutral $CP$-averaged

Table 71: Theory expectations for $\Delta S_{J/\psi K^0_S}$ or $\delta\phi_{J/\psi K^0_S}$, related via $\Delta S_{J/\psi K^0_S} \simeq \delta\phi_{J/\psi K^0_S} \cos\phi_1$. The values in large parentheses are not given in the corresponding reference, but have been calculated for convenience via eq. (307).

| Strategy | $\Delta S_{J/\psi K^0_S}$[%] | $\delta\phi_{J/\psi K^0_S}$[°] |
|---|---|---|
| QCDF/pQCD [659, 660] | $\vert\ \vert \lesssim 0.1$ | $(\vert\ \vert \lesssim 0.1)$ |
| OPE [662] | $(\vert\ \vert \lesssim 0.9)$ | $\vert\ \vert \lesssim 0.68$ |
| Broken $U$-spin [666, 668] | $0 \pm 2$ | $(0.0 \pm 1.6)$ |
| Broken $U$-spin [667] | $([-5, -0.5])$ | $[-2.0, -0.4]$ |
| $SU(3)$ at $\mathcal{O}(\varepsilon)$ [669] | $\vert\ \vert \lesssim 1$ | $\vert\ \vert \lesssim 0.8$ |
| Broken $SU(3)$ [665] | $(-(1.4^{+0.9}_{-1.1}))$ | $-(1.10^{+0.70}_{-0.85})$ |

rates,

$$\begin{aligned} \Delta_K &\equiv \frac{\bar\Gamma_{B_d\to J/\psi K^0} - \bar\Gamma_{B^+\to J/\psi K^+}}{\bar\Gamma_{B_d\to J/\psi K^0} + \bar\Gamma_{B^+\to J/\psi K^+}}\,, \\ \Delta_\pi &\equiv \frac{2\bar\Gamma_{B_d\to J/\psi \pi^0} - \bar\Gamma_{B^+\to J/\psi \pi^+}}{2\bar\Gamma_{B_d\to J/\psi \pi^0} + \bar\Gamma_{B^+\to J/\psi \pi^+}}\,. \end{aligned} \tag{311}$$

Determining $\Delta_{\pi,K}$ at the desired precision requires the control of potentially enhanced isospin-breaking effects in the $B^0/B^+$ production ratio from independent measurements, which is possible using data from Belle and Belle II, but requires a dedicated analysis [671].

*10.2.2. Experiment* $\sin 2\phi_1$ *from* $b \to c\bar{c}s$ *decay modes.*

$B^0 \to J/\psi K^0_S$ The $B^0 \to J/\psi K^0_S$ decay mode leads to an experimentally very clean signature. Moreover, it presents a relatively large branching fraction, so a large signal yield is expected. Even if the contribution of penguin diagrams with a different CKM phase is expected to be at less than 1% level, this effect can become appreciable at the end of the Belle II data taking and should be taken into account for the $\phi_1$ determination.

Belle has updated the time-dependent $CP$ asymmetry measurement using the whole data sample [80], obtaining:

$$\begin{aligned} S_{J/\psi K^0_S} &= +0.670 \pm 0.029(\text{stat}) \pm 0.013(\text{syst}), \\ A_{J/\psi K^0_S} &= -0.015 \pm 0.021(\text{stat}) \ {}^{+\,0.045}_{-\,0.023}(\text{syst}). \end{aligned} \tag{312}$$

While $S_{J/\psi K^0_S}$ is still dominated by the statistical error, the measurement of $A_{J/\psi K^0_S}$ is already dominated by the systematic uncertainties. For the extrapolation of the statistical errors, we assume the same $B\bar{B}$ vertex separation capability of Belle. We then scale the statistical error according to the square root of the integrated luminosity. Systematic errors include uncertainties in the wrong-tag fractions, a possible fit bias, uncertainties in the signal fractions, the background $\Delta t$ distribution, $\tau_{B^0}$ and $\Delta m_d$. All these depend on control samples or on Monte Carlo statistics and are expected to scale with square root of the integrated luminosity, as the statistical errors. The two remaining systematic errors, tag side interference and uncertainty due to the vertex reconstruction algorithm, do not scale with the integrated luminosity. A dedicated study is thus needed.

Table 72: Belle II expected sensitivity on the $CP$ parameters of $B^0 \to J/\psi K^0_S$. Expected statistical, reducible systematic and non reducible systematic uncertainties are shown. An integrated luminosity of 50 ab$^{-1}$ is assumed. Three cases are considered: 'No improvement', where Belle irreducible systematic uncertainties are assumed to not improve in Belle II; 'Vertex improvement', where an improvement of 50% is assumed for the systematic due to the vertex positions; 'Leptonic categories', where the analysis is performed using only the leptonic categories for flavour tagging.

| | No improvement | Vertex improvement | Leptonic categories |
|---|---|---|---|
| $S_{J/\psi K^0_S}$ (50 ab$^{-1}$) | | | |
| stat. | 0.0035 | 0.0035 | 0.0060 |
| syst. reducible | 0.0012 | 0.0012 | 0.0012 |
| syst. irreducible | 0.0082 | 0.0044 | 0.0040 |
| $A_{J/\psi K^0_S}$ (50 ab$^{-1}$) | | | |
| stat. | 0.0025 | 0.0025 | 0.0043 |
| syst. reducible | 0.0007 | 0.0007 | 0.0007 |
| syst. irreducible | $^{+0.043}_{-0.022}$ | $^{+0.042}_{-0.011}$ | 0.011 |

In the Belle measurement [80] the error due to the tag-side interference was evaluated by comparing the fit results with and without the tag-side interference term [672]. This term, however, is well defined in the case of the $B^0 \to D^{*-}l^+\nu$ decay mode. The systematic error from tag-side interference can then be reduced, in the measurement of the $CP$ parameters of $B^0 \to J/\psi K^0_S$, taking into account the tag-side interference term into the default fitter and assigning as error the tag-side interference term uncertainty. Another way of avoiding the effect of the tag-side interference is to use only leptonic categories for the flavour tagging. However, this would reduce the tagging efficiency by about a factor of three, which results in increase of statistical error.

The systematic uncertainty due to the vertex reconstruction had, at Belle, different causes: alignment of the vertex detector, vertex algorithms and vertex resolution. All these components are expected to be reduced at Belle II. The uncertainty on the alignment of the vertex detector (see Sec. 5.3.3) has two components: one that depends on the size of the control samples that will be used to perform the alignment (and thus will scale as the statistical uncertainties) and an irreducible component. The new vertex algorithm for the tag side removes the systematic effect coming from the selection of the tracks used for the vertex fit and improves, by almost a factor two, its resolution. The vertex resolution of the $CP$ side will improve by a factor two compared to Belle thanks to the new Pixel Vertex Detector. We assume, for this study, a factor two for the reduction of the systematic uncertainty due to the vertex reconstruction.

Table 72 shows expected Belle II sensitivity to the $B^0 \to J/\psi K^0_S$ $CP$ asymmetry parameters. The measurement is expected to be dominated by systematic errors. In the case of $A_{J/\psi K^0_S}$, the smallest total error is obtained when performing the analysis using only the leptonic categories for flavour tagging.

Table 73: Belle II expected sensitivity to the $B^0 \to J/\psi\pi^0$ $CP$ asymmetry parameters for an integrated luminosity of 50 ab$^{-1}$. See also Table 72.

| | No improvement | Vertex improvement | Leptonic categories |
|---|---|---|---|
| $S_{J/\psi\pi^0}$ (50 ab$^{-1}$) | | | |
| stat. | 0.027 | 0.027 | 0.047 |
| syst. reducible | 0.009 | 0.009 | 0.009 |
| syst. irreducible | 0.050 | 0.025 | 0.025 |
| $A_{J/\psi\pi^0}$ (50 ab$^{-1}$) | | | |
| stat. | 0.020 | 0.020 | 0.035 |
| syst. reducible | 0.004 | 0.004 | 0.004 |
| syst. irreducible | 0.045 | 0.042 | 0.017 |

*Expected sensitivity of the time-dependent asymmetries of $B^0 \to J/\psi\pi^0$.* The $B^0 \to J/\psi\pi^0$ decay mode, proceeding through $b \to c\bar{c}d$ transition, can be used to constrain theoretical uncertainties in $B^0 \to J/\psi K^0_S$ . Both BaBar [673] and Belle [674] have performed the time-dependent analysis of $B^0 \to J/\psi\pi^0$, the latter obtaining:

$$
\begin{aligned}
S_{J/\psi\pi^0} &= -0.65 \pm 0.21(\text{stat}) \pm 0.05(\text{syst}), \\
A_{J/\psi\pi^0} &= -0.08 \pm 0.16(\text{stat}) \pm 0.05(\text{syst}),
\end{aligned}
$$

largely dominated by statistical errors. A precise measurement will be possible using the high integrated luminosity collected by Belle II at the end of its data taking. Table 73 shows the expected sensitivity to the time-dependent $CP$ asymmetry parameters assuming an integrated luminosity of 50 ab$^{-1}$. The algorithm for calculating the expected statistical and systematic uncertainties is the same as in the previous section. A relative uncertainty of a few percent is expected. This will translate to an uncertainty of $\sim 0.1^\circ$ on $\phi_1$ from $B^0 \to J/\psi K^0_S$, when using eq. (310) to estimate the penguin pollution and ignoring for now the $\mathcal{O}(\epsilon^2)$ effects from $SU(3)$ breaking.

*Expected sensitivity from the combined analysis of $b \to c\bar{c}s$ decay modes.* We show next the projected sensitivity to $\phi_1$ for the combination of the most of the relevant $b \to c\bar{c}s$ decay modes. Belle has published a combined analysis of decay modes with $\eta_f = -1$ $CP$ eigenvalue, $B^0 \to J/\psi K^0_S$, $B^0 \to \psi(2S)K^0_S$, $B^0 \to \chi_{c1}K^0_S$, and with $\eta_f = +1$ $CP$ eigenvalue, $B^0 \to J/\psi K^0_L$ [675]. The averaged $CP$ asymmetry parameters,

$$
\begin{aligned}
S_{c\bar{c}s} &= 0.667 \pm 0.023(\text{stat}) \pm 0.012(\text{syst}), \\
A_{c\bar{c}s} &= 0.006 \pm 0.016(\text{stat}) \pm 0.012(\text{syst}),
\end{aligned}
\tag{313}
$$

are still dominated by the statistical errors.

The combination of $CP$-odd and $CP$-even final states returns a combined tag-side interference systematic that is smaller than the systematic in each individual mode. The tag-side interference term has the opposite sign for different $CP$ eigenstates which produces a partial cancellation in the combined result.

Table 74: Belle II expected sensitivity for the $CP$ asymmetry parameters in the combination of $b \to c\bar{c}s$ modes. An integrated luminosity of 50 ab$^{-1}$ is assumed.

| | No improvement | Vertex improvement | Leptonic categories |
|---|---|---|---|
| $S_{c\bar{c}s}$ (50 ab$^{-1}$) | | | |
| stat. | 0.0027 | 0.0027 | 0.0048 |
| syst. reducible | 0.0026 | 0.0026 | 0.0026 |
| syst. irreducible | 0.0070 | 0.0036 | 0.0035 |
| $A_{c\bar{c}s}$ (50 ab$^{-1}$) | | | |
| stat. | 0.0019 | 0.0019 | 0.0033 |
| syst. reducible | 0.0014 | 0.0014 | 0.0014 |
| syst. irreducible | 0.0106 | 0.0087 | 0.0035 |

Table 74 shows the expected sensitivity for the $CP$ asymmetry parameters in the combined $b \to c\bar{c}s$ analysis at Belle II . The $B^0 \to J/\psi K_{S,L}$ decay modes contribute the largest fraction of the yield. The addition of the remaining channels improves the statistical uncertainty by about 10%. While all the modes will be important for the measurements performed during the first years of data taking, the opportunity of including them in the final analysis performed using the full expected luminosity should be considered only if the systematic uncertainties will result better than those reported in Table 74. A precision better than 1% is expected, when setting aside the theoretical issue of penguin pollution. The strategies to deal with the latter differ from mode to mode as discussed above and may also be combined in a global analysis.

*10.2.3. Other $b \to c\bar{c}X$ decay modes.* The arguments above hold equally well for the other initial and final states, as long as the corresponding amplitude is dominated by the $b \to c\bar{c}X$ transition. The following three decay channels are the most relevant: (i) $B \to \psi(X)K_S^0$, *i.e.*, replacing the $J/\psi$ by other charmonia, (ii) $B \to J/\psi V$, *i.e.* replacing the pseudoscalar in the final state by a vector meson (or two pseudoscalars), specifically the decay $B_s \to J/\psi\phi$, (iii) $B \to D^{(*)}D^{(*)}$, *i.e.*, two charmed mesons in the final state. These options are briefly discussed in the remainder of this section.

(i) Replacing the $J/\psi$ by other charmonia makes little difference theoretically, but the experimentally convenient $\ell^+\ell^-$ final state becomes unavailable. The $SU(3)$ symmetry analysis for controlling penguin pollution proceeds analogously to the $J/\psi K_S^0$ case. The relevant $SU(3)$ related modes need to be, of course, re-measured with each charmonium state. For the $\psi(2S)K_S^0$ final state, the penguin pollution has been estimated in Ref. [662].

(ii) The main change when replacing the $K_S^0$ by a vector meson is that an angular analysis becomes necessary in order to disentangle the different polarisation amplitudes which transform differently under $CP$. Doing so provides presently the best extraction of $\phi_s$ via $B_s \to J/\psi\phi$ [230]. Regarding $\phi_d$, the fact that $K^*$ is not a $CP$ eigenstate complicates the extraction. However, the final state $K^* \to K_S^0\pi^0$ can be used. The penguin pollution in the $B \to J/\psi K^*(\to K_S^0\pi^0)$ mode has been estimated from a theoretical calculation to be of similar size as in $B \to J/\psi K_S^0$ [662].

A flavour $SU(3)$ symmetry analysis is again possible [665, 676], but complicated by the fact that the three polarisation amplitudes per channel require three independent sets of matrix elements, introducing a larger number of nuisance parameters into the analysis. Furthermore the fact that the $\phi$ meson is an admixture of flavour $SU(3)$ octet and singlet states has to be taken into account. When using $J/\psi\,\rho$ or $J/\psi\,K^*$ final states as control modes, only the octet amplitude can be restricted, while in order to control the singlet amplitude, data for $B_{d,s} \to J/\psi\,\omega$ and $B_d \to J/\psi\,\phi$ are necessary. So far the corresponding singlet amplitudes, stemming *e.g.* from exchange diagrams, have been neglected. This type of amplitude, however, has been shown in $B \to DD$ decays to be larger than naively expected [677, 678]. While this does not allow one to infer anything regarding $B \to J/\psi V$ decays, it demonstrates that experimental data should be used to control these amplitudes instead of theoretical assumptions.

Apart from $B_d \to J/\psi K^0_S$ and $B_s \to J/\psi\phi$, also the $B_s \to J/\psi f_0(980)$ decay has been proposed as a means to extract $\phi_s$ [679], $f_0(980)$ being the largest resonance in $B_s \to J/\psi\pi^+\pi^-$. Since it is a scalar meson, this mode does not require an angular analysis. Of concern in this case is the unclear hadronic nature of the $f_0(980)$, and its mixing with the $\sigma(f_0(500))$ resonance. This renders a symmetry analysis of the type described above more complicated [680]. It therefore seems hard to achieve the control of subleading contributions to a comparable level as in $B_s \to J/\psi\phi$.

(iii) The $B \to D^{(*)}D^{(*)}$ modes exhibit essentially the same features as $B \to J/\psi\,K^0_S$, only that they are not as straightforward to interpret. The corresponding "golden" channels are the tree-level $b \to c$ transitions into $CP$ eigenstates, specifically $B_d \to D^+D^-$, sensitive to $\phi_1$, and $B_s \to D_s^+D_s^-$, sensitive to $\phi_s$ [663]. This strategy has been extended to full $SU(3)$, model-independently including symmetry-breaking corrections [677]. The strategy requires precision measurements of many branching fractions and $CP$ asymmetries [677, 678]. Explicit calculation of penguin pollution in these modes is very challenging, since even in the formal limit $m_{b,c} \to \infty$ these decays do not factorise [244]. Some analyses have nonetheless employed this approximation [678].

## *10.3. Determination of $\phi_1$ in gluonic penguin modes*

### *10.3.1. Theory:* $\sin 2\phi_1$ *from* $b \to q\bar{q}s$, $q = u, d, s$. *Contributing author S. Jäger*

The penguin-dominated modes $b \to q\bar{q}s$ $(q = u, d, s)$ are interesting for at least the following three reasons:

(1) They probe the $B_d - \bar{B}_d$ mixing phase through different short-distance vertices than the tree-dominated $b \to c\bar{c}s$ decays.
(2) They are loop-dominated in the Standard Model, and hence may be more sensitive to new-physics effects than the tree-dominated modes.
(3) They comprise a large number of different final states, which can help in disentangling nonperturbative long-distance physics from short-distance information such as $\phi_1$ or beyond-Standard-Model (BSM) contributions to the weak Hamiltonian.

On general grounds, the charmless $b \to q\bar{q}s$ decay amplitude $\mathcal{A}_f \equiv \mathcal{A}(\bar{B} \to f)$ can be written as

$$\mathcal{A}_f = \lambda_c^s P_f + \lambda_u^s T_f + \mathcal{A}_f^{\rm NP}. \tag{314}$$

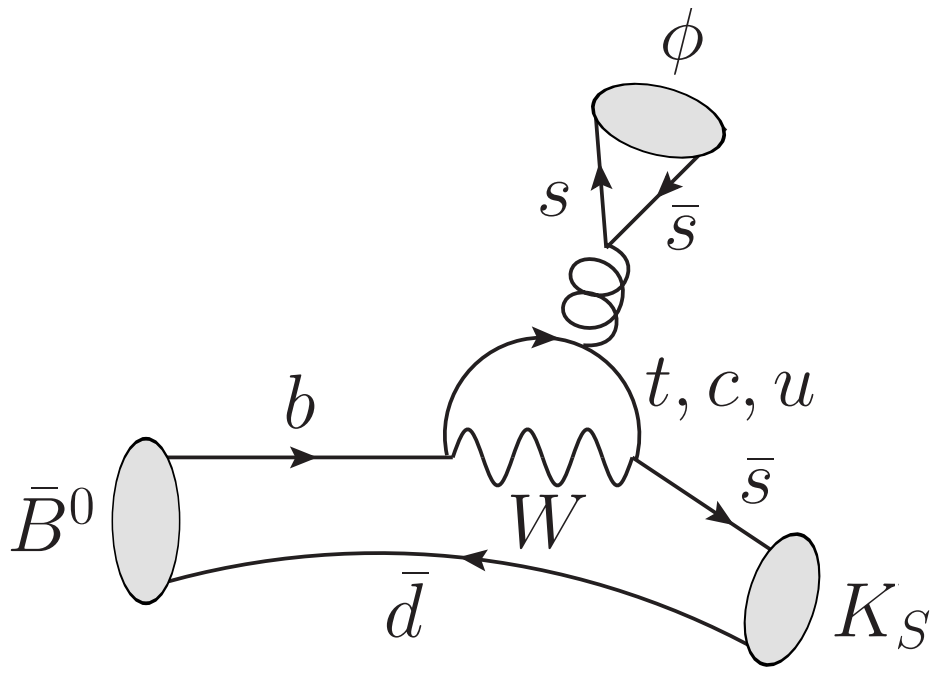

Fig. 100: The QCD penguin contributions to $B \to \phi K_s^0$ decay. The $T_f$ amplitude receives a contribution from $c-$quark in the loop, while $t-$ and $u-$quark contribute to both $T_f$ and $P_f$ in (314).

Note that compared to $b \to sc\bar{c}$ decays in eq. (305), here the "penguin amplitude" $P_f$ is multiplied by a large CKM factor, $\lambda_c^s$, while the "tree amplitude", $T_f$, is CKM suppressed. The $\mathcal{A}_f^{\rm NP}$ is a possible BSM contribution. The latter will, in general, have indefinite $CP$ properties, *i.e.* $|\bar{\mathcal{A}}^{\rm NP}| \neq |\mathcal{A}^{\rm NP}|$. It is worth noting that the "tree amplitude" $T_f$ as defined here contains effects not only from tree-level $W$ exchange (operators $Q_1$ and $Q_2$, including their 'up loop' contractions) but also part of the QCD and electroweak penguin operator contributions (top loops). These carry the combination of CKM matrix elements $\lambda_t^{(s)} = V_{ts}V_{tb}^* = -(1+\epsilon_{uc})\lambda_c^{(s)}$ (where we have defined $\epsilon_{uc} \equiv \lambda_u^{(s)}/\lambda_c^{(s)} = \mathcal{O}(\lambda^2)$), which differs slightly in weak phase from the penguin amplitude in eq. (314).

In the SM and when $\epsilon$ is neglected, the $b \to q\bar{q}s$ modes are pure penguin with the same weak phase (indeed, CKM combination) as the tree-dominated $b \to c\bar{c}s$ decays. As a consequence, direct $CP$ asymmetries vanish in this limit. If $f$ is a $CP$ eigenstate, the coefficient $S_f$ in the time-dependent $CP$-asymmetry then measures the same phase $\phi_1$ as in the $b \to c\bar{c}s$ decays. Departures from this limit may come from the tree amplitude $T_f$ (often called the "tree pollution") as well as from possible NP effects. Introducing the tree-to-penguin ratio $r_f^T = T_f/P_f$ and the BSM-to-SM ratio $r_f^{\rm NP} \equiv \mathcal{A}_f^{\rm NP}/(\lambda_c^{(s)}P_f)$, one can make the following statements:

- Branching ratios are affected at $\mathcal{O}(|\epsilon_{uc}r_f^T|, |r_f^{\rm NP}|)$.
- Direct $CP$ asymmetries in the SM are of $\mathcal{O}\big(|\epsilon_{uc}|{\rm Im}\, r_f^T\big)$. A possible BSM contribution to the direct $CP$ asymmetry will likewise require both a weak and a strong phase difference relative to the SM penguin.
- The sine-coefficients in time-dependent $CP$ asymmetry differ from $\sin 2\phi_1$ by $\Delta S_f$ (see eg [681]),

$$-\eta_f^{\rm CP} S_f = \sin 2\phi_1 + \Delta S_f, \tag{315}$$

where

$$\Delta S_f = 2\cos 2\phi_1 \sin\phi_3 |\epsilon_{uc}| {\rm Re}\, r_f^T + \Delta S_f^{\rm NP}. \tag{316}$$

The first term on the right-hand side is due to the SM tree pollution, while $\Delta S_f^{\rm NP}$ denotes the potential NP contribution.

Table 75: The predictions for $\Delta S_f$ (315), for charmless two-body final states listed in the first column, using different theoretical approaches, are listed in the second, third, and fourth column, while the experimental values ([230]) are given in the last column.

| Mode | QCDF [681] | QCDF (scan) [681] | $SU(3)$ | Data |
|---|---|---|---|---|
| $\pi^0 K_S^0$ | $0.07^{+0.05}_{-0.04}$ | $[0.02, 0.15]$ | $[-0.11, 0.12]$ [683] | $-0.11^{+0.17}_{-0.17}$ |
| $\rho^0 K_S^0$ | $-0.08^{+0.08}_{-0.12}$ | $[-0.29, 0.02]$ | | $-0.14^{+0.18}_{-0.21}$ |
| $\eta' K_S^0$ | $0.01^{+0.01}_{-0.01}$ | $[0.00, 0.03]$ | $(0 \pm 0.36) \times 2\cos(\phi_1)\sin\gamma$ [684] | $-0.05 \pm 0.06$ |
| $\eta K_S^0$ | $0.10^{+0.11}_{-0.07}$ | $[-1.67, 0.27]$ | | — |
| $\phi K_S^0$ | $0.02^{+0.01}_{-0.01}$ | $[0.01, 0.05]$ | $(0 \pm 0.25) \times 2\cos(\phi_1)\sin\gamma$ [684] | $0.06^{+0.11}_{-0.13}$ |
| $\omega K_S^0$ | $0.13^{+0.08}_{-0.08}$ | $[0.01, 0.21]$ | | $0.03^{+0.21}_{-0.21}$ |

To measure $\phi_1$ it is crucial to control the effects due to non-vanishing $r_f^T$. Since $\mathrm{Re}\, r_f^T \propto \cos\delta_f = 1 - \delta_f^2/2 + \ldots$, with $\delta_f = \arg r_f^T$ the strong phase difference, it is less sensitive to final-state rescattering effects. Small strong phases are predicted in the heavy-quark expansion, see below. In addition, a given NP scenario will affect different modes in different ways.

Theoretical information on $r_f^T$ comes from the use of flavour $SU(3)$ relations and the heavy-quark expansion. The methods based on $SU(3)$ relate the $b \to q\bar{q}s$ transitions to the $b \to q\bar{q}d$ transitions, such as $B \to \pi\pi$, $B \to \pi\eta'$, and so on *i.e.* the modes that are related by the $U$-spin subgroup of $SU(3)$ to $B \to \pi K, K\eta'$, ... In $b \to q\bar{q}d$ transitions the tree contributions are CKM enhanced compared to the penguins, making it possible to obtain experimental information on $r_f^T$.

The heavy-quark expansion gives rise to QCD factorisation (QCDF) of charmless hadronic $B \to M_1 M_2$ decay amplitudes [244, 415, 416] (see also [682]), allowing to calculate the decay amplitudes directly in terms of weak decay form factors and meson light-cone distribution amplitudes. Qualitatively, the QCD factorisation formula implies the following:

- Naïve factorisation holds up to perturbative corrections and power corrections. In particular, there is a limit—the heavy-quark limit—in which naïve factorisation holds.
- Imaginary parts of strong amplitudes (strong phases) are small, $\mathcal{O}(\alpha_s; \Lambda/m_B)$.
- Real parts of strong amplitudes, and thus also $\mathrm{Re}\, r_f^T$, are roughly approximated by their naïve-factorisation expressions. The exception are colour-suppressed tree amplitudes, for which there are strong cancellations at the naïve-factorisation level.

*$\Delta S_f$ phenomenology for $B_d \to \phi K_S^0, \eta' K_S^0$, etc.* Many $S_f$ measurements for charmless final states have been performed at the $B$-factories. A selection of them is shown in Table 75. The last column lists the experimental values for $\Delta S_f = -\eta_f^{\mathrm{CP}} S_f - (\sin\phi_1)_{c\bar{c}s}$, combining errors in quadrature. Here $(\sin\phi_1)_{c\bar{c}s}$ is the HFLAV average of $\sin\phi_1$ measurements using final states with charmonia [230], while $S_f$ are the HFLAV averages for each individual penguin ($q\bar{q}s$) mode. These data can be compared to theory predictions.

A systematic treatment in QCD factorisation has been given in [681] for each of the listed modes. The calculation constrains the QCD penguin amplitudes through the branching fraction measurements, see [681] for details. The second and third columns show two different error estimates, one combining individual errors in quadrature, the other scanning over them. One observes that deviations are mostly predicted to be small, notably in the $\eta' K_S^0$ and $\phi K_S^0$ final states. Typically they have a definite, mode-dependent, sign. The fourth column of Table 75, shows mode-specific predictions for $\Delta S_f$ obtained with the help of flavour $SU(3)$ to fix or constrain tree-to-penguin ratios from data, see *e.g.*, [683–686]. To obtain further control of the predictions one could use global fits of $SU(3)$ amplitudes, such as in, *e.g.*, [687], but it would be crucial to include $SU(3)$ breaking effects to control the systematic errors associated with them. A first step in this direction could be the NP tests using sum rules valid to higher orders in $SU(3)$ breaking [688].

QCDF generally predicts definite or preferred signs of the $\Delta S_f$ shift, similarly to predictions from naïve factorisation. This implies a definite pattern of shifts to be compared with data. At present there is no significant tension between these predictions and data. The theoretical errors are generally much smaller than the experimental ones. The measurements of $b \to q\bar{q}s$ time-dependent $CP$ asymmetries thus provide theoretically clean NP discovery modes for Belle II.

NP can give rise to peculiar patterns of shifts in certain scenarios. For instance, BSM scenarios with "right-handed currents" can enhance the weak Hamiltonian QCD and electroweak penguin operators with a right-handed strange quark field, $Q_i'$, well above their SM values. Their matrix elements are related to the "SM" operators with left-handed strange quark field, $Q_i$, through

$$\langle f|Q_i'|\bar{B}^0\rangle = \eta_{\rm P}\langle f|Q_i|\bar{B}^0\rangle,$$

where $f$ was assumed to have definite parity, and $\eta_P = \pm 1$. An example of such a scenario is low scale supersymmetry with substantial mixing between right-handed squark flavours [689] (for grand-unified scenarios where this naturally arises see, *e.g.*, [690, 691]). This modifies the QCD penguin amplitudes and can produce parity-dependent shifts $\Delta S_f$. (The parameter values in [689] are ruled out by the LHC SUSY searches and measurements of $CP$ violation in $B_s$ mixing. However, the correlation between $B_s$ mixing and $\Delta F = 1$ decays is model-dependent).

*$\Delta S$ versus $\Delta A$ in $B \to \pi^0 K_S^0$.* For $\pi^0 K_S^0$ it is possible, with very limited theory input, to relate $S_{\pi^0 K_S^0}$ and $A_{\pi^0 K_S^0}$ from the four measured branching ratios in the $B \to \pi K$ system [692, 693]. The reason why this is possible is that most of the required hadronic matrix elements can be obtained using isospin. The starting point is the isospin relation

$$\sqrt{2}\,\mathcal{A}(B^0 \to \pi^0 K^0) + \mathcal{A}(B^0 \to \pi^- K^+) = -\Big[(\hat{T}+\hat{C})e^{i\phi_3} + \hat{P}_{\rm ew}\Big] \equiv 3\mathcal{A}_{3/2}, \qquad (317)$$

in which the QCD penguin amplitudes cancel out on the left-hand side. The subscript of $\mathcal{A}_{3/2}$ reminds us that the $\pi K$ final state has isospin $I = 3/2$. A similar relation holds for the $CP$-conjugate amplitudes, with $\mathcal{A}_{3/2} \to \bar{\mathcal{A}}_{3/2}$ and $\phi_3 \to -\phi_3$. Here $\hat{T}$, $\hat{C}$ and $\hat{P}_{\rm ew}$ are, respectively, the colour-allowed tree, colour-suppressed tree, and electroweak penguin contributions.

The important point is that $\mathcal{A}_{3/2}$ can be obtained with good accuracy. The $SU(3)$ flavour symmetry relates $|\hat{T}+\hat{C}|$ to $\mathcal{B}(B \to \pi^+\pi^0)$, while $P_{\rm ew}/(\hat{T}+\hat{C})$ is given directly by $C_{9,10}$ Wilson coefficients and CKM elements [694, 695]. Measurements of $\mathcal{B}(B^0 \to \pi^0 K^0)$, $\mathcal{B}(B^0 \to \pi^+ K^-)$, and the corresponding $CP$ asymmetries then suffice to determine $S_{\pi^0 K_S^0}$.

It is instructive to remove $A_{\pi^0 K_S^0}$ from the input data set and view the construction as a mostly data-driven prediction of the relation between $A_{\pi^0 K_S^0}$ and $S_{\pi^0 K_S^0}$ (Figure 101 left).

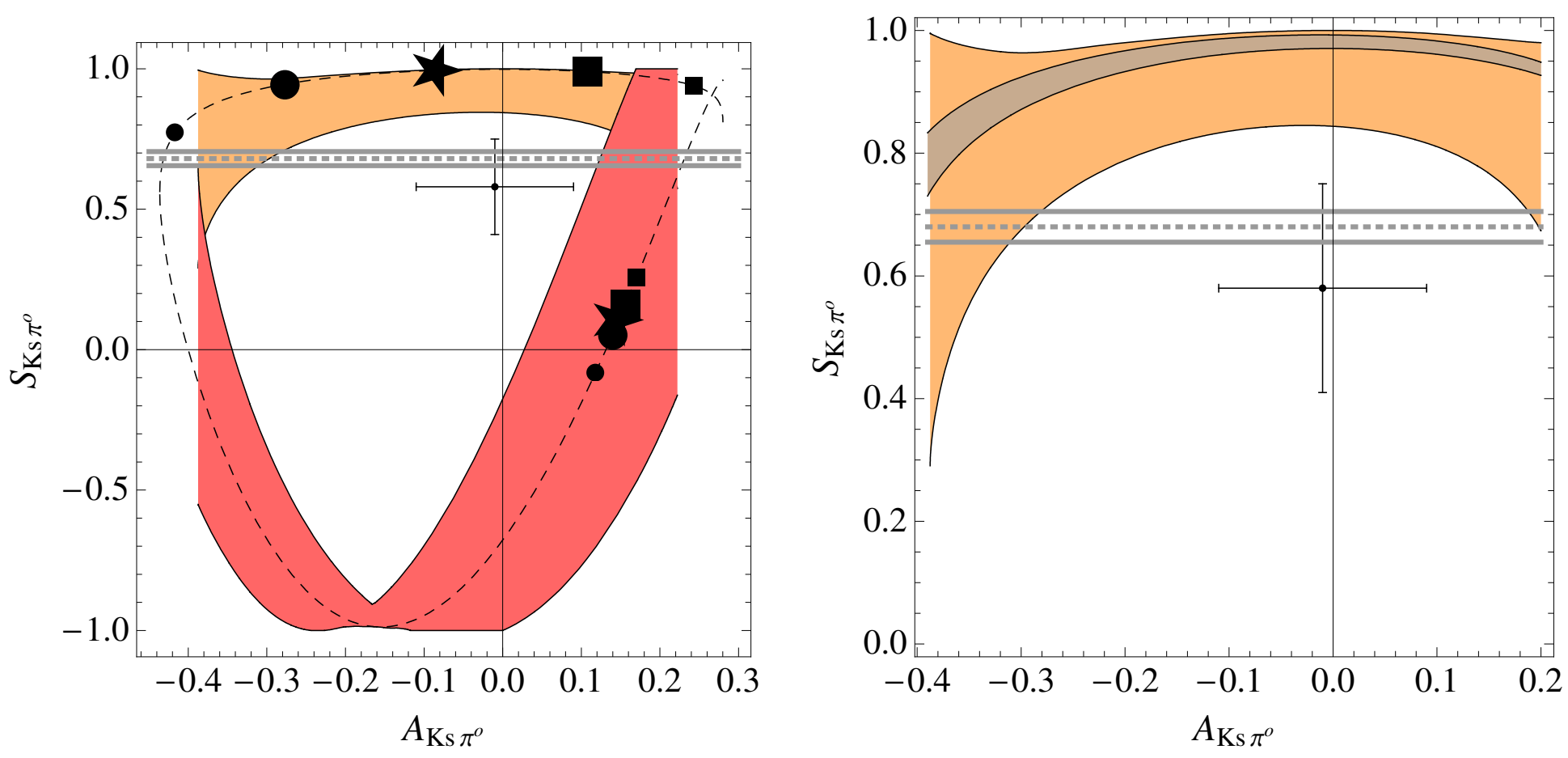

Fig. 101: Data-driven prediction for time-dependent $CP$ violation in $B \to \pi^0 K_S^0$. Figures from [692]. Left panel: Existing constraints. Only the top (orange, horizontal) band is consistent with $SU(3)$ and the heavy-quark limit. Right panel: Belle-II projection for about 10 ab$^{-1}$ of integrated luminosity (thin brown band). For details, see text.

Differently shaped points along the bands are distinguished by different values of the strong phase of the tree-to-penguin ratio $r_c \equiv (\hat{T} + \hat{C})/\hat{P}$. There is a four-fold ambiguity in the constuction, due to an ambiguity in determining the $I = 3/2$ amplitudes. The ambiguity is partly resolved by taking the strong phase of $r_c$ to be small, leading to a single closed "loop" of solutions in Figure 101. The smallness of the phase is implied by QCDF [696, 697] or alternatively from $SU(3)$ relations with $CP$ violation in $B \to \pi^+\pi^-$, see [692] for a detailed discussion. Of the remaining two solutions, one is again wildly inconsistent with both QCDF predictions and with $SU(3)$ relations with $B \to \pi\pi$ data. This leaves the top orange band in Figure 101. The experimental data on $A_{K_S^0\pi^0}, S_{K_S^0\pi^0}$ (cross) and $\sin\phi_1$ from $b \to c\bar{c}s$ (grey horizontal band) at the time of [692] are also displayed.

The data-driven nature of the method implies that Belle II measurements can be used to sharpen predictions considerably. This is illustrated in Figure 101 (right), where the 2008 situation is displayed in orange (broad band) and a projected uncertainty in brown (thin band), assuming unchanged central values of $B \to \pi K$ with 10 times larger data sets. The uncertainty (width of the brown band) is then determined by the $SU(3)$-breaking which were taken to be $\mathcal{O}(20\%)$ at the level of the amplitudes.

In summary, time-dependent $CP$ violation in $b \to q\bar{q}s$ penguin-dominated transitions provides complementary ways to access $\phi_1$ in the SM, and provides good sensitivity to BSM scenarios. The tree pollution in the $S_f$ coefficients is theoretically understood and small in several of the charmless two-body final states. In the special case of the $B \to \pi K$ system, an isospin analysis in conjunction with the use of the heavy-quark expansion provides a data-driven determination for $S_{\pi^0 K_S^0}$ at Belle II with percent-level accuracy.

*10.3.2. Experiment:* $\sin 2\phi_1$ *from* $b \to q\bar{q}s$, $q = u, d, s$. *Contributing authors: A. Gaz, S. Lacaprara*

In this section we present a complete sensitivity study for the time-dependent $CP$ violation parameters in the penguin dominated modes $B^0 \to \phi K^0$ and $\eta' K^0$ and an extrapolation of the sensitivity for $B^0 \to \omega K^0_S$.

For time-dependent $CP$ violation analysis the current implementation of the Belle II simulation and reconstruction software gives a realistic estimate of the $\Delta t$ resolution and the effective tagging efficiency. The reconstruction efficiencies, on the other hand, are most probably underestimated, especially for neutral particles. As more realistic estimates we thus use the values for the reconstruction efficiencies that were achieved by BaBar and Belle in the previous analyses (we also quote the efficiencies obtained with the current simulation).

The analyses of $b \to q\bar{q}s$ decays have several common features. The dominant background is due to random combinations of particles produced in continuum events ($e^+e^- \to q\bar{q}$, $q = u, d, s, c$) and, to a much smaller extent, from $B$-meson decays to charmed particles. To study it we simulated a large sample of $e^+e^- \to u\bar{u}, d\bar{d}, s\bar{s}, c\bar{c}$ events corresponding to $1-5$ ab$^{-1}$ of equivalent luminosity. The combinatorial background is without any peaking structure in the main analysis variables $M_{\rm bc}$ and $\Delta E$. It can be easily modelled from data, selecting sidebands of the $M_{\rm bc}$ and/or $\Delta E$ distributions.

The other significant background components are due to charmless $B$-decays with topologies similar to the decay under study. While much less frequent than the combinatorial background, they do require modelling of peaking structures in the main variables of the analysis, including $M_{\rm bc}$ and to a smaller extent $\Delta E$. This is more sensitive to extra or missing particles. Background charmless $B$ decays can also be $CP$ violating and can potentially bias the main measurement. We study these effects using a 5.0 ab$^{-1}$ equivalent-luminosity sample of simulated $B\bar{B}$ decays.

The $CP$ violating parameters $S$ and $A$ are extracted from an unbinned multi-dimensional maximum likelihood fit which includes the proper decay time difference $\Delta t$, and variables that discriminate against backgrounds: $M_{\rm bc}$, $\Delta E$, the output of the continuum suppression multivariate discriminator, the invariant masses of resonances, helicity angles, etc... We assume that all these variables are uncorrelated.

For some of the modes there will be competition from the LHCb experiment. Their plan is to measure $S_{\phi K^0}$ with uncertainty of 0.06 for an integrated luminosity of 50 fb$^{-1}$ and extensive upgrade in their hadronic trigger [698]. From this projection it is therefore reasonable to expect that LHCb will approach the sensitivity of BaBar and Belle at the end of Run2 in the year 2018. LHCb is expected to be less competitive in the $B^0 \to \eta' K^0$ and $\omega K^0_S$ channels.

The systematic uncertainties for $B^0 \to \eta' K^0$, the channel with higher yield, have been extrapolated from Belle results using the same assumption used for $B^0 \to J/\psi K^0$. They become comparable to statistical uncertainties for an integrated luminosity of about 10 ab$^{-1}$. The other $b \to q\bar{q}s$ decay channels will be statistically limited up to 50 fb$^{-1}$.

The equivalent integrated luminosity used for these studies is 5 ab$^{-1}$. With this dataset, the measurements will still be dominated by statistical uncertainties. We verified that the statistical sensitivity of the analyses scales well with $1/\sqrt{\mathcal{L}}$.

*$B_d \to \phi K^0$. Contributing author: A. Gaz*

BaBar and Belle extracted the $B_d \to \phi K^0$ $CP$ asymmetry parameters from time-dependent analysis of the $K^+K^-K^0$ final state [699, 700]. We perform a sensitivity study using a quasi-two body approach, taking into account the decay channels:

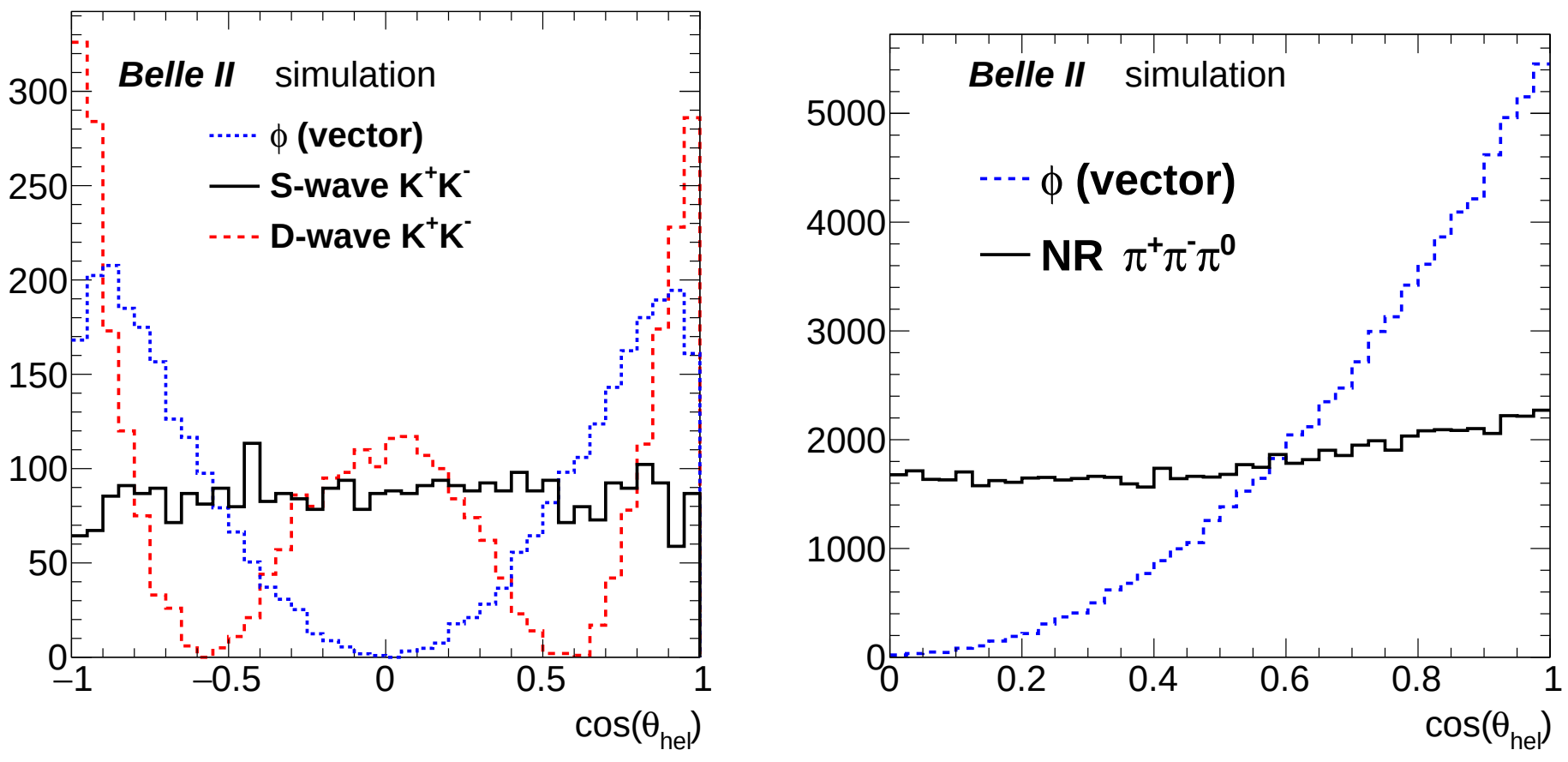

Fig. 102: Distributions of $\cos\theta_H$ for the $\phi \to K^+K^-$ case (left plot) and for the $\phi \to \pi^+\pi^-\pi^0$ case (right). The plots include effects due to the detector acceptance, easily visible at the edges of the distributions.

$\phi(K^+K^-)K^0_S(\pi^+\pi^-)$, $\phi(K^+K^-)K^0_S(\pi^0\pi^0)$, $\phi(\pi^+\pi^-\pi^0)K^0_S(\pi^+\pi^-)$, $\phi(K^+K^-)K^0_L$. The channel $\phi(\pi^+\pi^-\pi^0)K^0_S(\pi^+\pi^-)$ was never used before, due to its very low significance with small integrated luminosity. For the $K^0_L$ channel we extrapolate previous BaBar and Belle results and do not perform a simulation.

*Analysis strategy.* In the $B^0 \to \phi(K^+K^-)K^0$ decay there is a non-negligible non-resonant $s$-wave contribution, even when restricting the $K^+K^-$ invariant mass to a narrow range around the $\phi$ resonance. The $s$-wave has a $CP$-phase that differs from the resonant contribution and, if ignored, would lead to a significant bias in the measurement. This component could include also a contribution from $f_0(980) \to K^+K-$ decays which, in a more refined version of the analysis could be treated as an additional component.

The two components can be disentangled using a full Dalitz plot analysis. An alternative, much simpler, option is to include in the maximum likelihood fit the helicity angle, $\theta_H$, of the $\phi$ candidate. For $\phi \to K^+K^-$, $\theta_H$ is the angle between $K^+$ and $B$ flight directions in the $\phi$ rest frame, while for $\phi \to \pi^+\pi^-\pi^0$, $\theta_H$ is the angle between the normal to the plane formed by the three pions and the $B$ momentum in the $\phi$ rest frame. Figure 102 (left) shows the $\cos\theta_H$ distributions for the scalar, vector, and tensor components for $\phi \to K^+K^-$, while Figure 102 (right) shows the $\cos\theta_H$ distributions for the scalar and vector components for $\phi \to \pi^+\pi^-\pi^0$. Toy Monte Carlo studies show that this approach gives unbiased results for the $CP$ asymmetry parameters for both the vector and the scalar components, provided that each component has an adequate number of events, typically $\mathcal{O}(100)$.

*Event selection.* Table 76 summarises the main selection cuts applied for each of the investigated channels. Cuts are applied to the main discriminating variables $M_{\rm bc}$ and $\Delta E$, and to the invariant masses of the intermediate resonances; all these are intentionally set quite loose so that backgrounds can be fitted and modelled from the sidebands. Furthermore, requirements on the flight length significance (*flLenSig*, the ratio between the measured flight length and its estimated uncertainty) of the $K_S^0$ candidates, on the PID of the charged kaons, and on the probability that final state particles originate from a common vertex (PrbVtx) are set. The requirements on the distance of closest approach $d_0$, its $z$ coordinate $z_0$, and on the number of PXD hits associated to the tracks that originate from the $\phi$ decay, reject poorly reconstructed events that are expected to have significantly worse $\Delta t$ resolution.

Table 76: Main selection requirements for the channels used in the $\phi K_S^0$ sensitivity study. The selection efficiency $\varepsilon$ and candidate multiplicity for signal events are given at the bottom.

| Variable | $\phi(K^+K^-)K_S^0(\pi^+\pi^-)$ | $\phi(K^+K^-)K_S^0(\pi^0\pi^0)$ | $\phi(\pi^+\pi^-\pi^0)K_S^0(\pi^+\pi^-)$ |
|---|---|---|---|
| $M_{\rm bc}$ (GeV) | $> 5.25$ | $> 5.25$ | $> 5.25$ |
| $\Delta E$ (GeV) | in $[-0.2, 0.2]$ | in $[-0.1, 0.2]$ | in $[-0.1, 0.2]$ |
| $m(\pi^0)$ (GeV) | — | in $[0.10, 0.14]$ | in $[0.10, 0.14]$ |
| $E(\pi^0)$ (GeV) | — | — | $> 0.35$ |
| $m(\phi)$ (GeV) | in $[1.00, 1.05]$ | in $[1.00, 1.05]$ | in $[0.97, 1.04]$ |
| $m(K_S^0)$ (GeV) | in $[0.48, 0.52]$ | in $[0.44, 0.51]$ | in $[0.48, 0.52]$ |
| flLenSig($K_S^0$) | $> 5$ | — | $> 5$ |
| PIDk($K^\pm$) | $> 0.2$ | $> 0.2$ | — |
| PrbVtx($\phi$) | $> 10^{-4}$ | $> 10^{-4}$ | $> 10^{-4}$ |
| PrbVtx($K_S^0$) | $> 10^{-4}$ | — | $> 10^{-4}$ |
| PrbVtx($B$) | $> 10^{-4}$ | $> 10^{-4}$ | $> 10^{-4}$ |
| | for each track coming from the $\phi$ decay: | | |
| $d_0$ (cm) | $< 0.08$ | $< 0.08$ | $< 0.08$ |
| $z_0$ | $< 0.3$ | $< 0.3$ | $< 0.3$ |
| #PXDhits | $\geq 1$ | $\geq 1$ | $\geq 1$ |
| $\varepsilon$ | 31.1% | 14.2% | 17.4% |
| Cand. multiplicity | 1.0008 | 1.0701 | 1.0470 |

The selection efficiencies and candidate multiplicities for signal events are reported at the bottom of Table 76. While the efficiency for the $\phi(K^+K^-)K_S^0(\pi^+\pi^-)$ mode is comparable to what was achieved at BaBar and Belle, the modes with $\pi^0$ are performing significantly worse. We expect that the planned developments on the reconstruction software of the calorimeter and the selection of photon candidates will improve significantly the overall efficiency as we approach the start of data taking.

*$\Delta t$ resolution.* The resolution on the proper decay time difference $\Delta t$ is one of the critical aspects of time-dependent $CP$-violation analyses. The $\phi \to K^+K^-$ decay is particularly challenging, since the $\phi \to K^+K^-$ decay is just above the kinematical threshold. The $K^\pm$ momenta are soft in the $\phi$ rest frame, while the $\phi$ boost leads to quite unfavourable angle between the kaon momenta.

From a simple geometric vertex fit to the tracks originating from the signal $B$ decay vertex we thus expect a resolution that is much worse than for the $J/\psi \to \mu^+\mu^-$ decays. Fitting with three gaussians the $\Delta t_{gen} - \Delta t_{meas}$ distribution without any external constraint gives $\Delta t = 2.08$ps (1.18ps) resolution by for the $\phi \to K^+K^-$ ($\phi \to \pi^+\pi^-\pi^0$) decay.

Table 77: $\Delta t$ resolution obtained with the *iptube* constraint and, when applicable, the $K_S^0$ flight direction constraint. The value reported is the weighted average of the $\sigma$'s of the three Gaussians.

| Channel | $\Delta t$ resolution (ps) |
|---|---|
| $\phi(K^+K^-)K_S^0(\pi^+\pi^-)$ | 0.75 |
| $\phi(K^+K^-)K_S^0(\pi^0\pi^0)$ | 0.77 |
| $\phi(\pi^+\pi^-\pi^0)K_S^0(\pi^+\pi^-)$ | 0.78 |

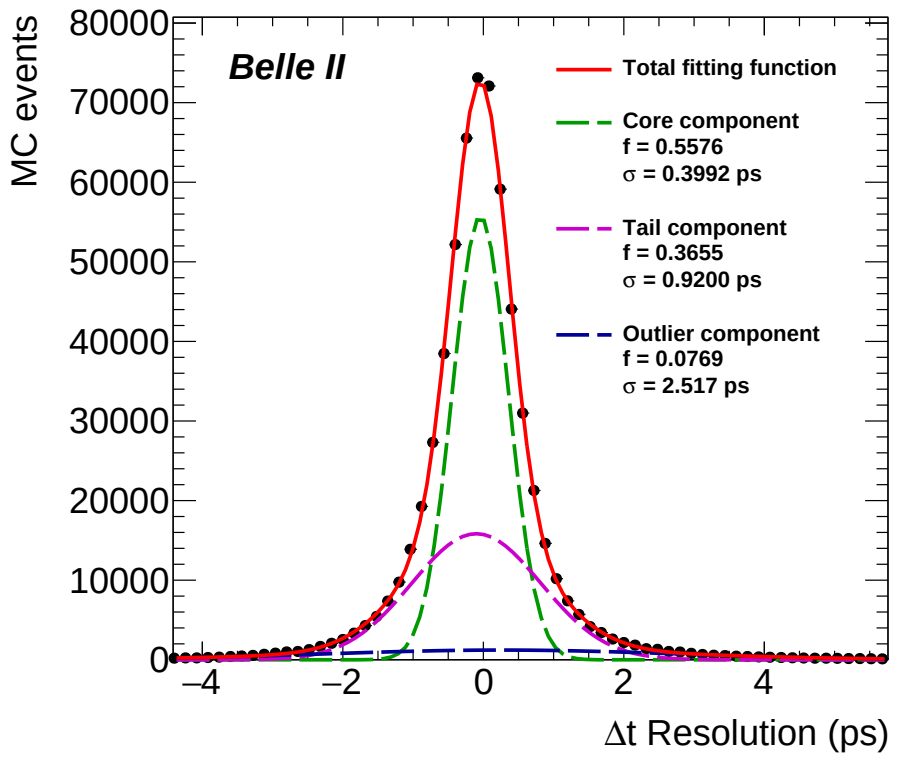

Fig. 103: Example of three-gaussian fit on the $\Delta t$ resolution. The (integral) fractions and the widths of the three gaussians components are reported on the figure.

External constraints greatly improve the $\Delta t$ resolution. One can apply the *iptube* constraint (see Section 6.2.3) and, in the case of $K_S^0 \to \pi^+\pi^-$, we can also add a constraint from the $K_S^0$ flight direction. Figure 103 shows an example of a fit to the $\Delta t$ resolution, while Table 77 summarises the best achievable resolutions for the three $B^0 \to \phi K_S^0$ channels investigated.

*Continuum background.* The discrimination between $B\bar{B}$ and continuum events relies on variables that are sensitive to the different event topologies (spherical for $B\bar{B}$, events, jet-like for continuum). We utilise a FastBDT [70] multivariate discriminator for efficient separation between signal $B \to \phi K_S^0$ and continuum events. The algorithm takes as input 30 variables that provide at least some discrimination power between the two categories. The most powerful among those are the cosine of the angle between the thrust axis of the $B$ candidate and the thrust axis of the rest of the event, the ratio between the second and zeroth Fox-Wolfram moments, the cosine of the angle between the $B$ candidate flight direction and the $z$-axis, the Cleo Cones and the Kakuno-Super-Fox-Wolfram (KSFW) moments (S. Sec. 6.4).

Figure 104 shows an example of the separation that is achievable with the FastBDT algorithm and the selected set of variables. Just to illustate the discrimination power of the method, by setting a cut on the output variable of the FastBDT discriminator that retains 95% of the signal events, we reject 82.7% of the continuum background in the $\phi(K^+K^-)K_S^0(\pi^+\pi^-)$ case, 85.5% in $\phi(K^+K^-)K_S^0(\pi^0\pi^0)$, and 75.9% in $\phi(\pi^+\pi^-\pi^0)K_S^0(\pi^+\pi^-)$.

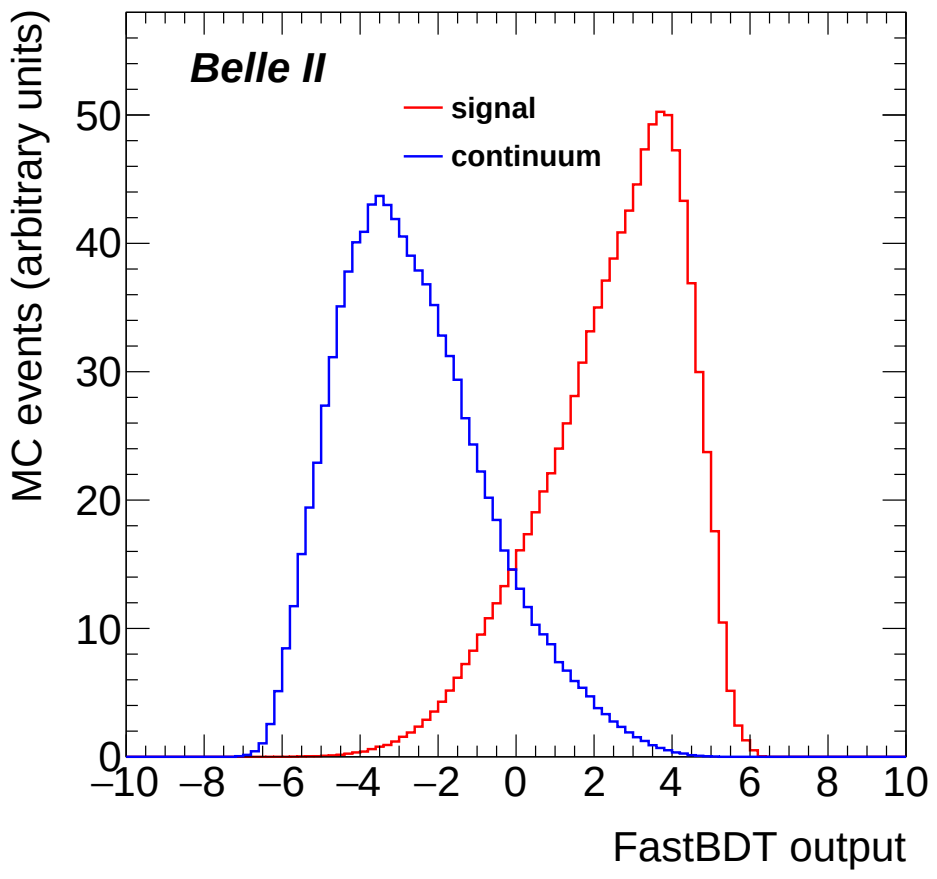

Fig. 104: Output of the FastBDT multivariate discriminator utilised for continuum suppression in the channel $\phi(K^+K^-)K^0_S(\pi^+\pi^-)$. The red (blue) histogram corresponds to the output of signal $B^0 \to \phi K^0_S$ (continuum) events.

Table 78: Number of generic $B\bar{B}$ events passing the selection (see Table 76) for the different channels. The equivalent luminosity of the generic MC sample used is 5 ab$^{-1}$. For comparison, also the expected signal yield is given.

| Channel | $B^+B^-$ | $B^0\bar{B}^0$ | expected signal |
|---|---|---|---|
| $\phi(K^+K^-)K^0_S(\pi^+\pi^-)$ | 43 | 97 | 2280 |
| $\phi(K^+K^-)K^0_S(\pi^0\pi^0)$ | 39 | 58 | 765 |
| $\phi(\pi^+\pi^-\pi^0)K^0_S(\pi^+\pi^-)$ | 627 | 1512 | 545 |

In order to keep the selection efficiency as high as possible for signal events, we do not apply cuts on the output of the FastBDT discriminator, but rather use it as a variable in the maximum likelihood fit when extracting the $S$ and $A$ parameters.

*$B\bar{B}$ background.* A preliminary estimate of the yield of $B\bar{B}$ backgrounds is obtained by checking how many events of generic $B^+B^-$ and $B^0\bar{B}^0$ Monte Carlo events pass the selection. For this estimate we use the full 5 ab$^{-1}$ available at the time of writing.

Table 78 summarises the number of generic $B\bar{B}$ background events passing the selection for different channels. The numbers are quite high for the $\phi(\pi^+\pi^-\pi^0)K^0_S(\pi^+\pi^-)$ channel, so further work is needed in order to optimise the selection and reject more background.

For the $\phi(K^+K^-)K^0_S(\pi^+\pi^-)$ channel, one of the modes that has highest probability of passing the selection is $B^0 \to \phi K^{*0}$, with $K^{*0} \to K^0_S\pi^0$ (with the $\pi^0$ not being reconstructed). For the real data analysis, further studies on this background mode will be needed, in order to avoid biases on the $S$ and $A$ parameters.

*Maximum Likelihood Fit.* In the multidimensional Maximum Likelihood Fit, the time-independent variables are $M_{\rm bc}$, $\Delta E$, the output of the FastBDT discriminator used for continuum suppression, $m(\phi)$, and $\cos\theta_H$. The first three variables are very powerful in discriminating between signal and continuum background events, while the latter two mostly separate the signal from real $\phi$ candidates from the non-resonant $K^+K^-$ (or $\pi^+\pi^-\pi^0$) events.

For each of the decay channels under study, we consider five components:

(1) Signal from real $\phi$ mesons;
(2) SXF: self cross-feed events originating from real signal events in which the reconstruction of the $B$ signal candidate utilised one or more particles from the decay of the other $B$ in the event;
(3) Non-resonant: events in which the selected $\phi$ candidate originates from non-resonant $K^+K^-$ or $\pi^+\pi^-\pi^0$ events;
(4) Combinatorial: mostly arising from continuum background;
(5) $B\bar{B}$ background.

*Estimate of sensitivity from pseudo-experiments studies.* To estimate the sensitivity of the analysis we performed an ensemble test with 1 ab$^{-1}$ and 5 ab$^{-1}$ equivalent Monte Carlo (MC) samples. For the signal, SXF and non-resonant events, the full MC simulation is used and combinatorial and $B\bar{B}$ background events are generated by the parameters of each corresponding pdf determined by the fully simulated generic MC datasets. To study the sensitivity on the signal and background yields, we perform a scan varying the number of injected signal and background events, covering approximately one order of magnitude about the expected values. For each mode and each of the investigated signal/background hypotheses we performed 1000 pseudo-experiments.

We confirm that the fit returns unbiased results about the signal for both scalar and vector components as well as background yields. To perform the test under an extreme condition, we studied the case with opposite values of $S$ for the signal (+0.7) and for the non-resonant component (-0.7). We find these two components are correctly separated and the obtained $CP$ violation parameters are consistent with inputs. We also see that the dependence of the uncertainty $\sigma(S)$ depends very mildly on the background yields, and that the dependence on the signal yield is, as expected, $1/\sqrt{N_{sig}}$.

Table 79 summarises the sensitivity estimates for the two integrated luminosity scenarios considered. We estimate the expected yield of $\phi K^0_L$ based on previous BaBar and Belle analyses (but use the same $\Delta t$ resolution we estimate in $\phi\to K^+K^-$ for Belle II).

$B_d\to\eta' K^0_S$*Contributing authors: S. Lacaprara, A. Mordà*

The $\eta' K^0$ decay channel shares many features with $\phi K^0$. The main differences are that the $\eta'$ is a pseudoscalar particle, its decay channels are more complex, and that the branching fraction is about 10 times larger [701].

The BaBar and Belle collaborations performed the $CP$-violation analyses for this channel using $467\cdot 10^6$ [702] and $772\cdot 10^6$ $B\bar{B}$ pairs [703], respectively. The published results for $S_{\eta' K^0_S}$ are still dominated by statistical uncertainties ($S_{\eta' K^0_S} = +0.57\pm 0.08\pm 0.02$(BaBar), $S_{\eta' K^0_S} = +0.68\pm 0.07\pm 0.03$(Belle)).

The $\eta'$ decay chains considered for this analysis are:

(1) $\eta'(\to\eta(\to\gamma\gamma)\pi^+\pi^-) : \eta'(\eta_{\gamma\gamma}\pi^{\pm})$
(2) $\eta'(\to\eta(\to\pi^+\pi^-\pi^0)\pi^+\pi^-)$: $\eta'(\eta_{3\pi}\pi^{\pm})$

Table 79: Sensitivity estimates for $S_{\phi K^0}$ and $A_{\phi K^0}$ parameters for 1 ab$^{-1}$ and 5 ab$^{-1}$ integrated luminosity. The efficiency $\varepsilon_{reco}$ used in this estimate has not been taken from the simulation, but is rather an estimate taking into account the expected improvements. Systematic uncertainties, negligible for these integrated luminosities, are not included.

| Channel | $\varepsilon_{reco}$ | Yield | $\sigma(S_{\phi K^0})$ | $\sigma(A_{\phi K^0})$ |
|---|---|---|---|---|
| 1 ab$^{-1}$ lumi.: | | | | |
| $\phi(K^+K^-)K_S^0(\pi^+\pi^-)$ | 35% | 456 | 0.174 | 0.123 |
| $\phi(K^+K^-)K_S^0(\pi^0\pi^0)$ | 25% | 153 | 0.295 | 0.215 |
| $\phi(\pi^+\pi^-\pi^0)K_S^0(\pi^+\pi^-)$ | 28% | 109 | 0.338 | 0.252 |
| $K_S^0$ modes combination | | 400 | 0.135 | 0.098 |
| $K_S^0 + K_L^0$ modes combination | | | 0.108 | 0.079 |
| 5 ab$^{-1}$ lumi.: | | | | |
| $\phi(K^+K^-)K_S^0(\pi^+\pi^-)$ | 35% | 2280 | 0.078 | 0.055 |
| $\phi(K^+K^-)K_S^0(\pi^0\pi^0)$ | 25% | 765 | 0.132 | 0.096 |
| $\phi(\pi^+\pi^-\pi^0)K_S^0(\pi^+\pi^-)$ | 28% | 545 | 0.151 | 0.113 |
| $K_S^0$ modes combination | | 2000 | 0.060 | 0.044 |
| $K_S^0 + K_L^0$ modes combination | | | 0.048 | 0.035 |

(3) $\eta'(\to \rho^0(\to \pi^+\pi^-)\gamma)$: $\eta'(\rho\gamma)$.

the $K^0$ can be a $K_S^0$, decaying into $K_S^0 \to \pi^+\pi^-$ ($K_S^{(\pm)}$) or $\pi^0\pi^0$ ($K_S^{(00)}$), or a $K_L^0$. At the time of writing the sensitivity study for the channel $\rho^0\gamma$ is not ready yet, and the modes with $K_L^0$ have not yet been studied. Among the four remaining channels we put more emphasis on the final states where the $K_S^0$ decays into charged pions. In particular, the channel with $\eta$ decaying into 3 $\pi$ and $K_S^0$ into a pair of $\pi^0$ has not been used by Belle and BaBar due to the very low reconstruction efficiency on signal events and large background yields.

Table 80: Selection requirements for the channels used in the $\eta'(\to \eta\pi^\pm)K_S^0$ sensitivity study.

| $\eta'(\eta_{\gamma\gamma}\pi^\pm)K_S^{(\pm)}$ | $\eta'(\eta_{\gamma\gamma}\pi^\pm)K_S^{(00)}$ | $\eta'(\eta_{3\pi}\pi^\pm)K_S^{(\pm)}$ | $\eta'(\eta_{3\pi}\pi^\pm)K_S^{(00)}$ |
|---|---|---|---|
| $M_{\rm bc} > 5.25$ GeV | | | |
| $-0.2 < \Delta E < 0.2$ GeV | $-0.15 < \Delta E < 0.2$ GeV | $-0.15 < \Delta E < 0.15$ GeV | $-0.15 < \Delta E < 0.25$ GeV |
| $0.06 < E_\gamma < 6$ GeV | | | |
| - | | $0.1 < m_{\pi^0} < 0.15$ GeV | |
| $0.52 < m_\eta < 0.57$ GeV | $0.48 < m_\eta < 0.57$ GeV | $0.52 < m_\eta < 0.57$ GeV | $0.52 < m_\eta < 0.57$ GeV |
| $0.93 < m_{\eta'} < 0.98$ GeV | $0.93 < m_{\eta'} < 0.98$ GeV | $0.93 < m_{\eta'} < 0.98$ GeV | |
| $0.48 < m_{K_S^0} < 0.52$ GeV | $0.42 < m_{K_S^0} < 0.52$ GeV | $0.48 < m_{K_S^0} < 0.52$ GeV | $0.40 < m_{K_S^0} < 0.52$ GeV |
| for each track coming from the $\eta'$ decay: | | | |
| $\Delta\log\mathcal{L}(\pi, K) > -10$ | | | |
| $d_0 < 0.16\,$cm | | | |
| $z_0 < 0.2\,$cm | | | |

*Signal reconstruction, backgrounds and selection.* For each final state, signal candidates are found by reconstructing the whole decay chains. This is done by reconstructing all the intermediate particles starting from the final state tracks, and proceeding back up to the head of the decay ($B^0$ or $\bar{B}^0$).

Selection criteria listed in table 80 are applied on each of the reconstructed particles in the decay chain. In particular the requirements on the invariant masses of the intermediate particles efficiently reduce the amount of candidates arising from random combination of tracks and photons in the event (combinatoric backgrounds). The selection criteria are in general looser for channels with neutral particles decaying to a pair of photons.

Once the signal $B^0$ decay chain has been reconstructed, the rest of the event is fed to the flavour tagger algorithm (S. Sec. 6.5) to determine the flavour of the $B$ on the tag side.

*Self cross feed and multiple candidates.* Given the complex final states considered, often more than one candidate per event fulfills the selection requirements, especially for the modes with $\eta \to \pi^+\pi^-\pi^0$. For signal events, the average number of candidates per event is 1.1 and 2.7 for the $\eta'(\eta_{\gamma\gamma}\pi^{\pm})K_S^{(\pm)}$ and $\eta'(\eta_{3\pi}\pi^{\pm})K_S^{(\pm)}$ channels respectively. For channels with $K_S^0$ decaying into a pair of $\pi^0$, preliminary studies show that the multiplicity is even higher: $\sim 5$ and $\sim 30$, for the final states with $\eta \to \gamma\gamma$ and $\eta \to \pi^+\pi^-\pi^0$ respectively.

Among the selected candidates there is usually the one reconstructed with the proper combination of tracks corresponding to the actual decay chain, together with others built with a wrong combination of the final state tracks. Those candidates will be referred in the following (according to the notation used in the previous section) as self cross-feed ("SXF")candidates.

The increased fraction of SXF candidates, compared to the previous analyses from Belle, is mostly due to the higher level of background arising from the beams interactions in the higher luminosity regime. Currently the tracking and photons reconstruction algorithm for Belle2 are still under development and, once optimised, will likely end up in an increased true signal purity of the sample of selected candidates; in the meanwhile a novel strategy has been designed to deal with the larger fraction of self cross-feed candidates.

In order to discriminate the true signal against the cross-feed selected candidates, a multivariate algorithm Boosted Decision Tree (BDT) [239] has been trained using kinematic and geometrical variables. These include invariant masses of intermediate particles, vertex $\chi^2$, impact parameters of the pion tracks, and the variables describing the reconstructed photons by the ECL. The BDT output variable, $BDT_{SXF}$ henceforth, is used as an additional input to the final fit. Indeed, it provides good separation between the true and cross-feed selected signal candidates, and, in addition, it improves the discrimination of background events (described in the next subsection) which behave like the cross-feed ones. The respective distributions are shown in Figure 105 for the $\eta(3\pi)$ channel.

Selected candidates are ranked according to the value of the $BDT_{SXF}$ and the following three strategies have been explored in order to deal with multiple candidates:

- Strategy A: for each event keep only the candidate with the highest $BDT_{SXF}$ value;
- Strategy B: for each event keep only the two candidates with the highest $BDT_{SXF}$ value;
- Strategy C: keep all the candidates in each event.

The advantage of strategy A is that it leads to only one candidate per event. The cost is the reduced signal efficiency. The other two strategies give higher signal efficiencies, but also increase the number of cross-feed candidates. These can still be separated in the maximum likelihood fit that uses $BDT_{SXF}$. Strategy B uses just two candidates, because in most cases the signal has the highest or second highest $BDT_{SXF}$. Including the third candidate does not increase the signal efficiency significantly.

Table 81: Selection efficiency $\varepsilon$ and fraction of signal cross feed candidates $\varepsilon_{SXF}$ for the $\eta'(\eta_{\gamma\gamma}\pi^{\pm})K_S^{(\pm)}$ and $\eta'(\eta_{3\pi}\pi^{\pm})K_S^{(\pm)}$ channels when selecting only one (A), two (B), or all (C) the candidates in the event. The selected strategy is labeled with $\star$.

| Channel | Strategy | $\varepsilon$ | $\varepsilon_{SXF}$ |
|---|---|---|---|
| $\eta'(\eta_{\gamma\gamma}\pi^{\pm})K_S^{(\pm)}$ | C$\star$ | 23.0 % | 3.8 % |
| | A | 6.7 % | 2.6% |
| $\eta'(\eta_{3\pi}\pi^{\pm})K_S^{(\pm)}$ | B$\star$ | 8.0 % | 6.0% |
| | C | 9.5 % | 28.6% |

For the $\eta'(\eta_{\gamma\gamma}\pi^{\pm})K_S^{(\pm)}$ final state the candidate multiplicity is very close to 1 and the three strategies are almost equivalent. The strategy C was chosen in order to keep the signal efficiency as high as possible. For the $\eta'(\eta_{3\pi}\pi^{\pm})K_S^{(\pm)}$ instead, the strategy B has been adopted, since it allows for a good signal efficiency while keeping the fraction of cross-feed events at a reasonable level, see Table 81. The goodness of such choice has also quantitatively been checked by computing (through ensemble tests) the expected statistical uncertainty and the bias on the $S_f$ parameters for each of the above mentioned scenarios: the strategy B has been found the one giving the smallest statistical uncertainty while keeping a negligible bias.

Despite the improvement in the $K_S^0$ reconstruction, the computed signal efficiencies of the modes with $K_S^0 \to \pi^+\pi^-$ decays are comparable with those achieved by Belle and BaBar; this is likely due to the low reconstruction efficiency of the intermediate $\eta$ and $\pi^0$ decaying into photons. Preliminary studies one modes with $K_S^0 \to \pi^0\pi^0$ decays show a significantly lower efficiency with respect to those estimated for the channels with $K_S^0$ decaying into charged pions. The reconstruction efficiencies for modes with $\eta$ and $\pi^0$ intermediate states are expected to improve with better reconstruction algorithms.

*Background suppression.* Background candidates originate from two sources: random combination of particles from continuum events, and from actual $B\bar{B}$ events (peaking).

Among the two sources of background, the former is relatively easy to model, by looking at side bands of $M_{\rm bc}$ and $\Delta E$ in the data, whose selection requirements are kept rather loose. At present, the expected yields of this contribution is estimated from a large MC production (0.7 ab$^{-1}$). As for the $\phi K_S^0$, the best discrimination between the continuum background and the signal is achieved by a multivariate algorithm sensitive to the event topology (spherical for $B\bar{B}$ events, jet-like for continuum). For this analysis, we used a BDT algorithm, using the same set of variables presented for the $\phi K^0$. These variables are explained in detail in Sec. 6.4. The output distribution of the BDT algorithm is shown in Figure 105 for the $\eta(\gamma\gamma)K_S^0(\pi^{\pm})$ channel.

This variable allows a good separation between background and signal, *e.g.*, it is possible to retain 95% of the signal and removing 50% of the background, or reject 97.5% of the background with a relative signal efficiency of 50%. Given the relative ease with which one can tell apart signal and continuum backgrounds events using $M_{\rm bc}$, $\Delta E$, and the BDT, an even better strategy is to include the BDT as discriminating variable in the multidimensional maximum likelihood fit.

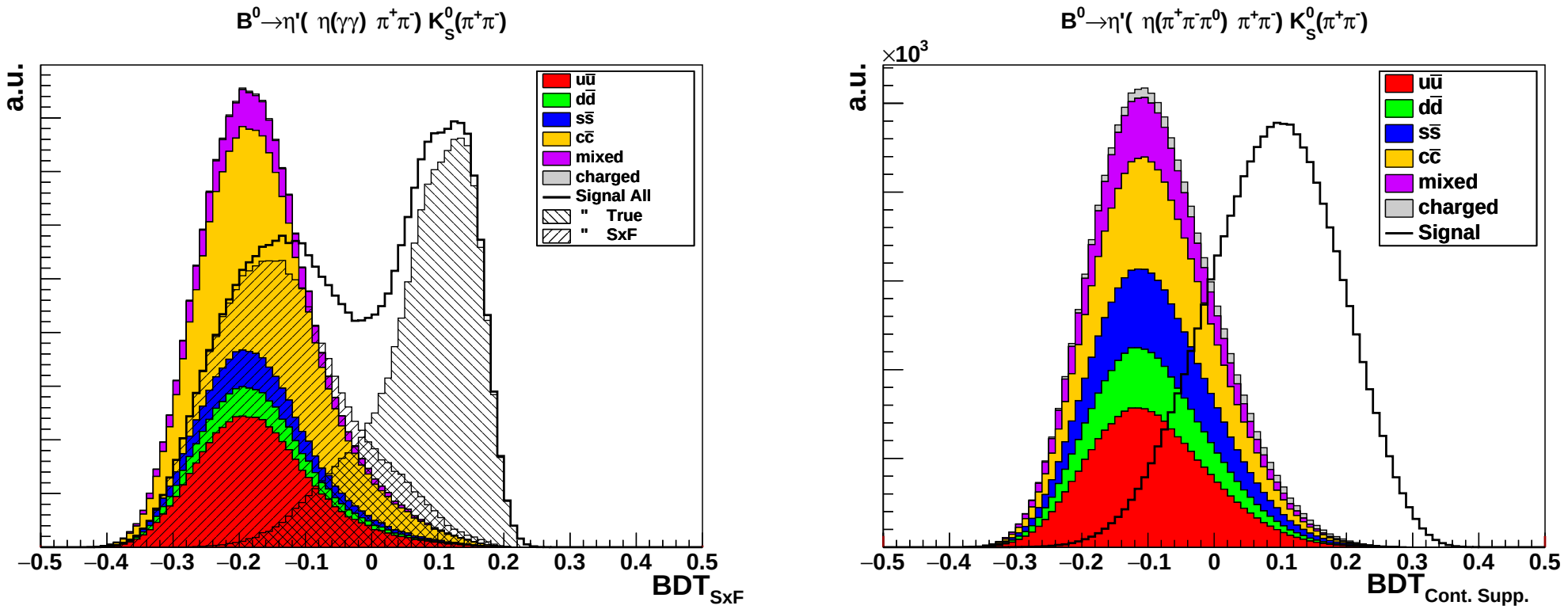

Fig. 105: Left: distribution of the $BDT_{SXF}$ variable for true signal events, self cross-feed candidates, and background events for the $\eta'(\eta_{\gamma\gamma}\pi^{\pm})K_S^{(\pm)}$ channel. Right: output of the BDT multivariate discriminator used for continuum suppression in the channel $\eta'(\eta_{3\pi}\pi^{\pm})K_S^0(\pi^{\pm})$ (”signal” distribution accounts for both the true and the SXF signal candidates).

Table 82: Expected yields of continuum and peaking ($B\bar{B}$) events passing the selection for the different channels. The equivalent luminosity of the generic MC sample used is 1 ab$^{-1}$. The continuum background yield is before any cut on the continuum suppression variable.

| Channel | Continuum | $B^0\bar{B}^0$ | $B^+B^-$ |
|---|---|---|---|
| $\eta'(\eta_{\gamma\gamma}\pi^{\pm})K_S^{(\pm)}$ | 16413 | 1834 | 57 |
| $\eta'(\eta_{3\pi}\pi^{\pm})K_S^{(\pm)}$ | 4508 | 304 | 13 |

The peaking background from $B\bar{B}$ events was estimated from Monte Carlo simulation, by analysing a data sample corresponding to $\sim 0.7$ ab$^{-1}$.

Table 82 summarises the amount of continuum and peaking background for different channels, for an integrated luminosity of 1 ab$^{-1}$. In general, the peaking background is much smaller than the continuum, and the continuum mostly comes from $c\bar{c}$ events and $u\bar{u}$ for $\eta \to 2\gamma$ channel.

$\Delta t$ *resolution.* The signal vertex precision ultimately determines the resolution on proper time difference $\Delta t$. In the case of decay $\eta' \to \eta(2\gamma)\pi^+\pi^-$, only the two charged pions can be used to build the $B$ decay vertex, whereas, for decay $\eta' \to \eta(3\pi)\pi^+\pi^-$, two additional charged pions are present and are used for vertex reconstruction. The decay topology is thus more favourable than the $\phi \to K^+K^-$ case described before, but not as good as the golden $J/\psi \to \mu^+\mu^-$ one, given the invariant mass of the parent particle $\eta'$.

The resolution on $\Delta t$ is estimated as for the $\phi K^0$ analysis, with a three-gaussian fit to the $\Delta t - \Delta t_{true}$ distribution, expressing the resolution as the weighted average of the $\sigma$ of each component. As for the case of $\phi K^0$, the vertex resolution can be improved by using the *iptube* constraint(see Sec. 6.2.3), as well as using the $K_S^0$ flight direction, for the $K_S^0 \to \pi^+\pi^-$ decays.

Table 83: The $\Delta t$ resolution for true, SXF and all selected candidates, for $\eta'(\eta_{\gamma\gamma}\pi^{\pm})K_S^{(\pm)}$ and $\eta'(\eta_{3\pi}\pi^{\pm})K_S^{(\pm)}$ channels (obtained by fitting the $B^0$ signal vertex with the *iptube* constraint and the information on the $K_S^0$ flight direction.

| Channel | True | SXF | All |
|---|---|---|---|
| $\eta'(\eta_{\gamma\gamma}\pi^{\pm})K_S^{(\pm)}$ | 1.22 ps | 2.87 ps | 1.45 ps |
| $\eta'(\eta_{3\pi}\pi^{\pm})K_S^{(\pm)}$ | 1.17 ps | 2.36 ps | 1.50 ps |

Without beam background, the resolution for the $\eta \to \gamma\gamma$ final state decreases from 1.89 ps without any constraints, to 1.62 ps including the constraints of $K_S^0$, and to 0.91 ps by adding the *iptube* constraint. In this last case the resolutions of the three gaussian components of the time resolution models are 0.49 ps, 1.14 ps and 2.97 ps, each of them accounting respectively for 56.5%, 36.2%, and 7.3% of the full model. Similarly, the $\Delta t$ resolution for channel with $\eta \to \pi^+\pi^-\pi^0$ improves from 1.25 ps to 0.88 ps with *iptube* and $K_S^0$ flight direction information (in this case the resolutions of the three gaussians are 0.45 ps, 1.07 ps and 2.88 ps respectively with weights 56.5%, 34.2%, and 9.3%). For both channels the larger improvement comes from the *iptube* constraint, while the one due to the $K_S^0$ direction is marginal.

Beam background causes a degradation of the $\Delta t$ resolution for signal events, plus tails due to the larger fraction of cross feed candidates. The values of the $\Delta t$ resolution for the true, cross feed and all candidates for channels $\eta(\gamma\gamma)K_S^0(\pi^{\pm})$ and $\eta(\to \pi^+\pi^-\pi^0)K_S^0(\pi^{\pm})$ are reported in Table 83.

*Estimate of sensitivity from ensemble tests studies.* To estimate the statistical uncertainties expected from this analysis, we performed a set of ensemble test studies. As input parameters we used $S_{\eta' K_S^0} = 0.7$ and $A_{\eta' K_S^0} = 0$. Using MC we performed the multivariate maximum likelihood fit and extracted pdf for the various distributions. For the time independent part we used $M_{\rm bc}$, $\Delta E$, the cross-feed discriminating BDT variable, and the continuum suppression BDT variables, both described above. The generated ensemble MC datasets, including signal, self cross-feed, continuum background, and peaking background, correspond to integrated luminosities of 1 and 5 ab$^{-1}$. The signal, as well as the signal cross feed, were obtained by extracting a random sub-sample from the full signal MC dataset. Both the backgrounds samples, instead, were randomly generated from the pdf fitted from the full sample.

The expected statistical uncertainties for extracted $S_{\eta' K_S^0}$ and $A_{\eta' K_S^0}$ are summarised in Table 84. A comparison with BaBar and Belle published results shows that these preliminary results are comparable for similar integrated luminosities.

No significant bias is observed for $S_{\eta' K_S^0}$, while a non-negligible bias arises in the estimation of $A_{\eta' K_S^0}$, which is likely due to a correlation between the $BDT_{SXF}$ and the Continuum Suppression BDT variables. Such effect can be mitigated by further optimizing the sets of input variables of each of the two classifiers, in order to reduce the correlation; another solution constists in taking into account such a correlation by implementing it in the likelihood fit model.

In the Table 84, also the channel $\eta'(\eta_{\gamma\gamma}\pi^{\pm})K_S^{(00)}$ is reported. The efficiency found in this sensitivity study is a factor of two lower, suffering from the poor $\pi^0$ reconstruction currently available. Improvements are expected before data taking, so we used instead an efficiency taken from similar study performed in Belle. The sensitivity for the decay channel with $\eta'(\rho\gamma)K_S^{(00)}$ has been estimated using the expected yield, based on Belle efficiency, and with the resolution found in the analysis for the $\eta'(\eta_{\gamma\gamma}\pi^{\pm})K_S^{(\pm)}$ channel. The $K_L^0$ modes were not analysed yet, so the values in Table 84 are obtained by extrapolation of the Belle measurement to the Belle II expected luminosity.

Table 84: The estimated resolutions (statistical uncertainties only) from ensemble test studies for $CP$-violating $S_f$ and $A_f$ parameters for a integrated luminosities of 1 and 5 ab$^{-1}$ for different channels.

| Channel | yield | $\sigma(S_f)$ | $\sigma(A_f)$ |
|---|---|---|---|
| | 1 ab$^{-1}$ | | |
| $\eta'(\eta_{\gamma\gamma}\pi^{\pm})K_S^{(\pm)}$ | 969 | 0.13 | 0.08 |
| $\eta'(\eta_{\gamma\gamma}\pi^{\pm})K_S^{(00)}$ | 215 | 0.27 | 0.17 |
| $\eta'(\eta_{3\pi}\pi^{\pm})K_S^{(\pm)}$ | 283 | 0.25 | 0.16 |
| $\eta'(\rho\gamma)K_S^{(\pm)}$ | 2100 | 0.09 | 0.05 |
| $\eta'(\rho\gamma)K_S^{(00)}$ | 320 | 0.22 | 0.14 |
| $K_S^0$ modes | 3891 | 0.065 | 0.040 |
| $K_L^0$ modes | 1546 | 0.17 | 0.11 |
| $K_S^0 + K_L^0$ modes | 5437 | 0.060 | 0.038 |
| | 5 $ab^{-1}$ | | |
| $\eta'(\eta_{\gamma\gamma}\pi^{\pm})K_S^{(\pm)}$ | 4840 | 0.06 | 0.04 |
| $\eta'(\eta_{\gamma\gamma}\pi^{\pm})K_S^{(00)}$ | 1070 | 0.12 | 0.09 |
| $\eta'(\eta_{3\pi}\pi^{\pm})K_S^{(\pm)}$ | 1415 | 0.11 | 0.08 |
| $\eta'(\rho\gamma)K_S^{(\pm)}$ | 10500 | 0.04 | 0.03 |
| $\eta'(\rho\gamma)K_S^{(00)}$ | 1600 | 0.10 | 0.07 |
| $K_S^0$ modes | 19500 | 0.028 | 0.021 |
| $K_L^0$ modes | 7730 | 0.08 | 0.05 |
| $K_S^0 + K_L^0$ modes | 27200 | 0.027 | 0.020 |

*Systematic Uncertainties.* A precise determination of the systematics uncertainties is not available at the time of this publication, so we estimate them following the guidelines described in Sec. 10.2.1. The current measurement by Belle [703] reports contributions of several sources of systematics uncertainty. Some are irreducible, such as vertex reconstruction ($\pm 0.014$) and tag-side interference ($\pm 0.001$), and some are reducible, like $\Delta t$ resolution, signal fraction, background $\Delta t$ pdf, flavour tagging, fit bias (accounting, summed in quadrature, to $\pm 0.038$). We can assume that the reducible systematics will scale with the luminosity, since they are evaluated via control samples and Monte Carlo simulated events. As in the $J/\psi K^0$ channel, the vertex related systematics are expected to be reduced by a factor two thanks to the new Pixel Vertex detector and improved tracking and alignment algorithms (see Sec. 10.2.2). As a conservative scenario we also consider the case when the vertex related systematics does not change.

The expected systematic uncertainties are summarised in Table 85, for two luminosities (5 and 50 $ab^{-1}$) and two scenarios: conservative, without scaling down the vertex related systematic, and optimistic, including that rescaling.

The measurement of $S$ with this channel will be affected by a systematic uncertainty similar to the statistical one at an integrated luminosity of $L = 10$ (20) $ab^{-1}$ in the conservative (optimistic) scenario.

Table 85: Estimated systematic uncertainties on $S_f$ for $B^0 \to \eta' K^0$ decay for two different luminosity and with two different hypothesis about the vertex related uncertainties: conservative and optimistic (in parenthesis)

| L ($ab^{-1}$) | stat. ($10^{-2}$) | syst. ($10^{-2}$) | total ($10^{-2}$) |
|---|---|---|---|
| 5 | 2.7 | 2.1 (1.7) | 3.4 (3.2) |
| 50 | 0.85 | 1.8 (1.3) | 2.0 (1.5) |

*Extrapolation of the $\omega K^0_S$ sensitivity.* Given the similarity of the decay channels of $\eta' \to \eta\pi^+\pi^-$ and $\omega \to \pi^0\pi^+\pi^-$, we extrapolate the Belle II sensitivity to the time-dependent $CP$ violation parameters on $B^0 \to \omega K^0_S$. We assume the $\Delta t$ resolution to be the same for these two modes. We thus rescale the uncertainties on $S_f$ and $A_f$ from the $\eta(2\gamma)K^0_S(\pi^\pm)$ channel by the expected $\omega K^0_S$ yields. For these we use a reconstruction efficiency of 21%, which is derived from the efficiency quoted in the latest BaBar paper [702], rescaled by the ratio of Belle II and BaBar efficiencies for the $\eta' K^0$ channel. The results are collected in Table 86.

### 10.4. Determination of $\phi_2$

#### 10.4.1. Theory: $\phi_2$ from $B \to \pi\pi$, $B \to \rho\rho$, and $B \to \rho\pi$. Contributing authors: Y. Grossman, M. Gronau

The theoretically most precise way of determining the phase $\phi_2$ is based on applying isospin symmetry to $B \to \pi\pi, \rho\rho$ decays [704]. The decays $B^+ \to \pi^+\pi^0$, $B^0 \to \pi^+\pi^-$, $B^0 \to \pi^0\pi^0$ including their charge-conjugates, and corresponding $B$ decays involving longitudinally polarised $\rho$ mesons, $\rho_L$, provide sufficient information to determine $\phi_2$. A complete study of these processes resolves discrete ambiguities in $\phi_2$. The discrete ambiguities are also resolved by a more complex study of $B \to \rho\pi$ that involves non-identical final state particles [705]. We describe the current status of the isospin method for determining $\phi_2$ in $B \to \pi\pi$ and $B \to \rho_L\rho_L$, pointing out its sensitivity to specific measurements and the potential of improving the precision at Belle II. This discussion follows largely a very recent study [706] where further details can be found.

*Determination of $\phi_2$ from $B \to \pi\pi$ and $B \to \rho\rho$.* Using the unitarity of the CKM matrix one can write in full generality for the decay amplitude (an equal formalism applies to $B \to \rho_L\rho_L$),

$$\mathcal{A}_{+-} \equiv \mathcal{A}(B^0 \to \pi^+\pi^-) = |T_{+-}|e^{i\phi_3} + |P_{+-}|e^{i\delta} , \tag{318}$$

Table 86: Extrapolated sensitivity for the $\omega K^0_S$ mode. The $\Delta t$ resolution is taken from the $\eta' K^0_S$ study, while we assume a reconstruction efficiency of 21%.

| $\omega(\pi^+\pi^-\pi^0)K^0_S(\pi^\pm)$ | | | |
|---|---|---|---|
| L (ab$^{-1}$) | yield | $\sigma(S)$ | $\sigma(A)$ |
| 1 | 334 | 0.17 | 0.14 |
| 5 | 1670 | 0.08 | 0.06 |
| 50 | 16700 | 0.024 | 0.020 |

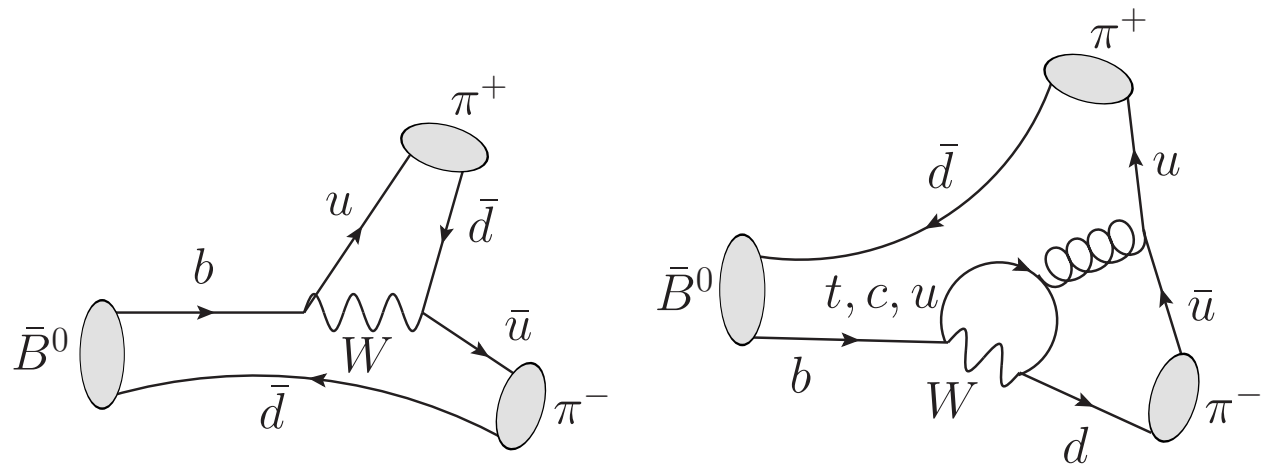

Fig. 106: Sample diagrams for the $T$ (left) and $P$ (right) amplitudes in (318) for the $\bar{B}^0 \to \pi^+\pi^-$ decay.

where $|T|$ is the magnitude of the tree amplitude with weak phase $\phi_3$, while $P$ is the magnitude of the penguin amplitude with strong phase $\delta$. The tree amplitude for $b \to u\bar{u}d$ transitions carries isospin 1/2 and 3/2, while the $b \to d$ penguin amplitude carries only isospin 1/2. The spinless two-pion state can only have isospin 0 and 2, so that the $B \to \pi\pi$ amplitudes, denoted by the pion charges, obey the relation

$$\mathcal{A}_{+-}/\sqrt{2} + \mathcal{A}_{00} = \mathcal{A}_{+0} \ . \tag{319}$$

The $\Delta I = 3/2$ amplitude $\mathcal{A}_{+0}$ has no penguin contribution and thus has the weak phase $\phi_3$, while $\bar{\mathcal{A}}_{-0}$ has weak phase $-\phi_3$. Defining $\tilde{\mathcal{A}} \equiv e^{2i\phi_3}\bar{\mathcal{A}}$, we have

$$\tilde{\mathcal{A}}_{+-}/\sqrt{2} + \tilde{\mathcal{A}}_{00} = \tilde{\mathcal{A}}_{-0} \ , \tag{320}$$

where the two triangles (319) and (320) have a common base, $\mathcal{A}_{+0} = \tilde{\mathcal{A}}_{-0}$. The sides of these two triangles are determined by decay rates and direct asymmetries. This fixes two angles $\theta_{+-} = \arg(\mathcal{A}_{+-}/\mathcal{A}_{+0}), \theta_{00} \equiv \arg(\mathcal{A}_{00}/\mathcal{A}_{+0})$ for $B$ decays and $\tilde{\theta}_{+-} = \arg(\tilde{\mathcal{A}}_{+-}/\tilde{\mathcal{A}}_{-0}), \tilde{\theta}_{00} \equiv \arg(\tilde{\mathcal{A}}_{00}/\tilde{\mathcal{A}}_{-0})$ for $\bar{B}$ decays. The two differences between these pairs of angles, $\Delta\theta_{+-} = \tilde{\theta}_{+-} - \theta_{+-}, \Delta\theta_{00} \equiv \tilde{\theta}_{00} - \theta_{00}$, then determine $\phi_2$ via the relations

$$\begin{aligned} \sin(2\phi_2 + \Delta\theta_{+-}) &= \frac{S_{+-}}{\sqrt{1-(A_{+-})^2}} \ , \\ \sin(2\phi_2 + \Delta\theta_{00}) &= \frac{S_{00}}{\sqrt{1-(A_{00})^2}} \ . \end{aligned} \tag{321}$$

A discrete ambiguity in $\Delta\theta_{+-}$ and $\Delta\theta_{00}$ may remain due to the possibility of flipping either triangle about its base.

The above determination of $\phi_2$ receives corrections from isospin breaking, either due to electroweak penguins, or due to mass difference of up and down quarks and their electric charges. We discuss the induced errors below.

*Determination of $\phi_2$ from $B \to \rho\pi$.* There is a major difference between $\phi_2$ determination from $B \to \rho\pi$ compared to that obtained from $B \to \pi\pi$, and $B \to \rho\rho$. From the overlaps of the resonances in the Dalitz plot of the time-dependent decay $B \to \rho\pi \to \pi^+\pi^-\pi^0$ one can now determine both the magnitudes and relative phases of the decay amplitudes $\mathcal{A}_{ij} \equiv \mathcal{A}(B^0 \to \rho^i\pi^j)$, $i, j = +, -, 0$, as well as their $CP$ conjugate counterparts. For instance, the time-dependent decay rate for initial $B^0$ is given by [705, 707, 708]

$$\begin{aligned}\Gamma(B^0 \to \pi^+\pi^0\pi^0(t)) \propto (|\mathcal{A}_{+-0}|^2 + |\bar{\mathcal{A}}_{-+0}|^2)\\ + (|\mathcal{A}_{+-0}|2 - |\bar{\mathcal{A}}_{-+0}|^2)\cos(\Delta mt) - 2\Im(e^{-2i\phi_1}\bar{\mathcal{A}}_{-+0}\mathcal{A}^*_{+-0})\sin(\Delta mt),\end{aligned} \tag{322}$$

where we shortened $\mathcal{A}_{+-0} \equiv \mathcal{A}(B^0 \to \pi^+\pi^-\pi^0)$, $\bar{\mathcal{A}}_{+-0} \equiv A(\bar{B}^0 \to \pi^+\pi^-\pi^0)$. Each of these decay amplitudes to the three body final state is a sum of the quasi-two body decay amplitudes, $\mathcal{A}_{ij}$, that overlap in the corners of the Dalitz plot.

For $B \to \rho\pi$ decays it is convenient to split the amplitudes into the tree and penguin amplitudes according to the so called "$t-$convention", see, *e.g.*, [708],

$$\begin{aligned}e^{i\phi_1}\mathcal{A}_a = e^{-i\phi_2}\mathcal{T}_a + \mathcal{P}_a,\\ e^{-i\phi_1}\bar{\mathcal{A}}_a = e^{i\phi_2}\mathcal{T}_a + \mathcal{P}_a,\end{aligned} \tag{323}$$

where $a = +-, -+, 00$. The relative phases between the two amplitudes on the left hand side are directly measured from the coefficient of $\sin(\Delta mt)$ in eq. (322). Note that $\mathcal{T}_{\pm,0}$ and $\mathcal{P}_{\pm,0}$ also contain the strong phases. There are thus eleven observables, 6 magnitudes and 5 relative phases between decay amplitudes. These are described by twelve unknown parameters: the weak phase $\phi_2$, 6 magnitudes of tree and penguin amplitudes, and 5 relative strong phases between these amplitudes. A complex isospin relation [705, 707]

$$\mathcal{P}_{-+} + \mathcal{P}_{+-} + 2\mathcal{P}_{00} = 0, \tag{324}$$

reduces the number of unknowns by two, leading to an over-constrained system. Unlike in $B \to \pi\pi$ and $B \to \rho\rho$ here the isospin was used only to relate the penguin amplitudes. Since these are smaller than the tree amplitudes the error induced by isospin breaking is expected to be smaller as a result [709].

*Current status of $B \to \pi\pi$.* Current $B \to \pi\pi$ measurements are summarised in Table 2 of [706] and include the six variables, $\mathcal{B}^{+0}_{\rm av}, \mathcal{B}^{+-}_{\rm av}$, $\mathcal{B}^{00}_{\rm av}, A_{+-}, A_{00}, S_{+-}$, where the subscript "av" denotes an average for the process and its $CP$ conjugate. Values of $\mathcal{B}^{00}_{\rm av}$ and $A_{00}$ in the table are based on preliminary Belle measurements [710] averaged with those measured by BaBar [711]. Final Belle results for $\mathcal{B}^{00}_{\rm av}$ and $A_{00}$ have recently been published in [712]. Solutions for $\phi_2$ were obtained in [706] using a Monte Carlo program generating the above six observables assuming they obey Gaussian distributions and calculating corresponding values for $\chi^2$. A minimum value $\chi^2_{\rm min} = 0.338$ occurs at four values of $\phi_2$, $\phi_2 = (95, 128.9, 141.1, 175)^\circ$. $\Delta\chi^2 \equiv \chi^2 - \chi^2_{\rm min} \leq 1$ is satisfied for $\phi_2$ in the range ([87,104],[120,150],[166,183])°, in agreement with ranges found by the CKMfitter Collaboration [657]. The error on $\phi_2$ scales roughly as the error on *all* six variables, while reducing the error on any individual variable does not significantly affect the error on $\phi_2$.

Future measurements of $S_{00}$ at Belle II using external photon conversion [713] will distinguish solutions near $\phi_2 = 129^\circ$ and $141^\circ$, yielding $S_{00} \simeq -0.70$ for the $\chi^2_{\rm min}$ solution, from those near $95^\circ$ and $175^\circ$, yielding $S_{00} \simeq 0.67$.

Table 87: Inputs to the determination of $\phi_2$ from an isospin analysis of $B \to \rho\rho$. Branching fractions are multiplied by longitudinal $\rho$ polarisation fractions [88, 230].

| Quantity | Value ($\times 10^{-6}$) | Quantity | Value |
|---|---|---|---|
| $f_L \mathcal{B}_{\rm av}(B^+ \to \rho^+\rho^0)$ | $21.18 \pm 1.71^a$ | $A_{+-}$ | $0.00 \pm 0.09$ |
| $f_L \mathcal{B}_{\rm av}(B^0 \to \rho^+\rho^-)$ | $27.42 \pm 1.95$ | $A_{00}$ | $0.20 \pm 0.85$ |
| $f_L \mathcal{B}_{\rm av}(B^0 \to \rho^0\rho^0)$ | $0.67 \pm 0.12^b$ | $S_{+-}$ | $-0.14 \pm 0.13$ |
| | | $S_{00}$ | $0.3 \pm 0.73$ |

[a]Branching ratio corrected by factor [88] $\tau(B^0)/\tau(B^+) = 0.929$.
[b]Averaged values of branching ratio and longitudinal fraction using also [714].

*Current status of $B \to \rho_L\rho_L$.* In order to perform a similar analysis for $B \to \rho_L\rho_L$ one uses branching ratios multiplied by fractions $f_L$ for decays leading to longitudinal $\rho$ polarisation. In addition to these three branching fractions and corresponding $CP$ asymmetries, $A_{+-}, A_{00}, S_{+-}$, the BaBar collaboration has measured also $S_{00}$ [715]. The seven averaged measured observables are listed in Table 87. The value of $\phi_2$ corresponding to $\Delta\chi^2 \leq 1$ is now $(92.0^{+4.7}_{-5.0})^\circ$. Since the $B$ and $\bar{B}$ isospin triangles are exactly flat for the solution of $\chi^2_{\rm min} = 0.202$, there exists merely a single second solution near $\phi_2 = 180^\circ$ which is ruled out by the measurement of $\phi_1$, assuming the unitarity of the CKM matrix. The precision in $\phi_2$ may be improved by reducing errors on $f_L\mathcal{B}_{\rm av}(B^0 \to \rho^+\rho^-)$ and $f_L\mathcal{B}_{\rm av}(B^+ \to \rho^+\rho^0)$, for which the Belle analysis [716] was based on only about ten percent of its $\Upsilon(4S)$ sample, and in particular by improving the current rather crude measurement of $S_{00}$. A data sample of $10^{10}$ $B\bar{B}$ pairs can reduce all current errors at least by a factor of two, for which one finds $\Delta\chi^2 \leq 1$ for $\phi_2 = (92.0 \pm 2.5)^\circ$.

*Current status of $B \to \rho\pi$.* Both BaBar [717] and Belle [718] performed time-dependent Dalitz plots analyses of the $B \to \rho\pi \to \pi^+\pi^-\pi^0$ decays. The result of these analyses are the 27 coefficients multiplying the bilinears of quasi-two body $\rho\pi$ decay form factors. These coefficients are directly related to the magnitudes and relative phases of the quasi-two body $B \to \rho\pi$ decay amplitudes. In addition to the isospin relation (324), the experiments also include the information from $B^+ \to \rho^+\pi^0, \rho^0\pi^+$ decays using an isospin pentagon relation [719] between the decay amplitudes.

From this Belle [718] obtained the constraint $68^\circ < \phi_2 < 95^\circ$ at 68.3% C.L. for the $\phi_2$ solution consistent with the SM. BaBar [717] did not attach a C.L. interval to the scan of $\phi_2$ that they presented, since their study indicated that the scan itself was not yet statistically robust.

*Electroweak penguin corrections.* Higher order electroweak penguin (EWP) operators contribute to $B \to \pi\pi$ and $B \to \rho\rho$. Neglecting EWP operators multiplied by tiny Wilson coefficients $(C_7, C_8)$, isospin symmetry relates dominant EWP operators to the $\Delta I = 3/2$ current-current operator in the effective Hamiltonian [720]. Hadronic matrix elements of the latter operator form the bases of the isospin triangles for $B \to \pi\pi$ and $B \to \rho_L\rho_L$.

Consequently the bases for $B$ and $\bar{B}$ form a small calculable relative angle [695, 721],

$$\mathrm{Arg}(\tilde{\mathcal{A}}_{-0}\mathcal{A}^*_{+0}) = -3\left(\frac{C_9(m_b)+C_{10}(m_b)}{C_1(m_b)+C_2(m_b)}\right)\frac{|V_{tb}V_{td}|}{|V_{ub}V_{ud}|}\sin\phi_2 \quad (325)$$

$$= 3.42\,\alpha(m_W)\,\frac{\sin(\phi_1+\phi_2)\sin\phi_2}{\sin\phi_1}\ . \quad (326)$$

Using $\alpha(m_W) = 1/129$, $\phi_1 = 22.6°$ [91] and $\phi_2 \simeq 90°$, obtained from $B \to \rho_L\rho_L$, one calculates a small negative shift in $\phi_2$ due to EWP corrections, $\Delta\phi_2(\mathrm{EWP}) = -\frac{1}{2}\mathrm{Arg}(\tilde{A}_{-0}A^*_{+0}) = -1.8°$. A shift of a similar size applies to $\phi_2$ determined from $B \to \rho\pi$ [709].

*Other isospin breaking effects.* Not all isospin breaking effects on $\phi_2$ can be included at present. We can judge the expected size of the bias in $\phi_2$ through the isospin breaking effects that we can estimate. One example is $\pi^0$-$\eta$-$\eta'$ mixing which introduces isospin breaking in $B \to \pi\pi$ through an additional $I = 1$ amplitude, while isospin-conserving terms obey the triangle relation (319) [709, 722]. We follow the discussion in [709], updating bounds on $B$ decays involving $\eta$ and $\eta'$. Mixing of $\pi^0, \eta$ and $\eta'$ adds a small isospin singlet component to the dominantly isotriplet neutral pion state, $|\pi^0\rangle = |\pi_3\rangle + \epsilon|\eta\rangle + \epsilon'|\eta'\rangle$, where $\epsilon = 0.017 \pm 0.003, \epsilon' = 0.004 \pm 0.001$ [723]. Applying flavour SU(3) to $B$ decays into pairs of non-strange pseudoscalar mesons [724] (thereby keeping an uncertainty at the level of 30% in isospin breaking terms), one reaches two conclusions:

(1) The isospin relation (319) becomes $\mathcal{A}_{+-}/\sqrt{2} + \mathcal{A}_{00} = \mathcal{A}_{+0}(1-\epsilon_0)$, where $\epsilon_0 = \sqrt{2/3}\,\epsilon + \sqrt{1/3}\,\epsilon' = 0.016 \pm 0.003$. This affects very slightly the current range in $\phi_2$, becoming $(94.5^{+9.2}_{-8.5})°$ instead of $(95^{+9}_{-8})°$.

(2) The amplitude $\mathcal{A}_{+0}$ can be written in terms of the pure $\Delta I = 3/2$ amplitude $\mathcal{A}_{+3}$ carrying a weak phase $\phi_3$, corrected by isospin breaking terms involving $\mathcal{A}_{0\eta}$ and $\mathcal{A}_{0\eta'}$, $\mathcal{A}_{+0} = (1+\epsilon_0)\mathcal{A}_{+3} + \sqrt{2}\,\epsilon\mathcal{A}_{0\eta} + \sqrt{2}\,\epsilon'\mathcal{A}_{0\eta'}$. This can be shown to imply an upper bound on the correction to $\phi_2$ [709]: $|\Delta\phi_2| \le \sqrt{2\tau(B^+)/\tau(B^0)}(\epsilon\sqrt{\mathcal{B}_{0\eta}/\mathcal{B}_{+0}} + \epsilon'\sqrt{\mathcal{B}_{0\eta'}/\mathcal{B}_{+0}})$. Using the above values of $\epsilon, \epsilon'$ and updated branching ratios [88] we find $|\Delta\phi_2| < 1.2°$.

The shift in $\phi_2$ from this effect is expected to be much smaller for $B \to \rho\pi$. The reason is that unlike for $B \to \pi\pi$ and $B \to \rho\rho$, here the isospin relations are used only to relate the penguin contributions. Thus the $\phi_2$ determination from $B \to \rho\pi$ are not affected by the isospin-breaking in tree amplitudes, but only in penguin amplitudes. Since penguin amplitudes are subleading to tree amplitudes, the isospin breaking effects on $\phi_2$ extraction are expected to be suppressed [709].

Another two effects that we can analyse are due to the finite $\rho$ width and due to the $\rho$-$\omega$ mixing on $\phi_2$ extraction from $B \to \rho_L\rho_L$. Amplitudes for $B \to \rho\rho$ depend on dipion invariant masses, $m^2_{12} \equiv (p_{\pi_1}+p_{\pi_2})^2, m^2_{34} \equiv (p_{\pi_3}+p_{\pi_4})^2$, for the two pion pairs forming $\rho$ mesons. The isospin method assumes equal $\rho$ masses, $m_{12} = m_{34}$, leading to the absence of an $I = 1$ final state for the two identical bosons. Choosing the two invariant masses to lie in a common mass range around the $\rho$ peak, *e.g.*, 400 MeV/$c^2 < m_{12}, m_{34} <$ 1150 MeV/$c^2$ [725], introduces an $I = 1$ amplitude for unequal masses [726]. The $I = 1$ amplitude, which is antisymmetric in $m_{12} \leftrightarrow m_{34}$, does not interfere in the decay rate with the symmetric $I = 0, 2$ amplitudes. Its contribution is expected to be of order $(\Gamma_\rho/m_\rho)^2$. Furthermore, isospin breaking leads to $\rho^0$-$\omega$ mixing of order a few percent. Its effect is to form a prominent peak at the $\omega$ mass followed by a sharp dip [709]. These effects have to be studied and resolved experimentally by a judicious choice of ranges for the two di-pion masses.

There are isospin effects that have not been captured in the above estimates. For instance, the reduced matrix elements of operators in the effective Hamiltonian between initial $B^0$ and $B^+$ states and final states involving $\pi^3, \pi^+$ were assumed to obey exact SU(2) relations. Furthermore, the $\Delta I = 5/2$ corrections were assumed to vanish. Further information could be gained from QCDF/SCET calculations, if the relevant isospin breaking in light-cone distribution amplitudes is available from the lattice [709].

*Formally going beyond leading order.* So far we discussed isospin breaking and ways to overcome and estimate it. Here we remark that in principle there are observables where the theoretical error is only of second order in isospin breaking and thus below the per-mill level. It may be impossible to measure them to the required level of accuracy, and thus these ideas are challenging experimentally. Nevertheless they may point to a way forward in tests for presence of NP.

The main point is the following. The $B \to n\pi$ amplitudes have a sum-rule that holds to $n-2$ level in isospin breaking. The proof of this statement is based on the following facts:

(1) There are $n+1$ different decays modes in $B \to n\pi$. In term of isospin these $n+1$ amplitudes have $\Delta I$ equal to

$$\frac{1}{2},\ \frac{3}{2},\ \ldots,\ \frac{2n+1}{2}. \tag{327}$$

(2) To leading order the weak Hamiltonian has only $\Delta I = 1/2, 3/2$.

(3) Isospin breaking effects are always $\Delta I = 1$.

Thus, in order to generate a non-zero $(2n+1)/2$ amplitudes, $n-1$ spurion insertions are needed. That is, the isospin sum-rules are completely broken at the $n-1$ level.

This is a known result in the $B \to \pi\pi$ where the sum-rule is broken at first order in isospin breaking. The nontrivial result is that with three pions we have a sum rule that is broken only at the second order in isospin breaking.

Having a sum rule that is broken only by 2nd order effects does not guarantee that we can get $\phi_2$ to the same accuracy. It is thus quite nontrivial that one can indeed determine $\phi_2$ from $B \to \pi\pi\pi$ even when including the first order isospin breaking. The problem is that the method requires measuring the time-dependent $CP$ asymmetry in the $B \to \pi^0\pi^0\pi^0$ to an accuracy of less than 1%. Such a measurement is very unlikely to be done at Belle II.

*10.4.2. Experiment: sensitivity study of the Branching fraction and $CP$ violation parameters in $B^0 \to \pi^0\pi^0$ and expected $\phi_2$ sensitivity from $B \to \pi\pi$, $B \to \rho\rho$, and $B \to \rho\pi$. Contributing authors: F. Abudinén, L. Li Gioi*

We estimate the Belle II sensitivity on $\phi_2$-angle with the isospin analysis of the decay modes $B \to \pi\pi$ and $B \to \rho\rho$. The input parameters of the isospin analysis are the branching fractions $\mathcal{B}_{+-}$, $\mathcal{B}_{00}$ and $\mathcal{B}_{+0}$, as well as the $CP$ violation parameters $A_{+-}$, $S_{+-}$, $A_{00}$ and $S_{00}$. In the case of $B \to \rho\rho$ the isospin analysis applies only for decays with longitudinally polarised $\rho$ mesons, therefore, the branching fractions are multiplied by the fraction $f_L$ of decays leading to longitudinal $\rho$ polarisation.

We present sensitivity studies of the $CP$ violation parameters and the branching fraction of $B^0 \to \pi^0\pi^0$. We estimate the sensitivity to $S_{\pi^0\pi^0}$ performing a feasibility study of the time-dependent $CP$-analysis using converted photons and $\pi^0$ Dalitz decays. The sensitivity to $A_{\pi^0\pi^0}$ and $\mathcal{B}_{\pi^0\pi^0}$ is estimated from a time-integrated $CP$-analysis considering only photons which are reconstructed as clusters in the calorimeter. For the other $B \to \pi\pi$ and for all the $B \to \rho\rho$ input parameters, we estimated the sensitivity through extrapolation of the Belle results.

*$B^0 \to \pi^0\pi^0$.* At present, there is not enough data to perform a time-dependent $CP$-analysis of the decay mode $B \to \pi^0\pi^0$. Neutral pions decay to about $(98.823 \pm 0.034)\%$ [88] into two photons, and, without external photon conversion $\gamma \to e^+ e^-$, they do not provide information to reconstruct the vertex of the $B^0$. Also the fraction of useful Dalitz decays $\pi^0 \to e^+e^-\gamma$ is very small $(1.174 \pm 0.035)\%$ [88]. Consequently, the isospin analysis of the $B \to \pi\pi$ decay mode has been performed without $S_{\pi^0\pi^0}$ leading to an eightfold ambiguity in the solution of $\phi_2$ in the range $[0, \pi]$. Only a measurement of $S_{\pi^0\pi^0}$ could reduce the number of possible solutions.

For this study, we generate and reconstruct signal MC events where $B_{\rm sig} \to \pi^0\pi^0$ and $B_{\rm tag} \to$ generic. Assuming that $\mathcal{B}_{\pi^0\pi^0} = 1.91 \cdot 10^{-6}$ [88], the total number of expected events in 50 ab$^{-1}$ is about 103,000. For the $CP$-analyses, the following three decay modes are considered as signal:

1. $B^0_{\rm sig} \to \pi^0_{\gamma\gamma}(\to \gamma\gamma)\ \pi^0_{\gamma\gamma}(\to \gamma\gamma)$,
2. $B^0_{\rm sig} \to \pi^0_{\rm dal}(\to e^+e^-\gamma)\ \pi^0_{\gamma\gamma}(\to \gamma\gamma)$,
3. $B^0_{\rm sig} \to \pi^0_{\gamma_c\gamma}(\to \gamma_c(\to e^+e^-)\gamma)\ \pi^0_{\gamma\gamma}(\to \gamma\gamma)$.

We found out that only about 3% of the generated events contain at least one photon that undergoes conversion within the Pixel Vertex Detector (PXD) volume. Of these, two thirds of the conversions take place in the beam pipe, and the remaining convert in the PXD material. Figure 107 shows the MC vertices of converted photons inside the PXD in the $xy$-plane. Additionally, about 6% of the generated events contain at least one converted photon in the Strip Vertex Detector (SVD) volume outside of the PXD.

*Signal reconstruction.* For each event, signal candidates are found by reconstructing the whole decay chain for the three considered signal decay modes. Photons and $\pi^0_{\gamma\gamma}$ are reconstructed in a similar way as in Belle [712]. Photons correspond to neutral clusters in the electromagnetic calorimeter (ECL); the energy of the clusters is required to be greater than 50 MeV in the barrel region, 100 MeV in the front-end cap and 150 MeV in the back-end cap of the ECL. The reconstruction of $\pi^0_{\gamma\gamma}$ is performed using pairs of photons with invariant masses in the range 105 MeV/$c^2$ $< m_{\gamma\gamma} <$ 165 MeV/$c^2$ corresponding to about $\pm 2.5\sigma$ around the nominal $\pi^0$ mass, where $\sigma$ is the current mass resolution shown in Fig. 108 (left). To reduce combinatorial background, $\pi^0_{\gamma\gamma}$ candidates with small helicity angles $(|\cos(\theta_H)| > 0.95)$ are rejected. The helicity angle $\theta_H$ is defined as the angle between the $\pi^0$ boost direction in the laboratory frame and the momentum of one of the $\gamma$ daughters in the $\pi^0$ rest frame.

Electron-positron pairs are reconstructed using pairs of oppositely charged tracks, where both tracks have an electron PID $\mathcal{L}(e : \pi) > 0.8$ (see eq. (5) in Sec. 5.5) and an impact parameter $d_0 < 5$ cm. For each electron-positron pair, we require that one or both tracks have at least one PXD hit. Events with electron-positron pairs without PXD hits, including also events with SVD hits, are not suitable for a time-dependent $CP$ analysis since their time resolution is at least a factor 3 worse than for the events used in this analysis.

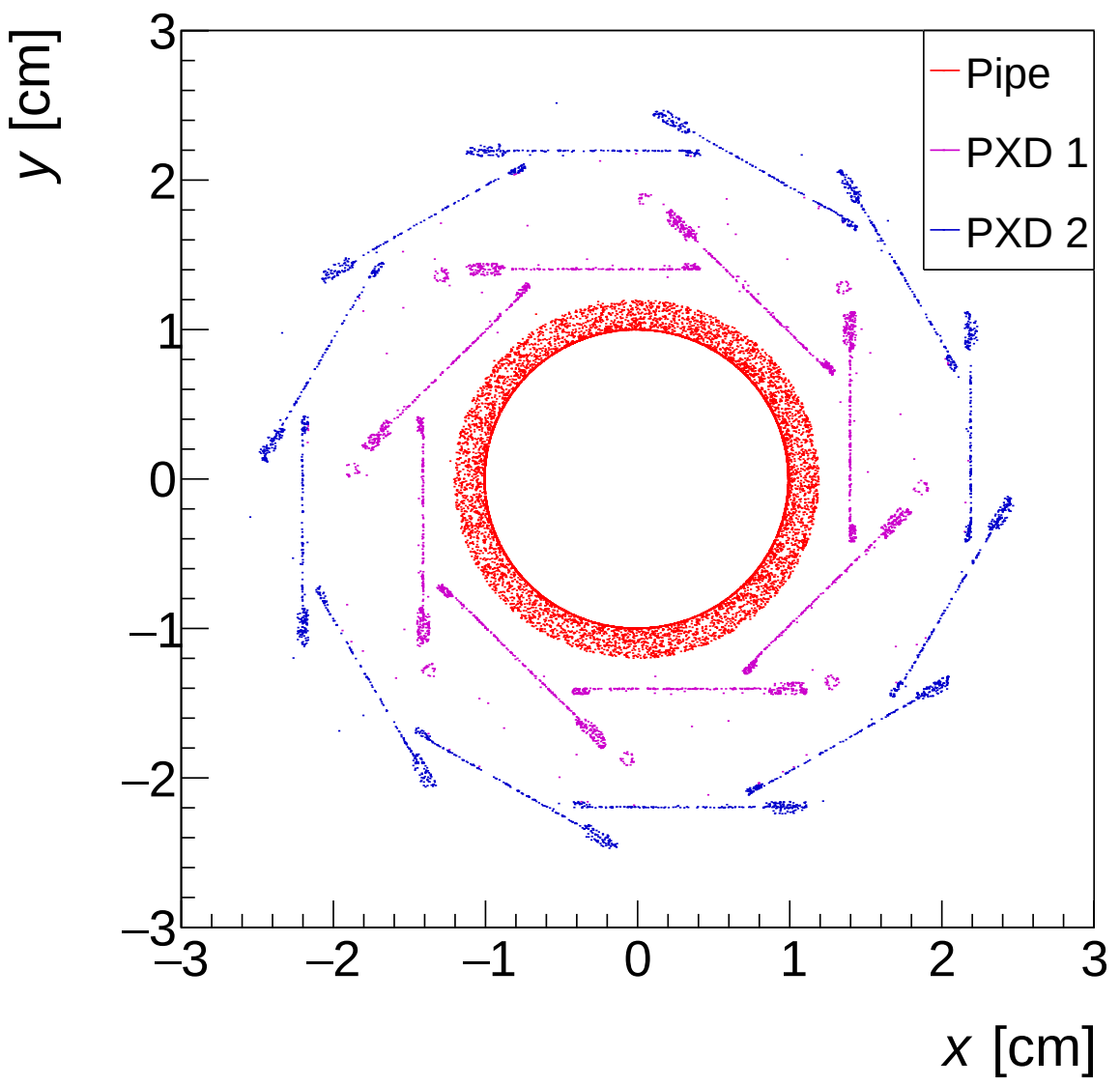

Fig. 107: Conversion vertices in the $xy$-plane inside the PXD. Conversions in the beam pipe as well as in the first and second layers of the PXD are shown in red, magenta and blue, respectively.

Converted photons $\gamma_c$ are reconstructed using $e^+e^-$ pairs with invariant masses $m_{e^+e^-} < 0.3$ GeV/$c^2$. The non-zero mass $m_{e^+e^-}$ results from the fact that the momentum of the tracks is determined at the point of closest approach to the interaction point (IP) and not at the conversion vertex. The $\gamma_c$ momentum is obtained by adding the momenta of the $e^+$ and the $e^-$ tracks.

The reconstruction of neutral pions with one converted photon $\pi^0_{\gamma_c\gamma}$ is performed using pairs of a reconstructed converted photon $\gamma_c$ and a photon corresponding to a neutral cluster. Dalitz pions $\pi^0_{\rm dal}$ are reconstructed using three particles for each candidate: an $e^+e^-$ pair and a photon corresponding to a neutral cluster.

All three kinds of pions are reconstructed within the same mass range chosen for $\pi^0_{\gamma\gamma}$ (105 MeV/$c^2$ $< m_{\pi^0} <$ 165 MeV/$c^2$). The reason is that the mass resolution for $\pi^0_{\gamma_c\gamma}$ ($\sigma \sim$ 12 MeV) and the mass resolution for $\pi^0_{\rm dal}$ ($\sigma \sim$ 10 MeV) differ only by about 1 MeV/$c^2$ from the mass resolution of $\pi^0_{\gamma\gamma}$ ($\sigma \sim$ 11 MeV). For $\pi^0_{\gamma_c\gamma}$ and $\pi^0_{\rm dal}$ candidates, the helicity angle $\theta_H$ is defined as the angle between the $\pi^0$ boost direction in the laboratory frame and the momentum of the neutral ECL cluster $\gamma$ in the $\pi^0$ rest frame. Candidates with helicity angles $|\cos(\theta_H)| > 0.95$ are rejected. Figure 108 (right) shows the $\cos(\theta_H)$ distributions for signal and background $\pi^0$ candidates. The asymmetry in the case of $\pi^0_{\gamma_c\gamma}$ and $\pi^0_{\rm dal}$ occurs because tracks are less affected by low momentum background in comparison with neutral clusters (the momentum threshold for tracks is higher than for neutral clusters).

At the end, $B^0_{\rm sig}$ candidates are reconstructed requiring that $M_{\rm bc} > 5.26$ GeV/$c^2$ and $-0.3$ GeV $< \Delta E <$ 0.2 GeV for the considered three signal decay modes. The same requirements for $M_{\rm bc}$ and $\Delta\ E$ were applied previously by Belle in the time-integrated analysis of $B^0 \to \pi^0\pi^0$ [712].

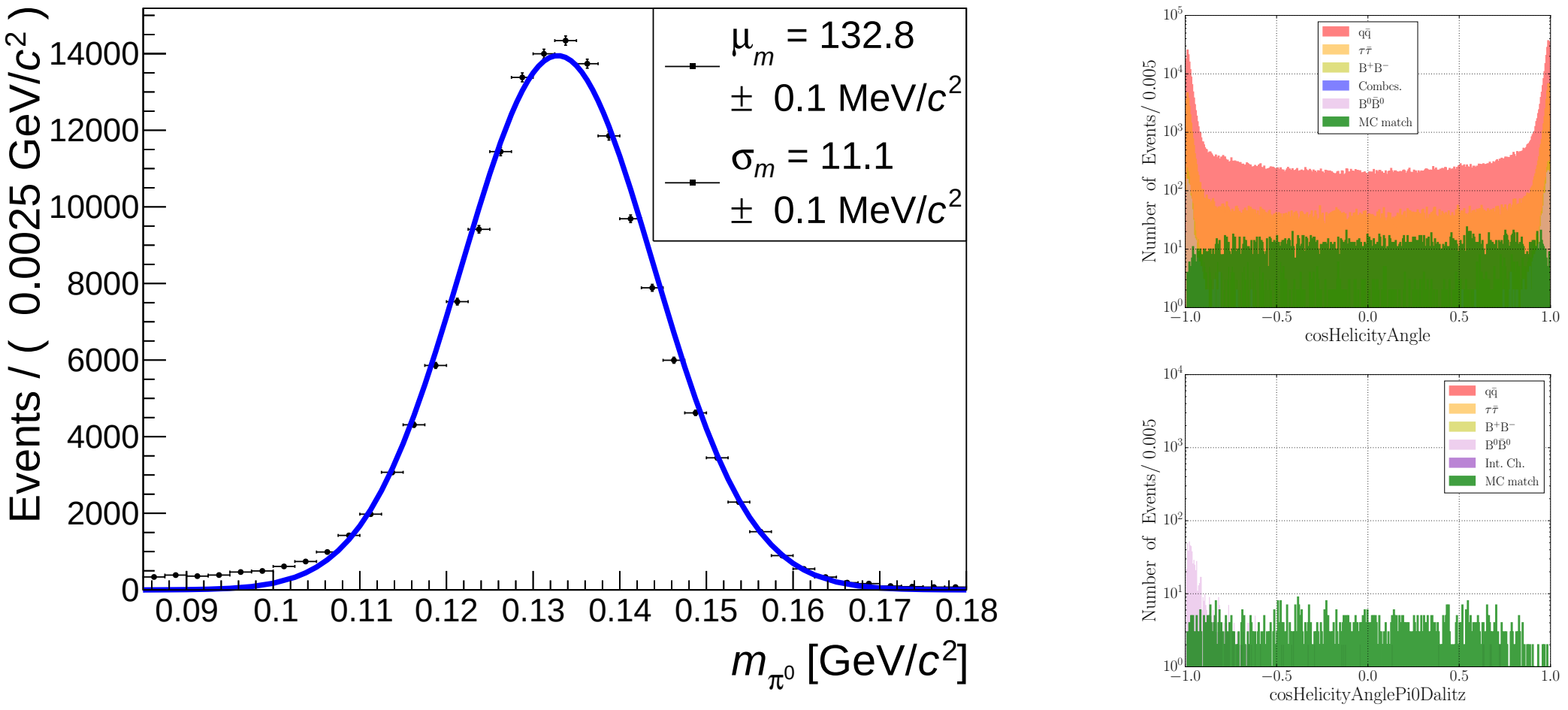

Fig. 108: Mass distribution of the reconstructed $\pi^0 \to \gamma\gamma$ (left). Distributions of $\cos\theta_H$ for the $\pi^0_{\gamma\gamma} \to \gamma\gamma$ case (right top) as well as for the $\pi^0_{\rm dal} \to e^+e^-\gamma$ and $\pi^0_{\gamma_{\rm c}\gamma}(\to \gamma_{\rm c}(\to e^+e^-)\gamma)$ cases (right bottom).

We found out that almost half of the reconstructed $B^0_{\rm sig}$ candidates with a reconstructed Dalitz $\pi^0_{\rm dal}$ are in reality generated $B^0$'s with a converted photon and vice versa. This occurs because the final state particles are in the same kinematic phase space and because the topology of the decay is very similar if at least one of the $e^+e^-$ tracks is required to have a PXD hit.

We select only one $B^0_{\rm sig}$ candidate per event. If an event contains both a $B^0_{\rm sig}$ candidate with a reconstructed converted photon and a $B^0_{\rm sig}$ candidate with a reconstructed Dalitz $\pi^0_{\rm dal}$, we select the candidate with the reconstructed Dalitz $\pi^0_{\rm dal}$. After reconstruction and final selection, about 270 events with $B^0_{\rm sig}$ candidates reconstructed with Dalitz $\pi^0_{\rm dal}$ and about 50 events with $B^0_{\rm sig}$ candidates reconstructed with converted photons remain in a signal MC sample corresponding to 50 ab$^{-1}$. The latter number of events is too small for a time-dependent $CP$ violation analysis. Since reconstructed $B^0_{\rm sig}$ candidates with converted photons have a worse $\Delta t$ resolution and a worse value of the figure of merit (definition in Continuum suppression subsection), they can not be added to the sample with reconstructed Dalitz $B^0_{\rm sig}$ candidates. Therefore, the sensitivity study for the time-dependent $CP$ violation is performed only for events with reconstructed Dalitz $B^0_{\rm sig}$ candidates. The $B^0_{\rm sig}$ candidates reconstructed from two $\pi^0_{\gamma\gamma}$ are used for time-integrated $CP$ violation study. There is no event overlap between events with $B^0_{\rm sig}$ candidates reconstructed from two $\pi^0_{\gamma\gamma}$ and events containing Dalitz decays or converted photons.

*$\Delta t$ Resolution.* For the time-dependent $CP$-analysis, we use only events reconstructed as Dalitz events $B^0_{\rm sig} \to \pi^0_{\rm dal}(\to e^+e^-\gamma)\ \pi^0_{\gamma\gamma}(\to \gamma\gamma)$. Using MC information, we find out that about 54% of these events correspond to signal events with true Dalitz $\pi^0_{\rm dal}$ decays and about 46% correspond to events with a converted photon. The vertex of the $B^0_{\rm sig}$ is reconstructed using the two tracks $(e^+e^-)$ together with the *iptube* constraint, an ellipsoid constraint whose transverse size corresponds to the beam size at the IP and is oriented along the boost direction (see Sec. 6.3.2).

The reconstruction of the $B^0_{\rm tag}$ vertex is performed as explained in Sec. 6.2. The time difference $\Delta t$ is calculated from the difference $\Delta l$ in the lab frame, which corresponds to the difference between the reconstructed decay vertices of $B^0_{\rm sig}$ and $B^0_{\rm tag}$ in boost direction.

Figure 109 shows the $\Delta t^{\rm rec} - \Delta t^{\rm gen}$ residuals for fully reconstructed events. The $\Delta t$ resolution $\sigma_{\Delta t}$ for events with a converted photon is about 0.3 ps larger than for true Dalitz events. Therefore, these two types of signal events have to be considered separately.

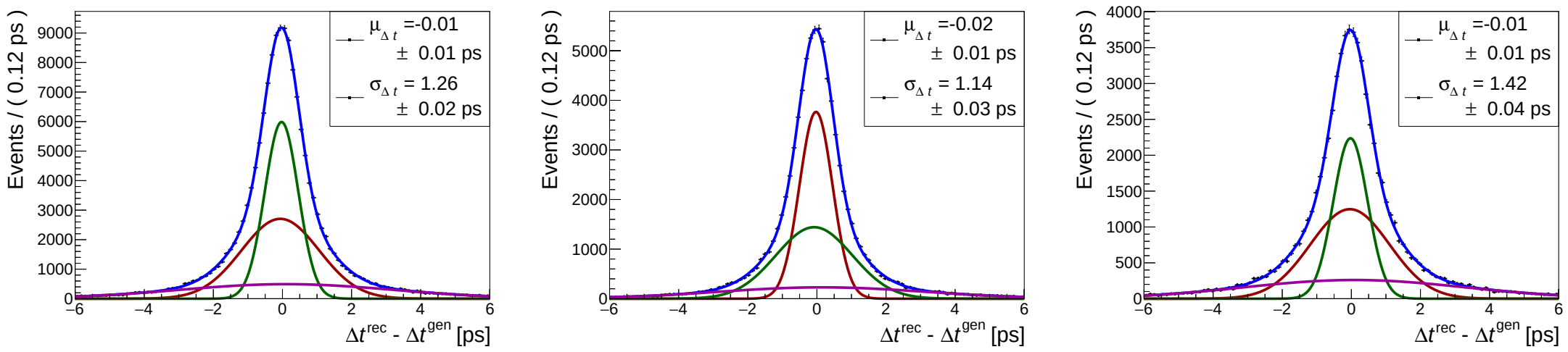

Fig. 109: $\Delta t$ residual distributions for events with reconstructed Dalitz $\pi^0_{\rm dal}$: for all the events (left), only for true Dalitz events (middle) and only for events with a converted photon (right). All events are reconstructed as Dalitz events. The fits are performed with three Gaussian distributions. The shift $\mu_{\Delta t}$ and the resolutions $\sigma_{\Delta t}$ are the weighted averages of the mean values and the standard deviations of the three Gaussian functions.

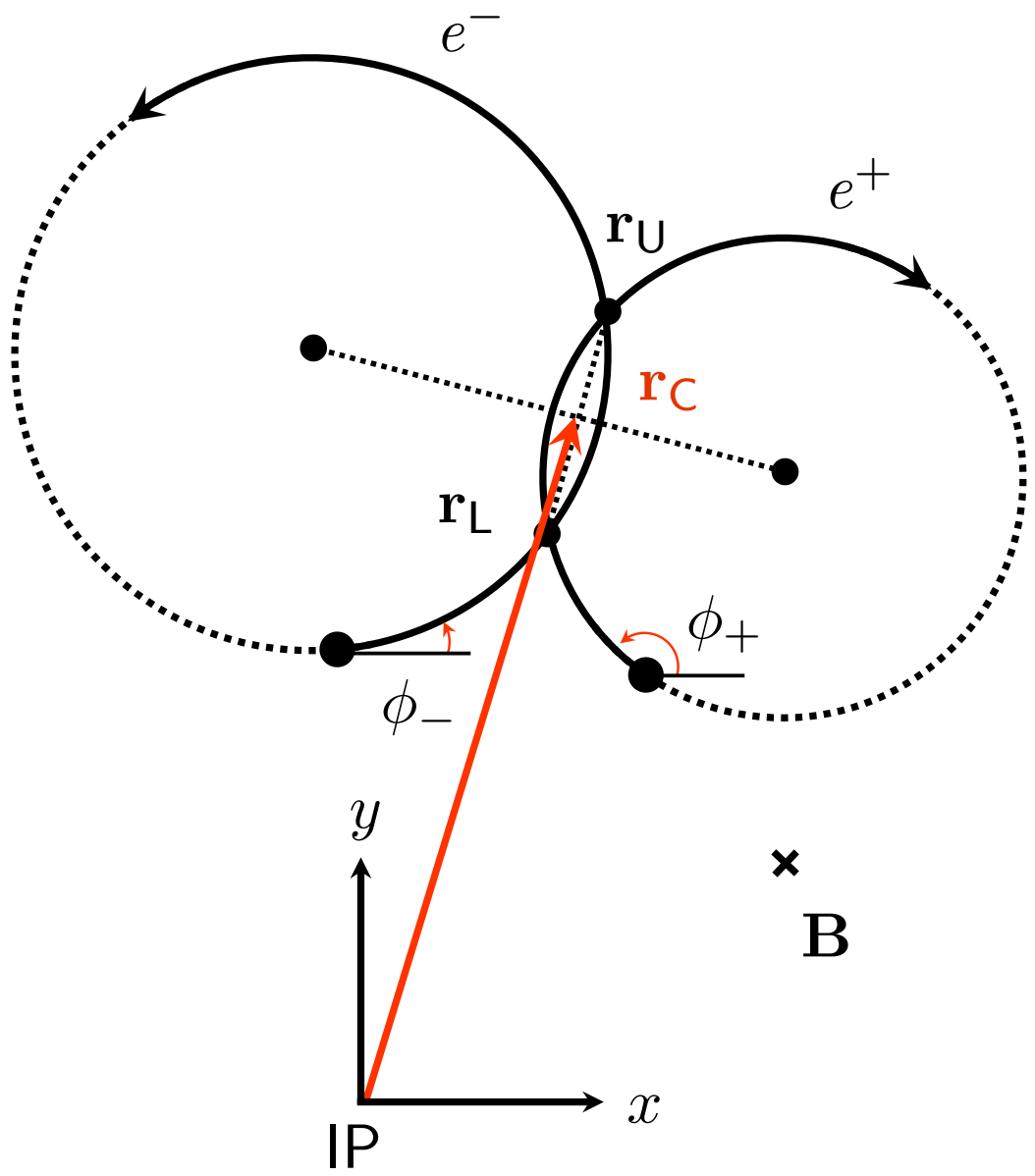

Fig. 110: Illustration of $\mathbf{r}_{\rm C}$, $\mathbf{r}_{\rm L}$, $\mathbf{r}_{\rm U}$, $\phi_{e^+}$ and $\phi_{e^-}$ on the $xy$-plane ($r\phi$-plane). The vector $\mathbf{B}$ corresponds to the magnetic field.

*Separation between $B^0_{sig}$ candidates with a true Dalitz $\pi^0_{dal}$ decay and with a converted photon from one of the two $\pi^0$'s.* In order to distinguish the two types of signal candidates used for the time-dependent $CP$-analysis, *i.e.* between $B^0_{\rm sig}$ candidates with a true Dalitz $\pi^0_{\rm dal}$ decay and $B^0_{\rm sig}$ candidates with a converted photon from one of the two $\pi^0$'s, a FastBDT [70] multivariate method was trained with four input variables that provide separation capacity. These input variables make use of the information related to the properties of the reconstructed $e^+e^-$ pair, for which at least one of the tracks is required to have a PXD hit. The first two variables, $r_{\rm L}$ and $r_{\rm U}$, correspond to the possible two solutions for the $e^+e^-$ vertex in the $xy$-plane ($r\phi$-plane): $r_{\rm L}$ is the solution closest to the IP and $r_{\rm U}$ the farthest one. For the calculation, we consider the $e^+e^-$ tracks as two circles with radii $r_1$ and $r_2$ and calculate the intersections between them. The line connecting the two intersection points crosses the line connecting the centers $\mathbf{c}_1$ and $\mathbf{c}_2$ of the two circles at the point

$$\mathbf{r}_{\rm C} = \mathbf{c}_1 + \frac{\mathbf{c}_2 - \mathbf{c}_1}{2}\left(1 + \frac{r_1^2 - r_2^2}{|\mathbf{c}_2 - \mathbf{c}_1|^2}\right). \tag{328}$$

One can write then

$$\mathbf{r}_{\rm U,L} = \mathbf{r}_{\rm C} \pm y \cdot \mathbf{n}, \tag{329}$$

where $\mathbf{n}$ is a unit vector perpendicular to the line connecting the centers of the two circles. If there are two intersections, $y$ is a real number such that $y > 0$. If there is only one intersection $y = 0$. In case that there is no intersection, we set $\mathbf{r}_{\rm L,U} = \mathbf{r}_{\rm C}$. An illustration of the two circle approach is shown in Figure 110.

The fourth and the fifth variables correspond to the angular differences $\Delta\theta_{e^+e^-} = \theta_{e^+} - \theta_{e^-}$ and $\Delta\phi_{e^+e^-} = \phi_{e^+} - \phi_{e^-}$, where the angles $\theta$ and $\phi$ for each track are calculated at the respective point of closest approach to the IP; within the Belle II software, the tracks are represented by the helix parameters calculated at the point of closest approach to the IP.

Figure 111 shows the distributions of the four input variables and the output $y_{\rm DC}$ of the FastBDT classifier for $B^0_{\rm sig}$ candidates with true Dalitz $\pi^0_{\rm dal}$ decays and for $B^0_{\rm sig}$ candidates with converted photons. In particular, the input variable $\Delta\phi_{e^+e^-}$ has a large separation power. For events with Dalitz $\pi^0_{\rm dal}$ decays, the $\Delta\phi_{e^+e^-}$ distribution is symmetric around zero degree. In this case, the $B^0$ decay vertex is only a few microns away from the IP in the $r\phi$-plane and then the $\pi^0_{\rm dal}$ decay follows immediately, therefore, the angles at the point of closest approach correspond to a good approximation to the true opening angles at the decay vertex. In contrast, the distribution of $\Delta\phi_{e^+e^-}$ for converted photons is asymmetric. For these events, the vertex is far from the interaction point (at least 1 cm). Thus, the $\phi$ angles at the point of closest approach to the IP are biased with respect to the true values at the conversion vertex. Because the curvature of reconstructed electrons has always the same sign, which is opposite to the sign of the curvature of reconstructed positrons, the angular difference $\Delta\phi_{e^+e^-}$ is always biased in the same direction.

*Continuum suppression and flavour tagging.* Continuum events ($e^+e^- \to q\bar{q}, q = u, d, s, c$) are the dominant source of background. The continuum background is studied using an available MC sample which corresponds to 2 ab$^{-1}$. We employ a FastBDT multivariate method to discriminate between $B^0 \to \pi^0\pi^0$ and continuum events. The method is trained with variables that have a separation power between spherical $B\bar{B}$ events and jet-like continuum events (see Sec. 6.4). From the available set of 30 continuum suppression variables,

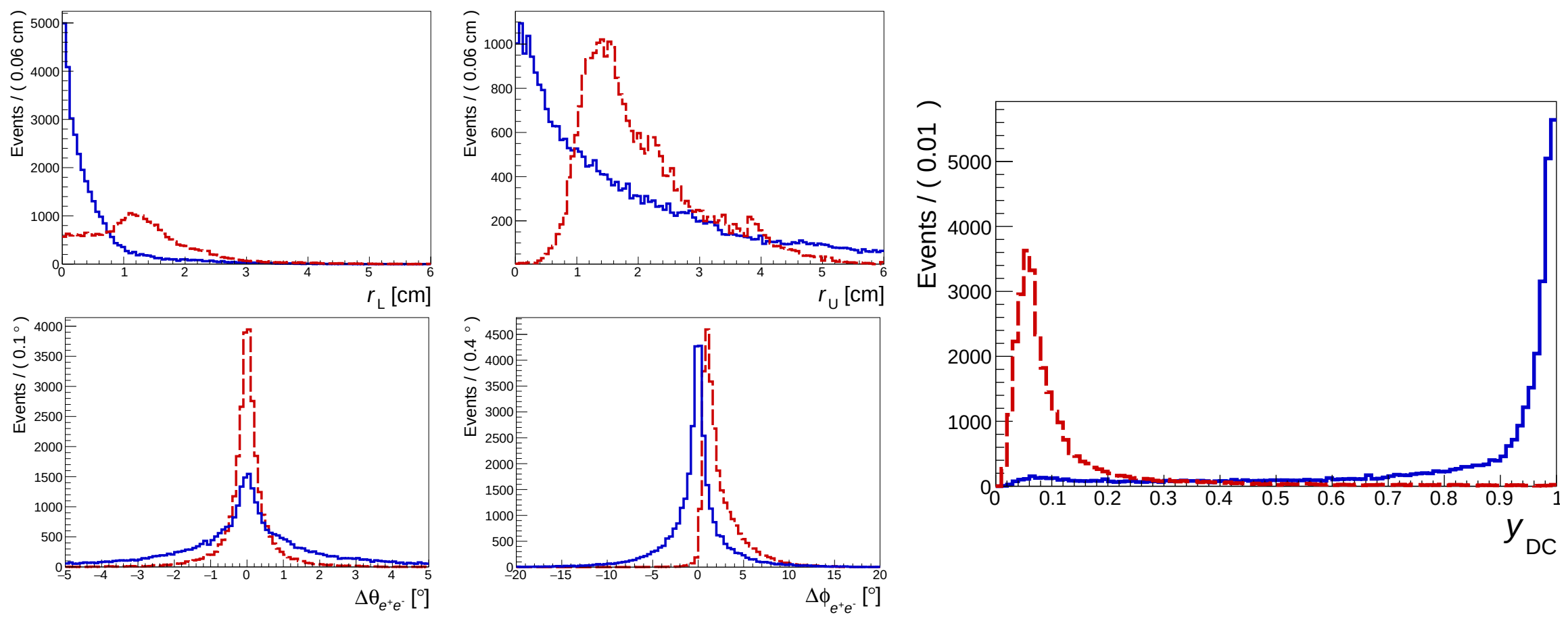

Fig. 111: The four input variables, $r_{\mathrm{L}}$ (left top), $r_{\mathrm{U}}$ (middle top), $\Delta\theta_{e^+e^-}$ (left bottom) and $\Delta\phi_{e^+e^-}$ (middle bottom), together with the output $y_{\mathrm{DC}}$ of the Dalitz-Conversion classifier (right). $B^0_{\mathrm{sig}}$ candidates with true Dalitz $\pi^0_{\mathrm{dal}}$ decays and with converted photons are shown by the solid blue and the long dashed red curves, respectively.

3 variables were discarded because of their strong correlation with $M_{\mathrm{bc}}$ and $\Delta E$: the magnitude of the $B_{\mathrm{sig}}$ thrust together with the first and the second Cleo Cones. The output of the multivariate method is normalised to the range $[0, 1]$. We select events where the output of the multivariate method is larger than 0.976, maximising the figure of merit

$$\mathrm{FoM} = \frac{n_{\mathrm{sig}}}{\sqrt{n_{\mathrm{sig}} + n_{\mathrm{cont}}}}, \tag{330}$$

where $n_{\mathrm{sig}}$ is the number of signal events and $n_{\mathrm{cont}}$ is the number of continuum events. The flavour of the remaining $B^0_{\mathrm{tag}}$ is determined using the flavour tagger algorithm explained in Sec. 6.5. The output flavour dilution $r \in [0, 1]$ peaks at 0 for continuum background and at 1 for signal events. In order to additionally remove some continuum background, events with $r < 0.1$ are rejected. In this way, only signal events which do not provide flavour separation power are discarded.

*$B\bar{B}$ background.* Sources of background from $B\bar{B}$ events are studied with a 4 ab$^{-1}$ MC sample. The largest contribution comes from $B^+ \to \rho^+(\to \pi^+\pi^0)\pi^0$ decays, where the $\pi^+$ is lost. Events where the remaining $\pi^0$ pair decays into four photons which arrive at the ECL are the main $B\bar{B}$ background for $B^0 \to \pi^0_{\gamma\gamma}\pi^0_{\gamma\gamma}$ candidates. Those events which contain a converted photon or a Dalitz $\pi^0$ are the main background $B\bar{B}$ source for $B^0 \to \pi^0_{\mathrm{dal}}\pi^0_{\gamma\gamma}$ candidates. This background peaks at the same value of $M_{\mathrm{bc}}$, but is shifted in $\Delta E$ towards negative values due to the missing $\pi^+$.

*Efficiencies.* Table 88 presents the absolute reconstruction efficiency and the efficiency after final selection including the requirements on flavour dilution and on continuum suppression for the different decay modes. We verified that the reconstruction efficiency is constant over the whole $\Delta t$ fit range. The events with converted photons and Dalitz $\pi^0$'s are reconstructed as $B^0_{\mathrm{sig}} \to \pi^0_{\mathrm{dal}}\ \pi^0_{\gamma\gamma}$.

Table 88: Fraction of generated events in the acceptance $n_{\rm gen}^{\rm acc}/n_{\rm gen}$, reconstruction efficiency $n_{\rm rec}/n_{\rm gen}^{\rm acc}$ and efficiency after final selection $n_{\rm rec}^{\rm FS}/n_{\rm gen}^{\rm acc}$ (the efficiencies are normalised to the number of generated events in the acceptance $n_{\rm gen}^{\rm acc}$). Events with converted photons and Dalitz $\pi^0$'s (first and second rows) were reconstructed as $B^0_{\rm sig} \to \pi^0_{\rm dal}\ \pi^0_{\gamma\gamma}$. The highlighted row corresponds to the whole set used for time-dependent $CP$-analysis.

| Decay. Channel | $n_{\rm gen}^{\rm acc}/n_{\rm gen}$ [%] | $n_{\rm rec}/n_{\rm gen}^{\rm acc}$ [%] | $n_{\rm rec}^{\rm FS}/n_{\rm gen}^{\rm acc}$ [%] |
|---|---|---|---|
| $B^0 \to\ \pi^0_{\rm dal}\ \pi^0_{\gamma\gamma}$ | 2.0 | 52.0 | 7.2 |
| $B^0 \to\ \pi^0_{\gamma_c\gamma}\ \pi^0_{\gamma\gamma}$ | 3.0 | 48.8 | 4.2 |
| Dal + Conv | 5.0 | 50.1 | 5.4 |
| $B^0 \to\ \pi^0_{\gamma\gamma}\ \pi^0_{\gamma\gamma}$ | 76.2 | 86.0 | 19.2 |

*Estimate of sensitivity from pseudo-experiments.* The expected statistical uncertainties are estimated performing sets of pseudo-experiments based on simulated experiments. A time-integrated $CP$-analysis is performed for events reconstructed as $B^0 \to \pi^0_{\gamma\gamma}(\to \gamma\gamma)\ \pi^0_{\gamma\gamma}(\to \gamma\gamma)$ and a time-dependent $CP$-analysis for events reconstructed as $B^0_{\rm sig} \to \pi^0_{\rm dal}(\to e^+e^-\gamma)\ \pi^0_{\gamma\gamma}(\to \gamma\gamma)$. We extracted pdf's for the distributions of the different components and performed an unbinned extended multi-dimensional maximum likelihood fit using MC. For the time-dependent and the time-integrated analyses we used $\Delta E$ and $M_{\rm bc}$ as fit variables. For the time-dependent analysis, the classifier output $y_{\rm DC}$ was used in addition. The fit was performed assuming no correlation between the fit variables.

The generated toy MC for each pseudo-experiment set corresponds to an integrated luminosity of 50 ab$^{-1}$ including signal, background from wrongly reconstructed signal events (WRS), $B\bar{B}$ background and continuum background. The signal and the background from wrongly reconstructed signal events were obtained by extracting random sub-samples from the generated signal MC. Wrongly reconstructed signal events correspond to signal events were the $B^0_{\rm sig}$ candidate was reconstructed by wrong combination of final state particles. Different input values of $A_{\pi^0\pi^0}$ and $S_{\pi^0\pi^0}$ were used to generate the signal MC: we considered the world averages $A_{\pi^0\pi^0} = 0.43$ (2016) [88], $A_{\pi^0\pi^0} = 0.34$ (2017) [650], the latest Belle measurement $A_{\pi^0\pi^0} = 0.14$ [712] and the predicted value $S_{\pi^0\pi^0} = 0.65$ [650]. The assumed branching fraction $\mathcal{B}_{\pi^0\pi^0} = 1.91 \cdot 10^{-6}$ [88] yields 15068 signal events for the time-integrated analysis and 271 for the time-dependent analysis. The latter number of events is composed of 147 signal events with Dalitz decays and 124 signal events with conversions. These two types of signal events are considered as two independent signal components in the maximum likelihood fit which can be distinguished by the $y_{\rm DC}$ fit variable. The purity

$$\text{Purity} = \frac{n_{\rm sig}}{n_{\rm sig} + n_{\rm WRS} + n_{B\bar{B}} + n_{\rm cont}} \tag{331}$$

and the fractions of wrongly reconstructed signal events are presented in Table 89. The continuum and the $B\bar{B}$ background events were generated from pdf's which were modelled using MC samples corresponding to 2 and 4 ab$^{-1}$, respectively. For each combination of input values $A_{\pi^0\pi^0}$ and $S_{\pi^0\pi^0}$, we generated 527 pseudo-experiments performing time-dependent and time-integrated analyses. Projections of the fit results for one example pseudo-experiment are shown in Figs. 112 and 113 for the time-dependent and the time-integrated analysis, respectively.

Table 89: Purity and fraction $\frac{n_{\mathrm{WRS}}}{n_{\mathrm{sig}}+n_{\mathrm{WRS}}}$ of wrongly reconstructed signal events after the final selection.

| Decay Channel | Purity [%] | $n_{\mathrm{WRS}}/n_{\mathrm{sig}}+n_{\mathrm{WRS}}$ [%] |
|---|---|---|
| Dal + Conv | 17.6 | 1.1 |
| $B^0 \to \pi^0_{\gamma\gamma}\ \pi^0_{\gamma\gamma}$ | 15.8 | 1.0 |

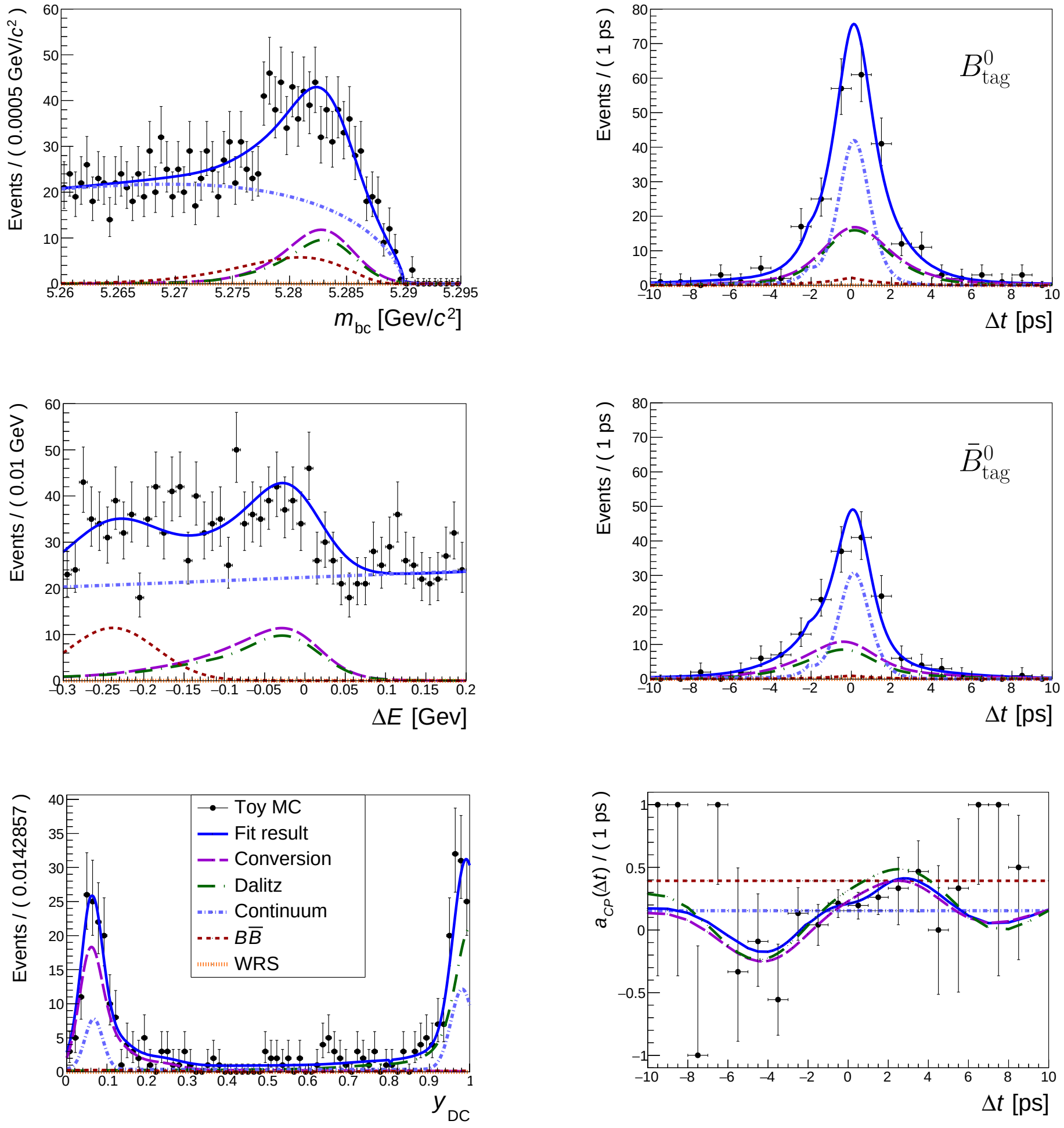

Fig. 112: Projections of the fit results for candidates reconstructed as $B^0 \to \pi^0\ (\to e^+e^-\gamma)\ \pi^0\,(\to \gamma\gamma)$. The projections for one example pseudo-experiment are shown onto $M_{\mathrm{bc}}$ (left top), $\Delta E$ (left middle), $y_{DC}$ (left bottom) and $\Delta t$ (right). The $\Delta t$ projection is shown for $B^0$ mesons tagged as $B^0$ (right top) and as $\bar{B}^0$ (right middle) together with the $CP$ asymmetry (right bottom). Points with error bars represent the toy MC sample. The full fit results are shown by the solid blue curves. Contributions from signal with Dalitz decays, signal with conversions, generic $B\bar{B}$, continuum background and background from wrongly reconstructed signal events are shown by the long dashed-dotted green, long dashed violet, short dashed red, dash-dotted blue and dotted orange curves, respectively. The input values used for this pseudo-experiment are $A_{\pi^0\pi^0} = 0.34$ and $S_{\pi^0\pi^0} = 0.65$.

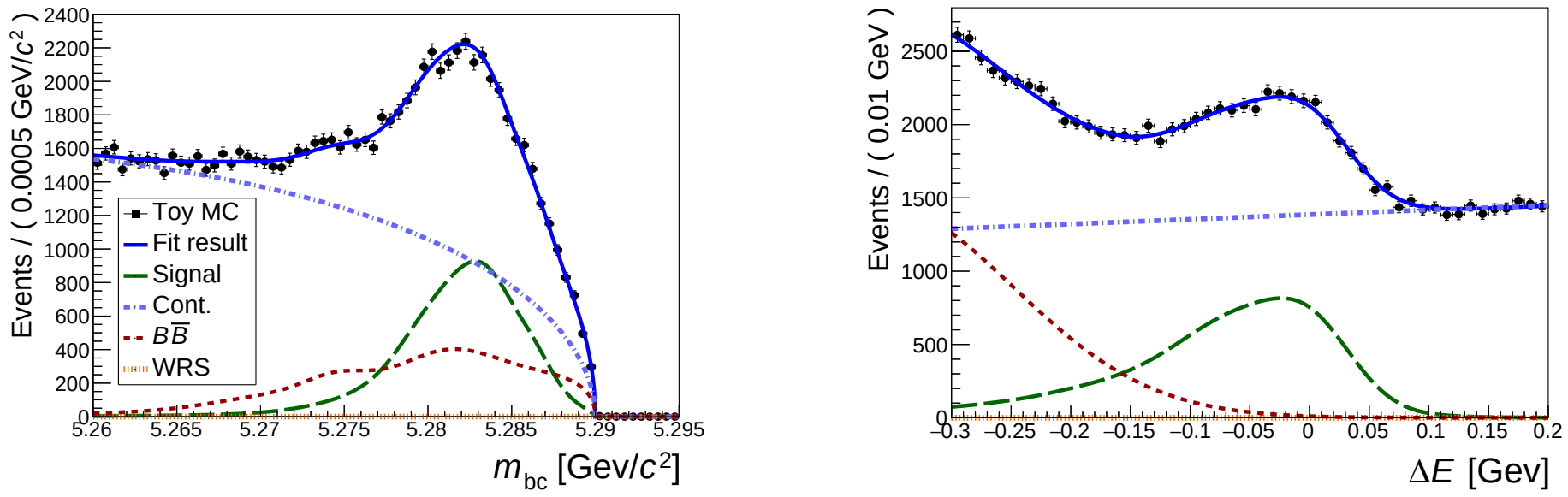

Fig. 113: Projections of the fit results for candidates reconstructed as $B^0 \to \pi^0\ (\to \gamma\ \gamma)\ \pi^0\ (\to \gamma\gamma)$. The projections for one example pseudo-experiment are shown onto $M_{\rm bc}$ (left) and $\Delta E$ (right). Points with error bars represent the toy sample. The full fit results are shown by the solid blue curves. Contributions from signal, generic $B\bar{B}$ events, continuum background and background from wrongly reconstructed signal events are shown by the long dashed green, short dashed red, dash-dotted blue and dotted orange curves, respectively. The input values used for this pseudo-experiment are $A_{\pi^0\pi^0} = 0.34$ and $S_{\pi^0\pi^0} = 0.65$.

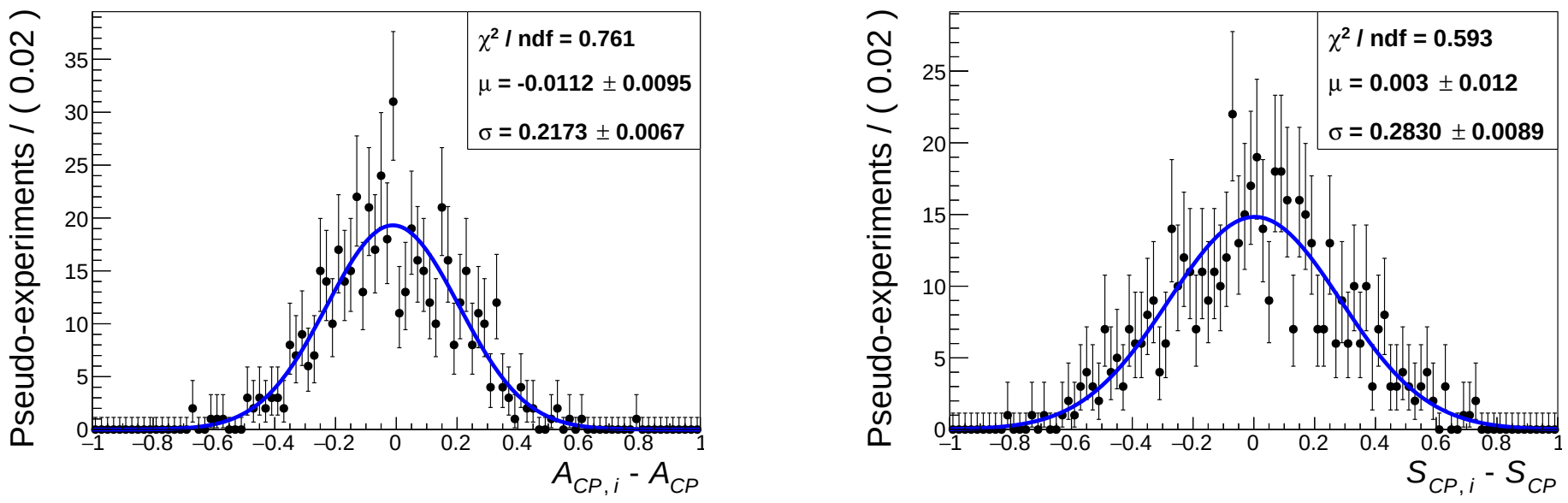

Fig. 114: Residuals distributions of $A_{CP}$ (left) and $S_{CP}$ (right) for the fit of $B^0 \to \pi^0\ (\to e^+\ e^-\ \gamma)\ \pi^0\ (\to \gamma\gamma)$. The input values used for these pseudo-experiments are $A_{\pi^0\pi^0} = 0.34$ and $S_{\pi^0\pi^0} = 0.65$.

We verify that the signal yields and the $CP$-violation parameters are determined without bias and without over- or underestimation of the error through examination of the fit pulls. Figure 114 shows the residuals distributions for the $CP$-violation parameters extracted from the time-dependent analysis. Figure 115 shows the residuals distributions for $n_{\rm sig}$ and $A_{\pi^0\pi^0}$ extracted from the time-integrated analysis. These distributions are fitted with a single Gaussian function and the value of $\sigma$ is taken as statistical uncertainty of the measured parameters. For the statistical uncertainty of the branching fraction we use $\Delta\mathcal{B}_{\pi^0\pi^0}/\mathcal{B}_{\pi^0\pi^0} = \sigma_{n_{\rm sig}}/n_{\rm sig}$. Table 90 shows the estimated statistical uncertainties of $A_{\pi^0\pi^0}$, $S_{\pi^0\pi^0}$ and $\Delta\mathcal{B}_{\pi^0\pi^0}/\mathcal{B}_{\pi^0\pi^0}$ for different values of the input parameters $A_{\pi^0\pi^0}$ and $S_{\pi^0\pi^0}$ used for the generation of the signal MC.

*Extrapolation of the $B^0 \to \pi\pi$ sensitivities.* The expected statistical uncertainties for $\mathcal{B}_{\pi^+\pi^-}$, $\mathcal{B}_{\pi^+\pi^0}$, $A_{\pi^+\pi^-}$ and $S_{\pi^+\pi^-}$ are estimated through extrapolation of Belle measurements at 0.8 ab$^{-1}$ assuming a final integrated luminosity of 50 ab$^{-1}$ at Belle II. For $\mathcal{B}_{\pi^0\pi^0}$, $A_{\pi^0\pi^0}$ and $S_{\pi^0\pi^0}$, the statistical uncertainties are taken from the sensitivity study that was performed using the current world averages for $A_{\pi^0\pi^0}$ and $S_{\pi^0\pi^0}$ as input values (first row of Table 90).

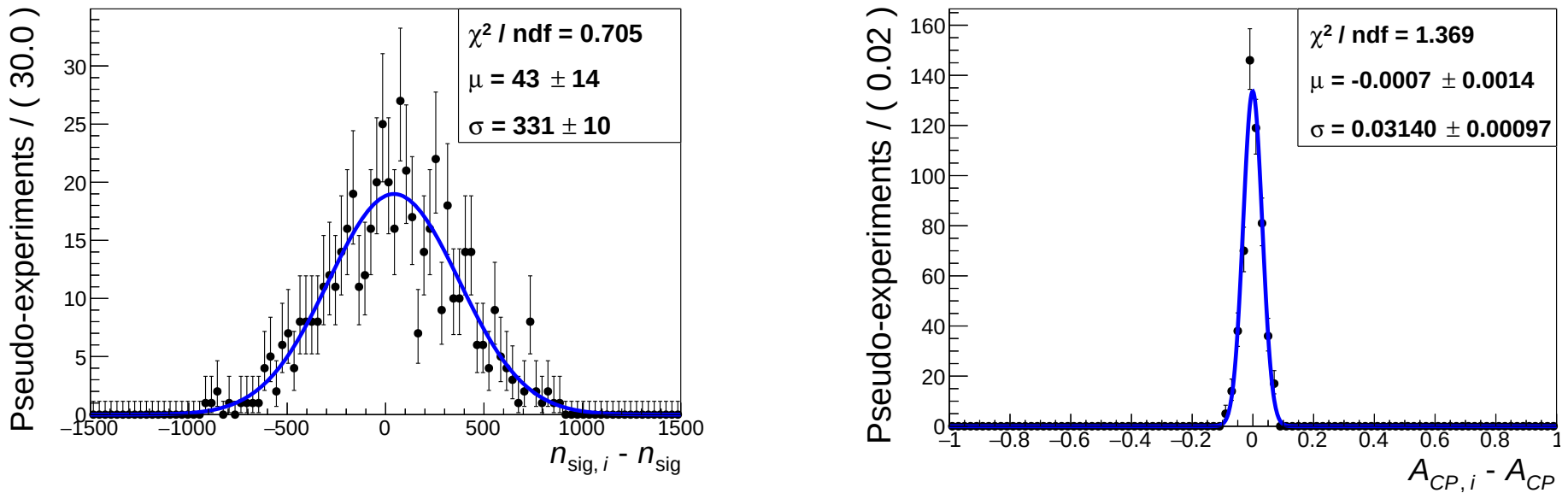

Fig. 115: Residuals distributions of the signal yield $n_{\text{sig}}$ (left) and the parameter $A_{CP}$ (right) for the fit of $B^0 \to \pi^0 (\to \gamma\,\gamma)\ \pi^0 (\to \gamma\gamma)$. The input values used for these pseudo-experiments are $A_{\pi^0\pi^0} = 0.34$ and $S_{\pi^0\pi^0} = 0.65$.

Table 90: Statistical uncertainties $\Delta A_{\pi^0\pi^0}$, $\Delta S_{\pi^0\pi^0}$ and $\Delta\mathcal{B}_{\pi^0\pi^0}/\mathcal{B}_{\pi^0\pi^0}$ for different input values of $A_{\pi^0\pi^0}$ and $S_{\pi^0\pi^0}$ used for the generation of signal MC.

| Input values | | Time-dependent | | Time-integrated | |
|---|---|---|---|---|---|
| $A_{\pi^0\pi^0}$ | $S_{\pi^0\pi^0}$ | $\Delta A_{\pi^0\pi^0}$ | $\Delta S_{\pi^0\pi^0}$ | $\Delta A_{\pi^0\pi^0}$ | $\Delta\mathcal{B}_{\pi^0\pi^0}/\mathcal{B}_{\pi^0\pi^0}$ [%] |
| 0.34 [650] | 0.65 [650] | 0.22 | 0.28 | 0.03 | 2.2 |
| 0.43 [88] | 0.79 | 0.23 | 0.29 | 0.03 | 2.2 |
| 0.14 [712] | 0.83 | 0.21 | 0.26 | 0.03 | 2.4 |
| 0.14 [712] | 0.40 | 0.20 | 0.29 | 0.03 | 2.3 |
| 0.14 [712] | −0.61 | 0.22 | 0.27 | 0.03 | 2.3 |
| 0.14 [712] | −0.94 | 0.22 | 0.28 | 0.03 | 2.4 |

An estimation of possible systematic uncertainties is performed following the guidelines in Sec. 10.2.1. We assume that reducible systematics will scale with the luminosity since they are evaluated with control samples and MC events. We sum in quadrature the irreducible and the extrapolated reducible systematic uncertainties. For $\mathcal{B}_{\pi^+\pi^-}$ and $\mathcal{B}_{\pi^+\pi^0}$, the list of sources of systematic uncertainties in Table II of [727] is considered, and, for $\mathcal{B}_{\pi^0\pi^0}$ and $A_{\pi^0\pi^0}$, the lists in [712]. We assume all sources in these lists to be reducible apart from the number of $B$ mesons (1.37% for $\mathcal{B}_{\pi^+\pi^-}$ and $\mathcal{B}_{\pi^+\pi^0}$ as well as 1.4% for $\mathcal{B}_{\pi^0\pi^0}$) and the contribution from the PHOTOS MC generator (0.8%). We add an additional reducible flavour tagging contribution of $\pm 0.0034$ to $A_{\pi^0\pi^0}$ considering Table VI of [711]. For $A_{\pi^+\pi^-}$ and $S_{\pi^+\pi^-}$ the systematic sources in Table II of [728] are considered. Apart from the tag side interference ($\pm 3.18$ for $A_{\pi^+\pi^-}$ and $\pm 0.17$ for $S_{\pi^+\pi^-}$) and the $\Delta t$ resolution ($\pm 0.42$ for $A_{\pi^+\pi^-}$ and $\pm 1.01$ for $S_{\pi^+\pi^-}$), we assume all sources in this list to be reducible. The $\Delta t$ resolution contribution is reduced by factor two considering an improvement thanks to the PXD and the new reconstruction algorithms (see Sec. 10.2.1). A summary of the Belle measurements and the extrapolated uncertainties is presented in Table 91.

Table 91: Branching fractions and $CP$ asymmetry parameters entering in the isospin analysis of the $B \to \pi\pi$ system: Belle measurements at 0.8 ab$^{-1}$ together with the expected Belle II sensitivity at 50 ab$^{-1}$.

| | Value | 0.8 ab$^{-1}$ | 50 ab$^{-1}$ |
|---|---|---|---|
| $\mathcal{B}_{\pi^+\pi^-}$ [10$^{-6}$] | 5.04 | $\pm0.21 \pm 0.18$ [727] | $\pm0.03 \pm 0.08$ |
| $\mathcal{B}_{\pi^0\pi^0}$ [10$^{-6}$] | 1.31 | $\pm0.19 \pm 0.19$ [712] | $\pm0.03 \pm 0.03$ |
| $\mathcal{B}_{\pi^+\pi^0}$ [10$^{-6}$] | 5.86 | $\pm0.26 \pm 0.38$ [727] | $\pm0.03 \pm 0.09$ |
| $A_{\pi^+\pi^-}$ | 0.33 | $\pm0.06 \pm 0.03$ [728] | $\pm0.01 \pm 0.03$ |
| $S_{\pi^+\pi^-}$ | $-0.64$ | $\pm0.08 \pm 0.03$ [728] | $\pm0.01 \pm 0.01$ |
| $A_{\pi^0\pi^0}$ | 0.14 | $\pm0.36 \pm 0.10$ [712] | $\pm0.03 \pm 0.01$ |

The systematic uncertainty for $S_{\pi^0\pi^0}$ is assumed to be about 10% of the statistical uncertainty and in the order of magnitude of the systematic uncertainties of $A_{\pi^+\pi^-}$ and $S_{\pi^+\pi^-}$. This gives in total $\Delta S_{\pi^0\pi^0} = \pm0.28 \pm 0.03$.

*Extrapolation of the $B^0 \to \rho\rho$ sensitivities.* The expected statistical uncertainties for all the parameters entering in the isospin analysis of the $B \to \rho\rho$ decay mode are estimated through extrapolation of Belle and BaBar measurements assuming a final integrated luminosity of 50 ab$^{-1}$ at Belle II. We consider BaBar measurements only for $A_{\rho^0\rho^0}$ and $S_{\rho^0\rho^0}$ since these measurements have not been performed by Belle.

An estimation of possible systematic uncertainties is performed in a similar way as in the previous section for the $B \to \pi\pi$ system: the irreducible and the extrapolated reducible systematic uncertainties are summed in quadrature. For $\mathcal{B}_{\rho^+\rho^-}$ and $f_{L,\rho^+\rho^-}$, the list of sources of systematic uncertainties in Table VIII of [725] is considered; for $\mathcal{B}_{\rho^0\rho^0}$ and $f_{L,\rho^0\rho^0}$, the list in Table VIII of [729]; and for $\mathcal{B}_{\rho^+\rho^0}$ and $f_{L,\rho^+\rho^0}$, the list of sources given in [716]. We assume all sources in these lists to be reducible apart from the number of $B$ mesons (1.4%). For $A_{\rho^+\rho^-}$ and $S_{\rho^+\rho^-}$, the list in Table VIII of [725] is taken into account. Apart from the tag side interference ($\pm1.02 \cdot 10^{-2}$ for $A_{\rho^+\rho^-}$ and $\pm0.08 \cdot 10^{-2}$ for $S_{\rho^+\rho^-}$), we assume all sources in this list to be reducible. Although the measurement of $A_{\rho^0\rho^0}$ and $S_{\rho^0\rho^0}$ was performed by BaBar, we consider for these parameters the sources of systematic uncertainties taken into account by Belle for the measurement of $A_{\rho^+\rho^-}$ and $S_{\rho^+\rho^-}$. A summary of the Belle and BaBar measurements together with the extrapolated uncertainties is presented in Table 92.

*$\phi_2$ expected sensitivity using isospin analysis.* We estimate the Belle II sensitivity to the $\phi_2$-angle performing the isospin analysis introduced in Sec. 10.4.1 for $B \to \pi\pi$ and $B \to \rho\rho$. We perform a scan of the confidence for $\phi_2$ from a $\chi^2$ distribution which is obtained by minimising $-2\log(\mathcal{L})$. The likelihood $\mathcal{L}$ has the form of a multivariate normal distribution

$$\chi^2 = -2\log\left[\frac{\exp\left(\frac{1}{2}\left(\mathbf{x}_{\rm data} - \mathbf{x}_{\rm theo}\right)^T \mathit{\Sigma}^{-1}\left(\mathbf{x}_{\rm data} - \mathbf{x}_{\rm theo}\right)\right)}{\sqrt{(2\pi)^n \det \mathit{\Sigma}}}\right]. \tag{332}$$

where $\mathbf{x}_{\rm data}$ and $\mathbf{x}_{\rm theo}$ are vectors containing respectively the measured values and the theoretical predictions of the parameters $\mathcal{B}_{+-}$, $\mathcal{B}_{00}$, $\mathcal{B}_{+0}$, $A_{+-}$, $S_{+-}$, $A_{00}$ and $S_{00}$. For the theoretical predictions, we adopt the alternative amplitude parametrisation proposed in [730].

Table 92: Branching fractions, fractions of longitudinally polarised events and $CP$ asymmetry parameters entering in the isospin analysis of the $B \to \rho\rho$ system: Belle measurements at 0.8 ab$^{-1}$ and 0.08 ab$^{-1}$, BaBar measurements at 0.5 ab$^{-1}$ and expected Belle II sensitivity at 50 ab$^{-1}$.

| | Value | 0.8 ab$^{-1}$ | 50 ab$^{-1}$ |
|---|---|---|---|
| $f_{L,\rho^+\rho^-}$ | 0.988 | $\pm 0.012 \pm 0.023$ [725] | $\pm 0.002 \pm 0.003$ |
| $f_{L,\rho^0\rho^0}$ | 0.21 | $\pm 0.20 \pm 0.15$ [729] | $\pm 0.03 \pm 0.02$ |
| $\mathcal{B}_{\rho^+\rho^-}$ [10-6] | 28.3 | $\pm 1.5 \pm 1.5$ [725] | $\pm 0.19 \pm 0.4$ |
| $\mathcal{B}_{\rho^0\rho^0}$ [10-6] | 1.02 | $\pm 0.30 \pm 0.15$ [729] | $\pm 0.04 \pm 0.02$ |
| $A_{\rho^+\rho^-}$ | 0.00 | $\pm 0.10 \pm 0.06$ [725] | $\pm 0.01 \pm 0.01$ |
| $S_{\rho^+\rho^-}$ | $-0.13$ | $\pm 0.15 \pm 0.05$ [725] | $\pm 0.02 \pm 0.01$ |
| | Value | 0.08 ab$^{-1}$ | 50 ab$^{-1}$ |
| $f_{L,\rho^+\rho^0}$ | 0.95 | $\pm 0.11 \pm 0.02$ [716] | $\pm 0.004 \pm 0.003$ |
| $\mathcal{B}_{\rho^+\rho^0}$ [10-6] | 31.7 | $\pm 7.1 \pm 5.3$ [716] | $\pm 0.3 \pm 0.5$ |
| | Value | 0.5 ab$^{-1}$ | 50 ab$^{-1}$ |
| $A_{\rho^0\rho^0}$ | $-0.2$ | $\pm 0.8 \pm 0.3$ [715] | $\pm 0.08 \pm 0.01$ |
| $S_{\rho^0\rho^0}$ | 0.3 | $\pm 0.7 \pm 0.2$ [715] | $\pm 0.07 \pm 0.01$ |

The covariance matrix $\Sigma$ contains the uncertainties in the diagonal and the correlations between the measured parameters in the non-diagonal part.

Figure 116 shows the results of the scan of the confidence for the $\phi_2$-angle performing the isospin analysis of $B \to \pi\pi$. We use the Belle measurements and the projection for Belle II summarised in Table 91 without and with $S_{\pi^0\pi^0}$ constraint. One can recognise the eight possible solutions and the improvement of rejection power at Belle II even without $S_{\pi^0\pi^0}$. The scan including the $S_{\pi^0\pi^0}$ constraint is performed for several $S_{\pi^0\pi^0}$ central values. Compatible $S_{\pi^0\pi^0}$ values are estimated by calculating the theoretical predictions. For these calculations, we used the fit parameters obtained at the solutions of the scan performed without the $S_{\pi^0\pi^0}$ constraint. The compatible values of $S_{\pi^0\pi^0}$ were used as input values for the sets of pseudo-experiments in Sec. 10.4.2 (S. Table 90). As it can be seen in Fig. 116 (right), the solutions for each value of $S_{\pi^0\pi^0}$ that is compatible with the scan performed without the $S_{\pi^0\pi^0}$ constraint overlap with two of the eight possible solutions, thus reducing the ambiguities in the determination of the $\phi_2$-angle by a factor 4.

Because of the experimental precision, it is possible that the value of $S_{\pi^0\pi^0}$ that will be measured at Belle II will not be compatible with any of the four predicted values obtained from the the scan without the $S_{\pi^0\pi^0}$ constraint. Figure 117 shows different possible scenarios: a value of $S_{\pi^0\pi^0}$ that is compatible with the solution around 88° and two values that are not compatible with any of the eight solutions. In both situations there is a large range that can be excluded at one $\sigma$. In the case of the compatible value $S_{\pi^0\pi^0} = 0.83$, the width of the solution around 88° corresponding to a confidence of one $\sigma$ is about 4° and thus $\Delta\phi_2 \approx 2°$.

Figure 118 (left) shows the results of the scan of the confidence for the $\phi_2$-angle performing the isospin analysis of $B \to \rho\rho$. The analysis was performed using the current Belle measurements without $S_{\rho^0\rho^0}$ constraint. The results of the scan are consistent with the Belle results presented in Fig. 7 of [729]. Since the $B \to \rho\rho$ system exhibits a two fold ambiguity, we focus on the range which is consistent with the current measurements of the unitarity triangle. Figure 118 (left) shows also projections for Belle II without and with the $S_{\rho^0\rho^0}$ constraint. The measurements and the projections are summarised in Table 92. The estimated sensitivity for Belle II at the one $\sigma$-level without the $S_{\rho^0\rho^0}$ constraint is found to be about $\Delta\phi_2 \sim 1°$. For the case with the $S_{\rho^0\rho^0}$ constraint, an estimation was performed using the central value $S_{\rho^0\rho^0} = -0.14$. This value was chosen such that the solutions of the scan using this value are compatible with the solutions of the scan performed without the $S_{\rho^0\rho^0}$ constraint. The improvement with $S_{\rho^0\rho^0}$ at the one $\sigma$ is about one third: $\Delta\phi_2 \sim 0.7°$.

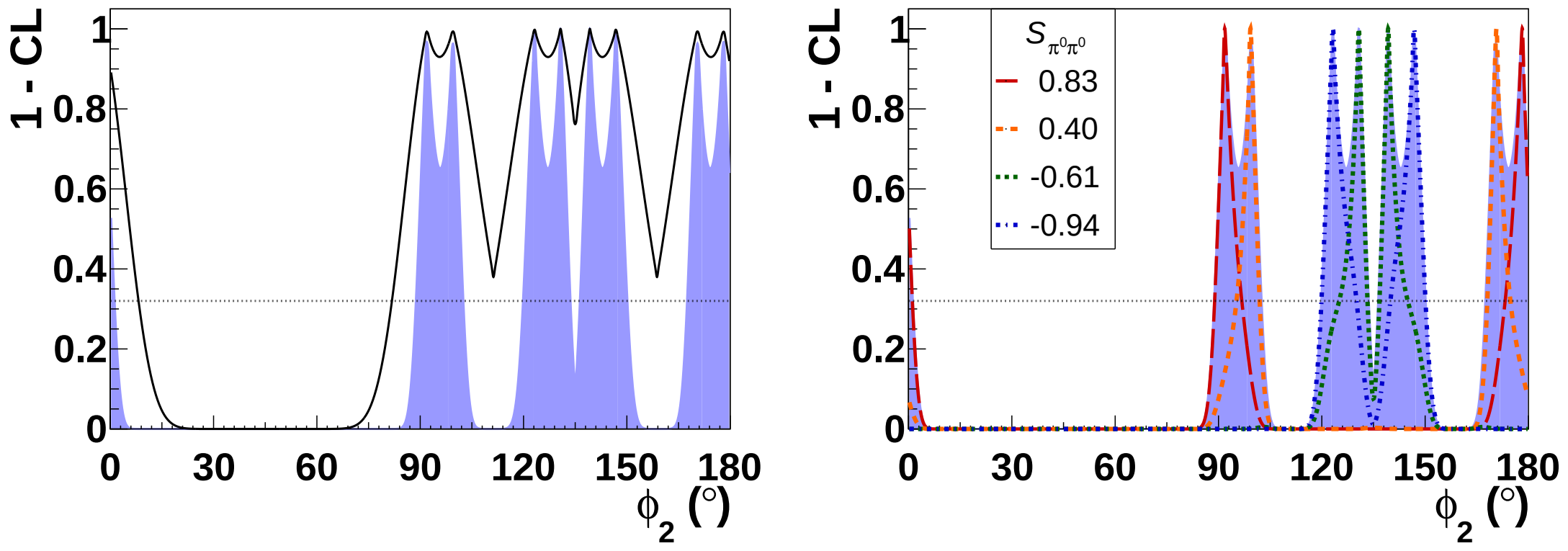

Fig. 116: Scan of the confidence for $\phi_2$ performing isospin analysis of the $B \to \pi\pi$ system. The black solid line (left) shows the result of the scan using data from Belle measurements (see Table 91). The blue shaded area in both plots shows the projection for Belle II. Results of the scan adding the $S_{\pi^0\pi^0}$ constraint (right): each line shows the result for a different $S_{\pi^0\pi^0}$ value. The dotted horizontal lines correspond to one $\sigma$.

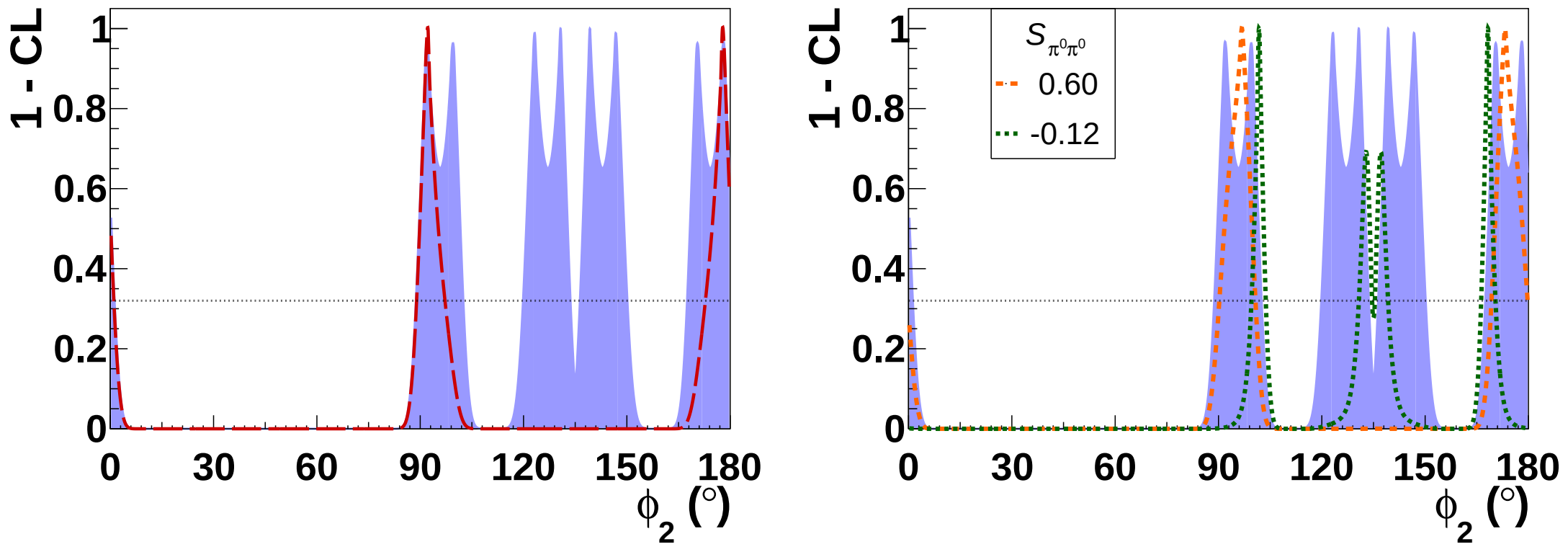

Fig. 117: Scan of the confidence for $\phi_2$ performing isospin analysis of the $B \to \pi\pi$ system. The blue shaded area in both plots shows the projection of the Belle measurements (S. Fig. 116) for Belle II. Results of the scan with additional $S_{\pi^0\pi^0}$ constraints are shown by dashed lines. Each line correspond to different input $S_{\pi^0\pi^0}$ values. The red long dashed line on the left figure shows the result for $S_{\pi^0\pi^0} = 0.83$. The dotted horizontal line correspond to one $\sigma$.

Figure 118 (right) shows the results of the scan of the confidence for $\phi_2$ combining the isospin analyses of $B \to \pi\pi$ and $B \to \rho\rho$. In order to have consistent central values of the input parameters for the study of the Belle II sensitivity, the value of $\mathcal{B}_{\pi^0\pi^0}$ was adjusted to be $1.27 \cdot 10^{-6}$. This adjustment, which is within one sigma of the measured value (S. Table 91), ensures that the solutions of the isospin analyses of $B \to \pi\pi$ and $B \to \rho\rho$ correspond to the same true value of the $\phi_2$. The $\phi_2$ scan using current Belle measurements was performed without $S_{00}$ constraints. The projections for Belle II were performed for both cases, without and with $S_{00}$ constraints. For the former case, the estimated sensitivity is found to be about $\Delta\phi_2 \sim 1°$. For the case with the $S_{00}$ constraints, the analysis is performed with central values of $S_{\rho^0\rho^0}$ and $S_{\pi^0\pi^0}$ which are compatible in terms of $\phi_2$ ($S_{\rho^0\rho^0} = -0.14$ and $S_{\pi^0\pi^0} = 0.75$). The improvement in the $\phi_2$-precision at the one $\sigma$ level with the $S_{00}$ constraints is about factor 2: from $\Delta\phi_2 \sim 1°$ to $\Delta\phi_2 \sim 0.6°$.

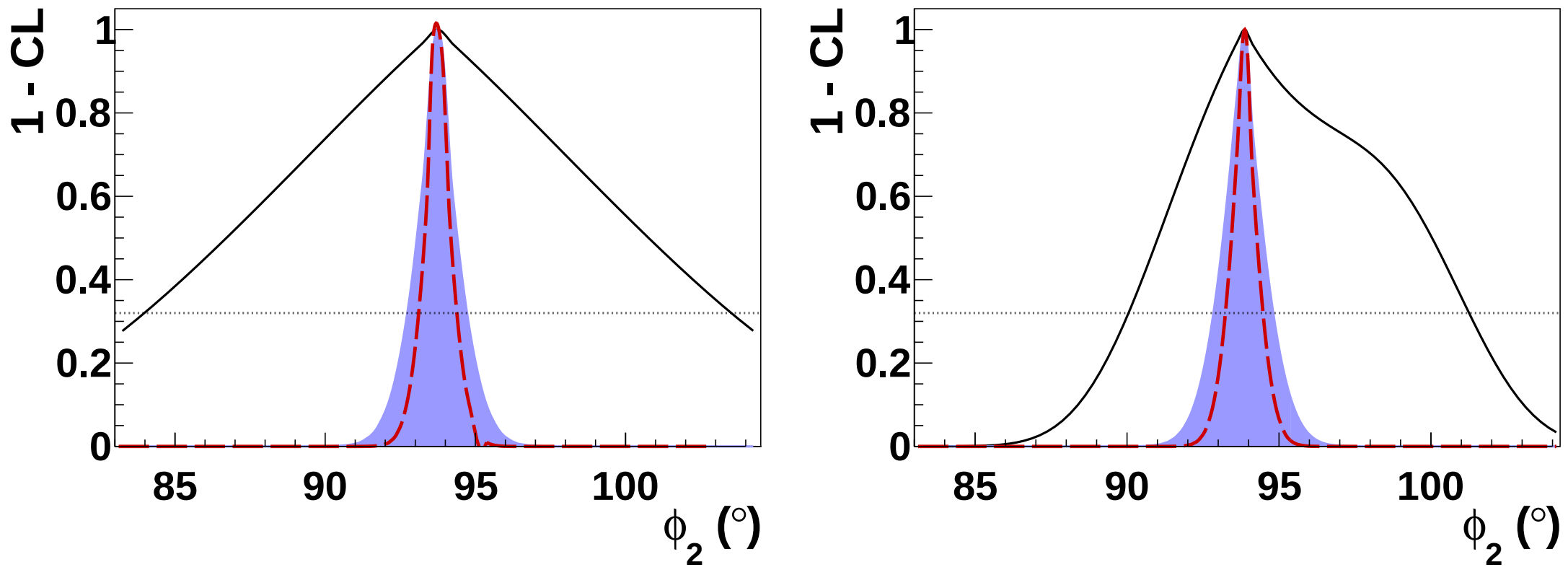

Fig. 118: Scans of the confidence for $\phi_2$ performing an isospin analysis of the $B \to \rho\rho$ system (left) and combining the isospin analyses of the $B \to \pi\pi$ and the $B \to \rho\rho$ systems (right). The black solid lines show the results of the scans using data from measurements at current precision (S. Tables 92 and 91). The blue shaded areas show the projections for Belle II. The red long dashed lines show the results of the scans adding the $S_{00}$ constraints: $S_{\rho^0\rho^0} = -0.14$ and $S_{\pi^0\pi^0} = 0.75$. The dotted horizontal lines correspond to one $\sigma$.

*$\phi_2$ from $B^0 \to \rho\pi$.* The measurement of $\phi_2$ from a Dalitz plot analysis of $B^0 \to \pi^+\pi^-\pi^0$ has been pioneered by Belle [718] and BaBar [717]. Both Collaborations succeeded in extracting meaningful information about $\phi_2$, however BaBar performed some studies on the robustness of the extraction of $\phi_2$ and discovered the existence of two *secondary solutions* on either side of the expected primary solution. These secondary solutions do not arise from ambiguities intrinsic on the method, but are rather artefacts that result from the limited statistics of the sample that was analysed and are expected to vanish with significantly larger datasets.

This strongly motivates the repetition of the analysis at Belle II, with a data sample of at least a few ab$^{-1}$. The dominant background will arise from random combinations of particles arising from continuum events and the model for the signal component will include the $\rho(770)$, $\rho(1450)$, and $\rho(1700)$ resonances.

For the reasons explained above, and due to the difficulty of realistically simulating the full Dalitz plot analysis on the Monte Carlo, we do not provide any prediction of the sensitivity attainable by Belle II.

## 10.5. Time dependent $CP$ violation analysis of $B^0 \to K^0_S \pi^0(\gamma)$

### 10.5.1. Theory: probing New Physics with $B^0 \to K^0_S \pi^0 \gamma$. Contributing authors: F. Bishara, A. Tayduganov

The radiative loop $b \to s\gamma$ processes have been extensively studied as a probe of NP beyond the SM. In the SM $b_L \to s_R \gamma_R$ is $m_s/m_b$ suppressed compared to $b_R \to s_L \gamma_L$, if QCD interactions are switched off. In order to use this as a probe of NP it is important to estimate reliably the QCD corrections to these expectations. We review the current status below.

The short distance contributions to $b \to s\gamma$ are given by

$$\mathcal{O}_7^{(\prime)} = \frac{e}{16\pi^2} m_b \bar{s}_{L(R)} \sigma^{\mu\nu} b_{R(L)} F_{\mu\nu}\,, \tag{333}$$

in the effective weak Hamiltonian. The operator $\mathcal{O}_7$ describes the $b_R \to s_L \gamma_L$, while $\mathcal{O}_7'$ describes the $b_L \to s_R \gamma_R$ process. Due to chiral suppression, in the SM $C_7'/C_7 \simeq m_s/m_b$. The $b \to s\gamma$ also receives long distance contributions, most notably from the "charm loop" contributions from insertion of $\mathcal{O}^c_{1,2}$ operators.

We focus on the $B \to K^*\gamma$ decay. An analysis of $1/m_b$ expansion based on Soft Collinear Effective Theory (SCET) in [506, 731] shows that the the right-handed helicity amplitude is suppressed. The largest contribution is expected to come from $\mathcal{O}_2 = (\bar{s}_L \gamma_\mu c_L)(\bar{c}_L \gamma^\mu b_L)$ operator, giving a parametric estimate:

$$\frac{\mathcal{M}(\overline{B} \to \overline{K}^* \gamma_R)}{\mathcal{M}(\overline{B} \to \overline{K}^* \gamma_L)} \sim \frac{(C_2/3)}{C_7} \frac{\Lambda_{\rm QCD}}{m_b} \sim 10\%\,. \tag{334}$$

The numerical value was obtained using naïve dimensional analysis for the relevant matrix element.

A light-cone sum rule based assessment of the matrix element shows, however, that this is further suppressed [491, 507] (for previous works see [490, 511, 732]). The matrix element $\mathcal{M}(\overline{B} \to \overline{K}^* \gamma_R)$ receives two types of contributions. The $m_b \to \infty$ contribution is perturbatively calculable and is $\mathcal{O}(m_s/m_b)$ suppressed compared to $\mathcal{M}(\overline{B} \to \overline{K}^* \gamma_L)$. The contributions to $\mathcal{M}(\overline{B} \to \overline{K}^* \gamma_L)$ from hard collinear gluon exchanges vanish, while the contributions from soft gluons are $1/m_b$ suppressed. This gives the estimate

$$\frac{\mathcal{M}(\overline{B} \to \overline{K}^* \gamma_R)}{\mathcal{M}(\overline{B} \to \overline{K}^* \gamma_L)} \sim \frac{(C_2/3)}{C_7} \frac{\Lambda^2_{\rm QCD}}{m_b^2} \sim \text{few}\%\,. \tag{335}$$

In conclusion, the charm loop effect could entail a theoretical uncertainty $\sim (2 \div 10)\%$ on the ratio of the right-handed polarisation amplitude over the left-handed one, though most likely on the lower end of this range. The NP effects can be clearly established only if the deviation from the SM is sufficiently large.

One way to measure the photon polarisation is to study the time-dependent $CP$-asymmetry in the radiative decays of the neutral $B$-mesons into the final hadronic self-conjugate state $f_{CP}$ [502]. The asymmetry arises from the interference between the $B \to f_{CP}\gamma$ and $B \to \overline{B} \to f_{CP}\gamma$ amplitudes (and similarly for $CP$ conjugated decays). Since the $B(\overline{B})$-meson decays predominantly into a photon with right(left)-handed helicity, the dominant amplitudes are $B \to f_{CP}\gamma_R$ and $B \to \overline{B} \to f_{CP}\gamma_L$, which cannot interfere. In the SM the time-dependent asymmetry is thus generated by suppressed amplitudes, of order $\mathcal{O}(m_s/m_b)$ or $\mathcal{O}(\Lambda^2/m_b^2)$ as discussed above. NP can induce much larger contribution to the "wrong" helicity amplitudes inducing a larger time-dependent $CP$ asymmetry.

For radiative decay $B(t) \to f_{CP}\gamma$, neglecting direct $CP$-violation and the small width difference between two $B$-mesons[30], the $CP$-asymmetry is given by [502]

$$\mathcal{A}_{CP}(t) \equiv \frac{\Gamma(\overline{B}(t) \to f_{CP}\gamma) - \Gamma(B(t) \to f_{CP}\gamma)}{\Gamma(\overline{B}(t) \to f_{CP}\gamma) + \Gamma(B(t) \to f_{CP}\gamma)} \approx S_{f_{CP}\gamma} \sin(\Delta m t)\,, \tag{336}$$

with

$$S_{f_{CP}\gamma} \equiv \eta_f \frac{2\Im[e^{-i2\phi_1}\,\mathcal{M}_L\mathcal{M}_R]}{|\mathcal{M}_L|^2 + |\mathcal{M}_R|^2} \simeq \eta_f \frac{2\Im[e^{-i2\phi_1}\,C_7C_7'|}{|C_7|^2 + |C_7'|^2}\,, \tag{337}$$

where $\mathcal{M}_{L(R)}$ are the amplitudes of $\overline{B} \to f_{CP}\gamma_{L(R)}$, $\eta_f = \pm 1$ is the $CP$-eigenvalue of $f_{CP}$, $\phi_1$ the phase in $B - \overline{B}$ mixing. The measurement of $\mathcal{A}_{CP}(t)$ allows us to determine the ratio of two amplitudes $\mathcal{M}_{L,R}$ together with the $CP$ violating phase $\phi_1$ but not each of them separately.

The decay $B \to K^*(\to K_S^0\pi^0)\gamma$ has the largest branching fraction and hence has the largest potential for the time-dependent $CP$ asymmetry search. The SM prediction for this asymmetry is [494]

$$S^{\rm SM}_{K_S^0\pi^0\gamma} \sim -2\frac{m_s}{m_b}\sin 2\phi_1 = -(2.3 \pm 1.6)\%\,, \tag{338}$$

which is to be compared with the current world average [230]

$$S^{\rm exp}_{K_S^0\pi^0\gamma} = -0.16 \pm 0.22\,. \tag{339}$$

Another method to search for the non-SM right-handed photon is to use $B \to K^*(\to K^+\pi^-)\gamma$ events with $\gamma \to e^+e^-$ conversions that occur in the detector [596]. If the photon helicity is mixed, the photon has an elliptical polarisation (or linear polarisation if the size of the left- and right-handed amplitudes are the same). Interference between the intermediate on-shell photon polarisations produces oscillations in the angular kinematic observable $\psi$ (relative twist between the $e^+e^-$ conversion plane and the $K - \pi$ decay plane) [516]. Measuring the amplitude and phase of these oscillations or equivalently two quadrant-type asymmetries would permit to extract the absolute value and the relative weak phase of the polarisation amplitudes ratio $\mathcal{M}_R/\mathcal{M}_L$.

This method requires high angular resolution in order to reconstruct the lepton kinematics after conversion. Moreover, a detector whose thickness is on the order of one radiation length or less is required to avoid multiple leptonic rescattering. All of the above makes this approach experimentally challenging.

*10.5.2. Experiment: sensitivity study of the time dependent CP asymmetries in $B^0 \to K_S^0\pi^0\gamma$. Contributing author: A. Martini*

In this section we use the Belle II simulation to estimate the sensitivity for measuring the time-dependent $CP$ asymmetry on the $B^0 \to K_S^0\pi^0\gamma$ decay:

$$\wp(\Delta t) = \frac{e^{-\frac{\Delta t}{\tau_{B^0}}}}{4\tau_{B^0}}(1 \pm S_{K_S^0\pi^0\gamma}\sin(\Delta m\Delta t) \pm A_{K_S^0\pi^0\gamma}\cos(\Delta m\Delta t)). \tag{340}$$

The experimental results from BaBar and Belle on the $CP$ violation parameters are respectively, $S_{K_S^0\pi^0\gamma} = -0.78 \pm 0.59 \pm 0.09$, $A_{K_S^0\pi^0\gamma} = 0.36 \pm 0.33 \pm 0.04$ [733] and $S_{K_S^0\pi^0\gamma} =$ $-0.10 \pm 0.31 \pm 0.07$, $A_{K_S^0\pi^0\gamma} = -0.20 \pm 0.20 \pm 0.06$ [519], all of which are still dominated by statistical errors.

[30] Non-negligible width difference $\Delta\Gamma_s$ in $B_s$-mesons leads to one more measurable quantity proportional to $\sinh(\Delta\Gamma t/2)$, also sensitive to the right-handed currents (see, e.g., Ref. [503]).

Table 93: Selection criteria for photon candidates in $B \to K_S^0 \pi^0 \gamma$ reconstruction.

| Detector region | forward | barrel | backward |
|---|---|---|---|
| Energy (MeV) | $> 103$ | $> 97$ | $> 72$ |
| E9/E25 | $> 0.80$ | $> 0.78$ | $> 0.71$ |
| minC2Hdist (cm) | 55 | 36 | 49 |

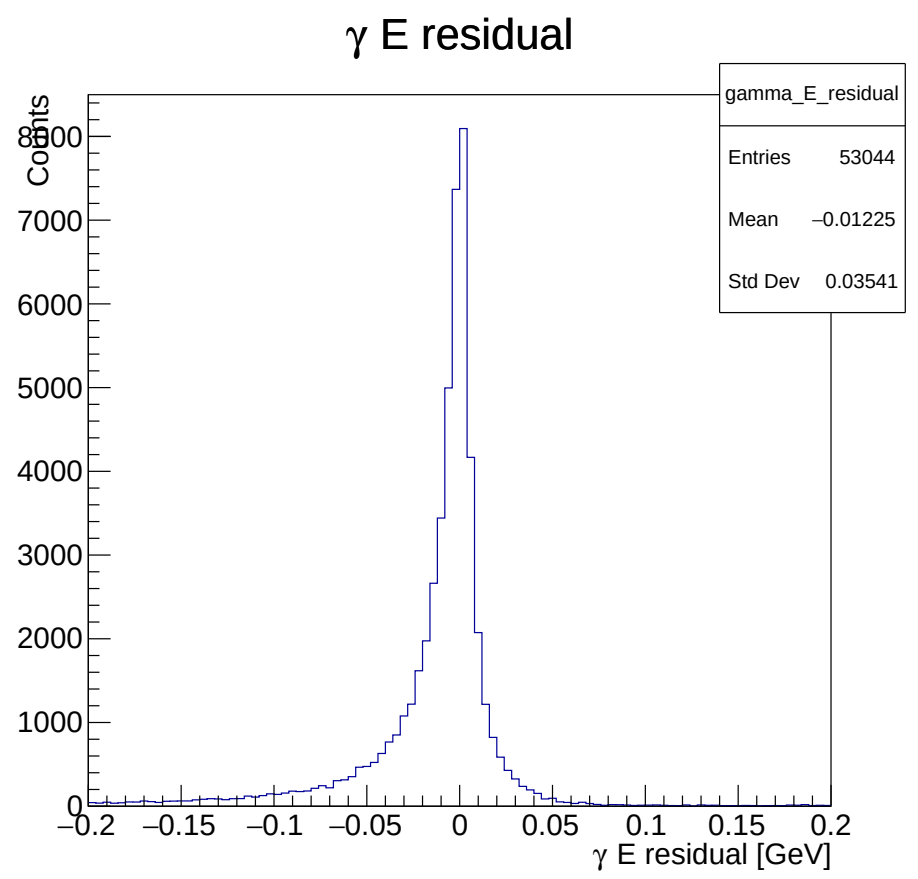

Fig. 119: Momentum residual distribution of all photons reconstructed in the events.

We focus in particular on the reconstruction of the signal side $B$ decay ($B^0_{sig}$) and on evaluating the $\Delta t$ resolution using a constraint on the beam spot size. Refined event selection strategies and the impact of beam backgrounds are not considered in this study.

*Signal reconstruction.* A Monte Carlo sample containing $10^4$ $B^0_{sig} \to K_S^0 \pi^0 \gamma$ decays, with the decay of the other $B$ meson in the event ($B^0_{tag}$) undergoing a generic decay has been generated and reconstructed. No beam background has been added to the simulation. To avoid effects due to possible incorrect assignment in the reconstruction of the intermediate resonances, all reconstructed particles are matched to the generated ones by using Monte Carlo truth information.

Photons are selected using four different variables, with selection criteria listed in Table 93. The $E9/E25$ variable is the ratio of deposited energies in the $3 \times 3$ and the $5 \times 5$ calorimeter cell blocks, while "minC2Hdist" is the minimum distance between the photon's estimated position and the closest track. Figure 119 shows the residuals of the momentum distribution of all the photons in the event. The long negative tail is due to calorimeter leakages and to interactions of photons with material in front of the calorimeter.

Neutral pions are reconstructed through the decay $\pi^0 \to \gamma\gamma$, with the invariant mass of the photon pair in the range [110, 150 MeV]. A kinematic mass fit, requiring a p-value $> 1‰$ (this is supposed to retain 99.9% of the true $\pi^0$ candidates) is then performed. The residuals distribution of the $\pi^0$ mass is shown in Figure 120 (left plot). The residuals are slightly negatively biased and the $\sigma$ value of the pull distribution (not shown here) is significantly larger than 1. This is due to an non optimised selection of photons, which will be improved with the next versions of the reconstruction algorithms.

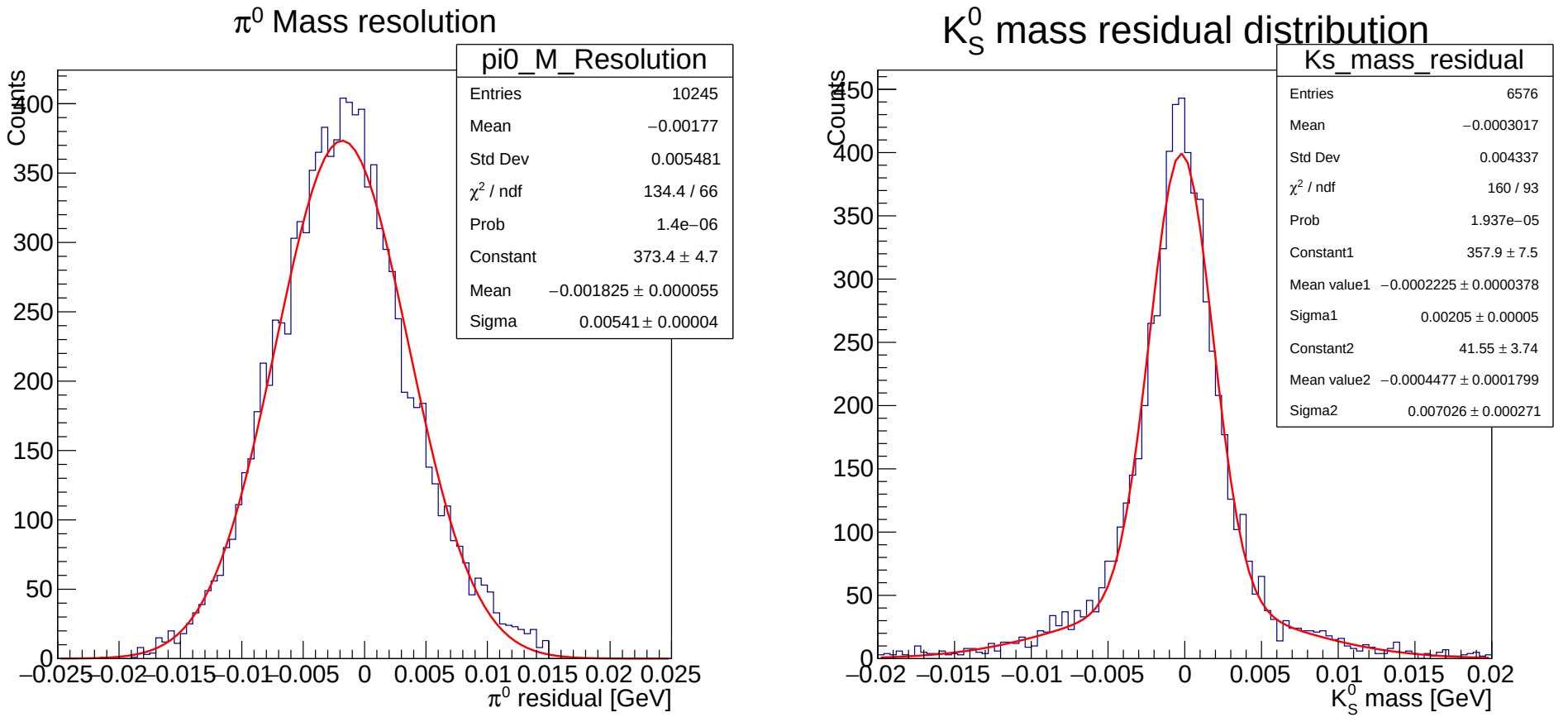

Fig. 120: Residuals of the reconstructed mass for the $\pi^0$ (left) and $K_S^0$ (right) candidates.

The reconstruction of $K_S^0$ is one of the crucial points of this analysis, because this is practically the only source of information to determine the vertex position of $B^0_{sig}$. For this reason $K_S^0$'s are reconstructed only through the decay into charged pions $K_S^0 \to \pi^+\pi^-$. Kaon candidates are selected in the invariant mass range between 400 MeV and 600 MeV; a vertex fit with a p-value greater than 1‰ is subsequently performed. The residuals distribution of the $K_S^0$ mass is shown in Figure 120 (right).

The $B^0_{sig}$ is reconstructed by requiring its mass to be between 5.0 GeV and 5.5 GeV and performing a vertex fit demanding p-value $>$ 1‰. The $K_S^0$ flight direction is extrapolated backwards and matched to the estimated region in which the $e^+e^-$ collisions take place. Given the non negligible flight length of the $B^0$ candidates and to avoid any biases in the time-dependent analysis, the beamspot constraint is applied only in the plane perpendicular to the boost direction, hence the name of *iptube* constraint. The ellipsoid in the transverse plane has semi-axes: $\sigma_x \simeq 6\ \mu$m, and $\sigma_y \simeq 42$ nm.

The $M_{\rm bc}$ (beam constrained mass) distribution and the $B^0_{sig}$ vertex residual distribution are reported in Figure 121 (left plot).

The distribution of the $B^0_{sig}$ vertex position along the boost axis (right plot of Fig. 121) is fitted with a combination of two gaussians with two very different widths. This is dominated by the resolution of the reconstruction of charged tracks. Kaons that decay outside the SVD (transverse flight distance $\rho > 3.8$ cm) are reconstructed from charged pions that are only seen by the CDC. Without SVD hit information, the precision on the $K_S^0$ vertices becomes significantly worse. The profile plot in Figure 122 shows the error on $B^0_{sig}$ vertex position with respect to the $K_S^0$ transverse flight distance (radius $\rho$).

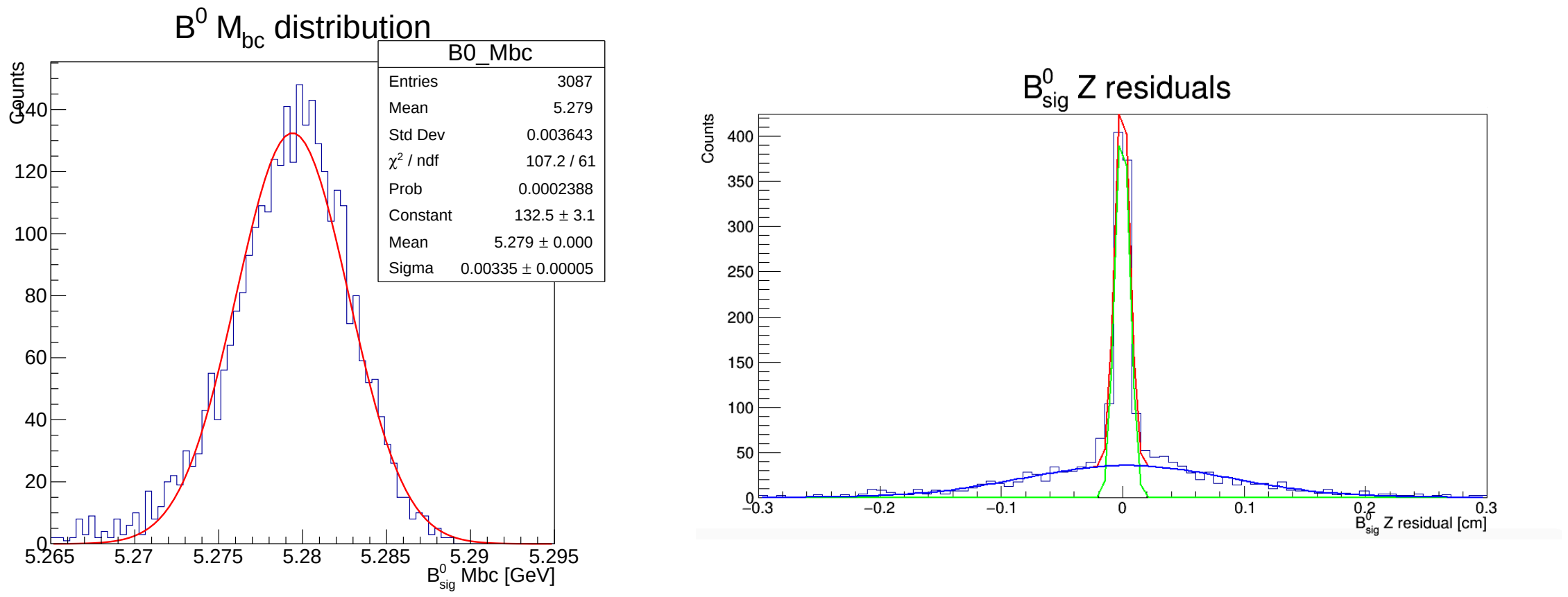

Fig. 121: Left: $M_{\rm bc}$ distribution of $B^0_{sig}$, fitted with a single gaussian giving $\sigma \sim 3.3$ MeV mass resolution. Right: residual distribution of the $B^0_{sig}$ vertex. The resolution provided by the core and the tail gaussian are respectively $\sim 58$ $\mu$m and $\sim 850$ $\mu$m.

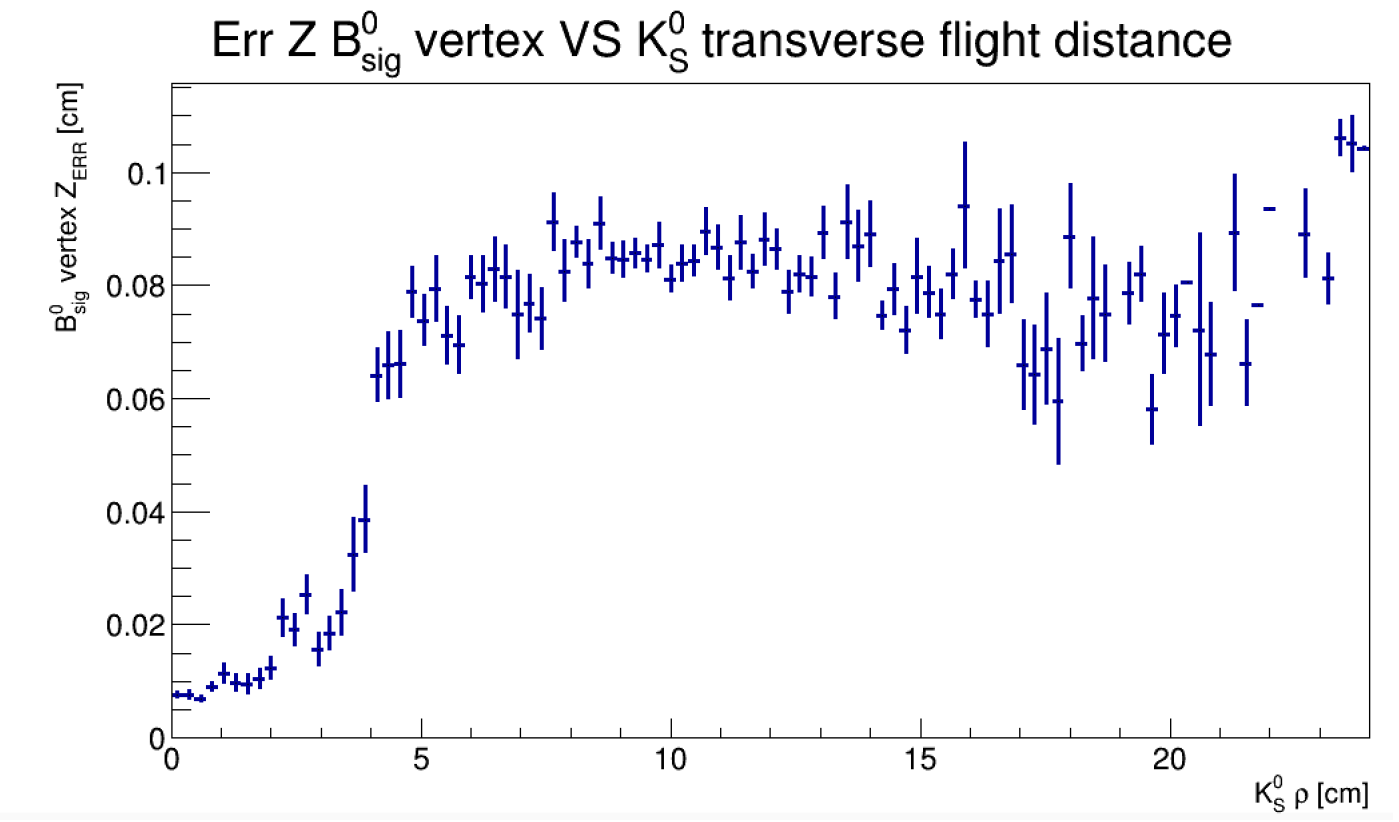

Fig. 122: Profile plot of the errors on the $z$ coordinate of the $B^0_{sig}$ vertex versus the $K^0_S$ transverse flight distance ($\rho$). The $K^0_S$'s decaying within the PXD volume have a significantly better resolution.

The step at around 3.8 cm, is visible, while there is no transition at 16 cm, corresponding to the inner radius of the CDC. This is due to the fact that tracking algorithms are still under development (we expect an improvement for $K^0_S$'s decaying within the SVD volume).

The other contribution to the uncertainty in $\Delta t$ measurement comes from the $B^0_{tag}$ vertex determination, which uses the standard algorithm that is common to the other time-dependent analyses. The resolution of the $B^0_{tag}$ vertex position along the boost axis is $\sim 42$ $\mu$m.

Table 94: Reconstruction efficiencies of particles involved in the decay of $B^0_{sig}$ (with the current version of the Belle II software).

| | $K^0_S$ | $\pi^0$ | $\gamma$ | $B^0$ |
|---|---|---|---|---|
| $\varepsilon^{reco}$ | 58.6 % | 53.7 % | 83.4 % | 26.2 % |

*Efficiency studies.* The reconstruction efficiency was studied using a Monte Carlo sample in which $B^0_{sig}$ is forced to decay to the $K^0_S\pi^0\gamma$ final state, while $B^0_{tag}$ decays to a $\nu\bar{\nu}$ pair. In this way any cross-feed between the two $B$ meson decays is avoided.

The obtained efficiencies are summarised in Table 94.

*$\Delta t$ resolution.* The distance of the $B^0_{sig}$ and $B^0_{tag}$ decay vertices along the boost direction ($\Delta z$) is related (with very good approximation) to $\Delta t$ via the formula: $\Delta z = \beta\gamma c\Delta t$, (with $\beta$ and $\gamma$ characterising the Lorentz boost of the $\Upsilon(4S)$ in the laboratory system). The $\Delta t$ resolution is measured from the residual distribution shown in Figure 123. The distribution is fitted with the sum of two gaussians where the narrow gaussian (core) has $\sigma \simeq 0.84$ ps while the wide one (tail gaussian) has $\sigma \simeq 9.2$ ps. The wide component can be suppressed by imposing that both the pions coming from $K^0_S$ have at least one SVD hit associated to their tracks. The $\Delta t$ residual distribution relative to the new sample of $K^0_S$ is shown in Figure 123.

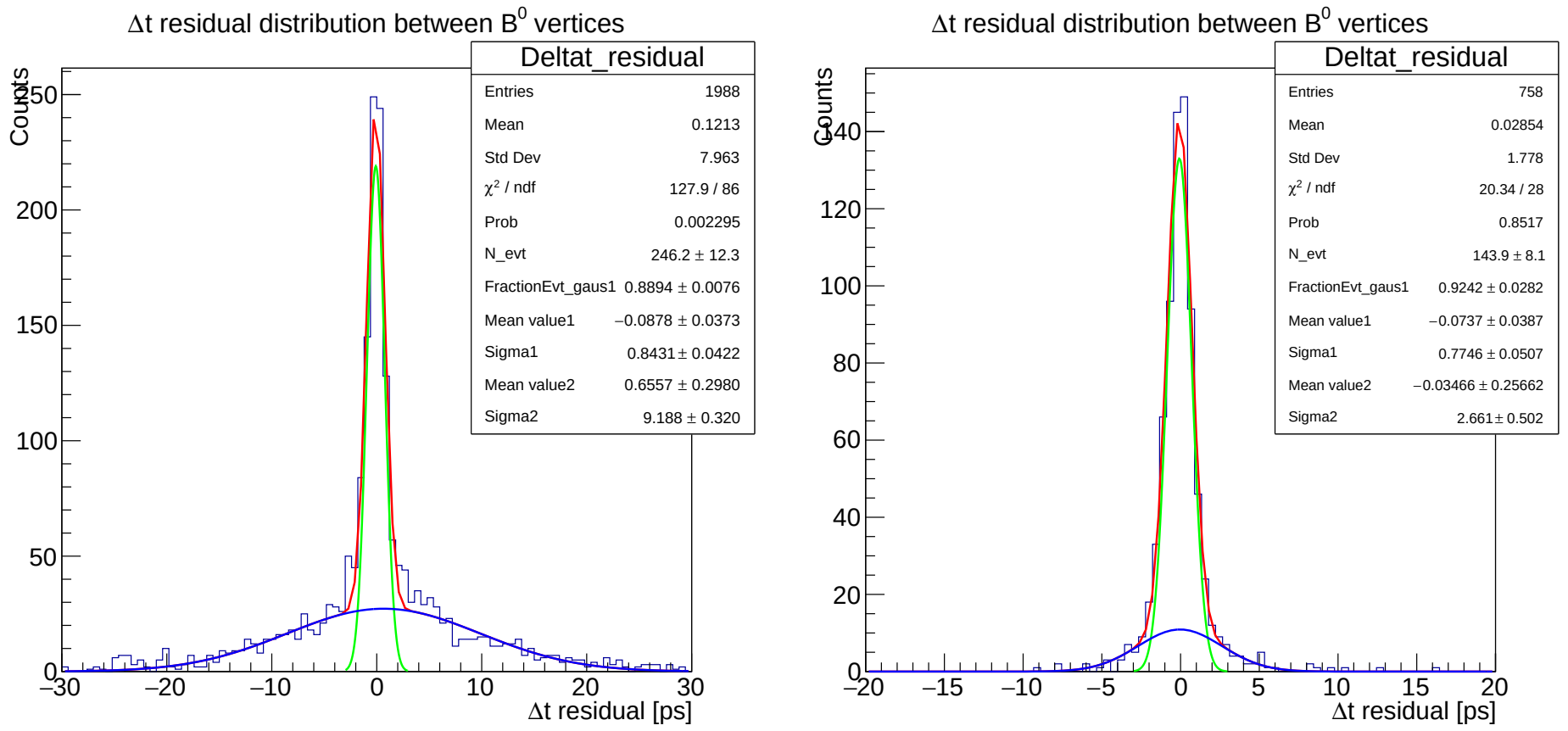

Fig. 123: $\Delta t$ residual distribution for $B \to K^0_S\pi^0\gamma$ without (left) and with (right) SVD hit requirements.

The results show a significant improvement but, selecting only the sub-samples of $K^0_S$ with better resolution, we can use only $\sim 35\%$ of the events. The resolution values are $\sim 0.77$ ps (from the core-gaussian) and $\sim 2.66$ ps (from the tail-gaussian). Requiring $M_{K\pi}$ <1.8 GeV (the region in which the Soft Collinear Effective Theory (SCET) framework can provide reliable predictions), the statistic drops by a further factor $\sim$6, but the residual distribution does not change significantly. Considering the resolution as the weighted average of the two $\sigma$ values, we obtain an average of $\simeq$0.94 ps.

Table 95: Sensitivity estimation on $S$ and $A$ parameters at different values of integrated luminosity.

| Int. Lum. $ab^{-1}$ | Stat(S) | Stat(A) |
|---|---|---|
| 2 | 0.15 | 0.10 |
| 10 | 0.07 | 0.05 |
| 50 | 0.031 | 0.021 |

Table 96: Extrapolated sensitivity for the $K_S^0\pi^0$ mode. The $\Delta t$ resolution is taken from the $K_S^0\pi^0\gamma$ study and we assume for this mode a reconstruction efficiency of 30%.

| Channel | Yield | $\sigma(S)$ | $\sigma(A)$ |
|---|---|---|---|
| | 1 $ab^{-1}$ | | |
| $K_S^0(\pi^\pm)\pi^0$ | 1140 | 0.20 | 0.13 |
| | 5 $ab^{-1}$ | | |
| $K_S^0(\pi^\pm)\pi^0$ | 5699 | 0.09 | 0.06 |

Table 97: Expected uncertainties on the $S$ and $A$ parameters for the channels sensitive to $\sin 2\phi_1$ discussed in this chapter for an integrated luminosity of 5 and 50 ab$^{-1}$. The present (2017) World Average [230] errors are also reported.

| | WA (2017) | | 5 ab$^{-1}$ | | 50 ab$^{-1}$ | |
|---|---|---|---|---|---|---|
| Channel | $\sigma(S)$ | $\sigma(A)$ | $\sigma(S)$ | $\sigma(A)$ | $\sigma(S)$ | $\sigma(A)$ |
| $J/\psi K^0$ | 0.022 | 0.021 | 0.012 | 0.011 | 0.0052 | 0.0090 |
| $\phi K^0$ | 0.12 | 0.14 | 0.048 | 0.035 | 0.020 | 0.011 |
| $\eta' K^0$ | 0.06 | 0.04 | 0.032 | 0.020 | 0.015 | 0.008 |
| $\omega K_S^0$ | 0.21 | 0.14 | 0.08 | 0.06 | 0.024 | 0.020 |
| $K_S^0\pi^0\gamma$ | 0.20 | 0.12 | 0.10 | 0.07 | 0.031 | 0.021 |
| $K_S^0\pi^0$ | 0.17 | 0.10 | 0.09 | 0.06 | 0.028 | 0.018 |

*Sensitivity studies.* An estimate of the sensitivity of Belle II on the $CP$ violation parameters $S$ and $A$ is obtained using a study based on pseudo-experiments, in which the expected $\Delta t$ resolution is used. The results, reported in Table 95, are very promising, especially considering that significant improvements are expected in the reconstruction software. On the other hand, the impact of physics and beam backgrounds still needs to be estimated.

*Extrapolation of the $K_S^0\pi^0$ sensitivity.* We estimate the sensitivity to the $S_{K_S^0\pi^0}$ and $A_{K_S^0\pi^0}$ parameters of the $K_S^0\pi^0$ mode analogously to what we have done in section 10.3.2. The vertex reconstruction position resolution is taken from the study of $K_S^0\pi^0\gamma$ presented above, and we assume a reconstruction efficiency of 30%, based on the performance of BaBar and Belle. The results are presented in Table 96.

### 10.6. Conclusions

We summarise in Table 97 the expected uncertainties to the $S$ and $A$ $CP$-violating parameters in the channels sensitive to $\sin 2\phi_1$ discussed in this chapter. For the $J/\psi K^0$ mode, we provide the estimate, dominated by systematic uncertainties, for the full 50 ab$^{-1}$ dataset. For the penguin dominated modes the estimates are based on an integrated luminosity of 5 ab$^{-1}$, for which we can safely assume that all the channels will still be dominated by the statistical uncertainties and the assumptions on which the current studies are based are valid. In the 5th and the last columns of Table 97 we also report the present HFLAV WA errors on each of the observables. For most of the penguin dominated modes Belle II is projected to reduce the WA errors by a factor of 2 to 3 already with 5 ab$^{-1}$.

We projected the uncertainty on the determination of $\phi_2$ considering the isospin analyses of $B \to \pi\pi$ and $B \to \rho\rho$. The $B \to \rho\pi$ system, which is usually considered together with $B \to \pi\pi$ and $B \to \rho\rho$, was not taken into account due to the difficulty of realistically simulating the full Dalitz plot analysis of $B^0 \to \pi^+\pi^-\pi^0$ in MC. The expected uncertainties on $\phi_2$ extracted via isospin analysis of $B \to \pi\pi$ and $B \to \rho\rho$ and via combined isospin analysis of these two decay systems are summarised in Table 98. The projections of the experimental errors and the central values of previous measurements that enter the isospin analysis of $B \to \pi\pi$ and $B \to \rho\rho$ are presented in Tables 91 and 92, respectively. Additionally, we performed a feasibility study for the novel time-dependent $CP$ analysis of the decay $B \to \pi^0\pi^0$. The uncertainty on the measurement of the time-dependent asymmetry $S_{\pi^0\pi^0}$ is estimated to be $\Delta S_{\pi^0\pi^0} = \pm 0.28 \pm 0.03$. Consequently, the current 8-fold ambiguity in the determination of $\phi_2$ performing the isospin analysis of $B \to \pi\pi$ will be reduced by factor 4 (see Fig. 116). It is also possible, that the values of $\phi_2$ extracted from the isospin analysis including $S_{\pi^0\pi^0}$ have a tension to the values expected within the SM (see Fig. 117). The sensitivity study of $B \to \pi^0\pi^0$ and the projections of previous measurements were performed for a total Belle II integrated luminosity of 50 ab$^{-1}$. So far, we did not consider isospin breaking effects on the projection of the sensitivity to $\phi_2$. Possible ways to extract the size of the bias in $\phi_2$ due to isospin breaking effects were discussed in Sec. 10.4. At present, isospin breaking effects can be only partially included. In principle, there are observables where the theoretical error is only of second order in isospin breaking and thus below the per-mill level. However, it may be impossible to measure them to the required level of accuracy.

Table 98: Current world average error [650] and expected uncertainties on the determination of $\phi_2$ performing isospin analyses of the decay systems $B \to \pi\pi$ and $B \to \rho\rho$ together with a combined isospin analysis of these two systems. For the current world average error, also the decay system $B \to \rho\pi$ was considered.

| Channel | $\Delta\phi_2$ [°] |
|---|---|
| Current world average | $^{+4.4}_{-4.0}$ |
| $B \to \pi\pi$ | 2.0 |
| $B \to \rho\rho$ | 0.7 |
| $B \to \pi\pi$ and $B \to \rho\rho$ Combined | 0.6 |

Finally, Table 99 summarizes the current and expected experimental and theoretical uncertainties on the CKM angles $\phi_1$ and $\phi_2$. We should stress once more that, as explained above, the theoretical uncertainties based on data-driven techniques cannot be precisely estimated at this stage.

Table 99: Summary of the current and expected sensitivities on the CKM angles $\phi_1$ and $\phi_2$. As explained in the text, theoretical uncertainties based on data-driven techniques (e.g. those releted to EWP amplitudes in the determination of $\phi_2$) can be given only with a large degree of uncertainty.

| | Current | 50 ab$^{-1}$ projection |
|---|---|---|
| $\phi_1$: | | |
| Experimental: | 0.7° | 0.2° |
| Theoretical - QCDF & pQCD | 0.1° | 0.1° |
| Theoretical - SU(3) | 1.7° | 0.8° |
| $\phi_2$: | | |
| Experimental: | 4.2° | 0.6° |
| Theoretical: | 1.2° | $< 1.0°$ |

# 11. Determination of the UT angle $\phi_3$

*Editors: M. Blanke, Y. Grossman, J. Libby*
*Additional section writers: J. Brod, G. Tetlalmatzi-Xolocotzi, I. Watson, J. Zupan*

## *11.1. Introduction*

This working group is dedicated to examining the Belle II potential to determine the unitarity triangle angle $\phi_3$ (also denoted as $\gamma$) defined as

$$\phi_3 \equiv -\arg\left(V_{ub}^* V_{ud} / V_{cb}^* V_{cd}\right), \tag{341}$$

in $B \to DK$, $B \to D\pi$, and related modes. In general when we talk about $B \to DK$ we refer to a family of related decays like $B^-$ decay into $DK$, $D^*K$, $DK^*$ and $D^*K^*$, as well as multi-body decays, as they are all sensitive to $\phi_3$ as well. Only the hadronic part of the amplitude is different.

The key feature of $B \to DK$ decays is that they arise solely from the interference of tree level diagrams of differing weak and strong phases. No $B$ mixing, nor penguin amplitudes, are involved. Here, $D$ represents a general superposition of $D^0$ and $\overline{D}^0$. The tree-level nature of the amplitudes involved in $B \to DK$ allows the theoretically clean extraction of $\phi_3$. Clearly, an improved knowledge of the unitarity triangle angle $\phi_3$ is very useful to further test the SM. The current precision on $\phi_3$ is an order of magnitude worse than that on $\phi_1$ [77] and it is the only measurement of the unitarity triangle that can be improved significantly by experimental advances alone.

We move to a very brief discussion of the main idea. Sensitivity to $\phi_3$ can be obtained by studying $CP$-violating observables in $B \to DK^-$ decays. There are two tree amplitudes contributing to $B^- \to DK^-$ decays: $B^- \to D^0 K^-$ and $B^- \to \overline{D}^0 K^-$. The amplitude for the second decay is both CKM and colour suppressed with respect to that for the first. The ratio of the suppressed to favoured amplitudes is written as

$$\frac{A(B^- \to \overline{D}^0 K^-)}{A(B^- \to D^0 K^-)} = r_B e^{i(\delta_B - \phi_3)} \ , \tag{342}$$

where $r_B \approx 0.1$ is the ratio of magnitudes and $\delta_B$ is the strong phase difference. The fact that the hadronic parameters $r_B$ and $\delta_B$ can be determined from data together with $\phi_3$ makes these measurements essentially free of theoretical uncertainties.

Several different types of $D$ decay are utilised to determine $\phi_3$. Examples of $D$ decays include $CP$-eigenstates, which was proposed by Gronau, London and Wyler [734, 735], Cabibbo-favoured (CF) and doubly-Cabibbo-suppressed (DCS) decays, which was proposed by Atwood, Dunietz and Soni [736, 737], self-conjugate modes, which was proposed by Giri, Grossman, Soffer and Zupan [738], and singly Cabibbo-suppressed (SCS) decays, which was proposed by Grossman, Ligeti and Soffer [739]. The different methods are known by their proponents' initials, which are given in Table 100, along with the $D$ final states that have so far been studied. Note that $K_S^0\phi$ has also been included in early GLW measurements but has been dropped from more recent analyses given that the same data forms part of the $K_S^0 K^+ K^-$ sample, which can be studied with the GGSZ method.

The remainder of this chapter is structured as follows. In Secs. 11.2 and 11.3 the theoretical limits on the accuracy of the measurements of $\phi_3$ and the scope for new physics to appear in these measurements is discussed, respectively. Section 11.4 provides a snapshot of the latest Belle II sensitivity studies related to $\phi_3$. In Sec. 11.5 we review the charm decay measurements that are required to reach the ultimate precision. Section 11.6 compares the current and future sensitivity of LHCb to that of Belle II, before the outlook and conclusions are given in Sec. 11.7.

Table 100: Methods and $D$ decay modes used in $B^- \to DK^-$ and $B^- \to D^*K^-$ measurements. Those in parentheses have not been published by Belle.

| Type of $D$ decay | Method name | $D$ final states studied |
|---|---|---|
| $CP$-eigenstates | GLW | $CP$-even: $K^+K^-$, $\pi^+\pi^-$; $CP$-odd $K_S^0\pi^0$, $K_S^0\eta$ |
| CF and DCS | ADS | $K^\pm\pi^\mp$, $K^\pm\pi^\mp\pi^0$, $(K^\pm\pi^\mp\pi^+\pi^-)$ |
| Self-conjugate | GGSZ | $K_S^0\pi^+\pi^-$, $(K_S^0K^+K^-)$, $(\pi^+\pi^-\pi^0)$, $(K^+K^-\pi^0)$, $(\pi^+\pi^-\pi^+\pi^-)$ |
| SCS | GLS | $(K_S^0K^\pm\pi^\mp)$ |

### *11.2. The ultimate precision*

In the original formulations of the methods for extracting $\phi_3$ from $B^- \to DK^-$ and $B_d \to DK_S^0$ the small effects due to $D-\bar{D}$ and $K-\bar{K}$ mixing, as well as due to $CP$ violation in the $D$ and $K$ sectors, were neglected. Several studies of the impact of mixing and direct $CP$ violation in charm decays have been made since then [740–749]. These studies show that $\phi_3$ can be extracted without bias as long as appropriate modifications of the formalism are made and the measured values of the mixing and direct $CP$ violation parameters are included as external inputs. Even if the effect of mixing is neglected the size of the induced bias is less than $1^\circ$ [748]. The inclusion of direct $CP$ violation in $D$ decays does require that at least one decay mode has no direct $CP$ violation. This breaks the phase shift reparametrisation symmetry which would otherwise prevent the model-independent determination of $\phi_3$ [745]. It is expected that Cabibbo allowed $D$ decays have vanishing small direct $CP$ asymmetry, a fact that can be checked experimentally.

Measurements of $\phi_3$ can also be made using the $B^- \to D\pi^-$ decay mode, which has sensitivity to $\phi_3$ in the same manner as $B^- \to DK^-$. However, the size of the direct $CP$ asymmetry is much smaller due to the ratio $r_B$ of the suppressed to favoured amplitudes being very small, approximately 0.005. The reduced sensitivity due to the smaller interference is compensated to some degree by the much larger branching fraction for $B^- \to D\pi^-$ compared to $B^- \to DK^-$ [77]. $D$ mixing and direct $CP$ violation must be accounted for carefully in $B^- \to D\pi^-$ measurements of $\phi_3$ because the bias on the extracted value of $\phi_3$ would be $\mathcal{O}(10^\circ)$ otherwise [748].

For both the $B \to DK$ and the $B \to D\pi$ modes, the irreducible theoretical uncertainty is due to diagrams of higher-order in the electroweak expansion. Second-order weak box diagrams have a dependence on the CKM parameters that differs from the tree diagrams. This can induce a shift, $\delta\phi_3$, in the extracted value of $\phi_3$. An effective-field-theory calculation including a resummation of the large logarithms $\log(m_b/m_W)$ in the corrections to the Wilson coefficients that gives $\delta\phi_3 \sim 2 \times 10^{-8}$ [750]. Long distance contributions are at most a factor of a few larger than the calculated short-distance contribution. For $B \to DK$, the relative shift in $\phi_3$ due to neglecting these weak-box diagrams is $\lesssim 10^{-7}$ [750]. This is many orders of magnitude below the present experimental precision as well as the one anticipated at Belle II. The estimate of the analogous uncertainty for the extraction of $\phi_3$ from $B \to D\pi$ decays suffers from a possible approximate cancellation in the leading term so that the relative shift can be up to $\delta\phi_3 < 10^{-4}$ [751]. More data can exclude the possibility of a large cancellation, so that the estimate of $\delta\phi_3$ can be made more precise also for the $B \to D\pi$ mode.

### *11.3. New physics in $\phi_3$*

The traditional way of testing for new physics (NP) contributions to the angle $\phi_3$ is to compare the value obtained from tree-level decays with the one obtained from penguin-dominated processes, and to look for deviations. This strategy relies on negligible NP contributions to SM tree-level processes.

Recent studies, however, have shown that, when state-of-the-art experimental measurements and theoretical determinations are taken into account, NP contributions of up to $\mathcal{O}(40\%)$ and $\mathcal{O}(20\%)$ to the tree-level Wilson coefficients $C_1$ and $C_2$ respectively are not excluded [752, 753]. Allowing for general complex contributions $\Delta C_{1(2)}$ to the tree-level Wilson coefficients $C_{1(2)}$ and using that $|C_1(m_b)/C_2(m_b)| \approx 0.22$, and $|\Delta C_1(m_b)/C_2(m_b)|$ and $|\Delta C_2(m_b)/C_2(m_b)|$ are small, Eq. (342) should be modified according to

$$r_B e^{i(\delta_B - \phi_3)} \to r_B e^{i(\delta_B - \phi_3)} \cdot \left[1 + (r_{A'} - r_A)\frac{\Delta C_1}{C_2}\right] , \tag{343}$$

with

$$r_{A(A')} = \frac{\langle \bar{D}^0 K^- | Q_1^{\bar{c}us}\big(Q_1^{\bar{u}cs}\big) | B^- \rangle}{\langle \bar{D}^0 K^- | Q_2^{\bar{c}us}\big(Q_2^{\bar{u}cs}\big) | B^- \rangle} , \tag{344}$$

and

$$\begin{aligned} Q_1^{\bar{u}_1 u_2 d_1} &= (\bar{u}_1^\alpha b^\beta)_{V-A}(\bar{d}_1^\beta u_2^\alpha)_{V-A} , \\ Q_2^{\bar{u}_1 u_2 d_1} &= (\bar{u}_1^\alpha b^\alpha)_{V-A}(\bar{d}_1^\beta u_2^\beta)_{V-A} . \end{aligned} \tag{345}$$

Here, $\alpha, \beta$ are colour indices and $u_1, u_2$ label the combinations of the up type quarks $u, c$.

A fit to the semileptonic asymmetries $a_{sl}^{d(s)}$, the $B_s$ decay width difference $\Delta\Gamma_s$, the branching ratio for the process $B \to X_s\gamma$ and the total life time of $B$-hadrons, as well as various observables associated with the hadronic decays $B \to \pi\pi$, $B \to \rho\rho$, $B \to \rho\pi$, $\bar{B}_d^0 \to D^{*+}\pi^-$ and $\bar{B}_d^0 \to D^{(*)0}h^0$ (with $h^0 = \pi^0, \eta, \omega$), such as indirect CP asymmetries and hadronic to semileptonic ratios, yields the allowed ranges for $\Delta C_1$ and $\Delta C_2$ quoted above (see [752] and [753] for details).

From Eq. (343) we can see that the shift in the CKM phase $\phi_3$ is sensitive mainly to new complex weak phases in $\Delta C_1$. The main impediment in evaluating the impact on $\phi_3$ quantitatively is the unknown hadronic matrix elements in Eq. (343). Naive colour counting gives $r_A \approx \mathcal{O}(1)$, $r_{A'} \approx \mathcal{O}(N_c = 3)$, while naive factorisation yields

$$r_A \approx \frac{f_D F_0^{B\to K}(0)}{f_K F_0^{B\to D}} \approx 0.4. \tag{346}$$

There are large uncertainties in this determination, but it is unlikely that the two ratios $r_A$ and $r_{A'}$ cancel accidentally. Using $r_A - r_{A'} \approx -0.6$ as a conservative (and rough) estimate, it follows that deviations in $\phi_3$ of $\mathcal{O}(4°)$ are consistent with the current experimental constraints [753]. A more detailed statistical study leads to similar conclusions [754]. It also reveals how the shifts in $\phi_3$ due to possible NP at tree level can be reduced if the theoretical and experimental status of different $B$-physics observables and hadronic parameters are improved, for instance the observable $\sin(2\beta)$, extracted from the transition $B \to J/\psi K_S^0$.

The possible deviations on $\phi_3$ induced by NP at tree level are close to the current precision achieved in the direct measurement. This result is a strong motivation for the $1^\circ$ precision being pursued by Belle II, and for further study of the NP contributions and the associated theoretical uncertainties.

### *11.4. Belle II sensitivity study*

In this section we summarise the status of the Belle II studies related to $\phi_3$. Sections 11.4.1 to 11.4.3 report a preliminary study of the determination of $\phi_3$ using the GGSZ analysis of $B^\pm \to [K_S^0\pi^+\pi^-]_D K^\pm$, which is the Golden Mode for Belle II. This study uses the Belle II simulation, though some aspects are not yet optimised. Section 11.4.4 then describes an extrapolation of the combination of the GGSZ measurements with the ADS and GLW measurements of $B \to D^{(*)}K$ based on Belle measurements to give an indication of the precision that can be reached by Belle II.

*11.4.1. Model-Independent Dalitz Analysis Overview.* The $B^\pm \to DK^\pm$ mode using the GGSZ method was the most sensitive channel to $\phi_3$ at Belle. Therefore, our initial efforts are concentrated on this decay. The sensitivity in this method arises when a $D$ decays to a self-conjugate three-body final state that allows a comparison of the Dalitz plot for $B^+$ and $B^-$ from which $\phi_3$, $r_B$ and $\delta_B$ can be determined from a single decay [738, 755]. This technique with $B^\pm \to [K_S^0\pi^+\pi^-]_D K^\pm$ has been seen to be the most sensitive single analysis at Belle because of the significant $D^0 \to K_S^0\pi^+\pi^-$ branching fraction of $(2.85 \pm 0.20)\%$ [77] and good $K_S^0$ reconstruction efficiency. Therefore, we perform a sensitivity analysis in this mode to understand how the $\phi_3$ measurement will evolve as a function of the amount of data collected at Belle II.

The decay rate for $B^\pm \to [K_S^0\pi^+\pi^-]_D K^\pm$ can be written as:

$$\begin{aligned} \mathrm{d}\Gamma_{B^-}(m_+^2, m_-^2) \propto \quad & |A_+|^2 + r_B^2|A_-|^2 + \\ & 2r_B|A_+||A_-|\left(\cos\delta_D\cos(\delta_B+\phi_3) - \sin\delta_D\sin(\delta_B+\phi_3)\right)\mathrm{d}p \,, \end{aligned} \quad (347)$$

where $m_+^2$ ($m_-^2$) is the invariant mass of the $K_S^0\pi^+$ ($K_S^0\pi^-$) from the $D$ decay, $A_+ = A_D(m_+^2, m_-^2)$ and $A_- = A_D(m_-^2, m_+^2)$ are the $D^0$ and $\bar{D}^0$ decay amplitudes, $\delta_D$ is the phase difference between $A_+$ and $A_-$, $\delta_B$ is the phase difference between $D^0$ and $\bar{D}^0$ diagrams in the $B^- \to DK^-$ decay and $\mathrm{d}p$ is an infinitesimal phase-space element. Similarly, the $B^+$ decay is found by substituting $\phi_3 \leftrightarrow -\phi_3$ and $A_+ \leftrightarrow A_-$.

Thus, in order to measure $\phi_3$ with these decays, we need to know the phase difference between $A_D(m_+^2, m_-^2)$ and $A_D(m_-^2, m_+^2)$ at each point in the Dalitz plot. This suggests that the strategy for the measurement is to, first, construct a model for $A_D(m_+^2, m_-^2)$. This can be done by reconstructing $D$ mesons produced in the decay $D^{*\pm} \to D\pi^\pm$, which gives a known flavour state for the $D$. An amplitude model can be fit to the flavour sample to determine $A_D(m_+^2, m_-^2)$. Then, one can use this model as input to fit a $B^\pm \to DK^\pm$ sample to the parameters $r_B, \delta_B$, and $\phi_3$. It has been found, though, that in order to eliminate bias due to the physical boundary $r_B = 0$, it is better to convert the physics parameters into Cartesian

coordinates for the fit:

$$(x_\pm, y_\pm) = r_B(\cos(\delta_B \pm \phi_3), \sin(\delta_B \pm \phi_3)) ,$$

which can then be reinterpreted in terms of $r_B, \delta_B$, and $\phi_3$.

This method has been followed at previous experiments, for example in a Belle analysis for $B^\pm \to DK^\pm$ and $B^\pm \to D^* K^\pm$ [756]. The analysis found, however, that the amplitude model leads to a systematic uncertainty of 8.9° on $\phi_3$. With the much larger data sample anticipated at Belle II, the statistical uncertainty on $\phi_3$ will reduce below 1°. Therefore, a method which eliminates the model systematic uncertainty, replacing it with a model-independent uncertainty measured from data, was proposed [738].

To remove the model-dependency and obtain degree-level precision, a binned approach is used. The Dalitz plot is divided into $2N$ bins numbered $-N$ to $N$, excluding 0, where interchanging bin $i$ with bin $-i$ corresponds to interchanging $m_-^2$ with $m_+^2$. Further, we choose positive bins to lie in the region $m_-^2 > m_+^2$. The binning used in the current analysis is shown in Fig. 124. The number of events expected in a given bin $i$ can then be found by integrating the amplitude over the phase space $\mathcal{D}_i$ of the Dalitz bin. For the flavour Dalitz plot in $D^{*\pm} \to D\pi^\pm$ the number of events $K_i$ is simply

$$K_i \propto \int_{\mathcal{D}_i} |A_+|^2 \mathrm{d}p ,$$

where $\mathcal{D}_i$ indicates the integration is over the $i^{\mathrm{th}}$ bin. The values of $K_i$ can be measured directly then used as inputs to the $B^\pm \to DK^\pm$ analysis. The number of $B^+ \to DK^+$ events in each bin is given by

$$\begin{aligned} N_i^+ &\propto \int_{\mathcal{D}_i} |A_-|^2 + r_B^2 |A_+|^2 + 2|A_+||A_-|(x_+ \cos\delta_D + y_+ \sin\delta_D)\mathrm{d}p , \\ &\propto K_{-i} + r_B^2 K_i + 2\sqrt{K_i K_{-i}}(x_+ c_i - y_+ s_i) . \end{aligned}$$

Here we introduced the amplitude averaged phase variations over the Dalitz plot bins

$$\begin{aligned} c_i = \quad c_{-i} &= \frac{\int_{\mathcal{D}_i} |A_+||A_-| \cos\delta_D \mathrm{d}p}{\sqrt{\int_{\mathcal{D}_i} |A_+|^2 \mathrm{d}p}\sqrt{\int_{\mathcal{D}_i} |A_-|^2 \mathrm{d}p}} , \\ s_i = -s_{-i} &= \frac{\int_{\mathcal{D}_i} |A_+||A_-| \sin\delta_D \mathrm{d}p}{\sqrt{\int_{\mathcal{D}_i} |A_+|^2 \mathrm{d}p}\sqrt{\int_{\mathcal{D}_i} |A_-|^2 \mathrm{d}p}} . \end{aligned}$$

A similar expression can be derived for the number of events $N_i^-$ expected in $B^- \to DK^-$ with replacements as before. If we also explicitly split the positive and negative Dalitz bins, and introduce an overall normalisation $h_B$ then the equations for the number of events expected in each of the Dalitz bins is given by the equations:

$$\begin{aligned} N_i^+ &\propto K_{-i} + r_B^2 K_i + 2\sqrt{K_i K_{-i}}(x_+ c_i + y_+ s_i) , \\ N_i^- &\propto K_i + r_B^2 K_i + 2\sqrt{K_i K_{-i}}(x_- c_i - y_- s_i) . \end{aligned} \tag{348}$$

This approach was used by Belle to measure $\phi_3$ in $B^\pm \to DK^\pm$ [757], $D \to K_S^0 \pi^+ \pi^-$, and by LHCb in $B^\pm \to DK^\pm$, $D \to K_S^0 \pi^+ \pi^-$ and $D \to K_S^0 K^+ K^-$ [758].

The phase-difference parameters have been measured by CLEO-c from quantum correlated $D\bar{D}$ decays of the $\psi(3770)$, which is discussed further in Sec. 11.5. The binning can be set to minimise the phase variation or to optimise sensitivity to $\phi_3$ including information about the $B$ decay yields by also taking into account the amplitude variation; the latter approach is adopted here. Setting the binning based on these criteria requires a model in order to divide the Dalitz plot. However, a model-dependency is not introduced by this procedure. Instead, if the model is incorrect, then the binning selected will simply not be optimal, since the direct measurement of the parameters is still valid and does not depend on the details of how the binning was derived.

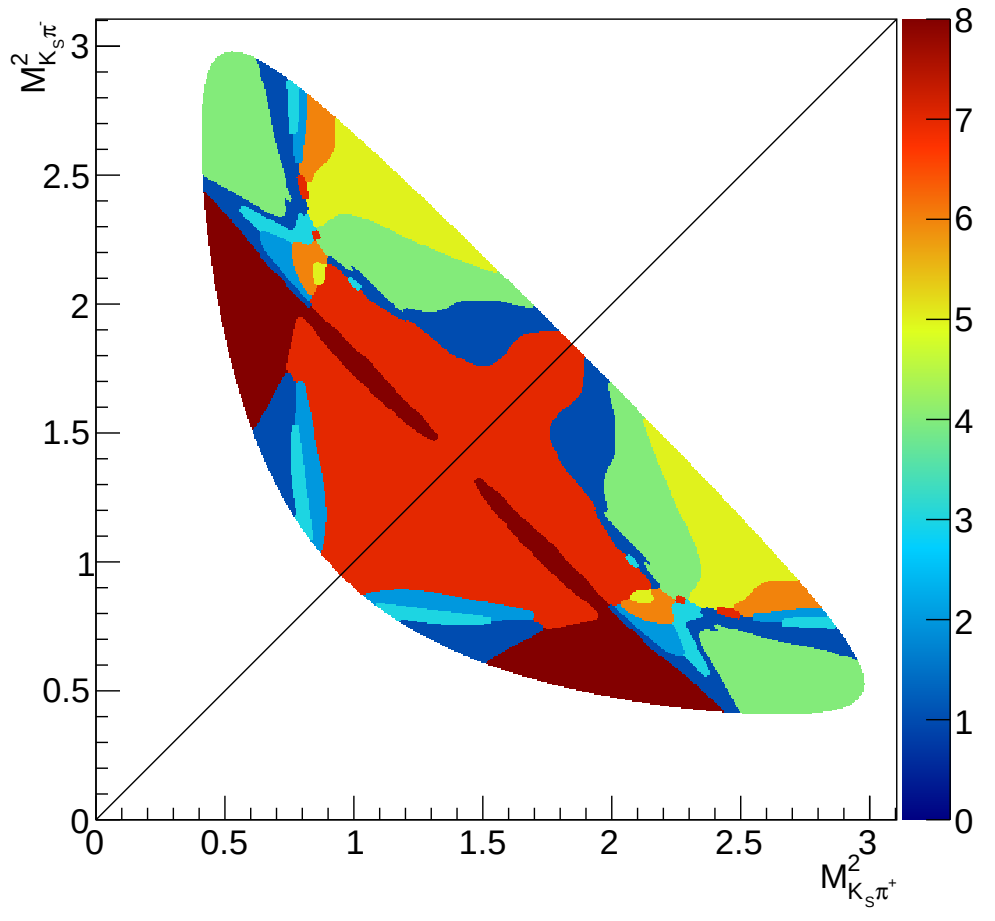

Fig. 124: Dalitz binning used for the $D \to K^0_S\pi^+\pi^-$ analyses.

*11.4.2. Signal reconstruction at Belle II.* The dataset used is the MC5 production dataset without beam background, and corresponds to integrated luminosity of 2 ab$^{-1}$. For purely hadronic modes which do not use calorimeter-based observables, like the one studied here, the beam background is expected to have little effect. The selection is not as refined as that of the Belle analysis due to the lack of multivariate continuum suppression and tuning of the $K^0_S$ reconstruction. However, the implementation of the whole analysis even with a non-optimal selection gives a conservative estimate of what can be achieved.

From the dataset, a sample for further study is selected based on a loose final state particle requirement, using the decay chains $D^{*\pm} \to \pi^{\pm}D$, $B^{\pm} \to D\pi^{\pm}$ and $B^{\pm} \to DK^{\pm}$ with $D \to K^0_S\pi^+\pi^-$ in all cases. The final state hadronic particles are selected with the Belle II framework's standard particle selection. The $K^0_S$ is selected with a BDT trained on the $K^0_S$ flight distance, mass, the minimum distance between the vertex and the electron beam direction, and vertex $p$-value, after an initial dipion selection satisfying $0.477 < M(\pi^+\pi^-) < 0.518$ GeV. After selecting $D$ candidates, a mass-vertex fit using RAVE is performed to improve the resolution of the Dalitz plot variables.

In the $D^{*\pm} \to D\pi^{\pm}$ channel, it is additionally required that $D$ candidates have $1.82 < M(D) < 1.9$ GeV and $\Delta M = M(D\pi^{\pm}) - M(D) < 0.155$ GeV and $p_D > 1.8$ GeV. The momentum selection is chosen so that the mean of the momentum spectrum of the $D^{*\pm}$ and $B$ signal events are the same; this partially eliminates any systematic effects due to the differing acceptance of the two samples over the Dalitz plot. In the $B^{\pm} \to DK^{\pm}$ channel, the additional requirements are $\Delta E < 0.15$ GeV, $M_{\rm bc} > 5.25$ GeV and $1.85 < M(D) < 1.88$ GeV.

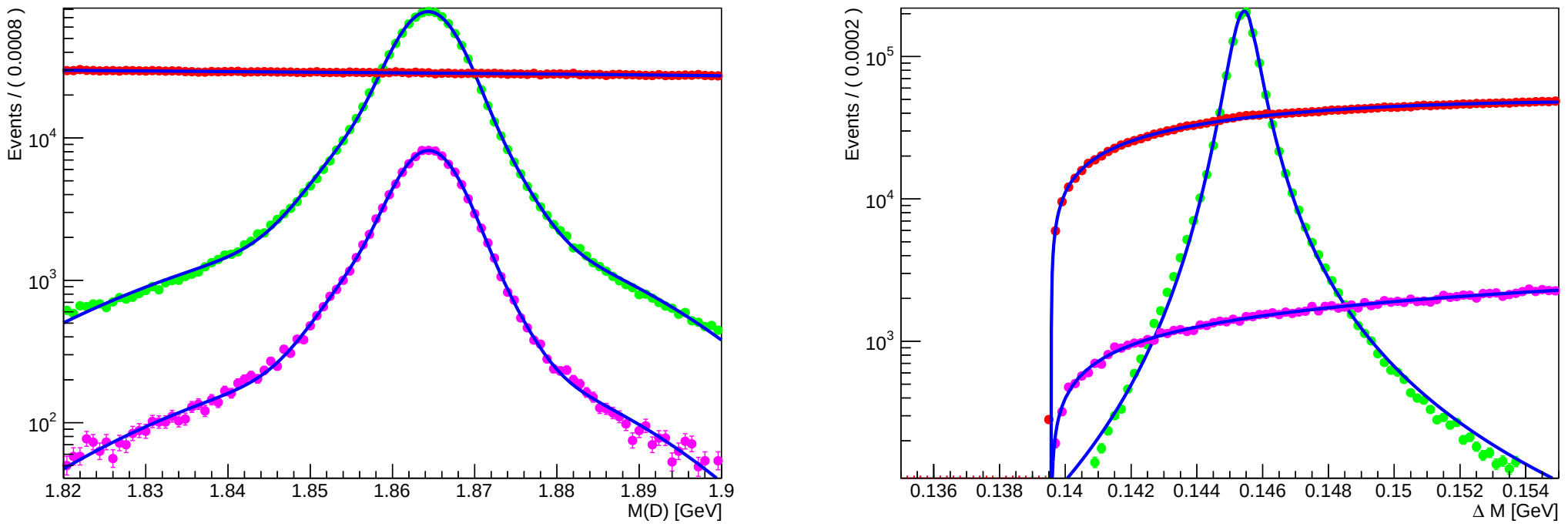

Fig. 125: Fits to the signal (green), fake slow pion background (blue) and combinatorial background (red) components of the $D^*$ in $M(D)$ (top) and $\Delta M$ (bottom) distributions.

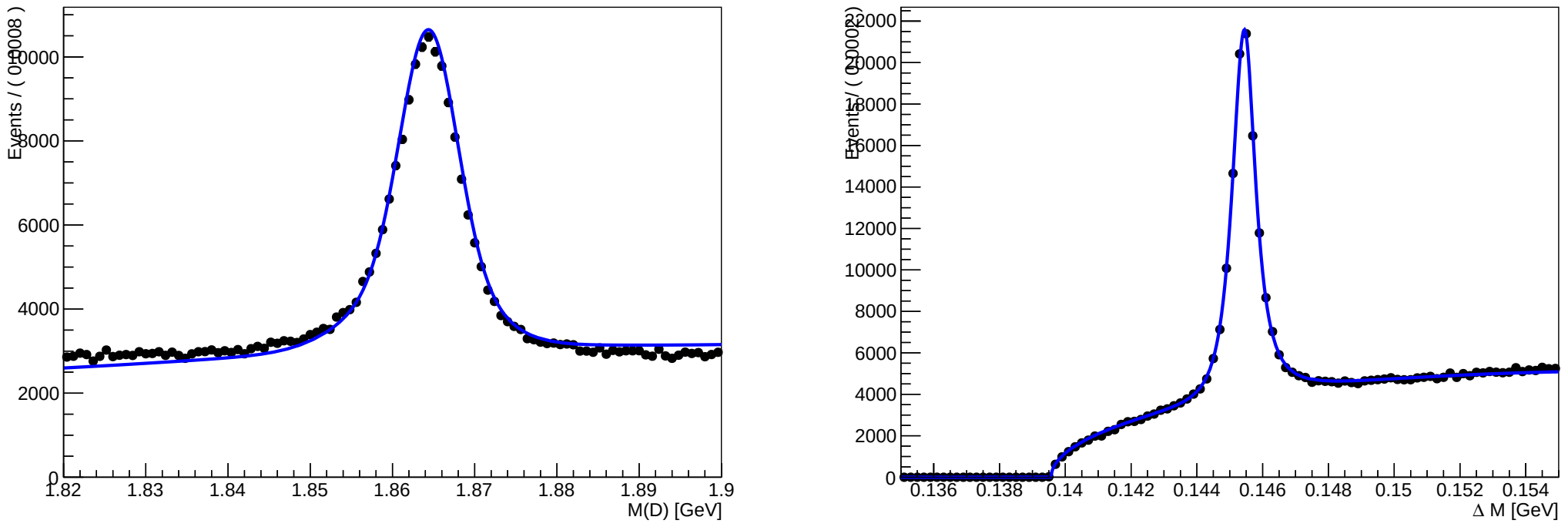

Fig. 126: Example fit to bin 4 of $D^{*\pm} \to D\pi^{\pm}$ in generic Belle II MC. Left shows the overall fit to $D$ mass, right shows $\Delta M = M(D\pi^{\pm}) - M(D)$.

Following the previous Belle analysis of the GGSZ mode, the $D^{*\pm}$ sample is fit in $M(D)$ and $\Delta M = M(D\pi^{\pm}) - M(D)$ to extract the number of signal events in each Dalitz bin. There are two background types: combinatorial and a correctly reconstruct $D$ candidate (true $D$) paired with a slow pion not originating from a $D^*$. The $D$ mass distribution is modelled by a sum of a Gaussian and two bifurcated Gaussians in the $D^*$ and true $D$ background component, and a third-order Chebyshev polynomial for the combinatorial background. The $\Delta M$ component is modelled by a bifurcated Student's $t$-function in the $D^{*\pm}$ component, and an Argus background function for the true $D$ and purely combinatorial components. These distributions are fixed by fitting to the full generic MC separated into the three components by matching the reconstructed $D^{*\pm}$ to the event generator information. Figure 125 shows the fits to the components.

The full generic MC is then divided into Dalitz bins and the number events for each component is obtained by fitting the sum of components. Figure 126 shows an example fit in one of the Dalitz bins.

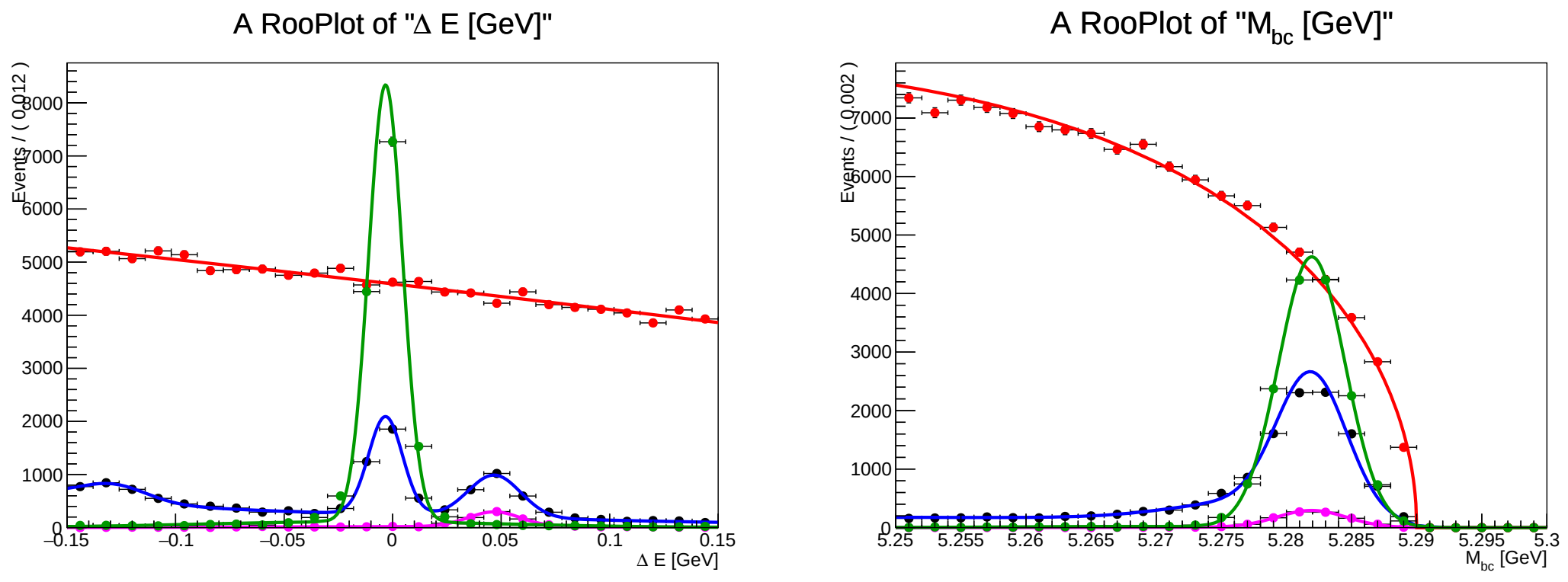

Fig. 127: Fits to the signal and background components of the Belle II Monte Carlo in $B^{\pm} \to [K_S^0\pi^+\pi^-]_D K^{\pm}$. Left shows $\Delta E$ and right shows $M_{\rm bc}$. The red component is from events reconstructed from $e^+e^-$ to $u$, $d$, $s$ or $c$ quarks pairs, the green component from a dedicated MC sample of signal $B^{\pm} \to [K_S^0\pi^+\pi^-]_D K^{\pm}$ events and the blue from arbitrary $B\bar{B}$ (not excluding signal) in the generic MC.

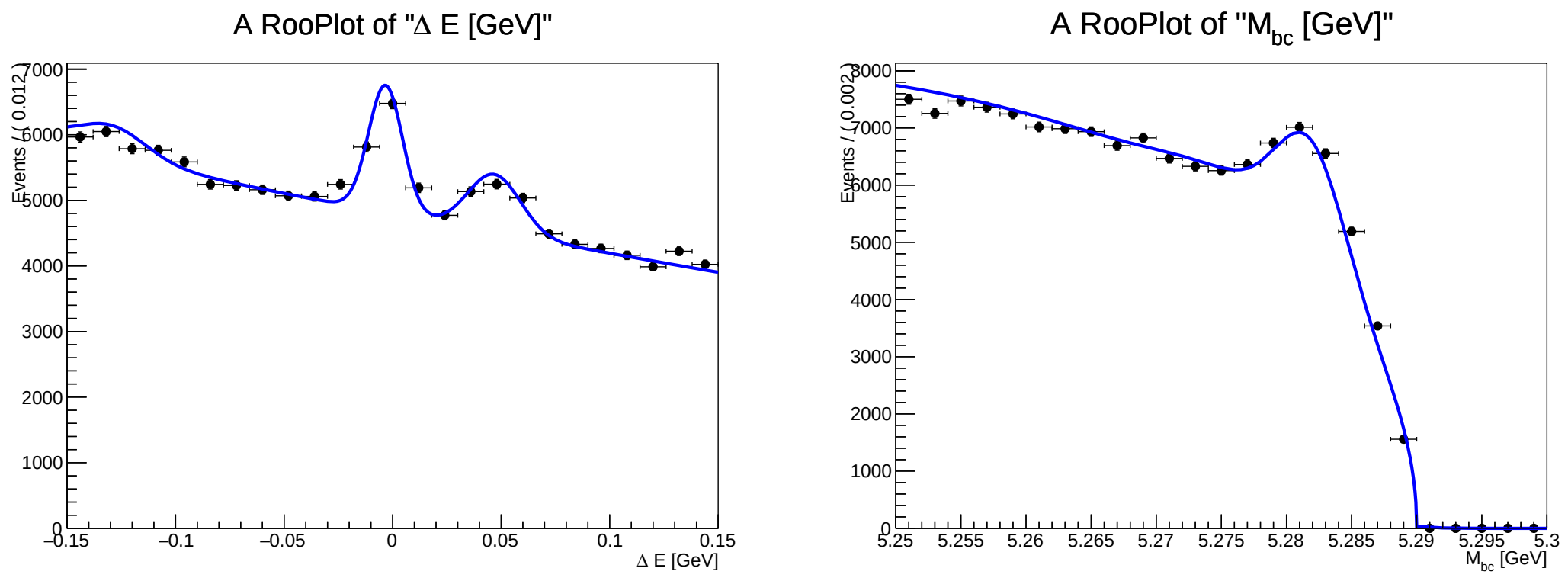

Fig. 128: Overall fit to the generic Monte Carlo in the $B^{\pm} \to [K_S^0\pi^+\pi^-]_D K^{\pm}$ mode. The fit is subsequently performed in Dalitz bins in order to measure the $(x, y)$ physics parameters which can be used extract $\phi_3$.

The signal and background components in $B^{\pm} \to DK^{\pm}$ are separated by fitting in two dimensions of $M_{\rm bc}$ and $\Delta E$. The signal $B^{\pm} \to DK^{\pm}$ component is modelled with a three Gaussians in $\Delta E$ and a correlated Gaussian in $M_{\rm bc}$, where the Gaussian mean of the $M_{\rm bc}$ is allowed to vary with $\Delta E$. The peaking $B^{\pm} \to D\pi^{\pm}$ background component, where the $\pi$ is misidentified as a $K$, is modelled with the sum of two Gaussians in $M_{\rm bc}$ and a sum of two Gaussian distributions in $\Delta E$, without correlation between the components. The generic $B\bar{B}$ component is modelled with a sum of a Gaussian and an exponential function in $\Delta E$ and in $M_{\rm bc}$ a sum of an Argus background and a Gaussian for the fully reconstructed component. The continuum background is modelled with a second-order Chebyshev in the $\Delta E$ and an Argus function in $M_{\rm bc}$. The fits to these components are shown in Fig. 127. A subsequent fit to the whole of the MC with fixed shapes for the components is presented in Fig. 128. In a more realistic analysis the abundant $B \to D\pi$ sample that has limited sensitivity to $\phi_3$ will act as a control sample on which to tune the probability density functions that will be used in the fit.

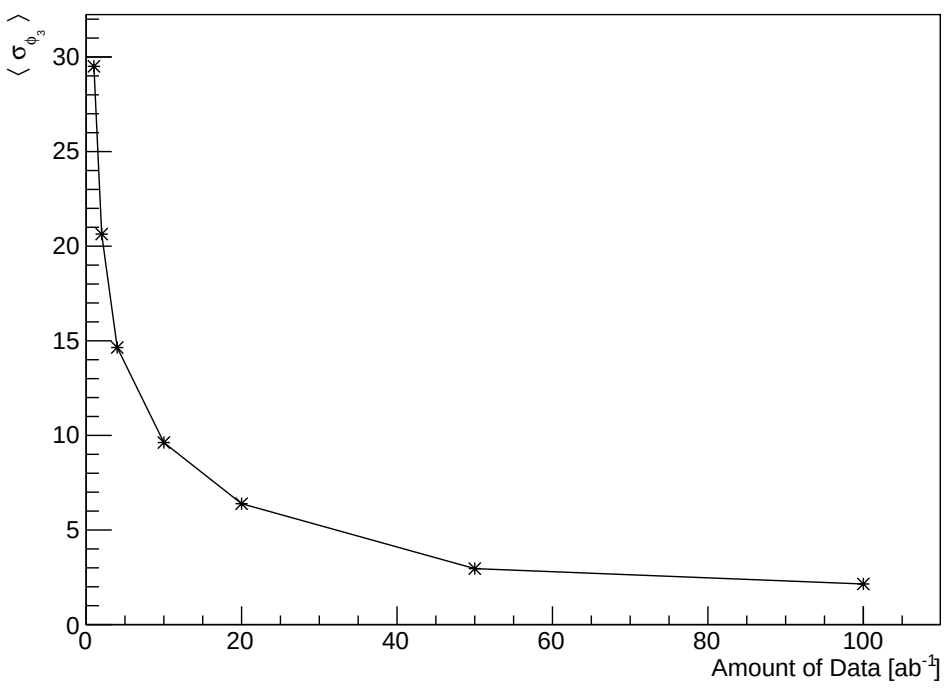

Fig. 129: The top figure shows the expected uncertainty (based on toy Monte Carlo studies) versus luminosity on $\phi_3$.

The whole MC sample is then divided into Dalitz bins, and the MC is fit again to measure the $(x, y)$ parameters. For the background components, the number of events of the component in each separate bin is allowed to vary freely. For the $B^{\pm} \to DK^{\pm}$ signal, the overall number of events is allowed to vary, while the number of events in the individual bins is governed by Eq. 348.

*11.4.3. Sensitivity to $\phi_3$ with increased luminosity.* To study the sensitivity to $\phi_3$ versus luminosity, toy Monte Carlo studies were performed. The signal and background components were each individually binned into an $M_{\rm bc} \times \Delta E$ grid, and an uncorrelated Dalitz bin array. The grid and Dalitz bin were then sampled the expected number of times for a given luminosity to produce a distribution for each individual component at that luminosity. These distributions were added together then the fit procedure of the previous section was performed to obtain an $(x, y)$ measurement. This procedure was repeated several hundred times per luminosity to build up an $(x, y)$ distribution expected at that luminosity.

The standard deviations of these toy distributions are taken as the expected uncertainty for a given luminosity. From these widths, and taking the current world average [759] *[UT fit average]* of $\phi_3$ parameters to fix the underlying true $(x, y)$, the $\phi_3$ resulting from these $(x, y)$ uncertainties is derived. This is done by generating pairs of $(x, y)$ for $B^+$ and $B^-$ using the $(x, y)$ derived from the UTFit parameters as the mean and the uncertainties from the toys as a width (assuming no correlation) and running this procedure several hundred times to generate a distribution for $\phi_3$ whose width we take as the expected $\phi_3$ uncertainty. Figure 129 shows how the expected uncertainty on $\phi_3$ scale with luminosity based on these toy Monte Carlo studies. It shows that the expected uncertainty with an integrated 50 ab$^{-1}$ is approximately $3^\circ$.

There are several possible future refinements of this study:

(1) including additional channels such as $D^0 \to K^0_S K^+ K^-$ and $B^+ \to D^{*0} K^+$;
(2) better signal-to-background separation by including a continuum suppression variable in the signal extraction fit; and

(3) derive the $\phi_3$ estimators from a likelihood profile fit of the $(x, y)$ that includes information about the correlation information for each of the several hundred toys.

However, this preliminary study clearly demonstrates the excellent capabilities of Belle II to determine $\phi_3$ from this mode.

*11.4.4. Extrapolation of the combination of measurements to Belle II luminosities.* The value of $\phi_3$ from a combination of Belle measurements alone is $(68\pm13)^\circ$ (see *e.g.* [760]) and is dominated by the GGSZ measurement of $B^- \to D^{(*)}(K_S^0\pi^+\pi^-)K^-$ [756], which should be considered the **Golden Mode** for Belle II. However, there have also been measurements using the ADS and GLW techniques [761–763] that have non-negligible weight in the combination. This includes an ADS/GLW analysis of $B^+ \to D^*(D\{\gamma, \pi^0\})K^+$ [762], which has only been measured at the $e^+e^-$ $B$ factories. Therefore, $\phi_3$ programme at Belle II must at least include all these modes and possibly others (see Sec. 11.7) to realise its full potential.

To be quantitative we will just restrict ourselves to the measurements so far made and extrapolate this combination to Belle II luminosities. Therefore, the expectation is a sensitivity of $3.6^\circ$ and $1.6^\circ$ with datasets corresponding to 10 ab$^{-1}$ and 50 ab$^{-1}$, respectively. The most important systematic uncertainties are related to the inputs from charm physics, which will be discussed in Sec. 11.5, the signal extraction fits and backgrounds from charmless $B$ decay. The latter two sources can be controlled using the $B^+ \to D\pi^+$ sample and sidebands of the $M_D$ distribution, respectively, so they should scale with the statistical uncertainty.

## *11.5. Auxiliary measurements*

The precise determination of $\phi_3$ using $B^- \to DK^-$ is reliant upon external inputs from the charm sector. The accurate determination of charm-mixing parameters [764] means that any bias from this source in the determination of $\phi_3$ can be corrected for as discussed in Sec. 11.2. In addition, $D$ meson branching fractions of both CF and DCS decays provide important inputs to ADS measurements [765, 766].

However, the most important auxiliary measurements are related to $D$ decay strong-phases, which are an essential input to interpret the measurements related to $\phi_3$. In principle these parameters could be extracted from the $B$ data along with $\phi_3$, $\delta_B$ and $r_B$, but the sensitivity to $\phi_3$ would be diluted. Therefore, measurements of the strong-phases are taken from elsewhere.

The strong-phase difference between the $D^0$ and $\overline{D}^0$ decays to $K^+\pi^-$ is required for the two-body ADS measurement and it is accurately determined using the combination of charm-mixing measurements [764]. For multibody ADS measurements two parameters must be determined due to the variation of the strong-phase difference over the allowed phase-space: the coherence factor $R$ and average strong-phase difference $\delta_D$. Recently there has been a new analysis to determine the $R$ and $\delta_D$ for $D \to K^-\pi^+\pi^0$ and $D \to K^-\pi^+\pi^+\pi^-$ [767], which uses quantum-correlated $D^0\overline{D}^0$ pairs produced at the $\psi(3770)$. (For a comprehensive review of quantum-correlated measurements relevant to $\phi_3$ see Ref. [768].) At the $\psi(3770)$ the $D$ decay of interest is tagged in events where the other $D$ decays to a $CP$-eigenstate, a state with a kaon of opposite or same-sign charge as the signal or $K_{S,L}^0\pi^+\pi^-$. The last of these tags is an addition since the first determination of $R$ and $\delta_D$ reported by the CLEO-c collaboration [769]. The updated results are used to perform the combinations reported elsewhere in these proceedings.

The model-independent GGSZ method requires two parameters related to the strong-phase difference to be determined for each bin of the Dalitz plot. Such measurements have been reported by the CLEO Collaboration [770] using a data sample corresponding to an integrated luminosity of 818 pb$^{-1}$. These measurements have been used by both the Belle [757] and LHCb [758] collaborations to determine $\phi_3$ from $B^- \to DK^-$ data. The systematic uncertainty on $\phi_3$ related to the statistical precision of the CLEO measurements is currently 2°, which is not dominant, but it will become much more significant with the future running of LHCb and Belle II. Therefore, improvements in the measurements of the strong phase parameters are desirable. BESIII has accumulated an integrated luminosity of 2.92 fb$^{-1}$ at the $\psi(3770)$ which is 3.5 times larger than that analysed by CLEO. Preliminary results for the $D \to K^0_S\pi^+\pi^-$ parameters using the same binning as CLEO have been reported [771], which give a significant improvement in the statistical uncertainty on the measurements. BESIII can accumulate around 4 fb$^{-1}$ of integrated luminosity per year of running at the $\psi(3770)$; therefore, a two year run at the $\psi(3770)$ by BESIII would reduce the uncertainty on $\phi_3$ from the determination of strong phases in the GGSZ method to a negligible level.

Quantum-correlated measurements are also opening up new pathways to determining $\phi_3$. A measurement of the $CP$ content of $D \to \pi^+\pi^-\pi^0$ and $D \to K^+K^-\pi^0$ [772] using the full CLEO-c $\psi(3770)$ data set has shown that $D \to \pi^+\pi^-\pi^0$ is $(96.8 \pm 1.7 \pm 0.6)\%$ $CP$-even. Therefore, this mode can be used as an additional GLW measurement to augment $D \to h^+h^-$, given it has a significantly larger branching fraction [77]. Most recently a preliminary study of using $D \to K^0_S\pi^+\pi^-\pi^0$ as a GLW and GGSZ mode has been reported [773]. This mode has a large branching fraction of 5.2% [77] and is largely $CP$-odd with a $CP$-even fraction of only $(0.246 \pm 0.018)$, which has been measured using the full CLEO-c data set. Therefore, this mode can be used in a GLW analysis. Furthermore, by binning the five-dimensional phase space, the values of $c_i$ and $s_i$ can be determined in the quantum correlated data, which then allows a GGSZ type measurement. This has been done with nine bins using the CLEO-c data. A toy simulation study based on these quantum-correlated measurements indicates a statistical uncertainty of 3.5° maybe possible with a Belle II sample of 50 ab$^{-1}$. There is no reliable amplitude model of this final state to guide the choice of binning to maximise $\phi_3$ sensitivity as there is for $D \to K^0_S\pi^+\pi^-$; this means there is scope to improve the sensitivity of this mode to $\phi_3$ if an amplitude model is developed.

### *11.6. Review of LHCb $B \to D^{(*)}K^{(*)}$ measurements*

LHCb have recently updated their $\phi_3$ average using the data collected at a centre-of-mass energy of 7 TeV and 8 TeV. The combination of $B \to DK$ modes gives $\phi_3 = (72.2^{+6.8}_{-7.3})^\circ$ [774] the most precise determination from a single experiment. The balance of the contributions to the average at LHCb is somewhat different due to the lower relative selection efficiency for $K^0_S$ in the forward hadronic environment. Here GGSZ and ADS/GLW, including the four-body final states $K^-\pi^+\pi^-\pi^+$ and $\pi^+\pi^-\pi^+\pi^-$, are on an almost equal footing in terms of sensitivity to $\phi_3$.

An extrapolation of these results under a variety of assumption about the size of the available BESIII data set has been performed [775]. These studies predict a precision of around 4° after Run 2 and a precision of around a 1° after the phase-1 upgrade. Therefore, the precision possible with an upgraded LHCb and Belle II is very similar and is a true area of competition between the two experiments. The future prospect of Belle II sensitivity to $\phi_3$ is plotted in Fig. 130.

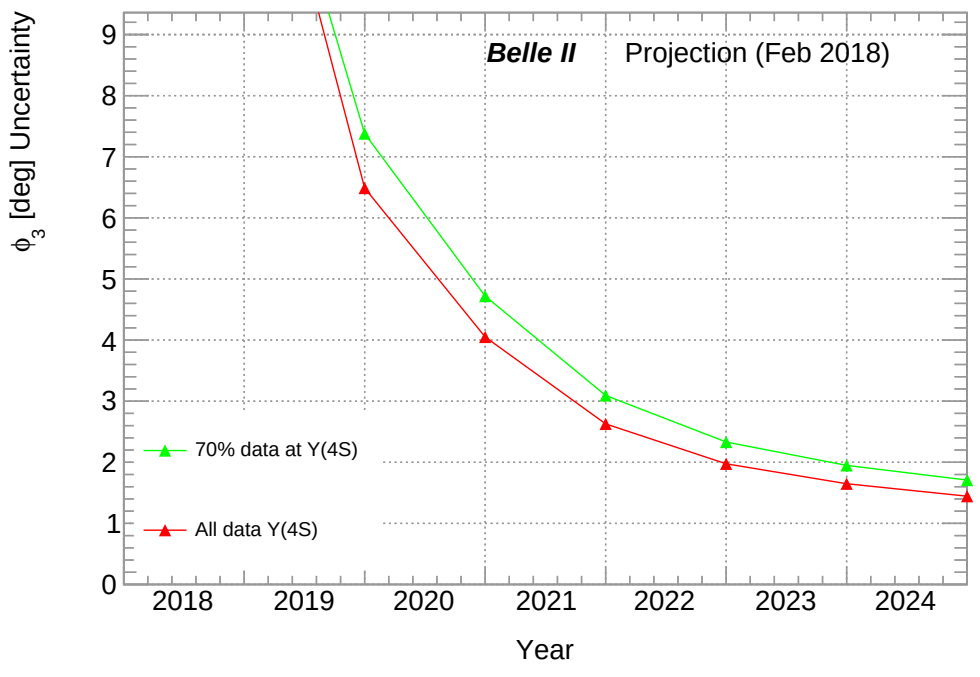

Fig. 130: The future prospect of Belle II sensitivity to $\phi_3$.

### *11.7. Outlook and conclusions*

We have reviewed the exquisite theoretical cleanliness of determining $\phi_3$ in $B \to DK$ decays, hence allowing these measurements to be a standard candle against which other Standard Model CKM measurements can be compared. Further, the current level of precision of the measurements of $\phi_3$ is such that there can still be NP contributions at the level of 4°. Both of these observations provide the motivation to produce a degree-level precision measurement at Belle II.

The first study of the sensitivity of Belle II based on the full simulation has been made for the GGSZ analysis of $B^+ \to (K^0_S\pi^+\pi^-)_D K^+$ using a generic MC sample corresponding to an integrated luminosity of 2 ab$^{-1}$. Based on these studies a 3° precision is anticipated from a 50 ab$^{-1}$ data set inline with naive expectations. Further, there is still much scope to refine the analysis further. However, the anticipated precision based on the combination of all Belle results, including GLW and ADS as well, is not completely dominated by the GGSZ measurement alone, such that once the full combination is extrapolated the uncertainty is expected to be 1.6° with a 50 ab$^{-1}$ dataset. A caveat is that the extrapolation is predicated on there being sufficient BESIII data collected at the $\psi(3770)$, approximately 10 fb$^{-1}$, to determine the strong-phase difference parameters required. If such a sample does not exist there will be a few degree systematic uncertainty that will limit the impact of the GGSZ measurements on the combination.

However, further improvements are possible as several $B \to DK$ modes have not been exploited at Belle. In particular there are several modes with significant branching fractions that have neutral particles $K^0_{S,L}$, $\pi^0$ and $\eta$ that are yet to be exploited:

- $CP$-even: $\pi^0\pi^0$, $K^0_L\pi^0$, $K^0_S\pi^0\pi^0$, $K^0_S\eta\pi^0$, $K^0_S K^0_S K^0_S$;
- $CP$-odd: $K^0_S K^0_S K^0_L$, $\eta\pi^0\pi^0$, $\eta'\pi^0\pi^0$, $K^0_S K^0_S\pi^0$, $K^0_S K^0_S\eta$; and
- Self-conjugate: $K^0_L\pi^+\pi^-$, $K^0_L K^+K^-$, $K^0_S\pi^+\pi^-\pi^0$, $\pi^+\pi^-\pi^0\pi^0$.

The improved particle identification, energy resolution in the electromagnetic calorimeter and in the continuum suppression algorithms at Belle II will all benefit the selection of these modes. The fully charged four-body modes $D \to K^-\pi^+\pi^+\pi^-$, $D \to \pi^+\pi^-\pi^+\pi^-$, $D \to K^-K^-\pi^+\pi^+$ are also of interest, but LHCb will reconstruct significantly larger samples because of the absence of neutral particles in the final state. Another type of measurement that appears to have excellent potential is the double-Dalitz analysis of $B^0 \to D(K_S^0\pi^+\pi^-)K^+\pi^-$ [776], which so far has received no attention at Belle or Belle II.

In summary, $\phi_3$ is the single place where a purely experimental improvement can be made in determining the Unitarity Triangle at Belle II, that in turn will allow for comparison with NP sensitive measurements. The sensitivity has been established using the Golden Mode $B^+ \to (K_S^0\pi^+\pi^-)_D K^+$ and extrapolating the Belle measurements. However, given the improvements in detector performance and the many modes that are yet to be explored there is scope to go beyond this baseline sensitivity.

# 12. Charmless Hadronic $B$ Decays and Direct $CP$ Violation

*Editors: M. Beneke, C-W. Chiang, P. Goldenzweig*
*Additional section writers: B. Pal, G. Bell, C. Bobeth, H-Y. Cheng, A. Datta, T. Feldmann, T. Huber, C-D. Lu, J. Virto*

## *12.1. Introduction*

Charmless hadronic final states in $B$ decays have branching fractions of order $10^{-5}$ or less, since either the final state is reached by the $b \to u$ transition, which is suppressed by the small CKM matrix element $|V_{ub}|$, or the transition is loop-suppressed. Charmless decays are a good place to observe direct $CP$ violation, since the smallness of the leading amplitude often implies that another amplitude with a different CKM factor is of similar size. If the two amplitudes also have a substantial (strong) phase difference, this leads to sizeable direct $CP$ violation, which has indeed been observed. There is a large number of potentially interesting decay modes. There are 130 (quasi) two-body final states, where the two mesons are from the ground-state pseudo-scalar or vector nonet alone. The number multiplies when more exotic mesons or three-body final states are considered. Belle II is expected to considerably extend the knowledge of such hadronic final states.

The theoretical description of hadronic $B$ decays starts from the effective weak-interaction Lagrangian for $\Delta B = 1$ transitions,

$$\mathcal{L}_{\text{eff}} = -\frac{G_F}{\sqrt{2}} \sum_{p=u,c} \lambda_p^{(D)} \sum_i C_i\, Q_i^p, \tag{349}$$

where $Q_i^p$ denotes the so-called tree, QCD and electroweak penguin, and dipole operators, and $\lambda_p^{(D)} \equiv V_{pb}V_{pD}^*$ ($p = u, c$, $D = d, s$). Any $B$ decay to a final state $f$ can then be expressed in the form

$$A(\bar{B} \to f) = \lambda_u^{(D)} A_f^u + \lambda_c^{(D)} A_f^c\,, \tag{350}$$

where $A_f^p$ are the matrix elements of the above Lagrangian. The Wilson coefficients $C_i$ include the physics from the highest scales, including $M_W$, down to the scale $m_b$, and their calculation is under complete theoretical control. $\mathcal{L}_{\text{eff}}$ above assumes the Standard Model (SM), and the convention that $\lambda_t^{(D)}$ is eliminated by the unitarity relation $\lambda_u^{(D)} + \lambda_c^{(D)} + \lambda_t^{(D)} = 0$. The structure of the operators $Q_i$, the values of their Wilson coefficients, and the flavour structures can be modified in extensions of the SM. The decay amplitude $A(\bar{B} \to f) = \langle f|\mathcal{L}_{\text{eff}}|\bar{B}\rangle$ then requires the computation of the hadronic matrix elements $\langle f|Q_i|\bar{B}\rangle$ of the local operators $Q_i$. When $f$ consists of two or more hadrons this is a difficult strong interaction problem, which cannot be solved with lattice QCD. Systematic expansions can be performed in the heavy quark mass, that is in $\Lambda/m_b$, where $\Lambda$ is the strong interaction scale, or in light quark masses, that is in $m_q/\Lambda$ ($q = u, d, s$). The corresponding theoretical approaches are referred to as the "factorisation" and "SU(3)" frameworks, respectively.

This chapter summarises recent developments in the field of charmless hadronic $B$ decays with possible relevance to Belle II physics, collected from contributions to the B2TIP workshop series (see Chapter 1 for details). It does not provide a comprehensive discussion of the field. A compact introduction to the theory of charmless decay and a summary of experimental results from BaBar and Belle can be found in the "Charmless $B$ decays" chapter of Ref. [2]. The present chapter provides only a few projections for Belle II results and uncertainties, since the large number of potential final states and observables, many of them not measured before, do not allow a more detailed study.

The chapter is divided into two main parts. The larger first part deals with aspects of two-body or quasi two-body final states, and starts with a discussion of global SU(3) analyses of charmless $B$ decays. This is followed by several contributions related to the factorisation framework, a section on $\pi K$ final states and the corresponding ones with vector mesons, a brief discussion on $CP$ violation in $B_s$ decays, specifically $B_s \to K_S^0 K_S^0$, and concludes with a focus on polarisation and angular distributions in vector-vector final states. The second part features two sections devoted to the theoretically less developed and experimentally less explored subject of three-body decays.

## *12.2. SU(3) analysis of two-body decays*

*[Contributing Author: Cheng-Wei Chiang]*

When the difference between the masses of the light quarks (up, down and strange) is neglected, QCD exhibits an exact SU(3) flavour symmetry. For charmless $B$ decays this implies that the hadronic decay amplitudes (with their CKM factors stripped off) of many different decays are related to a few reduced matrix elements multiplied by known Clebsch-Gordan coefficients. This approach to hadronic decays of heavy mesons was developed in the 1980's [777–780], originally for charmed meson decays, and has since been used extensively for hadronic $B$ decays [781] (and references therein). In practice, the SU(3) treatment amounts to an expansion in $m_s/\Lambda$. Since usually only the leading SU(3) symmetric term is considered, the approximation amounts to ignoring the effect of the strange quark mass on the long-distance dynamics.

Valuable knowledge about strong dynamics in various decay topologies has been obtained via this approach. With sufficient data it enables us to extract the (reduced) decay matrix elements directly from data without reference to further theoretical assumptions. The results include the effects of the strong interaction to all orders, including long-distance rescattering. This provides a good guide to the sizes and strong phases of certain amplitudes, such as the unexpectedly large colour-suppressed tree amplitude discussed below. On the other hand, being primarily data-driven through fits, the SU(3) approach does not by itself provide an explanation of such findings and the result depends on the quality of the experimental data.

Once the hadronic SU(3) amplitudes are determined from sufficient data, one can use the obtained information to make predictions for yet unmeasured observables. For example, the best-fit results from a fit to $B^0$ and $B^+$ decays have been used to predict the $B_s$ decays and obtained a fairly good agreement with those few observables that have been measured.

Within the limitations of its approximation, the quality of the SU(3) analysis depends on the quality of the data. Belle II is expected to collect a substantial amount of new information that should allow us to extend the SU(3) analysis to final states with two light vector mesons ("$VV$ modes"), and to attain a precision that requires the inclusion of sub-leading amplitudes, as well as SU(3)-breaking effects on the dominant amplitudes.

In the following we briefly discuss the formalism and the main results from the analysis of present pseudoscalar-pseudoscalar ($PP$) and pseudoscalar-vector ($PV$) final-state data. We then provide a short outlook on relevant issues for Belle II.

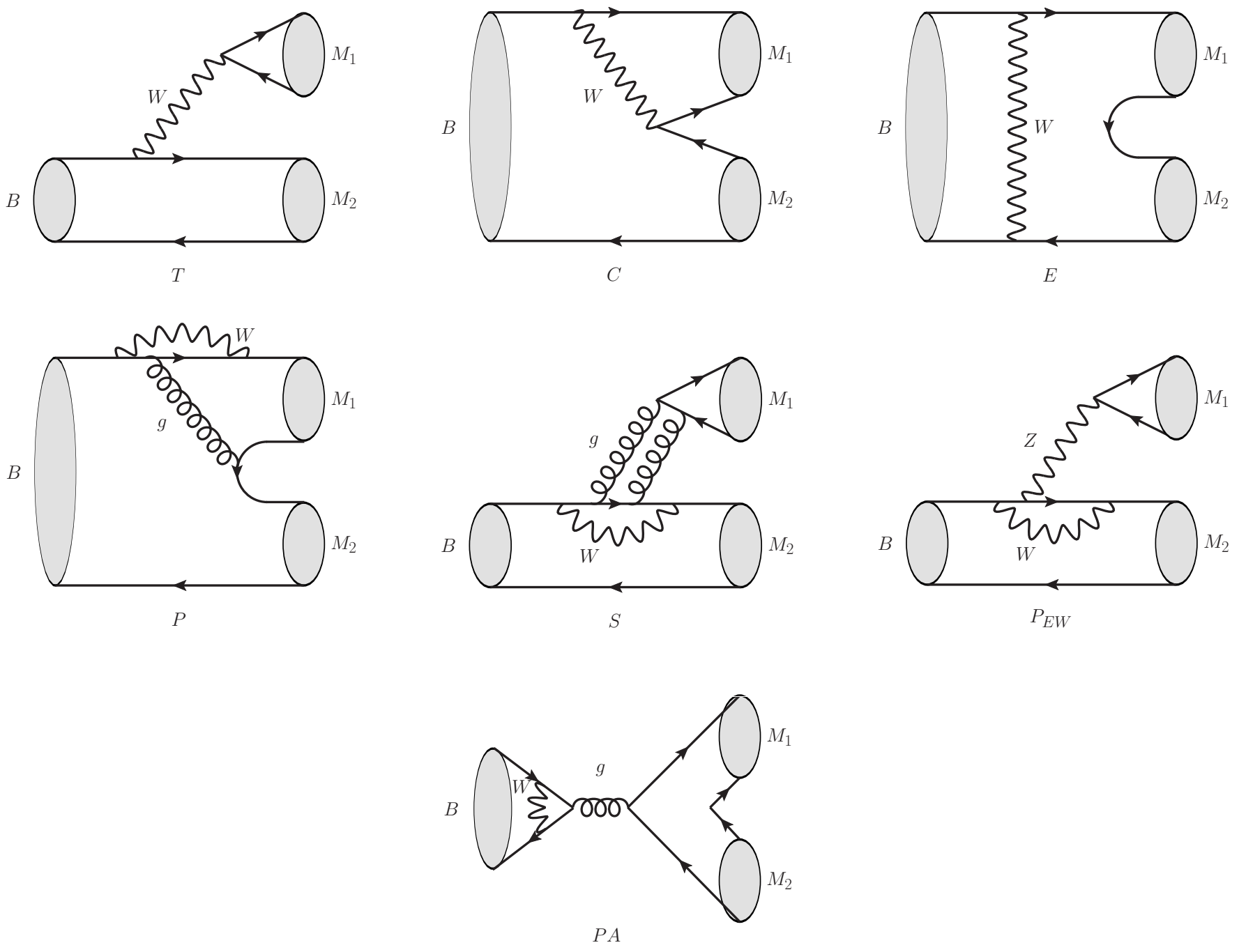

Fig. 131: Graphical representation of the most important SU(3) amplitudes of $B \to M_1 M_2$ decays in the flavour topology classification.

*SU(3) amplitudes.* In practice, instead of the group-theoretical reduced matrix elements, one uses an equivalent set of transition amplitudes for heavy meson decays categorised according to the topology of their flavour flow. Among these flavour diagrams, seven types have been identified to play an indispensable role in explaining the current data. Leaving out the CKM factors, they are:

(1) $T$, the colour-favoured tree diagram with an external $W$ emission;
(2) $C$, the colour-suppressed tree diagram with an internal $W$ emission;
(3) $E$, the $W$-exchange diagram;
(4) $P$, the QCD penguin diagram;
(5) $S$, the flavour-singlet QCD penguin diagram;
(6) $P_{EW}$, the electroweak (EW) penguin diagram;
(7) $PA$, the penguin-annihilation diagram.

The graphical representation of these amplitudes is shown in Figure 131.

The first three amplitudes $T$, $C$, $E$ are generated at tree level in the electroweak interaction. $T$ is the dominant amplitude, whereas $C$ is naively suppressed relative to $T$ by a colour factor of $N_c = 3$, and $E$ by helicity conservation and hadronic form factors. The remaining four amplitudes are induced only at the one-loop level. Compared to the first five amplitudes, the EW penguin amplitude is one order higher in the weak interaction and thus even smaller in magnitude. However, due to non-perturbative strong interaction dynamics, the hierarchy of the amplitudes is not seen as clearly in the current data as suggested by the above arguments.

The physical $\eta$ and $\eta'$ mesons are mixtures of the SU(3) singlet and octet states or, alternatively, of $\eta_q = \frac{1}{\sqrt{2}}(u\bar{u} + d\bar{d})$ and $\eta_s = s\bar{s}$ according to Ref. [782]

$$\begin{pmatrix} \eta \\ \eta' \end{pmatrix} = \begin{pmatrix} \cos\varphi & -\sin\varphi \\ \sin\varphi & \cos\varphi \end{pmatrix} \begin{pmatrix} \eta_q \\ \eta_s \end{pmatrix} , \tag{351}$$

where the mixing angle $\varphi$ is determined by lattice calculations as $\varphi \approx 46°$ [783]. For vector mesons, the $\phi$ meson is usually assumed to be a pure $s\bar{s}$ state. Since the mixing is a SU(3) breaking effect, including the $\eta$, $\eta'$, $\omega$ and $\phi$ by assuming a universal mixing angle for all amplitudes is an assumption that goes beyond the systematic SU(3) treatment, which however, greatly enhances the global fit, since it enlarges the set of final states at the expense of adding only one new topological amplitude, $S$. It is noted [781] that when the $\eta$-$\eta'$ mixing angle $\varphi$ is included as a free parameter in the fit, one obtains a value consistent with the lattice result quoted above.

Amplitude analyses using the current data show that SU(3) flavour symmetry is a satisfactory working assumption at the current level of experimental precision, meaning that the magnitude and strong phase of each flavour diagram can be taken to be the same for $\Delta S = 0$ and $|\Delta S| = 1$ transitions. In physical processes, the above-mentioned flavour amplitudes always appear in certain combinations, multiplied by appropriate CKM factors. In the case of strangeness-conserving ($\Delta S = 0$) transitions, we have

$$\begin{aligned} t &= \lambda_u^{(d)} T - \left(\lambda_u^{(d)} + \lambda_c^{(d)}\right) P_{EW}^C , \\ c &= \lambda_u^{(d)} C - \left(\lambda_u^{(d)} + \lambda_c^{(d)}\right) P_{EW} , \\ e &= \lambda_u^{(d)} E , \\ p &= -\left(\lambda_u^{(d)} + \lambda_c^{(d)}\right) \left(P - \frac{1}{3} P_{EW}^C\right) , \\ s &= -\left(\lambda_u^{(d)} + \lambda_c^{(d)}\right) \left(S - \frac{1}{3} P_{EW}\right) , \\ pa &= -\left(\lambda_u^{(d)} + \lambda_c^{(d)}\right) PA . \end{aligned} \tag{352}$$

In the SU(3) limit, the corresponding amplitudes for strangeness-changing ($|\Delta S| = 1$) transitions are obtained by replacing $d$ by $s$ in the CKM factors. Even though the colour-suppressed EW penguin diagram $P_{EW}^C$, which is both loop-suppressed and sub-leading in weak interactions, is not strongly required by the data (and was therefore not shown in Fig. 131), we include it in the above expressions for completeness. A full flavour amplitude decomposition of all $PP$ and $VP$ modes can be found, for example, in Ref. [781].

Whether a particular flavour amplitude is important phenomenologically is determined by the available data and their precision. Currently, the above seven flavour diagrams are sufficient to explain the observed data of $PP$ decays. In the case of the $PV$ modes, both the $E$ and $PA$ diagrams are not yet called for. Moreover, in this case one has to distinguish whether the spectator quark in the $B$ meson ends up in the vector or pseudoscalar meson. Therefore, the number of flavour amplitudes required for $PV$ modes is doubled, with the corresponding flavour amplitude denoted with a subscript $V$ or $P$. These two sets of amplitudes are different in both strength and strong phase *a priori*. Yet they can be related to each other under the assumption of (naive) factorisation. The SU(3) approach can also be applied to $VV$ final states, in which case one needs three parameters for each flavour diagram, one for every helicity amplitude. A global fit then requires polarisation data for every decay mode in the fit, which is not yet available.

In the following, we highlight some results of recent global analyses. We adopt the convention of fixing $T$ (in the case of $PP$ decays) and $T_P$ (in the case of $PV$ decays) to be real and positive. All other strong phases, denoted by $\delta_X$ for amplitude $X$, are then defined relative to these amplitudes. The experimental observables include the $CP$-averaged branching fractions and $CP$ asymmetries (direct and indirect). The former primarily determine the magnitude of each flavour amplitude, while the latter are more useful in fixing the associated strong phase.

*The PP Decays.* In the case of $PP$ final states, Ref. [781] shows that the magnitudes of the flavour amplitudes follow the hierarchy $|T| \gtrsim |C| > |P|, |E| > |S| > |P_{EW}| \sim |PA|$. The importance of the $E$ and $PA$ annihilation amplitudes mainly comes from the data on $B^0 \to K^+K^-$, $\pi^+\pi^-$ and $\pi^0\pi^0$ decays. The $E$ amplitude is seen to have a size about the same as the QCD penguin amplitude $P$ and a phase of $\sim -130°$. On the other hand, the $PA$ amplitude has a similar phase as $E$ but is one order of magnitude smaller in size than $P$.

An unexpected outcome is the value of the colour-suppressed tree amplitude. Not only does the $C$ amplitude have a nontrivial phase of about $-70°$, its magnitude is about 70% of $|T|$. The combination of both is at odds with QCD factorisation calculations [784, 785]. Large $|C|$ is not attributed only to the branching fractions of a small set of observables such as of $B^0 \to \pi^0\pi^0$ and/or $K^0\pi^0$, as might naively be expected. Rather, a large complex $C$ amplitude is a consequence of fitting to the observed direct $CP$ asymmetries in $B \to K\pi$ decays. In particular, it is required to explain the so-called $K\pi$ $CP$-puzzle, that is, the observation of the $CP$ asymmetry difference $\Delta A_{K\pi} \equiv A_{CP}^{K^+\pi^0} - A_{CP}^{K^+\pi^-} = 0.122 \pm 0.022$, by Standard Model hadronic physics. Moreover, it helps alleviate the rate deficit problem of the $B^0 \to \pi^0\pi^0$ decay.

The electroweak penguin amplitude $P_{EW}$ is found to have a strong phase of about $-80°$, and a similar phase is also found in the $P_{EW,V}$ amplitude for the $PV$ decays. Such a large phase is not only unexpected in the factorisation formalism, but difficult to understand from the basic structure of the weak effective Hamiltonian [720, 721, 786]. It is an open question whether better data from Belle II or a better theoretical understanding of SU(3) breaking effects in the dominant amplitudes can resolve this apparent contradiction in the subdominant $P_{EW}$ amplitude. Similarly, the status of the singlet amplitude $S$ has not been clarified. In the SU(3) fit it plays an essential role particularly in explaining the branching fractions of the $\eta' K$ decays [687]. It is found to be $\sim 60\%$ of the $P$ amplitude and $\sim 4$ times larger than the $P_{EW}$ amplitude and has a strong phase of about $-100°$. On the other hand, the explanation in terms of the interference pattern of standard QCD penguin amplitudes $P$ [787, 788] suggests a less important role of $S$.

The SU(3) determination of flavour amplitudes from $B_d$ and $B^\pm$ decays leads to predictions for $B_s$ decays. The $\eta\eta'$ and $\eta'\eta'$ modes are expected to have the largest decay rates among the $B_s$ decays.

*The PV Decays.* This sector shows the hierarchy $|T_{P,V}| > |C_{P,V}| > |P_{P,V}| \sim |P_{EW,V}| > |P_{EW,P}| \sim |S_{P,V}|$. Current data do not significantly constrain the magnitudes and phases of

the $E_{P,V}$ amplitudes. A global fit without the singlet $S_{P,V}$ amplitudes gives essentially the same values for most parameters except for $C_V$ and $P_{EW,V}$, but has a much worse quality, indicating a strong need for $S_{P,V}$ to describe data in the SU(3) framework. Unlike in the $PP$ sector, the singlet amplitudes are smaller than the electroweak penguin amplitudes $P_{EW,P,V}$.

The $C_V$ amplitude is about twice as large in magnitude than the $C_P$ amplitude, giving the ratios $|C_V/T_P| \sim 0.6$ and $|C_P/T_V| \sim 0.35$. Although with large errors, $C_P$ and $C_V$ have strong phases around $-30°$ and $-90°$, respectively. It thus appears that $C_V$ receives large corrections beyond factorisation, as in the $PP$ sector.

The QCD penguin amplitudes are about one order of magnitude smaller than the colour-allowed tree amplitudes, with $|P_P|$ slightly larger than $|P_V|$. It is noted that $P_P$ and $P_V$ are almost opposite in phase, in agreement with the proposal made in Ref. [787, 789] and the predictions from factorisation. This property results in constructive and destructive interference effects in the $\eta K^*$ and $\eta' K^*$ modes, respectively. Besides, $P_P$ has only a small strong phase of $\sim -20°$ relative to $T_P$, so that $P_V$ is almost opposite to both $T_P$ and $T_V$. This leads to a significant interference effect on modes involving $C_{P,V}$ and $P_{P,V}$. For example, among the $B_s$ decays to $PV$ final states, the $B_s \to \rho^+ K^-$ mode is predicted to have the largest branching fraction of order $15 \times 10^{-6}$.

A striking finding is that $|P_{EW,V}|$ is comparable to $|P_V|$. In contrast, in the $PP$ sector $|P_{EW}|$ is suppressed by one order of magnitude relative to $|P|$. This observation has some important implications for $CP$ violation in the $K^*\pi$ modes and for the branching fractions of $B_s \to \phi\pi^0$ (and $\phi\rho^0$). In the absence of the colour-suppressed amplitude, the $K^{*+}\pi^0$ and $K^{*+}\pi^-$ decays should have the same $CP$ asymmetry. Just as in the $B \to K\pi$ decays, a sign flip in $A_{CP}(K^{*+}\pi^0)$ will occur in the presence of a large complex $C_V$. This is in contradiction with the experimental observation that $CP$ asymmetries of $K^{*+}\pi^0$ and $K^{*+}\pi^-$ are of the same sign. This enigma can be resolved by noting that since $|\lambda_c^{(s)}| \gg |\lambda_u^{(s)}|$ and $|P_{EW,V}| \sim |P_V|$, $P_{EW,V}$ contributes substantially and renders $A_{CP}(K^{*+}\pi^0)$ the correct sign. In the $K\pi$ case, $P_{EW}$ is suppressed relative to $P$, only affecting the magnitude of $A_{CP}(K^+\pi^0)$ but not its sign.

The experimental status of the $K\pi$ and $K^*\pi$ systems is discussed in Section 12.4. For $B_s \to \phi\pi^0$ decays, there is currently no measurement. Theoretical predictions of its branching fraction within the SM are $(1.6^{+1.1}_{-0.3}) \times 10^{-7}$ in the framework of QCD factorisation [790] and $(1.94 \pm 1.14) \times 10^{-6}$ in the framework of flavour symmetry [781]. Preliminary studies at Belle [791] (unpublished) show that one can expect a signal yield of $0.5 - 1.14$ with the full 121 fb$^{-1}$ of $\Upsilon(5S)$ data, for this range of predicted branching fractions. In a theoretical analysis motivated by the $K\pi$ $CP$-puzzle (Section 12.3.4), it has been shown [790] that in models with modified or additional $Z$ bosons an increase of the branching fraction by an order of magnitude is possible without inconsistencies with other measurements. The $K\pi$ decays are dominated by isospin-conserving processes, but have a small contribution from isospin-violating penguin processes as well. In the isospin-violating process $B_s \to \phi\pi^0$ the penguin processes are expected to dominate, which means that potential NP contributions can have a much larger relative effect. If these contributions exist, an observation of the $B_s \to \phi\pi^0$ decay is possible. For $B_s \to \phi\rho^0$ decays, LHCb reports $4\sigma$ evidence with a branching fraction of $(2.7 \pm 0.7) \times 10^{-7}$ [792], which is lower than, but still consistent with the SM prediction of $(4.4^{+2.2}_{-0.7}) \times 10^{-7}$. While $B_s \to \phi\rho^0$ tests some of the same physics models as $B_s \to \phi\pi^0$, there are cases, *e.g.*, with parity-symmetric NP models, where only $B_s \to \phi\pi^0$ is sensitive.

*Further Discussion and Outlook.* Beyond the hierarchy of amplitudes the SU(3) approach makes many specific predictions for observables that are not yet well measured. For example, better or new measurements of branching fractions such as $Br(\pi^0 K^0)$, $Br(\eta K^0)$, $Br(\eta' K^0)$, as well as for the $\bar{K}^{*0}\pi^0$, $\rho^+ K^-$, $\bar{K}^{*0}\eta$, $\bar{K}^{*0}\eta'$, $\omega\bar{K}^0$, $\phi\bar{K}^0$, $\phi\pi^0$ and $\phi\eta'$ modes, the direct $CP$ asymmetries $A_{CP}^{\eta\pi^0}$, $A_{CP}^{\eta\eta'}$, $A_{CP}^{\eta' K^+}$, $A_{CP}^{\eta K_S^0}$ and the time-dependent $CP$ asymmetries $\mathcal{S}^{\pi^0 K_S^0}$, $\mathcal{S}^{\eta K_S^0}$, $\mathcal{S}^{\eta' K_S^0}$, $\mathcal{S}^{K^0 K^0}$, $\mathcal{S}^{\eta\pi^0}$, $\mathcal{S}^{\eta'\pi^0}$ at Belle II will be very useful in discriminating between different theoretical approaches. We refer to Ref. [781] for a detailed discussion of these and other specific examples.

With more and better data from Belle II, the flavour SU(3) symmetry approach will enable us to learn more about the role of each flavour amplitude in the $PP$ and $PV$ decays and inform us whether additional smaller amplitudes are called for. More insights can be obtained from applying the approach to the helicity amplitudes of $VV$ final states. At the same time, more precise data will allow us to address the question whether the SU(3) limit continues to be a satisfactory working assumption. A better understanding or parametrisation of SU(3) breaking is necessary to distinguish reliably the smaller amplitudes from potential SU(3) breaking effects in the dominant amplitudes. Also, it is known that decay constants and form factors, which appear in the factorisation framework exhibit sizeable SU(3) breaking. If there are significant corrections to perturbative factorisation as appears to be indicated for some amplitudes, this creates an ambiguity in the treatment of SU(3) breaking, which can only be resolved in terms of a complete parametrisation of SU(3) breaking of the charmless decay amplitudes. The likely lack of predictivity due to the increase in independent amplitudes may be compensated by the amount of data anticipated from Belle II or may motivate combinations of the SU(3) and factorisation approaches [793]. Due to its data-driven nature, the SU(3) approach will profit like no other theoretical approach from the input of Belle II and hence contribute to the possible discovery of NP in charmless $B$ meson decays.

## *12.3. Factorisation approach to two-body decays*

### *12.3.1. Introduction . [Contributing Author: M. Beneke]*

The notion of factorisation in $B$ physics originally referred to an approximation of the hadronic matrix elements $\langle f|Q_i|\bar{B}\rangle$ relevant to charmless two-body decays in terms of the product of a light-meson decay constant and a $B$ to light meson transition form factor [794]. In contrast "factorisation" or "QCD factorisation" now refers to a systematic separation of scales in $\langle f|Q_f|\bar{B}\rangle$. Contrary to the (useful but ad-hoc) approximation of "naive" factorisation, QCD factorisation implies an expansion of the matrix element in the small parameters $\alpha_s(\mu)$ and $\Lambda/m_b$, with $\mu = m_b$ or $\sqrt{m_b\Lambda}$, one of the perturbative scales. Since the $\alpha_s$ series can be calculated order by order (with increasing effort), but only the leading term in the $1/m_b$ expansion assumes a simple form, the generic accuracy of this approach is limited by power corrections of generic size $\Lambda/m_b \simeq (10-20)\%$ at the amplitude level. The actual importance of power corrections depends, however, on the specific amplitude and observable.

The QCD factorisation approach developed in Refs. [244, 415, 416] replaces the naive factorisation ansatz by a factorisation formula that includes radiative corrections and

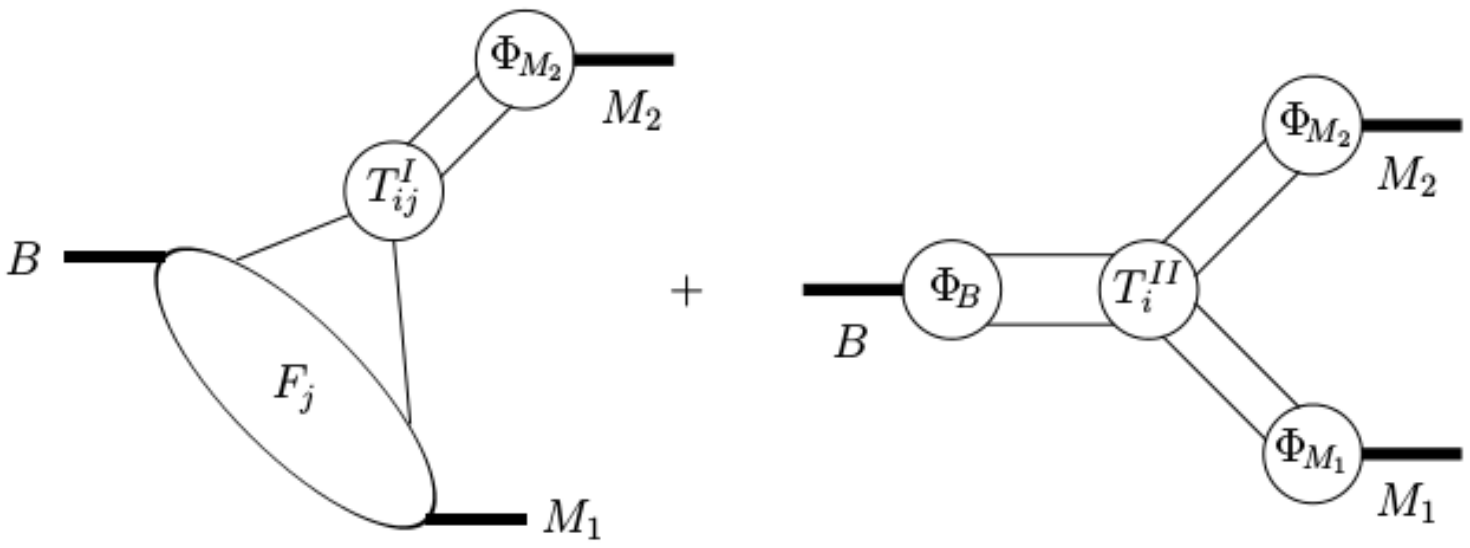

Fig. 132: Graphical representation of the factorisation formula given in Eq. (353). Figure from Ref. [244].

spectator-scattering effects. The basic formula for the hadronic matrix elements is

$$\begin{aligned} \langle M_1 M_2|Q_i|\bar{B}\rangle &= F^{BM_1}(0)\int_0^1 du\, T_i^I(u)\Phi_{M_2}(u) \\ &+ \int_0^\infty d\omega \int_0^1 du \int_0^1 dv\, T_i^{II}(\omega,u,v)\,\Phi_B(\omega)\Phi_{M_1}(v)\Phi_{M_2}(u) \\ &= F^{BM_1}\,T_i^I \star \Phi_{M_2} + \Phi_B \star [H_i^{II}\star J^{II}]\star\Phi_{M_1}\star\Phi_{M_2}\,, \end{aligned} \tag{353}$$

where $F^{BM_1}(0)$ is a (non-perturbative) $B$ to light-meson transition form factor, $\Phi_{M_i}$ and $\Phi_B$ are light-cone distribution amplitudes, and $T_i^{I,II}$ are perturbatively calculable hard-scattering kernels. $M_1$ is the meson that picks up the spectator quark from the $B$ meson.[31] The equation is illustrated in Fig. 132. The third line uses a short-hand notation $\star$ for convolutions and indicates that the spectator-scattering effect in the second line is a convolution of physics at the hard scale $m_b$, encoded in $H_i^{II}$, and the hard-collinear scale $\sqrt{m_b \Lambda}$, encoded in the jet function $J^{II}$. Eq. (353) shows that there is no long-distance interaction between the constituents of the meson $M_2$ and the $(BM_1)$ system at leading order in $1/m_b$. This is the precise meaning of "factorisation". Strong interaction scattering phases are generated at leading order in the heavy-quark expansion only by perturbative loop diagrams contributing to the kernels $T_i^I$ and $H_i^{II}$. Thus the strong phases and therefore direct $CP$ asymmetries are generically of order $\delta \sim \mathcal{O}(\alpha_s(m_b), \Lambda/m_b)$.

Factorisation as embodied by Eq. (353) is not expected to hold at sub-leading order in $1/m_b$. Some power corrections related to scalar currents are enhanced by factors such as $m_\pi^2/((m_u+m_d)\Lambda)$. Some corrections of this type, in particular those related to scalar penguin amplitudes, nevertheless appear to be calculable and turn out to be numerically important. On the other hand, attempts to compute sub-leading power corrections to hard spectator-scattering in perturbation theory usually result in infrared divergences, which signal the breakdown of factorisation. These effects are usually estimated and included into the error budget. All weak annihilation contributions belong to this class of effects and often constitute the dominant source of theoretical error, in particular for the direct $CP$ asymmetries. Factorisation as above applies to pseudoscalar flavour-non-singlet final states and to the longitudinal polarisation amplitudes for vector mesons. Final states with $\eta$ and $\eta'$ require

[31] The definition of $M_1$ and $M_2$ in Eq. 353 and Fig. 132 are opposite to what is shown in Fig. 131

additional considerations, but can be included [788]. The transverse helicity amplitudes for vector mesons are formally power-suppressed but can be sizeable, and do not factorise in a simple form [795, 796]. The description of polarisation is therefore more model-dependent than branching fractions and $CP$ asymmetries. QCD factorisation results are available for a variety of complete sets of final states. Refs. [696, 796] contain the theoretical predictions for pseudoscalar and vector meson final states. A similar analysis has been performed for final states with a scalar meson [797], axial-vector mesons [798, 799], and a tensor meson [800]. We refer to these papers for the present status of charmless $B$ decay calculations in the factorisation approach, and to Ref. [2] for an extended version of this very brief theoretical introduction.

Several variations of factorisation have been considered in the literature. In this chapter we shall also refer to the "perturbative QCD" (PQCD) framework [801, 802]. PQCD makes the stronger additional assumption that the $B$ meson transition form factors $F^{B\to M_1}(0)$ are also dominated by short-distance physics and factorise into light-cone distribution amplitudes. Both terms in Eq. (353) can then be combined to

$$\langle M_1 M_2 | Q_i | \bar{B} \rangle = \phi_B \star [T_i^{\rm PQCD} \star J^{\rm PQCD}] \star \phi_{M_1} \star \phi_{M_2}. \tag{354}$$

It should be mentioned that while the assumptions that lead to Eq. (353) are generally accepted and have been verified in the computation of radiative corrections, the additional assumption required for Eq. (354) has remained controversial, since it relies on regularising the infrared sensitivity by intrinsic transverse momentum, and the power counting in $1/m_b$ has not been clarified. The original PQCD factorisation formula Eq. (354) was revised due to infrared divergences in loop effects [803], which weakens its predictive power. Most phenomenological analyses predate this revision, with few exceptions [804].

Making first principle calculations of charmless $B$ decay amplitudes precise requires good knowledge of hadronic input parameters such as form factors and moments of distribution amplitudes, accurate perturbative computations, and some understanding of power corrections, either from theory or guided by data. While many issues involved are discussed in the references quoted above, the following summarises recent theoretical progress on each of them. In order to facilitate the comparison with the SU(3) terminology, we note that there is a one-to-one correspondence between the amplitudes notation $\alpha_i$, $\beta_i$ introduced in Ref. [696] and used below, and the topological flavour amplitudes, provided SU(3) breaking effects are neglected in the former. The correspondence is

$$T \leftrightarrow \alpha_1,\ C \leftrightarrow \alpha_2,\ E \leftrightarrow \beta_2,\ P \leftrightarrow \alpha_4 + \beta_3,\ S \leftrightarrow \alpha_3,\ P_{EW} \leftrightarrow \alpha_{3,\rm EW},\ PA \leftrightarrow \beta_4 \tag{355}$$

etc. A complete list can be found in Ref. [805].

*12.3.2. B meson light-cone distribution. [Contributing Author: T. Feldmann]*

The $B$ meson light-cone distribution amplitudes (LCDAs) constitute essential hadronic input parameters not only in the QCD factorisation formula Eq. (353) for exclusive charmless $B$ decays, but also for the computation of spectator corrections to heavy-to-light form factors and rare radiative decays. They also enter correlation functions in certain variants of the QCD sum-rule approach. The most important parameter is the inverse moment [415]

$$\frac{1}{\lambda_B} \equiv \int_0^\infty \frac{d\omega}{\omega}\, \Phi_B(\omega)\,, \tag{356}$$

which enters the overall size of spectator-scattering effects in Eq. (353) through the quantity

$$r_{\rm sp} \quad = \quad \frac{9 f_\pi f_B^{\rm HQET}}{m_b\, F^{B\to\pi}(0)}\, \frac{1}{\lambda_B}\,, \tag{357}$$

where $f_B^{\rm HQET}$ denotes the $B$ meson decay constant in heavy quark effective theory (HQET) and $F^{B\to\pi}(0)$ the $B\to\pi$ transition form factor at $q^2=0$. For example, the colour-suppressed tree amplitude $\alpha_2$ for $B\to\pi\pi$ decays has been calculated as [785]

$$\alpha_2(\pi\pi) = 0.220 - [0.179 + 0.077\, i]_{\rm NLO} + \left[\frac{r_{\rm sp}}{0.445}\right] \{[0.114]_{\rm LOsp} + [0.067]_{\rm tw3}\} + \ldots \tag{358}$$

It is to be noted that spectator-scattering effects tend to partly cancel the NLO vertex corrections which enhances the sensitivity on $\lambda_B$. The inverse moment $\lambda_B$ also enters the factorisation formula for partial rate of the $B\to\gamma\ell\nu$ decay with large photon energy $E_\gamma \gg \Lambda$ [806–808]. Since, at leading power,

$$\frac{d\Gamma}{dE_\gamma} \propto \frac{1}{\lambda_B^2}, \tag{359}$$

this process can be used to determine $\lambda_B$ experimentally, providing crucial input to the phenomenology of charmless decays.

The strongest experimental bound on $\lambda_B$ (from Belle) using the up-to-date theoretical results from Refs. [223, 246] are still rather weak [247]. Belle II is uniquely suited to improving the measurement of $\lambda_B$ with $B\to\ell\nu\gamma$ decays (Section 8.4.1). In the following, we therefore summarise recent theoretical progress in the understanding of the leading-twist $B$ meson light-cone distribution amplitude (LCDA).

The formal definition of the relevant LCDA[32] $\Phi_B(\omega)$ is given in terms of the Fourier transform of the hadronic matrix element of a light-cone operator in HQET [809],

$$m_B f_B^{\rm (HQET)}\, \phi_B^+(\omega) = \int \frac{d\tau}{2\pi}\, e^{i\omega\tau}\, \langle 0|\bar q(\tau n)\, [\tau n, 0]\, \not{n}\gamma_5\, h_v^{(b)}(0)|\bar B(m_B v)\rangle\,, \tag{360}$$

where $v^\mu$ is the heavy-quark velocity, $n^\mu$ a light-like ($n^2=0$) vector, and $[\tau n, 0]$ denotes a light-like Wilson line connecting the two field positions 0 and $\tau n$. The LCDA $\phi_B^+(\omega)$ can be interpreted as the probability amplitude for the distribution of the light antiquark's momentum $k$ in a two-particle Fock state of the $B$ meson, more precisely its light-cone projection $\omega \equiv n\cdot k$.

The LCDA $\phi_B^+$ is a scale-dependent quantity. The scale-dependence is controlled by a renormalisation-group (RG) equation [810]. Recent progress is related to the identification of the eigenfunctions of the one-loop RG kernel [811]:

$$f_{\omega'}(\omega) \equiv \sqrt{\frac{\omega}{\omega'}}\, J_1\left(2\sqrt{\frac{\omega}{\omega'}}\right) \tag{361}$$

with eigenvalues

$$\gamma_{\omega'} = -\left(\Gamma_{\rm cusp}\, \ln\frac{\mu}{\hat\omega'} + \gamma_+\right)\,. \tag{362}$$

Here $J_1(z)$ is a Bessel function. It has been noted that Eq. (362) can also be constructed from the symmetries of collinear conformal transformations [812]. The eigenfunctions are labelled by a continuous parameter $\omega'$, which can be viewed as a "dual momentum" (we also use the notation $\hat{\omega}' \equiv \omega' \, e^{-2\gamma_E}$ below). The spectrum in $\omega'$ defines the "dual" representation of the $B$ meson LCDA in terms of a function $\rho_B^+(\omega')$, related to the original LCDA via

$$\phi_B^+(\omega) = \int_0^\infty \frac{d\omega'}{\omega'} \, f_{\omega'}(\omega) \, \rho_B^+(\omega') \quad \Leftrightarrow \quad \rho_B^+(\omega') = \int_0^\infty \frac{d\omega}{\omega} \, f_{\omega'}(\omega) \, \phi_B^+(\omega) \, . \tag{363}$$

[32] Now denoted as $\phi_B^+(\omega)$. There is another Dirac structure, which defines a sub-leading twist LCDA, denoted as $\phi_B^-(\omega)$.

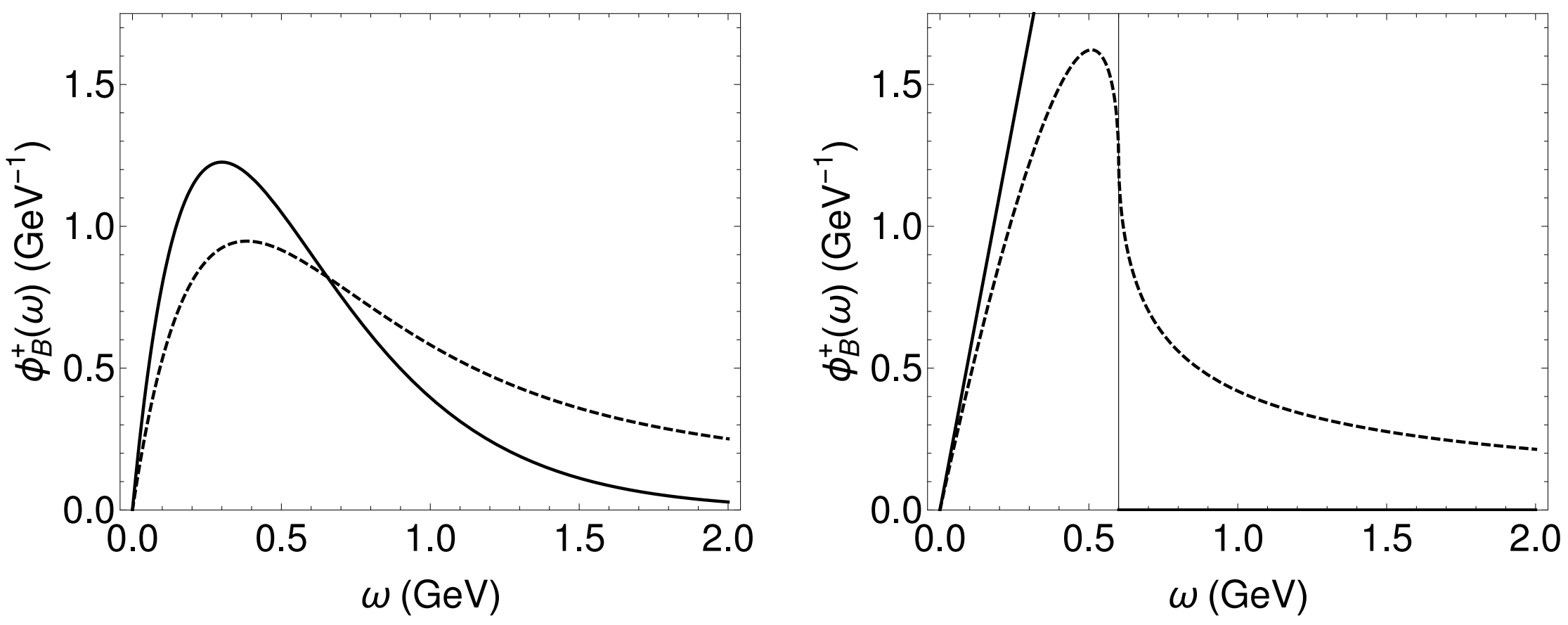

Fig. 133: Two models for $\phi_B^+(\omega)$ (solid lines): "Exponential" (left panel), and "Parton" (right panel) model, as defined in Eq. (364). The dashed lines illustrate the effect of RG evolution, see Ref. [811] for further details.

The key result is that the scale dependence of the dual function $\rho_B^+(\omega')$ is described by a simple multiplicative RG factor [811], which facilitates the construction and interpretation of models for the LCDA.

The effect of the RG evolution is illustrated in Fig. 133 for two models,

$$\begin{aligned} \text{"Exponential" [809]:} \quad \phi_B^+(\omega, \mu_0) &= \frac{\omega \, e^{-\omega/\omega_0}}{\omega_0^2} \quad \leftrightarrow \quad \rho_B^+(\omega', \mu_0) = \frac{e^{-\omega_0/\omega'}}{\omega'} \, , \\ \text{"Parton" [813]:} \quad \phi_B^+(\omega, \mu_0) &= \frac{\omega \, \theta(2\bar{\Lambda} - \omega)}{2\bar{\Lambda}^2} \quad \leftrightarrow \quad \rho_B^+(\omega', \mu_0) = \frac{1}{\bar{\Lambda}} \, J_2\!\left(2\sqrt{\frac{2\bar{\Lambda}}{\omega'}}\right) , \end{aligned} \tag{364}$$

where $\omega_0$ and $\bar{\Lambda} = M_B - m_b$ are the two parameters of the model, and the model form is supposed to hold at a particular reference scale $\mu_0$. One then observes from the figure that the RG evolution towards higher scales generates a "radiative tail" at large values of $\omega$.

In applications of QCD factorisation the jet function in the factorisation theorem takes the form of a polynomial in $\ln \omega$, or $\ln \omega'$ in dual space. One thus generically needs the

logarithmic moments[33]

$$L_n(\mu) \;\; \equiv \;\; \int_0^\infty \frac{d\omega'}{\omega'} \ln^n\left(\frac{\hat{\omega}'}{\mu}\right) \rho_B^+(\omega',\mu)\,, \tag{365}$$

for $n = 0, 1, 2, \ldots$. These obey the RG equation [811],

$$\frac{dL_n(\mu)}{d\ln\mu} = \Gamma_{\rm cusp}(\alpha_s)\, L_{n+1}(\mu) - \gamma_+(\alpha_s) L_n(\mu) - n\, L_{n-1}(\mu)\,, \tag{366}$$

which mixes neighbouring moments. For phenomenological applications one may consider – either in dual space or in original $\omega$ space – a truncated set $\{L_0, L_1, \ldots L_N\}$ of logarithmic moments, or a particular *model* for $\rho_B^+(\omega')$ or $\phi_B^+(\omega)$, respectively.

An advantage of the dual representation is that large and small values of $\omega'$ are clearly separated because they do not mix under renormalisation. This is useful, since large values of $\omega'$ can be described by perturbative dynamics which implements the QCD-improved parton model, subject to constraints from a local operator product expansion (OPE) in the context of HQET [814–816]. At fixed order in the strong coupling one finds the model-independent result [817]

$$\rho_B^+(\omega')_{\rm pert.} = C_0\, \frac{1}{\bar\Lambda}\, J_2\left(2\sqrt{\frac{2\bar\Lambda}{\omega'}}\right) + (C_0 - C_1)\, \frac{4}{\bar\Lambda}\, J_4\left(2\sqrt{\frac{2\bar\Lambda}{\omega'}}\right) + \ldots \tag{367}$$

for $\omega' \gg \bar\Lambda$ with matching coefficients

$$\begin{aligned} C_0 &= 1 + \frac{\alpha_s C_F}{4\pi}\left(-2L^2 + 2L - 2 - \frac{\pi^2}{12}\right) + \mathcal{O}(\alpha_s^2), \\ C_0 - C_1 &= \frac{\alpha_s C_F}{4\pi}\left(-\frac{13}{4}\right) + \mathcal{O}(\alpha_s^2), \end{aligned} \tag{368}$$

and $L = \ln\mu/\hat\omega'$. Equation (367) reduces to the free parton result in Eq. (364) for $\alpha_s \to 0$. Furthermore, the RG equations can be used to resum large logarithms $|L| \gg 1$. This implies that the function $\rho_B^+(\omega')$ falls off faster than $1/\omega'$ [817]. If one splits the logarithmic moments in Eq. (365) as $L_n = L_n^+ + L_n^-$ by separating the $\omega'$ integral at $\omega' = \mu$ into a large $\omega'$ part $(+)$ and small $\omega'$ part $(-)$, one concludes that

- $L_n^+$ are completely determined by the (RG-improved) OPE and contain the information on the HQET parameters $\bar\Lambda$ etc.
- $L_n^-$ depend on (non-local) IR dynamics and are *unrelated* to HQET parameters $\bar\Lambda$ etc.

Numerically, one typically finds that $L_n^- \gg L_n^+$, and thus the information from the LCDAs needed in phenomenological applications of the QCD factorisation approach in *exclusive* $B$ decays, is basically unrelated to the information on the HQET parameters entering the OPE analysis of *inclusive* $B$ decays. While the non-local effects entering $L_n^-$ are notoriously difficult to estimate with lattice-QCD simulations, a dedicated analysis within the QCD sum rule approach, using the dual representation of the $B$ meson LCDA, might improve the

[33] The relation to the convention used, for instance, in Ref. [814] reads

$$L_0 \equiv \frac{1}{\lambda_B(\mu)}\,, \qquad L_1 \equiv L_0\, \sigma_1(\mu)\,, \qquad L_2 \equiv L_0\, \sigma_2(\mu)\,.$$

situation on the theoretical side. Information on the first few moments $L_n^-$ can then be used in the future as a theory prior in a global analysis of exclusive radiative, semi-leptonic and charmless $B$ decays.

*12.3.3. Weak annihilation . [Contributing Author: C. Bobeth]*

Weak annihilation (WA) corresponds to parts of the decay amplitude where the constituent quarks of the decaying $B$ meson are annihilated by one of the local $|\Delta B| = 1$ four-quark operators in the weak effective Lagrangian in E q. (349), and two quarks of the final state are created by the operator. The remaining quark pair in the final state is created via QCD interactions.

The flavour amplitudes $E$ and $PA$ in Fig. 131 represent the subset of annihilation topologies that are most relevant for the SU(3) approach. In factorisation approaches to charmless two-body decays, WA is of subleading order in $1/m_b$, but is a potentially important contribution in all of those cases where the leading order amplitudes are small. This applies obviously to pure annihilation modes, but also to penguin-dominated transitions, especially when there is a vector meson in the final state.

Such $1/m_b$ corrections are not covered by the factorisation formula in Eq. (353). Technically, this manifests itself in so-called "end-point divergences" in convolutions of the hard scattering kernels with light-meson distribution amplitudes, if one attempts such a factorisation. In Refs. [416, 696] a parameterisation of the annihilation amplitudes has been introduced, which replaces the divergent expressions by hadronic parameters, $\rho_A$. There are WA amplitudes $b_{1,2}$ due to current-current operators ($Q_{1,2}$), $b^p_{3,4}$ due to QCD penguin operators ($Q_{3,4,5,6}$), and $b^p_{3,4,\mathrm{EW}}$ due to electroweak penguin operators ($Q_{7,8,9,10}$), which in QCD factorisation are parameterised as

$$\begin{aligned}
b_1 &\propto C_1 A_1^i, \\
b_2 &\propto C_2 A_1^i, \\
b_3^p &\propto C_3 A_1^i + C_5(A_3^i + A_3^f) + N_c C_6 A_3^f, \\
b_4^p &\propto C_4 A_1^i + C_6 A_2^i, \\
b_{3,\mathrm{EW}}^p &\propto C_9 A_1^i + C_7(A_3^i + A_3^f) + N_c C_8 A_3^f, \\
b_{4,\mathrm{EW}}^p &\propto C_{10} A_1^i + C_8 A_2^i.
\end{aligned} \tag{369}$$

As already mentioned, $E \leftrightarrow b_2$ and $PA \leftrightarrow b_4^p$ in terms of flavour amplitudes, while $b_3^p$ can always be absorbed into $P$. The $b_i$ coefficients are identical to the $\beta_i$ coefficients, which appear in Eq. (355), up to a proportionality factor defined in Ref. [696]. In the above equation $C_i$ denote the Wilson coefficients of operators $Q_i$ and $p = u, c$. The $A^i_{1,2,3}$ and $A^f_3$ can be regarded as non-perturbative objects[34] with strong phases. They are further expressed in terms of quantities $\rho^{i,f}_{A_{1,2,3}}$ [696], where the superscript indicates whether the gluon that creates the second quark-antiquark pair in the final state was radiated off the initial ($i$) or final ($f$) state (anti-) quarks. The sizes of the Wilson coefficients determine greatly the importance of the various WA amplitudes in $CP$-averaged observables.

[34] A proper factorisation theorem would establish a relation to matrix elements of well-defined operators. These matrix elements have to be either determined from data or calculated by non-perturbative means.

The theoretical uncertainties due to WA are estimated by varying the complex-valued $\rho_A$'s within ranges given by naive dimensional arguments for each observable separately. This conservative procedure yields large uncertainties, especially in $CP$ asymmetries and, of course, all pure annihilation modes, that allow for agreement with most of the data, in part because it allows the situation where different values of the $\rho_A$'s lead to agreement between predictions and measurements of different observables of one and the same decay mode.

Note that in the framework of light-cone sum rules WA contributions are free of endpoint divergences [818] due to different assumptions and approximations. This approach yields the same dependence on Wilson coefficients $C_i$ as given in Eq. (369) and the non-perturbative $A^{i,f}_{1,2,3}$ can be evaluated explicitly. We also mention that WA tree diagrams are calculated within the PQCD framework [802, 819].

The explicit dependence of amplitudes on WA contributions $b_i^{(p)}$ for charmless $B \to PP, PV, VP$ Refs. [416, 696] and $B \to VV$ [796, 820] can be quite different, such that certain groups of decays and/or certain observables have an enhanced sensitivity to a particular $A^{i,f}_{1,2,3}$. For example $b \to (s,d)q\bar{q}$ transitions dominated by QCD- and QED-penguin operators depend mainly on $b_3^p$, where $A_3^f$ is enhanced by a colour factor $N_c = 3$ such that $C_6 \approx 8C_5 \approx 3C_3$ in the SM at the renormalisation scale $\mu \sim m_b$. Further, the pure annihilation decays $B_d \to K^+K^-$, $B_s \to \pi^+\pi^-$ depend on $b_4^p$, $b_1$, being sensitive to $A^i_{1,2}$. It is therefore of utmost importance to improve and extend measurements for as many decay systems as possible to test these predictions and relations.

Belle II is the only experiment that can provide measurements of complete decay systems related by $u \leftrightarrow d$ quark exchange such as for example $B^+ \to K^+\pi^0, K^0\pi^+$ and $B^0 \to K^+\pi^-, K^0\pi^0$, due to its identification capabilities for charged *and* neutral particles. This enables Belle II to provide combinations of observables within such decay systems accounting for cancellations of common experimental systematic uncertainties. Prominent examples are ratios of branching fractions such as

$$R_n = \frac{1}{2}\frac{Br(B^0 \to K^+\pi^-)}{Br(B^0 \to K^0\pi^0)}, \tag{370}$$

or differences of $CP$ asymmetries such as $\Delta A_{K\pi} = A_{CP}^{K^+\pi^0} - A_{CP}^{K^+\pi^-}$ etc., which are less sensitive to theory uncertainties.

Some phenomenological studies supplement the leading-power QCDF predictions with WA contributions and infer the latter from data. The two main strategies can be classified as

(1) *either* fit whole WA amplitudes $b_i^{(p)}$, see Ref. [821] for the $B \to K\pi$ system;[35]
(2) *or* use short-distance Wilson coefficients from a given model and fit only long-distance parts $A^{i,f}_{1,2,3}$ [822–826].

The advantage of strategy 2 over 1 is that it consistently uses the Wilson coefficients of a given model, SM or extensions thereof, in both, the leading $1/m_b$ and WA contributions.

The relative size of WA contributions to leading amplitudes has been determined from recent data of $b \to (s,d)q\bar{q}$ transitions for the decay systems

- $B \to K + (\pi, \eta^{(\prime)}, K)$; $B_s \to \pi\pi, K\pi, KK$;
- $B \to K + (\rho, \phi, \omega)$; $B \to K^* + (\pi, \eta^{(\prime)})$;

[35] Ref. [821] fits subleading contributions in general, of which one is WA.

- $B \to K^* + (\rho, \phi, \omega, K^*)$; $B_s \to \phi\phi, K^*\phi, K^*K^*$;

following strategy 2 [823], assuming one universal $\rho_A$ per decay system. These systems depend primarily on the WA amplitude $b_3^p$ being dominated by $A_3^f$. The fits show that within the SM data do not require a huge $b_3^p$, but usually they are a sizeable fraction of the leading amplitude $\alpha_4^p$. For example at 68 % probability the minimal required fraction varies among the $B \to PP$ systems in the range $0 - 60$ %, and is larger ($40 - 80$ %) for $B \to PV$ systems as well as for $B \to VV$ systems ($20 - 90$ %). This is related to the larger absolute size of the QCD penguin coefficient $\alpha_4^p = a_4^p \pm r_\chi^{M_2} a_6^p$ for $PP$ compared to $PV$ and $VV$ final states (see the following subsection). Only in a few cases can the data be explained at 95 % probability without WA. Within current large experimental uncertainties, the goodness-of-fit is always excellent except for the $B \to K\pi$ system, where tensions of around $2 - 3\,\sigma$ are observed for observable combinations $R_n$ and $\Delta A_{K\pi}$. The potential underlying mechanism (WA or subleading hard-scattering contributions) of this so-called $K\pi$ $CP$-puzzle can be further scrutinised by improved measurements of direct and mixing-induced $CP$-asymmetries in $B^0 \to K^0\pi^0$ expected from Belle II.

Pure WA decays, such as the observed modes $B^0 \to K^+K^-$ and $B_s \to \pi^+\pi^-$, depend on $A_{1,2}^i$, but not on the hadronic quantity $\lambda_B$ and form factors. The fit of the preferred regions of a universal $\rho_A = \rho_A^i$ from branching fractions of $B^0 \to K^+K^-$ and $B_s \to \pi^+\pi^-$ shows a strong incompatibility [823, 827], but experimental uncertainties are still large. In the case $A_1^i \approx A_2^i$ only one strong phase would be present yielding tiny $CP$ asymmetries. It is important to measure the latter and to search for other pure WA decays, such as $B^0 \to K^-K^{*+}$, $K^{*-}K^+$, $K^{*-}K^{*+}$ and related $B_s$ decays.

In view of the large number of $\rho^{i,f}_{A_{1,2,3}}$ but limited set of observables per decay, the assumption that the dependence on $u$-, $d$- and $s$-quarks is small in initial- and final-state interactions would allow one to combine different decay systems to fit for universal $\rho_A^i$ and $\rho_A^f$ parameters. However, due to the aforementioned incompatibility of purely WA decays $B^0 \to K^+K^-$ and $B_s \to \pi^+\pi^-$, different $\rho_A^{i,f}$'s are often assumed in $B_{u,d}$ and $B_s$ decays. This has been done for $B_{u,d,s} \to \pi\pi, \pi K, KK$ in Ref. [822], $B_{u,d} \to PV$ in Ref. [824] and in combination with $B_s \to PV$ in Ref. [825], as well as $B_{u,d} \to VV$ in Ref. [826]. In these papers also $\lambda_B$ and in part subleading hard-scattering contributions have been included as fit parameters.

For example, in $B \to PP$ the $\rho_A^i$'s are constrained from the pure annihilation decays $B_d \to K^+K^-$ and $B_s \to \pi^+\pi^-$, where current data allow for similar size $\rho_A^i$'s in $B_{u,d}$ and $B_s$ decays, and the same has been tested for $\rho_A^f$. In the global fit the aforementioned tension between $B^0 \to K^+K^-$ and $B_s \to \pi^+\pi^-$ is less significant such that there are no indications of SU(3) flavour breaking within the current experimental accuracy. However, the data prefer $\rho_A^i \neq \rho_A^f$. These fits also prefer values of $\lambda_B \simeq 200$ MeV, similar to values inferred from data of tree-dominated decays $B \to \pi\pi$, $\pi\rho$, $\rho\rho$ [785]. Based on the stronger assumption of equal $\rho_A^i$ in $B_{u,d}$ and $B_s$ decays, and analogously for $\rho_A^f$, predictions for not yet measured $B_s \to P^0P^0$ ($P = \pi, K$) modes are given, which can be tested in the future.

In the case of $B \to PV$, the currently large experimental uncertainties in $B_s$ decays also allow for universal $\rho_A^i$ and $\rho_A^f$ in $B_{u,d}$ and $B_s$ decays, and again prefer $\rho_A^i \neq \rho_A^f$. Improved measurements of $B_s \to PV$ are necessary to investigate whether there is sizeable SU(3) breaking, demanding a dedicated $B_s$ physics run of Belle II.

Additional polarisation dependence enters through $A_h^{i,f}$ ($h = L, T$, with $T = \perp, \parallel$) in $B \to VV$ decays, which also allows for the assumption of polarisation-dependent $\rho_{A,h}^{i,f}$. The assumption of polarisation-independent $\rho_{A,L}^{i,f} = \rho_{A,T}^{i,f}$ leads to similar observations as in $B \to PP, PV$ decays. On the other hand it is found that the data of $B \to VV$ decays can be also described with polarisation-dependent but universal $\rho_{A,h}^{i} = \rho_{A,h}^{f}$ for initial and final state radiation.

The preferred regions of $\rho_A^f$ from $B \to PP, PV, VV$ decays are close to each other, which is not the case for the $\rho_A^i$.

*12.3.4. Direct CP asymmetries at NLO . [Contributing Authors: G. Bell, T. Huber]*

Direct $CP$ asymmetries require the interference of two decay amplitudes with different $CP$ ("weak") and rescattering ("strong") phases. As already observed, within QCD factorisation strong phases are generated only through loop effects proportional to $\alpha_s(m_b)$ or power corrections proportional to $\Lambda/m_b$. One therefore generically expects that direct $CP$ asymmetries are small, which is in qualitative agreement with experimental data. Larger strong phases and hence larger $CP$ asymmetries may arise whenever the leading-order term is suppressed, *e.g.* by colour factors or Wilson coefficients.

The dependence on the strong phases makes theoretical calculations of direct $CP$ asymmetries more involved than those of branching ratios or mixing-induced $CP$ asymmetries. A clear picture about the relative size and sign of direct $CP$ asymmetries requires, in particular, controlling subleading terms in the double expansion in $\alpha_s(m_b)$ and $\Lambda/m_b$. Whereas the former can be systematically computed using loop techniques, the latter cannot be calculated at present and their modelling introduces sizeable theoretical uncertainties.

The various contributions to the decay amplitudes are typically classified according to their topological structure into tree, QCD penguin, electroweak penguin and annihilation topologies. In the notation of Ref. [696], the $\bar{B} \to \pi \bar{K}$ amplitudes, which play an important role in the following discussion, are parametrised as

$$\begin{aligned}
\mathcal{A}_{B^- \to \pi^- \bar{K}^0} &= \lambda_p \, A_{\pi\bar{K}} \Big[ \hat{\alpha}_4^p - \tfrac{1}{2} \alpha_{4,\mathrm{EW}}^p \Big] \, , \\
\sqrt{2}\, \mathcal{A}_{B^- \to \pi^0 K^-} &= \lambda_p \, A_{\pi\bar{K}} \Big[ \delta_{pu} \, \alpha_1 + \hat{\alpha}_4^p + \alpha_{4,\mathrm{EW}}^p \Big] + \lambda_p \, A_{\bar{K}\pi} \Big[ \delta_{pu} \, \alpha_2 + \tfrac{3}{2} \alpha_{3,\mathrm{EW}}^p \Big] \, , \\
\mathcal{A}_{\bar{B}^0 \to \pi^+ K^-} &= \lambda_p \, A_{\pi\bar{K}} \Big[ \delta_{pu} \, \alpha_1 + \hat{\alpha}_4^p + \alpha_{4,\mathrm{EW}}^p \Big] , \\
\sqrt{2}\, \mathcal{A}_{\bar{B}^0 \to \pi^0 \bar{K}^0} &= \lambda_p \, A_{\pi\bar{K}} \Big[ - \hat{\alpha}_4^p + \tfrac{1}{2} \alpha_{4,\mathrm{EW}}^p \Big] + \lambda_p \, A_{\bar{K}\pi} \Big[ \delta_{pu} \, \alpha_2 + \tfrac{3}{2} \alpha_{3,\mathrm{EW}}^p \Big] \, , \qquad (371)
\end{aligned}$$

up to power-suppressed annihilation topologies, which are not shown for simplicity (the exact expressions can be found in Ref. [696]). The corresponding amplitudes with $\pi \to \rho$ or/and $K \to K^*$ take the same form with the appropriate meson substitution. Here $\lambda_p = V_{pb} V_{ps}^*$ and the terms must be summed over $p = u, c$. The prefactors $A_{M_1 M_2} \propto f_{M_2} F^{BM_1}(M_2^2)$ reflect the factorised structure of the hadronic matrix elements in terms of a form factor and a decay constant.

The above $\Delta S = 1$ amplitudes are dominated by the charm penguin topology $\hat{\alpha}_4^c$. A non-zero direct $CP$ asymmetry is then generated via its interference with the contribution $\propto \lambda_u$. If this is the colour-allowed tree topology $\alpha_1$ – and if one neglects the other topologies for the moment – one obtains $\Delta A_{K\pi} = A_{CP}^{K^+\pi^0} - A_{CP}^{K^+\pi^-} = 0$. The observed value $\Delta A_{K\pi} = (12.2 \pm 2.2)\%$ constitutes the so-called $B \to \pi K$ $CP$-puzzle, which has attracted a lot of attention in the past, since it could hint at a NP contribution to the electroweak penguin amplitude $\alpha^c_{3,\mathrm{EW}}$. This interpretation is, however, flawed by the fact that the remaining topologies cannot be neglected. It is equally possible to explain the $B \to \pi K$ $CP$-puzzle by purely hadronic effects, if the colour-suppressed tree amplitude $\alpha_2$ and its phase are larger than naively expected.

In order to better understand the pattern of direct $CP$ asymmetries, perturbative corrections to the QCD factorisation framework have been worked out to next-to-next-to-leading order (NNLO), *i.e.* $\mathcal{O}(\alpha_s^2)$, accuracy.[36] According to the factorisation formula in Eq. (353), this includes two sets of hard-scattering kernels – vertex corrections ($T_i^I$) and spectator-scattering contributions ($T_i^{II}$) – for each topological amplitude.

Both types of $\mathcal{O}(\alpha_s^2)$ corrections have been worked out for the tree topologies [784, 785, 828–831]. Using the input parameters specified in Ref. [785], the colour-allowed tree amplitude for the $\pi\pi$ final states becomes (see also Ref. [831, 832])

$$\alpha_1(\pi\pi) = 1.000^{+0.029}_{-0.069} + (0.011^{+0.023}_{-0.050})i\,, \tag{372}$$

which is close to its leading-order value 1.009. As the amplitude is stable under radiative corrections, the theoretical uncertainties are small and the strong phase is negligible.

The situation is quite different for the colour-suppressed tree amplitude. In order to understand why the respective uncertainties are much larger, it is instructive to disentangle the various perturbative contributions. Extending Eq. (358) to NNLO [785, 831], the expression for $\alpha_2(\pi\pi)$ reads

$$\begin{aligned}\alpha_2(\pi\pi) &= 0.220 - [0.179 + 0.077\,i]_{\mathrm{NLO}} - [0.031 + 0.050\,i]_{\mathrm{NNLO}} \\ &\quad + \left[\frac{r_{\mathrm{sp}}}{0.445}\right]\Big\{[0.114]_{\mathrm{LOsp}} + [0.049 + 0.051i\,]_{\mathrm{NLOsp}} + [0.067]_{\mathrm{tw3}}\Big\} \\ &= 0.240^{+0.217}_{-0.125} + (-0.077^{+0.115}_{-0.078})i\,. \end{aligned} \tag{373}$$

Here the first term is the leading-order result, and the next two terms represent corrections to the vertex kernel $T_i^I$. Note that the real part almost cancels in this sum, which makes this amplitude particularly sensitive to the spectator-scattering mechanism ($T_i^{II}$). Unfortunately, the normalisation of this contribution – encoded in $r_{\mathrm{sp}}$ – is currently only poorly constrained, which is mainly related to the $B$-meson light-cone distribution amplitude ($r_{\mathrm{sp}} \propto 1/\lambda_B$, see Section 12.3.2). It is therefore possible to enhance the colour-suppressed tree amplitude by tuning the hadronic parameters, but the relative strong phase between $\alpha_1$ and $\alpha_2$ is stable under this variation, and predicted to be small. Thus, the $B \to \pi K$ $CP$-puzzle cannot be explained by perturbative corrections to $\alpha_2$.

The NNLO calculation of the penguin topologies is incomplete to date. Whereas the spectator-scattering contributions are known [786], only the current-current [833] and magnetic dipole [834] operator insertions to the kernels $T_i^I$ have been computed so far. The new $\mathcal{O}(\alpha_s^2)$ corrections are particularly important for the imaginary part of the QCD penguin amplitudes

[36] As direct $CP$ asymmetries first arise at $\mathcal{O}(\alpha_s)$, the counting of the perturbative orders is shifted by one unit, and the $\alpha_s^2$ correction represents an NLO effect in this case.

[833, 835]. The partial NNLO result reads

$$a_4^u(\pi\bar{K})/10^{-2} = (-2.46^{+0.49}_{-0.24}) + (-1.94^{+0.32}_{-0.20})i\,,$$
$$a_4^c(\pi\bar{K})/10^{-2} = (-3.34^{+0.43}_{-0.27}) + (-1.05^{+0.45}_{-0.36})i\,. \tag{374}$$

The most recent numbers for the electroweak penguin amplitudes can be found in Ref. [786].

The full QCD penguin amplitude $\hat{\alpha}_4^p$ in Eq. (371) is a combination of three terms,

$$\hat{\alpha}_4^p = a_4^p \pm r_\chi^{M_2} a_6^p + \beta_3^p\,, \tag{375}$$

where $a_4^p$ is the leading-power contribution from above, $r_\chi^{M_2} a_6^p$ is a power-suppressed scalar penguin amplitude (currently known to NLO [416]), and $\beta_3^p$ is the penguin annihilation amplitude. The plus (minus) sign applies to decays where the meson $M_1$, which picks up the spectator quark, is a pseudoscalar (vector) meson, see Figure 132. Eq. (375) has two important implications. First, as the second term depends on the spins of the final-state mesons, QCD factorisation predicts a specific hierarchy of the penguin amplitudes for final states with pseudoscalar and vector mesons. This pattern is clearly reflected in the experimental data [833]. Second, although the first term in Eq. (375) is the only leading-power contribution, all terms may numerically be of similar magnitude. The NNLO correction to $a_4^p$ is therefore diluted in the full QCD penguin amplitude $\hat{\alpha}_4^p$.

These features are essential for understanding the theoretical predictions for direct $CP$ asymmetries. As an example, Table 101 shows (partial) NNLO numbers for $\bar{B} \to \pi\bar{K}^{(*)}$ and $\bar{B} \to \rho\bar{K}$ decays [833].[37] First, one notes that the predicted $CP$ asymmetries are generically larger for $\pi\bar{K}^*$ and $\rho\bar{K}$ final states and have larger uncertainties than for $\pi\bar{K}$ final states. The reason is that the $a_6^p$ term, which exceeds the formally leading term $a_4^p$ and adds up constructively for $\pi\bar{K}$, is practically absent for $\pi\bar{K}^*$ and adds destructively for $\rho\bar{K}$. The charm penguin amplitude $\hat{\alpha}_4^c$ is therefore smaller and more uncertain, and the interference with the tree amplitudes is more important for $\pi\bar{K}^*$ and $\rho\bar{K}$. For the same reason, the NNLO corrections to $a_4^p$ are more pronounced for the $\pi\bar{K}^*$ and $\rho\bar{K}$ direct $CP$ asymmetries. If one adds the weak-annihilation term $\beta_3^p$ to the short-distance contribution, one is left with the column NNLO+LD (long distance). The weak annihilation has a large impact on the direct $CP$ asymmetries, and the parameterisation in Eq. (369) from Ref. [416] introduces sizeable theoretical errors.

The table also shows the direct $CP$ asymmetry difference $\Delta A_{K\pi}$ and the corresponding quantity for the $PV$ final states, in which the colour-allowed tree amplitude cancels out to good approximation, and the value of the asymmetry sum rule parameter

$$I_{K\pi} = A_{CP}^{K^+\pi^-} + A_{CP}^{K^0\pi^+}\frac{Br(K^0\pi^+)}{Br(K^+\pi^-)}\frac{\tau_{B^0}}{\tau_{B^+}} - 2A_{CP}^{K^+\pi^0}\frac{Br(K^+\pi^0)}{Br(K^+\pi^-)}\frac{\tau_{B^0}}{\tau_{B^+}} - 2A_{CP}^{K^0\pi^0}\frac{Br(K^0\pi^0)}{Br(K^+\pi^-)}, \tag{376}$$

where the colour-suppressed tree amplitude cancels out. The parameter $I_{K\pi}$ is expected to be small on general grounds [836], but all $K\pi$ $CP$ asymmetries must be measured to high precision.

[37] The table does not provide results for the $\rho\bar{K}^*$ final states, because (partial) NNLO accuracy is available only for the longitudinal polarisation amplitude, while the transverse amplitudes are much more uncertain, see Section 12.6. Polarisation effects in the $\rho\bar{K}^*$ final states are discussed in Section 12.6.3 and Table 103 quotes results for the direct $CP$ asymmetries.

Table 101: Direct $CP$ asymmetries $A_{CP}$, difference of direct CP asymmetry $\Delta A$, and the isospin breaking parameter $I$ (all in percent) for the $\pi K$, $\pi K^*$, and $\rho K$ final states. The theoretical values are taken from Ref. [833]. The column NNLO+LD (long distance) includes an estimate of non-factorisable annihilation contributions. The theoretical errors are due to CKM and hadronic parameters, respectively. The errors on the experimental values of $\Delta A$ and $I$ are computed from those of the individual observables appearing in Eq. (??) for $\pi K$ (and analogous sum rules for $\pi K^*$ and $\rho K$,) ignoring possible correlations. The fourth column contains the experimental world average (WA) values from Ref. [230]. The last column includes the precision on $I$ determined by fitting Eq. (??), using the complete set of measurements from Belle for $I_{K\pi}$, and from BaBar for $I_{K^*\pi}$ and $I_{K\rho}$ (Section 12.4). The first and second errors in parentheses are obtained by repeating the fit with the errors on the branching fractions and $A_{CP}$ scaled to the expected results with 5 and $50\,\mathrm{ab}^{-1}$ of Belle II data, respectively.

| | NLO | NNLO | NNLO + LD | Exp (WA) | Exp (Belle II) |
|---|---|---|---|---|---|
| $A_{CP}^{\pi^-\bar{K}^0}$ | $0.71^{+0.13+0.21}_{-0.14-0.19}$ | $0.77^{+0.14+0.23}_{-0.15-0.22}$ | $0.10^{+0.02+1.24}_{-0.02-0.27}$ | $-1.7\pm1.6$ | |
| $A_{CP}^{\pi^0K^-}$ | $9.42^{+1.77+1.87}_{-1.76-1.88}$ | $10.18^{+1.91+2.03}_{-1.90-2.62}$ | $-1.17^{+0.22+20.00}_{-0.22-6.62}$ | $4.0\pm2.1$ | |
| $A_{CP}^{\pi^+K^-}$ | $7.25^{+1.36+2.13}_{-1.36-2.58}$ | $8.08^{+1.52+2.52}_{-1.51-2.65}$ | $-3.23^{+0.61+19.17}_{-0.61-3.36}$ | $-8.2\pm0.6$ | |
| $A_{CP}^{\pi^0\bar{K}^0}$ | $-4.27^{+0.83+1.48}_{-0.77-2.23}$ | $-4.33^{+0.84+3.29}_{-0.78-2.32}$ | $-1.41^{+0.27+5.54}_{-0.25-6.10}$ | $1\pm10$ | $-14\pm13$ |
| $\Delta A_{K\pi}$ | $2.17^{+0.40+1.39}_{-0.40-0.74}$ | $2.10^{+0.39+1.40}_{-0.39-2.86}$ | $2.07^{+0.39+2.76}_{-0.39-4.55}$ | $12.2\pm2.2$ | |
| $I_{K\pi}$ | $-1.15^{+0.21+0.55}_{-0.22-0.84}$ | $-0.88^{+0.16+1.31}_{-0.17-0.91}$ | $-0.48^{+0.09+1.09}_{-0.09-1.15}$ | $-14\pm11$ | $-27\pm14(7)(3)$ |
| $A_{CP}^{\pi^-\bar{K}^{*0}}$ | $1.36^{+0.25+0.60}_{-0.26-0.47}$ | $1.49^{+0.27+0.69}_{-0.29-0.56}$ | $0.27^{+0.05+3.18}_{-0.05-0.67}$ | $-3.8\pm4.2$ | |
| $A_{CP}^{\pi^0K^{*-}}$ | $13.85^{+2.40+5.84}_{-2.70-5.86}$ | $18.16^{+3.11+7.79}_{-3.52-10.57}$ | $-15.81^{+3.01+69.35}_{-2.83-15.39}$ | $-6\pm24$ | $-6\pm24$ |
| $A_{CP}^{\pi^+K^{*-}}$ | $11.18^{+2.00+9.75}_{-2.15-10.62}$ | $19.70^{+3.37+10.54}_{-3.80-11.42}$ | $-23.07^{+4.35+86.20}_{-4.05-20.64}$ | $-23\pm6$ | |
| $A_{CP}^{\pi^0\bar{K}^{*0}}$ | $-17.23^{+3.33+7.59}_{-3.00-12.57}$ | $-15.11^{+2.93+12.34}_{-2.65-10.64}$ | $2.16^{+0.39+17.53}_{-0.42-36.80}$ | $-15\pm13$ | |
| $\Delta A_{CP}^{K^*\pi}$ | $2.68^{+0.72+5.44}_{-0.67-4.30}$ | $-1.54^{+0.45+4.60}_{-0.58-9.19}$ | $7.26^{+1.21+12.78}_{-1.34-20.65}$ | $17\pm25$ | |
| $I_{K^*\pi}$ | $-7.18^{+1.38+3.38}_{-1.28-5.35}$ | $-3.45^{+0.67+9.48}_{-0.59-4.95}$ | $-1.02^{+0.19+4.32}_{-0.18-7.86}$ | $-5\pm45$ | $69\pm32(15)(6)$ |
| $A_{CP}^{\rho^-\bar{K}^0}$ | $0.38^{+0.07+0.16}_{-0.07-0.27}$ | $0.22^{+0.04+0.19}_{-0.04-0.17}$ | $0.30^{+0.06+2.28}_{-0.06-2.39}$ | $-12\pm17$ | |
| $A_{CP}^{\rho^0K^-}$ | $-19.31^{+3.42+13.95}_{-3.61-8.96}$ | $-4.17^{+0.75+19.26}_{-0.80-19.52}$ | $43.73^{+7.07+44.00}_{-7.62-137.77}$ | $37\pm11$ | |
| $A_{CP}^{\rho^+K^-}$ | $-5.13^{+0.95+6.38}_{-0.97-4.02}$ | $1.50^{+0.29+8.69}_{-0.27-10.36}$ | $25.93^{+4.43+25.40}_{-4.90-75.63}$ | $20\pm11$ | |
| $A_{CP}^{\rho^0\bar{K}^0}$ | $8.63^{+1.59+2.31}_{-1.65-1.69}$ | $8.99^{+1.66+3.60}_{-1.71-7.44}$ | $-0.42^{+0.08+19.49}_{-0.08-8.78}$ | $6\pm20$ | $5\pm26$ |
| $\Delta A_{K\rho}$ | $-14.17^{+2.80+7.98}_{-2.96-5.39}$ | $-5.67^{+0.96+10.86}_{-1.01-9.79}$ | $17.80^{+3.15+19.51}_{-3.01-62.44}$ | $17\pm16$ | |
| $I_{K\rho}$ | $-8.75^{+1.62+4.78}_{-1.66-6.48}$ | $-10.84^{+1.98+11.67}_{-2.09-9.09}$ | $-2.43^{+0.46+4.60}_{-0.42-19.43}$ | $-37\pm37$ | $-44\pm49(25)(11)$ |

The experimental uncertainty of the asymmetry sum rule is currently dominated by the $\pi^0\bar{K}^0$ direct $CP$ asymmetry, which will be one of the key measurements at Belle II (see the following subsection). The related $\pi\bar{K}^*$, $\rho\bar{K}$ and $\rho\bar{K}^*$ channels provide additional insights on the pattern of direct $CP$ asymmetries in penguin-dominated $\Delta S=1$ transitions. As discussed above, all interference effects, and also the theoretical uncertainties are expected to be enhanced in these channels, which are therefore of significant interest for both NP searches and theory testing.

Table 102: Branching fractions (top) and $A_{CP}$ (bottom) measurements for $B \to K\pi$ decays from Ref. [230].

| $\mathcal{B}(10^{-6})$ | | |
|---|---|---|
| Mode | BaBar | Belle |
| $K^+\pi^-$ | $19.1 \pm 0.6 \pm 0.6$ | $20.0 \pm 0.34 \pm 0.60$ |
| $K^+\pi^0$ | $13.6 \pm 0.6 \pm 0.7$ | $12.62 \pm 0.31 \pm 0.56$ |
| $K^0\pi^+$ | $23.9 \pm 1.1 \pm 1.0$ | $23.97 \pm 0.53 \pm 0.71$ |
| $K^0\pi^0$ | $10.1 \pm 0.6 \pm 0.4$ | $9.68 \pm 0.46 \pm 0.50$ |

| $A_{CP}$ | | | |
|---|---|---|---|
| Mode | BaBar | Belle | LHCb |
| $K^+\pi^-$ | $-0.107 \pm 0.016^{+0.006}_{-0.004}$ | $-0.069 \pm 0.014 \pm 0.007$ | $-0.080 \pm 0.007 \pm 0.003$ |
| $K^+\pi^0$ | $0.030 \pm 0.039 \pm 0.010$ | $0.043 \pm 0.024 \pm 0.002$ | |
| $K^0\pi^+$ | $-0.029 \pm 0.039 \pm 0.010$ | $-0.011 \pm 0.021 \pm 0.006$ | $-0.022 \pm 0.025 \pm 0.010$ |
| $K^0\pi^0$ | $-0.13 \pm 0.13 \pm 0.03$ | $0.14 \pm 0.13 \pm 0.06$ | |

The theoretical predictions shown in Table 101 can be further improved in the future. On the one hand, this requires completing the NNLO calculation of the leading-power penguin amplitude $a_4^p$. In view of its phenomenological relevance, one should also consider computing the scalar penguin amplitude $a_6^p$ to the same precision. In addition, one should attempt to improve the modelling of the weak-annihilation amplitudes, *e.g.* through a data-driven approach (see previous section).

While this short review focussed on the $\pi K$ channels and their relatives, many more direct $CP$ asymmetries can be measured in charmless two-body decays. In general, one expects that the same theoretical methods hold for the leading amplitudes in the heavy-quark limit (in the case of $B \to VV$ decays, this applies then only to the longitudinal amplitude).

## *12.4. Experimental status of $B \to \pi K^{(*)}$ and $\rho K^{(*)}$ decays*

*[Contributing Author: P. Goldenzweig]*

The experimental status of the branching fraction and $A_{CP}$ measurements of the $K\pi$ system are displayed in Table 102. Both Belle and BaBar report a complete set of measurements of the eight observables, while LHCb only reports the values for $A_{CP}^{K^+\pi^-}$ and $A_{CP}^{K^0\pi^+}$. The most demanding of these measurements is the all-neutral final state $K^0\pi^0$. It requires vertex reconstruction of the charged pions from the neutral kaon decays and depends crucially on a vertex detector with a large radial acceptance. Belle measures $A_{CP}^{K^0\pi^0} = +0.14 \pm 0.13 \pm 0.06$ with a data sample of approximately 600 fb$^{-1}$ [837]. The main systematic uncertainty contributions are ordered from largest to smallest as follows: tag side interference ($\pm 0.054$), vertex reconstruction ($\pm 0.022$), background fraction ($\pm 0.022$), and potential fit biases ($\pm 0.020$). These are expected to improve with the larger data set, particularly since similar systematic uncertainties in the analyses of the other $K\pi$ modes, which all have more signal events, are all substantially smaller.

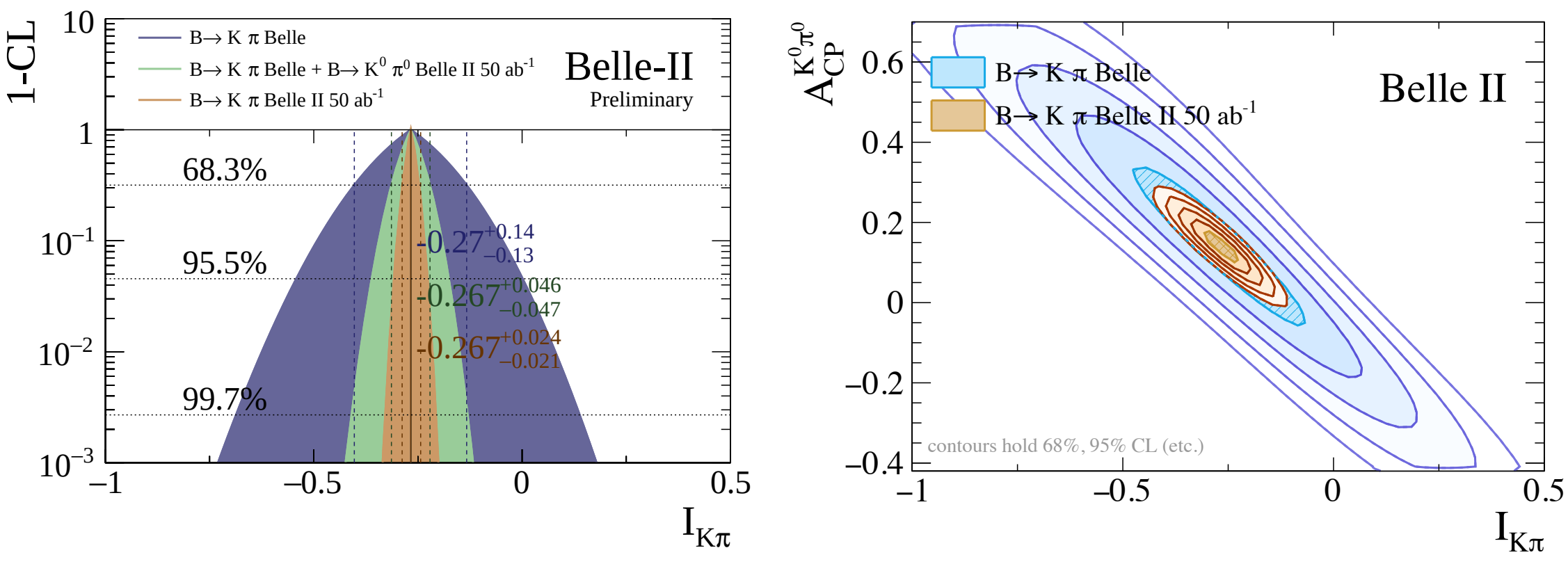

Fig. 134: Precision of $I_{K\pi}$ with: current Belle results; $K^0\pi^0$ with $50\,\text{ab}^{-1}$; all channels with $50\,\text{ab}^{-1}$ (left). 2D comparison of $I_{K\pi}$ vs. $A_{CP}^{K^0\pi^0}$ with current Belle results and all channels with $50\,\text{ab}^{-1}$ (right).

Belle has found the value of the isospin breaking identity parameter, $I_{K\pi}$ as defined in Eq.(376), to be $-0.270 \pm 0.132 \pm 0.060$ [727]. To determine the effect on the precision of $I_{K\pi}$ with Belle II data, the errors on Belle's measurements of the branching fractions and $A_{CP}$ are scaled to the expectations at 5 and $50\,\text{ab}^{-1}$, and fits are performed with the GammaCombo fit package [838] to extract $I_{K\pi}$. The only possible correlated errors for the $A_{CP}$ measurements are detector bias errors, which are estimated with different methods for each channel; thus the bias errors are assumed to be uncorrelated. Additionally, the systematic uncertainties are conservatively estimated and they are still smaller than the statistical errors. With the large Belle II dataset, the correlations will need to be taken into account. The precision with 5 (50) $\text{ab}^{-1}$ is found to be 0.07 (0.03). These results are shown in the first horizontal block of Table 101, alongside the NLO, NNLO, and NNLO+LD predictions (described in detail in Section 12.3.4), all in %. To isolate the effect of the all neutral mode, an additional fit is performed where only the $K^0\pi^0$ measurements are scaled to the expectation at $50\,\text{ab}^{-1}$. Clearly the precision is limited by $K^0\pi^0$, as displayed in Fig. 134 (left). The dependence on the precision of $I_{K\pi}$ is further demonstrated by a simplistic 2D comparison of $I_{K\pi}$ vs. $A_{CP}^{K^0\pi^0}$ shown in Fig. 134 (right).

The experimental results for the branching fractions and $A_{CP}$ measurements for the $K^*\pi$, $K\rho$ and $K^*\rho$ systems are tabulated in Ref. [230]. To determine the effect on the precision of the isospin breaking parameters $I_{K^*\pi}$, $I_{K\rho}$ and $I_{K^*\rho}$ with Belle II data, the errors on the branching fractions and $A_{CP}$ measurements are scaled to the expectations at 5 and $50\,\text{ab}^{-1}$, and fits are performed to extract the corresponding $I$ (analogous to the $K\pi$ system). The results of the fits to $I_{K^*\pi}$ and $I_{K\rho}$ are listed in the second and third blocks of Table 101, respectively, alongside the theoretical predictions. Here, the inputs to the fits are from BaBar's complete set of branching fraction and $A_{CP}$ measurements, as Belle does not yet have results for all observables. The vector-vector decay $K^*\rho$ is discussed in detail in Sec. 12.6.3. Here, the comparison of NNLO results to experiment is presently not possible, as the longitudinal $A_{CP}$ for $K^{*0}\,\rho^+$ has not been measured. Furthermore, the NNLO computation is not possible for transverse amplitudes, as they are power-suppressed and there is no complete QCD factorisation theorem for them. The results for the fit to the $K^*\rho$ system

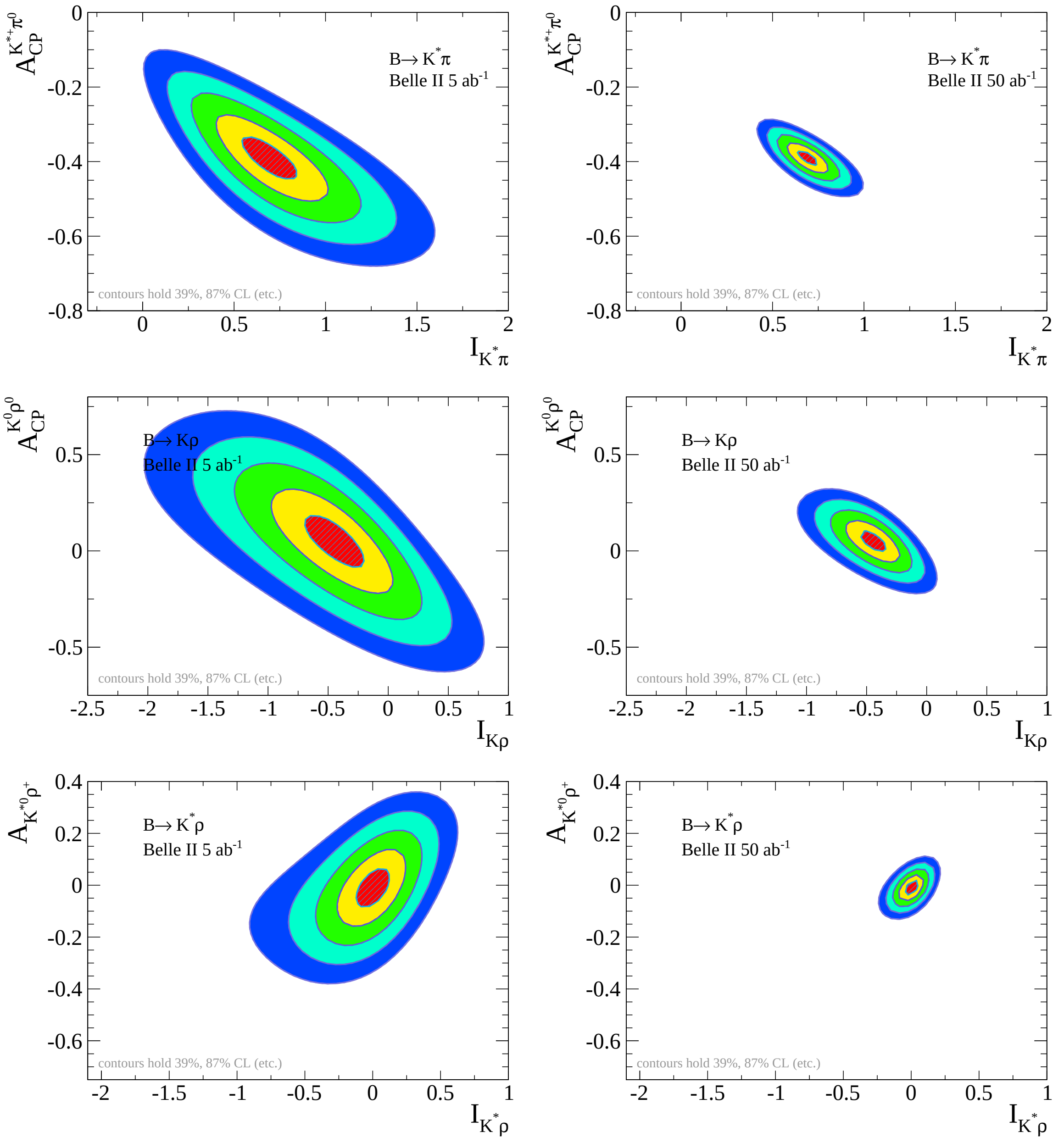

Fig. 135: Demonstration of the limitations on the measurments of the isospin sum rules due to the precision of $A_{CP}^{K^{*+}\pi^0}$ for $I_{K^*\pi}$ (top), $A_{CP}^{K^0\rho^0}$ for $I_{K\rho}$ (middle), and $A_{CP}^{K^{*0}\rho^+}$ for $I_{K^*\rho}$ (bottom) for $5\,\mathrm{ab}^{-1}$ (top) and $50\,\mathrm{ab}^{-1}$ (bottom) of Belle II data. The results for $I$ are listed in Table 101.

(also using BaBar's complete set of branching fractions) are $I_{K^*\rho} = 0.4 \pm 26.4(12.4)(4.4)\%$, where the first and second errors in parentheses are obtained by repeating the fit with the errors on the branching fractions and $A_{CP}$ scaled to the expected results with 5 and $50\,\mathrm{ab}^{-1}$ of Belle II data, respectively. Analogous to the $K\pi$ system, 2D contours are plotted for the isospin breaking parameters vs. the channel with the largest error in $A_{CP}$ (Fig. 135): $K^{*+}\pi^0$, $K^0\rho^0$ and $K^{*0}\rho^+$ for the $K^*\pi$, $K\rho$ and $K^*\rho$ systems, respectively.

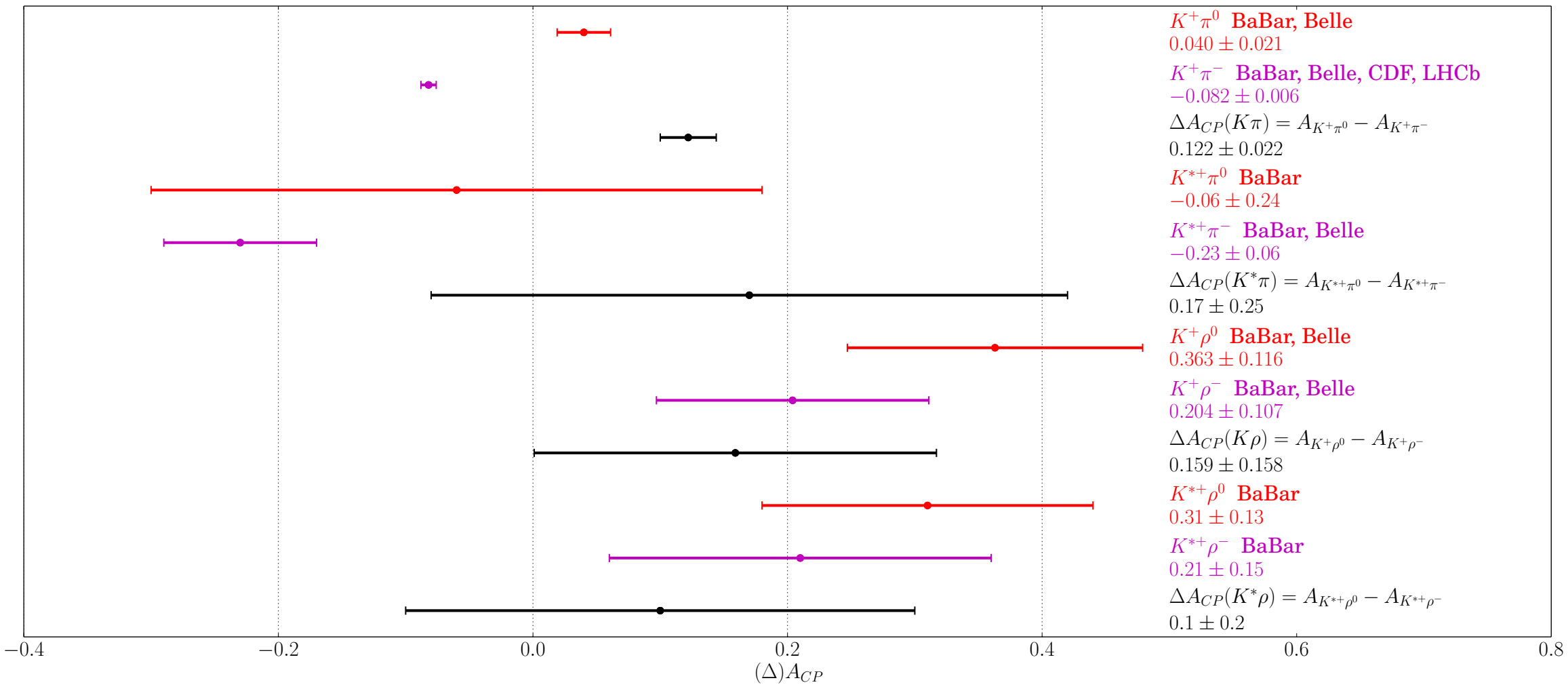

Fig. 136: World averages of $A_{CP}$ and $\Delta A_{CP}$ for the $K\pi$, $K^*\pi$, $K\rho$ and $K^*\rho$ systems.

A summary of the world average results for $A_{CP}$ and $\Delta A_{CP}$ for all four systems is provided in Fig. 136 and Tab. 101. While the uncertainty has improved greatly in $K\pi$, it is still too large in the $PV$ and $VV$ systems to be conclusive and thus requires high-precision measurements from Belle II.

### *12.5. CP violation in $B_s^0$ decays and $B_s^0 \to K^0\bar{K}^0$*

*[Contributing Author: B. Pal]*

The observation of the decay $B_s^0 \to K^0\bar{K}^0$ (candidate $K^0$ mesons are reconstructed via the decay $K_S^0 \to \pi^+\pi^-$) by the Belle Collaboration [839] is the first observation of a charmless two-body $B_s^0$ decay involving only neutral hadrons in the final state. In the SM, this decay proceeds mainly via a $b \to s$ penguin transition, and thus is sensitive to any NP that propagates in the internal loop.

With the full Belle $\Upsilon(5S)$ data set of 121.4 fb$^{-1}$, a total of $29.0^{+8.5}_{-7.6}$ signal candidates are observed with a significance of $5.1\sigma$. The measured branching fraction is $\mathcal{B}(B_s^0 \to K^0\bar{K}^0) = (19.6^{+5.8}_{-5.1} \pm 1.0 \pm 2.0) \times 10^{-6}$, where the first uncertainty is statistical, the second is systematic, and the third reflects the uncertainty due to the total number of $B_s^0\bar{B}_s^0$ pairs. This result is in good agreement with SM predictions [697, 781, 822, 827, 840–843]. This branching fraction implies that Belle II will reconstruct over 1000 of these decays with $5\,\text{ab}^{-1}$ of $\Upsilon(5S)$ data (assuming similar reconstruction efficiency to Belle). Results of a Toy MC study are shown in Fig. 137, where the signal region projections are identical to Ref. [839]. Such a sample would allow for a much higher sensitivity search for NP in this $b \to s$ penguin-dominated decay. In particular, the SM prediction for $CP$ violation in $B_s^0 \to K^0\bar{K}^0$ is very small, and thus any $CP$ asymmetry observed could be an indication of NP. It has been argued in Refs. [844, 845] that the direct $CP$ asymmetry of the decay $B_s^0 \to K^0\bar{K}^0$ is a very promising observable to search for the effects of NP. It was shown that the direct $CP$ asymmetry, which is $\lesssim 1\%$ in the SM, can be an order of magnitude larger in the presence of SUSY, while the branching fraction remains unaffected.

Measuring $CP$ violation in $B_s^0$ decays at $e^+e^-$ $B$ factories, however, cannot be performed using conventional time-dependent techniques, which require the two $B_s^0$ vertices be reconstructed with an accuracy of about $\sim 10\ \mu m$. The current Belle II detector design will not achieve $\sim 10\ \mu m$ resolution. Additionally, determining the flavour of the initial state $B_s^0$ will be difficult, due to the rapid $B_s^0$ oscillations. However, it can be studied in a manner similar to $CP$ violation studies in $K^0$ decays [846, 847]. In the SM, the lifetime distribution of $B_s^0$ decays into a fixed $CP$ eigenstate, $f_{CP}$ (*e.g.*, the $K_S^0 K_S^0$ final state is a $CP$-even eigenstate) is governed by a single exponential. $CP$ violation would be established if a second exponential component were observed in the decay, *i.e.*, the $CP$-eigenstate would not be a mass-eigenstate. The Belle II experiment will be able to perform this study and will clarify the presence of NP in the decay $B_s^0 \to K^0 \bar{K}^0$.

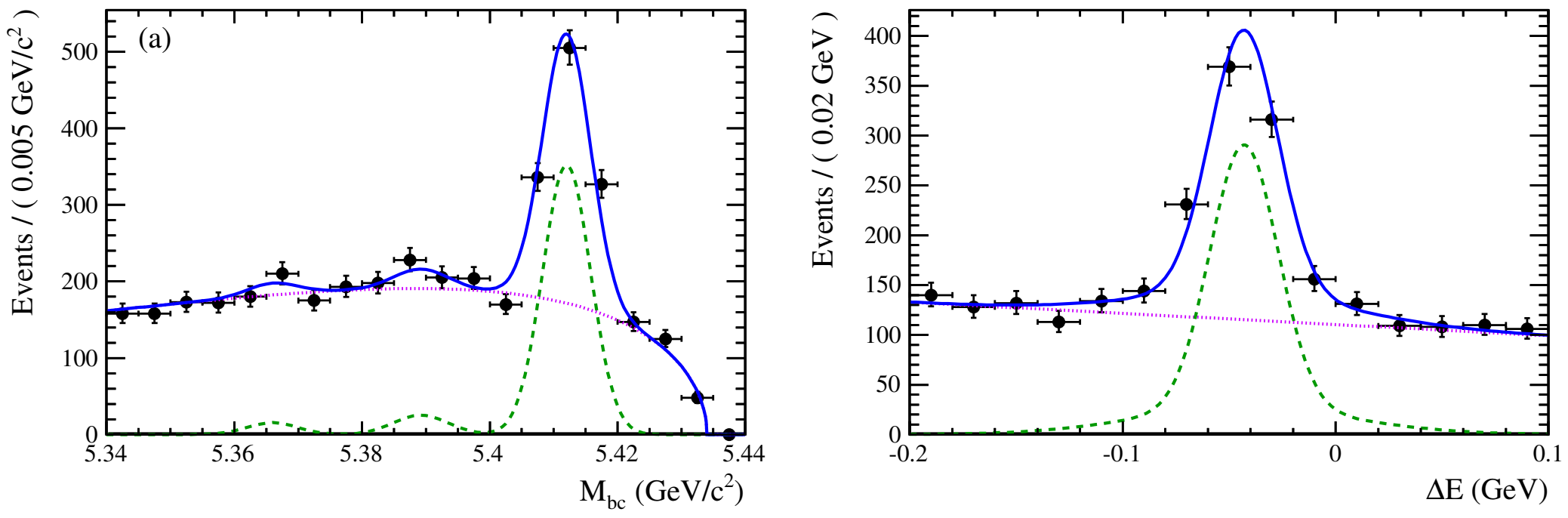

Fig. 137: Projections for $B_s \to K^0 \overline{K}^0$ decay by a toy MC study for $5\,\mathrm{ab}^{-1}$ of Belle II $\Upsilon(5\mathrm{S})$ data, based on the Belle measurement [839]: (a) $M_{\rm bc}$ in $\Delta E \in (-0.11, 0.02)$ GeV; and (b) $\Delta E$ in $M_{\rm bc} \in (5.405, 5.427)$ GeV$/c^2$. Both projections contain a cut on the continuum suppression network output variable of $C'_{\rm NN} > 0.5$. The points with error bars are data, the (green) dashed curves show the signal, (magenta) dotted curves show the continuum background, and (blue) solid curves show the total. The three peaks in $M_{\rm bc}$ arise from $\Upsilon(5S) \to B_s^0 \bar{B}_s^0, B_s^{*0} \bar{B}_s^0 + B_s^0 \bar{B}_s^{*0}$, and $B_s^{*0} \bar{B}_s^{*0}$ decays.

## *12.6. $B_{(s)} \to VV$ decays*

Theoretically, $B_{(s)} \to VV$ decays are two-body final states, however, experimentally they are at least four-body decays, since the vector mesons decay via the strong interaction with a non-negligible width. Vector mesons can be produced in three polarisation states, corresponding to the longitudinal and the two helicity $\pm 1$ amplitudes. The fraction of a given polarisation state is an interesting observable, as well as other observables constructed from the helicity amplitudes, in addition to the branching fractions. The phenomenology of $B \to VV$ decays offers rich opportunities for our understanding of the mechanism for hadronic weak decays and their $CP$ asymmetry, and the search for physics beyond the SM. The following three subsections provide first an overview of the theoretical and experimental status of $VV$ final states, then a discussion of triple-product observables, and finally, of polarisation in the system of the $\rho K^*$ final states.

### *12.6.1. Polarisation . [Contributing Authors: M. Beneke, C.-D. Lu]*

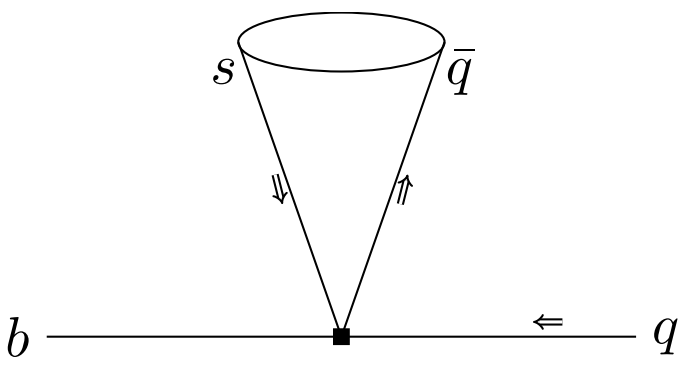

Fig. 138: Naive quark helicities in the charmless $B_{(s)} \to VV$ decays.

In the SM, all charmless $B$ decays occur through the $V - A$ weak interaction. This implies that the outgoing light quark in the current containing the $b$-quark is left-handed, while the anti-quark from the other current is right-handed. This makes one of the final-state vector mesons naturally longitudinally polarised as in Fig. 138. The other must then also be longitudinal for pseudoscalar $B$ meson decay. To form a negatively polarised vector meson, one has to flip the spin of the energetic anti-quark, which results in a negative helicity and $1/m_b$ suppression [795]. To obtain a vector meson in the positive helicity state, one has to flip the spin orientation of two energetic quarks, which is even further suppressed. Therefore, it is naively expected that the helicity amplitudes $A_i$ in heavy meson decay satisfy the power-counting hierarchy [848]

$$A_0 : A_- : A_+ = 1 : \frac{\Lambda}{m_b} : \left(\frac{\Lambda}{m_b}\right)^2 , \tag{377}$$

where $\Lambda$ denotes the strong-interaction scale. In the naive factorisation approach, longitudinal polarisation dominates the branching fractions of $B \to VV$ decays [849]. In the QCD factorisation (QCDF) approach this continues to hold formally in the leading-power approximation [850, 851], but is violated numerically by several large power-suppressed effects [795, 796].

The hierarchy in Eq. (377) is indeed verified in the experimental measurements of tree-dominated final states such as $B \to \rho^+\rho^-$, etc. In sharp contrast to the expectations, however, a large transverse polarisation fraction (around 50%) was observed in $B \to K^*\phi$ decays by Belle [852] and BaBar [853]. Large transverse polarisation of order 50% has subsequently also been observed for $B \to K^*\rho$, $B_s \to K^*\phi$ and $B_s \to \phi\phi$ decays. The fact that the scaling behaviour shown in Eq. (377) is apparently violated — at least numerically — in penguin-dominated $B$ decays has triggered considerable theoretical interest both in the QCD factorisation approach [795, 796, 799, 854–856] and in the perturbative QCD (PQCD) approach [841, 857–860].

Several mechanisms have been proposed to explain the observation of large transverse polarisation in penguin-dominated decays. The most convincing appears to be a large annihilation contribution from the scalar $(S - P) \times (S + P)$ penguin operator $Q_6$ in the weak effective Hamiltonian, originally introduced in Ref. [795] and further analysed in the QCD factorisation framework in Ref. [796]. This operator is already known to contribute significantly to final states with pseudoscalar mesons [802, 819]. From Fig. 139, one can see that one quark spin needs to be flipped to obtain longitudinal or[1] negative polarisations. As a result, although power-suppressed, the contribution from this diagram is of the same order for the longitudinal and negative helicity amplitude. Since the annihilation contribution to the longitudinal amplitude is already sizeable for longitudinally polarised vector mesons, whose factorisable penguin amplitude is suppressed, it is plausible the polarisation fractions satisfy

$$f_L \approx f_{\parallel} \approx f_{\perp}. \tag{378}$$

The basic picture is confirmed in the perturbative QCD approach [841], that is, both QCDF and PQCD invoke penguin annihilation to explain the observed large transverse polarisation fraction in the penguin-dominated $B \to VV$ decays $B \to K^*\phi$, $B \to K^*\rho$. Recent updates of the respective results of branching ratios, longitudinal and transverse polarisation fractions, relative strong phases and the $CP$ asymmetry variables in $B \to VV$ decays can be found in Refs. [855, 856, 861]. Other explanations of the large transverse polarisation include its attribution to the charm-penguin amplitude, final-state interactions [862, 863], form-factor tuning [864], and even NP [865, 866], some of which have already been ruled out by experiment, since they cannot produce the relation $f_{\parallel} \approx f_{\perp}$ and the observed relative strong phases between different polarisation states.

We note that, theoretically, the longitudinal amplitude for $VV$ final states is similar to the single decay amplitude in $PV$ and $VP$ final states. However, the helicity $\pm 1$ amplitudes are power-suppressed and theoretically not as well understood. As a consequence the prediction of polarisation-related observables in $B \to VV$ modes are not on the same footing as the calculations of branching fractions and even $CP$ asymmetries for the $PP$, $PV$ and $VP$ final states. When transverse polarisation is sizeable as for penguin-dominated final states, even the calculation of branching fractions and $CP$ asymmetries from first principles is on less solid grounds. This should be kept in mind when comparing observations to theoretical predictions. As an aside we mention that the hierarchy in Eq. (377) is not respected by electromagnetic effects, which generate a transverse electroweak penguin amplitude, which instead of Eq. (377) satisfies [854]

$$A_0 : A_- : A_+ = 1 : \frac{\alpha_{\rm em} m_b}{\Lambda} : \alpha_{\rm em}\,. \tag{379}$$

The measurement and theoretical interpretation of polarisation in colour-suppressed $B$ decays is also not completely settled. The branching fraction $B^0 \to \rho^0\rho^0$ was measured by BaBar and Belle in 2008 as $(0.9 \pm 0.32 \pm 0.14) \times 10^{-6}$ [715] and $(0.4 \pm 0.4^{+0.2}_{-0.3}) \times 10^{-6}$ [867], respectively. With such a small decay rate of $B^0 \to \rho^0\rho^0$, by isospin symmetry the rate for the decay $B^0 \to \rho^+\rho^-$ ought to be twice that of $B^+ \to \rho^+\rho^0$. Experimentally, however, the first is only slightly larger than the second, indicating significant isospin violation. In 2012, Belle updated its $B^0 \to \rho^0\rho^0$ to $1.02 \pm 0.30 \pm 0.15$ [729], in better agreement with the isospin triangle expectation. However, the increase comes from the transverse polarisation, which results in a very small longitudinal polarisation fraction $f_L = 0.21^{+0.18}_{-0.22} \pm 0.13$ [729] in the Belle measurement, which is in conflict with the BaBar measurement and in particular the LHCb result [714] $f_L = 0.745^{+0.048}_{-0.058} \pm 0.034$, which dominates the world average. The theoretical interpretation also remains somewhat ambiguous. Leading-order PQCD calculations find the longitudinal polarisation fraction for the $B^0 \to \rho^0\rho^0$ decay to be as small as 12% [861]. This results from a large cancellation of two hard-scattering emission diagrams and the annihilation diagram in the longitudinal polarisation amplitude, while the chirally enhanced annihilation and hard-scattering emission diagrams provide a sizeable transverse amplitude. On the other hand, in the QCDF approach Ref. [796] finds $f_L = (90^{+3+8}_{-4-63})\%$, which supports a larger value, although with a very large uncertainty, which arises from non-factorisable spectator-scattering contributions. The current situation leaves much room for better measurements of the branching fractions and the longitudinal polarisation fractions of colour-suppressed $VV$ final states and their theoretical understanding. The existing $f_L$ measurements in $B$ decays are summarised in Fig. 140. Besides the currently measured channels, other modes such as $B^0 \to \rho^0\omega$, $B^0 \to \omega\omega$, $B^0 \to K^{*+}K^{*-}$, $B^- \to \phi\rho^-$, $B^0 \to \phi\rho^0$ and $B^0 \to \phi\omega$ will provide further insight into the QCD dynamics that governs the different helicity amplitudes.

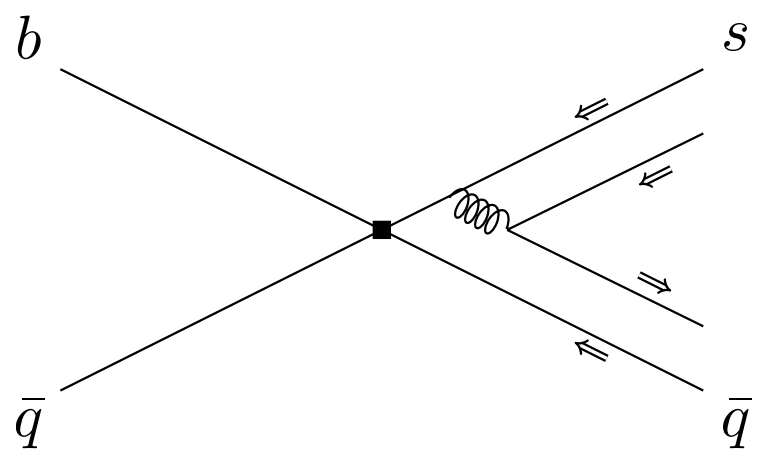

Fig. 139: Quark helicities in the penguin annihilation diagram of charmless $B \to VV$ decays.

The experimental effort to study the charmless $B_s \to VV$ decays has already started, for example with the measurements of branching ratios and polarisation fractions of decays $B_s \to \bar{K}^{*0}K^{*0}$, $B_s \to \phi\bar{K}^{*0}$ and $B_s \to \phi\phi$. However, most of the $B_s \to VV$ decays have not yet been measured, leaving much room for the Belle II experiment. Many of the branching ratios and direct $CP$ observables need the input from Belle II, since it is more difficult for the LHCb experiment to measure absolute branching fractions. The results of the existing $f_L$ measurements for $B_s$ decays are also summarised in Fig. 140.

Theoretically, the prediction of $B_s \to VV$ decays follows similar patterns as $B \to VV$, but there is a larger uncertainty for the non-perturbative input parameters, such as the form factors and light-cone wave functions/distribution amplitudes. Results for all $B_s \to VV$ modes can be found in Refs. [796, 856] and Ref. [861] in the QCDF and PQCD approach, respectively, and show agreements and disagreements depending on whether there is agreement on the underlying dynamical mechanism governing a particular decay. In general, the QCDF approach adopts a more conservative approach on theoretical uncertainties due to power-suppressed effects, including all helicity amplitudes. The above mentioned references also contain a comprehensive coverage of direct $CP$ asymmetries and the parameters $f_\perp$, $\phi_\parallel = \arg(A_\parallel/A_0)$, and $\phi_\perp = \arg(A_\perp/A_0)$, which enter the complete angular analysis. These parameters (and the corresponding $CP$ asymmetries) are in part related to the positive-helicity amplitude, which is expected to be strongly suppressed in the SM, but is otherwise poorly understood theoretically. Many of the existing polarisation results refer only to $f_L$, while the perpendicular polarisation fraction, $f_\perp$, the relative phases $\phi_\parallel$, $\phi_\perp$, and the helicity specific $CP$ asymmetry parameters $A^0_{CP}$ and $A^\perp_{CP}$ have been measured only in five channels: $B^0(B^+) \to K^*(K^{*+})\phi$ and $B_s \to \bar{K}^{*0}\phi$, $B_s \to \bar{K}^{*0}K^{*0}$ and $B_s \to \phi\phi$. Complete angular analyses of penguin-dominated and colour-suppressed $VV$ final states are expected to provide further insight on the surprisingly complex dynamics of these decays, and are, at least in principle (see Ref. [868] for an example), sensitive to NP with right-handed currents.

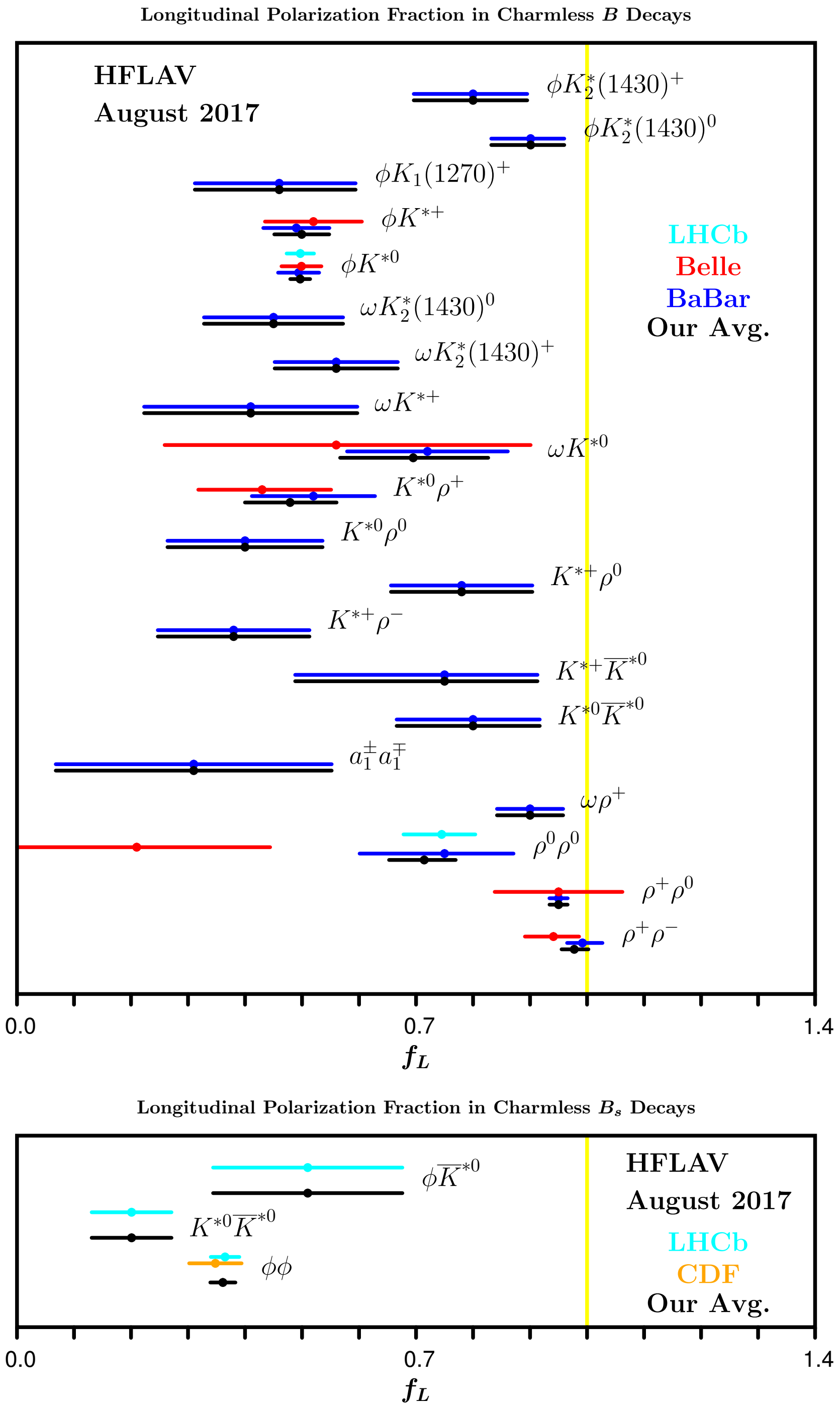

Fig. 140: Longitudinal polarisation fraction in charmless $B$ and $B_s$ decays from Ref. [230].

*12.6.2. Triple product asymmetries . [Contributing Author: A. Datta]*

In addition to the standard polarisation observables, one can measure the so-called triple product asymmetries (TPAs) [869, 870] in the angular distribution of $B \to V_1 V_2$ decays. As any $CP$-violating quantity, a non-vanishing TPA needs the interference of at least two amplitudes with a weak phase difference $\Delta\phi$. However, while direct $CP$ asymmetries are proportional to $\sin\Delta\phi\sin\Delta\delta$ and therefore also require a strong phase difference $\Delta\delta$, TPAs go as $\sin\Delta\phi\cos\Delta\delta$. Hence direct $CP$ violation and TPAs complement each other. If the strong phases are small, then the TPA is maximal. Even in the absence of $CP$-violating effects, T-odd triple products (also called "fake" TPAs), which go as $\cos\Delta\phi\sin\Delta\delta$, can provide useful complementary information on NP [871].

*Basic definitions.* The general amplitude for $B(p) \to V_1(k_1, \varepsilon)V_2(k_2, \eta)$ is

$$A_{\lambda_1,\lambda_2} = a\,\varepsilon^*_{\lambda_1}\cdot\eta^*_{\lambda_2} + \frac{b}{m_B^2}(p\cdot\varepsilon^*_{\lambda_1})(p\cdot\eta^*_{\lambda_2}) + i\frac{c}{m_B^2}\epsilon_{\mu\nu\rho\sigma}p^\mu q^\nu \varepsilon^{*\rho}_{\lambda_1}\eta^{*\sigma}_{\lambda_2}, \tag{380}$$

where $q \equiv k_1 - k_2$. The amplitude $c$ has $L = 1$ and is parity-odd while the amplitudes $a$ and $b$ are combinations of the $L = 0$ and $L = 2$ partial waves. We note that in the $B$ meson rest frame the last term takes the form of a triple product $TP \equiv \vec{q}\cdot(\vec{\epsilon}^* \times \vec{\eta}^*)$. The TPAs are related to the interference of this amplitude with the other two, $\mathrm{Im}\,(ac^*)$ and $\mathrm{Im}\,(bc^*)$.

The polarisation vectors can be transverse $(\epsilon_\pm, \eta_\pm)$ or longitudinal $(\epsilon_0, \eta_0)$. Helicity conservation allows $A_{+,+}, A_{-,-}, A_{0,0}$ which we will denote as $A_+, A_-, A_0$. The amplitudes in the transversity bases are related to $a$, $b$ and $c$ above by

$$\begin{aligned}
A_0 &= -ax - \frac{m_1 m_2}{m_B^2}b(x^2-1) \approx -(2a+b)\frac{m_B^2}{4m_1m_2},\\
A_\parallel &= \sqrt{2}a,\\
A_\perp &= 2\sqrt{2}\,\frac{m_1m_2}{m_B^2}c\sqrt{x^2-1} \approx \sqrt{2}c,
\end{aligned} \tag{381}$$

where $x = k_1\cdot k_2/(m_1m_2) \approx m_B^2/(2m_1m_2)$ and the approximation holds for $m_B \gg m_1, m_2$. The relation between the transversity and helicity amplitudes are

$$A_+ = (A_\parallel + A_\perp)/\sqrt{2}, \qquad A_- = (A_\parallel - A_\perp)/\sqrt{2}. \tag{382}$$

The asymmetry

$$A_T = \frac{\Gamma[TP>0] - \Gamma[TP<0]}{\Gamma[TP>0] + \Gamma[TP<0]}$$

is T-odd. In terms of the transversity amplitudes that appear in the $B \to V_1V_2$ angular distribution, we can define

$$A_T^{(1)} \equiv \frac{\mathrm{Im}(A_\perp A_0^*)}{A_0^2 + A_\parallel^2 + A_\perp^2}, \qquad A_T^{(2)} \equiv \frac{\mathrm{Im}(A_\perp A_\parallel^*)}{A_0^2 + A_\parallel^2 + A_\perp^2}, \tag{383}$$

and

$$\bar{A}_T^{(1)} \equiv -\frac{\mathrm{Im}(\bar{A}_\perp \bar{A}_0^*)}{\bar{A}_0^2 + \bar{A}_\parallel^2 + \bar{A}_\perp^2}, \qquad \bar{A}_T^{(2)} \equiv -\frac{\mathrm{Im}(\bar{A}_\perp \bar{A}_\parallel^*)}{\bar{A}_0^2 + \bar{A}_\parallel^2 + \bar{A}_\perp^2} \tag{384}$$

for the $CP$-conjugate decay. While all these observables are T-odd, they do not by themselves violate time-reversal invariance. It is evident from the definition that they can be generated by a strong phase difference between the transversity amplitudes alone. Assuming $CPT$ invariance, a $CP$-violating observable is obtained from comparing $A_T$ and $\bar{A}_T$. One therefore

constructs

$$A_{TP}^{true,1,2} = \frac{1}{2}\left(A_T^{(1,2)} + \bar{A}_T^{(1,2)}\right) \propto \sin\Delta\phi\cos\Delta\delta, \tag{385}$$

$$A_{TP}^{fake,1,2} = \frac{1}{2}\left(A_T^{(1,2)} - \bar{A}_T^{(1,2)}\right) \propto \cos\Delta\phi\sin\Delta\delta. \tag{386}$$

The first quantity is the $CP$-violating TPA; the second, which is non-zero even for $\Delta\phi = 0$ if there is a strong phase difference, is referred to as "fake TPA".

As is the case with rate asymmetries and the full angular distribution, when the $V_1V_2$ final state be reached by both $B$ and $\bar{B}$ mesons, such as in $B_d \to K^*\bar{K}^*$ and $B_s \to J/\psi\phi, \phi\phi$ etc., mixing effects have to be included and the measurement of TPAs becomes a time-dependent problem.

When the final-state particles can be reached through a scalar background (resonant or non-resonant) — for example, $B \to V_1V_2 \to f$ and $B \to V_1S \to f$ — one has to include the interference effects. In particular, when a neutral vector meson is detected via its decay $V \to PP'$ ($P, P'$ are pseudoscalars), there is usually a background from the decay of a scalar resonance $S \to PP'$, or from the scalar non-resonant $PP'$ production [679, 872]. Then it is necessary to add another (scalar) helicity to the angular analysis of $B_s \to V_1(\to P_1P_1')V_2(\to P_2P_2')$ in presence of the scalar background [873]. The most general amplitude contains six helicities: $h = VV$ (3), $VS$, $SV$, and $SS$, each with a corresponding amplitude $A_h$. After squaring the amplitude the general angular analysis contains 21 terms. Allowing for time dependence due to $B_s$-$\bar{B}_s$ mixing, the angular distribution can be written as

$$\frac{d^4\Gamma(t)}{dtd\cos\theta_1 d\cos\theta_2 d\phi} = \frac{9}{8\pi}\sum_{i=1}^{21} K_i(t)X_i(\theta_1,\theta_2,\phi), \tag{387}$$

where $\theta_1, \theta_2$ and $\phi$ are the helicity angles in Fig. 141. We can express

$$K_i(t) = \left[a_i\cosh{(\Delta\Gamma/2)t} + b_i\sinh{(\Delta\Gamma/2)t} + c_i\cos\Delta mt + d_i\sin\Delta mt\right], \tag{388}$$

where the individual functions $a_i, b_i, c_i$, and $d_i$ for $i = 1,\ldots,21$ are time independent. The expressions for these coefficients in terms of the helicity amplitudes and mixing phase can be found in Ref. [873]. Various $CP$-violating quantities including TPAs are related to these coefficients as follows:

(1) Direct $CP$ asymmetries are represented by $c_i$ ($i =$ 1-4, 7, 13-16, 18, 20, 21), $a_i$ ($i =$ 8-11);
(2) The indirect $CP$ asymmetries are: $d_i$ ($i =$ 1-4, 7, 13-16, 18, 20, 21), $b_i$ ($i =$ 8-11);
(3) The triple products are: $a_i$ ($i = 5, 6, 17, 19$), $c_{12}$;
(4) The mixing-induced triple products: $b_i$ ($i = 5, 6, 17, 19$), $d_{12}$.

*SM expectations and observations.* The TPAs $A_T^{(1,2)}$ both involve the transverse polarisation amplitudes $A_\perp$, $A_\parallel$. For the reasons discussed in the general overview on $VV$ final states this makes it difficult to make reliable theoretical predictions from first principles. Nevertheless a few general observations can be made, based on the amplitude hierarchy in Eq. (377) as well as the existing observations of longitudinal polarisation fractions $f_L$ and direct $CP$ asymmetries.

- Due to the left-handedness of the the weak interaction $A_+ \ll A_-$ is expected. Within uncertainties, there is no experimental evidence of a violation of this hierarchy, which implies $A_{\|} \approx A_{\perp}$, hence $A_T^{(2)}$ is power-suppressed by $\Lambda/m_B$ relative to $A_T^{(1)}$. An observation of $A_T^{(2)} \sim A_T^{(1)}$ in any pure vector-vector decay mode would indicate a departure from pure left-handedness.

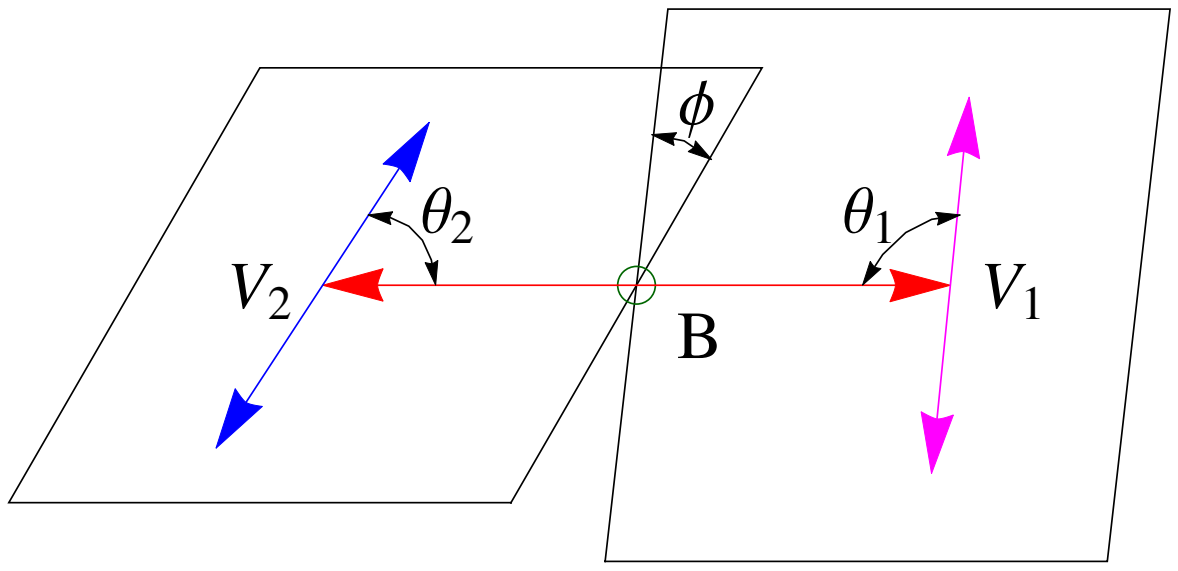

Fig. 141: Definition of the helicity angles in Eq. (387).

- The hierarchy in Eq. (377) also implies that $A_T^{(1)}$ is itself a power-suppressed quantity. However, as discussed in the previous section, this hierarchy is numerically not respected by penguin-dominated decays and is possibly also violated in colour-suppressed decay modes. Thus, final states with large transverse amplitude fractions are favourable for the measurement of TPAs and can then provide valuable complementary information on $CP$ violation without requiring the generation of a sizeable strong phase difference.
- Strangeness-changing penguin-dominated decays are dominated by an amplitude with a single weak phase, hence one does not expect large "true" TPAs $A_{TP}^{true,1,2}$. They are therefore especially sensitive to NP [865, 874]. In addition, the "fake" TPAs are of own interest, since they provide information on the helicity structure of NP interactions, see above.

TPAs have already been measured in some $B$ decay final states at BaBar, Belle and at hadron machines by CDF and LHCb [875–878]. These measurements have in turn provided strong constraints on various NP models [871]. TPAs can also be probed in $b$ baryon decays [879–881], as well as in semileptonic $B$ decays [882–884]. In $B_{s,d}^0$ decays, where the final state can be reached by $B_{s,d}^0$ and $\bar{B}_{s,d}^0$ decays, the TPAs appear in the time-integrated untagged angular distribution [873, 874].

Analyses including amplitudes for scalar backgrounds have been performed by Babar [853] and LHCb [885] in their studies of the decays $B_s \to J/\psi\phi$ [886] and $B_s \to \phi\phi$. In both cases the $\phi$ is detected through its decay to $K^+K^-$, and there is a resonant ($f_0$) or non-resonant scalar background. The angular analyses were performed with four and five helicities, respectively. TPAs have been measured in $B_s \to \phi\phi$ [887] with results consistent with the SM expectation of no CP violation. In addition, LHCb has studied the decay $B_s \to K^{*0}(892)\bar{K}^{*0}(892)$ [888], and found that each of these vector mesons has a background coming from the scalar resonance $K_0^{*0}(1430)$. In this case, as one does not have identical particles in the final state (in contrast to $B_s \to \phi\phi$), all six helicities and the general angular distribution in Eq. (387) must be considered. Some of these asymmetries in the $K_i(t)$ appear in the untagged distribution and have been measured in Ref. [888]. They were found to be consistent with SM predictions with the present precision of the measurements. Recently, a flavour-tagged decay-time-dependent amplitude analysis of $B_s \to (K^+\pi^-)(K^+\pi^-)$ decays was presented in Ref. [889] where the $K\pi$ combinations come from intermediate $K^*$ resonances with spin 0, 1 and 2. Many TPAs can be obtained from the interference of the various helicity amplitudes in this decay.

New physics if present in $B$ decays is more likely to be observed in rare decays where it can compete with the SM contribution. For a number of years, there has been a certain inconsistency among the measurements of the branching ratios and $CP$ asymmetries of the four $B \to \pi K$ decays as discussed earlier in this chapter. If the data is interpreted in terms of NP then it points to a new contribution to the electroweak penguin amplitude, which may come from models with a new neutral gauge boson ($Z'$). These models would also produce TPAs in vector-vector final states that proceed through the $b \to s\bar{q}q$ transitions where $q = u, d, s$. More precise measurements of TPAs in decays where TPAs have been measured as well TPA measurements in new decays like the $B \to \rho K^*$ decays (see also the following subsection) would be very interesting.

*12.6.3. Electroweak penguins in $B \to \rho K^*$ decays . [Contributing Author: M. Beneke]*

The system of the four $B \to \rho K^*$ decays and their $CP$ conjugates deserves special mentioning here, since it represents the $VV$ cousin of the much discussed $\pi K$ final states. The particular interest in the $\rho K^*$ final states is due to the facts that

- the dominant QCD penguin amplitude is at least a factor of two smaller than for the $\pi K$ system, while the tree and electroweak penguin amplitudes are of similar size; hence all interference effects are enhanced for the $\rho K^*$ final states, possibly providing a clue to the "$\pi K$ puzzles";
- the polarisation degree of freedom can provide additional clues. Furthermore, there is an electromagnetic penguin effect in the transverse amplitudes, which modifies the electroweak penguin amplitude. In the SM, it appears only in negative helicity amplitude [854].

It is therefore of great interest to measure the full angular distributions for all four $\rho K^*$ final states, which should be feasible at Belle II owing to its high statistics and good particle identification.

We briefly elaborate on these two facts and refer to Refs. [796, 854] for further details. The amplitude decomposition of $B \to \rho K^*$ decays similar to Eq. (371) for $\pi K$ reads

$$\begin{aligned}
A_h(\rho^- \bar{K}^{*0}) &= P_h \\
\sqrt{2}\, A_h(\rho^0 K^{*-}) &= [P_h + P_h^{EW}] + \epsilon_{\rm KM} e^{-i\gamma}\, [T_h + C_h] \\
A_h(\rho^+ K^{*-}) &= P_h + \epsilon_{\rm KM} e^{-i\gamma}\, T_h \\
-\sqrt{2}\, A_h(\rho^0 \bar{K}^{*0}) &= [P_h - P_h^{EW}] + \epsilon_{\rm KM} e^{-i\gamma}\, [-C_h]. \qquad (389)
\end{aligned}$$

Here we used the topological amplitude notation,[38] and neglected the colour-suppressed electroweak penguin amplitude $\alpha^p_{4,\rm EW}$. The subscript $h = 0, \pm 1$ denotes the helicity amplitudes and $\epsilon_{\rm KM} = |V_{ub}V_{us}^*|/|V_{cb}V_{cs}^*| \sim 0.025$ implies the CKM suppression of the tree amplitudes. The system is ideally suited to probe the electroweak penguin amplitudes, which enter in three different combinations $P_h + k \cdot P_h^{EW}$, or, in factorisation notation,

$$A_{\rho\bar{K}^*}\,\hat{\alpha}_4^{p,h} + k \cdot A_{\bar{K}^*\rho}\,\frac{3}{2}\alpha_{3,\mathrm{EW}}^{p,h}, \qquad k = 1, 0, -1, \tag{390}$$

allowing various kinds of interferences. The smallness of $P_h$ implies that $|P_h^{\mathrm{EW}}/P_h|$ is sizeable and the interference can be large. Similarly, one expects sizeable direct $CP$ asymmetries in the final states with charged $K^*$ mesons due to the enhanced interference with the colour-allowed tree amplitude. A comparison of theoretical predictions and present experimental results for the $CP$-averaged branching fractions and direct $CP$ asymmetries summed over all helicity states is shown in Table 103. The theoretical results represent an update of the QCD factorisation results [796], where the QCD penguin amplitude $P_h$ is determined from the $\phi K^*$ angular distribution rather than from the theoretical calculation. The qualitative pattern of the branching fractions and especially the $CP$ asymmetries is in good agreement with observations, given uncertainties, but a helicity-specific analysis would provide interesting further insights.

[38] For the relation to the factorisation notation, see the end of Section 12.3.1.

Table 103: $CP$-averaged branching fractions and direct $CP$ asymmetries of the $\rho K^*$ final states. Theoretical values correspond to an update of Ref. [796]. Experimental values are taken from the HFLAV compilation [230] (August 2017 web update) except for the $B^- \to \bar{K}^{*0}\rho^-$ branching fraction, which is the average of the BaBar and Belle measurements reported in Refs. [890, 891].

| Mode | BrAv/$10^{-6}$ | | $A_{CP}$ / percent | |
|---|---|---|---|---|
| | Theory | Experiment | Theory | Experiment |
| $B^- \to \bar{K}^{*0}\rho^-$ | $6.6^{+0.3+3.3+0.3}_{-0.3-1.4-0.7}$ | $9.2 \pm 1.5$ | $1^{+0+1+1}_{-0-1-2}$ | $-1 \pm 16$ |
| $B^- \to K^{*-}\rho^0$ | $5.1^{+1.6+2.1+0.5}_{-1.4-1.0-0.8}$ | $4.6 \pm 1.1$ | $18^{+5+9+31}_{-5-9-22}$ | $31 \pm 13$ |
| $\bar{B}^0 \to K^{*-}\rho^+$ | $6.1^{+1.8+2.9+0.5}_{-1.6-1.3-0.6}$ | $10.3 \pm 2.6$ | $10^{+3+8+38}_{-2-7-27}$ | $21 \pm 15$ |
| $\bar{B}^0 \to \bar{K}^{*0}\rho^0$ | $2.4^{+0.2+1.1+0.2}_{-0.2-0.6-0.1}$ | $3.9 \pm 0.8$ | $-16^{+4+16+10}_{-5-14-6}$ | $-6 \pm 9$ |

Table 104: Longitudinal polarisation fraction of the $\rho K^*$ final states. Theoretical values correspond to an update of Ref. [796]. Experimental values are taken from the HFLAV compilation [230] (August 2017 web update).

| Mode | $f_L$ / percent | |
|---|---|---|
| | Theory | Experiment |
| $B^- \to \bar{K}^{*0}\rho^-$ | $67^{+0+14+0}_{-0-10-3}$ | $48 \pm 8$ |
| $B^- \to K^{*-}\rho^0$ | $89^{+1+6+1}_{-2-5-3}$ | $78 \pm 12$ |
| $\bar{B}^0 \to K^{*-}\rho^+$ | $70^{+3+13+1}_{-4-10-6}$ | $38 \pm 13$ |
| $\bar{B}^0 \to \bar{K}^{*0}\rho^0$ | $34^{+3+23+2}_{-3-14-0}$ | $40 \pm 14$ |

A full angular analysis is presently not available for any of the $\rho K^*$ final states. Table 104 summarises the averages of the existing longitudinal polarisation fraction measurements and the theoretical prediction [796]. The difficulties with calculating the transverse QCD penguin amplitudes reliably have been discussed in Section 12.6.1. One notes that the agreement is quite satisfactory for the final states with the neutral $\rho$ meson, especially what concerns the largely different $f_L$. The charged $\rho$ final states, however, show a discrepancy, even with uncertainties, which appears surprising given their simpler amplitude structure in Eq. (389).

A full angular analysis of the $\rho K^*$ final states is especially interesting in view of the fact that there is a contribution from the electromagnetic dipole operator $Q_{7\gamma}$ to the transverse polarisation amplitudes, which dramatically changes the power counting in the heavy quark limit. Comparing Eqs. (377, 379) in Section 12.6.1, one notes that the negative helicity amplitude is enhanced by a factor $(m_B/\Lambda)^2$ compared to the counting in the absence of the electromagnetic dipole effect. This arises due to the transition $b \to s\gamma^\star$ to a photon with virtuality $m_\rho^2$, which then converts to a $\rho$ meson. This process modifies the electroweak penguin amplitude as

$$\alpha_{3,\mathrm{EW}}^{p,-} = \alpha_{3,\mathrm{EW}|\mathrm{no}\ C_{7\gamma}}^{p,-} - \frac{2\alpha_{\mathrm{em}}}{3\pi} C_{7\gamma}^{\mathrm{eff}} \frac{m_B m_b}{m_\rho^2}, \tag{391}$$

where the double power enhancement is evident in the additional contribution. It changes the real part from $-0.010^{+0.002}_{-0.002}$ to the value $+0.015^{+0.004}_{-0.003}$. Due to the change in sign, the pattern of interference between the electroweak penguin and QCD penguin amplitude is now opposite for the longitudinal and negative-helicity amplitude.

Due to the left-handed nature of the weak interaction, $Q_{7\gamma}$ contributes through the above effect only to the negative-helicity amplitude. Since the term proportional to the Wilson coefficient of the electromagnetic dipole operator, $C_{7\gamma}^{\mathrm{eff}}$, is the largest contribution to the negative-helicity electroweak penguin amplitude, the interference patterns in Eq. (390) are sensitive to possible anomalous contributions to $C_{7\gamma}^{\mathrm{eff}}$, including its phase. An anomalous right-handed component would lead to a corresponding enhancement of the positive-helicity amplitude, which otherwise is strongly suppressed in the SM.

The numerical effect of the electromagnetic dipole contribution to the branching fractions, direct $CP$ asymmetries and longitudinal polarisation fractions of the final states containing $\rho^0$ is shown in Table 105 by including and excluding the second term on the right-hand side of Eq. (391) in the theoretical prediction.[39] Already in the "excluded" results, the longitudinal polarisation fractions of the $\rho K^*$ final states are predicted to differ such that $f_L(K^{*-}\rho^0) > f_L(\bar{K}^{*0}\rho^-) > f_L(\bar{K}^{*0}\rho^0)$. This follows from the large longitudinal electroweak penguin contribution. The transverse electromagnetic dipole effect amplifies the hierarchy among the three $f_L$ predictions.

The present situation is inconclusive. Observables more sensitive to the electroweak penguin amplitudes can be defined by taking the helicity-specific $CP$-averaged decay rate ratios [854]

$$S_h \equiv \frac{2\bar{\Gamma}_h(\rho^0 \bar{K}^{*0})}{\bar{\Gamma}_h(\rho^- \bar{K}^{*0})}, \qquad S_h' \equiv \frac{2\bar{\Gamma}_h(\rho^0 K^{*-})}{\bar{\Gamma}_h(\rho^- K^{*0})}, \tag{392}$$

[39] These theoretical numbers are *not* updated relative to Ref. [796] and therefore differ from those in the previous tables. The difference is small except for the longitudinal polarisation fraction of the $\bar{K}^{*0}\rho^0$ final state. The main purpose of this table is to show the difference of the two results.

Table 105: Predicted branching fraction, longitudinal polarisation, and direct $CP$ asymmetry of the two $\rho K^*$ final states sensitive to the electroweak penguin amplitude with the power-enhanced transverse contribution proportional to $C_{7\gamma}$ included or excluded. Experimental results for comparison (exp.).

| | $B^- \to K^{*-}\rho^0$ | | | $\bar{B}^0 \to \bar{K}^{*0}\rho^0$ | | |
|---|---|---|---|---|---|---|
| | incl. | excl. | exp. | incl. | excl. | exp. |
| BrAv/$10^{-6}$ | 4.5 | 5.4 | $4.6 \pm 1.1$ | 2.4 | 1.4 | $3.9 \pm 0.8$ |
| $f_L$ / % | 84 | 70 | $78 \pm 12$ | 22 | 37 | $40 \pm 14$ |
| $A_{CP}$ / % | 16 | 14 | $31 \pm 13$ | $-15$ | $-24$ | $-6 \pm 9$ |

and $S_h^{''} \equiv S_h/S_h'$. In particular $S_h^{''}$ differs by up to a factor of four whether or not the electromagnetic dipole contribution is included [796]. It would be very interesting to detect this effect in the complete angular distribution, which is essentially equivalent to a measurement of photon polarisation in the radiative decay $B \to K^*\gamma$. It should be emphasised that the $CP$-average of helicity-specific decay rates is not the same as the $CP$-average of polarisation fractions $f_h$. When the standard variables are used, the relation involves $CP$ asymmetries. The $S$-observables defined above are better suited to an investigation of helicity-specific effects. Experimentally they can be determined from the same data as the standard observables, thus avoiding unfolding complicated correlations in the errors of $CP$ asymmetries, branching and polarisation fractions.

### *12.7. Three-body charmless B decays*

In the final section of this chapter we focus on the relatively new and much less developed subject of three-body charmless $B$ decays. Two subsections approach the topic from the general theoretical and the phenomenological points of view. Particular interest in the subject arises from the observation of large local $CP$ asymmetries in Dalitz plot analyses by LHCb [892–894].

#### *12.7.1. Theoretical framework . [Contributing Author: J. Virto]*

In complete analogy with two-body decays (see Section 12.1), three-body $B$-decay amplitudes can be decomposed according to their CKM structure:

$$A(\bar{B} \to f) = \lambda_u^{(D)} A_f^u + \lambda_c^{(D)} A_f^c \tag{393}$$

where now $f = M_a M_b M_c$ is a three-body charmless final state and $A_f^p$ are given by the corresponding matrix elements of dimension-six operators in the weak effective Lagrangian in Eq. (349),

$$A_f^p = -\frac{G_F}{\sqrt{2}} \sum_{i=1\ldots 6,8} C_i(\mu) \langle f | Q_i^p(\mu) | \bar{B} \rangle \ . \tag{394}$$

As in two-body decays, the theory challenge is to calculate the matrix elements $\langle M_a M_b M_c | Q_i(\mu) | \bar{B} \rangle$ from first principles in QCD, or else to establish rigorous relationships between various of these matrix elements that can be exploited phenomenologically. In

this respect three-body decays are considerably more challenging than two-body decays (and correspondingly less well understood), but provide a number of theoretical and phenomenological advantages:

- The number of different three-body final states is about ten times larger than the number of two-body decays. In addition, each final state has a non-trivial kinematic multiplicity (a two-dimensional phase space) as opposed to two-body decays where the kinematics is fixed by the masses. This leads to a much richer phenomenology.
- "Quasi-two-body" decays $\bar{B} \to M_a M(\to M_b M_c)$, where $M$ decays strongly, are only well defined in the context of the three-body decay, in the narrow width approximation, and neglecting any "non-resonant" background (*e.g.* the overlap with other nearby or very wide resonances in the $M_b M_c$ channel). Therefore, the full understanding of the three-body decay provides corrections to the quasi-two-body approximation. In addition, this allows for performing spectroscopy by looking for resonant structures in the kinematic distributions, and to measure their spin.
- Factorisation properties of three-body decays depend continuously on two kinematic invariants, thus allowing for more detailed data-driven studies of factorisation and power corrections in $B$ decays.
- Strong phases in two-body decays are either perturbative $[\mathcal{O}(\alpha_s(m_b))]$ or power suppressed $[\mathcal{O}(\Lambda/m_b)]$. Therefore, the corresponding $CP$ asymmetries are predicted to be suppressed correspondingly, and leading-power predictions are on a less solid footing, since $\alpha_s(m_b)/\pi \sim \Lambda/m_b$. On the contrary, strong phases in three-body decays arise non-perturbatively already at the leading power, through complex phases in matrix elements such as $F_\pi \sim \langle 0|j|\pi\pi\rangle$ and $F^{B\pi\pi} \sim \langle \pi\pi|j|\bar{B}\rangle$. These matrix elements and their phases can in principle be obtained from data from other, unrelated decay modes. Localised direct $CP$ asymmetries can therefore be large, potentially leading to improved extraction of CKM angles from direct $CP$ violation.

The theory of three-body non-leptonic decays is still in an early stage of development. Here we provide a brief overview of the subject (see Ref. [895] for an extended version).

*Kinematics.* We consider a decay

$$\bar{B}(p_B) \to M_a(p_1) M_b(p_2) M_c(p_3) \ . \tag{395}$$

Fixing the masses of the initial and final hadrons, the kinematics is completely specified by two invariant masses of two pairs of final state particles (*e.g.* $s_{ab}$ and $s_{ac}$, with $s_{ab} \equiv 2(p_1 \cdot p_2)/m_B^2$, etc.) All physical kinematic configurations thus define a two-dimensional region in the $s_{ab}$-$s_{ac}$ plane, which in the limit where all final particles are massless is a triangle defined by $s_{ab} > 0$, $s_{ac} > 0$, $s_{ab} + s_{ac} < 1$ (see Fig. 142). The amplitude of the process is a function of these two invariant masses $\mathcal{A}(s_{ab}, s_{ac})$, and the density plot of the differential decay rate

$$\frac{d^2\Gamma}{ds_{ab}\, ds_{ac}} = \frac{m_B}{32(2\pi)^3} |\mathcal{A}(s_{ab}, s_{ac})|^2 \tag{396}$$

in that region is called the Dalitz plot. Labeling particles by their momenta removes any ambiguity related to identical particles, but reduces the relative physical phase space (Dalitz plot) to one half, or one sixth (see Fig. 142).

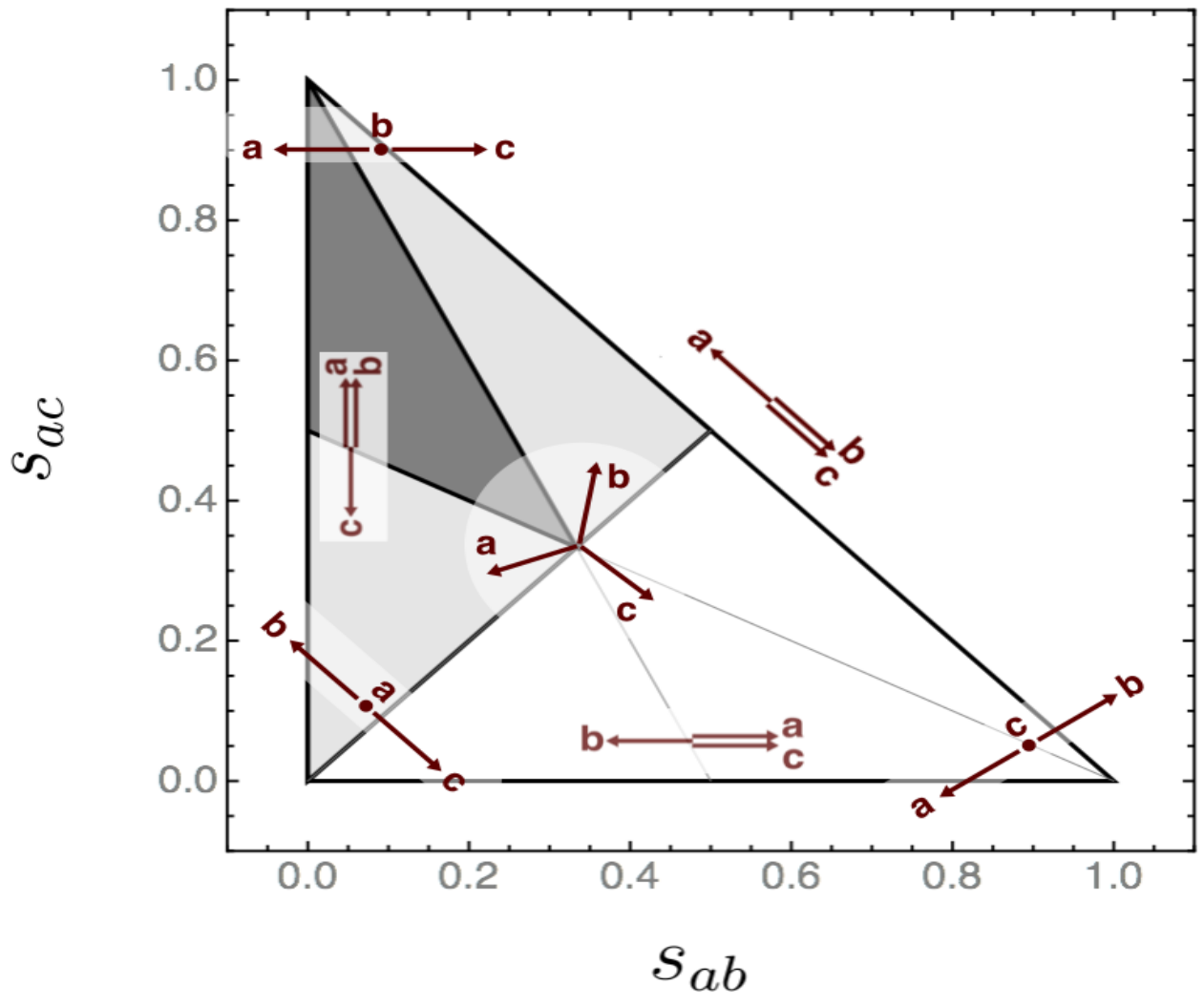

Fig. 142: Phase space of the three-body decay $B \to M_a M_b M_c$ in terms of the normalised invariants $s_{ab}, s_{ac}$. Special kinematic configurations are indicated. If $M_b = M_c$ or $M_a = M_b = M_c$ then $s_{ab} \to s_{ab}^{\text{low}}$ and $s_{ac} \to s_{ab}^{\text{high}}$, and the phase space is reduced to the light-gray and dark-gray regions respectively.

The Dalitz plot contains different regions with "special" kinematics (Fig. 142). The central region corresponds to the case in which all three final particles fly apart at $\sim 120^{\rm o}$ angles with large energy ($E \sim m_B/3$). The corners correspond to the case in which one final-state particle is approximately at rest (*i.e.* soft), and the other two fly back-to-back with large energy ($E \sim m_B/2$). The central part of the edges correspond to the case in which two particles move collinearly with large energy and the other particle recoils back. The significance of these special kinematic configurations is that different theoretical approaches may be applicable in these different regions, as will be discussed below.

*Partial-wave expansions and isobars.* The Dalitz plot is typically dominated by resonant quasi-two-body contributions along the edges. Therefore, a first-order approximation is to regard the three-body decay as a coherent sum of quasi-two-body decays $\bar{B} \to R_{ij}^{(\ell)}(\to M_i M_j) M_k$, where $R_{ij}^{(\ell)}$ denotes a resonance in the $(ij)$ channel with spin $\ell$. This resonance contributes to the region $s_{ij} \sim (m_{R_{ij}} \pm \Gamma_{R_{ij}})^2/m_B^2$, where $m_{R_{ij}}, \Gamma_{R_{ij}}$ are the mass and width respectively, and the profile of this contribution in the other Dalitz-plot variable $s_{ik}$ is specified by the spin $\ell$. In each channel it is thus convenient to expand the amplitude in partial waves. For example, one may trade the variable $s_{ac}$ by the angle $\theta_c$ between the momenta $\vec{p}_3$ and $\vec{p}_B$ in the $(M_a M_b)$ rest frame, which in the massless limit is given by $(1 - s_{ab}) \cos\theta_c = s_{ab} + 2 s_{ac} - 1$. The amplitude $\mathcal{A}(s_{ab}, s_{ac})$ can then be expanded in Legendre polynomials:

$$\mathcal{A}(s_{ab}, s_{ac}) = \sum_{\ell=0}^{\infty} (2\ell + 1) \mathcal{A}^{(\ell)}(s_{ab}) P_\ell(\cos\theta_c) \ , \tag{397}$$

and spin $\ell$ resonances in the $(ab)$ channel contribute only to the corresponding partial wave. Note however that truncating the wave expansion to any finite order makes the right-hand side of Eq. (397) algebraic in $s_{ac}$, while the l.h.s typically contains singularities in the physical region in the $s_{ac}$ (and $s_{bc}$) channels [896]. Therefore, the series cannot converge. A popular solution to this issue is provided by the *isobar* model, where the amplitude is modeled by a finite set of partial waves simultaneously in all three channels:

$$\mathcal{A}(s_{ab}, s_{ac}) = \sum_{\ell=0}^{\ell_{\max}} (2\ell+1)\, a_\ell^{ab}(s_{ab}) P_\ell(\cos\theta_c) + (abc \to bca) + (abc \to cab) \ . \tag{398}$$

Typically, the isobaric amplitudes $a_\ell(s)$ are modeled by energy-dependent Breit-Wigner amplitudes. In addition, a "non-resonant" (smooth) component can be added to the amplitude, but the exact kinematic dependence of this component is rather arbitrary.

*Final-state interactions and $CPT$ constraint.* Final-state interactions are often invoked as a possible source of non-perturbative strong phases, leading to large localised $CP$ asymmetries in three-body decays. However, the isobar model does not include coupled channel effects (beyond resonance interference) or three-body rescattering. These effects may be modeled separately, or analysed by means of dispersive methods (*e.g.* Refs. [118, 896–898]).

Additional constraints may be obtained by combining $CPT$ invariance and unitarity, which imply that [899]

$$\sum_f \left[\Gamma(B \to f) - \Gamma(\bar{B} \to \bar{f})\right] = 0 \ , \tag{399}$$

where the sum runs over all states $f$ with the same flavour quantum numbers. The individual exclusive rates need not be equal, as there might be direct $CP$ violation in exclusive modes; but all $CP$ asymmetries of exclusive decays to same-flavour final states must sum to zero. Due to the large phase space available in $B$ meson decays, the multiplicity of such final states is very large, and so the constraint in Eq. (399) is by itself of little use. However, this constraint may be imposed on simple models with a few coupled channels, leading to some insight on the importance of final-state interactions, resonance interference, and a qualitative understanding of the patterns of $CP$ asymmetries in different modes. For example, a two-channel model with coupled $S$-wave $(\pi^+\pi^-)$ and $(K^+K^-)$ states satisfying the $CPT$ constraint [900] shows good qualitative agreement for the observed $CP$ asymmetries in $B^\pm \to K^\pm\pi^+\pi^-$ and $B^\pm \to K^\pm K^+K^-$ in the region $1\,\mathrm{GeV}^2 \lesssim m^2_{\pi\pi,KK} \lesssim 2.2\,\mathrm{GeV}^2$, where $S$-wave $\pi^+\pi^- \leftrightarrow K^+K^-$ scattering is expected to be important, and explains qualitatively why these asymmetries (properly weighted by the branching ratios) are equal and of opposite sign. The same pattern is observed in $B^\pm \to \pi^\pm\pi^+\pi^-$ and $B^\pm \to \pi^\pm K^+K^-$. More complicated models including resonant contributions from $\rho(770)$ and $f_0(980)$ have also been studied in this context [901].

*Flavour symmetries and SU(3) relations.* The approximate SU(3) flavour symmetry of QCD has been used extensively to study two-body charmless decays (see Section 12.2), and it is equally useful in the case of three-body decays. By arranging all three-body final states and effective operators into SU(3) representations, the matrix elements $\langle M_a M_b M_c|Q_i(\mu)|\bar{B}\rangle$ can be expressed in terms of reduced matrix elements and Clebsch-Gordan coefficients. Some relationships can be established between observables where reduced matrix elements cancel exactly or approximately, or global fits to data can be performed to determine the reduced matrix elements and CKM parameters. In the remainder of this section we discuss the methods based on factorisation.

*Naive factorisation.* For two-body charmless $B$ decays, naive factorisation is a prediction of QCD in the heavy-quark limit and at the leading order in $\alpha_s(m_b)$ [415], and perturbative "non-factorisable" corrections can be computed consistently (see Section 12.3.1). While such a theory has not been fully developed in the three-body case, many phenomenological analyses have been performed *assuming* that "naive factorisation plus $\mathcal{O}(\alpha_s)$ corrections" is a good approach to three-body decays, too. It is very likely that this is the case in the kinematic regions where one invariant mass is small and the other two are large (near the edges of the Dalitz plot). Indeed, these regions contain "quasi-two-body" configurations corresponding to two-body decays with one strong resonance in the final state (such as $B \to \rho\pi$), to which the QCD factorisation formula applies [696].

Considering the kinematic region where $s_{bc} \ll 1$, and denoting the two-meson system with small invariant mass by $[M_b M_c]$, the naive factorisation formula for the amplitude $A^p_{M_a[M_b M_c]}$ in Eq. (394) is given by Ref. [902]

$$A^p_{M_a[M_bM_c]} = \sum_k \left[\alpha_k^p(M_a,[M_bM_c])A^k_{M_a,[M_bM_c]} + \alpha_k^p([M_bM_c],M_a)A^k_{[M_bM_c],M_a}\right], \quad (400)$$

where

$$A^k_{M_a,[M_bM_c]} = -\frac{G_F}{\sqrt{2}}\langle M_a|j_k^1|\bar{B}\rangle\langle [M_bM_c]|j_k^2|0\rangle, \quad (401)$$

$$A^k_{[M_bM_c],M_a} = -\frac{G_F}{\sqrt{2}}\langle [M_bM_c]|j_k^1|\bar{B}\rangle\langle M_a|j_k^2|0\rangle \; . \quad (402)$$

Here $j_k^{1,2}$ are local bilinear colour-singlet currents and $a_k^p$ are the usual coefficients in QCD factorisation [696]. Annihilation contributions as well as hard-scattering corrections are typically neglected. A simple way to make sense of NLO vertex corrections and penguin contractions in $a_k^p(M_a,[M_bM_c])$, which would involve a light-cone distribution amplitude of the system $[M_bM_c]$, is to adopt a multi-resonance model [902]. This requires partial-wave decomposition in the $(M_bM_c)$ channel, which immediately involves all values of $s_{ab}$, including the kinematic regions where either $M_b$ or $M_c$ are soft (the corners of the Dalitz plot). The resonance model is also used for the matrix elements $\langle [M_bM_c]|j_k^1|\bar{B}\rangle$ and $\langle [M_bM_c]|j_k^2|0\rangle$.

A even more aggressive approach is to extend this factorisation formula to the whole Dalitz plot [903, 904]. This provides a more complete set of predictions but a loss of theoretical justification. These phenomenological analyses (see following subsection) include also an estimate of non-resonant contributions, by calculating the $B \to M_bM_c$ form factors $\langle [M_bM_c]|j_k^1|\bar{B}\rangle$ within the heavy-meson chiral perturbation theory at an unphysical kinematic point where the two mesons are soft, and then using an exponential one-parameter ansatz to extrapolate to the physical region. This parameter is assumed universal, is fitted to the "non-resonant" component of $B^- \to \pi^-\pi^+\pi^-$ provided by the $B$ factories, and used to predict non-resonant contributions in other modes. These (model-dependent) predictions are in reasonable agreement with data for $B^- \to K^-K^+K^-$ and $B^- \to K^-\pi^+\pi^-$ branching fractions [904], but the significance of this agreement is not always easy to interpret.

*QCD factorisation.* Different forms of factorisation theorems may be conjectured depending on the scaling of the two kinematic invariants with $m_b$ [905–907].

In the central region of the Dalitz plot, where all invariant masses are of order $m_B$ ($s_{ab} \sim s_{ac} \sim 1/3$), the following factorisation formula can be proposed [907]:

$$\langle M_a M_b M_c|Q_i|\bar{B}\rangle_{\text{center}} = F^{B\to M_a}\, T_i^I \star \Phi_b \star \Phi_c + T_i^{II} \star \Phi_B \star \Phi_a \star \Phi_b \star \Phi_c \ , \quad (403)$$

where the convolutions of hard-scattering kernels and distribution amplitudes are written schematically, as in Eq. 353. The hard kernels $T_i^{I,II}$ can be computed perturbatively in QCD. To the lowest order (at order $\alpha_s$), only $T_i^I$ contributes, and arises from diagrams with an insertion of the operator $Q_i$ and all possible insertions of a hard gluon which splits into a quark-antiquark pair with large invariant mass. The convolutions of the resulting perturbative kernels $T_i^I$ with the pion light-cone distributions can be computed without encountering end-point singularities, thus providing a check of the factorisation formula. This check is non-trivial since the kernels $T_i^I(u,v)$ already depend on the momentum fraction of the quarks at the leading order, making the convolutions non-trivial.

At certain edges of the Dalitz plot, where one invariant mass becomes small, the gluon propagator in some of the diagrams becomes soft, leading to a $1/s$ behaviour in the amplitude. This behaviour is related to non-perturbative dynamics that results, for example, in the formation of resonances. This is the case for *e.g.* $B^\pm \to \pi^\pm\pi^-\pi^+$ in the region where $m_{\pi^+\pi^-} \sim m_\rho$. The decay thus looks very much like a two-body decay, and one expects a similar factorisation formula [905, 907]:

$$\begin{aligned}\langle M_a M_b M_c|Q_i|B\rangle_{s_{bc}\ll 1} &= F^{B\to\pi^a}\, T_a^I \star \Phi_{bc} + F^{B\to M^b M^c}\, T_{bc}^I \star \Phi_a \\ &\quad + T^{II} \otimes \Phi_B \star \Phi_a \star \Phi_{bc} \ . \end{aligned} \quad (404)$$

Here $\Phi_{bc}$ denotes a two-meson distribution amplitude (2MLCDA), and $F^{B\to M_b M_c}$ denotes a $B \to M_b M_c$ form factor. Conceptually, this factorisation formula is at the same level of theoretical rigour as the factorisation formula for two-body decays to unstable particles (*e.g.* $B \to \rho\pi$), but requires more complicated hadronic input (discussed below). This is the cost of generalizing quasi-two-body decays beyond the narrow-width approximation.

The three-body amplitude in the central region is power- and $\mathcal{O}(\alpha_s)$-suppressed with respect to the amplitude at the edge [905, 906]. The interpolation between one region and the other can be understood by noting that some parts of the central region amplitude arise from factorisation of 2MLCDAs or $B \to M_b M_c$ form factors at large $s_{bc}$, and one can check analytically the correspondence of such parts of the amplitudes [907]. Numerically, it is found that, in the case of $B^- \to \pi^-\pi^+\pi^-$, a good matching of the 2MLCDA part of the amplitude between the center and the edge happens only for $m_B \gtrsim 20$ GeV, but not for physical values ($m_B \sim 5$ GeV), suggesting that power corrections to Eq. (403) are too large in reality, and preclude a description of the central region in terms of single pion states.

We finish this section summarizing a few facts about two-pion distribution amplitudes and $B \to \pi\pi$ form factors, relevant for three-body decays with two collinear pions.

*Generalised distribution amplitudes.* An example of a $2\pi$LCDA in Eq. (404) is given by the matrix element [907, 908]

$$\Phi_{\pi\pi}(z,\zeta,k_{12}^2) = \int \frac{dx^-}{2\pi} e^{iz(k_{12}^+ x^-)} \langle \pi^+(k_1)\pi^0(k_2)|\bar{u}(x^- n_-)\, \not{n}_+ d(0)|0\rangle \ ,$$

where $k_{12}^\mu = k_1^\mu + k_2^\mu \simeq (k_{12}^+/2)n_+^\mu$, $\zeta = k_{12}^+/k_1^+$, and we have suppressed a Wilson line that makes the non-local quark current gauge invariant. At the leading order the kernel $T_a^I$ in Eq. (404) does not depend on $z$, and only the normalisation for $\Phi_{\pi\pi}$ is needed [905]:

$$\int dz\, \Phi^q_{\pi\pi}(z,\zeta,s) = (2\zeta - 1)F_\pi(s) \ , \tag{405}$$

where $F_\pi(s)$ is the pion vector form factor. The absolute value of the pion form factor is well known experimentally in a wide range of energies (see Fig. 5 in Ref. [907]). Higher moments of the $2\pi$LCDA are needed at higher orders, but these are not well-known.

*$B \to \pi\pi$ form factors.* $B \to \pi\pi$ form factors are accessible from measurements of $B \to \pi\pi\ell\nu$ observables [909]. At low dipion masses and at large recoil of the dipion, these form factors can be studied by means of light-cone sum rules. One may consider light-cone sum rules with two-pion distribution amplitudes [910] or with $B$-meson distribution amplitudes [911]. In the first case one arrives at a closed expression for the form factors in terms of moments of the $2\pi$LCDAs:

$$F^{B\to\pi\pi}(k_{12}^2,\zeta) \sim \frac{1}{f_B}\int du\, f(u,k_{12}^2)\ \Phi_{\pi\pi}(u,\zeta,k_{12}^2) \ . \tag{406}$$

The disadvantage of this method is that moments of $2\pi$LCDAs are not well-known.

In the second case, one obtains sum-rules that depend on weighted convolutions of the form factors with the pion form factor $F_\pi(s)$ [911]:

$$\int ds\ g(s)\, F_\pi^\star(s)\, F^{B\to\pi\pi}(s,\zeta) \sim f_B \int d\omega\ h(\omega)\ \phi_B^+(\omega) \tag{407}$$

and depend on moments of the $B$ meson LCDA $\phi_B^+$ discussed in Section 12.3.2. These sum rules allow for testing of models for the $B \to \pi\pi$ form factors. In the limit where the pion form factor is dominated by an infinitely narrow $\rho$ meson, the sum rules reduce analytically to the known sum-rules for the $B \to \rho$ form factors [912].

A factorisation formula for $B \to \pi\pi$ form factors at large dipion masses has also been proven at NLO recently [913]. This also proves part of the factorisation formula in Eq. (403) at NLO.

*12.7.2. Phenomenological analysis . [Contributing Author: H-Y. Cheng]*

Evidence of inclusive integrated direct $CP$ asymmetries $A_{CP}^{\rm incl}$ in charmless three-body decays of charged $B$ mesons, $B^+ \to \pi^+\pi^+\pi^-$ ($4.2\sigma$), $B^+ \to K^+K^+K^-$ ($4.3\sigma$) and $B^+ \to K^+K^-\pi^+$ ($5.6\sigma$), has been found by LHCb [892–894]. LHCb has also observed very large asymmetries $A_{CP}^{\rm low}$ of order 60-70% in some small invariant-mass regions of phase space. For example, $A_{CP}^{\rm low}(K^-\pi^+\pi^-) = 0.678 \pm 0.085$ for $m^2_{K^-\pi^+\ \rm high} < 15$ GeV$^2$ and $0.08 < m^2_{\pi^+\pi^-\ \rm low} < 0.66$ GeV$^2$.

As is evident from the previous subsection, three-body decays of heavy mesons are much more complicated than the two-body ones, in particular as they receive both resonant and non-resonant contributions. Contrary to three-body $D$ decays where the non-resonant signal is usually rather small and less than 10% [914], non-resonant contributions play an essential role in penguin-dominated three-body $B$ decays. For example, the non-resonant fraction of $KKK$ modes is of order (70-90)%. It follows that non-resonant contributions to the penguin-dominated modes should be also dominated by the penguin mechanism. The relevance and importance of non-resonant effects are often not appreciated in the literature. Resonant effects are conventionally described within the isobar model in terms of the usual Breit-Wigner formalism. For charmless three-body decays of $B$ mesons into three pseudoscalar mesons, there exist vector and scalar resonances.

$CP$ violation in three-body decays is also more intricate than in the two-body case. While $CP$ violation is just a number in the latter case, it is the distribution of the $CP$ asymmetry in the Dalitz plot that is measured in three-body decays. Hence, the Dalitz-plot analysis of $A_{CP}$ distributions can reveal very rich information about $CP$ violation. Besides the integrated $CP$ asymmetry, the local asymmetry can be large and positive in some region and negative in another. A successful model must explain not only the inclusive asymmetry but also regional $CP$ violation. Therefore, the measured $CP$-asymmetry Dalitz distributions put stringent constraints on the models.

The following discussion is based on the model and results of Refs. [904, 915], which examined $CP$ violation in three-body decays and stressed the crucial role played by the non-resonant contributions. Indeed, if the non-resonant term is essential to account for the total rate, it should play a role for $CP$ violation, too.

*Decay Rates.* Unlike hadronic two-body $B$ decays, established frameworks such as such as QCD factorisation (QCDF) [415] or perturbative QCD (PQCD) [801, 802] are not yet on the same footing for three-body decays (see the previous subsection and the original papers Refs. [905–907] and Refs. [916, 917]). Hence, Refs. [903, 915] take the factorisation approximation Eqs. (400)–(402) of the three-body decay amplitudes as a working hypothesis rather than starting from first principles.

*Non-Resonant contributions.* In general, the decay amplitude is the coherent sum of resonant contributions together with the non-resonant background

$$A = \sum_R A_R + A_{\rm NR}. \tag{408}$$

Consider the non-resonant contributions induced by the $b \to u$ transition to the tree-dominated $B^- \to K^+K^-\pi^-$ and $B^- \to \pi^+\pi^-\pi^-$ decays. The non-resonant amplitude induced by the $b \to u$ transition process reads

$$\begin{aligned} A^{\rm HMChPT}_{\rm transition} &\equiv \langle P_3(p_3)|(\bar{q}u)_{V-A}|0\rangle \, \langle P_1(p_1)P_2(p_2)|(\bar{u}b)_{V-A}|B\rangle_{\rm NR} \\ &= -\frac{f_{P_3}}{2}\left[2m_3^2 r + (m_B^2 - s_{12} - m_3^2)\omega_+ + (s_{23} - s_{13} - m_2^2 + m_1^2)\omega_-\right], \end{aligned} \tag{409}$$

where $(\bar{q}_1 q_2)_{V-A} = \bar{q}_1\gamma_\mu(1-\gamma_5)q_2$. The form factors $r$, $\omega_\pm$ and $h$ can be evaluated in the framework of heavy-meson chiral perturbation theory (HMChPT) [918]. However, as pointed out in Refs. [903, 915], the predicted non-resonant rates based on HMChPT are then too large for tree-dominated decays. The branching fractions of non-resonant $B^- \to \pi^+\pi^-\pi^-$ and $B^- \to K^+K^-\pi^-$ are found to be of order $75 \times 10^{-6}$ and $33 \times 10^{-6}$, respectively, one order of magnitude larger than the corresponding measured total branching fractions of $15.2 \times 10^{-6}$ and $5.0 \times 10^{-6}$. The issue has to do with the applicability of HMChPT. In order to apply this approach, the two final-state pseudoscalars in the $B \to P_1P_2$ transition matrix element have to be soft, which is not generally the case. Hence, an ansatz for the momentum dependence of non-resonant amplitudes in an exponential form,

$$A_{\rm transition} = A^{\rm HMChPT}_{\rm transition}\, e^{-\alpha_{\rm NR} p_B\cdot(p_1+p_2)} e^{i\phi_{12}}, \tag{410}$$

is assumed, so that the HMChPT results are recovered in the soft meson limit of $p_1, p_2 \to 0$.

For penguin-dominated decays $B \to KKK$ and $B \to K\pi\pi$, the non-resonant background induced by the $b \to u$ transition process is small compared to experiment due to the large CKM suppression $|V_{ub}V^*_{us}| \ll |V_{cb}V^*_{cs}| \approx |V_{tb}V^*_{ts}|$ associated with the $b \to u$ tree transition relative to the $b \to s$ penguin process. This implies that the two-body matrix element of scalar densities such as $\langle K\bar{K}|\bar{s}s|0\rangle$ induced from the penguin diagram should have a large non-resonant component. The measured kaon electromagnetic form factors can be used to extract $\langle K\bar{K}|\bar{q}\gamma_\mu q'|0\rangle^{\rm NR}$ and $\langle K\bar{K}|\bar{s}s|0\rangle^{\rm NR}$ first, then SU(3) flavour symmetry is applied to relate them to other two-body matrix elements [903]. The non-resonant component of the matrix element of scalar density is given by Ref. [903]

$$\langle K^+(p_2)K^-(p_3)|\bar{s}s|0\rangle_{\rm NR} = \frac{v}{3}\,(3F_{NR} + 2F'_{NR}) + \sigma_{\rm NR} e^{-\alpha s_{23}}. \tag{411}$$

with $v = m^2_{K^+}/(m_u + m_s) = (m^2_K - m^2_\pi)/(m_s - m_d)$.

*Resonant contributions.* In general, vector and scalar resonances contribute to the two-body matrix elements $\langle P_1P_2|V_\mu|0\rangle$ and $\langle P_1P_2|S|0\rangle$, respectively. The intermediate vector meson contributions to three-body decays are identified through the vector current, while the scalar resonances are mainly associated with the scalar density. Both scalar and vector resonances can contribute to the three-body matrix element $\langle P_1P_2|J_\mu|B\rangle$. The intermediate resonances are described by a coherent sum of Breit-Wigner expressions

$$\begin{aligned}
\langle P_1P_2|\bar{q}_1\gamma_\mu q_2|0\rangle^R &= \sum_i \langle P_1P_2|V_i\rangle\langle V_i|\bar{q}_1\gamma_\mu q_2|0\rangle \times \frac{1}{s_{12} - m^2_{V_i} + i m_{V_i}\Gamma_{V_i}}, \\
&+ \sum_i \langle P_1P_2|S_i\rangle\langle S_i|\bar{q}_1\gamma_\mu q_2|0\rangle \times \frac{-1}{s_{12} - m^2_{S_i} + i m_{S_i}\Gamma_{S_i}}, \\
\langle P_1P_2|\bar{q}_1 q_2|0\rangle^R &= \sum_i \langle P_1P_2|S_i\rangle\langle S_i|\bar{q}_1 q_2|0\rangle \times \frac{-1}{s_{12} - m^2_{S_i} + i m_{S_i}\Gamma_{S_i}},
\end{aligned} \tag{412}$$

where $V_i = \phi, \rho, \omega, \cdots$ and $S_i = f_0(980), f_0(1370),\, f_0(1500), \cdots$ for $P_1P_2 = \pi^+\pi^-$, and $V_i = K^*(892), K^*(1410),\, K^*(1680), \cdots$ and $S_i = K^*_0(1430), \cdots$ for $P_1P_2 = K^\pm\pi^\mp$.

*Branching fractions.* Table 106 summarises the calculated branching fractions of resonant and non-resonant components in penguin-dominated decays $B^- \to K^+K^-K^-$ and $K^-\pi^+\pi^-$ decays. It is known that the predicted rates for penguin-dominated channels $K^-\phi$ in $B^- \to K^+K^-K^-$ decays, and $K^*\pi$, $K^*_0(1430)\pi$ and $\rho K$ in $B^- \to K^-\pi^+\pi^-$ within the factorisation approach are substantially smaller than the data. To overcome this problem, the penguin-annihilation induced power corrections calculated in Ref. [855] have been used. Regarding the quasi-two-body mode $B^- \to \bar{K}^{*0}_0(1430)\pi^-$, BaBar has measured the three-body decay $B^- \to K^0_S\pi^-\pi^0$ and obtained $Br(B^- \to \bar{K}^{*0}_0(1430)\pi^- \to K^-\pi^+\pi^-) = (31.0 \pm 3.0 \pm 3.8^{+1.6}_{-1.6}) \times 10^{-6}$ [921], in good agreement with the Belle's result $(32.0 \pm 1.0 \pm 2.4^{+1.1}_{-1.9}) \times 10^{-6}$ [920]. Hence, the rate predicted by naive factorisation is too small by a factor of 3. This is still an unresolved puzzle in both, the QCD factorisation and PQCD approaches [922, 923].

Table 106: Branching fractions (in units of $10^{-6}$) of resonant and non-resonant (NR) contributions to $B^- \to K^+K^-K^-, K^-\pi^+\pi^-$ [904].

| $B^- \to K^+K^-K^-$ | | | |
|---|---|---|---|
| Decay mode | BaBar [699] | Belle [919] | Theory |
| $\phi K^-$ | $4.48 \pm 0.22^{+0.33}_{-0.24}$ | $4.72 \pm 0.45 \pm 0.35^{+0.39}_{-0.22}$ | $4.4^{+0.0+0.8+0.0}_{-0.0-0.7-0.0}$ |
| $f_0(980)K^-$ | $9.4 \pm 1.6 \pm 2.8$ | $< 2.9$ | $11.2^{+0.0+2.7+0.0}_{-0.0-2.1-0.0}$ |
| $f_0(1500)K^-$ | $0.74 \pm 0.18 \pm 0.52$ | | $0.63^{+0.0+0.11+0.0}_{-0.0-0.10-0.0}$ |
| $f_0(1710)K^-$ | $1.12 \pm 0.25 \pm 0.50$ | | $1.2^{+0+0.2+0}_{-0-0.2-0}$ |
| $f_2'(1525)K^-$ | $0.69 \pm 0.16 \pm 0.13$ | | |
| NR | $22.8 \pm 2.7 \pm 7.6$ | $24.0 \pm 1.5 \pm 1.8^{+1.9}_{-5.7}$ | $21.1^{+0.8+7.2+0.1}_{-1.1-5.7-0.1}$ |
| Total | $33.4 \pm 0.5 \pm 0.9$ | $30.6 \pm 1.2 \pm 2.3$ | $28.8^{+0.5+7.9+0.1}_{-0.6-6.4-0.1}$ |
| $B^- \to K^-\pi^+\pi^-$ | | | |
| Decay mode | BaBar [919] | Belle [920] | Theory |
| $\bar{K}^{*0}\pi^-$ | $7.2 \pm 0.4 \pm 0.7^{+0.3}_{-0.5}$ | $6.45 \pm 0.43 \pm 0.48^{+0.25}_{-0.35}$ | $8.4^{+0.0+2.1+0.0}_{-0.0-1.9-0.0}$ |
| $\bar{K}_0^{*0}(1430)\pi^-$ | $19.8 \pm 0.7 \pm 1.7^{+5.6}_{-0.9} \pm 3.2$ | $32.0 \pm 1.0 \pm 2.4^{+1.1}_{-1.9}$ | $11.5^{+0.0+3.3+0.0}_{-0.0-2.8-0.0}$ |
| $\rho^0 K^-$ | $3.56 \pm 0.45 \pm 0.43^{+0.38}_{-0.15}$ | $3.89 \pm 0.47 \pm 0.29^{+0.32}_{-0.29}$ | $2.9^{+0.0+0.7+0.0}_{-0.0-0.2-0.0}$ |
| $f_0(980)K^-$ | $10.3 \pm 0.5 \pm 1.3^{+1.5}_{-0.4}$ | $8.78 \pm 0.82 \pm 0.65^{+0.55}_{-1.64}$ | $6.7^{+0.0+1.6+0.0}_{-0.0-1.3-0.0}$ |
| NR | $9.3 \pm 1.0 \pm 1.2^{+6.7}_{-0.4} \pm 1.2$ | $16.9 \pm 1.3 \pm 1.3^{+1.1}_{-0.9}$ | $15.7^{+0.0+8.1+0.0}_{-0.0-5.2-0.0}$ |
| Total | $54.4 \pm 1.1 \pm 4.6$ | $48.8 \pm 1.1 \pm 3.6$ | $42.2^{+0.2+16.1+0.1}_{-0.1-10.7-0.1}$ |

The non-resonant component of $B \to KKK$ is governed by the $K\bar{K}$ matrix element of scalar density $\langle K\bar{K}|\bar{s}s|0\rangle$. By the same token, the non-resonant contribution to the penguin-dominated $B \to K\pi\pi$ decays should also be dominated by the $K\pi$ matrix element $\langle K\pi|\bar{s}q|0\rangle$ of the scalar density. Applying the SU(3) symmetry relation, $\langle K^-(p_1)\pi^+(p_2)|\bar{s}d|0\rangle_{\rm NR} = \langle K^+(p_1)K^-(p_2)|\bar{s}s|0\rangle_{\rm NR}$, one finds too large non-resonant and total branching fractions, namely $Br(B^- \to K^-\pi^+\pi^-)_{\rm NR} \sim 29.7 \times 10^{-6}$ and $Br(B^- \to K^-\pi^+\pi^-)_{\rm tot} \sim 68.5 \times 10^{-6}$. It also leads to negative asymmetries $A_{CP}^{\rm incl}(B^- \to K^-\pi^+\pi^-) \sim -0.8\%$ and $A_{CP}^{\rm resc}(B^- \to K^-\pi^+\pi^-) \sim -6.4\%$ opposite in sign compared with the data. To accommodate the rates, as argued in Ref. [915], some sort of power corrections such as final-state interactions are assumed to give a large strong phase $\delta$ to the non-resonant component of $\langle K^-\pi^+|\bar{s}d|0\rangle$, parametrised as

$$\langle K^-(p_1)\pi^+(p_2)|\bar{s}d|0\rangle_{\rm NR} \quad = \quad \frac{v}{3}\left(3F_{\rm NR} + 2F'_{\rm NR}\right) + \sigma_{\rm NR}e^{-\alpha s_{12}}e^{i\delta}. \tag{413}$$

It is then found that $\delta \approx \pm\pi$ accommodates both, the non-resonant branching fractions and the $CP$ asymmetry for $B^- \to K^-\pi^+\pi^-$. Yet, it should be stressed again that the predicted total rate of $B^- \to K^-\pi^+\pi^-$ is smaller than the measurements of both BaBar and Belle. This is ascribed to the fact that the calculated $K_0^*(1430)\pi^-$ rate in naive factorisation is too small by a factor of 3.

*Direct $CP$ violation.* In Refs. [904, 915], there are three sources of strong phases: from effective Wilson coefficients, from propagators of resonances, and from the matrix element of the scalar density $\langle M_1M_2|\bar{q}_1q_2|0\rangle$. There are two sources for the phase in the penguin matrix element of scalar densities: $\sigma_{\rm NR}$ and $\delta$ for $K\pi$-vacuum matrix elements, see Eq. (413).

Table 107: Predicted inclusive and regional $CP$ asymmetries (in %) for various charmless three-body $B$ decays [904]. Two local regions of interest for regional $CP$ asymmetries are the low-mass regions specified in Refs. [892, 893] for $A_{CP}^{\rm low}$ and the rescattering region of $m_{\pi\pi}$ and $m_{K\bar{K}}$ between 1.0 and 1.5 GeV for $A_{CP}^{\rm resc}$. Resonant (RES) and non-resonant (NR) contributions to direct $CP$ asymmetries are considered.

| | $\pi^-\pi^+\pi^-$ | $K^+K^-\pi^-$ | $K^-\pi^+\pi^-$ | $K^+K^-K^-$ |
|---|---|---|---|---|
| $(A_{CP}^{\rm incl})_{\rm NR}$ | $25.0^{+4.4+2.1+0.0}_{-2.7-3.1-0.1}$ | $-25.6^{+2.2+1.7+0.2}_{-3.0-1.1-0.1}$ | $9.1^{+1.3+2.2+0.1}_{-1.8-2.0-0.1}$ | $-7.8^{+1.4+1.3+0.1}_{-0.9-1.5-0.1}$ |
| $(A_{CP}^{\rm incl})_{\rm RES}$ | $5.3^{+0.0+1.6+0.0}_{-0.0-1.3-0.0}$ | $-16.3^{+0.0+0.9+0.1}_{-0.0-0.8-0.1}$ | $6.9^{+0.0+2.1+0.1}_{-0.0-1.8-0.1}$ | $1.2^{+0.0+0.0+0.0}_{-0.0-0.0-0.0}$ |
| $(A_{CP}^{\rm incl})_{\rm NR+RES}$ | $8.3^{+0.5+1.6+0.0}_{-1.1-1.5-0.0}$ | $-10.2^{+1.6+1.5+0.1}_{-2.5-1.4-0.1}$ | $7.3^{+0.2+2.1+0.1}_{-0.2-2.0-0.1}$ | $-6.0^{+1.8+0.8+0.1}_{-1.2-0.9-0.1}$ |
| $(A_{CP}^{\rm incl})_{\rm expt}$ | $5.8 \pm 2.4$ | $-12.3 \pm 2.2$ | $2.5 \pm 0.9$ | $-3.6 \pm 0.8$ |
| $(A_{CP}^{\rm low})_{\rm NR}$ | $58.3^{+3.6+2.6+0.8}_{-3.7-4.0-0.8}$ | $-25.0^{+2.8+2.7+0.3}_{-5.4-2.5-0.3}$ | $48.9^{+\ 7.0+7.6+0.3}_{-10.5-8.2-0.3}$ | $-13.0^{+2.0+2.8+0.2}_{-1.2-3.2-0.2}$ |
| $(A_{CP}^{\rm low})_{\rm RES}$ | $4.5^{+0.0+1.6+0.0}_{-0.0-1.2-0.0}$ | $-4.9^{+0.0+0.5+0.0}_{-0.0-0.4-0.0}$ | $57.1^{+0.0+\ 7.9+0.9}_{-0.0-16.6-0.9}$ | $1.6^{+0.0+0.1+0.0}_{-0.0-0.1-0.0}$ |
| $(A_{CP}^{\rm low})_{\rm NR+RES}$ | $21.9^{+0.5+3.0+0.0}_{-0.4-3.3-0.1}$ | $-17.5^{+0.6+1.7+0.1}_{-0.9-1.5-0.1}$ | $49.4^{+0.7+\ 9.4+0.8}_{-1.0-14.2-0.8}$ | $-16.8^{+3.5+2.8+0.2}_{-2.3-3.2-0.2}$ |
| $(A_{CP}^{\rm low})_{\rm expt}$ | $58.4 \pm 9.7$ | $-64.8 \pm 7.2$ | $67.8 \pm 8.5$ | $-22.6 \pm 2.2$ |
| $(A_{CP}^{\rm resc})_{\rm NR}$ | $36.7^{+6.2+3.2+0.1}_{-3.7-4.6-0.2}$ | $-27.7^{+3.1+3.0+0.4}_{-5.9-2.7-0.4}$ | $31.8^{+4.6+4.6+0.3}_{-6.7-4.5-0.3}$ | $-10.8^{+1.8+2.2+0.2}_{-1.2-2.5-0.2}$ |
| $(A_{CP}^{\rm resc})_{\rm RES}$ | $7.0^{+0.0+1.8+0.0}_{-0.0-1.5-0.0}$ | $-5.6^{+0.0+0.5+0.0}_{-0.0-0.4-0.0}$ | $1.1^{+0.0+0.6+0.0}_{-0.0-0.5-0.0}$ | $0.96^{+0.00+0.02+0.01}_{-0.00-0.02-0.01}$ |
| $(A_{CP}^{\rm resc})_{\rm NR+RES}$ | $13.4^{+0.5+2.0+0.0}_{-1.1-2.1-0.0}$ | $-20.4^{+1.2+2.0+0.2}_{-1.8-1.8-0.2}$ | $4.1^{+0.2+0.9+0.0}_{-0.3-0.9-0.0}$ | $-3.8^{+1.5+0.5+0.1}_{-1.0-0.5-0.1}$ |
| $(A_{CP}^{\rm resc})_{\rm expt}$ | $17.2 \pm 2.7$ | $-32.8 \pm 4.1$ | $12.1 \pm 2.2$ | $-21.1 \pm 1.4$ |

The LHCb data indicate that decays involving a $K^+K^-$ pair have a larger $CP$ asymmetry ($A_{CP}^{\rm incl}$ or $A_{CP}^{\rm resc}$) than their partner channels. The asymmetries are positive for channels with a $\pi^+\pi^-$ pair and negative for those with a $K^+K^-$ pair. In other words, when $K^+K^-$ is replaced by $\pi^+\pi^-$, the $CP$ asymmetry flips its sign. This can be understood in terms of U-spin symmetry, which leads to the relation [924, 925]

$$R_1 \equiv \frac{A_{CP}(B^- \to \pi^-\pi^+\pi^-)}{A_{CP}(B^- \to K^-K^+K^-)} = -\frac{\Gamma(B^- \to K^-K^+K^-)}{\Gamma(B^- \to \pi^-\pi^+\pi^-)}, \tag{414}$$

and

$$R_2 \equiv \frac{A_{CP}(B^- \to \pi^-K^+K^-)}{A_{CP}(B^- \to K^-\pi^+\pi^-)} = -\frac{\Gamma(B^- \to K^-\pi^+\pi^-)}{\Gamma(B^- \to \pi^-K^+K^-)}. \tag{415}$$

The predicted signs of the ratios $R_1$ and $R_2$ are confirmed by experiment. However, because of the momentum dependence of three-body decay amplitudes, U-spin or flavour SU(3) symmetry do not lead to any testable relations between $A_{CP}(\pi^-K^+K^-)$ and $A_{CP}(\pi^-\pi^+\pi^-)$ and between $A_{CP}(K^-\pi^+\pi^-)$ and $A_{CP}(K^+K^-K^-)$. That is, symmetry arguments alone do not give hints at the relative sign of the $CP$ asymmetries in the pair of $\Delta S = 0(1)$ decays.

Following the framework of Refs. [903, 915] we present in Table 107 the calculated inclusive and regional $CP$ asymmetries in the adopted model, including both resonant and non-resonant mechanisms and their interference. For the non-resonant contributions, direct $CP$ violation arises solely from the interference of tree and penguin non-resonant amplitudes. For example, in the absence of resonances, the $CP$ asymmetry in $B^- \to K^-\pi^+\pi^-$ stems mainly from the interference of the non-resonant tree amplitude $\langle \pi^+\pi^-|(\bar{u}b)_{V-A}|B^-\rangle_{\rm NR}\langle K^-|(\bar{s}u)_{V-A}|0\rangle$ with the non-resonant penguin amplitude $\langle \pi^-|\bar{d}b|B^-\rangle\langle K^-\pi^+|\bar{s}d|0\rangle_{\rm NR}$.

It is clear from Table 107 that non-resonant $CP$ violation is usually much larger than resonant $CP$ violation and that the interference effect is generally quite significant. If non-resonant contributions are turned off in the $K^+K^-K^-$ mode, the predicted asymmetries will be wrong in sign when compared with experiment. The main contributions to $(A_{CP}^{\rm incl})_{\rm RES}$ arise from $\phi K^-, f_0(1500)K^-, f_0(1710)K^-$, all giving positive contributions. This is not a surprise because the mode $B^- \to K^+K^-K^-$ is dominated by the non-resonant background. Hence, the magnitude and the sign of its $CP$ asymmetry should also be governed by the non-resonant term. The observed negative $A_{CP}^{\rm incl}(K^+K^-K^-)$ is a strong indication of the importance of non-resonant effects.

From Table 107, it is also evident that except for the $K^+K^-K^-$ mode, the resonant contributions to integrated inclusive $CP$ asymmetries are of the same sign and similar magnitudes as $A_{CP}^{\rm incl}$. On the other hand, the predicted $(A_{CP}^{\rm low})_{\rm RES}$ and $(A_{CP}^{\rm resc})_{\rm RES}$ by resonances alone for other modes are usually too small compared to the data, especially for the former.

The LHCb data indicate that the $CP$ asymmetries are positive for channels with a $\pi^+\pi^-$ pair and negative for those with a $K^+K^-$ pair, as discussed above. This observation appears to imply that final-state rescattering may play an important role for direct $CP$ violation. Based on the constraint of $CPT$ invariance on final-state interactions, the authors of Refs. [897, 900] have studied $CP$ violation in charmless three-body charged $B$ decays. They assume that only the two channels $\alpha = \pi^+\pi^-P^-$ and $\beta = K^+K^-P^-$ $(P = \pi, K)$ in $B^-$ decays are strongly coupled through strong interactions and treat the third meson $P$ as a bachelor. Applying the $CPT$ relation to describe the $CP$-asymmetry distribution in $B^- \to K^+K^-P^-$ decays after fitting the model to the $B^- \to \pi^+\pi^-P^-$ channels, they find that final-state rescattering of $\pi^+\pi^- \leftrightarrow K^+K^-$ dominates the asymmetry in the mass region between 1 and 1.5 GeV. In Refs. [904, 915], the partial rates and $CP$ asymmetries are calculated in the model based on naive factorisation as discussed above, without taking into account final-state interactions explicitly and without data fitting. While the calculated direct $CP$ asymmetries for $K^+K^-K^-$ and $\pi^+\pi^-\pi^-$ modes are in good agreement with experiment in both magnitude and sign, the predicted asymmetries in $B^- \to \pi^-K^+K^-$ and $B^- \to K^-\pi^+\pi^-$ are wrong in sign compared to experiment. In order to accommodate the non-resonant branching fraction and $CP$ asymmetry observed in $B^- \to K^-\pi^+\pi^-$, the matrix element $\langle K\pi|\bar{s}q|0\rangle$ is modified by extra strong phase $\delta$ of order $\pm\pi$ in addition to the phase characterised by the parameter $\sigma_{\rm NR}$, as mentioned above. The phase $\delta$ may arise from final-state interactions.

In the study of $B^- \to \pi^-\pi^+\pi^-$, Refs. [903, 915] encountered a conflict between theory and experiment for the $CP$ asymmetry $A_{CP}^{\rho^0\pi^-}$. Both BaBar [926] and LHCb [894] measurements of $B^- \to \pi^+\pi^-\pi^-$ indicate a positive $CP$ asymmetry in the $m(\pi^+\pi^-)$ region peaked at $m_\rho$. On the other hand, all theories predict large and negative $CP$ violation in $B^- \to \rho^0\pi^-$. Therefore, the issue with $CP$ violation in $B^- \to \rho^0\pi^-$ needs to be resolved.

As mentioned at the beginning, the magnitude and sign of $CP$ asymmetries in the Dalitz plot vary from region to region. The $CP$-asymmetry Dalitz distributions in some (large) invariant mass regions have been studied in the factorisation model [904], finding qualitative agreement with experiment for $K^+K^-K^-$ and $\pi^+\pi^-\pi^-$ modes and the correct sign for $K^-\pi^+\pi^-$. However, it appears that the phase $\delta$ needs to vanish in the large invariant mass region for $K^+K^-\pi^-$ in order to accommodate the observations. Thus it is possible that the phase $\delta$ must be allowed to be energy dependent. This issue needs to be investigated.

The large Belle II dataset will enable the study of additional three-body channels with neutral final state particles which may exhibit large local $CP$ asymmetries, such as $B^0 \to K^+K^-K^0_S$, $B^0 \to K^+K^-\pi^0$, $B^0 \to K^+\pi^0\pi^0$, $B^+ \to K^0_S\pi^+\pi^0$, $B^+ \to K^0_SK^0_SK^+$ and $B^+ \to K^0_SK^0_S\pi^+$.

## *12.8. Conclusions*

The large datasets collected by Belle, BaBar, and LHCb have enabled the study of many charmless hadronic $B_{(s)}$ decays and have allowed for a detailed comparison with theoretical predictions and models. Some tantalizing questions have emerged and await the large dataset of Belle II to be further understood. The expected precision in $B \to K^0\pi^0$ with $50\,\mathrm{ab}^{-1}$ of data will be sufficient for NP studies and may resolve the $K\pi$ $CP$-puzzle. The analogous isospin sum rules for the multi-body $\pi K^*$ and $\rho K^{(*)}$ decays are also promising avenues to resolve this puzzle, but are statistically limited and must be measured with high precision to reveal whether an anomalous pattern of direct $CP$ violation is emerging. The study of $B \to VV$ decays requires large statistics to perform full angular analyses, and thus there remains enormous potential. While the majority of analyses at Belle and BaBar were limited to only measuring the longitudinal polarisation fraction, full angular analyses will be possible for many $VV$ channels at Belle II. Of particular interest are $\rho K^*$ decays, where a polarisation analysis will reveal if there is an enhanced contribution proportional to electromagnetic penguins. Belle II will also be uniquely suited to search for $CP$ asymmetries in $B \to 3h$ decays with multiple neutral particles in the final state, which will serve to complement related searches at LHCb, where the observation of large local $CP$ asymmetries in multiple channels has generated enormous interest from the theoretical and phenomenological communities. A sizeable $B_s$ dataset will also be necessary to study rare decays such as the penguin dominated $B_s \to \phi\pi^0$, where an excess above the SM prediction would be a clear indication of NP, and, *e.g.*, the recently observed $B^0_s \to K^0\bar{K}^0$ decay, where Belle II expects to reconstruct $\mathcal{O}(1000)$ events with $5\,\mathrm{ab}^{-1}$ which will enable a $CP$ violation study and will serve to clarify the presence of NP in the decay. There are countless additional charmless hadronic $B_{(s)}$ decays which will be within the reach of Belle II. This will open up a new era of discovery and complementarity with other experiments.

# 13. Charm Physics

*Editors: G. Casarosa, A. L. Kagan, A. A. Petrov, A. J. Schwartz,*
*Additional section writers: J. Bennett, R. Briere, G. De Pietro, S. Fajfer, M. Jung, L. Li, T. Nanut, U. Nierste, S. Schacht, S. Sharpe*

## *13.1. Introduction*

Studies of transitions involving the charm quark play an important role in both searches for New Physics (NP) and in understanding QCD. The large yields of charmed mesons and baryons that will be produced at Belle II makes searches for NP in charm transitions a vibrant avenue for research. At the $\Upsilon(4S)$ resonance, the charm production cross section is approximately 2.7 nb: 1.6 nb from prompt $c\bar{c}$ production, and 1.1 nb from secondary $\Upsilon(4S) \to B \to D$ production. Thus the number of $D$ ($D_s$) mesons produced is expected to be $> 10^9$ ($> 10^8$) per ab$^{-1}$ of data. Decays of charmed mesons and baryons probe a variety of NP scenarios, *e.g.*, couplings to intermediate charged Higgs states, and decays to light dark matter particles.

Searches for NP in charm decays fall into three categories: *(1)* searches for processes that are forbidden in the Standard Model (SM); *(2)* studies of processes that are forbidden at tree level in the SM; and *(3)* studies of processes that are allowed at tree level. The first category probes violations of principles upon which modern quantum field theories are based, such as locality, unitarity, gauge invariance, and Lorentz invariance. The second category includes processes that occur via higher-order electroweak diagrams, such as flavour-changing neutral currents (FCNC). However, the relatively small mass of the intermediate state $b$ quark and tiny values of the Cabbibo-Kobayashi-Maskawa (CKM) matrix elements $V_{cb}, V_{ub}$ make the short-distance SM amplitudes very small. Thus, these processes tend to be long-distance dominated, and SM predictions of most FCNC $|\Delta C| = 1$ and $|\Delta C| = 2$ processes have significant uncertainties. Finally, charm transitions allowed at tree level also test the SM. For example, measurements of leptonic and semileptonic decays can be directly compared to lattice QCD calculations, which have greatly improved in precision over the past few years. In addition, SM sum rules and symmetry relations among decay amplitudes based on $SU(2)$ and $SU(3)$ can be tested experimentally, *e.g.*, by measuring branching fractions. Violations of such relations would indicate the presence of NP.

Experimentally, an $e^+e^-$ collider experiment is ideal for studying charm decays, including those that are very rare or forbidden. Backgrounds are much lower than at a hadron machine, trigger efficiencies are much higher, and acceptances tend to be flat across Dalitz plots. There are usually numerous control samples available with which to study backgrounds and estimate systematic uncertainties. Because the recorded luminosity can be determined by measuring Bhabha scattering, absolute (in addition to relative) branching fractions can be measured. Because the initial state is known, unknown particles can be searched for via energy-momentum conservation: one calculates the difference between the 4-momentum of the initial state and the sum of the 4-momenta of all visible particles, squares that "missing 4-momentum" to get a "missing mass," and looks for structure in the missing mass spectrum. Finally, photons, $\pi^0$'s, and final state particles decaying to $\pi^0$'s such as $\eta$, $\eta'$, $\rho^+$, $\omega$, and $K^{*+}$ are much easier to reconstruct in an $e^+e^-$ experiment than in a hadron collider experiment.

This chapter is organised as follows. We first review experimental techniques such as flavour tagging and partial reconstruction, and then we discuss some highlights from the Belle II charm physics program. The latter is not a complete discussion but rather focuses on several topics of high interest that the Belle II detector is well-suited to address. These topics are divided into the following categories: leptonic and semileptonic decays; rare and radiative decays; mixing and indirect $CP$ violation; and direct $CP$ violation. Within each category there is a theory discussion followed by a discussion of experimental sensitivity. A dedicated section on lattice QCD calculations is also included. Finally, we conclude with a listing of "Golden Modes," *i.e.*, those decay modes that Belle II should measure well and also have especially good sensitivity to NP.

## *13.2. Experimental Techniques*

*Authors: G. Casarosa, G. De Pietro*

The Belle II detector will offer improved performance in the reconstruction of charm events with respect to the previous generation of $B$-factories. Before presenting the physics reach of Belle II, we discuss charm flavour-tagging techniques, and expected improvements in decay vertex resolution and reconstruction efficiency.

*13.2.1. Flavour-Tagging Methods.* In order to measure CPV it is crucial to determine the flavour of the $D^0$ or $\overline{D}^0$ at production. At $B$-factories this was achieved selecting the $D^0$ coming from the $D^{*+} \to D^0\pi^+$ with the charge of the pion determining the charm-quark flavour of the neutral meson. The $D^0$ mesons coming from $B$ decays were excluded[40] in order to have a better measurement of the decay proper time, therefore only $D^0$ from $D^{*+}$ in $c\bar{c}$ events were used. After a brief summary of the expected performance of the $D^*$ method at Belle II, we present a new flavour-tagging method, the ROE method, that could potentially increase statistics and also provide useful control samples for our measurements. We will also comment on the possibility of exploiting $D^0$ mesons from a partial reconstruction of $B$ decays for time-integrated measurements. In Table 108 we report a summary of efficiencies and mistagging rates for the methods presented.

*$D^*$ method* This is the "golden" flavour-tagging method: it provides a clean sample of flavour-tagged $D^0$ and it has been used extensively at B-Factories. The primary purpose of the method is to identify the flavor of the $D^0$ meson at its production through the charge of the pion emitted with the $D^0$ in the $D^{*+}$ decay. The low Q value of the $D^{*+} \to D^0\pi^+$ decay allows for a powerful criterion to be applied to the reconstructed difference of $D^{*+}$ and $D^0$ masses, which eliminates a considerable fraction of the combinatorial background, as shown in Fig. 143. At Belle II we achieve a resolution on $\Delta m$ of $\sim 180$ keV/$c^2$ (estimated for $D^*$-tagged $D^0 \to K\pi$ candidates), a factor of two better than that achieved by Belle and BaBar; this will increase the background rejection power.

The typical reconstruction efficiency at BaBar was around $\epsilon_{D^*} = 80\%$, with a mistagging rate of approximately $\omega_{D^*} = 0.2\%$. Studies on simulated events show that Belle II will have a similar reconstruction efficiency.

*ROE method*

[40] To remove $D^0$ candidates from $B$ decays, at Belle and BaBar $D^0$ mesons were usually required to have the momentum in the CM frame greater than $\sim 2.5$ GeV/$c$.

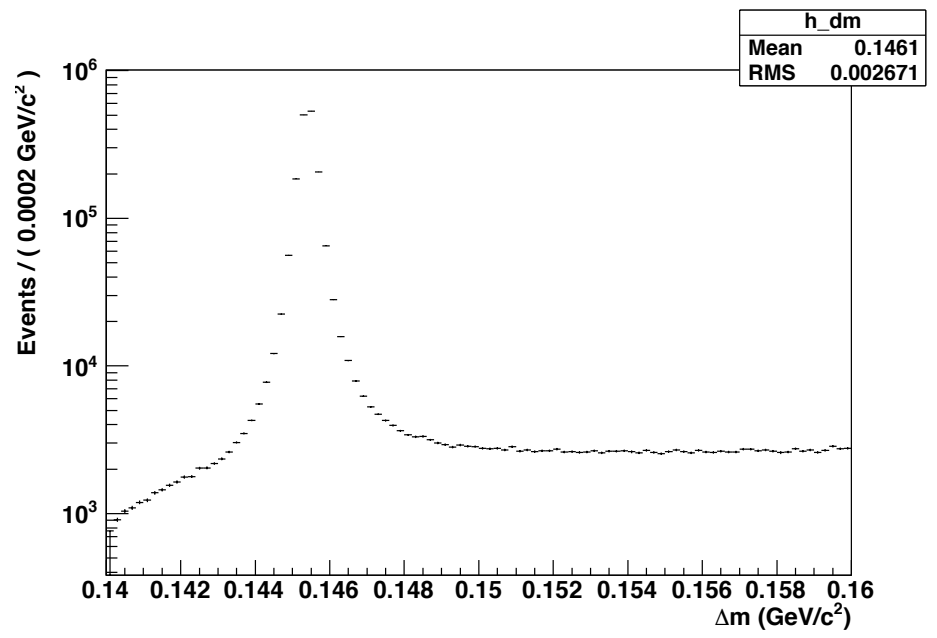

Fig. 143: $\Delta m = m(D^{*+}) - m(D^0)$ distribution for reconstructed $D^*$-tagged $D^0 \to K\pi$.

We present a new flavour-tagging method with the goal of increasing the size of the sample of tagged $D^0$ candidates. This is achieved by adding $D^0$ mesons produced in $c\bar{c}$ events that are not coming from $D^{*+}$ decays.

This new method consists in looking at the so-called rest of the event (ROE) with respect to the neutral $D$ meson whose flavour we want to tag. The principle of the ROE method is shown in Fig. 144. Suppose a $c$ quark hadronizes into a $D^0$ meson, and $\bar{c}$ hadronizes into an anti-charmed meson or an anti-charmed baryon. Since the Cabibbo-favoured transition for an anti-charm quark is $\bar{c} \to \bar{s}$, we expect to find at least one hadron containing an $\bar{s}$-quark in the ROE, namely a $K^+$ $(u\bar{s})$ or a $K^0$ $(d\bar{s})$.

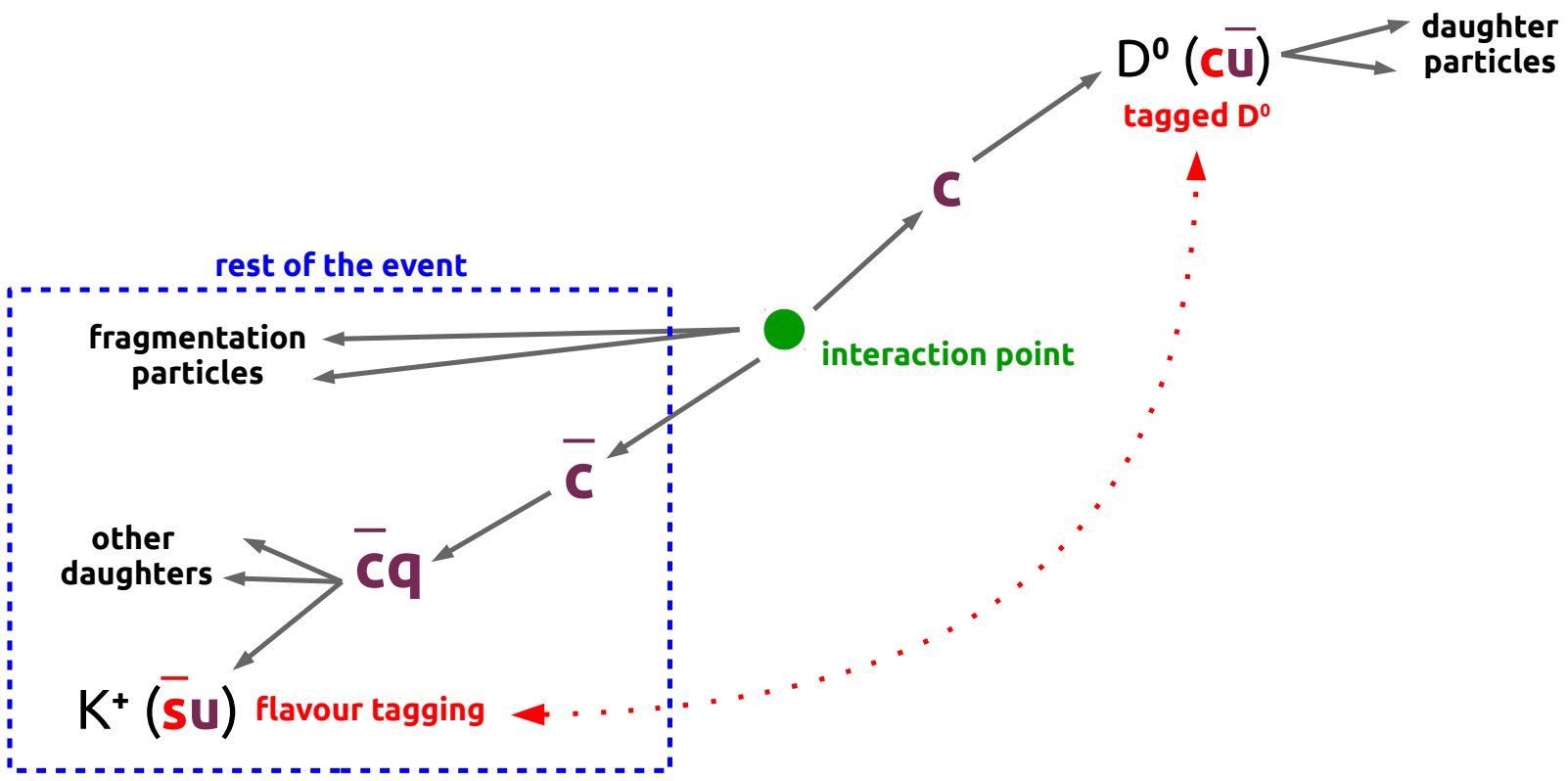

Fig. 144: The principle on which the ROE flavour-tagging method is based. The events with only one $K^\pm$ in the ROE are selected; the flavour of the neutral $D$ meson is determined by the charge of the kaon.

The flavour-tagging is performed by selecting events with only one $K^\pm$ in the ROE and using the charge of the kaon to determine the flavour of the other $D^0$ at the time of its production. A correctly identified $K^\pm$ produced by a Cabibbo-Favoured (CF) decay ( $c \to su\bar{d}$) of a charmed hadron is labeled as "signal $K^\pm$".

Of course, not every event with a single $K^\pm$ in the ROE correctly determines the flavour of the neutral $D$. The main source of mistagging is given by kaons produced from $s$ or $\bar{s}$ quarks in the primary fragmentation. If the $K^\pm$ in the ROE is generated from the hadronisation of an $s$ $(\bar{s})$ quark instead of a CF decay, the charge of the kaon is not correlated with the flavour of the neutral $D$ meson; *i.e.*, the $K^{\pm}$ in the ROE will randomly tag the flavour of the neutral $D$ meson. This type of background is labelled as "$K^{\pm}$ from $c\bar{c}s\bar{s}$". There are two other minor sources of mistagging: a charged kaon in the ROE produced by a Doubly-Cabibbo-Suppressed (DCS) decay ($c \to du\bar{s}$) labelled as "$K^{\pm}$ from DCS decay", and a charged kaon produced by a CF decay of a $D^0$ that has undergone mixing (labelled as "$K^{\pm}$ from mixing"). This last type of background is heavily suppressed, since the time-integrated probability for $D^0$-$\bar{D}^0$ oscillations is measured to be very small. Thus this background is neglected in the following.

Other sources of mistagging arise from the reconstruction of the charged kaons in the ROE. Applying a soft selection, we risk contaminating the list of $K^{\pm}$ candidates with tracks produced by other charged particles, mainly charged pions and protons (background from "fake $K^{\pm}$"). On the other hand, if the selection is too tight, we risk missing some $K^{\pm}$ candidates in the ROE and miscounting the number of $K^{\pm}$ (background from "missing $K^{\pm}$").

The crucial part of this new method consists of the selection of the $K^{\pm}$ candidates in the ROE, and this is performed using a multivariate classification. The chosen classifier method is the Fast Boosted Decision Tree, FBDT [70]. To reject poorly measured candidates, a preselection of the tracks based on PID and track fit probability is applied before the multivariate classification. The FBDT makes use of the following variables: track momentum in the lab frame, cosine of the track polar angle, track impact parameters, track fit probability, number of hits in PXD, SVD, CDC, and PID selectors for $K$, $\mu$, $e$, $p$. A two-step selection based on the FBDT output variable is applied: first a soft cut is applied to correctly count the number of charged kaons in the ROE, then a tighter cut is applied in order to remove the fake charged kaons from the list. The resulting tagging efficiency is $\epsilon = 26.7\%$ and the mistagging rate is $\omega = 13.3\%$. This is referred to as "criteria A" in Table 108. With this selection, 87.8% of the selected events have a single $K^{\pm}$ at the generator level ($\rho_{1K}$).

More than half of the mistagged events are due to $K^{\pm}$ from $c\bar{c}s\bar{s}$. A veto on the $K^0_S$ (reconstructed in the $\pi^+ \pi^-$ final state) in the ROE is applied after the FBDT selection in order to reduce the mistagging due to $c\bar{c}s\bar{s}$ events. Since the two charm quarks are produced back-to-back, a signal $K^{\pm}$ in the ROE tends to be produced in the opposite direction with respect to the neutral $D$ meson. A cut on the relative angle ($\theta^*_{\rm rel}$) between the direction of the charged kaon and the direction of the neutral $D$ meson in the center-of-mass frame further helps in reducing the mistagging rate. Applying these two post-FBDT selection criteria, the performance of the flavour-tagging method becomes $\epsilon = 16.8\%$, $\omega = 9.8\%$, and $\rho_{1K} = 90.9\%$. This is referred to as "criteria B" in Table 108.

Since this analysis has been made with the Belle II offline software (BASF2) version released at the time of writing this manuscript, some tools that will be available with the final version of the software were missing. In particular, the $K^0_L$ reconstruction was missing. Moreover, some improvements in the $K^0_S$ reconstruction and in the PID are expected. In order to evaluate how much the performance of this new flavour-tagging could change with further versions of the software, we studied the limiting case in which all generated $K^0_S \to \pi^+\pi^-$ and $K^0_L$ decays in the ROE are vetoed. With this special veto and with the same cut on $\theta^*_{\rm rel}$ as before, we obtain the following performance for the ROE flavour-tagging method: $\epsilon = 15.9\%$, $\omega = 4.9\%$, and $\rho_{1K} = 93.3\%$. This is referred to as "criteria C" in Table 108.

The final evaluation of the efficiency and mistagging rate can be performed directly with data events that are double-tagged, *i.e.*, tagged with *both* the $D^{*+}$ and ROE methods. We estimate that Belle II can measure the mistagging rate of the ROE method with a statistical uncertainty of $\sim$1% using an integrated luminosity of 13 fb$^{-1}$.

It should be noted that the number of $D^0$ mesons that are taggable using the ROE method, *i.e.*, produced via $e^+e^- \to D^0\bar{D}X$, is similar to the number of mesons taggable using the $D^*$ method, *i.e.*, produced via $e^+e^- \to D^{*+}\bar{D}X$.

*Partial Reconstruction of $B$ decays* Most $B$ mesons decay into a charmed hadron plus other particles. Charm measurements at $B$-factories have not fully exploited this large sample of charmed hadrons. Here we briefly present a reconstruction technique used to measure the absolute branching fraction of the $D^0 \to K^-\pi^+$ channel [927].

The technique consists of partially reconstructing the semileptonic decay $B^0 \to D^{*+}\ell^-\nu$ with $D^{*+} \to D^0\pi^+$. Experimentally, only two tracks are required to be reconstructed: the charged lepton and the low momentum (or "slow") pion of the $D^{*+}$ decay. These two oppositely charged tracks are geometrically fitted to a common vertex. Two assumptions has to be made in order to reconstruct the decay tree:

(1) the momentum of the $B^0$ in the the $\Upsilon(4S)$ reference frame is neglected since it is small compared to the momenta of the reconstructed tracks;
(2) the momentum vector of the $D^{*+}$ is estimated by rescaling the momentum of the slow pion: the low Q-value of the $D^{*+}$ decay allows the approximation that the pion is at rest in the $D^{*+}$ rest frame.

With these assumptions, the missing mass squared ($M_\nu^2$) of events can be calculated via

$$M_\nu^2 \;=\; (\sqrt{s}/2 - E_{D^{*+}} - E_\ell)^2 - (\vec{p}_{D^{*+}} + \vec{p}_\ell)^2\,, \tag{416}$$

where $E$ and $\vec{p}$ are the energy and momentum of the subscript particle, and $\sqrt{s}$ is the center-of-mass energy. The $M_\nu^2$ distribution peaks at zero for signal events and can is fitted to obtain the signal yield.

We note that although the $D^0$ is not reconstructed, its presence is indicated by a small value of $M_\nu^2$, and its flavour is identified by both the charge of the slow pion and the charge of the lepton. Although this reconstruction technique has a high efficiency of around 65% and a low mistagging rate, it suffers from the low branching fraction of the $B^0$ semileptonic decay. As listed in Table 108, considering $\ell = e, \mu$, only 13 $D^0$ are reconstructed for every 100 reconstructed with the $D^{*+}$ technique. The application of this technique is therefore limited by the low $D^0$ yield. However, it can be used to improve the measurement of branching fractions and also to search for rare processes where the uniqueness of a $B\bar{B}$ event environment can be exploited. Further studies are needed to understand the effective power of this reconstruction technique at Belle II.

*13.2.2. D Proper Time Resolution.* The Belle II vertex detector allows one to reconstruct the $D^0$ decay vertex with a precision of $\sim 40\ \mu$m, a significant improvement with respect to Belle and BaBar. This ability is due to the reduced distance between the first pixel layer and the interaction point. The resolution on the $D^0$ decay time is improved by a factor two, and this should greatly improve the precision of time-depending measurements of $D^0$-$\bar{D}^0$ mixing and searches for $CP$ violation.

Table 108: Flavour-tagging summary. The number of $D^0$ mesons produced is relative to the number originating from $D^*$ decays. The numbers regarding partial $B$ reconstruction are estimated from a BaBar analysis [927], while the others are extracted from Belle II simulations using a cut on the $D^0$ center-of-mass momentum ($p^* > 2.5$ GeV/c). Criteria A, B, C are described in the text.

| Flavour-tagging Method | Produced $D^0$ $N_{D^0}$ | Mistagging $\omega$ | Efficiency $\epsilon$ | $Q = \epsilon\,(1-2\omega)^2$ |
|---|---|---|---|---|
| $D^*$ | 1 | 0.2% | 80% | 79.7% |
| ROE - criteria A | 3 | 13.3% | 26.7% | 20.1% |
| ROE - criteria B | 3 | 9.8% | 16.8% | 13.7% |
| ROE - criteria C | 3 | 4.9% | 15.9% | 15.7% |
| partial $B$ reconstruction | 0.13 | $< 1\%$ | 65% | $\sim 62\%$ |

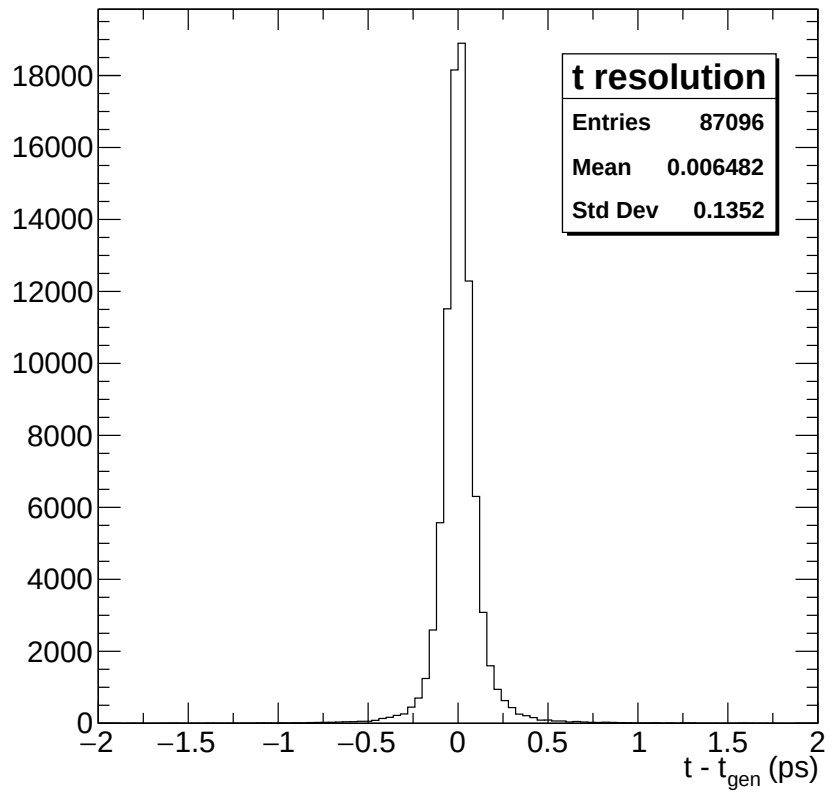

Fig. 145: $D^0$ proper time residuals for $D^*$-tagged $D^0 \to K^+K^-$ reconstructed decays.

Table 109: Typical mean and RMS of proper time resolutions. These values are obtained for candidates reconstructed in the $D^*$-tagged $D^0 \to K^+K^-$ channel.

| Experiment | $t$ resolution Mean | RMS |
|---|---|---|
| Belle II | 6.5 fs | 135 fs |
| BaBar | -0.48 fs | 271 fs |

Figure 145 shows the residuals of the proper decay time $t = (\vec{p}\cdot\vec{d})/|\vec{p}|$, where $\vec{p}$ is the reconstructed momentum and $\vec{d}$ is the vector connecting the $D^0$ production and decay vertices. For comparison, Table 109 lists typical resolutions of proper decay time and errors on decay time for Belle II and BaBar. The average proper time error is a factor of three smaller at Belle II. The same improvement in resolution is also achieved for prompt $D^0$ production, *i.e.*, $D^0$ mesons that do not originate from $D^{*+}$ decays.

## 13.3. Leptonic and semileptonic decays

Leptonic and semileptonic decays of charm involve both well-understood weak interactions physics and non-perturbative strong-interaction effects. Leptonic decays of charm mesons are used to extract the product of a decay constant and a CKM matrix element: $|V_{cd}|f_D$ or $|V_{cs}|f_{D_s}$. Semileptonic decays are used to extract the product of a form-factor normalisation at $q^2 = m^2_{\ell\nu} = 0$ and a CKM matrix element: $|V_{cd}|f^{\pi}_{+}(0)$ or $|V_{cs}|f^{K}_{+}(0)$. In each case, there is a factor parametrising strong-interaction effects due to the fact that the quarks are bound in mesons. These factors can now be calculated with good precision via lattice QCD (LQCD). One typically uses the experimental data in three ways:

- inputting CKM matrix elements to yield a measurement of decay constants or form factors; comparing these to theoretical calculations tests lattice QCD.
- taking ratios of branching fractions such that CKM matrix elements cancel; this can provide a high precision test of lattice QCD.
- inputting lattice QCD values for decay constants or form factors to yield a measurement of CKM matrix elements $|V_{cd}|$ and $|V_{cs}|$.

### 13.3.1. Theory. Author: S. Fajfer

Searching for new physics (NP) at the LHC is the most efficient way to see the effects of NP at energies larger than 1 TeV. The alternative way to search for NP is via high precision measurements at low energies. For example, measuring flavour-changing neutral-current processes are often considered to be a promising way to detect NP. However, indications of a difference between the measured branching fraction for $B \to D^{(*)}\tau\nu_\tau$ and the theoretical predictions (see *e.g.* [291]) have stimulated discussions on the presence of NP in *charged current* processes. The $c \to s\ell\nu_\ell$ transition within charm mesons offers interesting tests of the SM as well as non-perturbative QCD dynamics. Precise values of the decay constants for $D$ and $D_s$ mesons are now known from unquenched lattice QCD calculations that include the effects of dynamical up, down, strange and charm quarks [928, 929]. The shapes of the semileptonic form factors $f_{+,0}(q^2)$ for the process $D \to K\ell\nu$ over the whole physical $q^2$ region were also recently calculated using lattice QCD [929, 930]. In order to extract the $|V_{cs}|$ and $|V_{cd}|$ elements of the Cabibbo-Kobayashi-Maskawa (CKM) matrix, the theoretical predictions performed within the SM can be compared to the experimental values of the total or differential branching fractions. Alternatively, constraints on the effects of NP can be derived by fixing the value of the CKM matrix element using another independent source. The relevant NP states are usually assumed to be much heavier than the typical hadronic energy scale, in which case they can be integrated out together with the $W$ boson. The result is that NP appears as non-standard higher dimensional operators in the low energy effective description of $c \to s\ell\nu_\ell$ transitions. In general the effective Lagrangian can be written [931]:

$$\mathcal{L}_{eff} = -\frac{4G_F}{\sqrt{2}} V_{cs} \sum_{\ell=e,\mu,\tau} \sum_i c_i^{(\ell)} \mathcal{O}_i^{(\ell)} + \text{h.c.} \tag{417}$$

The usual four-fermion operator is $\mathcal{O}^{(\ell)}_{SM} = \left(\bar{s}\gamma_\mu P_L c\right)\left(\bar{\nu}_\ell \gamma^\mu P_L \ell\right)$ with the coefficient $c^{(\ell)}_{SM} = 1$. The non-SM effective operators that involve only the (pseudo)scalar quark and lepton densities and keeping only the SM neutrinos are:

$$\mathcal{O}^{(\ell)}_{L(R)} = \left(\bar{s} P_{L(R)} c\right)\left(\bar{\nu}_\ell P_R \ell\right). \tag{418}$$

These operators might be induced by integrating out the new non-SM charged scalar boson at the tree level. Such a boson can arise in a two-Higgs doublet model (THDM), *i.e.*, the extension of the SM with an additional scalar doublet [932]. In the approach of Ref. [931], the coefficients $c^{(\ell)}_{S,R(L)}$ are complex-valued and depend on the flavour of the charged lepton. The additional dependence (besides the factor of $m_\ell$) on the charged lepton's flavour is present in the THDM of the type-III [279] or in the aligned THDM [280, 933].The tensor operator $\left(\bar{s}\sigma_{\mu\nu}P_R c\right)\left(\bar{\nu}_\ell \sigma^{\mu\nu} P_R \ell\right)$ could also appear together with the (pseudo)scalar operators, after integrating out a scalar leptoquark at the tree level. Such contributions are ignored due to the lack of reliable information on the tensor form factors. First the constraints on the linear combination of the Wilson coefficients $c^{(\ell)}_{L(R)}$ from the measured branching fractions of the purely leptonic $D_s \to \ell\nu$ decay mode can be determined. The hadronic matrix element of the corresponding axial vector current is parametrised by the decay constant $f_{D_s}$ via $\langle 0|\bar{s}\gamma_\mu\gamma_5|D_s(k)\rangle = f_{D_s}\, k_\mu$. Following the procedure described in [931], the branching fraction is then modified to

$$\mathcal{B}(D_s \to \ell\nu_\ell) = \tau_{Ds}\frac{m_{Ds}}{8\pi} f^2_{Ds}\left(1 - \frac{m^2_\ell}{m^2_{Ds}}\right)^2 G^2_F \ \times (1+\delta^{(\ell)}_{em})|V_{cs}|^2 m^2_\ell \left|1 - c^{(\ell)}_P \frac{m^2_{Ds}}{(m_c+m_s)m_\ell}\right|^2, \tag{419}$$

where the pseudoscalar combination of the couplings is $c^{(\ell)}_P \equiv c^{(\ell)}_R - c^{(\ell)}_L$. In evaluating these constraints on $c_P$, the authors of Ref. [931] used the most recent theoretical value of the decay constant $f_{D_s} = 249.0(0.3)(^{+1.1}_{-1.5})$ MeV, as calculated with sub-percent precision by the Fermilab Lattice and MILC collaborations [928]. The allowed regions for the real and imaginary parts of $c_P$ are shown in Fig. 146.

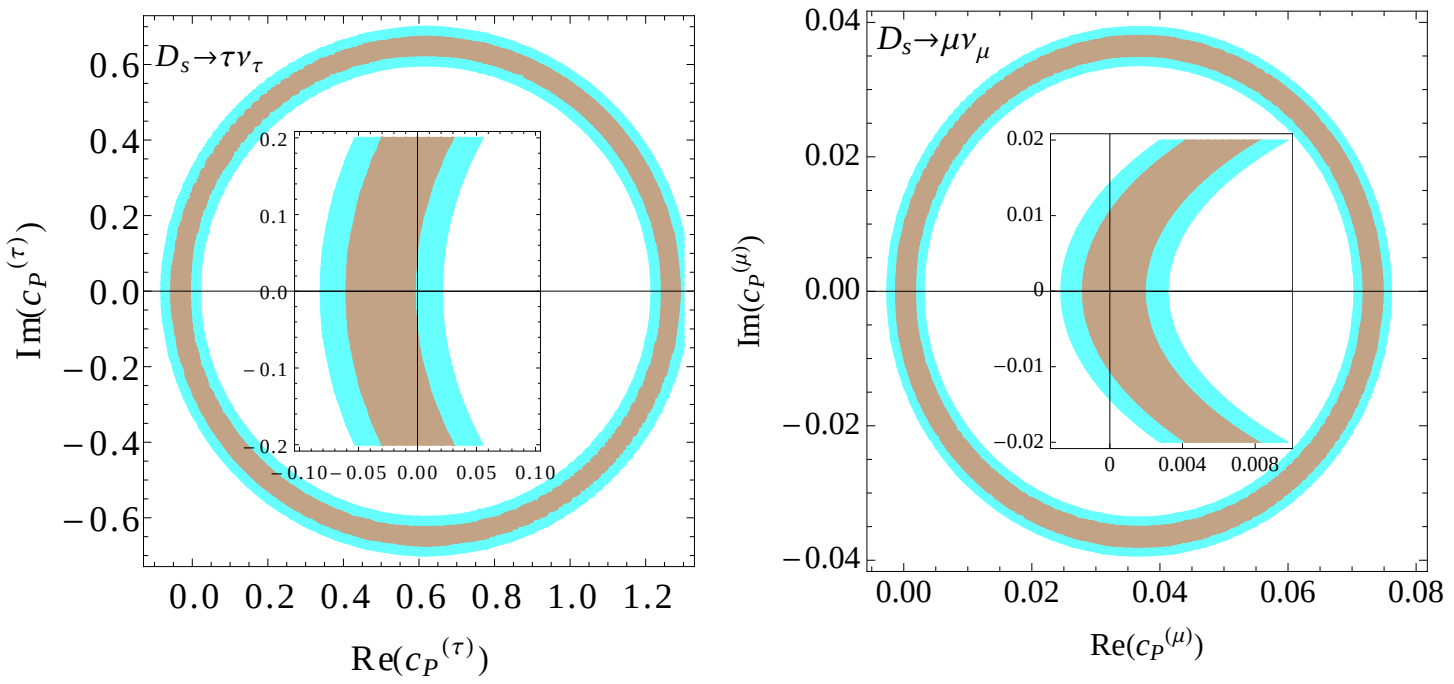

Fig. 146: Allowed regions of the effective coupling $c^{(\tau)}_P$ (left panel) and $c^{(\mu)}_P$ (right panel), extracted from the branching fraction of the decay mode $D_s \to \tau(\mu)\nu$, respectively. The 68% (95%) C.L. regions of the parameters are shown in darker (lighter) shades.

The leptonic branching fractions for $D^+_s \to \tau^+(\mu^+)\nu$ have been measured by the Belle Collaboration [934]. Belle also sets an upper limit $\mathcal{B}(D^-_s \to e^-\nu) < 1.0\times 10^{-4}$ (95% CL) [934], which leads to the constraint $|c^{(e)}_P| < 0.005$ [931]. This calculation uses the value $|V_{cs}| = 0.97317^{+0.00053}_{-0.00059}$ resulting from a global fit of the unitary CKM matrix as performed by the CKMFitter Collaboration [91].

One can consider the ratio of the branching fractions, *i.e.*, $R_{\tau/\mu} = \mathcal{B}(D_s \to \tau\nu)/\mathcal{B}(D_s \to \mu\nu)$ in order to test the lepton flavour universality of the charged current. The pseudoscalar

Wilson coefficient $c_P^{(\ell)}$ appears also in the semileptonic decays of the pseudoscalar to vector mesons, which then offer a larger number of observables than the two-body leptonic decays due to the existence of the non-trivial angular distributions as described in [931]. The information about the helicity-suppressed contribution can be extracted experimentally by comparing the decays that involve electrons and muons in the final state. The helicity-suppressed contributions [931] are subdominant; this results in the small sensitivity of $D \to K^* \ell \nu_\ell$ and $D_s \to \phi \ell \nu_\ell$ to the coefficient $c_P^{(\ell)}$ when compared to that of purely leptonic decays. Also, the knowledge of the form factors in these transitions is currently less precise. Information about the decay mode $D \to K^* \ell \nu$ is obtained from $D \to K\pi\ell\nu$ decays, in which the dominant vector intermediate state interferes with the scalar $K\pi$ amplitude and also, to a smaller extent, with higher waves [935]. The extraction of the possible NP effects from the angular analysis thus requires careful disentangling of such resonant (and also other non-resonant) contributions. Lattice calculations would provide easier access to the form factors for $D_s \to \phi\ell\nu$, in which neither of the two mesons contain light valence quarks.

In order to analyse NP effects in $D \to K^* \ell\nu$, one needs to know the behaviour of the form factors. Most of the experimental approaches to the form factors assume single pole dominance, in which the main contribution arises from the lowest pole outside the physically allowed region. An analysis of $D \to K\pi\ell\nu$ decays was performed by BaBar [936] and more recently by BESIII [937]. Both groups used the simple pole parametrisation of form factors and extracted the ratios of form factors for the $D \to K^*$ transition at a single kinematic point. In Ref. [931] the ratio of the decay widths of the longitudinally and transversally polarised $K^*$ fractions $R_{L/T}$ was considered as an observable which is sensitive to $c_P^{(\ell)}$. However, the resulting constraint is much weaker than that shown in Fig. 146.

The scalar combination of Wilson coefficients $c_S^{(\ell)} = c_R^{(\ell)} + c_L^{(\ell)}$ enters the amplitude for semileptonic $D \to K\ell\nu$ decays. In this case there are lattice evaluation of the form factors, performed by the HPQCD collaboration in Ref. [930] and there are measured values of the branching ratios [77]. These can then constrain the values of $c_S^{(\ell)}$, $\ell = e, \mu$. Using lattice QCD results [930] and the measured decay rates [77], one can derive a constraint on the Wilson coefficient $c_S^{(\mu)} \equiv c_R^{(\mu)} + c_L^{(\mu)}$, *e.g.*, $|c_S^{(e)}| < 0.2$ at 95% C.L. [931].

The most interesting observables in which to search for NP are the forward-backward asymmetry and the $CP$-violating transverse muon polarisation in decays involving muons in the final state. Deviations from the SM in these observables have not been excluded. In Ref. [931] it was found that the differential forward-backward asymmetry for low $q^2$ can be about 10%. It was also found that the ratio $R_{\mu/e}(q^2) \equiv (d\Gamma^{(\mu)}/dq^2)/(d\Gamma^{(e)}/dq^2)$ can be used to test lepton flavour universality [931]. By allowing the first generation of leptons to interact as in the SM, and NP to affect the second generation, it was found that this ratio can deviate from the SM value by (10–20)%.

*13.3.2. Experiment. Authors: J. Bennett, R. Briere, A. J. Schwartz*

*Leptonic decays $D^+ \to \ell^+\nu$.* The low backgrounds of an $e^+e^-$ experiment allow one to study purely leptonic $D^-_{(s)} \to \ell^-\bar{\nu}$ decays. Belle has measured the branching fractions for $D_s^- \to \mu^-\bar{\nu}$ and $D_s^- \to \tau^-\bar{\nu}$ [934], and inputting the value of $f_{D_s}$ as calculated from lattice QCD [140] results in the world's most precise determination of $|V_{cs}|$. The $D_s^- \to \ell^-\bar{\nu}$ event sample for Belle II will be significantly larger than that for Belle, and this will allow for a more precise determination of $|V_{cs}|$. In addition, Belle II should measure $D^- \to \mu^- \bar{\nu}$ decays, and, from the branching fraction, determine $|V_{cd}|$ with an uncertainty of $< 2\%$.

The method used by Belle to reconstruct $D_s^- \to \mu^- \bar{\nu}$ decays is as follows [934]. First, a "tag-side" $D^0$, $D^+$, or $\Lambda_c^+$ is reconstructed, nominally recoiling against the signal $D_s^-$. The decay modes used for this are listed in Table 110. In addition, tag-side $D^0$ and $D^+$ mesons can be paired with a $\pi^+$, $\pi^0$, or $\gamma$ candidate to make a tag-side $D^{*+} \to D^0\pi^+$, $D^{*+} \to D^+\pi^0$, $D^{*0} \to D^0\pi^0$, or $D^{*0} \to D^0\gamma$ candidate. The remaining pions, kaons, and protons in the event are then grouped together into what is referred to as the "fragmentation system" $X_{\rm frag}$. The particle combinations allowed for $X_{\rm frag}$ are also listed in Table 110. Because the signal decay is a $D_s^-$, $X_{\rm frag}$ must include a $K^+$ or $K_S^0$ in order to conserve strangeness. If the tag side were a $\Lambda_c^+$, then $X_{\rm frag}$ must include a $\bar{p}$ in order to conserve baryon number. After $X_{\rm frag}$ is identified, a photon recoiling against $D_{\rm tag}$ and having $p > 120$ MeV/$c$ is required. This photon is consistered to originate from $D_s^{*-} \to D_s^- \gamma$: the missing mass squared, $M^2_{\rm miss} = (P_{CM} - P_{\rm tag} - P_{X_{\rm frag}} - P_\gamma)^2$, is required to be within a narrow window centered around $M^2_{D_s}$. The event is subsequently required to have a $\mu^-$ candidate, presumably originating from $D_s^- \to \mu^- \bar{\nu}$. The signal yield is obtained by fitting the distribution of "neutrino" missing mass $M^2_\nu = (P_{CM} - P_{\rm tag} - P_{X_{\rm frag}} - P_\gamma - P_{\mu^-})^2$, which should peak at zero. The Belle signal yield, and the much larger yield expected for Belle II, are listed in Table 111.

The above method can also be used to search for $D^- \to \mu^- \bar{\nu}$ and $D^0 \to \nu\bar{\nu}$ decays [938]. In the latter case, the $D^0$ is required to originate from $D^{*+} \to D^0\pi^+$, and the daughter $\pi^+$ momentum is used when calculating the missing mass. Requiring that $M_{\rm miss}$ lie within a narrow window centred around $M_{D^0}$ results in an inclusive sample of $D^0$ decays. The Belle yield for this sample, and the expected Belle II yield, are also listed in Table 111. The Belle II yield would allow for a $7\times$ more sensitive search for $D^0 \to \nu\bar{\nu}$ (or any invisible final state) than that achieved by Belle.

Table 110: List of tag modes and $X_{\rm frag}$ used for analysis of $D_s^- \to \ell^- \nu$ decays at Belle [934].

| Tag side: | $D^0$ | $D^+$ | $\Lambda_c^+$ |
|---|---|---|---|
| Final state: | $K^-\pi^+$ | $K^-\pi^+\pi^+$ | $pK^-\pi^+$ |
| | $K^-\pi^+\pi^0$ | $K^-\pi^+\pi^+\pi^0$ | $pK^-\pi^+\pi^0$ |
| | $K^-\pi^+\pi^+\pi^-$ | $K_S^0\,\pi^+$ | $pK_S^0$ |
| | $K^-\pi^+\pi^+\pi^-\pi^0$ | $K_S^0\,\pi^+\pi^0$ | $\Lambda\pi^+$ |
| | $K_S^0\,\pi^+\pi^-$ | $K_S^0\,\pi^+\pi^+\pi^-$ | $\Lambda\pi^+\pi^0$ |
| | $K_S^0\,\pi^+\pi^-\pi^0$ | $K^+K^-\pi^+$ | $\Lambda\pi^+\pi^+\pi^-$ |
| $X_{\rm frag}$ : | $K_S^0\pi^+$ | $K_S^0$ | same as for $D^+$ tag $+\ \bar{p}$ |
| | $K_S^0\pi^+\pi^0$ | $K_S^0\,\pi^0$ | |
| | $K_S^0\pi^+\pi^+\pi^-$ | $K_S^0\,\pi^+\pi^-$ | |
| | $K^+$ | $K_S^0\,\pi^+\pi^-\pi^0$ | |
| | $K^+\,\pi^0$ | $K^+\,\pi^-$ | |
| | $K^+\,\pi^+\pi^-$ | $K^+\,\pi^-\pi^0$ | |
| | $K^+\,\pi^+\pi^-\pi^0$ | $K^+\,\pi^-\pi^+\pi^-$ | |

Table 111: Belle's $D_s^- \to \mu^-\bar{\nu}$ [934] and inclusive $D^0$ [939] signal yields, and the yields expected for Belle II. The latter are obtained by either scaling the Belle results or from MC simulation studies.

| Mode | Belle<br>(0.91, 0.92 $\mathrm{ab}^{-1}$) | Belle II<br>(50 $\mathrm{ab}^{-1}$) |
|---|---|---|
| $D_s^- \to \mu^-\bar{\nu}$ | $492 \pm 26$ | 27000 |
| $D^- \to \mu^-\bar{\nu}$ | – | 1250 |
| inclusive $D^0 \to$ anything | $(695 \pm 2) \times 10^3$ | $38 \times 10^6$ |

*Semileptonic decays $D \to h\ell^+\nu$.* Both Belle and BaBar have measured semileptonic $D$ decays. An early Belle analysis used 280 $\mathrm{fb}^{-1}$ of data to reconstruct $126 \pm 12$ ($106 \pm 12$) $\pi e\nu$ ($\pi\mu\nu$) decays with an average purity of $S/(S+B) = 79\%$ [940]. A more efficient BaBar analysis used 380 $\mathrm{fb}^{-1}$ of data to reconstruct $5303 \pm 121$ $\pi e\nu$ decays, but with more background: $S/(S+B) = 53\%$ [941]. However, the systematic error on the branching fraction for the BaBar result was in fact less than that of Belle. Scaling the BaBar result to the expected Belle II integrated luminosity, one predicts for Belle II a very large sample of $7.0 \times 10^5$ $\pi e\nu$ decays in 50 $\mathrm{ab}^{-1}$ of data.

As a feasibility study, semileptonic charm decays have been studied using the 1 $\mathrm{ab}^{-1}$ sample of $c\bar{c}$ MC. Events are reconstructed according to the reaction $e^+e^- \to c\bar{c} \to D_{\mathrm{tag}}^{0/+} D^{*-} X_{\mathrm{frag}}^{+/0}$, where $D^{*-} \to D_{\mathrm{sig}}^0 \pi^-$ (charge conjugation is assumed throughout). Finally, the $D_{\mathrm{sig}}^0$ decays to the $hl\nu$ final state, where $h = K, \pi$ and $l = e, \mu$. The $D_{\mathrm{tag}}$ can be either a $D^0$ or $D^+$ reconstructed in several decay modes. The number and charge of fragmentation particles depends on the charge of the $D_{\mathrm{tag}}$. A preliminary list of tag and fragmentation modes to be implemented is given in Table 112.

Table 112: List of tag modes and $X_{\mathrm{frag}}$ used for analysis of $D^0$ semileptonic decays at Belle II.

| Tag side: | $D^0$ | $D^+$ |
|---|---|---|
| Final state: | $K^-\pi^+$ | $K^-\pi^+\pi^+$ |
| | $K^-\pi^+\pi^0$ | $K^-\pi^+\pi^+\pi^0$ |
| | $K^-\pi^+\pi^+\pi^-$ | $K_S^0\pi^+$ |
| | $K^-\pi^+\pi^+\pi^-\pi^0$ | $K_S^0\pi^+\pi^0$ |
| | $K_S^0\pi^+\pi^-$ | $K_S^0\pi^+\pi^+\pi^-$ |
| | $K_S^0\pi^+\pi^-\pi^0$ | $K^+K^-\pi^+$ |
| $X_{\mathrm{frag}}$ : | $\pi^+$ | none |
| | $\pi^+\pi^0$ | $\pi^0$ |
| | $\pi^+\pi^+\pi^-$ | $\pi^+\pi^-$ |
| | | $\pi^+\pi^-\pi^0$ |

The details of the missing neutrino are determined using the recoil reconstruction method, as described above for leptonic $D^+ \to \ell^+\nu$ decays. For semileptonic decays at Belle II, the reconstruction proceeds in two steps. First, the signal $D^*$ is reconstructed using the recoil against the $D_{\text{tag}}X_{\text{frag}}$ system. Next, the search for semileptonic decays of the $D^0_{\text{sig}}$ is conducted by reconstructing the mass recoiling against the $D_{\text{tag}}X_{\text{frag}}\pi_s^- hl$ system, where $\pi_s$ is the slow pion from the $D^*$ decay. The four-momentum of the missing neutrino is determined by the equation

$$P_{\text{miss}} = P_{e^+} + P_{e^-} - P_{D_{\text{tag}}} - P_{X_{\text{frag}}} - P_h - P_l. \tag{420}$$

Then, the missing mass is constructed as $M_{\text{miss}} = \sqrt{P^2_{\text{miss}}}$, or, alternatively, the missing energy as $U_{\text{miss}} = E_{\text{miss}} - |\vec{p}_{\text{miss}}|$. For correctly reconstructed events, both $M_{\text{miss}}$ and $U_{\text{miss}}$ peak at zero. Due to its superior resolution, $U_{\text{miss}}$ is used in this analysis.

With relatively few selection criteria, it is possible to get a clean $U_{\text{miss}}$ spectrum. In each event, the reconstructed charged tracks must originate at the interaction point ($|z_0| < 4$ cm, $|d_0| < 2$ cm) and survive a loose cut on the track fit quality. Hadrons must satisfy a standard requirement on the particle identification likelihood $\mathcal{L}$ ($\mathcal{L}(K) > 0.50$, $\mathcal{L}(\pi) > 0.50$), while leptons must satisfy only a loose requirement ($\mathcal{L}(\mu) > 0.10$, $\mathcal{L}(e) > 0.10$). Each $K^0_S$ is reconstructed from $\pi^+\pi^-$ pairs, subjected to a vertex fit, and required to have an invariant mass within 20 MeV/c$^2$ of the nominal $K^0_S$ mass [77].

Each $D_{\text{tag}}$ candidate within a 30 MeV/c$^2$ mass window of the nominal $D$ mass is subjected to a vertex fit. The mass recoiling against the $D_{\text{tag}}$ candidate and fragmentation particles must fall within a 500 MeV/c$^2$ window of the nominal $D^*$ mass. Finally, the difference in recoil masses for the $D_{\text{tag}}X_{\text{frag}}$ and $D_{\text{tag}}X_{\text{frag}}\pi_s$ system (equivalent to the difference in invariant mass of the $D^*$ and $D_0$ candidates) must be less than 0.15 GeV/c$^2$.

Additional selection criteria are still under investigation but will possibly include restrictions on the PID likelihood ratio for leptons, lepton momenta, the number of extra tracks in the event, and the unassociated ECL energy in the event. Imposing loose restrictions on these values, one obtains the $M_{\text{miss}}$ and $U_{\text{miss}}$ distributions shown in Fig. 147. The missing mass resolution is comparable to that of the most recent Belle analysis [942]. A similar analysis of Belle II MC light quark continuum samples yields no events, indicating that the continuum background for this analysis will be small.

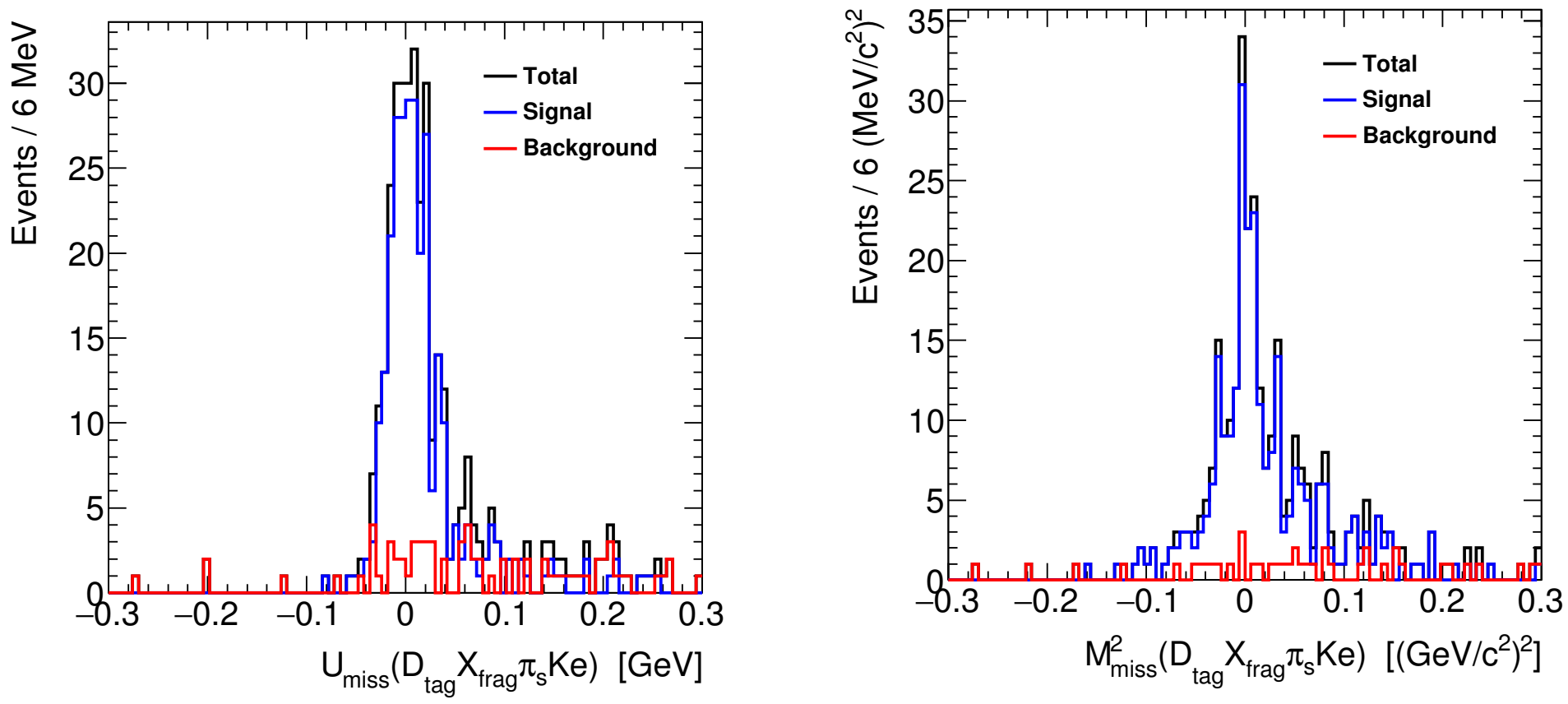

Fig. 147: $U_{\text{miss}}$ (top) and missing mass squared (bottom) for semileptonic charm decays reconstructed using a 1 ab$^{-1}$ sample of generic $c\bar{c}$ events using basf2 release-00-07-00. Only a single tag mode, $D^0_{\text{tag}} \to K^-\pi^+$ and a single fragmentation pion, is reconstructed.

*Discussion.* As branching fraction measurements of leptonic and semileptonic decays do not require measuring decay times, they can be performed with high statistics at BESIII. Thus we compare the sensitivity of Belle II to that of BESIII.

All decays of interest involve unobservable neutrinos. With charm threshold data, one has simple initial states, either $e^+e^- \to \psi(3770) \to D^0\bar{D}^0, D^+D^-$, or $e^+e^- \to D_s^{*+}D_s^- +$ c.c. near 4170 MeV. The lack of additional fragmentation particles allows one to fully reconstruct ("tag") one $D_{(s)}$ decay to a hadronic final state, and then study the other decay. The neutrino may be inferred via energy-momentum conservation, leading to rather clean signal peaks. With continuum charm data at $B$ factory energies, one produces more complex final states. Not only are there an unknown number of fragmentation particles, but the $c$ and $\bar{c}$ quarks may appear in many different pairings of charm hadrons (*e.g.*, $D^0D_s^-$, $\Lambda_c\bar{D}^0$). Belle overcomes this by fully reconstructing a tag-side $D$ or $\Lambda$ decay. To obtain enough efficiency, Belle sums over many different exclusive states. For both experiments there is sufficient phase space (and reduced helicity suppression) for the $\tau\nu$ mode to be important for the $D_s$ but not the $D$ decay.

For $D_s^+ \to \tau^+\nu, \mu^+\nu$ decays, we note that the error of 1.4% on the $D_s$ lifetime contributes a significant 0.7% to the systematic error on $f_{D_s}$. As discussed above Belle used 0.91 ab$^{-1}$ to measure $f_{D_s} = (255.5 \pm 4.2 \pm 5.1)$ MeV [934]; this has higher precision than a similar measurement using 3 fb$^{-1}$ by BESIII. While BESIII may take more data, they cannot match the factor of 50 increase in data that Belle II will have over Belle. The Belle result is systematics limited, with the largest contributions to the systematic uncertainty being due to normalisation, efficiency, and particle identification. These uncertainties are determined by studying control samples, and thus the uncertainties should be reduced with increased luminosity.

For $D^+ \to \mu^+\nu$ decays, both CLEOc (0.82 fb$^{-1}$) [943] and BESIII (2.9 fb$^{-1}$)[944] have measured $|V_{cd}|f_D$. For the latter, the statistical error on the leptonic branching fraction is 5%. While Belle has not yet published a measurement, Belle II is expected to reconstruct over 1200 of these decays, giving a statistical error of 2.8% and a resulting uncertainty on $|V_{cd}|f_D$ of half of this: 1.4%. This precision is similar to what BESIII would achieve with its planned final data set of 15 fb$^{-1}$. For both experiments, the measurement should be statistics-dominated.

For semileptonic $D \to (K/\pi)\ell\nu$ decays, those for the charged meson, $D^+$, provide identical information as that for the neutral meson, $D^0$, but are reconstructed less efficiently due to the presence of neutral hadrons $K_S, \pi^0$. This disadvantage is partially offset by the 2.5-times-longer $D^+$ lifetime, *i.e.*, the semileptonic branching fractions are 2.5 times larger. In addition, studying a mode such as $D^+ \to K^{*0}\ell^+\nu$, $K^{*0} \to K^-\pi^+$ avoids having to reconstruct a neutral hadron with a cost of reconstructing an additional track relative to $h^-\ell^+\nu$ modes. However, the production of $D^{*+}$ mesons with the subsequent decay $D^{*+} \to D^0\pi^+$ will produce larger samples of $D^0$ mesons than $D^+$ at Belle II. The yield of $D^0 \to \pi^-e^+\nu$ decays estimated for Belle II, $7.0 \times 10^5$ in 50 ab$^{-1}$ of data, can be compared to that for BESIII. The latter experiment reconstructed $6297 \pm 87$ events in 2.9 fb$^{-1}$ of data [945], implying that the final BESIII data sample of 15 fb$^{-1}$ would yield 33000 $D^0 \to \pi^-e^+\nu$ decays. This is a large sample but an order of magnitude less than that of the full Belle II data set.

### 13.4. *Rare decays*

#### 13.4.1. *Theory. Author: A. A. Petrov*

In general, rare decays of $D$ mesons are mediated by quark-level flavor-changing neutral current (FCNC) transitions $c \to u\ell^+\ell^-$ and $c \to u\gamma^*$ (followed by $\gamma^* \to \ell^+\ell^-$). Both these decays and $D^0$-$\overline{D}^0$ mixing proceed only at one loop in the SM, and these amplitudes are highly suppressed by the GIM mechanism.

*Rare decays with charged leptons.* The simplest rare decay is purely leptonic: $D^0 \to \ell^+\ell^-$. This transition has a very small SM contribution, so it can serve as a clean probe of amplitudes due to NP. Other rare decays such as $D \to \rho\gamma$ receive significant SM contributions, which are often difficult to calculate [946–949]. There exist several experimental constraints on $D^0 \to \ell_1^+\ell_2^-$ branching fractions [230, 950–952]:

$$\begin{aligned}
\mathcal{B}(D^0 \to \mu^+\mu^-) &< 7.6 \times 10^{-9}, \\
\mathcal{B}(D^0 \to e^+e^-) &< 7.9 \times 10^{-8}, \\
\mathcal{B}(D^0 \to \mu^\pm e^\mp) &< 1.3 \times 10^{-8}.
\end{aligned} \tag{421}$$

Theoretically, all NP contributions to $c \to u\ell^+\ell^-$ transitions (and also to $D^0$-$\overline{D}^0$ mixing) can be parameterised in terms of an effective Hamiltonian:

$$\mathcal{H}_{NP}^{\rm rare} = \sum_{i=1}^{10} \frac{\widetilde{\rm C}_i(\mu)}{\Lambda^2}\, \widetilde{Q}_i, \tag{422}$$

where $\widetilde{\rm C}_i$ are Wilson coefficients, $\widetilde{Q}_i$ are the effective operators, and $\Lambda$ represents the energy scale of NP interactions that generate $\widetilde{Q}_i$'s. There are only ten of these operators with canonical dimension six:

$$\begin{aligned}
\widetilde{Q}_1 &= (\overline{\ell}_L\gamma_\mu\ell_L)(\overline{u}_L\gamma^\mu c_L)\ , \\
\widetilde{Q}_2 &= (\overline{\ell}_L\gamma_\mu\ell_L)(\overline{u}_R\gamma^\mu c_R)\ , \\
\widetilde{Q}_3 &= (\overline{\ell}_L\ell_R)\ (\overline{u}_R c_L)\ , \\
\widetilde{Q}_4 &= (\overline{\ell}_R\ell_L)(\overline{u}_R c_L)\ , \\
\widetilde{Q}_5 &= (\overline{\ell}_R\sigma_{\mu\nu}\ell_L)(\overline{u}_R\sigma^{\mu\nu} c_L)\ ,
\end{aligned} \tag{423}$$

and five additional operators $\widetilde{Q}_6, \ldots, \widetilde{Q}_{10}$ obtained from those in Eq. (423) by interchanging $L \leftrightarrow R$, *e.g.* $\widetilde{Q}_6 = (\overline{\ell}_R\gamma_\mu\ell_R)(\overline{u}_R\gamma^\mu c_R)$, $\widetilde{Q}_7 = (\alpha/4)(\overline{\ell}_R\gamma_\mu\ell_R)(\overline{u}_L\gamma^\mu c_L)$, etc.

The Hamiltonian of Eq. (422) is quite general, and thus it also contains the SM contribution usually denoted by the operators $Q_9 = (\alpha/4)(\widetilde{Q}_1 + \widetilde{Q}_7)$ and $Q_{10} = (\alpha/4)(\widetilde{Q}_7 - \widetilde{Q}_1)$ (together with a substitution $\Lambda \to \sqrt{G_F^{-1}}$). It is worth noting that matrix elements of several operators or their linear combinations vanish in the calculation of $\mathcal{B}(D^0 \to \ell^+\ell^-)$: $\langle \ell^+\ell^-|\widetilde{Q}_5|D^0\rangle = \langle \ell^+\ell^-|\widetilde{Q}_{10}|D^0\rangle = 0$ (identically), $\langle \ell^+\ell^-|Q_9|D^0\rangle \equiv (\alpha/4)\langle \ell^+\ell^-|(\widetilde{Q}_1 + \widetilde{Q}_7)|D^0\rangle = 0$ (vector current conservation), etc. The most general $D^0 \to \ell^+\ell^-$ decay amplitude can be written

$$\mathcal{M}(D^0 \to \ell^+\ell^-) = \overline{u}(p_-, s_-)\,[\ A + \gamma_5 B\ ]\, v(p_+, s_+)\,. \tag{424}$$

Any NP contribution described by the operators of Eq. (422) gives for the amplitudes $A$ and $B$:

$$|A| = \frac{f_D M_D^2}{4\Lambda^2 m_c}\left[\widetilde{C}_{3-8}+\widetilde{C}_{4-9}\right] \tag{425}$$

$$|B| = \frac{f_D}{4\Lambda^2}\left[2m_\ell\left(\widetilde{C}_{1-2}+\widetilde{C}_{6-7}\right)+\frac{M_D^2}{m_c}\left(\widetilde{C}_{4-3}+\widetilde{C}_{9-8}\right)\right], \tag{426}$$

with $\widetilde{C}_{i-k} \equiv \widetilde{C}_i - \widetilde{C}_k$. The amplitude of Eq. (424) results in the following branching fractions for the lepton flavour-diagonal and off-diagonal decays:

$$\mathcal{B}(D^0 \to \ell^+\ell^-) = \frac{M_D}{8\pi\Gamma_D}\sqrt{1-\frac{4m_\ell^2}{M_D^2}} \times \left[\left(1-\frac{4m_\ell^2}{M_D^2}\right)|A|^2+|B|^2\right] \tag{427}$$

$$\mathcal{B}(D^0 \to \mu^+ e^-) = \frac{M_D}{8\pi\Gamma_D}\left(1-\frac{m_\mu^2}{M_D^2}\right)^2 \times \left[|A|^2+|B|^2\right]. \tag{428}$$

In the latter expression, the electron mass is safely neglected. Any NP model that contributes to $D^0 \to \ell^+\ell^-$ can be constrained by bounds on the Wilson coefficients appearing in Eqs. (425) and (426). We note that, because of helicity suppression, studies of $D^0 \to e^+e^-$ (and consequently analyses of lepton universality using this channel) are experimentally challenging. Experimental limits on $\mathcal{B}(D^0 \to \mu^+e^-)$ give constraints on lepton-flavour-violating interactions via Eq. (428). Similar limits can also be obtained from two-body charmed quarkonium decays [953].

Table 113: Predictions for $D^0 \to \mu^+\mu^-$ branching fraction from correlations of rare decays and $D^0$-$\overline{D}^0$ mixing for $x_D \sim 1\%$ (from [954]). Notice that experimental constraints are beginning to probe charm sector of R-parity violating SUSY models.

| Model | $\mathcal{B}(D^0 \to \mu^+\mu^-)$ |
|---|---|
| Stand. Model (LD) | $\sim$ several $\times 10^{-13}$ |
| $Q = +2/3$ Singlet | $4.3 \times 10^{-11}$ |
| $Q = -1/3$ Singlet | $1 \times 10^{-11}$ $(m_S/500\ \text{GeV})^2$ |
| 4th Family | $1 \times 10^{-11}$ $(m_S/500\ \text{GeV})^2$ |
| $Z'$ Model | $2.4 \times 10^{-12}/(M_{Z'}(\text{TeV}))^2$ |
| Family Symmetry | $0.7 \times 10^{-18}$ (Case A) |
| RPV-SUSY | $4.8 \times 10^{-9}$ $(300\ \text{GeV}/m_{\tilde{d}_k})^2$ |
| Experiment | $\leq 7.6 \times 10^{-9}$ |

In studying NP contributions to rare decays in charm, it can be advantageous to study *correlations* of various processes, for example $D^0$-$\overline{D}^0$ mixing and rare decays [954]. In general, one cannot predict the rare decay rate by knowing just the mixing rate, even if both $x_D$ and $\mathcal{B}(D^0 \to \ell^+\ell^-)$ are dominated by a single operator contribution. It is, however, possible to do so for a restricted subset of NP models [954]; these results are presented in Table 113.

*Rare charm decays with missing energy.* High-luminosity $e^+e^-$ flavour factories such as Belle II provide a perfect opportunity to search for rare processes that require high purity of the final states. In particular, searches for $D$-decays to final states that contain neutrinos, such as $D \to \pi(\rho)\nu\overline{\nu}$, are possible at those machines due to the fact that pairs of $D$-mesons are produced in a charge-correlated state. The SM predicts extremely small branching fractions for $D$-decay processes with neutrinos in the final state, *i.e.* $\mathcal{B}(D^0 \to \nu\overline{\nu}) \simeq 1 \times 10^{-30}$, and $\mathcal{B}(D^0 \to \nu\overline{\nu}\gamma) \simeq 3 \times 10^{-14}$ [955]. Thus, any detection of decays of $D$ states into channels with missing energy in the current round of experiments would indicate NP. It is important to note that these NP models could be substantially different from models described in previous sections: experimentally, it is impossible to say if the missing energy signature were generated by a neutrino or by some other weakly-interacting particle.

Recently, a variety of models with light ($\sim$ MeV) dark matter (DM) particles have been proposed to explain the null results of experiments indirectly searching for dark matter (see, *e.g.*, Refs. [956, 957]). Such models predict couplings between quarks and DM particles that can be described using effective field theory (EFT) methods [958]. These models can be tested at $e^+e^-$ flavour factories by studying $D$ (or $B$) mesons decaying into a pair of light dark matter particles or a pair of DM particles and a photon. The latter process eliminates helicity suppression of the final state [955]. It is conceivable that searches for light DM in heavy meson decays could even be more sensitive than direct detection and other experiments, as DM couplings to heavy quarks could be enhanced (*e.g.*, the "Higgs portal" model of Ref. [956]).

Branching fractions for the heavy meson states decaying into $\chi_s\overline{\chi}_s$ and $\chi_s\overline{\chi}_s\gamma$, where $\chi_s$ is a DM particle of spin $s$, can be calculated in the EFT framework. Since production of scalar $\chi_0$ states avoids helicity suppression, these are discussed here. For the cases $s = 1/2$ and $s = 1$, see Ref. [955].

A generic effective Hamiltonian for scalar DM interactions has the simple form

$$\mathcal{H}_{\text{eff}} = 2\sum_i \frac{C_i}{\Lambda^2} O_i, \tag{429}$$

where $\Lambda$ is the energy scale associated with the particle(s) mediating interactions between the SM and DM fields, and $C_i$ are the Wilson coefficients. The effective operators $O_i$ are

$$\begin{aligned}
O_1 &= m_c(\overline{u}_R c_L)(\chi_0^*\chi_0), \\
O_2 &= m_c(\overline{u}_L c_R)(\chi_0^*\chi_0), \\
O_3 &= (\overline{u}_L\gamma^\mu c_L)(\chi_0^* \overleftrightarrow{\partial}_\mu \chi_0), \\
O_4 &= (\overline{u}_R\gamma^\mu c_R)(\chi_0^* \overleftrightarrow{\partial}_\mu \chi_0),
\end{aligned} \tag{430}$$

where $\overleftrightarrow{\partial} = (\overrightarrow{\partial} - \overleftarrow{\partial})/2$ and the DM *anti-particle* $\overline{\chi}_0$ may or may not coincide with $\chi_0$. The branching fraction for the two- body decay $D^0 \to \chi_0\chi_0$ is

$$\mathcal{B}(D^0 \to \chi_0\chi_0) = \frac{(C_1 - C_2)^2}{4\pi M_D \Gamma_{D^0}} \left[\frac{f_D M_D^2 m_c}{\Lambda^2(m_c + m_q)}\right]^2 \times \sqrt{1 - 4x_\chi^2}\,, \tag{431}$$

where $x_\chi = m_\chi/M_{D^0}$ is a rescaled DM mass. This rate is not helicity-suppressed, so it could allow one to study DM properties at an $e^+e^-$ flavour factory.

Using the formalism above, the photon energy distribution and the decay width of the radiative transition $D^0 \to \chi_0\, \chi_0\, \gamma$ can be calculated:

$$\frac{d\Gamma}{dE_\gamma}(D^0 \to \chi_0\chi_0\gamma) = \frac{f_D^2 \alpha C_3 C_4}{3\Lambda^4}\left(\frac{F_D}{4\pi}\right)^2 \times \frac{2M_D^2 E_\gamma (M_D(1-4x_\chi^2) - 2E_\gamma)^{3/2}}{\sqrt{M_D - 2E_\gamma}} \tag{432}$$

$$\mathcal{B}(D^0 \to \chi_0\chi_0\gamma) = \frac{f_D^2 \alpha C_3 C_4 M_D^5}{6\Lambda^4 \Gamma_{D^0}}\left(\frac{F_D}{4\pi}\right)^2 \times \left(\frac{1}{6}\sqrt{1-4x_\chi^2}(1-16x_\chi^2 - 12x_\chi^4) - 12x_\chi^4 \log\frac{2x_\chi}{1+\sqrt{1-4x_\chi^2}}\right). \tag{433}$$

We observe that Eqs. (432) and (433) are independent of $C_{1,2}$; this is due to the fact that $D \to \gamma$ form factors of scalar and pseudoscalar currents vanish. In this manner, studies of $D^0 \to$ (missing energy) and $D^0 \to (\gamma + \text{missing energy})$ processes probe complementary operators in the effective Hamiltonian of Eq. (429). Similar conclusions hold for decays of $B$ mesons into final states with missing energy.

*13.4.2. Experiment.*

*Radiative Modes $D \to V\gamma$. Author: T. Nanut*

Radiative decays $D^0 \to V\gamma$ are a promising probe of NP, as theoretical studies [959, 960] predict that NP contributions can enhance the time-integrated $CP$ asymmetry

$$A_{CP} \equiv \frac{\Gamma(D^0 \to f) - \Gamma(\overline{D}^0 \to \bar{f})}{\Gamma(D^0 \to f) + \Gamma(\overline{D}^0 \to \bar{f})} \tag{434}$$

up to an order of magnitude relative to the SM expectation, which is $\sim 10^{-3}$. As these decays are dominated by long distance contributions, measurement of their branching fractions can test non-perturbative QCD calculations [961, 962].

Belle has recently made the first measurements of $A_{CP}$ in $D^0 \to V\gamma$ decays, where $V = \phi, \overline{K}^{*0}$, or $\rho^0$. The results are

$$\begin{aligned} A_{CP}(D^0 \to \rho^0\gamma) &= +0.056 \pm 0.152 \pm 0.006 \\ A_{CP}(D^0 \to \phi\gamma) &= -0.094 \pm 0.066 \pm 0.001 \\ A_{CP}(D^0 \to \overline{K}^{*0}\gamma) &= -0.003 \pm 0.020 \pm 0.000\,, \end{aligned}$$

where the first error is statistical and the second is systematic. These results are consistent with no $CP$ violation. The dominant error is by far the statistical one, which implies that the precision will be significantly improved at Belle II, with its much larger data set.

The extraction of signal and the corresponding statistical uncertainty depend heavily on backgrounds. The dominant background arises from $D^0 \to X\pi^0$ decays in which one of the photons from the subsequent $\pi^0 \to \gamma\gamma$ decay is undetected. For example, $D^0 \to \phi\pi^0$ is background to $D^0 \to \phi\gamma$; $D^0 \to K^-\rho^+$, $\rho^+ \to \pi^+\pi^0$ is background to $D^0 \to \overline{K}^{*0}\gamma$; and $D^0 \to \rho^+\pi^-$, $\rho^+ \to \pi^+\pi^0$ is background to $D^0 \to \rho^0\gamma$. As these backgrounds have the same charged final state particles and only one photon is missed in the reconstruction, the distribution of the reconstructed $D^0$ mass is shifted towards lower values but still overlaps with the signal peak, as shown in Fig. 148. Since there are many such decays of this type and their branching fractions can exceed that of the signal by several orders of magnitude, it is crucial to suppress this type of background as much as possible.

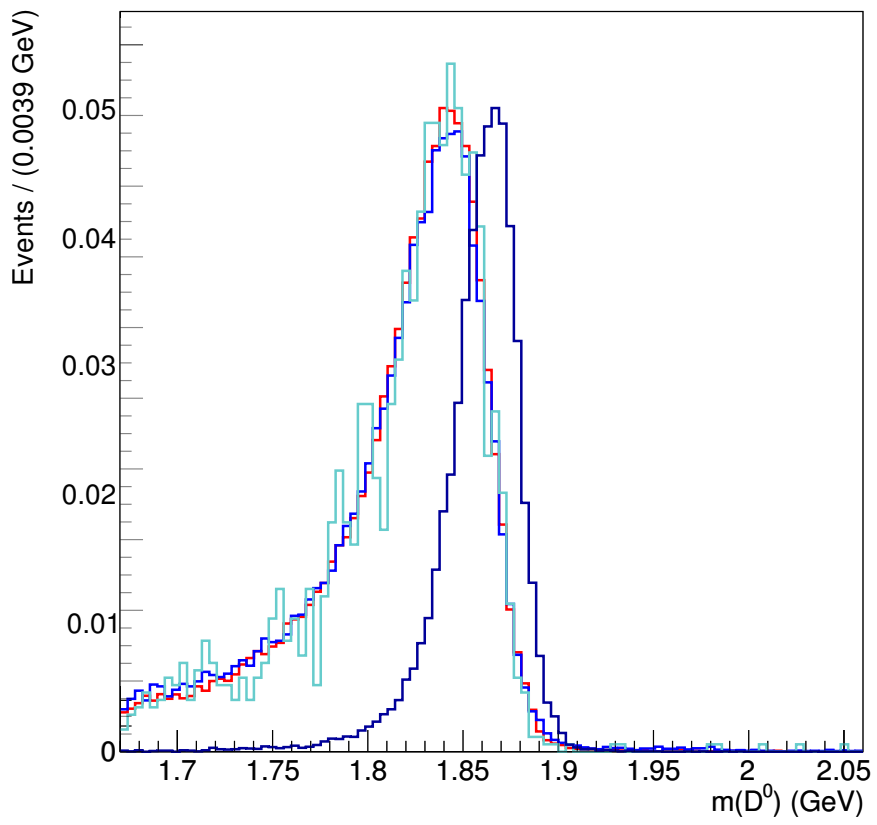

Fig. 148: Comparison of normalised distributions for $m(D^0)$ of signal (right-most distribution) and various $\pi^0$-type backgrounds, for $D^0 \to \overline{K}^{*0}\gamma$ decays at Belle.

The separation power between signal and $\pi^0$ background, which is reflected in the statistical error, is governed by the $D^0$ mass resolution of signal and background (*i.e.*, the extent of overlap of the peaks), and the signal-to-background ratio. We study these two aspects using Monte Carlo (MC) simulation. As the three signal decays studied by Belle share the same kinematics and have similar backgrounds, we investigate the Belle II sensitivity to only one signal mode and one $\pi^0$-type background: $D^0 \to \phi\gamma$ signal and $D^0 \to \phi\pi^0$ background. We generate Belle II MC samples of both signal and background events and reconstruct $D^0 \to \phi\gamma$ candidates in both.

To reduce the substantial $\pi^0$ background for this type of analysis, Belle developed a dedicated $\pi^0$ veto. This veto employs a neural network utilising two mass veto variables. Each variable is obtained by pairing the signal candidate photon with all other photons in the event whose energy exceeds a specific minimum value. The di-photon mass combination that lies closest to the mass of the $\pi^0$ is recorded and assigned to the signal candidate photon. To reject background with minimal signal loss, the energy of the second photon is required to be $>75$ MeV for the first veto variable and $>30$ MeV for the second. This minimum energy requirement is effective as there is a high multiplicity of soft photons, which increases the chance that a random combination of photons will have a diphoton mass close to that of the $\pi^0$. This is demonstrated in Figure 149, which shows that, even for MC signal events in which there is *no* $\pi^0$ background, there is a peak in the diphoton mass distribution near $M(\pi^0)$ when the energy requirement on the second photon is low. This false peak can cause a signal inefficiency.

The performance of the $\pi^0$ veto in Belle II has been studied using MC simulation, and the results are shown in Fig. 150. This figure plots $D^0 \to \phi\pi^0$ background rejection as a function of $D^0 \to \phi\gamma$ signal efficiency for both Belle and Belle II. Comparing the two curves indicates that the performance is very similar; in fact the Belle II results indicate slightly higher background rejection for a given signal efficiency.

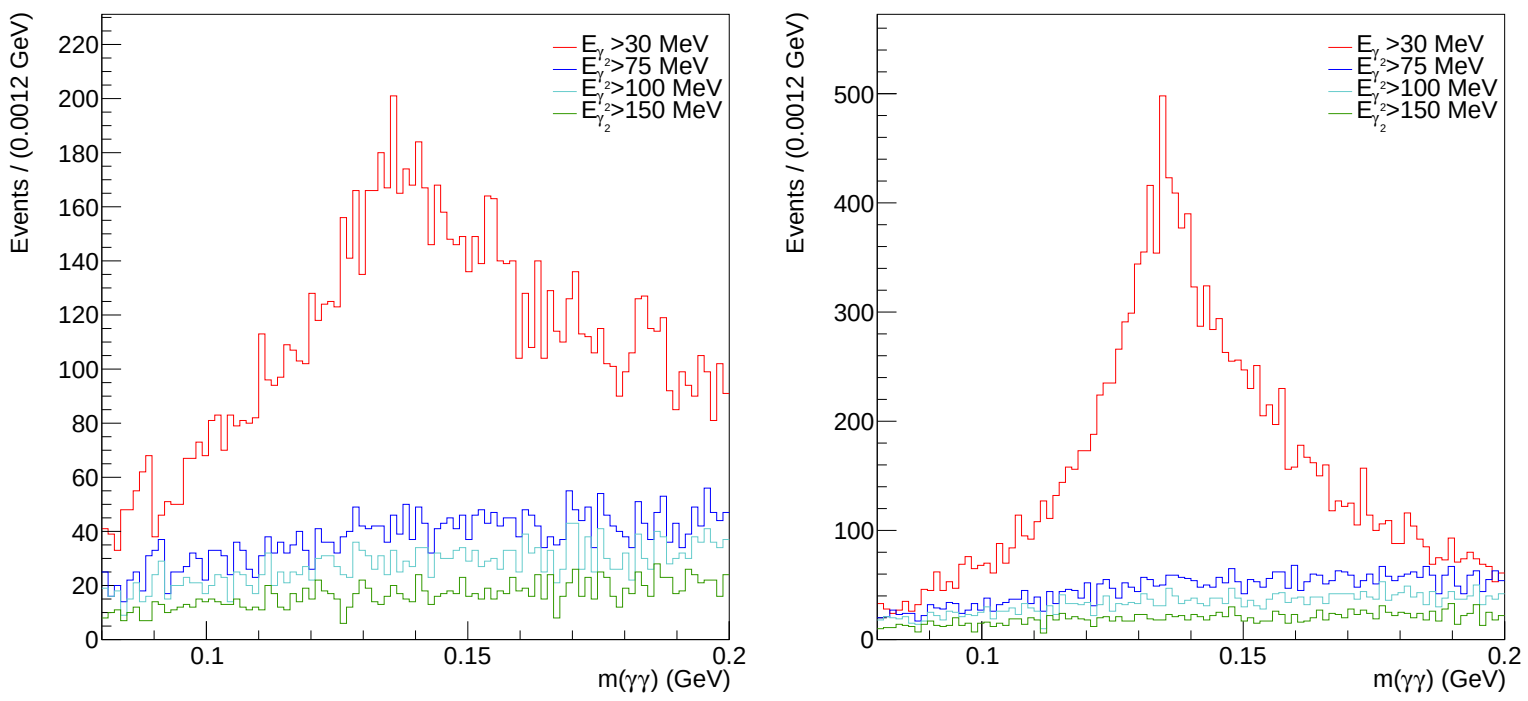

Fig. 149: Comparison between the diphoton invariant mass distributions with different requirements on the energy of the second photon for $D^0 \to \phi\gamma$ signal MC events, for Belle (left) and Belle II (right).

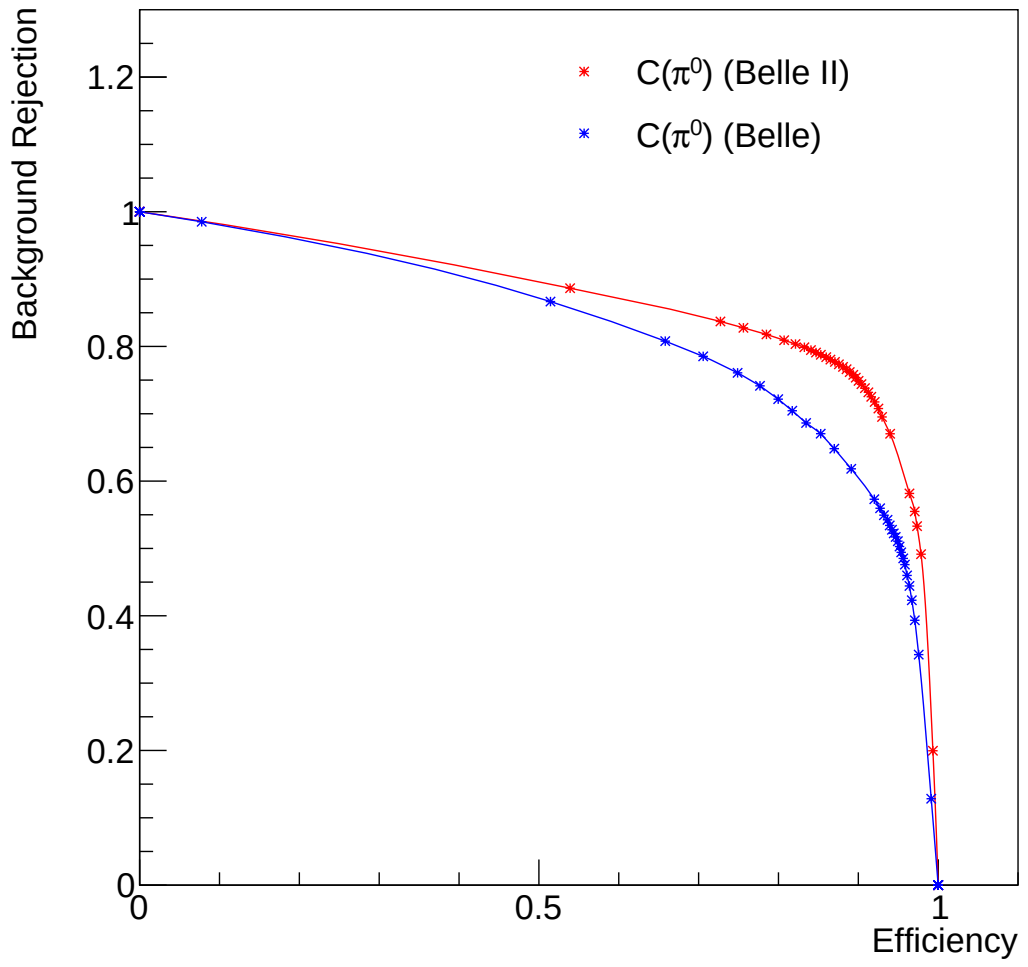

Fig. 150: Performance of the $\pi^0$ veto for Belle and Belle II: $D^0 \to \phi\pi^0$ background rejection as a function of $D^0 \to \phi\gamma$ signal efficiency.

The variables that are used in the fit to measure the signal yield are the reconstructed $D^0$ mass and the cosine of the helicity angle, which is the angle between the $D^0$ and daughter particles of the vector meson in the rest frame of the vector meson. The comparison of $D^0$ mass distributions between Belle and Belle II MC samples for both signal and background is shown in Fig. 151. The fitted widths for the Belle and Belle II signal samples are $0.0122 \pm 0.0001$ GeV and $0.0164 \pm 0.0002$ GeV, respectively. For the background samples, the widths are $0.0162 \pm 0.0004$ GeV and $0.0194 \pm 0.0003$ GeV, respectively. The Belle II resolution is comparable to, but slightly worse than, that of Belle. The means of the Belle and Belle II mass distributions are very similar. The decrease in the resolution of Belle II is attributed to slightly worse energy resolution of the electromagnetic calorimeter (ECL). However, the ECL reconstruction software is still under development, and the Belle II resolution is expected to improve, eventually reaching the level of that of Belle. Figure 152 shows analogous plots for the second fit variable, $\cos\theta_H$. The distributions for Belle and Belle II samples match well.

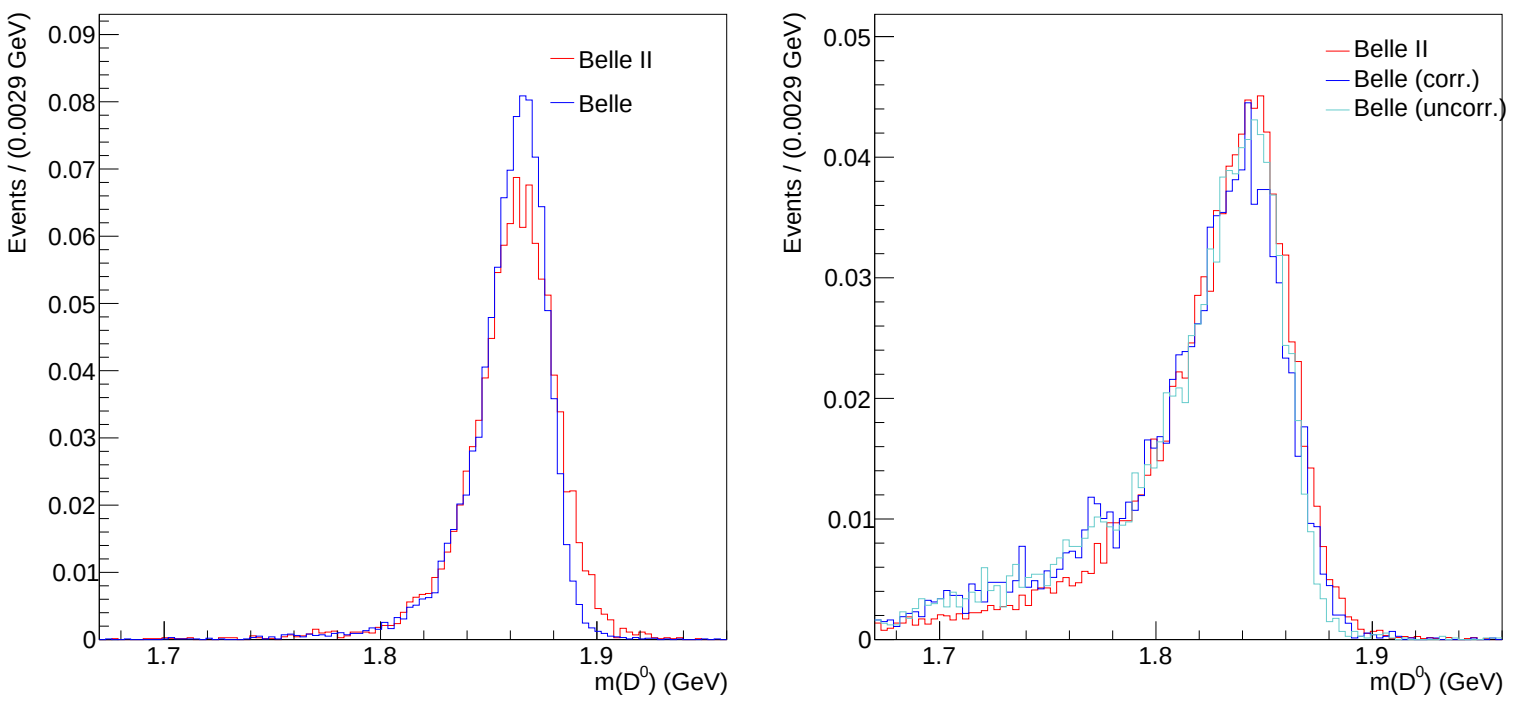

Fig. 151: Comparison of $m(D^0)$ distributions between Belle and Belle II for signal $D^0 \to \phi\gamma$ decays (left), and for background events reconstructed as $D^0 \to \phi\gamma$ (right), from MC simulation. For Belle, a corrected distribution is also shown; this distribution was calibrated to match that of the data.

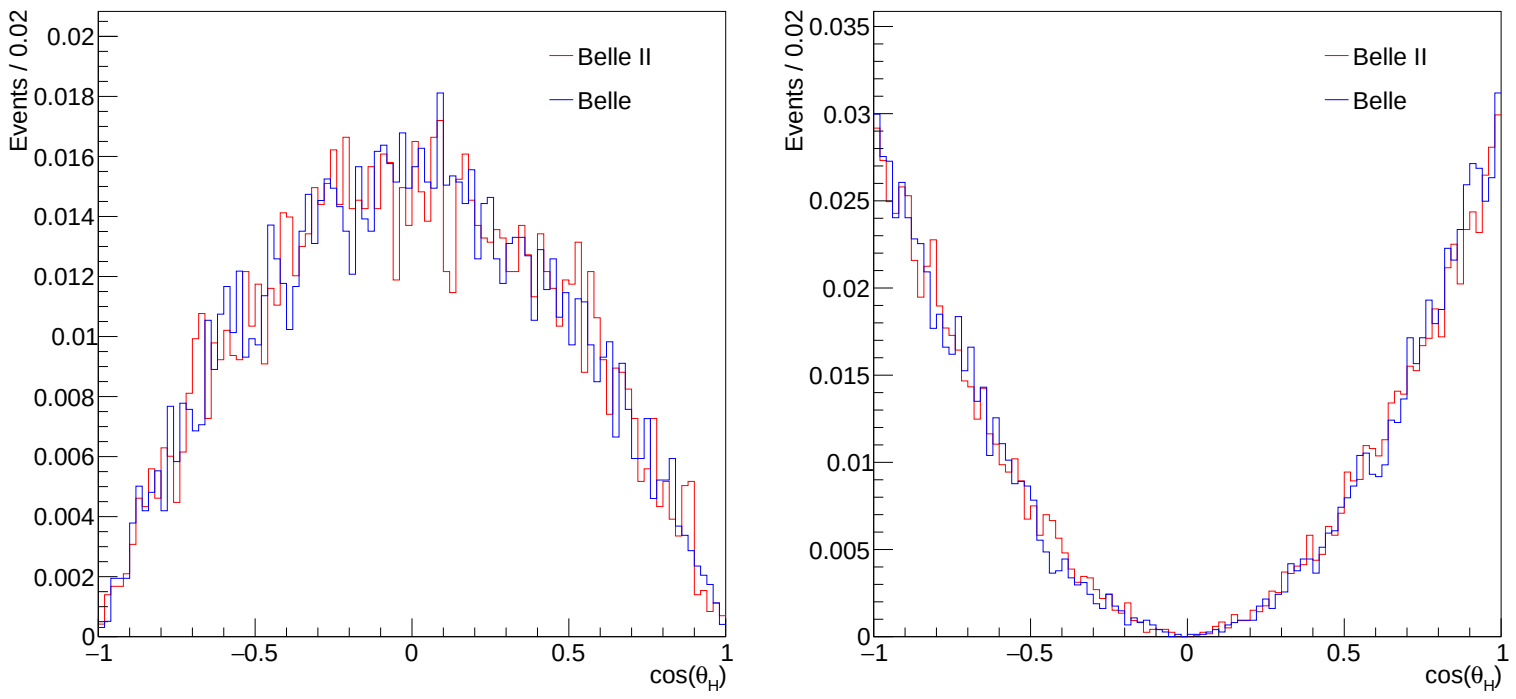

Fig. 152: Comparison of $\cos\theta_H$ distributions between Belle and Belle II for signal $D^0 \to \phi\gamma$ decays (left), and for background events reconstructed as $D^0 \to \phi\gamma$ (right), from MC simulation.

As the resolution of both fitted variables is very similar for Belle and Belle II, and the performance of the $\pi^0$ veto is similar, we conclude that the ratio of signal to background will be similar for the two experiments. Thus we estimate the Belle II sensitivity for $A_{CP}$ of $D^0 \to V\gamma$ decays by scaling (reducing) the Belle statistical uncertainty by the ratio of integrated luminosities. The resulting sensitivities for 5, 15, and 50 ab$^{-1}$ of data are listed in Table 114. The table shows that the statistical error should be reduced to the level of 1-2% for the full Belle II data set. As these $A_{CP}$ measurements are relative, *i.e.*, $A_{CP}(\phi\gamma)$ is measured relative to $A_{CP}(D^0 \to K^+K^-)$ (which has a similar final state), most systematic uncertainties should cancel. Thus the overall systematic error for Belle II should be similar to that for Belle, and the statistical error will remain the dominant one. We conclude that the final results for Belle II will provide an order of magnitude greater sensitivity to NP than that achieved by Belle.

## *13.5. Charm mixing*

### *13.5.1. CP Violation Theory. Author: A. L. Kagan*

Table 114: $A_{CP}$ results of the Belle study and extrapolation of the statistical uncertainty to Belle II, for different values of integrated luminosity.

| | Int. luminosity | $A_{CP}(D^0 \to \rho^0\gamma)$ | | |
|---|---|---|---|---|
| Belle result | 1 ab$^{-1}$ | +0.056 | ±0.152 | ±0.006 |
| | 5 ab$^{-1}$ | | ±0.07 | |
| Belle II statistical error | 15 ab$^{-1}$ | | ±0.04 | |
| | 50 ab$^{-1}$ | | ±0.02 | |
| | | $A_{CP}(D^0 \to \phi\gamma)$ | | |
| Belle result | 1 ab$^{-1}$ | −0.094 | ±0.066 | ±0.001 |
| | 5 ab$^{-1}$ | | ±0.03 | |
| Belle II statistical error | 15 ab$^{-1}$ | | ±0.02 | |
| | 50 ab$^{-1}$ | | ±0.01 | |
| | | $A_{CP}(D^0 \to \overline{K}^{*0}\gamma)$ | | |
| Belle result | 1 ab$^{-1}$ | −0.003 | ±0.020 | ±0.000 |
| | 5 ab$^{-1}$ | | ±0.01 | |
| Belle II statistical error | 15 ab$^{-1}$ | | ±0.005 | |
| | 50 ab$^{-1}$ | | ±0.003 | |

In the SM, $CP$ violation in mixing enters at $O(|V_{cb}V_{ub}/V_{cs}V_{us}|) \sim 10^{-3}$. What is the resulting theoretical uncertainty on the indirect CP violation observables? How large is the current window for New Physics (NP)? What is an appropriate parameterisation for indirect CP violating effects, given the expected sensitivity in the LHCb/Belle-II era? These points are addressed below, based on work to appear in [963] (also see [964, 965]).

We begin with an introduction to the formalism for treating CP violation in mixing. The transition amplitudes for $D^0$-$\overline{D}^0$ mixing are written as

$$\langle D^0|H|\overline{D}^0\rangle \;=\; M_{12} - \frac{i}{2}\Gamma_{12}\,, \tag{435}$$

where $M_{12}$ is the dispersive mixing amplitude. In the SM it is dominated by long-distance contributions of off-shell intermediate states. A significant short distance effect would be due to new physics (NP). $\Gamma_{12}$ is the absorptive mixing amplitude, and is due to long-distance contributions of on-shell intermediate states. The $D$ meson mass eigenstates, obtained by diagonalizing the $2\times 2$ Hamiltonian $H = M - i\Gamma/2$, are $|D_{1,2}\rangle = p|D^0\rangle \pm q|\overline{D}^0\rangle$. The differences between their eigenvalues are parameterized as $x \equiv (m_2 - m_1)/\Gamma$ and $y \equiv (\Gamma_2 - \Gamma_1)/2\Gamma$. The subscripts label the masses and widths of the two mass eigenstates; by convention the "2" state usually corresponds to the CP-even state in the absence of CP violation. The parameters $x, y$ give rise to $D^0$-$\overline{D}^0$ mixing and can be measured.

We define the following three underlying theoretical parameters: $x_{12}$, $y_{12}$, $\phi_{12}$. The first two are CP-conserving:

$$x_{12} \;\equiv\; \frac{2|M_{12}|}{\Gamma} \qquad \text{and} \qquad y_{12} \;\equiv\; \frac{|\Gamma_{12}|}{\Gamma}\,, \tag{436}$$

while the last is a phase difference that gives rise to CP violation in mixing:

$$\phi_{12} \;\equiv\; \arg\left(\frac{M_{12}}{\Gamma_{12}}\right)\,. \tag{437}$$

It can be shown that $x \approx x_{12}$ and $y \approx y_{12}$ up to small corrections quadratic in the amount of CP violation. CP violation in mixing occurs due to subleading $O(V_{cb}V_{ub})$ suppressed SM decay amplitudes (containing the CKM phase $\gamma$), and possible NP short distance mixing amplitudes and decay amplitudes containing new weak phases.

There are in fact two types of CP violation due to mixing; both are referred to as "indirect" CP violation. The first is CP violation in the mixing ("CPVMIX"), which arises when $\phi_{12} \neq 0$, and is due to interference between the dispersive and absorptive mixing amplitudes. CPVMIX can be directly measured via the semileptonic CP asymmetry

$$A_{\rm SL} \equiv \frac{\Gamma(D^0 \to K^+\ell^-\nu) - \Gamma(\overline{D}^0 \to K^-\ell^+\nu)}{\Gamma(D^0 \to K^+\ell^-\nu) + \Gamma(\overline{D}^0 \to K^-\ell^+\nu)} = \frac{|q/p|^4 - 1}{|q/p|^4 + 1} = \frac{2x_{12}y_{12}}{x_{12}^2 + y_{12}^2} \sin\phi_{12} \,. \quad (438)$$

The second type of CP violation is due to interference between a direct decay amplitude and a "mixed" amplitude followed by decay ("CPVINT"), *i.e.*, interference between $D^0 \to f$ and $D^0 \to \overline{D}^0 \to f$. For decays to a CP eigenstate final state, there are two CPVINT observables (introduced in [966]),

$$\lambda_f^M \equiv \frac{M_{12}}{|M_{12}|}\frac{A_f}{\overline{A}_f} = \eta_f^{CP}\left|\frac{A_f}{\overline{A}_f}\right| e^{i\phi_f^M}\,, \quad \lambda_f^\Gamma \equiv \frac{\Gamma_{12}}{|\Gamma_{12}|}\frac{A_f}{\overline{A}_f} = \eta_f^{CP}\left|\frac{A_f}{\overline{A}_f}\right| e^{i\phi_f^\Gamma}\,, \quad (439)$$

that parameterise the interference for a dispersive mixing amplitude and an absorptive mixing amplitude, respectively. Here, $\phi_f^M$ and $\phi_f^\Gamma$ are the corresponding weak phases, $A_f = \langle f|H|D^0\rangle$ and $\overline{A}_f = \langle f|H|\overline{D}^0\rangle$ are the decay amplitudes, and $\eta_f^{CP} = +(-)$ for CP even (odd) final states. For decays to a non-CP-eigenstate final state $f$, and its CP-conjugate $\bar{f}$, there are two pairs of observables,

$$\lambda_f^M \equiv \frac{M_{12}}{|M_{12}|}\frac{A_f}{\overline{A}_f} = \left|\frac{A_f}{\overline{A}_f}\right| e^{i(\phi_f^M - \Delta_f)}\,, \quad \lambda_f^\Gamma \equiv \frac{\Gamma_{12}}{|\Gamma_{12}|}\frac{A_f}{\overline{A}_f} = \left|\frac{A_f}{\overline{A}_f}\right| e^{i(\phi_f^\Gamma - \Delta_f)}\,, \quad (440)$$

and $\lambda_{\bar{f}}^M$, $\lambda_{\bar{f}}^\Gamma$, obtained by substituting $f \to \bar{f}$ and $\Delta_f \to -\Delta_f$ in (440), where $\Delta_f$ is the strong phase difference between the decay amplitudes. Note that the absorptive and dispersive phases are related to the pure mixing phase $\phi_{12}$ as

$$\phi_{12} = \phi_f^M - \phi_f^\Gamma\,. \quad (441)$$

In general, the weak phases $\phi_f^M$ and $\phi_f^\Gamma$ are final-state specific due to "non-universal" weak and strong phases entering SM CKM-suppressed contributions and possible NP contributions to the subleading decay amplitudes. However, in the case of the Cabibbo-Favored/Doubly-Cabibbo-Suppressed (CF/DCS) decays in the SM, these phases are universal. More generally, NP phases entering the CF/DCS amplitudes would need to be very exotic in origin, or tuned, to evade the $\epsilon_K$ constraint [967]. Thus it is a well-motivated assumption to take $\phi_f^M$ and $\phi_f^\Gamma$ to be final state independent, in general, for CF/DCS decays.

Non-vanishing $\phi_f^M$ and $\phi_f^\Gamma$ cause *time-dependent* CP asymmetries. For example, in SCS decays to the CP-eigenstates $f = K^+K^-$ and $f = \pi^+\pi^-$, the effective lifetimes $\hat{\tau}$ (or inverse lifetimes $\hat{\Gamma} = 1/\hat{\tau}$) for $D^0$ and $\overline{D}^0$ decays will differ:

$$\Delta Y_f \equiv \frac{\hat{\Gamma}_{\overline{D}^0 \to f} - \hat{\Gamma}_{D^0 \to f}}{2\Gamma_D} = -x_{12}\sin\phi_f^M + a_f^d\, y_{12}\,. \quad (442)$$

The second term on the RHS is the direct CP violating contribution, where the direct CP asymmetry is defined as

$$a_f^d = 1 - \left|\bar{A}_f/A_f\right|\,. \quad (443)$$

It can, in principle, be disentangled experimentally from the dispersive CPVINT contribution with the help of time integrated CP violation measurements, in which $a_f^d$ also enters without a mixing suppression [968],

$$A_{CP} \;\equiv\; \frac{\Gamma(D^0 \to f) - \Gamma(\overline{D}^0 \to \bar{f})}{\Gamma(D^0 \to f) + \Gamma(\overline{D}^0 \to \bar{f})} \;=\; \frac{\langle t \rangle}{\tau_D} \cdot \Delta Y_f + \; a_f^d \,. \tag{444}$$

At Belle II, the factor $\langle t \rangle/\tau_D$ is very close to unity, whereas at LHCb this factor is close to two[969]. Examples of time-dependent CP asymmetries in decays to non-CP eigenstates include the SCS final states $f = K^* K$ or $f = \rho\pi$, and CF/DCS final states $f = K^\pm \pi^\mp$. These asymmetries generally depend on both $\phi_f^M$ and $\phi_f^\Gamma$ because of the additional strong phase $\Delta_f$ [963].

Finally, we relate the dispersive and absorptive observables to the more familiar parameterisation of indirect CP violation currently in use; see, *e.g.*, Ref. [970]. The latter consists of the CPVMIX parameter $|q/p| - 1$, and the CPVINT observables

$$\lambda_f \equiv \frac{q}{p}\frac{\bar{A}_f}{A_f} = -\eta_f^{CP}\, |\lambda_f|\, e^{i\,\phi_{\lambda_f}} \tag{445}$$

for CP eigenstate final states, and their generalisation to pairs of observables $\lambda_f$, $\lambda_{\bar{f}}$ for non-CP eigenstate final states, with arguments $\phi_{\lambda_f} \pm \Delta_f$. The relation between $|q/p| - 1$ and $\phi_f^M$, $\phi_f^\Gamma$ follows from (438), (441), while $\phi_{\lambda_f}$ is given by [963],

$$\tan 2\phi_{\lambda_f} = -\,\frac{x_{12}^2 \sin 2\phi_f^M + y_{12}^2 \sin 2\phi_f^\Gamma}{x_{12}^2 \cos 2\phi_f^M + y_{12}^2 \cos 2\phi_f^\Gamma}\,. \tag{446}$$

Indirect CP violation can be equivalently described in terms of the parameters $\phi_f^M$, $\phi_f^\Gamma$, $x_{12}$, $y_{12}$ emphasized in this report, or the more familiar ones $|q/p|$, $\phi_{\lambda_f}$, $x$, $y$. The same number of independent parameters is employed in each case.

*The superweak limit.* Until recently, fits to measurements of indirect CP violation have been sensitive to values of $\phi_{12}$ down to the 100 mrad level. This level of precision probes large short-distance NP effects. In particular, the effects of weak phases in the subleading decay amplitudes can be safely neglected in the indirect CP violation observables. In this limit, referred to as the superweak limit, a non-vanishing $\phi_{12}$ would be entirely due to short-distance NP in $M_{12}$. Thus, the dispersive and absorptive weak phases satisfy

$$\phi_f^M = \phi_{12}, \qquad \phi_f^\Gamma = 0\,, \tag{447}$$

and the $\phi_{\lambda_f}$ reduce to the familiar "universal" CPVINT phase $\phi$ entering current fits. Note that the phase $\phi_{12}$ would be the only source of indirect CP violation. Therefore, CPVMIX and CPVINT would be related as [971–973],

$$\tan 2\phi \approx -\frac{x_{12}^2}{x_{12}^2 + y_{12}^2} \sin 2\phi_{12}\,, \qquad \tan\phi \approx \left(1 - \left|\frac{q}{p}\right|\right)\,\frac{x}{y}\,, \tag{448}$$

where the first relation is the superweak limit of (446). However, in the superweak limit, the effects of weak phases in the SCS decay amplitudes are kept in the *direct* CP violation observables (where they are not suppressed by $x_{12}$, $y_{12}$). For example, (442) reduces to $\Delta Y_f = -x_{12} \sin\phi_{12}$, while the second term on the RHS of (444) is kept.

With only one phase $\phi_{12}$ controlling all indirect CP violation, the superweak fits to CP violation data are highly constrained. In particular, the Heavy Flavour Averaging Group (HFLAV) [230] and the UTfit Collaboration [965] obtain the following $1\sigma$ and 95% CL fit results (in radians):

$$
\begin{aligned}
\text{HFLAV} : \phi_{12} &= 0.00 \pm 0.03, \quad [-0.09, +0.08] \\
\text{UTfit} : \phi_{12} &= 0.01 \pm 0.05, \quad [-0.10, +0.15] \,.
\end{aligned}
\tag{449}
$$

The HFLAV fit uses all available charm mixing and CP violation data. The HFLAV superweak results for $\phi$ and $|q/p|$ are:

$$
\phi = 0.00 \pm 0.01 \text{ [rad]}, \quad |q/p| = 0.999 \pm 0.014 \,. \tag{450}
$$

*Approximate Universality.* With the continuing improvement in experimental sensitivity expected from Belle II and LHCb, achieving $O(10$ mrad$)$ precision for $\phi_{12}$ may be possible. Thus, we must consider possible deviations from the superweak limit due to the subleading decay amplitudes, and how best to parameterize such deviations. In particular, we need to estimate the size of the final state dependence in $\phi_f^M$ and $\phi_f^\Gamma$. We accomplish this via a $U$-spin flavour symmetry decomposition of the $D^0$-$\overline{D}^0$ mixing amplitudes. Crucially, this also yields order of magnitude estimates of indirect CP violating effects in the SM [963].

Employing CKM unitarity, the $U$-spin decomposition of the SM mixing amplitude $\Gamma_{12}$ can be written as $(\lambda_i \equiv V_{ci}V_{ui}^*)$,

$$
\Gamma_{12} = \frac{(\lambda_s - \lambda_d)^2}{4}\,\Gamma_2 + \frac{(\lambda_s - \lambda_d)\lambda_b}{2}\,\Gamma_1 + \frac{\lambda_b^2}{4}\,\Gamma_0, \tag{451}
$$

and similarly for $M_{12}$, with substitutions $\Gamma_i \to M_i$. The $U$-spin amplitudes $\Gamma_{2,1,0}$ and $M_{2,1,0}$ are the $\Delta U_3 = 0$ elements of $\Delta U$= 2, 1, 0 (5-plet, triplet, and singlet) multiplets, respectively. The $\Delta U = 2, 1, 0$ amplitudes enter at $O(\epsilon^2)$, $O(\epsilon)$ and $O(1)$, respectively, in $SU(3)_F$ flavour symmetry breaking. The expansion parameter $\epsilon$ characterizes the size of the symmetry breaking. Although $M_2$, $\Gamma_2$ enter at $O(\epsilon^2)$, they dominate due to their large CKM factors, and yield the mass and lifetime differences, *i.e.* $x_{12}$ and $y_{12}$. CP violation in the SM is due to $M_1$, $\Gamma_1$, and arises at $O(\epsilon)$ via the CKM phase $\gamma = \arg(\lambda_b)$ entering the SCS decays. The effects of $M_0$, $\Gamma_0$ are of $O(\lambda_b^2)$, and therefore negligible.

We define a pair of theoretical absorptive and dispersive CP violation phases, $\phi_2^\Gamma$ and $\phi_2^M$, respectively, with respect to the $\Delta U = 2$ direction in the mixing amplitude complex plane, proportional to $(\lambda_s - \lambda_d)^2$, i.e the direction of the $\Gamma_2$ and $M_2$ contributions, cf. (451):

$$
\phi_2^\Gamma \equiv \arg\left(\frac{\Gamma_{12}}{\Gamma_{12}^{\Delta U=2}}\right) \approx \text{Im}\left(\frac{2\lambda_b}{\lambda_s - \lambda_d}\frac{\Gamma_1}{\Gamma_2}\right) \sim \left|\frac{\lambda_b}{\theta_c}\right| \sin\gamma \times \frac{1}{\epsilon}, \tag{452}
$$

and similarly for $\phi_2^M$, with the substitution $\Gamma \to M$ everywhere in (452). The second relation in (452) is obtained from the ratio of $\Gamma_1$ to $\Gamma_3$ contributions in (451), while $\Gamma_1/\Gamma_3 = O(1/\epsilon)$ is used in the last relation. In addition, $\phi_{12} = \phi_2^M - \phi_2^\Gamma$. Taking the nominal value $\epsilon \sim 0.2$ for $U$-spin breaking in (452), we arrive at the rough SM estimates

$$
\phi_{12} \sim \phi_2^\Gamma \sim \phi_2^M \sim 3 \times 10^{-3} \,. \tag{453}
$$

Thus, values for these phases as large as $\sim 10$ mrad are certainly plausible.

The phases $\phi_2^M$ and $\phi_2^\Gamma$ are the theoretical analogs of the final state dependent phases $\phi_f^M$ and $\phi_f^\Gamma$, respectively. Another useful theoretical phase defined with respect to the $\Delta U = 2$ direction is the theoretical analog of the final state dependent phases $\phi_{\lambda_f}$. It is given by

$$\phi_2 \equiv \arg\left(\frac{q}{p}\frac{1}{\Gamma_{12}^{\Delta U=2}}\right) . \tag{454}$$

The estimate, $\phi_2 \sim 3\times 10^{-3}$, follows from (453) and the substitutions $\phi_f^{\Gamma(M)} \to \phi_2^{\Gamma(M)}$, $\phi_{\lambda_f} \to \phi_2$ in (453).

Next, we assess the deviations of the final state dependent phases $\phi_f^M$, $\phi_f^\Gamma$, and $\phi_{\lambda_f}$, cf. (439), (440), (445), from their theoretical counterparts, in order to arrive at the appropriate minimal parameterisation of indirect CP violating effects in the LHC-b/Belle-II era. The misalignments between these phases, for given final state $f$, satisfy

$$\delta\phi_f \equiv \phi_f^\Gamma - \phi_2^\Gamma = \phi_f^M - \phi_2^M = \phi_2 - \phi_{\lambda_f} . \tag{455}$$

We can characterise the magnitude of the misalignment in the SM as follows: (i) For CF/DCS decays it is precisely known and negligible, *i.e.* $\delta\phi_f = O(\lambda_b^2/\theta_c^2)$, implying that to excellent approximation, $\phi_f^{\Gamma(M)} = \phi_2^{\Gamma(M)}$, and $\phi_{\lambda_f} = \phi_2$; (ii) In SCS decays, $\delta\phi_f$ is related to direct CP violation as $\delta\phi_f = a_f^d \cot\delta$ (via the $U$-spin decomposition of the decay amplitudes [974]), where a strong phase $\delta = O(1)$ is expected due to large rescattering at the charm mass scale. Thus, for $f = \pi^+\pi^-$, $K^+K^-$, the experimental bounds $a_f^d \lesssim O(10^{-3})$ imply that $\delta\phi_f \lesssim O(10^{-3})$; (iii) In SCS decays, $\delta\phi_f = O(\lambda_b \sin\gamma/\theta_c)\times\cot\delta$, *i.e.* it is $O(1)$ in $SU(3)_F$ breaking. Thus, (452) yields $\delta\phi_f/\phi_2^\Gamma = O(\epsilon)$, implying an order of magnitude suppression of the misalignment.

We conclude that in the SM, the deviations of the final state dependent phases from the theoretical phases are entirely negligible for CF/DCS decays, whereas for SCS decays they yield $\sim 10\%$ corrections. Thus, in the LHCb/Belle-II era, with a potential sensitivity of 10 mrad, a single pair of dispersive and absorptive phases suffices to parametrise all indirect CP violating effects, which we can identify with our theoretical phases $\phi_2^M$ and $\phi_5^\Gamma$, respectively. We refer to this fortunate circumstance as *approximate universality*. Moreover, approximate universality generalises beyond the SM under the following conservative assumptions about NP decay amplitudes containing new weak phases: (i) they can be neglected in CF/DCS decays [967], (ii) in SCS decays their magnitudes are similar to, or smaller than the SM QCD penguin amplitudes, as already hinted at by the experimental bounds on the direct CP asymmetries $a^d_{K^+K^-}$, $a^d_{\pi^+\pi^-}$. These assumptions can ultimately be tested by future direct CP violating measurements.

Under approximate universality, the final state dependent CPVINT phases $\phi_f^M$, $\phi_f^\Gamma$ are replaced with the final state independent phases $\phi_2^M$, $\phi_2^\Gamma$ in the expressions for the time-dependent CP asymmetries. For example, $\Delta Y_f = -x_{12}\sin\phi_2^M + y_{12}\,a_f^d$, cf. (442). A global fit to the CP violation data with any two of the three phases $\phi^M$, $\phi^\Gamma$, $\phi_{12}$ is equivalent to the traditional two-parameter fit for the parameters $|q/p|$ and $\phi = \phi_2$. The relations

$$\left|\frac{q}{p}\right| - 1 \approx \frac{|x||y|}{x^2+y^2}\sin\phi_{12}, \qquad \tan 2(\phi_2 + \phi^\Gamma) \approx -\frac{x_{12}^2}{x_{12}^2+y_{12}^2}\sin 2\phi_{12} , \tag{456}$$

together with $\phi_{12} = \phi_2^M - \phi_2^\Gamma$, allow one to translate between $(\phi_2, |q/p|)$ and $(\phi_2^M, \phi_2^\Gamma)$ (the second relation in (456) follows from (446) in the approximate universality limit). In this manner it is possible to separately determine the dispersive and absorptive CP violation phases. Large short distance NP contributions, which would reside in the former, could therefore be isolated.

*Theory Summary.* We have described indirect CP violating effects in terms of the (final state dependent) dispersive and absorptive CP violating weak phases. This description allowed us to estimate the size of indirect CP violation in the SM, and to arrive at a minimal parameterisation appropriate for the LHCb/Belle-II era. Up until recently, the sensitivity of indirect CP violation measurements has been sufficient to probe for large short distance NP contributions. Fits to the CP violation data carried out in the superweak limit have therefore been appropriate. In the superweak limit the only source of indirect CP violation is the mixing phase $\phi_{12}$. The UTfit and HFLAV fits yield $\phi_{12} \lesssim 0.10$ [rad] at the 95% CL. A $U$-spin based decomposition of the mixing amplitudes implies that $\phi_{12}$ could be as large as $\sim 10$ mrad in the SM, cf. (453). Hence, we concluded that there is currently an $O(10)$ window for NP in indirect CP violation.

As the sensitivity of indirect CP violation measurements improves towards 10 mrad in the indirect CP violation phases, it becomes necessary to take into account the effects of weak phases in subleading decay amplitudes. We have argued that their contributions to the indirect CP violation observables can be accounted for in the SM, and under conservative assumptions for NP, with only two theoretical dispersive and absorptive phases $\phi_2^M$ and $\phi_2^\Gamma$ (denoted as approximate universality), cf. (452). These phases could be as large as $\sim 10$ mrad in the SM, while the final state dependent corrections are an order of magnitude smaller .

The parameterisation in terms of $\phi_2^M$ and $\phi_2^\Gamma$ is equivalent to the traditional two-parameter fits to $\phi$ and $|q/p|$, where $\phi$ is identified with the theoretical phase $\phi_2$ in (454), with direct CP asymmetries in the CF/DCS decays set to zero. The translation between them is given in (456). In the second relation, the LHS contains the combination $\phi_2 + \phi_2^\Gamma$. However, our $U$-spin based estimates for these phases in the SM are not far from the $1\sigma$ error on $\phi$ in the superweak fit, cf. (450), (453). Going forward, this confirms that we must move beyond the superweak limit (in which $\phi_2^\Gamma = 0$), and fit for the independent parameters $\phi_2^M$ and $\phi_2^\Gamma$. The two-parameter fit yields much larger errors on $\phi = \phi_2$ and $|q/p|$. For example, the current HFLAV errors increase by $O(10)$ compared to the superweak fit results. However, this should ultimately be overcome by the improved statistics at Belle-II and LHCb.

*13.5.2. Lattice Calculations. Author: S. Sharpe*

To search for new physics using charmed mesons, it is obviously crucial to accurately predict the standard model contributions to the decay and mixing amplitudes. The issue addressed in this section is the extent to which Lattice QCD (LQCD) can, over the next few years, provide such predictions.

LQCD provides is a method for determining the strong-interaction contributions to certain types of hadronic amplitudes, using numerical simulations of the partition function of QCD. All approximations that are made (finite lattice spacing, finite volume, etc.) can be systematically removed, so that results with fully controlled errors are possible. For those quantities that are presently accessible, LQCD results have now achieved percent level precision or better. These quantities include the light hadron spectrum, decay constants (including those of the $D$ and $D_s$), semileptonic form factors (including those for $D \to K$ and $D \to \pi$ decays), and mixing matrix elements (such as $B_K$ and $B_B$). For a recent review of the status of such calculations see Ref. [140]. The results confirm that QCD indeed describes the strong interactions in the non-perturbative regime, and provide predictions that play a crucial role in the search for new physics by looking for inconsistencies in unitarity triangle analyses.

At present, however, results with high precision are only available for processes involving single hadrons and a single insertion of a weak operator. For charmed mesons, the only high precision quantities are thus the above-mentioned decay constants and semileptonic form factors, as well as the short-distance part of the $D^0 - \overline{D^0}$ mixing amplitude. The decay amplitudes (*e.g.* those for $D \to \pi\pi$ and $D \to K\bar{K}$), as well as the long-distance part of the mixing amplitude, are more challenging quantities because they involve multiple-particle states. The progress towards LQCD calculations of these quantities will be discussed below.

Before doing so we give an update on results for the short-distance contributions to $D^0 - \overline{D^0}$ mixing. We recall that, in the Standard Model, these arise only from loops involving $b$-quarks, and contribute only a small part of the mixing amplitude. The largest contribution involves intermediate light quarks and is long-distance dominated. This holds also for the CP-violating part of the mixing amplitude, which is, in any case, expected to be very small in the Standard Model due to the CKM factors and lack of enhancement from the top-quark loop. In light of these considerations, $D^0$ mixing is an excellent place to look for contributions from new physics. Integrating out the heavy particles in generic Beyond-the-Standard-Model (BSM) theories leads to Lorentz scalar $|\Delta C| = 2$ operators with all possible chiral structures, eight in all. Their matrix elements can be parameterised in terms of five $B$-parameters, generalisations of the single $B$-parameter that is needed for the standard model operator composed of left-handed currents. Explicit forms of these operators and the definitions of the corresponding $B$-parameters are given, for example, in Ref. [193]. Analogous matrix elements are needed to parameterise the BSM contributions to kaon and $B$ mixing.

Last year, the first fully controlled results for the $D$-mixing $B$-parameters have become available [193]. These were obtained by the ETM collaboration using simulations with dynamical up, down, strange and charm quarks (with up and down degenerate). The quarks were discretised with the twisted-mass action. The results have errors ranging from 3-8%, larger than the state-of-the-art quantities mentioned above, but sufficiently small for most phenomenological purposes. The values of the five $B$-parameters range from 0.65 up to 0.97.

There are also preliminary results from the Fermilab Lattice plus MILC collaborations [975]. These use staggered light quarks and the Fermilab heavy quark action for the charm quark, with up, down and strange sea quarks. The results have larger errors than those of Ref. [193], but have similar central values [976]. Final results are expected soon.

We now return to the more challenging, and more interesting, case of the $D$ decay amplitudes. To understand the challenges facing a lattice calculation of, say, the $D \to \pi\pi$ amplitude, it is useful to begin by discussing the simpler case of $K \to \pi\pi$ decays. For the latter decays, a LQCD calculation is now possible, and indeed first results have been obtained. For the $\Delta I = 3/2$ transition, these have fully controlled errors [178], while for the $\Delta I = 1/2$ case a complete result is available only at a single lattice spacing [179, 977]. Nevertheless, the successful calculation of both real and imaginary (CP-violating) parts of the $\Delta I = 1/2$ amplitude is a *tour do force*, and substantially moves the boundaries of what is possible from LQCD.

There are three main technical challenges for $K \to \pi\pi$ calculations: (i) the fact that one necessarily works in finite volume so the states are not asymptotic two-particle states; (ii) the need to calculate Wick contractions (such as the penguin-type contractions) that involve gluonic intermediate states in some channels, and (iii) the presence of an effective weak Hamiltonian with many operators, some of which can mix in lattice regularisation with lower-dimension operators. The former challenge was solved some time ago [177]. It has taken many years, however, for methods, algorithms and computational power to improve to the point that the numerical aspects of all three challenges can be overcome. Fully controlled results for all amplitudes, and in particular for the Standard Model prediction for $\epsilon'$, are expected in the next few years.

To extend these results to the charm case one faces two additional challenges. The first is that, to calculate a decay matrix element, one must use final states with energy $E = M_D$, and these are highly excited compared to the strong interaction ground states. For example in the $I = S = 0$ sector, with total momentum $\vec{P} = 0$, the lightest state consists of two pions at rest, with $E \approx 2M_\pi$ (up to small finite-volume corrections). In a correlation function, the contribution of this lightest state will dominate over that from states with $E = M_D$ by an exponential factor, $e^{(M_D - 2M_\pi)\tau}$ (where $\tau$ is Euclidean time). This is relevant because, in a LQCD calculation, the weak Hamiltonian is able to connect states having different energies. Thus if one simply considers a correlator in which one first creates a $D$ meson, then inserts the weak Hamiltonian, and finally destroys two pions, the dominant contribution will be from the unphysical (and uninteresting) amplitude for $D \to (\pi\pi)\big|_{\rm rest}$. In order to obtain the desired physical amplitude one must therefore project the final state onto one having with $E = M_D$. This can be done in principle by using appropriate final state operators, tuned to avoid couplings to lighter states. While this will certainly be challenging, it is encouraging that, over the last five years, there have been tremendous advances in the methodology for extracting excited state energies, using the methods described in Refs. [978, 979]. These have allowed the first calculations of resonance properties in several channels (as reviewed, for example, in Ref. [980]. Thus we expect this challenge to be solvable using adaptations of existing methods.

The second, and more difficult, challenge is as follows. Even when one has fixed the quantum numbers of a final state, say to $I = S = 0$, the strong interactions necessarily mix together all kinematically allowed states having these quantum numbers. When $E = M_D$, this means that, even if one creates a state with a two-pion operator, it will mix with states containing $K\overline{K}$, $\eta\eta$, $4\pi$, $6\pi$, $\ldots$. This is an inevitable feature of working in finite volume, where the particles do not lie in asymptotic states and repeatedly scatter off one another. It means that, even if one could calculate the matrix element between a $D$ meson and one of these finite-volume states, this would not be related to the desired infinite-volume amplitudes. This is the generalisation of the problem solved by Lellouch and Lüscher for a single two-particle channel. We stress that this is not an issue introduced by discretisation of space-time, but instead by the computational necessity of working with a finite system.

There has been significant progress towards a solution to this theoretical problem in the last few years. The first step, taken in Ref. [189], was to solve the analogous problem for a particle that can decay to any number of two-particle channels (and assuming that the scattering is dominantly s-wave). Thus if the $D$ decayed only to $\pi\pi$, $K\bar{K}$ and $\eta\eta$, for example, then the problem is solved in principle.

The next major step was to determine the "quantisation condition" for a finite-volume three-particle system, *e.g.* three pions in a box [981, 982]. This relates the finite-volume energies to the two- and three-particle infinite-volume scattering amplitudes, and generalizes to three particles the two-particle formalism of Lüscher. The result is quite complicated, but has recently been checked by comparing the threshold expansion to results obtained using non-relativistic quantum mechanics and perturbation theory [983]. Other checks are underway. The derivation also makes various simplifying assumptions (*e.g.* not allowing any $2 \leftrightarrow 3$ transitions) and work is actively underway to relax these. We expect that a fully general three-particle quantisation condition will be available within a year, and that the next step, a generalisation of the Lellouch-Lüscher method to decays involving two and three particles, will follow shortly after.

The extension of the theoretical work to four or more particles will be the next challenge. No work has been done on this to date, and it is difficult to give a time scale for expected progress. We do note, however, that there is great interest in the development of these methods in the hadron spectroscopy community.

Finally, we comment briefly on the possibility of calculating long-distance contributions to $D^0 - \overline{D^0}$ mixing using lattice methods. Here the challenge is that there are two insertions of the weak Hamiltonian, with many allowed states propagating between them. Significant progress has been made recently on the corresponding problem for kaons [984, 985], but the $D^0$ system is much more challenging. The main problem is that there are many strong-interaction channels having $E < m_D$, and these lead to exponentially enhanced contributions that must be subtracted and corrected for. Further theoretical work is needed to develop a practical method.

*13.5.3. Experiment.* The experimental status of $D^0\overline{D}^0$ mixing and CP violation measurements is summarized in Table 115. The first evidence for mixing was obtained by Belle in $D^0 \to h^+h^-$ decays [986] and by BaBar in $D^0 \to K^+\pi^-$ decays [987]. The first observations of mixing with more than $5\sigma$ significance were made by LHCb [988] and CDF [989]. To-date, no evidence of $CP$ violation in $D^0$ decays has been obtained by any experiment.

*Wrong-sign decays $D^0 \to K^+\pi^-$.* *Author: A. Schwartz*

Given the much larger samples of flavour-tagged DCS $D^0 \to K^+\pi^-$ decays Belle II will collect over those collected by Belle and BaBar, and also the improved decay time resolution, one expects that Belle II will have significantly greater sensitivity to mixing and $CP$ violation in $D^0$ decays than that of the first generation of $B$ factories. To study this, we have performed a "toy" MC study in which "wrong-sign" $D^0 \to K^+\pi^-$ and $\overline{D}^0 \to K^-\pi^+$ decays are generated, their decay times smeared by the expected decay time resolution of Belle II, and the resulting decays times fitted for mixing parameters $x$, $y$ and $CP$-violating parameters $|q/p|$, $\phi$. The fit results are compared with the generated (true) values and the residuals plotted. The RMS of these distributions are taken as the precision Belle II should achieve for these parameters. Below we provide details of this study and the results.

Table 115: The experimental status of $D^0\overline{D}^0$ mixing and $CP$ violation in different decays.

| **Decay Type** | **Final State** | **LHCb** | **Belle** | **BaBar** | **CDF** | **CLEO** | **BES III** |
|---|---|---|---|---|---|---|---|
| DCS 2-body(WS) | $K^+\pi^-$ | ★ | ★ | ● | ★ | ✓ | |
| DCS 3-body(WS) | $K^+\pi^-\pi^0$ | | ✓ $A_{CP}$ | ● | | ✓ $A_{CP}$ | |
| CP-eigenstate | $K^+K^-$, $\pi^+\pi^-$ | ●$^{(a)}_{A_{CP}}$ | ● | ● | ✓ $A_{CP}$ | ✓ | |
| Self-conjugated 3-body decay | $K^0_S\pi^+\pi^-$ | ✓ | ✓ | ✓ $A_{CP}$ | ✓ | | |
| | $K^0_S K^+K^-$ | | ✓$^{(b)}$ | ✓ | | | |
| Self-conjugated SCS 3-body decay | $\pi^+\pi^-\pi^0$ | ✓ $A_{CP}$ | ✓ $A_{CP}$ | ✓$^{mixing}_{A_{CP}}$ | | | |
| | $K^+K^-\pi^0$ | | | ✓ $A_{CP}$ | | | |
| SCS 3-body | $K^0_S K^\pm\pi^\mp$ | ✓ $\delta_{K^0_S K\pi}$ | | | | ✓ $\delta_{K^0_S K\pi}$ | |
| Semileptonic decay | $K^+\ell^-\nu_\ell$ | | ✓ | ✓ | | ✓ | |
| Multi-body(n≥4) | $\pi^+\pi^-\pi^+\pi^-$ | ✓ $A_{CP}$ | | | | | |
| | $K^+\pi^-\pi^+\pi^-$ | ★ | ✓ $A_{CP}$ | ✓ | | | |
| | $K^+K^-\pi^+\pi^-$ | ✓$^{(c)}_{A_{CP}}$ | | ✓ $A_T$ | | ✓ $A_{CP}$ | |
| $\psi(3770) \to D^0\overline{D}$ via correlations | | | | | | ✓ $\delta_{K\pi}$ | ✓ $y_{CP}$ |

★ for observation ($> 5\sigma$); ● for evidence ($> 3\sigma$); ✓ for measurement.

(a) LHCb measured the indirect CP asymmetry in Phys. Rev. Lett. **112**, 041801 (2014).

(b) Belle measured $y_{CP}$ in $D^0 \to K^0_S\phi$ in Phys. Rev. D **80**, 052006 (2009).

(c) LHCb searched for CP violation using T-odd correlations in JHEP **10** (2014) 005.

The yield of flavour-tagged $D^0 \to K^+\pi^-$ decays collected by Belle was $4024 \pm 88$ in 400 fb$^{-1}$ of data and $11478 \pm 177$ in 976 fb$^{-1}$ of data, corresponding to two independent analyses. Scaling the latter result, which has higher statistics, by luminosity gives the expected Belle II signal yields listed in Table 116 for 5 ab$^{-1}$, 20 ab$^{-1}$, and 50 ab$^{-1}$ of data. Thus, for our MC study we generate samples of $D^{*+} \to D^0\pi^+$, $D^0 \to K^+\pi^-$ and $D^{*-} \to \overline{D}{}^0\pi^-$, $\overline{D}{}^0 \to K^-\pi^+$ decays corresponding to these integrated luminosities. It is expected that it will take Belle II approximately two years, five years, and 10 years, respectively, to collect these samples.

Table 116: Flavor-tagged $D^{*+} \to D^0\pi^+$, $D^0 \to K^+\pi^-$ signal yields (including charge-conjugates) obtained by Belle and expected for Belle II.

| Luminosity (ab$^{-1}$) | Belle | Belle II |
|---|---|---|
| 0.400 | 4024 | |
| 0.976 | 11478 | |
| 5.0 | | 58800 |
| 20 | | 235200 |
| 50 | | 588000 |

The $D^0 \to K^+\pi^-$ decay times are generated according to the following probability density functions (PDFs):

$$\frac{N(D^0\to f)}{dt} \propto e^{-\overline{\Gamma}\,t}\left\{R_D + \left|\frac{q}{p}\right|\sqrt{R_D}(y'\cos\phi - x'\sin\phi)(\overline{\Gamma}t) + \left|\frac{q}{p}\right|^2\frac{(x'^2+y'^2)}{4}(\overline{\Gamma}\,t)^2\right\}$$
$$\frac{N(\overline{D}^0\to \bar{f})}{dt} \propto e^{-\overline{\Gamma}\,t}\left\{\overline{R}_D + \left|\frac{p}{q}\right|\sqrt{\overline{R}_D}(y'\cos\phi + x'\sin\phi)(\overline{\Gamma}t) + \left|\frac{p}{q}\right|^2\frac{(x'^2+y'^2)}{4}(\overline{\Gamma}\,t)^2\right\}, \tag{457}$$

where $x' = x\cos\delta + y\sin\delta$, $y' = y\cos\delta - x\sin\delta$, and $\delta$ is the strong phase difference between $D^0 \to K^-\pi^+$ and $\overline{D}^0 \to K^-\pi^+$ amplitudes. The parameter $R_D$ is the squared ratio of the magnitudes of $D^0 \to K^+\pi^-$ to $D^0 \to K^-\pi^+$ amplitudes, and $\overline{R}_D$ is the squared ratio of the magnitudes of $\overline{D}^0 \to K^-\pi^+$ to $\overline{D}^0 \to K^+\pi^-$ amplitudes. We subsequently smear these decay times using a Gaussian resolution function with a mean of zero and a width set equal to the expected decay time resolution of Belle II: 135 fs. Finally, we fit these distributions with the PDF's of Eq. (457) convolved with the Gaussian resolution function. This convolution is done analytically, resulting in a PDF consisting of Error functions [Erf(x)]:

$$\begin{aligned}\frac{N(D^0\to f)}{dt} \propto\ & R_D\,\sigma\sqrt{\frac{\pi}{2}}\,e^{\sigma^2/(2\tau^2)}\,[1+\mathrm{Erf}(x)]\,e^{-t/\tau}\ + \\ & \left|\frac{q}{p}\right|\sqrt{R_D}\,(y'\cos\phi - x'\sin\phi)\sigma^2 e^{-t^2/(2\sigma^2)}\ + \\ & \frac{\sigma}{\tau}\sqrt{\frac{\pi}{2}}\,e^{\sigma^2/(2\tau^2)}(t\tau-\sigma^2)\,[1+\mathrm{Erf}(x)]\,e^{-t/\tau}\ + \\ & \left|\frac{q}{p}\right|^2\frac{(x'^2+y'^2)}{4}\left(\sigma^2 t - \frac{\sigma^4}{\tau}\right)e^{-t^2/(2\sigma^2)}\ + \\ & \frac{\sigma}{\tau^2}\sqrt{\frac{\pi}{2}}\,e^{\sigma^2/(2\tau^2)}(\tau t-\sigma^2)\left[\sigma^4 - 2\sigma^2\tau t + \tau^2(t^2+\sigma^2)\right][1+\mathrm{Erf}(x)]\,e^{-t/\tau}\,,\end{aligned}$$

where $\sigma = 135$ fs, $\tau = \tau_D = 410$ fs, and $x = t/(\sigma\sqrt{2}) - \sigma/(\tau\sqrt{2})$. The corresponding smeared PDF for $\overline{D}^0$ decays is the same as that above but with the substitutions $R_D \to \overline{R}_D$, $\phi \to -\phi$, and $|q/p| \to |p/q|$.

Two distinct generation + fitting procedures are performed: assuming $CP$ conservation and allowing for $CP$ violation. For the first procedure we set $R_D = \overline{R}_D$, $|q/p| = 1$, and $\phi = 0$. In this case the PDFs for $D^0$ and $\overline{D}^0$ decays are identical [see Eqs. (457)], and the fitted parameters are $R_D$, $x'^2$, and $y'$. For the second fit we float all six parameters: $R_D$, $\overline{R}_D$, $x'$, $y'$, $|q/p|$, and $\phi$. Typical unsmeared and smeared decay time distributions are shown in Fig. 153, along with the fit result, for a typical fit assuming $CP$ conservation. For this fit the generated values were $R_D = 0.335\%$, $x'^2 = (0.01)^2$, and $y' = 0.01$. For fits allowing for $CP$ violation, the samples were generated with $|q/p| = 0.90$ and $\phi = \pi/4$.

The generation + fitting procedure is repeated for an ensemble of 1000 experiments, and the differences between the fit results and the true (generated) values are plotted; see Fig. 154. The RMS of these distributions of residuals are taken as the uncertainties on these parameters and are listed in Table 117. We note that the residuals for $x'$ and $\phi$ for the $CP$-allowed fits exhibit a bifurcated structure. This is due to nonlinearities in the PDF, which cause the fitter to occasionally converge to a local (rather than global) minimum. This problem can be alleviated by fitting for a better-behaved set of parameters, for example, $\alpha \equiv x' \sin\phi$, $\beta \equiv y' \cos\phi$, and $\gamma \equiv \tan\phi$. However, we do not study such transformations here.

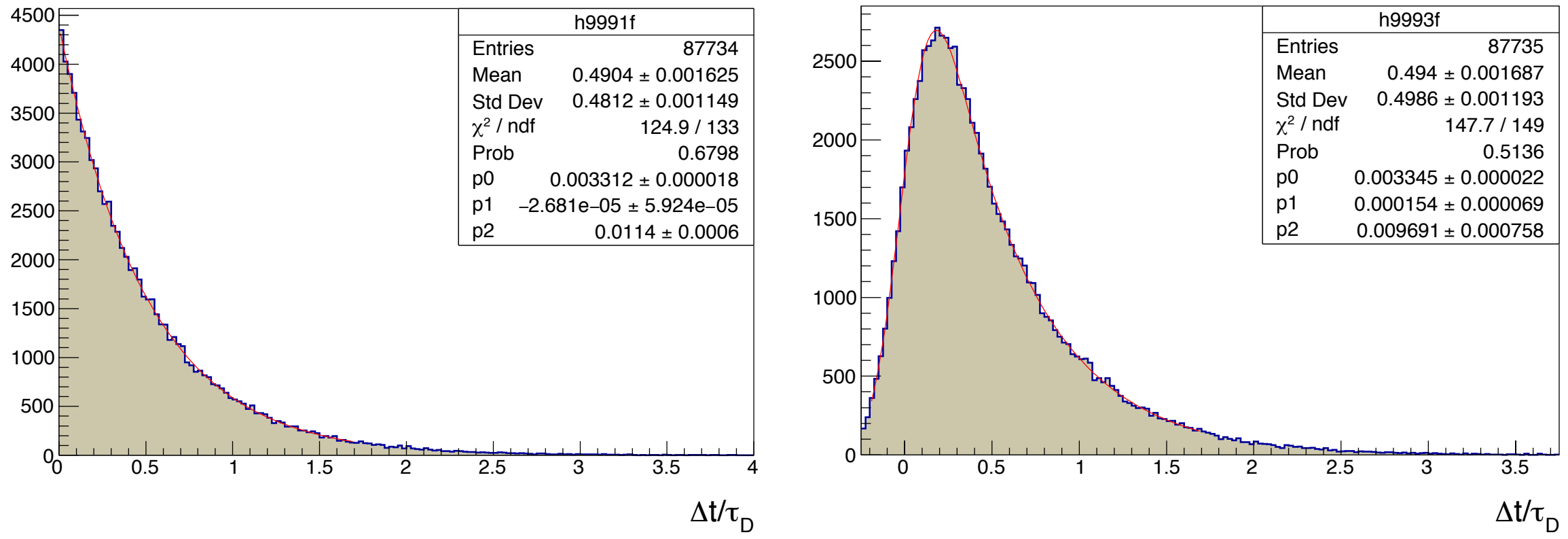

Fig. 153: Unsmeared (left) and smeared (right) decay time distributions for $D^{*+} \to D^0\pi^+$, $D^0 \to K^+\pi^-$ decays, and projections of the fit result, for 20 ab$^{-1}$ of Belle II data. The fit shown is for no $CP$ violation; the fitted parameters are $p_0 = R_D$, $p_1 = x'^2$, and $p_2 = y'$. The generated values are $R_D = 0.335\%$, $x'^2 = (0.01)^2$, and $y' = 0.01$.

Table 117: Uncertainties on mixing parameters $x$, $y$ and $CP$-violating parameters $|q/p|$, $\phi$ resulting from fitting Belle II samples of flavor-tagged $D^0 \to K^+\pi^-$ and $\overline{D}{}^0 \to K^-\pi^+$ decays. The first two rows correspond to no $CP$ violation; the last four rows correspond to allowing for $CP$ violation.

| Parameter | 5 ab$^{-1}$ | 20 ab$^{-1}$ | 50 ab$^{-1}$ |
|---|---|---|---|
| $\delta x'^2$ $(10^{-5})$ | 6.2 | 3.2 | 2.0 |
| $\delta y'$ (%) | 0.093 | 0.047 | 0.029 |
| $\delta x'$ (%) | 0.32 | 0.22 | 0.13 |
| $\delta y'$ (%) | 0.23 | 0.15 | 0.097 |
| $\delta\lvert q/p\rvert$ | 0.174 | 0.073 | 0.043 |
| $\delta\phi$ (°) | 13.2 | 8.4 | 5.4 |

We also plot the MINOS errors for these parameters as returned by the fits; see Fig. 155. The mean values of these errors should nominally match the RMS of the residuals distributions of Fig. 154. This is confirmed for the case of no $CP$ violation, but for the $CP$-allowed case the situation is more complicated: the mean of the MINOS errors for $|q/p|$ matches the RMS of the residuals distribution, but the mean for $x'$ is somewhat larger, and those for $y'$ and $\phi$ are about twice as large.

This study does not yet include backgrounds, which are expected to be modest. A preliminary look at the effect of backgrounds indicates that the fitted errors on the mixing and $CP$ violation parameters increase by $\sim 20\%$. We have also neglected systematic uncertainties, which are expected to be small (for the Belle analysis including $CP$ violation, the systematic uncertainties increased the errors on the fitted parameters by 12% [990]).

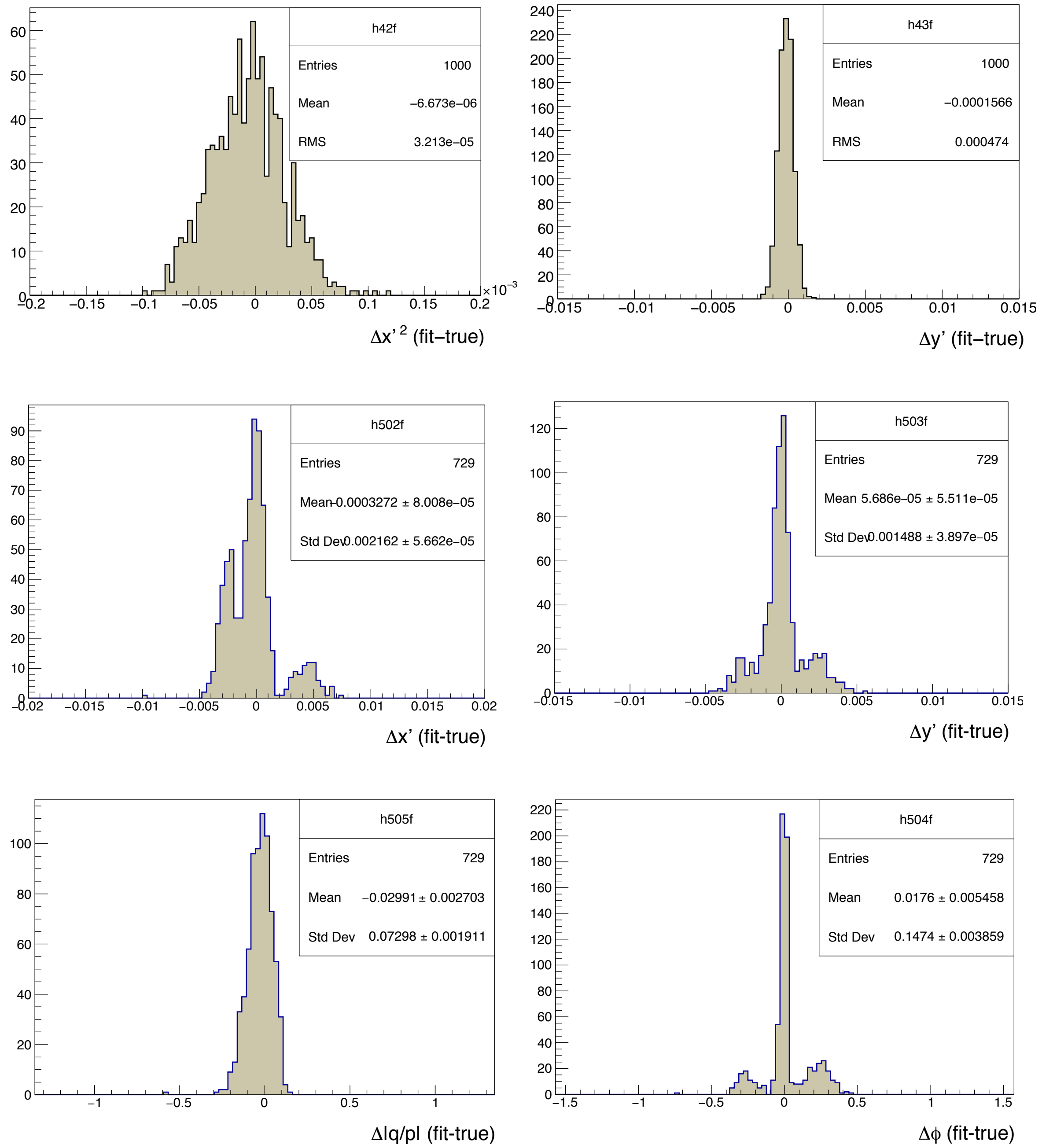

Fig. 154: Residual distributions obtained by simultaneously fitting the decay time distributions of $D^0 \to K^+\pi^-$ and $\overline{D}^0 \to K^-\pi^+$ decays, for an ensemble of 1000 experiments (see text). Each experiment corresponds to 20 ab$^{-1}$ of data. The top row corresponds to no $CP$ violation; the middle and bottom rows correspond to allowing for $CP$ violation. The units for $\Delta\phi$ are radians.

*Wrong-sign decays $D^0 \to K^+\pi^-\pi^0$.* *Authors: L. Li, A. Schwartz*

The "wrong-sign" (WS) decay $D^0 \to K^+\pi^-\pi^0$ proceeds via the same weak amplitudes as does the WS decay $D^0 \to K^+\pi^-$: directly via a doubly Cabibbo-suppressed decay, and indirectly via mixing followed by a Cabbibo-favoured decay. The latter amplitude provides sensitivity to mixing and indirect $CP$ violation. However, fitting for mixing parameters requires understanding the decay amplitude, which typically contains numerous intermediate resonances. The different magnitudes and phases of the intermediate states necessitates performing a time-dependent Dalitz plot analysis to measure mixing.

Both Belle and BaBar have studied $D^0 \to K^+\pi^-\pi^0$ decays. Belle has measured the ratio of rates for WS decays to "right-sign" $D^0 \to K^-\pi^+\pi^0$ decays: $R_{WS} = (0.229 \pm 0.015)\%$ [991]. BaBar has performed a time-dependent fit to the $(m^2_{K^+\pi^-}, m^2_{\pi^-\pi^0})$ Dalitz plot to measure the effective mixing parameters $x'' = x \cdot \cos\delta_{K\pi\pi^0} + y \cdot \sin\delta_{K\pi\pi^0}$ and $y'' = y \cdot \cos\delta_{K\pi\pi^0} - x \cdot \sin\delta_{K\pi\pi^0}$, where $\delta_{K\pi\pi^0}$ is the strong phase difference between the amplitudes for $D^0 \to K^+\rho^-$ and $\overline{D}{}^0 \to K^+\rho^-$. The results are $x'' = (2.61^{+0.57}_{-0.68} \pm 0.39)\%$ and $y'' = (-0.06^{+0.55}_{-0.64} \pm 0.34)\%$ [992].

Assuming Belle II has the same $D^0 \to K^+\pi^-\pi^0$ reconstruction efficiency as that of BaBar, we estimate the Belle II signal yield by scaling the BaBar yield by the ratio of luminosities. The result is 225 000 flavor-tagged signal decays in 50 ab$^{-1}$ of data. We estimate the Belle II sensitivity to $x''$ and $y''$ by performing an MC simulation study, generating 10 independent data sets of 225 000 WS events each. We also generate corresponding samples of RS events, which are needed to determine the ratio of the magnitudes of the WS and RS decay amplitudes. To generate WS decays, we use a seven-resonance decay model as measured by BaBar [992], whereas to generate RS decays we use an 11-resonance decay model as measured by Belle [993] [41]

For both samples the decay times are smeared by an assumed resolution of $\sigma = 140$ fs. The mixing parameters used for the event generation are $x'' = 2.58\%$, $y'' = 0.39\%$, and $\delta_{K\pi\pi^0} = 10°$. After generation, we fit the samples for parameters $x''$ and $y''$ as well as for the magnitudes and phases of the intermediate states. A typical time-dependent fit to the Dalitz plot is shown in Fig. 156.

The resulting fit residuals for the 10 experiments are plotted in Fig. 157, and the RMS of these distributions is taken as the expected precision of Belle II for these parameters. This precision is:

$$\begin{aligned}\sigma_{x''} &= 0.057\% \quad (\text{stat. error only},\ 50\ \text{ab}^{-1}) \\ \sigma_{y''} &= 0.049\% \quad (\text{stat. error only},\ 50\ \text{ab}^{-1})\,. \end{aligned} \tag{458}$$

These uncertainties are an order of magnitude smaller than the errors obtained by BaBar. These errors do not include systematic uncertainties, which are expected to be small and are discussed below. In addition, these errors do not include the effect of backgrounds. From a study with Belle data, we find that the presence of backgrounds increases the fitted errors on $x''$ and $y''$ by approximately 40%. Applying this scaling to the values of Eq. (458) gives errors of $\sigma_{x''} = 0.080\%$ and $\sigma_{y''} = 0.070\%$. These estimates are probably conservative, as backgrounds should be smaller at Belle II than at Belle due to improved vertex resolution, improved mass resolution, and improved particle identification.

[41] For this decay there are thirteen possible resonances. However, the $\rho(1450)$ and $\rho(1700)$ have masses whose peaks are outside the Dalitz plot boundary but whose widths are sufficiently wide such that their tails extend into the region of interest. As their phases differ by close to 180°, it is difficult for the fitter to distinguish them, and therefore we keep only the $\rho(1700)$. We also remove the $K^*(1680)^0$, whose fitted fraction is negligibly small ($< 0.1\%$).

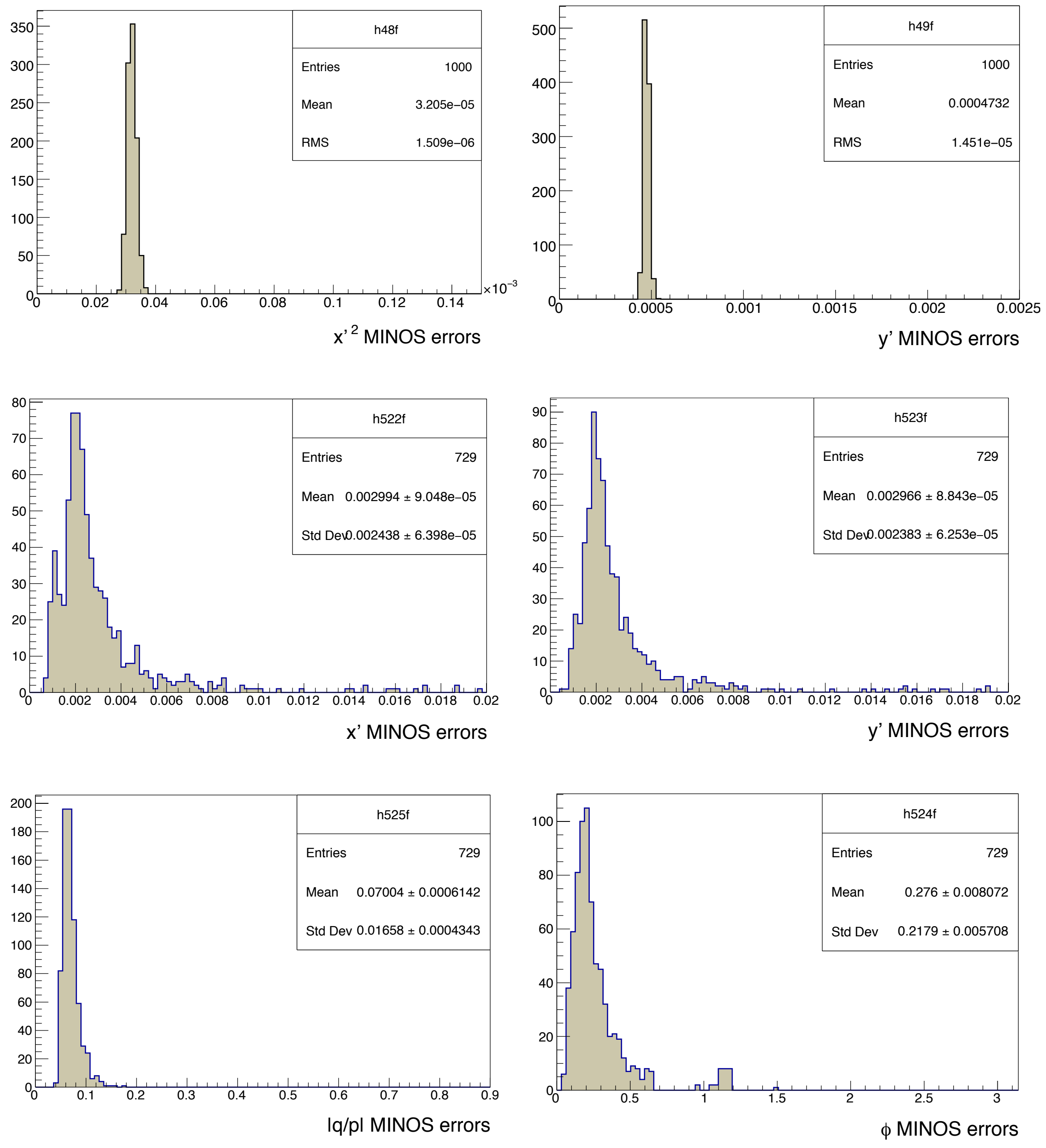

Fig. 155: MINOS errors obtained by simultaneously fitting the decay time distributions of $D^0 \to K^+\pi^-$ and $\overline{D}^0 \to K^-\pi^+$ decays, for an ensemble of 1000 experiments (see text). Each experiment corresponds to 20 ab$^{-1}$ of data. The top row corresponds to no $CP$ violation; the middle and bottom rows correspond to allowing for $CP$ violation. The units for $\phi$ are radians.

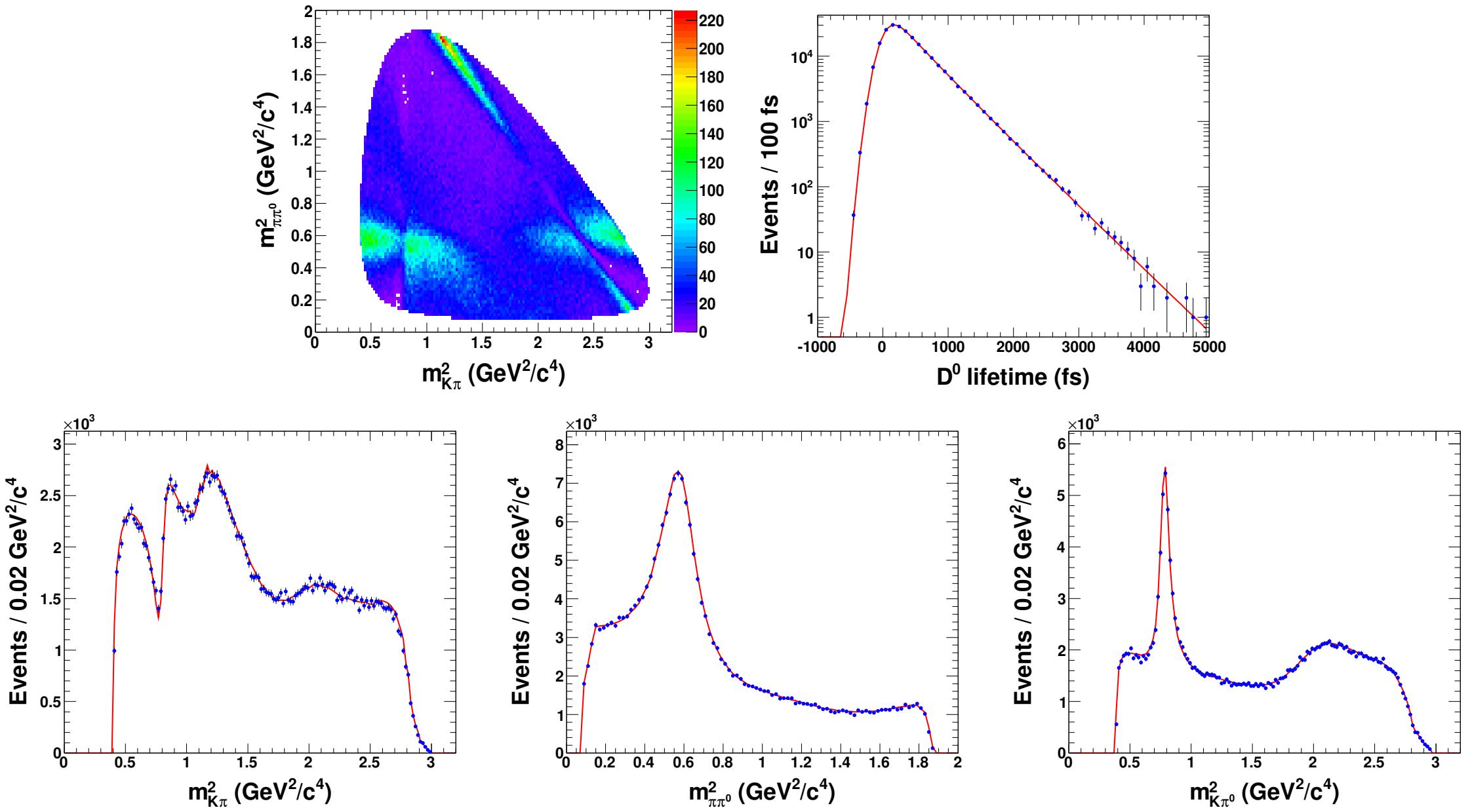

Fig. 156: Time-dependent fit to the Dalitz plot of WS $D^0 \to K^+\pi^-\pi^0$ decays. The decay times are smeared by the expected Belle II decay-time resolution of 140 fs. The second row shows projections of the fitted Dalitz variables $m^2_{K^+\pi^-}$ (left), $m^2_{\pi^-\pi^0}$ (middle), and $m^2_{K^+\pi^0}$ (right).

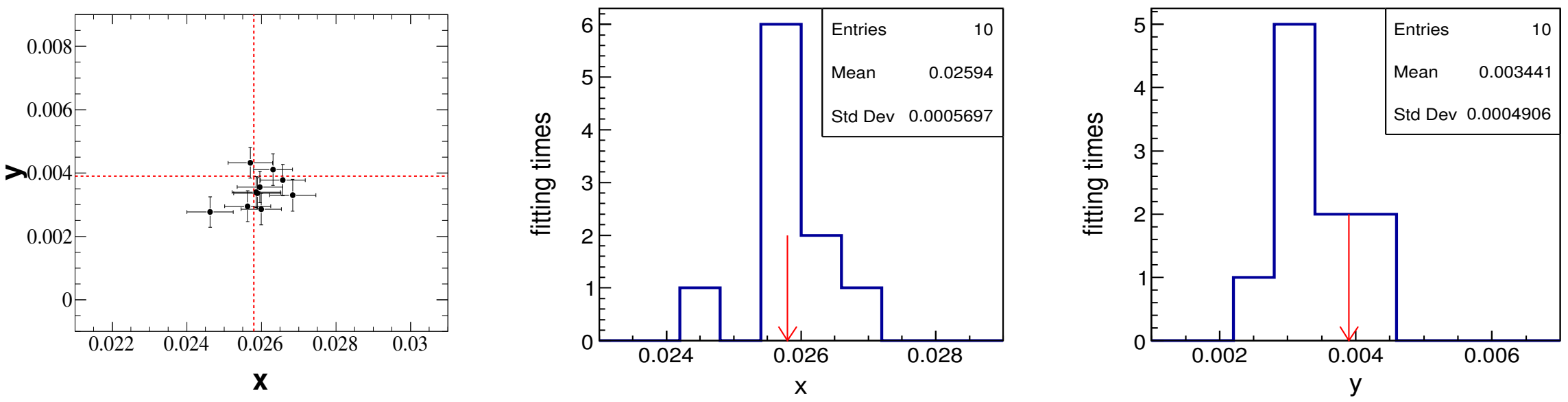

Fig. 157: Residuals resulting from fitting the time-dependent Dalitz plot of WS $D^0 \to K^+\pi^-\pi^0$ decays, for 10 experiments corresponding to 50 ab$^{-1}$ of data (see text). The red lines and arrows show the input (true) values of $x''$ and $y''$.

The systematic errors in this measurement can be classified as "reducible," *i.e.*, those that decrease with increasing data sample size, and "irreducible," *i.e.*, those that do not. The reducible systematic errors in the BaBar analysis [992] were dominated by uncertainty in the RS decay model (0.22 times the statistical uncertainty), uncertainty in the WS decay model (0.22 times the statistical uncertainty), and uncertainties due to the decay time resolution (0.10 times the statistical uncertainty). The irreducible systematic errors were dominated by the choice of fitted decay time range (0.3-0.5 times the statistical uncertainty). This last uncertainty should be notably smaller at Belle II, as the decay time resolution is only half that at BaBar.

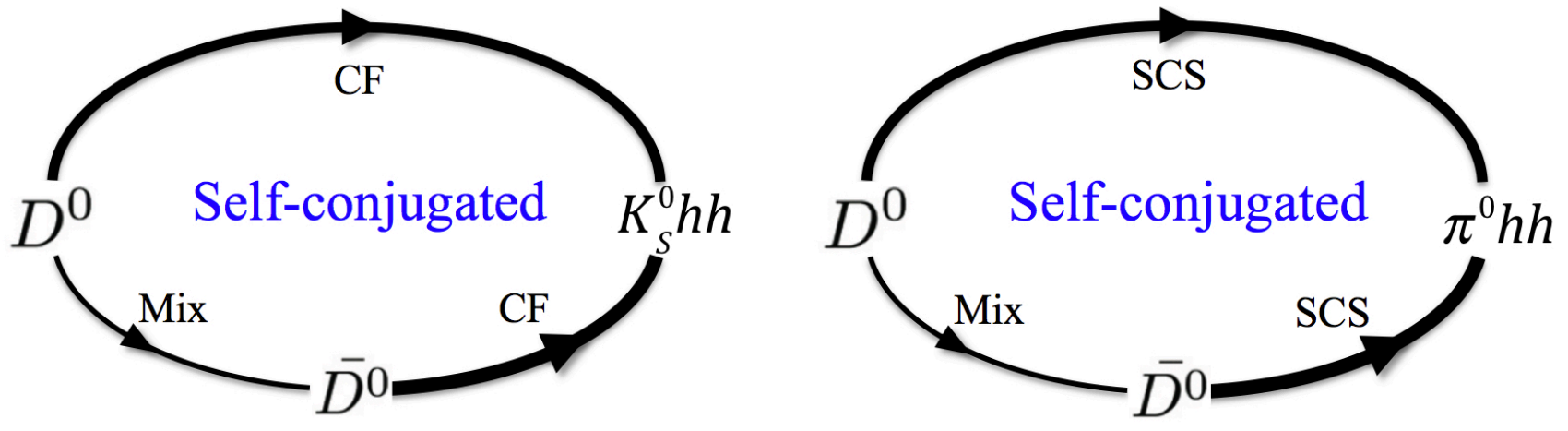

Fig. 158: Interfering amplitudes for $D^0 \to K_S^0\pi^+\pi^-$ (left) and $D^0 \to \pi^0\pi^+\pi^-$ (right).

In summary, Belle II can measure the mixing parameters $x''$ and $y''$ by fitting the time-dependent Dalitz plot of $D^0 \to K^+\pi^-\pi^0$ decays, and the resulting uncertainties should be almost an order-of-magnitude smaller than those obtained previously by BaBar. As the strong phase $\delta_{K\pi\pi^0}$ can be independently measured by BESIII using double-tagged events recorded at the $\psi(3770)$ resonance, this measurement can provide strong constraints on the underlying mixing parameters $x$ and $y$.

*Self-conjugate decays $D^0 \to K_S^0\pi^+\pi^-$.* *Author: L. Li, A. Schwartz*

The self-conjugate decays $D^0 \to K_S^0\pi^+\pi^-$ and $D^0 \to \pi^0\pi^+\pi^-$ proceed via the Cabibbo-favored ($K_S^0$ daughter) and singly-Cabibbo-suppressed ($\pi^0$ daughter) interfering amplitudes shown in Fig. 158. This interference gives sensitivity to $D^0$-$\overline{D}^0$ mixing. By separately fitting samples of $D^*$-flavor-tagged $D^0$ and $\overline{D}^0$ decays, one can also measure indirect $CP$-violating parameters $|q/p|$ and $\phi$.

Denoting the two interfering amplitudes as $A_1$ and $A_2$, the overall decay amplitudes squared have the form

$$
\begin{aligned}
|\mathcal{M}_f|^2 &= \left\{|A_1|^2 e^{-y\Gamma t} + |A_2|^2 e^{y\Gamma t} + 2\Re[A_1 A_2^*]\cos(x\Gamma t) + 2\Im[A_1 A_2^*]\sin(x\Gamma t)\right\} e^{-\Gamma t} \\
|\bar{\mathcal{M}}_f|^2 &= \left\{|\bar{A}_1|^2 e^{-y\Gamma t} + |\bar{A}_2|^2 e^{y\Gamma t} + 2\Re[\bar{A}_1 \bar{A}_2^*]\cos(x\Gamma t) + 2\Im[\bar{A}_1 \bar{A}_2^*]\sin(x\Gamma t)\right\} e^{-\Gamma t}\,.
\end{aligned}
$$

These expresssions show that self-conjugate decays provide sensitivity directly to $x$ and $y$, i.e., without being "rotated" by a strong phase difference between $D^0$ and $\overline{D}^0$ decays. The current status of mixing and $CP$ violation measurements in self-conjugate 3-body decays is summarized in Table 118.

We estimate the sensitivity of Belle II to mixing parameters $x$, $y$, and $CP$-violating parameters $|q/p|$, $\phi$ from $D^0 \to K_S^0\pi^+\pi^-$ decays by scaling from a recent Belle measurement. The Belle analysis reconstructed $1.23 \times 10^6$ $D^*$-flavor-tagged decays in 0.921 ab$^{-1}$ of data. Assuming Belle II has the same reconstruction efficiency, we estimate the Belle II signal yield by scaling by the ratio of luminosities. The result is $67 \times 10^6$ signal decays in 50 ab$^{-1}$ of data. The Belle errors on $x$ and $y$ were $\sigma_x = 0.19\%$ and $\sigma_y = 0.15\%$. We divide these errors into three parts: statistical uncertainty, reducible systematic uncertainty, and irreducible systematic uncertainty. We scale the first two uncertainties by the ratio of luminosities between Belle and Belle II, and then add the result in quadrature to the irreducible systematic uncertainty. The resulting errors are listed in Table. 119. These estimates are conservative, as they do not account for the improved decay-time resolution of Belle II over that of Belle.

Table 118: Measurements of mixing and $CP$ violation in self-conjugate 3-body decays. The errors listed are statistical, systematic, and systematic due to the decay amplitude model.

| | | |
|---|---|---|
| No CPV | $D^0 \to K^0_S\pi^+\pi^-$ (Belle 921 fb$^{-1}$) | $x = \left(+0.56 \pm 0.19\,^{+0.03}_{-0.09}\,^{+0.06}_{-0.09}\right)\%$ <br> $y = \left(+0.30 \pm 0.15\,^{+0.04}_{-0.05}\,^{+0.03}_{-0.06}\right)\%$ |
| | $D^0 \to K^0_S\pi^+\pi^-,\ K^0_S K^+K^-$ (BaBar 469 fb$^{-1}$) | $x = (+0.16 \pm 0.23 \pm 0.12 \pm 0.08)\%$ <br> $y = (+0.57 \pm 0.20 \pm 0.13 \pm 0.07)\%$ |
| | $D^0 \to K^0_S\pi^+\pi^-$ (LHCb 1.0 fb$^{-1}$) | $x = (-0.86 \pm 0.53 \pm 0.17)\%$ <br> $y = (+0.03 \pm 0.46 \pm 0.13)\%$ |
| | $D^0 \to \pi^0\pi^+\pi^-$ (BaBar 468 fb$^{-1}$) | $x = (+1.5 \pm 1.2 \pm 0.6)\%$ <br> $y = (+0.2 \pm 0.9 \pm 0.5)\%$ |
| Indirect CPV | $D^0 \to K^0_S\pi^+\pi^-$ (Belle 921 fb$^{-1}$) | $x = \left(+0.56 \pm 0.19\,^{+0.04}_{-0.08}\,^{+0.06}_{-0.08}\right)\%$ <br> $y = \left(+0.30 \pm 0.15\,^{+0.04}_{-0.05}\,^{+0.03}_{-0.07}\right)\%$ <br> $\lvert q/p\rvert = 0.90\,^{+0.16}_{-0.15}\,^{+0.05}_{-0.04}\,^{+0.06}_{-0.05}$ <br> $\arg(q/p) = \left(-6 \pm 11 \pm 3\,^{+3}_{-4}\right)^\circ$ |

With the high statistics of Belle II, the systematic uncertainty due to the $D^0$ decay model is expected to become the dominant uncertainty. This can be avoided by using strong phase differences measured experimentally, i.e., by BESIII using double-tagged events recorded on the $\psi(3770)$ resonance. Using this method, the authors of Ref. [994] estimate the resulting precision for $x$, $y$. The resulting statistical errors for a sample of $100 \times 10^6$ reconstructed $D^0 \to K^0_S\pi^+\pi^-$ decays are $\sigma_x = 0.017\%$ and $\sigma_y = 0.019\%$. The systematic errors are estimated by propagating the errors on the binned strong phases as measured by CLEOc [770]; the results are $\sigma_x(\text{syst}) = 0.076\%$ and $\sigma_y(\text{syst}) = 0.087\%$. These systematic errors, while larger than the statistical errors, constitute upper bounds, as more precise measurements of the binned strong phases are expected from BESIII.

*Singly Cabibbo-suppressed decays $D^0 \to K^0_S K^\pm\pi^\mp$.* *Author: L. Li, A. Schwartz*

Whereas the CF and DCS amplitudes producing a $K^+\pi^-$ final state give rise to branching fractions that differ by a factor of $\sim 300$, for singly Cabibbo-suppressed (SCS) final states there is approximate equality between the two branching fractions, i.e., $B(D^0 \to K^0_S K^-\pi^+) = (0.35 \pm 0.05)\%$ and $B(D^0 \to K^0_S K^+\pi^-) = (0.26 \pm 0.05)\%$. This similarity implies that the decay amplitudes have similar magnitudes, which in turn gives greater interference between the amplitudes and thus greater sensitivity to mixing and indirect $CP$ violation.

Table 119: Expected precision for mixing parameters $x$, $y$, and $CP$-violating parameters $|q/p|$, $\phi$ from $D^0 \to K_S^0\pi^+\pi^-$ decays.

| **Data** | **stat.** | **syst.** | | **Total** | **stat.** | **syst.** | | **Total** |
|---|---|---|---|---|---|---|---|---|
| | | red. | irred. | | | red. | irred. | |
| | | $\sigma_x$ $(10^{-2})$ | | | | $\sigma_y$ $(10^{-2})$ | | |
| 976 $\text{fb}^{-1}$ | 0.19 | 0.06 | 0.11 | 0.20 | 0.15 | 0.06 | 0.04 | 0.16 |
| 5 $\text{ab}^{-1}$ | 0.08 | 0.03 | 0.11 | 0.14 | 0.06 | 0.03 | 0.04 | 0.08 |
| 50 $\text{ab}^{-1}$ | 0.03 | 0.01 | 0.11 | 0.11 | 0.02 | 0.01 | 0.04 | 0.05 |
| | | $\lvert q/p\rvert$ $(10^{-2})$ | | | | $\phi$ $(^\circ)$ | | |
| 976 $\text{fb}^{-1}$ | 15.5 | 5.2-5.6 | 7.0-6.7 | 17.8 | 10.7 | 4.4-4.5 | 3.8-3.7 | 12.2 |
| 5 $\text{ab}^{-1}$ | 6.9 | 2.3-2.5 | 7.0-6.7 | 9.9-10.1 | 4.7 | 1.9-2.0 | 3.8-3.7 | 6.3-6.4 |
| 50 $\text{ab}^{-1}$ | 2.2 | 0.7-0.8 | 7.0-6.7 | 7.0-7.4 | 1.5 | 0.6 | 3.8-3.7 | 4.0-4.2 |

Experimentally, SCS $D^0 \to K_S^0 K^\pm\pi^\mp$ decays should have greater purity than DCS $D^0 \to K^+\pi^-$ decays due to the larger branching fraction. These SCS decays have been studied by both CLEO [995] and LHCb [996]. An MC study [997] indicates that the precision obtainable for $y$ should be similar to that obtained for $y'$ using $D^0 \to K^+\pi^-$ decays.

## *13.6. $CP$ asymmetries of $D \to PP'$ decays*

### *13.6.1. Theory. Authors: M. Jung, U. Nierste, S. Schacht*

$CP$ asymmetries in non-leptonic $D$ decays have long been considered a quasi-null test of the Standard Model (SM), since they vanish for Cabibbo-allowed and doubly-suppressed modes, and are very small for singly-suppressed decays;[42] the latter is mainly due to the fact that they are strongly CKM-suppressed by the factor $\text{Im}[(V_{cb}^* V_{ub})/(V_{cs}^* V_{us})] \sim 10^{-3}$. However, recent experimental progress – see Table 120 – changed this situation: several measurements achieved sensitivity down to the SM level, such that a potential significant measurement cannot easily be considered NP anymore, apart from exceptional channels like $D^+ \to \pi^+\pi^0$ [998, 999]. The main challenge at this level is therefore the distinction between NP and the SM, taking into account finite contributions from the SM; this has been a main focus in recent theory analyses, see, *e.g.*, Refs. [968, 974, 1000–1022].

Determining the SM contributions precisely turns out to be extremely difficult. The reason is that there is so far no reliable method to determine the corresponding hadronic matrix elements (ME's), related to the fact that the charm quark is neither very heavy nor light compared to a typical QCD scale $\Lambda_{\text{QCD}} \sim 300 - 500$ MeV. This is in contrast to the situation in $B$ or $K$ decays. So far there are also no lattice calculations available for the relevant three-body ME's, see, however, Ref. [189] for recent progress in that direction. Direct calculation can be avoided when employing symmetry methods, specifically the $\text{SU}(3)_F$ and isospin flavour symmetries. Instead of calculating the ME's in question, symmetries *relate* them and thereby allow one to determine them from data. The main concern in such analyses becomes symmetry-breaking contributions. These can be systematically included but yield additional degrees of freedom, complicating the determination of the ME's, as will be discussed below. Nevertheless, they allow for the identification of NP contributions in the presence of a sizable SM background, in the form of sum rules and patterns that hold in the SM but can be violated by NP. This type of analysis relies on experimental input not only of (many) $CP$ asymmetries, but also of branching fractions, and the results improve with more precise inputs.

[42] We do not consider $CP$-violation in the kaon system, which affects decays that produce $K^0$ or $\overline{K}^0$.

Table 120: Most precise measurements of $A_{CP}$ in singly Cabibbo-suppressed $D \to PP'$ decays. For older measurements and world averages, see Ref. [230].

| Mode | $A_{CP}$ [%] | Ref. |
|---|---|---|
| $D^0 \to K^+K^-$ | $0.04 \pm 0.12 \pm 0.10$ | [1023] |
| $D^0 \to \pi^+\pi^-$ | $0.07 \pm 0.14 \pm 0.11$ | [1023] |
| $D^0 \to K^0_S K^0_S$ | $-0.02 \pm 1.53 \pm 0.17$ | [1024] |
| $D^0 \to \pi^0\pi^0$ | $-0.03 \pm 0.64 \pm 0.10$ | [1025] |
| $D^+ \to K^0_S K^+$ | $0.03 \pm 0.17 \pm 0.14$ | [1026] |
| $D^+_s \to K^0_S \pi^+$ | $-0.36 \pm 0.09 \pm 0.07$ | [1027] |

In order to discuss the future key impact of Belle II in this context, we provide an overview of two theoretical frameworks for $D \to PP'$ decays, which are both based on the approximate $\mathrm{SU}(3)_F$ symmetry of QCD. After that we give our predictions for key measurements at Belle II.

*Theoretical framework.* The $\mathrm{SU}(3)_F$-symmetry approach can be implemented in two different ways, whose close connection has been realised from the start [777, 1028, 1029]. In the "plain group theory" approach (see recent Refs. [1002, 1012, 1017–1019]), one uses directly the Wigner-Eckart theorem in order to obtain a decomposition of the amplitudes in terms of reduced $\mathrm{SU}(3)_F$ matrix elements and Clebsch-Gordan coefficients [1030–1032]. In the "dynamical" approach (see recent Refs. [1003, 1004, 1020–1022]), one uses a decomposition of the decay amplitudes in terms of topological diagrams. These diagrams, which are defined by their flavour-flow, are meant to include all-order QCD effects. In both approaches one can include $\mathrm{SU}(3)_F$-breaking effects in a systematic way. These come into play through the different masses of the light quarks. The mass terms can be written as

$$m_u \bar{u}u + m_d \bar{d}d + m_s \bar{s}s \;=\; \mathcal{H}_1 + \mathcal{H}_8^{\not{I}} + \mathcal{H}_8^{\Delta I=0}\,. \tag{459}$$

In the limit $m_u = m_d = m_s$, the $\mathrm{SU}(3)_F$ limit is restored; *i.e.*, only the $\mathrm{SU}(3)_F$-singlet operator $\mathcal{H}_1$ survives, which does not break $\mathrm{SU}(3)_F$. The second operator, $\mathcal{H}_8^{\not{I}} \sim m_u - m_d$, breaks isospin and can be neglected to very good approximation, leaving $\mathcal{H}_8^{\Delta I=0} \sim m_s$ to determine the major part of $\mathrm{SU}(3)_F$-breaking. In order to apply the Wigner-Eckart theorem, initial and final states as well as the relevant Hamiltonian are classified according to their $\mathrm{SU}(3)_F$ representations. The initial states $(D^0, D^+, D_s^+)$ transform as a $\bar{\mathbf{3}}$. The two-body final states of kaons and pions are symmetrised due to a Bose symmetry and transform as

$$(\mathbf{8} \otimes \mathbf{8})_S = \mathbf{1} \oplus \mathbf{8} \oplus \mathbf{27}\,. \tag{460}$$

The SM operators have the flavour structure $\bar{u}c\bar{q}q'$ with $q, q' = d, s$, and correspond to the $\mathrm{SU}(3)_F$ representations

$$\mathbf{3} \otimes \bar{\mathbf{3}} \otimes \mathbf{3} = \mathbf{3}_1 \oplus \mathbf{3}_2 \oplus \bar{\mathbf{6}} \oplus \mathbf{15}\,. \tag{461}$$

Note that this decomposition is only sensitive to the flavour structure of a given effective operator and not to its Dirac structure. The first-order $\mathrm{SU}(3)_F$-breaking representations can be obtained from the tensor products of the $\mathrm{SU}(3)_F$ limit representations with the perturbation operator $\mathcal{H}_8^{\Delta I=0}$. The CKM-leading part of the amplitude of a decay $d$ can then be written as [1018]

$$\mathcal{A}^{\mathrm{C}}(d) = \lambda_C \left( \sum_{i,k} c_{d;ik} A_i^k + \sum_{i,j} c_{d;ij} B_i^j \right)\,, \tag{462}$$

with C=SCS, CF, DCS, and $\lambda_{\mathrm{CF}} \equiv V_{cs}^* V_{ud}$, $\lambda_{\mathrm{DCS}} \equiv V_{cd}^* V_{us}$, and

$$\lambda_{\mathrm{SCS}} \equiv \frac{V_{cs}^* V_{us} - V_{cd}^* V_{ud}}{2}\,. \tag{463}$$

The $A_i^k$ and $B_i^j$ denote the reduced $\mathrm{SU}(3)_F$-limit and -breaking matrix elements for the final state representation $i$ and the operator representation $k, j$, respectively. The Clebsch-Gordan coefficients $c_{d,ik}$ can be found in Ref. [1018]. To give an example, the CKM-leading $\mathrm{SU}(3)_F$-limit decay amplitude of the decay channel $D_s^+ \to K^+\pi^0$ can be written as [1018]

$$\mathcal{A}^{\mathrm{SCS}}(D_s^+ \to K^+\pi^0)/\lambda_{\mathrm{SCS}} = -\frac{1}{5} A_{27}^{15} + \frac{1}{5} A_8^{15} + \frac{1}{\sqrt{10}} A_8^{\bar{6}}\,. \tag{464}$$

Note that, due to linear dependences of the parametrisation, the number of matrix elements can be reduced by redefinitions; see Ref. [1018] for details. The same holds for the parametrisation in terms of topological diagrams [1020]. $\mathcal{A}_X(d)$ is the part of the amplitudes that is obtained when setting all $A_i^k$ in Eq. (462) to zero. In order to measure the maximal $\mathrm{SU}(3)_F$ breaking present in the system of all 17 $D \to P_8 P_8'$ decays, we define

$$\delta_X \equiv \frac{\max_{ij} |B_i^j|}{\max\left( |A_{27}^{15}|, |A_8^{\bar{6}}|, |A_8^{15}| \right)}\,, \text{ and} \tag{465}$$

$$\delta_X' \equiv \max_d \left| \frac{\mathcal{A}_X(d)}{\mathcal{A}(d)} \right|\,, \tag{466}$$

which are complementary measures regarding interference effects between different $\mathrm{SU}(3)_F$-breaking ME's. By performing a global fit to the available data, one can test whether the fit worsens significantly for given values of the measures for maximal $\mathrm{SU}(3)_F$ breaking, $\delta_X^{(\prime)}$, compared to the null hypothesis, which does not constrain the amount of $\mathrm{SU}(3)_F$ breaking. The result is shown in Fig. 159. One finds that the data can be described by $\mathrm{SU}(3)_F$ breaking of $\sim 30\%$, *i.e.* the perturbative expansion is consistent with the data. This of course does not exclude larger values for $\mathrm{SU}(3)_F$ breaking.

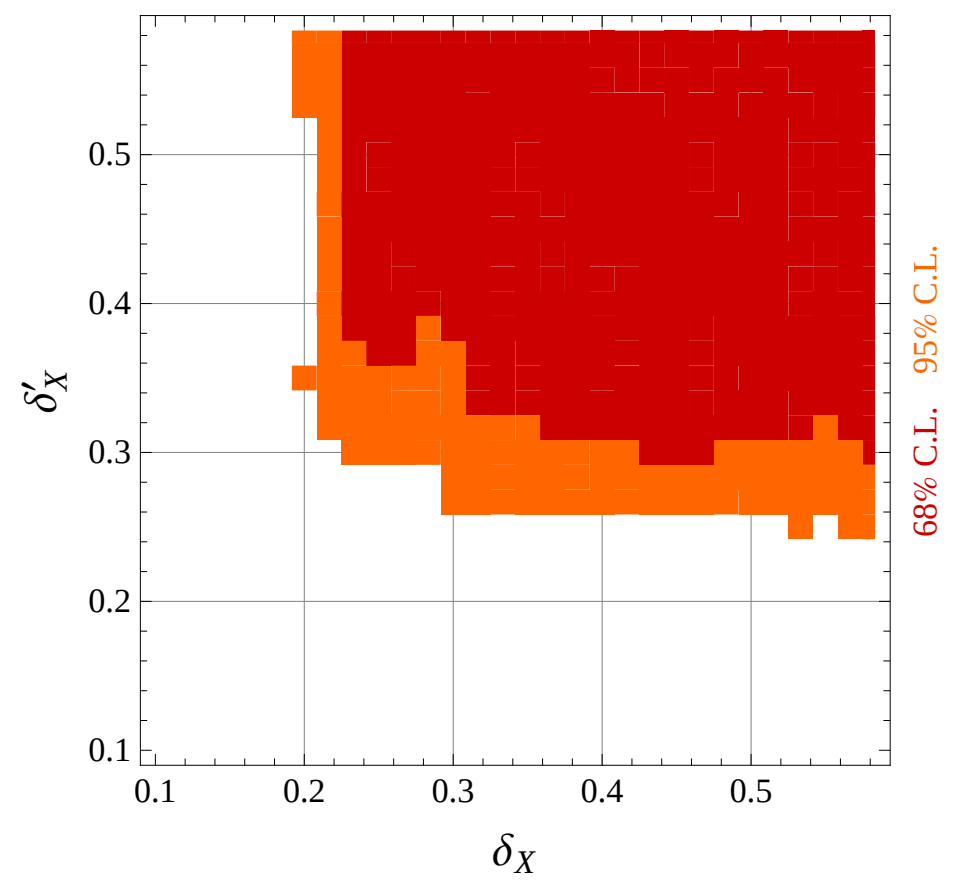

Fig. 159: Allowed regions for $\delta_X$ and $\delta'_X$ at 68% CL (red) and 95% CL (orange) with respect to the global minimum of a fit where the amount of $\mathrm{SU}(3)_F$ breaking is unrestricted. Figure taken from Ref. [1018].

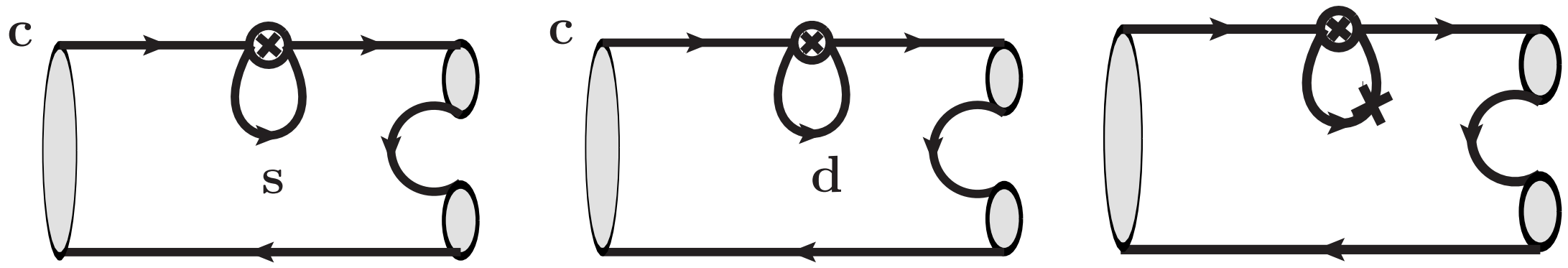

(a) $P_s$ (b) $P_d$ (c) $P_{\mathrm{break}} \equiv P_s - P_d$

Fig. 160: The relevant penguin diagrams. Figures taken from Ref. [1020].

In the language of topological diagrams, the CKM-leading amplitude of the example decay channel $D_s^+ \to K^+\pi^0$ [see Eq. (464)] has the decomposition

$$\frac{\mathcal{A}(D_s^+ \to K^+\pi^0)}{\lambda_{\mathrm{SCS}}} = -\frac{1}{\sqrt{2}}\left(A + A_1^{(1)} + A_2^{(1)}\right) - \frac{1}{\sqrt{2}}\left(C + C_3^{(1)}\right) - \frac{1}{\sqrt{2}}P_{\mathrm{break}}\,, \quad (467)$$

which includes $\mathrm{SU}(3)_F$-breaking topologies denoted by a superscript "$^{(1)}$" and also the $\mathrm{SU}(3)_F$-breaking difference of penguin diagrams [974]

$$P_{\mathrm{break}} \equiv P_s - P_d\,, \quad (468)$$

as shown in Fig. 160. The $\mathrm{SU}(3)_F$-limit topological diagrams are shown in Fig. 161. As a Feynman rule for the perturbation $H_{\not{SU(3)}_F} = (m_s - m_d)\bar{s}s$, we write a cross on the corresponding quark line [1033]. Also, in the topological approach, one can define measures of $\mathrm{SU}(3)_F$ breaking analogous to Eqs. (465) and (466); see Ref. [1020] for details.

The reduced matrix elements of the "plain group theory" approach and the topological diagrams of the "dynamical" approach can be mapped onto each other. For the explicit mapping including linear $\mathrm{SU}(3)_F$ breaking, see Appendix B and Table XVII in Ref. [1020]. In both cases the dependence on the parameters has the same algebraic properties. The rank of the parametrisation is identical, and so are the sum rules between the amplitudes [1012]. Consequently, as long as no dynamical input is assumed for the topological amplitudes, the two frameworks are equivalent to each other. Note, however, that assuming a certain amount of $SU(3)_F$ breaking in the form of ratios of $SU(3)_F$ matrix elements, and in the form of ratios of topological amplitudes, are in general not equivalent [1020]. This is a consequence of the fact that reduced matrix elements are a linear combination of several topological amplitudes, and vice versa.

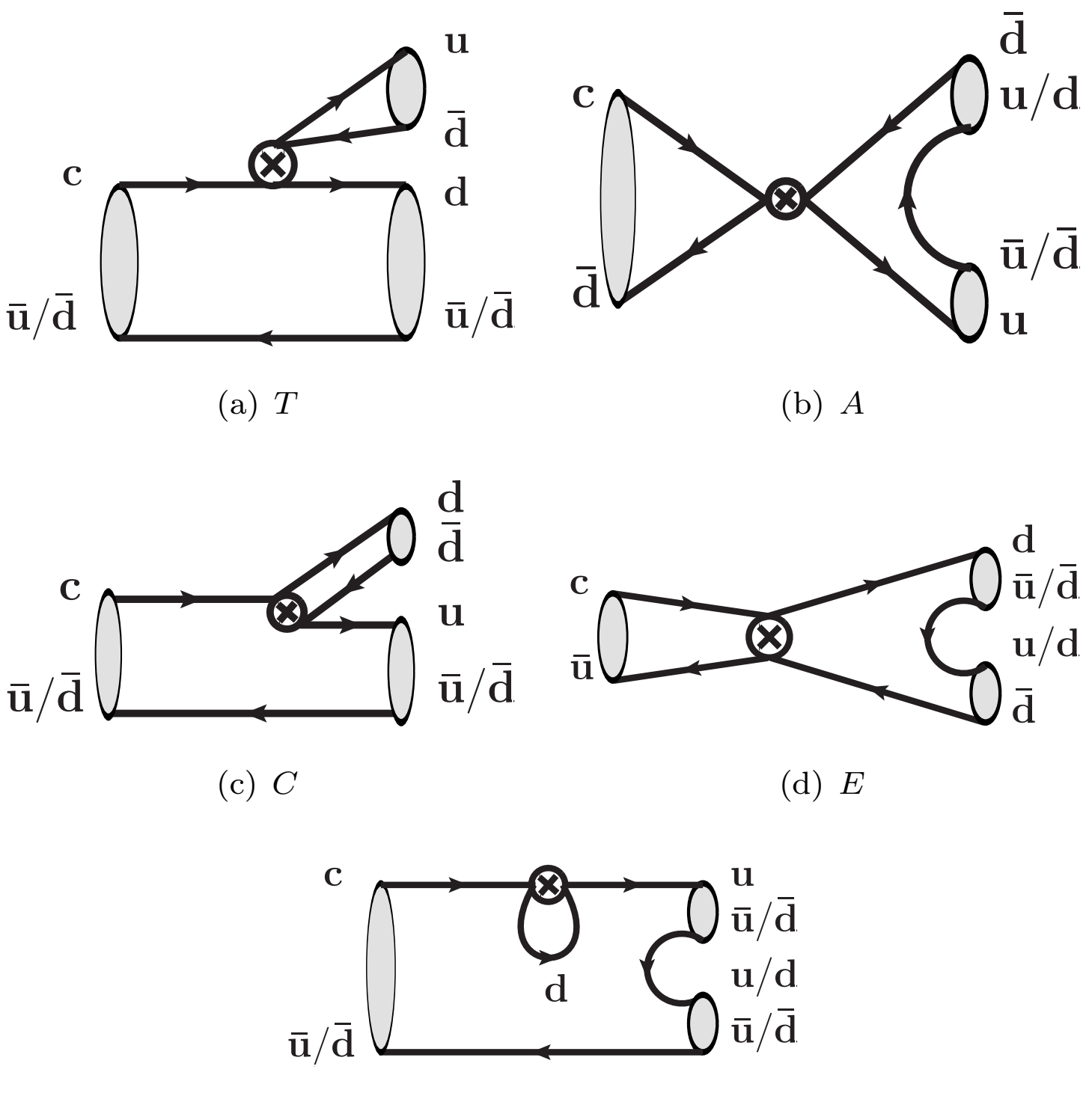

(e) $P_d$

Fig. 161: $SU(3)_F$-limit topological amplitudes. Figures taken from Ref. [1020].

For further dynamical input, one can utilise the $1/N_C$ expansion [1034, 1035]. This is used in Refs. [1020, 1021] for calculating tree and annihilation diagrams, and to estimate upper limits for exchange and colour-suppressed diagrams.

Sum rules between amplitudes as given in Ref. [1012, 1018] have the disadvantage that in most cases the corresponding relative strong phases are unknown. However, there are also sum rules between decay rates [1012] and between $CP$ asymmetries [688, 1021]. These are further discussed in the next section.

*Predictions for key measurements at Belle II .* In this section we discuss several key measurements that Belle II can perform in the field of non-leptonic charm decays. The following discussion is summarized in Table 121.

The difficulty in predicting individual $CP$ asymmetries is due to their dependence on independent combinations of hadronic matrix elements (sums of penguin diagrams), which are not constrained by a fit to branching fractions. The branching fractions contain only differences of these penguin diagrams, up to some heavily CKM-suppressed (*i.e.*, negligible) corrections. Therefore, the corresponding predictions depend strongly on the assumptions regarding these penguin diagrams. The way out of this situation is to consider correlations of $CP$ asymmetries that are determined by sum rules. These sum rules depend only on topologies that can be extracted from a global fit to branching fractions, and thus they eliminate the dependence on the sum of penguins.

Table 121: Key measurements that Belle II can improve. †Our average, statistical and systematic error added in quadrature. *Our estimate from experimental data in Refs. [77, 934, 1036], see text before Eq. (478).

| Observable | Current Measurement | Phenomenological impact |
|---|---|---|
| $A_{CP}(D^0 \to \pi^0\pi^0)$ | $-0.0003 \pm 0.0064$ [230, 1025, 1037] | SM test with $A_{CP}$ sum rule I |
| $A_{CP}(D_s^+ \to K^+\pi^0)$ | $-0.266 \pm 0.238 \pm 0.009$ [1038] | SM test with $A_{CP}$ sum rule II |
| $A_{CP}(D^+ \to \pi^+\pi^0)$ | $0.023 \pm 0.012 \pm 0.002$ [1039] | SM null test |
| $A_{CP}(D^0 \to K_S^0 K_S^0)$ | $-0.004 \pm 0.015$ †[1024, 1037, 1040] | Possible near future observation channel of CP violation |
| $R(D_s^+)$ | $0.02 \pm 0.35$ * | Distinguishing different theoretical treatments |

There are two exceptions to this general observation. First, in the isospin limit we have [998]

$$A_{CP}(D^+ \to \pi^+\pi^0) = 0\,, \tag{469}$$

up to corrections that are expected to be tiny: $\mathcal{O}(\%)$ relative to other $CP$ asymmetries [999]. Second, generically one expects that $A_{CP}(D^0 \to K_S^0 K_S^0)$ is enhanced [1011, 1018, 1022] with respect to other modes for the following reason: the CKM-leading part of the amplitude ($\propto \lambda_{\rm SCS}$) vanishes in the $\mathrm{SU}(3)_F$ limit, while the CKM-suppressed part ($\propto V_{cb}^* V_{ub}$) does not. Furthermore, $A_{CP}(D^0 \to K_S^0 K_S^0)$ receives contributions from a large exchange diagram. Estimating penguin annihilation contributions ($PA$) through a perturbative calculation, the authors of Ref. [1022] find

$$\left|A_{CP}(D^0 \to K_S^0 K_S^0)\right| \;\leq\; 1.1\% \quad (95\%\ \mathrm{CL})\,, \tag{470}$$

*i.e.*, this asymmetry could be as large as one percent. This is of similar size as recent measurements [1024, 1040], which makes this mode very promising for discovering $CP$ violation in the up-quark sector. The observables $A_{CP}(D^+ \to \pi^+\pi^0)$ and $A_{CP}(D^0 \to K_S^0 K_S^0)$ are especially well-suited for Belle II because of the neutral particles in the final state.

Regarding correlations of $CP$ asymmetries, in the $\mathrm{SU}(3)_F$ limit we have

$$a_{CP}^{\rm dir}(D^0 \to K^+K^-) \;=\; -a_{CP}^{\rm dir}(D^0 \to \pi^+\pi^-)\,, \tag{471}$$

$$a_{CP}^{\rm dir}(D^+ \to K_S^0 K^+) \;=\; -a_{CP}^{\rm dir}(D_s^+ \to K^0\pi^+)\,. \tag{472}$$

Taking into account $\mathrm{SU}(3)_F$-breaking corrections, these sum rules can be generalised to contain three direct $CP$ asymmetries each:

I. sum rule among $a_{CP}^{\rm dir}(D^0 \to K^+K^-)$, $a_{CP}^{\rm dir}(D^0 \to \pi^+\pi^-)$, and $a_{CP}^{\rm dir}(D^0 \to \pi^0\pi^0)$;
II. sum rule among $a_{CP}^{\rm dir}(D^+ \to K_S^0 K^+)$, $a_{CP}^{\rm dir}(D_s^+ \to K^0\pi^+)$, and $a_{CP}^{\rm dir}(D_s^+ \to K^+\pi^0)$.

In these sum rules, the coefficients contain only topological amplitudes that can be extracted from the branching fractions. Thus the unknown penguin combination is eliminated. Details are given in Ref. [1021]. To fully benefit from the sum rules, one would perform a global fit to the $D \to PP'$ data. Note that the generalised sum rules do not include $\mathrm{SU}(3)_F$-breaking corrections to the penguin contributions. Consequently, they have an inherent uncertainty of the order of the $\mathrm{SU}(3)_F$ breaking, approximately 30% of the size of the $CP$ asymmetry.

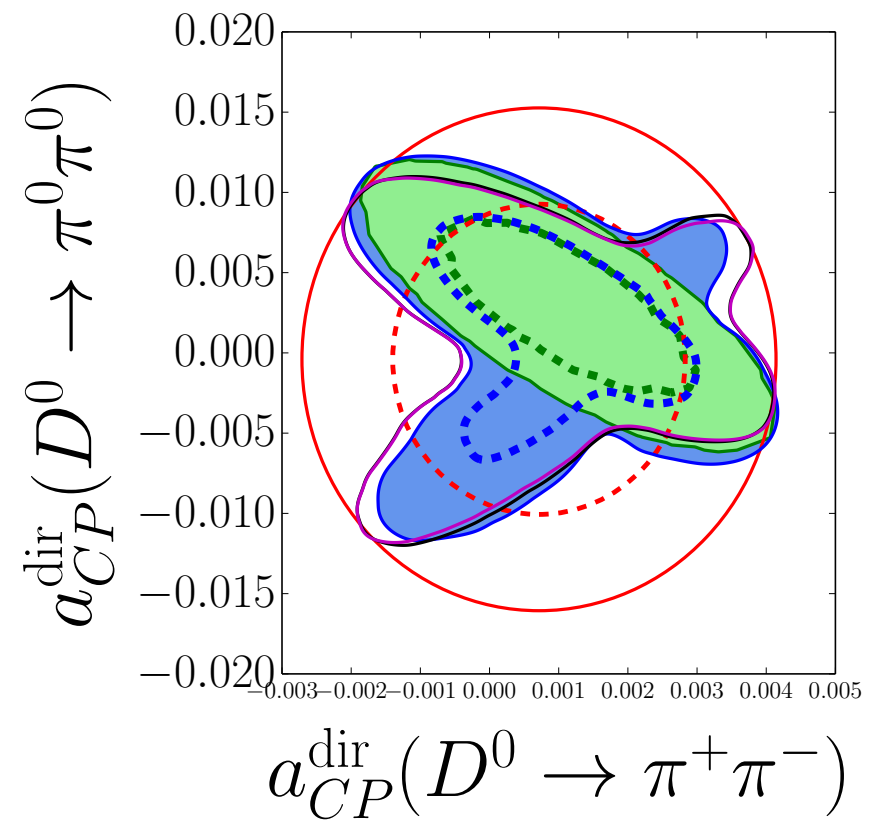

Fig. 162: Correlation between $a_{CP}^{\rm dir}(D^0 \to \pi^0\pi^0)$ and $a_{CP}^{\rm dir}(D^0 \to \pi^+\pi^-)$ from sum rule I. The solid (dashed) lines show allowed 95% (68%) CL regions. In red we show the current experimental data at the date of publication of Ref. [1021]. In blue we show the current fit result, including $1/N_c$ counting for the topological amplitudes. In green we show the same for a future data scenario where the errors of all branching ratios are improved by a factor $\sqrt{50}$. The solid black and magenta curve show current and future data scenario, respectively, without including the $1/N_c$ counting. Note that the magenta and black curve lie partially on top of each other. Figure taken from Ref. [1021].

The correlation between $a_{CP}^{\rm dir}(D^0 \to \pi^+\pi^-)$ and $a_{CP}^{\rm dir}(D^0 \to \pi^0\pi^0)$ resulting from sum rule I is shown in Fig. 162. Also shown are contours corresponding to a future scenario in which uncertainties on the branching fractions are divided by a factor of $\sqrt{50}$. These latter contours illustrate that more precise measurements of branching fractions will significantly improve our knowledge of the uncalculable ME's, and this in turn improves our predictions for $CP$ asymmetries. A similar correlation between direct $CP$ asymmetries appearing in sum rule II can be plotted; however, to test this sum rule to an interesting level, a better measurement of $A_{CP}(D_s^+ \to K^+\pi^0)$ is needed.

Better measurements of branching fractions can also test the theoretical description of the data. This is especially true for DCS modes. A good example is the asymmetry

$$R(D_s^+) \;=\; \frac{\mathcal{B}(D_s^+ \to K_S^0 K^+) - \mathcal{B}(D_s^+ \to K_L^0 K^+)}{\mathcal{B}(D_s^+ \to K_S^0 K^+) + \mathcal{B}(D_s^+ \to K_L^0 K^+)}\,. \tag{473}$$

A fit with the topological-amplitudes method results in [1020]

$$R(D_s^+) = 0.11^{+0.04}_{-0.14} \quad (1\sigma)\,. \tag{474}$$

In the literature, several other predictions for this observable are available. In the $\mathrm{SU}(3)_F$ limit, the authors of Ref. [1041] find

$$R(D_s^+)^{\mathrm{SU}(3)_F\ \mathrm{limit}} = -0.0022 \pm 0.0087\,. \tag{475}$$

Partially including $\mathrm{SU}(3)_F$-breaking effects, the authors of Ref. [1042] find

$$R(D_s^+)^{\text{partial}\,\cancel{\mathrm{SU}(3)_F}} = -0.008 \pm 0.007\,. \tag{476}$$

This result is identical to a result found using QCD factorisation [1043]. (QCD factorisation involves an expansion in $2\Lambda_{\mathrm{QCD}}/m_c$, which, however, is close to one.) Note that the prediction of Eq. (474) has quite large uncertainties; these reflect the conservative treatment of $\mathrm{SU}(3)_F$ breaking, which is allowed to be as large as 50% in the fit.

Belle has measured the combination [934]

$$\mathcal{B}(D_s^+ \to K_S^0 K^+) + \mathcal{B}(D_s^+ \to K_L^0 K^+) \;=\; 0.0295 \pm 0.0011 \pm 0.0009\,, \tag{477}$$

where the $K_S^0$ or $K_L^0$ is identified only by its missing mass. From Eq. (477) and the measurement $\mathcal{B}(D_s^+ \to K_S^0 K^+) = (1.50 \pm 0.05)\%$ [77, 1036], one obtains

$$R(D_s^+)^{\text{exp}} = 0.02 \pm 0.35\,. \tag{478}$$

In the future, more precise data on $R(D_s^+)$ could allow one to quantify $\mathrm{SU}(3)_F$ breaking in this mode.

Deviations from the SM for the observables in Table 121 can occur in several new physics models. For example, the operator with flavour structure $\bar{u}c\bar{u}u$ studied in Ref. [1044] gives the additional $\mathrm{SU}(3)_F$ representations [1018]

$$\mathcal{H}^{\mathrm{NP},\bar{u}c\bar{u}u} = -\frac{V_{cb}^* V_{ub}}{2}\left(\mathbf{15}_{3/2}^{\mathrm{NP}} + \frac{1}{\sqrt{2}}\mathbf{15}_{1/2}^{\mathrm{NP}} + \sqrt{\frac{3}{2}}\mathbf{3}_{1/2}^{\mathrm{NP}}\right). \tag{479}$$

The new quantity in Eq. (479) is the additional $\mathbf{15}_{3/2}^{\mathrm{NP}}$ representation, which comes with a weak phase relative to the $\mathbf{15}_{3/2}$ already present in the SM. This leads to the "smoking-gun" signal for $\Delta I = 3/2$ NP models:

$$A_{CP}(D^+ \to \pi^+\pi^0) \neq 0\,. \tag{480}$$

As another example, some models [1005] give rise to an operator with flavour structure $\bar{s}c\bar{u}s$ without a corresponding operator $\bar{d}c\bar{u}d$, breaking the discrete $U$ spin symmetry of $\mathcal{H}$. This results in the representations

$$\mathcal{H}^{\mathrm{NP},\bar{s}c\bar{u}s} = -\frac{V_{cb}^* V_{ub}}{2}\left(\sqrt{\frac{3}{2}}\mathbf{15}_{1/2}^{\mathrm{NP}} - \bar{\mathbf{6}}_{1/2}^{\mathrm{NP}} - \sqrt{\frac{3}{2}}\mathbf{3}_{1/2}^{\mathrm{NP}}\right). \tag{481}$$

A sign of this model would be violation of the $CP$ asymmetry sum rules, for example, a deviation from the SM prediction in Fig. (162). A proof of principle for distinguishing the above NP models from the SM has been given in Ref. [1018].

*Conclusions.* We have highlighted key measurements of non-leptonic charm decays where Belle II is expected to have a large impact. The individual $CP$ asymmetries $A_{CP}(D^+ \to \pi^+\pi^0)$ and $A_{CP}(D^0 \to K_S^0 K_S^0)$ can be used on their own as a test of the SM and as a discovery channel for $CP$ violation in the up-quark sector, respectively. Due to the difficulty of estimating penguin diagrams, the best way to study further $CP$ asymmetries is to look at their correlations. In order to test the $CP$ asymmetry sum rules, further improvement of $A_{CP}(D^0 \to \pi^0\pi^0)$ and the poorly measured $A_{CP}(D_s^+ \to K^+\pi^0)$ is particularly important. Also, time-dependent measurements of $CP$ asymmetries for $D^0$ decays would be helpful [1022, 1045]. In addition, future measurements of the asymmetry $R(D_s^+)$ could distinguish different theoretical approaches to SM predictions for charm decays.

*13.6.2. Experiment. Authors: G. Casarosa, A. J. Schwartz*

The Belle II experiment is ideal for searching for time-integrated $CP$-violating effects in a variety of final states, and will reach precisions of the order of 0.01% level. The LHCb experiment has provided extremely precise measurements of $CP$ asymmetries in decays with charged particles in the final state. The excellent $\gamma$ and $\pi^0$ reconstruction (and thus $\eta$, $\eta'$, and $\rho^+$ reconstruction) will allow Belle II to search for $CP$ violation in complementary final states that contain neutral particles. The high flavour-tagging efficiency with low dilution, the numerous control samples with which to study systematics, in addition to the excellent reconstruction of charged particle will allow Belle II to compete with LHCb in measurements of time-integrated $CP$ violation.

*Extrapolations from Belle Measurements.* A listing of $D^0$, $D^+$, and $D_s^+$ decay modes that Belle has studied is given in Table 122. The table lists the $CP$ asymmetry $A_{CP} = [\Gamma(D^0 \to f) - \Gamma(\overline{D}^0 \to \bar{f})]/[\Gamma(D^0 \to f) + \Gamma(\overline{D}^0 \to \bar{f})]$ measured by Belle, and the precision expected for Belle II. The latter is estimated by scaling the Belle statistical error ($\sigma_{\rm stat}$) by the ratio of integrated luminosities, and by dividing the systematic error into those that scale with luminosity such as background shapes measured with control samples ($\sigma_{\rm syst}$), and those that do not scale with luminosity such as decay time resolution due to detector misalignment ($\sigma_{\rm irred}$). The overall error estimate is calculated as

$$\sigma_{\rm Belle\ II} = \sqrt{(\sigma^2_{\rm stat} + \sigma^2_{\rm syst}) \cdot (\mathcal{L}_{\rm Belle}/50\ {\rm ab}^{-1}) + \sigma^2_{\rm irred}}.$$

For most of the decay modes listed, the expected uncertainty on $A_{CP}$ is $\lesssim 0.10\%$.

It is important to note that the estimates in this table do not take into account expected improvements in reconstruction at Belle II with respect to Belle, and the additional flavour-tagging techniques that we have presented above. Also, the different background conditions are not included in the estimation.

*Systematic Uncertainties.* In this section we briefly discuss the main sources of systematic errors and we classify them as *reducible* or *irreducible* as they have been treated in Table 122.

Detection Asymmetry The determination of the flavour or charge of the $D$ meson is achieved by looking at the charge of a final state pion or kaon. Any asymmetry in the detection and reconstruction between positive and negative tracks will induce a fake asymmetry in the final measurement. The detection asymmetry is measured in data using control samples, with different techniques employed by Belle [1046] and BaBar [1047], and a correction is applied to the number of $D$ and $\overline{D}$ candidates, usually by weighting events. This correction depends on the direction and momentum of the track and can be as large as 3%. The associated systematic error depends on the precision to which the correction is determined, which in turn depends on the statistics of the control sample used. Hence it is a *reducible* systematic error.

Forward-Backward Asymmetry The interference between the photon and the $Z$ bosons in the $e^+e^-$ interaction induces a forward-backward asymmetry ($A_{\rm FB}$) in the direction of the $c$ (and consequently $\bar{c}$) quark. A fake $CP$ asymmetry is induced by a non-null $A_{\rm FB}$ coupled with the Belle II detector acceptance. In order to decouple $A_{\rm FB}$ from $A_{CP}$, the asymmetry measurement is performed in bins of the cosine of the polar angle in the centre-of-mass of the charmed meson ($\cos\theta^*$). This angle is a good approximation for the direction of the charmed quark. The forward-backward asymmetry is an odd function of $\cos\theta^*$, while the $CP$ asymmetry is of course independent; thus the value of $A_{CP}$ can be extracted with a fit with a constant to the averages of the number-of-candidates asymmetry in symmetric bins of $\cos\theta^*$. This is also clearly a *reducible* systematic error since with more data the determination of $A_{\rm FB}$ will be more precise and less dependent on the binning in $\cos\theta^*$ and the impact on the determination of the $CP$ asymmetry will be smaller.

Table 122: Time-integrated $CP$ asymmetries measured by Belle, and the precision expected for Belle II in 50 ab$^{-1}$ of data.

| Mode | $\mathcal{L}$ (fb$^{-1}$) | $A_{CP}$ (%) | Belle II 50 ab$^{-1}$ |
|---|---|---|---|
| $D^0 \to K^+K^-$ | 976 | $-0.32 \pm 0.21 \pm 0.09$ | $\pm 0.03$ |
| $D^0 \to \pi^+\pi^-$ | 976 | $+0.55 \pm 0.36 \pm 0.09$ | $\pm 0.05$ |
| $D^0 \to \pi^0\pi^0$ | 966 | $-0.03 \pm 0.64 \pm 0.10$ | $\pm 0.09$ |
| $D^0 \to K_S^0\pi^0$ | 966 | $-0.21 \pm 0.16 \pm 0.07$ | $\pm 0.02$ |
| $D^0 \to K_S^0 K_S^0$ | 921 | $-0.02 \pm 1.53 \pm 0.02 \pm 0.17$ | $\pm 0.23$ |
| $D^0 \to K_S^0\eta$ | 791 | $+0.54 \pm 0.51 \pm 0.16$ | $\pm 0.07$ |
| $D^0 \to K_S^0\eta'$ | 791 | $+0.98 \pm 0.67 \pm 0.14$ | $\pm 0.09$ |
| $D^0 \to \pi^+\pi^-\pi^0$ | 532 | $+0.43 \pm 1.30$ | $\pm 0.13$ |
| $D^0 \to K^+\pi^-\pi^0$ | 281 | $-0.60 \pm 5.30$ | $\pm 0.40$ |
| $D^0 \to K^+\pi^-\pi^+\pi^-$ | 281 | $-1.80 \pm 4.40$ | $\pm 0.33$ |
| $D^+ \to \phi\pi^+$ | 955 | $+0.51 \pm 0.28 \pm 0.05$ | $\pm 0.04$ |
| $D^+ \to \pi^+\pi^0$ | 921 | $+2.31 \pm 1.24 \pm 0.23$ | $\pm 0.17$ |
| $D^+ \to \eta\pi^+$ | 791 | $+1.74 \pm 1.13 \pm 0.19$ | $\pm 0.14$ |
| $D^+ \to \eta'\pi^+$ | 791 | $-0.12 \pm 1.12 \pm 0.17$ | $\pm 0.14$ |
| $D^+ \to K_S^0\pi^+$ | 977 | $-0.36 \pm 0.09 \pm 0.07$ | $\pm 0.02$ |
| $D^+ \to K_S^0 K^+$ | 977 | $-0.25 \pm 0.28 \pm 0.14$ | $\pm 0.04$ |
| $D_s^+ \to K_S^0\pi^+$ | 673 | $+5.45 \pm 2.50 \pm 0.33$ | $\pm 0.29$ |
| $D_s^+ \to K_S^0 K^+$ | 673 | $+0.12 \pm 0.36 \pm 0.22$ | $\pm 0.05$ |

$K_S^0$ in the final state  Many interesting channels have a $K_S^0$ in the final state. Particular attention must be paid in these cases because the final $CP$ asymmetry will contain a contribution due to $CP$ violation in the $K^0\overline{K}^0$ system. This contribution can be calculated, *e.g.*, for the Belle $D^+ \to K_S^0 K^+$ analysis [1048] it is $-(0.328 \pm 0.006)\%$.

Moreover additional contributions to the measured asymmetry arise from:

(1) $K^0\overline{K}^0$ regeneration effect in the material;
(2) $K_S^0 - K_L^0$ interference.

These are conservatively considered *irreducible* systematics. However, for Belle II it may be possible to reduce these with further simulation studies, as the physics underlying them is well-understood. (For Belle and BaBar analyses, it was not necessary to reduce these as the statistical error dominated.)

Other Contributions A few small systematic contributions related to the selection of events and the selection of the control sample used to correct for the detection asymmetry are also included. Other systematic contributions are related to the fitting of invariant mass distribution used to extract the number of D candidates, *e.g.* the estimation of the PDF of the background events from the sidebands, and extraction of the signal PDF resolution parameters. These contributions are usually all reducible, and they do not dominate the total error. With a much larger data sample their impact should be re-evaluated.

*Expected Improvements.* As described in section 13.2, we expect that the reconstruction efficiency for charm will be higher at Belle II than it was at Belle or BaBar. The improved vertex resolution should result in improved rejection of random combinations of tracks. The improved tracking and the extended radius of the drift chamber should improve the $K^0_S$ reconstruction efficiency, and also that of converted photons. The improved discrimination between $K^\pm$ and $\pi^\pm$ will reduce cross-feed backgrounds among topologically similar decay modes. Moreover, the measurements will benefit from the improved photon and $\pi^0$ reconstruction. However, the overall increase of backgrounds complicates simply extrapolating the overall reconstruction improvements of Belle II.

The high statistics that Belle II is expected to accumulate will allow to estimate the $CP$ asymmetry in bins of proper time. This technique has already been used at LHCb and is expected be available at Belle II both for $D^*$-tagged $D^0$ decaying to at least two charged tracks (or one track and a $K^0_S$) and for prompt charmed mesons decaying to at least one charged track, or one $K^0_S$.

Impact of the ROE flavour-tagging method Beside the usual $D^{*+}$-tagged sample of $D^0$ mesons, and additional sample will be available, exploiting the ROE flavour-tagging method. The overlap of the two samples is estimated to be $< 3\%$, and the overlapping events are assigned to only one of the two samples so that the samples are independent. We can conservatively estimate the reduction of the statistical error on $A_{CP}$ assuming two independent $A_{CP}$ measurements: one performed on the $D^{*+}$-tagged sample with a statistical error $\sigma^*_{A_{CP}}$, and the other one performed on the ROE-tagged sample with a statistical error $\sigma^0_{A_{CP}}$. We can then evaluate the error of the combination of the two. The ratio of statistical errors of the two independent measurements is:

$$\frac{\sigma^0_{A_{CP}}}{\sigma^*_{A_{CP}}} = \sqrt{\frac{1}{3}\cdot\frac{Q^*}{Q^0}\cdot\frac{\rho^*_{\rm reco}}{\rho^0_{\rm reco}}} \equiv \alpha\,, \tag{482}$$

where $Q = \epsilon_{\rm tag}(1-2\omega)^2$ is the effective tagging efficiency, $\rho_{\rm reco}$ is the purity of the reconstructed sample, the symbol 0 ($*$) denotes the sample tagged with the ROE ($D^{*+}$) method, and the factor of three in the denominator accounts for the different number of generated prompt $D^0$ versus $D^{*+}$. Values of $Q$ are listed in Table 108.

If $\sigma^c_{A_{CP}}$ is the statistical uncertainty associated to $A_{CP}$ obtained combining the two independent measurements, we have

$$\frac{\sigma^c_{A_{CP}}}{\sigma^*_{A_{CP}}} = \frac{\alpha}{\sqrt{1+\alpha^2}} \quad \text{, with } \alpha \equiv \alpha(\rho^*_{\rm reco}/\rho^0_{\rm reco})$$

The value of $\rho^*_{\rm reco}/\rho^0_{\rm reco}$ depends on the reconstruction performance of the detector and also on the final state. We conservatively estimate the ratio $\rho^*_{\rm reco}/\rho^0_{\rm reco} = 1.4$ from a BaBar analysis [1049]. The plots of the two ratios $\sigma^0_{A_{CP}}/\sigma^*_{A_{CP}}$ and $\sigma^c_{A_{CP}}/\sigma^*_{A_{CP}}$ as functions of $\alpha$ are shown in Fig. 163, for different values of $Q^0$. The best scenario is obtained for criteria A, since the tagging efficiency $\epsilon^0_{\rm tag}$ is significantly higher than that for the other two cases (see Table 108). We estimate that the error on $A_{CP}$ would be reduced by $\sim 15\%$ if, in addition to using $D^*$ decays for flavour tagging, one also used the ROE method for flavour tagging. Since $\sigma_{A_{CP}} \propto 1/\sqrt{\mathcal{L}}$, where $\mathcal{L}$ is the integrated luminosity, one can interpret this reduction of the statistical uncertainty as being equivalent to an effective increase in integrated luminosity of $(1/0.85)^2 - 1 = 38\%$.

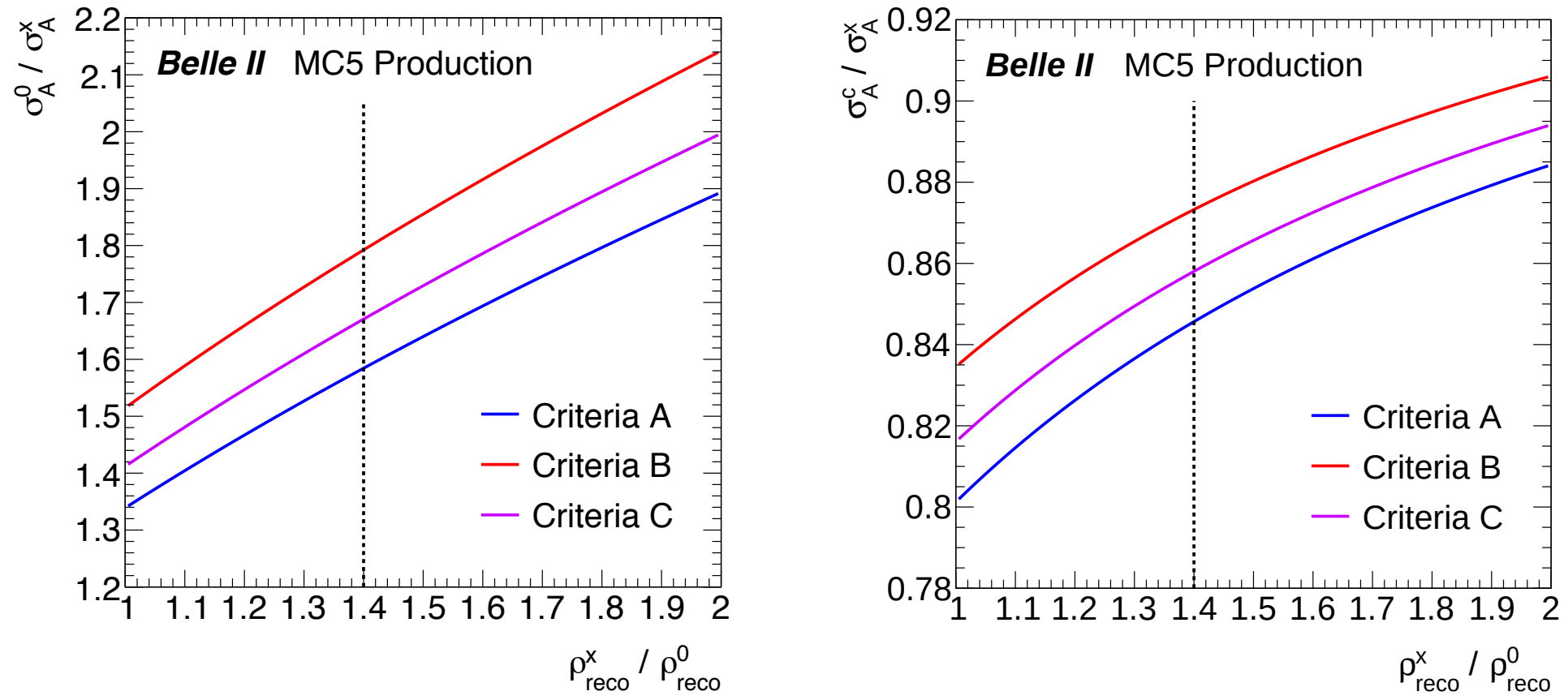

Fig. 163: Left: trend of $\sigma^0_{A_{CP}}/\sigma^*_{A_{CP}}$ as a function of $\rho^*_{\rm reco}/\rho^0_{\rm reco}$. Right: trend of $\sigma^c_{A_{CP}}/\sigma^*_{A_{CP}}$ as a function of $\rho^*_{\rm reco}/\rho^0_{\rm reco}$. In both cases the vertical dotted line is drawn at the value $\rho^*_{\rm reco}/\rho^0_{\rm reco} = 1.4$. Criteria A-C are described in Section 13.2.1 and Table 108.

A further improvement in the precision can be achieved with a simultaneous fit to the two separately tagged $D^0$ samples, instead of the *a posteriori* combination of two measurements. A quantitative estimation is not straightforward, but previous analyses [1049] have shown that a reduction of the statistical error of up to 5% is possible.

*Estimate of $\delta A_{CP}$ for $D^0 \to K^0_S K^0_S$ and $D^+ \to \pi^+\pi^0$.* In this section we discuss two channels that have generated much theoretical interest:

(1) $D^0 \to K^0_S K^0_S$: $CP$ violation may be as large as 1% in the SM;
(2) $D^+ \to \pi^+\pi^0$: no $CP$ violation is expected in the SM.

The first decay mode is a promising channel for obtaining first evidence of $CP$ violation in the charm sector. The second decay mode is of particular interest in the search for NP contributions.

Both Belle [1024] and LHCb [1040] have measured $A_{CP}$ for $D^0 \to K^0_S K^0_S$ decays using their full or current datasets ($921\,{\rm fb}^{-1}$ and $3\,{\rm fb}^{-1}$ respectively). The statistical and systematic errors of the Belle measurement are significantly smaller than those of LHCb, and they will improve further at Belle II. The Belle result is $A_{CP}(D^0 \to K^0_S K^0_S) = (-0.02 \pm 1.53 \pm 0.17)\%$. The systematic uncertainty is very small because the measurement is normalized to the asymmetry in the $D^0 \to K^0_S \pi^0$ channel:

$$A_{CP}(D^0 \to K^0_S K^0_S) \;=\; A_{\rm raw}(K^0_S K^0_S) - A_{\rm raw}(K^0_S \pi^0) + A_{CP}(K^0_S \pi^0)\,, \qquad (483)$$

where $A_{\rm raw}(f) = [N(D^0 \to f) - N(\overline{D}^0 \to f)]/[N(D^0 \to f) + N(\overline{D}^0 \to f)]$ is the raw asymmetry. Note that the asymmetry that arises from the difference in strong interactions in material of $K^0$ and $\overline{K}^0$ is null in this channel, as the final state is $K^0\overline{K}^0$. The systematic error is almost completely reducible, as it is dominated by the error on $A_{CP}(D^0 \to K_S^0\pi^0)$ (which will itself be improved at Belle II). Scaling the statistical error with luminosity and adding in quadrature the expected error on $A_{CP}(D^0 \to K_S^0\pi^0)$, we obtain an uncertainty for $50\,\text{ab}^{-1}$ of data of 0.23%. This precision is notably smaller than the allowed window for $CP$ violation in this decay, Eq. (470).

For $D^+ \to \pi^+\pi^0$ decays, to estimate the Belle II precision for $A_{CP}$ we perform an MC simulation study. To reduce backgrounds, we require that the $D^+$ originate from $D^{*+} \to D^+\pi^0$ decays. We then apply a set of preliminary selection criteria in order to reduce the main sources of background. The resulting distribution of energy released $Q \equiv M(D^+\pi^0) - M(D^+) - M_{\pi^0}$, where $M(D^+\pi^0)$ is the $\pi^+\pi^0\pi^0$ invariant mass and $M(D^+)$ is the $\pi^+\pi^0$ invariant mass, is plotted in Fig. 164. Both signal and background events plotted correspond to $50\,\text{ab}^{-1}$ of data; the signal curve is scaled by a factor of 10 for greater visibility. Misreconstructed signal candidates are included in the signal distribution. At this stage, a selection efficiency of approximately 30% and a background rejection of 96-99% are achieved. This performance is good, but, as this study was done with an early version of the Belle II reconstruction code (release-00-06-00), we expect further improvements as the recontruction code is developed. As the signal-to-background ratio shown in Fig. 164 is similar to that achieved at Belle, we estimate the sensitivity of Belle II by simply scaling the Belle errors by the square root of the ratio of luminosities. The result is an uncertainty for $50\,\text{ab}^{-1}$ of data of 0.17%.

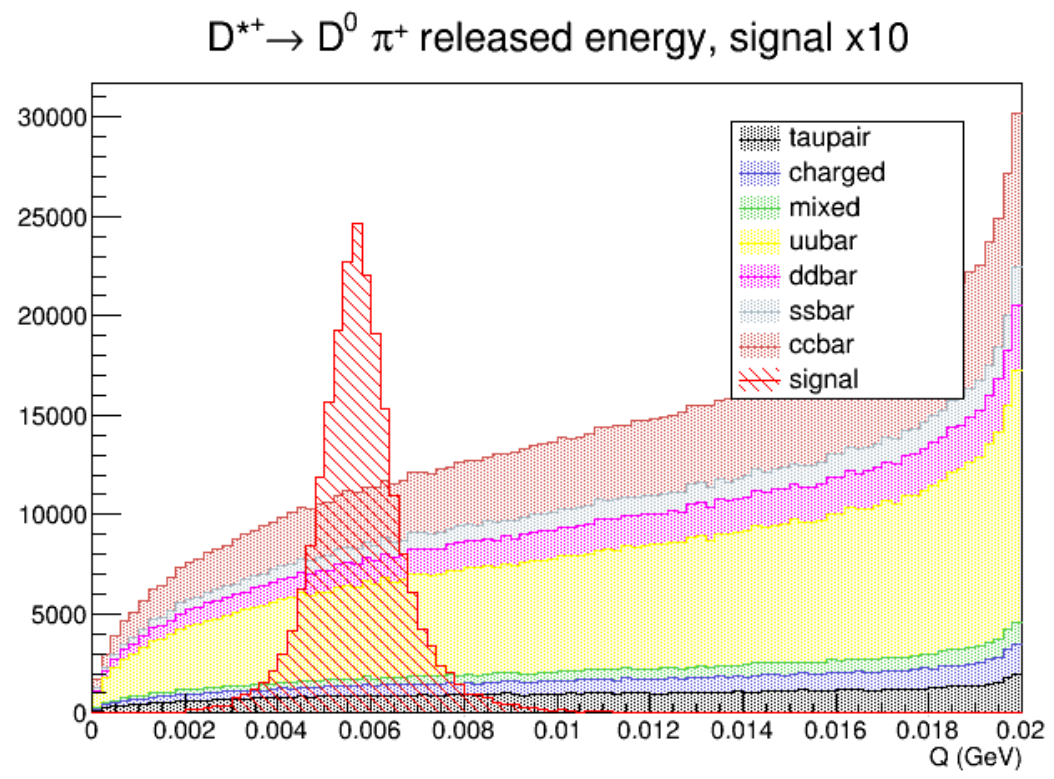

Fig. 164: Energy released $Q \equiv M(D^+\pi^0) - M(D^+) - M_{\pi^0}$ , for $D^{*+} \to D^+\pi^0$, $D^+ \to \pi^+\pi^0$ decays after a preliminary selection. The signal distribution (in red) is multiplied by 10. The overall background distribution consists of stacked contributions from different event types.

## *13.7. The Golden Channels*

In summary, the charm physics program of Belle II is quite broad, covering mixing, both indirect and direct $CP$ violation, semileptonic decays, hadronic decays, rare and forbidden decays, etc. There is substantial discovery potential. We list some of the most promising channels and measurements for discovering NP in Table 123, and refer to this listing as the "golden channels."

Table 123: The "golden channels" for charm physics.

| **Channel** | **Observable** | **Belle/BaBar Measurement** | | **Scaled** | |
|---|---|---|---|---|---|
| | | $\mathcal{L}$ [ab$^{-1}$] | Value | 5 ab$^{-1}$ | 50 ab$^{-1}$ |
| | | **Leptonic Decays** | | | |
| $D_s^+ \to \ell^+\nu$ | $\mu^+$ events | 0.913 | $492 \pm 26$ | 2.7k | 27k |
| | $\tau^+$ events | | $2217 \pm 83$ | 12.1k | 121k |
| | $f_{D_s}$ | | 2.5% | 1.1% | 0.34% |
| $D^+ \to \ell^+\nu$ | $\mu^+$ events | - | - | 125 | 1250 |
| | $f_D$ | - | - | 6.4% | 2.0% |
| | | **Rare and Radiative Decays** | | | |
| $D^0 \to \rho^0 \gamma$ | $A_{CP}$ | 0.943 | $+0.056 \pm 0.152 \pm 0.006$ | $\pm 0.07$ | $\pm 0.02$ |
| $D^0 \to \phi \gamma$ | $A_{CP}$ | | $-0.094 \pm 0.066 \pm 0.001$ | $\pm 0.03$ | $\pm 0.01$ |
| $D^0 \to \overline{K}^{*0} \gamma$ | $A_{CP}$ | | $-0.003 \pm 0.020 \pm 0.000$ | $\pm 0.01$ | $\pm 0.003$ |
| | | **Mixing and Indirect (time-dependent) $CP$ Violation** | | | |
| $D^0 \to K^+\pi^-$ | $x'^2$ (%) | 0.976 | $0.009 \pm 0.022$ | $\pm 0.0075$ | $\pm 0.0023$ |
| (no $CPV$) | $y'$ (%) | | $0.46 \pm 0.34$ | $\pm 0.11$ | $\pm 0.035$ |
| ($CPV$ allowed) | $\|q/p\|$ | World Avg. [230] | $0.89\,^{+0.08}_{-0.07}$ | $\pm 0.20$ | $\pm 0.05$ |
| | $\phi\,(^\circ)$ | with LHCb | $-12.9\,^{+9.9}_{-8.7}$ | $\pm 16^\circ$ | $\pm 5.7^\circ$ |
| $D^0 \to K^+\pi^-\pi^0$ | $x''$ | 0.384 | $2.61\,^{+0.57}_{-0.68} \pm 0.39$ | - | $\pm 0.080$ |
| | $y''$ | | $-0.06\,^{+0.55}_{-0.64} \pm 0.34$ | - | $\pm 0.070$ |
| $D^0 \to K^0_S\pi^+\pi^-$ | $x$ (%) | 0.921 | $0.56 \pm 0.19\,^{+0.04}_{-0.08}\,^{+0.06}_{-0.08}$ | $\pm 0.16$ | $\pm 0.11$ |
| | $y$ (%) | | $0.30 \pm 0.15\,^{+0.04}_{-0.05}\,^{+0.03}_{-0.07}$ | $\pm 0.10$ | $\pm 0.05$ |
| | $\|q/p\|$ | | $0.90\,^{+0.16}_{-0.15}\,^{+0.05}_{-0.04}\,^{+0.06}_{-0.05}$ | $\pm 0.12$ | $\pm 0.07$ |
| | $\phi\,(^\circ)$ | | $-6 \pm 11 \pm 3\,^{+3}_{-4}$ | $\pm 8$ | $\pm 4$ |
| | | **Direct (time-integrated) $CP$ Violation in %** | | | |
| $D^0 \to K^+K^-$ | $A_{CP}$ | 0.976 | $-0.32 \pm 0.21 \pm 0.09$ | $\pm 0.10$ | $\pm 0.03$ |
| $D^0 \to \pi^+\pi^-$ | $A_{CP}$ | 0.976 | $+0.55 \pm 0.36 \pm 0.09$ | $\pm 0.16$ | $\pm 0.05$ |
| $D^0 \to \pi^0\pi^0$ | $A_{CP}$ | 0.966 | $-0.03 \pm 0.64 \pm 0.10$ | $\pm 0.28$ | $\pm 0.09$ |
| $D^0 \to K^0_S\,\pi^0$ | $A_{CP}$ | 0.966 | $-0.21 \pm 0.16 \pm 0.07$ | $\pm 0.08$ | $\pm 0.02$ |
| $D^0 \to K^0_S\,K^0_S$ | $A_{CP}$ | 0.921 | $-0.02 \pm 1.53 \pm 0.17$ | $\pm 0.66$ | $\pm 0.23$ |
| $D^0 \to K^0_S\,\eta$ | $A_{CP}$ | 0.791 | $+0.54 \pm 0.51 \pm 0.16$ | $\pm 0.21$ | $\pm 0.07$ |
| $D^0 \to K^0_S\,\eta'$ | $A_{CP}$ | 0.791 | $+0.98 \pm 0.67 \pm 0.14$ | $\pm 0.27$ | $\pm 0.09$ |
| $D^0 \to \pi^+\pi^-\pi^0$ | $A_{CP}$ | 0.532 | $+0.43 \pm 1.30$ | $\pm 0.42$ | $\pm 0.13$ |
| $D^0 \to K^+\pi^-\pi^0$ | $A_{CP}$ | 0.281 | $-0.60 \pm 5.30$ | $\pm 1.26$ | $\pm 0.40$ |
| $D^0 \to K^+\pi^-\pi^+\pi^-$ | $A_{CP}$ | 0.281 | $-1.80 \pm 4.40$ | $\pm 1.04$ | $\pm 0.33$ |
| $D^+ \to \phi\pi^+$ | $A_{CP}$ | 0.955 | $+0.51 \pm 0.28 \pm 0.05$ | $\pm 0.12$ | $\pm 0.04$ |
| $D^+ \to \pi^+\pi^0$ | $A_{CP}$ | 0.921 | $+2.31 \pm 1.24 \pm 0.23$ | $\pm 0.54$ | $\pm 0.17$ |
| $D^+ \to \eta\pi^+$ | $A_{CP}$ | 0.791 | $+1.74 \pm 1.13 \pm 0.19$ | $\pm 0.46$ | $\pm 0.14$ |
| $D^+ \to \eta'\pi^+$ | $A_{CP}$ | 0.791 | $-0.12 \pm 1.12 \pm 0.17$ | $\pm 0.45$ | $\pm 0.14$ |
| $D^+ \to K^0_S\,\pi^+$ | $A_{CP}$ | 0.977 | $-0.36 \pm 0.09 \pm 0.07$ | $\pm 0.05$ | $\pm 0.02$ |
| $D^+ \to K^0_S\,K^+$ | $A_{CP}$ | 0.977 | $-0.25 \pm 0.28 \pm 0.14$ | $\pm 0.14$ | $\pm 0.04$ |
| $D_s^+ \to K^0_S\,\pi^+$ | $A_{CP}$ | 0.673 | $+5.45 \pm 2.50 \pm 0.33$ | $\pm 0.93$ | $\pm 0.29$ |
| $D_s^+ \to K^0_S\,K^+$ | $A_{CP}$ | 0.673 | $+0.12 \pm 0.36 \pm 0.22$ | $\pm 0.15$ | $\pm 0.05$ |

## 14. Quarkonium(like) Physics

*Editors: N. Brambilla, B. Fulsom, C. Hanhart, Y. Kiyo, R. Mizuk, R. Mussa, A. Polosa, S. Prelovsek, C. P. Shen*

*Additional section writers: S. Godfrey, F.-K. Guo, K. Moats, A. Nefediev, S. Olsen, P. Pakhlov, U. Tamponi, A. Vairo*

### *14.1. Introduction*

[N. Brambilla, A. Vairo]

This chapter is devoted to quarkonium(like) states at Belle II. Quarkonium is a bound state of a heavy quark and a heavy antiquark: *i.e.* bound states of the type $c\bar{c}, b\bar{b}$, and $b\bar{c}$. $t\bar{t}$ cannot give rise to a quarkonium state, because the $t$ quark decays weakly before a proper bound state can be formed. We use "Quarkonium(like)" as title for the chapter, since we also include the (potentially) exotic — non-$\bar{Q}Q$ — $X$, $Y$, $Z$ states that have been observed at Belle and at other accelerator based experiments. Very likely they involve more degrees of freedom besides the heavy quark and the heavy antiquark. Quarkonium is a system of great physical interest in general and of great interest for the Belle experiment in particular. Indeed, the most cited paper of the Belle collaboration [1050] concerns the first observation of an exotic quarkonium candidate (the famous $X(3872)$).

In this chapter we will report about the current status of our theory understanding of quarkonium and the present status of its experimental investigation within a discussion focused on the Belle II program.

The chapter is structured as follows. Section 14.2 discusses the physics of quarkonium, its impact and relevance. Section 14.3 introduces the theoretical methods to study quarkonia within QCD, *i.e.* nonrelativistic effective field theories and lattice. Then we present concrete theoretical descriptions of quarkonium processes and corresponding results and predictions. We separate the presentation of the theoretical concepts into Section 14.4, where quarkonia below the threshold for the decay into two open flavour mesons are discussed, and Section 14.5, where quarkonia and quarkonium-like states above this threshold are the subject. At this threshold many additional degrees of freedom become dynamical and should be considered in theory. Accordingly the properties of quarkonium states appear to differ below and above this threshold and all of the new exotic $X$, $Y$, $Z$ states have been found at or above this strong decay threshold. In particular, Section 14.4 presents predictions for masses, decay widths, and transitions. We also discuss how to use high order perturbative calculations inside nonrelativistic effective field theories together with lattice or experimental data on quarkonium observables to obtain precise extractions of Standard Model parameters like the heavy quark masses and the strong coupling constant $\alpha_s$. To have precise values for these parameters is relevant to the search for beyond Standard Model (BSM) physics that is discussed later in this document: the uncertainties in these quantities jeopardise the search for new physics. Section 14.5 explains the physics of quarkonium at or above the open flavour threshold and compares several approaches to describe the $X$, $Y$, $Z$ states and their predictions. The lattice and QCD based effective field theory descriptions of states at and above threshold are currently being developed. We summarise the latest developments and progress that will likely take place in the near future as well as their potential impact on prospects for Belle II.

Sections 14.6 and 14.7 present recent experimental results and the perspectives for Belle II for charmonium and bottomonium respectively, discussed in relation to the theoretical descriptions presented before.

Finally, Section 14.8 presents the early physics program of Belle II in relation to quarkonium together with an analysis of competing experiments. A part is dedicated to a quantitive analysis of the states that could be observed at Belle II and the relevant processes. Action items for experiment and theory are listed in Section 14.9 emphasising modes of greatest importance at Belle II. Conclusions are presented in Section 14.10.

## *14.2. Heavy Quarkonium: physical picture*

[N. Brambilla, A. Vairo]

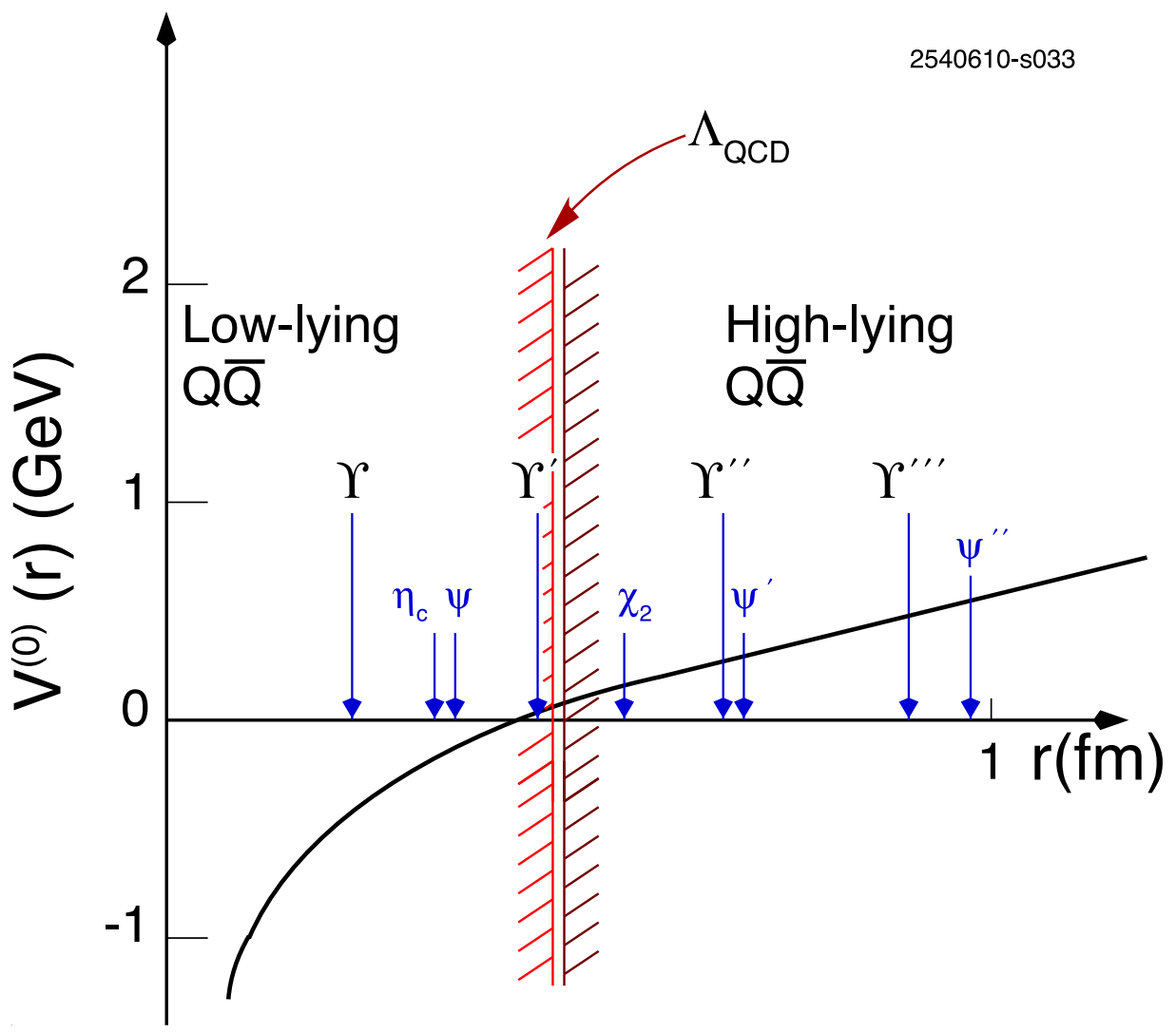

Fig. 165: The static $Q\bar{Q}$ potential plotted as a function of the $Q\bar{Q}$ distance $r$ in comparison to typical quarkonia radii. The static potential has a Coulombic behaviour (asymptotic freedom) for small $r$ and a linearly rising behaviour for large $r$ (confinement). Low-lying quarkonia are quarkonia states with a typical radius smaller that the inverse of $\Lambda_{\rm QCD}$, the scale at which nonperturbative effect become dominant. High-lying quarkonia are quarkonia states with typical radius bigger than $\Lambda_{\rm QCD}$.

The contemporary interest into heavy quarkoniua stems from the fact that on one hand a solid description of quarkonium has been developed from QCD, based on the recent advances in QCD nonrelativistic effective field theories and lattice QCD [1051–1054] at least for what concerns states below and away from the open-flavour strong decay threshold. On the other hand a wealth of data on discoveries of new states, measurements of new processes and increased statistics and precision have been accumulated thanks to the $B$-factories [2], the tau-charm factories and, at present, the experiments at LHC [1051, 1052].

The theoretical interest in quarkonium states below the open flavour threshold stems from the hierarchy of the physical scales that characterize them. Heavy quarks have a mass $m$ larger than the QCD scale $\Lambda_{\mathrm{QCD}}$, which is the scale at which nonperturbative effects become dominant (let's say a few hundred MeV). This implies that processes happening at the scale of the heavy quark mass can be described by perturbative QCD while nonperturbative effects are suppressed by powers of $\Lambda_{\mathrm{QCD}}/m$. This is also true for heavy-light mesons. However, in the case of a bound state of two heavy quarks the situation becomes more interesting. Because the system is nonrelativistic, quarkonium is characterised by the heavy-quark bound-state velocity, $v \ll 1$, ( $v^2 \sim 0.3$ for $c\bar{c}$, $v^2 \sim 0.1$ for $b\bar{b}$ with $E = mv^2 \sim 500$ GeV) and by a hierarchy of energy scales: the mass $m$ (hard scale), the relative momentum $p \sim mv$ (soft scale), and the binding energy $E \sim mv^2$ (ultrasoft scale). For energy scales close to $\Lambda_{\mathrm{QCD}}$, perturbation theory breaks down and one has to rely on nonperturbative methods. Regardless, the nonrelativistic hierarchy of scales,

$$m \gg p \sim 1/r \sim mv \gg E \sim mv^2\,, \tag{484}$$

where $r \sim 1/(mv)$ is the typical distance between the heavy quark and the heavy antiquark, also persists below the scale $\Lambda_{\mathrm{QCD}}$, since $m \gg \Lambda_{\mathrm{QCD}}$, $\alpha_{\mathrm{s}}(m) \ll 1$, and phenomena occurring at the scale $m$ can be always treated perturbatively. The coupling may also be small if $mv \gg \Lambda_{\mathrm{QCD}}$ and $mv^2 \gg \Lambda_{\mathrm{QCD}}$, in which case $\alpha_{\mathrm{s}}(mv) \ll 1$ and $\alpha_{\mathrm{s}}(mv^2) \ll 1$, respectively. This is likely to happen only for the lowest charmonium and bottomonium states as for the excited states, $r$ becomes bigger and $r \ll 1/\Lambda_{\mathrm{QCD}}$ is no longer true. Direct information on the radius of the quarkonia systems is not available, and thus the attribution of some of the lowest bottomonia and charmonia states to the perturbative or the nonperturbative soft regime is at the moment still ambiguous.

Annihilation and production processes take place at the scale $m$, binding takes place at the scale $mv$, while very low-energy gluons and light quarks are sufficiently long-lived that a bound state has time to form and therefore are sensitive to the scale $mv^2$. Ultrasoft gluons are responsible for retardation (*i.e.* nonpotential) effects analogous to the Lamb shift. The existence of several scales makes QCD calculations for quarkonium difficult. In perturbative calculations of loop diagrams the different scales get entangled, challenging our abilities to perform higher-order calculations. In lattice QCD, the existence of several scales for quarkonium sets requirements on the lattice spacing ($a < 1/m$) and the overall size of the lattice ($L > 1/(mv^2)$) that are challenging to be meet.

However, it is precisely this rich structure of separated energy scales that makes heavy quarkonium an ideal laboratory to test the interplay between perturbative and nonperturbative QCD within a controlled environment. Quarkonia with different radii have varying sensitivities to the Coulombic and confining interactions, as shown in Fig. 165. Additionally, the large mass of quarkonium makes it also a suitable system to probe some part of the parameter space of BSM models in decays.

The greatest excitement in the field came in the last decade from the discovery of a large number of states close to and above the open flavour strong decay threshold showing exotics properties: the $X$, $Y$, $Z$ states, see Tables 125 and 126. Starting from the discovery of the $X(3872)$ in 2003 [1050], more than 20 candidates have been reported by various experiments with properties at odds with the expectations from quark models, which had been until then very successful. An interesting feature is that many of these exotics states have a comparatively small width. Most prominent among them are the charged states found in both the charmonium and bottomonium mass ranges. These states have to contain at least 4 quarks: they are explicitly exotic. Other states of explicit exotic nature, two pentaquark states in the charmonium mass region, have been recently discovered by the LHCb Collaboration [1055].

With these states we have for the first time the possibility to explore the nonstandard configurations that a non-Abelian theory like QCD can generate [1051] and have been long-conjectured: hybrids, multiquark configurations, pentaquarks. Belle II will play a role of paramount importance for the characterisation of these states, taking an important step towards a solution of what is among the most challenging problems in contemporary particle physics. The theory to describe such states from QCD lags behind, however. At present most of the studies are carried out at the level of models, which focus on a limited number of degrees of freedom assumed to be prominent. Initial applications of effective field theories based either on quark–gluon degrees of freedom or on hadronic degrees of freedom, and of lattice QCD to study these states have begun, and the field is likely to develop very rapidly.

### *14.3. Theory Methods*

[N. Brambilla, A. Vairo]

*14.3.1. Introduction.* The modern approach to heavy quarkonium is provided by lattice QCD and nonrelativistic effective field theories (NR EFTs) [1054].

Lattice QCD is a reliable non-perturbative method to study hadron properties based directly on QCD. It relies on numerical path integrations in finite and discretized Euclidean space-time.

On the other hand, taking advantage of the existence of a hierarchy of scales, one can substitute simpler but equivalent NR EFTs for QCD [1054]. A hierarchy of EFTs may be constructed by systematically integrating out modes associated with high-energy scales not relevant for a particular quarkonium system. Such integration is performed as part of a matching procedure that enforces the equivalence between QCD and the EFT at a given order of the expansion in the velocity $v$ of the heavy quark in the bound state. The EFT realises a factorisation between the high-energy contributions carried by the matching coefficients and the low-energy contributions carried by the degrees of freedom left as dynamical in the EFT Lagrangian. The Poincaré symmetry remains intact at the level of the NR EFT in a nonlinear realisation that imposes exact relations among the EFT matching coefficients [1056].

*14.3.2. Nonrelativistic Effective Field Theories.*

*Physics at the scale m: NRQCD.* Quarkonium annihilation and production occur at the scale $m$. The suitable EFT is Nonrelativistic QCD [1057, 1058], which follows from QCD by integrating out the scale $m$. As a consequence, the effective Lagrangian is organised as an expansion in $1/m$ and $\alpha_s(m)$:

$$\mathcal{L}_{\rm NRQCD} = \sum_n \frac{c_n(\alpha_s(m), \mu)}{m^n} \times O_n(\mu, mv, mv^2, ...), \tag{485}$$

where $O_n$ are the operators of nonrelativistic QCD (NRQCD) that are important at the low-energy scales $mv$ and $mv^2$, $\mu$ is the NRQCD factorisation scale, and $c_n$ are the Wilson coefficients of the EFT that encode the contributions from the scale $m$ and are nonanalytic in $m$. Only the upper (lower) components of the Dirac fields matter for quarks (antiquarks) at energies lower than $m$. The low-energy operators $O_n$ are constructed out of two or four heavy quark/antiquark fields plus gluons. Four-fermion operators have to be added. Matrix elements of $O_n$ depend on the scales $\mu$, $mv$, $mv^2$ and $\Lambda_{\rm QCD}$. Thus operators are counted in powers of $v$. The imaginary part of the coefficients of the four-fermion operators contains the information on heavy quarkonium annihilation. The NRQCD heavy quarkonium Fock state is given by a series of terms, where the leading term is a $Q\overline{Q}$ in a colour-singlet state, and the first correction, suppressed in $v$, comes from a $Q\overline{Q}$ in an octet state plus a gluon. Higher order terms are subleading in increasing powers of $v$. NRQCD is suitable for spectroscopy studies on the lattice and it is used especially to calculate properties of bottomonium since the large mass of the $b$ quark would put stringent requirement on the lattice step. The latest lattice results on charmonium and bottomonium spectroscopy are reported in Section 14.4.3. Inclusive and exclusive quarkonium decays can be calculated in NRQCD [1054, 1058] at higher order in the expansion in $v$ and in $\alpha_s$, the main problem being the proliferation of low energy nonperturbative matrix elements at higher order in the velocity expansion [1059, 1060] that should be still calculated on the lattice. See [1061] for an example of such lattice calculations.

NRQCD is also the theory used to study quarkonium production see [1051, 1062]. One of the most interesting production processes for Belle is double quarkonium production, which turned out to be also a discovery tool for new states, see Section 14.6.4.

*Physics at the scale $mv, mv^2$: pNRQCD.* To study the physics at the scales $mv$, $mv^2$, and bound state formation, it is convenient to integrate out also the scale of the momentum transfer obtaining EFTs called potential NRQCD (pNRQCD) [1063, 1064]. If the typical radius of the quarkonium state is smaller than the inverse of $\Lambda_{\rm QCD}$ the dynamical degrees of freedom are quark-antiquak pairs in either the colour singlet or the colour octet configuration as well as low energy gluons (higher energy gluons are integrated out). In this case the matching can be done in perturbation theory. The corresponding EFT is called 'weakly coupled pNRQCD'. Otherwise, the residual dynamical degree of freedom is only the colour singlet and the matching is nonperturbative, *i.e.* the matching coefficients cannot be calculated via a perturbative expansion in $\alpha_s$. However, the gauge-invariant gluon correlators (generalised Wilson loops) in terms of which the matching coefficients are written can be calculated using lattice QCD or in QCD vacuum models. This EFT is called 'strongly-coupled pNRQCD'. In both cases the zeroth order problem is a Schrödinger equation and the matching coefficients are the potentials which are directly obtained from QCD. We will present details of this and applications in the next section. In the case of strongly coupled pNRQCD, the singlet is the only degree of freedom (apart from ultrasoft pions). In this situation the non-relativistic EFT is similar to a quark potential model, however, with a few important differences: The potentials are not modeled but obtained directly from QCD [1054, 1065, 1066], all the scales of the problem are considered systematically and each observable is calculated within a well defined power counting scheme.

When we are close to or above the open flavour threshold things become radically different. A description of quarkonium based on potentials (obtained from QCD using strongly-coupled pNRQCD) emerges only below the strong decay threshold. At and above the the open heavy-flavour threshold, new degrees of freedom become relevant containing explicitly light quarks and excited gluons and many new states are expected. NRQCD is still a good EFT for states close to and just above threshold, at least as long as their binding energies remain much smaller than the heavy quark mass. The heavy quarks move slowly in these states, and the static limit should remain a good starting point.

To construct a quark–gluon based EFT description is difficult, because it entails identifying the dynamical degrees of freedom, symmetries and an appropriate and small expansion parameter. Lattice *ab initio* calculations are also difficult because they require dealing with a plethora of excited states and considering also scattering states. Still we will present pioneering results in both directions.

In the next section we will deal with quarkonia below threshold. For the case of charmonium, all such states have been observed. For bottomonium most $S$-wave and $P$-wave states have been observed with exception of the $\eta_b(3S)$ and most $3P$ states. Only one bottomonium $1D$-wave state below $B\bar{B}$ threshold has been observed and no $2D$ states. Observation of $2D$ and other excited states is a challenging task from experimental point of view; various search strategies are discussed in Section 14.7.

### *14.3.3. Lattice.* [N. Brambilla, S. Prelovsek]

Lattice QCD obtains an expectation value of a desired quantity $C$ via numerical path integration $\int \mathcal{D}G\mathcal{D}q_i\mathcal{D}\bar{q}_i e^{-S_{QCD}} C$ formulated on a discretised and finite Euclidean space-time. The parameters of the lattice action $S_{QCD}$ are the quark masses $m_{q_i}$ and the strong coupling $g_s$. The quantity $C$ that gives information on the quarkonium(-like) masses is the correlator $C(t) = \langle \Omega|\mathcal{O}(t)\mathcal{O}^\dagger(0)|\Omega\rangle$. Here operators $\mathcal{O} \simeq \bar{Q}Q$ or $\bar{Q}q\bar{q}Q$ create/annihilate the system with quantum numbers of interest and $|\Omega\rangle$ is the vacuum. The correlator renders energies of QCD eigenstates and therefore hadron masses via Eq. 493 below. For a pedagogical introduction we refer to the excellent books [1067] and [1068]. Some recent reviews on spectroscopy including quarkonium(-like) states are given in [1069–1071].

Two approaches concerning the heavy quark $Q$ that enters the correlator $C$ are employed: (i) $Q$ can be a moving (*i.e.* non-static) quark, where particular care is needed in discretisation due to the non-zero lattice spacing: the $c$ quark is often treated with a proper relativistic formulation, while the heavier $b$ is approached with $\mathcal{L}_{NRQCD}$ or with improved discretisations of $\mathcal{L}_{QCD}$. (ii) The heavy quark, particularly $Q = b$, can alternatively be treated as static in the 0th order of strongly coupled pNRQCD (Sec. 14.4.2). The employed operator $\mathcal{O} = \bar{Q}(0)Q(r)$ or $\bar{Q}(0)q\bar{q}Q(r)$ keeps $Q$ and $\bar{Q}$ at fixed distance $r$ and the correlator leads to the potential $V(r)$. Quarkonium masses are then extracted from the Schrödinger equation.

The matrix elements for electromagnetic and weak transitions are extracted from an analogous path integral with $C = \langle \Omega | \mathcal{O}_f(t_f)\ J^\mu(Q,t)\ \mathcal{O}_i(t_i) | \Omega \rangle$ which involves the transition matrix element $H_i \to H_f$ as detailed in Eq. (494).

The described procedure straightforwardly leads to hadron masses and transitions only for hadrons that are stable with respect to strong decay and are away from thresholds. This (approximately) applies for quarkonium below $D\bar{D}$ and $B\bar{B}$ thresholds. The properties of the resonances $R \to H_1 H_2$ and states near $m_{H_1} + m_{H_2}$ threshold have to be inferred from simulating the $H_1 H_2$ scattering on the lattice and extracting the scattering matrix. The pioneering work concerning quarkonium(like) states has recently employed this strategy, and is discussed in Sec. 14.5.4.

## *14.4. Theory for heavy quarkonium states below open flavour threshold*

In this section we summarise the theory for states below threshold and corresponding applications to spectra, transitions and decays.

### *14.4.1. Weakly coupled pNRQCD.* [N. Brambilla, Y. Kiyo, A. Vairo]

For systems with a small radius ($mv \gg \Lambda_{\rm QCD}$), the effective Lagrangian is organised as an expansion in $1/m$ and $\alpha_{\rm s}(m)$, inherited from NRQCD, and an expansion in $r$ (being $r$ the quark-antiquark distance, $r \sim (mv)^{-1}$) [1063]:

$$\mathcal{L}_{\rm pNRQCD} = \int d^3r \sum_n \sum_k \frac{c_n(\alpha_{\rm s}(m),\mu)}{m^n} \times V_{n,k}(r,\mu',\mu)\, r^k \times O_k(\mu', mv^2, ...), \tag{486}$$

where $O_k$ are the operators of pNRQCD that are dominant at the low-energy scale $mv^2$, $\mu'$ is the pNRQCD factorisation scale and $V_{n,k}$ are the Wilson coefficients of the EFT that encode the contributions from the scale $r$ and are nonanalytic in $r$: they are the potentials. The degrees of freedom that make up the operators $O_k$ are $Q\overline{Q}$ states, colour-singlet $S$, colour-octet $O_a T^a$, and (ultrasoft) gluons. The operators are defined in a multipole expansion. In the equations of motion of pNRQCD, we may identify $V_{n,0} = V_n$ with the $1/m^n$ potentials that enter the Schrödinger equation and $V_{n,k\neq 0}$ with the couplings of the ultrasoft degrees of freedom that provide nonpotential corrections to the Schrödinger equation. Since the degrees of freedom that enter the Schrödinger description are, in this case, both $Q\overline{Q}$ colour singlet and $Q\overline{Q}$ colour octets, both singlet and octet potentials exist. The bulk of the interaction is contained in potential-like terms, but non-potential interactions, associated with the propagation of low-energy degrees of freedom are, in general, present as well and start to contribute at next-to-leading (NLO) in the multipole expansion. They are typically related to nonperturbative effects.

If the quarkonium system is small, the soft scale is perturbative and the potentials can be *entirely* calculated in perturbation theory. They are renormalisable, develop a scale dependence, and satisfy renormalisation group equations that eventually allow resummation of potentially large logarithms.

The singlet potential $V_S$ is a quantity of great relevance and it has been studied since the inception of QCD. When the soft scale is perturbative, it can be calculated in perturbation theory and it is given by the sum of the static singlet potential $V_0$ and $1/m$ and $1/m^2$ spin-dependent and velocity dependent singlet potentials [1072, 1073]

$$V_S = V_0(r) + V_{1/m}(r) + V_{1/m^2}(r) + \cdots \tag{487}$$

In particular the static potential is $V_0(r) = -C_F\alpha_V(r)/r$ where $\alpha_V(r)$ represent $\alpha_s$ in the $V$ scheme and parametrises the strength of interquark force is presently fully known at three loops [1074, 1075]. The static potential develops a logarithmic divergence that is compensated in the static energy by the presence of low energy gluons [1076].

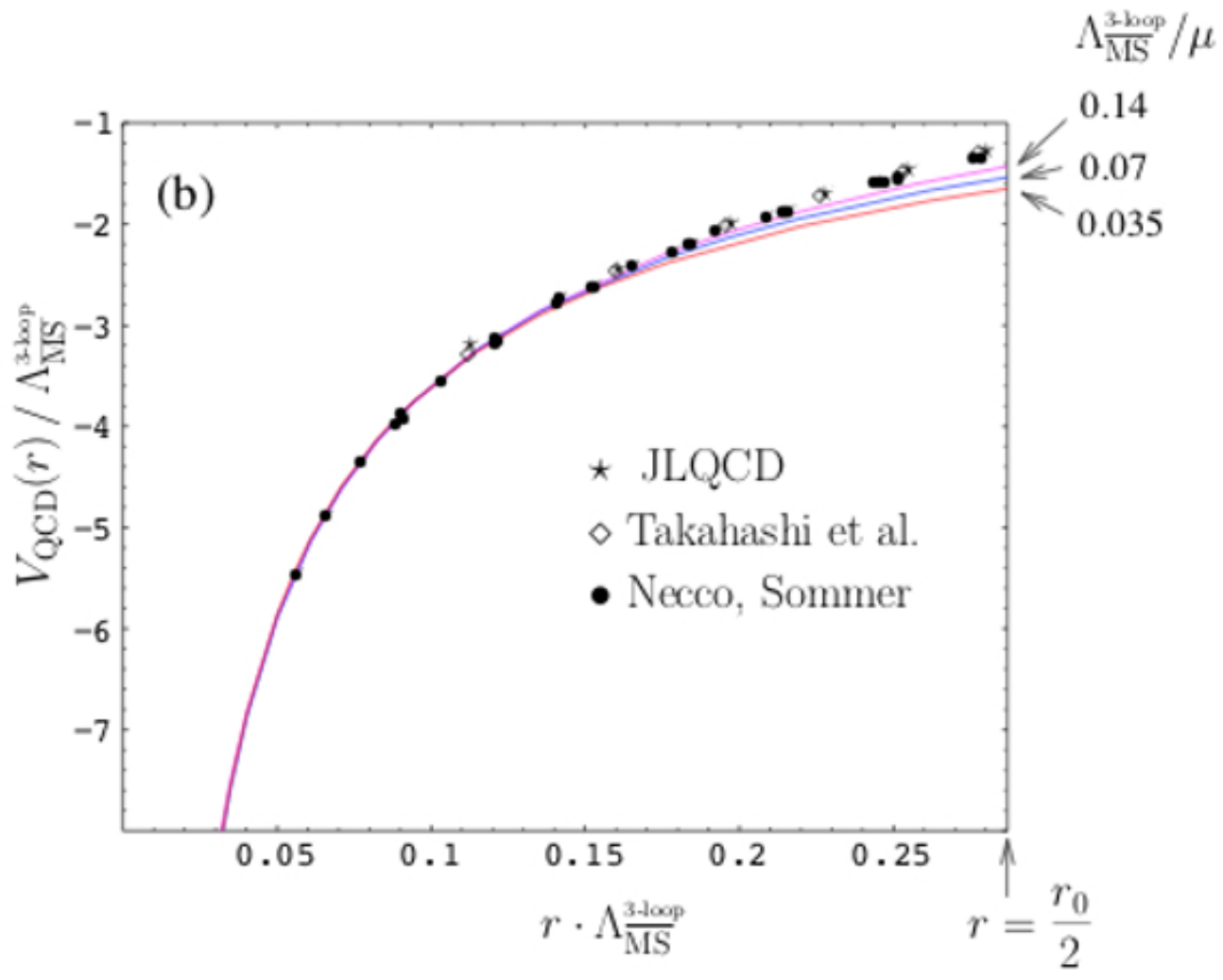

Fig. 166: Comparison in [1075] between the potential in weak coupling expansion in pNRQCD and lattice simulation of the static energy.

The quark-antiquark static energy is given in perturbation theory by the sum of the static singlet potential and the ultrasoft correction (coming from the low energy gluons) $E(r) = V_0(r) + \delta E_{us}$ and it is a function of the quark-antiquark distance $r$. It is known at next-to-next-to-next-to leading logarithmic accuracy (NNNLL) [1077]. It is a physical quantity, to be identified with the QCD static energy, and its perturbative determination can be compared to the lattice determination.

In Fig.166 such a comparison is shown [1075, 1078, 1079] in a short distance region, where the higher order perturbative computation and the lattice determination agrees well: the agreement persists up to about 0.25 fm. The quark-antiquark distance $r$ is given in terms of the lattice Sommer scale $r_0 \sim 0.5$ fm.

*Spectroscopy and precise determination of SM parameters.* [N. Brambilla, Y. Kiyo, A. Vairo]

Using the pNRQCD Lagrangian Eq.(486), quarkonium energy levels have been calculated for general quantum number $n^{2s+1}L_j$ to NNLO [1080, 1081] and NNNLO [1072, 1073, 1082]. To perform a phenomenological analysis it is mandatory to cure a problem of the perturbative series (technically one should enforce the infrared renormalon cancellation [1083, 1084] between the static potential $V_0(r)$ and heavy quark pole mass $m_q$). This can be conveniently done by using a variety of heavy quark short distance masses. For the quarkonium spectroscopy such an analysis was performed in Refs. [1085] and [1086] at NNLO and NNNLO, respectively, using the $\overline{\mathrm{MS}}$ mass. The bottomonium spectrum below $B\bar{B}$ threshold predicted at NNLO and NNNLO and the corresponding experimental values are summarised in Fig. 167. As input values $\alpha_s^{(5)}(M_z) = 0.1184, m_b^{\overline{\mathrm{MS}}}(m_b^{\overline{\mathrm{MS}}}) = 4.20\,\mathrm{GeV}$ are used, where $m_b^{\overline{\mathrm{MS}}}$ is adjusted to reproduce the experimentally measured mass $M_{\Upsilon(1S)} = 9.460\,\mathrm{GeV}$.

Although the overall structures are well explained by the fixed order perturbative predictions, some of the level splittings are far from the experimental value. This is to be expected for excited states for which the typical radius is no longer smaller than the confinement scales, and for which nonperturbative effects become dominant at the level of the potential.

Dedicated studies of the ground state hyperfine splitting have been performed for bottomonium [1087, 1088] with the result $M(\Upsilon(1S)) - M(\eta_b(1S) = 41 \pm 14$ MeV and for $B_c$ with the result $M(B_c^*) - M(B_c) = 50 \pm 17(th)^{+15}_{-12}(\delta\alpha_\mathrm{s})$ MeV [1089], where $th$ represents uncertainty due to high-order perturbative corrections and nonperturbative effects, and $\delta\alpha_\mathrm{s}$ the uncertainty in $\alpha_s(M_Z)$.

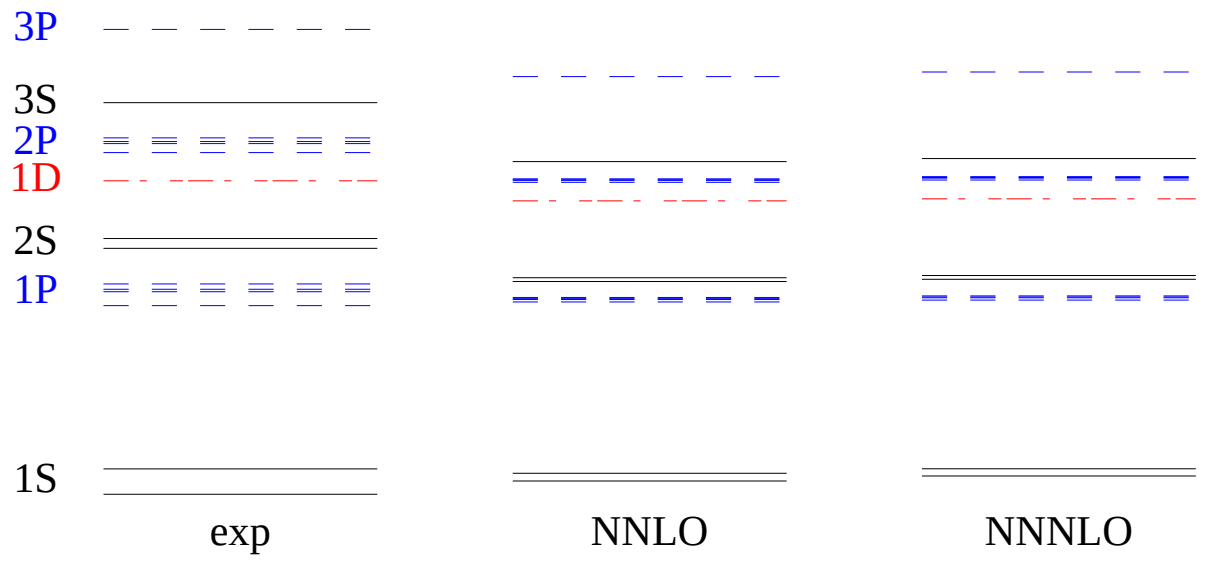

Fig. 167: Experimental and perturbative QCD results [1086] for bottomonium spectrum. The black/blue/red lines correspond to S-wave($\Upsilon(nS), \eta_b(nS)$)/P-wave($\chi_b(nP), h_b(nP)$)/D-wave ($\Upsilon(1D; J^{PC} = 2^{--})$). For $n = 3$ states, only the spin triplet states $\Upsilon(3S)$(black line) and $J$-averaged $\chi_b(3P)$ (dashed blue line) are shown.

For most of the phenomenological analyses for bottomonia the charm quark mass effect is neglected, but there exist some theory predictions [1090–1093] for the $1S$ energy levels that include the effect of finite charm mass. Such an analysis becomes important for precision bottom quark mass extraction from experimentally measured $1S$ bottomonium masses. In Fig.168 $1S$ bottomonium masses versus the $\overline{\mathrm{MS}}$ bottom quark mass is shown. The extracted bottom quark masses are given by $m_b^{\overline{\mathrm{MS}}}(m_b^{\overline{\mathrm{MS}}}) = 4.207$ and $4.187\,\mathrm{GeV}$, which reproduce the experimental masses of $\Upsilon(1S)$ and $\eta_b(1S)$, respectively. Combining these values leads to one of the most precise bottom quark mass determination from a perturbation approach

$$\overline{m}_b^{\mathrm{ave}} = (4197 \pm 2(d_3) \pm 6(\alpha_s) \pm 18(\mathrm{h.o.}) \pm 5(m_c))\,\mathrm{MeV}, \tag{488}$$

where the errors correspond to uncertainties on a constant called $d_3$, in the potential, on $\alpha_s$, higher order terms ($h.o.$), and the input $m_c$ value, respectively [1092]. This shows the importance of systematic computations in the effective field theory approach.

Although one can predict the charmonium masses with exactly the same formula used for bottomonium replacing $(m_b, n_l = 4)$ by $(m_c, n_l = 3)$, most of the charmonium states have too large a radius and therefore a soft scale lying outside the perturbative regime. For instance the minimal sensitivity scale for which the perturbative prediction is stable against the $\mu$- variation is given by $\mu \sim 2\mathrm{GeV}(\alpha_s \sim 0.3)$ for the low-lying states $\eta_c(1S), J/\Psi(1S)$, while for excited charmonium states $\mu \leq 0.8{\rm GeV}(\alpha_s \geq 0.6)$ and perturbative expansion in $\alpha_s$ fails to converge. As a result, except for the low-lying $\eta_c(1S)$ and $J/\psi(1S)$, the perturbative series for the energy levels does not converge well. For instance, the perturbative series with $m_c^{\overline{\rm MS}} = 1226, 1266\,{\rm MeV}$ gives $M^{pert}_{\eta_c(1S)} = (2452 + 242 + 162 + 103 + 24)\,{\rm MeV}$, $M^{pert}_{J/\psi(1S)} = (2532 + 263 + 170 + 109 + 23)\,{\rm MeV}$, respectively, where the last terms represent the $\mathcal{O}(m_c\alpha_s^5)$ binding energy corrections. The values are obtained by evaluating the perturbative series in $\alpha_s(\mu) = 0.269, 0.288$ with $\mu = 2.45, 2.13{\rm GeV}$ for $\eta_c(1S), J/\Psi(1S)$, respectively. In average the charm quark $\overline{\rm MS}$ mass is

$$\overline{m}_c^{\rm ave} = (1246 \pm 2(d_3) \pm 4(\alpha_s) \pm 23({\rm h.o.}))\,{\rm MeV}. \tag{489}$$

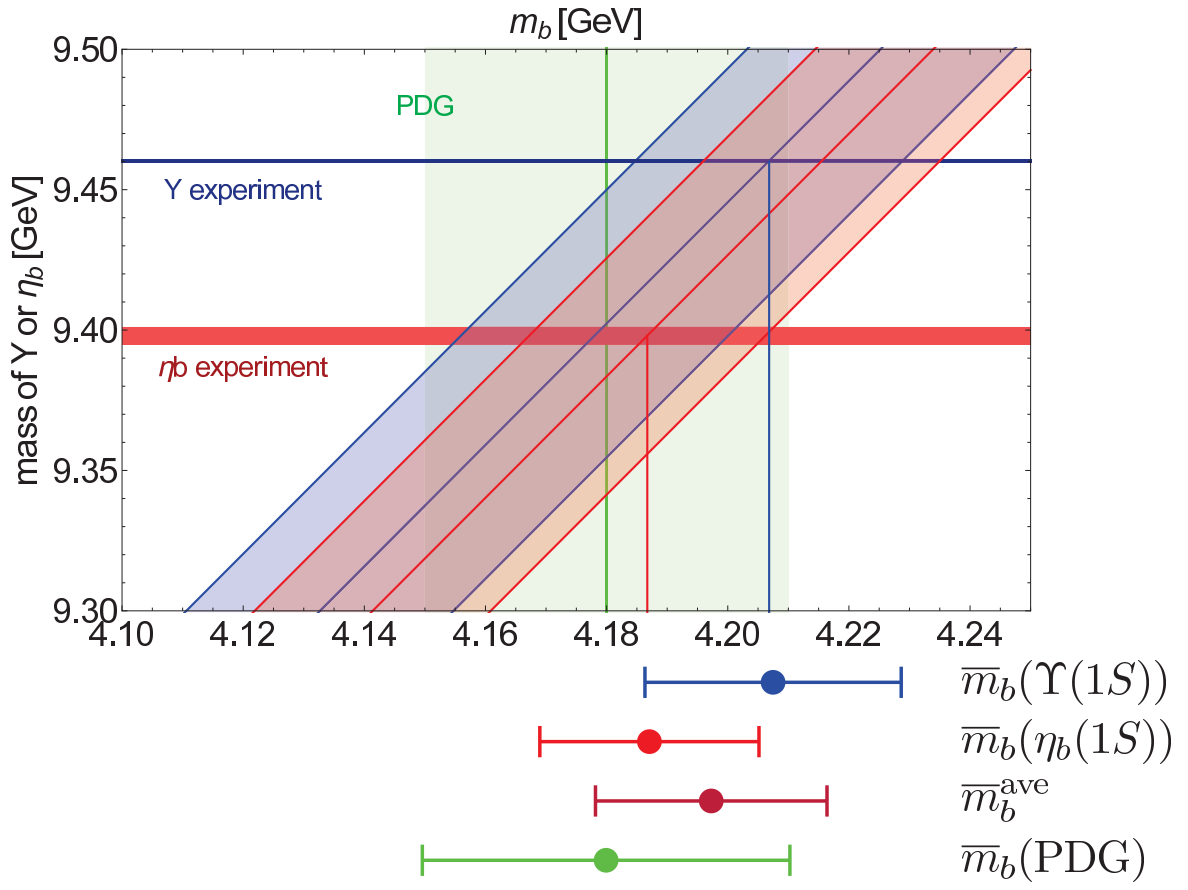

Fig. 168: $\varUpsilon(1S), \eta_b(1S)$ masses v.s. bottom quark $\overline{\rm MS}$ mass taking into account finite charm quark mass [1092]. The blue/red region indicates the theory ambiguity, for $\varUpsilon(1S)/\eta_b(1S)$, respectively, for a given $m_b^{\overline{\rm MS}}$.

By comparing the lattice calculation of the static energy and the static potential at NNNLL in perturbation theory it was possible to extract a precise determination of $\alpha_{\rm s}$: $\alpha_{\rm s}(M_Z) = 0.116^{+0.0012}_{-0.0008}$ [1078, 1079]. Precise lattice data at smaller quark-antiquark distance will allow a more precise extraction. Such extractions of the strong coupling constant from quarkonium are important because they are independent from other determinations and are done at high order in perturbation theory.

*Decays and transitions.* [N. Brambilla, Y. Kiyo, A. Vairo]

While the quarkonium energy levels are evaluated by computing expectation values of the nonrelativistic Hamiltonian, the quarkonium decay widths depend on the square of the wavefunctions, for example, leptonic widths are determined by the wavefunction at the origin $|\psi(0)|^2$. Reliable predictions for the leptonic decays enable another useful check of the QCD dynamics. The leptonic decay width $\varUpsilon(1S) \to \ell^+\ell^-$ for the bottomonium ground state is calculated to NNNLO QCD [1094]

$$\begin{aligned}\Gamma(\varUpsilon(1S) \to \ell^+\ell^-) &= \frac{32}{243}\alpha^2\alpha_s^3 m_b^{\rm PS}(1 + 0.37 + 0.95 - 0.04) \\ &= 1.08 \pm 0.05(\alpha_s)^{+0.01}_{-0.20}(\mu)\,{\rm keV},\end{aligned} \tag{490}$$

to be compared with the measurement $\Gamma(\Upsilon(1S) \to e^+e^-) = 1.340$keV. In Eq.(490) the uncertainties are due to uncertainty of $\alpha_s(M_z)$ and the scale variation (3GeV $< \mu <$ 10GeV). Here the theory prediction was evaluated using a potential subtracted mass $m_b^{\rm PS} = 4.484\,$GeV, which corresponds to $m_b^{\overline{\rm MS}} = 4.163\,$GeV. This means that even at NNNLO the theory prediction lacks roughly a 30% contribution, which remains substantial even if the theoretical uncertainty is taken into account. There can be nonperturbative effects related to nonlocal gluon condensates [1054, 1063] or local gluon condensate corrections [1095, 1096], which are not well known and therefore not included nor estimated in the above theory prediction. It would be important to develop estimates of such nonperturbative condensates corrections. The decays of charmonium are more challenging in perturbative QCD because of the bad convergence of the perturbation series.

The two-photon decay width $\eta_b(1S) \to \gamma\gamma$ and the leptonic decay width are proportional to the same wavefunction to NLO due to the spin symmetry of heavy quarks. This suggests that the decay ratio $\Gamma(n^3S_1 \to e^+e^-)/\Gamma(n^1S_0) \to \gamma\gamma)$ [1097, 1098] is more appropriate to obtain reliable results that are stable against the renormalisation scale variation. Using $\Gamma(\Upsilon(1S) \to e^+e^-) = 1.340 \pm 0.018\,$keV as an input the spin ratio provides a prediction for the spin-singlet decay width [1098]

$$\Gamma(\eta_b(1S) \to \gamma\gamma) = (0.54 \pm 0.15)\,\text{keV}. \tag{491}$$

Table 124: Widths for the magnetic dipole transitions in pNRQCD for bottomonium and charmonium in eV units. Note that different values for the photon energy $k_\gamma = (M_i^2 - M_f^2)/(2M_i)$ are used for some theory predictions.

| Decay | Ref. [1099] | Ref. [1100] |
|---|---|---|
| $\Upsilon(1S) \to \eta_b(1S)\gamma$ | 3.6(2.9) | 15.18(51) |
| $h_b(1P) \to \chi_{b0}(1P)\gamma$ | 1.0(2) | 0.962(35) |
| $h_b(1P) \to \chi_{b1}(1P)\gamma$ | 17(4) | $8.99(55) \times 10^{-3}$ |
| $\chi_{b2}(1P) \to h_b(1P)\gamma$ | 90(20) | 0.118(6) |
| $\Upsilon(2S) \to \eta_b(2S)\gamma$ | | 0.668(60) |
| $\Upsilon(2S) \to \eta_b(1S)\gamma$ | | $6^{+26}_{-06}$ |
| $\eta_b(2S) \to \Upsilon(1S)\gamma$ | | |
| $J/\psi(1S) \to \eta_c(1S)\gamma$ | $1.5(1.0) \times 10^3$ | $2.12(40) \times 10^3$ |

Among quarkonium transitions, electromagnetic transitions [1101] are theoretically clean and rather straightforward compared to the hadronic ones. The pNRQCD description of magnetic dipole (M1) and electric dipole (E1) radiative transitions have been developed in Refs. [1099, 1102], and the precision was raised to $k_\gamma^3/m^2 \times \mathcal{O}(\alpha_s^2, v^2)$ and $k_\gamma^3/m^2 \times \mathcal{O}(v^4)$ ($k_\gamma$ emitted photon energy) for the allowed and hindered magnetic dipole transitions, respectively, in Ref. [1100]. In this latter work several improvements have been obtained taking into account the higher order corrections of the heavy quark static potential, $\mathcal{O}(\alpha_s^2)$ correction to the heavy quark anomalous magnetic moment, and the LL resummation of large logarithms for the NRQCD interactions. The results are summarised in Table 124. While there are no experimental data for $2^1S_0 \to 1^3S_1\gamma$, the radiative transition $2^3S_1 \to 1^1S_0\gamma$ is available and the theory prediction $\Gamma(\Upsilon(2S) \to \eta_b(1S)\gamma) = 6^{+26}_{-6}\,\mathrm{eV}$ agrees within uncertainties with the experimental value $\Gamma_{\rm exp} = 12.5(4.8)\,\mathrm{eV}$ [88].

The radiative transition for $J/\psi$ is also available and $\Gamma_{\rm exp}(J/\psi \to \eta_c(1S)\gamma) = (1.58 \pm 0.37)\,\mathrm{keV}$ is consistent with the theory predictions.

Calculations of higher order contribution in E1 transitions and related applications to the phenomenology are under way [1103].

The study of electromagnetic radiative transition in the EFT approach enables model independent analyses of the decays including photon energy line-shapes [1104]. All this can provide additional insights into bound state dynamics, *e.g.* from the electromagnetic radius $\langle r^2 \rangle$ and expectation value $\langle p^2 \rangle$ of quarkonia based on first principles of QCD.

*14.4.2. Strongly coupled pNRQCD.* [N. Brambilla, A. Vairo]

For systems with a radius $r^{-1} \sim mv \sim \Lambda_{\rm QCD}$ the soft scale is nonperturbative. Only colour singlet quark-antiquark operators can appear in the EFT at the soft scale. The matching to NRQCD is organised as an expansion in $1/m$ but no expansion in $\alpha_s$ can be made at the soft scale. Consequently the matching coefficients of pNRQCD (the potentials) are nonperturbative objects defined through QCD averages of gauge invariant nonlocal operators called generalised Wilson loops.

*Potentials, energy levels, inclusive decays.* [N. Brambilla, Y. Kiyo, A. Vairo]

When $mv \sim \Lambda_{\rm QCD}$ the colour singlet quark-antiquark pair $S$ is the only dynamical degree of freedom and the pNRQCD Lagrangian is written as [1054, 1065, 1066]:

$$\mathcal{L}_{\rm pNRQCD} = S^\dagger \left( i\partial_0 - \frac{\mathbf{p}^2}{2m} - V_S(r) \right) S\,. \tag{492}$$

The singlet potential $V_S(r)$ can be expanded in powers of the inverse of the quark mass; static, $1/m$ and $1/m^2$ terms were calculated long ago [1065, 1066]. The form of these potentials is summarised in [1054]. They involve NRQCD matching coefficients (containing the contribution from the hard scale) and low-energy nonperturbative parts given in terms of static Wilson loops and field-strength insertions in the static Wilson loop (containing the contribution from the soft scale). In this regime of pNRQCD, we recover the quark potential singlet model. However, here the potentials are calculated from QCD by nonperturbative matching. These generalised Wilson loops have been calculated on the lattice (for the most precise recent determination see [1105, 1106]) and in QCD vacuum models [1107–1109].

Then, away from and below open-flavour thresholds, all the heavy quarkonium masses can be obtained by solving the Schrödinger equation with such potentials [1107, 1110].

An example for the application of this method is the mass of the $h_b$. The lattice data show a vanishing long-range component of the spin-spin potential. Thus the potential appears to be entirely dominated by its short-range, delta-like, part, suggesting that the $^1P_1$ state should be close to the centre-of-gravity of the $^3P_J$ system. Indeed, this is consistent with measurements [1111, 1112].

If we explicitly consider light quarks, each quarkonium state may develop a width due to decay through pion emission. Pions in this situation act as ultrasoft degrees of freedom. The heavy-light states develop a mass gap of order $\Lambda_{\rm QCD}$ with respect to quarkonium [1113]. The imaginary part of the potentials give the quarkonium inclusive decay widths [1113]. In particular the NRQCD low energy matrix elements can be rewritten in pNRQCD in terms of quarkonium wave functions and few universal purely gluonic correlators, reducing in this way the number of unknowns. Still lattice calculations of these gluonic correlators have not yet been performed.

*14.4.3. Lattice QCD.* [S. Prelovsek]

Quarkonia $\bar{Q}Q$ below the open-flavour threshold are discussed in this section; those are (almost) stable in the approximation when heavy-quark annihilation is omitted. This simplifying approximation of omitting the Wick contraction related to $\bar{Q}Q$ annihilation is undertaken in all lattice studies presented below. We present also results on higher-lying charmonia that are treated as stable, *i.e.* their strong decay to a pair of charmed hadrons is ignored. Resonances that are considered as unstable and near-threshold states are discussed in Sec. 14.5.4.

The mass of a stable hadron $m_n = E_n(P=0)$ is obtained from the energy $E_n$ of the lattice QCD eigenstate $|n\rangle$ at zero momentum. The eigen-energies $E_n$ are extracted from the time-dependence of the correlation functions

$$C_{ij}(t) = \langle \Omega | \mathcal{O}_i(t) \mathcal{O}_j^\dagger(0) | \Omega \rangle = \sum_n Z_i^n Z_j^{n*} \; e^{-E_n t} \; , \tag{493}$$

where $Z_i^n \equiv \langle \Omega | \mathcal{O}_i | n \rangle$ and $|\Omega\rangle$ denotes the vacuum. The $\mathcal{O}_i$ are the interpolating fields that create/annihilate the physical system with given quantum numbers $J^{PC}$ and only operators of type $\mathcal{O} \simeq \bar{Q}Q$ ($Q = c, b$) are used throughout this Section. In fact, the continuum rotation group is reduced on the lattice to a discrete one, and operators correspond to irreducible representations of the discrete group. Several $J^P$ contribute to a given irreducible representation and careful studies are needed to identify the quantum numbers of the eigenstates.

*Spectra below open-flavour threshold .* The masses of charmonia $m_n = E_n(P = 0)$ obtained from the correlation functions (Eq. 493) are extrapolated to continuum, infinite volume and physical quark masses. The simulations involve light dynamical quarks and non-static heavy quarks with variety of heavy-quark discretisations. Recent precision spectra [1115–1119] are in impressive agreement with the experimental masses, and there are no major open issues. The main remaining uncertainty is due to the omission of $c\bar{c}$ annihilation.

The spectrum of bottomonia below $B\bar{B}$ contains many more states. The recent lattice spectrum in Fig. 169 [1114] is based on non-static $b$ quarks within NRQCD and presents a valuable guidance for states that have not been found in experiment yet.

The other approach considers strongly-coupled potential NRQCD introduced in Section 14.4.2, which determines $\bar{Q}Q$ potentials $V(r)$ at orders $1/m^0$, $1/m$ and $1/m^2$ in the heavy quark expansion [1065, 1066]. The most precise lattice determination of those potentials has been done in the quenched approximation [1105, 1106, 1110], where the spin-independent $1/m^2$ corrections are still missing. The spectra of charmonia and bottomonia based on these potentials are shown in Figs. 2 and 3 of [1110]. The analogous dynamical results suffer from large errors at present, but work is underway.

*Charmonia within a single-hadron approach.* The most extensive spectra of the excited charmonia have been calculated within the so-called single-hadron approach by the Hadron

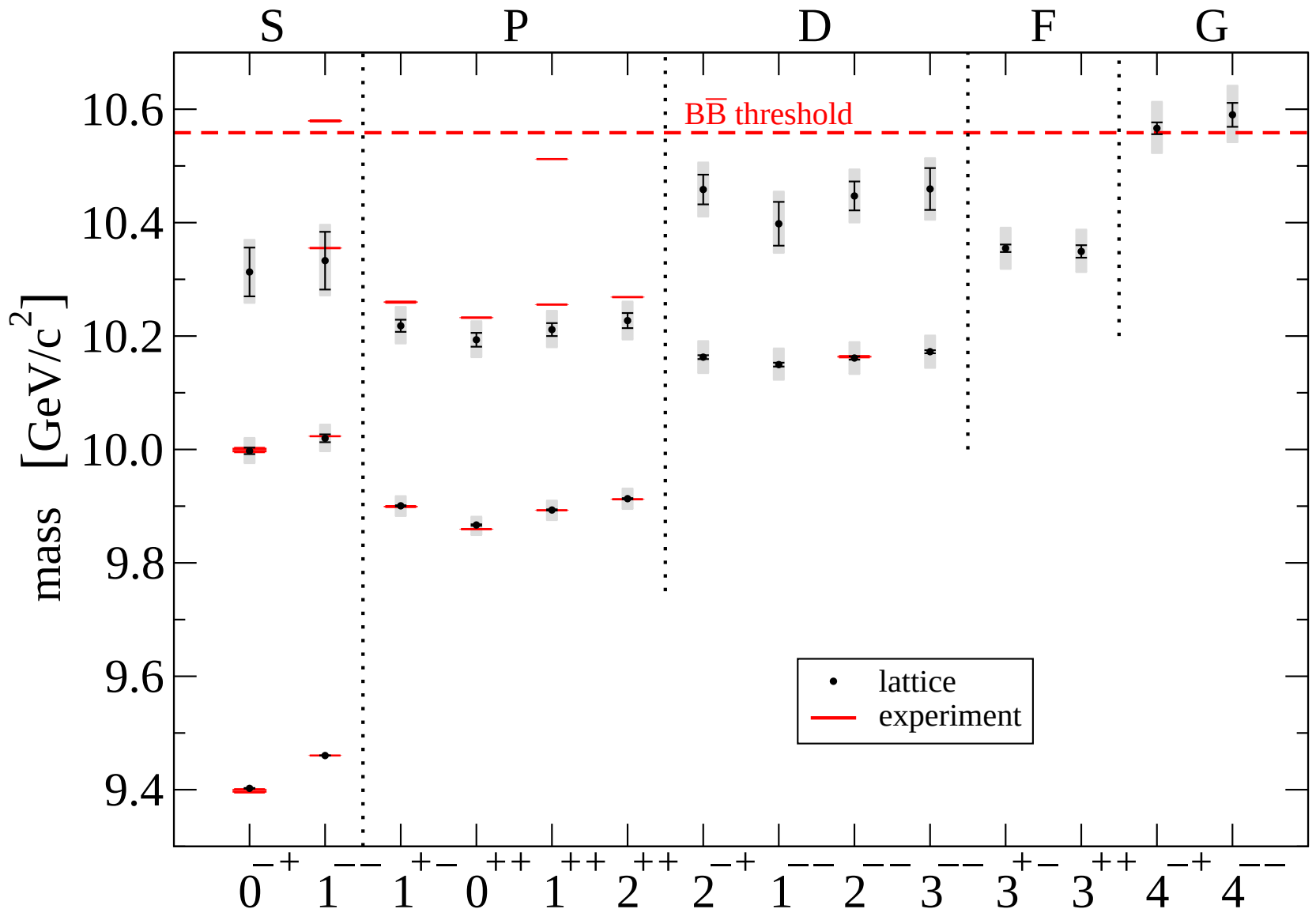

Fig. 169: Lattice spectra of bottomonia below $B\bar{B}$ threshold from [1114]. The $J^{PC}$ and $L$ of the $\bar{b}b$ multiplet are shown.

Spectrum Collaboration. Several complete quark-antiquark multiplets $nL$ were found in a recent simulation with $m_\pi \simeq 240$ MeV [195] (green boxes in Fig. 170). The comparison of earlier spectra at $m_\pi \simeq 400$ MeV [1120] and those at $m_\pi \simeq 240$ MeV shows only mild light-quark mass dependence (Fig. 6 of Ref. [195]).[43] Multiplets of hybrid states were also found (shown in red and blue) and some of them carry exotic $J^{PC}$. Spectra of excited $D$ and $D_s$ mesons were also presented in [195]. The single-hadron treatment ignores strong decays of resonances and threshold effects. It provides valuable reference spectra, but can not lead to reliable conclusions on the existence of near-threshold exotic states, for example.

*Radiative transitions and leptonic widths.* Certain transitions were investigated between quarkonia that lie below the strong decay threshold. The transition $\langle H_f|J^\mu(Q)|H_i\rangle$ is typically determined from correlators of the type

$$\langle \Omega|\mathcal{O}_f(t_f)\; J^\mu(Q,t)\; \mathcal{O}_i(t_i)|\Omega\rangle \propto \sum_{H_i,H_f} \langle \Omega|\mathcal{O}_f|H_f\rangle \cdot e^{-E_{H_f}(t_f-t)}\langle H_f|J^\mu(Q,t)|H_i\rangle e^{-E_{H_i}(t-t_i)}\langle H_i|\mathcal{O}_i|\Omega\rangle \;, \tag{494}$$

where $\mathcal{O}_i$ ($\mathcal{O}_f$) are interpolators that create a tower of initial (final) hadrons with certain quantum numbers. The leptonic decay constants of vector mesons $V \to l^+l^-$ are obtained by extracting $\langle 0|J^\mu|V\rangle$ from correlators.

[43] For excited charmonia close to $S$-wave open-charm thresholds, a nonanalytic pion mass dependence is expected which could be sizeable [1121].

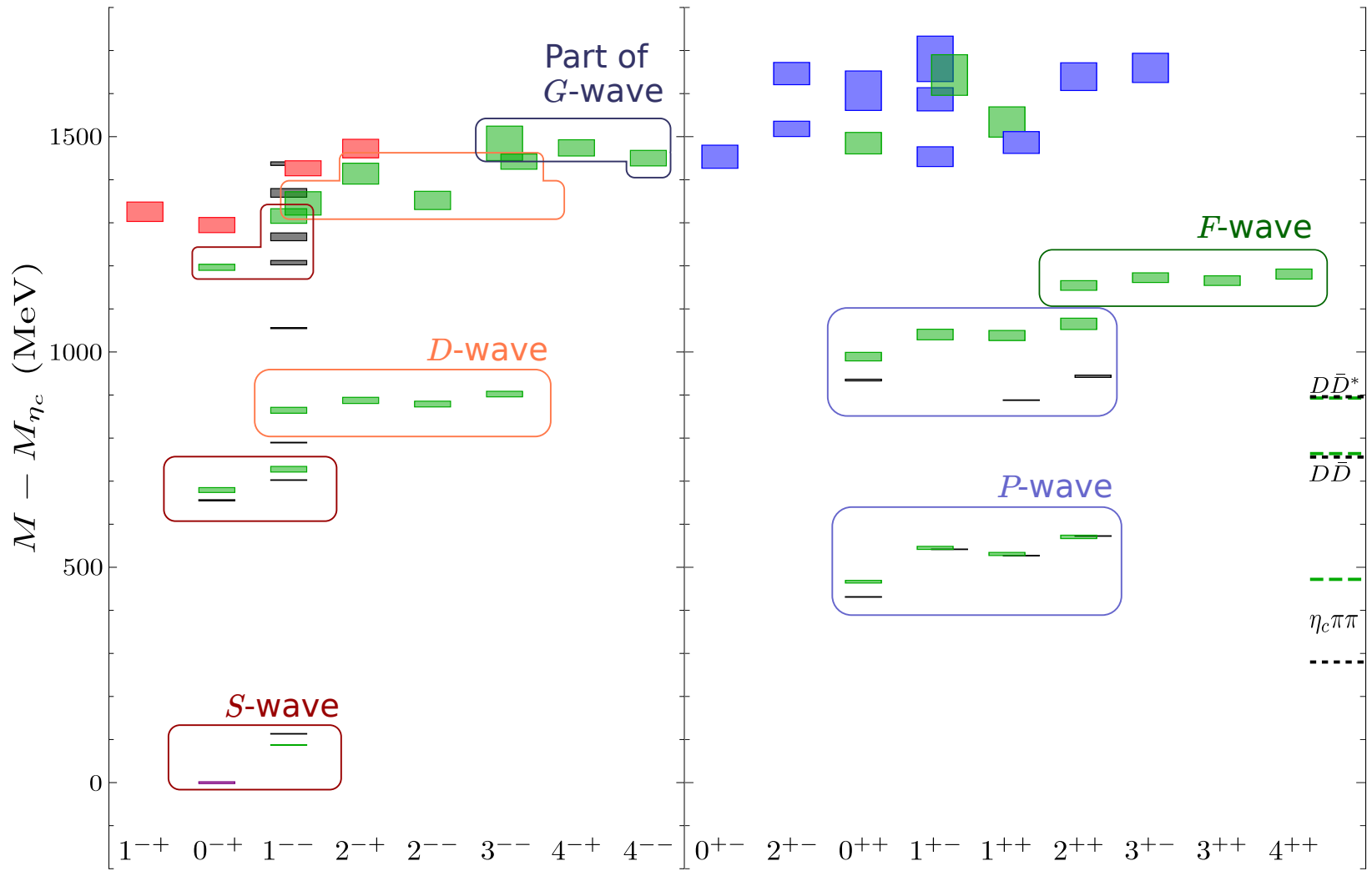

Fig. 170: Lattice spectra of excited charmonia within the single-hadron approximation at $m_\pi \simeq 240$ MeV [195]. Lattice results are shown in green (conventional $c\bar{c}$ multiplets $nL$), red and blue (hybrid candidates from lightest and first-excited supermultiplets, respectively). Experimental masses are shown by black lines. The multiplets are identified based on overlaps $\langle \Omega|\mathcal{O}_i|n\rangle$ (Eq. 493).

An extensive study by the HPQCD collaboration with full error budget leads to $\Gamma^{lat}[\Upsilon(2S) \to \eta_b(1S)\gamma] = 1.72(55) \cdot 10^{-2}$ keV using the Lattice NRQCD [1122]. This hindered M1 transition would have zero rate in the extreme non-relativistic limit due to the orthogonality of the radial wavefunctions. So its matrix element is sensitive to a multitude of effects such as relativistic corrections to the leading order current and to the wavefunctions, particularly those which affect the hyperfine splitting. Such delicate effects need to be considered also when fitting the photon spectra from the experimental data.

The leptonic widths for $\Upsilon(1S, 2S, 3S)$ and the matrix elements for radiative transitions $\Upsilon(1S, 2S, 3S) \to \gamma\eta_b(1S)$, $\Upsilon(2S) \to \gamma\eta_b(2S)$, $\eta_b(2S, 3S) \to \gamma\Upsilon(1S)$ were determined with lattice NRQCD in [1123]. The subsequent higher-order corrections $O(v^6)$ [1124] lead to better agreement with the observed widths $\Upsilon(2S, 3S) \to \gamma\eta_b(1S)$. It would be interesting to compare the rates of $\Upsilon(2S) \to \gamma\eta_b(1S)$ and $\eta_b(2S) \to \gamma\Upsilon(1S)$, since the difference of the rates arises solely from the spin dependent interactions

The quenched study of radiative transitions between charmonia [1125] considered ground states as well as excited states, states of high spin and hybrids. The excited charmonia above open charm threshold were treated in a simplified single-hadron approach discussed above. The dynamical simulation that considered also the extrapolation to the continuum limit, rendered $\Gamma(J/\psi \to \eta_c\gamma) = 2.64(11)(3)$ keV that is within $2\sigma$ from the experimental value, and predicted $\Gamma(\eta_c(2S) \to J/\psi\gamma) = (15.7 \pm 5.7)$ keV and $\Gamma(h_c \to \eta_c\gamma) = 0.72(5)(2)$ MeV [1126, 1127]. The transitions between ground state charmonia were determined also in [1128] and favourable comparison with experiment is provided in Table IV of [1128]. A comparison of quenched and dynamical calculations can provide important information for the role of light quarks in heavy quarkonium systems [1121], which may lead to new insights into the $XYZ$ structures to be discussed below.

Hadronic transitions between quarkonia (via $\pi$, $\eta$, $\pi\pi$, etc.) have not been considered in lattice QCD to our knowledge. The transitions via a single pion could potentially be addressed by studying the matrix element $\langle H_2|u\gamma^m u\gamma_5 d|H_1\rangle$. The general hadronic transitions and related non-leptonic decays are challenging for lattice QCD study. The framework and prospects to study non-leptonic decays on the lattice are discussed in the theory overview of this report.

*14.4.4. Summary.* In summary the theory for quarkonium states below threshold has been constructed and highly developed. Thanks to this quarkonium can be described in QCD and becomes an important system to probe strong interactions. Lattice calculation of masses and electro-weak matrix elements is generally straightforward for such states. On the EFT side, the soft scale is perturbative for systems characterised by a small radius (lowest states). Nonperturbative corrections appear in the form of local and nonlocal condensates: in this case high order resummed perturbation theory and lattice can be compared to experiment and quarkonium becomes a tool for precision determinations. For systems characterised by a larger radius (higher states) the soft scale is nonperturbative, still quarkonium properties can be calculated either with the EFT potentials or directly from the lattice, and quarkonium becomes a tool for the investigation of confinement. The alliance of EFTs and lattice will be important for further progress.

## *14.5. Theory and theory predictions for quarkonia at and above the open flavour threshold*

*14.5.1. Introduction.* [N. Brambilla, A. Vairo]

When we consider the region close to or above the lowest open flavour threshold (see Tables 125 and 126), things change substantially compared to the physics of the previous section. First, let us consider the simplified case without light quarks, in which case the degrees of freedom are heavy quarkonium, heavy hybrids (*i.e.* bound states of heavy quark, heavy antiquark and gluonic excitations) and glueballs. In the static limit, at and above the $\Lambda_{\rm QCD}$ threshold, a tower of hybrid static energies (*i.e.* of gluonic excitations) should be considered on top of the $Q\overline{Q}$ static singlet energy. The spectrum has been thoroughly studied on the lattice [1129]. At short quark-antiquark distance, the spectrum of the hybrid static energies is described in the leading multipole expansion of pNRQCD by the octet potential plus a nonperturbative mass scale, which is called *gluelump mass* [1063, 1130]. At large distances the static energies resemble a string pattern. Some of these states may develop a width if decays to lower states with glueball emission (such as hybrid→glueball + quarkonium) are allowed.

Then, once light fermions have been incorporated into the spectrum to describe the realistic situation, new gauge-invariant states appear beside the heavy quarkonia, hybrids, and glueballs. In fact close to threshold, there is no longer a mass gap between the heavy quarkonium and the creation of a $\overline{Q}q - Q\overline{q}$ pair. Thus, for a study of near-threshold heavy quarkonia, these degrees of freedom must be included in the spectrum and in the effective field theory Lagrangian. States made of $Q\bar{Q}$ and light quarks and anti-quarks include those built on pairs of heavy-light mesons ($D\bar{D}$, $B\bar{B}$, ...) and pairs of heavy-light baryons, like hadronic molecular states; states composed of the usual quarkonium states (built on the static potential) and light hadrons (hadro-quarkonium); tetraquark states; and likely many others. Moreover how these different kinds of states "talk to each other" becomes an important issue. This explains why, from the QCD point of view, so many states of a new nature may appear in this region of the spectrum. However, a systematic QCD description of these states has not yet been developed and it is not yet possible to derive from QCD what the dominant degrees of freedom and their interactions are.

The states are typically searched as poles in the hadron-hadron scattering matrix. First steps towards the extraction of this scattering matrix from lattice have been done for $D^{(*)}\bar{D}^{(*)}$, possibly coupled with a scattering of charmonium and a light meson.

On the parallel front, models are developed in order to obtain more detailed information on these systems. Exceptional cases, *e.g.* those for which the state is extremely close to a threshold (like the $X(3872)$), allow for a kind of "universal" effective field theory treatment [1131–1133] largely inspired by EFT treatments for the nucleon–nucleon interaction [1134]. The models, as detailed in the next sections, are based on a special choice of degrees of freedom assumed to be dominant. The resulting models are not equivalent because different dynamics are attributed to different configurations. Due to the absence of further theoretical input from QCD, many tetraquark studies at the moment rely just on phenomenological forms for the tetraquark interaction. This will change in the near future following the first pioneering lattice calculations of tetraquark static energies [1135, 1136] and further explorations of the special hierarchy of dynamical scales on top of the nonrelativistic and perturbative expansions discussed so far [211, 1137].

In the next sections we will summarise the observed exotic states, introduce various approaches and their predictions. After that we will turn to the lattice QCD and QCD based effective field theory results existing at the moment.

*14.5.2. Observed states.* [C. Hanhart, R. Mizuk]

All hadrons containing $c\bar{c}$ or $b\bar{b}$ quarks with masses above the $D\bar{D}$ or $B\bar{B}$ thresholds are presented in Tables 125 and 126. The names of the recently observed states are not well established yet. We partly use the convention of PDG [77], however, vector and isovector states are denoted $Y$ and $Z$, respectively. The Tables give mass and width values, $J^{PC}$ quantum numbers, a list of the processes in which the state is seen, corresponding references and significances (or "np" for "not provided"), discovery year and the status for each production and decay channel (here "NC!" stands for "need confirmation"). For isovector states the $C$-parity is given for the neutral member of the isotriplet.

Table 125 shows the states that have masses very near the lowest open flavour thresholds, *e.g.*, the $X(3872)$ near $D^0\bar{D}^{*0}$, the $Z_b(10610)$ and $Z_b(10650)$ near $B\bar{B}^*$ and $B^*\bar{B}^*$, respectively. The threshold proximity is often interpreted as a signature for a hadronic molecule, however, also within the tetraquark scenario the appearance of very near threshold states could be natural — for a detailed discussion of these issues we refer to *Molecules* and *Tetraquarks* subsections of Section 14.5.3, respectively. All states have properties that make them distinct from quarkonia. The $X(3872)$ decays into $\rho J/\psi$ and $\omega J/\psi$ with comparable probabilities, which corresponds to strong violation of isospin symmetry and is unexpected for quarkonium. The $Z_b$ states are isovectors, thus in addition to a $b\bar{b}$ pair they should contain light quarks. Another interesting property of $Z_b$ is that they decay with comparable probabilities into spin-triplet and spin-singlet bottomonia [$\Upsilon(nS)$ $(n = 1, 2, 3)$ and $h_b(mP)$ $(m = 1, 2)$]. This would correspond to a strong violation of Heavy Quark Spin

Table 125: Quarkonium-like states at the lowest open flavour thresholds.

| State | $M$, MeV | $\Gamma$, MeV | $J^{PC}$ | Process (mode) | Experiment (#$\sigma$) | Year | Status |
|---|---|---|---|---|---|---|---|
| $X(3872)$ | $3871.69 \pm 0.17$ | $< 1.2$ | $1^{++}$ | $B \to K(\pi^+\pi^- J/\psi)$ | Belle [1050, 1138] (>10), BaBar [1139] (8.6) | 2003 | Ok |
| | | | | $p\bar{p} \to (\pi^+\pi^- J/\psi) \ldots$ | CDF [1140–1142] (11.6), D0 [1143] (5.2) | 2003 | Ok |
| | | | | $pp \to (\pi^+\pi^- J/\psi) \ldots$ | LHCb [1144–1146] (np), CMS [1147] (np) | 2012 | Ok |
| | | | | $Y(4260) \to \gamma(\pi^+\pi^- J/\psi)$ | BESIII [1148] (6.3) | 2013 | NC! |
| | | | | $B \to K(\omega J/\psi)$ | Belle [1149] (4.3), BaBar [1150] (4.0) | 2005 | NC! |
| | | | | $B \to K(\gamma J/\psi)$ | Belle [1149, 1151] (5.5), BaBar [1152, 1153] (3.6), LHCb [1154] (> 10) | 2005 | Ok |
| | | | | $B \to K(\gamma \psi(2S))$ | BaBar [1153] (3.5), Belle [1151] (0.2), LHCb [1154] (4.4) | 2008 | NC! |
| | | | | $B \to K(D^0\bar{D}^{*0})$ | Belle [1155, 1156] (6.4), BaBar [1157] (4.9) | 2006 | NC! |
| $Z_c(3900)^+$ | $3891.2 \pm 3.3$ | $40 \pm 8$ | $1^{+-}$ | $Y(4260) \to \pi^-(\pi^+ J/\psi)$ | BESIII [1158] (>8), Belle [1159] (5.2), CLEO data [1160] (>5) | 2013 | Ok |
| | | | | $Y(4260, 4360) \to \pi^0(\pi^0 J/\psi)$ | CLEO data [1160] (3.5), BESIII [1161] (10.4) | 2013 | Ok |
| | | | | $Y(4260, 4390) \to \pi^-(\pi^+ h_c)$ | BESIII [1162] (2.1) | 2013 | NC! |
| | | | | $Y(4260) \to \pi^-(D\bar{D}^*)^+$ | BESIII [1163, 1164] (18) | 2013 | Ok |
| | | | | $Y(4260) \to \pi^0(D\bar{D}^*)^0$ | BESIII [1165] (>10) | 2015 | Ok |
| $Z_c(4020)^+$ | $4022.9 \pm 2.8$ | $7.9 \pm 3.7$ | $?^{?-}$ | $Y(4260, 4390) \to \pi^-(\pi^+ h_c)$ | BESIII [1162] (8.9) | 2013 | NC! |
| | | | | $Y(4260, 4390) \to \pi^0(\pi^0 h_c)$ | BESIII [1166] (>5) | 2014 | NC! |
| | | | | $Y(4360) \to \pi^-(\pi^+ \psi(2S))$ | Belle [1167] (3.5), BESIII [1168] (9.2) | 2014 | NC! |
| | | | | $Y(4260) \to \pi^-(D^*\bar{D}^*)^+$ | BESIII [1169] (10) | 2013 | NC! |
| | | | | $Y(4260) \to \pi^0(D^*\bar{D}^*)^0$ | BESIII [1170] (5.9) | 2015 | NC! |
| $Z_b(10610)^+$ | $10607.2 \pm 2.0$ | $18.4 \pm 2.4$ | $1^{+-}$ | $\Upsilon(10860) \to \pi^-(\pi^+\Upsilon(1S, 2S, 3S))$ | Belle [1171–1173] (>10) | 2011 | Ok |
| | | | | $\Upsilon(10860) \to \pi^0(\pi^0\Upsilon(2S, 3S))$ | Belle [1174] (6.5) | 2013 | NC! |
| | | | | $\Upsilon(10860) \to \pi^-(\pi^+ h_b(1P, 2P))$ | Belle [1171, 1172] (16) | 2011 | Ok |
| | | | | $\Upsilon(10860) \to \pi^-(B\bar{B}^*)^+$ | Belle [1175, 1176] (9.3) | 2012 | NC! |
| $Z_b(10650)^+$ | $10652.2 \pm 1.5$ | $11.5 \pm 2.2$ | $1^{+-}$ | $\Upsilon(10860) \to \pi^-(\pi^+\Upsilon(1S, 2S, 3S))$ | Belle [1171–1173] (>10) | 2011 | Ok |
| | | | | $\Upsilon(10860) \to \pi^-(\pi^+ h_b(1P, 2P))$ | Belle [1171, 1172] (16) | 2011 | Ok |
| | | | | $\Upsilon(11020) \to \pi^-(\pi^+ h_b(1P))$ | Belle [1177] (3.3) | 2015 | NC! |
| | | | | $\Upsilon(10860) \to \pi^-(B^*\bar{B}^*)^+$ | Belle [1175, 1176] (8.1) | 2012 | NC! |

Symmetry (HQSS) for pure quarkonium, however, can be explained naturally within both the tetraquark picture [1238] as well as the molecular scenario [1239]. The isovector states in the charmonium sector, the $Z_c(3900)$ and $Z_c(4020)$, seem to be close relatives of the $Z_b$ states. Their masses as reported in the current literature are located somewhat above the $D\bar{D}^*$ and $D^*\bar{D}^*$ thresholds, which is disastrous for the molecular interpretation. However, those mass determinations neither take into account the interference between the $Z_c$ signals and the non-resonant background, which could shift the peak position by as much as $\Gamma/2$, nor the proper analytic structure of the amplitudes.

Recently BESIII studied the process $e^+e^- \to \pi^+\pi^-\psi(2S)$ and reported observation of an intermediate charged state [1168]. Previously, Belle had found evidence for this state [1167].

Table 126: Quarkonium(-like) states above the open flavour thresholds.

| State | $M$, MeV | $\Gamma$, MeV | $J^{PC}$ | Process (mode) | Experiment (#$\sigma$) | Year | Status |
|---|---|---|---|---|---|---|---|
| $\psi(3770)$ | $3773.13 \pm 0.35$ | $27.2 \pm 1.0$ | $1^{--}$ | $e^+e^- \to (D\bar{D})$ | PDG [77] | 1977 | Ok |
| | | | | $B \to K(D\bar{D})$ | Belle [1178, 1179] (5.5), BaBar [1157] (6.4) | 2003 | Ok |
| | | | | $e^+e^- \to (\pi^+\pi^- J/\psi)$ | BES [1180] (3), CLEO [1181] (11.6) | 2003 | Ok |
| | | | | $e^+e^- \to (\pi^0\pi^0 J/\psi)$ | CLEO [1181] (3.4) | 2005 | NC! |
| | | | | $e^+e^- \to (\eta\, J/\psi)$ | CLEO [1181] (3.5) | 2005 | NC! |
| | | | | $e^+e^- \to (\phi\, \eta)$ | CLEO [1182] (5) | 2005 | NC! |
| | | | | $e^+e^- \to (\gamma\, \chi_{c0,1})$ | PDG [77] | 2005 | Ok |
| $\psi_2(3823)$ | $3822.2 \pm 1.2$ | $< 16$ | $2^{--}$ | $B \to K(\gamma\chi_{c1})$ | Belle [1183] (3.8) | 2013 | NC! |
| *or* $X(3823)$ | | | | $e^+e^- \to \pi^+\pi^-(\gamma\chi_{c1})$ | BESIII [1184] (6.2) | 2015 | NC! |
| $X(3860)$ | $3862^{+48}_{-35}$ | $201^{+177}_{-106}$ | $0/2^{++}$ | $e^+e^- \to J/\psi\,(D\bar{D})$ | Belle [1185] (6.5) | 2017 | NC! |
| $X(3915)$ | $3918.4 \pm 1.9$ | $20 \pm 5$ | $0/2^{?+}$ | $B \to K(\omega J/\psi)$ | Belle [1186] (8), BaBar [1150, 1187] (19) | 2004 | Ok |
| *or* $Y(3940)$ | | | | $e^+e^- \to e^+e^-(\omega J/\psi)$ | Belle [1188] (7.7), BaBar [1189] (7.6) | 2009 | Ok |
| $\chi_{c2}(2P)$ | $3927.2 \pm 2.6$ | $24 \pm 6$ | $2^{++}$ | $e^+e^- \to e^+e^-(D\bar{D})$ | Belle [1190] (5.3), BaBar [1191] (5.8) | 2005 | Ok |
| $X(3940)$ | $3942^{+9}_{-8}$ | $37^{+27}_{-17}$ | $?^{?+}$ | $e^+e^- \to J/\psi\,(D\bar{D}^*)$ | Belle [1192, 1193] (6) | 2005 | NC! |
| $\psi(4040)$ | $4039 \pm 1$ | $80 \pm 10$ | $1^{--}$ | $e^+e^- \to$ (hadrons) | PDG [77] | 1978 | Ok |
| | | | | $e^+e^- \to (\eta J/\psi)$ | BESIII [1194] (>10), Belle [1195] (6.0) | 2012 | NC! |
| $Z(4050)^+$ | $4051^{+24}_{-43}$ | $82^{+51}_{-55}$ | $?^{?+}$ | $\bar{B}^0 \to K^-(\pi^+\chi_{c1})$ | Belle [1196] (5.0), BaBar [1197] (1.1) | 2008 | NC! |
| $X(4140)$ | $4146.5^{+6.4}_{-5.3}$ | $83^{+30}_{-25}$ | $1^{++}$ | $B^+ \to K^+(\phi J/\psi)$ | CDF [1198, 1199] (5.0), Belle [1200] (1.9), | 2009 | Ok |
| *or* $Y(4140)$ | | | | | LHCb [1201] (1.4), CMS [1202] (>5), | | |
| | | | | | D0 [1203] (3.1), BaBar [1204] (1.6), | | |
| | | | | | LHCb [1205, 1206] (8.4) | | |
| | | | | $p\bar{p} \to (\phi J/\psi)\ldots$ | D0 [1207] (4.7) | 2015 | NC! |
| $\psi(4160)$ | $4153 \pm 3$ | $103 \pm 8$ | $1^{--}$ | $e^+e^- \to$ (hadrons) | PDG [77] | 1978 | Ok |
| | | | | $e^+e^- \to (\eta J/\psi)$ | Belle [1195] (6.5), BESIII [1208] (>5) | 2013 | NC! |
| $X(4160)$ | $4156^{+29}_{-25}$ | $139^{+113}_{-65}$ | $?^{?+}$ | $e^+e^- \to J/\psi\,(D^*\bar{D}^*)$ | Belle [1193] (5.5) | 2007 | NC! |
| $Z(4200)^+$ | $4196^{+35}_{-32}$ | $370^{+99}_{-149}$ | $1^{+-}$ | $\bar{B}^0 \to K^-(\pi^+ J/\psi)$ | Belle [1209] (6.2) | 2014 | NC! |
| $Z(4250)^+$ | $4248^{+185}_{-45}$ | $177^{+321}_{-72}$ | $?^{?+}$ | $\bar{B}^0 \to K^-(\pi^+\chi_{c1})$ | Belle [1196] (5.0), BaBar [1197] (2.0) | 2008 | NC! |
| $Y(4260)$ | $4221.1 \pm 2.5$ | $47.7 \pm 4.0$ | $1^{--}$ | $e^+e^- \to (\pi^+\pi^- J/\psi)$ | BaBar [1210, 1211] (8), CLEO [1212, 1213] (11), | 2005 | Ok |
| | | | | | Belle [1159, 1214] (15), BESIII [1158, 1215] (np) | | |
| | | | | $e^+e^- \to (\pi^0\pi^0 J/\psi)$ | CLEO [1212] (5.1), BESIII [1161] (np) | 2006 | Ok |
| | | | | $e^+e^- \to (K^+K^- J/\psi)$ | CLEO [1212] (3.7) | 2006 | NC! |
| | | | | $e^+e^- \to (f_0(980) J/\psi)$ | BaBar [1211] (np), Belle [1159] (np) | 2012 | Ok |
| | | | | $e^+e^- \to (\pi^+\pi^- h_c)$ | BESIII [1162, 1216] (10) | 2013 | NC! |
| | | | | $e^+e^- \to (\pi^0\pi^0 h_c)$ | BESIII [1166] (np) | 2014 | NC! |
| | | | | $e^+e^- \to (\omega\chi_{c0})$ | BESIII [1217] (>9) | 2014 | NC! |
| | | | | $e^+e^- \to (\gamma\, X(3872))$ | BESIII [1148] (6.3) | 2013 | NC! |
| | | | | $e^+e^- \to (\pi^- Z_c(3900)^+)$ | BESIII [1158, 1164] (>8), Belle [1159] (5.2) | 2013 | Ok |
| | | | | $e^+e^- \to (\pi^0 Z_c(3900)^0)$ | BESIII [1161, 1165] (10.4) | 2015 | Ok |
| | | | | $e^+e^- \to (\pi^{\mp,0} Z_c(4020)^{\pm,0})$ | BESIII [1162, 1166, 1169, 1170] (>10) | 2013 | Ok |

Table 126: (continued)

| State | $M$, MeV | $\Gamma$, MeV | $J^{PC}$ | Process (mode) | Experiment (#$\sigma$) | Year | Status |
|---|---|---|---|---|---|---|---|
| $X(4274)$ | $4273.3^{+19.1}_{-9.0}$ | $56.2^{+13.8}_{-15.6}$ | $1^{++}$ | $B^+ \to K^+(\phi J/\psi)$ | CDF [1199] (3.1), LHCb [1201] (1.0), | 2011 | NC! |
| *or* $Y(4274)$ | | | | | CMS [1202] (>3), D0 [1203] (np), | | |
| | | | | | LHCb [1205, 1206] (6.0) | | |
| $X(4350)$ | $4350.6^{+4.6}_{-5.1}$ | $13^{+18}_{-10}$ | $0/2^{?+}$ | $e^+e^- \to e^+e^-(\phi J/\psi)$ | Belle [1218] (3.2) | 2009 | NC! |
| $Y(4360)$ | $4341.2 \pm 5.4$ | $101.9 \pm 9.3$ | $1^{--}$ | $e^+e^- \to (\pi^+\pi^-\psi(2S))$ | Belle [1167, 1219] (8), BaBar [1220] (np) | 2007 | Ok |
| | | | | $e^+e^- \to (\pi^+\pi^- J/\psi)$ | BESIII [1215] (7.6) | 2016 | NC! |
| | | | | $e^+e^- \to (\pi^+\pi^-\psi_2(3823))$ | BESIII [1184] (np) | 2015 | NC! |
| | | | | $e^+e^- \to (\pi^0 Z_c(3900)^0)$ | BESIII [1161] (np) | 2015 | NC! |
| | | | | $e^+e^- \to (\pi^- Z_c(4020)^+)$ | Belle [1167] (3.5), BESIII [1168] (9.2) | 2014 | NC! |
| $Y(4390)$ | $4391.6 \pm 6.4$ | $139.5 \pm 16.1$ | $1^{--}$ | $e^+e^- \to (\pi^+\pi^- h_c)$ | BESIII [1216] (10) | 2016 | NC! |
| | | | | $e^+e^- \to (\pi^{\mp,0} Z_c(4020)^{\pm,0})$ | BESIII [1162, 1166] (np) | 2013 | NC! |
| $\psi(4415)$ | $4421 \pm 4$ | $62 \pm 20$ | $1^{--}$ | $e^+e^- \to$ (hadrons) | PDG [77] | 1976 | Ok |
| | | | | $e^+e^- \to (\eta J/\psi)$ | Belle [1195] (np), BESIII [1208] (>5) | 2013 | NC! |
| | | | | $e^+e^- \to (\omega\chi_{c2})$ | BESIII [1221] (10.4) | 2015 | NC! |
| | | | | $e^+e^- \to (D\bar{D}_2^*(2460))$ | Belle [1222] (10) | 2007 | NC! |
| $Z(4430)^+$ | $4478^{+15}_{-18}$ | $181 \pm 31$ | $1^{+-}$ | $\bar{B}^0 \to K^-(\pi^+\psi(2S))$ | Belle [1223–1225] (6.4), | 2007 | Ok |
| | | | | | BaBar [1226] (2.4), LHCb [1227, 1228] (13.9) | | |
| | | | | $\bar{B}^0 \to K^-(\pi^+ J/\psi)$ | Belle [1209] (4.0) | 2014 | NC! |
| $X(4500)$ | $4506^{+16}_{-19}$ | $92^{+30}_{-29}$ | $0^{++}$ | $B^+ \to K^+(\phi J/\psi)$ | LHCb [1205, 1206] (6.1) | 2016 | NC! |
| $Y(4660)$ | $4643 \pm 9$ | $72 \pm 11$ | $1^{--}$ | $e^+e^- \to (\pi^+\pi^-\psi(2S))$ | Belle [1167, 1219] (5.8), BaBar [1220] (5) | 2007 | Ok |
| | | | | $e^+e^- \to (\Lambda_c^+\bar{\Lambda}_c^-)$ | Belle [1229] (8.2) | 2007 | NC! |
| $X(4700)$ | $4704^{+17}_{-26}$ | $120^{+52}_{-45}$ | $0^{++}$ | $B^+ \to K^+(\phi J/\psi)$ | LHCb [1205, 1206] (5.6) | 2016 | NC! |
| $\Upsilon(4S)$ | $10579.4 \pm 1.2$ | $20.5 \pm 2.5$ | $1^{--}$ | $e^+e^- \to$ (hadrons) | PDG [77] | 1985 | Ok |
| | | | | $e^+e^- \to (\pi^+\pi^-\Upsilon(1S,2S))$ | BaBar [1230, 1231] (>10), | 2006 | Ok |
| | | | | | Belle [1232, 1233] (11.2) | | |
| | | | | $e^+e^- \to (\eta\,\Upsilon(1S))$ | BaBar [1231] (>11) | 2008 | Ok |
| | | | | $e^+e^- \to (\eta\, h_b(1P))$ | Belle [1111] (11) | 2015 | Ok |
| $\Upsilon(10860)$ | $10891 \pm 4$ | $54 \pm 7$ | $1^{--}$ | $e^+e^- \to$ (hadrons) | PDG [77] | 1985 | Ok |
| | | | | $e^+e^- \to (\pi^+\pi^-\Upsilon(1S,2S,3S))$ | Belle [1172, 1173, 1234] (>10) | 2007 | Ok |
| | | | | $e^+e^- \to (\pi^0\pi^0\Upsilon(1S,2S,3S))$ | Belle [1174] (np) | 2013 | Ok |
| | | | | $e^+e^- \to (f_0(980)\,\Upsilon(1S))$ | Belle [1172–1174] (>8) | 2011 | Ok |
| | | | | $e^+e^- \to (f_2(1275)\,\Upsilon(1S))$ | Belle [1172–1174] (np) | 2011 | NC! |
| | | | | $e^+e^- \to (\eta\,\Upsilon(1S,2S))$ | Belle (10) | 2012 | NC! |
| | | | | $e^+e^- \to (K^+K^-\Upsilon(1S))$ | Belle [1234] (4.9) | 2007 | NC! |
| | | | | $e^+e^- \to (\omega\,\chi_{b1,2}(1P))$ | Belle [1235] (12) | 2014 | Ok |
| | | | | $e^+e^- \to ((\pi^+\pi^-\pi^0)_{\text{non-}\omega}\,\chi_{b1,2}(1P))$ | Belle [1235] (4.9) | 2014 | NC! |
| | | | | $e^+e^- \to (\pi^+\pi^-\Upsilon_J(1D))$ | Belle (9) | 2012 | NC! |
| | | | | $e^+e^- \to (\eta\,\Upsilon_J(1D))$ | Belle (np) | 2014 | NC! |
| | | | | $e^+e^- \to (\pi\, Z_b(10610, 10650))$ | Belle [1172, 1174] (>10) | 2011 | Ok |
| | | | | $e^+e^- \to (B_s^*\bar{B}_s^*)$ | Belle [1236] (np) | 2016 | NC! |
| $\Upsilon(11020)$ | $10987.5^{+11.0}_{-3.4}$ | $61^{+9}_{-28}$ | $1^{--}$ | $e^+e^- \to$ (hadrons) | PDG [77] | 1985 | Ok |
| | | | | $e^+e^- \to (\pi^+\pi^-\Upsilon(1S,2S,3S))$ | Belle [1237] (np) | 2015 | NC! |
| | | | | $e^+e^- \to (\pi^{\mp} Z_b(10610, 10650)^{\pm})$ | Belle [1177] (5.3) | 2015 | NC! |

The state is only $2.8\,\sigma$ away from the $Z_c(4020)$[44] and likely corresponds to a new decay channel of that state.

As we move even higher above the lowest open flavour thresholds (see Table 126) the interpretation of the states becomes even more difficult. The number of states in the charmonium region is especially large. All states, except $\psi_2(3823)$ and $X(3860)$, possess properties unexpected for $c\bar{c}$ levels. Most of the states have hadronic transitions to lower charmonia with anomalously high rates. Not only do the recently observed $XYZ$ states have this property, but it is shared by the $\psi(4040)$, $\psi(4160)$ and $\psi(4415)$ known since 1970s. Indeed, Belle observed the above $\psi$ states in the energy dependence of the $e^+e^- \to J/\psi\,\eta$ cross section using the initial state radiation (ISR) process [1195], and BESIII confirmed this measurement at several energies [1208]. BESIII has also found that the $e^+e^- \to \chi_{c2}\,\omega$ cross section peaks near $\psi(4415)$ [1221]. The only charmonium-like states for which hadronic transitions are not known yet are $X(3940)$ and $X(4160)$. However, their masses are quite far from the expectations derived from conventional quark models, thus they also have unexpected properties.

The rate of the $\psi(3770)$ decay to $J/\psi\pi^+\pi^-$ is not anomalously high. However, the decay to $J/\psi\eta$ is not strongly suppressed, which corresponds to violation of HQSS. The rates of the decays into charged and neutral $D\bar{D}$ pairs are substantially different, thus violating isospin conservation. These properties point to a multiquark admixture in the $\psi(3770)$ [1240].

Recently Belle observed a new state, $X(3860)$, produced via $e^+e^- \to J/\psi(D\bar{D})$ [1185]. The mass and width of the state are $M = (3862^{+48}_{-35})\,\mathrm{MeV}/c^2$ and $(\Gamma = 201^{+177}_{-106})\,\mathrm{MeV}$, respectively. The spin-parity hypothesis $0^{++}$ is favoured over the $2^{++}$ at the $2.5\,\sigma$ level. The properties of $X(3860)$ agree well with expectations for the charmonium level $\chi_{c0}(2P)$.

Even before the observation of $X(3860)$, the authors of the phenomenological paper [1241] interpreted the near-threshold enhancement in the $\gamma\gamma \to D\bar{D}$ cross section as a signal of the $\chi_{c0}(2P)$. The mass and width estimated in Ref. [1241] are consistent with the measurement by Belle.

It is puzzling that a $0^{++}$ state is already known in this mass region: the $X(3915)$, with $M = (3918.4 \pm 1.9)\,\mathrm{MeV}/c^2$ and $\Gamma = (20 \pm 5)\,\mathrm{MeV}$. Properties of $X(3915)$ do not fit expectations for the $\chi_{c0}(2P)$ [1241, 1242]. The fact that this state is $190\,\mathrm{MeV}/c^2$ above the $S$ wave $D\bar{D}$ threshold but is only $20\,\mathrm{MeV}$ wide, is especially unusual.

Alternatively, the authors of Ref. [1243] proposed that the spin-parity of the $X(3915)$ is in fact $2^{++}$. Indeed, the $2^{++}$ assignment was found to be disfavoured relative to the $0^{++}$ assignment by the analysis of one-dimensional angular distributions in the $\gamma\gamma \to X(3915) \to J/\psi\,\omega$ process under the assumption that the $2^{++}$ state is produced only with helicities $\pm 2$ [1189], as expected for pure charmonium. It is pointed out [1243] that if $X(3915)$ has a non-$c\bar{c}$ admixture then the suppression of helicity 0 could be lifted, and with this contribution allowed, the $2^{++}$ hypothesis is no longer excluded. In this case the $X(3915)$ could correspond to a new decay channel of the $\chi_{c2}(2P)$ state [1243]. The large helicity 0 component would call for a prominent exotic component in the $\chi_{c2}(2P)$ state — in Ref. [1244] it is discussed to what extent this pattern is consistent with a molecular nature of this state. Further experimental studies, to be performed both in $B$ decays and in two-photon production, are needed to clarify the puzzle of the states near $3.9\,\mathrm{GeV}/c^2$.

[44] This number is possibly even smaller, since the errors in masses do not include systematic uncertainties due to neglecting interferences.

Recently BESIII observed that the signal of the vector state $Y(4260)$ is in fact a sum of signals of two structures, an enhancement that may be traced to $Y(4360)$ and a state with mass near $4.22\,\mathrm{GeV}$ [1215]. Note that already the original data peaked at this energy, however, since the distribution is highly asymmetric, this feature was diminished in the experimental analysis using a symmetric Breit-Wigner distribution — note that the molecular picture for the $Y(4260)$ naturally leads to an asymmetric lineshape as discussed in Sec. 14.5.3. This lower mass state is still called $Y(4260)$ by BESIII, though now its parameters have changed considerably (see Table 126). The new state decays to both spin-triplet and spin-singlet charmonia, which corresponds to a violation of HQSS and is unexpected for quarkonium. In addition, there are now a lot more vector states than expected $c\bar{c}$ levels in the considered mass region — and there might be even more than shown in the table. For example here we list the decay $Y(4660) \to \Lambda_c\bar{\Lambda}_c$ following Refs. [1245, 1246], although the distribution peaks at $4630\,\mathrm{MeV}$ and might in fact point to an additional state. The high mass $\psi$ states decay predominantly into $D^{(*(*))}\bar{D}^{(*)}$ channels, while no open flavour decays were found for the vector $Y$ states, which is also puzzling for the charmonium assignment.

In the bottomonium sector there are only three states in the region above the open flavour thresholds, the $\Upsilon(10580)$, $\Upsilon(10860)$ and $\Upsilon(11020)$ (see Table 126). For brevity, we will refer to them as $\Upsilon(4S)$, $\Upsilon(5S)$ and $\Upsilon(6S)$ according to the potential model assignment. However, they all show properties unexpected for pure $b\bar{b}$ pairs. The mass splitting between the $\Upsilon(4S)$ and $\Upsilon(5S)$ is larger by $(73 \pm 11)\,\mathrm{MeV}/c^2$ than that between the $\Upsilon(3S)$ and $\Upsilon(4S)$, while for a pure $b\bar{b}$ system it is expected to be smaller by about $40\,\mathrm{MeV}/c^2$ [1247]. The rates of $\Upsilon(5S) \to \Upsilon(nS)\pi^+\pi^-$ and $\Upsilon(6S) \to \Upsilon(nS)\pi^+\pi^-$ transitions are two orders of magnitude higher than expected for a pure bottomonium [1234, 1237]. The $\eta$ transitions, that in a pure bottomonium involve the spin flip of heavy quark and are suppressed by three orders of magnitude relative to $\pi^+\pi^-$ transitions, are not strongly suppressed in case of the $\Upsilon(5S)$ and are even enhanced in case of the $\Upsilon(4S)$ [1231]. In addition, the open-bottom two-body decays of the $\Upsilon(5S)$ show a sizable breaking of HQSS, see *e.g.* [1248, 1249], which is expected to be a very good approximation for bottomonia.

Thus practically all known hadrons containing $c\bar{c}$ or $b\bar{b}$ quarks with masses above corresponding open flavour thresholds have properties unexpected for a pure $Q\bar{Q}$ state; their structure is possibly more complicated. Theoretical interpretations of these states are discussed in the next subsections.

In 2015, the LHCb Collaboration reported the observation of two pentaquark-like structures in the $\Lambda_b \to P_c K^-(P_c \to J/\psi p)$ channel [1055], with opposite parities. Similar models as the one discussed above have been proposed to explain these new resonances, *e.g.* the (compact) pentaquark [1250], or the meson-baryon molecule [1251, 1252]. Ref. [1253] also considers the possibility for the narrower peak to be due to a kinematical singularity. Belle II can search for these (and similar) states in $c\bar{c}p$ both inclusively and in association with an antiproton.

*14.5.3. Models.* [C. Hanhart]

As described below in more detail most of the models for exotic states can be classified according to their clustering of the quarks and the relevant degrees of freedom:

- ◦ If the heavy quark-antiquark pair forms a compact quarkonium-like core surrounded by light quarks and anti-quarks, the state is called **hadroquarkonium**.
- ◦ If the light and the heavy quark as well as the light and the heavy anti–quark combine to form compact diquark and anti-diquark substructures (diquarkonium), respectively, one speaks of **tetraquarks** (note, this applies to the most prominent tetraquark model, but there are also tetraquark models that do not assume any diquark clustering).
- ◦ If the quarks and anti-quarks combine to form a pair of heavy hadrons, the object is called a **hadronic molecule**. When located close to the threshold of the molecular constituents the molecules can become quite extended — a feature that is crucial for a well-defined hadronic molecule and is claimed to lead to observable consequences.

In addition to those possibilities, there are exotics expected with gluons as active degrees of freedom:

- ◦ **Glueballs**: They are bound systems of gluons and do not carry any valence quarks. A lattice calculation based on quenched QCD, in which case glueballs do not mix with ordinary mesons, revealed glueball masses up to almost 5 GeV, with the lightest vector state having a mass of almost 4 GeV [1254].
- ◦ **Hybrid states**: In those both the gluons and the quarks act as active degrees of freedom and contribute to the quantum numbers [1255].

Many of the exotic states are located near thresholds. This led various authors to claim that they are simply kinematical effects that find their origin in the non-analyticity of any $S$–wave meson loop when crossing a threshold [1256–1263]. However, as stressed in Ref. [1264], if this were correct, those resonances should not show up as pronounced signals in the elastic channel (= the channel close to whose threshold the actual state is located) [45]. Based on this reasoning basically none of the near-threshold states found can be purely of kinematic origin; in other words: for all pronounced signals in elastic channels there must be a pole of the $S$-matrix nearby — they all classify as states.

In reality all the physical wave functions might contain some fraction of all of the mentioned configurations and at least the neutral ones even an admixture of regular quarkonium. Till now only a few studies investigate the interplay of quark model poles and exotics in the quarkonium mass range, see, *e.g.*, Refs. [1266–1271] for the quarkonium effect on hadronic molecules.

At present it is the main focus of research in the field to identify the most prominent component in the wave function of some given state. Already this enterprise calls for refined theories that allow one to relate observables to the underlying substructures in a controlled way as well as experiments of sufficient quality and quantity to be decisive.

*Tetraquarks.* [A. D. Polosa]

The constituent quark model has been by far the most successful tool for the classification and interpretation of hadrons. Despite its obvious limitations, the systematic search of $SU(3)$ multiplets provides the most reliable guideline in hadron spectroscopy. Exotic states with non-minimal quark content were forecasted by Gell-Mann in the very first paper on the quark model [1272]. The proposal of diquarks as effective degrees of freedom inside baryons came out in the late 60s: it is based on the observation that a $qq$ pair in the antisymmetric colour configuration binds according to the tree-level calculation (one gluon exchange). Some phenomena, like the $x \to 1$ of the ratio of proton and neutron PDFs, or the ratio of fragmentation functions into $\Sigma$ and $\Lambda$, can be qualitatively understood assuming the existence of these coloured objects. Also, some evidence of a scalar diquark was found in lattice QCD [1273].

[45] In Ref. [1265] the reasoning of Ref. [1264] is questioned. However, it should be stressed that in that work the mechanism to produce structures in the elastic and in the inelastic channels is very different and appears to be somewhat implausible.

Diquarks can be the constituent bricks of a new, rich multiquark spectroscopy. In 2003, Jaffe and Wilczek [1274] proposed a diquark-diquark-antiquark explanation for the positive-strangeness $\Theta^+$ baryon (whose existence was later contested by an higher statistics analysis). Soon after, Maiani *et al.* [1275, 1276] interpreted the light scalar sector in terms of diquark-antidiquark states.

For the Hamiltonian of the multiquark system we take [1277]

$$H = \sum_i m_i + 2\sum_{i<j} \kappa_{ij} \mathbf{S}_i \cdot \mathbf{S}_j T_i^a T_j^a, \tag{495}$$

$T$ being the $SU(3)$ generators and $\mathbf{S}_i$ the spins of the constituent quarks. The spin-spin interaction is local (proportional to $\delta(\mathbf{r})$). The spin-one diquarks are heavier and less likely to be produced. The parity of $S$-wave tetraquarks is positive.

Taking heavy quarks into account one can produce spin-one diquarks as well. Hence, the $S$-wave states carry $0^{++}$, $1^{++}$, $1^{+-}$ and $2^{++}$ quantum numbers [1278]. For each of them, the full $SU(3)$ nonet is in principle expected. The natural assignment for the $X(3872)$ is the $1^{++}$ member of the multiplet. This allows one to fix the unknown diquark mass and get predictions for the masses of the other states. In the first version of the model, the chromomagnetic couplings $\kappa$ were estimated from the splittings in the ordinary mesons and baryons spectra; however, this picture does not fit with the observed $Z_c'(4020)$ state.

On the other hand, if one thinks of the diquarks as pointlike particles separated in space [1238], the only nonzero contribution is due to the $\kappa_{cq}$ coupling inside the diquark itself; the Hamiltonian is diagonal in the diquark masses, and the resulting spectrum is compatible with the experiment.

An effective description of the tetraquark [1279]. can be given in terms of a double well potential segregating the the two diquarks apart: a system with two length scales, namely, the size of the diquark and that of the whole hadron. The tunneling amplitude of a heavy quark through the barrier separating the diquarks is exponentially suppressed with respect to the switching amplitude of the the two light quarks to produce a pair of open charm/beauty mesons.

One can observe that if the ratio of the two tetraquark length scales is chosen appropriately (and for very reasonable values of it), the two neutral and charged $X_{u,d}, X^\pm$ states are expected to be all quasi-degenerate. They will preferably decay into open charm mesons and, with smaller rates, into charmonia, as observed. The charged components, however, are forced to decay only into the suppressed charmonia modes, because of the heavier $D^\pm D^{*0}$ thresholds.

The quasi-degenerate $X_u$ and $X_d$ particles will get mixed. One can show that there are mixing angle regions allowing to explain, at the same time, the observed isospin breaking pattern of the decays into $\psi\,\omega$ and $\psi\,\rho$ and to keep the $\psi\,\rho^{\pm}$ modes well below the neutral one in $B$ decays [1279].

We notice that this approach can answer satisfactorily long standing questions challenging the diquark-antidiquark model of exotic resonances. The tetraquark description of $X$ and $Z$ resonances is shown to be compatible with present limits on the non-observation of charged partners $X^{\pm}$ of the $X(3872)$ and the absence of a hyperfine splitting between the two different neutral states. The $Z_c(3900)$ can be described as the partner of the $X(3872)$ with reversed charge conjugation [1238]. In the same picture, $Z_c$ and $Z_b$ particles are expected to form complete isospin triplets plus singlets. It is also explained why the decay rate into final states including quarkonia are suppressed with respect to those having open charm/beauty states.

The generalisation to $J = 1$ and $L = 1$ excitations reproduces the spectrum of vector states; the extra contribution to the Hamiltonian is

$$\Delta H_{L=1,J=1} = B_c \frac{\mathbf{L}^2}{2} - 2a\mathbf{L}\cdot\mathbf{S}. \tag{496}$$

For a complete treatment of this problem, see [1280] where tensor interactions are included.

The hadronic decay matrix elements depend on the details of the dynamics (see for example [1281]), and the most reliable predictions on the ratios of branching fractions are due to the fulfillment of heavy quark spin symmetry [1238, 1282]. Predictions on radiative decays can also be achieved.

Other properties of diquark-antidiquark mesons were forecasted in the late 70s in the context of the so-called baryonium in dual theories [1283–1285]. The isospin violation was predicted to happen in heavy baryonia because of the smallness of $\alpha_s(m_Q)$, with $Q = c, b$: this quenches the light quark annihilations and leads the eigenstates to align with the flavour basis. Charmed baryonia have more recently been considered in [1245].

LHCb recently observed a rich structure in the $B^+ \to XK^+(X \to J/\psi\phi)$ channel, and confirmed the $X(4140)$ seen at Tevatron and CMS, albeit with much broader width. Were these states confirmed, they naturally constitute the candidates for a $[cs][\bar{c}\bar{s}]$ tetraquark multiplet [1286].

In the bottomonium sector, tetraquark interpretations of the axial states were also proposed by Ali *et al.*. [1282, 1287].

The main drawback of the tetraquark model is the experimental absence of many of the predicted states, most notably of the charged partners of the $X(3872)$ and of its bottomonium analogous; see the discussion above and [1279]. Moreover, in the original models, the presence of several close meson-meson thresholds is ignored for it was supposed they had naturally to occur right below the mass of the observed states (since diquarks are less bound than a colour singlet).

On the other hand the tetraquark model predicted the presence of charged states as the $Z^+(4430)$ and the $Z^{\pm}_{c,b}$ resonances. With present data, all of them are *above* the respective meson-meson threshold with the corresponding quantum numbers. Since diquarks are slightly less bound than colour singlets this explains why compact tetraquarks should be observed close, but above the corresponding meson-meson thresholds.

A tentative solution to the problem of charged states was proposed also in [1288], following some ideas presented in Refs. [1131, 1289, 1290]. In this picture the observed states are neither pure tetraquarks nor meson molecules, but rather the result of an hybridisation between the first ones and the two-meson states. According to the Feshbach formalism, the scattering length for an unbound pair of particles (open channel) is dramatically enhanced whenever a discrete level of the same quantum numbers (closed channel) happens to be close and *above* the onset of the continuous spectrum. This hybridisation is in contrast to the formation of a bound state, in which case the discrete level must be below threshold.

In this model, the two-meson spectrum corresponds to the open channel, while the closed one is provided by the compact tetraquark previously described. The hybridisation between the two consists in an inelastic scattering that temporarily rearranges the internal structure of the four-quark system.

The phenomenon described here induces a resonant enhancement in the production of tetraquarks and would be compatible (as any compact tetraquark model) with their production in high energy and high $p_T$ proton-(anti)proton collisions, as opposed to what expected for real loosely bound molecules, as discussed in the literature [1291, 1292] [46].

The enhancement in the scattering length together with the fact that the energy of the pair must be smaller than some $E_{\max}$, also instruct us on the total width of the state. This is now expected to be $\Gamma \sim \sqrt{\delta}$, where $\delta$ is computed with respect to the *closest* threshold from below, as already explained [47]. This prediction correctly fits many of the observed $X$ and $Z$ states [1288].

The analysis does not straightforwardly generalise to excited orbital and radial states and to better understand the absence of the isospin partners of the $X(3872)$ we need to turn to the more complete description introduced in [1279] and sketched at the beginning of this section — inspired in part from the Feshbach resonance idea.

The picture in [1279] strongly suggests some experimental tests to be done: *i*) improve on the bounds on $X^{\pm}$ in $J/\psi\rho^{\pm}$ decays by at least one order of magnitude *ii*) improve on the precision on the mass measurement of the neutral component of the $X(3872)$ in various decay channels *iii*) determine wether the $Z_c(3900)$ is produced or not in $B$ decays and in prompt $pp$ collisions — search all of $X, Z$ states both in $B$ decays and in prompt hadron collisions, and measure their production cross sections.

*Hybrid states.* [A. V. Nefediev]

Hybrid hadrons (or simply hybrids) are states where not only quarks and antiquarks but also gluons contribute to the quantum numbers and other properties of the system. Indeed, due to the non-Abelian nature of the interaction mediated by gluons, the latter can play a role of extra constituents of hadrons or even form a new type of compounds — glueballs made entirely of two, three, or more gluons. In particular, in the language of the potential quark models, conventional mesons are described as the radial and orbital excitations of the quark-antiquark pair connected by a confining potential or, in more sophisticated models, by a string-like object usually referred to as the flux-tube or the QCD string. Hybrid excitations correspond to the vibrational modes of this object formed by gluons. In the simplest realisation, the hybrid meson is a quark-antiquark pair accompanied by a single excitation quantum of the glue. There exists a vast literature on theoretical approaches to hybrids. The corresponding predictions can be found, for example, in [1297, 1298] (bag model), [1255] (flux-tube model), [1299–1301] (Coulomb-gauge QCD approach), [1137] (nonrelativistic QCD (NRQCD) approach), [1302] (potential quark model), [1303–1305] (constituent gluon model), [1306–1314] (QCD string approach). Predictions of different models for the hybrids may slightly differ from each other. For example, in the flux tube model [1315], the hybrid excitations are visualised as phonon-type objects while in the constituent gluon model [1303–1305] the latter carry colour and spin. This can lead to different predictions for the quantum numbers and other properties of the hybrids. Nevertheless, there is a consensus about the most general and most important features of hybrids which are discussed below.

[46] Note that Refs. [1293–1296] come to a different conclusion about the production of shallow bound states.

[47] The fact that $\sqrt{\delta}$ is referred to the smallest detuning rather than to the larger available ones — those related to other decay modes — shows that this is not just a simple phase space effect.

One of the straightforward consequences of the presence of an extra degree of freedom in the system is a richer set of quantum numbers available for the hybrids. For example, while the quantum numbers $0^{+-}$, $1^{-+}$ and so on are not accessible in the standard $J^{PC}$ scheme for the conventional quark-antiquark mesons, such options are allowed for hybrids. Thus, the experimental observation of a state containing a heavy quark-antiquark pair which at the same time has exotic quantum numbers would provide a strong candidate for a hybrid meson.

Another consequence of the excited glue in hybrids is a higher mass of the latter compared to the conventional mesons. Indeed, whatever model for the hybrid is used, its ground state is expected to have a mass, roughly, $2m_Q + 1$ GeV, where $m_Q$ is the mass of the heavy quark and the extra 1 GeV comes from the gluons. Therefore, in the spectrum of charmonium and bottomonium, it is natural to expect the lightest hybrid to have a mass slightly above 4 GeV and around 11 GeV, respectively. Indeed, the eight lowest $c\bar{c}g$ hybrids predicted in the flux-tube model [1255] reside around 4.1-4.2 GeV, with the $1^{--}$ and $1^{-+}$ states being among those. Other models give similar predictions. To mention just a few, in a potential model, with the $c\bar{c}$ pair considered as a colour-octet source, the tensor hybrid is predicted at 4.12 GeV [1302]; calculations within the QCD string model give for the mass for the exotic $1^{-+}$ charmonium hybrid $4.2 \pm 0.2$ GeV [1316] or $4.3 - 4.4$ GeV [1313, 1314]. The sibling states with a magnetic gluon and with the quantum numbers $0^{-+}$, $1^{--}$, and $2^{-+}$ are found in [1313, 1314] to lie within the range $4.3 - 4.5$ GeV; the nonrelativistic effective field theory approach of [1137] predicts multiple hybrid states with different quantum numbers, including the total spin $J$ as large as 3 or 4, in the range $4.0 - 4.7$ GeV — this approach is discussed in some detail in Sec. 14.5.5. It should be noticed that the existence of the sibling hybrid states with different quantum numbers lying close to each other is yet another natural consequence of the extra degree of freedom introduced in the system. In the flux tube model, the spin splittings between such hybrids are due to the long-range Thomas precession and they were found to be small [1317]. In the QCD string model, these splittings are much larger, and they mostly come from the perturbative short-ranged forces [1313, 1314]. In the EFT of [1137] they stem from different gluonic excitation operators which appear in the multipole expansion in perturbative NRQCD.

The situation with hybrids in the spectrum of bottomonium looks similar. Indeed, the most recent calculations place the lowest bottomonium hybrids around 11 GeV, in agreement with the simple estimate made above, see Refs. [1137, 1314, 1318] for more details.

An independent source of information about the masses and splittings of the hybrid mesons is provided by the results of the lattice calculations which are collected in *Hybrids* in Section 14.5.4.

To summarise, both theory and numerical simulations indicate that hybrids with different quantum numbers, including those with exotic ones, may exist in the mass region $4-5$ GeV and around 11 GeV in the spectrum of charmonium and bottomonium, respectively.

However, it remains difficult to identify hybrid states in the experimental spectra. Therefore, it is important to establish selection rules which would allow one to disentangle the conventional mesons from the hybrids.

For the leptonic decays, a straightforward selection rule of this kind follows from the fact that the quark-antiquark pair in the conventional vector meson can easily annihilate into lepton pairs, but a similar decay is forbidden for hybrids which contain the $Q\bar{Q}$ pair in the colour octet.

Open-flavour decays also provide a set of selection rules and criteria for hybrids. In particular, it was found long ago that, due to the symmetry of the wave function, a selection rule exists which forbids the decay of the vector hybrid with a magnetic gluon into a pair of $S$-wave open-flavour heavy-light mesons in the final state [1304, 1305, 1319–1321] . On the contrary, hybrids with electric gluons couple quite strongly to such $S$-wave pairs and, as a result, they are very broad and not observable experimentally. At the same time, for the allowed decays, for example, into one $S$-wave and one $P$-wave open-flavour meson, the relative strength of such decays encoded in the corresponding coupling constants depends strongly on the pair creation mechanism for the light-quark pair ($\bar{q}q$): it is created with the quantum numbers of the vacuum, for conventional mesons, however it is coupled to the gluon for the hybrid, and as such it carries the quantum number of the vector. Thus, the recoupling coefficients for the decays into the $(\bar{Q}q)(\bar{q}Q)$ final state differ substantially for the $\bar{Q}Q$ mesons and for the $Q\bar{Q}g$ hybrids — see examples in [1313, 1322].

It has to be noticed that, in the strict heavy-quark limit, the quadruplet of the $P$-wave heavy-light mesons turns to a pair of doubly degenerate states $P_j$ corresponding to a particular total momentum of the light quark, $j=1/2$ or $j=3/2$. Since the $P_{1/2}$ and the $P_{3/2}$ mesons decay via pion emission to the lower-lying $S$-wave heavy-light mesons in the $S$-wave and in the $D$-wave, respectively, then the $P_{1/2}$ mesons appear to be much broader than the $P_{3/2}$ ones. This makes it hardly feasible to identify experimentally either of the two $P_{1/2}$ quadruplet members in the final state.

In the meantime, production of a heavy-light meson from the $P_{3/2}$ doublet accompanied by an $S$-wave meson is only possible if the produced light-quark pair has the total angular momentum equal to 1. This condition is not fulfilled for the vector bottomonium where $j_{q\bar{q}}=0$ and, therefore, the amplitude for its decay into such a final state vanishes in the heavy-quark limit [1323, 1324]. Meanwhile, as mentioned above, open-flavour decays of $Q\bar{Q}g$ hybrids proceed through gluon conversion into a light quark-antiquark pair which carries the quantum numbers of the vector, $j_{q\bar{q}}=1$. This implies that there is no suppression for the amplitude of the vector hybrid decay into a pair of one $S$-wave and one $P_{3/2}$ open-bottom meson. Therefore, in the strict heavy-quark limit, the decays to the open-flavour final states containing the members of the positive-parity quadruplet of heavy-light mesons could be used as test modes for the hybrid [1314].[48]

Beyond the strict heavy-quark limit, corrections of two types have to be taken into account. On the one hand, there exist corrections which are controlled by the parameter $\Lambda_{\rm QCD}/m_Q$. Such corrections are sizeable in the charm sector, so that the heavy-quark symmetry can provide only qualitative predictions for charmonia. In the meantime, since $m_b \gg \Lambda_{\rm QCD}$, the heavy-quark symmetry constraints are typically very well met in bottomonium systems. On the other hand, for a finite $m_Q$, the physical $P$-level heavy-light mesons come as certain mixtures of the $P_{1/2}$ and $P_{3/2}$ states governed by the mixing angle $\theta$. Then, the probability of the decay of the genuine quarkonium $\bar{Q}Q$ (of the $Q\bar{Q}g$ hybrid) into the final state containing one $S$-level and one narrow $P$-level open-flavour meson is proportional to $\sin^2\theta$ ($\cos^2\theta$). Recent estimates of the mixing angle $\theta$ in the charmed and bottomed systems demonstrate that, at least in the $b$-sector, the given selection rule may allow one to distinguish the hybrid from the conventional meson [1314].

Also, the final-state momentum distributions in the open-flavour decay channels can provide an additional valuable information about the nature of the decaying state [1300, 1325]. The method is reminiscent of the Franck-Condon factorisation principle in molecular physics which is based on the so-called "velocity superselection rule." The rule states that the heavy quark does not change its velocity upon emitting or interacting with the light degrees of freedom, such as light quarks, gluons, pions, and so on, with a momentum of the order of the typical QCD scale $\Lambda_{\rm QCD}$. This entails that the momentum distribution of the heavy mesons in the open-flavour decays should be proportional to the momentum distribution of the heavy quarks inside the parent hadrons, thus giving a window to their internal structure. As was mentioned above, excited glue brings a large contribution to the energy of the hybrid, so that the low-lying hybrids have the quark-antiquark pair in the ground state. Meanwhile, for a conventional meson, one needs to excite the radial motion of the quark-antiquark pair to arrive at the state with the same mass and the same $J^{PC}$ quantum numbers. For example, in the framework of potential quark models, the vectors $\Upsilon(10860)$ and $\Upsilon(11020)$ lying in the mass range around 11 GeV correspond to the fourth and the fifth radial excitation of the $b\bar{b}$ pair, respectively, conventionally denoted as $\Upsilon(5S)$ and $\Upsilon(6S)$. Their wave functions possess 4 and 5 nodes, respectively, which is to be confronted with the nodeless wave function of the $b\bar{b}g$ hybrid. Since, for a given total energy, the relative momentum of the mesons in the final state is fixed by the energy conservation law then different two- and three-body open-flavour final states probe different parts of the wave function of the decaying state — see [1300, 1325] for further details. Therefore, studying the two- and three-body open-bottom final states, it should be possible to make a conclusion concerning the source of the $B$ meson pair: a smooth distribution would identify the source as the hybrid while the distribution with residual (after smearing due to the quark recoil in $B$ mesons) structures would indicate a conventional (highly radially excited) quarkonium as the corresponding source.

For hidden-charm decays of the charmonium hybrids two different types of transitions can be identified: with and without conventional charmonium production in the final state. In the former case, the decay proceeds through the $c\bar{c}$ pair conversion from colour octet to colour singlet via emission of a single gluon and with a consequent annihilation of the gluons into light hadrons [1326],

$$\psi_g(c\bar{c}g) \to (c\bar{c})(gg) \to \psi(c\bar{c}) + \text{light hadrons}, \tag{497}$$

where $\psi_g$ is a hybrid while $\psi$ is a conventional charmonium. Such decays populate final states with hidden charm which may provide a clear experimental signal if the charmonium $\psi(c\bar{c})$ is the $J/\psi$ or a higher charmonium which decays into $J/\psi$ through a cascade. On the contrary, in the latter case, the $c\bar{c}$ pair in the hybrid annihilates into gluons which then convert into light hadrons,

$$\psi_g(c\bar{c}g) \to (ng) \to \text{light hadrons}, \quad n \geqslant 2, \tag{498}$$

so that such decays enhance no-charm final states [1327]. See also [1328] for further details of the experimental signatures and search strategies for the charmonium hybrids in the $B$ meson decays.

[48] If the state is close to the threshold of the $S$-wave and $P$-wave heavy mesons, the strong coupling would also imply a sizeable hadronic molecule component to be discussed below.

Additional information on hybrids can be obtained from the production reactions where hybrids are created in line with the conventional mesons. For $B$-factories, the most relevant production reaction of this kind is a decay of the $B$ meson of the form $B \to \psi_g + X$, where $X$ stands for the rest of the products of the decay. It has to be noticed that, although both types of hybrids with a heavy quark-antiquark pair and with one heavy and one light quark can be produced in such $B$ meson decays, only hybrids containing a $c\bar{c}$ pair are eigenstates of the $C$-parity operator and as such they can possess exotic quantum numbers to be regarded as a smoking-gun-like signature of an exotic state. Then the underlying weak transition is governed by the CKM-favoured decay $b \to c\bar{c}s$. Since the $c\bar{c}$ pair in such a transition should not be in the colour singlet, it is produced through the colour octet intermediate state — the corresponding theoretical estimates can be found, for example, in [1328, 1329]. In particular, in [1329], the branching fraction for the decay $B \to \psi_g(0^{+-}) + X$ is estimated at the level of $10^{-3}$ while it is argued in [1328] that $\mathcal{B}[B \to \psi_g + X]$ can be as large as $\simeq 1\%$ for any quantum numbers of the hybrid, including the exotic ones $1^{-+}$. Therefore, hybrids are produced in the $B$ meson decays with the probability comparable with that for the conventional charmonia.

The experimental status of hybrids is obscure because up to now not a single hybrid is identified beyond doubt. One of the most prominent candidates for a hybrid state is $Y(4260)$ (see Table 126) which demonstrates some features expected for a charmonium hybrid. In particular, it has a mass close to the phenomenological and lattice predictions discussed above and, what is more important, it has a decay pattern (small electronic width and not seen open-charm decays of a particular type) that is not typical for conventional mesons but that is specific for hybrids. Thus, this state could be a hybrid charmonium with a spin-1 [1137, 1330] or spin-0 $c\bar{c}$ core [1313, 1320]. However, further studies of the open-charm decays of this state [1331] question its hybrid nature — alternative scenarios for the $Y(4260)$ are discussed below. In addition, if the resonant spectrum of transition $e^+e^- \to h_c(1P)\,\pi\pi$ around 4.22 GeV [1216] is found to be dominated by the $Y(4260)$ that also decays into $J/\psi\pi\pi$, the data might call for a mixture of two structures, since the appearance of both transitions would otherwise violate heavy quark spin symmetry.

To summarise, hybrids in the spectrum of charmonia and bottomonia are expected to possess a few specific features which are expected to allow one to disentangle them from conventional quarkonia. Among those are

- the exotic quantum numbers not accessible for the quark-antiquark system;
- the co-existence in the same mass region with sibling states with different quantum numbers;
- abnormally small leptonic width;
- peculiar decay pattern into open-flavour meson pairs;
- smooth momentum distribution in the two- and three-body open-flavour decays.

*Hadroquarkonia.* [C. Hanhart]

Triggered by the observation that a large number of exotic candidates decay into a quarkonium accompanied by one or more pions, these candidates were proposed in Refs. [1332, 1333] to consist of a core provided by a heavy quarkonium surrounded by an excited state of light–quark matter. In this picture the mentioned decays are understood as setting free the quarkonium core in the process of de-exciting the light–quark cloud into one or more pions.

It is expected that the dominant decay modes of hadroquarkonia are given by light quarks in combination with the core quarkonium. In particular, since in heavy quark systems the spins of the heavy quarks and the total angular momentum of the light quarks are conserved individually, any given state should decay either into a spin 1 or a spin 0 quarkonium, but not into both. However, this spin symmetry selection rule can be evaded by mixing [1334]. Following this idea $Y(4260)$ (potentially seen not only in the final state $J/\psi\pi\pi$ but also in $h_c\pi\pi$) and $Y(4360)$ (seen in $\psi'\pi\pi$) could be mixtures of two hadrocharmonia with spin triplet and spin singlet heavy quarkonium cores. The same kind of mixing could also operate for hybrids.

The mixing scenario of Ref. [1334] opens an interesting opportunity: using the proposed scenario for $Y(4260)$ and $Y(4360)$ as input one can use spin symmetry to predict in total 4 spin partners of the mentioned states — most special amongst them is a pseudoscalar state, which is significantly lighter than $Y(4260)$ [1335]. Although this state appears at a similar mass as the second radial excitation predicted within typical quark models, it could still be identified via its prominent decay into $\eta_c^{(\prime)}\pi\pi$, while a decay into $D^*\bar{D}$ should not occur. Testing these predictions provides crucial tests for the hadroquarkonium model.

Since in leading order the light quark cloud does not feel the flavour of the quarkonium core the masses of the bottomonium partners of hadrochamonia can be found simply by adding the mass difference of the assumed core state and its bottomonium partner to the hadrocharmonium mass. Although this picture can get distorted to some extent via the interactions with neighbouring states, it should be clear that a lot can be learned from a comparison of the bottomonium spectrum and the charmonium spectrum.

*Hadronic Molecules.* [F.-K. Guo, C. Hanhart]

In contrast to the compact tetraquarks discussed above that are formed from coloured (anti)-diquarks, hadronic molecules are understood as bound states of two colour neutral hadrons. This results in a different analytic structure of the corresponding amplitudes that leads to observable consequences, if the corresponding states are located close to the relevant threshold. For a recent review, see [1336]. To get a quantum mechanical understanding of this statement one may think of the wave function of a physical state to be made of two components: a two–hadron and a compact component. In the modern literature the former is often called the molecular component. Already in 1963 Weinberg showed that the probability to find the molecular component inside the physical wave function, $(1-\lambda^2)$, is related to the physical coupling of the state to the continuum channel via [1337]

$$\frac{g_{\rm eff}^2}{4\pi} = 4M_{\rm thr.}^2(1-\lambda^2)\sqrt{\frac{2\epsilon}{\mu}} \leq 4M_{\rm thr.}^2\sqrt{\frac{2\epsilon}{\mu}} \ , \tag{499}$$

where $M_{\rm thr.} = m_1 + m_2$ denotes the threshold mass, and $m_{1,2}$ and $\mu$ denote the masses of the individual constituents and their reduced mass, respectively. The binding energy, $\epsilon$, is defined with respect to the continuum threshold via

$$M = m_1 + m_2 - \epsilon \ , \tag{500}$$

where $M$ denotes the mass of the state considered. Eq. (499) is correct up to corrections of the order of $\gamma R$, where $\gamma = \sqrt{2\mu\epsilon}$ denotes the binding momentum and $R$ the range of forces. Since $g_{\rm eff}^2$ is nothing but the residue at the pole for the state considered, via Eq. (499) the amount of molecular component in a wave function becomes an observable, *e.g.* the scattering length scales as [1337]

$$a = -\left(\frac{1-\lambda^2}{1-\lambda^2/2}\right)\frac{1}{\gamma} \ . \tag{501}$$

For binding momenta $\gamma$ much smaller than any intrinsic scale of the system considered and $\lambda^2 \to 0$, all physics gets controlled by the single scale $\gamma$. In particular the scattering length gets unnaturally large. This gives rise to various universal phenomena as detailed in Ref. [1338] as well as allows for a construction of effective field theories based on hadronic degrees of freedom [1131–1133] largely inspired by EFT treatments for the nucleon–nucleon interaction [1134].

The derivation of Eq. (499) involves the normalisation of a bound state wave function and it therefore holds rigorously only for stable bound states. However, it was shown that it can be generalised to states coupling to remote inelastic channels [96]. In addition, in order to keep the corrections small, the considered bound systems should be shallow. Generalisations of the Weinberg approach to coupled channels as well as resonances can be found in Refs. [1339–1342]. In any case, as soon as one adopts the physical picture also for somewhat more deeply bound systems, namely that the coupling of a state gets large when it has a sizeable molecular component, quite significant observable consequences emerge, like highly asymmetric line shapes, as was shown, *e.g.*, in Refs. [1343, 1344] on the example of the $Y(4260)$ as a $D_1\bar{D}$–molecular structure (for a recent discussion, which also contains the most recent BESIII data of relevance for the $Y(4260)$, we refer to Ref. [1345]).

While detailed predictions for new states within a molecular picture are difficult, since they require a detailed dynamical modelling analogous to that necessary to describe few-nucleon systems, some general statements are still possible. For instance, molecules should form (predominantly) in $S$–waves since the centrifugal barrier is always repulsive. Therefore the quantum numbers of the constituents already define the molecules they can (most easily) form. In addition, only narrow states can form hadronic molecules, since a shallow bound state that contains a broad building block would be very short lived or might not even have the time to be formed before the constituent decays [1346, 1347].

In addition, it appears natural to expect that the one-pion exchange plays an important role in the formation of the bound states [1348], which after all is also understood as a crucial ingredient to the nuclear force. In this context it is important to acknowledge that the strength of the one-pion exchange changes by a factor $-1/3$ when switching from an isoscalar to an isovector channel. Thus, if the pion exchange provides a crucial contribution to the binding of the isoscalar $X(3872)$ one might be tempted to claim that there should be no charged molecules. However, there is an additional change in sign, when switching to systems of opposite $C$-parity. As a result of this one should expect that, if there is an isoscalar molecule of a given $C$-parity, the isovector partner, if it exists, should have opposite $C$-parity [1335]. This is in contrast to the tetraquark picture where for each $J^{PC}$ there should always be both an isoscalar and an isovector state, but in line with experiment at least for the states near the $D\bar{D}^*$ threshold, since the $X(3872)$ has positive $C$ parity while the $Z_c(3900)^+$ has negative $C$ parity. Following this logic one might also expect an isoscalar partner of the $Z_c(4020)^+$ near the $D^*\bar{D}^*$ threshold with $J^{PC} = 1^{++}$, which, however, does not exist since an $S$-wave $D^*\bar{D}^*$ state with spin 1 has negative $C$-parity [1349].

In heavy quark systems spin symmetry can provide an important diagnostic for the study of the structure of hadrons [1335]. Detailed studies for the implications of the heavy quark spin symmetry on both states in the charmonium and bottomonium sector can be found in Refs. [1239, 1350–1354]. While the pattern does not get destroyed by the inclusion of the one-pion exchange, spin symmetry violations driven by the mass difference in the open flavour mesons might distort the pattern severely [1355, 1356]. This becomes apparent, *e.g.*, when looking at the spin 2 partner of the $X(3872)$. This state was predicted to be a shallow bound state of $D^*\bar{D}^*$ with the same binding energy as $X(3872)$ [1349, 1354] and a narrow width [1357]. However, a crucial part of the one-pion exchange is its tensor force that induces sizable transitions from $S$- to $D$-waves. In case of the spin 2 partner of the $X(3872)$ these transitions allow for a decay of the state to $\bar{D}D$ in $D$-wave which might result not only in a significant mass shift but also in a sizeable width for this state [1355]. A recent study of the lineshapes of $Z_b(10610)$ and $Z_b(10650)$ revealed, that a proper renormalisation of the full one-pion exchange contribution is possible only, if a naively order $p^2$ counter term that induces $S-D$ transitions is promoted to leading order. The effect of this is that a significant fraction of the tensor force contribution gets canceled [1358]. What impact this counter term can have in the charmonium systems remains to be seen. Predictions based solely on interactions of a heavy meson pair could also be distorted due to the interplay with the preexisting charmonium states [1271] — for a more general discussion of this scenario we refer to Refs. [1266, 1267]. In this context a detailed knowledge of the bottonomium spectrum would be extremely valuable since the mentioned violations of spin symmetry should be suppressed significantly in those heavier systems. Moreover, the location of the open flavour thresholds relative to the quarkonia is expected to be different in the bottom sector compared to the charm sector.

Even without any detailed calculation it should be clear that especially by comparing charmonium and bottomonium sectors a great deal of physics may be revealed, *e.g.* molecular states are located at a binding energy where the kinetic energy matches the potential energy. Accordingly the binding momentum, $\gamma = \sqrt{2\mu\epsilon}$, is an important characteristic of a two-hadron bound state. In heavy-light two-hadron molecular states the reduced mass is close to the mass of the light hadron and as such the binding energies of, *e.g.*, the $KD^{(*)}$ and the $K\bar{B}^{(*)}$ system should be similar. Therefore, if indeed $D_{s0}^*(2317)$ and $D_{s1}(2460)$ are bound states of $KD$ and $KD^*$, respectively, as claimed in Refs. [1359–1363], which can naturally explain the otherwise puzzling fact $M_{D_{s1}(2460)} - M_{D_{s0}^*(2317)} \simeq M_{D^*} - M_D$, the actual masses of the corresponding bottom states should reveal important information on the flavour dependence of the binding potential. In Ref. [1364], the radiative decays of these scalar and axial vector states are identified as the most promising discovery modes of these predicted bottom states (isospin violating decay modes are estimated in Refs. [1364] within the molecular picture and in Refs. [1365, 1366] using heavy meson chiral perturbation theory). In the scenario that the $D_{s0}^*(2317)$ and $D_{s1}(2460)$ are hadronic molecules, some quantitative predictions have been confronted with both lattice QCD and experimental data. Using the parameters fixed in Ref. [1367], which leads to an $\sim 70\%$ $DK$ component in the $D_{s0}^*(2317)$, the finite-volume energy-levels in the scalar isoscalar $D\pi, D\eta$ and $D_s\bar{K}$ coupled-channel system [1368] were found to be in a remarkable agreement with the lattice results by the Hadron Spectrum Collaboration [174]. The same parameters also result in a good description [1369] of the precise measurements of the $D^+\pi^-$ and $\bar{D}^0K^-$ angular moments in the $B^- \to D^+\pi^-\pi^-$ [336] and $B_s^0 \to \bar{D}^0K^-\pi^+$ [1370] processes. These agreements may be regarded as a strong support of the hadronic molecular scenario. Nevertheless, these comparisons are only in the scalar sector, since the quality of the data in the $1^+$ sector [327] is not good enough. The expected much better data from Belle-II will be crucial to allow for more firm statements.

In contrast to the heavy-light systems, the flavour dependence of the kinetic energy of heavy-heavy systems is much stronger, since here the reduced mass is of the order of the heavy meson mass. Thus, for heavy-heavy two-hadron molecular systems one should expect significant differences in the bottomonium and charmonium spectra, as exemplified by the large binding energy difference between the $X(3872)$ and $X_b$ in Ref. [1354].

Before closing this subsection we would like to give a few examples of molecular candidates in the heavy hadron spectrum. One of the prime candidates for a molecular state is $X(3872)$. Its mass lies extremely close to the $D^0\bar{D}^{*0}$ threshold and therefore a natural explanation for this state might be a $1^{++}$ $D\bar{D}^*$ molecule [1371]. As a consequence of the separation to the $D^+D^{*-}$ channel of only 8 MeV, strong isospin breaking is predicted in this scenario [1371, 1372]. The comparable rates in the $\omega J/\psi$ and $\rho^0 J/\psi$ channels appear to be consistent with an interpretation of $X(3872)$ as an isoscalar $D\bar{D}^*$ molecule when the different widths of $\rho$ and $\omega$ as well as the mass difference between the $D\bar{D}^*$ thresholds are taken into account [1373]. Also in Ref. [1374] it becomes apparent that only an isoscalar structure is consistent with the decay properties of $X(3872)$. The copious production of $X(3872)$ at very high $p_T$ in $pp$ and $p\bar{p}$ collisions was claimed to be in conflict with a pure molecular assignment [1290, 1291], however, the role of rescattering could be crucial in enhancing the production cross sections [1293–1295]. Recently the production of shallow molecules was revisited from a different angle in Ref. [1296] again showing that the observed production rates are not in conflict with expectations. However, it is fair to say that up to date this issue is not resolved in the literature. The comparison with light nuclei production at high $p_T$ proposed in [1292] might shed further light on the molecular assignment. However, there is a crucial difference between a light nucleus and the $X(3872)$, namely that the former does not allow for a $q\bar{q}$ component, which can prevent a decisive conclusion from being drawn from this comparison [1336].

Other quarkonium-like states very close to open-flavour thresholds include the charged $Z_b^{\pm}(10610, 10650)$ and $Z_c(3900, 4020)^{\pm}$. For these, also, a molecular interpretation was proposed shortly after their discovery: $B\bar{B}^*$ and $B^*\bar{B}^*$ for the $Z_b$ states [1239], and $D\bar{D}^*$, $D^*\bar{D}^*$ for the $Z_c$ states [1354, 1375]. Measured spin and parity of $J^P = 1^+$ for $Z_b(10610)$ and $Z_b(10650)$ [1173] and for $Z_c(3900)$ [1163, 1164] correspond to heavy-light mesons in the $S$-wave, in line with the molecular interpretation. The experimental fact that, *e.g.*, the $Z_b$ states decay predominantly into the open flavour channels although being located in mass very close to their thresholds, is claimed to be a "smoking gun" of a molecular structure [114, 1239, 1376]. A recent combined analysis of the $B^{(*)}\bar{B}^*$ and $h_b(mP)\pi$ channels using amplitudes consistent with unitarity and analyticity indicates that $Z_b(10610)$ and $Z_b(10650)$ may in fact be virtual molecular states with poles within 2 MeV from the corresponding thresholds [114]. A similar conclusion for the $Z_c(3900)$ was made in Ref. [1377], where an above-threshold resonance solution was also found in addition to the virtual state one. To fully pin down the pole locations of the $Z_b$ and $Z_c$ states additional data of better statistics appear to be necessary. Those studies are really important: if the masses of all the charged states mentioned in this paragraph were indeed located above the corresponding thresholds, it would challenge the molecular interpretation (and might support the tetraquark interpretation — see discussion in the tetraquark subsection above) for molecular structures naturally appear either below threshold or are broad [1378] since the coupling of the given state to the channel that forms is large (see Eq. (499)). To further establish if indeed the $Z_b$ states are of molecular nature or consist predominantly of more compact components like tetraquarks in Ref. [1376] several decay ratios are given based on the molecular picture. In Ref. [1379] similar relations are derived for the $Z_c$ states both within the molecular and the tetraquark picture. Indeed, given the heavy flavour symmetry of QCD, a detailed comparison of the charmonium and the bottomonium spectra should provide deep insights into the way how nature forms hadrons.

The above examples concern candidates of hadronic molecules formed by two ground state heavy mesons. Molecules formed of other heavy hadrons are possible as well. The most discussed example is the $Y(4260)$ which was conjectured to be predominantly a $D_1(2420)\bar{D}$ molecule [1375]. If this were the case, the $D_1\bar{D}$ decay should play a significant role in the $Y(4260)$ physics, but data are still inconclusive on that [1163, 1343]. The resulting binding energy of 70 MeV pushes this state out of the validity range of Weinberg theorem [1337], and makes the $Y(4260)$ predictions more model-dependent. However, this assignment not only provides a natural mechanism for the production of a $D\bar{D}^*$ molecule, $Z_c(3900)^+$, but also allows subsequently for the prediction of a copious production of $X(3872)$, also assumed to be a $D\bar{D}^*$ molecular state, in $Y(4260)$ radiative decays [1380]. This prediction was confirmed shortly after at BESIII [1148]. The same radiative transitions naturally occur, if the $Y(4260)$ is identified as the orbital excitation of the $X(3872)$ tetraquark [1238]. The $Y(4360)$ with a large $D_1\bar{D}^*$ component could be the spin partner of the $Y(4260)$ [1381–1383], but a detailed microscopic calculation to make this connection solid is still lacking. The new value of the $Y(4260)$ mass of about 4.22 GeV (see Table 126) agrees better than the old one with the expectations for the molecule [1343, 1344]. In addition, the mass difference between $Y(4260)$ and $Y(4360)$ is now closer to that between $D$ and $D^*$ mesons, as expected in the molecular picture.

The $\Upsilon(6S)$ is situated near the $B_1(5721)\bar{B}$ threshold, where $B_1(5721)$ is a narrow $P$-wave excitation with the spin-parity of the light degrees of freedom $j^P = 3/2^+$, and can be the bottomonium analogue of the $Y(4260)$. A contribution of the $B_1(5721)\bar{B}$ pairs to the $\Upsilon(6S)$ decays[49] has a very clear experimental signature: the $Z_b(10610)\pi$ final state should be produced, while the $Z_b(10650)\pi$ should not [1384], in full analogy to $Y(4260)$. This prediction is distinct from the observations at $\Upsilon(5S)$, where both $Z_b$ states are produced in roughly equal proportion. Present data provide only a very loose constraint on the relative yields of $Z_b(10610)$ and $Z_b(10650)$, and do not exclude the single $Z_b(10610)$ hypothesis at a high confidence level [1177]. The observation of a possible analogy of $Y(4260)$ and $\Upsilon(6S)$ also makes one suggest that the radiative decay of $\Upsilon(6S)$ to the $X_b$, the to-be-found bottomonium partner of $X(3872)$, should be sizeable and well be an ideal discovery channel (*e.g.* via $\Upsilon(6S) \to \gamma X_b \to \gamma[\Upsilon(1S)\omega]$).

*Effect of continuum channels on quarkonia.* [C. Hanhart, R. Mizuk]

For hadronic molecules to be formed it is necessary that the scattering potential of two heavy mesons is sufficiently strong that its resummation leads to a pole in the scattering matrix. For this mechanism to be convincing one expects that the two-meson continuum also influences at least some properties of states that are believed to have a pronounced $Q\bar{Q}$ component. In the literature, the continuum contribution is sometimes called coupled-channel effects or molecular admixture.

For most heavy quarkonium transitions, we notice that $M_{Q\bar{Q}} - 2M_{Q\bar{q}} \ll M_{Q\bar{q}}$, where $M_{Q\bar{Q}}$ and $M_{Q\bar{q}}$ are the masses of the heavy quarkonium and an open-flavour heavy meson, respectively. As a result, the intermediate heavy mesons are nonrelativistic with a small velocity

$$v \sim \sqrt{|M_{Q\bar{Q}} - 2m_{Q\bar{q}}|/m_{Q\bar{q}}} \ll 1 \tag{502}$$

and the coupled-channel effects in the transitions can be investigated using a non-relativistic effective field theory. A systematic analysis reveals that certain transitions acquire a $1/v$ enhancement compared to the transition captured by the quark model. This is especially the case if the quark model transition is suppressed either by a violation of flavour selection rules as in $\psi' \to J/\psi\pi$ and $\psi' \to J/\psi\eta$ [1385] or by small wave function overlaps as in hindered M1 transitions between two $P$-wave $Q\bar{Q}$ states [1386, 1387] (a detailed discussion on the power counting and additional examples see Ref. [1388]). On the other hand, based on the same power counting rules, it was argued that the transitions $\Upsilon(4S) \to h_b\pi^0/\eta$ have only a small pollution from the bottom-meson loops. They are dominated by a short-distance contribution proportional to the light quark mass difference [1389] and could be used for the extraction of light quark mass ratio. Furthermore, the prediction, made before the discovery of the $h_b(1P)$, on the branching fraction of the order of $10^{-3}$ for the decay $\Upsilon(4S) \to h_b\eta$ was verified by the Belle measurement, $(2.18 \pm 0.11 \pm 0.18) \times 10^{-3}$ [1111]. In addition, in Ref. [1121], it is pointed out that coupled-channel effects can even introduce sizeable and nonanalytic pion mass dependence in heavy quarkonium systems which couple to open-flavour heavy meson pairs in an $S$-wave.

[49] The coupling of a pair of $3/2^+$ and $1/2^-$ bottom mesons to an $S$-wave $b\bar{b}$ system, however, requires a sizeable HQSS breaking to be present [1323, 1324].

A possible explanation for the unusual features of $\Upsilon(4S)$, $\Upsilon(5S)$, and $\Upsilon(6S)$ listed in Sec. 14.5.2 is the presence of heavy mesons in their wave functions. In this picture the $\Upsilon(4S)$, $\Upsilon(5S)$ and $\Upsilon(6S)$ states are mixtures of the $b\bar{b}$ and $B_{(s)}^{((*)*)}\bar{B}_{(s)}^{((*)*)}$ pairs [with $B_{(s)}^{**}$ we denote the $P$-wave excitations of $B$ or $B_s$ mesons]. It was realised at the time of observation of the $\Upsilon(4S)$, $\Upsilon(5S)$ and $\Upsilon(6S)$ in 1980s [1390, 1391] that the too large splitting between the $\Upsilon(4S)$ and $\Upsilon(5S)$ states is due to the contribution of hadron loops [1392], which is another language to discuss the molecular admixture. Enhanced transitions into hidden flavour final states are due to rescattering of the on-shell heavy mesons [1393, 1394]. Finally, the molecular admixture is also responsible for the violation of HQSS. Indeed, an admixture of a specific $B\bar{B}$, $B\bar{B}^*$ or $B^*\bar{B}^*$ meson pair is not an eigenstate of the $b\bar{b}$ total spin. Table 127 presents the decomposition of the $P$-wave $B^{(*)}\bar{B}^{(*)}$ pairs with $J^{PC} = 1^{--}$ into the $b\bar{b}$ spin eigenstates $\psi_{ij}$, where $i$ is the total spin of the $b\bar{b}$ pair and $j$ is the total angular momentum contributed by all other degrees of freedom, including both the spin of light quarks and the orbital angular momentum $L = 1$ [1395, 1396]. Various $\psi_{ij}$ components give rise to transitions that are forbidden by HQSS for pure $b\bar{b}$ states. Experimental signatures for $\psi_{ij}$ components are presented in Table 128.

Table 127: The decomposition of the $P$-wave $B^{(*)}\bar{B}^{(*)}$ pairs with $J^{PC} = 1^{--}$ into the $b\bar{b}$ spin eigenstates [1397].

| State | Decomposition into $b\bar{b}$ spin eigenstates |
|---|---|
| $B\bar{B}$ | $\frac{1}{2\sqrt{3}}\,\psi_{10} + \frac{1}{2}\,\psi_{11} + \frac{\sqrt{5}}{2\sqrt{3}}\,\psi_{12} + \frac{1}{2}\,\psi_{01}$ |
| $B\bar{B}^*$ | $\frac{1}{\sqrt{3}}\,\psi_{10} + \frac{1}{2}\,\psi_{11} - \frac{\sqrt{5}}{2\sqrt{3}}\,\psi_{12}$ |
| $(B^*\bar{B}^*)_{S=0}$ | $-\frac{1}{6}\,\psi_{10} - \frac{1}{2\sqrt{3}}\,\psi_{11} - \frac{\sqrt{5}}{6}\,\psi_{12} + \frac{\sqrt{3}}{2}\,\psi_{01}$ |
| $(B^*\bar{B}^*)_{S=2}$ | $\frac{\sqrt{5}}{3}\,\psi_{10} - \frac{\sqrt{5}}{2\sqrt{3}}\,\psi_{11} + \frac{1}{6}\,\psi_{12}$ |

Table 128: Experimental signatures for the $b\bar{b}$ spin eigenstates $\psi_{ij}$ [1397].

| Spin eigenstate | Expected decays |
|---|---|
| $\psi_{10}$ | $\Upsilon(nS)\,\pi^+\pi^-$, $\Upsilon(nS)\,K^+K^-$ in $S$ wave |
| $\psi_{11}$ | $\Upsilon(nS)\,\eta$, $\Upsilon(nS)\,\eta'$ |
| $\psi_{11}$, $\psi_{12}$ | $\Upsilon(nS)\,\pi^+\pi^-$, $\Upsilon(nS)\,K^+K^-$ in $D$ wave |
| $\psi_{01}$ | $\eta_b(nS)\,\omega$, $\eta_b(nS)\,\phi$, $h_b(nP)\,\eta$, $h_b(nP)\,\eta'$ |

One can expect that the closer is the physical state to the threshold, the larger is the admixture of the corresponding meson pairs. Therefore $B\bar{B}$ should be the dominant admixture in the $\Upsilon(4S)$ wave function, while in case of the $\Upsilon(5S)$ the dominant HQSS-violating contribution could be the $B_s^*\bar{B}_s^*$ component [1396]. This picture gives successful description of the observed decay pattern, while Tables 127 and 128 give guidance for further experimental searches.

The above analysis considers only the ground state $S$ wave bottom mesons. Contributions of excited bottom mesons were recently discussed for the $\Upsilon(nS)$ states in a quark model [1398]. The $\Upsilon(6S)$ situated near the $B_1(5721)\bar{B}$ threshold, may even qualify as a hadronic molecule, which was discussed above in the hadronic molecular subsection.

The $\Upsilon(5S)$ and $\Upsilon(6S)$ states are not good candidates for compact tetraquarks, since their signals are seen in the total hadronic cross section, and thus the open-flavour channels are likely to give a dominant contribution to the $\Upsilon(5S)$ and $\Upsilon(6S)$ peaks. The $\Upsilon(5S)$ is also not a good candidate for hadrobottomonia, since it decays into various final states with bottomonia, instead of a single one.

In this sense it is of paramount importance that we get better information on the hadron spectrum in the bottomonium sector above the $\bar{B}B$ threshold. Various possible strategies that could be followed at Belle II for those discoveries also for quantum numbers different from $1^{--}$ are listed in Sec. 14.7.2. Clearly, the same decay chains are also well suited to discover new exotic candidates.

We are convinced that the spectrum of bottomonium(-like) states above the $\bar{B}B$ threshold will provide crucial information on the inner workings of QCD and that Belle II can and will be a key player in this field in the years to come.

*14.5.4. Lattice QCD.* [S. Prelovsek]

In the energy region near or above threshold, the masses of bound-states and resonances have to be inferred from the infinite-volume scattering matrix $S(E)$ of the one-channel (elastic) or multiple-channel (inelastic) scattering of two-hadrons. This has been done for QCD in practice only recently and the approaches currently used or proposed are briefly summarised below. The poles of the resulting $S(E)$ provide the masses of resonances and bound states.

(i) The most rigorous and the most widely used way to extract $S(E)$ is Lüscher's method. He has shown that the energy of eigenstate $E$ in finite volume gives the scattering matrix $S(E)$ at that energy in infinite volume [196]. This leads to $S_l(E) = e^{2i\delta_l(E)}$ for partial wave $l$ only for specific values of $E$ since the spectrum in a finite volume is discrete. The energies of lattice eigenstates are extracted from the correlation matrix (Eq. 493), where interpolators preferably span the most important Fock-components. For quarkonium(-like) states one takes for example $\mathcal{O} \simeq \bar{Q}\Gamma Q$, two-meson interpolators $\mathcal{O} = (\bar{Q}\Gamma_1 q)(\bar{q}\Gamma_2 Q)$, $(\bar{Q}\Gamma_1 Q)(\bar{q}\Gamma_2 q)$ and $\mathcal{O} = [\bar{Q}\Gamma_1\bar{q}][Q\Gamma_2 q]$ with $Q = c, b$ and $q = u, d, s$. The energy eigenstates $|n\rangle$ are predominantly "one-meson" states (*e.g.* $\chi_{c0}$) or predominantly "two-meson" states (*e.g.* $D\bar{D}$) - in interacting theory they are mixtures of those. The Lüscher approach has been thoroughly verified on the conventional resonances like $\rho$. The generalisation of this approach to the multichannel scattering is discussed *e.g.*, in Refs. [185, 1399]. The scattering of the particles with integer/half-integer spin in generic moving frames is considered, *e.g.*, in Ref. [187, 1400].

A modification of the Lüscher approach has been proposed, based on the use of the Unitary Chiral Perturbation Theory in a finite volume [1401–1404]. Within this approach, the free parameters are directly fitted to the energy levels measured on the lattice. Using the same values of parameters, the infinite-volume formalism allows one to calculate the scattering amplitudes and determine the resonance pole positions in the complex plane.

(ii) Another possibility to extract $S(E)$ is the HALQCD approach [207], which starts by determining the two-hadron Bethe-Salpeter wave function and two-hadron potential $V(r)$ from lattice. The phase shift $\delta(E)$ and $S(E)$ are then obtained using the Schrödinger equation for given $V(r)$. This approach has not been verified on conventional resonances yet. There are ongoing discussions as to whether this approach is as rigorous as Lüscher's approach.

(iii) The Born-Oppenheimer approach may be applied for the systems with heavy quarks $Q$, where the static heavy-quark sources are surrounded by the light degrees of freedom. The static energy is calculated as a function of distance $r$ between static pair $Q(0)Q(r)$ or $Q(0)\bar{Q}(r)$ in presence of the light degrees of freedom. In lattice sections 14.4.3 and 14.5.4 we refer to this static energy as the potential $V(r)$ as commonly used in the literature; there is a subtle difference[50] between static energy and the potential mentioned shortly after Eq. (487).

Related potentials were mentioned in the previous sections. The potential is then used in the Schrödinger equation to determine the properties of the system.

(iv) Recently, a novel method was suggested that allows to directly extract the real and imaginary part of the optical potential from lattice data [1405]. The method relies on the analytic continuation into the complex plane and is applicable, even if the intermediate states contain more than two particles.

All presented simulations of heavy quarkonia omit Wick contractions with $\bar{Q}Q$ annihilation, while all other Wick contractions are taken into account.

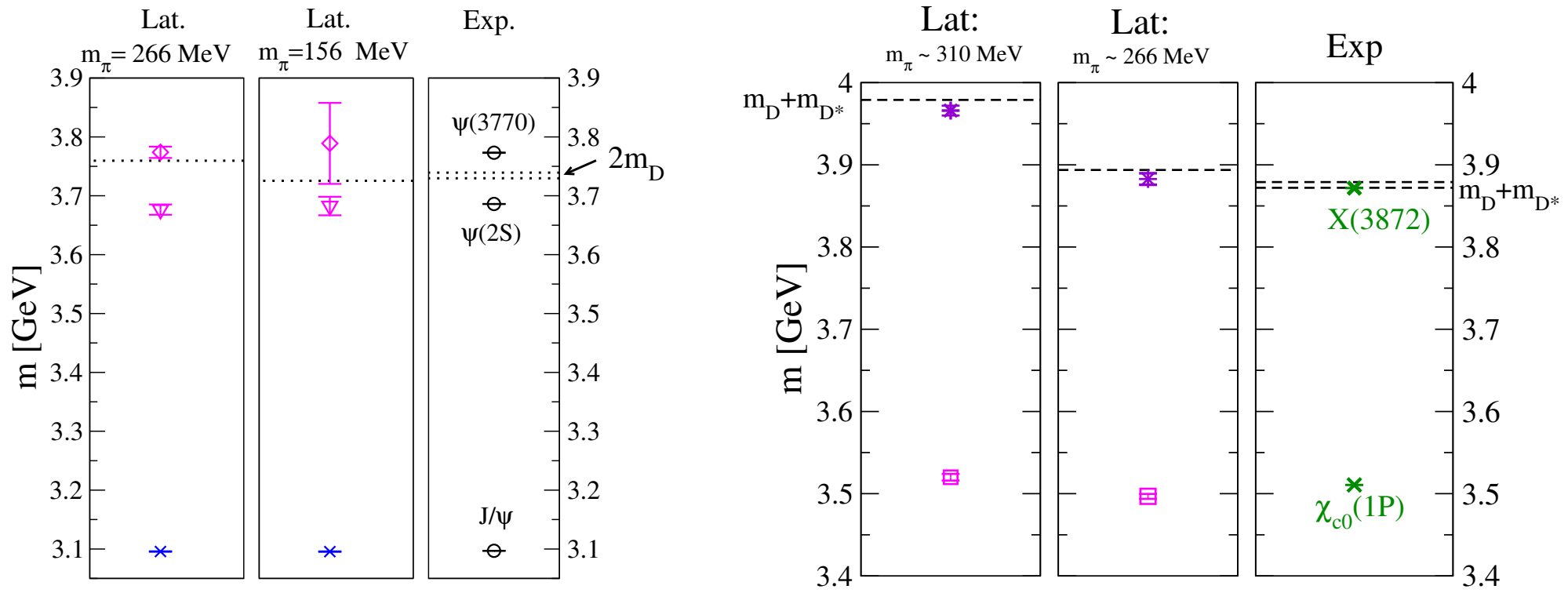

Fig. 171: Left: The spectrum of the vector charmonia from [197]: the diamond denotes the resonance mass of $\psi(3770)$, while the triangle denotes the pole mass of the bound state $\psi(2S)$; both are obtained from $D\bar{D}$ scattering matrix. Right: The location of $X(3872)$ with $I=0$ which emerges as shallow bound state in $D\bar{D}^*$ scattering at $m_\pi \simeq 266$ MeV [198] and $m_\pi \simeq 310$ MeV [1406] (update of [199]).

*Vector and scalar resonances .* Until recently, all quarkonia above $D\bar{D}$ ($B\bar{B}$) threshold were treated ignoring the strong decay to a pair of charmed (beauty) mesons. The pioneering investigation of mixing between $c\bar{c}$ and $D\bar{D}$ was presented in [1407]. The first simulation that determined $m_R$ and $\Gamma_R$ of such resonances using Lüscher's approach considered vector and scalar charmonia [197]. The Breit-Wigner-type fit of the $D\bar{D}$ scattering matrix $S(E)$ in $P$-wave leads to the resonance mass and width of $\psi(3770)$, which agree with experiment within sizeable errors (see Fig. 171)[51]. The $\psi(2S)$ appears as a bound state pole below threshold.

[50] They differ in perturbative QCD at three-loops for ultrasoft corrections. In the nonperturbative regime such corrections are not there.

In the scalar channel, only the ground state $\chi_{c0}(1P)$ is understood and there was no commonly accepted candidate for its first excitation $\chi_{c0}(2P)$ until recently. Some identifed $X(3915)$ with this state, but this was seriously questioned in [1241–1243] as discussed in Section 14.5.2. The alternative candidate $X^*(3860)$ was observed by Belle in 2017 [1185] as a rather broad resonance in $D\bar{D}$ invariant mass. The scattering matrix for $D\bar{D}$ in $S$-wave was extracted from lattice [197] and provide an indication for a rather narrow resonance slightly below 4 GeV with a width $\Gamma[\chi'_{c0} \to D\bar{D}] < 100$ MeV, which is compatible with the experimental $X^*(3860)$ at a $2.7\sigma$ level [1185]. Further experimental and lattice QCD efforts are required to map out the $D\bar{D}$ and $J/\psi\omega$ scattering in more detail.

*Charmonium-like $X(3872)$ and $X(4140)$.* The $X(3872)$ lies experimentally on the $D^0\bar{D}^{0*}$ threshold and its existence on the lattice can not be established without taking into account the effect of this threshold. This was first done by simulating $D\bar{D}^*$ scattering in [198], using $D\bar{D}^*$ and $c\bar{c}$ operators. The $D\bar{D}^*$ scattering matrix in $I(J^{PC}) = 0(1^{++})$ channel was determined using Lüscher's approach. The pole was found just below the threshold (Fig. 171) and it was associated with a bound state $X(3872)$. The more recent simulation using HISQ action confirms the existence of such a pole [199, 1406]. The finite-volume corrections for such shallow bound states fall as exponentials of $pL$, where $p$ is the binding momentum [1408]. For a discussion of the quark mass dependence of the $X(3872)$ we refer to Refs. [1409, 1410].

The lattice study [1411] investigated the overlaps of the different operators needed to make the $X(3872)$ with $I = 0$ visible on the lattice. The energy eigenstate related to $X(3872)$ appears in the simulation only if $D\bar{D}^*$ as well as $c\bar{c}$ interpolating fields are employed. The $X(3872)$ does not appear in absence of $c\bar{c}$ interpolators, even if (localised) interpolators $[\bar{c}\bar{q}]_{3_c}[cq]_{\bar{3}_c}$ or $[\bar{c}\bar{q}]_{6_c}[cq]_{\bar{6}_c}$ are in the interpolator basis. Although the overlaps are scheme and scale dependent, and no theoretically strict conclusion can be driven from them, this might nevertheless suggest that the $c\bar{c}$ Fock component is most likely more essential for creating the $X(3872)$ than the diquark-antidiquark one.

A charged $X(3872)$ with $J^{PC} = 1^{++}$ was not found in [1411] although the diquark-antidiquark interpolators were incorporated. The reliable search for the neutral $I = 1$ state would need to incorporate isospin breaking effects [1412], but that has not been performed on the lattice yet.

The experimental candidate $X(4140)$ with hidden strangeness and $J^{PC} = 1^{++}$ was recently confirmed by the high-statistics LHCb study [1205, 1206]. The lattice search for it was performed with $J/\psi\phi$, $D_s\bar{D}_s^*$ and $[\bar{c}\bar{s}][cs]$ interpolators [1411]. If $X(4140)$ was an elastic resonance in $J/\psi\phi$ or $D_s\bar{D}_s^*$ where both channels are decoupled, one would expect an additional lattice eigenstate near $m_X$. Such an eigenstate was not found, so $X(4140)$ as a dominantly elastic resonance (on the unphysical sheet) is not supported. This may point to a coupled-channel origin of this structure, which may still allow for the presence of a state as a pole in the infinite-volume coupled-channel $S$-matrix. The $S$-wave and $P$-wave $J/\psi\,\phi$ scattering matrices from the simulations of Ref. [1413], which omit $\bar{s}s$ annihilation, also do not support the elastic resonance $X(4140)$.

[51] Actually, the $\psi \to D\bar{D}$ coupling ($g$) is compared to the experiment, rather than the width $\Gamma = g^2p^3/(6\pi m_R^2)$.

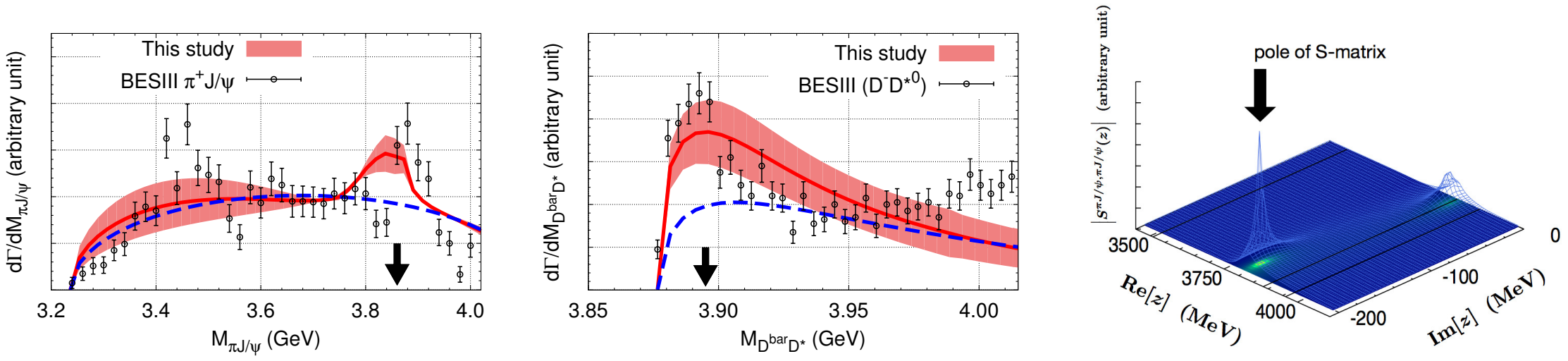

Fig. 172: Simulation of $Z_c^+(3900)$ based on the HALQCD method [208]. Left two panes: red lines present the lattice QCD results for $Y(4260) \to J/\psi\pi\pi$ and $Y(4260) \to D\bar{D}^*\pi$ differential rates, while blue lines present results if $J/\psi\pi$-$D\bar{D}^*$ coupling is turned off by setting $V_{J/\psi\pi,D\bar{D}^*} = 0$. Right pane: Poles of the scattering matrix in the complex energy plane.

*Charged $Z_{c,b}^+$ and $B\bar{B}$ potentials.* The lattice search for the manifestly exotic states $Z_c^+$ with flavour content $\bar{c}c\bar{d}u$ and $I^G(J^{PC}) = 1^+(1^{+-})$ is challenging since the experimental candidates lie above several thresholds and can decay to several final states via strong interaction. A reliable treatment requires the simulation of coupled channels and the extraction of the coupled channel scattering matrix.

The only simulation that determined the coupled-channel S(E) for such systems applied the HALQCD approach [208]. The potential $V_{\pi J/\psi,\pi J/\psi}(r)$ related to Nambu-Bethe-Salpeter equation is determined between the $J/\psi$ and $\pi$ as a function of their separation $r$. The other diagonal and off-diagonal potentials $V_{\alpha,\beta}(r)$ for three channels $\alpha, \beta = J/\psi\pi^+,\ D\bar{D}^*,\ \eta_c\rho$ were also determined [208] according to the formalism of Ref. [1414]. The off-diagonal potential between channels $\pi J/\psi$ and $D\bar{D}^*$ is larger than other ones, which seems to indicate a sizeable coupled channel effect near $D\bar{D}^*$ threshold. The potentials render a $3\times 3$ scattering matrix for three coupled two-meson channels [208]. This is then used to determine the three body decay $Y(4260) \to J/\psi\pi\pi$ and $Y(4260) \to D\bar{D}^*\pi$ in a semi-phenomenological way. The differential rate indeed shows a peak around $Z_c$ mass (red line in Fig. 172). If the potential between $J/\psi\pi$ and $D\bar{D}^*$ is turned off, the peak disappears (blue dashed line). It indicates that the coupling of $J/\psi\pi$ and $D\bar{D}^*$ channels seems to be crucial for the existence of $Z_c$. This needs to be verified also by the Lüscher method. Notice that a virtual state pole deep in the complex plane was found in [208]. This is consistent with the fact that the obtained peaks in Fig. 172 are not as narrow as the observed ones. However, a phenomenological fit to the BESIII data for the $Z_c(3900)$ reveals a pole, as either a virtual state or a resonance, much closer to the threshold [1377].

The $S$-wave and $P$-wave scattering matrices near $D\bar{D}^*$ threshold were determined using only $D\bar{D}^*$ interpolating fields in [1415], which may not be reliable since the ground state of the system is $J/\psi\pi$. The authors conclude that no evidence for $Z_c(3900)^+$ is found.

A simplified search for $Z_c^+$–states extracted the energies of eigenstates [199, 1416] without determining $S(E)$. The simulation of the coupled channels (with a large number of meson-meson and diquark-antidiquark interpolators) renders eigen-energies close to energies of the non-interacting two-meson states, *e.g.* $J/\psi\pi^+$, $D\bar{D}^*$. A scenario with elastic resonance poles[52] on the unphysical sheet predicts an additional eigenstate, but such eigenstate was not found in the actual simulations [199, 1416][53]. This could indicate that the effect of coupled channels is significant for experimental $Z_c$, in line with the HALQCD result [208]. Notice that the analytic study of coupled channels [1417], which renders the experimental differential rates [1377] and at the same time quantitatively agrees with the lattice energy levels [1416], allows for scenarios where $Z_c$ can be either a virtual state with a pole below $D\bar{D}^*$ threshold [54].

The lattice search for a pair of $Z_b^+$ from Belle [1171] is challenging as each of them can decay to at least 5 two-meson final states. The additional difficulty is that the discrete two-meson energies[55] are much denser than for the $Z_c$ channel. The only exploratory study considered $B(0)\bar{B}^{(*)}(r)$ potentials within the Born-Oppenheimer approach [210]. The challenge is that the $\Upsilon(nS)\pi$ represents the ground state of the $\bar{Q}(0)Q(r)\bar{d}u$ system, which leads roughly to the energy $V_{Q\bar{Q}}(r) + m_\pi$. The $B(0)\bar{B}^*(r)$ potential had to be extracted from the first excited eigenstate, and the Schrödinger equation for this potential rendered a possible indication of a $\bar{b}b\bar{d}u$ bound state with a binding energy of $(58 \pm 71)$ MeV and $I(J^P) = 1(1^+)$ [210]. The effects of the heavy-quark spin have been considered analytically and it was found that the bound state with binding energy $59\,^{+30}_{-38}$ MeV persists after their inclusion [1135]. Exploratory studies of related $B\bar{B}$ potentials for $I = 0$ channel [1418, 1419] followed the discovery of $Q\bar{Q}$ string breaking [1420]. Most of the interesting spectroscopic aspects related to $\bar{b}b$ and $B\bar{B}$ mixing are still unexplored on the lattice.

*$\bar{q}\bar{q}QQ$ states, $BB^{(*)}$ and $DD^{(*)}$ potentials .* The $\bar{q}\bar{q}QQ$ system was considered with static quarks $Q$ within Born-Oppenheimer approximation [1419, 1421–1423] and also with non-static $Q$ [1136, 1424–1426]. On the lattice, this open heavy-flavour channel is simpler than the closed heavy-flavour one since the ground state is $BB^{(*)}$ and leads to the potential/interaction of the $BB^{(*)}$ (for $Q = b$). Most lattice studies explored whether a bound state could exist in the $\bar{q}\bar{q}QQ$ system. Among all the investigated channels (including $q = u, d, s$, $Q = b, c$, $J = 0, 1, 2$), the channel $\bar{u}\bar{d}bb$ with $I(J^P) = 0(1^+)$ gives an indication for a bound state below $BB^*$ in several studies [1421, 1422, 1426]. This is manifested as an eigenstate about 200 MeV below $BB^*$ threshold [1426] and the same study favours also $\bar{u}\bar{s}bb$ with $J^P = 1^+$ as a bound state. The $DD^{(*)}$ and $\bar{K}D$ ($cs\bar{u}\bar{d}$) systems are typically not found to

[52] Resonance scenario [1417] does not predict an additional eigenstate since the phase shift does not pass through $\pi/2$ at resonance there.

[53] An extra eigenstate (in addition to expected two-meson states) has been found for all resonances and all bound states that have been established on the lattice up to now.

[54] or a resonance above threshold but the phase shift does not pass through $\pi/2$ at resonance in this case [1417].

[55] Non-interacting energies $E \simeq 2\sqrt{m_B^2 + n(2\pi/L)^2}$ for $B\bar{B}$ are denser than for $D\bar{D}$.

form bound states on the lattice [1424, 1425]. However, certain phenomenological studies favour $DD^*$ bound by only few MeV [1427, 1428], which prompts further lattice investigation to achieve such a precision. A search for $DD^*$ bound states in double-charm production at Belle II is an exciting possibility, where the search strategies have been investigated in [1427, 1428].

Possible existence of resonances in such systems remain yet to be explored.

*Hadro-quarkonium: $Q\bar{Q}$ and light hadrons.* The hadro-quarkonium picture [1333] consists of a colour-singlet $Q\bar{Q}$ pair to which a colour-singlet light hadron $H$ is bound by residual QCD forces — see Sec. 14.5.3. This scenario was tested in the Born-Oppenheimer approach by determining the modification of the potential between a static quark-antiquark pair induced by the presence of a light hadron [1429]. The modification $\Delta V_H(r) = V_H(r) - V_0(r)$ was found negative $\Delta V_H(r) < 0$ for all investigated hadrons $H = \pi,\ K,\ \rho,\ K^*,\ N,\ \Sigma,\ \Lambda,\ \Delta,\ \Sigma^*,\ \Omega,\ \Xi,\ \Xi^*$. The main effect was as a reduction of the linear slope of the potential. At a distance of 0.5 fm the potential was lowered by only about $2-3$ MeV for all these hadrons. The Schrödinger equation for these potentials yields quarkonia that are more tightly bound by $1-7$ MeV in presence of the light hadrons.

*Hybrids and other exotics.* Hybrid states $\bar{Q}gQ$ have been generally discussed in the *Hybrid states* subsection of Sec. 14.5.3 and in Sec. 14.5.5. They have been addressed on the lattice, for example, in [209, 1120, 1430–1436]. The $\bar{Q}gQ$ are unstable resonances above $(\bar{Q}q)(\bar{q}Q)$ threshold in dynamical QCD, while practically all previous lattice simulations neglected their strong decays. This challenge is the main reason why there has been little progress on this front recently. Quenched QCD is somewhat simpler in this respect as the threshold for the decay $\bar{Q}gQ \to (\bar{Q}Q) + glueball$ lies higher. Most studies consider $J^{PC} = 1^{-+}$, which are much easier to disentangle from the conventional mesons.

The recent results for the masses of $\bar{c}gc$ hybrids with relativistic quarks [195] are shown by red and blue in Fig. 170 above. The $1^{-+}$ state is obtained at 4.31(2) GeV [195], which is close to the other determinations, for example 4.4 GeV in Ref. [1431] and $4.405 \pm 0.038$ GeV in Ref. [1432]. The $1^{--}$ candidate resides at 4.41(2) GeV [195], close to $4.379 \pm 0.149$ GeV from [1433] (the authors claim that they have found a radially excited vector hybrid, however such an interpretation was criticised in [1434]).

The lightest bottomonium hybrid was found to be $1^{-+}$ at 10.9 GeV in a quenched simulation with relativistic $b$ quarks [1430]. The other approach is within the Born-Oppenheimer approximation, where hybrid static energies for excited static $Q(0)\bar{Q}(r)$ are calculated first. This has been done long ago [209] only in quenched approximation so far. The lightest hybrid $\bar{b}gb$ was also found at about 10.9 GeV and the masses for lowest few hybrids show reasonable agreement with the results based on quenched NRQCD [209].

An experimental search for hybrids, particularly those with exotic quantum numbers $0^{+-}$ and $1^{-+}$, is of high importance.

*$Q\bar{Q}$ potentials for other exotic states* The $\bar{Q}(0)Q(r)$ potentials related to hybrids, mentioned in the previous section, represent just one of the possible types of interest. The potentials related to light adjoint mesons in static octet-colour sources $(Q\bar{Q})$ were calculated in the quenched approximation [1437]. An analytic study proposed to calculate many interesting Born-Oppenheimer potentials related to $\bar{Q}(0)Q(r)$, accompanied with glue and/or light quarks with various angular momenta and flavour quantum numbers [211]. Most of these have not been calculated on the lattice yet, but recent dynamical simulations discussed above [210, 1429] investigate along these lines.

*Pentaquarks* The NPLQCD collaboration finds an interesting indication for an $\eta_c N$ bound state approximately 20 MeV below $\eta_c N$ threshold ($N$ denotes a nucleon) [1438]. This is the only pentaquark candidate containing $c\bar{c}$ from lattice studies up to now. As the simulation is done at $SU(3)$ flavour symmetric point ($m_\pi \simeq 800$ MeV), it is not clear yet whether this bound state persists to physical $m_\pi$. A Belle II search for such a bound-state pentaquark and its siblings is of prime interest. A $\eta_c p$ ($p$=proton) resonance or a bound state could be searched in inclusive decays $Y(nS) \to \eta_c p X$ by considering the invariant mass $M(\eta_c p)$. Here $\eta_c$ is reconstructed through its deacy to $K^* K\pi$, $2(\pi^+\pi^-)$, ... and $X$ is anything or something with a reconstructed antiproton. An $\eta_c p$ structure would be found for $M(K^* K\pi\ p) < m_{\eta_c} + m_p$. Another production mechanism could be through $\Lambda_b$ decays if $E_{cm} > 2m_{\Lambda_b}$ would be reached at Belle II.

The simulation of two pentaquarks $P_c$ discovered by LHCb [1055] will be more difficult as they are located 0.4 GeV above $J/\psi p$ threshold and they have several open decay channels. Such lattice results are not available yet. Some light on such systems might be inferred from [1429] where $\bar{Q}Q$ potential $V(r)$ in presence of baryons is considered within the hadroquarkonium picture.

*Lattice Outlook.* Lattice QCD spectra of charmonia and bottomonia below open-flavour threshold are precise, under control and in reasonable agreement with experiment. The decay constants and radiative transitions between some of these states have been determined, and the remaining ones are tractable. $\bar{Q}Q$ annihilation is omitted in all simulations presented here, and this presents the main remaining uncertainly for quarkonia below open-flavour threshold. This annihilation would invoke decays to light hadrons and presents a considerable challenge.

Information on the states above or slightly below threshold have to be inferred from the scattering matrix extracted on the lattice. The hadronic resonances that can strongly decay only to one two-hadron final state can be treated rigorously by simulating the scattering in this channel. The same applies for stable hadrons that are situated slightly below one threshold. Reasonable results can be obtained also for the hadrons where a coupling to one channel is dominant, while coupling to others may be neglected. Such an approximation is not reliable for exotic candidates that could involve important coupled-channel effects. The radiative and weak transitions $\langle H_1|J^\mu|H_2\rangle$, where the initial or final hadron decay strongly via one channel, could be studied along the lines of a recent pioneering simulation [204] in the light sector.

The states that can decay into two or three different two-hadron final states are challenging, but manageable in principle. The scattering matrix for few-coupled channels has been recently extracted via Lüscher-type method for other systems [174, 206, 1439].[56] Some analogous results relevant to quarkonium(like) spectroscopy at Belle II can be expected by the time it starts operating.

| | $l$ | $J^{PC}\{s=0,s=1\}$ | $E_n^{(0)}$ |
|---|---|---|---|
| $H_1$ | 1 | $\{1^{--},(0,1,2)^{-+}\}$ | $\Sigma_u^-,\Pi_u$ |
| $H_2$ | 1 | $\{1^{++},(0,1,2)^{+-}\}$ | $\Pi_u$ |
| $H_3$ | 0 | $\{0^{++},1^{+-}\}$ | $\Sigma_u^-$ |
| $H_4$ | 2 | $\{2^{++},(1,2,3)^{+-}\}$ | $\Sigma_u^-,\Pi_u$ |
| $H_5$ | 2 | $\{2^{--},(1,2,3)^{-+}\}$ | $\Pi_u$ |

Table 129: $J^{PC}$ multiplets with $l \leq 2$ for the $\Sigma_u^-$ and $\Pi_u$ gluonic states. We follow the naming notation $H_i$ used in [209, 211], which orders the multiplets from lower to higher mass. The last column shows the gluonic static energies that appear in the Schrödinger equation of the respective multiplet.

Many interesting aspects related to systems containing $\bar{b}b$ or $bb$ will be obtained by further investigations based on Born-Oppenheimer potentials. Other interesting future directions for lattice studies of quarkonium-like states were recently put forward in [1440, 1441].

*14.5.5. Exotics from QCD with EFTs.* [N. Brambilla, A. Vairo]

In Ref. [1137] an effective field theory description of hybrid states has been constructed on the basis of pNRQCD. Lattice calculation of the two lowest static hybrid energy curves ($\Sigma_u$ and $\Pi_u$) is provided in the quenched approximation [209] (see the *Hybrids and other exotics* subsection of Sec. 14.5.4). The main uncertainty comes from the poor lattice knowledge of a nonperturbative parameter entering the calculation called gluelump mass. Up to now the EFT does not include spin but manages to obtain coupled Schrödinger equations at leading order that cause an effect on the energy level called $\Lambda$ doubling already known from molecular physics. The obtained multiplets are given in Table 129 and compared to the neutral heavy quark mesons above open flavour threshold in Fig. 173 and with a recent direct lattice calculation of the hybrid masses multiplets [195] in Fig. 174. It is interesting to compare these results to those obtained in other approaches.

The Born-Oppenheimer approximation for hybrids [211] produces spin-symmetry multiplets with the same $J^{PC}$ constituents as the $H_i$ multiplets in Table 129. However, in all the existing Born-Oppenheimer papers the mixing terms have been neglected and therefore the masses of opposite parity states are degenerate, in contrast with the EFT approach in which the degeneracy between opposite parity states is broken ($\Lambda$ doubling).

In the constituent gluon picture hybrids are assumed to be composed of a gluonic excitation bound to a heavy quark-antiquark pair - see the *Hybrid states* of Sec. 14.5.3. The gluons are assumed to appear in $J^{PC}$ representations unlike the case of pNRQCD or Born-Oppenheimer descriptions, in which the gluonic states appear in $\Lambda_\eta^\sigma$ representations. The quantum numbers of the resulting hybrid are obtained by adding those of the gluon and those of the heavy quark–antiquark pair using the standard rules for addition of angular momenta. In this way one gets the same $J^{PC}$ quantum numbers but arranged in larger multiplets. If in the constituent gluon picture we add the gluon quantum numbers $1^{+-}$ to a $S$-wave heavy-quark antiquark pair in a spin singlet $\{0^{-+}\}$ or spin triplet $\{1^{--}\}$ state, then we get exactly the quantum numbers of $H_1$. Similarly, for $P$-wave quarkonia with quantum numbers $\{1^{+-}, (0, 1, 2)^{++}\}$ (corresponding to different spin states) we get $H_2 \cup H_3 \cup H_4$. $H_5$ would then be included in the combination with the next quarkonium quantum numbers.

[56] Special attention is neecssary to get rid of possible model-dependence in the parameterisations of the coupled-channel $T$-matrix. Sometimes, even the number of poles in the relevant energy region could be different [174, 1368].

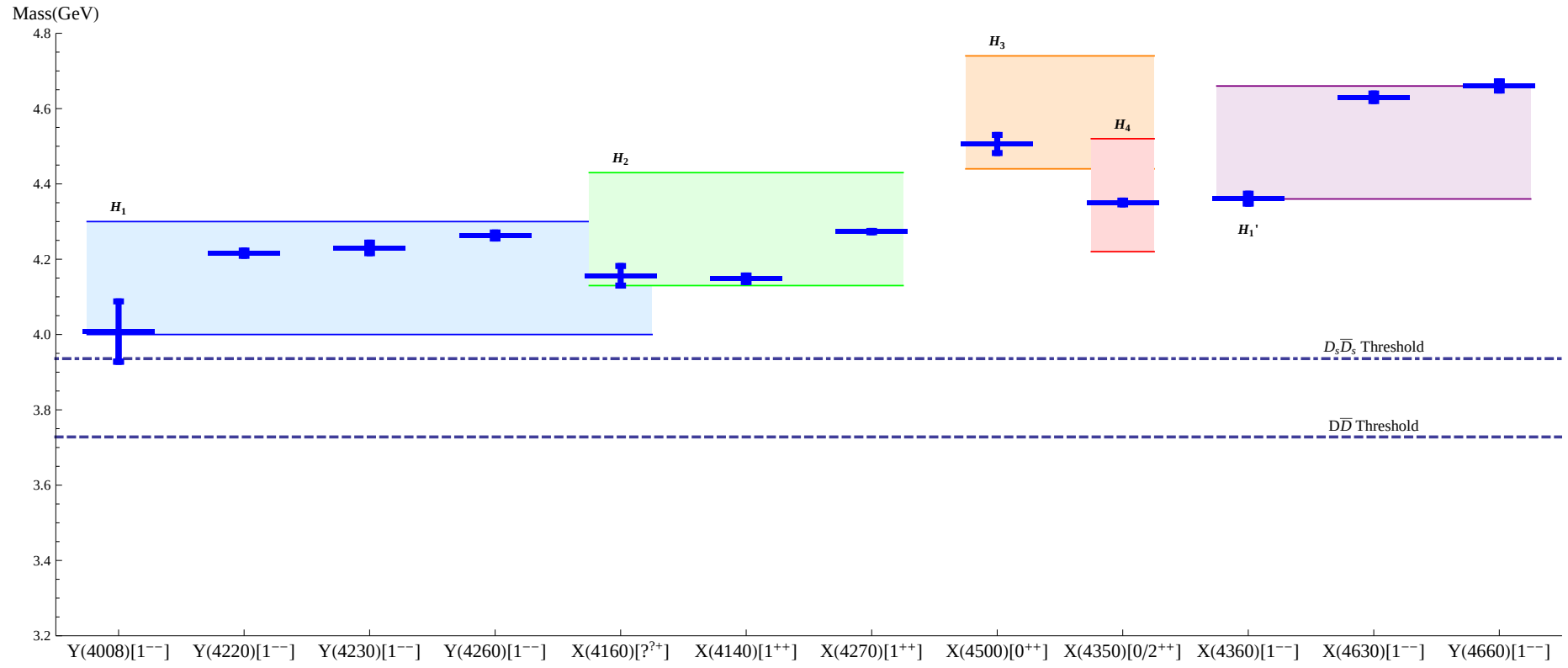

Fig. 173: Comparison of the experimental candidate masses for the charmonium sector with the results of [1137] for the hybrid masses. The experimental states are plotted in solid blue lines with error bars corresponding to the average of the lower and upper mass uncertainties. The results for the $H_1$, $H_2$, $H_3$ $H_4$ and $H_1'$ multiplets have been plotted in error bands corresponding to the gluelump mass uncertainty of $\pm 0.15$ GeV.

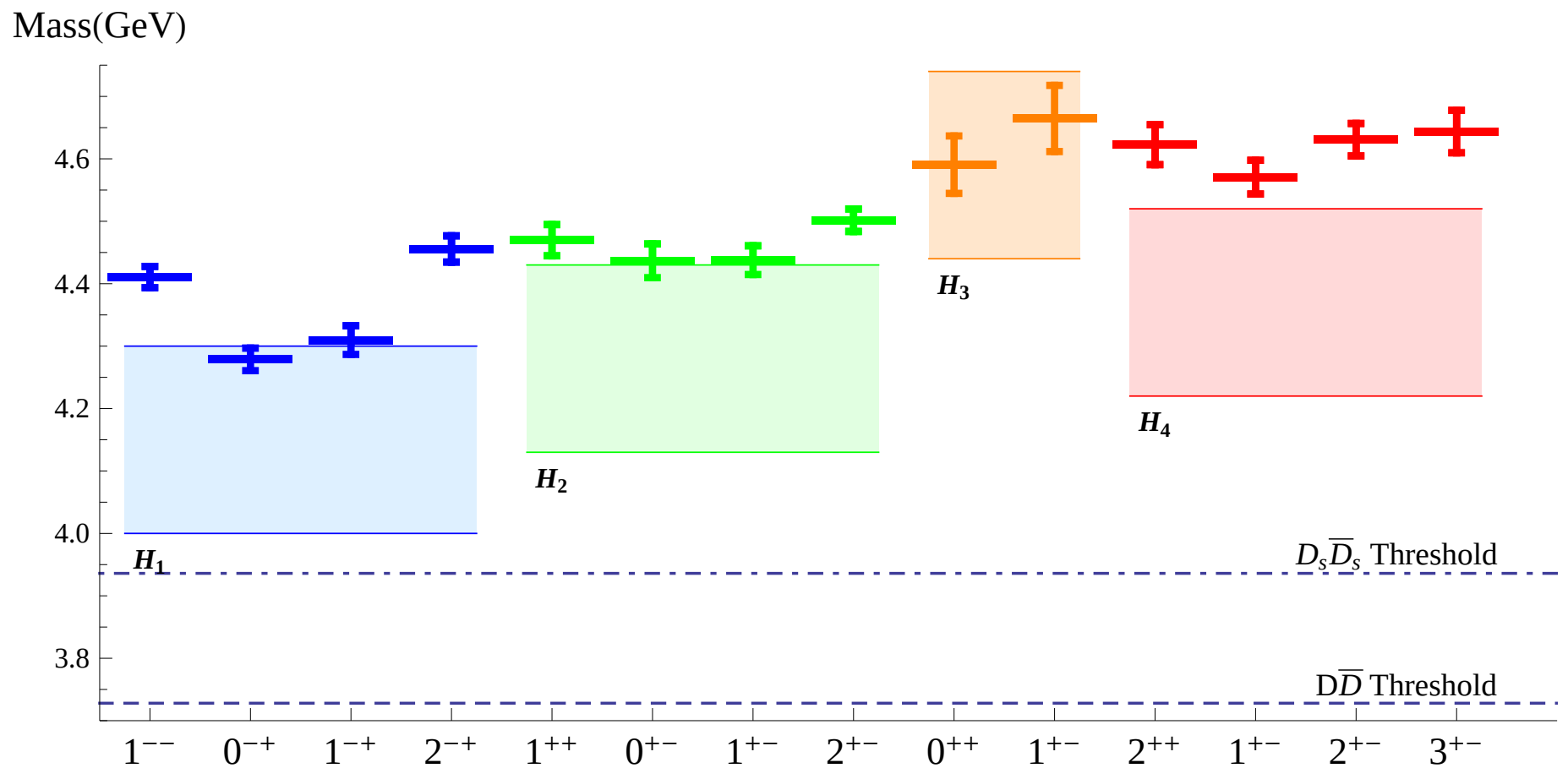

Fig. 174: Comparison of the results from direct lattice computations of the masses for charmonium hybrids [195] with the results of [1137] for the hybrid masses. The direct lattice mass predictions are plotted in solid lines with error bars corresponding to the mass uncertainties. The results for the $H_1$, $H_2$, $H_3$ $H_4$ multiplets have been plotted in error bands corresponding to the gluelump mass uncertainty of $\pm 0.15$ GeV.

Since for pNRQCD in the limit $r \to 0$ we recover spherical symmetry, we can see the constituent gluon picture as the short distance limit of the pNRQCD. Furthermore, one can interpret the finer multiplet structure of pNRQCD with respect to the constituent gluon picture as the effect of the finite distance $r$ between the heavy-quark pair.

The flux tube model [1315] arises from the idea that for QCD in the strong-coupling regime one can think of the gluonic degrees of freedom as having condensed into a collective string-like flux tube. In this picture the spectrum of gluonic static energies can be interpreted as the vibrational excitation levels of the string. The lowest excitations of such a string will correspond to nonrelativistic, small, transverse displacement oscillations and as such should be well described by the Hamiltonian of a continuous string. The eigenstates of such a Hamiltonian are characterised by the phonon occupation number and their polarisations, while the spectrum eigenenergies correspond to the different phonon occupation numbers. The hybrid quantum numbers are constructed by specifying the gluonic states via phonon operators. The value of $\Lambda$ corresponds to the number of phonons with clockwise polarisation minus the number of phonons with anticlockwise polarisation. From here one can construct the $J^{PC}$ quantum numbers of the hybrid states in an analogous way to the Born-Oppenheimer picture. The first excited energy level is a one-phonon state, which necessarily corresponds to a $\Lambda = 1$ state, unlike in the pNRQCD case, where the first excited energy level can be $\Lambda = 0$, 1. Thus, the pattern of the spin-symmetry multiplets emerging from the flux tube model in the case of the first excited static energy is the one in Table 129 except for the nonexistence of $H_3$.

In order to interpret the data it is necessary to include spin and enlarge the description to decay and transitions [1442]. Work in this direction is currently in progress. Moreover by adding light quarks with isospin number into the picture [211, 1137, 1318] it will be possible eventually to describe tetraquarks and molecular states directly from QCD, combining EFTs and lattice QCD. For a discussion of EFTs for molecular states based on hadronic degrees of freedom we refer to Sec. 14.5.3.

In [1443] the van der Waals forces between two $\Upsilon(1S)$ or two $\eta_b$ states have been studied to investigate the possibility of the formation of a bound state of two quarkonia. Van der Waals QCD forces may also play a role for the binding of the charmonium pentaquark.

*14.5.6. Summary.* In summary, the energy region at and above the strong decay threshold is very interesting for future investigation holding the promise to unveil new interactions and new binding mechanisms of the strong interactions. At the moment a plethora of new $X$, $Y$, $Z$ states have been observed and in part paired by an enormous theoretical activity with a number of phenomenological models. A theory description inside EFTs and lattice is being currently developed and new results will be soon available to match the experimental development.

### *14.6. Belle II prospects for Charmonium(-like) states*

Presently more than twenty charmonium-like states are known (see Tables 125 and 126). Their properties do not agree with the expectations for pure $c\bar{c}$ levels, which indicates that their structure is exotic. Theoretical approaches to the description of these states are presented in the previous Section. For each state several variants of interpretation exist (tetraquark, hybrid, hadroquarkonium, molecule, ...), and thus the structures of the state remains not established. Existing experimental data are not precise enough or are not sufficiently full to discriminate various interpretations. Thus further experimental input is crucial for understanding of the quarkonium-like states. This Section describes future studies on charmonium and charmonium-like states that can be performed at Belle II. The states are grouped according to the processes in which they are produced: $B$ decays, Initial State Radiation, Two Photon Process and Double Charmonium Production. All the considered states are situated above the $D\bar{D}$ threshold.

*14.6.1. B decays.* [R. Mizuk, S.L. Olsen]

Charmonium(-like) states are produced in $B$ meson decays in association with a kaon: $B \to K\, X_{c\bar{c}}$. Such processes are CKM favoured and therefore have relatively large branching fractions, typically $10^{-4} - 10^{-3}$. Decays of $B$ mesons produced at the $\Upsilon(4S)$ have very useful properties for studies of charmonium(-like) states:

- Both the $B$ and the kaon are spinless, therefore the state $X_{c\bar{c}}$ is produced polarised (with $J_Z = 0$ relative to kaon path). This helps to discriminate various spin and parity hypotheses for the $X_{c\bar{c}}$.
- Using hadronically tagged events (*i.e.,* with a fully reconstructed second $B$) one can measure absolute branching fractions of the charmonium(-like) states, as discussed below.

In $B$ decays, Belle observed two narrow charmonium levels: the $\eta_c(2S)$ reconstructed in its $K^0_S K^\pm \pi^\mp$ channel [1444], and the $\psi_2(1D)$ reconstructed via its radiative transition $\psi_2(1D) \to \gamma\, \chi_{c1}$ [1183]. The $\eta_c(2S)$ is a radial excitation of the spin-singlet $S$-wave state, and its mass is below the $D\bar{D}$ threshold. By now all charmonium levels below the $D\bar{D}$ threshold are known. The $\psi_2(1D)$ is a spin-triplet member of the $1D$ multiplet with $J^{PC} = 2^{--}$. It is above the $D\bar{D}$ threshold but, due to unnatural spin-parity, can not decay to this channel. Since the $\psi_2(1D)$ is below the $D\bar{D}^*$ threshold, it is very narrow [1445, 1446]. Only one more narrow charmonium level remains unobserved: the $\eta_{c2}(1D)$, a spin-singlet $1D$ state with $J^{PC} = 2^{-+}$. It is expected to reside between the $D\bar{D}$ and $D\bar{D}^*$ thresholds and, similarly to $\psi_2(1D)$, can not decay to $D\bar{D}$ due to unnatural spin-parity [1445, 1446]. A promising search channel might be $B \to K\,(h_c\gamma)$ [1445]. The lattice QCD predictions for the mass of this state are presented in Fig. 169 and Sec. 14.4.3.

$B$ decays have also been a rich source of charmonium-like mesons. Of the 24 known states 10 have been observed in $B$ decays (see Tables 125 and 126). In addition to the well known $X(3872)$ observed in the $\pi^+\pi^- J/\psi$ channel, these include the: $X(3915)$ seen in $\omega\, J/\psi$; $Z(4050)^+$ and $Z(4250)$ in $\pi^+\chi_{c1}$; $Z(4200)^+$ in $\pi^+ J/\psi$; $Z(4430)^+$ in $\pi^+\psi(2S)$ and in $\pi^+ J/\psi$; $X(4140)$, $X(4274)$, $X(4500)$ and $X(4700)$ in $\phi\, J/\psi$. With the possible exception of the $X(3872)$, the experimental information on all other states is very incomplete: the $Z(4050)^+$, $Z(4200)^+$, $Z(4250)^+$ and high mass $X$ states were seen in one experiment only and therefore need confirmation; the spin-parities of the $X(3915)$, $Z(4050)$ and $Z(4250)$ are still not determined; for most states only one decay channel is known. Belle II can considerably improve this situation.

Belle and LHCb performed full amplitude analyses of the $B^0 \to K^\pm\pi^\mp J/\psi$, $B^0 \to K^\pm\pi^\mp\psi(2S)$ and $B^0 \to K\phi\, J/\psi$ decays, which permitted $J^P$ quantum numbers determinations of the corresponding intermediate charmonium-like states. It is very important to apply full amplitude analyses to $B \to K\,\omega\,J/\psi$ and $B \to K\,\pi\,\chi_{c1}$ decays to determine the spin-parities of the $X(3915)$, $Z(4050)$ and $Z(4250)$.

While the $X(3872)$ has been seen to decay into $D^0\bar{D}^{*0}$ final states, open-flavour decays of the other states have yet to be seen. These can be investigated by comprehensive studies of the three-body decay processes $B \to K(D\bar{D})$, $B \to K(D\bar{D}^*)$ and $B \to K(D^*\bar{D}^*)$, for $D^{(*)}\bar{D}^{(*)}$ final states with both zero and non-zero total charge. The addition of one more pion will provide access to the final states with the $P$-wave $D$ mesons: $K(D\bar{D}^{**})$ and $K(D^*\bar{D}^{**})$. Since the states are broad, full amplitude analyses will be necessary in all cases. It will also be of considerable interest to study decays that include $D_s$ mesons.

Systematic investigations of charmonium plus light hadron final states, $B \to K\,(c\bar{c} + h)$, will be useful both for uncovering new decay channels of known charmonium-like mesons and for new charmonium-like meson searches. In this case, one should consider all narrow charmonium states: $\eta_c$; $\eta_c(2S)$; $J/\psi$; $\psi(2S)$; $h_c$; $\chi_{cJ}$ and $\psi_2(1D)$, and light hadron systems such as: $\pi^0$; $\pi^\pm$; $\pi^+\pi^-$; $\eta$; $\omega$; and $\phi$. Recently Belle performed a systematic search for $B$ decays into final states with an $\eta_c$ meson: $B^\pm \to K^\pm\pi^0\eta_c$; $B^\pm \to K^\pm\pi^+\pi^-\eta_c$; $B^\pm \to K^\pm\eta\,\eta_c$; and $B^\pm \to K^\pm\omega\,\eta_c$ [1447], but no signals were seen for any of these channels with the available data sample. Belle also studied the decays $B \to K\,\eta J/\psi$ [1448], $B \to K\pi\chi_{c2}$ and $B \to K\pi\pi\chi_{c1,2}$ [1449]; possibly there is a hint of the $Y(4140)$ in $\pi^+\pi^-\chi_{c1}$ [1450]. There is interest to search for radiative transitions of new states, in which case a charmonium should be combined with a photon. Belle studied the $B \to K\,\gamma\,\chi_{c1,2}$ decays and this resulted in the first evidence for $\psi_2(1D)$ [1183]. Investigations of these channels with a charmonium replaced by $X(3872)$ might also be interesting; recently Belle observed the first decay of this kind, $B \to K\pi X(3872)$ [1451].

Apart from $X(3872)$, which likely has a dominant $D^0\bar{D}^{*0}$ molecule-like component, the interpretations of the other states produced in $B$ decays remain unclear. An interesting possibility for the $X(3915)$ is that it is not a new state, but a new decay channel of the established $\chi_{c2}(2P)$ meson [1243] (see Section 14.5.2). To test this hypothesis, one should determine the $X(3915)$ spin-parity using an amplitude analysis of the $B \to K\,\omega J/\psi$ decays, and search for an intermediate $\chi_{c2}(2P)$ in the $B \to KD\bar{D}$ decays. Additional experimental information on the charmonium-like states, such as their spin-parities, open flavour and new hidden flavour decay channels, will facilitate their interpretation and help to discriminate between various models. More precise measurements of masses and widths of charmonium(-like) states are also important.

Determinations of absolute branching fractions of the $XYZ$ states (and, thus, partial decay widths) are essential. This can be done at Belle II by identifying their inclusive production in $B \to KX$ decays via the missing mass recoiling against the kaon. For this, one needs to know the initial momentum vector of the $B$ meson with good precision, which can be determined in $B\bar{B}$ events where the accompanying $\bar{B}$ meson is fully reconstructed (*i.e.,* using techniques similar to those described in Ref. [1452]).

For many of the above-described measurements, Belle II will have competition from the LHCb experiment. However, for absolute branching fraction measurements and for studies of final states that include neutral particles, Belle II will have considerably lower background.

*14.6.2. Initial State Radiation.* [C. P. Shen]

The idea of utilising initial-state radiation (ISR) from a high-mass state to explore electron-positron processes at all energies below that state was outlined in Ref. [1453]. The possibility of exploiting such processes in high luminosity $\phi-$ and $B$-factories was discussed in Refs. [1454–1456] and motivates the hadronic cross section measurement. States with $J^{PC} = 1^{--}$ can be studied with ISR technology using the huge Belle II data sample.

The ISR cross section for a hadronic final state $f$ is related to the corresponding $e^+e^-$ cross section $\sigma_f(s)$ by:

$$\frac{\sigma_f(s,x)}{dx} = W(s,x)\sigma_f(s(1-x)),$$

where $x = \frac{2E_\gamma}{\sqrt{s}}$; $E_\gamma$ is the energy of the ISR photon in the nominal $e^+e^-$ centre-of-mass frame; $\sqrt{s} = E_{c.m.}$ is the nominal $e^+e^-$ centre-of-mass energy; and $\sqrt{s(1-x)}$ is the effective centre-of-mass energy at which the final state $f$ is produced. The function $W(s,x)$ is calculated with an accuracy better than 1% and describes the probability density function for ISR photon emission [1053].

Although dramatic progress has been made on the study of the $XYZ$ states and the conventional charmonium states, there are still many questions to be studied in more detail. For example: Are the $X(3872)$ and $\psi_2(1D)$ in $e^+e^- \to \gamma\pi^+\pi^- J/\psi$ coming from resonance decays or continuum production? Are there other similar $X$ states in similar processes such as $\chi_{c2}(2P)$, $X(3915)$, $X(4140)$, and $X(4350)$? Is there a $Z_{cs}$ state decaying into $K^\pm J/\psi$ or $D_s^- D^{*0} + c.c.$, $D_s^{*-} D^0 + c.c.$? Can the $Z_c$ states decay into light hadrons?

More data are necessary for ISR studies. Belle II will accumulate 10 ab$^{-1}$ (50 ab$^{-1}$) data at around $\Upsilon(4S)$ by 2020 (2024). Compared to the current BESIII experiment, with ISR events the whole hadron spectrum is visible so that the line shape of the resonance and fine structures can be investigated. The disadvantage is the effective luminosity and detection efficiency are relatively low. Figure 175 shows the effective luminosity from 3 to 5 GeV in the Belle II data samples. We can see that, for 10 ab$^{-1}$ Belle II data, we have about 400–500 pb$^{-1}$ of data for every 10 MeV in the range 4–5 GeV. Of course, the ISR analyses have a lower efficiency than in direct $e^+e^-$ collisions because of the extra ISR photons and the boost given to events along the beam direction. Even taking these effects into account, the full Belle II data sample, which corresponds to about 2,000–2,300 pb$^{-1}$ data for every 10 MeV from 4–5 GeV, will result in similar statistics as BESIII for modes like $e^+e^- \to \pi^+\pi^- J/\psi$. Belle II has the advantage that data at different energies are expected to be accumulated at the same time, making the analysis much simpler than at BESIII at 60 data points. In addition, Belle II gets access to events above 4.6 GeV, which is currently the maximum energy of BEPCII. Very interesting in this context would be the search for the predicted pseudoscalar spin partner of $Y(4660)$ that should have a mass of 4616 MeV [1350] and could be produced in radiative decays of $Y(4660)$. This state should exist, if indeed $Y(4660)$ has a prominent $f_0(980)\psi(2S)$ component as claimed in Ref. [1457].

With a data sample larger than 10 ab$^{-1}$ at Belle II, ISR processes $e^+e^- \to \pi^+\pi^- J/\psi$, $\pi^+\pi^-\psi(2S)$, $K^+K^- J/\psi$, $K^+K^-\psi(2S)$, $\gamma X(3872)$, $\pi^+\pi^-\psi_2(1D)$, $\pi^+\pi^- h_c$, $\pi^+\pi^- h_c(2P)$, $\omega\chi_{cJ}$, $\phi\chi_{cJ}$, $\eta J/\psi$, $\eta' J/\psi$, $\eta\psi(2S)$, $\eta h_c$, $(D^*\bar{D}^*)^\pm\pi^\mp$, and so on can be studied. Some golden modes are:

- $e^+e^- \to \pi^+\pi^- J/\psi$: The $Y(4260)$ state was observed and confirmed by BaBar [1210], CLEO [1213] and Belle experiments [1458]. Besides the $Y(4260)$, Belle also observed a broad excess near 4 GeV, called $Y(4008)$ [1214]. With the full BaBar data sample of 454 fb$^{-1}$, the $Y(4008)$ structure was not confirmed [1211]. The difference between the BaBar and Belle interpretation around 4.01 GeV is large. Recently, BESIII reported a precise measurement of the $e^+e^- \to \pi^+\pi^- J/\psi$ cross section from 3.77 to 4.60 GeV using data samples with an integrated luminosity of 9 fb$^{-1}$ [1215]. While the nature of the events at around 4 GeV is still ambiguous, the dominant resonant structure, the so called $Y(4260)$, was found to have a mass of $(4222.0 \pm 3.1 \pm 1.4)$ MeV/$c^2$ and a width of $(44.1 \pm 4.3 \pm 2.0)$ MeV. In addition, a new resonance with a mass of around 4.32 GeV/$c^2$ is needed to describe the high precision data. With a 10 ab$^{-1}$ (50 ab$^{-1}$) data sample at Belle II, the expected statistical error on the $e^+e^- \to \pi^+\pi^- J/\psi$ cross section will be 7.5% (3.0%) at 4.23 GeV/$c^2$. The questions on the existence of the $Y(4008)$ and if there are more structures around 4.26 GeV/$c^2$ will be answered — note that such a light exotic vector state seems incompatible with a molecular interpretation while, *e.g.*, the tetraquark picture calls for light vectors — see discussions in Sec. 14.5.3. Belle also observed a charged charmonium-like state $Z_c(3900)$ in the $M_{\max}(\pi^\pm J/\psi)$ distributions [1159]. Its properties will be measured with much improved precision.

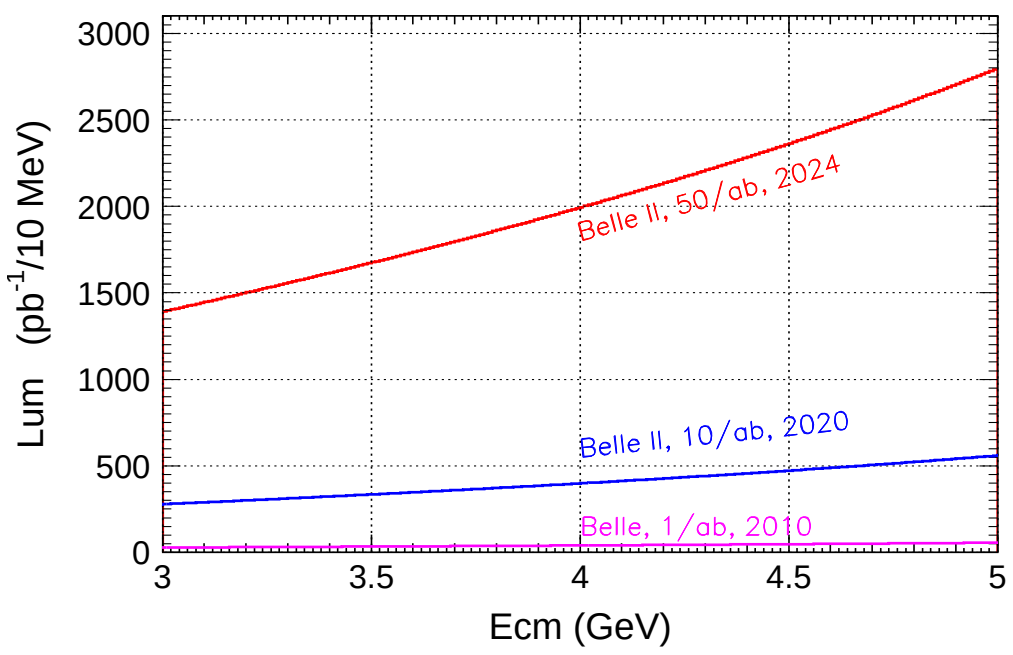

Fig. 175: Effective luminosity at low energy in the Belle and Belle II $\Upsilon(4S)$ data samples.

- $e^+e^- \to \pi^+\pi^-\psi(2S)$: The $Y(4360)$ was discovered by BaBar [1459], while Belle observed two resonant structures at 4.36 and 4.66 GeV/$c^2$, denoted the $Y(4360)$ and $Y(4660)$ [1219]. BaBar confirmed the existence of the $Y(4660)$ state later [1220]. Later the $Y(4360)$ and $Y(4660)$ parameters were measured with improved precision with the full 980 fb$^{-1}$ data sample of Belle [1167]. Belle also noticed a few events in the vicinity of the $Y(4260)$ mass, but the signal significance of the $Y(4260)$ was only 2.4$\sigma$. Evidence for a charged charmonium-like structure at 4.05 GeV/$c^2$, denoted the $Z_c(4050)$, was observed in the $\pi^\pm\psi(2S)$ intermediate state in $Y(4360)$ decays, which might be the excited state of the $Z_c(3900)$. With a 10 ab$^{-1}$ (50 ab$^{-1}$) data sample at Belle II, the expected statistical error on the $e^+e^- \to \pi^+\pi^-\psi(2S)$ cross section will be 12% (5.0%) at 4.36 GeV/$c^2$. The questions on the presence of the $Y(4260)$ in this channel and the existence of $Z_c(4050)$ will be answered.
- $e^+e^- \to K^+K^- J/\psi$: In the updated analysis, a strange partner of the $Z_c(3900)^\pm$, called the $Z_{cs}$, was searched for in $K^\pm J/\psi$ system [1460]. There are clear $K^+K^- J/\psi$ signal events and the cross sections are measured from threshold to 6.0 GeV. Rich structures with large statistical errors are observed; different fits were tried with poor fit quality. No obvious structures were observed in the $K^\pm J/\psi$ system. With a 10 ab$^{-1}$ (50 ab$^{-1}$) data sample at Belle II, the expected statistical error on the $e^+e^- \to K^+K^-J/\psi$ cross section is 15% (6.5%) at 4.53 GeV/$c^2$. Possible resonance structures to $K^+K^-J/\psi$ and $Z_{cs}$ to $K^\pm J/\psi$ can be searched for.

- $e^+e^- \to \pi^+\pi^-h_c$: BESIII measured $e^+e^- \to \pi^+\pi^-h_c$ cross sections at 13 energies of 3.90–4.42 GeV [1162]. There is a clear evidence for an exotic charmonium-like structure with a greater than $8.9\sigma$ statistical significance in the $\pi^\pm h_c$ system at 4.02 GeV/$c^2$, referred to as $Z_c(4020)$, and there are also some events at around 3.9 GeV/$c^2$, which could be $Z_c(3900)$. Adding $Z_c(3900)$ with fixed mass and width to the fit results in a statistical significance of $2.1\sigma$. BESIII also studied $e^+e^- \to (D^*\bar{D}^*)^\pm\pi^\mp$ at a centre-of-mass energy of 4.26 GeV using a 827 pb$^{-1}$ data sample. A structure near the $(D^*\bar{D}^*)^\pm$ threshold, denoted the $Z_c(4025)^\pm$, was observed [1169]. Very recently, the cross sections of $e^+e^- \to \pi^+\pi^-h_c$ at centre-of-mass energies from 3.896 to 4.600 GeV were measured. Two structures are observed around 4.22 and 4.39 GeV/$c^2$, which are called $Y(4220)$ and $Y(4390)$, respectively [1216]. With a 10 ab$^{-1}$ (50 ab$^{-1}$) data sample at Belle II, the expected statistical error on the $e^+e^- \to \pi^+\pi^-h_c$ cross section is 15% (6.5%) at 4.23 GeV/$c^2$. The $Y(4220)$, $Y(4390)$, and $Z_c(4020)/Z_c(4025)$ are expected to be confirmed or denied.
- $e^+e^- \to \omega\chi_{c0}$: Based on data samples collected at 9 centre-of-mass energies from 4.21 to 4.42 GeV, BESIII searched for the production of $e^+e^- \to \omega\chi_{c0}$ [1217]. Assuming the $\omega\chi_{c0}$ events come from a single resonance, the fitted mass and width of the resonance are $(4230 \pm 8 \pm 6)$ MeV/$c^2$ and $(38 \pm 12 \pm 2)$ MeV, respectively. The position of this resonance is consistent with the $Y(4220)$ state observed in the cross section of $e^+e^- \to \pi^+\pi^-h_c$ [1461]. It also indicates that the $Y(4260)$ signals observed in $e^+e^- \to \pi^+\pi^-J/\psi$ may have fine structures, and the lower mass structure at about 4230 MeV/$c^2$ has a sizeable coupling to the $\omega\chi_{c0}$ channel. With a 10 ab$^{-1}$ (50 ab$^{-1}$) data sample at Belle II, the expected statistical error on the $e^+e^- \to \omega\chi_{c0}$ cross section is 35% (15%) at 4.23 GeV/$c^2$. The question on the existence of the $Y(4220)$ will be further investigated.

Table 130 shows the estimated statistical errors on the cross section measurements at some typical centre-of-mass points for some golden ISR reactions with the total luminosity of 10 ab$^{-1}$ (50 ab$^{-1}$) at Belle II and the involved $XYZ$ states.

The PHOKHARA event generator is used at Belle II to simulate the ISR process at next-to-leading order accuracy. This includes virtual and soft photon corrections to one photon emission events and the emission of two real hard photons. At the moment, version 9.1 has been transferred into the Belle II environment [1462], where only a few modes, such as $e^+e^- \to \pi^+\pi^-$, $\pi^+\pi^-\pi^0$, $p\bar{p}$, and $n\bar{n}$, are available. More modes with a charmonium final state have also been added by Belle II. Figure 176 shows the simulated $\pi^+\pi^-$ invariant mass distribution from $e^+e^- \to \pi^+\pi^-$ MC sample simulated with the PHOKHARA generator in the Belle II environment.

*14.6.3. Two Photon Collisions.* [C. P. Shen]

At $e^+e^-$ colliders, two-photon interactions are studied via the process $e^+e^- \to e^+e^-\gamma^*\gamma^* \to e^+e^-R$. Almost all of the beam energy is kept by the scattered electron and positron and usually those are not detected (non-tagged events). If the scattering angle is sufficiently large, they can be detected in the forward region (tagged events). This process gives access to the resonances with $J^{PC} = 0^{++}, 0^{-+}, 2^{++}, 2^{-+}, ...$

Table 130: Estimated statistical errors on the cross section measurements at some typical centre-of-mass points for some golden ISR reactions with the total luminosity of 10 $ab^{-1}$ (50 $ab^{-1}$) at Belle II. Here the largest values of the cross sections from the latest measurements [1167, 1215–1217, 1460] are taken and the signal efficiency at Belle II is assumed to be around 22% of that at BESIII according to the measurements of $e^+e^- \to \pi^+\pi^- J/\psi$ at BESIII and Belle [1159, 1215].

| Golden Channels | $E_{c.m.}$ (GeV) | Statistical error (%) | Related $XYZ$ states |
|---|---|---|---|
| $\pi^+\pi^- J/\psi$ | 4.23 | 7.5 (3.0) | $Y(4008)$, $Y(4260)$, $Z_c(3900)$ |
| $\pi^+\pi^- \psi(2S)$ | 4.36 | 12 (5.0) | $Y(4260)$, $Y(4360)$, $Y(4660)$, $Z_c(4050)$ |
| $K^+K^- J/\psi$ | 4.53 | 15 (6.5) | $Z_{cs}$ |
| $\pi^+\pi^- h_c$ | 4.23 | 15 (6.5) | $Y(4220)$, $Y(4390)$, $Z_c(4020)$, $Z_c(4025)$ |
| $\omega\chi_{c0}$ | 4.23 | 35 (15) | $Y(4220)$ |

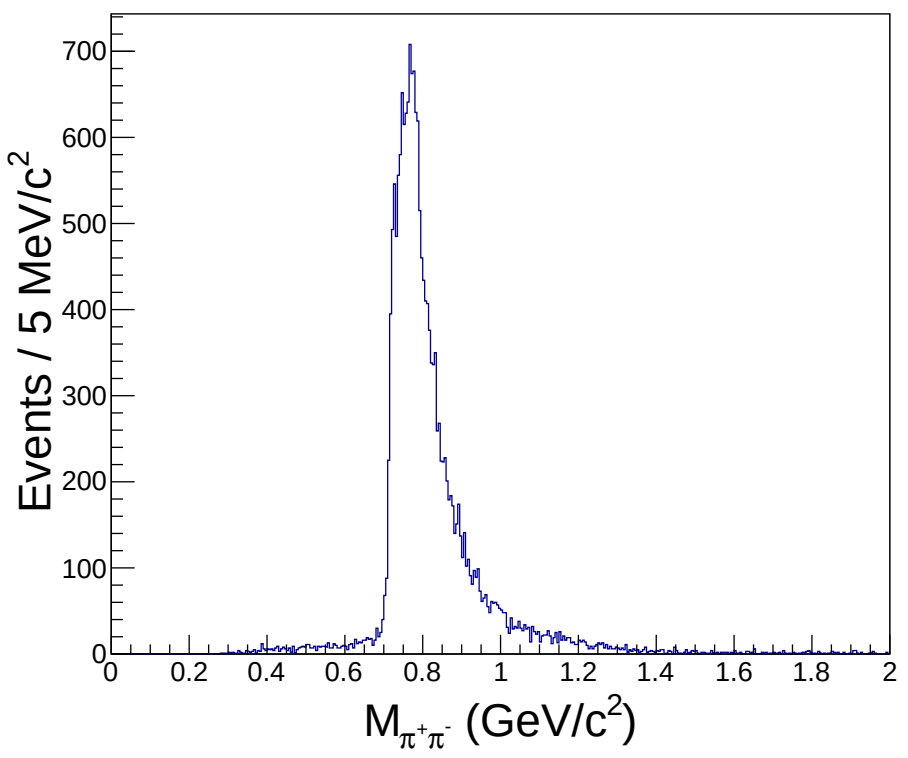

Fig. 176: The $\pi^+\pi^-$ invariant mass distribution from $e^+e^- \to \pi^+\pi^-$ MC sample simulated with the PHOKHARA generator in the Belle II environment.

The total production cross section of $e^+e^- \to e^+e^- R$ is given by

$$\begin{aligned} \sigma(e^+e^- \to e^+e^- R) &= \int \sigma(\gamma\gamma \to R) \frac{d\mathcal{L}_{\gamma\gamma}}{dW_{\gamma\gamma}} dW_{\gamma\gamma} \\ &= (2J+1) \cdot \mathcal{K} \cdot \Gamma_{\gamma\gamma}, \end{aligned}$$

where $\frac{d\mathcal{L}_{\gamma\gamma}(\mathcal{W}_{\gamma\gamma})}{dW_{\gamma\gamma}}$ is the transverse two-photon luminosity function [1463], $W_{\gamma\gamma}$ is the effective center-of-mass energy of the two-photon system and the factor $\mathcal{K}$ could be obtained from Monte Carlo integration.

$$\begin{aligned}\mathcal{K} &= \int 8\pi^2 \frac{d\mathcal{L}_{\gamma\gamma}(\mathcal{W}_{\gamma\gamma})}{dW_{\gamma\gamma}} \delta(W_{\gamma\gamma}^2 - m_R^2) \frac{1}{m_R} dW_{\gamma\gamma} \\ &= 4\pi^2 \frac{1}{m_R^2} \frac{d\mathcal{L}_{\gamma\gamma}(\mathcal{W}_{\gamma\gamma})}{dW_{\gamma\gamma}}.\end{aligned}$$

Here $\int \delta(W_{\gamma\gamma}^2 - m_R^2) dW_{\gamma\gamma} = \frac{1}{2m_R}$ is used.

Experimental studies for two-photon physics at Belle II have merits since all the data at any energy point can be used to investigate the lower invariant mass region. Physics at higher invariant mass regions, $W_{\gamma\gamma} > 5$ GeV, is not suitable because the luminosity function for two-photon collisions steeply decreases with increasing $W_{\gamma\gamma}$ and the backgrounds from single photon annihilation processes are considerable. With the total integrated luminosity larger than 10 ab$^{-1}$ at Belle II, the two-photon processes listed below are our priorities.

A state at 3930 MeV was discovered by Belle in $\gamma\gamma \to D\bar{D}$ [1190], and confirmed by BaBar later [1191]. The experimental results on the mass, angular distributions, and $\Gamma(X(3930) \to \gamma\gamma)\mathcal{B}(X(3930) \to D\bar{D})$ are all consistent with the expectation for the $\chi_{c2}(2P)$. So far, this is the only unambiguously identified radially excited $P$-wave charmonium state in experiment. The $X(3915)$ was discovered by Belle in $\gamma\gamma \to \omega J/\psi$ [1188]. Later, a spin-parity analysis was performed for $X(3915) \to \omega J/\psi$ by BaBar [1189], and the results suggest that the quantum numbers of this state are $J^P = 0^+$. It was therefore identified as the $\chi_{c0}(2P)$. However, assigning the $X(3915)$ as the $\chi_{c0}(2P)$ state faces many problems as discussed in Sec. 14.5.2. At Belle II with higher statistics, the $\gamma\gamma \to D\bar{D}$ process needs to be analysed carefully to give more precise parameters of the $\chi_{c2}(2P)$. The angular distribution for the broad bump below the narrow peak of the $\chi_{c2}(2P)$ should be determined to obtain the $J^P$ value and check if it could be assigned as the $\chi_{c0}(2P)$. For $\gamma\gamma \to D\bar{D}$, we expect obvious contributions from $\gamma\gamma \to D\bar{D}^* \to D\bar{D}\pi$ and $\gamma\gamma \to D\bar{D}(n)\gamma$. All of these processes are cross contaminated. Their cross sections can be measured, however, by applying iterative methods. The lattice QCD masses of $\chi_{c0,c2}(2P)$ within an approach that neglects their strong decays are given in Fig. 170. The results of an exploratory lattice simulation which takes into account their $D\bar{D}$ strong decays was discussed in *Scalar resonances* of Section 14.5.4.

Another important two-photon process is $\gamma\gamma \to \phi J/\psi$. With the full amplitude analysis of $B^+ \to K^+\phi J/\psi$ performed by LHCb, four $\phi J/\psi$ structures, $X(4140)$, $X(4274)$, $X(4500)$ and $X(4700)$, are observed [1205]. The structures in the $\phi J/\psi$ mass spectrum seem very rich. The $M(\phi J/\psi)$ distribution at Belle II needs to be revisited to confirm or deny the existence of the $X(4350)$ and search for more exotic states. Note here due to the Landau-Yang theorem [1464], the $X(4140)$ and $X(4274)$ with $J^{PC} = 1^{++}$ can not be produced in two-photon process $\gamma\gamma \to \phi J/\psi$.

With somewhat smaller boost at Belle II, the efficiency in two-photon process may be a little higher compared to that at Belle. With higher luminosity and detection efficiency, we expect interesting results from $\gamma\gamma \to \phi J/\psi$.

#### *14.6.4. Double Charmonium Production.* [P. Pakhlov]

The discovery of a number of double-charmonium production processes in $e^+e^-$ annihilation at $B$-factories was initiated by the observation of $e^+e^- \to J/\psi X$, where $X$ is $\eta_c$, $\chi_{c0}$, $\eta_c(2S)$ by Belle [1465]. This production mechanism provides a powerful tool for an understanding of the interplay between perturbative QCD (pQCD) (and its expansions beyond the leading order) and non-perturbative effects, in particular with application of the light-cone approximation and the nonrelativistic QCD (NRQCD) factorisation approaches. The first calculations using NRQCD within the leading order pQCD for the $e^+e^- \to J/\psi\,\eta_c$ cross-section gave a value, which was an order of magnitude smaller than the measured cross section [1466]. The importance of relativistic corrections was realised in Ref. [1467, 1468]; the authors, using the light-cone approximation to take into account the relative momentum of heavy quarks in the charmonium states, managed to calculate the cross section which is close to the experimental value. Some authors have been able to reproduce the experimental result using next-to-leading (NLO) corrections [1469, 1470]. The present variety of different alternative approaches that explain the experimental result points to the need to check the suggested models with new data.

For the moment the production of $J/\psi$ and $\psi(2S)$ with spin-0 charmonia is established with a very high significance [1471, 1472]. The processes $e^+e^- \to J/\psi\,X$ are identified from peaks in the mass spectrum of the system recoiling against the reconstructed $J/\psi$ in inclusive $e^+e^- \to J/\psi\,X$ events. Belle also reported strong evidence for the process $e^+e^- \to J/\psi\chi_{c1}$ [1473]. Furthermore, the later process was reconstructed from both sides: with a fully reconstructed $J/\psi$ and $\chi_{c1}$ signal seen in the recoil spectrum and vice versa with a $J/\psi$ peak seen recoiling against the reconstructed $\chi_{c1}$. A hint of the process $e^+e^- \to J/\psi\chi_{c2}$ was also seen. At Belle II it is likely that the full list of possible charmonium pairs can be measured with a good accuracy, which can be used to verify the charmonium production models. Another important mission for Belle II is to perform angular analyses (*e.g.* to measure the $J/\psi$ production and $J/\psi$ helicity angle distributions for $e^+e^- \to J/\psi X$) that gives access to the ratio of different orbital angular momentum contributions in the two body process, which also allows to check the consistency of the models with the experimental data. Finally, Belle II with unprecedentedly high statistics and data taken at different energies from the $\Upsilon(1S)$ to $\Upsilon(6S)$ resonances is capable of measuring the $\sqrt{s}$-dependence of the double charmonium cross sections.

This process can also serve as an efficient tool to study the charmonium decays, in particular to measure their absolute branching fractions. The double charmonium production mechanisms provide an efficient and clean tagging for charmonia states, produced recoiling against the $J/\psi$. At Belle the statistics of tagged $\eta_c$ and $\eta_c(2S)$ from the process $e^+e^- \to J/\psi X$ was about 1000 and 700 events, respectively [1473]. With a 50 times higher data set Belle II can measure the absolute branching fractions for $\eta_c$, $\eta_c(2S) \to K^0_S K\pi$ with a $\sim 1\%$ accuracy.

On the other hand, the double-charmonium production mechanism offers a unique opportunity to search for and study new $C$-even charmonium states, produced in association with the effectively reconstructed $C$-odd charmonia such as $J/\psi$ or $\psi(2S)$. New states can be exclusively revealed as peaks in the recoil mass spectrum against the $J/\psi$ or $\psi(2S)$, using two-body kinematics. Moreover, the decay channels of those new charmonia states can be also studied. Known examples from Belle [1192, 1193] are $X(3940) \to D\bar{D}^*$ and $X(4160) \to D^*\bar{D}^*$, which are tagged by the $\bar{D}^*$ peak in the spectrum recoiling against reconstructed $J/\psi D$ or $J/\psi D^*$ combinations, respectively. These two states are the only charmonium-like states for which hidden flavour decay channels are not known. As mostly spin-zero ordinary charmonia are produced in the process $e^+e^- \to J/\psi\, X$ and production of the radial excitations are not suppressed, the $X(3940)$ and $X(4160)$ could be naturally interpreted as the $\eta_c(3S)$ and $\eta_c(4S)$ states. As these assignments are not well accommodated by the potential models [1051, 1052], it is important to perform a full amplitude analysis at Belle II to measure spin-parities and finally identify these states. New states can appear at Belle II with much higher statistics in the processes which have been already studied at Belle (such as $e^+e^- \to J/\psi\, D\bar{D}$, $J/\psi\, D\bar{D}^*$, $J/\psi\, D^*\bar{D}^*$ [1185, 1193]), or with new final states, which were not accessible at Belle because of low reconstruction efficiencies. They include charmonia other than $J/\psi$: $e^+e^- \to \eta_c X$, $e^+e^- \to h_c X$, $e^+e^- \to \chi_{cJ} X$, $e^+e^- \to \eta_c(2S)X$, $e^+e^- \to \psi(2S)X$, or other final states for new charmonium states (*e.g.* charmonium+light hadrons, or $D_s^{(*)}\bar{D}_s^{(*)}$, charmed baryons final states, or $D^{(*)}\bar{D}^{(*)}\pi$).

Interesting but experimentally challenging topics related to the $e^+e^- \to c\bar{c}c\bar{c}$ process are searches for doubly charmed molecules $DD^{(*)}$ and doubly charmed baryons. The former are discussed in Section 14.5.4 "$\bar{q}\bar{q}QQ$ states, $BB^{(*)}$ and $DD^{(*)}$ potentials"; they can be formed in double $2(c\bar{c})$ production because the phase space in this process is limited. The $DD^*$ molecular state can be searched for in the $DD^*$, $DD\pi$ or $DD\gamma$ channels. Experimental difficulties arise due to low efficiency of the full reconstruction of the final states and due to large expected combinatorial background from single $(c\bar{c})$ production. Production of doubly charmed baryons in $e^+e^-$ annihilation is discussed in Refs. [1474, 1475]. It is predicted that the $c\bar{c}c\bar{c}$ events hadronise predominantly into the doubly charmed baryons. However, their reconstruction is also difficult because of small expected branching fractions into the final states convenient for reconstruction, and high background. Thus far, the obtained upper limit for the cross section of the production of weakly decaying doubly charmed baryons, the $\Xi_{cc}$, by Belle [1476] is at least an order of magnitude higher than the theoretical expectation.

*14.6.5. Summary.* The only missing very narrow charmonium level, the spin-singlet $\eta_{c2}(1D)$ with $J^{PC} = 2^{-+}$, can be searched for at Belle II in $B$ decays.

Experimental information about the charmonium-like states above the $D\bar{D}$ threshold is very incomplete. With the significant increase of statistics compared to Belle, the Belle II experiment can measure more precisely the line shapes of the states, determine their spin-parities, search for new decay channels (in particular, for the open flavour channels which are not known for most of the states but are very important for their interpretation), and search for new states expected in various approaches. More precise and detailed experimental information about the states should facilitate their interpretation and help to discriminate between various models.

## *14.7. Belle II prospects for Bottomonium(-like) states*

### *14.7.1. Bottomonia below $B\bar{B}$ threshold.* [B. Fulsom, R. Mussa, U. Tamponi]

At Belle II, there are generally three ways to access bottomonia below the $B\bar{B}$ threshold: via decays of higher mass states (*e.g.* $\Upsilon(4S, 5S, 6S)$)), production of $1^{--}$ states via initial-state radiation, or by direct production via operation at a lower centre-of-mass energy. From the previous generation of $B$-factories, we learned that the spin-flipping transitions via charged bottomonium-like states are key to reach the spin singlet sector of the spectrum (para-bottomonia). These studies can be performed in Belle II using the high statistics data samples that will be integrated at $\Upsilon(4S)$ and $\Upsilon(5S)$ peak energies. High precision studies of the rare hindered transitions, which will allow to measure the small but sizeable relativistic effects in bottomonium, will in turn require one dedicated period of data taking on the peak of the $\Upsilon(3S)$ resonance.

The lattice QCD results for masses and radiative transitions of bottomonia were given in Sec. 14.4.3. In particular, the almost complete spectrum of states below $\bar{B}B$ threshold from lattice NRQCD is presented in Fig. 169. The calculated masses are generally in good agreement with masses of already observed states. Hadronic transitions are challenging for rigorous lattice treatment and have not been considered yet. Potential model predictions are collected in Table 132.

Below the $\Upsilon(4S)$ threshold, there are several predicted bottomonium states that have yet to be positively identified: separation of the $\chi_b(3P)$ triplet, the $\Upsilon(2D_3)$ triplet, $\eta_b(3S)$, $\Upsilon(1D_{1,3})$, $\eta_b(1D)$, and the $F$-wave states. Evidence for $\eta_b(2S)$ is below the $5\sigma$ threshold [1112]. Of the known states there are several important parameters that either need to be measured or have conflicting experimental results in need of resolution. Examples include the masses and widths of the $\eta_b$ states, $\chi_{b0}$ widths, and the mass splitting of the $\Upsilon(1D)$ states.

Experimentally, few facts should be taken into account to understand the Belle II potential. Radiative and hadronic transitions can be identified either inclusively, by studying the $\eta$, $\omega$, $\gamma$ or $\pi^+\pi^-$ recoil spectrum in hadronic events, or exclusively when the final state of a decay chain can be fully reconstructed. In bottomonium, this restricts us to the $\Upsilon(nS) \to e^+e^-, \mu^+\mu^-$ modes, since the hadronic annihilations are known to result in many multi-body final states with small branching fraction. While the former technique grants a high-efficiency reconstruction almost insensitive to the quantum numbers of the final state, its power is limited by a small signal-versus-background ratio, due to the large combinatorial background arising by the final state annihilation or the underlying *continuum* process $e^+e^- \to q\bar{q}$. This latter process is dominant in the datasets collected above the $\Upsilon(4S)$ energy: Monte Carlo simulation of the $\Upsilon(6S) \to \eta h_b(1P)$ process show that more than 90% of the combinatorial background in the signal region is due to the continuum, even after applying a topological selection to suppress it based on the Fox-Wolfram moments. A strong commitment by the experimental community in finding better tools to further suppress the continuum contribution is therefore desirable and needed to improve the physics results on hadronic transitions of any run at $\Upsilon(4S)$, $\Upsilon(5S)$ and $\Upsilon(6S)$ energies.

Inclusive radiative transitions, on the other hand, are even more problematic. Even though the combinatorial background is in some cases smaller and the reconstruction efficiency much higher than any hadronic transition, the activity in the calorimeter arising from the beam background, which scales approximately with the instantaneous luminosity, limits the photon energy resolution and adds a large background. This makes it quite difficult to detect any transitions emitting a photon with energy in the centre of mass frame lower than 50 MeV. While the improved Belle II tracking grants better resolution and much large tracking efficiency in the low transverse momentum region compared to Belle, no striking improvement in the photon detection performance is expected.

*$\eta_b$ measurements.* The combined BaBar [1477, 1478] and CLEO [1479] results from $\Upsilon(mS) \to \gamma\eta_b(1S)$ decays give $m_{\eta_b(1S)} = (9391.1 \pm 2.9)$ MeV, while the combined Belle [1112, 1480] measurements via $h_b(nP) \to \gamma\eta_b(1S)$ find $m_{\eta_b(1S)} = (9403.4 \pm 1.9)$ MeV. These measurements disagree at the $\sim$3.5$\sigma$ level. Combining the two existing measurements of the $\eta_b(1S)$ width [1112, 1480] still has an appreciable uncertainty, $\Gamma_{\eta_b(1S)} = 10.8^{+3.5}_{-3.0}$ MeV. As for $\eta_b(1S)$, Belle [1112] measured $m_{\eta_b(2S)} = 9999.0 \pm 3.5^{+2.8}_{-1.9}$ MeV in the decay $h_b(2P) \to \gamma\eta_b(2S)$, after disconfirming [1481] an independent analysis of the CLEO dataset which found $m_{\eta_b(2S)} = (9974.6 \pm 2.3 \pm 2.1)$ MeV [1482] .

Ultimately, the best way to produce $\eta_b$ states at Belle II could be via $\Upsilon(4S)$ decays. With 50 ab$^{-1}$ of $\Upsilon(4S)$ data and $\mathcal{B}(\Upsilon(4S) \to \eta h_b(1P)) = (1.83 \pm 0.16 \pm 0.17) \times 10^{-3}$, $\sim$ 4M $\eta_b(1S)$ can be produced and reconstructed (accounting for efficiency and $\eta$ branching ratios). An extended run of $\sim$2 ab$^{-1}$ at $\Upsilon(5S)$ energy offers access to $h_b(1P, 2P)$, potentially resulting in 0.25M of both $\eta_b(1S, 2S)$ after efficiency and reconstruction. Given $\mathcal{B}(\Upsilon(3S) \to \gamma\eta_b(1S)) = (5.1 \pm 0.7) \times 10^{-4}$ [77], one expects roughly 0.6M $\eta_b(1S)$ events if 300 fb$^{-1}$ is collected at the $\Upsilon(3S)$. The best limit for the suppressed $M1$ transition, $\mathcal{B}(\Upsilon(3S) \to \gamma\eta_b(2S)) < 6.2 \times 10^{-4}$ [1483], implies less than 0.5M $\eta_b(2S)$ events from a similar dataset. This last analysis would require a detailed understanding of nearby peaking contributions from $\chi_b$, $\Upsilon(1D)$, and initial-state radiation photon lines, and extraction of a signal above a large inclusive photon background. The $\pi\pi$ transitions from the $\Upsilon(5S)$ and the $\eta$ transition from the $\Upsilon(4S)$ will allow the study of the $\eta_b \to \gamma\gamma$ transition with very little background in exclusive mode ($\pi\pi + 5\gamma$ or $5\gamma$ from the $\Upsilon(4S)$, and $\pi\pi + 3\gamma$ from the $\Upsilon(5S)$). Very precise prediction is available in NRQCD [1097, 1484] for the two photon decays of both $\eta_b(1S)$ and $\eta_b(2S)$, as discussed in the theory section.

Another potential unexplored pathway well-suited to the initial running conditions could be via $\Upsilon(3S) \to \gamma\chi_{b0}(2P) \to \gamma\eta\eta_b(1S)$. The $\chi_b(2P) \to \eta\eta_b(1S)$ decay could have a branching fraction as large as $10^{-3}$ [1485]. When combined with $\mathcal{B}(\Upsilon(3S) \to \gamma\chi_{b0}(2P)) = (5.9 \pm 0.6)\%$ [77] and an estimated efficiency of 5%, one might expect $\sim$12 events per fb$^{-1}$, or a few thousand events with a 300 fb$^{-1}$ $\Upsilon(3S)$ sample.

*$\Upsilon(m^3D_J)$ measurements.* Although much progress has been made on understanding of the $\Upsilon(1^3D_J)$ triplet by CLEO, BaBar, and Belle, isolation and identification of the individual $\Upsilon(1^3D_J)$ states remains elusive.

Belle observed $\Upsilon(5S) \to \eta\Upsilon(1D)$ decays with a branching fraction of $(2.8 \pm 0.7 \pm 0.4) \times 10^{-3}$, corresponding to a cross section $\sigma(e^+e^- \to \eta\Upsilon(1D)) \approx 400$ fb, in inclusive $\eta$ transitions. Further preliminary studies investigated $\Upsilon(1D)$ decays to exclusive $\gamma\chi_b(1P) \to \gamma\Upsilon(1S)$ final states. The current level of statistics was insufficient to resolve the various $J$ components.

From $\Upsilon(3S)$, two decay pathways have been employed: the four-photon radiative decay cascade [1486, 1487], and radiative decays to $\Upsilon(1^3D_J)$ followed by a dipion decay to $\Upsilon(1S)$ [1488]. With an integrated luminosity of 250 fb$^{-1}$, thousands of $1^3D_J$ will be produced in radiative transitions for each $J$ value, which should be sufficient to identify each of the $1^3D_J$ states and measure their masses precisely enough to test theoretical predictions for their splittings. The $D$-wave states may also be accessed via dipion transitions from $\Upsilon(5S)$ [1480].

In addition, the spin singlet state $\eta_b(1D)$ can only be reached via E1 radiative transitions from $h_b(2P)$, which can be reached only from $\Upsilon(5,6S)$.

Based on the BaBar results [1488], a factor of $7-9$ increase in statistics is needed to resolve and observe previous hints of $J=1,3$ $\Upsilon(1D)\to\pi^+\pi^-\Upsilon(1S)$ peaks, given a similar reconstruction efficiency (and if they are indeed significant). For the exclusive four-photon cascade, approximately 4, 15, and 2 events are expected per $\text{fb}^{-1}$ for the $J=1$, 2, and 3 states, respectively (assuming an efficiency of $\sim 15\%$). Good photon energy resolution is critical for disentangling overlapping photon energies in this decay mode.

Another possibility could be to search for the $\Upsilon(1D_1)$ state via the inclusive $\Upsilon(3S)$ photon spectrum ($\Upsilon(1D_1)\to\gamma\chi_{b0}(1P)$ produces a photon of $E^*_{CM}\sim 288$ MeV). This is the highest of the six $\Upsilon(1^3D_J)\to\gamma\chi_{bJ}(1P)$ photon transition energies, and may possibly be identified as a lone peak apart from a Gaussian encompassing several other photon transition lines. Energy resolution and higher statistics of running directly at $\Upsilon(3S)$ is necessary for such an analysis.

In addition to the above noted $\Upsilon(1^3D_J)$ strategies, there could also be the chance to produce the $J=1$ states directly via a beam energy scan [1489]. $\Gamma_{e^+e^-}$ widths of the $\Upsilon(D)$ states are predicted to be very small ($<2$ eV) [1490, 1491], but a dedicated scan with the high instantaneous luminosity of Belle II could cover this in a relatively short time. The production cross section is proportional to the BR to $e^+e^-$ which is roughly 3 orders of magnitude smaller than that for the $S$-wave states. The small number of signal events will also make it challenging to see the $n^3D_1$ states above backgrounds. With these caveats we estimate the number of $n^3D_1$ produced by multiplying the raito of $nD/2S$ BRs to $e^+e^-$ times the $2^3S_1$ cross section. This gives $\sim 13$ pb for the $1^3D_1$ which results in $\sim 1.3\times10^6$ $1^3D_1$ for 100 $\text{fb}^{-1}$ of integrated luminosity yielding $\sim 100$ events in $1^3D_1\to\mu^+\mu^-$ but many more in radiative decay chains via intermediate $1P$ states.

Similarly we estimate $\sigma(e^+e^-\to 2^3D_1)\sim 18$ pb yielding $\sim 2\times10^6$ $2^3D_1$ for 100 $\text{fb}^{-1}$ of integrated luminosity. It should thus be possible to observe the $2^3D_1$ state, which is predicted to have mass around 10.45 GeV/c$^2$, above the $\Upsilon(3S)$. No known way to access these states with hadronic transitions is known, even if the observation of $\Upsilon(5S)\to\eta\Upsilon(1D)$ hints to a possible production of $\Upsilon(2D)$ via $\eta$ transitions from $\Upsilon(6S)$. Preliminary MC simulations show that the cross section $\sigma[e^+e^-\to\eta\Upsilon(2D)]$ at the $\Upsilon(6S)$ energy must exceed 3.5 pb (0.5 pb) in order to observe this process with an integrated luminosity of 50 $\text{fb}^{-1}$ (250 $\text{fb}^{-1}$) collected at the $\Upsilon(6S)$ energy, using the inclusive $\eta$ meson recoil technique. These results can be largely improved by a better rejection of the continuum background.

*$h_b$ measurements.* The $h_b$ has been discovered in the process $\Upsilon(5S,6S)\to\pi^+\pi^-h_b(1P,2P)$ (via intermediate $Z_b^\pm$) [1112], and $\Upsilon(4S)\to\eta h_b(1P)$. The controversial evidence of $h_b(1P)$ in decays of $\Upsilon(3S)\to\pi^0h_b(1P)$ [1492] deserves further studies. No evidence for $\Upsilon(3S)\to\pi^+\pi^-h_b(1P)$ was found [1493].

$\eta$ and dipion transitions from $\Upsilon(4S)$ and $\Upsilon(5S)$ to $h_b(nP)$ are the best production source for these states at Belle II, with branching fractions of $\sim 2\times10^{-3}$, in agreement with the prediction in [1389], and $(4-6)\times10^{-3}$, respectively. Considering efficiency, this equates to approximately 11M $h_b(1P)$ events in the full Belle II $\Upsilon(4S)$ data set. An additional 0.8 (1.4)M $n=1(2)$ could be reconstructed from 2ab$^{-1}$ of $\Upsilon(5S)$ data.

For direct running at $\Upsilon(3S)$, the BaBar measurement of $\Upsilon(3S) \to \pi^0 h_b(1P)$), efficiency of $\sim 15\%$ implies about 250 events per $\mathrm{fb}^{-1}$, though the background level for the $\pi^0$ transition is very high. A factor of $\sim$3 increase in statistics would be necessary to convert their evidence into an observation. Besides providing an alternative measurement of the masses of $h_b(1P)$ and $\eta_b(1S)$, this process is the only isospin violating transition between narrow bottomonia, in striking contrast with the non-observation of the transition $\Upsilon(3S) \to \eta\Upsilon(1S)$. This process may also provide comparative information on the $\eta_b(1S)$ from a separate production method.

The dipion transition sets a branching fraction limit for $\Upsilon(3S) \to \pi^+\pi^- h_b(1P)$ of less than $1.2 \times 10^{-4}$, challenging most theoretical models, the lowest of which was on the order of $\mathcal{O}(10^{-4})$ [1494–1496].

*$\chi_b(3P)$ bottomonia.* ATLAS, D0, and LHCb have observed radiative $\chi_b(3P)$ decays to $\Upsilon(1S, 2S, 3S)$. At Belle II, this could potentially be accessed by $\Upsilon(4S) \to \gamma\chi_b(3P)$ decays, though the rate is expected to be low. It is expected that several $\mathrm{ab}^{-1}$ of integrated luminosity will be accumulated at the $\Upsilon(4S)$. Assuming 10 $\mathrm{ab}^{-1}$ of integrated luminosity roughly $10^{10}$ $\Upsilon(4S)$ will be produced so that sufficient numbers of events will be produced in decay chains proceeding via the $3^3P_2$ and $3^3P_1$ that these states should be observed in radiative decays of the $\Upsilon(4S)$. We do not expect that the $3^3P_0$ will be observed in this manner due to its larger width and consequently smaller branching ratios for radiative transitions. Another interesting possibility for studying the $3P$ states via radiative transitions from the $\Upsilon(4S)$ utilizes hadronic transitions from the $\Upsilon(3S)$ and $\Upsilon(2S)$ or $\Upsilon(1S)$ in the decay chain. In the full Belle IIdataset, this would yield some tens of events for the $3^3P_2$ and $3^3P_1$ intermediate states but only $\mathcal{O}(1)$ event for the $3^3P_0$. This might be sufficient to resolve these states.

Lattice QCD guidance on where $^3P$ states are located is not available yet (see Fig. 169). This is an even more challenging task as it would need to take into account also the effect of $\bar{B}B^{(*)}$ threshold.

*Search for F-wave bottomonia.* The lattice QCD predictions for these states are given in Fig. 169. These states are likely best accessed via radiative decays of $\Upsilon(2^3D_J)$ (also yet to be discovered). Reaching $\Upsilon(2^3D_J)$ could potentially be accomplished by two-photon transitions from $\Upsilon(4S)$, a dipion transition from $\Upsilon(6S)$, or direct production from initial state radiation or operation at the appropriate energy. The $F$-wave states decay dominantly to $\gamma\Upsilon(1^3D_J)$, which also typically decay radiatively to $\chi_b(1P)$ states, or via rare dipion transitions to $\Upsilon(1S)$. Overall, this is a difficult experimental challenge. The production rate of these states is expected to be very low (indeed, the first step of reaching $\Upsilon(2^3D_J)$ has already never been realised). Assuming a beam energy scan to discover and directly produce $\Upsilon(2^3D_1)$, the mostly promising search method could be via the inclusive photon spectrum for a $\sim$ 90 MeV photon corresponding to $\Upsilon(2D_1) \to \gamma\chi_{b2}(1F)$. This low-energy photon search would face high backgrounds, and overlapping transition lines from other bottomonium and initial-state radiation processes.

*Hadronic transitions and decays.* CLEO and Belle have searched for or identified dozens of exclusive final states for $\chi_b$ and $\Upsilon$ [1497–1499] decays (generally at the level of $< \mathcal{O}(10^{-4})$. Two examples of interesting measurements with this approach are $\Upsilon \to \gamma\eta_b(nS) \to \gamma +$ *hadrons* [1481, 1482], or using the ratio of $\Upsilon(2S) \to \gamma\chi_b(1P) \to \gamma + hadrons$ to $\Upsilon(3S) \to \gamma\chi_b(1P) \to \gamma + hadrons$ to access $\Upsilon(3S) \to \gamma\chi_b(1P)$ [1497].

There are several allowed dipion transitions amongst bottomonium states that could be studied in detail, for example $\Upsilon(3S) \to \pi^+\pi^-\Upsilon(1S, 2S)$ and $\chi_b(2P) \to \pi^+\pi^-\chi_b(1P)$, and their neutral pion counterparts. While the dipion transitions between $\Upsilon$ states and their atypical $m_{\pi\pi}$ structure have been studied extensively [1500], the transitions between $\chi_b$ states have yet to be fully exploited and understood [1493, 1501].

If the $\pi\pi$ invariant mass distributions for the dipion transitions were measured precisely, additional interesting physics aspects can be extracted. For instance, the $\pi\pi$ $S$-wave scattering length $a_0 - a_2$ can be measured precisely with an uncertainty of $\sim 0.5\%$ using the cusp effect at the $\pi^+\pi^-$ threshold with about 4M events for $\Upsilon(3S) \to \Upsilon(2S)\pi^0\pi^0$ [1502]. A fine measurement of the dipion mass spectrum at around 1 GeV in the transition $\Upsilon(4S) \to \Upsilon(1S)\pi^+\pi^-$ is necessary to see clearly the nontrivial behaviour caused by the $f_0(980)$ [1503, 1504] which is invisible in the present analysis [1233].

*Radiative decays.* The lattice results on radiative decays $\Upsilon(1S, 2S, 3S) \to \eta_b(1S, 2S)\gamma$ and $\eta_b(2S, 3S) \to \Upsilon(1S)\gamma$ were summarised in Section 14.4.3. The transition $\Upsilon(2S, 3S) \to \eta_b(1S)\gamma$, for example, would be absent in the strict non-relativistic limit and its rate crucially depends on a multitude of interesting relativistic effects. Results on other, yet unobserved, transitions discussed below can be expected in the years to come.

Radiative decays between $\Upsilon$ and $\chi_b$ have been extensively studied by several experiments in the past, most recently by BaBar [1505, 1506]. Two outstanding questions regarding these transitions are the observation of $\chi_{b0}(2P) \to \gamma\Upsilon(1S)$, and the understanding of $\Upsilon(3S) \to \gamma\chi_{bJ}(1P)$ decays. The former has a single claim [1507], while the recent high-statistics BaBar measurement could only reach a significance of $2.2\sigma$, putting this observation within reach with approximately 5 times the $\Upsilon(3S)$ dataset. These hindered E1 transitions are both experimentally difficult to measure due to overlapping photon transition energies, and theoretically difficult to calculate due to the effects of higher-order corrections. While they have been measured by both CLEO [1483, 1508] and BaBar [1505], there remains some disagreement over the suppressed $\Upsilon(3S) \to \gamma\chi_{b1}(1P)$ transition.

The hindered M1 transitions $\chi_{bJ}(nP) \to \gamma h_b(mP)$ and $h_b(nP) \to \gamma\chi_{bJ}(mP)$ with $n > m$ are even more difficult to be measured. However, any observation of such transitions would be a clear hint at nontrivial coupled-channel effects due to light quarks [1387], and thus deserve to be searched for. The most promising processes are $h_b(2P) \to \gamma\chi_{bJ}(1P)$ and $\chi_{bJ}(2P) \to \gamma h_b(1P)$ [1387]. The former can be studied as part of the exclusive decay chain $\Upsilon(5S) \to \pi^+\pi^- h_b(2P) \to \pi^+\pi^-\gamma\chi_{bJ}(1P) \to \pi^+\pi^-\gamma\gamma\Upsilon(1S)$, with $\Upsilon(1S)$ reconstructed in di-muon or di-electron decay. We expect roughly $\mathcal{O}(10)$ reconstructed events/ab$^{-1}$ at the $\Upsilon(5S)$ energy, with a small or negligible background coming from radiative QED processes or other transitions among bottomonia. The latter transition can be studied using a large dataset collected at the $\Upsilon(3S)$ energy. In this case the analysis should be done inclusively, studying single and multi-photon recoil mass spectra in hadronic events, in search for either the $\Upsilon(3S) \to \gamma\chi_{bJ}(2P) \to \gamma\gamma h_b(1P)$ or the $\Upsilon(3S) \to \gamma\chi_{bJ}(2P) \to \gamma\gamma h_b(1P) \to \gamma\gamma\gamma\eta_b(1S)$ cascade decays.

Another class of radiative transitions which should become statistically within reach at Belle II are the $b\bar{b} \to \gamma c\bar{c}$ transitions. This process is predicted in NRQCD [1509] and is expected to be enhanced by the interference of QCD and QED amplitudes. Weak evidence of these transitions were observed at Belle, but the final comparison with theory predictions will be feasible only at Belle II.

*14.7.2. Dedicated runs above $B\bar{B}$ threshold at Belle II.* [R. Mizuk]

Above the $B\bar{B}$ threshold there are five hadrons containing the $b\bar{b}$ quarks. Properties of all of them are inconsistent with their structure being a pure $b\bar{b}$ pair. Unlike in the charmonium sector, there are clear winners in the interpretation of the bottomonium-like states. The vector states $\Upsilon(10580)$, $\Upsilon(10860)$ and $\Upsilon(11020)$ [or $\Upsilon(4S)$, $\Upsilon(5S)$ and $\Upsilon(6S)$] are likely mixtures of the $b\bar{b}$ pair and molecular component of $B\bar{B}$ or $B_s\bar{B}_s$ mesons in ground or excited states. The isospin-one states $Z_b(10610)$ and $Z_b(10650)$ likely have purely molecular structures of $B\bar{B}^*$ and $B^*\bar{B}^*$, respectively.

Anomalous properties of $\Upsilon(4S)$ and $\Upsilon(5S)$ were established somewhat occasionally using on-resonance data samples that were collected to study $B$ and $B_s$ mesons. Subsequently Belle performed a high statistics energy scan with the total luminosity of $27\,\text{fb}^{-1}$ that played a crucial role in interpreting the $\Upsilon(5S)$ and $\Upsilon(6S)$. Further studies at Belle II should proceed along the same lines: in addition to increased data samples at the $\Upsilon(4S)$ and $\Upsilon(5S)$, it is useful to perform an energy scan with improved statistics, and to collect data at the $\Upsilon(6S)$ and at higher mass states if they are found in the scan.

*Energy scan.* ***Coupled-channel analysis of exclusive cross sections.*** The total $b\bar{b}$ cross section above the $B\bar{B}$ threshold has several features: the $\Upsilon(4S)$, $\Upsilon(5S)$, and $\Upsilon(6S)$ peaks, and dips near the $B\bar{B}^*$, $B^*\bar{B}^*$ and $B_s^*\bar{B}_s{}^*$ thresholds [1237, 1510]. The exclusive cross sections for open flavour final states, such as $B^{(*)}\bar{B}^{(*)}$, $B^{(*)}\bar{B}^{(*)}\pi$ or $B_s^{(*)}\bar{B}_s^{(*)}$, that almost saturate the total $b\bar{b}$ cross section in this energy region are expected to have much more features. The Unitarised Quark Model [1511] predicts several maxima and zeros in each exclusive cross section, which are shifted in different final states, producing a relatively featureless sum. The oscillatory behaviour of exclusive cross sections is related to the nodes of the $\Upsilon(5S)$ and $\Upsilon(6S)$ wave functions. Exclusive cross sections contain complete information about the corresponding energy region. Combined coupled-channel analysis of all exclusive cross sections will allow to determine the pole positions of the $\Upsilon$ states, their electronic widths, and couplings to various channels. Thus, the ability of the currently favoured interpretation to describe the data will be put to test, and all relevant parameters will be determined.

Recently Belle measured the $B_s^{(*)}\bar{B}_s^{(*)}$ cross sections from corresponding thresholds up to 11.02 GeV [1236]. Results for $B\bar{B}$, $B\bar{B}^*$, $B^*\bar{B}^*$ and $B^{(*)}\bar{B}^{(*)}\pi$ are still expected. These measurements will provide the first attempt of a coupled-channel analysis.

***Search for new vector states.*** The final states with bottomonia, such as $\Upsilon(nS)\pi^+\pi^-$, $\Upsilon(nS)\eta$ or $h_b(nP)\pi^+\pi^-$, contribute only at a few percent to the total $b\bar{b}$ cross section. They can be used in the coupled-channel analysis and play an important role in searching for new states. A "smoking gun" of the compact tetraquarks and hadrobottomonia are suppressed decays to the open flavour channels. Thus, hidden flavour cross sections provide a unique way to search for such states. Even for molecular states, for which open flavour channels dominate, the channels with bottomonia could have higher sensitivity because they usually have higher reconstruction efficiency and no non-resonant continuum contribution.

Recently Belle measured the $\Upsilon(nS)\pi^+\pi^-$ $(n = 1, 2, 3)$ and $h_b(mP)\pi^+\pi^-$ $(m = 1, 2)$ cross sections [1177, 1237]. They exhibit clear $\Upsilon(5S)$ and $\Upsilon(6S)$ peaks. With the available statistics no new significant structures are found.

***Promising energy regions.*** The states with molecular admixture are naturally located near the corresponding threshold. The positions of the thresholds in the region above 11.0 GeV where no high-statistics data are available are listed in Table 131. We consider only pairs of narrow $S$- and $P$-wave mesons and baryons. The present energy limit of the SuperKEKB accelerator of 11.24 GeV will allow to investigate the $B^{(*)}\bar{B}^{**}$ and $B_s^{(*)}\bar{B}_s^{**}$ threshold regions. Increase of maximal energy by at least 100 MeV will allow to explore the $\Lambda_b\,\bar{\Lambda}_b$ threshold and to search for baryon-antibaryon molecular states. Presence of potentially interesting dynamics in a heavy baryon-antibaryon channel is strongly suggested by the data [1229] in the charmonium sector near the $\Lambda_c\bar{\Lambda}_c$ threshold. The region of promising thresholds extends up to 12 GeV. The energy region 11.5 – 11.6 GeV is of special importance to search for partners of $Z_b(10610)$ and $Z_b(10650)$, as discussed in the next section.

Table 131: Thresholds of narrow $S$ and $P$ wave mesons and baryons.

| Particles | Threshold, GeV/$c^2$ |
|---|---|
| $B^{(*)}\bar{B}^{**}$ | 11.00 – 11.07 |
| $B_s^{(*)}\bar{B}_s^{**}$ | 11.13 – 11.26 |
| $\Lambda_b\,\bar{\Lambda}_b$ | 11.24 |
| $B^{**}\bar{B}^{**}$ | 11.44 – 11.49 |
| $B_s^{**}\bar{B}_s^{**}$ | 11.48 – 11.68 |
| $\Lambda_b\,\bar{\Lambda}_b^{**}$ | 11.53 – 11.54 |
| $\Sigma_b^{(*)}\,\bar{\Sigma}_b^{(*)}$ | 11.62 – 11.67 |
| $\Lambda_b^{**}\,\bar{\Lambda}_b^{**}$ | 11.82 – 11.84 |

In the high statistics energy scan Belle collected about 1 fb$^{-1}$ per point, and the statistical uncertainty in measured cross sections was quite high [1177, 1236, 1237]. Thus at Belle II it is useful to collect about 10 fb$^{-1}$ per scan point. Since expected energy smearing at Belle II is similar to that at Belle – close to 5 MeV, no narrow peak will be missed if the step of the scan is 10 MeV.

*On resonance data at $\Upsilon(6S)$ and at higher mass states.* Once a new state is found it is of interest to collect about 500 fb$^{-1}$ at its peak. Searches for new transitions from vector states, searches for missing bottomonia and for molecular states in such transitions, and searches for excited $B$ and $B_s$ mesons are among the topics to be addressed with these data.

***Mechanism of hadronic transitions and structure of vector bottomonium-like states.*** In Section 14.5.3 it is shown that the rates of hadronic transitions are sensitive to the structure of the parent state. Final states that have already been seen at $\Upsilon(4S)$ or $\Upsilon(5S)$ include $\Upsilon(nS)\pi^+\pi^-$, $\Upsilon(nS)\eta$, $\Upsilon(1S)K^+K^-$, $h_b(nP)\pi^+\pi^-$, $h_b(1P)\eta$, $\chi_{bJ}(1P)\omega$, $\Upsilon_J(1D)\pi^+\pi^-$, $\Upsilon_J(1D)\eta$ and $Z_b\pi$ (see the full list in Ref. [1397]). Further final states to be investigated are given in Table 128. These studies will benefit greatly from increased $\Upsilon(4S)$ and $\Upsilon(5S)$ data samples at Belle II. The final states that have already been investigated at the $\Upsilon(6S)$ energy are limited to the most prominent channels $\Upsilon(nS)\pi^+\pi^-$ and $h_b(nP)\pi^+\pi^-$ because only a small amount of scan data with effective luminosity of $3\,\text{fb}^{-1}$ is available. It is of interest to compare the transitions from $\Upsilon(6S)$ and $\Upsilon(5S)$ since the two states are relatively close in energy and the differences should be due to their different structures. The comparison requires an increase of statistics at the $\Upsilon(6S)$ peak which is a good topic for initial data taking at Belle II (see Section 14.8).

***Search for missing conventional bottomonia below the $B\bar{B}$ threshold.*** The $121\,\text{fb}^{-1}$ data sample at the $\Upsilon(5S)$ was highly instrumental in finding missing bottomonium levels, *e.g.* first observation of $h_b(1P)$ and $h_b(2P)$, first evidence for $\eta_b(2S)$, first precise measurement of the $\eta_b(1S)$ mass, and first measurement of its width [1112, 1512]. Possible production channels for still missing bottomonium levels and their thresholds are shown in Table 132. Many thresholds are above the $\Upsilon(6S)$, which motivates investigation of higher mass region and increase of the SuperKEKB $e^+e^-$ collision energy. Various kinematic effects in the production of excited states are discussed in Ref. [1397].

Table 132: Missing bottomonium levels below the $B\bar{B}$ threshold, their quantum numbers, potential model predictions for masses [1247], light hadrons emitted in the transitions from vector bottomonium-like states to the considered bottomonia and thresholds of these transitions [1397].

| Name | $L$ | $S$ | $J^{PC}$ | Mass, MeV/$c^2$ | Emitted hadrons [Threshold, GeV/$c^2$] |
|---|---|---|---|---|---|
| $\eta_b(3S)$ | 0 | 0 | $0^{-+}$ | 10336 | $\omega$ [11.12], $\phi$ [11.36] |
| $h_b(3P)$ | 1 | 0 | $1^{+-}$ | 10541 | $\pi^+\pi^-$ [10.82], $\eta$ [11.09], $\eta'$ [11.50] |
| $\eta_{b2}(1D)$ | 2 | 0 | $2^{-+}$ | 10148 | $\omega$ [10.93], $\phi$ [11.17] |
| $\eta_{b2}(2D)$ | 2 | 0 | $2^{-+}$ | 10450 | $\omega$ [11.23], $\phi$ [11.47] |
| $\Upsilon_J(2D)$ | 2 | 1 | $(1,2,3)^{--}$ | 10441 – 10455 | $\pi^+\pi^-$ [10.73], $\eta$ [11.00], $\eta'$ [11.41] |
| $h_{b3}(1F)$ | 3 | 0 | $3^{+-}$ | 10355 | $\pi^+\pi^-$ [10.63], $\eta$ [10.90], $\eta'$ [11.31] |
| $\chi_{bJ}(1F)$ | 3 | 1 | $(2,3,4)^{++}$ | 10350 – 10358 | $\omega$ [11.14], $\phi$ [11.38] |
| $\eta_{b4}(1G)$ | 4 | 0 | $4^{-+}$ | 10530 | $\omega$ [11.31], $\phi$ [11.55] |
| $\Upsilon_J(1G)$ | 4 | 1 | $(3,4,5)^{--}$ | 10529 – 10532 | $\pi^+\pi^-$ [10.81], $\eta$ [11.08], $\eta'$ [11.49] |

The transitions listed in Table 132 can be reconstructed inclusively using missing mass of the emitted light hadrons. In case of the spin-triplet states there is also a possibility of exclusive reconstruction. The dominant transitions between the bottomonia below the $B\bar{B}$ threshold are radiative $E1$ transitions. Thus the chain $\Upsilon_J(1G) \to \gamma\chi_{bJ}(1F) \to \gamma\gamma\Upsilon_J(1D) \to \gamma\gamma\gamma\chi_{bJ}(1P) \to \gamma\gamma\gamma\gamma\Upsilon(1S)$ corresponds to dominant transitions and can be used for exclusive reconstruction, with $\Upsilon(1S) \to e^-e^-$ or $\mu^+\mu^-$. More details on the bottomonium decays can be found in *e.g.* Ref. [1247].

***Search for molecular states near $B^{(*)}_{(s)}\bar{B}^{(*)}_{(s)}$ thresholds*** The $Z_b(10610)$ and $Z_b(10650)$ states situated near the $B\bar{B}^*$ and $B^*\bar{B}^*$ thresholds, respectively, were observed in pionic transitions from the $\Upsilon(5S)$ and $\Upsilon(6S)$: $\Upsilon(nS) \to Z_b\pi$ (see Table 125). It is expected [1352] that there are other molecular states near the $B\bar{B}$, $B\bar{B}^*$ and $B^*\bar{B}^*$ thresholds (see Table 133), however there are difficulties with producing them at the $\Upsilon(5S)$ or $\Upsilon(6S)$. Indeed, all other than $Z_b^{(\prime)}$ isovector states have negative $G$-parity, therefore they can not be produced with the emission of a single pion, but require emission of a $\rho$ meson. Production of isosinglet states requires emission of $\eta$ or $\omega$ mesons. All these processes have higher thresholds that are in the range $11.15 - 11.43\,\text{GeV}$ (see Table 133). To search for them a new vector state has to be found above this range, and thus the SuperKEKB energy would need to be increased up to $11.5 - 11.6\,\text{GeV}$. Most of the states can be reached also via radiative transitions that have much lower thresholds, however, corresponding rates carry the suppression factor of $\alpha_{\rm QED}$.

Table 133: Expected molecular states with the structure $B\bar{B}$, $B\bar{B}^*$ and $B^*\bar{B}^*$ [1397].

| $I^G(J^P)$ | Name | Content | Co-produced particles [Threshold, GeV/$c^2$] | Decay channels |
|---|---|---|---|---|
| $1^+(1^+)$ | $Z_b$ | $B\bar{B}^*$ | $\pi$ [10.75] | $\Upsilon(nS)\pi$, $h_b(nP)\pi$, $\eta_b(nS)\rho$ |
| $1^+(1^+)$ | $Z_b'$ | $B^*\bar{B}^*$ | $\pi$ [10.79] | $\Upsilon(nS)\pi$, $h_b(nP)\pi$, $\eta_b(nS)\rho$ |
| $1^-(0^+)$ | $W_{b0}$ | $B\bar{B}$ | $\rho$ [11.34], $\gamma$ [10.56] | $\Upsilon(nS)\rho$, $\eta_b(nS)\pi$ |
| $1^-(0^+)$ | $W_{b0}'$ | $B^*\bar{B}^*$ | $\rho$ [11.43], $\gamma$ [10.65] | $\Upsilon(nS)\rho$, $\eta_b(nS)\pi$ |
| $1^-(1^+)$ | $W_{b1}$ | $B\bar{B}^*$ | $\rho$ [11.38], $\gamma$ [10.61] | $\Upsilon(nS)\rho$ |
| $1^-(2^+)$ | $W_{b2}$ | $B^*\bar{B}^*$ | $\rho$ [11.43], $\gamma$ [10.65] | $\Upsilon(nS)\rho$ |
| $0^-(1^+)$ | $X_{b1}$ | $B\bar{B}^*$ | $\eta$ [11.15] | $\Upsilon(nS)\eta$, $\eta_b(nS)\omega$ |
| $0^-(1^+)$ | $X_{b1}'$ | $B^*\bar{B}^*$ | $\eta$ [11.20] | $\Upsilon(nS)\eta$, $\eta_b(nS)\omega$ |
| $0^+(0^+)$ | $X_{b0}$ | $B\bar{B}$ | $\omega$ [11.34], $\gamma$ [10.56] | $\Upsilon(nS)\omega$, $\chi_{bJ}(nP)\pi^+\pi^-$, $\eta_b(nS)\eta$ |
| $0^+(0^+)$ | $X_{b0}'$ | $B^*\bar{B}^*$ | $\omega$ [11.43], $\gamma$ [10.65] | $\Upsilon(nS)\omega$, $\chi_{bJ}(nP)\pi^+\pi^-$, $\eta_b(nS)\eta$ |
| $0^+(1^+)$ | $X_b$ | $B\bar{B}^*$ | $\omega$ [11.39], $\gamma$ [10.61] | $\Upsilon(nS)\omega$, $\chi_{bJ}(nP)\pi^+\pi^-$ |
| $0^+(2^+)$ | $X_{b2}$ | $B^*\bar{B}^*$ | $\omega$ [11.43], $\gamma$ [10.65] | $\Upsilon(nS)\omega$, $\chi_{bJ}(nP)\pi^+\pi^-$ |

The state $X_b$ near the $B\bar{B}^*$ threshold (see Table 133) would be a bottom partner of the $X(3872)$. However, the properties of the $X_b$ are expected to differ vastly from the $X(3872)$: it should be dominantly an isoscalar state and the decays into $\Upsilon(nS)\pi\pi$, breaking isospin symmetry, should be strongly suppressed [1354]. Therefore, it should be searched for in the final states such as $\Upsilon(nS)\pi\pi\pi$, $\chi_{bJ}\pi\pi$ and $\Upsilon(nS)\gamma$ [1295, 1354]. A search for the $\Upsilon(1S)\omega$ final state using the Belle data set at $\sqrt{s} = 10.867$ GeV/$c^2$ was negative [1235].

One can search also for the $Z_{bs}$ states with the structure of $B_s^*\bar{B}$, $B_s\bar{B}^*$ and $B_s^*\bar{B}^*$ [1397]. They can be produced in association with a kaon: $e^-e^- \to Z_{bs}K$, at energies above 11.20 GeV. These resonances would decay into the states of bottomonium plus a kaon, and also to heavy meson pairs with one $B$ meson being either $B_s$ or $B_s^*$.

***Spectroscopy of $B$ and $B_s$ mesons.*** Taking data at high energies potentially gives access to excited $B$ and $B_s$ mesons. The missing $B_s$ mesons with $J^P = 0^+$ and $1^+$ are expected not far below $BK$ and $B^*K$ thresholds, respectively. The effect of tresholds was taken into account in the lattice study [1513], where the bound-state poles in the scattering matrices of $BK$ and $B^*K$ were found below thresholds. The predicted masses are $m_{B_{s0}} = 5.711(13)(19)$ GeV and $m_{B_{s1}} = 5.750(17)(19)$ GeV for $J^P = 0^+$ and $1^+$ states, respectively. Therefore they should be narrow, similarly to the $D_{sJ}$ case. They could be produced near the $B_s^{(*)}\bar{B}_s^{**}$ thresholds [1514] that are within the current reach of SuperKEKB.

*$D_s$ and $B_s$ excited states .* The scalar $B_{s0}$ and axial $B_{s1}$ states have not been found experimentally yet and Belle II has potential to search for them. They are expected not far below $BK$ and $B^*K$ thresholds. The analogous study of $D_{s0}(2317)$ rendered it 37(17) MeV below $DK$, while $D_{s1}(2460)$ was found 44(10) MeV below $DK$ [1515, 1516], both in agreement with experiment.

***Conclusion*** Dedicated data taking is needed to establish the current interpretation of known bottomonium-like states, check its predictions, and search for new states. An energy scan from the $B\bar{B}$ threshold up to the highest possible energy with $\sim 10$ fb$^{-1}$ per point and $\sim 10$ MeV steps is of high interest. Measured exclusive open- and hidden-flavour cross sections will be used in coupled-channel analyses to establish the nature of the vector states and in the searches for new states. At each new vector state it is useful to collect $\sim 500$ fb$^{-1}$ to perform a detailed study of corresponding transitions, to search for missing conventional bottomonia, excited $B_{(s)}$ mesons and molecular states – partners of $X(3872)$, $Z_b(10610)$ and $Z_b(10650)$. It is of high interest to increase the maximal energy of the SuperKEKB collider, which is currently 11.24 GeV. The increase to about 11.35 GeV will allow to cover the $\Lambda_b\bar{\Lambda}_b$ threshold region. Further increase to 11.5 – 11.6 GeV is of particular importance for searches of molecular states – partners of $X(3872)$ and $Z_b$. Studies of the whole resonance region require an increase up to 12 GeV. More detailed discussion of this subject can be found in Refs. [1248, 1397].

## *14.8. Early Physics Program at Belle II*

[B. Fulsom]

The Belle II Experiment is scheduled to begin its first “physics” run in early 2019. As a prelude to this, there were two commissioning periods known as “Phase 1” (early 2016) and “Phase 2” (early 2018) where a varied collection of smaller detectors are deployed for measuring background rates and operating conditions. During Phase 1, beams were circulated, but the solenoid was inactive, no collisions took place, and the Belle II detector was not yet installed. For Phase 2 all detector subsystems except for the vertex detector were fully deployed to study colliding beam events. A total of 0.5 fb$^{-1}$ of collision data was collected for commissioning. The first physics run (“Phase 3”) in early 2019 will involve the entire Belle II detector, with the machine expected to operate with an instantaneous luminosity of at least $1 \times 10^{35}$ cm$^{-2}$ s$^{-1}$. In addition to data collected at the nominal $\Upsilon(4S)$ energy for commissioning purposes, data collected at different centre-of-mass energies during Phase 3 represent an important opportunity for the Belle II experiment to have an early scientific impact. These opportunities largely lie in the realm of quarkonium and “new states” physics, as described previously in this chapter.

*14.8.1. Potential operating points.* Table 134 summarises recent data collected at the $\Upsilon$ resonances. Since existing $\Upsilon(4S)$ and $\Upsilon(5S)$ datasets cannot be matched during the early periods, this leaves the possibility for quick acquisition of uniquely large samples at $\Upsilon(1S)$, $\Upsilon(2S)$, $\Upsilon(3S)$, $\Upsilon(6S)$, off-resonance, and $E_{\rm CM}$ scan points if sufficiently justified. One of the primary drivers of the physics will be the amount of integrated luminosity available during these early periods.

Based on the expected operating conditions and physics prospects, collecting data above the $\Upsilon(4S)$ offers the best physics opportunities during the early stages of the experiment.

Table 134: Existing $\Upsilon$-related datasets.

| Experiment | Scans Off. Res. | $\Upsilon(6S)$ fb$^{-1}$ | $\Upsilon(5S)$ fb$^{-1}$ | $\Upsilon(5S)$ $10^6$ | $\Upsilon(4S)$ fb$^{-1}$ | $\Upsilon(4S)$ $10^6$ | $\Upsilon(3S)$ fb$^{-1}$ | $\Upsilon(3S)$ $10^6$ | $\Upsilon(2S)$ fb$^{-1}$ | $\Upsilon(2S)$ $10^6$ | $\Upsilon(1S)$ fb$^{-1}$ | $\Upsilon(1S)$ $10^6$ |
|---|---|---|---|---|---|---|---|---|---|---|---|---|
| CLEO | 17.1 | - | 0.1 | 0.4 | 16 | 17.1 | 1.2 | 5 | 1.2 | 10 | 1.2 | 21 |
| BaBar | 54 | $R_b$ scan | | | 433 | 471 | 30 | 122 | 14 | 99 | – | |
| Belle | 100 | $\sim 5.5$ | 36 | 121 | 711 | 772 | 3 | 12 | 25 | 158 | 6 | 102 |

The $\Upsilon(6S)$ energy region ($\sim$ 11020 MeV) is particularly interesting, both because only $<$ 5.6fb$^{-1}$ of data have been collected there previously, and also because of the discoveries of multiquark $Z_b$ states in its midst [1177]. At $\sim$ 11 GeV, 20 fb$^{-1}$ could be used to understand $\Upsilon(6S) \to \pi^{\pm} Z_b^{\mp}$ decays, *e.g.* the relative production of $m_{Z_b} = 10610$ MeV/$c^2$ versus 10650 MeV/$c^2$, in decays to $h_b(1P, 2P)$ and $\Upsilon(1S, 2S, 3S)$. It may also be possible to search for $Z_b$ partners [1239] in decays $\Upsilon(6S) \to \gamma W_b$ and $\Upsilon(6S) \to \pi^+\pi^- W_b$, and in analogy to $\Upsilon(5S)$ decays, study bottomonium transitions with sufficient phase space for $h_b(3P)$, $\Upsilon(2D)$, and F-wave discovery, as discussed in the previous section. Other Belle results for decays to $\pi\pi\Upsilon$ may point to cross section enhancements indicative of these intermediate states for energies in the range of 10.7 - 10.8 GeV, where only $\sim$ 2fb$^{-1}$ of data have been collected [1237].

Energies below $\Upsilon(4S)$ are useful for both the study of bottomonium states and their transitions, and physics Beyond the Standard Model in searches for the dark sector and light Higgs. Datasets in the $>$ 200 fb$^{-1}$ range during Phase 3 offer a chance to reach this type of physics from the $\Upsilon(3S)$. Another strategy could be an $E_{cm}$ scan of the expected $\Upsilon(1^3D_1)$ and $\Upsilon(2^3D_1)$ mass regions (10160 MeV and 10450 MeV, respectively) to discover these states directly in $e^+e^-$ collisions.

*14.8.2. Operating conditions.* The majority of Phase 2 focused on accelerator commissioning, ultimately reaching an instantaneous luminosity of $\sim 0.5 \times 10^{34}$ cm$^{-2}$s$^{-1}$. A total of 500 pb$^{-1}$ of data was collected. It was found that the beam energy spread was near the expected value of $\sim$ 5 MeV, and this is promising for physics studies in phase 3. The first full physics run of Belle II is expected to be with nominal operating conditions. During this "Phase 3", data will be collected at $\Upsilon(4S)$, with options for exploring other energy values once a suitable $B\overline{B}$ sample has been collected for validation, commissioning, and other early physics studies.

## *14.9. Action Items*

Experiment:

- It is important to perform an energy scan from the $B\bar{B}$ threshold up to the highest possible energy with about 10 fb$^{-1}$ per point, and to measure energy dependence of exclusive open flavour ($B\bar{B}$, $B\bar{B}^*$, $B^*\bar{B}^*$, $B^{(*)}\bar{B}^{(*)}\pi$, $B_s\bar{B}_s$ etc) and hidden flavour ($\Upsilon(nS)\pi^+\pi^-$, $h_b(nP)\pi^+\pi^-$, $\Upsilon(nS)\eta$, etc.) cross sections. This information is crucial for understanding vector bottomonium-like states.
- Collect data at $\Upsilon(6S)$ and at any new peak observed in the energy scan. These data will allow investigation of the decay mechanism of bottomonium-like states, search for missing conventional bottomonia, predicted bottomonium-like states and missing $P$-wave excitations of $B_s$ mesons.

- Maximal energy of SuperKEKB is expected to be 11.24 GeV. The region above this energy is previously unexplored and it is of paramount importance to increase the energy. There is a $\Lambda_b \bar{\Lambda}_b$ threshold at 11.24 GeV with potentially interesting baryon-antibaryon dynamics and more promising thresholds all the way up to 12 GeV. Transitions from new vector states provide possibly a unique way to produce partners of the $X(3872)$, $Z_b(10610)$ and $Z_b(10650)$. Most of relevant transitions are kinematically allowed if the mass of the vector state is above 11.5 GeV.
- Spin and parity of the quarkonium-like state are very important to discriminate various interpretations. For many states this information is missing. One should perform full amplitude analyses of the corresponding production processes to measure $J^P$.
- All quarkonium-like states are above the open flavour thresholds, and decay pattern into open flavour channels and corresponding line shapes are crucial for understanding of them. One should systematically search for open flavour decays of all quarkonium-like states.

Theory:

- Action items for phenomenological approaches
  - Within all approaches to QCD exotics predictions should be provided for states with quantum numbers not yet observed. This could be done, *e.g.*, employing the heavy quark spin symmetry. At Belle those could be searched for in the decay chains of heavy vector states.
  - For all predicted states quantitative statements about partial decay widths or at least branching ratios should be provided, not only allowing one to identify potential discovery modes but also as stringent test of the assumed dynamics.
  - Predictions also for the bottom sector are necessary again for various quantum numbers.
  - The mixing of exotic states with regular quarkonia needs to be investigated.
- Action items for lattice QCD
  - Calculate scattering matrices for $D_{(s)}^{(*)} \bar{D}_{(s)}^{(*)}$ and for $charmonium\ +\ light{-}hadron$ with non-static heavy quarks. First in the one-channel approximation and then taking as many coupled channels as possible. Determine the position of poles in the scattering matrix and try to connect poles with the experimentally observed states.
  - Approach bottomonium states close to $B^{(*)} \bar{B}^{(*)}$ threshold with non-static b-quarks and make an effort to take into account the effect of this threshold.
  - Determine yet undetermined Born-Oppenheimer potentials for static heavy quarks using lattice QCD, for example those related to $Z_b$.
  - Calculate yet undetermined radiative transitions between quarkonia below threshold. Try to make a step towards a rigorous treatment of this problem also for states above open-charm threshold.
  - Consider effects of $\bar{Q}Q$ annihilation in lattice simulations.
  - Consider effects of isospin breaking in lattice QCD for channels where it might be important, for example $X(3872)$.

### *14.10. Conclusions*

Since the turn of the century a large number of states that can not be explained by the until then very successful quark model was discovered experimentally. These discoveries lead to a renaissance of hadron spectroscopy both with respect to experimental as well as theoretical activities. On the theory side there are important developments in three branches: Effective field theories, both based on quark-gluon dynamics nested in QCD as well as based on hadron-hadron dynamics, model building and lattice QCD studies.

There are firm theoretical predictions for the spectrum of quarkonia below open-flavour threshold by means of lattice QCD and NRQCD, for example. All such charmonia have already been discovered experimentally and they agree with calculated masses well. The bottomonium spectrum below $\bar{B}B$ contains many more states and some of the predicted ones have not been discovered yet. The predicted spectra of higly-excited charmonia should serve as a valuable guidline for experimental searches, although most of these lattice simulations ignore strong decays of these states. Meanwhile, first lattice attempts have been made to treat quarkonia above open-flavour thresholds as strongly decaying resonances.

Most of the discovered exotic hadrons lie near some threshold and they are strongly decaying states. Suggested structures for the new states are hybrids ($\bar{Q}Q$ supplemented with an active gluon degree of freedom), tetraquarks (bound systems of heavy diquarks and anti-diquarks), hadroquarkonia (compact $\bar{Q}Q$ cores surrounded by a light quark cloud) and hadronic molecules (bound states of colour-neutral hadrons in analogy to nuclei). As of today there is no consensus yet achieved within the community which of those structures is the most relevant — it is even neither excluded that there are groups of different nature, nor that contributions of all kinds are significant for various states simultaneously. Currently most effort goes into generating more predictions within the different pictures individually but mixing scenarios need to be necessarily on the agenda in the not too far future both amongst exotic structures as well as between exotics and $\bar{Q}Q$ states.

In the chapter it was demonstrated that different assumed structures for each exotic state lead to different predictions for decay branching ratios. Moreover the location of spin partner states that necessarily exist as a consequence of heavy-quark spin symmetry is known to be sensitive to the intrinsic dynamics of the states. Therefore there is a lot of work waiting for Belle II, not only to complete our experimental knowledge of new states in the charm sector but especially to map out exotics in the bottom sector where so far only two exotic states are identified unambiguously.

The experimental studies at Belle II should proceed along the same lines as at Belle: search for missing quarkonia and for expected partners of exotic states, search for new decay channels of known states, and detailed measurement of all accessible properties, including spin-parities, absolute branching fractions, line-shapes, and so on. All this should be possible given the expected significant increase in luminosity at Belle II. With only modest additional effort and time dedicated to operating at energies other than the $\Upsilon(4S)$ resonance, it is possible to make important scientific gains in this area. More detailed experimental information will help to resolve many puzzles currently present in the field of heavy quarkonia. The high precision of recent and especially future data require analysis employing sound theoretical tools. This appears to be realised most efficiently within close collaborations of experimenters and theoreticians.

# 15. Tau and low multiplicity physics

*Editors: T. Ferber, K. Hayasaka, E. Passemar, J. Hisano*
*Additional section writers: H. Aihara, I. Bigi, V. Braun, G. Colangelo, H. Czyz, S. Eidelman, D. Epifanov, M. Hoferichter, M. Jamin, T. Konno, E. Kou, K. Maltman, B. Moussallam, D. Nomura, N. Offen, A. Pich, M. Procura, P. Roig, M. Roney, J. Sasaki, N. Shimizu, Y. Shimizu, B. Shwartz, P. Stoffer, F. Tenchini, T. Teubner, S. Uehara, Z. Was*

## *15.1. Introduction*

The enormous amount of $e^+e^-$ collisions that are expected from the Belle II experiment features a unique environment for electroweak and QED studies: about 45 billion of both $\tau\tau$ and $\mu\mu$ pairs are expected in the full dataset. The Belle II experiment will therefore offer fantastic possibilities to study $\tau$ physics and low multiplicity final states with high precision.

The $\tau$ lepton is an extremely convenient probe to search for new physics (NP) beyond the Standard Model (BSM) because of the well-understood mechanisms that govern its production and decay in electroweak interactions. With its large mass, it is the only lepton that can decay into hadrons, thus providing a clean laboratory to study QCD effects in the 1 GeV energy region.

The Belle II experiment will be well suited to study $\tau$ physics, in fact since the decays of $\tau$ leptons involve neutrinos in the final state their study is very difficult at hadron colliders such as LHC.

Two experiments at $e^+e^-$ colliders capable of producing tau leptons, BES III at IHEP, Beijing and KEDR at BINP, Novosibirsk, are statistically limited with respect to Belle II and therefore have a $\tau$ physics program basically limited to measuring its mass. Note that there are a few proposals of tau-charm factories for the future. $\tau$ decays offer a whole range of possible studies, from understanding strong interactions, to precise tests of electroweak interactions, and potential discoveries of New Physics with lepton flavour violation (LFV) and lepton universality violation.

Non–$\tau$ physics such as initial-state radiation (ISR), two–photon physics, and dark sector searches will profit from both the significantly larger statistics compared to Belle or BaBar and also from triggers specifically designed to collect data for these analyses.

## *15.2. Charged Lepton Flavour Violation in $\tau$ decays*

*(Contributing author: K. Hayasaka, H. Hisano, T. Konno, E. Passemar, Y. Shimizu, F. Tenchini)*

In the Standard Model, the presence of only left-handed neutrinos implies that lepton flavour is conserved and that neutrinos are massless. From the experimental observation of neutrino oscillations, we now know that lepton flavour is violated in the neutrino sector. However, this alone does not necessarily mean that charged lepton flavour is violated and that charged LFV processes will be observed in near future experiments. Even if we extend the Standard Model to include neutrino masses, generated by the Higgs mechanism, these processes are suppressed by the fourth power of the neutrino masses such that their branching ratios are too small to be observed [1517, 1518] (*e.g.* $< 10^{-54}$ for $\mu \to e\gamma$ and $< 10^{-53}$ for $\tau \to \mu\gamma$, with the latest averages of the measured neutrino mass and mixing).

Lepton flavour symmetries are not exact in nature, rather they are only accidental. Many BSMs at the TeV scale predict charged LFV interactions at a level reachable in the near future experiments: such as supersymmetric standard models [1519–1523], little Higgs models [1524], low-scale seesaw models [1525], leptoquark models [1526], $Z'$ models [1527], and extended Higgs models [1528–1532].

Let us consider charged LFV transitions in tau lepton decays. Stringent bounds already exist for $\mu$-$e$ transitions, for example the latest result of the MEG experiment is $\mathrm{Br}(\mu^+ \to e^+\gamma) < 4.2 \times 10^{-13}$ (90% CL) [1533], and gives strong constraints on BSMs. On the other hand, the bounds on the $\tau$-$\mu$ or $\tau$-$e$ transitions are much weaker. Some new physics scenarios, such as the SUSY seesaw model [1519], may have enhanced LFV couplings for tau leptons. Moreover, the CMS hint of Higgs LFV coupling suggested a $\tau$-$\mu$ coupling at the 1% level [1534] which triggered many new theoretical activities. While it seems that the anomaly was not confirmed by more recent measurements at CMS and ATLAS [1535, 1536], models that could explain such an anomaly have been presented. They typically show very interesting correlations between $H \to \mu\tau$ and LFV tau lepton decays. Constraining LFV from tau decays offers therefore a very interesting complementarity with energy frontier collider constraints.

Studying LFV processes in tau decays offers several advantages compared to muon decays. Since the tau lepton is much heavier than the muon, many more types of LFV processes can be studied: $\tau \to \mu/e + \gamma$ and $\tau \to \mu/e + l^+l^-$ $(l = \mu/e)$, the counterparts of $\mu \to e\gamma$ and $\mu \to 3e$, respectively. In addition, the tau lepton has semileptonic LFV channels whose final states have one or two mesons (even more) of isospin zero or one. These final states allow us to test the LFV couplings between quarks and leptons. If charged LFV is discovered, we can identify fundamental LFV interactions by matching the pattern of the branching ratios to the predictions in BSMs. Furthermore, tau leptons can have more exotic LFV decay processes, such as $\tau^+ \to \mu^- e^+ e^+$ (all lepton flavour symmetries are not conserved) and $\tau \to \Lambda\pi^-$ (baryon number is not conserved).

We choose $\tau \to 3\mu$ and $\tau \to \mu\gamma$ as the golden modes for studying charged LFV. Firstly the $\tau \to 3\mu$ channel, with its purely leptonic final state, is expected to be free of background. This allows us to scale the experimental uncertainties linearly with the luminosity. Thus, we naturally expect at least a 50 times increase in discovery potential at SuperKEKB. We will briefly discuss the Belle II prospects as well as a comparison to the LHCb experiment at the end of the following section.

The $\tau \to \mu\gamma$ decay, has the largest LFV branching ratio in models where the LFV processes are induced by one-loop diagrams including heavy particles, such as in supersymmetric models. For example, if $\tau \to 3\mu$ is induced by photon-penguin diagrams, the ratio of $\mathrm{Br}(\tau \to 3\mu)$ and $\mathrm{Br}(\tau \to \mu\gamma)$ is $2.2 \times 10^{-2}$. However, a search for $\tau \to \mu\gamma$ may suffer from background due to $\tau \to \mu\nu\bar{\nu}$ with radiated photons, or radiative di-muon events, such that the scaling of the sensitivity is non-trivial. At Belle II, the higher beam background will make the search more difficult but at the same time, its high luminosity will allow us to impose a more stringent experimental criteria compared to Belle. In the following section, we show the results of a Belle II sensitivity study.

We should also note the complementarity of semileptonic LFV transitions in $\tau$ or $B$ decays, such as $\tau \to \mu h$ ($h$ being hadrons), $B \to K^{(*)}\tau\mu$ and $\tau \to 3\mu$ channels. If the LFV processes are induced by a tree-level exchange of Higgs bosons or $Z'$ bosons, the branching ratio of $\tau \to 3\mu$ may provide information for the normalisation for the LFV couplings, and the ratios between the branching ratios of $\tau \to 3\mu$ and the semileptonic LFV processes allow us to discriminate between models. If $\tau \to 3\mu$ is not discovered, while the semileptonic LFV processes are, that would give us the indication that LFV couplings are generated from more exotic models such as those that contain leptoquarks.

### *15.2.1. Theory.* [57]

Table 135: Relations between LFV tau decay modes and effective operators. Here, $I$ stands for isospin of the final states. Table adapted from [124].

| | $\tau \to \mu\gamma$ | $\tau \to 3\mu$ | $\tau \to \mu\pi^+\pi^-$ | $\tau \to \mu K\bar{K}$ | $\tau \to \mu\pi$ | $\tau \to \mu\eta^{(\prime)}$ |
|---|---|---|---|---|---|---|
| $C_{DL,R}$ | $\surd$ | $\surd$ | $\surd$ | $\surd$ | – | – |
| $C_{SLL,RR}$ | – | $\surd$ | – | – | – | – |
| $C_{VLL,RR}$ | – | $\surd$ | – | – | – | – |
| $C_{VLR,RL}$ | – | $\surd$ | – | – | – | – |
| $C^q_{VL,R}$ | – | – | $\surd$ $(I=1)$ | $\surd$ $(I=0,1)$ | – | – |
| $C^q_{SL,R}$ | – | – | $\surd$ $(I=0)$ | $\surd$ $(I=0,1)$ | – | – |
| $C_{GL,R}$ | – | – | $\surd$ | $\surd$ | – | – |
| $C^q_{AL,R}$ | – | – | – | – | $\surd$ $(I=1)$ | $\surd$ $(I=0)$ |
| $C^q_{PL,R}$ | – | – | – | – | $\surd$ $(I=1)$ | $\surd$ $(I=0)$ |
| $C_{\tilde{G}L,R}$ | – | – | – | – | – | $\surd$ |

The low-energy effective operators for a LFV $\tau$–$\mu$ transition are the following (those for a LFV $\tau$–$e$ transition are derived by exchanging $\mu$ and $e$) [124],

$$\mathcal{L}_{eff} = \mathcal{L}^{(D)}_{eff} + \mathcal{L}^{(4l)}_{eff} + \mathcal{L}^{(lq)}_{eff} + \mathcal{L}^{(G)}_{eff} + \cdots ,$$

[57] J. Hisano, E. Passemar and Y. Shimizu

where

$$
\begin{aligned}
\mathcal{L}_{eff}^{(D)} &= -\frac{m_\tau}{\Lambda^2}\left[\left(C_{DL}\bar{\mu}\sigma^{\rho\sigma}P_L\tau + C_{DR}\bar{\mu}\sigma^{\rho\sigma}P_R\tau\right)F_{\rho\sigma} + h.c.\right], \\
\mathcal{L}_{eff}^{(4l)} &= -\frac{1}{\Lambda^2}\left[C_{SLL}(\bar{\mu}P_L\tau)(\bar{\mu}P_L\mu) + C_{SRR}(\bar{\mu}P_R\tau)(\bar{\mu}P_R\mu) + C_{VLL}(\bar{\mu}\gamma^\mu P_L\tau)(\bar{\mu}\gamma_\mu P_L\mu)\right. \\
&\quad + C_{VRR}(\bar{\mu}\gamma^\mu P_R\tau)(\bar{\mu}\gamma_\mu P_R\mu) + C_{VRL}(\bar{\mu}\gamma^\mu P_R\tau)(\bar{\mu}\gamma_\mu P_L\mu) \\
&\quad \left. + C_{VLR}(\bar{\mu}\gamma^\mu P_L\tau)(\bar{\mu}\gamma_\mu P_R\mu) + h.c.\right], \\
\mathcal{L}_{eff}^{(lq)} &= -\frac{1}{\Lambda^2}\sum_{q=u,d,s}\left[\left(C_{VL}^q\bar{\mu}\gamma^\rho P_L\tau + C_{VR}^q\bar{\mu}\gamma^\rho P_R\tau\right)\bar{q}\gamma_\rho q\right. \\
&\quad + \left(C_{AL}^q\bar{\mu}\gamma^\rho P_L\tau + C_{AR}^q\bar{\mu}\gamma^\rho P_R\tau\right)\bar{q}\gamma_\rho\gamma_5 q + G_F m_\tau m_q\left(C_{SL}^q\bar{\mu}P_R\tau + C_{SR}^q\bar{\mu}P_L\tau\right)\bar{q}q \\
&\quad + G_F m_\tau m_q\left(C_{PL}^q\bar{\mu}P_R\tau + C_{PR}^q\bar{\mu}P_L\tau\right)\bar{q}\gamma_5 q \\
&\quad \left. + G_F m_\tau m_q\left(C_{TL}^q\bar{\mu}\sigma^{\rho\sigma}P_R\tau + C_{TR}^q\bar{\mu}\sigma^{\rho\sigma}P_L\tau\right)\bar{q}\sigma_{\rho\sigma}q + h.c.\right], \\
\mathcal{L}_{eff}^{(G)} &= \frac{G_F m_\tau}{\Lambda^2}\frac{9\alpha_s}{8\pi}\left[\left(C_{GL}\bar{\mu}P_R\tau + C_{GR}\bar{\mu}P_L\tau\right)G_{\rho\sigma}^a G^{a\rho\sigma}\right. \\
&\quad \left. + \left(C_{\tilde{G}L}\bar{\mu}P_R\tau + C_{\tilde{G}R}\bar{\mu}P_L\tau\right)G_{\rho\sigma}^a\tilde{G}^{a\rho\sigma} + h.c.\right].
\end{aligned}
\tag{503}
$$

Here, $F_{\rho\sigma}$ and $G_{\rho\sigma}$ are field-strength tensors for photons and gluons, respectively. For simplicity we show pure leptonic four-Fermi operators for $\tau \to 3\mu$ in $\mathcal{L}_{eff}^{(4l)}$, though the inclusion of those for $\tau^- \to \mu^- e^+ e^-$ is straightforward.

The Wilson coefficients in Eq. (503) depend on the choice of UV model. The LFV dipole operators in $\mathcal{L}_{eff}^{(D)}$ are induced by loop diagrams in the renormalisable models. In the SUSY SMs, slepton mass terms are sources of lepton-flavour violation, and the integration of a SUSY particle generates the LFV dipole operators via the loop diagrams. The LFV dipole operators induce $\tau \to \mu/e\gamma$ and other LFV processes via the penguin diagrams. The Higgs or $Z'$ boson can generate LFV four-Fermi operators in $\mathcal{L}_{eff}^{(4l)}$ and $\mathcal{L}_{eff}^{(lq)}$ if they have LFV interactions. In Eq. (503) heavy quarks (top, bottom and charm quarks) are integrated out. The gluonic operators in $\mathcal{L}_{eff}^{(G)}$ come from loop diagrams of those heavy quarks or unknown coloured particles if they have LFV scalar or pseudoscalar couplings.

Table 135 shows relations between LFV tau decay modes and effective operators. (The branching ratios for the LFV tau lepton decay modes are summarised as functions of the Wilson coefficients in [124].) When the LFV dipole operators are non-vanishing, the pure leptonic and semileptonic LFV tau lepton decay modes are predicted to proceed via penguin diagrams, in addition to $\tau \to \mu\gamma$. Their branching ratios are suppressed by $\alpha$ compared with $Br(\tau \to \mu\gamma)$, such as $Br(\tau^- \to \mu^-\mu^+\mu^-)/Br(\tau^- \to \mu^-\gamma) \simeq 2.2 \times 10^{-2}$, as mentioned above.

The specific decay modes of semileptonic LFV tau decays depend on the Wilson coefficients of the gluon and quark LFV operators. If the coefficients of the four-Fermi operators correspond to vector or scalar couplings, the final states will include at least two mesons. If the coefficients correspond to a pseudoscalar or axial-vector coupling, the final states may include only one meson. The final states of semileptonic LFV tau lepton decays will have isospin zero or one, depending on the Wilson coefficients. On the other hand, purely leptonic tau lepton decays, such as $\tau \to 3\mu$, are induced by any type of purely leptonic four-Fermi operators. Thus, from the branching ratios of the LFV tau lepton decay, we can identify the Wilson coefficients and the UV physics. If the golden mode $\tau \to 3\mu$ is discovered, it will provide the normalisation for the LFV couplings, and we can discriminate models by comparing it with the branching ratios of other decay types.

We will discuss some predictions for the LFV tau lepton decay modes, assuming the SUSY extensions of the SM as benchmark models. In the SUSY SMs, the SUSY breaking mass terms for the left-handed and/or right-handed sleptons may be lepton-flavour violating such that the LFV dipole operators induce $\tau \to \mu/e\gamma$ and other LFV processes via penguin diagrams. In Fig. 177 we show the branching ratio for $\tau \to \mu\gamma$ by introducing the left-handed (right-handed) stau and smuon mixing mass term with black (red) dots. The branching ratio for the golden mode $\tau \to 3\mu$ is $2.2 \times 10^{-2}$ times smaller than that for $\tau \to \mu\gamma$. Here the bino, wino, and Higgsino masses are 250 GeV, 500 GeV, and 1 TeV, respectively, while $\tan\beta$ is 30. In the case of left-handed (right-handed) slepton mixing, the smuon and stau masses are taken from 500 GeV to 2 TeV, while right-handed (left-handed) sleptons masses are 5 TeV. We exclude the points where the sleptons are lighter than the second lightest neutralino or the light chargino, since the LHC results give constraints on these points. The left-handed sleptons interact as $\mathrm{SU}(2)_L$, and the branching ratio is larger when the left-handed sleptons have non-zero mixing mass terms. It is found that the SUSY SMs could even exceed the current experimental bounds on the branching ratios.

In Fig. 178 we show the branching ratios of $\tau \to \mu\gamma$ and $\tau \to e\gamma$ in the SUSY seesaw model. Here the CMSSM boundary conditions for the SUSY breaking parameters are assumed, that is, $0.5 < m_0$, $m_{1/2} < 10$ TeV, $|A_0| < 3$, $\mu > 0$, while $\tan\beta = 30$. The mixing mass terms for left-handed sleptons are generated by renormalisation group effects. The SUSY seesaw model is reconstructed from the observed oscillation parameters by assuming a specific set of Yukawa couplings that suppress $\mu \to e\gamma$. The procedure is detailed in Ref. [1537]. For the blue (red) points the normal (inverted) hierarchy is assumed for the neutrino mass ordering. The branching ratios in Fig. 178 are smaller than in Fig. 177. This is mainly from the observed Higgs mass, since it requires the SUSY particle masses to be heavier. These results demonstrate that the target of the Belle II experiment with $\tau \to \mu\gamma$ is a more generic flavour structure rather than the CMSSM-like structure.

*15.2.2. Experiment.* [58] In this subsection, we describe the experimental techniques to search for LFV signals, based on Belle analyses. Belle performed searches for 46 lepton-flavour-violating $\tau$ decay modes using nearly the entire data sample of approximately 1000 fb$^{-1}$. No evidence for LFV decays was observed in any of the modes studied, where 90% C.L. ULs on the branching fractions at the $O(10^{-8})$ level were set. At Belle II a sensitivity at the $O(10^{-9})$–$O(10^{-10})$ level is expected allowing one to explore a wider region of parameter space in various NP scenarios.

*$\tau \to \mu\gamma$.* In LFV analyses, in order to evaluate the signal yield, two independent variables are typically used: the reconstructed mass of the signal and the difference between the sum of energies of the signal $\tau$ daughters and the beam energy ($\Delta E$) in the CM frame. In the

[58] K. Hayasaka, T. Konno, F. Tenchini

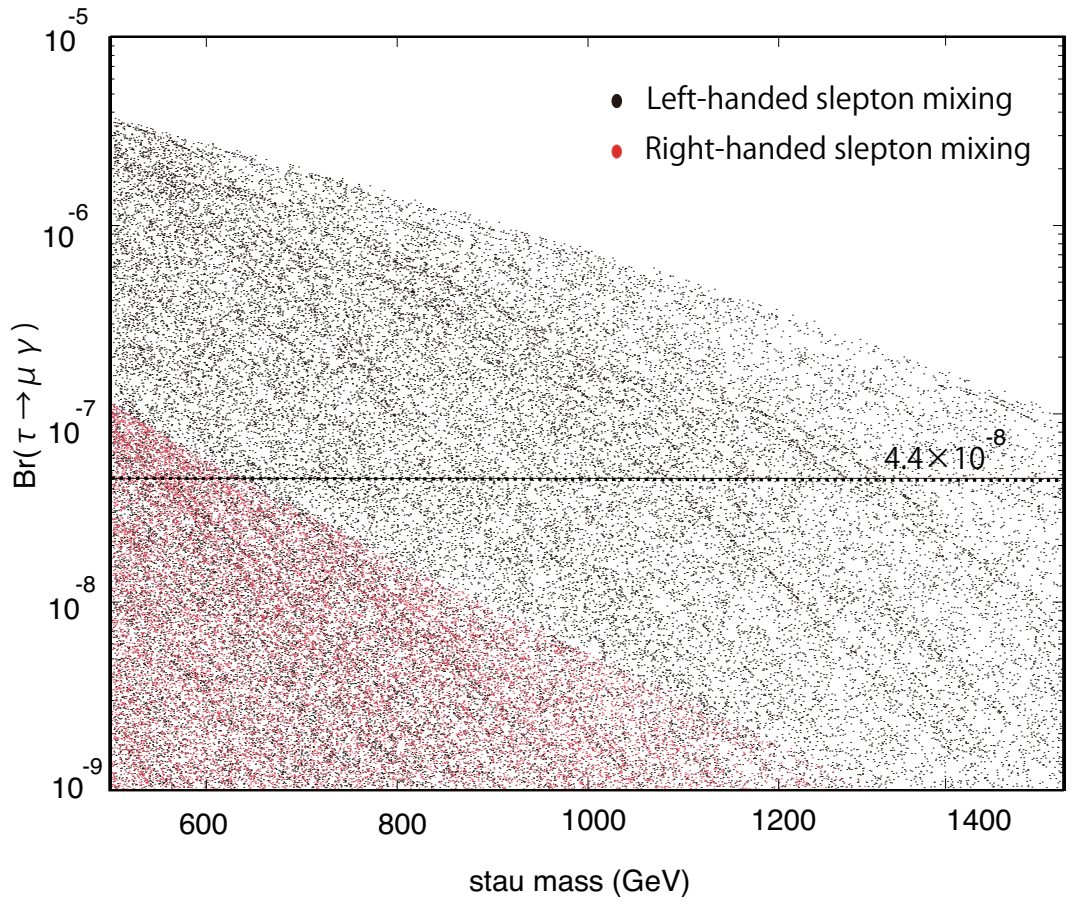

Fig. 177: Branching ratio of $\tau \to \mu\gamma$ in the SUSY SMs with left-handed (black) and right-handed (red) smuon-stau mixing mass terms. The dashed line indicates the current experimental bound. See text for details on the input parameters.

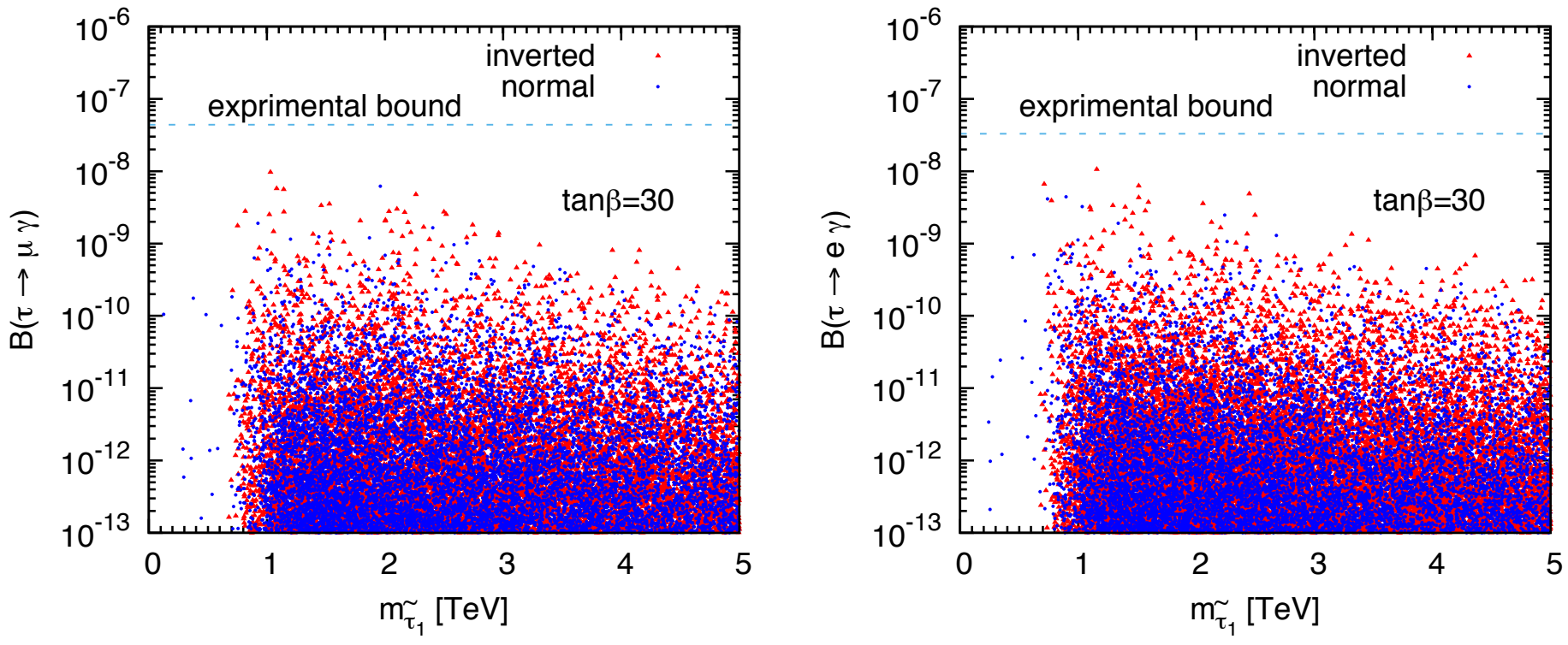

Fig. 178: Branching ratios of $\tau \to \mu\gamma$ and $\tau \to e\gamma$ in the SUSY seesaw model under assumption of specific textures of Yukawa couplings which suppress $\mu \to e\gamma$. For the blue (red) points normal (inverted) hierarchy is assumed for neutrino mass ordering. Here, CMSSM boundary conditions for SUSY breaking parameters are assumed. See text for details on the input parameters.

$\tau \to \mu\gamma$ case, these variables are defined as

$$M_{\mu\gamma} = \sqrt{E_{\mu\gamma}^2 - P_{\mu\gamma}^2}, \tag{504}$$

$$\Delta E = E_{\mu\gamma}^{\rm CM} - E_{\rm beam}^{\rm CM}, \tag{505}$$

where $E_{\mu\gamma}$ and $P_{\mu\gamma}$ are the sum of the energies and the magnitude of the vector sum of the momenta for the $\mu$ and the $\gamma$, respectively. The superscript CM indicates that the variable is defined in the CM frame, *e.g.* $E_{\rm beam}^{\rm CM}$ is the beam energy in the CM frame. For signal, $M_{\mu\gamma}$ and $\Delta E$ should peak at $M_{\mu\gamma} \sim m_\tau$ and $\Delta E \sim 0$ (GeV), while for the background, $M_{\mu\gamma}$ and $\Delta E$ will smoothly vary without peaking structure.

Taking into account the resolution of the detector and the correlation between $M_{\mu\gamma}$ and $\Delta E$, we use an elliptical signal region. To avoid biases, we typically perform blind analyses, where the data in the signal region are blinded when determining the selection criteria and the systematic uncertainties.

The observed $M_{\ell\gamma}$–$\Delta E$ distributions at Belle (based on a sample of $4.9 \times 10^8$ $\tau^+\tau^-$ pairs [1538]) are shown in Fig. 179 for $\tau \to \mu\gamma$ and $\tau \to e\gamma$. The signal yield is evaluated from an extended unbinned maximum-likelihood fit to the $M_{\ell\gamma}$–$\Delta E$ distributions. The main background (BG) is from $\tau \to \ell\nu_\ell\nu_\tau$ with radiated photons, radiative di-muon (for $\mu\gamma$), and Bhabha (for $e\gamma$) events. Event selection criteria used to reduce background are discussed further in the context of the Belle II sensitivity study shown later in this section.

The upper limit obtained from this analysis yields $Br(\tau \to \mu\gamma\ (e\gamma)) = 4.5 \times 10^{-8}$ $(1.2 \times 10^{-7})$ at 90% C.L.

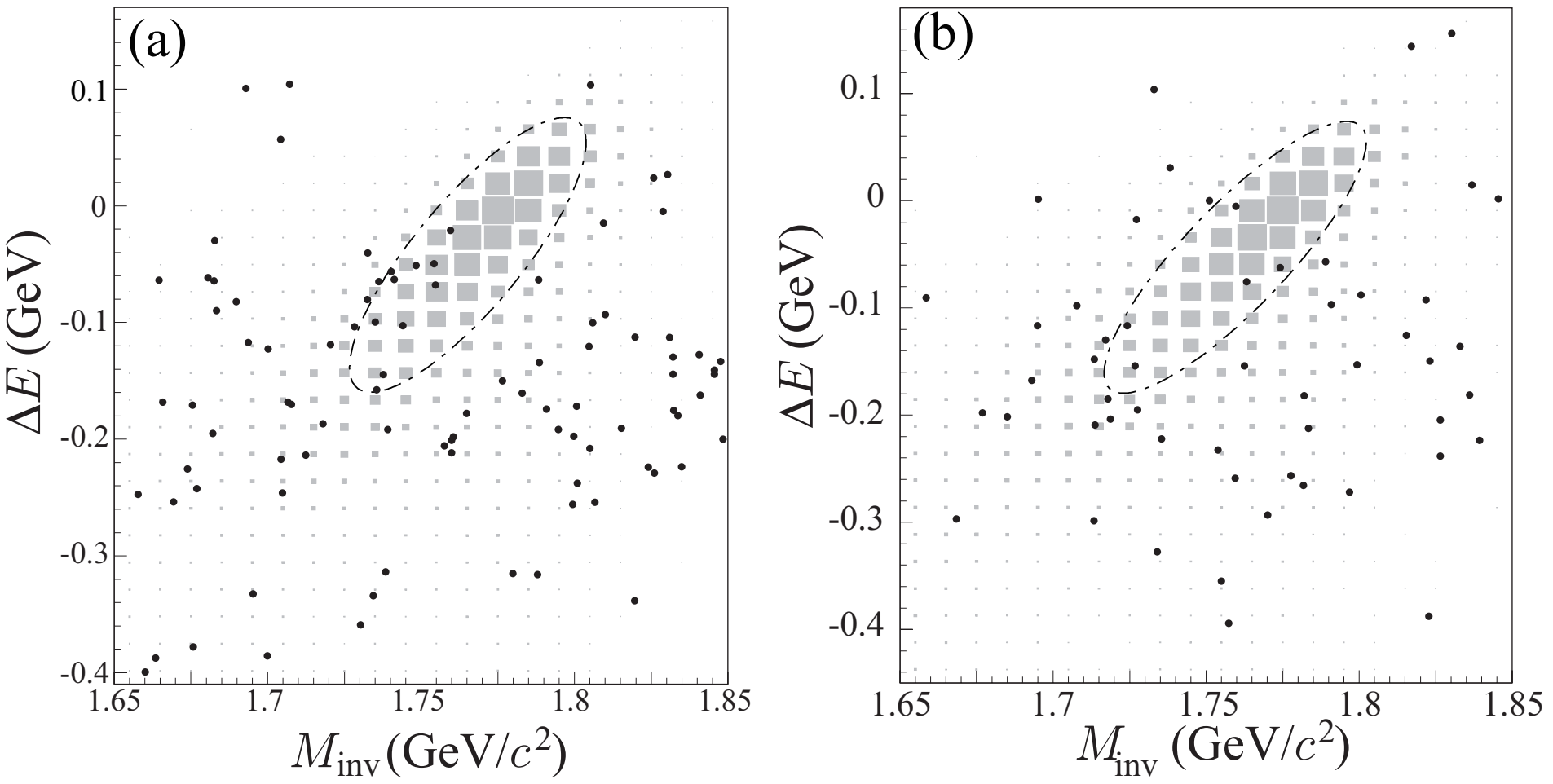

Fig. 179: $M_{\ell\gamma}$–$\Delta E$ distributions from the Belle search for (a) $\tau \to \mu\gamma$ and (b) $\tau \to e\gamma$ [1538]. The black dots and shaded boxes show the data and signal MC, respectively, and the ellipse is the $2\sigma$ signal region.

*Beam background studies.* At Belle II, the beam background to the $\tau$ LFV searches is potentially a more serious concern than at Belle, as the ultimate goal is to achieve background free signal windows. However, beam background is likely to produce only low energy photons, which would be removed by typical selection criteria. A preliminary Belle II $\tau \to \mu\gamma$ study with beam background was performed using MC samples, in order to determine the feasibility of $\tau$ LFV analyses in this more contaminated environment.

We first studied generic SM-decaying $\tau^+\tau^-$ pairs generated with nominal beam background (denoted BGx1) and without background (denoted BGx0) in order to study its impact on the distributions of various physics observables and develop background reduction techniques. As a result, we introduced the following basic selection criteria:

- For photon clusters:
  - $E_\gamma > 0.100(\text{forward endcap}), 0.090(\text{barrel}), 0.160(\text{backwards endcap})$ GeV;
  - $|\Delta t_{cluster}| < 50$ ns.

- For charged particles:
  - Track fit p-value $> 0.01$;
  - Track distance from interaction point (along the beam axis) $|dz| < 0.5$ cm;
  - $P_t > 0.08$ GeV.

The distributions of each of the above variables, except for the p-value, are shown in Fig. 180.

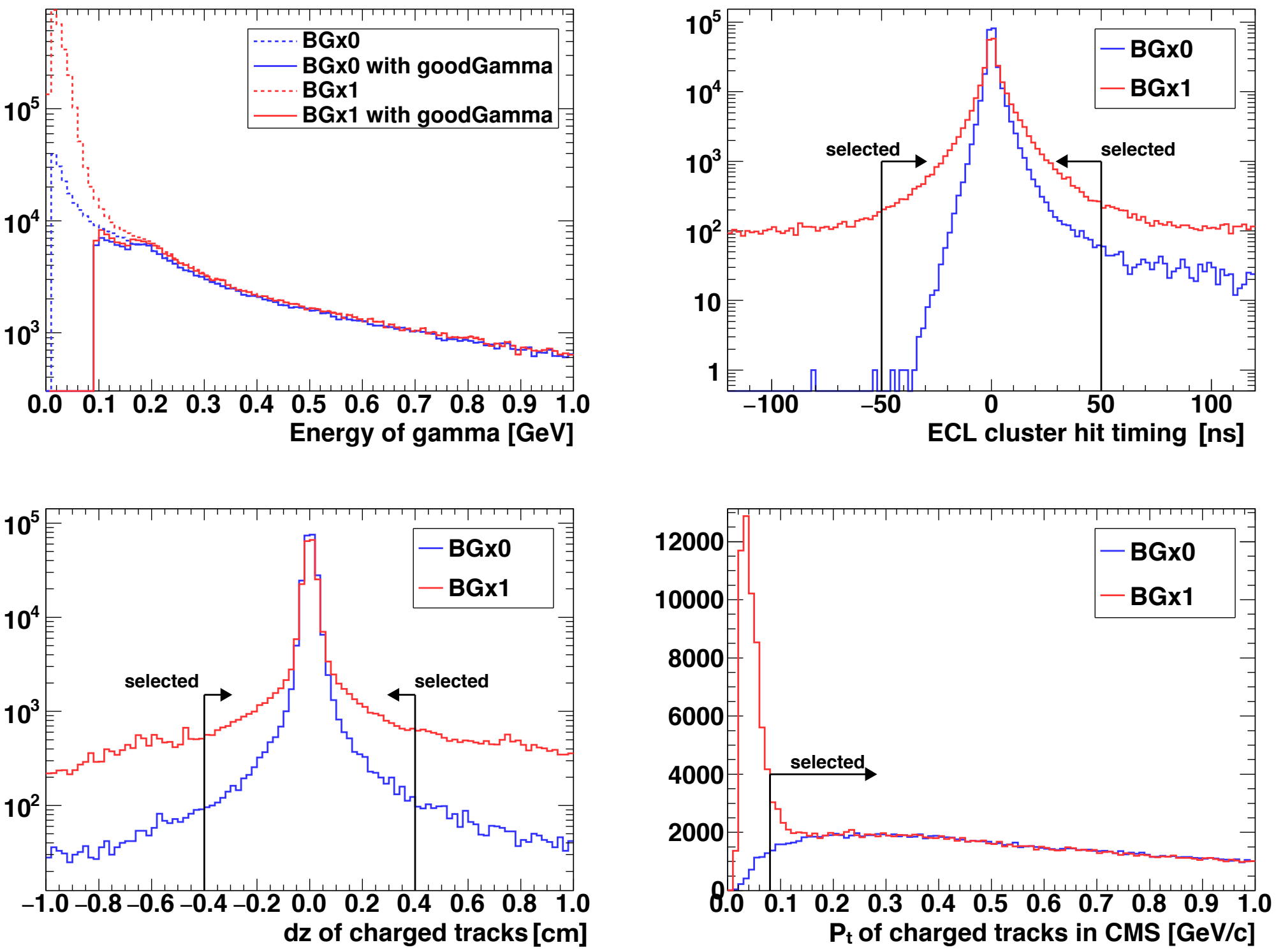

Fig. 180: Distribution of $\gamma$ energy (top-left), distribution of ECL cluster timing (top-right), distribution of track distance from the interaction point along the beam, $dz$ (defined in Sec.5), (bottom-left) and the $P_t$ distribution of charged particle tracks (bottom-right). The bold lines show the distributions after the Belle II quality cut, which rejects low energy photons. The black arrows show the selection criteria adopted for background rejection.

We examined the $\tau$ LFV decay mode $\tau \to \mu\gamma$, by loosely following the corresponding Belle analysis and reconstructing $\tau$ MC signal samples generated both with and without beam background. In addition to the signal side, a single charged particle in the accompanying (tag) side of the $\tau$ pair was also required for successful candidate selection.

The final state particle selection criteria was verified to largely mitigate the effect of beam background, as demonstrated by the distribution of the reconstructed $\tau$ energy in the centre of mass frame (Fig. 181 (left)). The phase space in the invariant mass and the beam energy difference of the signal $\tau$ can also be seen in Fig. **??** (right), where 29.6% of BGx1 events and 35.1% of BGx0 events passed this selection; we therefore can estimate a 16% decrease in signal due to background.

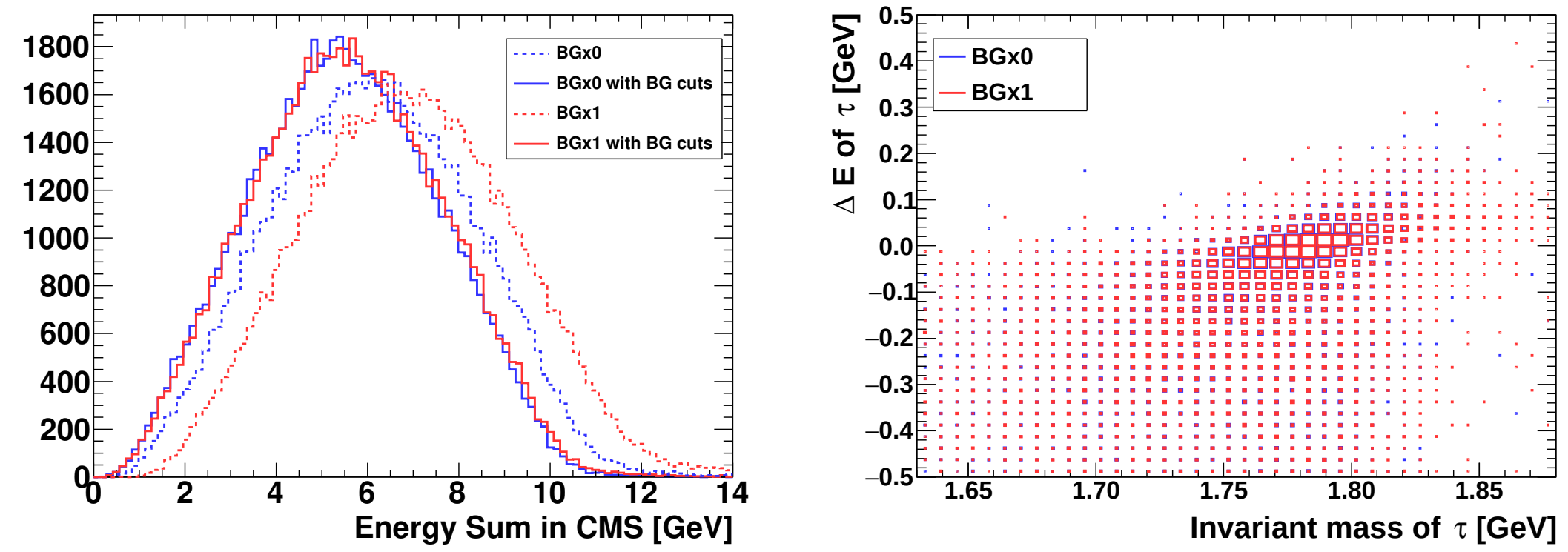

Fig. 181: (Left) Energy sum distribution in the CMS frame: dashed lines show histograms without any selection, while bold lines show histograms with all selection criteria applied. (Right) Correlation between invariant mass and energy difference, $\Delta E$, of the signal $\tau$ after $\tau \to \mu\gamma$ pre-selection.

*Sensitivity Studies.* We perform a sensitivity study for a $\tau \to \mu\gamma$ analysis in Belle II, in order to validate the stated sensitivity projections, demonstrate measurement feasibility with the new beam conditions, and to investigate new observables that could potentially improve the separation power for future analyses.

The expected $\Delta E$ resolution in Belle II is expected to be superior to Belle when neutral particles are part of the reconstructed decay. This is due to improvements in photon position measurement, and hence 4-vector derivation. Overall, this means that better background separation is in principle achievable in order to achieve a background-free signal region.

The analysis approach mimics the previous Belle analysis in [1539], with several additional criteria to further reduce the background. We reconstruct the signal in the mode $\tau \to \mu\gamma$, while the tag side is reconstructed from one non-$\mu$ charged track which is assumed to originate from a SM tau decay. The study is performed on $3 \times 10^6$ signal MC events and $10^8 \div 10^9$ events for each process acting as a background to the analysis: $\tau \to \mu\nu\overline{\nu}$, $\tau \to e\nu\overline{\nu}$, $\tau \to \pi\nu$, $e^+e^- \to \mu^+\mu^-(\gamma)$, $e^+e^- \to e^+e^-(\gamma)$, $e^+e^- \to q\overline{q}$ (for each $q = u, d, s, c$), $e^+e^- \to B^+B^-$ and $e^+e^- \to B^0\overline{B}^0$. The channels are subsequently rescaled to a luminosity of 1 ab$^{-1}$, equivalent to the full Belle dataset, for ease of comparison. The selection criteria applied are similar to the Belle analysis, with some changes due to differences in the Belle II detector. The criteria described in the preceding section on beam background studies are applied as a preselection. In addition, several new criteria are implemented. We discuss the most significant ones below and show their impact against the dominant background processes, due to $e^+e^- \to \mu^+\mu^-(\gamma)$, $\tau \to \pi\nu$ or $\tau \to \mu\nu\overline{\nu}$ with random photons, as well as hadronic continuum where relevant. $e^+e^- \to \Upsilon(4S)$ is omitted from most plots to avoid clutter, as it is easily suppressed through basic event shape requirements.

*Energy, momentum, timing.* We expect tag side tracks from $\mu$-pair backgrounds to peak at $p_{CM}^{tag} \sim 5.5\,\mathrm{GeV}/c$ due to momentum conservation; we thus require $p_{CM}^{tag} < 2.5\,\mathrm{GeV}/c$. As a significant portion of background process is reconstructed from random combinations of charged tracks with low-energy photons, we require the total energy collected in the ECL from neutral clusters to be $2 < E_{ECL} < 6\,\text{GeV}$. This is shown in Figure 182.

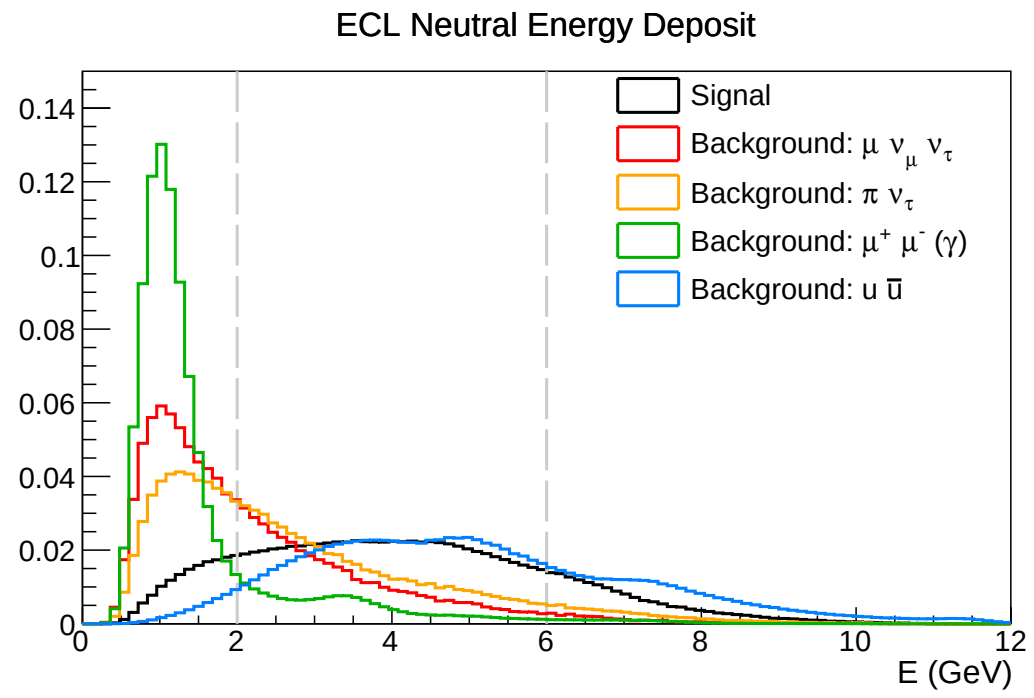

Fig. 182: Total energy deposit in the ECL by neutral particles. The dashed gray lines indicate the selection criteria.

We expect the ECL energy cluster ratio E9oE25 to be consistent with that of a EM shower (see Sec. 5.6.1) and peak close to 1. We require E9oE25$>$ 0.95 (see Figure 183 (left)). In order to select time-coincident $\tau \to \mu\gamma$ decays, we require the value of cluster timing to be within $\pm 1$ ns (see Figure 183 (right)).

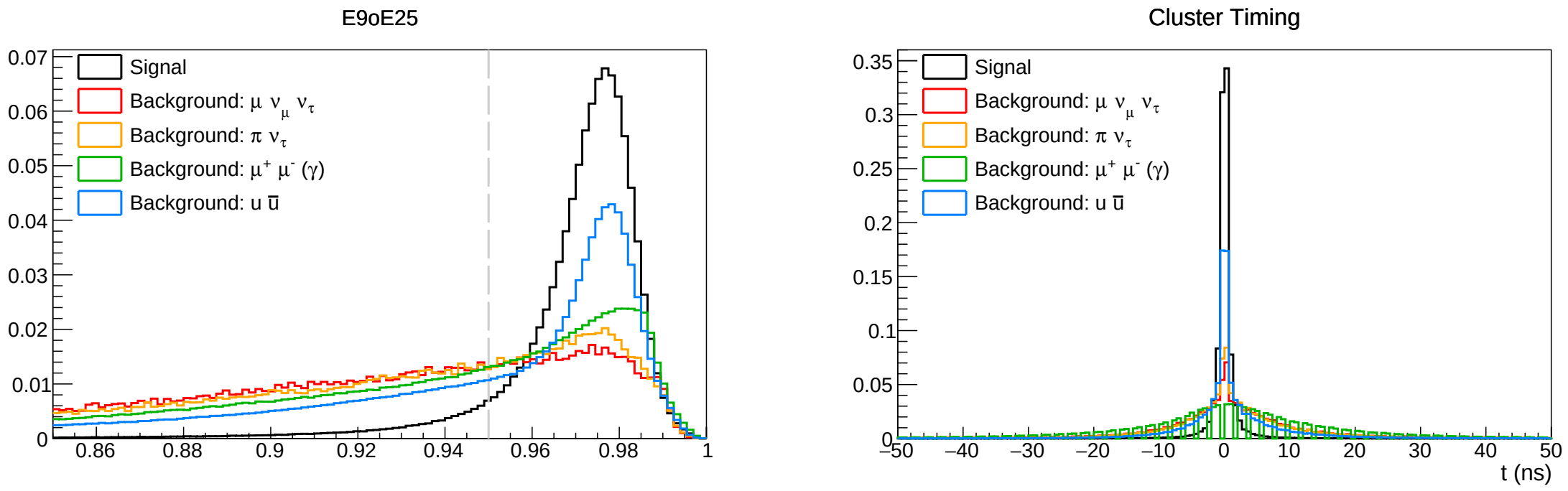

Fig. 183: Left: E9oE25 cluster ratio distributions. The dashed gray line indicates the selection criterion. Right: Cluster timing distributions.

*Event thrust.* We define the thrust scalar $T$ for a collection of $N$ particles as the value

$$T = \frac{\sum_{i=1}^{N} |\mathbf{T} \cdot \mathbf{p_i}|}{\sum_{i=1}^{N} |\mathbf{p_i}|}, \tag{506}$$

where the thrust axis $\mathbf{T}$ is the unit vector along which the total projection of the particle momenta is maximised. We calculate the thrust scalars for both the signal side and the rest of the event, *i.e.* the collection of particles (including the tag) belonging to the event but not used to build the signal side. The thrust distribution for the signal is shown in Fig. 184 (left). Signal $\mu\gamma$ pairs have similar momenta with a small opening angle between them; therefore the signal-side thrust shows a clear peak around 0.942. For most background events, instead, the signal side is reconstructed from a high energy track paired with a low-energy photon from bremsstrahlung, beam background or other processes. As the track momentum alone dominates the thrust calculation, the distribution produces a peak at 1. We therefore require for the signal thrust to be in the range of 0.936 to 0.944. A similar criterion is also used in the Belle analysis, but in Belle II the improved photon momentum measurements allow for a tighter requirement. An additional criterion can be required on the angle between the signal thrust and the beam axis to be $\cos\theta_{T,Bz} < 0.8$, shown in Fig. 184 (right).

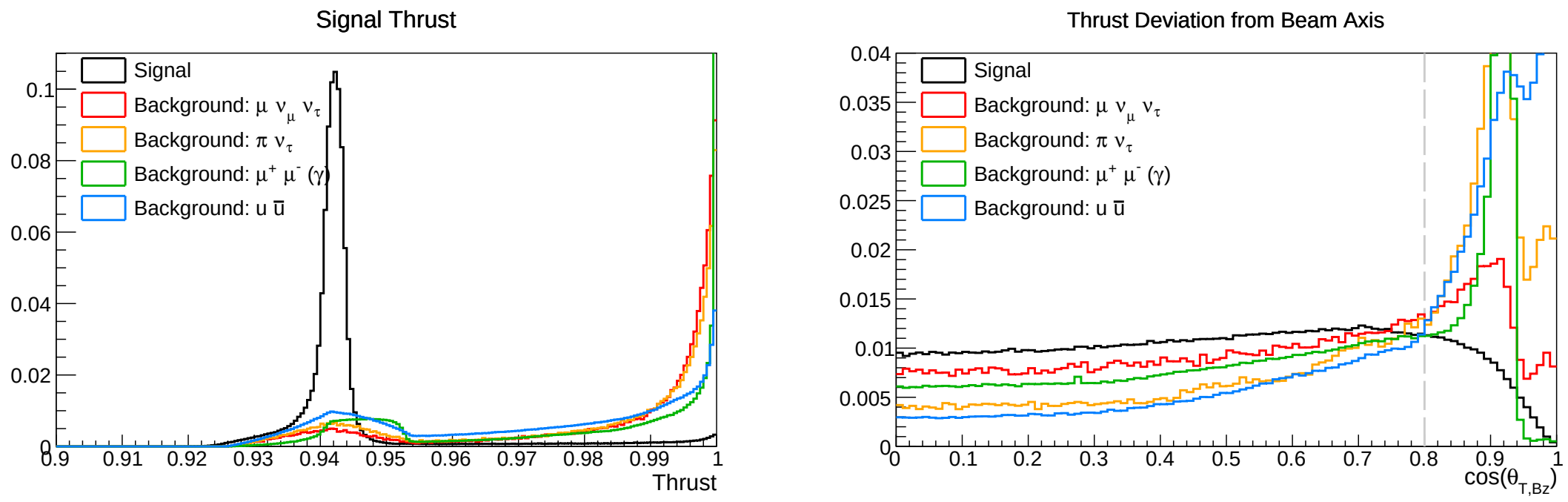

Fig. 184: (Left) Signal-side thrust scalar distributions. (Right) Angle between the signal thrust axis and the beam axis. The dashed gray line indicates the selection criterion.

Due to the large number of tracks and photons produced in hadronic processes, misreconstructed $e^+e^- \to q\bar{q}$ and $e^+e^- \to B\overline{B}$ events have on average a lower rest-of-event thrust than leptonic channels. We can thus also select on the magnitude of the rest-of-event thrust vector to exclude $B\overline{B}$ events and strongly suppress $q\bar{q}$continuum. This is shown in Fig. 185.

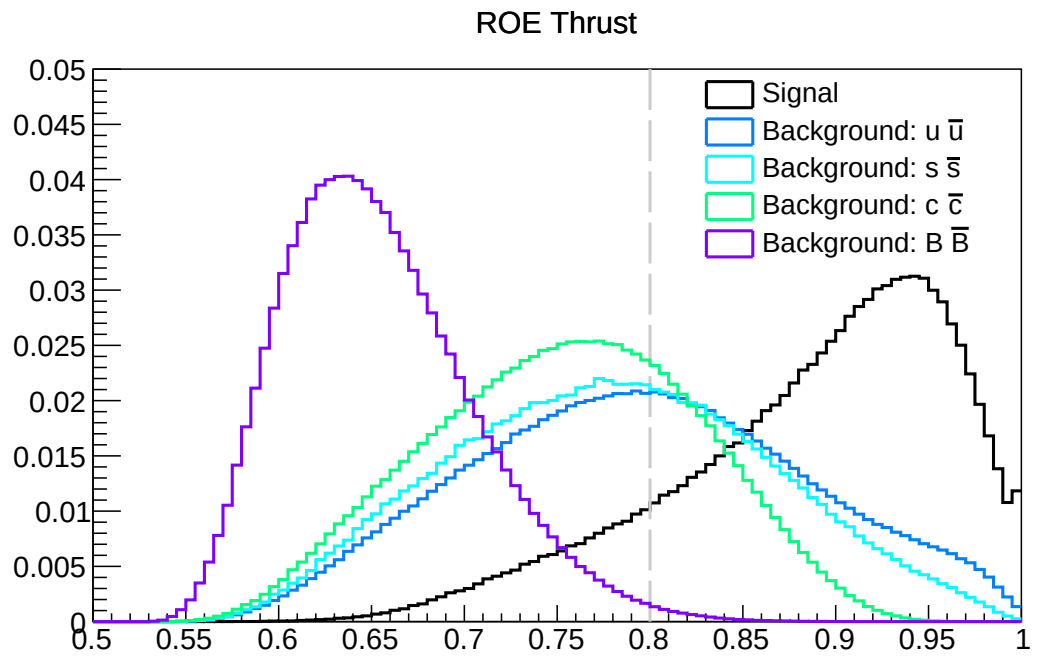

Fig. 185: Rest-of-event thrust scalar distributions. The dashed gray line indicates the selection criterion.

*Event Shape.* Further suppression can be achieved by selecting on higher level event shape variables. The first such are CLEO cones [74], which are defined by binning the space around the signal thrust axis in nine 10° polar angle regions and then measuring the forward and backwards momentum flow through those regions in the lab frame. Distributions for the first and second CLEO cone are shown in Fig. 186 together with sample selections used for this study. The full set of selection criteria is given in Table 136. Event shape Fox-Wolfram moments can also provide strong separation power. Sample criteria are shown in Table 137. These are especially effective to reject surviving continuum background events.

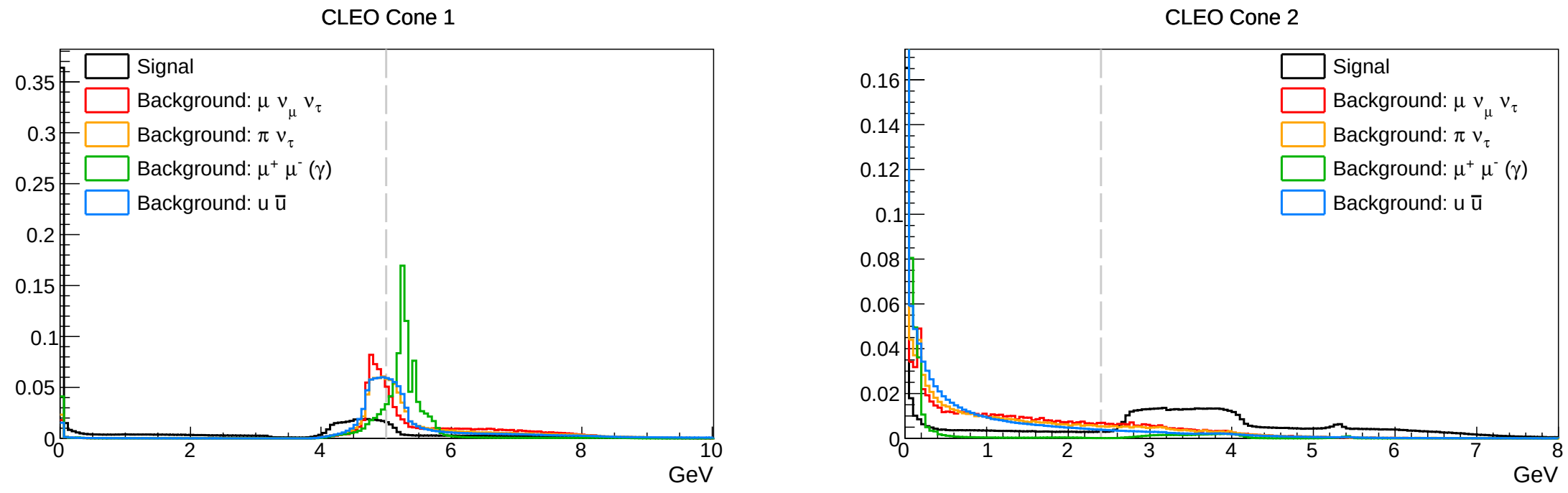

Fig. 186: Momentum flow distributions for the first (left) and second (right) CLEO cones. The dashed gray lines indicate the selection criteria.

| CLEO cone | lower | upper |
|---|---|---|
| cc1 | – | 5 |
| cc2 | 2.4 | – |
| cc3 | – | – |
| cc4 | – | 1.7 |
| cc5 | – | 0.9 |
| cc6 | – | 0.7 |
| cc7 | – | 0.5 |
| cc8 | – | – |
| cc9 | – | 0.4 |

Table 136: Sample selection criteria for CLEO cones.

Table 138 lists the remaining events in the extended signal region ($-0.4 < \Delta E < 0.2\,\mathrm{GeV}$, $1.65 < M_{inv} < 1.85\,\mathrm{GeV}/c^2$) after the event selection. As can be seen the expected background rate is vastly reduced. This selection has a 7.23% signal efficiency; the remaining signal distribution is fit with asymmetric gaussians yielding $\langle \Delta E \rangle \approx 47\,\mathrm{MeV}$, $\langle M_{inv} \rangle \approx 1.79\,\mathrm{GeV}/c^2$, which is consistent with our expectation of $\langle \Delta E \rangle \sim 0$ and $\langle M_{inv} \rangle \sim m_\tau$.

To determine an upper limit on the sensitivity to the branching fraction, we perform a 72° phase space rotation and select a smaller signal region centered on the means of the gaussian fits. This is shown in Fig. 187. Selecting this subregion reduces the signal efficiency to 4.59%; however, no MC background events are present. Assuming the background distribution follows Poisson statistics, the sensitivity can be estimated without the need of likelihood fits, which is acceptable for a basic sensitivity study. Based on the observation of 0 events, we can set an upper limit of $n_{events} < 2.3$ at 90% CL and convert it into an upper limit to the branching ratio under the hypothesis of no signal, for 1 ab$^{-1}$ luminosity and a $e^+e^- \to \tau\tau$

| FWM | lower | upper |
|---|---|---|
| Hso(0,0) | 0.05 | 1 |
| Hso(0,1) | -0.05 | 0.3 |
| Hso(0,2) | – | 0.48 |
| Hso(0,3) | -0.1 | 0.25 |
| Hso(2,0) | -0.1 | 1 |
| Roo(1) | -0.018 | 0.08 |
| Roo(3) | -0.01 | 0.007 |
| $R_2$ | 0.4 | – |

Table 137: Sample selection criteria for Super- and reduced Fox-Wolfram moments.

| Channel | Events in sample (scaled to 1 $ab^{-1}$) | Events after selection |
|---|---|---|
| $\tau \to \mu\nu\overline{\nu}$ | $1.60 \times 10^8$ | 163 |
| $\tau \to \pi\nu$ | $3.34 \times 10^8$ | 40 |
| $\tau \to e\nu\overline{\nu}$ | $1.64 \times 10^8$ | 0 |
| $e^+e^- \to \mu^+\mu^-(\gamma)$ | $1.148 \times 10^9$ | 15 |
| $e^+e^- \to e^+e^-(\gamma)$ | $3 \times 10^{11}$ | 0 |
| $e^+e^- \to B^+B^-$ | $5.50 \times 10^8$ | 0 |
| $e^+e^- \to B^0\overline{B}^0$ | $5.50 \times 10^8$ | 0 |
| $e^+e^- \to q\overline{q}$ | $3.69 \times 10^9$ | $9(u\overline{u})+3(d\overline{d})+3(c\overline{c})$ |

Table 138: Events remaining in the extended signal region after selection.

production cross section of 0.919 nb:

$$B(\tau \to \mu\gamma) < \frac{n_{events}}{\epsilon_{signal} \times n_\tau} = \frac{2.3}{4.59 \times 10^{-2} \times 2 \times 9.19 \times 10^8} = 2.7 \times 10^{-8}(90\%CL) \quad (507)$$

This compares favourably to the Belle result of $B(\tau \to \mu\gamma) < 4.5 \times 10^{-8}$ using 535 $fb^{-1}$. Although this is only a preliminary study, if a background-free selection were to be successfully achieved and maintained over the full Belle II dataset this would have a great impact, as the limit scales as $1/n_\tau$ in the zero-background hypothesis. This would lead us a factor 100 improvement compared to the limit obtained by Belle.

*Other modes.* Here we review some of the other $\tau$ LFV measurements performed at Belle, including the golden $\tau \to 3\mu$ mode. All of these channels are largely background-free and therefore a sensitivity improvement of about two orders of magnitude can be expected with Belle II due to the increase in luminosity.

*$\tau \to \ell\ell'\ell''$.* For the decays $\tau \to \ell\ell'\ell''$, $M_{\ell\ell\ell}$ and $\Delta E$ are similarly used to identify signal events. Figures 188(a) and (b) show the three-lepton invariant mass versus $\Delta E$ ($M_{\ell\ell\ell}$–$\Delta E$) distributions, respectively, for the $\tau^- \to e^-e^+e^-$ and $\tau^- \to \mu^-\mu^+\mu^-$ candidates after selection at Belle with nearly the entire data sample ($7.2 \times 10^8$ $\tau^+\tau^-$ pairs) [1539]. No events in the signal region have been found in any of the six modes. The 90% C.L. upper limits on the branching fractions in units of $10^{-8}$ are given in Table 139.

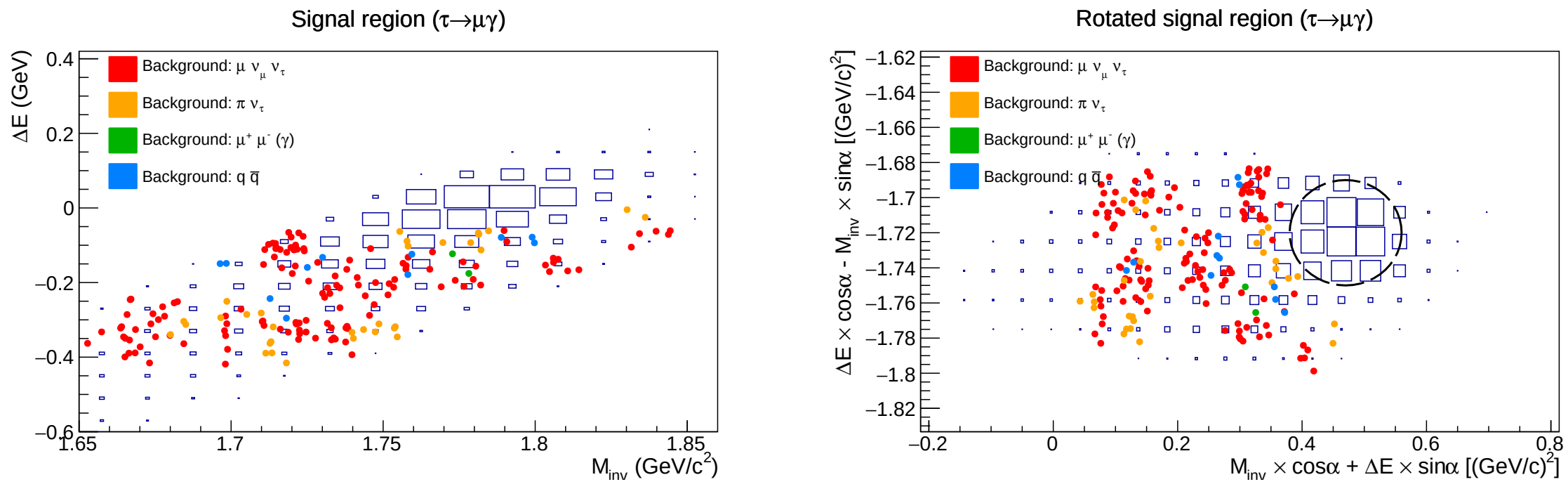

Fig. 187: $M_{inv}$–$\Delta E$ extended signal region (*top*), and the same region rotated by $\alpha = 72°$ (*bottom*). Shaded boxes indicate the event distribution for $\tau \to \mu\gamma$; dots are unscaled background events coloured according to the legend. The dotted ellipse represents the final, background-free selection.

LHC experiments have also studied $\tau \to 3\mu$ where the current upper limit obtained by LHCb is $4.6 \times 10^{-8}$ using 1.0 fb$^{-1}$ data sample at 7 TeV and 2.0 fb$^{-1}$ data sample at 8 TeV. LHCb has a plan to accumulate a 50 fb$^{-1}$ data sample at 14 TeV by 2029, and to improve the tau trigger efficiency through various upgrades. Since the LHCb analysis has a large background contribution, the sensitivity is expected to primarily run proportionally to the inverse of the square root of the data increase. Therefore, the expected upper limit of the branching fraction for $\tau \to 3\mu$ will be approximately $4.6 \times 10^{-8}/\sqrt{50/3 \times 2 \times 2} = 5.6 \times 10^{-9}$. So while the LHC experiments will be good competitors, Belle II is expected to have higher sensitivity owing to the background-free environment. The expected Belle II upper limit is $2.1 \times 10^{-8}/(50/0.8) = 3.3 \times 10^{-10}$ with the 50 ab$^{-1}$ data sample.

Table 139: Summary of the efficiency (Eff.), the expected number of BG events ($N_{BG}^{\rm exp}$), and the upper limit on the branching fraction (UL) at 90% C.L. for $\tau^- \to \ell^- \ell'^+ \ell''^-$ at Belle.

| Mode | Eff.(%) | $N_{BG}^{\rm exp}$ | UL ($10^{-8}$) |
|---|---|---|---|
| $e^-e^+e^-$ | 6.0 | $0.21 \pm 0.15$ | 2.7 |
| $e^-\mu^+\mu^-$ | 6.1 | $0.10 \pm 0.04$ | 2.7 |
| $e^-e^+\mu^-$ | 9.3 | $0.04 \pm 0.04$ | 1.8 |
| $\mu^-e^+\mu^-$ | 10.1 | $0.02 \pm 0.02$ | 1.7 |
| $e^-\mu^+e^-$ | 11.5 | $0.01 \pm 0.01$ | 1.5 |
| $\mu^-\mu^+\mu^-$ | 7.6 | $0.13 \pm 0.06$ | 2.1 |

*$\tau \to \ell P^0$ ($P^0 = \pi^0, \eta, \eta'$).* The results for $\tau$ decays into a lepton and a neutral pseudoscalar ($\pi^0, \eta, \eta'$) at Belle [1540] are summarised in Table 140. A single event is found in $\tau \to e\eta(\to \gamma\gamma)$, consistent with the expected background, while no events are observed in other modes. The obtained 90% C.L. ULs on the branching fraction are in the range $(2.2 - 4.4) \times 10^{-8}$.

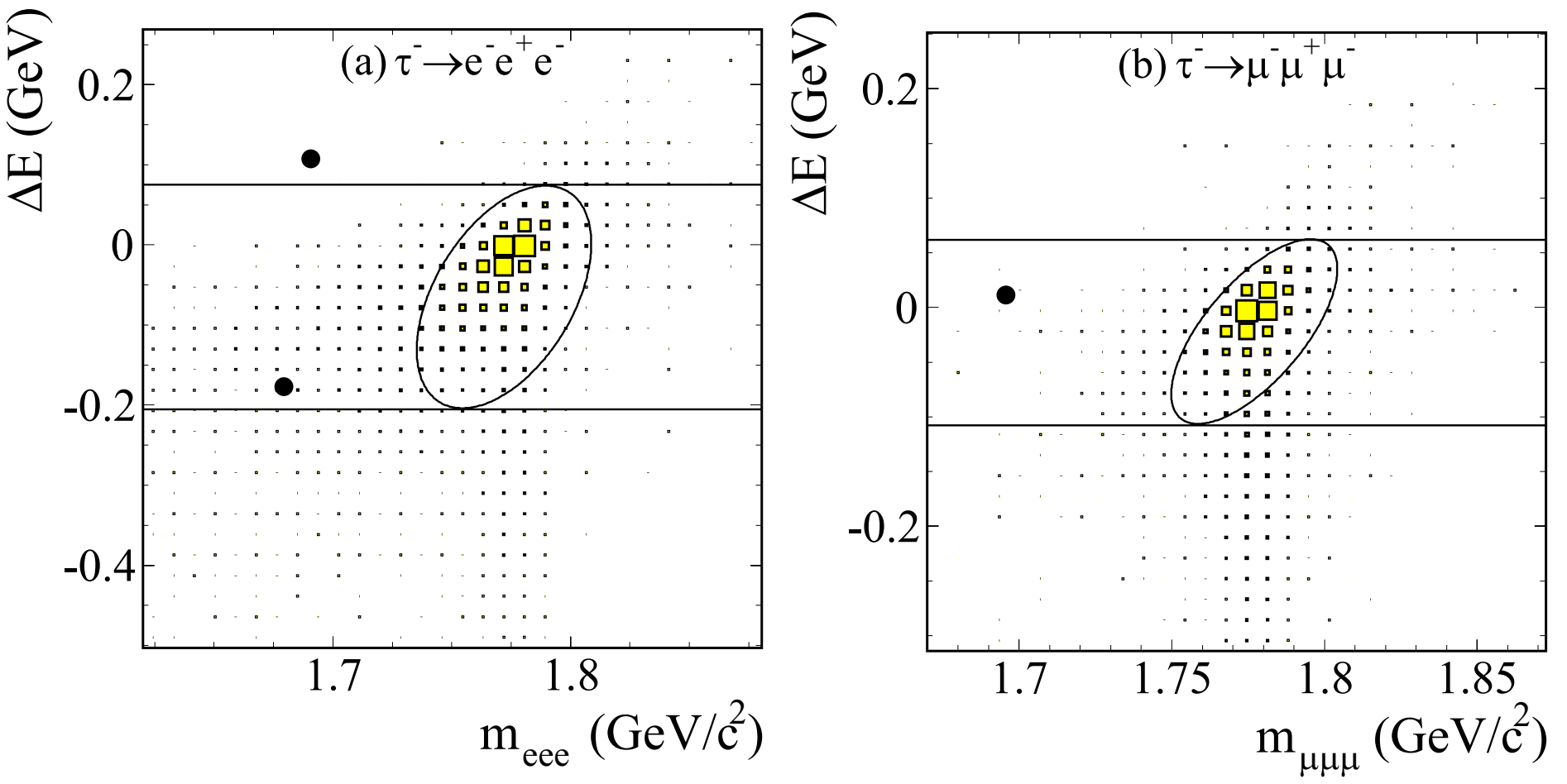

Fig. 188: $M_{\ell\ell\ell}$–$\Delta E$ distributions for the Belle analyses of (a) $\tau^- \to e^-e^+e^-$ and (b) $\tau^- \to \mu^-\mu^+\mu^-$ modes [1539]. The black dots and shaded boxes show the data and signal MC, respectively. The ellipse is the signal region. The region formed by the two parallel lines, excluding the signal ellipse region, is the sideband region used to evaluate the expected number of background events in the signal region.

Table 140: Summary of the efficiency (Eff.), the expected number of BG events ($N_{BG}^{\rm exp}$), and the upper limit on the branching fraction (UL) for $\tau \to \ell P^0$ at Belle, where (comb.) means the combined result from sub-decay modes.

| Mode | Eff.(%) | $N_{BG}^{\rm exp}$ | UL ($10^{-8}$) |
|---|---|---|---|
| $\mu\eta(\to \gamma\gamma)$ | 8.2 | $0.63 \pm 0.37$ | 3.6 |
| $e\eta(\to \gamma\gamma)$ | 7.0 | $0.66 \pm 0.38$ | 8.2 |
| $\mu\eta(\to \pi\pi\pi^0)$ | 6.9 | $0.23 \pm 0.23$ | 8.6 |
| $e\eta(\to \pi\pi\pi^0)$ | 6.3 | $0.69 \pm 0.40$ | 8.1 |
| $\mu\eta$(comb.) | | | 2.3 |
| $e\eta$(comb.) | | | 4.4 |
| $\mu\eta'(\to \pi\pi\eta)$ | 8.1 | $0.00^{+0.16}_{-0.00}$ | 10.0 |
| $e\eta'(\to \pi\pi\eta)$ | 7.3 | $0.63 \pm 0.45$ | 9.4 |
| $\mu\eta'(\to \gamma\rho^0)$ | 6.2 | $0.59 \pm 0.41$ | 6.6 |
| $e\eta'(\to \gamma\rho^0)$ | 7.5 | $0.29 \pm 0.29$ | 6.8 |
| $\mu\eta'$(comb.) | | | 3.8 |
| $e\eta'$(comb.) | | | 3.6 |
| $\mu\pi^0$ | 4.2 | $0.64 \pm 0.32$ | 2.7 |
| $e\pi^0$ | 4.7 | $0.89 \pm 0.40$ | 2.2 |

*Summary of LFV projections.* We provide a naive set of 50 ab$^{-1}$projections for 90% C.L. upper limits based on zero-background scenarios in Fig. 189. In this case, all limits follow linearly with the increase in integrated luminosity, with a reasonable assumption that the Belle II analyses will have the same efficiency as that of Belle. As shown earlier in this section, it may be feasible to reach zero-background in many channels, particularly those with all charged particle final states. A simulation study of a challenging case was presented, that of $\tau \to \mu\gamma$, which we found could conceivably be isolated from background with optimised analysis techniques. It would be realistic to assume that some modes with neutrals may eventually contain some irreducible background, in which case the improvement on the limits will be approximately one order of magnitude, rather than two in the zero background case presented in the figure.

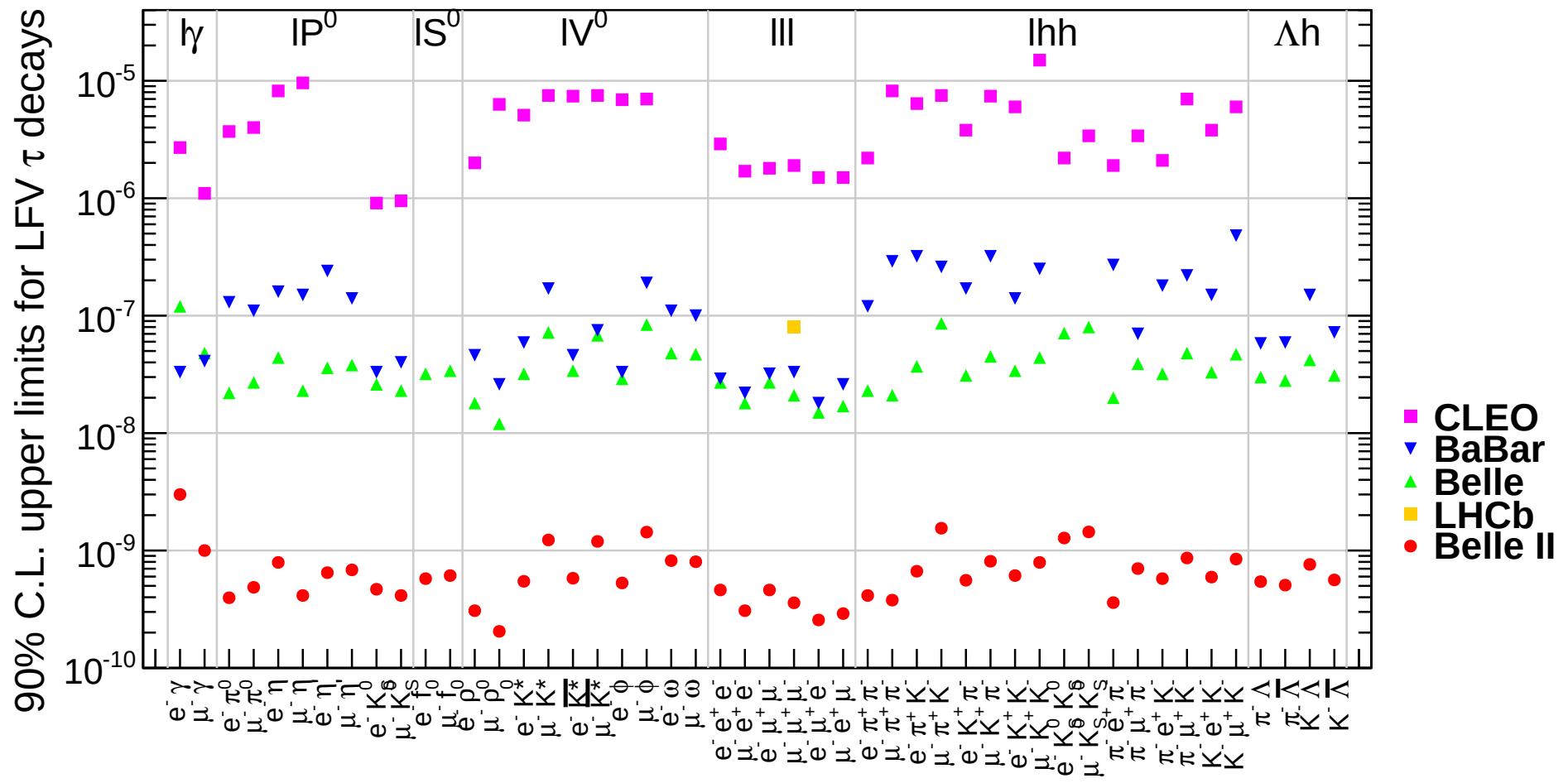

Fig. 189: Current 90% C.L. upper limits for the branching fraction of $\tau$ LFV decays obtained in the CLEO, BaBar, and Belle experiments. Purple boxes, blue inverted triangles, green triangles and yellow boxes show CLEO, BaBar, Belle and LHCb results, respectively, while red circles express the Belle II future prospects, where they are extrapolated from Belle results assuming the integrated luminosity of 50 ab$^{-1}$.

## 15.3. *CP violation in $\tau$ decays*

*(Contributing authors: I. Bigi, K. Hayasaka, E. Kou, B. Moussallam, E. Passemar)*

In the three-generation SM, the violation of $CP$ is explained by the Kobayashi-Maskawa mechanism. It predicts the $CP$ violation in the quark sector as well as an absence of $CP$ violation in the lepton sector. Therefore, the study of $CP$ violation in semi-hadronic $\tau$ decays offers a unique search of physics beyond the SM, namely a new source of $CP$ violation beyond the Kobayashi-Maskawa mechanism. If $\tau$ $CP$ violation is observed, it implies that there exists a new $CP$ violating coupling in the $\tau - \nu_\tau$ and/or $d - u$ or $s - u$ current in addition to the one induced in the SM by the $K^0 - \overline{K}^0$ mixing.

The first $CP$ asymmetry measurement in $\tau$ decays was performed using the decay rate difference between $\tau^+ \to \pi^+ K_S^0 \bar{\nu}_\tau$ and $\tau^- \to \pi^- K_S^0 \nu_\tau$:

$$\mathcal{A}_\tau = \frac{\Gamma(\tau^+ \to \pi^+ K_S^0 \bar{\nu}_\tau) - \Gamma(\tau^- \to \pi^- K_S^0 \nu_\tau)}{\Gamma(\tau^+ \to \pi^+ K_S^0 \bar{\nu}_\tau) + \Gamma(\tau^- \to \pi^- K_S^0 \nu_\tau)} \tag{508}$$

This $CP$ asymmetry is non-zero in SM due to the $K^0-\overline{K}^0$ mixing. Assuming the $CPT$ invariance, we can write the $CP$ asymmetry in terms of the kaon mixing parameter $\epsilon$. With a first order approximation, and neglecting the effect from $\epsilon'/\epsilon$, the SM prediction yields [1541–1543]

$$\mathcal{A}_\tau^{\rm SM} \simeq 2{\rm Re}(\epsilon) \simeq (0.36 \pm 0.01)\% \tag{509}$$

where we assume that the experimental results are obtained by time integration between a time much smaller than $K_S^0$ life time and time much longer than the $K_L^0$ life time (see [1543] for the discussion on time-dependence effects). Note that we are discussing the simplest case with two hadrons in the final state but we would obtain the same result adding more pions. Thus, more generally, [1541]

$$\begin{aligned} & A_{\mathbf{CP}}(\tau^- \to \nu K^0 X'_{S=0}) \\ = \; & A_{\mathbf{CP}}(\tau^- \to \nu K_S^0 X'_{S=0}) + A_{\mathbf{CP}}(\tau^- \to \nu K_L^0 X'_{S=0}) + A_{\mathbf{CP}}(\tau^- \to \nu [K_S^0/K_L^0] X'_{S=0}) \\ = \; & 2\,{\rm Re}\,\epsilon_K + 2\,{\rm Re}\,\epsilon_K - 4\,{\rm Re}\,\epsilon_K = 0 \end{aligned} \tag{510}$$

where $X' = \pi, \pi\pi, \pi\pi\pi \cdots$ with corresponding charge and $[K_S^0/K_L^0]$ represents the $[K_S^0 - K_L^0]$ interference term. The SM prediction in Eq. (509) must be compared to the result obtained by the BaBar collaboration [1544]:

$$\mathcal{A}_\tau = (-0.36 \pm 0.23 \pm 0.11)\% \ , \tag{511}$$

which is 2.8 $\sigma$ away from the SM expectation in Eq. (509). A similar level of $CP$ violation, due to kaon mixing, should be observed in the $D$ meson decay, $\mathcal{A}_D = \frac{\Gamma(D^+\to\pi^+K_S^0)-\Gamma(D^-\to\pi^-K_S^0)}{\Gamma(D^+\to\pi^+K_S^0)+\Gamma(D^-\to\pi^-K_S^0)}$. These asymmetries are related to the $\tau$ $CP$ asymmetry as $A_\tau = -A_D$ [1543]. The experimental average of $D$ meson $CP$ asymmetry is found to be $A_D = (-0.41 \pm 0.09)\%$, much more precise than $\mathcal{A}_\tau$. Thus, an improved measurement of $\mathcal{A}_\tau$ is certainly the first priority at Belle II.

This intriguing result motivates us to further investigate $CP$ violation measurements in hadronic $\tau$ decays. The $CP$ asymmetry $\mathcal{A}_\tau$ discussed above is an angular integrated observable (parity even) and is sensitive to a particular type of new physics couplings. For example, in [1545], it is pointed out that a tensor coupling induced by a new physics model may interfere with the SM vector coupling and produce such $CP$ violation. However, in a recent study [1546] it has been shown that this tensor coupling can not explain such a large effect. Moreover, they also show that such a large asymmetry is incompatible with constraints coming from the neutron EDM and $D^0 - \overline{D}^0$ mixing. An interesting new physics search in the $D^+ \to K_S^0\pi^+$ process is proposed in [1547]. The prospect of the relevant measurement at Belle II is discussed in detail in the charm physics chapter, Chapter 13. If this discrepancy persists at Belle II, it would therefore point towards the existence of light BSM physics evading these constraints. On the other hand, in order to probe scalar or pseudoscalar couplings, which can be induced by charged Higgs bosons, a parity odd angular observable is needed. In the next subsections, we discuss some details on the $CP$ violation measurement using the angular observables.

We note an interesting $CPT$ test that Belle II may be able to perform:

$$A_{CP}(\tau^- \to \nu K^- X'_{S=0}) + A_{CP}(\tau^- \to \nu K^0 X'_{S=0}) = 0. \tag{512}$$

where $X' = \pi, \pi\pi, \pi\pi\pi$ etc., with corresponding charges. This relation can be derived from the $CPT$ invariance relation

$$\Gamma(\tau^- \to \nu X^-_{S=0}) = \Gamma(\tau^+ \to \bar{\nu} X^+_{S=0}), \quad \Gamma(\tau^- \to \nu X^-_{S=-1}) = \Gamma(\tau^+ \to \bar{\nu} X^+_{S=+1}). \tag{513}$$

*15.3.1. $CP$ violation in angular observables.* The general angular formalism of hadronic $\tau$ decays contain 16 structure functions [1548]. Not all of them are measurable as the final state neutrino direction is lost. Nevertheless, it has been shown in Ref. [1548] that 13 out of 16 are measurable if the information on the azimuthal and polar angles of the hadronic system in the laboratory frame, $\phi, \beta$[59] in addition to the two angles, $\theta$ and $\psi$, which characterise the relative orientation of the laboratory, hadron and the tau rest-frames, (find the details below) are known. We do not use the angle $\theta$ in the following discussion since it is useful only when the $\tau$ polarisation is known [1550], which is not the case for Belle II. Various $CP$ violation measurements have been proposed for $\tau \to 2\pi\nu$ [1551], $\tau \to K\pi\nu$ [1549], $\tau \to 3\pi\nu$ [1552], $\tau \to K\pi\pi\nu, KK\pi\nu$ [1553–1555] decay channels.

Let us start with the two hadron final state, $\tau \to P_1 P_2 \nu$ $(P_1 P_2 = \pi\pi, \pi\eta, \pi K, \eta K)$. In [1549], it has been shown that the $CP$ violating angular observable can be composed even without knowing the $\tau$ direction as follows. Let us work in the $P_1 P_2$ momentum rest frame. In this frame, we first define the laboratory direction as $+z$ and then the direction of $P_1$ by the polar angle $\beta$ and the azimuthal angle $\phi$ (see Fig. 190). The direction of the $\tau$ can not be measured due to the missing neutrino, however, the polar angle $\psi$ can be computed with the following formula [1549]:

$$\cos\psi = \frac{x(m_\tau^2 + Q^2) - 2Q^2}{(m_\tau^2 - Q^2)\sqrt{x^2 - 4Q^2/s}}, \tag{514}$$

where $x = \frac{2E_{P_1}}{\sqrt{Q^2}}$ and $E_{P_1}$ is the energy of the hadron $P_1$ in the laboratory frame. $Q$ is the invariant mass of $P_1, P_2$. We fix the relative azimuthal angle such that $\tau$ is on the $y - z$ plane to remove an unnecessary angle. As a result we can write the angle between $P_1$ and $\tau$ in the hadronic rest frame as

$$\cos\alpha = \cos\beta\cos\psi + \sin\beta\sin\psi\cos\phi \tag{515}$$

We are left with one angle, $\phi$, undetermined though[60] this does not cause any problem since the azimuthal angle does not carry additional information. Therefore we can integrate it out, which cancels the second term in Eq. (515) (*i.e.* integration along the cone in Fig. (190)). As we will see in the next subsection, using this term $\cos\beta\cos\psi$, we can construct one $CP$ violating observable for the two hadron final state case, that can be measured at Belle II.

Before closing this introduction, we briefly discuss three hadron final states. While the $CP$ violation search in $\tau \to P_1 P_2 \nu$ is sensitive to a new (pseudo-)scalar contribution in the intermediate state, the $CP$ violation search in $\tau \to P_1 P_2 P_3 \nu$ is sensitive to a new pseudoscalar contribution (see *e.g.* [1553–1555]) since $P_1P_2P_3$ proceeds mainly through axial-vector resonances. For the three hadron final states, we define the following angles $\tau^-$ : the $\beta$ angle, which is an angle between the direction of the momentum for $P_1P_2P_3$ system in the lab frame and the perpendicular direction of the plane which includes the momenta of $P_1$, $P_2$ and $P_3$ in the $P_1P_2P_3$ rest frame, and the $\gamma$ angle, which is the angle between the $P_1$ direction in the $P_1P_2P_3$ rest frame and the plane made from two directions constructing $\beta$. Note that for $\tau^+$, the charge conjugate must be considered (see Fig.1 in [1554]). Using these angles, it is found that three $CP$ asymmetries, two T-even and one T-odd, can be constructed [1554]. Note that the T-even observable requires a strong phase to measure $CP$ violation while the T-odd one does not. Extracting the hadronic form factors in the three hadron final states is more difficult comparing to the two hadron case. Therefore we do not go into the detail here. Nevertheless, we emphasise that three hadron final state can offer an excellent null-test: if non-zero $CP$ violation is observed, it can be immediately interpreted as a signal of new physics.

[59] In [1548], the azimuthal angle is defined as $\alpha$ instead of $\phi$. Note also that [1549] assumes that the laboratory frame coincides with the $e^+e^-$ CM frame, which is not the case at Belle II.

[60] The $\phi$ angle could be determined with the two-fold ambiguity by using the other $\tau$ produced together from $e^+e^-$ annihilation.

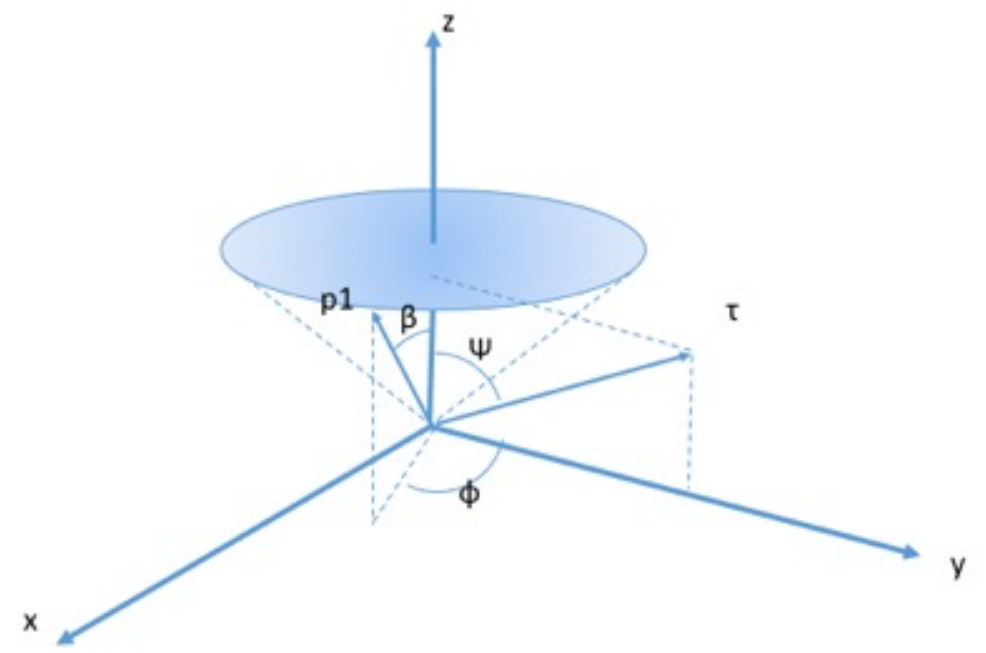

Fig. 190: Kinematics of $\tau \to K\pi\nu$

*15.3.2. $CP$ violation measurement in $\tau \to P_1P_2\nu$ and determination of the hadronic form factors .* As mentioned in the previous section, $CP$ violation in two hadron final states occurs due to the interference between vector and (pseudo-)scalar currents that are accompanied by a strong phase. This amplitude is theoretically very well described by the form factors, which can be extracted from the same measurement. In this section we give a short account of the properties of these amplitudes and of the potential progress which could be achieved at Belle II.

Decays into two light pseudoscalar mesons $\tau^- \to P_1^- P_2^0 \nu_\tau$ with $P_i = \pi, K, \eta$ represent approximately 27% of the hadronic tau decays. Up to electromagnetic corrections, the decay amplitudes have the Fermi form

$$\mathcal{T} = -\frac{G_F}{\sqrt{2}} V_{uq}\, \bar{u}_\tau \gamma^\mu (1-\gamma^5) u_\nu \langle P_1^-(q_1) P_2^0(q_2) |\, \bar{\psi}_q \gamma_\mu \psi_u |0\rangle \tag{516}$$

where $q$ is the $d$ or $s$ quark and $V_{uq}$ the corresponding CKM matrix element. The four-momenta $q_1$, $q_2$ are those of $P_1$ and $P_2$ in the laboratory frame[61].

[61] Note that the formulae given in Ref. [1549] assume that the laboratory frame coincides with the $e^+e^-$ CM frame, which is not the case at Belle II.

The hadronic matrix element involves the vector current only due to parity conservation in QCD and can be expressed in terms of two form factors,

$$\langle P_1(q_1)P_2(q_2)|j_\mu^{qu}|0\rangle = C^{12}\big\{(q_1+q_2)_\mu\, f_-^{12}(Q^2) + (q_1-q_2)_\mu\, f_+^{12}(Q^2)\big\} \tag{517}$$

where $Q^2=(q_1+q_2)^2$ and the overall normalisation factor $C^{12}$ reads: $C^{\pi^-\pi^0}=\sqrt{2}$, $C^{\pi^-\bar{K}^0}=-1$, $C^{\pi^-\eta}=\sqrt{2}$, $C^{K^-\pi^0}=1/\sqrt{2}$, $C^{K^-\eta}=\sqrt{3/2}$, $C^{K^-\bar{K}^0}=-1$. It is convenient to introduce the scalar form factor $f_0^{12}(s)$ as the following combination

$$f_0^{12}(Q^2) = f_+^{12}(Q^2) + \frac{Q^2}{\Delta_{12}} f_-^{12}(Q^2)\ , \quad \Delta_{12} = M_1^2 - M_2^2 \tag{518}$$

such that at $Q^2=0$, $f_0^{12}(0)=f_+^{12}(0)$. Expressing the two kinematical factors in eq. (517) in the centre-of-mass frame of the meson pair one easily sees that $f_+^{12}(s)$ is associated with a pair in a $J=1$ angular-momentum state, it is thus the vector form factor while the scalar form factor $f_0^{12}$ is associated with pair in a $J=0$ state.

The differential $\tau^-$ decay width reads

$$\begin{aligned}\frac{d\Gamma^{\tau^-}}{d\cos\alpha dW} &= \frac{(C^{12}G_F V_{uq})^2}{128\pi^3}\left(\frac{m_\tau^2}{Q^2}-1\right)^2 \frac{q_{cm}}{m_\tau} \\ &\times \Bigg\{|f_0^{12}(Q^2)|^2\Delta_{12}^2 + 4|f_+^{12}(Q^2)|^2 q_{cm}^2 Q^2\left(\frac{Q^2}{m_\tau^2}+(1-\frac{Q^2}{m_\tau^2})\cos^2\alpha\right) \\ &\quad -4\mathrm{Re}[f_0^{12}(Q^2)f_+^{12}(Q^2)^*]\Delta_{12}\, q_{cm}\sqrt{Q^2}\cos\alpha\Bigg\}\end{aligned} \tag{519}$$

where $W=\sqrt{Q^2}$ and $q_{cm}$ is the momentum in the hadronic center of mass system, $q_{cm}^2 = ((Q^2)^2 - 2Q^2(M_1^2+M_2^2)+\Delta_{12}^2)/4Q^2$. The decay widths integrated over $\cos\alpha$ are usually dominated by the vector form factor so that little is known at present on the scalar form factors from an experimental point of view. If one could measure the dependence on $\cos\alpha$, we would obtain more precise values of the two form factors and furthermore, their relative phase which is of particular interest in relation to the $CP$ violation as shown below.

As has been pointed out in Ref. [1556], the interference of the vector and scalar form factor may be obtained using the forward-backward asymmetry:

$$\begin{aligned}A_{\rm FB}^{\tau^-}(Q^2) &= \frac{\int_0^1 \left[d\Gamma^{\tau^-}(\cos\alpha) - d\Gamma^{\tau^-}(-\cos\alpha)\right] d\cos\alpha}{\int_0^1 \left[d\Gamma^{\tau^-}(\cos\alpha) + d\Gamma^{\tau^-}(-\cos\alpha)\right] d\cos\alpha} \\ &= \frac{-2\mathrm{Re}[f_0^{12}(Q^2)f_+^{12}(Q^2)^*]\Delta_{12}\, q_{cm}\sqrt{Q^2}}{|f_0^{12}(Q^2)|^2\Delta_{12}^2 + 4|f_+^{12}(Q^2)|^2 q_{cm}^2 Q^2\left(\frac{Q^2}{m_\tau^2}+(1-\frac{Q^2}{m_\tau^2})\frac{1}{3}\right)}\end{aligned} \tag{520}$$

where $\phi$ in $\cos\alpha$ can be integrated (see Eq. (515)) so that in practice, $\cos\alpha = \cos\beta\cos\psi$. Note that the $A_{\rm FB}^{\tau^-}$ is related to the average weighted by $\cos\alpha$, $\langle\cos\alpha\rangle_{\tau^-}$,

$$\begin{aligned}\langle\cos\alpha\rangle_{\tau^-} &\equiv \frac{\int_{-1}^1 \cos\alpha d\Gamma^{\tau^-}(\cos\alpha)d\cos\alpha}{\int_{-1}^1 d\Gamma^{\tau^-}(\cos\alpha)d\cos\alpha} \\ &= \frac{2}{3}A_{\rm FB}^{\tau^-}(Q^2)\end{aligned} \tag{521}$$

In the presence of a charged Higgs boson exchange, the QCD scalar form factor is modified in the following way,

$$f_0^{12}(Q^2) \to f_0^{12}(Q^2)\left(1+\eta_s\frac{Q^2}{m_H^2}\right) . \tag{522}$$

As we expect $\left|\eta_s\frac{Q^2}{m_H^2}\right| \ll 1$, the phase difference between the vector and the scalar form factor can still be obtained by $A_{\rm FB}$ within a good approximation.

A new physics contribution as in Eq. (522) can induce $CP$ violation, which can be measured with the $CP$ violating observable discussed in the introduction. It can be given as (see *e.g.* [1557, 1558])

$$\begin{aligned} A_{\rm CP}(Q^2) &= A_{\rm FB}^{\tau^-}(Q^2) - A_{\rm FB}^{\tau^+}(Q^2) \\ &= \frac{3}{2}(\langle\cos\alpha\rangle_{\tau^-} - \langle\cos\alpha\rangle_{\tau^+}) \end{aligned} \tag{523}$$

The decay rate for $\tau^+$ can be obtained by $\eta_s \to \eta_s^*$ in Eq. (519). Thus, we find

$$A_{\rm CP}(Q^2) = \frac{4{\rm Im}[f_0^{12}(Q^2)f_+^{12}(Q^2)^*]{\rm Im}[\eta_s\frac{Q^2}{m_H^2}]\Delta_{12}\, q_{cm}\sqrt{Q^2}}{|f_0^{12}(Q^2)|^2\Delta_{12}^2 + 4|f_+^{12}(Q^2)|^2 q_{cm}^2 Q^2\left(\frac{Q^2}{m_\tau^2} + (1-\frac{Q^2}{m_\tau^2})\frac{1}{3}\right)} \tag{524}$$

where we neglected the subdominant charged Higgs contribution in the denominator. We can see that having the hadronic form factor information, we can determine the $CP$ violating parameter $\eta_s\frac{Q^2}{m_H^2}$ from this formula. Thus, obtaining the vector and scalar form factors is important.

It should be noted that $A_{\rm CP}(Q^2)$ is a T-even observable so that the weak phase can be observed only when there is a strong phase. Fortunately, the strong phase in $\tau \to P_1P_2\nu$ process is expected to be relatively large and it can be well defined theoretically. Recently, a model-independent parametrisation of the form factors for this decay based on dispersion relations have been introduced (see *e.g.* Ref. [1559]). They take into account final-state interactions, as has been done in the case of the kaon decays [1560]. We will review this method in the following text.

*Vector form factors.* In the following we recall some properties of the form factors in QCD. Of particular importance is the property of analyticity, which follows from confinement. The $\pi\pi$ vector form factor, in particular, is of great interest in connection with a precise evaluation of the hadronic contributions to the $g-2$ of the muon. Analyticity-based descriptions provide improved extrapolations of the experimental data below the $\pi\pi$ threshold region. Both the vector and scalar form factors can be defined as analytic functions of the energy variable with a right-hand cut along the real axis and they have the property of real analyticity, *i.e.* $f^{12}(z^*) = f^{12*}(z)$ (see *e.g.* [1561] for a review and also Sec. 7.4). This implies that one can express the form factors as a phase dispersive representation,

$$f^{12}(Q^2) = P_N(Q^2)\Omega^{12}(Q^2), \tag{525}$$

with

$$\Omega^{12}(Q^2) = \exp\left[\frac{Q^2}{\pi}\int_{s_0}^{\infty} ds'\,\frac{\phi^{12}(s')}{s'(s'-Q^2)}\right] \tag{526}$$

where $\phi^{12}(s')$ is the phase of the form factor and $P_N$ is a polynomial. In QCD, we expect the form factors to vanish at infinity (see Ref. [1562] for a review)

$$f^{12}(Q^2) \sim \alpha_s(Q^2)/Q^2, \tag{527}$$

which constrains the degree $N$ of the polynomial and the value of the phase at infinity to satisfy

$$N = \frac{\phi^{12}(\infty)}{\pi} - 1 \,. \tag{528}$$

The phase representation is effective for the $\pi\pi$ or $\pi K$ form factors as Watson's theorem relates the phase of the form factor to the scattering phase shift of $\pi\pi$ or $\pi K$ in a finite energy range where the scattering is elastic. As an illustration of this, a description of $f_+^{\pi\pi}$ of the following form was proposed [1563]

$$f_+^{\pi\pi}(Q^2) = \Omega_1(Q^2)\, \Omega_{in}(Q^2) \tag{529}$$

where $\Omega_1$ is the Omnès function associated with $I = J = 1$ $\pi\pi$ phase-shift $\delta_1^1$ (recent parametrisations of $\delta_1^1$, constrained by the Roy equations can be found in [116, 117]). The second term, $\Omega_{in}$, takes into account the effect of the effective onset of inelasticity close to the $\omega\pi$ threshold via a simple polynomial of a conformally mapped variable (see [1563]). Similar types of parameterisations can be used for the $\pi K$ vector form factor: see, *e.g.*, Ref. [1564] where such analytic representations are used for combining $\tau \to K\pi\nu$ and $K_{l3}$ data in order to derive and improve determination of $V_{us} f_+^{K\pi}(0)$.

The $\tau \to \pi\eta\nu$ mode belongs to a category of so-called "second class current" processes, which are suppressed by isospin breaking effects. The values of the form factors at $Q^2 = 0$ are proportional to the isospin breaking quark mass ratio,

$$f_+^{\eta\pi}(0) = f_0^{\eta\pi}(0) = \frac{\sqrt{3}(m_d - m_u)}{4(m_s - (m_d + m_u)/2)}(1 + O(m_q)) \tag{530}$$

where the NLO chiral corrections can be found in Ref. [1565]. In the region of the $\rho(770)$ resonance, the discontinuity of the $\eta\pi$ vector form factor is dominated by the $\eta\pi \to \pi\pi$ amplitude which is well constrained by a number of recent experiments on $\eta \to 3\pi$ decays. This information was used in Ref. [1566] to provide a quantitative estimate of the form factor (see Ref. [1567] for an update and a list of references).

*Scalar form factors.* The Omnès phase representation is particularly useful for the scalar form factors because meson-meson interactions with $J = 0$ may contain broad resonances (like the "$\kappa$") or no resonances at all. This is the case for the $\pi^-\pi^0$ scalar form factor, which we first consider:

**a) $\pi^-\pi^0$:**

$\pi^-\pi^0$ scattering in the $S$-wave corresponds to the $I = 2$ isospin. Ignoring inelasticity effects in this channel we can identify the phase of the form factor with the scattering phase $\delta_0^2$ and use the following Omnès representation for the form factor

$$f_0^{\pi\pi}(Q^2) = \exp\left[\frac{Q^2}{\pi}\int_{4m_\pi^2}^{\infty} ds' \frac{\delta_0^2(s')}{s'(s' - Q^2)}\right] \,. \tag{531}$$

Parameterisations for the phase-shift $\delta_0^2$ can be found in Refs. [116, 117]. Note that $f_0^{\pi\pi}$ is not particularly suppressed but its influence on the decay width is, due to the multiplicative $M_{\pi^+}^2 - M_{\pi^0}^2$ factor. Equation 531 should provide a better estimate for $f_0^{\pi\pi}$ than that used in Ref. [1568] in their search for $CP$ violation in the $\tau \to \pi\pi\nu$ channel. The parameter $f_0^{\pi\pi}$ induces a forward-backward asymmetry, which is visible in a small energy range close to the threshold, see Ref. [1567].

**b)** $K^-\pi^0$, $\bar{K}^0\pi^-$

The phase of the $K\pi$ scalar form factors corresponds to $K\pi$ scattering with $I = 1/2$ in the elastic region. Experimental measurements of the scattering phase-shift (*e.g.* [1569] and references therein) have been performed. The phase representation encodes the influence of the broad $K_0^*(800)$ resonance and that of the more conventional $K_0^*(1430)$. Furthermore, the $K\pi$ scalar form factor obeys the Dashen-Weinstein chiral constraint [1570],

$$f_0^{K\pi}(m_K^2 - m_\pi^2) = \frac{F_K}{F_\pi} + O\left(\frac{m_\pi^2}{\Lambda^2}\right) \tag{532}$$

where the leading $O(m_\pi^2)$ corrective term was computed in Ref. [1571]. Results for $f_0^{K\pi}$ at $s = 0$ are available from lattice QCD (see [140] ). Both constraints can be encoded in the dispersive representation. Inelastic scattering for $J = 0$ $K\pi$ was shown to remain small below the $K\eta'$ threshold and can be approximated by a two-channel $T$-matrix in a range suitable for $\tau$ decay. This was used in Ref. [1572] to derive the $K\pi$ scalar form factor from an Omnès matrix. This model should be valid over most of the $\tau$ decay range. At present, some evidence for $f_0^{K\pi}$ was observed at Belle [1573] below 800 MeV. Keeping track of the $\cos\beta\cos\psi$ dependence, substantial improvements can be achieved at Belle II.

**c)** $\pi^-\eta$

The $S$-wave $\pi\eta$ scattering phase shift has not been measured but a qualitative picture emerges for its behaviour around 1 GeV from Flatté-type parameterisations of a number of production processes (*e.g.* [1574]) while the behaviour at low energy is constrained by chiral symmetry [1575]. Some progress in determining the phase shift from lattice simulations has also been achieved recently [1439]. The influence of the scalar form factor on the $\tau \to \pi\eta\nu$ branching fraction is usually estimated to be of comparable size to that of the vector form factor, or even larger, (*e.g.* [1567] and references therein). Measuring this mode with a number of energy bins would shed new light on the properties of the $\eta\pi$ interaction as well as the quark content properties of the $a_0(980)$ scalar resonance (via the size of its coupling to the $\bar{u}d$ operator).

*15.3.3. CP violation measurements with angular observables.* Belle searched for $CP$ violation, $A_{\rm CP}$, in angular observables of the mode $\tau^\pm \to K_S^0\pi^\pm\nu_\tau$ using a 699 fb$^{-1}$ data sample [1576]. The $K_S^0\pi^\pm$ invariant mass distribution of the $3.2 \times 10^5$ $\tau^\pm \to K_S^0\pi^\pm\nu_\tau$ signal candidates is shown in Fig. 191 (left). We see clearly additional resonance structures on top of the $K^*(890)$. It is most important to first understand these resonance structures.

The measured $CP$ asymmetry $A^{\rm CP}$ is shown in Fig. 191-(right) as a function of $K_S^0\pi^\pm$ invariant mass after correcting the known detector effects. At Belle, almost all contributions to the systematic uncertainty come from the detector bias on $A^{CP}$ (See Table II in Ref. [1576]), was evaluated using a control sample in which one tau decays into three charged pions and a neutrino, and the pions are not daughters of a $K_S^0$. Since this source of the systematic uncertainty depends on the statistics of the control sample, it is expected that the systematic uncertainty of the Belle II analysis will follow the integrated luminosity of the data sample. Thus, with an 50 ab$^{-1}$ data sample, we can expect around $\sqrt{70}$ times improvement for both the statistical and systematic uncertainties, *i.e.*, $|A^{CP}| < (0.4 - 2.6) \times 10^{-4}$ at 90% confidence level depending on $M_{K\pi}$ under the assumption that the central value of $|A^{CP}|$ is zero. It should be emphasised that at such a high precision, the extraction of the $CP$ parameter $\eta_s$ has to be done in conjunction with the form factor determination.

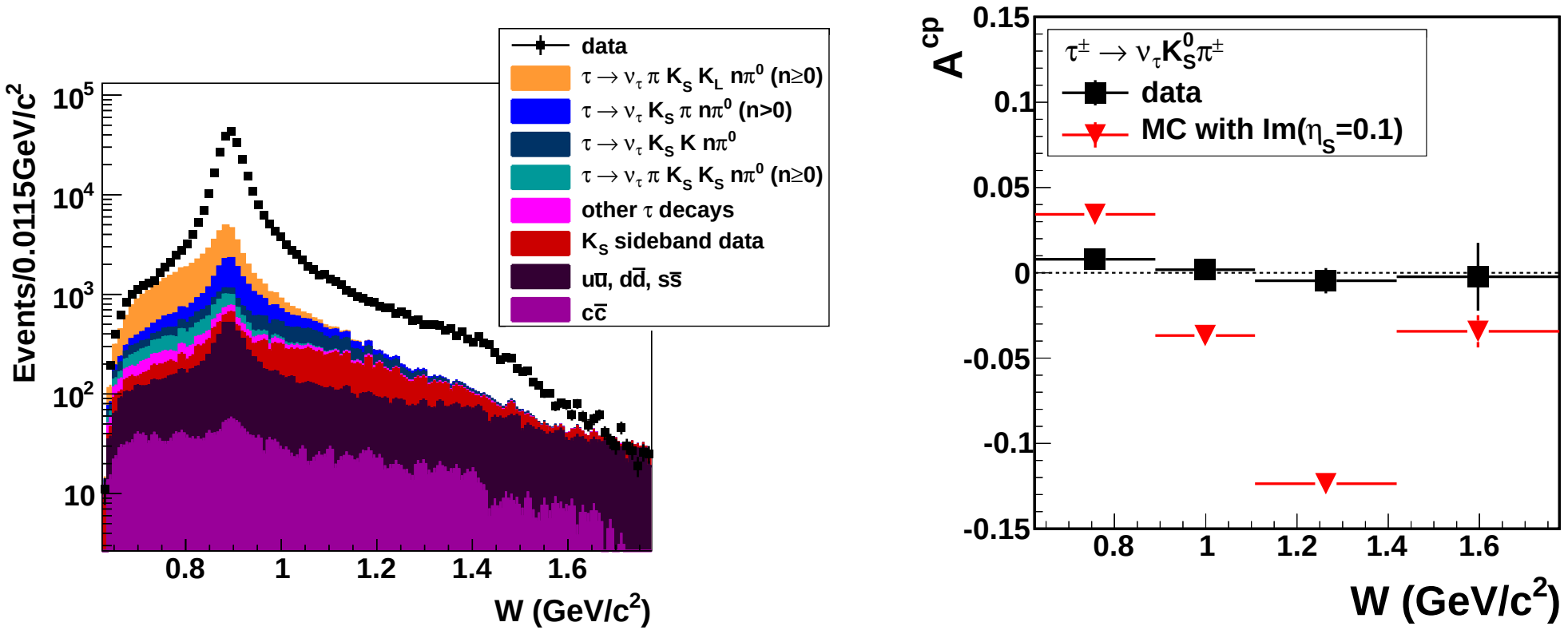

Fig. 191: Left: Invariant mass spectrum of the $K_S^0\pi^\pm$ system in $\tau \to K_S^0\pi^\pm\nu_\tau$ candidates [1576]. Right: Measured $CP$-violating asymmetry $A^{\rm CP}$ as a function of the $K_S^0\pi^\pm$ invariant mass $W$ after subtraction of background (black squares) [1576]. The inverted red triangles show the expected asymmetry when $\Im(\eta_S) = 0.1$ (to obtain this prediction, a linear combination of Breit-Wigner shapes of the vector resonances $K^*(890)$ and $K^*(1410)$ and the scalar resonances $K_0^*(800)$ and $K_0^*(1430)$ are used [1576]). At Belle II, this $CP$ asymmetry is expected to be improved by nearly a factor of $\sqrt{70} = 8.4$.

Finally, we comment on the prospect of the $CP$ violation measurement with an angular analysis of the three hadron final state. The most promising channel is $\tau \to K\pi\pi\nu$. The $CP$ violation search can be performed by using the three observables, $A^{\rm CP}_{(i)}$ $(i = 1, 2, 3)$ (see [1554] for detail). Note that this analysis was not done at Belle. To maximise the sensitivity of $A^{\rm CP}_{(i)}$ $(i = 1, 2, 3)$, the $M_{K\pi\pi}$, $M_{K\pi}$ and $M_{\pi\pi}$ distributions should all be measured. Similarly to the analysis of $\tau \to K_S^0\pi\nu$, the $\tau \to \pi\pi\pi\nu$ channel can be used as a controlled sample to evaluate the detector bias by assuming the CP violation in $\tau \to \pi\pi\pi\nu$ is much smaller than that in $\tau \to K\pi\pi\nu$. In this analysis, $\tau \to \pi\pi\pi\nu$ is also a major background since it has an almost 30 times larger branching fraction than that of $\tau \to K\pi\pi\nu$ and, due to the incompleteness of the particle identification, $\tau \to \pi\pi\pi\nu$ may enter into the $\tau \to K\pi\pi\nu$ sample. Since at Belle II the probability of misidentifying a $\pi$ as a $K$ will be reduced to around a half of that at Belle, we can expect a higher purity $\tau \to K\pi\pi\nu$ sample.

## *15.4. Other $\tau$ measurements*

*15.4.1. Leptonic $\tau$ decays: Michel parameter determination.* *(Contributing author: H. Aihara, D. Epifanov, J. Sasaki, N. Shimizu)* In the Standard Model (SM), $\tau$ decays proceed via the charged weak interaction, which is described by the exchange of a $W^\pm$ with a pure vector coupling to only left-handed fermions. Decays of $\tau$ leptons, such as $\tau^- \to \ell^-\bar{\nu}_\ell\nu_\tau$, $\tau^- \to \ell^-\bar{\nu}_\ell\nu_\tau\gamma$ and $\tau^- \to \ell^-\ell'^+\ell'^-\bar{\nu}_\ell\nu_\tau$ $(\ell, \ell' = e, \mu)$, are of special interest as electroweak couplings in these decays can be probed without disturbance from the strong interaction.

This makes them an ideal system to study the Lorentz structure of the charged weak current. Assuming left-handed neutrinos, the most general, Lorentz invariant, derivative-free and lepton-number-conserving four-lepton point interaction Lagrangian can be written as [1577]:

$$\mathcal{L} = \frac{4G_F}{\sqrt{2}} \sum_{\substack{N=S,V,T \\ i,j=L,R}} g_{ij}^N \Big[\bar{\Psi}_i(l)\Gamma^N \Psi_n(\nu_l)\Big] \Big[\bar{\Psi}_m(\nu_\tau)\Gamma_N \Psi_j(\tau)\Big]; \tag{533}$$

$$\Gamma^S = 1,\ \Gamma^V = \gamma^\mu,\ \Gamma^T = \frac{i}{2\sqrt{2}}(\gamma^\mu\gamma^\nu - \gamma^\nu\gamma^\mu).$$

The $\Gamma_N$ matrices ($\gamma^\nu$ are Dirac matrices) define the properties of the two currents under a Lorentz transformation with $N = S, V, T$ for scalar, vector and tensor interactions, respectively. The indices $i$ and $j$ label the right- or left-handedness (R, L) of the charged leptons. Ten non-trivial terms are characterised by ten complex coupling constants $g_{ij}^N$; those with $g_{RR}^T$ and $g_{LL}^T$ are identically zero. In the SM, the only non-zero coupling constant is $g_{LL}^V = 1$, this property is also known as (V-A)⊗(V-A) Lorentz structure of the matrix element. As the couplings can be complex, with arbitrary overall phase, there are 19 independent parameters. The total strength is determined by the Fermi constant $G_F$, which constrains the coupling constants to be $|g_{ij}^S| \leq 2$, $|g_{ij}^V| \leq 1$ and $|g_{ij}^T| \leq 1/\sqrt{3}$.

In leptonic $\tau$ decay, $\tau^- \to \ell^-\bar{\nu}_\ell\nu_\tau$, where neutrinos are not detected and the spin of the outgoing charged lepton is not determined, only four Michel parameters $\rho$, $\eta$, $\xi$ and $\delta$ are experimentally accessible. They are bilinear combinations of the $g_{ij}^N$ coupling constants [1578] and appear in the predicted energy spectrum of the charged lepton. In the $\tau$ rest frame, neglecting radiative corrections, this spectrum is given by [914]:

$$\frac{d\Gamma(\tau^\mp)}{dx\, d\Omega_\ell} = \frac{4G_F^2 m_\tau E_{\max}^4}{(2\pi)^4}\sqrt{x^2 - x_0^2}\Bigg(x(1-x) + \frac{2}{9}\rho(4x^2 - 3x - x_0^2) + \eta x_0(1-x)$$
$$\mp \frac{1}{3}P_\tau \cos\theta_\ell \xi \sqrt{x^2 - x_0^2}\Big[1 - x + \frac{2}{3}\delta\big(4x - 4 + \sqrt{1 - x_0^2}\big)\Big]\Bigg),$$
$$x = E_\ell/E_{\max},\quad E_{\max} = m_\tau\big(1 + m_\ell^2/m_\tau^2\big)/2,\quad x_0 = m_\ell/E_{\max},$$

where $P_\tau$ is the $\tau$ polarisation, $\Omega_\ell$ is the solid angle of the outgoing lepton and $\theta_\ell$ is the angle between the $\tau$ spin and the lepton momentum. In the SM, the (V-A) charged weak current is characterised by $\rho = 3/4$, $\eta = 0$, $\xi = 1$ and $\delta = 3/4$. In the radiative leptonic decay, $\tau^- \to \ell^-\bar{\nu}_\ell\nu_\tau\gamma$, three additional parameters, $\bar{\eta}$, $\eta''$ and $\xi\kappa$ can be extracted [1579]. Michel formalism for the five-lepton $\tau$ decay, $\tau^- \to \ell^-\ell'^+\ell'^-\bar{\nu}_\ell\nu_\tau$, can be found in the recent paper [1580].

Measurement of $\xi$, $\delta$ and $\xi\kappa$ parameters requires knowledge of the $\tau$ spin direction. In experiments at $e^+e^-$ colliders with unpolarised $e^\pm$ beams, the average polarisation of a single $\tau$ is zero. However, spin-spin correlations between the $\tau^+$ and $\tau^-$ produced in the reaction $e^+e^- \to \tau^+\tau^-$ can be exploited [1581]. The main idea of the method is to consider events where both taus decay to the selected final states. One tau decays to the signal mode while the opposite tau, which decays via $\tau^+ \to \pi^+\pi^0\bar{\nu}_\tau$, serves as a spin analyser. We choose the $\tau^+ \to \pi^+\pi^0\bar{\nu}_\tau$ (it is characterised by $\xi_\rho$ parameter, in the SM $\xi_\rho = 1$) decay mode because it has the largest branching fraction as well as properly studied dynamics. To write the total differential cross section we follow the approach developed in Refs. [1582, 1583]. The differential cross section is used to construct the probability density function (PDF), and Michel parameters are extracted in the unbinned maximum likelihood fit of the selected events.

Further, we describe the construction of the PDF for ordinary leptonic decay. The total differential cross section for the $(\tau^- \to \ell^- \bar{\nu}_\ell \nu_\tau;\ \tau^+ \to \pi^+\pi^0\bar{\nu}_\tau)$ (or, briefly, $(\ell^-;\ \rho^+)$) events in the full nine-dimensional phase space, $\frac{d\sigma}{d\vec{z}}(\vec{\Theta})$ ($\vec{\Theta} = \{1,\ \rho,\ \eta,\ \xi_\rho\xi,\ \xi_\rho\xi\delta\}$, $\vec{\Theta}_{\rm SM} = \{1,\ 3/4,\ 0,\ 1,\ 3/4\}$), is used to construct the PDF for the measurement vector $\vec{z} = \{p_\ell,\ \cos\theta_\ell,\ \phi_\ell,\ p_\rho,\ \cos\theta_\rho,\ \phi_\rho,\ m_{\pi\pi},\ \cos\tilde{\theta}_\pi,\ \tilde{\phi}_\pi\}$. The main background processes, $(\ell^-;\ \pi^+\pi^0\pi^0)$, $(\pi^-;\ \rho^+)$ and $(\rho^-;\ \rho^+)$ with the fractions, $\lambda_{3\pi}$, $\lambda_\pi$ and $\lambda_\rho$, respectively, are included in the PDF analytically. The remaining background with the fraction $\lambda_{\rm other}$ is described by the $\mathcal{P}^{\rm MC}_{\rm bg}(\vec{z})$ PDF, which is evaluated from the large Monte Carlo sample. The total PDF for the $(\ell^-;\ \rho^+)$ events is written as:

$$\begin{aligned}\mathcal{P}(\vec{z}) = \quad & \frac{\varepsilon(\vec{z})}{\bar{\varepsilon}}\Bigg((1-\lambda_{3\pi}-\lambda_\pi-\lambda_\rho-\lambda_{\rm other})\frac{S(\vec{z}\,|\,\vec{\Theta})}{\int\frac{\varepsilon(\vec{z})}{\bar{\varepsilon}}S(\vec{z}\,|\,\vec{\Theta})d\vec{z}} \\ & +\lambda_{3\pi}\frac{B_{3\pi}(\vec{z}\,|\,\vec{\Theta})}{\int\frac{\varepsilon(\vec{z})}{\bar{\varepsilon}}B_{3\pi}(\vec{z}\,|\,\vec{\Theta})d\vec{z}}+\lambda_\pi\frac{B_\pi(\vec{z})}{\int\frac{\varepsilon(\vec{z})}{\bar{\varepsilon}}B_\pi(\vec{z})d\vec{z}}+\lambda_\rho\frac{B_\rho(\vec{z})}{\int\frac{\varepsilon(\vec{z})}{\bar{\varepsilon}}B_\rho(\vec{z})d\vec{z}}+\lambda_{\rm other}\mathcal{P}^{\rm MC}_{\rm bg}(\vec{z})\Bigg),\end{aligned} \tag{534}$$

where $S(\vec{z}\,|\,\vec{\Theta})$, $B_{3\pi}(\vec{z}\,|\,\vec{\Theta})$, $B_\pi(\vec{z})$ and $B_\rho(\vec{z})$ are the cross sections for the $(\ell^-;\ \rho^+)$, $(\ell^-;\ \pi^+\pi^0\pi^0)$, $(\pi^-;\ \rho^+)$ and $(\rho^-;\ \rho^+)$ events, respectively; $\varepsilon(\vec{z})$ is the detection efficiency for signal events in the full phase space; and $\bar{\varepsilon} = \int\varepsilon(\vec{z})S(\vec{z}\,|\,\vec{\Theta}_{\rm SM})d\vec{z}/\int S(\vec{z}\,|\,\vec{\Theta}_{\rm SM})d\vec{z}$ is an average signal detection efficiency. There are several corrections that must be incorporated into the procedure to take into account real experimental situation. Physics corrections include electroweak higher-order corrections to the $e^+e^- \to \tau^+\tau^-$ cross section [1584, 1585]. Apparatus corrections include the effect of the finite detection efficiency and resolution, the effect of the external bremsstrahlung for $(e^-;\ \rho^+)$ events, and the $e^\pm$ beam energy spread.

A study of Michel parameters $\rho$, $\eta$, $\xi_\rho\xi$ and $\xi_\rho\xi\delta$ in leptonic $\tau$ decays using a 485 fb$^{-1}$ data sample collected at Belle showed that the statistical uncertainties of Michel parameters are already of the order of $10^{-3}$, see Table 141 [1586]. Although systematic uncertainties coming from the physical and apparatus corrections as well as from the normalisation are below 1%, there are still relatively large systematic uncertainties from the experimental corrections to the detection efficiency. Currently, the largest contribution, $(1 \div 3)$%, comes from the trigger efficiency correction. The expected statistical uncertainties of Michel parameters at Belle-II (with the total planned luminosity integral of 50 ab$^{-1}$) are already of the order of $10^{-4}$. At

Table 141: Statistical uncertainties of Michel parameters, $\rho$, $\eta$, $\xi_\rho\xi$, $\xi_\rho\xi\delta$, $\bar{\eta}$ and $\xi_\rho\xi\kappa$ in ordinary and radiative leptonic $\tau$ decays [1587] at Belle (with a 485 fb$^{-1}$ (703 fb$^{-1}$) data sample for ordinary (radiative) leptonic decays).

| Exp. | $(\Delta\rho)_{\rm stat}$, $10^{-4}$ | $(\Delta\eta)_{\rm stat}$, $10^{-4}$ | $(\Delta\xi_\rho\xi)_{\rm stat}$, $10^{-4}$ | $(\Delta\xi_\rho\xi\delta)_{\rm stat}$, $10^{-4}$ | $(\Delta\bar{\eta})_{\rm stat}$ | $(\Delta\xi_\rho\xi\kappa)_{\rm stat}$ |
|---|---|---|---|---|---|---|
| Belle | 13 | 62 | 39 | 25 | 1.5 | 0.4 |

Belle II the systematic uncertainties will be the dominant ones. To improve them, a high- and uniform-efficiency two-track trigger is needed.

*15.4.2. Searches for second class currents in $\tau$ decays.* *(Contributing authors: P. Roig, S. Eidelman)*

*Theory.* Hadronic currents can be classified according to their spin, parity and $G$-parity quantum numbers ($J^{PG}$) as [1588]: first class currents, with the quantum numbers $J^{PG} = 0^{++}(\sigma), 0^{--}(\pi), 1^{+-}(a_1), 1^{-+}(\rho)$; and second class currents (SCC), which have $J^{PG} = 0^{+-}(a_0), 0^{-+}(\eta), 1^{++}(b_1), 1^{--}(\omega)$, yet to be discovered. Mesons in brackets share $J^{PG}$ with the preceding current, yielding easily the simplest meson systems for a given class current.

$G$-parity combines charge and isospin symmetries. The latter is broken both by $m_u \neq m_d$ and $q_u \neq q_d$. Since these violations are small, $G$-parity is a good approximate symmetry of the strong interactions. Thus, within the SM and for definite $J^P$, hadron systems with $G$-parity corresponding to the weak left-handed (light-)quark current are allowed and easily produced. Those with the 'wrong' $G$-parity are suppressed and have SCC quantum numbers. Within the SM, a small violation of $G$-parity is induced by isospin breaking, giving rise to *induced* SCCs. In addition to this suppressed effect one may have *genuine* weak SCCs from unknown new physics, which may show up either in rates above the expectations coming from isospin violation or distinguished from the calculable SM background.

In principle SCCs could also be discovered in nuclear processes or in $\Sigma^{\pm}$ semileptonic decays [1588]. However, both face the challenge of separating possible violations of CVC from SCC effects [1589, 1590].

The discovery of either of the decays $\tau^- \to b_1^- \nu_\tau$ or $\tau^- \to a_0^- \nu_\tau$ would be an unambiguous signature of SCCs [1591]. Since $b_1$ decays dominantly to $\omega\pi$ and this final state can also be produced via ordinary first class current at a rate of $\sim 2\%$, angular analyses of the pions is needed to disentangle both types of currents. Resulting upper limits on SCCs are $Br \sim 1.4 \cdot 10^{-4}$ [1592], while Ref. [1593] roughly estimates $Br \sim 2.5 \cdot 10^{-5}$ based on spin-one meson dominance.

SCCs can also be searched through the $\tau^- \to \pi^- \eta \nu_\tau$ decays (not necessarily proceeding through $a_0$ exchange). In the SM, their suppressed amplitude can be understood in terms of the $\pi^0 - \eta$ mixing parameter $\epsilon_{\pi\eta}$ given by the value of Eq. (530) neglecting subleading $m_q$ corrections. Since $\epsilon_{\pi\eta} \sim 10^{-2}$, $Br(\tau^- \to \pi^- \eta \nu_\tau) \sim 10^{-5}$ is expected in the SM [1594].

Both form factors will contribute sizeably to the $\tau^- \to \pi^- \eta \nu_\tau$ decays ($m_\eta^2 >> m_\pi^2$). While their low-energy behaviour is determined by Chiral Perturbation Theory [1565], resonance dynamics is needed to describe them appropriately throughout the available phase space. Recently, the vector form factor contribution was estimated using $\eta \to 3\pi$ decay data $BR \sim 0.36 \cdot 10^{-5}$ [1566]. This was done using a dispersive approach [1567]. According to Ref. [1595], $\pi^-\eta$ vector form factor can be related to the very precisely measured $2\pi$ vector form factor [1596], which results in a negligible error in the corresponding prediction, $Br = (0.26 \pm 0.02) \cdot 10^{-5}$ [1595], in agreement with $[0.1, 0.4] \cdot 10^{-5}$ [1567].

The scalar form factor contribution is more involved theoretically. A phase dispersive representation of this form factor is supplemented with a sum rule constraint for the inelastic region and a realistic model for the phase-shift [1567]. The corresponding estimate for this contribution is $[0.1, 0.6] \cdot 10^{-5}$, versus $\sim 1.0 \cdot 10^{-5}$ [1566]. A coupled-channel dispersive analyses of the $\pi\eta - K\bar{K} - \pi\eta'$ channels within $U(3)$ Chiral Perturbation Theory with resonances [1597] determines $(1.41 \pm 0.09) \cdot 10^{-5}$ [1595]. Recent COMPASS data on the partial waves of the $\pi\eta^{(\prime)}$ system [1598] will help to check this uncertainty.

$Br(\tau^- \to \pi^-\eta'\nu_\tau) \in [0.2, 1.4] \cdot 10^{-6}$ [1599], $Br \in [10^{-7}, 10^{-6}]$ [1595] conclude scalar form factor dominance with an associated order-of-magnitude error in $Br$. Although BaBar fixed the impressive upper bound $Br < 4.0 \cdot 10^{-6}$ on these decays, the inaccuracy of the theory predictions does not allow to conclude if the first measurement of SCCs will correspond to $\tau^- \to \pi^-\eta'\nu_\tau$ decays.

With $Br(\tau^- \to \pi^-\eta\nu_\tau) \sim 1 \cdot 10^{-5}$, SCCs were not measured by BaBar or Belle because of the difficulty in controlling the associated backgrounds [2]. BaBar was able to set the upper limit $Br < 9.9 \cdot 10^{-5}$ [1600] while Belle determined the bound $Br < 7.3 \cdot 10^{-5}$ [1601]. Scaling the previous upper limits on both $Br(\tau^- \to \pi^-\eta^{(\prime)}\nu_\tau)$ according to Belle-II statistics should warrant the discovery of SCCs at Belle-II. New Physics can manifest through abnormally large branching fractions in either of them, but only $\tau^- \to \pi^-\eta\nu_\tau$ is predicted with enough accuracy to allow setting competitive restrictions [1602] on a possible charged Higgs exchange if the $BR$ is known with at least 20% accuracy [1567].

*Experimental status.* The most frequently discussed SCC decay mode is $\tau^- \to \eta\pi^-\nu_\tau$ for which theory predicts the branching ratio to be in the range $10^{-5} - 10^{-6}$. The smallness of the branching ratio makes its search very sensitive to various background processes, such as *e.g.*, that from $\tau^- \to \eta\pi^-\pi^0\nu_\tau$, which has a branching fraction $\sim 10^{-3}$, so that a missing $\pi^0$ mimics completely the decay looked for and thus produces the background very difficult to suppress. To understand such backgrounds better, Belle performed a high-statistics study of various exclusive decays that include an $\eta$ meson [1603].

In the BaBar search that used the $\eta \to \pi^+\pi^-\pi^0$ decay mode the above mentioned background dominates, however, other processes ($q\bar{q} + c\bar{c}$, $\eta K^0\pi^-\nu_\tau$, $\eta K^-\nu_\tau$) also give significant contributions, which in total are even larger than the first one. As a result, BaBar, with its much larger data sample than CLEO set an upper limit of $< 9.9 \times 10^{-5}$ [1600] only slightly improving an upper limit of $< 1.4 \times 10^{-4}$ from CLEO [1604], which used both $\eta \to \pi^+\pi^-\pi^0$ and $\eta \to \gamma\gamma$ decay modes. The latter decay mode looks more promising for future searches although serious backgrounds are still expected from $\tau^- \to \eta\pi^-\pi^0\nu_\tau$ and $\tau^- \to \pi^-\pi^0\nu_\tau$.

For the process $\tau^- \to \eta'\pi^-\nu_\tau$ theory predicts the branching ratio at the level of $10^{-6}$ [1565]. The background situation is better than for the previous decay and BaBar set an upper limit of $< 7.2 \times 10^{-6}$ [1605] improving by an order of magnitude that of $< 7.4 \times 10^{-5}$ from CLEO [1606].

The decay $\tau^- \to \omega\pi^-\nu_\tau$ is expected to proceed through the hadronic vector current mediated by the $\rho$, $\rho'$, $\rho''$ and higher excitations. If, however, second-class currents violating $G$-parity contribute to this decay, it can also proceed through a hadronic axial-vector current mediated, *e.g.*, by the $b_1(1235)$ resonance. The difference in spin-parity assignments for each of these states is reflected in different polarisations of the $\omega$ spin and hence in different expected angular distributions of $cos\chi$. The angle $\chi$ is defined as the angle between the normal to the $\omega$ decay plane and the direction of the fourth pion measured in the $\omega$ rest frame, and $l$ is the orbital angular momentum of the $\omega\pi$ system. The expected forms of the $cos\chi$ distribution are listed in the Table [1607].

*15.4.3. Measurement of the $\tau$ lepton mass.* *(Contributing author: S. Eidelman)* Mass is one of the most fundamental parameters of any particle and thus should be measured as accurately as possible. For the $\tau$ lepton this is particularly important since its width is proportional to the mass to the fifth power, so that any tests of Standard Model, *e.g.*, of leptonic universality, crucially depend on the mass value and its accuracy [1608].

Two methods of $\tau$ lepton mass measurement exist. In the threshold method one studies the energy dependence of the $\tau^+\tau^-$ production cross section in the energy range close to threshold: $\sqrt{s} - 2m_\tau \leq 200$ MeV. One can reach very high accuracy even with a limited data sample where the current world most precise result on the $\tau$ lepton mass belonging to BESIII has been obtained with about 1000 events only [1609]. Measurements of the $\tau$ lepton mass at higher accuracy with this method are limited by statistics, and the systematic uncertainties are due to the energy scale calibration and knowledge of the beam energy spread.

The $B$ factories, on the contrary, can collect much large data samples and have systematic uncertainties that differ from the threshold method. They are based on the so called pseudo-mass determination in which mass is estimated from the edge of the spectrum of invariant mass based on four-momenta of the detected hadrons - products of $\tau$ decay [1610]. This method allows for a separate determination of the mass for positive and negative $\tau$ leptons to be performed thus providing a test of $CPT$ invariance first realised by the OPAL Collaboration [1611]. Table 142 below summarised the current status of $\tau$ lepton mass measurements.

Table 142: Summary of $\tau$ lepton mass measurements.

| Group | $\sqrt{s}$, GeV | $N_{\text{ev}}$ | $m_\tau$, MeV |
|---|---|---|---|
| DELCO, 1978 | $3.1 - 7.4$ | 692 | $1783^{+3}_{-4}$ |
| ARGUS, 1992 | $9.4 - 10.6$ | $11k$ | $1776.3 \pm 2.4 \pm 1.4$ |
| BES, 1996 | $3.54 - 3.57$ | 65 | $1776.96^{+0.18+0.25}_{-0.21-0.17}$ |
| CLEO, 1997 | 10.6 | $98.5k$ | $1778.2 \pm 0.8 \pm 1.2$ |
| OPAL, 2000 | $\sim 90$ | $13.3k$ | $1775.1 \pm 1.6 \pm 1.0$ |
| KEDR, 2007 | $3.54 - 3.78$ | 81 | $1776.81^{+0.25}_{-0.23} \pm 0.15$ |
| Belle, 2007 | 10.6 | $\sim 400k$ | $1776.61 \pm 0.13 \pm 0.35$ |
| BaBar, 2009 | 10.6 | $\sim 682k$ | $1776.68 \pm 0.12 \pm 0.41$ |
| BESIII, 2015 | 3.54-3.60 | 1171 | $1776.91 \pm 0.12^{+0.10}_{-0.13}$ |
| PDG, 2016 | – | – | $1776.86 \pm 0.12$ |

In the Belle measurement the three most important sources of the systematic uncertainty were: beam energy and tracking system calibration (0.26 MeV), parameterisation of the spectrum edge (0.18 MeV) and limited MC statistics (0.14 MeV) [1612]. One expects that the two latter values will be greatly improved upon: the *ad hoc* parameterisation of the spectrum edge will be replaced with a theoretical spectrum directly following from the high-statistics measurement of the $\tau$ decay into the corresponding final state (usually $\tau^- \to \pi^-\pi^+\pi^-\nu_\tau$), which will be also used for the MC generators. The tracking system calibration should benefit from the larger data set of Belle II, whereas beam energy determination will improve following the progress achieved in the $B$ meson mass and $\Upsilon(4S)$ width measurements [88]. Very optimistically, one can hope to reach a level of (0.15-0.20) MeV for the total uncertainty on the mass making new $\tau$ lepton mass measurements an attractive independent test of threshold measurements and Standard Model in general.

*15.4.4. Electric Dipole Moment of the $\tau$. (Contributing authors: K. Hayasaka, E. Kou)* The current limit for the $\tau$ electric dipole moment (EDM) ($d_\tau$) is several orders of magnitude less restrictive than that for the electron, muon, or neutron. The difficulty of the $\tau$ EDM measurement comes from its short life time. Therefore, the $\tau$ EDM can not be measured in an electrostatic field. At a $e^+e^-$ collider, however, the $\tau$ EDM can be measured by using the correlation of decay product momenta in the process $e^+e^- \to \tau^+\tau^-$.

The matrix element for the process $e^+e^- \to \tau^+\tau^-$ is given by the sum of the SM term, $\mathcal{M}^2_{\rm SM}$, the EDM term, $|d_\tau|^2\mathcal{M}^2_{d^2}$, and the interference between them:

$$\mathcal{M}^2 = \mathcal{M}^2_{\rm SM} + {\rm Re}(d_\tau)\mathcal{M}^2_{\rm Re} + {\rm Im}(d_\tau)\mathcal{M}^2_{\rm Im} + |d_\tau|^2\mathcal{M}^2_{d^2}, \tag{535}$$

where ${\rm Re}(d_\tau)$ ( ${\rm Im}(d_\tau)$ ) is the real (imaginary) part of the EDM. These interference terms $\mathcal{M}^2_{\rm Re/Im}$ contain the following combination of spin-momentum correlations:

$$\begin{aligned}
\mathcal{M}^2_{\rm Re} &\propto (\boldsymbol{S}_+ \times \boldsymbol{S}_-)\cdot\hat{\boldsymbol{k}}, \quad (\boldsymbol{S}_+ \times \boldsymbol{S}_-)\cdot\hat{\boldsymbol{p}}, \\
\mathcal{M}^2_{\rm Im} &\propto (\boldsymbol{S}_+ - \boldsymbol{S}_-)\cdot\hat{\boldsymbol{k}}, \quad (\boldsymbol{S}_+ - \boldsymbol{S}_-)\cdot\hat{\boldsymbol{p}},
\end{aligned} \tag{536}$$

where $\boldsymbol{S}_\pm$ is a $\tau^\pm$ spin vector, and $\hat{\boldsymbol{k}}$ and $\hat{\boldsymbol{p}}$ are the unit vectors of the $\tau^-$ and $e^-$ momenta in the CM system, respectively. These terms are $CP$-odd since they change sign under a $CP$ transformation.

One could evaluate the value of the matrix elements if the values of $\boldsymbol{S}_\pm$ and $\hat{\boldsymbol{k}}$ could be measured on an event-by-event basis from the $\tau$-decay products. Although one can not know them completely due to missing neutrinos from $\tau$ decays, one can obtain the most probable values of $\boldsymbol{S}_\pm$ and $\hat{\boldsymbol{k}}$ by calculating approximate averages from measurements of the momenta of $\tau$ decay products. In the analysis of Belle, the method of optimal observables [1613] is employed. In this method, the observables $\mathcal{O}_{\rm Re}$ and $\mathcal{O}_{\rm Im}$

$$\mathcal{O}_{\rm Re} = \frac{\mathcal{M}^2_{\rm Re}}{\mathcal{M}^2_{\rm SM}}, \quad \mathcal{O}_{\rm Im} = \frac{\mathcal{M}^2_{\rm Im}}{\mathcal{M}^2_{\rm SM}}, \tag{537}$$

are evaluated using the most probable values of $\boldsymbol{S}_\pm$ and $\hat{\boldsymbol{k}}$. The means of $\mathcal{O}_{\rm Re}$, $\mathcal{O}_{\rm Im}$ are proportional to the EDM value and have maximal sensitivity. In order to obtain the maximal sensitivity, we measure as many modes as possible. For example in the Belle analysis with a 29.5 fb$^{-1}$ data sample [1614], the following 8 modes are used: $\tau^+\tau^- \to (e\nu_e\nu_\tau)(\mu\nu_\mu\nu_\tau)$, $(e\nu_e\nu_\tau)(\pi\nu_\tau)$, $(\mu\nu_\mu\nu_\tau)(\pi\nu_\tau)$, $(e\nu_e\nu_\tau)(\rho\nu_\tau)$, $(\mu\nu_\mu\nu_\tau)(\rho\nu_\tau)$, $(\pi\nu_\tau)(\pi\nu_\tau)$, $(\pi\nu_\tau)(\rho\nu_\tau)$, and $(\rho\nu_\tau)(\rho\nu_\tau)$.

The current mean values for ${\rm Re}(d_\tau)$ and ${\rm Im}(d_\tau)$ have been obtained by taking the weighted mean of 8 modes to be

$${\rm Re}(d_\tau) = (1.15 \pm 1.70) \times 10^{-17} e{\rm cm} \tag{538}$$

$${\rm Im}(d_\tau) = (-0.83 \pm 0.86) \times 10^{-17} e{\rm cm}. \tag{539}$$

The 95% C.L. intervals are

$$-2.2 \times 10^{-17} < \mathrm{Re}(d_\tau) < 4.5 \times 10^{-17} e\mathrm{cm} \quad , \tag{540}$$

$$-2.5 \times 10^{-17} < \mathrm{Im}(d_\tau) < 0.8 \times 10^{-17} e\mathrm{cm} \quad . \tag{541}$$

These limits are ten times more restrictive than previous experiments.

Now let us discuss the prospect for EDM and $g-2$ searches in $\tau$ decays at Belle II. In the $\tau$ EDM analysis, the statistical errors for $\mathrm{Re}(d_\tau)$ and $\mathrm{Im}(d_\tau)$ are expected to be proportional to an inverse of the square root of the integrated luminosity while the systematic error strongly depends on the understanding of MC and data samples since the dependence of the optimal observable for $d_\tau$ is evaluated with MC samples. In particular, the understanding of low momentum tracking is a large and important systematic uncertainty. Trigger, track-finding and PID efficiency systematics are, while slightly smaller, strongly correlated, demanding a nuanced approach to their evaluation. Their uncertainties are mostly data driven, and will continue to improve with more data. Therefore, we expect approximately a 40 times gain from the current result, *i.e.*, $|\mathrm{Re}, \mathrm{Im}(d_\tau)| < 10^{-18} - 10^{-19}$.

The tau $g-2$ can be evaluated in a way similar to that of the tau EDM, by giving $\frac{1}{2}\bar{\psi}\sigma^{\mu\nu}\psi\frac{ea_\tau}{2m_\tau}F_{\mu\nu}$ as a $g-2$ interaction term instead of $-\frac{1}{2}\bar{\psi}\sigma^{\mu\nu}\psi\frac{e\tilde{a}_\tau}{2m_\tau}\tilde{F}_{\mu\nu}$ in the case of tau EDM into the Lagrangian. However, it is expected that the sensitivity to tau $g-2$ will be worse than that of the tau EDM.

*15.4.5. Inclusive $\tau$ decays: $V_{us}$ and $\alpha_s$.* *(Contributing authors: M. Jamin, K. Maltman, E. Passemar, A. Pich)* Hadronic $\tau$ decays constitute a very interesting tool for studying Quantum Chromodynamics (QCD) and performing precise extractions of some of the fundamental parameters of the Standard Model. The most famous example is the determination of $\alpha_s(m_\tau)$, the strong coupling constant at the tau mass, and the test of the running of $\alpha_s$ from the tau mass $m_\tau$ to the $Z$ mass $M_Z$. Another example is the determination of the Cabibbo Kobayashi Maskawa (CKM) matrix element $V_{us}$ and the test of the unitarity of the first row of the CKM matrix. This was rendered possible by the measurement of not only the branching fractions of $\tau \to hadrons$, see figure 192, but also the experimental differential distributions with respect to the invariant squared-mass of the hadronic system, which generate information on the so-called spectral functions. The inclusive isovector, vector (V) and axial vector (A) spectral functions, and, with lower statistics, the inclusive flavour $us$ V+A spectral function sum, have been measured by ALEPH [1615–1618] and OPAL [1619, 1620], but not yet at the B factories. The ALEPH isovector distribution results are shown in figure 192. These results have triggered intense theoretical activities. Central observables for inclusive hadronic $\tau$ decays are the so-called $R_\tau$ ratio, and its differential version, $dR_\tau/ds$, with $s$ the invariant mass-squared of the hadronic system. $R_\tau$ is defined by

$$R_\tau = \frac{\Gamma\left(\tau^- \to \nu_\tau \text{ hadrons}^- (\gamma)\right)}{\Gamma\left(\tau^- \to \nu_\tau e^- \bar{\nu}_e (\gamma)\right)} \ . \tag{542}$$

The central theoretical object is the appropriate two-point correlation function of the colour-singlet vector $V^\mu_{ij} \equiv \bar{\psi}_j\gamma^\mu\psi_i$ or axial vector $A^\mu_{ij} \equiv \bar{\psi}_j\gamma^\mu\gamma_5\psi_i$ quark currents with $i, j = u, d, s$:

$$\Pi^{\mu\nu}_{ij}(q) = i\int d^4x\ e^{iqx}\left\langle 0 \left| T\left(\mathcal{J}^\mu_{ij}(x)\mathcal{J}^\nu_{ij}(0)^\dagger\right)\right| 0\right\rangle \ , \tag{543}$$

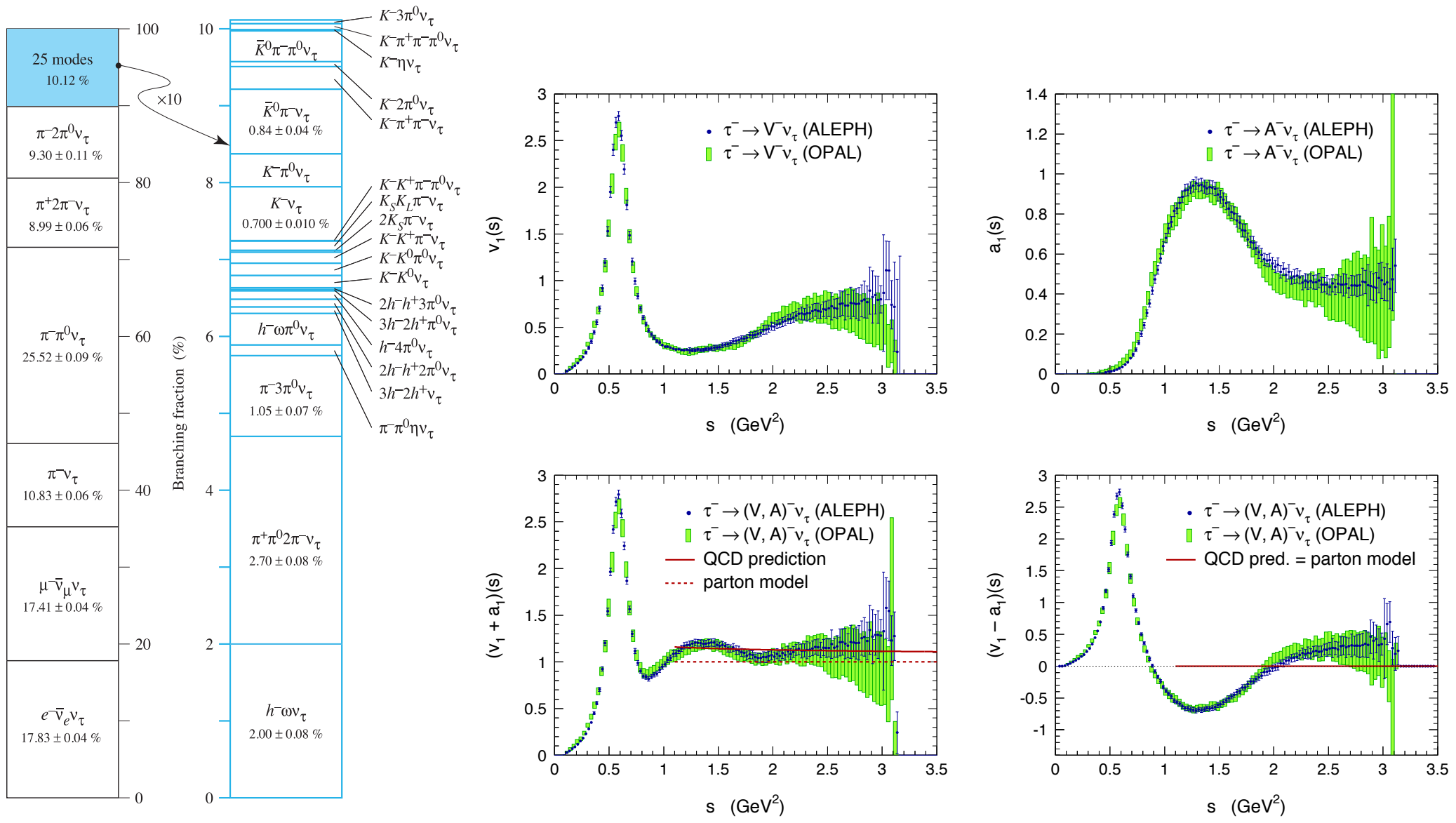

Fig. 192: Left: Measured tau branching fractions from the PDG [914]. Right: Spectral functions measured by ALEPH (blue data points [1615] and OPAL (green rectangular blocks) [1619] in the vector (upper-left) and axial (upper-right) non-strange channels. The lower plots show the sum (left) and difference (right) of the vector and axial spectral functions, together with the perturbative QCD predictions (continuous red lines). Figures taken from Ref. [1616]. A recent update for the ALEPH collaboration has been given in Ref. [1617].

with the current $\mathcal{J} = V, A$. The correlator has the Lorentz decomposition:

$$\Pi^{\mu\nu}_{ij,\mathcal{J}}(q) = \left(-g^{\mu\nu} q^2 + q^\mu q^\nu\right) \Pi^{(1)}_{ij,\mathcal{J}} + q^\mu q^\nu \Pi^{(0)}_{ij,\mathcal{J}} \ , \tag{544}$$

with $\Pi^{(0)}_{ij,\mathcal{J}}$ the longitudinal component and $\Pi^{(1)}_{ij,\mathcal{J}}$ the transverse component in the hadronic rest frame. Using the optical theorem, the width for the decay to hadrons mediated by the flavour $ij$ V or A current can be related to the imaginary part of the corresponding correlator, as shown schematically in figure 193. Explicitly, in the Standard Model, one has [1581]

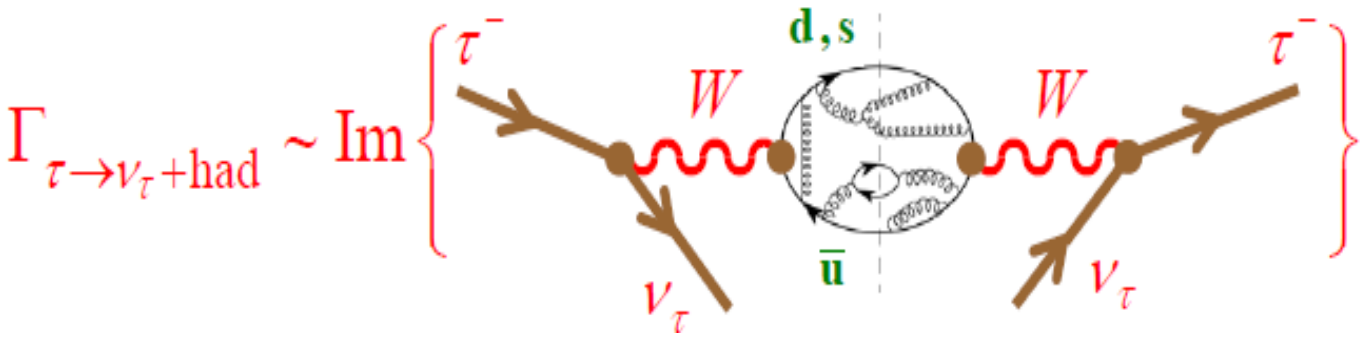

Fig. 193: Optical Theorem, figure from [1621].

$$R_\tau = 12\pi S_{EW} \int_0^{m_\tau^2} \frac{ds}{m_\tau^2} \left(1 - \frac{s}{m_\tau^2}\right)^2 \left[\left(1 + 2\frac{s}{m_\tau^2}\right) \text{Im}\Pi^{(1)}(\text{s}) + \text{Im}\Pi^{(0)}(\text{s})\right] \ , \tag{545}$$

where $S_{EW} = 1.0201 \pm 0.0003$ [1622] is the short-distance electroweak correction. The appropriate combination of correlators entering this equation is:

$$\Pi^{(J)}(s) = |V_{ud}|^2 \left(\Pi^{(J)}_{ud,V}(s) + \Pi^{(J)}_{ud,A}(s)\right) + |V_{us}|^2 \left(\Pi^{(J)}_{us,V}(s) + \Pi^{(J)}_{us,A}(s)\right) \ . \tag{546}$$

We can separate the inclusive contributions associated with specific quark currents as follows:

$$R_\tau = R_{\tau,V} + R_{\tau,A} + R_{\tau,S}\,, \tag{547}$$

where $R_{\tau,V}$ and $R_{\tau,A}$ correspond to the first two terms in Eq. (546), while $R_{\tau,S}$ contains the remaining Cabibbo-suppressed contributions. Non-strange vector and axial-vector hadronic $\tau$ decays can be distinguished experimentally, for the dominant n-pion decay modes, by counting the number of pions, with vector decays ($R_{\tau,V}$) producing an even number and axial vector decays ($R_{\tau,A}$) an odd number. Strange decays ($R_{\tau,S}$) are identified by the presence of an odd number of kaons in the final state. In principle, we need to calculate the correlator in Eq. (545) from $s = 0$ to $m_\tau^2$. Unfortunately, this is an energy region where QCD is non-perturbative, displaying clear resonances, as can been seen on figure 192, and a calculation is at present not possible. Nevertheless, the integral itself can be calculated systematically by exploiting the analytic properties of the correlators $\Pi^{(0+1)}(s)$ and $s\Pi^{(0)}(s)$, which are analytic functions of $s$ except along the positive real $s$ axis, where their imaginary parts have discontinuities. Using the closed contour in Fig. 194, $R_\tau$ can then be expressed as a contour integral in the complex $s$ plane running counter-clockwise around the circle $|s| = m_\tau^2$ [1623–1625]:

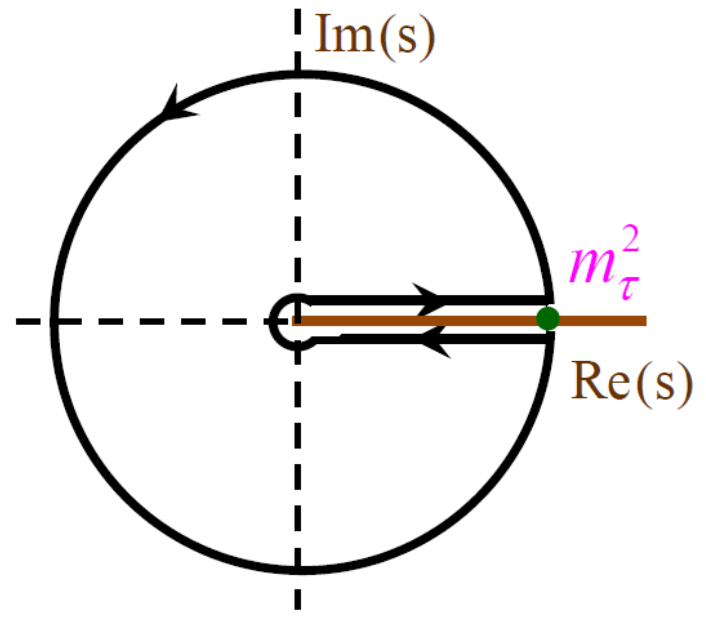

Fig. 194: Integration contour in the complex $s$ plane, used to obtain Eq. (548), figure taken from [1621].

$$R_\tau = 6\pi i S_{EW} \oint_{|s|=m_\tau^2} \frac{ds}{m_\tau^2}\left(1 - \frac{s}{m_\tau^2}\right)^2 \left[\left(1 + 2\frac{s}{m_\tau^2}\right)\Pi^{(0+1)}(s) - 2\frac{s}{m_\tau^2}\Pi^{(0)}(s)\right]\,, \tag{548}$$

The advantage of writing $R_\tau$ using Eq. (548) rather than Eq. (545) is that we are at sufficiently high energy on the circle ($|s| = m_\tau^2$) that we can use Operator Product Expansion (OPE) to calculate the correlator on the contour. The OPE relates the QCD quark-gluon dynamics to the inclusive hadron distributions actually observed in hadronic tau decays. This is only justified for integrated quantities such as Eq. (548) (global quark-hadron duality). Local violations of quark-hadron duality can be expected from the integration region near the real axis, where the OPE is not valid. They are fortunately reduced by the presence of the kinematic factor $\left(1 - \frac{s}{m_\tau^2}\right)^2$ which provides a double zero at $s = m_\tau^2$, suppressing contribution from the region near the branch cut. Whether this suppression is sufficient to make duality violating contributions negligible is the subject of intense theoretical debate, see for instance [1626].

The short-distance OPE can be used to organise the perturbative and non-perturbative contributions to the correlators into a systematic expansion in powers of $1/s$ [1627]

$$\Pi^{(J)}(s) = \sum_{D=2n} \sum_{\mathrm{dim}\mathcal{O}=D} \frac{C^{(J)}(s,\mu)\langle\mathcal{O}(\mu)\rangle}{(-s)^{D/2}} \ , \tag{549}$$

where the inner sum is over local gauge-invariant scalar operators of dimension $D = 0, 2, 4.....$ The parameter $\mu$ is an arbitrary factorisation scale, which separates long-distance non-perturbative effects, which are absorbed into the vacuum matrix elements $\langle\mathcal{O}(\mu)\rangle$, from short-distance effects, which are included in the Wilson coefficients $C^{(J)}(s,\mu)$. Inserting $\Pi^{(J)}(s)$ from Eq. (549) into the contour integral representation of $R_\tau$ in Eq. (548) we obtain:

$$\begin{aligned} R_{\tau,V/A} &= \frac{N_c}{2}|V_{ud}|^2 S_{EW}\left(1+\delta_P+\sum_{D=2,4,\ldots}\delta^{(D)}_{ud,V/A}\right) \\ R_{\tau,S} &= N_c|V_{us}|^2 S_{EW}\left(1+\delta_P+\sum_{D=2,4,\ldots}\delta^{(D)}_{us}\right) \end{aligned} \tag{550}$$

where $\delta^{(D)} = \left(\delta^{(D)}_V + \delta^{(D)}_A\right)/2$ is the average of the vector and axial-vector corrections and $N_C = 3$ the number of colours. We have several contributions:

- $\delta_P \equiv \delta^{(D=0)}$: this term is the purely perturbative QCD correction, neglecting quark masses, which is the same for all the components of $R_\tau$. This contribution is numerically dominant. It has been calculated up to $\mathcal{O}(\alpha_s^4)$ [1628]. The main uncertainties for this part come from the treatment of higher-order corrections, through the use of different renormalisation-group improvement prescriptions in the integration [1629–1632]. Its evaluation is subject to intense theoretical discussions [1626].
- $\delta^{(2)}$: this term represents the perturbative mass corrections. It is negligible for the non-strange part, where leading contributions are proportional to $m^2_{u,d}$, but not for $R_{\tau,S}$, which has leading $D = 2$ contributions proportional to $m_s^2$ [1633].
- $\delta^{(D\geq 4)}$: these are the non-perturbative contributions which involve $D = 4$ terms proportional to the gluon and quark condensates, and yet higher dimension condensate terms. We do not know in general how to calculate the non-perturbative condensates, but can attempt to fit them to data using $s_0$-dependent weighted integrals (moments) of the measured invariant-mass distribution, involving alternate (typically polynomial) weights, $w(s/s_0)$. These moments, denoted $R^w_{\tau,V/A}(s_0)$ and $R^w_{\tau,S}(s_0)$, are the $w$-reweighted analogues of the kinematically-weighted integrals $R_{\tau,V/A}(s_0)$ and $R_{\tau,S}(s_0)$ which correspond to a generalisation of $R_\tau$ in Eq. (545). $R_\tau(s_0)$ are obtained by restricting the integral in Eq. (545) to $0 < s < s_0$ for any $s_0 \leq m_\tau^2$ and reweighting the kinematic prefactors. $R_\tau(s_0)$ has a contour integral representation analogous to that of $R_\tau$, obtained by replacing in Eq. (548) the contour $|s| = m_\tau^2$ with $|s| = s_0$ and substituting $m_\tau^2$ by $s_0$ in the kinematic prefactors.

*$\alpha_s$ determination.* By comparing the theoretical prediction of $R_{\tau,V+A}$, Eq. (550) in the non-strange sector to the measured $\tau$ branching fractions as well as its moments $R^w_{\tau,V+A}(s_0)$ one can determine $\alpha_s$ [1634, 1635]. The determination of $\alpha_s$ from these finite-energy sum rule (FESR) analyses of inclusive non-strange hadronic $\tau$ decay data [1625] has the lowest scale amongst the various current precision determinations and hence provides the strongest test of the running predicted by QCD. In addition, as a result of the decrease in relative error generated by running to higher scales, the $\tau$ decay result provides the second most precise determination at the $Z$ scale[62]. The general situation for the determination of $\alpha_s$ remains somewhat unsettled, with determinations from shape observables, for example, lying significantly lower than those from $\tau$ decay and the lattice. Improvements of the current situation are required to take advantage of the precision determinations of Higgs branching fractions anticipated at a future ILC in searching for BSM physics [1636].

Because of the relatively low $\tau$ mass scale, non-perturbative effects are not totally negligible in the $\tau$-based $\alpha_s$ analysis. Moreover, even if one considers the inclusive non-strange experimental distribution, which sums vector (V) and axial vector (A) channel contributions and reduces the oscillatory behaviour of the spectral distribution, one would like to estimate the uncertainty induced by duality violations in the upper part of the decay distribution. This necessitates some modelling of DV contributions [1635, 1637], even for the so-called pinched weights used in the FESR analyses, which suppress the relative role of such DV contributions. The ability to test and constrain this modelling, and hence to obtain sensible estimates of systematic errors associated with the presence of residual DV contributions, is currently limited by the size of the errors in the ALEPH differential distributions [1617] in the region above $s \sim 2$ GeV$^2$. In the V channel, it would be very useful to take advantage of the $> 1000$ increase in statistics of the B-factories relative to ALEPH and OPAL to reduce the errors on the differential distributions for the $4\pi$ exclusive modes, which dominate the V spectral function in this region. Improvements of the $\tau \to 4\pi\nu_\tau$ results are also motivated by the discrepancy remaining between expectations based on the ALEPH $\tau \to 4\pi\nu_\tau$ distributions and recent SND and BaBar $e^+e^- \to \pi^+\pi^-\pi^0\pi^0$ cross-section results [1638, 1639]. In the longer term, improved B-factory determinations of the fully inclusive $ud$ V and A spectral distributions are also highly desirable for improving the determination of $\alpha_s$.

*$V_{us}$ determination.* Comparing the strange $|\Delta S| = 1$ and non-strange $|\Delta S| = 0$ tau decay widths gives the possibility to determine the CKM matrix element $V_{us}$. A determination of $V_{us}$ using hadronic $\tau$ decay data is of interest both for providing an additional independent determination in the scenario where BSM contributions are negligible and in the context of the recently observed discrepancy between experimental results and Standard Model expectations for the $B$ decay ratios $R(D)$ and $R(D^*)$, which suggest the possibility of BSM contributions coupled more strongly to the third generation [1640].

If quark masses are neglected, or in the $SU(3)$ limit, the experimental ratio of the strange to non-strange decay widths provides a direct measurement of $|V_{us}/V_{ud}|^2$. Away from this limit, one needs to take into account the small $SU(3)$-breaking contributions induced by the strange quark mass.

The original idea for extracting $|V_{us}|$ using hadronic $\tau$ decay data [1641, 1642] involved the construction of the flavour-breaking (FB) combination $\delta R_\tau$, defined by

$$\delta R_\tau \equiv \frac{R_{\tau,V+A}}{|V_{ud}|^2} - \frac{R_{\tau,S}}{|V_{us}|^2} , \qquad (551)$$

[62] For an overview of the various $\alpha_s$ determinations, see the PDG QCD review section at pdg.lbl.gov/2016/reviews/rpp2016-rev-qcd.pdf

where $R_{\tau,V+A}$ and $R_{\tau,S}$ are given in Eq. (547). In the $SU(3)$ limit, $\delta R_\tau = 0$. The idea was to solve Eq. (551) for $|V_{us}|$, using the contour integral representation of the left-hand side, and, on the right-hand side, external input for $|V_{ud}|$ [1643], and experimental input for $R_{\tau,V+A}$ and $R_{\tau,S}$ [1641, 1642, 1644–1654]. The result is

$$|V_{us}| = \left( \frac{R_{\tau,S}}{\frac{R_{\tau,V+A}}{|V_{ud}|^2} - \delta R_{\tau,th}} \right)^{1/2} . \qquad (552)$$

This expression represents the conventional implementation of the flavour-breaking, inclusive finite-energy sum rule (FESR) approach to determining $|V_{us}|$ from inclusive hadronic $\tau$ decay data [1641, 1642].

The FESR approach can be formulated more generally using an arbitrary polynomial weight, $w$, and the analogues, $R^w_{\tau,V+A}(s_0)$ and $R^w_{\tau,S}(s_0)$, of $R_{\tau,V+A}$ and $R_{\tau,S}$, obtained by reweighting, using the weight $w$ rather than the Standard Model kinematic weight, $w_\tau \equiv w(s/m_\tau^2)$ and integrating up to any kinematically allowed value, $s = s_0 \le m_\tau^2$, rather than all the way to $s = m_\tau^2$. Generalized versions, $\delta R^w_\tau(s_0)$, of $\delta R_\tau$, can then be constructed and more general, $w$- and $s_0$-dependent analogues of Eq. (552) obtained.

The conventional implementation of this general FB, inclusive FESR approach, represented by Eq. (552) [1641, 1642], employs, in this language, $s_0 = m_\tau^2$ and unreweighted versions of the experimental spectral integrals (corresponding to the choice $w = w_\tau$). With these choices, the spectral integrals entering the FESR are fixed by the inclusive non-strange and strange hadronic branching fractions. The cost of this experimental simplification is the presence, in the corresponding $w_\tau$-weighted OPE integral, of in-general-unsuppressed dimension $D = 6$ and 8 OPE contributions whose values are not known from external sources. Past analyses dealt with this problem by using the crude vacuum saturation approximation (VSA) for the $D = 6$ contribution and neglecting $D = 8$ contributions on the grounds that the $D = 6$ VSA estimate was small. The result of these analyses with new inputs on branching ratios from $B$-factories is a value of $V_{us}$ more than $3\sigma$ below that implied by three-family unitarity[63] and the super-allowed nuclear $\beta$ decay result for $V_{ud}$ [1643]. This is mainly due to the fact that the branching ratios measured by BaBar and Belle are smaller than previous world averages, which translates into smaller results for $R_{\tau,S}$ and $|V_{us}|$. Replacing the three largest branching ratios results $\left(\mathrm{Br}\left(\tau^- \to \nu_\tau \mathrm{K}^-\right), \mathrm{Br}\left(\tau^- \to \nu_\tau \bar{\mathrm{K}}^0 \pi^-\right) \text{ and } \mathrm{Br}\left(\tau^- \to \nu_\tau \mathrm{K}^- \pi^0\right)\right)$ by information from leptonic kaon decays ($K^- \to \mu^- \bar{\nu}_\mu$) and the combination of the measured spectra in $\tau^- \to \nu_\tau (K\pi)^-$ decays with $K_{\ell 3}$ ($K \to \pi \ell \bar{\nu}_\ell$) data one gets a result for $|V_{us}|$ in better agreement with CKM unitarity, see Ref. [1655].

The $V_{us}$ value, however, can also be obtained from FESRs with $s_0 < m_\tau^2$ and weights, $w$, other than $w_\tau$. Varying $s_0$ and $w$, one finds a highly significant unphysical $s_0$ and $w$ dependence [1656–1659]. These are eliminated when not just $V_{us}$ but also the higher dimension $D = 6$ and 8 OPE contributions are fitted to the data [1659]. Lattice data can also be used to obtain complementary information on the relevant OPE contributions [1659]. On the experimental side, with spectral integrals required over a range of $s_0$, inclusive branching fraction input no longer suffices; the full differential distributions are needed.

[63] See, *e.g.*, the HFLAV-Tau Spring 2017 report *www.slac.stanford.edu/xorg/hfag/tau/spring-2017*

At present, the total $V_{us}$ error is strongly dominated by the uncertainties in the weighted flavour $us$ spectral integrals. Contributions to these errors from the exclusive $K\pi$ and $K\pi\pi$ modes studied by BaBar and Belle, are at present dominated by uncertainties on the exclusive branching fractions which multiply the unit-normalised experimental distributions. Significantly reduced $V_{us}$ errors should thus be possible through improvements of the low-multiplicity strange mode branching fractions. Thus, experimental efforts at Belle II are particularly important.

An additional inclusive hadronic $\tau$ decay method for determining $V_{us}$ was also recently proposed in Ref. [1658**?** ]. A weighted dispersion relation is employed which (1) allows lattice input rather than the OPE to be used on the theory side and (2) involves only the flavour $us$ inclusive differential distribution on the experimental side. The weights used can be tuned to enhance relative contributions from the lower-multiplicity region of the $us$ distribution without unduly inflating the associated lattice errors. The latest results in [**?** ] show errors on $V_{us}$ significantly reduced compared to those obtained from the FB FESR approach and the $V_{us}$ value is higher. It suggests the method has the potential to become competitive with $K_{\ell 3}$ and $\Gamma[K_{\mu 2}]/\Gamma[\pi_{\mu 2}]$ determinations in future, making improvements of lower-multiplicity exclusive $us$ experimental data highly desirable.

To summarise, the following experimental results would be useful for improving the FB FESR and new lattice-based $us$ determinations of $V_{us}$:

- Improved $K^-\pi^0$ and $\bar{K}^0\pi^-$ branching fractions.
- Fully unfolded exclusive mode unit-normalised $K^-\pi^0$ and $\bar{K}^0\pi^-$ differential distributions, similar to Belle $\bar{K}^0\pi^-$ results, ideally including covariances, though improved overall normalisation is most immediately useful.
- Improved $K\pi\pi$ branching fractions, if possible including the smaller $K^-\pi^0\pi^0$ mode.
- A fully unfolded unit-normalised $\bar{K}^0\pi^-\pi^0$ distribution, analogous to that for $K^-\pi^+\pi^-$ reported by BaBar [1660], ideally including covariances, but, again, with improved overall normalisation of most immediate utility.
- If possible, a first B-factory version of the unit-normalised $K^-\pi^0\pi^0$ distribution.

## *15.5. MC event generators for $\tau$ physics*

*(Contributing authors: Z. Was, D. Epifanov)*

*15.5.1.* `KKMC` *for the $\tau$ lepton pair production .* Here we review status of the $\tau$ lepton production Monte Carlo generator `KKMC`. The main purpose of the program is the simulation of $f\bar{f} \to f'\bar{f}'$ processes at high energies. To achieve per-mille level precision a substantial effort was required. In this section we concentrate on those effects which are necessary for high precision to be achieved at a centre–of–mass energy of about 10 GeV. In such an energy regime many effects related to high energy electroweak interactions can be neglected. On the other hand, effects due to masses of outgoing fermions, as well as electromagnetic vacuum polarisation have to be considered. The precise modelling of subsequent $\tau$ decays as well as radiative corrections in decays are important for precision measurements. Lessons learned from fits to BaBar and Belle data are discussed below.

The Monte Carlo program `KKMC` for $e^+e^- \to f'\bar{f}'n\gamma$ was developed and tested for centre-of-mass energies above those necessary for Belle II (See Refs. [23, 24, 1661] and its recent upgrade for LHC applications in [1662]). It features a second-order matrix element for initial-and final-state QED effects, one loop electroweak corrections including line-shape corrections, and longitudinal and transverse spin effects of incoming electrons and outgoing fermions ($\tau$-leptons). Beamsstrahlung effects can be included in the simulation as well. A precision of 0.1% was achieved. For Belle II, applications of the effects mentioned above are limited, while specific effects for $B$-factories discussed in Ref. [37] were not included and added ad hoc later by BaBar. The achieved precision of 2-3 per-mille was considered to be sufficient, though more theoretical work is required to greatly improve on this limit.

One of the important features of KKMC is the possibility to generate $\tau$ lepton decays with all spin effects treated in the production process. The TAUOLA package [1663–1666] can then be used for the simulation of $\tau$-lepton decays, and PHOTOS [1667–1669] for simulation of QED radiative corrections in decays. Additional lepton pairs in the final state can be also generated with the help of the PHOTOS algorithm, described below, while the effect of both initial and final state pair emission can be simulated with the help of KORALW Monte Carlo [1670].

*15.5.2. New currents in the* TAUOLA *Package.* In this subsection, we will concentrate on physics extensions and novel applications to the TAUOLA package. We will stress importance of the three aspects of the work: (i) construction and implementation of hadronic currents for $\tau$ decay currents obtained from models (evaluated from QCD), (ii) presentation of experimental data in a form suitable to fits, (iii) preparations of algorithms and determination of distributions useful for fits.

We have prepared two new sets of currents, the first based mainly on theoretical considerations, the second on an effort from the BaBar collaboration. In Refs. [1671, 1672] it was shown how the Resonance Chiral Lagrangian approach was used for calculations of hadronic currents. This will be adapted for TAUOLA. In [1671] it was stressed that details such as additional resonances, more specifically the $f_2(1270)$, $f_0(1370)$ and $a_1(1640)$ observed by CLEO [1673] can not be introduced if fits to one-dimensional invariant mass spectra of two- and three-pion systems are used. In Ref. [1673] two-dimensional mass scattergrams were used as input for a parameterisation of TAUOLA currents (CLEO parameterisation [64] [1666]). This should be considered as a minimum for comparisons with the present day data. In fact, CLEO used a more detailed representation of the data in Ref. [1674]. It may be of interest to repeat such a data analysis, with the help of observables presented in Ref. [1548], adapted to the case of relativistic tau-pair production in Belle II.

Physics of $\tau$ lepton decays requires sophisticated strategies for the confrontation of phenomenological models with experimental data, owing to the high precision of experimental data. Changing the parameterisation for one channel may affect the background modelling of another. This demands simultaneous analysis of many decay channels. One has to keep in mind that the models used to obtain distributions in the fits may require refinement or even substantial rebuilding as a consequence of comparison with data. The topic was covered in detail in Ref. [1675].

[64] Note that for this parametrisation, differences between hadronic currents of $\tau \to \pi^+\pi^-\pi^-\nu$ and $\tau \to \pi^-\pi^0\pi^0\nu$ were ignored and isospin symmetry was imposed ($\rho\pi$ dominance). A version of the current without this constraint is nonetheless distributed with TAUOLA (all versions), but as a non-active option.

One may wish to calculate for each generated event (separately for the decay of $\tau^+$ and/or $\tau^-$) alternative weights; the ratios of the matrix element squared obtained with new currents, and the ones used in generation. A vector of weights can then be obtained and used in fits. Such a solution can be easily installed. For practical reasons, the use of semi-analytical distributions is much easier. It enables much faster calculation of errors for fit parameters including correlations, but experimental distributions must be available unfolded . This was important for fits of $3\pi$ currents obtained in Ref. [1676]. Modifications of the currents were necessary to obtain the results in Ref. [1677]. It is not clear, if such fitting, without additional help of observables as in [1548], can be used for the $KK\pi\nu_\tau$ and $K\pi\pi\nu_\tau$ $\tau$ decay channels, even if two-dimensional scattergrams are available. If experimental data are available as one- or at most two-dimensional histograms, then the associated currents still rely on models. With the present day experimental precision, even use of the Resonance Chiral Lagrangian should not be expected to have sufficient predictive power to describe multidimensional distributions from the constraints of fits to one- or two-dimensional histograms [1678, 1679]. This limitation is clearly visible in results for $4\pi$ currents [1680].

Currents have been developed for `TAUOLA` based on Refs. [1596, 1677, 1680, 1681], respectively, for two-, three-, four- and five-pion final states. This is now available in `FORTRAN` and `C++`, with the option that users can introduce their own C++ currents. Note also that the parameterisation for TAUOLA, equivalent to the one used by the BaBar collaboration for the default simulations is given in [1679].

*15.5.3. Status of the implementation of tau decays in* `TAUOLA`*.* Tau decays into the final states with leptons, which are implemented currently in TAUOLA, include ordinary leptonic decay, $\tau^- \to \ell^- \bar{\nu}_\ell \nu_\tau$ ($\ell = e, \mu$) and radiative leptonic decay, $\tau^- \to \ell^- \bar{\nu}_\ell \nu_\tau \gamma$. These decays are simulated together and they are separated according to the gamma energy threshold (its value is set in the TAUOLA). Also, the complete $\mathcal{O}(\alpha)$ QED corrections are implemented for the $\tau^- \to \ell^- \bar{\nu}_\ell \nu_\tau$ decay (leading order (LO) matrix element, virtual and soft photon corrections) [1664], while the $\tau^- \to \ell^- \bar{\nu}_\ell \nu_\tau \gamma$ decay is generated according to the LO matrix element only. Hence, the accuracy of the simulation of the $\tau^- \to \ell^- \bar{\nu}_\ell \nu_\tau$ decay, estimated to be about $10^{-3}$, is determined by the uncertainty of the theoretical formalism, i.e. contributions from various higher order electroweak corrections, like [1682]. Recently, the next-to-leading order (NLO) correction has been calculated for the $\tau^- \to \ell^- \bar{\nu}_\ell \nu_\tau \gamma$ decay, the corresponding corrections to the branching ratios were found to be about 3% for the $\tau^- \to \mu^- \bar{\nu}_\ell \nu_\tau \gamma$, and about 10% for the $\tau^- \to e^- \bar{\nu}_\ell \nu_\tau \gamma$ mode [1683]. These corrections have not been implemented in TAUOLA yet, hence the accuracy of the simulation of the radiative leptonic decays is only $(3 \div 10)$%. It should be mentioned that neither doubly radiative leptonic decay, $\tau^- \to \ell^- \bar{\nu}_\ell \nu_\tau \gamma\gamma$ (which is important for the precision studies of the radiative leptonic decay), nor 5-body leptonic decays, $\tau^- \to \ell^- \ell'^+ \ell'^- \bar{\nu}_\ell \nu_\tau$ are implemented in the standard TAUOLA package. For all decay channels, configurations like $\tau^- \to l^- \gamma\gamma$ with additional photons and/or $\tau^- \to l^- l'^+ l'^-$ with additional lepton pair, can be introduced into TAUOLA sample with the help of PHOTOS, mentioned below. Nevertheless, the code within the TAUOLA package for the simulation of the 5-body leptonic tau decays, according to Ref. [1580], have been developed at Belle and can be easily embedded in the official version of TAUOLA [1684].

The multipion hadronic tau decays ($\tau \to (2\pi,\, 3\pi,\, 4\pi,\, 5\pi)\nu$) have been studied with high statistics in several experiments, for some of these modes optimal parameterisations of the hadronic currents (spectral functions) were obtained. The most precise description of the hadronic current in the $\tau^- \to \pi^-\pi^0\nu_\tau$ decay was achieved at Belle from the fit of the experimental $\pi^-\pi^0$ invariant mass distribution [1596]. The parameterisation of the hadronic current in the $\tau^- \to \pi^-\pi^0\pi^0\nu_\tau$ decay was established by CLEO in their unbinned analysis of the $e^+e^- \to (\tau^- \to \pi^-\pi^0\pi^0\nu_\tau,\, \tau^+ \to \ell^+\nu_\ell\bar{\nu}_\tau)$ process in the full phase space [1673]. Up to now, this is the most sophisticated and precise study of the dynamics of hadronic tau decay, and such kind of analyses allows one to avoid the disadvantages of the studies of one- or two-dimensional distributions for tau decays into $\geq$3 pions mentioned above. It was found that the CLEO $\pi^-\pi^0\pi^0$ hadronic current fits well also the $\tau^- \to \pi^-\pi^+\pi^-\nu_\tau$ decay. For the $\tau^- \to (\pi^-\pi^+\pi^-\pi^0,\, \pi^-\pi^0\pi^0\pi^0)\nu_\tau$ decays the hadronic currents are written based on the experimentally measured cross sections of the reactions $e^+e^- \to \pi^+\pi^-\pi^0\pi^0$, $\pi^+\pi^-\pi^+\pi^-$ [1680] and the conserved vector current (CVC) theorem. Basically, such an approach allows one to describe the dynamics of the four-pion production with small uncertainty determined by the degree of the CVC theorem violation (of the order of 1%). The hadronic currents in the $\tau^- \to (\pi^-\pi^+\pi^-\pi^+\pi^-,\, \pi^-\pi^+\pi^-\pi^0\pi^0)\nu_\tau$ decays are described by the model from Ref. [1681]. To choose the appropriate model, a high statistics study of these decays in multidimensional phasespace should be performed at Belle II.

*15.5.4.* `PHOTOS` *Monte Carlo for bremsstrahlung and its systematic uncertainties.* While PHOTOS is described in detail elsewhere [34], two aspects of recent development should be noted. Firstly the emission of additional lepton pairs was introduced, which contributes through final-state bremsstrahlung. Secondly the package is now fully written in C++ [34]. Recent work on numerical tests and new applications, especially in the domain of LHC, have been performed with precision better than $10^{-3}$ [1685–1687]. Note that Photos algorithm features matrix-element phase-space separation. This is the case for the multi-photon mode of the operation as well. That is why the decay channel dependent electromagnetic form factors can be implemented into matrix element used for the decays of smaller multiplicity, see *e.g.* [1688].

## *15.6.* $e^+e^- \to \pi^+\pi^-$ *cross section for* $(g-2)_\mu$

*(Contributing authors: H. Czyz, T. Ferber, D. Nomura, M. Roney, B. Shwartz, T. Teubner)* The discrepancy between the measurement and the Standard Model calculations for the anomalous magnetic moment of the muon $(g-2)_\mu$ is close to $4\sigma$ . With the upcoming experiments at Fermilab [1689] and J–PARC [1690] we can expect a factor four improvement on the accuracy in each of them over the existing result [1691], reducing the experimental uncertainty to about $1.6 \times 10^{-10}$. The current theoretical uncertainty of $4.9 \times 10^{-10}$ (see [1692–1695] for recent reviews) is dominated from experimental input for the calculation of the leading order hadronic contribution. Of these experimental inputs, the largest contribution and also the largest uncertainty come from the two charged pion channel in the mass region around the $\omega - \rho$ interference. Therefore, without significant improvements in the measurement of the $e^+e^- \to \pi^+\pi^-$ cross section, there is no hope to improve the error coming from the SM calculations. The situation here is inconclusive: The three recent most accurate experimental measurements of the cross section of the reaction $e^+e^- \to \pi^+\pi^-$ by BaBar [1696], BESIII [1697] and KLOE [1698–1700] show some tension (see Fig. 195). The spread between KLOE and BaBar, not accounted within their quoted uncertainties, is bridged by the BESIII results. The CMD–2 [1701] and SND [1702] results are not helping in sorting out this issue.

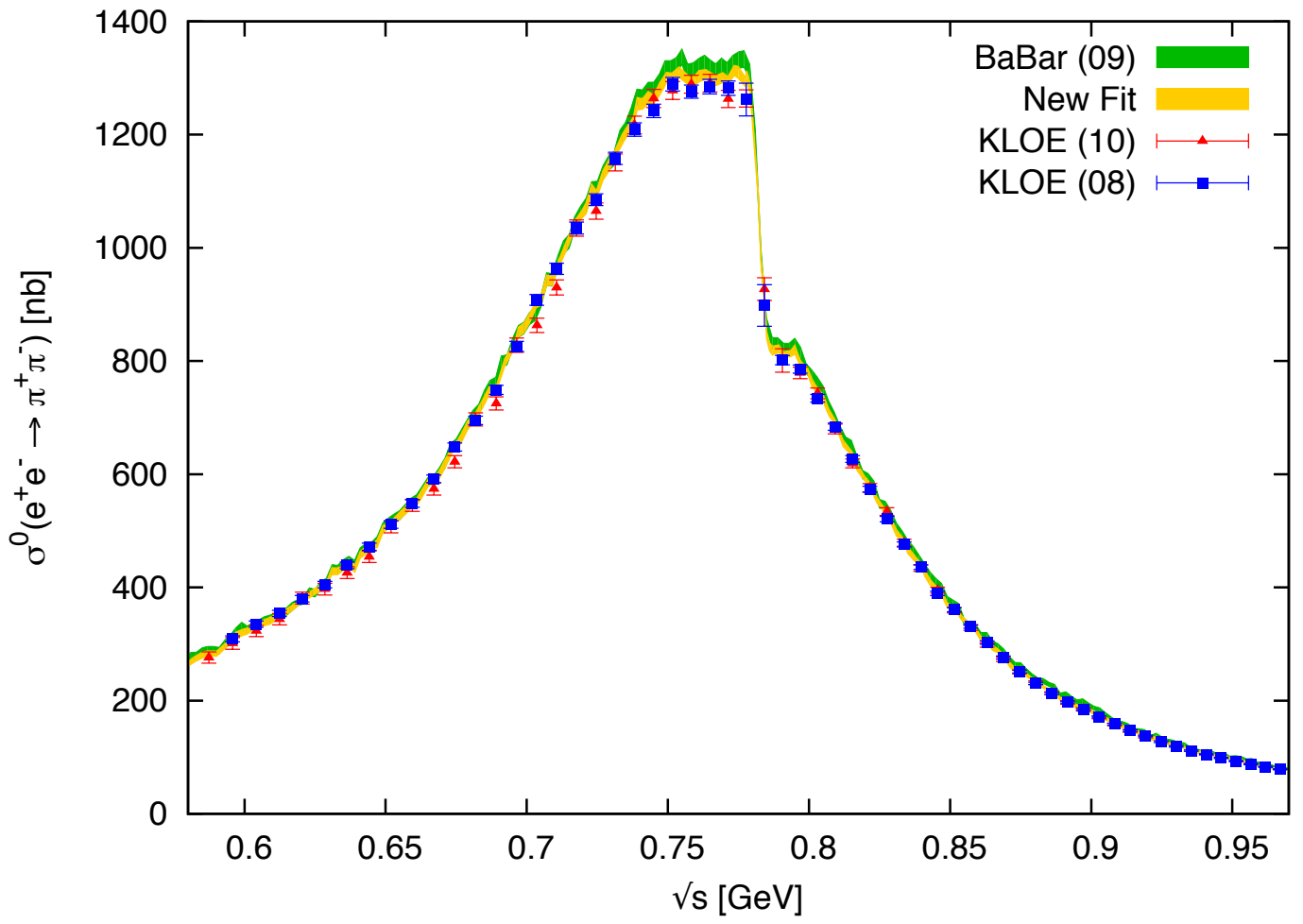

Fig. 195: Fit with all data in the $2\pi$ channel (light (yellow) band). Radiative return data from BaBar [1696] are shown by the darker (green) band, whereas the KLOE [1698–1700] data are displayed by the markers as indicated in the plot (reproduced from [1703]).

All experimental groups made a significant effort for the control of systematic errors, yet the difference is not understood at all and new experiments are needed. Moreover, in order to reduce the error on $(g-2)_\mu$ significantly, the goal for the final accuracy including both statistical and systematic uncertainties is to be 0.5% or lower.

*15.6.1. Experiment.* A measurement of $e^+e^- \to \pi^+\pi^-(\gamma)$ from threshold to approximately 3 GeV can be made with Belle II using the initial state radiation method [1453–1456]. The methodology published by BaBar [1696] can be used to estimate the precision reach of the Belle II dataset. The reduced centre-of-mass energy ($\sqrt{s'}$) spectrum of $e^+e^- \to X\gamma_{ISR}$ events gives the cross section for the process $e^+e^- \to X$ via

$$\frac{dN_{X\gamma_{ISR}}}{d\sqrt{s'}} = \frac{dL_{ISR}^{eff}}{d\sqrt{s'}} \epsilon_{X\gamma_{ISR}}(\sqrt{s'})\sigma_X^0(\sqrt{s'}) \tag{553}$$

where $\epsilon_{X\gamma_{ISR}}$ is the detection efficiency estimated initially with Monte Carlo (MC) simulation, subsequently corrected with control sample studies using data, $\sigma_X^0(\sqrt{s'})$ is the 'bare' cross-section that excludes vacuum polarisation, and $\frac{dL_{ISR}^{eff}}{d\sqrt{s'}}$ is the effective initial state radiation (ISR) luminosity. This ISR luminosity is obtained using $e^+e^- \to \mu\mu(\gamma)\gamma_{ISR}$ events and Equation 553, where the measured $\frac{dN_{\mu\mu\gamma_{ISR}}}{d\sqrt{s'}}$ distribution, the bare cross section $\sigma^0_{\mu\mu(\gamma)}(\sqrt{s'})$ calculated using QED, and $\epsilon_{\mu\mu(\gamma)\gamma_{ISR}}$ are used as input. With $\frac{dL_{ISR}^{eff}}{d\sqrt{s'}}$ in hand from these $e^+e^- \to \mu^+\mu^-(\gamma)\gamma_{ISR}$ events, the measurement of $\sigma^0(e^+e^- \to \pi^+\pi^-(\gamma))$ is obtained using

Equation 553, the measured $\frac{dN_{\pi\pi\gamma_{ISR}}}{d\sqrt{s'}}$ distribution and $\epsilon_{\pi\pi(\gamma)\gamma_{ISR}}$.

In essence, this is a precision measurement of the ratio $\sigma(e^+e^- \to \pi^+\pi^-(\gamma))/\sigma(e^+e^- \to \mu^+\mu^-(\gamma))$ as a function of the $\pi^+\pi^-$ and $\mu^+\mu^-$ invariant masses. The advantage is that it removes large components of the systematic effects related to the detection of the initial state radiation and most higher-order theoretical uncertainties. Such cancellations are required as every systematic uncertainty must be kept at a few-per-mil level.

The published BaBar measurement [1696] was performed using a sample of 232 fb$^{-1}$, which is approximately half of the entire BaBar dataset. That analysis selected $e^+e^- \to \pi^+\pi^-(\gamma)\gamma_{ISR}$ and $e^+e^- \to \mu^+\mu^-(\gamma)\gamma_{ISR}$ events where the muon and pion samples were separated using $\mu/\pi$ charged particle identification (PID). Backgrounds were suppressed using the $\chi^2$ of kinematic fits to the $e^+e^- \to X\gamma_{ISR}$ signal process that can allow for unmeasured photons radiated at small angles to the initial state electron and/or positron.

In Ref. [1696] the systematic error on the $\pi^+\pi^-$ cross section in the region of the $\rho$ ($0.6\ GeV < m_{\pi\pi} < 0.9\ GeV$), where most of the signal lies, is $\pm 0.5\%$, but is significantly larger below and above this region around the $\rho$ resonance. The systematic error is dominated by $\pi$-ID ($\pm 0.24\%$) and ISR luminosity from $\mu$-pairs ($\pm 0.34\%$). The ISR luminosity error itself, 0.34%, is dominated by the 0.29% systematic uncertainties on the $\mu$-ID with smaller contributions from trigger, tracking and acceptance uncertainties. The statistical error of the raw spectrum is 1.35% at the $\rho$ mass, which includes the statistical error of the measured efficiency corrections ($4.7 \times 10^{-3}$ at the $\rho$). Although larger outside the $\rho$ region, the systematic uncertainties did not exceed statistical errors over the full spectrum.

The lowest-order contribution of the $\pi\pi(\gamma)$ state to the muon magnetic anomaly is given by:

$$a_{mu}^{\pi\pi(\gamma),LO} = \frac{1}{4\pi^3} \int_{4m_\pi^2}^{\infty} ds' K(s') \sigma^0_{\pi\pi(\gamma)}(s'), \tag{554}$$

where $K(s')$ is a known kernel [1704]. With 232 fb$^{-1}$ of data, the BaBar integrated measurement from threshold to 1.8 GeV was $a_\mu = (514.1 \pm 2.2 \pm 3.1) \times 10^{-10}$, representing a statistical error of 0.4% and systematic error of 0.6%. Note that the systematic errors, though smaller in each bin of mass, are correlated across mass bins and therefore in the evaluation of the error on $a_\mu$ the systematic error dominates.

Nonetheless, significant improvements can be expected with 1 ab$^{-1}$ of Belle II data. Assuming similar selection approaches and that Belle II has a trigger for the $e^+e^- \to \pi^+\pi^-(\gamma)\gamma_{ISR}$ and $e^+e^- \to \mu^+\mu^-(\gamma)\gamma_{ISR}$ events that is at least as efficient as the BaBar trigger was, then one can expect a statistical error of 0.1%, or three times smaller than the error on $a_\mu$ expected from the next generation experiments at FERMILAB and J-PARC. Consequently, the focus will be on reducing the systematic errors. As BaBar included all the statistical components of the systematic errors in the statistical error, there is no trivial projection of the potential systematic error reach for Belle II. Experience has shown, however, that with a significantly larger data sample potential opportunities to identify and reduce the systematic errors in dedicated studies could arise.

The BaBar analysis employed the AfkQED MC generator (based on Ref. [1705]) to compare the $\mu$-pair cross sections to NLO QED and provide MC efficiencies for both $\pi^+\pi^-$ and $\mu^+\mu^-$ channels. The more accurate PHOKHARA generator (see Sec. 15.6.2) was used to study effects of additional ISR photons. With the anticipated improvements to PHOKHARA we anticipate not having to rely on AfkQED for Belle II.

As with the BaBar analysis, the efficiencies should be obtained from data-driven corrections to the MC. The PID efficiencies can be determined using the $x^+x^-\gamma_{ISR}$ sample itself, where one of the final state charged $x$ particles ($x = \mu, \pi, K$) is tagged with stringent PID criteria, and the second ('opposite') track identification is probed ('tag-and-probe' method). Such a sample was used in Ref. [1696] and set the level of the systematic errors associated with $\pi$-ID and $\mu$-ID, the dominant components of the systematic errors. One can augment that tag-and-probe sample with high statistics, pure samples of pions and kaons in low multiplicity events from a high purity sample of $\tau^-$ decays to three charged particles using the fact that $\tau^- \to K^+\pi^-\pi^-$ is forbidden. In that sample a clean sample of charged pions is obtained by making stringent requirements to select a $\pi^-\pi^-$ pair which forces the third charged particle to be $\pi^+$. The pure sample of kaons is obtained by making stringent requirements selecting a $K^+\pi^-$ pair in these $\tau$ decays, thereby forcing the third charged particle to be a $K^-$. In addition, a Belle II analysis can augment the tag-and-probe sample of muons using a dedicated sample of muons from $e^+e^- \to e^+e^-\mu^+\mu^-$ two-photon events where muons are selected in low total transverse momentum events with an electron tag and an oppositely charged, strictly identified muon.

An important cross-check on the sensitivity to higher order radiative effects was provided by comparing the muon absolute cross section dependence on the reduced centre-of-mass energy. This check was limited by the 1.1% error on the BaBar luminosity available at the time of that publication. It will be valuable for Belle II to have a few-per-mil level uncertainty on the absolute luminosity as this comparison can provide a means to reduce other systematic uncertainties in the analysis.

It is planned to have a purely neutral trigger based on a single, high-energetic photon at Belle II. In order to keep the rate of this trigger sufficiently low, a 3D veto for Bhabha events based on tracks and calorimeter clusters will be used. Preliminary studies using the Belle II trigger simulation of a 2 GeV photon trigger show a 100 % efficiency for all $e^+e^- \to \pi^+\pi^-(\gamma_{ISR})$ with the ISR photon in the barrel calorimeter. Additional triggers based on charged tracks only will allow a precision study of the trigger efficiencies using fully orthogonal triggers.

As mentioned, in the region of the $\rho$ the largest systematic errors in the BaBar analysis arose from PID. If those errors were to be removed, the systematic error would drop by 1/3. A Belle II analysis that has a negligible PID error with 1 ab$^{-1}$ of data would therefore have a total error on $a_\mu$ of $2 \times 10^{-10}$, or 0.4%, which is approaching the 0.3% error on $a_\mu$ expected from the next generation experiments at FERMILAB and J-PARC. One way to remove the PID uncertainties [1706] is to avoid using PID to separate the $e^+e^- \to \pi^+\pi^-(\gamma)$ and $e^+e^- \to \mu^+\mu^-(\gamma)$ event samples by exploiting the fact that they have different angular distributions in the centre-of-mass system of the charge pair because of the different spins of the muon and pion: $e^+e^- \to \mu^+\mu^-(\gamma)$ events have a $1 + cos^2\theta^*$ distribution and $e^+e^- \to \pi^+\pi^-(\gamma)$ events have a $sin^2\theta^*$ distribution. Such an analysis conducted with a large Belle II dataset with a focus on further reducing the other classes of systematic errors reported by BaBar [1696] has a reasonable chance of reducing the total hadron vacuum polarisation error to below the experimental error on $a_\mu$ expected from the FERMILAB and J-PARC experiments.

*15.6.2. Monte Carlo generator.* The measurement of the $e^+e^- \to \pi^+\pi^-$ cross section requires very precise Monte Carlo Event generators. The Monte Carlo generator PHOKHARA, the most accurate generator to date, has an accuracy of the ISR radiator function of 0.5% [1707]. This should be improved. The PHOKHARA group is planning to include NNLO corrections to ISR emissions in the leading logarithmic approximation. This should, according to estimates, allow for a reduction of the error originating from this source to the level of (0.1–0.2) %. Including the complete NNLO radiative corrections to ISR emissions is more demanding. Another issue to be addressed is to add the missing NLO corrections in the calculation of the cross section for $e^+e^- \to \pi^+\pi^-\gamma$. The complete NLO corrections were added already for the process $e^+e^- \to \mu^+\mu^-\gamma$ [1708], where it was found that the corrections, which were not included in PHOKHARA before, are small for all event selections used at experiments, which used the radiative return method. For $e^+e^- \to \pi^+\pi^-\gamma$, if scalar QED is used for modelling of the photon–pion interactions, one expects that the results will be similar. However, when including the pion form factor effects, it is difficult to predict the results and the answer will be known only after simulating the process with the complete corrections and with realistic event selections cuts used at experiments.

### *15.7. Two photon physics*

*15.7.1. $\pi^0$ and $\eta^{(\prime)}$ transition form factors. (Contributing authors: V. Braun, N. Offen, S. Uehara)*

*Theory.* The $\gamma^*\gamma \to \pi, \eta, \eta'$ transition form factors (FFs) are widely recognised as golden modes that allow one to access meson wave functions at small transverse separations, usually referred to as distribution amplitudes (DAs). The standard theory framework is based on collinear factorisation [1709–1711] complemented by estimates of soft end-point contributions using a simplified version of the light-cone sum rules [1712, 1713] as suggested first in [1714]. Such calculations have reached a high degree of maturity [1715–1719].

An alternative approach to the calculation of transition form factors makes use of transverse momentum dependent (TMD) meson wave functions (TMD- or $k_T$-factorisation [1720]). This is a viable technique that has been advanced recently to NLO, see *e.g.* Ref. [1721, 1722] for the electromagnetic pion form factor and $\gamma^*\gamma \to \pi^0$, and which can be applied to the $\gamma^*\gamma \to \eta, \eta'$ transitions as well. Because of a more complicated nonperturbative input, interpretation of these results is not straightforward.

The recent measurements of the $\gamma^*\gamma \to \pi^0$ FF at space-like momentum transfers in the interval $4 - 40$ GeV$^2$ by the BaBar and Belle collaborations [1723, 1724] caused much excitement and stimulated a flurry of theoretical activity. A strong scaling violation in the $Q^2 = 10 - 20$ GeV$^2$ range observed by BaBar [1723], see Fig. 196, necessitates a very large soft correction to the FF and a significant enhancement of the pion DA close to the end points. This would have profound implications for the studies of $B$ decays to final states involving energetic pions using QCD factorisation and/or LCSRs. The Belle data [1724] indicated a much softer scaling violation that is more consistent with common wisdom, although a certain enhancement of the end-point behaviour of the pion DA as compared to models based on truncated Gegenbauer expansion is favoured in this case as well [1716]. A clarification of this discrepancy is extremely important.

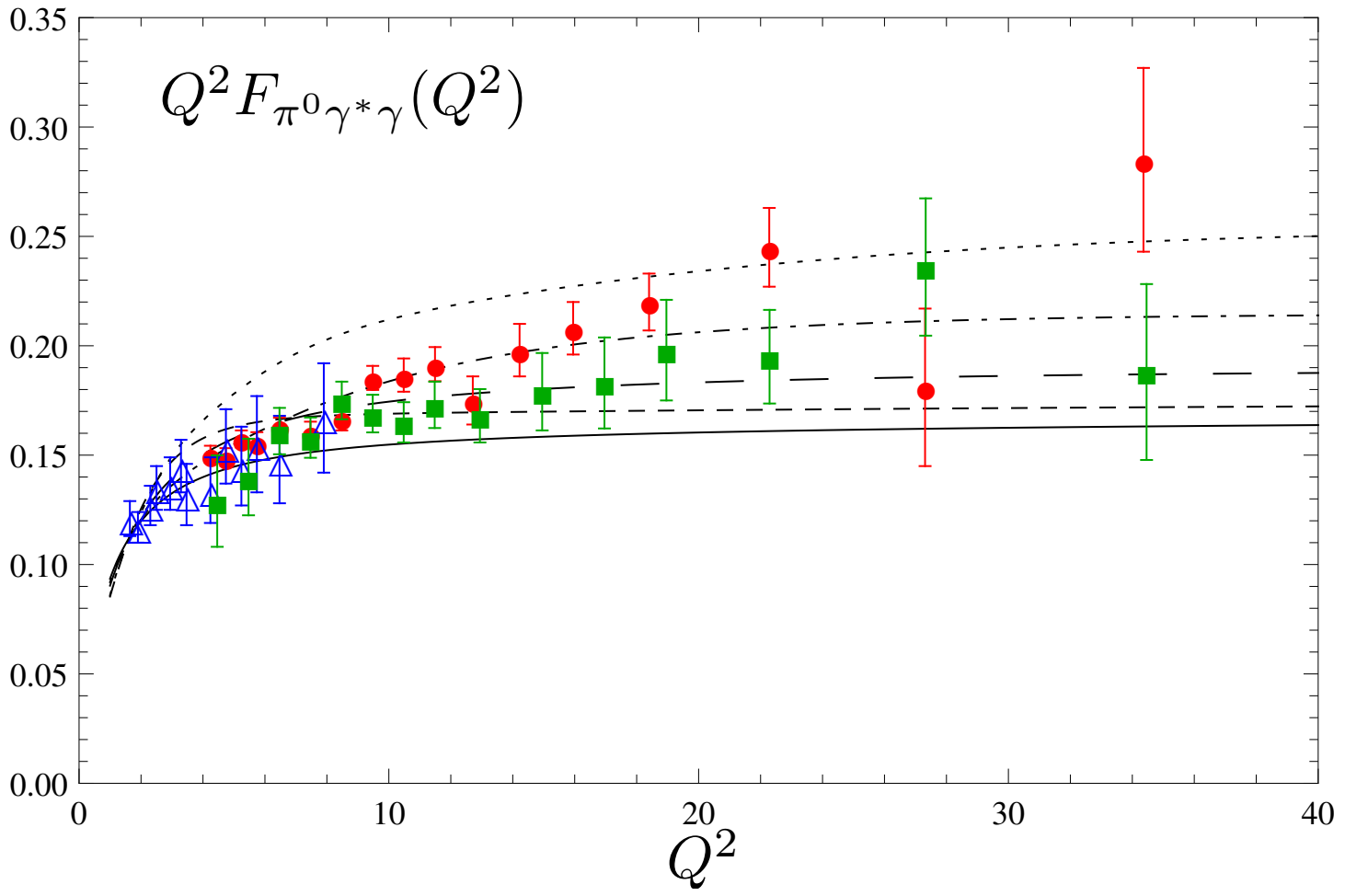

Fig. 196: The pion transition form factor for the "asymptotic" (solid line), "BMS"[1717] (short dashes), "holographic" (long dashes), "model II" of Ref. [1715] (dash-dotted) and "flat" (dots) pion DA. The experimental data are from BaBar [1723] (circles), Belle [1724] (squares), and CLEO [1725] (open triangles).

The question at stake is whether hard exclusive hadronic reactions are under theoretical control, which is highly relevant for all future high-intensity, medium energy experiments. Due to better pion identification and much higher statistics the Belle II experiment will be able to measure the $\gamma^*\gamma \to \pi^0$ form factor with unprecedented precision in the whole $Q^2$ range. This effort will be complemented on the theory side by high-precision lattice calculations of the second moment of the pion DA [1726–1728] and the NNLO calculation of the leading-twist contribution [1729, 1730].

The theory of $\gamma^*\gamma \to \eta, \eta'$ decays is similar to $\gamma^*\gamma \to \pi^0$ apart from a few technical elements. The most important question is whether the usual approach to $\eta, \eta'$ based on the concept of state mixing (see *e.g.* [782] and references therein) is adequate for the description of hard processes. Another issue is that $\eta$ mesons, different to the $\pi$, can contain a significant admixture of the two-gluon state at low scales, and alias a comparably large two-gluon DA. Several different reactions were considered in an effort to extract or at least constrain these contributions. Non-leptonic exclusive isosinglet decays [1731] and central exclusive production [1732] act as prominent probes for the gluonic Fock-state since the gluon production diagram enters already at LO. Exclusive semi-leptonic decays of heavy mesons were calculated in the framework of LCSRs [1733, 1734] and $k_T$-factorisation [1735]. These decays are simpler but the interesting gluon contribution enters only at NLO. Numerically it was shown that the gluonic contributions to $\eta$ production are negligible while they can reach a few percent in the $\eta'$-channel. Up to now experimental data are not conclusive in all these decays, with a vanishing gluonic DA being possible at a low scale. On the other hand, a large gluon contribution was advocated in [1736] from the analysis of $B_d \to J/\Psi\eta'$ transitions (see also [1737]).

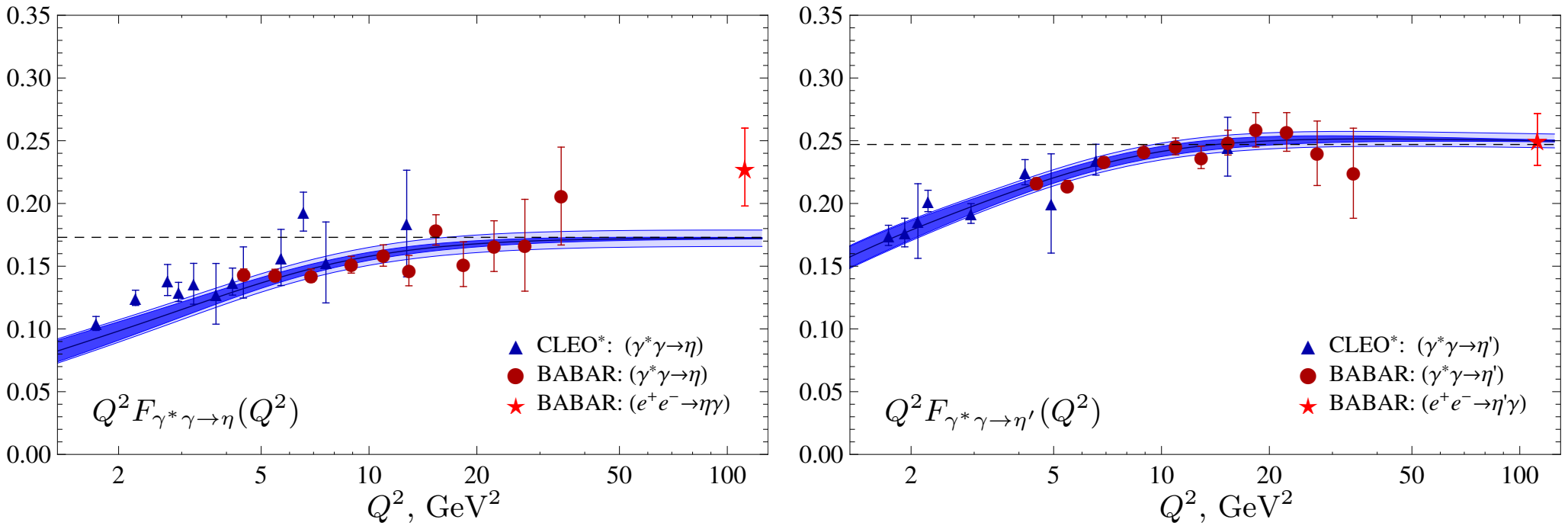

Fig. 197: Transition form factors $\gamma^*\gamma \to \eta$ (left panels) and $\gamma^*\gamma \to \eta'$ (right panels) [1725, 1738] compared to the LCSR calculation [1718]. Asymptotic values at large photon virtualities are shown by the horizontal dashed lines. The dark blue shaded areas correspond to uncertainty of the calculation due to the choice of various parameters. The light blue areas are obtained by adding the uncertainties in the mixing angles. The time-like data point [1739] at $|Q^2| = 112$ GeV$^2$ is shown by red stars for comparison.

The space-like FFs $\gamma^*\gamma \to \eta$ and $\gamma^*\gamma \to \eta'$ in the interval $4 - 40$ GeV$^2$ were measured by BaBar [1738]. In addition, in Ref. [1739] the processes $e^+e^- \to \gamma^* \to (\eta, \eta')\gamma$ were studied at a centre of mass energy of $\sqrt{s} = 10.58$ GeV. The measurements can be interpreted in terms of the $\gamma^*\gamma \to \eta, \eta'$ FFs at remarkably high time-like photon virtuality $Q^2 = -112$ GeV$^2$: Note that the time-like FFs are complex numbers, whereas only the absolute value is measured.

NLO QCD calculations for the FFs at Euclidean virtualities [1718] are, in general, in good agreement with data [1738], see Fig. 197, although the present statistical accuracy of the measurements is insufficient to distinguish between different models of the DAs and, in particular, determine the two-gluon DA. The most important effect of the NLO improvement is due to the finite renormalisation of the flavour-singlet axial current which results in a 20% reduction of the asymptotic value of the $\gamma^*\gamma \to \eta'$ form factor at large photon virtualities. It is interesting that the experimental result for $\gamma^*\gamma \to \eta'$ at $Q^2 = -112\,\text{GeV}^2$ [1739] is very close to the contribution of the asymptotic $\eta'$ meson DA, whereas the asymptotic contribution to $\gamma^*\gamma \to \eta$ is almost 50% below the data. This result urgently needs verification. If correct, it points to a much larger soft contribution alias a much broader DA of the $\eta$ meson as compared to $\eta'$, which would be in conflict with the state mixing approximation.

The Belle II experiment will be able to decrease the errors significantly. In addition, some of the parameters, most importantly the decay constants $f_\eta$, $f_{\eta'}$, will be calculated with high precision on the lattice. In this way the comparison of the QCD calculations with experimental results will allow one to study the structure of $\eta$, $\eta'$ mesons at short interquark separations, encoded in the DAs, on a quantitative level. This, in turn, will benefit theory studies of $B$ decays in final states involving $\eta$ and $\eta'$ mesons. The transition FF studies at time-like momentum transfers will eventually be complemented by studies of very rare exclusive decays of electroweak gauge bosons, *e.g.* $Z \to \eta\gamma$, in the high luminosity run at the LHC or, later, at a future lepton collider [1740, 1741].

Last but not least, in recent years there has been increasing interest in hard exclusive production of tensor mesons such as $f_2(1270)$, $K_2^*(1430)$ etc. by virtual photons or in heavy meson decays. One motivation is that having three different polarisations of tensor mesons in weak $B$ meson decays can shed light on the helicity structure of the underlying electroweak interactions. A different symmetry of the wave function and hence a different hierarchy of the leading contributions for the tensor mesons as compared to the vector mesons can lead to the situation that the colour-allowed amplitude is suppressed and becomes comparable to the colour-suppressed one. This feature can give an additional handle on penguin contributions. These studies are, comparatively, at their infancy, but the first results [1742] on the $\gamma\gamma^* \to f_2(1270)$ transition FFs are quite encouraging and agree well with the theory predictions [1743]. Also in this case Belle II has a potential to provide one with high quality data.

To summarise, studies of electromagnetic transition form factors at Belle II will result in stringent tests of the QCD factorisation formalism for hard exclusive reactions, provide one with quantitative information on the soft end-point contributions and in the long run enable novel searches for new physics.

*Experiment.* The $\pi^0$ transition form factor in $\gamma^*\gamma \to \pi^0$ has been measured in Belle [1724] where one of beam particles scatters into the acceptance of the detector by which the momentum transfer $Q^2$ of the virtual photon can be calculated, called a "Single-tag measurement". Therefore, this event has an electron (or a positron) and 2 photons. Usually, such events in Belle are strongly suppressed by the trigger to veto Bhabha events. Due to this Bhabha veto, a complicated selection condition for the polar-angle combinations of the electron and the two-photon system were imposed to reduce the uncertainty of the trigger inefficiency. As a result, it has turned out that the trigger efficiency is at the 10% level. The dominant sources for the systematic uncertainties are the extraction of the $\pi^0$ yield with the fit and the uncertainty of the trigger efficiency and the total systematic uncertainty for the combined cross section is between 8% and 14% (and between 4% and 7% for the form factor), depending on the $Q^2$ region. Figure 198 (black dots with error bar) shows the $Q^2$ dependence of the form factor, $Q^2|F(Q^2)|$. It is found that the form factor approaches asymptotically $0.209 \pm 0.016$ GeV [1724], which is slightly higher but consistent with the pQCD prediction of $\sim 0.185$ GeV [1744]. On the other hand, BaBar's measurement [1723] shows rapid growth with $Q^2$ in higher $Q^2$ region. Belle II results will draw great attention whether they reproduce these results or not.

In Belle II, since the trigger system will be designed taking into account this analysis, the previous restrictions on statistical power by the trigger, and the systematic uncertainty on the trigger efficiency will be improved upon. The statistical uncertainty is expected to reduce a factor of 8 (from 759 fb$^{-1}$ to 50 ab$^{-1}$) times 2.5 (expected improvement to the trigger efficiency) while the total systematic uncertainty should be at least 2 times smaller than that at Belle (mostly due to improvements on the trigger efficiency uncertainty). As a result, a factor 3 to 5 times more precise measurement is possible for the high $Q^2$ above 20 GeV$^2$, as shown in Fig. 198.

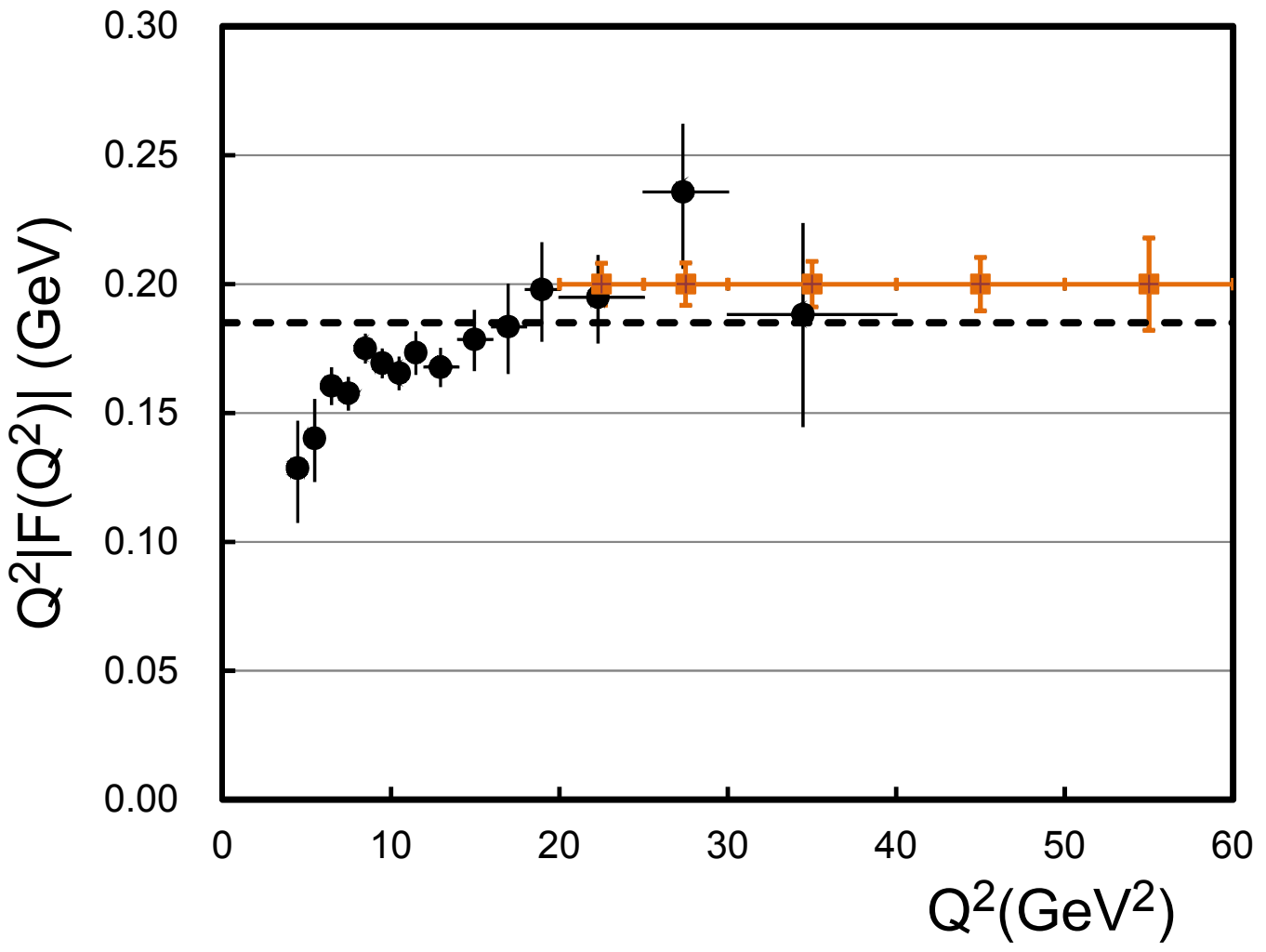

Fig. 198: Distribution of $Q^2|F(Q^2)|$. Black dots show the Belle result and the error bars for the red boxes show the Belle II expectation: The error bars are obtained from the quadratic sum of the statistical and systematic uncertainties. For the latter, the integrated luminosity is assumed to be 50 ab$^{-1}$. The central value for the red boxes are arbitrary. The dashed line is the asymptotic value of the TFF.

*15.7.2. Inputs for the determination of the hadronic contribution to light-by-light scattering in $(g-2)_\mu$. (Contributing authors: G. Colangelo, M. Hoferichter, M. Procura, P. Stoffer, S. Uehara)*

*Theory.* $\gamma\gamma$ physics allows one to constrain important input quantities needed for a data-driven analysis of the hadronic-light-by-light (HLbL) contribution to the anomalous magnetic moment of the muon $(g-2)_\mu$, a relation that can be studied in a systematic way within dispersion theory [1745–1749], where the HLbL amplitude is reconstructed in terms of its analytic properties. Expanding in terms of the mass of intermediate states, the dominant contribution at low energies originates from pseudoscalar poles, $\pi^0$, $\eta$, $\eta'$, followed by cuts generated by two-meson states, $\pi\pi$, $K\bar{K}$, and higher contributions e.g. from multi-pion intermediate states are further suppressed. This expansion scheme, illustrated in Fig. 199, ensures that all building blocks correspond to on-shell particles and are thus observable quantities, in the case of the pseudoscalar poles doubly-virtual transition form factors, for the two-meson cuts doubly-virtual helicity partial waves, and in principle similarly for higher intermediate states. Due to the suppression from phase space and energy thresholds the contributions from heavier states become more and more suppressed in a dispersive reconstruction of the amplitude, which together with the convolution with photon and muon propagators in the $(g-2)_\mu$ integral makes the low-energy region dominant. In practice, an explicit description in terms of individual channels is feasible for centre-of-mass energies $\lesssim 1.5$GeV, and information on the virtuality dependence is most critical in the same region. However, information on larger virtualities can still be useful to assess the asymptotic behaviour, with the pion transition form factor one prime example.

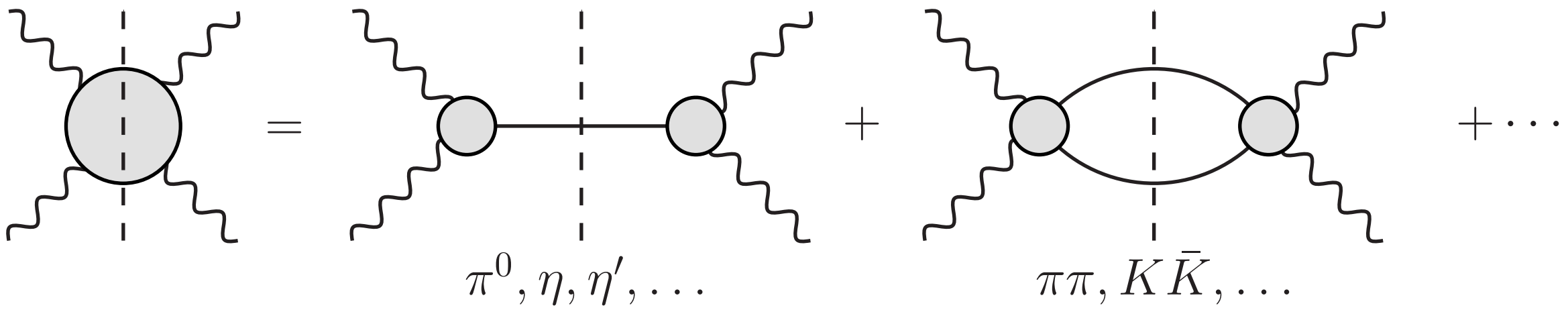

Fig. 199: Singularities of the HLbL amplitude. Solid lines denote meson intermediate states, wiggly lines the (virtual) photons, the gray blobs hadronic amplitudes, and the dashed lines indicate intermediate states taken on-shell.

While the $(g-2)_\mu$ integral requires only space-like virtualities, measurements for time-like photons can provide additional invaluable information, see [1746] for a comprehensive list of processes that can help constrain one- and two-pion intermediate states, and Fig. 200 for the corresponding space-like and time-like processes in $e^+e^-$ scattering. In fact, the singularity structure of the amplitudes is often dominated by (time-like) kinematics where the virtualities coincide with the mass of a vector meson, which can then be used to constrain the space-like amplitude. For instance, the pion vector form factor in the space-like region is predicted very accurately from the combination of time-like data and dispersion relations. Even though a direct measurement of the full virtuality dependence is often unrealistic, strategies along those lines can help constrain the full amplitudes from experimentally accessible quantities. For $\pi^0$ intermediate states this includes $\gamma\pi \to \pi\pi$ [1750, 1751], $\omega, \phi \to \pi^0\gamma^*$ [129], $\omega, \phi \to 3\pi$ [130], $e^+e^- \to 3\pi$ [1752], and similarly for $\eta$ and $\eta'$ in relation to $\eta, \eta' \to \pi\pi\gamma$ as well as crossed reactions [106, 107, 1753, 1754]. A complete analysis along these lines has been carried out for the $\pi^0$ pole [1755, 1756], indicating the impact of the various input quantities. In fact, a major component of the uncertainty estimate traces back to the asymptotic behaviour of the singly-virtual pion transition form factor, which could be clarified at Belle-II. In the same way, data for the on-shell process $\gamma\gamma \to \pi\pi$ (and $\gamma\gamma \to K\bar{K}$) have become sufficiently precise to allow for a detailed understanding on the level of partial waves [1757–1759], while a partial-wave analysis in the whole space of relevant virtualities is clearly beyond reach. However, already less comprehensive virtual data provide valuable constraints on the doubly-virtual amplitude [1760, 1761], as do several of the aforementioned processes by constraining the left-hand cut.

Finally, one way to estimate the impact of contributions beyond one- and two-meson intermediate states relies on the assumption that the dominant such terms are generated by resonances in the multi-meson systems [1762]. Information on their transition form factors crucial for this strategy can be partially reconstructed from light-by-light sum rules [1763, 1764], but estimates along these lines could be improved with more information on the electromagnetic form factors of axial-vector, scalar, and tensor resonances between 1 and 2GeV.

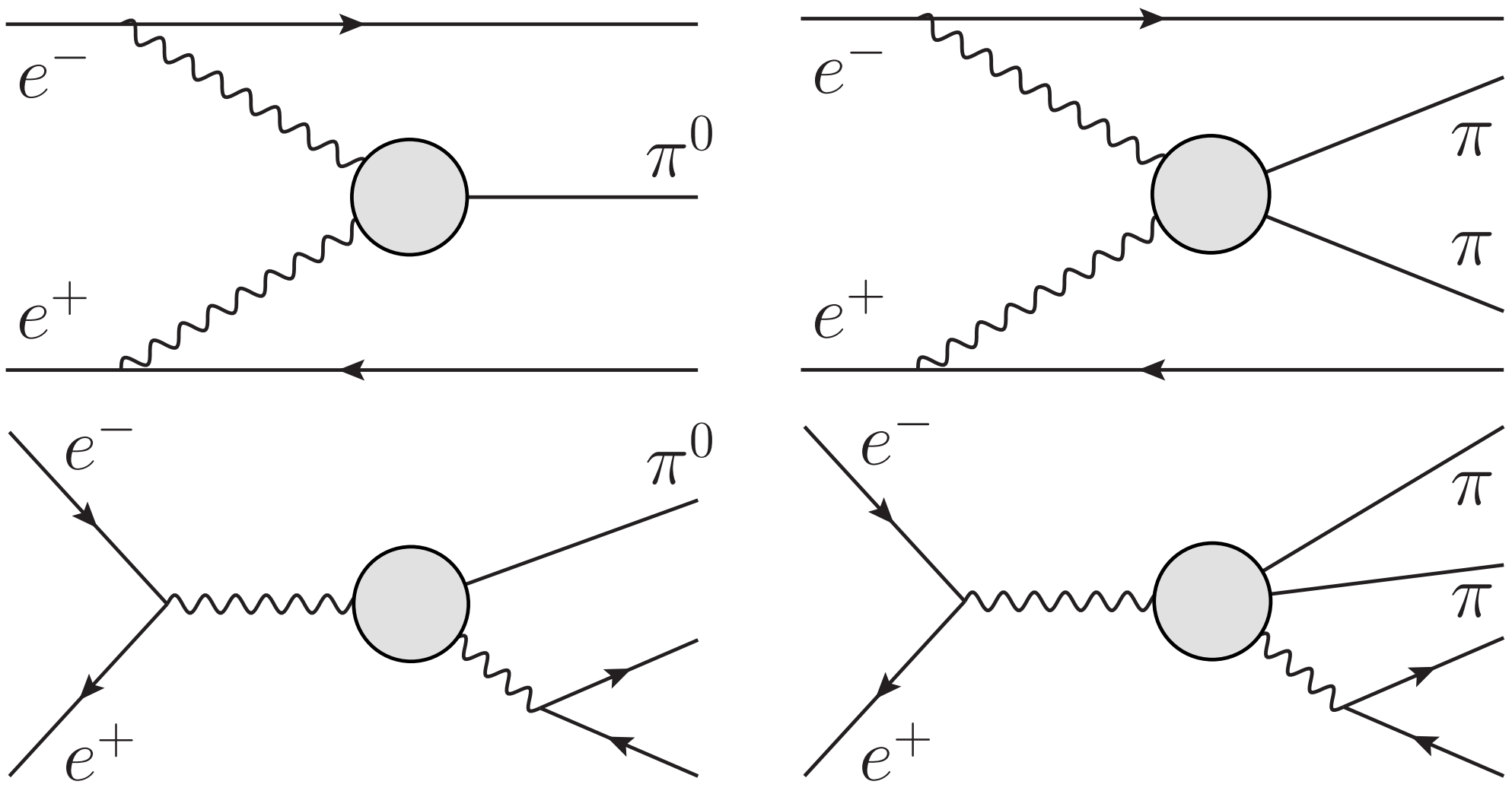

Fig. 200: $e^+e^- \to e^+e^-\pi^0$ and $e^+e^- \to e^+e^-\pi\pi$ in space-like (top) and time-like (bottom) kinematics.

*Experiment.* Collisions of two space-like photons are measured through double-tagged two-photon processes at $e^+e^-$ colliders. The reactions such as $\gamma^*\gamma^* \to \pi^0$, $\eta$, $\pi\pi$ *etc.* have a recoiling electron and positron, and a hadronic system in the final state. We "tag" both the electron and positron to measure the $Q^2$'s of the two virtual photons, where $Q^2 \equiv -q^2$ is the negative of the four-momentum squared of the virtual photon. In general, we need to measure the final-state particles exclusively, including the hadronic system, as the double-tag production has a small cross section, and background rejection requires clear identification of the final state. No experimental measurement of double-tag production of a hadronic system below $W < 5$ GeV has been reported to date, where $W$ is the centre-of-mass energy of the two-photon collision system and is the same as the invariant mass of the hadronic system.

The double-tag process, like Bhabha scattering, is detected via the presence of a high-energy electron-positron pair. We typically apply a Bhabha veto to reduce the trigger rate in the low-level trigger which should not reject the signals. It should be possible to keep good efficiency for this purpose by examining the angular correlation of the electron and positron and measuring activity originating from the hadronic system. In the case that hadronic system is decays with a high final state multiplicity, the trigger and measurement is substantially easier.

The $Q^2$ range over which is covered by the Belle II detector (with ECL) is typically above $Q^2 > 2$ GeV$^2$ for both electron- and positron-tag sides. The sensitivity to the high end of the $Q^2$ distribution will be limited by signal cross sections that steeply decrease with $Q^2$. As the accelerator energy and detector acceptance are asymmetric in the two beam directions, the dependence of the efficiency in $Q^2$ differs between the two tags. This feature can provide a method for a consistency check of efficiency evaluations, by swapping $Q^2$'s of the electron and positron sides ($Q^2_-$ and $Q^2_+$, respectively) based on $C$ symmetry of the differential cross section, $d^3\sigma/dW dQ^2_- dQ^2_+$. Pseudoscalar-meson pair production below $W < 2$ GeV, dominated by tensor- or scalar-meson resonance formation processes, have been measured with zero-tag and single-tag two-photon events [2, 1742].

The coupling of two time-like photons to a hadron (or a hadronic system) can be measured through an $e^+e^-$ annihilation process to hadron(s) accompanied by a virtual photon, which converts to an electron or muon pair. In the case of the electron pair, the final-state particle combination is just the same as that of the double-tag process. However, measurement methods of photon $q^2$ ($Q^2$) are different from the double-tag process case, and we have to distinguish them with the invariant mass of the electron pair.

We note that a hadronic system consisting of $\pi^+\pi^-$, $K^+K^-$ *etc.* can have a $C$-odd component that has converted from a single virtual photon, in addition to the targeted two-photon diagram. In some cases, the $C$-odd component can be larger than the $C$-even component and interfere with the latter, making it difficult to interpret the experimental results. Pure $C$ eigenstate such as $\pi^0\pi^0$ and $K^0_S K^0_S$ should have an advantage on this point.

## *15.8. Conclusions*

The high luminosity data at Belle II will provide us with a unique opportunity to investigate the physics program of the tau lepton and low multiplicity processes. The wealth of data will enable a great advance in this field in the coming years.

LFV processes, such as $\tau \to \mu(e)\gamma$, $\tau \to \mu(e)h$ ($h$ = hadrons) and $\tau \to \mu\mu\mu$ (or processes with $\mu$ replaced by $e$), are known to be very sensitive to new physics. In particular we choose $\tau \to \mu\gamma$ and $\tau \to \mu\mu\mu$ as the Belle II golden channels and have discussed these processes in detail from both a theoretical and an experimental perspective. $\tau \to \mu\gamma$ is known to be sensitive to SUSY GUT models. In this report, we have discussed the status of this class of models after taking into account the recent results from the LHC (ATLAS and CMS). From these direct particle searches a large parameter space of such models has been excluded in recent years. Not having some obvious new physics benchmark models at hand, we also discussed LFV from an Effective Field Theory point of view. On the experimental side, we have presented a beam background impact study based on Belle II simulation. We showed that the sensitivity to LFV processes will naturally improve as the integrated luminosity increases. At the same time, the high luminosity will allow us to impose further experimental criteria, which should help to reduce some systematic uncertainties.

$CP$ violation in $\tau$ decays is another golden measurement chosen in this chapter. The observed hint of a deviation from the SM expectation in the $\tau^\pm \to \pi^\pm K_S \nu_\tau$ channel is extremely intriguing. If it is confirmed, this will be the first observation of $CP$ violation in the leptonic sector, thus leading to a discovery of new physics. We also reviewed some $CP$ violation measurements involving angular observables with hadronic two-body and three-body final states. In the two-hadron final states, $CP$ violation can occur from the interference between the vector and the scalar currents, when, for example, a scalar particle, such as a new Higgs particle interacting with a coupling that is $CP$ violating. The hadronic part of these processes is described by form factors. The form factor parameterisation, which takes into account important final state interactions, has been discussed in detail. In particular, it was pointed out that the forward-backward asymmetry is one of the most interesting observables whose measurement will be largely improved in the era of Belle II.

In contrast to the other leptons, the tau lepton can decay into various leptonic and hadronic final states, which allow us to perform precise measurements of its properties. The Michel parameters have the advantage to allow us to investigate the $V - A$ interaction of the SM for leptons without hadronic uncertainties. The study of the second class currents provides a stringent test of the G-parity conservation in the SM. The Belle II measurement of the tau mass, which is a fundamental parameter of the SM, will offer an important test with respect to the threshold methods that are currently used, *e.g.* by BESIII and KEDR.

The g-2 and EDM of the tau lepton can be also measured: by knowing the initial $e^+e^-$ energy, we can construct so-called optimal observables. The hadronic tau lepton decays can also be used to determine the CKM matrix element $|V_{us}|$ which constitutes an important input for the global CKM matrix parameter fits. Moreover, the study of the hadronic tau decays is crucial to determine the strong coupling constant $\alpha_s$, input for various new physics search at collider experiments, including the Higgs mass. For this purpose, the spectral function measurements of the inclusive tau decays are needed. These studies at the $B$ factories have been very limited and the theory community is strongly wishing to have new data from Belle II.

All the investigations using tau leptons require a specific Monte Carlo generator, called KKMC for the production and TAUOLA for the decays. A mini-review of the status of KKMC and TAUOLA is presented.

The Belle II experiment will also contribute to the clarification of one of the most significant deviations from the SM (at $4\sigma$ level) observed in the measurement of the muon anomalous magnetic moment, $(g-2)_\mu$. As the experimental measurement of $(g-2)_\mu$ at Fermilab and J-PARC will be soon improved and eventually by a factor of 4, it is an urgent matter to improve the theoretical uncertainty, which is currently dominated by the uncertainty on the hadronic contributions. This can be achieved by measuring the cross section of $e^+e^- \to \pi^+\pi^-$ with the ISR technique. Belle was not able to perform such a measurement due to the lack of specific triggers. We reviewed the $e^+e^- \to \pi^+\pi^-$ cross section measurements following the BaBar analysis and a first sensitivity estimate at Belle II was provided. The so-called two photon processes, $e^+e^- \to \gamma\gamma e^+e^-$, provide information on the two photon coupling to $\pi$, $\eta$ and $\eta'$. There are two types of experimental signals: single-tag where either $e^+$ or $e^-$ is detected and double-tag where both $e^+$ and $e^-$ are detected. Both signals will be studied at Belle II. The former will provide information on the space-like pion form factor, which is a fundamental input for many of the theoretical computations of the B meson decays. The latter can be related and will improve the determination of the so-called light-by-light contribution to the theoretical prediction of the anomalous magnetic moment of the muon ($(g-2)_\mu$), which constitutes the most important theoretical uncertainty after the hadronic contribution mentioned above.

As presented in this section, the investigations of tau and low-multiplicity processes are very unique to Belle II. Belle II will therefore play a major role for searching for new physics signals either directly via LFV processes or $CP$ violating processes or indirectly by providing crucial information for testing the SM such as for $(g-2)_\mu$, $|V_{us}|$ extractions or $\alpha_s$ determination.

## 16. Dark Sectors and Light Higgs

*Editors: T. Ferber, K. Hayasaka, E. Passemar, J. Hisano*

*Additional section writers: M. Dolan, S. Godfrey, C. Hearty, G. Inguglia, F. Kahlhoefer, H. Logan, J. Pradler, K. Schmidt-Hoberg*

## *16.1. Theory*

*(Contributing authors: M. Dolan, F. Kahlhoefer, J. Pradler, K. Schmidt–Hoberg*)*

*Motivation and Context* In the past decades, an extensive experimental search program has been devoted to dark matter (DM) particles with mass and interaction strength comparable to the electroweak scale, the so-called WIMP paradigm. In recent years, however, the possibility that both DM and the particles mediating its interactions to the Standard Model (SM) have a mass at or below the GeV–scale has gained much traction; see, *e.g.* [1765] and references therein.

The interest in light DM and/or light mediators is not new [1766–1769]. Most notable is the frenzy of activity that set in once it was realised the harder-than-expected cosmic ray positron fluxes [1770–1772] may be tied to DM and light mediators that leptonically couple to the SM; see [1773–1776] among many following works.

Another recent interest in light mediators, possibly (much) lighter than the DM particle, concerns DM self-interactions [1777–1780]. The ensuing heat transfer from outer to inner parts of DM halos may potentially explain the discrepancies between $N$-body simulations of collisionless cold DM and observations on small scales, see [1781] and references therein.

When the DM mass is at or below the GeV-scale, cosmology provides another motivation for a richer dark sector, since achieving the correct relic abundance requires a light spin-1 vector ($V$) or spin-0 scalar particle $S$ facilitating the annihilation of DM via $\chi\chi \to V^*, S^* \to$ SM or $\chi\chi \to VV$ or $SS$. In the latter case, $V$ or $S$ is required to be lighter than DM and offers the possibility to seclude it from the SM [956, 1782]. At the same time, the mediator coupling to the SM cannot be made too weak, since it will eventually interfere with primordial nucleosynthesis [1783] (although the precise coupling values and lifetimes are model-dependent).

Independently of all such motivations, New Physics below the GeV-scale is a perfectly viable option *per se*, and therefore needs to be explored with all the experimental and observational tools available. The direct detection of sub-GeV DM particles is hampered by small energy depositions and finite detector thresholds, although it is still possible to set limits [1784, 1785].

The efforts to detect dark matter in the sub-GeV mass bracket as well as the particle mediating the interactions with SM states have therefore been actively pursued at fixed target experiments and with high intensity, low-energy colliders. Searches at previous experiments such as Babar [1786, 1787] have been the subject of theoretical interest and recasting [1788, 1789], while neutrino experiments such as MiniBoone [1790] and the proposed DUNE facility [1791] also have sensitivity in this parameter space. Limits have also been derived using existing results from beam-dump experiments [1789, 1792] and there are also proposals for multiple future experiments oriented towards low-mass dark sectors [1793, 1794].

*Light dark sectors* There are only a small number of ways to couple a new light SM gauge singlet to the SM in a renormalisable way [1795]. A new vector or pseudo-vector particle $V^\mu$ can couple to a SM current $J^\mu_{\rm SM}$ via the *vector portal*

$$\mathcal{L} \supset \epsilon V_\mu J^\mu_{\rm SM} \,. \tag{555}$$

The most frequently studied example is when $J^{\mu}_{\rm SM}$ is the electromagnetic current and the coupling $\epsilon = \kappa e/\cos\theta_W$ arises from kinetic mixing of the hypercharge ($Y$) and the vector field strengths, $(\kappa/2)V_{\mu\nu}F^{\mu\nu}_{Y}$ [1796], with $\kappa$ the kinetic mixing parameter. In this case, $V$ is then often called a "dark photon".

A new scalar particle $S$ can couple to the SM Higgs field $H$ via the *Higgs portal*

$$\mathcal{L} \supset \lambda S^2 (H^\dagger H) \; . \tag{556}$$

If this scalar acquires a vacuum expectation value, it will mix with the SM Higgs, leading to couplings to SM fermions of the form $\sin\theta \, y_q \, \bar{q}q \, S$ with $\theta$ the Higgs-singlet mixing angle and $y_q$ the Yukawa coupling of $q$.

Finally[65], for a pseudoscalar $P$, couplings to SM fermions can arise from the dimension-5 *axion portal* [1797]

$$\mathcal{L} \supset \frac{\partial_\mu P}{f_A} \bar{f}\gamma^\mu\gamma^5 f \; . \tag{557}$$

This term is obtained for example from the spontaneous breaking of a global symmetry, where $f_A$ is the scale at which the symmetry is broken. If the scale $f_A$ is sufficiently large, the pseudoscalar naturally obtains a small mass and small couplings. While this operator is higher-dimensional, there are very simple UV completions which give rise to such a term. For instance such a term naturally arises in extended Higgs sectors such as in the two Higgs doublet model (2HDM), encountered in the context of supersymmetry. In particular if an additional singlet is present as *e.g.* in the NMSSM there is a limit in which the singlet-like pseudoscalar is naturally at the GeV scale, precisely the region which will be probed by Belle II.

In addition to fermionic couplings, the axion portal generically induces couplings of the pseudoscalar to SM gauge bosons:

$$\mathcal{L} \supset -\sum_i \frac{\alpha_i}{8\pi} \frac{C_i}{f_A} F^b_{(i)\,\mu\nu} \tilde{F}^{b\,\mu\nu}_{(i)} \, P \; , \tag{558}$$

where $i = \{Y, 2, 3\}$ labels the different gauge groups of the SM, $F^{\mu\nu}_{(i)}$ denotes the corresponding field strength tensor and we have furthermore defined the dual field strength tensors $\tilde{F}^{\mu\nu}_{(i)} = \frac{1}{2}\epsilon^{\mu\nu\rho\sigma}F_{(i)\,\rho\sigma}$. Similar interactions are expected to arise from string compactifications [1798], with $f_A \sim M_{\rm string}$ and coefficients $C_i$ that are typically of order unity. Of particular interest is the pseudoscalar-photon coupling

$$\mathcal{L} \supset -\frac{g_{\gamma\gamma}}{4} F_{\mu\nu}\tilde{F}^{\mu\nu} \, P \; , \tag{559}$$

where we have introduced the effective coupling

$$g_{\gamma\gamma} \equiv \frac{\alpha_i}{2\pi} \frac{C_Y \cos^2\theta_{\rm W} + C_2 \sin^2\theta_{\rm W}}{f_A} . \tag{560}$$

Light pseudoscalars with couplings to SM gauge bosons can play an interesting role in cosmology [1799] and can potentially explain the anomalous muon magnetic dipole moment

[65] For completeness, we also mention the *neutrino portal*, $N(LH)$, where $N$ a sterile neutrino. There is indeed ample chance that this portal is realised in nature, since $N$ can be a (heavy) right handed neutrino that is being invoked for generating a SM neutrino mass term. Nevertheless, we will not discuss this portal further in the present context.

$(g-2)_\mu$ via Barr-Zee diagrams and light-by-light scattering [1800]. Relevant constraints come from collider experiments [1801–1803], fixed-target experiments [1804] and searches for rare decays [1805]. The expected Belle II sensitivity is discussed in Sec. 16.2.2.

Each of the new particles introduced above can be a viable DM candidate [636], provided it is sufficiently light and its couplings to the SM are so small that it is stable on cosmological scales (or, in the case of the Higgs portal, if the $Z_2$ symmetry remains unbroken). Alternatively, the particles can be unstable and act as the *mediator* between the SM and another (stable) particle in the dark sector (which is itself a SM singlet and may be either a boson or a fermion). The role of the mediator may then be to keep the DM particle in thermal equilibrium with the SM in the early Universe and provide the annihilation processes that are necessary to reproduce the observed DM relic abundance via thermal freeze-out. Depending on the details of the model, the mediator can couple to both quarks and leptons, only to quarks (leptophobic) or only to leptons (leptophilic). In principle, it can also couple with different strength to down-type quarks and to up-type quarks. The scalar and the pseudoscalar mediator are furthermore expected to couple to fermions proportional to their mass.

The crucial point is that – in contrast to DM candidates from portal interactions – mediators from portal interactions may have sizeable couplings to the SM, which can potentially be probed in particle physics experiments. In the following section we discuss the general ways how such a mediator can be produced in $e^+e^-$ colliders and what experimental signatures would result from its various decay modes.

*Production and decay modes* There are three fundamentally different ways to produce light mediators ($M = S, P, V$) at B factories [616, 1769, 1806–1808]:

(1) Direct (i.e. non-resonant) production from the annihilation of an electron-positron pair: $e^+e^- \to M + X$
(2) Resonant production from a tree-level decay, for example $e^+e^- \to \Upsilon(nS) \to M + X$.
(3) Resonant production from a loop-level rare decay, for example $e^+e^- \to B + X \to K + M + X$.

Once produced, the mediator can have three different types of decays:

(1) Invisible decays
(2) Leptonic decays
(3) Hadronic decays

If the DM mass is less than half of the mediator mass, the first decay mode is expected to be the dominant one and the production of the mediator will lead to missing momentum in the detector. If invisible decays are kinematically forbidden, there will typically be both leptonic and hadronic decays (unless of course the mediator is either leptophilic or leptophobic). For scalar and pseudoscalar mediators the leptonic decay modes will be dominated by the heaviest lepton that is kinematically accessible, while vector mediators decay equally into all available leptonic channels. Calculating the hadronic branching fractions for mediators in the GeV range is a difficult problem due to the onset of non-perturbative effects [1809, 1810], so we will limit ourselves to the discussion of general signatures.

The first production mechanism relies on the coupling of the mediator to electrons, so it is most relevant for vector mediators. Depending on the decay modes of the mediator, it may be difficult to reliably trigger on this production. If the mediator decays invisibly, it is essential that additional visible particles are produced in association with the mediator. The most promising channel then is to search for the production of a single photon in association with missing energy [1788]. Provided the photon energy is sufficiently large, backgrounds can be efficiently suppressed and a high trigger acceptance can be achieved. For leptonic decays, the events are easier to identify. A possible way to enhance the trigger acceptance, although at the cost of a smaller cross section, would be to insist that the $\gamma$ converts: $e^+e^- \to \gamma^* + M \to e^+e^- + M$ [1788]. For a (pseudo)scalar coupling to leptons proportional to their mass it may even be advantageous to consider $e^+e^- \to \gamma^* \to \tau^+\tau^-$ followed by final state radiation of the mediator from a tau lepton [1811].

A very efficient way to trigger on the second production mechanism is to focus on the case that the mediator is produced in the decay of an $\Upsilon(1S)$, which in turn is produced from the decay of $\Upsilon(2S)$ or $\Upsilon(3S)$: $e^+e^- \to \Upsilon(2S, 3S) \to \Upsilon(1S) + \pi\pi \to M + X + \pi\pi$. For example, for a spin-0 mediator or a spin-1 mediator with axial couplings one can study the radiative decay $\Upsilon(1S) \to M + \gamma$ (for a spin-1 mediator with purely vectorial couplings this process is forbidden by $C$ conservation [1769]). We note that, although Belle II will take most of the data at higher centre-of-mass energy (corresponding to the mass of $\Upsilon(4S)$ and $\Upsilon(5S)$), a sizeable number of $\Upsilon(2S)$ and $\Upsilon(3S)$ may still be produced if some of the centre-of-mass energy is taken away by a photon from initial state radiation. In principle the mediator could also be produced directly in the radiative decay of a heavier resonance, *e.g.* $\Upsilon(3S)$, in which case the signature would resemble the mono-photon signal discussed above [1786].

If a new mediator is produced in radiative decays of an $\Upsilon(1S)$ resonance, this will lead to a bump in the photon energy, which can be readily identified experimentally. If in addition the mediator decays leptonically, one also obtains a peak in the invariant mass of the dilepton pair, leading to very efficient suppression of backgrounds [1812]. Even if the mediator decays into $\tau^+\tau^-$ [1813], gluons [1814] or hadrons [1815] it may be possible to search for features in the invariant mass of the decay products, providing the most promising search strategy for these channels.

The third production mode relies on loop-induced flavour-changing processes (involving *e.g.* penguin diagrams with internal $W$-bosons and top-quarks) [616]. The rate for these processes may depend on the details of the model in the UV as well as on uncertain hadronic form factors [1797, 1810, 1816, 1817]. Nevertheless, the resulting experimental signatures are clear: For example, for an invisibly decaying mediator the process $B \to K + M$ will lead to $B \to K + \text{invisible}$, which is experimentally very similar to the rare decay $B \to K\nu\bar{\nu}$ [1810, 1818]. A measurement of this rare SM process can therefore be translated into a bound on the couplings of the new mediator, although the kaon momentum distribution may differ from the Standard Model prediction, thus motivating a rather inclusive search. For a visibly decaying mediator the process $B \to K\mu^+\mu^-$ is of particular importance, as it is often leads to the strongest limit on the SM coupling in the case of a scalar or pseudoscalar mediator [1797, 1816]. Rare B decays may also have a unique sensitivity to photonic decays of the mediator, which can be relevant for (pseudo)scalar mediators with a mass comparable to the muon threshold: $B \to K + M \to K + \gamma\gamma$ [1810].

Finally, we consider a somewhat more complicated set-up and assume that decays of the mediator into DM are kinematically forbidden and couplings to Standard Model particles are so small that the mediator decay length becomes sizeable. Such a scenario is in fact quite likely given the strong current bounds on the SM-mediator couplings (see *e.g.* [1810]). The resulting signature would then be a leptonic or hadronic decay of the mediator from a displaced vertex. The most promising production mode in this case would be direct production of the mediator in association with a photon, so that there is no activity at all at the collision point and the displaced vertex can be readily identified. Searching for displaced vertices in the rare decay $B \to K\mu^+\mu^-$ may also place strong constraints on such models [1819].

Alternatively, it may be possible to produce such a mediator off-shell, such that decays into a pair of DM particles are allowed [1788]. This process can for example be searched for in radiative $\Upsilon$ decays, taking into account that the photon energy is now continuous rather than having a bump: $\Upsilon(1S) \to \gamma + M^* \to \gamma +$ invisible. For a vector mediator one can instead study the case that the $\Upsilon(1S)$ decays fully invisibly, such that the event is only visible due to the pions from the decay of the heavier $\Upsilon$ resonance: $\Upsilon(3S) \to \Upsilon(1S) + \pi\pi \to \pi\pi + \mathrm{inv}$ [1820]. These searches for non-resonant invisible decays may also allow to constrain mediators with a mass above the centre-of-mass energy of the collider, provided the DM mass is small enough [1821, 1822]. For CP-even scalar mediators, an analogous search can be performed in the decays of scalar bottomium $\chi_b$ [1823]. These searches can be used to constrain the interactions of DM via heavy mediators in a model-independent effective operator approach.

To conclude this discussion, we note that it is also conceivable that there is more than one new mediator. For example, the mass for a vector mediator $V$ could arise from a dark Higgs boson $H' = (h' + v')/\sqrt{v'}$ giving interactions such as $(m_V^2/v')h'V_\mu^2$ and $(m_V/v')^2 h'^2 V_\mu^2$, while $H'$ couples to the SM via the Higgs portal. In such a scenario the dark Higgs may be produced via dark Higgsstrahlung from the vector mediator [1808], which can lead to striking signatures such as $e^+e^- \to 3\ell^+3\ell^-$ [1824, 1825].

*Specific examples* Let us discuss a few specific examples cases to elucidate the projected sensitivity of Belle-II. These examples are far from comprehensive, but they cover most of the search channels discussed above.

The first example is the search for invisibly–decaying Dark Photons $A$ into light DM $\chi$ , where $A$ couples to the SM via the kinetic mixing parameter $\epsilon$. The projected sensitivity and expected backgrounds are discussed in detail in Sec.16.2.1. Existing limits and projections of future experiments searching for invisibly–decaying Dark Photons are shown in Fig. 201. In case of a discovery the energy distribution of the single photon may even allow to perform a dark sector 'spectroscopy' [1828].

The second example, taken from [1811], considers a new scalar mediator coupling exclusively to leptons with coupling strength $g_\ell = \xi_\ell^S m_\ell/v$, with $\xi_\ell^S$ a mixing parameter. The presence of a light DM particle is not assumed, so the mediator decays dominantly into the heaviest lepton that is kinematically accessible. The left panel of figure 202 shows model-independent bounds, which only consider tree-level processes. In this case the leading constraints from Belle II result from processes where the scalar mediator is radiated from a tau lepton in the final state (orange dashed). If the mass of the scalar is below the muon threshold, its decay length can become comparable to the size of the detector, leading to constraints that become weaker for smaller mediator masses.

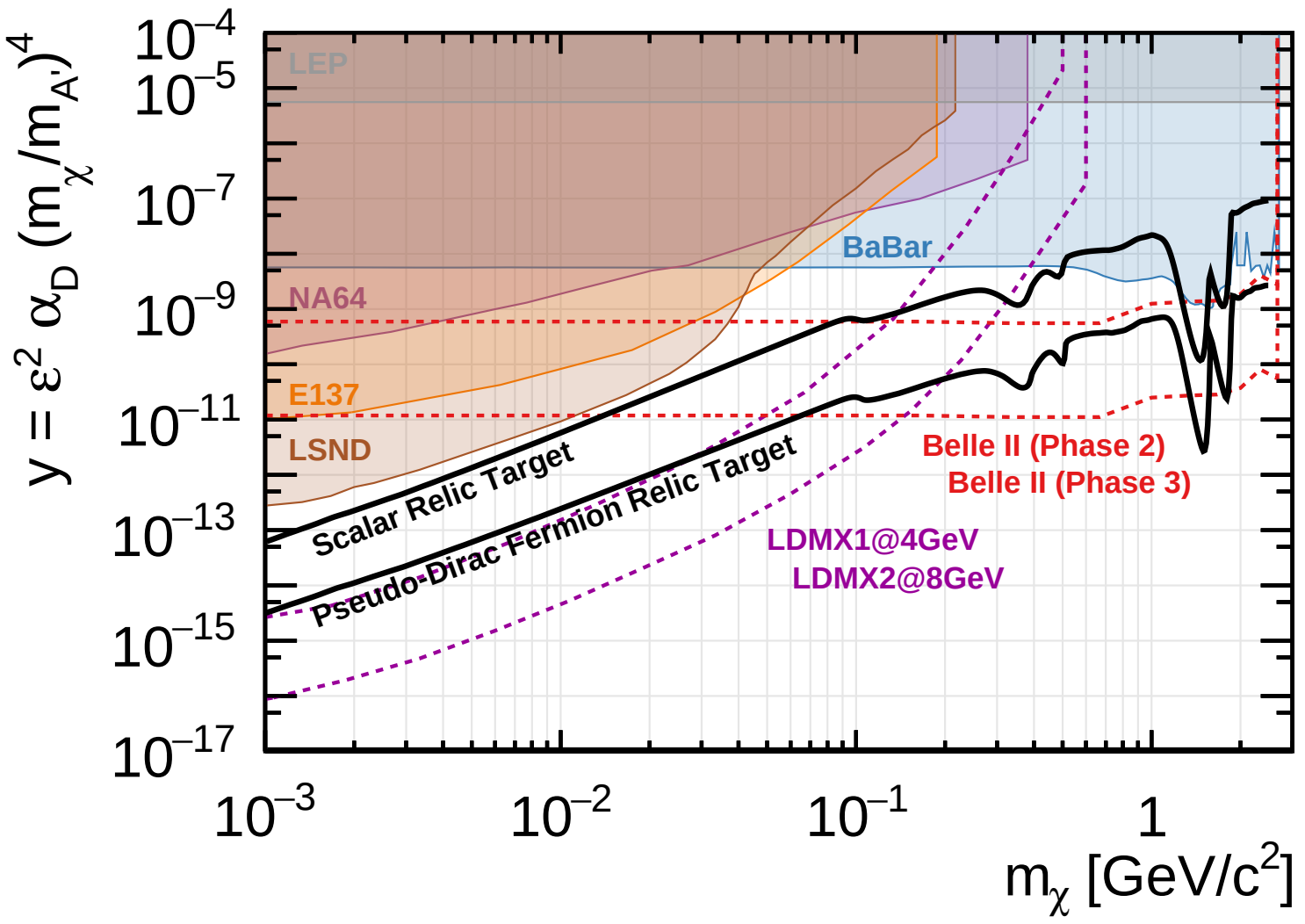

Fig. 201: Combined projections (LDMX, Belle II) and constraints, encapsulating direct production LDM constraints in the context of a kinetically mixed Dark Photon coupled to a LDM state that scatters elastically (or nearly elastically) at beam–dump, missing energy, and missing momentum experiments (Dark Photon mass $m_{A'} = 3m_\chi$ and coupling of the Dark Photon to Dark Matter $g_\chi = 0.5$ where applicable) [1826, 1827]. The Belle II projection for Phase 3 is extrapolated from the limit for Phase 2 (see Sec.16.2.1). Note that the relic density lines assume a standard cosmological history and that there is only a single component of dark matter, which only interacts via Dark Photon exchange.

The right panel considers a specific UV-completion in terms of a Leptonic Two-Higgs Doublet Model. In this case, it is possible to calculate the rate for loop-induced rare decays, such as $B \to K\mu^+\mu^-$ or $B_s \to \mu^+\mu^-$. The corresponding searches from LHCb are found to give very strong constraints, which may further be improved by Belle II by searching for displaced vertices in B meson decays. We note that the Leptonic Two-Higgs Doublet Model also predicts additional tree-level processes, such as $h \to SS$ that can be constrained by low-energy experiments. We refer to [1811] for details.

For the third example, we consider a pseudoscalar coupling exclusively to quarks with coupling strength $g_q = g_f m_q/v$. Again, the pseudoscalar is assumed to decay visibly, so it can be observed in radiative $\Upsilon$ decays, *e.g.* $\Upsilon(2S) \to \gamma +$ hadrons. In figure 203 we show in blue the bound on this process from BaBar [1814], calculating the $\Upsilon(2S)$ branching ratio following Ref. [1810]. For the Belle II projection (purple dotted), we assume that the sensitivity scales proportional to the square root of the number of $\Upsilon(2S)$ and $\Upsilon(3S)$ produced, which we take to be a factor of 16 larger than in BaBar. The grey shaded region indicates the pseudoscalar masses for which the process $B \to K + P$ is kinematically allowed and additional (model-dependent) constraints may be obtained from rare decays such as $B \to K\gamma\gamma$.

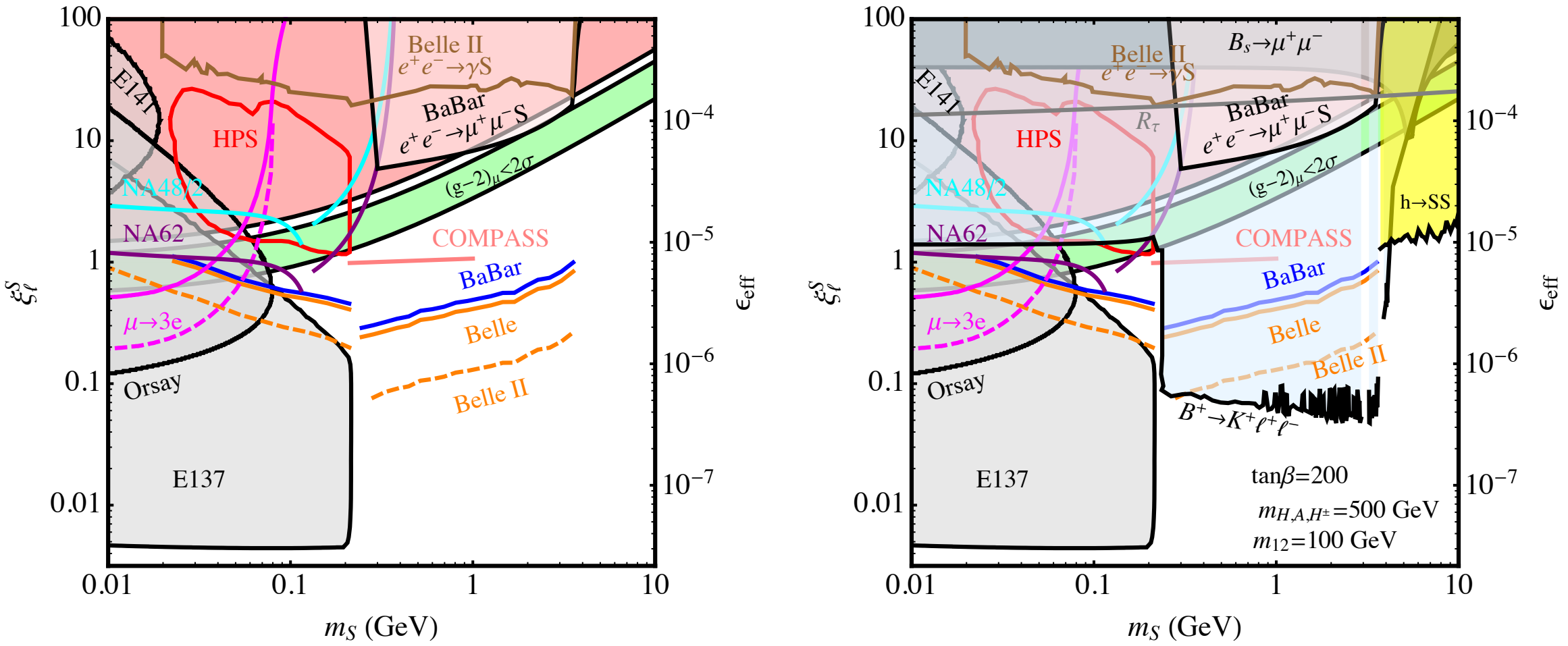

Fig. 202: Model–independent constraints on the effective coupling to leptons (left panel) and specific constraints for the Leptonic Two-Higgs Doublet Model (L2HDM+$\varphi$) UV completion (right panel) as a function of the scalar mass, reproduced from [1811].

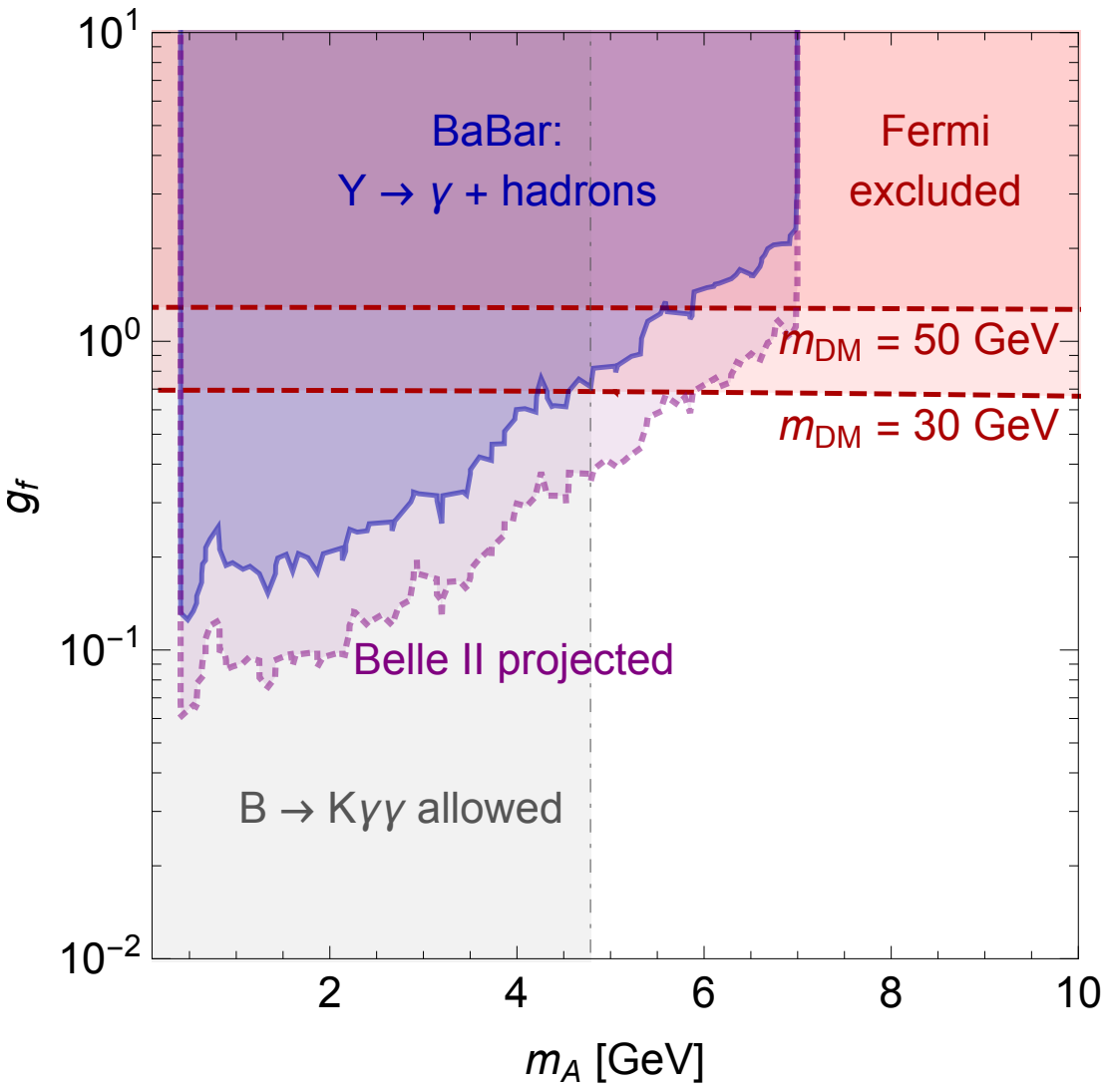

Fig. 203: Constraints on a leptophobic pseudoscalar from radiative $\Upsilon$ decays compared to bounds from Fermi-LAT. See [1810] for details.

It is intriguing to assume that such a light pseudoscalar mediator is responsible for setting the DM relic abundance via the processes $\chi\chi\to PP$ and $\chi\chi\to q\bar{q}$. Indeed, direct detection experiments are almost completely insensitive to DM particles interacting with quarks via pseudoscalar exchange, making it very difficult to experimentally test this scenario [1829]. For concreteness, we consider Dirac fermion DM and fix the coupling of the DM particle to the pseudoscalar mediator (for given masses and coupling to quarks) by the requirement to reproduce the observed relic abundance. For a fixed DM mass, we can then show bounds on the process $\chi\chi\to b\bar{b}$ from Fermi-LAT observations of dwarf spheroidal galaxies [1830] in the same parameter plane (red dashed). We find Belle II to be competitive with these constraints, in particular for $m_{\rm DM} > 50$ GeV.

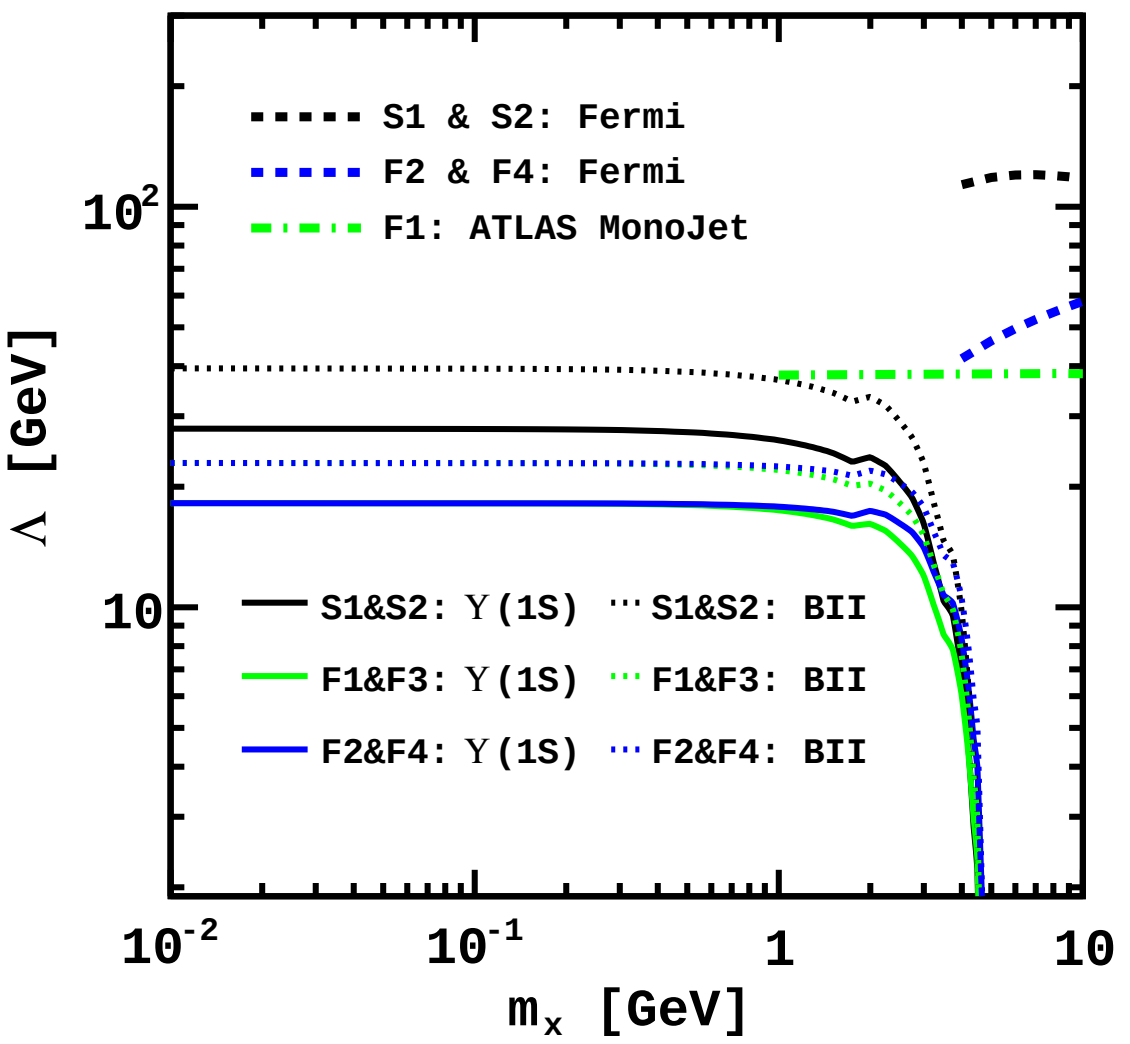

Fig. 204: Constraints on effective DM-quarks interactions from radiative $\Upsilon(1S)$ decays followed by an invisible decay of the off-shell mediator as a function of the DM mass, reproduced from [1822]. In the caption BII stands for the expected sensitivity of Belle II.

As a final example, taken from [1822], we consider the case that the mediator is so heavy that it cannot be produced on-shell in B factories, but the DM particle is kinematically accessible. In this case, the presence of a vector mediator can induce invisible $\Upsilon(1S)$ decays, while for a spin-0 mediator (or a spin-1 mediator with axial couplings) the $\Upsilon$ can decay into a photon and missing energy. Experimental bounds on these decays can then be translated to the suppression scale of the effective operator parametrising the interactions of DM with quarks. The estimated bounds which can be set by Belle II (which have been obtained by scaling those in [1822] by a factor of 4), shown in figure 204 for spin-0 mediators, depend on whether the DM particle is a scalar or a fermion as well as on the $CP$ properties of the mediator. Again, these constraints can be compared to bounds from Fermi-LAT and LHC monojet searches.

## *16.2. Experiment: Scattering Processes*

*(Contributing authors: T. Ferber*, S. Godfrey, C. Hearty, G. Inguglia, H. Logan)*

*16.2.1. Search for Dark Photons decaying into Light Dark Matter ("Single–photon search").* A significant number of experiments have recently published results for $A'$ searches where the $A'$ decays visibly into charged lepton pairs. Several other dedicated experiments will proceed over the next several years. A recent search by BaBar for the radiative production of the $A'$ in the $e^+e^-$ and $\mu^+\mu^-$ final states used 514 fb$^{-1}$ of data [1831]. The SM rates for $e^+e^- \to \gamma e^+e^-$ and $e^+e^- \to \gamma\mu^+\mu^-$ are large, and the search for the $A'$ consists of a search for a narrow peak in the dilepton invariant mass spectrum on top of a large background. The cross section for this process is proportional to $\varepsilon^2\alpha^2/E^2_{CM}$ [1807]. The decay branching fractions of the $A'$ are the same as a virtual SM photon of mass $M_{A'}$.

If the $A'$ is not the lightest dark sector particle, it will dominantly decay into light dark matter via $A' \to \chi\bar{\chi}$. Since the interaction probability of dark matter with the detector is negligible, the experimental signature of such a decay will be a mono–energetic ISR photon $\gamma_{ISR}$ with energy $E_\gamma = (E_{CM}^2 - M_{A'}^2)/(2E_{CM})$. This search requires a L1 trigger that is sensitive to single photons which was not available at Belle and was only partially available at BaBar. BaBar recorded 53 $\text{fb}^{-1}$ of data with a single photon trigger, primarily at the $\Upsilon(2S)$ and $\Upsilon(3S)$ — resonances. The effective threshold was 1.8 GeV in the centre of mass frame, resulting in limits on $\varepsilon$ for dark photon masses up to 8 GeV/$c^2$ [1827]. The analysis had significant backgrounds at low masses (high photon energy) from $e^+e^- \to \gamma\gamma$ in which a photon was not detected in the calorimeter due to gaps between crystals, and which was also missed by the muon system. A subset of this data was used to produce a preliminary result with limits on invisible decays of a light Higgs produced in radiative decays of the $\Upsilon(3S)$ [1786].

*Monte Carlo Simulation.* Signal MC events ($e^+e^- \to \gamma A', A' \to \chi\bar{\chi}$) are generated using MadGraph [1832] and a model based on [1833] that includes a dark photon $A'$ and fermionic dark matter $\chi$. Each signal sample is generated using a fixed dark photon mass $m_{A'}$ = [0.1, 0.5, 1.0, 2.0, 3.0, 4.0, 5.0, 6.0, 7.0, 8.0, 8.5, 8.75, 9.0, 9.25, 9.5, 9.75] GeV and contains 50000 events for each mass hypothesis. Events are generated for a maximal photon pseudo–rapidity of $\eta^*_\gamma < 1.681$, which corresponds to $|\cos(\theta^*_\gamma)| = 0.933$. The beam energy is set to $E^* = 10.58$ GeV. We assume a dark matter mass $m_\chi = 1\,\text{MeV/c}^2$, and we set the coupling to $g_\chi = g_e$. The decay width of the dark photon is set to the tree–level width which increases slowly with $m_{A'}$ and is of $\mathcal{O}(\text{MeV/c}^2)$. We assume that all decays of the $A'$ are into $\chi\bar{\chi}$ and set the kinetic mixing parameter to $\varepsilon = 1$. The resulting cross section, including vacuum polarisation corrections (up to about 10 %), is shown in Fig. 205.

The background in this analysis is dominated by high cross section QED processes $e^+e^- \to e^+e^-\gamma(\gamma)$ and $e^+e^- \to \gamma\gamma(\gamma)$ that produce one or more photons in the final states. If the charged tracks or additional photons are not detected or are out of detector acceptance, they can fake a single photon event. $e^+e^- \to \gamma\gamma(\gamma)$ events are simulated using BABAYAGA.NLO (see Sec.4.3) without any cut on the photon polar angles, and a minimum photon energy of 0.01 GeV which results in an effective cross section of $\sigma_{\gamma\gamma}$=25.2 nb. The phase space for radiative Bhabha events is split into three different regions: the one where both electrons are above $\theta^* > 1°$ (A), the one where one electron is below $\theta^* < 1°$ (B) and the one where both electrons are below $\theta^* < 1°$ (C). The latter case C has a negligible cross section after event selection (see Sec.16.2.1), and is not included in the full simulation.

Case A is simulated using BHWIDE [1834] with a cut on the minimal electron energy of $E_e$ >0.1 GeV. BHWIDE generates multi–photon initial state radiation, final state radiation and the interference of initial and final state radiation based on Yennie–Frautschi–Suura exponentiation (YFS) and exact NLO matrix elements. The main advantage over BABAYAGA.NLO for this particular configuration is a much higher generator speed. Some potential shortcoming of BHWIDE, like non–optimal vacuum polarisation corrections and missing beam energy spread in the basf2 implementation, are negligible for this analysis. The effective cross section for the BHWIDE sample is 30,950 nb which is, by far, the largest contribution before further preselection (see below).

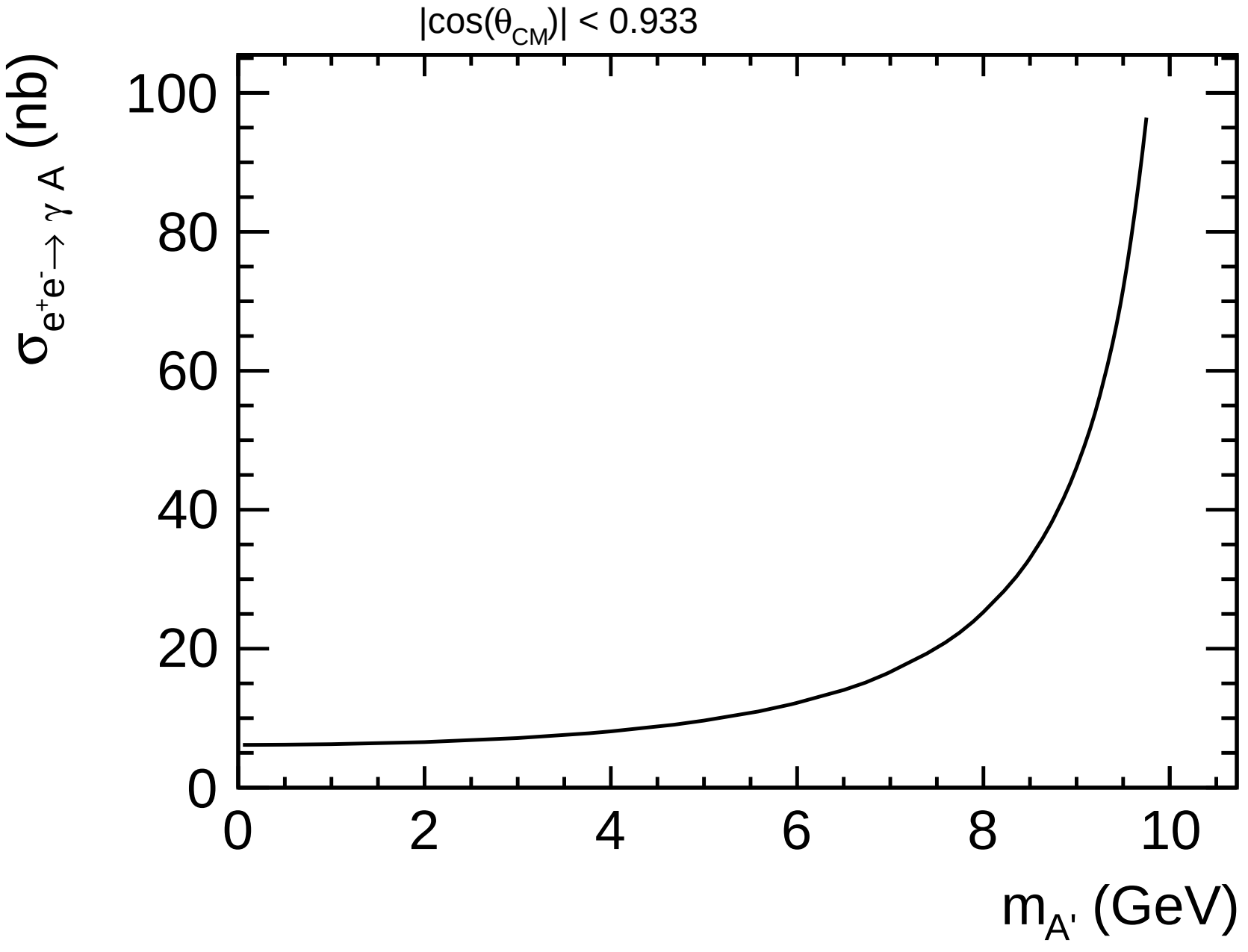

Fig. 205: Cross section for $e^+e^- \to \gamma A'$ as function of dark photon mass $m_{A'}$ calculated using MadGraph.

The contribution for case B is simulated using three different modes of TEEGG [1835]: $\mathcal{O}(\alpha^3)$ (single hard photon emission), $\mathcal{O}(\alpha^4)$ with soft corrections and $\mathcal{O}(\alpha^4)$ with hard corrections (double hard photon emission). The dominant diagram for these configurations is the $t$–channel amplitude and the processes are characterised by very small momentum transfers $Q^2$ where $Z$–exchange is negligible and ignored here. The $\mathcal{O}(\alpha^3)$ calculation is exact and the $\mathcal{O}(\alpha^4)$ corrections are included in the equivalent photon approximation. Additional cuts are applied for the TEEG sample: We require at least one photon with $\theta^* > 10°$ and an energy of $E^* > 1.4$ GeV and no other photons above 0.1 GeV. The effective cross sections are 16.90 nb, 12.80 nb, and 4.90 nb . All background samples are preselected before the events are passed to the detector simulations. We require no charged particle with $p_T >$0.15 GeV/c in $17° < \theta^{lab} < 150°$ and one photon with $E^{lab} > 1.4$ GeV/c and $17° < \theta^{lab} < 140°$. We use a 0.1 fb$^{-1}$ equivalent sample of all background processes with a reduced energy cut $E^{lab} > 0.7$ GeV/c and a 5.0 fb$^{-1}$ equivalent sample for BABAYAGA.NLO and TEEGG. The BHWIDE background has a very similar shape to the TEEGG samples and is scaled approximately according to the cross section.

The detector simulation and reconstruction use the Phase 2 BEAST 2 geometry as of release–00–07–01. All luminosity dependent beam backgrounds are scaled to 0.025 of the nominal background and all other backgrounds are scaled to 0.10 of the nominal background.

Two–photon background is not included. For technical reasons the backgrounds in the PXD and SVD octant are ignored.

*Trigger Efficiency.* The trigger efficiency has been evaluated using the L1emulator tool, which simulates the L1 trigger response using reconstructed quantities. These studies will need to be repeated using the full trigger simulation when it becomes available. We assume that the high level trigger efficiency is high.

There are two single–photon triggers for physics use (*i.e.*, not prescaled). Both look for an energy deposit in an ECL trigger tower, excluding the ring of towers closest to the beam line in each endcap. These innermost towers have very high rates of background from Bhabha showers in the VXD and CDC outside of the detector acceptance and depositing energy in the ECL with no accompanying charged track. The angular coverage of the triggers is $18.5^\circ < \theta < 139.2^\circ$. Prescaled versions will cover the full angular range of the ECL, $12^\circ < \theta < 157^\circ$.

The first trigger requires an energy deposition $E^* > 2\,\mathrm{GeV}$, where $E^*$ is the centre of mass (COM) energy. The event must not satisfy the Bhabha or $e^+e^- \to \gamma\gamma$ vetoes. An event is labeled as a Bhabha event if the highest momentum track has $p^* > 3\,\mathrm{GeV}/c$, the second highest has $p^* > 1\,\mathrm{GeV}/c$, they are separated by at least $143^\circ$ in the COM frame, and at least one of the two is associated with an ECL cluster with $E^* > 3\,\mathrm{GeV}$. An event is labeled a $\gamma\gamma$ event if the two most energetic ECL clusters have $E^* > 2\,\mathrm{GeV}$ and are separated by at least $150^\circ$ in the COM frame, and the event contains no tracks with $p_T > 300\,\mathrm{MeV}/c$ in the laboratory frame. Note that this is not just a single photon trigger. No requirement is placed on the number of charged tracks or additional clusters in the event, so that the trigger will be efficient for ISR production of $\pi^+\pi^-$ and similar final states. The effective cross section is estimated to be 4 nb, dominated by radiative Bhabha events.

The second trigger has a lower threshold, $E^* > 1\,\mathrm{GeV}$, and requires that the second cluster in the event be less than $0.2\,\mathrm{GeV}$. There are no other vetoes applied. The cross section is estimated to be 2.5 nb, largely due to radiative Bhabha events in which there really is only a single photon in the acceptance of the detector.

The efficiency for signal MC, as a function of $A'$ mass after combining the two triggers, is shown in Fig. 206. The loss of efficiency is primarily due to detector acceptance. The trigger efficiency for high energy photons within the angular region $18.5^\circ < \theta < 139.2^\circ$ is 95% (not shown in Fig. 206).

*Event Selection.* The basic event selection requires a calorimeter (ECL) cluster with COM energy $E^* > 1.8$ GeV, no other ECL clusters with $E^* > 0.1$ GeV, and no tracks with $p_T > 0.2$ GeV/c in the COM frame. The cuts on the second cluster and track momentum have not been optimised, but the results shown here are not sensitive to the specific values. These criteria have high efficiency for signal; the remaining criteria are designed to suppress

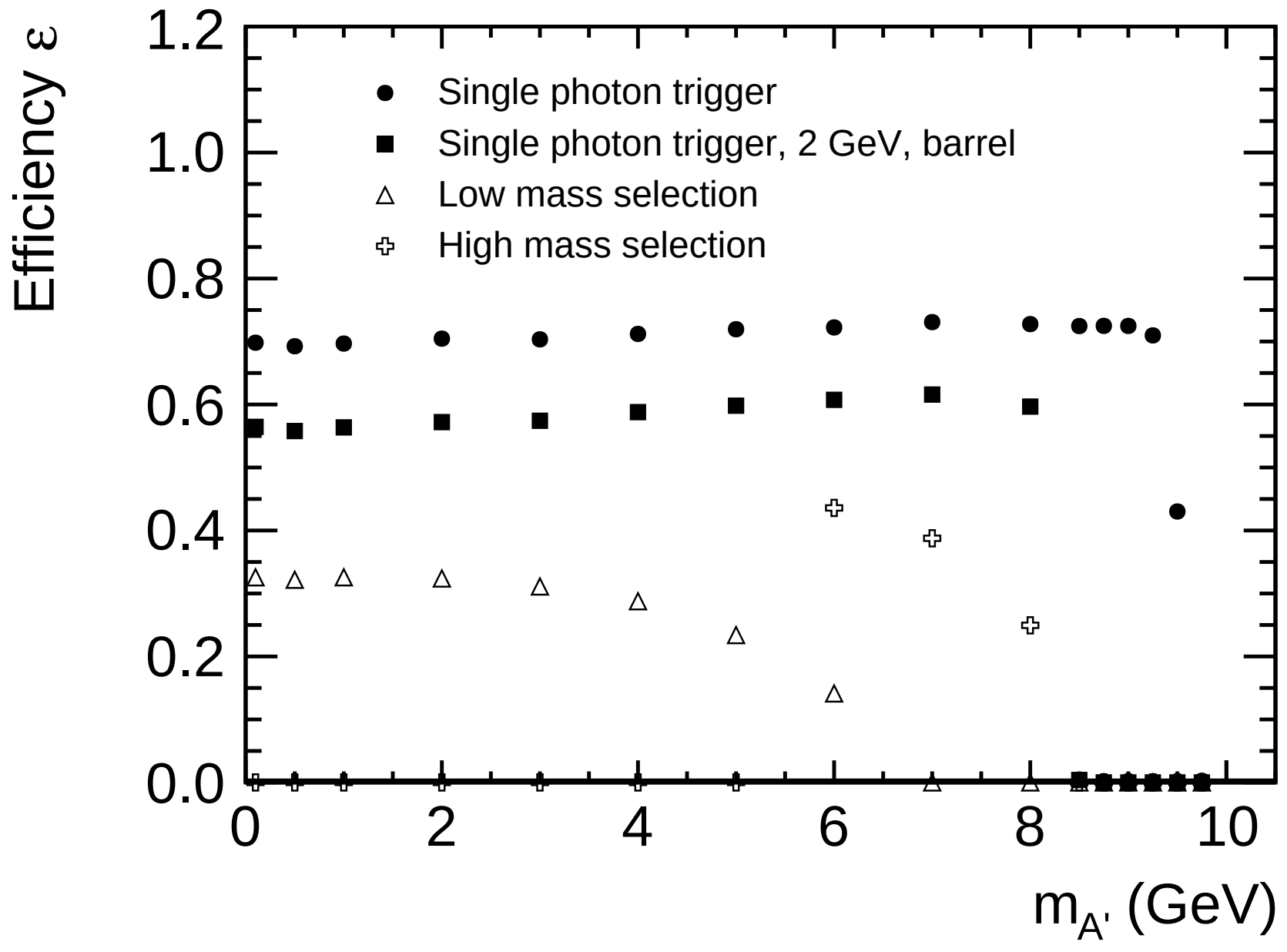

Fig. 206: Trigger efficiency for signal MC as a function of $A'$ mass (filled circles). The filled squares show the efficiency if the acceptance is reduced to the ECL barrel, $E^* > 2\,\text{GeV}$, a selection more relevant for the subsequent event selection. The open symbols show the overall analysis efficiency for the low mass and high mass selections discussed in Sec. 16.2.1.

physics backgrounds.

Backgrounds fall into two general categories. Irreducible backgrounds are those in which the final state includes one photon and no other particles in the acceptance of the detector, which for this purpose we consider to be the full coverage of the ECL, $12^\circ < \theta < 157^\circ$. In practice, efficient photon reconstruction is only available in the acceptance of the CDC, $17^\circ < \theta < 150^\circ$.

Simulation predicts approximately two million events of this type in 20 $\text{fb}^{-1}$ with $E^* > 1.8$ GeV and $22^\circ < \theta < 139^\circ$, 85% due to radiative Bhabhas, and the remainder due to radiative photon events $e^+e^- \to \gamma\gamma(\gamma)$ with at least three photons in the final state. Radiative muon pairs and other QED processes have small cross sections compared to these two sources. The kinematics of requiring all other particles to be outside of the detector acceptance produces a strong correlation between the maximum COM energy of the photon and $\theta$ (Fig. 207).

The dominant background at higher energies are those in which there is a second photon within the detector fiducial volume that is not detected. The event may also contain a third (or more) photon outside of the acceptance. Photon detection inefficiency in the ECL is due to the following, in order of importance:

(1) gap between the ECL barrel and the backwards endcap;

(2) gap between the ECL barrel and the forward endcap;
(3) $200\,\mu$m gaps between endcap crystals that are projective to the interaction point, plus 16 larger gaps for mechanical structure;
(4) a 1–1.5 mm gap for mechanical structure in the barrel at $\theta = 90°$;
(5) photons not converting in the crystal, which occurs with a probability of $3 \times 10^{-6}$.

Note that the gaps between crystals in the barrel do not project to the beam spot in either polar or azimuthal angle.

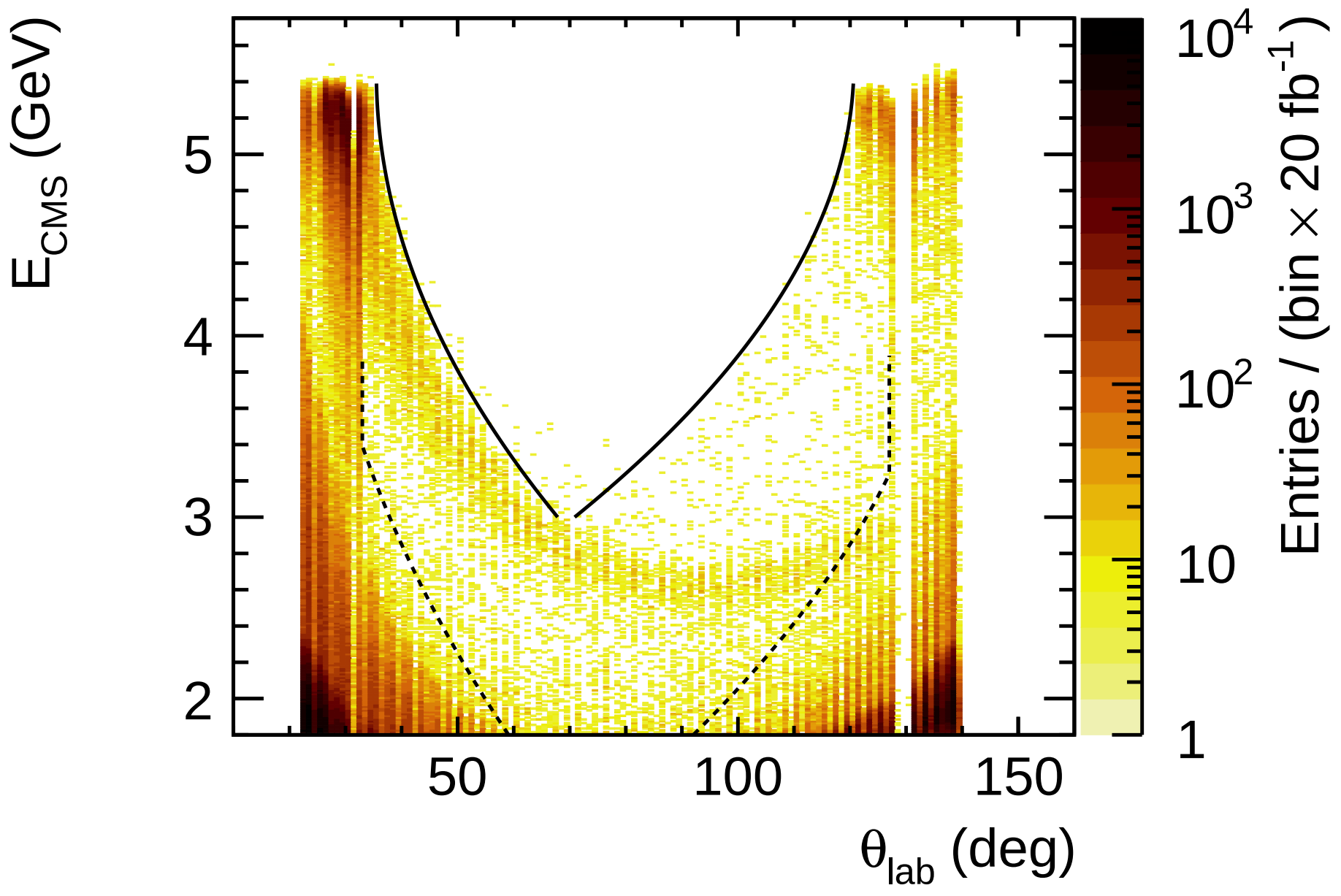

Fig. 207: COM Energy $E^*$ versus $\theta$ for background events satisfying all selection criteria other than the final cut on $\theta$. The large population of events at low energies and wide angles is due to irreducible background processes. The beam-energy events near $\theta$ of 30° and 130° are due to $e^+e^- \to \gamma\gamma$ where a photon goes undetected in a gap between the ECL barrel and an endcap. The band at intermediate energies (*e.g.* 2.6 GeV at $\theta = 90°$ arises from three photon final states in which a near-beam-energy photon is undetected in the backward barrel–endcap gap and a second (radiative) photon is near $\theta^* = 0$. The solid lines mark the fiducial region for $m_{A'} \leq 6\,\mathrm{GeV}/c^2$; the dashed lines are for higher masses.

The KLM can also be used to detect photons. Studies with signal MC indicate that most KLM clusters in such events are within 25° (3D, COM) of the signal photon. To suppress backgrounds due to ECL photon inefficiency, we require that there be no KLM clusters outside of this 25° cone. At the background levels expected for Phase 2 running, 3.6% of signal MC fail this selection.

The KLM also has regions of inefficiency, primarily at the transitions from the barrel to the endcaps, at the location of the solenoid cryogenics chimney (located near the barrel / backwards endcap transition), between octants in the barrel KLM, and between sectors of the endcap KLM. The chimney and the backward transition overlap with the backwards gap in the ECL, and produce the majority of the non-irreducible backgrounds.

The kinematic distribution of background events passing the basic selection criteria plus the additional KLM requirement is shown in Fig. 207. The number of background events used to extract the final sensitivity is summarized in Table 143. Note that we have not yet used azimuthal information. Since KLM inefficiency is concentrated at specific values of $\phi$, including this information—for example, as part of a neural net or BDT—will improve the final analysis.

Table 143: Expected number of background events after final selection, scaled to an integrated luminosity of $20^{-1}$ fb (see text for details).

| Dark Photon mass $m_A$ [GeV/c$^2$] | low mass selection | high mass selection |
|---|---|---|
| 0.5 | 4.3 | 15335 |
| 1 | 4.3 | 16012 |
| 2 | 4.3 | 14399 |
| 3 | 38.3 | 7931 |
| 4 | 42.6 | 5693 |
| 5 | 46.8 | 4659 |
| 6 | 51.1 | 3587 |
| 7 | – | 4251 |
| 8 | – | 1869 |

The final step of the selection is an energy dependent cut on $\theta$, which rejects the vast majority of events in Fig. 207. Due to the limited background MC statistics in some regions of this 2D histogram, a fully automatic process was not used to select the cuts, but rather a combination of an optimisation with manual intervention. For dark photon masses $m_{A'} \leq 6\,\text{GeV}/c^2$ (roughly $E^* > 3$ GeV), the $\theta$ selection produces a low background region, marked with the solid lines in Fig. 207. This region is selected to reject the band in Fig. 207 corresponding to a three–photon final state, where one photon is lost in the backwards ECL gap, and another is at $\theta^* = 0$. Simulation predicts 300 events per 20 fb$^{-1}$ with $E^* > 3$ GeV between the solid lines, all due to $e^+e^- \to \gamma\gamma\gamma(\gamma)$.

At higher masses $6 < m_{A'} \leq 8\,\text{GeV}/c^2$ (lower photon energy), a much wider $\theta$ region is used. It is chosen to almost completely reject the irreducible background, and to avoid the ECL endcaps, which have significantly lower efficiency. Simulation predicts that 25,000 background events within this $\theta$ region with $1.8 < E^* < 3.9\,\text{GeV}$, almost all due to $e^+e^- \to \gamma\gamma\gamma(\gamma)$.

Figure 208 shows the resulting background recoil mass distribution for the higher $A'$ mass region, along with an example of a signal distribution.

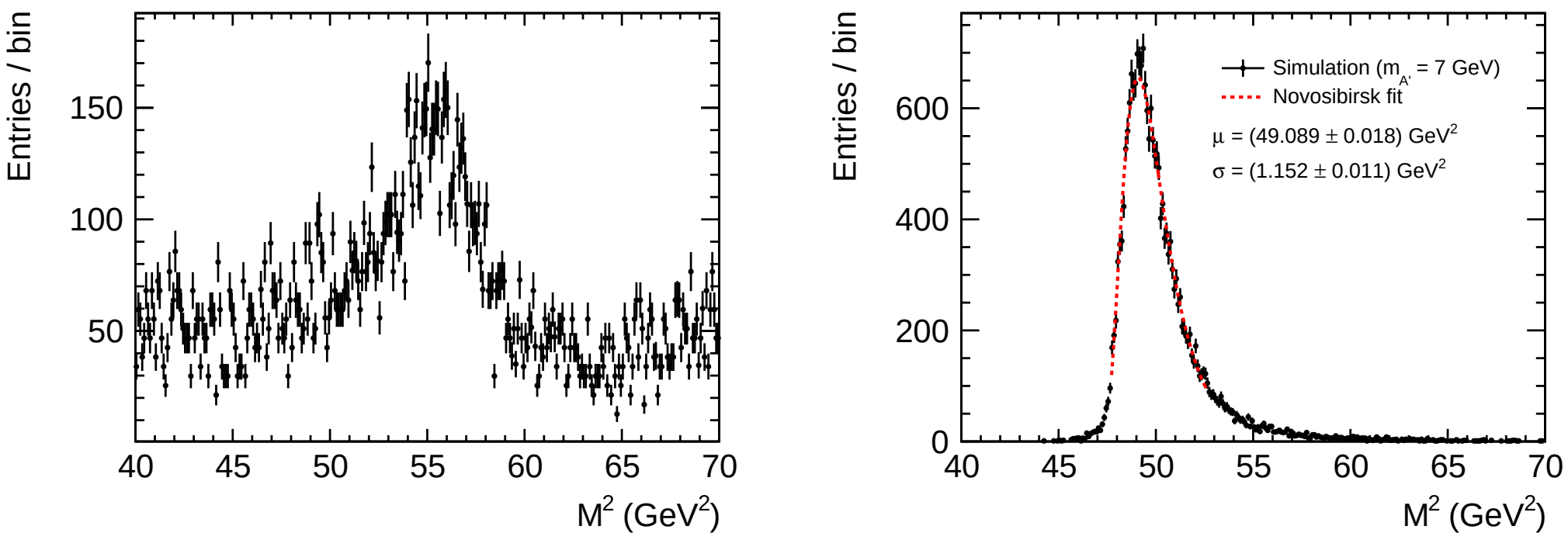

Fig. 208: (a) Recoil mass distribution squared for background events satisfying the selection criteria for the high $A'$ mass region. (b) Distribution for a 7 GeV/$c^2$ $A'$, fit with a Novosibirsk function [1836].

*Signal Extraction.* The final analysis will fit the measured recoil mass squared distribution with a Novosibirsk function [1836] of appropriate width to measure the $A'$ signal and an as-yet unspecified function for the background. The kinematic features present in the background distribution will make this quite challenging in the high mass region, and a simpler process—not suitable for the final analysis, since it uses MC truth information—has been used for this study. The procedure uses the photon energy distribution, rather than the equivalent recoil mass. The expected upper limit has been obtained for different $A'$ masses with $m_{A'} \leq 8\,\mathrm{GeV}/c^2$.

For each mass, the reconstructed photon COM energy has been fitted with a Novosibirsk function. The signal region is taken to be the photon energy range $[\mu_E - 3\sigma_E, \mu_E + 1.5\sigma_E]$, where $\mu_E$ and $\sigma_E$ are the Novosibirsk fit parameters for that mass. This range contains between 83% and 88% of the signal.

For each mass, we obtain the expected 90% CL upper limit on the observed number of signal events $\mu_S$ from the expected number of background events $\mu_B$. $\mu_B$ is the number of events in the signal region predicted by the generated background MC samples, scaled to $20\,\mathrm{fb}^{-1}$. We assume that N, the number of events observed, is the integer closest to $\mu_B$. $\mu_S$ is selected such that the Poisson probability of observing $\leq N$ events when expecting $\mu_B + \mu_S$ events is 0.1.

The upper limit on the cross section for $e^+e^- \to \gamma A'$, $A' \to$ invisible is $\sigma = \mu_S/\epsilon_S\mathcal{L}$, where $\epsilon_S$ is the signal efficiency (Fig. 206) and $\mathcal{L} = 20\,\mathrm{fb}^{-1}$ is the integrated luminosity. The equivalent limit on $\varepsilon$ is the square root of this cross section divided by the cross section calculated for $\varepsilon = 1$ (Fig. 205). Projected upper limits on $\varepsilon$ are summarised as a function of $A'$ mass in Fig. 209. The results are projected to be significantly better than BaBar due to the better hermeticity of the calorimeter and the efficiency of the KLM.

*Systematic Uncertainties.* We expect that the systematic uncertainties will be dominated by uncertainties in the predicted number and kinematic properties of background events.

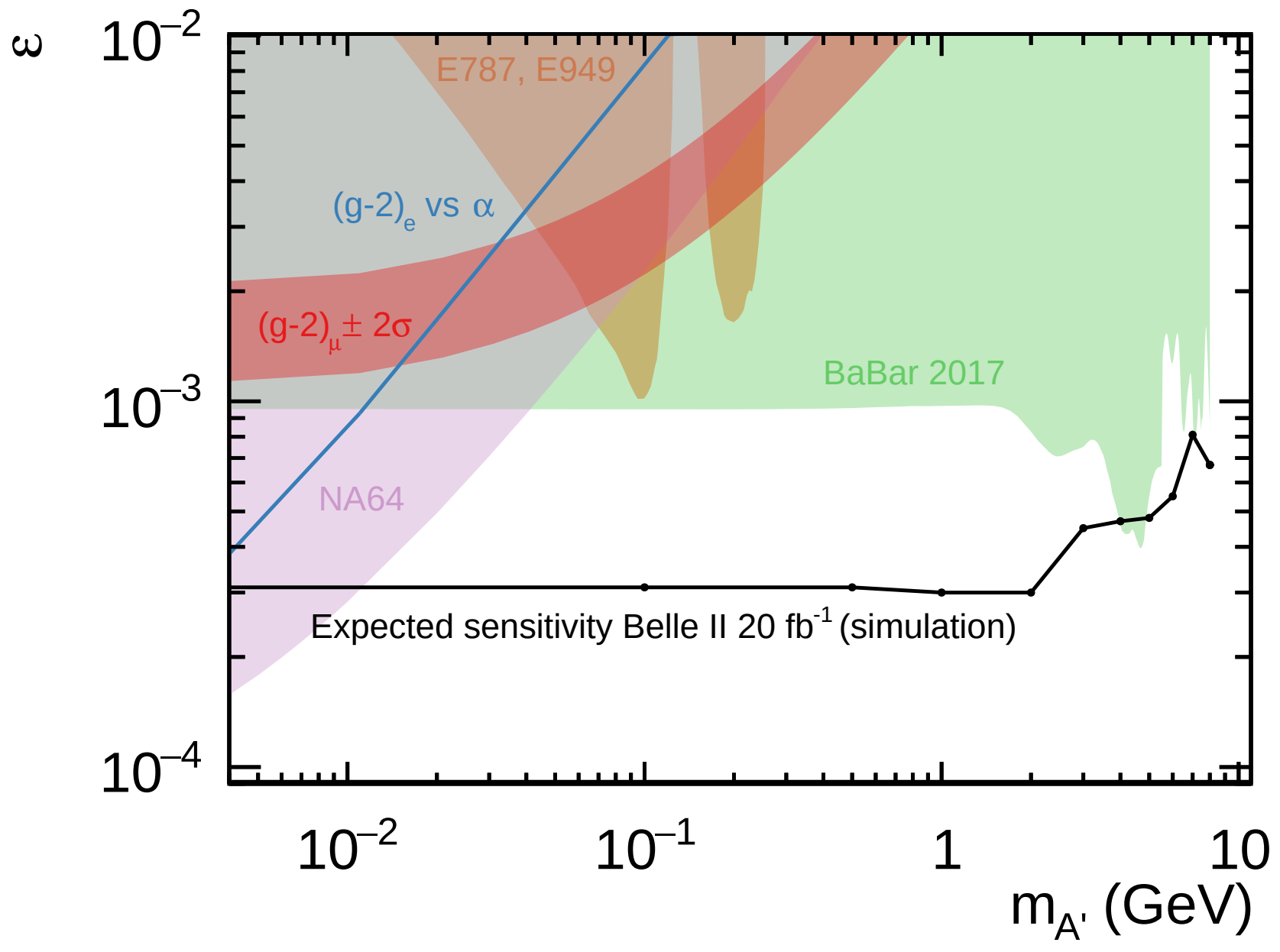

Fig. 209: Projected upper limits on $\varepsilon$ for the process $e^+e^- \to \gamma A'$, $A' \to$ invisible, for a $20\,\mathrm{fb}^{-1}$ Belle II data set (solid black curve).

At low $A'$ masses, we need to quantify the residual beam-energy photon backgrounds from $e^+e^- \to \gamma\gamma$. This will require photon control samples, such as kinematically fit radiative muon pairs, or $e^+e^- \to \gamma\gamma$ events in which one photon is reconstructed at full energy and the other has low energy, corresponding to a late conversion in the ECL crystal. The backgrounds for high $A'$ masses are dominated by events with one photon in the backwards barrel/endcap gap and a second near $\theta^* = 0$. The kinematically fit muon pair sample will be used to map the photon efficiency across this gap.

*16.2.2. Search for Axion-like particles.* Axions were originally motivated by the strong CP problem and have a fixed relation between coupling strength and mass. While the axion and its parameters are related to QCD, the coupling and mass of axion–like particles (ALPs) is taken to be independent and can appear in a variety of extensions to the SM. ALPs are pseudo–scalars ($J^P = 0^-$) with couplings to the different gauge bosons. The simplest search for an ALP at Belle II is via its coupling to $\gamma\gamma$ (Fig. 210) [1837]. Depending on the ALP mass $m_a$ and the coupling constant $g_{a\gamma\gamma}$, the ALP is long lived, producing a single photon final state, or decays in the detector to $\gamma\gamma$, producing a three photon final state. A wide range of ALP coupling to photons and ALP masses has been ruled by previous experiments, but two regions in the $g_{a\gamma\gamma}$–$m_a$ plane are of particular interest for a Belle II analysis. The first region are light ALPs with $m_a \approx 1\,\mathrm{MeV}$, $g_{a\gamma\gamma} \approx (10^{-5} - 10^{-6})\,\mathrm{GeV}^{-1}$ which is out of reach for beam–dump experiments and only disfavoured by model–dependent limits from cosmology. The second region are heavier ALPs with $0.1\,\mathrm{GeV} \lesssim m_a \lesssim 10\,\mathrm{GeV}$. Hypercharge couplings are excluded for $g_{aBB} \approx (10^{-2} - 10^{-3})$ $\mathrm{GeV}^{-1}$ by re–analysed LEP data, where the weaker limit is for coupling to photons only.

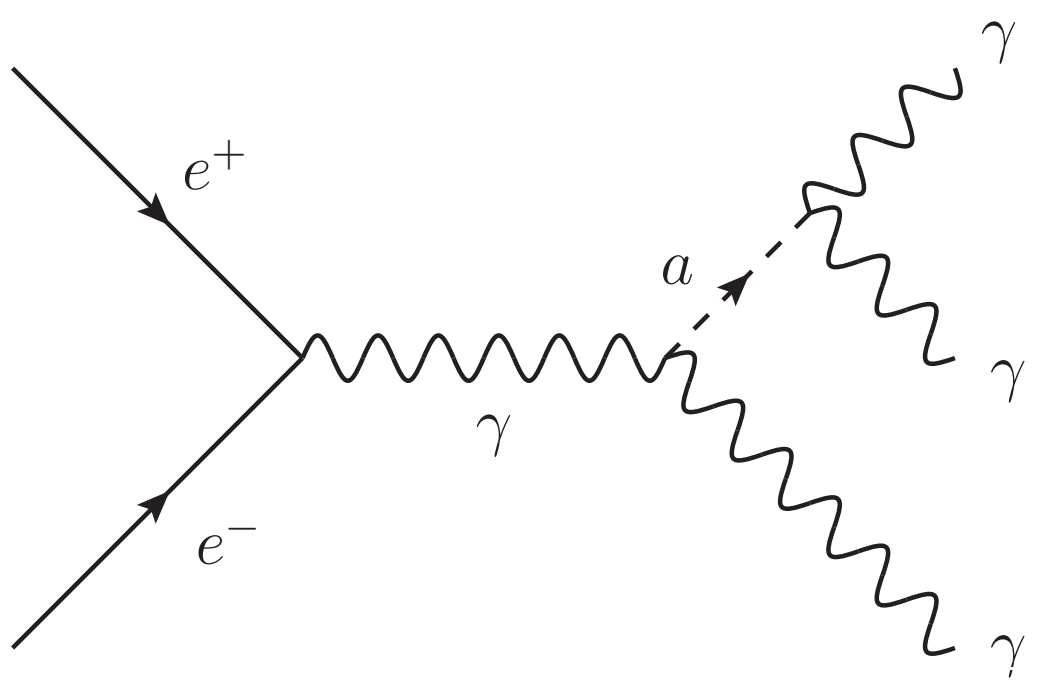

Fig. 210: Production of an Axion–like particle in association with a photon.

The three photon search will provide access to parameter space that is not addressed by any current measurements. At higher masses, $m_a \gtrsim 200\,\mathrm{MeV}/c^2$, the three photons are well separated. At lower masses, the ALP is sufficiently boosted that the two decay photons overlap in the calorimeter or, at the lowest masses, form a cluster that is reconstructed with only a single local maximum. The calorimeter group is developing software to reconstruct merged $\pi^0$ mesons that can be adapted for this analysis. However, the low mass region will also be a challenge for the trigger system, particularly at level 1, where the signal is indistinguishable from $e^+e^- \to \gamma\gamma$. The problem is not that these signals are difficult to trigger on, but rather that the plan is to prescale the $\gamma\gamma$ final state to reduce the throughput to the high level trigger, and to reduce the rate of events stored to disk. The preferred solution would be to delay the decision to reject $\gamma\gamma$ events to the HLT and apply no prescale at level 1 for the $\gamma\gamma$ trigger. The Belle II sensitivity for visible and invisible ALPs has been studied in detail in [1838].

It has been suggested [1805] to search for the ALP in the flavour-changing neutral current decay $B^+ \to K^{(*)+}a$, which is governed by the coupling $g_{aWW}$ of the ALP to $W^+W^-$. The decay $a \to \gamma\gamma$ produces a final state similar to $K^+\pi^0$. Both BaBar and Belle have measured this branching fraction with uncertainties of a few $\times 10^{-7}$ [727, 1839]. Similar uncertainties in an ALP search would provide significant constraints on ALP coupling to heavy charged bosons. Extrapolating to the full Belle II data set requires work, as the existing analyses are systematics dominated.

The $a \to$ invisible case produces a monoenergetic $K^{(*)+}$ in the $B^+$ rest frame. Such searches use the fully-reconstructed $B$ sample, and would be similar to the BaBar search for $B^+ \to K^{(*)+}J/\psi$, $J/\psi \to$ invisible [624]. This search was statistically limited, and could exploit the much larger data set that will be available to Belle II.

*16.2.3. Search for Dark Photons decaying into charged leptons and charged hadrons.* If there are no kinematically accessible dark sector final states available, dark photons produced via the ISR reaction $e^+e^- \to \gamma_{ISR}A'$ will decay to Standard Model particles, with branching fractions equal to a virtual photon of that mass. Particularly useful final states include $\mu^+\mu^-$, $e^+e^-$, and $h^+h^-$, where $h$ is a charged pion or kaon. BaBar has searched for prompt decays to $e^+e^-$ and $\mu^+\mu^-$ [1831], setting upper limits on the kinematic mixing parameter of $5 \times 10^{-4}$–$10^{-3}$, depending on mass. A search is ongoing at Belle for prompt decays to leptonic and hadronic final states, and for displaced decays to lepton pairs. With the large amount of data expected to be collected by the Belle II detector (about two orders of magnitude larger than that available at BaBar), one can expect to observe an excess of events due to a dark photon decays to charged leptons or charged hadrons with a mixing parameter of order of few $\times 10^{-4}$. This search requires the implementation of an efficient L1 two track triggers and it will also profit from photon triggers due to the presence of a single high energetic ISR photon. In order to maintain a high L1 trigger efficiency for $A' \to e^+e^-$, the unavoidable prescale factor for radiative Bhabha events is ideally implemented as function of track charge and polar angle.

One can extrapolate the existing BaBar limits of Dark Photon decays into charged particles to Belle II. The larger drift chamber radius of Belle II will yield an improved invariant mass resolution ($\sim$ factor 2) and better trigger efficiency for both muons ($\sim$ factor 1.1) and electrons ($\sim$ factor 2) is expected. The projected upper limits for different values of integrated luminosity are shown in Fig. 211.

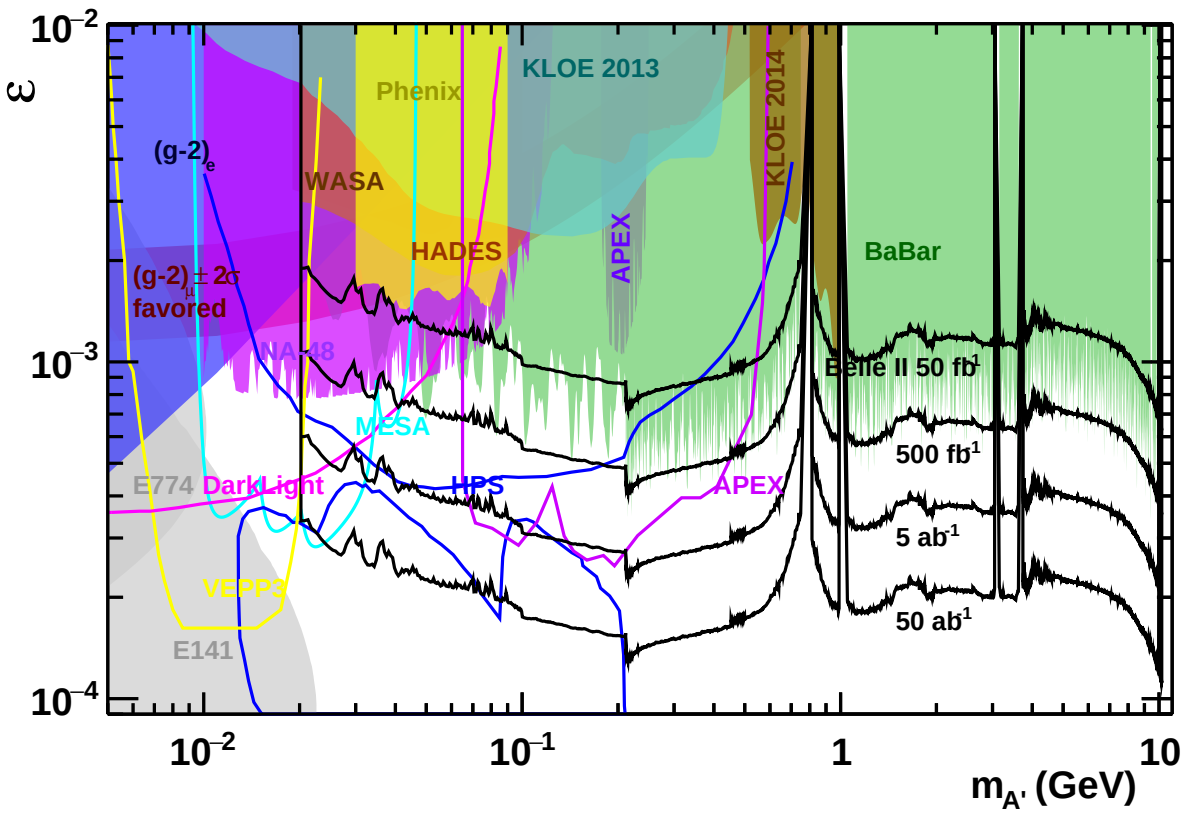

Fig. 211: Existing exclusion regions (90% CL) on the dark photon mixing parameter $\varepsilon$ and mass $M_{A'}$ (solid regions) for $A' \to \ell\ell$, with projected limits for Belle II and other future experiments (lines) (Figure reproduced from [1840]).

*16.2.4. Search for Dark Photons decaying into Light Dark Matter in $e^+e^- \to A'\ell^+\ell^-$.*
Dark photons can also be searched for in the reaction $e^+e^- \to A'\mu^+\mu^-$, with subsequent decays of the dark photon (also called a $Z'$ in this context) into a variety of final states [1841, 1842], including invisible ones. BaBar has performed this search for dark photon decays to muonic final states [1843], and the same analysis is in preparation at the Belle experiment. For the invisible case, a kinematic fit of the muons can be used to select events in which the missing energy is pointing into the barrel calorimeter, which has the best hermiticity. The trigger for this final state is the muon pair, which may be sensitive to higher $A'$ masses than the single photon trigger. A sensitivity to the mixing parameter at the level of $10^{-4}$–$10^{-3}$ can be expected in this channel.

## *16.3. Experiment: Quarkonium Decay*

*16.3.1. Searches for BSM physics in invisible $\Upsilon(1S)$ decays.* In the SM, invisible decays of $\Upsilon(1S)$ involve neutrinos in the final state are produced by $b\bar{b}$ annihilation with $BR[\Upsilon(1S) \to \nu\bar{\nu}] \simeq 10^{-5}$. Low mass dark matter (*i.e.* with a mass smaller than the mass of the $b-$quark), if it exists, should enhance this $BR$ [1821, 1844]. The ARGUS, CLEO, Belle and BABAR experiments have studied this channel with limited data providing upper limits to $BR[\Upsilon(1S) \to \text{invisible}] < 3.0 \times 10^{-4}$ at the 90% confidence level (C.L.) [1845–1848]. Low mass dark matter can also be probed in radiative $\Upsilon(1S)$ decays such as $\Upsilon(1S) \to \gamma + \text{invisible}$. The Next-to-Minimal supersymmetric extension of the SM (NMSSM) allows for the existence of low mass (GeV/c$^2$ scale) dark matter and of a low mass $CP$-odd Higgs boson ($A^0$), therefore if $M_{\Upsilon(1S)} > M_{A^0}$ and $M_{A^0} > 2M_\chi$ one would be able to observe the Wilczek production of $A^0$ ($\Upsilon(1S) \to \gamma A^0$) followed by the decay into dark matter $A^0 \to \chi\bar{\chi}$ [1821, 1844, 1849, 1850].
The main limitations for the study of transitions involving invisible decays of the $\Upsilon(1S)$ is that one has to find a tagging method that can be used to unambiguously infer the presence of the $\Upsilon(1S)$ even though its decay products are not seen in the detector. Generally one uses transitions such as $\Upsilon(2,3S) \to \pi^+\pi^-\Upsilon(1S)$ (with $BR[\Upsilon(3S) \to \pi^+\pi^-\Upsilon(1S)] \approx 4.5\%$ and $BR[\Upsilon(2S) \to \pi^+\pi^-\Upsilon(1S)] \approx 18.1\%$) followed by $\Upsilon(1S) \to \text{invisible}$; in this case one has to trigger and reconstruct final states in which only two low momentum pions are seen in the detector, trying to avoid to collect too many background events and at the same time maintaining a high trigger efficiency.

During Belle operations a special dedicated trigger was implemented to allow the search of $\Upsilon(1S) \to \text{invisible}$ from di-pion transition of the $\Upsilon(3S)$ in $e^+e^-$ collisions at the $\Upsilon(3S)$ peak, but no special triggers were implemented for the $\Upsilon(2S)$ data taking. Based on the Belle experiment experience and on simulation studies performed with the Level 1 Trigger Emulator Package (L1TriggerEmulator) of the Belle II Analysis Software (BASF2), it is found that the special trigger conditions such as a *single track trigger* with $p_t > 200$ MeV/c (long track trigger) and opening angle between the tracks in the $r - \phi$ plane larger than 30° would allow a trigger efficiency comparable to that of Belle (*i.e.* 85-90 %) in the process $\Upsilon(3S) \to \pi^+\pi^-\Upsilon(1S)$ and an efficiency of 30% for $\Upsilon(2S) \to \pi^+\pi^-\Upsilon(1S)$, $\Upsilon(1S) \to \text{invisible}$. Simulation studies have shown that trigger efficiencies for $\Upsilon(2S) \to \pi^+\pi^-\Upsilon(1S)$ can still be improved if a second dedicated trigger is implemented for this case (for example realising the $p_t$ threshold of the single track trigger).

If $s$ (=10.0233, 10.3552 GeV/c$^2$) is the centre-of-mass energy of the collision, $M_{\pi^+\pi^-}$ the invariant mass of the two-pion system, and $E^{CMS}_{\pi^+\pi^-}$ the energy of the two-pion system in the centre-of-mass reference frame, one defines a recoil mass for the two-pion system as $M_r^2 = s + M^2_{\pi^+\pi^-} - 2\sqrt{s}E^{CMS}_{\pi^+\pi^-}$. A signal of the decay $\Upsilon(1S) \to \text{invisible}$ is an excess of events in the $M_r$ distribution at a mass equivalent to that of the $\Upsilon(1S)$ (9.460 GeV/c$^2$). Another possibility to produce and tag $\Upsilon(1S)$ consists in the search for the production of higher spin resonances (such as $\Upsilon(2S)$ and $\Upsilon(3S)$) in ISR processes followed by their di-pion transition to $\Upsilon(1S)$. While the electron and the positron approaches before a head-on collision, either one or the other can radiate a photon, reducing the energy of the collision. As outlined and shown in [1851–1854], given the $e^+e^-$ collision rate, it is possible to precisely calculate the ISR cross section $\sigma_f(s)$ for the production of any final state $f$. The cross section for untagged ISR production (ISR photon not reconstructed) of $\Upsilon(3S)$ ($\Upsilon(2S)$) is 29 pb (17 pb). This technique, that was already widely used by the Belle Collaboration, provides an additional source of $\Upsilon(nS)$ resonances. The search for $\Upsilon \to$ invisible is characterised by the presence of different backgrounds in the $M_r$ distribution that will limit the final sensitivity: combinatorial and peaking backgrounds. The combinatorial background consists mostly of two-photon processes $e^+e^- \to e^+e^-X$ where the initial $e^+e^-$ proceed along the beam pipe escaping detection and $X \to \pi^+\pi^-$,$\pi^+\pi^-\pi^0$, $\mu^+\mu^-$; this background can be easily parametrised by a first-order polynomial function. The peaking background is due to two-body decays of the $\Upsilon(1S)$ with both the decay products travelling outside the detector acceptance. Since the $BR$ for $\Upsilon(1S)$ decays to lepton pairs are all of the order of 2.5%, this leads to a relatively large peaking background that needs to be carefully evaluated and that will constitute the main limitation to the measurement. An additional background source would be the reaction $e^+e^- \to \pi^+\pi^-\nu\bar{\nu}$, however due to the low cross section of the order of $10^{-6}$ pb, it would produce about 5 events out of 50 ab$^{-1}$ of data collected at the $\Upsilon(4S)$ peak and can therefore be neglected in this study.

In order to extract the expected sensitivities we consider a scenario in which data are collected at different energies and different luminosities as shown in Table 144. Taking into account the expected contribution from the peaking background, one can expect a sensitivity at the 90% confidence level (statistical only) of $1.3 \times 10^{-5}$ to $BR[\Upsilon(1S) \to \text{invisible}]$ combining the various channels; this value is comparable to the SM prediction $BR[\Upsilon(1S) \to \nu\bar{\nu}] \approx 1.0 \times 10^{-5}$. For the search of the process $\Upsilon(1S) \to \gamma + \text{invisible}$, it can be anticipated that taking into account the large data samples to be used and the expected total reconstruction efficiencies as outlined above, the Belle II experiment has the possibility to discover an excess of events at the 90% confidence level either if $BR[\Upsilon(1S) \to \gamma A^0] \times BR[A^0 \to \text{invisible}] > 5 \times 10^{-7}$, or $BR[\Upsilon(1S) \to \gamma\chi\bar{\chi}] > 5 \times 10^{-6}$ for $0 < m_\chi < 4.5$ GeV/c$^2$.

Table 144: Expected yields for various $\Upsilon(1S)$ tagging techniques where $L_{int}$ is the integrated luminosity considered for the extrapolation of the yields, $\epsilon$ is the expected total efficiency, $N(\Upsilon(1S))$ is the number of $\Upsilon(1S)$ produced in the process, and $N_{\Upsilon(1S)\to\nu\bar{\nu}}$ and $N_{NP}$ are the expected number of observed $\Upsilon(1S) \to$ invisible events assuming Standard Model ($1 \times 10^{-5}$) and new physics ($3 \times 10^{-4}$) rates, respectively.

| Process | $L_{int}(ab^{-1})$ | $\epsilon$ | $N(\Upsilon(1S))$ | $N_{\Upsilon(1S)\to\nu\bar{\nu}}$ | $N_{NP}$ |
|---|---|---|---|---|---|
| $\Upsilon(2S) \to \pi^+\pi^-\Upsilon(1S)$ | $0.2, \Upsilon(2S)$ | 0.1-0.2 | $2.3 \times 10^8$ | 232-464 | 6960-13920 |
| $\Upsilon(3S) \to \pi^+\pi^-\Upsilon(1S)$ | $0.2, \Upsilon(3S)$ | 0.1-0.2 | $3.2 \times 10^7$ | 32-64 | 945-1890 |
| $\Upsilon(4S) \to \pi^+\pi^-\Upsilon(1S)$ | $50.0, \Upsilon(4S)$ | 0.1-0.2 | $5.5 \times 10^6$ | 5.5-11 | 165-310 |
| $\Upsilon(5S) \to \pi^+\pi^-\Upsilon(1S)$ | $5.0, \Upsilon(5S)$ | 0.1-0.2 | $7.6 \times 10^6$ | 7.6-15.2 | 228-456 |
| $\gamma\Upsilon(2S) \to (\gamma)\pi^+\pi^-\Upsilon(1S)$ | $50.0, \Upsilon(4S)$ | 0.1-0.2 | $1.5 \times 10^8$ | 150-300 | 4500-9000 |
| $\gamma\Upsilon(3S) \to (\gamma)\pi^+\pi^-\Upsilon(1S)$ | $50.0, \Upsilon(4S)$ | 0.1-0.2 | $6.5 \times 10^7$ | 65-130 | 1950-3900 |

*16.3.2. Probe of new light CP even Higgs bosons from bottomonium $\chi_{b0}$ decay.* The decay of scalar bottomonium $\chi_{b0} \to \tau^+\tau^-$, can be sensitive to $s$-channel exchange of $CP$-even neutral Higgs bosons via the process $\Upsilon(3S) \to \gamma\chi_{b0}(2P) \to \gamma\tau^+\tau^-$ [1823]. Although the event rate for SM Higgs exchange is a few orders of magnitude too small to be observed, this process can put significant constraints on the parameters of the type-II two-Higgs-doublet model [932] when the discovered 125 GeV Higgs boson is the heavier of the two charge parity $CP$-even scalars. In this model the scalar couplings to $b$ quarks and $\tau$ leptons can be simultaneously enhanced for large values of the parameter $\tan\beta$, which is defined as the ratio of vacuum expectation values of the two Higgs doublets. The model contains two $CP$-even neutral scalars, which we call $H_{125}$ and $H_{\rm new}$. We identify $H_{125}$ with the discovered Higgs boson at 125 GeV. In this model the Standard Model expression for the partial width given by

$$\Gamma^H(\chi_0 \to \ell^+\ell^-) = \frac{M_{\chi_0}}{8\pi}\left[1 - \frac{4m_\ell^2}{M_{\chi_0}^2}\right]^{3/2}\left(\frac{m_q m_\ell}{v^2 M_H^2}\right)^2 f_{\chi_0}^2. \tag{561}$$

is modified by the presence of the second Higgs resonance;

$$\left(\frac{m_b m_\tau}{v^2 M_H^2}\right)^2 \to \left[\frac{m_b m_\tau}{v^2}\left(\frac{\kappa_b^{125}\kappa_\tau^{125}}{M_H^2} + \frac{\kappa_b^{\rm new}\kappa_\tau^{\rm new}}{M_{\rm new}^2 - M_{\chi_{b0}}^2}\right)\right]^2, \tag{562}$$

where in the above equations $v^2 = 1/\sqrt{2}G_F$ is the SM Higgs vacuum expectation value, $M_H$ is the Higgs mass, we have neglected $M_{\chi_0}^2$ relative to $M_H^2$ in the propagator and $f_{\chi_0}$ is the $\chi_{b0}$ decay constant. $M_{\rm new}$ is the mass of the second scalar $H_{\rm new}$ and the $\kappa$ factors represent the couplings of the two scalars to $b$ quarks or $\tau$ leptons normalised to the corresponding coupling of the SM Higgs boson [1855]. We have kept the $p^2 = M_{\chi_{b0}}^2$ dependence in the second term in Eq. (562) corresponding to $H_{\rm new}$ exchange because we will be interested in low $M_{\rm new}$. There is also a contribution to the $\chi_0 \to \ell^+\ell^-$ decay through a two-photon intermediate state which we estimate to be $\mathrm{BR}^{2\gamma}(\chi_{b0}(2P) \to \tau^+\tau^-) \simeq 6 \times 10^{-9}$ [1856].

To estimate the BR for $\chi_{b0} \to \tau^+\tau^-$ via SM Higgs exchange we estimate the $\chi_{b0}(2P)$ total width using the measured BR for $\chi_{b0} \to \gamma\Upsilon(1S)$ and the predicted partial with for this transition to obtain

$$\mathrm{BR}^H(\chi_{b0}(2P) \to \tau^+\tau^-) = (1.9 \pm 0.5) \times 10^{-12}. \tag{563}$$

In the 2HDM we set the couplings of the 125 GeV Higgs boson equal to their SM values (*i.e.*, working in the *alignment limit* [1857]) and the branching ratios in Eq. (563) are modified by the multiplicative factor

$$\left[1 + \frac{M_H^2}{M_{\rm new}^2 - M_{\chi_{b0}}^2}\tan^2\beta\right]^2. \tag{564}$$

The number of signal events grows with increasing $\tan\beta$ and decreasing $M_{\rm new}$. For a large enough enhancement, the $H_{\rm new}$-exchange contribution will dominate over the SM two-photon intermediate state process.

There is also a continuum signal from $\Upsilon \to \gamma H_{\rm new}^* \to \gamma\tau^+\tau^-$, in which the photon is not monoenergetic. On the $\Upsilon(3S)$ the total continuum signal rate is only about 0.5% of the resonant rates through the $\chi_{b0}(2P)$ and $\chi_{b0}(1P)$ and is spread over a large photon energy range so we neglect it in our results.

The resonant signal is a single photon, monoenergetic in the parent $\Upsilon$ rest frame, with the remainder of the collision energy taken up by the $\tau^+\tau^-$ pair. This must be discriminated from the reducible background $\Upsilon \to \gamma\chi_{b0}$ with $\chi_{b0}$ decaying to anything other than $\tau^+\tau^-$, as well as from the irreducible continuum background $e^+e^- \to \gamma\tau^+\tau^-$. We assume that the $\tau^+\tau^-$ identification purity will be good enough that the reducible backgrounds can be ignored. To estimate the sensitivity to the irreducible background we take as background the total number of $e^+e^- \to \gamma\tau^+\tau^-$ events with a photon energy within a window of width $2\delta E_\gamma$ around the characteristic photon energy.

In Fig. 212 we show the resulting $5\sigma$ discovery reach and 95% confidence level (CL) exclusion reach from 250 fb$^{-1}$ of data on the $\Upsilon(3S)$. We plot the sensitivity reach as a function of $M_{\rm new}$ and $\tan\beta$, assuming that the couplings of the 125 GeV Higgs boson take their SM values. We also show, using dotted lines, the parameter region allowed by direct searches, as computed using HiggsBounds 4.2.0 [1858]. $H_{\rm new}$ masses below about 10 GeV are generally excluded by searches for $\Upsilon \to \gamma H_{\rm new}$ [1849, 1859], which have not been included in HiggsBounds. We note that the 95% CL exclusion line for the $\Upsilon(3S)$-initiated process corresponds to 130 signal events on top of about 4300 background events, so that more sophisticated kinematic cuts could improve signal to background substantially and even more so for the $5\sigma$ discovery curves.

On the $\Upsilon(3S)$, the sensitivity comes almost entirely from decays to $\chi_{b0}(2P)$; the signal rate from $\chi_{b0}(1P)$ is more than one hundred times smaller but with comparable background. This process has the potential to probe a large region of the Type-II 2HDM parameter space with $H_{\rm new}$ masses below 80 GeV and moderate to large $\tan\beta$ that is currently unconstrained by existing searches. On the $\Upsilon(2S)$ the sensitivity is not as good due to a combination of lower signal rate and a larger photon linewidth, resulting in more background.

*16.3.3. Search for a $CP-$odd Higgs boson in radiative $\Upsilon(3S)$ decays.* Radiative decays of vector bottomonium states can be used in the search of a low mass $CP-$odd Higgs boson, $A^0$. Such a possibility has been already discussed in the case of invisible decays of $A^0$ to low mass dark matter, but if dark matter is such that $2M_\chi > M_{A^0}$ then decays of $CP-$odd Higgs boson to dark matter would be kinematically forbidden and $A^0$ would decay to SM final states, in particular one would expect for example $A^0 \to h^+h^-, l^+l^-$ ($h = \pi, K$ and $l = e, \mu, \tau$). One reaction to be studied in this sense is then , for example, $e^+e^- \to \Upsilon(3S)$ followed by $\Upsilon(3S) \to A^0\gamma$ with $A^0 \to l^+l^-$ or $A^0 \to h^+h^-$. This reaction is characterised by the presence of one photon and two oppositely charged leptons/hadrons (same flavour) in the final state and a peak in the photon energy spectrum is expected in correspondence of a peak in the leptons/hadrons invariant mass distribution for a signal $A^0$ production and decays. Searches for $A^0$ decays to final states including $l^+l^-$ ($l = \mu, \tau$) or hadrons have been performed by the BABAR experiment [1814, 1860, 1861]. No excess of events have been observed and 90% C.L. upper limits have been set as follow:

- BR[$\Upsilon(3S) \to \gamma A^0$] $\times$ BR[$A^0 \to \mu^+\mu^-$]$< 5.5 \times 10^{-6}$ for $0.2 < M_{A^0} < 9.3$ GeV/c$^2$ [1860],

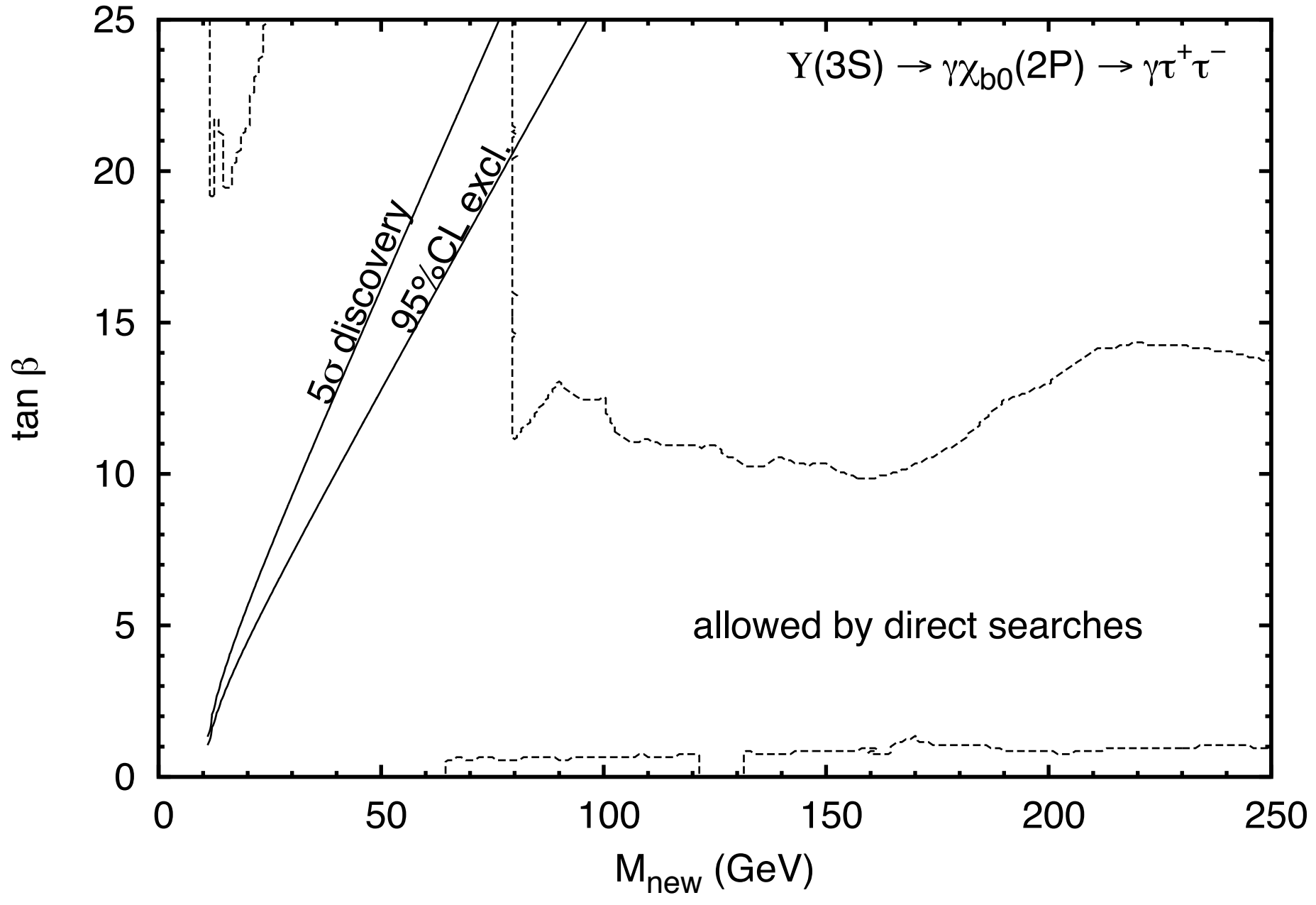

Fig. 212: $5\sigma$ discovery and 95% confidence level (CL) exclusion reach in the Type-II 2HDM from 250 fb$^{-1}$ of data on the $\Upsilon(3S)$. The sensitivity is to the regions to the left of the solid curves. We have set the couplings of the 125 GeV Higgs boson equal to their SM values. The dashed lines indicate the parameter regions still allowed by direct searches for $H_{\rm new}$, computed using HiggsBounds 4.2.0 [1858].

- BR[$\Upsilon(3S) \to \gamma A^0$] $\times$ BR[$A^0 \to \tau^+\tau^-$]$< 1.6 \times 10^{-4}$ for $4.3 < M_{A^0} < 9.5$ and $9.6 < M_{A^0} < 10.1$ GeV/c$^2$ [1861],
- BR[$\Upsilon(3S) \to \gamma A^0$] $\times$ BR[$A^0 \to h^+h^-$]$< 8 \times 10^{-5}$ for $0.3 < M_{A^0} < 7$ GeV/c$^2$ [1814].

These results have been obtained analysing a sample of $122 \times 10^6$ $\Upsilon(3S)$ resonances equivalent to an integrated luminosity of 28 fb$^{-1}$. At Belle II with improved detector performance and an integrated luminosity larger by a factor 7-10 one can expect large improvements also in the search for $CP-$odd Higgs boson production and decays.

*16.3.4. Prospects for lepton universality tests in $\Upsilon(1S)$ decays.* Leptonic decays of $\Upsilon(1S)$ can be used to test lepton universality [1862, 1863]. While the measured values for the BR[$\Upsilon(1S) \to l^+l^-$] ($l = e, \mu, \tau$) are consistent with each other within statistical precision [77]

$$BR[\Upsilon(1S) \to e^+e^-] = 2.38 \pm 0.11 \tag{565}$$

$$BR[\Upsilon(1S) \to \mu^+\mu^-] = 2.48 \pm 0.05 \tag{566}$$

$$BR[\Upsilon(1S) \to \tau^+\tau^-] = 2.60 \pm 0.10 \tag{567}$$

the central values might be hiding a tendency to increase as function of $m_l$; this might be a hint of lepton flavour non universality. The values reported in Eqs. (565)-(567) are based on the average of results that are over a decade old, and the most recent results used in those averages are the following:

$$BR[\Upsilon(1S) \to e^+e^-] = 2.29 \pm 0.08_{stat.} \pm 0.11_{sys.}, \tag{568}$$

$$BR[\Upsilon(1S) \to \mu^+\mu^-] = 2.49 \pm 0.02_{stat.} \pm 0.07_{sys.}, \tag{569}$$

$$BR[\Upsilon(1S) \to \tau^+\tau^-] = 2.53 \pm 0.13_{stat.} \pm 0.05_{sys.}. \tag{570}$$

Equations 568-570 show the values of the BR as obtained in the most recent measurements [1864–1866] together with their statistics and systematic errors. It is important to notice that the values shown in Eqs. (568)-(570) are based on very limited data sample of the order of few $\mathrm{fb}^{-1}$, consequently the statistical precision will be largely improved with the data sample size expected at Belle II experiment. The main systematic effects are due to the determination of the number of produced $\Upsilon(2S)$ resonances in the reaction $e^+e^- \to \Upsilon(2S) \to \pi^+\pi^-\Upsilon(1S)$ with $\Upsilon(1S) \to e^+e^-$ [1864], to the efficiency determination (dominated by uncertainties in the detector simulation) at the level of 1.8% and scale factor between on-resonance and off-resonance data in the determination of $\Upsilon(1S) \to \mu^+\mu^-$ with data collected at different centre-of-mass energies [1865], to $\tau, \mu$ selection criteria at the level of 3% and $\Upsilon \to \tau^+\tau^-$ at the level of 2% in the search for $\Upsilon(1S) \to \tau^+\tau^-$ with data collected at different centre-of-mass energies [1866]. Lepton universality implies that BR[$\Upsilon(1S) \to e^+e^-$]= BR[$\Upsilon(1S) \to \mu^+\mu^-$]= BR[$\Upsilon(1S) \to \tau^+\tau^-$] up to small corrections due to different available phase space, and consequently

$$R_{ll'} = \frac{BR[\Upsilon(1S) \to l^+l^-]}{BR[\Upsilon(1S) \to l'^+l'^-]} = 1 \tag{571}$$

where $l, l' = e, \mu, \tau$ and with the largest correction expected to appear in the $\tau\mu$ ratio with $R_{\tau\mu} = 0.992$. The observed $R_{ll'}$ values are all consistent with unity within 1-2 $\sigma$, so it is interesting to see how this will evolve as more precise measurements of BR[$\Upsilon(1S) \to l^+l^-$] becomes available. While detailed studies for these decay channels aren't available, the Belle II experiment is expected to achieve a better control of systematic effects allowing an improved determination of BR[$\Upsilon(1S) \to l^+l^-$] and of the $R_{ll'}$ ($l, l' = e, \mu, \tau$) ratios.

### *16.4. Conclusions*

Belle II offers exciting opportunities to explore dark sector physics both in scattering processes and in decays of $\Upsilon$ mesons in the MeV to GeV range. Unique triggers will be available already during the early running of Belle II that will allow to collect events with a single photon in the final state. The better hermiticity of the Belle II detector compared to BaBar make these searches competitive with even smaller datasets. With the final dataset the sensitivity of existing limits can be improved by almost an order of magnitude since these searches are generally not limited by systematics.

# 17. Physics beyond the Standard Model

*Editors: F. Bernlochner, R. Itoh, J. Kamenik, V. Lubicz, U. Nierste, Y. Sato, L. Silvestrini*

*Additional section authors: A. Buras, M. Blanke, T. Deppisch, F. De Fazio, J. Jones, L. Hofer, W-S. Hou N. Kosnik, J. Schwichtenberg, C. Smith D. Straub*

## *17.1. Introduction*

This chapter describes new physics models with interesting imprints on flavour-changing transitions, specifically those testable at Belle II.

Flavour physics probes virtual effects of heavy particles with masses far above the reach of the high-$p_T$ experiments ATLAS and CMS. In the SM all flavour-changing transitions originate from the Yukawa sector and are governed by very small numbers. Particularly sensitive to new physics are flavour-changing neutral current (FCNC) processes, which involve fermions of different generations but the same electric charge. Belle II can probe quark FCNC transitions of the type $b \to s$, $b \to d$, $c \to u$, and $s \to d$. Important observables to measure these processes are the meson-antimeson mixing amplitudes, which are called $|\Delta F| = 2$ processes as the flavour quantum number $F = B, S, C$ changes by two units. Equally interesting are FCNC decays that belong to the class of $|\Delta F| = 1$ transitions. In the SM FCNC amplitudes are tiny as they are governed by small CKM elements and are forbidden at tree-level, proceeding instead through an electroweak loop diagram. In addition, the CKM-favoured contribution to $s \to d$ and $c \to u$ transitions are GIM suppressed, with suppression factors of $(m_c^2 - m_u^2)/M_W^2$ and $(m_s^2 - m_d^2)/M_W^2$. The GIM suppression is most spectacular in FCNC decays of charged leptons (such as $\tau \to \mu\gamma$), which are suppressed by a factor of $\Delta m_\nu^2/M_W^2$ where $\Delta m_\nu^2$ is a difference of squared neutrino masses. Furthermore, SM flavour-changing transitions involve only left-handed fermion fields. This feature leads to a chirality suppression of leptonic and radiative decays. For example, the chirality flips in the decay rates of $B \to \mu^+\mu^-$ and $B \to X_s\gamma$ come with factors of $m_\mu^2/m_b^2$ and $m_b^2/M_W^2$, respectively.

Models of physics beyond the Standard Model (BSM) need not involve any of the above-mentioned suppression factors. For example, some of the models discussed in this chapter permit FCNC transitions at tree level. An important category in the classification of BSM theories is the property of *minimal flavour violation (MFV)*[621, 1867] . In MFV theories the only sources of flavour violation are the Yukawa matrices of the SM and enter the amplitudes in such a way that flavour-changing transitions involve the same CKM elements as the corresponding SM contribution. MFV theories may still have CP phases in addition to the Kobayashi-Maskawa phase. Usually the MFV property is an arbitrary add-on to a given model of new physics, for example the Minimal Supersymmetric Standard Model (MSSM) can be studied with or without MFV conditions.

New physics will modify measurements of effective Wilson coefficients (see Ch. 7) away from their SM expectation values. New operators can also be induced, meaning that Wilson coefficients that are negligible in the SM acquire non-zero values. In general, observables of different types may depend on the same Wilson operators, so that new physics will appear in different measurements in a correlated way. This is crucial for fully elucidating the nature of new physics. In this chapter we also discuss correlations with other experiments and observables outside the scope of Belle II, wherever appropriate. However, it is important to note that there are sectors of BSM flavour physics which are exclusively probed at Belle II.

Tables 145-149 provide a summary of the experimental signatures accessible by Belle II that are sensitive to probe the new physics models described in this chapter.

Table 145: A snapshot of the discovery potential of the selected new physics model for Belle II observables: charmless hadronic $B$ decays including time-dependent CP asymmetry. It illustrates, with the number of stars, how well each model can accommodate a potential deviation from the SM in a given observable. The more stars the models have, the more they allow new physics contributions. The "×" means the models does not permit significant new physics contributions for those observables. The "□" indicates that there is no specific study available. The stars are given mainly based on the text in the given sections. The *Experimental Signatures* encodes the competitiveness of Belle II: "★★★" meaning "superior to LHCb", "★★" meaning "competitive to LHCb", and "★" meaning "Belle II can contribute to the measurement, albeit withless sensitivity than LHCb".

| Observables | Experimental Sensitivity | Multi-Higgs Models (§17.2) | generic SUSY | MFV (§17.3) | $Z'$ models (§17.6.1) | gauged flavour (§17.6.2) | 3-3-1 (§17.6.3) | left-right (§17.6.4) | leptoquarks (§17.6.5) | compositeness (§17.7) | dark sector (§16.1) |
|---|---|---|---|---|---|---|---|---|---|---|---|
| $b \to s$ gluonic penguins: | | | | | | | | | | | |
| $S_{CP}(B_d^0 \to \eta^{(\prime)} K_S^0)$ | ★★★ | ★ | ★ | ★ | ★ | ★ | ★ | ★★ | × | □ | × |
| $S_{CP}(B_d^0 \to K_S^0 K_S^0 K_S^0)$ | ★★ | × | ★ | ★ | ★ | ★ | ★ | ★★ | × | □ | × |
| $S_{CP}(B_d^0 \to K_S^0 \pi^0)$ | ★★★ | × | ★ | ★ | ★★ | ★ | ★★ | ★★ | × | □ | × |
| $b \to d$ gluonic penguins: | | | | | | | | | | | |
| $S_{CP}(B_d^0 \to K_S^0 K_S^0)$ | ★ | × | ★ | ★ | ★ | ★ | ★ | ★★ | × | □ | × |
| $b \to s$ EW penguins: | | | | | | | | | | | |
| $\Delta A_{CP}(B \to K^{(*)}\pi)$ | ★★★ | × | ★ | × | ★★★ | ★ | ★★★ | □ | × | □ | × |
| $\mathcal{B}(B_s \to \phi \pi^0)$ | ★★ | × | ★ | × | ★★★ | ★ | ★★★ | □ | × | □ | × |
| $\mathcal{B}(B_s \to \phi \rho^0)$ | ★ | × | ★ | × | ★★★ | ★ | ★★★ | □ | × | □ | × |
| $b \to d$ EW penguins: | | | | | | | | | | | |
| $S_{CP}(B \to \pi\pi)$ | ★★★ | × | × | × | ★★★ | ★ | ★★★ | □ | □ | □ | × |
| $S_{CP}(B \to \rho\pi)$ | ★★ | × | × | × | ★★★ | ★ | ★★★ | □ | □ | □ | × |
| $S_{CP}(B \to \rho\rho)$ | ★★★ | × | × | × | ★★★ | ★ | ★★★ | □ | □ | □ | × |

## *17.2. Two Higgs-doublet models*

*Contributing author: Wei-Shu Hou, Ryosuke Itoh*

Table 146: A snapshot of the discovery potential of the selected new physics model for Belle II observables: semi-leptonic and leptonic $B$ decays. See caption of Table 145 for what the symbols stand for.

| Observables | Experimental Sensitivity | Multi-Higgs Models (§17.2) | generic SUSY | MFV (§17.3) | $Z'$ models (§17.6.1) | gauged flavour (§17.6.2) | 3-3-1 (§17.6.3) | left-right (§17.6.4) | leptoquarks (§18.3.1) | compositeness (§17.7) | dark sector (§16.1) |
|---|---|---|---|---|---|---|---|---|---|---|---|
| Inclusive Semileptonic B decays: | | | | | | | | | | | |
| $A_{CP}(B_d^0 \to X\ell\bar{\nu})$ | $\star\star\star$ | $\star$ | $\star\star\star$ | $\star$ | $\star\star\star$ | $\times$ | $\star\star\star$ | $\star$ | $\times$ | $\square$ | $\star$ |
| $A_{CP}(B_s^0 \to X\ell\bar{\nu})$ | $\star\star\star$ | $\star$ | $\star\star\star$ | $\star$ | $\star\star\star$ | $\times$ | $\star\star\star$ | $\star$ | $\times$ | $\square$ | $\times$ |
| $B \to D^{(*)}\tau\bar{\nu}$: | | | | | | | | | | | |
| Branching ratio | $\star\star$ | $\star\star$ | $\times$ | $\times$ | $\times$ | $\times$ | $\times$ | $\star$ | $\star\star\star$ | $\star$ | $\star$ |
| $q^2$ | $\star\star$ | $\star\star\star$ | $\times$ | $\times$ | $\times$ | $\times$ | $\times$ | $\star\star$ | $\star\star\star$ | $\star$ | $\star$ |
| $\tau$ properties | $\star\star\star$ | $\star\star\star$ | $\times$ | $\times$ | $\times$ | $\times$ | $\times$ | $\star\star$ | $\star\star\star$ | $\star$ | $\star$ |
| $B \to \pi\tau\bar{\nu}$: | | | | | | | | | | | |
| Branching ratio | $\star\star$ | $\star\star$ | $\times$ | $\times$ | $\times$ | $\times$ | $\times$ | $\star$ | $\star\star\star$ | $\square$ | $\star$ |
| $q^2$ | $\star\star$ | $\star\star\star$ | $\times$ | $\times$ | $\times$ | $\times$ | $\times$ | $\star\star$ | $\star\star\star$ | $\square$ | $\star$ |
| $\tau$ properties | $\star\star\star$ | $\star\star\star$ | $\times$ | $\times$ | $\times$ | $\times$ | $\times$ | $\star\star$ | $\star\star\star$ | $\square$ | $\star$ |
| Leptonic B Decays: | | | | | | | | | | | |
| $\mathcal{B}(B^+ \to \tau^+\nu)$ | $\star\star\star$ | $\star\star\star$ | $\times$ | $\star$ | $\times$ | $\times$ | $\times$ | $\star$ | $\star\star$ | $\square$ | $\star\star$ |
| $\mathcal{B}(B^+ \to \mu^+\nu)$ | $\star\star\star$ | $\star\star$ | $\times$ | $\star$ | $\times$ | $\times$ | $\times$ | $\star$ | $\star\star\star$ | $\times$ | $\star\star\star$ |
| $\mathcal{B}(B_{d,s}^0 \to \tau\tau)$ | $\star\star\star$ | $\star\star$ | $\star\star$ | $\star$ | $\star$ | $\times$ | $\star$ | $\times$ | $\star\star\star$ | $\square$ | $\times$ |
| $\mathcal{B}(B_{d,s}^0 \to \tau^\pm\ell^\mp)$ | $\square$ | $\star$ | $\star$ | $\times$ | $\star$ | $\times$ | $\star$ | $\times$ | $\star\star\star$ | $\square$ | $\times$ |

In this section, we discuss the two Higgs doublet models (2HDM), in which the SM is extended by an additional Higgs boson doublet. Given the discovery of the 125 GeV boson $h^0$, it is mandatory to find out which Higgs sector is realised in nature. We consider only two 2HDMs: Model II, or 2HDM-II, which coincides with the tree-level Higgs sector of the Minimal Supersymmetric Standard Model (MSSM), and the general 2HDM (G2HDM, also called 2HDM-III), i.e. without discrete $Z_2$ symmetry.

The charged Higgs boson $(H^+)$ coupling to quarks in 2HDM-II is

$$\mathcal{L}_Y = \bar{u}_i \left[\cot\beta\, \lambda_i^u V_{ij} L + \tan\beta\, V_{ij}\lambda_j^d R\right] d_j\, H^+ + \text{h.c.} \tag{572}$$

where $V$ is the CKM matrix, $\tan\beta$ is the ratio of VEVs of the two doublets, and $\lambda_i \equiv \sqrt{2}m_i/v$ are the diagonalised Yukawa couplings related to the quark mass $m_i$ and the Higgs vacuum expectation value $v = 246\,\text{GeV}$. In the 2HDM-II the charged leptons couple analogously to $d$-type quarks.

Table 147: A snapshot of the discovery potential of the selected new physics model for Belle II observables: electroweak penguin and radiative $B$ decays, including lepton flavour violating channels. See caption of Table 145 for what the symbols stand for.

| Observables | Experimental Sensitivity | Multi-Higgs Models (§17.2) | generic SUSY | MFV (§17.3) | $Z'$ models (§17.6.1) | gauged flavour (§17.6.2) | 3-3-1 (§17.6.3) | left-right (§17.6.4) | leptoquarks (§18.3.1) | compositeness (§17.7) | dark sector (§16.1) |
|---|---|---|---|---|---|---|---|---|---|---|---|
| Semileptonic $b \to s$ Penguin Decays: | | | | | | | | | | | |
| $B \to K^{(*)}\ell\ell$ angular | ⋆⋆ | × | × | ⋆⋆ | ⋆⋆ | × | ⋆⋆ | × | ⋆⋆⋆ | ⋆⋆ | × |
| $R(K^*)$, $R(K)$ | ⋆⋆ | × | × | × | ⋆⋆ | × | ⋆⋆ | × | ⋆⋆⋆ | ⋆⋆ | × |
| $\mathcal{B}(B \to X_s\ell\ell)$ | ⋆⋆⋆ | × | × | ⋆⋆⋆ | ⋆⋆ | × | ⋆⋆ | × | ⋆⋆⋆ | ⋆⋆ | × |
| $R(X_s)$ | ⋆⋆⋆ | × | × | × | ⋆⋆ | × | ⋆⋆ | × | ⋆⋆⋆ | ⋆⋆ | × |
| $\mathcal{B}(B \to K^{(*)}\tau\tau)$ | ⋆⋆⋆ | ⋆⋆⋆ | × | ⋆ | ⋆ | × | ⋆ | × | ⋆⋆⋆ | ⋆ | × |
| $\mathcal{B}(B \to X_s\tau\tau)$ | □ | ⋆⋆⋆ | × | ⋆ | ⋆ | × | ⋆ | × | ⋆⋆⋆ | ⋆ | × |
| $\mathcal{B}(B \to K^{(*)}\nu\nu)$ | ⋆⋆⋆ | × | × | ⋆ | ⋆ | × | ⋆ | × | ⋆⋆⋆ | ⋆ | × |
| $\mathcal{B}(B \to X_s\nu\nu)$ | □ | × | × | ⋆ | ⋆ | × | ⋆ | × | ⋆⋆⋆ | ⋆ | × |
| Semileptonic $b \to d$ Penguin Decays: | | | | | | | | | | | |
| $B \to \pi\ell\ell$ angular | ⋆⋆ | × | × | ⋆⋆ | ⋆⋆ | × | ⋆⋆ | × | ⋆⋆⋆ | ⋆ | × |
| $R(\rho)$, $R(\pi)$ | ⋆⋆ | × | × | × | ⋆⋆ | × | ⋆⋆ | × | ⋆⋆⋆ | ⋆ | × |
| $\mathcal{B}(B \to X_d\ell\ell)$ | ⋆⋆⋆ | × | × | ⋆⋆⋆ | ⋆⋆ | × | ⋆⋆ | × | ⋆⋆⋆ | ⋆ | × |
| $R(X_d)$ | ⋆⋆⋆ | × | × | × | ⋆⋆ | × | ⋆⋆ | × | ⋆⋆⋆ | ⋆ | × |
| $\mathcal{B}(B \to \pi\tau\tau)$ | □ | ⋆⋆⋆ | × | ⋆ | ⋆ | × | ⋆ | ⋆ | ⋆⋆⋆ | ⋆ | × |
| $\mathcal{B}(B \to \pi\nu\nu)$ | ⋆⋆⋆ | × | × | ⋆ | ⋆ | × | ⋆ | × | ⋆⋆⋆ | ⋆ | × |
| Semileptonic LFV B Decays: | | | | | | | | | | | |
| $\mathcal{B}(B \to Xe^{\pm}\mu^{\mp})$ | ⋆⋆⋆ | ⋆ | ⋆ | × | ⋆ | × | ⋆ | × | ⋆⋆⋆ | □ | × |
| $\mathcal{B}(B \to K^{(*)}\tau\ell)$ | ⋆ | ⋆ | ⋆ | × | ⋆ | × | ⋆ | × | ⋆⋆⋆ | □ | × |
| $\mathcal{B}(B \to \pi\tau\ell)$ | ⋆ | ⋆ | ⋆ | × | ⋆ | × | ⋆ | × | ⋆⋆⋆ | □ | × |
| Radiative Penguins: | | | | | | | | | | | |
| $\mathcal{B}(B \to X_s\,\gamma)$ | ⋆⋆⋆ | ⋆⋆⋆ | ⋆⋆⋆ | ⋆⋆⋆ | ⋆ | ⋆⋆ | ⋆ | ⋆ | ⋆ | ⋆⋆⋆ | × |
| $A_{CP}(B \to X_{s+d}\,\gamma)$ | ⋆⋆⋆ | ⋆⋆⋆ | ⋆⋆⋆ | × | ⋆ | ⋆ | ⋆ | ⋆⋆ | ⋆ | ⋆ | × |
| $S_{CP}(B^0_d \to K^0_S\pi^0\,\gamma)$ | ⋆⋆⋆ | ⋆⋆⋆ | ⋆⋆⋆ | ⋆⋆⋆ | ⋆ | ⋆ | ⋆ | ⋆⋆ | ⋆ | ⋆⋆⋆ | × |
| $S_{CP}(B^0_d \to \rho\,\gamma)$ | ⋆⋆⋆ | ⋆⋆⋆ | ⋆⋆⋆ | ⋆⋆⋆ | ⋆ | ⋆ | ⋆ | ⋆⋆ | ⋆ | □ | × |
| $B^0_s \to \eta^{(\prime)}\gamma$ lifetime | ⋆⋆⋆ | ⋆⋆⋆ | ⋆⋆⋆ | ⋆⋆⋆ | ⋆ | ⋆ | ⋆ | ⋆⋆ | ⋆ | □ | × |

Due to a power suppression of the leading SM contribution to the effective $bs\gamma$ coupling, it was found [1868] in the late 1980s that logarithmic corrections from $H^+$ loop can have a significant impact. In addition, due to the dipole or $\sigma_{\mu\nu}m_bR$ form of the $bs\gamma$ coupling, one could have a $\cot\beta$ factor from the coupling to top at one side of the loop compensating a $\tan\beta$ factor at the other side needed for $m_b$ factor. There is thus a $\tan\beta$-independent $H^+$ effect that turns out to be constructive with the SM contribution, which makes $B \to X_s\gamma$ a powerful tool to constrain $m_{H^+}$. Of course, QCD corrections and other sophisticated effects have to be taken into account, which have seen a dramatic progress over the past two decades, as discussed briefly in Sec. 9.2.1. The recent Belle update [425] of $B \to X_s\gamma$ is slightly lower than the SM expectation in central value, giving rise to the stringent bound $m_{H^+} > 570$ GeV [1869].

Table 148: A snapshot of the discovery potential of the selected new physics model for Belle II observables: $\tau$ decays. See caption of Table 145 for what the symbols stand for.

| Observables | Experimental Sensitivity | Multi-Higgs Models (§17.2) | generic SUSY | MFV (§17.3) | $Z'$ models (§17.6.1) | gauged flavour (§17.6.2) | 3-3-1 (§17.6.3) | left-right (§17.6.4) | leptoquarks (§18.3.1) | compositeness (§17.7) | dark sector (§16.1) |
|---|---|---|---|---|---|---|---|---|---|---|---|
| $\tau$ tree decays: | | | | | | | | | | | |
| $\mathcal{B}(\tau \to K\nu)/\mathcal{B}(\tau \to \pi\nu)$ | ⋆⋆⋆ | ⋆⋆ | × | × | × | × | × | ⋆ | ⋆⋆⋆ | □ | ⋆⋆ |
| $\mathcal{B}(\tau \to K^*\nu)/\mathcal{B}(\tau \to \rho\nu)$ | ⋆⋆⋆ | × | × | × | × | × | × | ⋆ | ⋆⋆⋆ | □ | ⋆⋆ |
| $\tau \to \mu$ decays: | | | | | | | | | | | |
| $\tau \to \mu\gamma$ | ⋆⋆⋆ | ⋆ | ⋆⋆⋆ | ⋆ | ⋆ | ⋆ | ⋆ | × | ⋆ | ⋆⋆⋆ | □ |
| $\tau \to \mu\pi^0$ | ⋆⋆⋆ | ⋆ | ⋆⋆ | × | ⋆⋆⋆ | × | ⋆⋆⋆ | × | ⋆⋆⋆ | □ | □ |
| $\tau \to \mu K^0_S$ | ⋆⋆⋆ | ⋆ | ⋆ | × | ⋆ | × | ⋆ | × | ⋆⋆⋆ | □ | □ |
| $\tau \to \mu\rho^0$ | ⋆⋆⋆ | × | ⋆⋆ | × | ⋆⋆⋆ | × | ⋆⋆⋆ | × | ⋆⋆⋆ | □ | □ |
| $\tau \to \mu K^{0*}$ | ⋆⋆⋆ | × | ⋆ | × | ⋆ | × | ⋆ | × | ⋆⋆⋆ | □ | □ |
| $\tau^- \to \mu^-\ell^-\ell^+$ | ⋆⋆ | ⋆⋆ | ⋆ | × | ⋆⋆⋆ | ⋆⋆⋆ | ⋆⋆⋆ | × | ⋆ | ⋆⋆⋆ | □ |
| $\tau^- \to \mu^-\mu^-e^+$ | ⋆⋆ | ⋆ | × | × | ⋆ | ⋆⋆⋆ | ⋆ | × | × | ⋆⋆⋆ | □ |
| $\tau \to e$ decays: | | | | | | | | | | | |
| $\tau \to e\gamma$ | ⋆⋆⋆ | ⋆ | ⋆⋆⋆ | ⋆ | ⋆ | ⋆ | ⋆ | × | ⋆ | ⋆⋆⋆ | □ |
| $\tau \to e\pi^0$ | ⋆⋆⋆ | ⋆ | ⋆⋆ | × | ⋆⋆⋆ | × | ⋆⋆⋆ | × | ⋆⋆⋆ | □ | □ |
| $\tau \to eK^0_S$ | ⋆⋆⋆ | ⋆ | ⋆ | × | ⋆ | × | ⋆ | × | ⋆⋆⋆ | □ | □ |
| $\tau \to e\rho^0$ | ⋆⋆⋆ | × | ⋆⋆ | × | ⋆⋆⋆ | × | ⋆⋆⋆ | × | ⋆⋆⋆ | □ | □ |
| $\tau \to eK^{0*}$ | ⋆⋆⋆ | × | ⋆ | × | ⋆ | × | ⋆ | × | ⋆⋆⋆ | □ | □ |
| $\tau^- \to e^-\ell^-\ell^+$ | ⋆⋆ | ⋆⋆ | ⋆ | × | ⋆⋆⋆ | ⋆⋆⋆ | ⋆⋆⋆ | × | ⋆ | ⋆⋆⋆ | □ |
| $\tau^- \to e^-e^-\mu^+$ | ⋆⋆ | ⋆ | × | × | ⋆ | ⋆⋆⋆ | ⋆ | × | × | ⋆⋆⋆ | □ |
| $\tau$ CP violation: | | | | | | | | | | | |
| $\tau$ EDM | ⋆⋆⋆ | □ | □ | × | □ | □ | □ | □ | □ | □ | □ |
| $A_{CP}(\tau \to K^0_S\pi\nu)$ | ⋆⋆⋆ | ⋆ | ⋆ | × | × | × | × | ⋆ | ⋆⋆⋆ | □ | □ |

Table 149: A snapshot of the discovery potential of the selected new physics model for Belle II observables: $D$ decays and Dark Sector. See caption of Table 145 for what the symbols stand for.

| Observables | Experimental Sensitivity | Multi-Higgs Models (§17.2) | generic SUSY | MFV (§17.3) | $Z'$ models (§17.6.1) | gauged flavour (§17.6.2) | 3-3-1 (§17.6.3) | left-right (§17.6.4) | leptoquarks (§18.3.1) | compositeness (§17.7) | dark sector (§16.1) |
|---|---|---|---|---|---|---|---|---|---|---|---|
| Charm tree Decays | | | | | | | | | | | |
| $\mathcal{B}(D^+ \to \ell\nu)/\mathcal{B}(D_s^+ \to \ell\nu)$ | $\star\star\star$ | $\star$ | $\times$ | $\times$ | $\times$ | $\times$ | $\times$ | $\times$ | $\star\star\star$ | $\square$ | $\star$ |
| $\mathcal{B}(D_s^+ \to \tau\nu)$ | $\star\star\star$ | $\star\star$ | $\times$ | $\star$ | $\times$ | $\times$ | $\times$ | $\times$ | $\star\star\star$ | $\square$ | $\star$ |
| $\mathcal{B}(D^+ \to \tau\nu)/\mathcal{B}(D_s^+ \to \tau\nu)$ | $\star\star\star$ | $\star\star$ | $\times$ | $\times$ | $\times$ | $\times$ | $\times$ | $\times$ | $\star\star\star$ | $\square$ | $\star$ |
| $A_{CP}(D^+ \to \pi^+\pi^0)$ | $\star\star\star$ | $\star$ | $\star$ | $\times$ | $\star\star$ | $\times$ | $\star\star$ | $\star\star$ | $\times$ | $\square$ | $\times$ |
| $A_{CP}(D^0 \to \pi^0\pi^0)$ | $\star\star\star$ | $\times$ | $\star$ | $\times$ | $\star$ | $\times$ | $\star$ | $\star$ | $\times$ | $\square$ | $\times$ |
| Charm FCNC Decays | | | | | | | | | | | |
| $D^0 \to \gamma\gamma$ | $\star\star\star$ | $\star$ | $\star$ | $\times$ | $\star$ | $\times$ | $\star$ | $\times$ | $\star$ | $\square$ | $\times$ |
| $D^0 \to \mu^+\mu^-$ | $\star\star\star$ | $\star$ | $\star$ | $\times$ | $\star$ | $\times$ | $\star$ | $\times$ | $\star\star$ | $\square$ | $\times$ |
| $D^0 \to e^+e^-$ | $\star\star\star$ | $\star$ | $\star$ | $\times$ | $\star$ | $\times$ | $\star$ | $\times$ | $\star\star$ | $\square$ | $\times$ |
| $D^0 \to$ invisible | $\star\star\star$ | $\star$ | $\star$ | $\times$ | $\star$ | $\times$ | $\star$ | $\times$ | $\star\star$ | $\square$ | $\times$ |
| Dark Sector (boson $A'$, fermion $\chi$): | | | | | | | | | | | |
| $e^+e^- \to A' \to$ invisible | $\star\star\star$ | $\times$ | $\times$ | $\square$ | $\times$ | $\times$ | $\times$ | $\times$ | $\times$ | $\times$ | $\star\star\star$ |
| $e^+e^- \to A' \to \ell\ell$ | $\star\star\star$ | $\star$ | $\times$ | $\square$ | $\star$ | $\times$ | $\star$ | $\times$ | $\times$ | $\times$ | $\star\star\star$ |
| $e^+e^- \to A'\gamma$ | $\star\star\star$ | $\star$ | $\times$ | $\square$ | $\star$ | $\times$ | $\star$ | $\times$ | $\times$ | $\times$ | $\star\star\star$ |
| $B \to$ invisible | $\star\star\star$ | $\times$ | $\times$ | $\square$ | $\star$ | $\times$ | $\star$ | $\times$ | $\star\star\star$ | $\times$ | $\star\star\star$ |
| $B \to KA'$ | $\star\star\star$ | $\times$ | $\times$ | $\square$ | $\times$ | $\times$ | $\times$ | $\times$ | $\times$ | $\times$ | $\star\star\star$ |
| $B \to \pi A'$ | $\star\star\star$ | $\times$ | $\times$ | $\square$ | $\times$ | $\times$ | $\times$ | $\times$ | $\times$ | $\times$ | $\star\star\star$ |
| $B^+ \to \mu^+\chi$ | $\star\star\star$ | $\times$ | $\times$ | $\square$ | $\times$ | $\times$ | $\times$ | $\times$ | $\times$ | $\times$ | $\star\star\star$ |
| $B^+ \to \mu^+\nu A'$ | $\star\star\star$ | $\times$ | $\times$ | $\square$ | $\times$ | $\times$ | $\times$ | $\times$ | $\times$ | $\times$ | $\star\star\star$ |
| $\Upsilon(3S) \to \gamma A'$ | $\star\star\star$ | $\times$ | $\times$ | $\square$ | $\times$ | $\times$ | $\times$ | $\times$ | $\times$ | $\times$ | $\star\star\star$ |

A second powerful constraint on $H^+$ comes from a tree level effect in $B^+ \to \tau^+\nu_\tau$, as discussed in Sec. 8.3.1. It is rather interesting that [238] $\mathcal{B}^{\rm 2HDM-II}/\mathcal{B}^{\rm SM}[B^+ \to \tau^+\nu_\tau] = r_H$, where

$$r_H = \left(1 - \tan^2 m_{B^+}^2/m_{H^+}^2\right), \tag{573}$$

involves only physical parameters, with no dependence on hadronic quantities. Measurements of $B \to \tau\nu$ came much later than $B \to X_s\gamma$ and provided another strong constraint on $m_{H^+}$ and $\tan\beta$, as already discussed in Sec. 8.3, which has been a main driver for the Belle II upgrade, especially in earlier years.

In 2012, however, BaBar announced their measurements of $R_D \equiv B(B \to D\tau\nu)/B(B \to D\ell\nu)$ and the analogously defined $R_{D^*}$, and claimed to rule out 2HDM-II, albeit with a low statistical significance. The main feature of these data, until today, are branching ratios which are *higher* than the SM prediction, while the 2HDM-II predicts *lower* branching ratios, unless the 2HDM contribution is so large that it overcompensates the SM piece. While several measurements by Belle are closer to the SM than the BaBar result, the 2015 result of LHCb for $R_{D^*}$ complied with the BaBar data, which elevated the interest from the broader community. This was for $R_{D^*}$ using $\tau \to \mu\nu\nu$. In 2017, however, LHCb announced a second measurement of $R_{D^*}$, now via 3-prong decay of $\tau$, and the result is more consistent with SM (and Belle). But the $R_{D^{(*)}}$ anomaly is far from gone and is a main target for Belle II, as discussed in Sec. 8.4.1. Combining various inputs, the effect of a scalar boson is not the most favoured explanation. But given the volatility of the experimental situation, it only makes the experimental clarification more imperative.

One impact of the $R_{D^{(*)}}$ anomaly is the gain in interest in G2HDM, i.e. 2HDM without a discrete $Z_2$ symmetry. As pointed out [1870] by Glashow and Weinberg in 1977, having two Yukawa matrices contributing to the mass matrix of each type of charged fermion would result in flavour-changing Higgs couplings. They proposed the Natural Flavour Conservation (or NFC) condition, that there can be only one Yukawa matrix per mass matrix, hence these matrices can be simultaneously diagonalised. This is usually implemented by a $Z_2$ symmetry in 2HDM, such as $u$- and $d$-type quarks receive mass each from its own Higgs doublet, resulting in Eq. (572), where there is no additional free parameter other than $\tan\beta$. With the advent of the $R_{D^{(*)}}$ anomaly, models without NFC were proposed [254–256, 279], utilizing the extra Yukawa couplings coming from the second Higgs doublet. Without any discrete $Z_2$ symmetry to implement NFC, this is in fact the "general" 2HDM, and was earlier called [1871] Model III, to distinguish from the two types (if one discounts the freedom on lepton side) of 2HDM with $Z_2$ .

In G2HDM, which possesses flavour-changing neutral Higgs (FCNH) couplings, the full Yukawa couplings are [1872], in matrix notation

$$\begin{aligned}\mathcal{L}_Y = \; & -\sum_{f=}^{u,\,d} \bar{\mathbf{f}} \Bigg[ \left( \frac{\mathbf{M}^f}{v} h^0 - \frac{\boldsymbol{\rho}^f}{\sqrt{2}} H^0 \right) \sin(\beta-\alpha) + \left( \frac{\boldsymbol{\rho}^f}{\sqrt{2}} h^0 + \frac{\mathbf{M}^f}{v} H^0 \right) \cos(\beta-\alpha) \\ & - i\,\mathrm{sgn}\,(Q_f) \frac{\boldsymbol{\rho}^f}{\sqrt{2}} A^0 \Bigg] \mathbf{f} \; - \Big[ \bar{\mathbf{u}} \big( \boldsymbol{\rho}^{u\dagger} \mathbf{V} L - \mathbf{V} \boldsymbol{\rho}^d R \big) \mathbf{d}\, H^+ + \text{ h.c.} \Big], \end{aligned} \qquad (574)$$

where $\mathbf{M}^f$ is the diagonal mass matrix for $f = u$- or $d$-type quarks, whereas $\boldsymbol{\rho}^f$ is likewise the Yukawa matrix for the doublet that is not responsible for mass generation. Keeping the convention of the 2HDM-II, $\cos(\beta-\alpha)$ is the $h^0$–$H^0$ mixing angle. But since $\tan\beta$ is unphysical when there is no $Z_2$ to distinguish the two doublets, a better notation is to call it $\cos\gamma$. Note that, in $\cos(\beta-\alpha) \equiv \cos\gamma \to 0$ limit, $h^0$ couplings become diagonal and would be equal to that of the SM Higgs boson, while $H^0$ and $A^0$ can have exotic, new Yukawa couplings. From the fact that we see no deviations so far [1873] in $h^0$ properties from SM Higgs, we seem to be either close to this "alignment" limit or close to the limit of a diagonal $\rho^f$. For a realisation of the second possibility cf. the Cheng-Sher ansatz [1874].

Comparing Eq. (574) with Eq. (572), the $\boldsymbol{\rho}^f$ couplings modulate the CKM matrix of the charged Higgs couplings, which various authors have utilised [254–256, 279] to account for the $R_{D^{(*)}}$ anomaly, and the role of Belle II is to clarify the experimental situation. It is certainly more complicated than the $\cot\beta$, $\tan\beta$ and SM Yukawa factors of Eq. (572). It also means that the aforementioned bound [1869] on $m_{H^+}$ for 2HDM-II no longer holds. *G2HDM brings in new flavour (exotic Yukawa couplings) parameters* into the game, which should be welcome news for Belle II in terms of potential measurables. So, just what are the $\boldsymbol{\rho}^u$ and $\boldsymbol{\rho}^d$ (and likewise $\boldsymbol{\rho}^\ell$) matrices? Since one has two Higgs doublets, one combination of the two Yukawa matrices gives the mass matrix, $\mathbf{M}^f$, which is diagonalised in the usual way. An orthogonal combination of the two Yukawa matrices gives rise to $\boldsymbol{\rho}^f$.

The possibility of new Yukawa couplings $\rho^f_{ij}$ should interest Belle II practitioners directly. Besides the $R_{D^{(*)}}$ anomaly, we mention two other examples. The ratio of $B^+ \to \tau^+\nu_\tau$ and $B^+ \to \mu^+\nu_\mu$, or $R_{\rm pl}$ is fixed kinematically for both SM and 2HDM-II. In G2HDM, $\rho^\ell_{ij}$ is nontrivial and modulated by both $\lambda_\mu$ and $\lambda_\tau$, so both $\tau$ or $\mu$ channels can pick up NP effects. This makes the precision measurement of $B^+ \to \mu^+\nu_\mu$ of interest in its own, where the recent hint [242] from the untagged analysis of the full Belle data set is encouraging. Second, we have already mentioned the link between $\tau \to \mu\gamma$ with $h^0 \to \tau\mu$, but the link is in fact more subtle. Naively, the link is through a one-loop diagram involving $\rho_{\tau\mu}$. However, even if this parameter is small, a two-loop mechanism [1875, 1876] connects with $\rho_{tt}$! Thus, there are two sources which can generate $\tau \to \mu\gamma$, and can be probed at Belle II.

In 2HDM-II, the alignment phenomenon of $\cos(\beta - \alpha) \to 0$ arises automatically in the decoupling limit of very heavy non-standard Higgs bosons [1857]. In scenarios with lighter $H^+$, $A^0$, $H^0$ bosons the alignment limit can only be realised via fine-tuning. This is because $\alpha$ depends on details of the Higgs potential, and there is no reason for $\beta$ and $\alpha$ to differ by $\pm\pi/2$.

To this end, a discovery scenario is studied for the charged Higgs with a mass of 800 GeV$/c^2$ at $\tan\beta = 40$. Fig. 213 shows the result for 50 ab$^{-1}$ in a projection fit given by the NP-Japan group. The figure shows a region with 1-$\sigma$ signal confidence level constraint. The constraint is obtained with a global fit to the predicted branching fractions of $B \to \tau\nu$, $B \to X_s\gamma$, and $K \to \mu\nu$ in a 50 ab$^{-1}$ projected data sample assuming the 2HDM of type II. The central values of the predictions are obtained using *SuperIso*[1877] with the given parameter values in the type-II 2HDM model. The errors on branching fraction measurements of $B \to \tau\nu$ and $B \to X_s\gamma$ are as described in the previous chapters. As for the $K \to \mu\nu$, the uncertainty on the lattice calculation is assumed to improve by a factor of 3 from the present while the experimental error is assumed to be unchanged from the value quoted in HFLAV 2016 [230]. In the plot, the upper limit obtained by the ATLAS experiment in 2016 [1878] is overlaid.

## *17.3. Minimal Flavour Violation (MFV)*

*(Contributing author: Christopher Smith)*

Phenomenologically, FCNC observables can be described as driven by effective operators. For example, $B \to X_{d,s}\nu\bar{\nu}$ and $B_{s,d} \to \mu^+\mu^-$ attract contributions from

$$\mathcal{H}_{eff} = \left[\frac{\mathcal{C}^{IJ}_{SM}}{M_W^2} + \frac{\mathcal{C}^{IJ}_{NP}}{\Lambda^2}\right](\bar{Q}^I\gamma_\mu Q^J)H^\dagger D_\mu H\ , \tag{575}$$

with $I, J = 1, 2, 3$, $Q$ the left quark doublet, $H$ the Higgs doublet, $D^\mu$ the SM covariant derivative, and the NP scale $\Lambda$ presumably higher than $M_W$. What makes FCNC so interesting is that the SM Wilson coefficients $\mathcal{C}^{IJ}_{SM}$, corresponding to the $Z$ penguin [1879], are severely suppressed by small CKM matrix elements. Since this is up to now in line with experimental results, the NP contributions tuned by $\mathcal{C}^{IJ}_{NP}$ have to be at least as suppressed if $\Lambda$ is to be only slightly above the electroweak scale, as required to prevent a strong hierarchy problem.

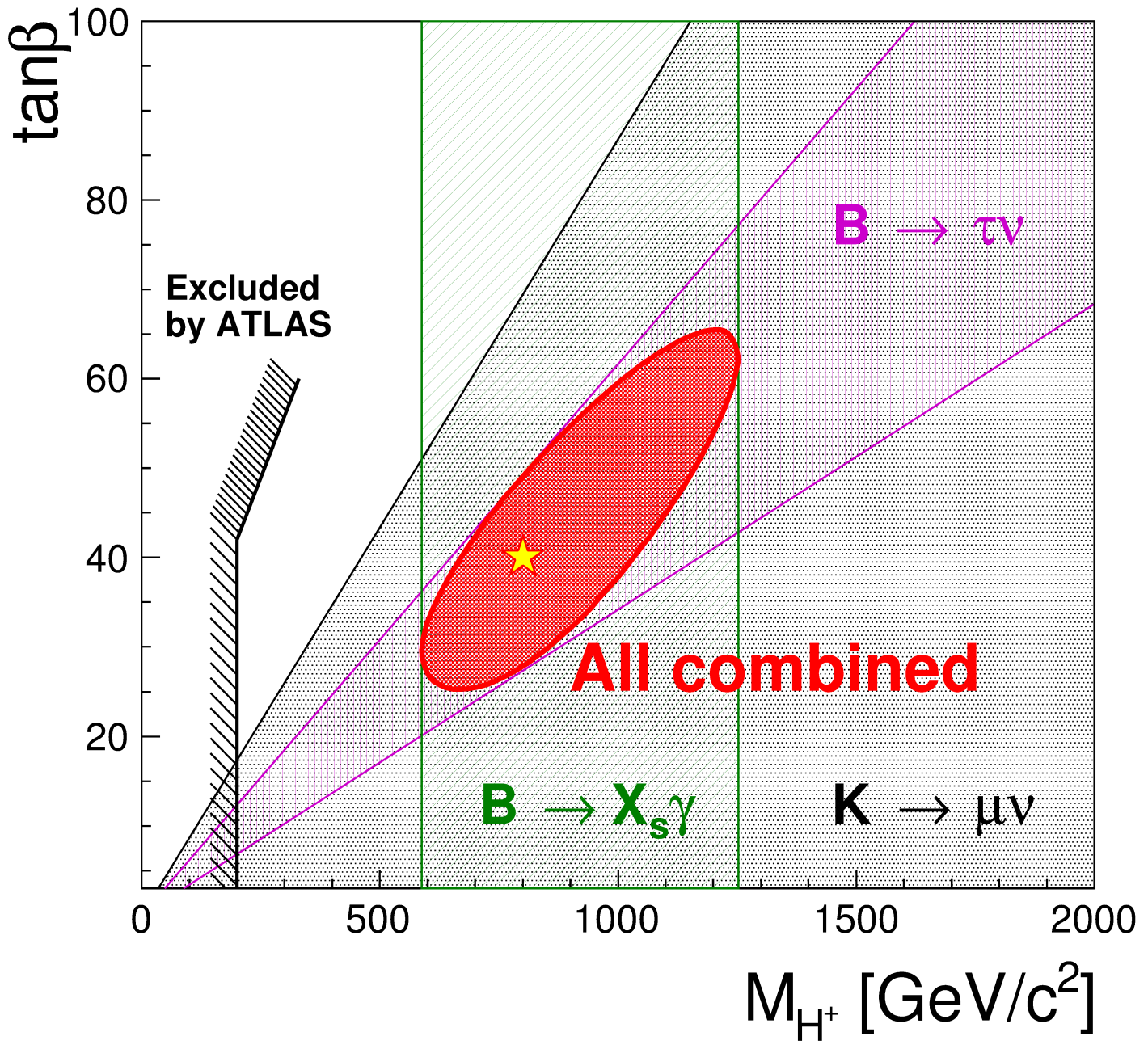

Fig. 213: $\tan\beta$ vs. the charged-Higgs mass $M_{H^+}$ for a discovery scenario with $M_H$=800 GeV and $\tan\beta = 40$ in the 2HDM of type II. Note that current data on $B \to D^{(*)}\tau\nu$ are incompatible with this model and are not included in this figure.

In practice there is no way to tell whether these suppressions are natural or not. Indeed, the naive definition of naturality –Lagrangian parameters should be $\mathcal{O}(1)$– makes no sense in the flavour sector, where the known Yukawa couplings $\mathbf{Y}_{u,d,e}$ are already highly non-natural. So, the best strategy to define a meaningful naturality principle for the NP flavour couplings is to compare them with the Yukawa couplings. There would be no flavour puzzle if the hierarchies of the NP flavour couplings required to pass the experimental constraints are similar to those observed for the quark and lepton masses and mixing parameters.

*Introducing Minimal flavour Violation.* To proceed, this similarity statement must be made precise. We ground it on a symmetry principle and deem natural those NP flavour couplings that respect minimal flavour violation [621]. This hypothesis can be defined in two steps [1880]: the first specifies how the flavour couplings are to be constructed, and the second requires the free parameters to be natural [1881, 1882].

*Construction principle.* The first condition for MFV is expressed straightforwardly in the spurion language: all the flavour couplings are required to be invariant under the flavour symmetry $G_F = U(3)^5$ exhibited by the flavour-independent SM gauge interactions [1867, 1883], but only the spurions required to account for the fermion masses and mixings are allowed. By this we mean that $G_F$ can be formally restored at the level of the whole SM if the Yukawa couplings $\mathbf{Y}_{u,d,e}$ are given definite $G_F$ transformation rules, *i.e.*, are promoted to spurions[66]. This purely formal manipulation provides us with a very useful tool. As soon as the SM Lagrangian becomes invariant under $G_F$, the SM amplitude for any possible process must also be expressible as manifestly $G_F$-invariant. Crucially, this invariance may require inserting Yukawa spurions in a very specific way in the amplitude. Its flavour structure can thus be established quite precisely without embarking into any computation.

In the presence of NP, allowing for only $\mathbf{Y}_{u,d,e}$ is clearly the minimal spurion content, since anything less would be insufficient to reproduce the well-known fermionic flavour structures. Typically, this does not forbid NP from introducing new flavour couplings, but forces them to be expressed as polynomials in the allowed spurions, that is, as functions of the Yukawa couplings $\mathbf{Y}_{u,d,e}$.

To illustrate this, let us take the effective operator (575). Both Wilson coefficients $\mathcal{C}_{SM}^{IJ}$ and $\mathcal{C}_{NP}^{IJ}$ are three-by-three matrices of complex numbers in flavour space. They explicitly break $G_F$ whenever $\mathcal{C}_{SM}, \mathcal{C}_{NP} \neq \mathbf{1}$, in which case their entries depend on the basis chosen for the quark fields. To formally restore the $G_F$ invariance, $\mathcal{C}_{SM}$ and $\mathcal{C}_{NP}$ must transform contragradiantly to the fields. This can be achieved thanks to the presence of the spurions. There are infinitely many combinations of spurions transforming like $\mathcal{C}_{SM,NP}$, so in full generality they are written as expansions

$$\begin{aligned}\mathcal{C}_i &= z_1^i \mathbf{1} + z_2^i \mathbf{Y}_u^\dagger \mathbf{Y}_u + z_3^i \mathbf{Y}_d^\dagger \mathbf{Y}_d \\ &+ z_4^i \{\mathbf{Y}_d^\dagger \mathbf{Y}_d, \mathbf{Y}_u^\dagger \mathbf{Y}_u\} + \ldots \,, \end{aligned} \tag{576}$$

for some a priori complex coefficients $z_j^i$.

Once this expansion is written down, the spurions have to be frozen back to their physical values in some basis. For example, we can now set $v\mathbf{Y}_u \sim \mathbf{m}_u V_{CKM}$, $v\mathbf{Y}_d \sim \mathbf{m}_d$ with $\mathbf{m}_{u,d}$ the diagonal quark mass matrices, and $v$ the Higgs vacuum expectation value. But, without any constraint on the $z_i$, any coupling can be expressed in this way [1884]. The infinite series of powers of $\mathbf{Y}_u^\dagger \mathbf{Y}_u$ and $\mathbf{Y}_d^\dagger \mathbf{Y}_d$ form a complete basis for the space of complex three-by-three matrices. So, the flavour couplings $\mathcal{C}_i$ can take on any value, they could even have all their entries of $\mathcal{O}(1)$.

*Naturality principle.* The spurion expansions are not entirely void of physical content. The numerical value of a flavour coupling like $\mathcal{C}_{SM,NP}^{IJ}$ depends on the basis chosen for the quark fields, and this renders any assertion on the size of the NP flavour couplings ambiguous. On the other hand, by construction, the coefficients occurring in the spurion expansions are basis independent. In particular, the experimental information drawn from flavour observables can be unambiguously translated into values or bounds for the coefficients [1880]. Three situations can arise:

- *MFV flavour structure*: The second condition for MFV is for all coefficients to be natural, $z_i \sim \mathcal{O}(1)$. In that case, all the flavour couplings inherit the peculiar numerical

[66] To force a Lagrangian parameter to transform under some symmetry, it is implicitly promoted to a non-propagating scalar field called a spurion, whose vacuum expectation value matches the parameter's physical value.

hierarchies of the spurions. For example, the leading non-diagonal effects for $\mathcal{C}_i$ arise from

$$\mathcal{C}_i^{I\neq J} \approx z_2^i \, (\mathbf{Y}_u^\dagger \mathbf{Y}_u)^{IJ} \approx z_2^i \, (m_t^2/v^2) \, V_{3I}^* V_{3J} \; , \tag{577}$$

in the down-quark mass basis. This perfectly reproduces the CKM coefficients occurring for the SM $Z$ penguin, with $z_2^{SM} \sim \alpha/4\pi$. If these suppressions of $\mathcal{C}_{NP}^{I\neq J}$ are necessary and sufficient for all FCNC processes, MFV solves the flavour puzzles.

- *Fine-tuned flavour structure*: If some coefficients are still required to be very small, $z_i \ll 1$, the suppression brought in by MFV is not sufficient and one needs either a NP scale $\Lambda$ much higher than the TeV, or a complementary/alternative fine-tuning mechanism for that specific flavour coupling. At present, this is partly the case for flavour-blind $CP$-violating effects, both at the NP and SM levels (where it is known as the strong $CP$ problem).
- *Generic flavour structure*: Some coefficients are required to be very large, $z_i \gg 1$, for instance if some FCNC processes are found to significantly deviate from their SM values. This would signal NP, of course, but also the presence of a new flavour structure within its dynamics. Indeed, though the terms of the expansions (577) form a complete basis, they barely do so; they nearly live in a lower-dimensional subspace. Therefore, a flavour structure not sufficiently aligned with $\mathbf{Y}_{u,d,e}$ generates huge coefficients when projected onto the expansions (577).

MFV thus offers an unambiguous test of naturality. It permits to precisely characterise the flavour puzzles and to identify non-standard flavour structures.

*Why should we trust MFV?.* In view of the severe restrictions MFV imposes on the NP flavour structures, it would seem appropriate to ground it on some full-fledged dynamical mechanism. The problem is that this may be far too ambitious. Its origin may lie in the physics responsible for the observed patterns of quark and lepton masses and mixings, in which case MFV may be explained only once a comprehensive solution to all the flavour issues is found. A second point is that it is actually not necessary to explain the origin of MFV or the internal structures of the spurions dynamically to interpret the MFV hypothesis in very meaningful and universal ways. Let us discuss three such phenomenological interpretations.

*Utilitarian interpretation.* MFV is at the very least a convenient tool. First, it offers an improved parametrisation. Instead of working with the ambiguous values of the couplings in some basis, one deals with the value of the coefficients of the expansions. There are as many free parameters in both descriptions [1884]. Second, the numerical size of the expansion coefficients is the only meaningful measure of the naturality of the NP couplings. It would not be consistent to say that a NP flavour coupling is unnatural if it is no more fine-tuned than those of the SM. So, MFV could be viewed as an **improved dimensional analysis** tool, designed to tackle the highly hierarchical flavour sector.

*Pragmatic interpretation.* Let us assume that some NP exists, whose dynamics is blind to the flavour of the fields. Its flavour sector is thus trivial and the only $G_F$-breaking term in the whole SM plus NP Lagrangian are the usual Yukawa couplings only.

In practice, such a flavour-blind NP setting is not tenable because the SM is not flavour blind. The non-trivial SM flavour mixings will spill onto the NP flavour sector through radiative corrections and/or RG evolution [1884–1886]. At least at the loop level, the flavour-blind NP dynamics combined with the SM flavour mixings will generate new contributions to the FCNC. This is where MFV enters since all the flavour transitions remain tuned by the Yukawa couplings. MFV is not a hypothesis in this case; it is strictly valid. So, MFV emerges as **the least acceptable flavour violation** for the NP sector.

*Redundancy interpretation.* MFV can be understood as statement about the mechanism at the origin of the flavour structures. To illustrate this, imagine a low-energy theory with two elementary flavour couplings $\mathbf{Y}$ and $\mathbf{A}$, which can be thought of as the Yukawa and NP couplings. At the very high scale, some flavour dynamics is active and introduces a single explicit breaking of $G_F$, which we call $\mathbf{X}$. The two low-energy flavour couplings are induced by this elementary flavour breaking, so it is possible to express them as:

$$\begin{cases} \mathbf{Y} = x_1^Y \mathbf{1} + x_2^Y \mathbf{X} + x_3^Y \mathbf{X}^2 \ , \\ \mathbf{A} = x_1^A \mathbf{1} + x_2^A \mathbf{X} + x_3^A \mathbf{X}^2 \ . \end{cases} \tag{578}$$

If the flavour dynamics was known, these coefficients could be computed explicitly. Lacking this, we simply assume they are natural. Also, for these expansions to make sense, powers of $\mathbf{X}$ must not grow unchecked. A sufficient condition is $\langle \mathbf{X} \rangle \lesssim 1$, since then all $\mathbf{X}^{n>2}$ can be eliminated in terms of $\mathbf{1}$, $\mathbf{X}$, and $\mathbf{X}^2$ without upsetting $x_i \sim \mathcal{O}(1)$. Under this condition, from Eq. (578), we can get rid of the unknown high-energy spurion $\mathbf{X}$ and derive the low-energy MFV expansions

$$\begin{cases} \mathbf{A} = y_1 \mathbf{1} + y_2 \mathbf{Y} + y_3 \mathbf{Y}^2 \ , \\ \mathbf{Y} = a_1 \mathbf{1} + a_2 \mathbf{A} + a_3 \mathbf{A}^2 \ , \end{cases} \tag{579}$$

for some $y_i, a_i$ coefficients. Naturality is preserved since $y_i, a_i \sim \mathcal{O}(1)$ when $x_i \sim \mathcal{O}(1)$. So, in this interpretation, neither the Yukawa $\mathbf{Y}$ nor the NP coupling $\mathbf{A}$ are fundamental, and the MFV expansions are understood as the only low-energy observable consequences of their **intrinsic redundancy**.

*Some MFV frameworks and expectations.* MFV strongly constrain the NP flavour structures, but not the rest of the dynamics. Typically, MFV is very effective at relating different observables since their scaling essentially derives from that of the CKM coefficients, but not so much at predicting their overall size which depends essentially on the masses of the NP particles. Let us give a few examples.

*Model-independent MFV:.* Let us take again the operator (575). The cleanest constraints on $\mathcal{C}_{NP}^{IJ}$ come from leptonic and semileptonic processes because the hadronic matrix elements are well-controlled theoretically. The golden modes are the $B_{d,s} \to \mu^+\mu^-$ decays, along with $K^+ \to \pi^+\nu\bar{\nu}$ (in principle, the $B_{d,s} \to (K, \pi, ...)\nu\bar{\nu}$ or $B_{d,s} \to (K, \pi, ...)\ell^+\ell^-$ processes could also be used). Assuming the NP contribution is at most saturating the experimental measurements, we find the values quoted in Table 150. For generic Wilson coefficients, the strongest constraints on $\Lambda$ come from the kaon sector. Indeed, the experimental results are in good agreement with the SM, so they roughly scale like the corresponding SM contributions. The kaon sector is the most suppressed by the CKM scaling, hence it is the one leaving the least room for NP. On the other hand, once MFV scalings are enforced, $\mathcal{C}_{NP}^{sd}$ is so suppressed that $\Lambda$ is allowed to be much lower, and $B$ physics takes the lead. Note, though, that this also means in practice that the NP scale should be very low to induce experimentally visible deviations, especially if the NP dynamics prevents tree-level FCNC.

Table 150: Effective New Physics scales derived under various hypotheses for the Wilson coefficient $\mathcal{C}_{NP}^{IJ}$, assuming its contribution at most totally saturates the experimental measurements. The last two columns correspond to MFV, with the CKM matrix elements $|V_{tI}^* V_{tJ}| \approx 4 \cdot 10^{-2},\ 8 \cdot 10^{-3},\ 3 \cdot 10^{-4}$ for $(I,J) = (b,s),(b,d),(s,d)$, respectively. The last column further assumes a loop level NP contribution, in which case $\mathcal{C}_{NP}$ essentially scales like $\mathcal{C}_{SM}$ and $\Lambda$ ends up close to the electroweak scale.

| $\mathcal{C}_{NP}^{IJ}$ | $\mathcal{O}(1)$ | $\mathcal{O}(g^2/4\pi)$ | $\mathcal{O}(\lvert V_{tI}^* V_{tJ}\rvert)$ | $\mathcal{O}(\lvert V_{tI}^* V_{tJ}\rvert \times g^2/4\pi)$ |
|---|---|---|---|---|
| $B_s \to \mu^+\mu^-$ | $\Lambda \gtrsim 12\,\mathrm{TeV}$ | $\Lambda \gtrsim 2.2\,\mathrm{TeV}$ | $\Lambda \gtrsim 2.5\,\mathrm{TeV}$ | $\Lambda \gtrsim 0.45\,\mathrm{TeV}$ |
| $B_d \to \mu^+\mu^-$ | $\Lambda \gtrsim 17\,\mathrm{TeV}$ | $\Lambda \gtrsim 3\,\mathrm{TeV}$ | $\Lambda \gtrsim 1.5\,\mathrm{TeV}$ | $\Lambda \gtrsim 0.27\,\mathrm{TeV}$ |
| $K^+ \to \pi^+\nu\bar{\nu}$ | $\Lambda \gtrsim 100\,\mathrm{TeV}$ | $\Lambda \gtrsim 18\,\mathrm{TeV}$ | $\Lambda \gtrsim 1.8\,\mathrm{TeV}$ | $\Lambda \gtrsim 0.33\,\mathrm{TeV}$ |

Table 151: Same as in Table 150, but for the meson mixing operator $\mathcal{Q}_{WW}^{IJ} \equiv (\bar{Q}^I \gamma_\mu Q^J)(\bar{Q}^I \gamma^\mu Q^J)$, induced in the SM by the W box diagram.

| $(\mathcal{C}_{WW}^{IJ})_{\mathrm{NP}}$ | $\mathcal{O}(1)$ | $\mathcal{O}((g^2/4\pi)^2)$ | $\mathcal{O}(\lvert V_{tI}^* V_{tJ}\rvert^2)$ | $\mathcal{O}(\lvert V_{tI}^* V_{tJ}\rvert^2 \times (g^2/4\pi)^2)$ |
|---|---|---|---|---|
| $B_s^0 - \bar{B}_s^0$ | $\Lambda \gtrsim 130\,\mathrm{TeV}$ | $\Lambda \gtrsim 4\,\mathrm{TeV}$ | $\Lambda \gtrsim 5\,\mathrm{TeV}$ | $\Lambda \gtrsim 0.17\,\mathrm{TeV}$ |
| $B_d^0 - \bar{B}_d^0$ | $\Lambda \gtrsim 650\,\mathrm{TeV}$ | $\Lambda \gtrsim 21\,\mathrm{TeV}$ | $\Lambda \gtrsim 5\,\mathrm{TeV}$ | $\Lambda \gtrsim 0.16\,\mathrm{TeV}$ |
| $K^0 - \bar{K}^0$ | $\Lambda \gtrsim 24000\,\mathrm{TeV}$ | $\Lambda \gtrsim 800\,\mathrm{TeV}$ | $\Lambda \gtrsim 8\,\mathrm{TeV}$ | $\Lambda \gtrsim 0.25\,\mathrm{TeV}$ |

Table 152: Same as in Table 150, but for the magnetic operator $\mathcal{Q}_\gamma^{IJ} \equiv \bar{D}^I \sigma_{\mu\nu} Q^J F^{\mu\nu} H^C$. In this case though, the SM contribution is not neglected, and we use the bounds set in Refs. [427], [431], and [1887].

| $(\mathcal{C}_\gamma^{I\neq J})_{\mathrm{NP}}$ | $\mathcal{O}(1)$ | $\mathcal{O}(m_{b,s}/v)$ | $\mathcal{O}(\lvert V_{tI}^* V_{tJ}\rvert \times m_{b,s}/v)$ | $\mathcal{O}(\lvert V_{tI}^* V_{tJ}\rvert \times g^2/4\pi \times m_{b,s}/v)$ |
|---|---|---|---|---|
| $b \to s\gamma$ | $\Lambda \gtrsim 220$ TeV | $\Lambda \gtrsim 34$ TeV | $\Lambda \gtrsim 7\,\mathrm{TeV}$ | $\Lambda \gtrsim 1.2\,\mathrm{TeV}$ |
| $b \to d\gamma$ | $\Lambda \gtrsim 56$ TeV | $\Lambda \gtrsim 9$ TeV | $\Lambda \gtrsim 0.8\,\mathrm{TeV}$ | $\Lambda \gtrsim 0.14\,\mathrm{TeV}$ |
| $s \to d\gamma$ | $\Lambda \gtrsim 220$ TeV | $\Lambda \gtrsim 5$ TeV | $\Lambda \gtrsim 0.1\,\mathrm{TeV}$ | $\Lambda \gtrsim 0.02\,\mathrm{TeV}$ |

Looking at Table 150, it is clear though that these $B$ and $K$ decay modes probe similar scales when MFV is active. In that case, and with the prospect of further experimental results on $K^+ \to \pi^+\nu\bar{\nu}$ in the near future, these $B$ decay modes are not the best place to look for new physics. Let us thus consider other observables, and take the operator $\mathcal{Q}_{WW}^{IJ} \equiv (\bar{Q}^I \gamma_\mu Q^J)(\bar{Q}^I \gamma^\mu Q^J)$, relevant for meson-antimeson mixing, and $\mathcal{Q}_\gamma^{IJ} \equiv \bar{D}^I \sigma_{\mu\nu} Q^J F^{\mu\nu} H^C$, for $d^I \to d^J\gamma$ transitions, with $Q$ the quark doublet, $D$ the down-type quark singlet, and $F^{\mu\nu}$ the QED field strength. The corresponding scales are given in Tables 151 and 152. From the MFV point of view, meson mixing does not look very promising to go well beyond the scales probed with $K^0 - \bar{K}^0$ (and this gets even worse in the presence of non-standard operators like $\left(\bar{D}^I Q^J\right)\left(\bar{Q}^I D^J\right)$ to which $K^0 - \bar{K}^0$ is particularly sensitive). On the other hand, $b \to s\gamma$ and to a lesser extent $b \to d\gamma$ have the unique ability to probe the magnetic operator, to which $K$ physics is essentially blind in the MFV case [1887]. The only caveat worth keeping in mind is the assumption that $\bar{D}^I \sigma_{\mu\nu} Q^J F^{\mu\nu} H^C$ is not accompanied by $\bar{D}^I \sigma_{\mu\nu} T^a Q^J G_a^{\mu\nu} H^C$, because the latter is already tightly bounded by $\varepsilon'_K$.

*Supersymmetric MFV.* Current constraints from direct searches at the LHC push the mass of supersymmetric particles quite far from the EW scale. So much so that if MFV is enforced, the prospect of observing any deviation in $B$ physics look dire (see Tables 150, 151, and 152).

There is however a caveat. While MFV only affects flavour couplings, this can influence the expected dynamics once in a supersymmetric setting. There are two interesting consequences. First, the squark masses derive nearly entirely from purely supersymmetric flavour couplings (the soft-breaking terms). So MFV restricts the squark mass spectrum, by requiring for example that the left squark mass matrix expresses itself as $\mathbf{M}_{LL}^2 = m_1^{LL}\mathbf{1} + m_2^{LL}\mathbf{Y}_u^\dagger \mathbf{Y}_u + ...$ [1888]. One of its predictions is that while in most cases the squarks should be quasi degenerate, MFV nevertheless permits to decouple the stop. A so-called natural SUSY-like spectrum arises when $m_2^{LL} \approx -m_1^{LL}/tr(\mathbf{Y}_u^\dagger \mathbf{Y}_u)$, which respects MFV naturality thanks to the large top quark Yukawa coupling [1889]. Thus, the stop could be much lighter than the other squarks, and could play a significant role in FCNC [1890–1892].

Second, the direct searches for supersymmetric particles usually assume that the so-called R parity is enforced. Typical signature then involve significant missing energy, that carried away by the stable lightest sparticle. But once MFV is present, this is not compulsory to satisfy proton decay bounds [1893–1896]. A perfectly viable supersymmetric setting with MFV then emerges, where baryon number violation would be significant when involving the stop. Current bounds on squark and gaugino masses would be evaded [1897]. With in addition a rather light stop from natural-SUSY like soft-breaking terms, significant supersymmetric effects in FCNC could still occur.

*Lepton-flavour violation under MFV.* To deal with lepton flavour violation, let us assume a seesaw mechanism is present. Two new spurions are then relevant at low energy [1898]: the tiny neutrino mass, $v^2 \mathbf{Y}_\nu^T (\mathbf{M}_R^{-1}) \mathbf{Y}_\nu = U^* \mathbf{m}_\nu U^\dagger$, where $\mathbf{m}_\nu$ is the diagonal left-handed neutrino mass matrix, $\mathbf{M}_R$ is the heavy $\nu_R$ Majorana mass matrix, $\mathbf{Y}_\nu$ is the neutrino Yukawa coupling, $U$ the PMNS neutrino mixing matrix, and more interestingly, an unsuppressed spurion $\mathbf{Y}_\nu^\dagger \mathbf{Y}_\nu$ (which cannot be fully reconstructed out of the available data on $\mathbf{m}_\nu$ and $U$ [1899]).

Let us now consider the $P \to P' \nu^I \bar{\nu}^J$ and $P \to P' \ell^I \bar{\ell}^J$ decay modes with $I \neq J$. When enforcing MFV, operators involving right-handed fermions are heavily suppressed by light fermion masses. The least suppressed operators able to induce these transitions is then of the form $\bar{Q}^I (\mathbf{Y}_u^\dagger \mathbf{Y}_u)^{IJ} Q^J \otimes \bar{L}^K (\mathbf{Y}_\nu^\dagger \mathbf{Y}_\nu)^{KL} L^L$, with $L$ the left-lepton doublet so from which $\mathcal{B}(P \to P' \nu^I \bar{\nu}^J) \sim \mathcal{B}(P \to P' \ell^I \bar{\ell}^J)$.

Because the quark and lepton flavour sectors are completely disconnected under MFV (the flavour group factorises), the hadronic current shows the same suppression as before. With a NP scale only slightly above the EW scale, the $\mathbf{Y}_u^\dagger \mathbf{Y}_u$ insertion alone naturally brings the NP contributions at most at around the SM ones. Then there remain the leptonic currents. Since $\mathbf{m}_\nu \sim \mathbf{Y}_\nu^T (\mathbf{M}_R^{-1}) \mathbf{Y}_\nu$, it would appear that taking $\mathbf{M}_R$ sufficiently large would ensure $\mathbf{Y}_\nu \sim \mathcal{O}(1)$, leading to $P \to P'\nu^I\bar{\nu}^J \sim P \to P'\nu^I\bar{\nu}^I$. However, such large $\mathbf{Y}_\nu$ are forbidden by $\ell^I \to \ell^J\gamma$, tuned by the same spurion insertion $E\mathbf{Y}_e(\mathbf{Y}_\nu^\dagger\mathbf{Y}_\nu)\sigma^{\mu\nu}LH^\dagger F_{\mu\nu}$. Conservatively, we can at most get $(\mathbf{Y}_\nu^\dagger\mathbf{Y}_\nu)^{I\neq J} \sim 1\%$, so that for $I \neq J$, $L = \nu, \ell$, $\mathcal{B}^{MFV}(K \to \pi L^I \bar{L}^J) \lesssim 10^{-15}$ and $\mathcal{B}^{MFV}(B \to (\pi, K)L^I\bar{L}^J) \lesssim 10^{-10}$, well beyond experimental reach. These modes are thus very powerful checks for the presence of new flavour structures, and would in particular react strongly to any NP spurion directly connecting the lepton and quark sectors.

## *17.4. Models with Lepton Flavour Violation (LFV)*

*(Contributing author: Nejc Košnik)*

Lepton flavour is exactly conserved in the SM due to the pattern of the lepton flavour group breaking due to Yukawa couplings, $U(3)_L \otimes U(3)_{e_R} \to U(1)_e \otimes U(1)_\mu \otimes U(1)_\tau$ that keeps individual leptonic flavours conserved. To account for the observed neutrino mass differences one can simply introduce a singlet right-handed neutrino(s) with appropriate Yukawa term that breaks leptonic flavour down to lepton number $U(1)_{e+\mu+\tau}$. LFV processes with charged leptons at low energies [67] are then induced by flavour mixing of virtual neutrinos, where only an extremely small GIM-violating effect of order $m_\nu^2/m_W^2$ survives, making such a framework effectively lepton flavour conserving [1900]. Less theoretically ad-hoc frameworks that induce neutrino masses possibly enhance charged LFV (CLFV) processes up to experimentally observable levels. The $U(1)_e \otimes U(1)_\mu \otimes U(1)_\tau$ symmetry is accidental in the SM and is in general expected to be broken in NP models, which makes searches for CLFV processes a very interesting null test of the SM. From the above reasoning it is also evident that NP models can in general be expected to violate lepton flavour. Recent hints of lepton flavour non-universality in $B \to K\ell\ell$ decays and intriguing deviations from the SM predictions in $B \to K^{(*)}\mu\mu$ spectra, if true, generally imply accompanying LFV processes [623]. In this section, we first introduce the effective Lagrangian for LFV, testable at Belle II, and provide a brief summary of the level of LFV in representative NP models.

*Effective Lagrangian for LFV at Belle II.* Narrowing our focus, for the time being, onto baryon and lepton number conserving processes involving two (or four) charged leptons we can set up a model independent parameterisation of heavy NP in terms of effective Lagrangian of mass-dimension 6 at the electroweak scale [619, 620, 1901]. There are in total 19 LFV-mediating operators that contain either 2 or 4 leptonic fields. This effective theory (SM-EFT) is a sensible starting point in studies of low-energy phenomenology of any NP model with degrees of freedom heavier than the weak scale. More suited to the processes to be studied at Belle II is the effective Lagrangian at scale $m_B$ that is matched onto to the SM-EFT through renormalisation group (RG) running due to the full SM group above the electroweak scale [1901] and due to strong and electromagnetic RG effects below the electroweak scale [396, 1902].

The $m_B$-scale Lagrangian of dimension-6 can be systematically broken down to

$$\mathcal{L}_{\text{eff}} = \mathcal{L}_{\text{eff}}^{(D)} + \mathcal{L}_{\text{eff}}^{(4\ell)} + \mathcal{L}_{\text{eff}}^{(\ell q)} + \mathcal{L}_{\text{eff}}^{(G)}, \tag{580}$$

[67] Since the outgoing neutrino flavours are not accessible experimentally we need at least two charged leptons in the asymptotic states to tag LFV.

where the definition of the above operators is given in Sec. 14.2.1. Here $\mathcal{L}_{\rm eff}^{4\ell}$ is the Lagrangian containing four leptons and is mainly responsible for the purely leptonic LFV decays $\tau^- \to \ell'^- \ell'^{\mp} \ell''^{\pm}$, that are among the Golden channels of Belle II. The semileptonic part of the LFV Lagrangian $\mathcal{L}_{\rm eff}^{\ell q}$ composed of $(\bar{q}q)(\bar{\ell}\ell)$ fields mediates LFV meson (M) decays, $M \to \ell\ell'$, $M \to M'\ell\ell'$, baryon (N) decays $N \to N'\ell\ell'$, and semileptonic $\tau$ decays, $\tau \to M\ell$ and $\tau \to PP\ell$. A very useful compilation of constraints on operators present in $\mathcal{L}_{\rm eff}^{\ell q}$ was presented in [1903]. The dipole Lagrangian

$$\mathcal{L}_{\rm eff}^{(D)} = -\frac{m_\tau}{\Lambda^2} \left[ C_{DL} \bar{\mu} \sigma^{\mu\nu} P_L \tau + C_{DR} \bar{\mu} \sigma^{\mu\nu} P_R \tau \right] F_{\mu\nu} \tag{581}$$

triggers the radiative LFV decays that are accessible at Belle II, e.g., $\tau \to \mu\gamma, e\gamma$. Phenomenology of leptonic and semileptonic meson decays as induced by operators contained within $\mathcal{L}_{\rm eff}^{(\ell q)}$ has been studied in [1904], whereas the operator basis of $\mathcal{L}_{\rm eff}^{(4\ell)}$ adapted to the $\mu \to 3e$ decay can be found in Eq. (112) of ref. [1905]. Refs. [124, 1906, 1907] studied model-independent and model-discriminating aspects of $\tau \to \ell\ell\ell'$ decay. For the role of effective gluonic operators $\mathcal{L}_{\rm eff}^{(G)}$ in $\tau \to \mu\eta^{(\prime)}$ see [124, 1908]. Finally, Belle II prospects for charged LFV in $\tau$ decays have been discussed in Sec. 14.2.

*Model Case Studies for LFV.* LFV processes involving four-fermion vertices $(\bar{q}q)(\bar{\ell}\ell)$, $(\bar{\ell}\ell)(\bar{\ell}\ell)$ can be mediated by tree-level amplitudes with renormalisable couplings in models with neutral mediators such as additional Higgs or $Z'$ gauge boson. In the two Higgs doublet model (2HDM) with generic Yukawa couplings (also known as Type III 2HDM) leptonic decays of mesons $P \to \ell\ell'$ and purely leptonic decays $\mu \to eee$, $\tau \to \ell\ell\ell'$ are induced by tree-level exchanges of neutral scalars. For moderately large $\tan\beta$ the $\mathcal{B}(\tau \to \mu\gamma)/\mathcal{B}(\tau \to \mu\mu\mu)$ and $\mathcal{B}(\tau \to e\gamma)/\mathcal{B}(\tau \to e\mu\mu)$ could be between 0.1 and $\sim 1$, and taking into account better Belle II sensitivity to decays with three final-state leptons makes the searches for $\tau \to 3\ell$ more promising [279]. For a study of LFV in the $\mu$-$\tau$ sector of the Minimal supersymmetric SM cf. [1909].

Non-diagonal $Z$ couplings to leptons can be a signature of vector-like leptons where the most promising modes to search for at Belle II involve $\tau$: $B_{(s)} \to \tau\mu, \tau e$, and $\tau \to \mu\phi$ can be of the order $10^{-10}$ and $10^{-8}$, respectively [1910]. In models with additional $U(1)'$ gauge symmetries LFV couplings of a lepton pair to $Z$ or $Z'$ boson are induced at tree-level [1911]. In the $Z'$ models LFV is induced due to off-diagonal gauge couplings of $Z^{(\prime)}$ to leptons. It was demonstrated that LFV $B$ meson decays are correlated to another Golden channel, $B \to K^{(*)}\nu\bar{\nu}$ [1912]. Models with very light $Z'$ contribute dynamically at distances comparable to $m_B^{-1}$ and their effects are not caught by the effective Lagrangian (580). A light $Z'$, with a mass above 2 MeV in order to avoid bounds on the number of relativistic degrees of freedom in the early universe, was considered as an explanation of the anomalous muon magnetic moment. However, in such cases $\tau \to \mu Z'$ with $Z'$ subsequently decaying to neutrinos gives too strong a constraint if $m_{Z'} < m_\tau$. For heavier masses, $m_{Z'} > m_\tau$, the scenario can explain $(g-2)_\mu$ provided that $Z'$ couples predominantly to right-handed fermions [1913].

Another mechanism for generating LFV processes is to introduce scalar or vector leptoquarks (LQs) close to the weak scale which are colour triplet states and can be singlets, doublets, or triplets under $SU(2)_L$ in order to be able to induce tree-level LFV in $\mathcal{L}_{\rm eff}^{(\ell q)}$, while contributions to $\mathcal{L}_{\rm eff}^{(4\ell)}$ and $\mathcal{L}_{\rm eff}^{(D)}$ are loop induced [1914]. In the leptoquark scenarios it is usually assumed that a single LQ multiplet with well defined SM quantum numbers is present at a time. Their typical UV embedding are the Grand Unified Theories, however composite scenarios with LQs can also be constructed [1914]. Tree-level LQ exchanges contribute to $\mathcal{L}_{\text{eff}}^{(\ell q)}$ and can be thus most efficiently probed in LFV decays of hadrons or in $\tau \to \ell M$, where $M$ denotes a meson [1915]. The decay $B \to K\mu\tau$ is an important test of the leptoquark scenarios designed to address the lepton flavour universality puzzle $R_K$ and related anomalies in $B \to K^{(*)}\mu\mu$ spectra [295]. Upper bounds on the Golden channel $\tau \to \mu\gamma$ constrain the scalar leptoquark contribution in explaining $R_{D^{(*)}}$ [273] as well as prohibit the attempts at explaining the $h \to \tau\mu$ puzzle [1916]. At one-loop level, LQs contribute to the Belle II Golden channel $\tau \to 3\ell$ [1526, 1914].

Seesaw models of type I, II, and III could be distinguished by the imprint of heavy fermionic/scalar mediators on the dimension-6 operator basis that leads to CLFV processes. These dimension-6 effective operators, which are responsible for CLFV processes, are suppressed with $M^{-2}$, where $M$ is the high scale linked to small neutrino masses [1917, 1918]. For the vector quarkonia decays with LFV it was demonstrated in [1904] that in the inverse seesaw neutrino mass realisation and in the sterile neutrino framework the quarkonia decays $\phi \to \ell\ell'$, $\psi \to \ell\ell'$, $\Upsilon \to \ell\ell'$ can be slightly enhanced, albeit no branching fraction can climb above the $10^{-12}$ level. Type I+III seesaw the semileptonic decays $\tau \to Pe$ and leptonic $\tau \to 3\ell$ present constraints that are slightly weaker compared to the constraints from $\mu$-$e$ sector [1919]. In supersymmetry versions of seesaw mechanisms the $\tau \to \mu\gamma$ can be close to the current upper bound [1920], however LFV meson decay $B \to K\mu\tau$ is suppressed below $10^{-10}$ [1520]. In versions of minimal flavour violation framework in the lepton sector it is well established that with current value of the neutrino mixing angle, $\sin\theta_{13}$, the $\tau \to \mu\gamma$ decay is beyond the reach of Belle II [1898]. The gauged lepton flavour in the framework of Pati-Salam model requires at least $3-4$ orders of magnitude improvement in $\tau \to \mu\gamma$ and $\tau \to 3\mu$ experimental upper bounds in order to potentially test this model [1921].

In Randall-Sundrum models the Golden channel $\tau \to \mu\gamma$ is reachable at Belle II. Another Golden channel, $\tau \to 3\mu$, is currently constrained to lie below $\lesssim 10^{-8}$ [1922], and is thus a sensible probe of such models at Belle II.

*Lepton Number violation.* In the effective theory approach to the SM extensions the lowest dimension operator is the dimension-5 Weinberg operator [637] that violates leptonic number by 2 units ($\Delta L = 2$) and leads to Majorana mass term for neutrinos. For heavy Majorana neutrinos, in addition to the LFV phenomena described above, Belle II could probe $\Delta L = 2$ processes that are resonantly enhanced [1923]. The authors point out semileptonic $\Delta L = 2$ processes $\tau^- \to \ell^+ M^- M'^-$ and $B^- \to \ell^- \ell'^- M'^+$ involving two charged leptons and two charged mesons, which are suitable targets of study in Belle II. Better limits on branching fraction of $\tau^- \to \ell^+ M^- M'^-$ would lead to stricter constraints on the mixing combination $|V_{\ell 4} V_{\tau 4}|$ in the mass range 0.1–1 GeV for the heavy Majorana neutrino. $B \to \ell^- \ell'^- M^+$ decays with $e$ and $\mu$ in the final states are uniquely sensitive to mixing angles $|V_{e4}|$, $|V_{\mu 4}|$ in the mass range 2–5 GeV.

## *17.5. Minimal Super Symmetric Model (MSSM) with $\boldsymbol{U(2)^5}$ symmetry*

*(Contributing author: Joel Jones)*

Supersymmetric models with generic flavour structure (sfermion soft SUSY breaking masses and trilinear couplings) suffer from severe flavour and $CP$ violation constraints. These require that the sfermion states lie in the PeV range or above [1924], exacerbating the EW little hierarchy problem. The strongest constraints come from the measurement of $\epsilon_K$, whose smallness in the SM is due to an approximate $U(2)$ flavour symmetry respected by the quarks of the first two generations and which results in the efficient GIM suppression of all FCNCs among the light quarks. Similarly, LFV $\mu \to e$ transitions, whose null searches put severe constraints on the slepton sector of SUSY models are absent in the $U(2)$ symmetric limit, when the flavour group contains $U(1)_\mu \times U(1)_e$. This motivates to consider NP models respecting approximate $U(2)$ flavour symmetries acting on the light SM fermion generations.

In fact, with the improvement of theoretical input, a tension within $CP$ violating observables was first pointed out in Refs. [1925, 1926]. This consisted of an incompatibility in the determination of $\epsilon_K$, $S_{\psi K^0_S}$ and $\Delta M_d/\Delta M_s$, namely, with two of these observables one can predict the third, and this prediction would be in tension with the current experimental measurements [1927]. A recent analysis [1928] with updated theoretical data confirms that the problem is still present, at a level above $2\sigma$.

A $U(2)^3$ flavour symmetry was proposed in [1929] as a way of solving the tension. This came naturally when considering the virtues of both MFV [621, 1867] and $U(2)$ flavour symmetry models [1930, 1931]. The main idea consists of imposing a $U(2)_Q \otimes U(2)_u \otimes U(2)_d$ symmetry acting on quarks, such that the first two generations transform as doublets, while the third generation remains a singlet. In order to reproduce the observed quark masses and mixings, one would need to introduce spurion fields transforming appropriately under the symmetries:

$$\Delta Y_u \sim (2,\,2,\,1) \qquad \Delta Y_d \sim (2,\,1,\,2) \qquad V_q \sim (2,\,1,\,1) \tag{582}$$

With these spurions, the Yukawa matrices would acquire a structure following a definite pattern:

$$Y_f \sim \left(\begin{array}{c|c} \Delta Y_f & V_q \\ \hline 0 & y_f \end{array}\right) \tag{583}$$

where $y_f$ should be of $\mathcal{O}(1)$[68]. Here, everything above the horizontal line has two rows, and everything to the left of the vertical lines has two columns. The parameters within the spurions are then fixed by requiring them to reproduce the quark masses and mixings. In the following we use $|V_{us}|$, $|V_{cb}|$, $|V_{ub}|$ and $\phi_3$ to build the CKM matrix, and vary them in the following range [77, 85, 143, 230, 353, 1932]:

$$|V_{us}| \in (0.2245,\, 0.2261) \tag{584}$$

$$|V_{cb}| \in (3.97,\, 4.30) \times 10^{-2} \tag{585}$$

$$|V_{ub}| \in (3.56, 4.65) \times 10^{-3} \tag{586}$$

$$\phi_3 \in (63.8^\circ,\, 78.0^\circ) \tag{587}$$

[68] For small values of $\tan\beta$, the suppression in $y_b$ would be justified by the addition of an extra $U(1)_b$ flavour symmetry.

As in MFV, the squark soft masses and trilinears are assumed to acquire a flavour structure based on the same spurions as in the Yukawas. One finds:

$$m^2_{\tilde{Q}} = \begin{pmatrix} I+\Delta_{LL} & x_Q^* V_q^* \\ x_Q V_q^T & m^2_{\text{light}}/m^2_{\text{heavy}} \end{pmatrix} m^2_{\text{heavy}} \quad (588)$$

$$m^2_{\tilde{U}} = \begin{pmatrix} I+\Delta^u_{RR} & x_U^* \Delta Y_u^T V_q^* \\ x_U \Delta Y_u^* V_q^T & m^2_{\text{light}}/m^2_{\text{heavy}} \end{pmatrix} m^2_{\text{heavy}} \quad (589)$$

$$m^2_{\tilde{D}} = \begin{pmatrix} I+\Delta^d_{RR} & x_D^* \Delta Y_d^T V_q^* \\ x_D \Delta Y_d^* V_q^T & m^2_{\text{light}}/m^2_{\text{heavy}} \end{pmatrix} m^2_{\text{heavy}} \quad (590)$$

where $\Delta_{LL} \sim V_q^* V^T + \Delta Y_u^* \Delta Y_u^T + \Delta Y_d^* \Delta Y_d^T$, $\Delta^u_{RR} \sim \Delta Y_u^T \Delta Y_u^*$ and $\Delta^d_{RR} \sim \Delta Y_d^T \Delta Y_d^*$. The trilinears follow the exact same structure of the Yukawas, proportional to $a_0\, m_{\text{heavy}}$, with different $\mathcal{O}(1)$ parameters.

The analyses carried out in [1929, 1933–1936] considered the first two generation of squarks to be completely decoupled from the theory, as in Effective Supersymmetry [1937, 1938]. In addition, no left-right mixing was considered, and only loops with gluinos were taken into account. In this limit, the SUSY contributions would modify $\epsilon_K$, $S_{\psi K^0_S}$ and $\Delta M_d/\Delta M_s$, such that:

$$\epsilon_K = \epsilon_K^{\text{SM(tt)}} \times \left(1 + x^2 F_0\right) + \epsilon_K^{\text{SM(tc+cc)}} \quad (591)$$

$$S_{\psi K^0_S} = \sin\left(2\phi_1 + \arg\left(1 + x F_0 e^{-2i\gamma}\right)\right) \quad (592)$$

$$\frac{\Delta M_d}{\Delta M_s} = \left(\frac{\Delta M_d}{\Delta M_s}\right)^{SM} \quad (593)$$

where $x$ is a combination of $\mathcal{O}(1)$ constants, expected to be smaller than 10, $F_0$ is a loop function depending on $m^2_{\tilde{g}}$ and $m^2_{\tilde{Q}_3}$, and $\gamma$ is an effective NP phase.

The results of [1929, 1934, 1936] confirmed that the new contributions could modify $\epsilon_K$ and $S_{\psi K^0_S}$ in the correct direction, such that the tension would be removed. In addition, a new contribution to $S_{\psi\phi}$ would be induced:

$$S_{\psi\phi} = \sin\left(2|\beta_s| - \arg\left(1 + x F_0 e^{-2i\gamma}\right)\right) \quad (594)$$

where one can see the same NP phase $\gamma$ appearing. This means that the NP contributions to $S_{\psi K^0_S}$ and $S_{\psi\phi}$ would be correlated. In fact, this was used later in [1936] to demonstrate a more complete correlation between the latter two observables and the value of $|V_{ub}|$. Nevertheless, it was found in [1939, 1940] that once the heavy squarks are included, non-negligible corrections appear. These corrections are related to the existence of a super-GIM mechanism, to contributions coming from the off-diagonal elements in the $1-2$ block, and to left-right mixing.

In addition, with the increasing bounds on the gluino mass, it is necessary to consider loops with lighter charginos or neutralinos [1940]. Even though the latter interact through smaller couplings, the bounds on their masses are not so stringent, so it is possible for them to give non-vanishing effects. The neutralino contribution should follow the same flavour structure as for the gluinos. In contrast, charginos and charged Higgs give an MFV-like contribution. All of these effects are expected to be small, but non-negligible.

In order to take all of these points into consideration, the framework has been implemented into `SPheno 3.3.8` [1941, 1942].The appropriate Wilson coefficients have been generated

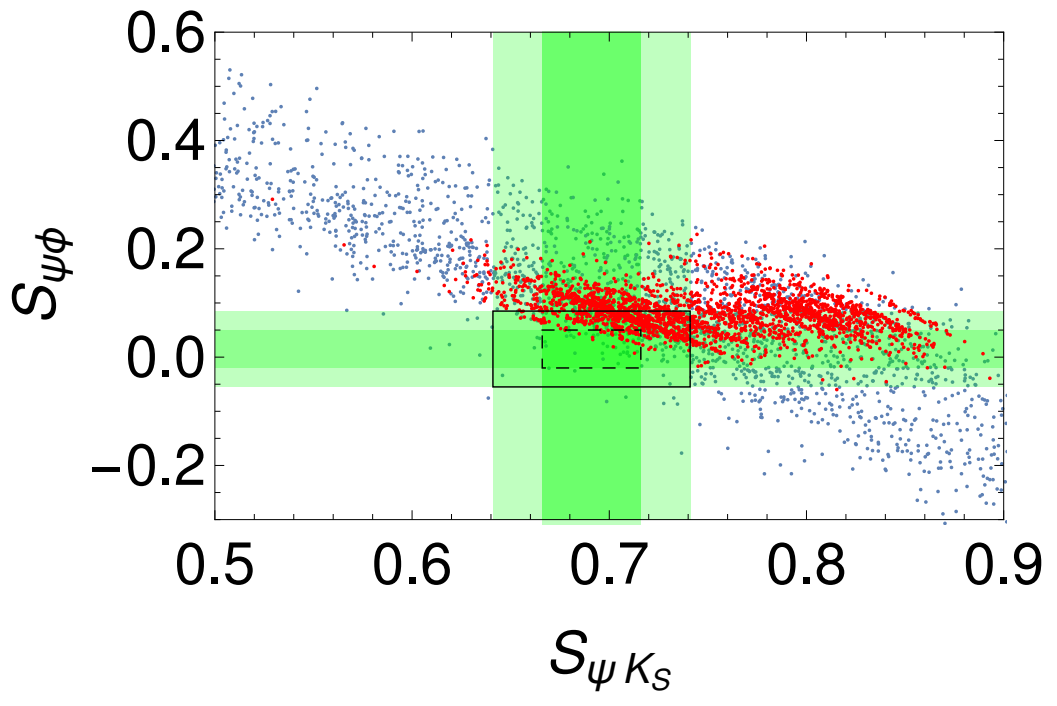

Fig. 214: Correlation between $S_{\psi K_S^0}$ and $S_{\psi\phi}$ in $U(2)^3$. Red points satisfy LHC mass constraints, and shaded green regions show bounds of experimental measurements at 1 and $2\sigma$. The dashed (solid) rectangles show the region where both bounds are satisfied at $1\sigma$ ($2\sigma$) using current world average values. [88, 230]

using `SARAH-FlavorKit 4.10.0` [1943–1945], which are used to generate our observables of interest. For the quark sector, these coefficients are processed with `flavio 0.19.0` [618].

Using this implementation, a scan of the SUSY parameter space is performed in the parameter regions

$$m_{\text{heavy}} \in (10,\ 30)\ \text{TeV} \qquad |a_0| \in (0,\ m_{\text{heavy}})$$
$$M_3 \in (1.8,\ 3)\ \text{TeV} \qquad \mu \in (0.1,\ 1)\ \text{TeV}$$
$$M_{1,2} \in (0.1,\ 1)\ \text{TeV} \qquad \tan\beta \in (2,\ 10)$$
$$M_A \in (1,\ 5)\ \text{TeV}$$

While $m^2_{\text{light}}$ is set to appropriate values such that tachyons are absent, and so that the latest bounds on gluino, stop and sbottom masses are satisfied. One also needs to check that the LSP is neutral.

We first demonstrate the correlation between $S_{\psi K_S^0}$ and $S_{\psi\phi}$. In Fig. 214, we show how both of these are modified in this framework. Points in red (blue) do (do not) satisfy the LHC constraints above. In addition, in regions shaded in green $S_{\psi K_S^0}$ and $S_{\psi\phi}$ are satisfied up to $2\sigma$. From the figure, we see that a smaller $S_{\psi K_S^0}$ will imply larger values of $S_{\psi\phi}$. Moreover, the current LHC bounds are not in conflict with the region where both observables satisfy their bounds. However, we also see it is very difficult to have $S_{\psi K_S^0}$ on its lower end and pass simultaneously the bounds on $S_{\psi\phi}$. It is interesting to note that many red points have very large values of $S_{\psi K_S^0}$. As we shall see, most of these cases are due to the value of $V_{ub}$, and not to SUSY effects. In fact, the current bounds on squark and gluino masses have a strong impact on the SUSY contributions to $\epsilon_K$ and $S_{\psi K_S^0}$. The bounds force $F_0$ to be small, which then requires $x$ to be large in order to give a large enough effect in $S_{\psi K_S^0}$. However, since $x$ appears quadratically in $\epsilon_K$, this can cause the latter to overshoot. This can be seen in Fig. 215. Comparison with between the SM-only points (orange) and those with SUSY contributions (blue) help understand how $U(2)^3$ can reduce the tension. Although this is still possible, it is also common to have a too large contribution to $\epsilon_K$. It is interesting that, as with $S_{\psi\phi}$, it is not favourable to have a small $S_{\psi K_S^0}$.

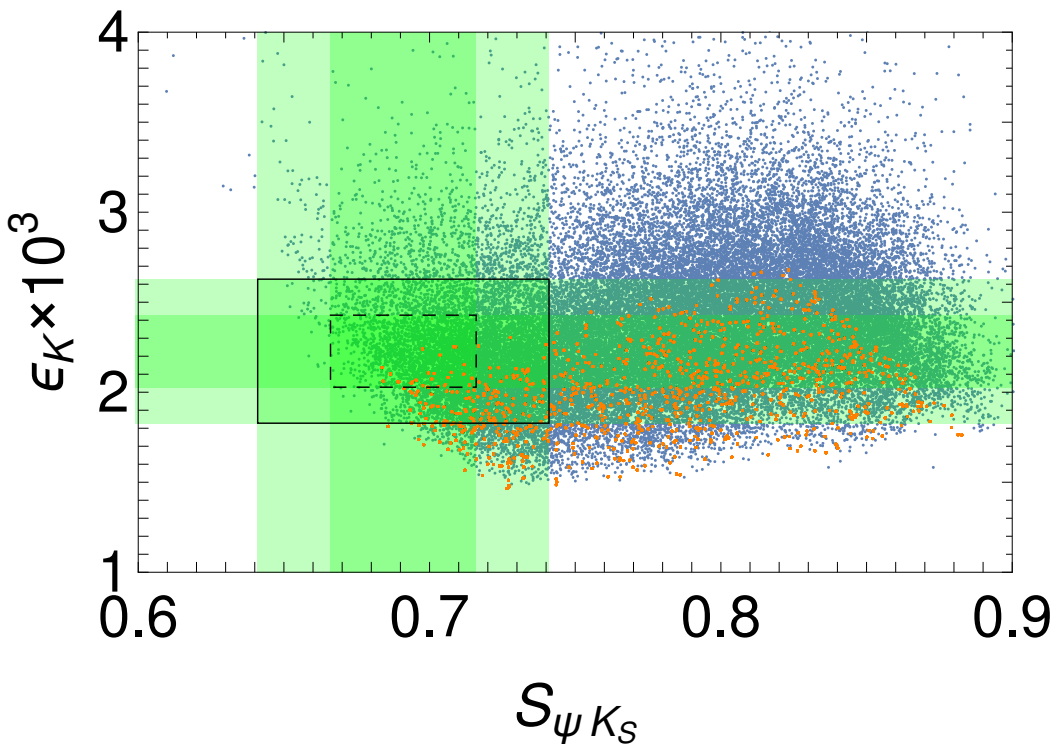

Fig. 215: Correlation between $S_{\psi K_S^0}$ and $\epsilon_K$ in $U(2)^3$. All points satisfy LHC mass constraints, and shaded green regions show bounds at 1 and $2\sigma$. The dashed (solid) rectangles show the region where both bounds are satisfied at $1\sigma$ $(2\sigma)$ using current world average values [88]. Orange points show the prediction without SUSY.

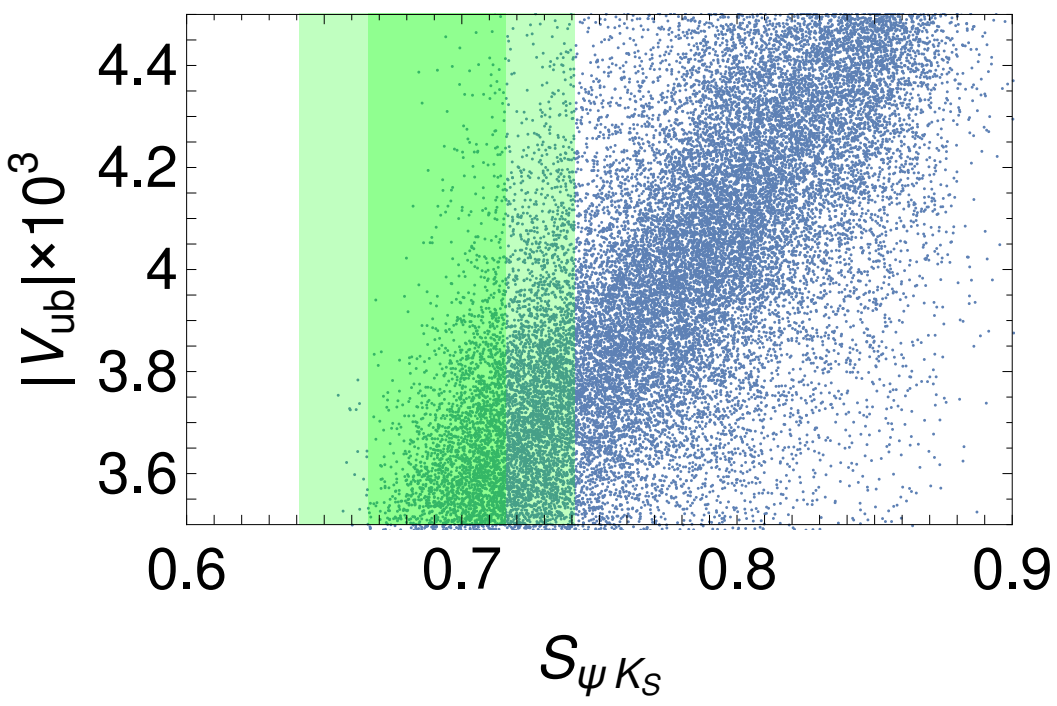

Fig. 216: Correlation between $S_{\psi K_S^0}$ and $|V_{ub}|$ in $U(2)^3$. All points satisfy LHC mass constraints, and have $\epsilon_K$ satisfying its bounds at $2\sigma$. Shaded green regions show bounds at 1 and $2\sigma$ using current world average values [88, 143, 353].

One attains further insight on the situation by selecting those points where the $\epsilon_K$ bounds are satisfied. One can then plot $|V_{ub}|$ as a function of $S_{\psi K_S^0}$, as is done in Fig. 216. The plots show that values of $|V_{ub}|$ close to its exclusive determination are favoured. This is consistent to what we mentioned before. If we take $|V_{ub}|$ close to its inclusive value, we find that the SUSY contribution to $S_{\psi K_S^0}$ must be much larger. This, however, is correlated to the contribution to $\epsilon_K$ through the parameter $x$, which causes the former to exceed its bounds.

Thus, for this framework to solve the flavour tension would imply a future measurement of $S_{\psi K_S^0}$ to favour its current upper range, and a future measurement of $|V_{ub}|$ to approach its exclusive value. For $\Delta F = 1$ decays, one does not find strong distinctions with respect to MFV. However, $CP$-asymmetries in $\Delta F = 1$ decays have been studied in detail in [1933]. For the lepton sector, there exist two possible approaches. The first one is to expand the $U(2)^3$ symmetry into $U(2)^5$ into the lepton sector by analogy, and generate LFV processes directly [1934]. The second approach, which is the one we discuss in detail below, is to modify the symmetry in order to generate neutrino masses and mixing, and then consider how that affects LFV.

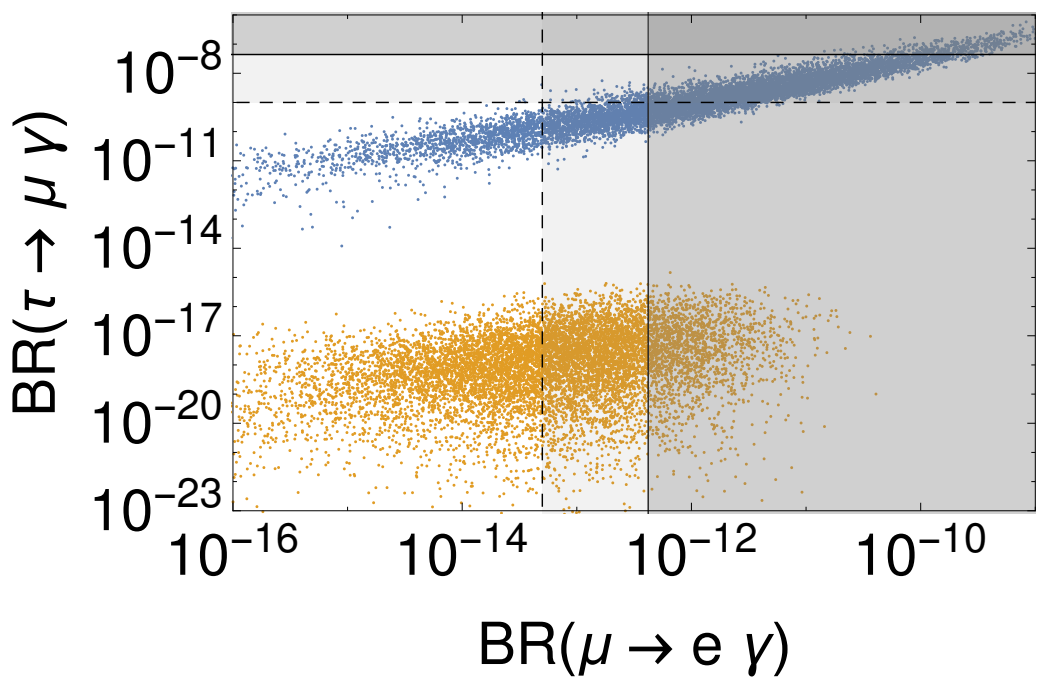

Fig. 217: Correlation between $\mu \to e\gamma$ and $\tau \to \mu\gamma$ in leptonic extension to $U(2)^3$. Blue points correspond to $U(3)^5_{\text{broken}}$ scenario, while yellow points correspond to $U(2)^5_{\text{shifted}}$. Solid lines indicate current constraints [1533, 1948], dashed lines indicate future prospects [1949].

In this second approach, there exist two realisations so far. The first one [1946] starts with a $U(3)^5$ symmetry, which is broken into $U(2)^5$ for the Yukawa couplings and $O(3)$ for the Majorana masses (denoted $U(3)^5_{\text{broken}}$). In this case, the neutrino masses are almost degenerate, such that one would expect an observation of neutrinoless double beta decay very soon. However, for the SUSY masses, one finds it more difficult to obtain a split spectrum, requiring a slight tuning in order to do so.

Another possibility [1947] is to start with $U(2)^5$, but then shift the $U(2)^2$ symmetry in the lepton sector towards the last two generations (denoted $U(2)^5_{\text{shifted}}$). This is achieved by the addition of two new $U(1)$ symmetries for the lepton sector, with their own spurions. Here, one favours a normal ordering for the neutrino masses, with no possibility of observing neutrinoless double beta decay. An important feature of this case is that one would expect the selectron to be the lightest slepton, instead of the stau.

The main SUSY signature of both realisations is LFV decays. As one can see in Fig. 217 both scenarios have comparable $\mu \to e\gamma$ rates, and can be severely constrained by the MEG II upgrade. However, if MEG does observe a signal, there is a small chance to differentiate $U(3)^5_{\text{broken}}$ from $U(2)^5_{\text{shifted}}$ by observing $\tau \to \mu\gamma$ decay. Needless to say, if one observes the latter, and does not see $\mu \to e\gamma$ in MEG II, then both frameworks would be simultaneously excluded. The predicted $\tau \to e\gamma$ branching ratio in both realisations is unfortunately too small to be observed in the near future.

### *17.6. Models with extended gauge sector*

#### *17.6.1.* $\mathbf{Z}'$ *models and modified* $\mathbf{Z}$ *couplings.* *(Contributing author: Lars Hofer)*

The simplest extension of the SM gauge sector is obtained by adding an extra $U(1)'$ symmetry. This modification augments the particle content by an additional gauge boson, $Z'$, coupling to those fermions that carry a non-vanishing $U(1)'$ charge. After $U(1)'$ and EW symmetry breaking, the corresponding part of the Langrangian takes the most general form

$$\begin{aligned}\mathcal{L}_{Z'} \;=\; & \Gamma^L_{uu'}\bar{u}\gamma^\mu P_L u' Z'_\mu + \Gamma^R_{uu'}\bar{u}\gamma^\mu P_R u' Z'_\mu + \Gamma^L_{dd'}\bar{d}\gamma^\mu P_L d' Z'_\mu + \Gamma^R_{dd'}\bar{d}\gamma^\mu P_R d' Z'_\mu + \\ & \Gamma^L_{\ell\ell'}\bar{\ell}\gamma^\mu P_L \ell' Z'_\mu + \Gamma^R_{\ell\ell'}\bar{\ell}\gamma^\mu P_R \ell' Z'_\mu + \Gamma^L_{\nu_\ell\nu_{\ell'}}\bar{\nu}_\ell\gamma^\mu P_L \nu_{\ell'} Z'_\mu + \text{h.c.},\end{aligned} \tag{595}$$

where $u = (u, c, t)$, $d = (d, s, b)$, $\ell = (e, \mu, \tau)$ and $\nu_\ell = (\nu_e, \nu_\mu, \nu_\tau)$ denote the SM fermion fields in the mass eigenbasis. If the scale associated with the $Z'$ is assumed to be well above the one of the electroweak interactions, its couplings to left-handed fermions have to preserve $SU(2)_L$ invariance implying the model-independent relations

$$\Gamma^L_{uu'} = V_{ud}\Gamma^L_{dd'}V^\dagger_{u'd'}, \qquad \Gamma^L_{\ell\ell'} = \Gamma^L_{\nu_\ell\nu_{\ell'}}, \tag{596}$$

with $V$ denoting the CKM matrix.

Any further constraints on the couplings $\Gamma^{L,R}_{ij}$ depend on the $U(1)'$ charges assigned to the SM fermions and on a potential embedding of the $U(1)'$ model in a more fundamental theory (see e.g. the models described in the following sections). Of interest for quark flavour physics are $Z'$ scenarios featuring flavour off-diagonal quark-couplings $\Gamma^{L,R}_{bs}$, $\Gamma^{L,R}_{bd}$ or $\Gamma^{L,R}_{uc}$. Such couplings can be obtained at tree-level with a family non-universal assignment of $U(1)'$ charges, e.g. [1950, 1951], requiring an extension of the Higgs sector in order to comply with the experimentally observed fermion masses and mixings. Alternatively, flavour off-diagonal couplings can be generated as effective couplings in an underlying more fundamental theory. A typical mechanism to generate these effective couplings involves heavy vector-like quarks that are charged under the $U(1)'$ symmetry and that mix with the SM fermions [599, 648, 1952, 1953].

In the lepton sector the most popular class of $Z'$ models is based on gauging $L_\tau - L_\mu$ lepton number [599, 648, 1950, 1952, 1954]. Such models are anomaly free [1955–1957] and phenomenologically appealing as they lead to a PMNS matrix that provides a good tree-level approximation for the measured pattern of neutrino mixing angles [1958–1960]. The vanishing coupling of the $Z'$ to electrons further allows to explain the present tensions in the LHCb measurement [391] of the LFUV observable $R_K = \mathrm{Br}(B \to K\mu^+\mu^-)/\mathrm{Br}(B \to Ke^+e^-)$ and helps to avoid LEP bounds on the $Z'$ mass $M_{Z'}$.

Tree-level exchange of the $Z'$ boson contributes to various $\Delta F = 1$ and $\Delta F = 2$ FCNC processes. The contributions to the respective Wilson coefficients are quadratic in the reduced $Z'$ couplings,

$$\widetilde{\Gamma}^{L,R}_{ij} \equiv \Gamma^{L,R}_{ij}/M_{Z'}, \tag{597}$$

and the absence of a loop suppression compared to the SM contribution boosts the sensitivity of FCNC observables to high $Z'$ masses beyond the reach of direct searches. In the following we discuss in a model-independent way, based on the general Lagragian (595), the $Z'$ phenomenology in $B$ physics.

*$B_{d,s} - \bar{B}_{d,s}$ mixing.* The observables in $B_q - \bar{B}_q$ mixing ($q = d, s$) probe the reduced $Z'$ couplings $\widetilde{\Gamma}^L_{bq}$ and $\widetilde{\Gamma}^R_{bq}$. They constrain all $B_{d,s}$ decay modes discussed in the following as a non-vanishing $Z'$ contribution to any of them necessarily also implies a contribution to $B_q - \bar{B}_q$ mixing. In the absence of an additional $CP$ phase, *i.e.* for $\arg(\Gamma^{L,R}_{bq}) = \arg(V^*_{tq}V_{tb})$, the $Z'$ contribution leads to an enhancement of $\Delta M_{B_q}$ for purely left-handed, purely right-handed or axial $Z'bs$ couplings, while it leads to a decrease for a vectorial $Z'bs$ coupling [1912, 1961].

*(Semi-)leptonic $B$ decays.* (Semi-)leptonic $B$ decays are the natural place to search for $Z'$ bosons in quark flavour physics due to their partial protection from polluting QCD effects. They probe the four products of couplings $\widetilde{\Gamma}^{L,R}_{bq}\widetilde{\Gamma}^{L,R}_{\ell\ell}$ generating the four Wilson coefficients $C^{(\prime)}_{9,10}$. Since only three out of the four products $\widetilde{\Gamma}^{L,R}_{bq}\widetilde{\Gamma}^{L,R}_{\ell\ell}$ are independent, the relation

$$C_9 \cdot C'_{10} \;=\; C'_9 \cdot C_{10} \tag{598}$$

is valid in models with a single $Z'$ boson, with the ratio $C'_9/C_9 = C'_{10}/C_{10}$ being free from the leptonic couplings and fixed from $B_q - \bar{B}_q$ mixing [1962].

Global fits to current $b \to s\ell^+\ell^-$ data including (among other modes) $B_s \to \mu^+\mu^-$, $B \to K^{(*)}\ell^+\ell^-$, $B_s \to \phi\mu^+\mu^-$, show some significant tensions pointing towards New Physics in these coefficients, in particular a negative contribution to the coefficient $C_9$ [497, 524, 612, 647, 1963]. A $Z'$ boson with a left-handed coupling to quarks and a vectorial coupling to leptons precisely generates this coefficient, whereas a $Z'$ boson with purely left-handed couplings to both quarks and leptons generates the pattern $C_9 = -C_{10}$. A solution of the present tensions in exclusive semi-leptonic $B_{(s)}$ decays via the latter scenario implies a reduced branching ratio for the purely leptonic mode $B_s \to \mu^+\mu^-$, consistent with current LHC data. An unambiguous discrimination between a $Z'$ model and non-perturbative QCD contributions as origin of the observed tensions requires an increased resolution in the invariant dilepton mass in the above-mentioned exclusive semi-leptonic channels, as well as the exploitation of neutrino modes like $B \to K^{(*)}\nu\bar{\nu}$ and additional more exotic modes like the baryonic $\Lambda_b \to \Lambda\mu^+\mu^-$[162, 1964, 1965] or inclusive $B \to X_s\ell^+\ell^-$.

If the $Z'$ mediates $b \to s$ transitions, it can be expected that it also generates $b \to d$ transitions. In a scenario with Minimal Flavour Violation, for instance, the corresponding couplings would be related as $|\Gamma^{L,R}_{bd}/\Gamma^{L,R}_{bs}| = |V_{td}/V_{ts}|$. Therefore also the corresponding $b \to d\ell\ell$ modes $B_d \to \ell\ell$, $B \to \pi\ell\ell$ and $B \to \rho\ell\ell$ need to be studied in detail. Belle II should be able to provide complementary input on $b \to d$ transitions through radiative decay measurements.

*LFUV and LFV $B_{d,s}$ decays.* $Z'$ models based on gauged $L_\tau - L_\mu$ lepton number introduce lepton-flavour universality violation (LFUV), a possibility suggested also by the current LHCb data [391] on the ratio $R_K = \mathrm{Br}(B \to K\mu^+\mu^-)/\mathrm{Br}(B \to Ke^+e^-)$. The hypothesis of LFUV can be tested by measuring a certain $b \to q\ell^+\ell^-$ decay for different lepton families $\ell = e, \mu, \tau$ and considering the ratio of the respective branching ratios, a theoretically clean observable due to the cancellation of hadronic uncertainties. For decays into vector mesons also more elaborate observables can be constructed from the full angular distribution [600].

It has been proposed to search for lepton-flavour violating (LFV) $B$ decay modes as well [623]. In $L_\tau - L_\mu$ models, for example, the symmetry breaking, which is needed for a realistic neutrino phenomenology, can induce such decays. Combined bounds from $B_s - \bar{B}_s$ mixing, $\tau \to 3\mu$ and $\tau \to \mu\nu\bar{\nu}$ constrain branching ratios of $b \to s\tau^\pm\mu^\mp$ modes to $\mathcal{O}(10^{-8}) - \mathcal{O}(10^{-6})$, depending on the amount of fine-tuning allowed in $B_s - \bar{B}_s$ mixing [1912, 1966]. For $b \to s\mu^\pm e^\mp$ sizeable branching ratios (up to $\mathcal{O}(10^{-7})$ if a certain fine-tuning in $B_s - \bar{B}_s$ is permitted) are only possible in a region disfavoured by current $b \to s\mu^+\mu^-$ data.

*Hadronic $B_{(s)}$ decays.* To fully scrutinise the $Z'$ model it is not sufficient to examine (semi-)leptonic $B$ decays; also purely hadronic decays $B_{(s)} \to M_1M_2$ of the $B_{(s)}$ meson into two light mesons $M_1$, $M_1$, mediated by quark-level transitions $b \to q\bar{q}'q'$ ($q = d, s$, $q' = u, d, s, c$), should be explored. Governed by the products of couplings $\widetilde{\Gamma}^{L,R}_{bq}\widetilde{\Gamma}^{L,R}_{q'q'}$, these decays are in general independent of the (semi-)leptonic ones and open a portal to probe the $Z'$ boson even in leptophobic models.

While the plethora of hadronic channels allows for over-constraining measurements, the fact that these modes are typically dominated by QCD penguin topologies hampers the sensitivity to high-scale New Physics. An exception occurs for isospin-violating $Z'$ models, *i.e.* for scenarios in which the $Z'$ couples in a different way to $u$- and $d$-quarks, leading to observable effects that cannot be mimicked by the isospin-conserving QCD penguins. As a consequence of the $SU(2)_L$ relation (596), isospin violation is CKM-suppressed in the left-handed $Z'$ couplings $\Gamma^L_{uu} \approx \Gamma^L_{dd}$, but can be introduced via the right-handed couplings $\Gamma^R_{uu} \neq \Gamma^R_{dd}$.

The two $B_s$ decay modes $B_s \to \phi\rho^0$ and $B_s \to \phi\pi^0$ are purely $\Delta I = 1$ decays and thus golden modes to search for an isospin-violating $Z'$ boson [790]. The absence of a QCD penguin amplitude renders their branching ratios particularly small within the SM, $\text{Br}(B_s \to \phi\rho^0) = 4.4^{+2.7}_{-0.7} \times 10^{-7}$ and $\text{Br}(B_s \to \phi\pi^0) = 1.6^{+1.1}_{-0.3} \times 10^{-7}$.

Also the decay $B \to K\pi$ is sensitive to isospin-violating $Z'$ models. Its transition amplitude decomposes into an $\Delta I = 0$ part (dominated by QCD penguins) and an $\Delta I = 1$ part (free from QCD penguins). By combining the four different modes $B^- \to \bar{K}^0\pi^-$, $B^- \to K^-\pi^0$, $\bar{B}^0 \to K^-\pi^+$ and $\bar{B}^0 \to \bar{K}^0\pi^0$ it is possible to construct observables that project out the $\Delta I = 1$ component. In the last decade some discrepancies in such observables lead to speculations about a "$B \to K\pi$ puzzle" [1967–1970], but in the meantime measurements have fluctuated towards the SM predictions reducing the "$B \to K\pi$ puzzle" to a $\gtrsim 2\sigma$ tension [1971] in the difference of direct $CP$ asymmetries $\Delta A_{CP} = A_{CP}(B^- \to K^-\pi^0) - A_{CP}(\bar{B}^0 \to K^-\pi^+)$.

*Modified Z couplings.* A scenario where high-scale new physics generates effective FCNC couplings for the SM $Z$ boson is phenomenologically similar to a $Z'$ model and can thus be probed by the same processes. Differences compared to the $Z'$ case arise from the fact that the $Z$ boson mass as well as its flavour-diagonal lepton couplings are known precisely from the LEP measurements. This significantly lowers the potential for observable departures from SM predictions in $B$ decays, in particular in $b \to s$ transition, where the observed $b \to s\ell^+\ell^-$ anomalies cannot be accommodated. In $b \to d$ transitions measurable imprints are still possible [1961, 1962].

*17.6.2. Gauged flavour models.* *(Contributing author: Jernej F. Kamenik)*

The idea of assuming horizontal (flavour) symmetries to be true (gauged) symmetries of nature has a long history (see Ref. [1972] for a review). Unfortunately, contrary to the (global) symmetry arguments underlying MFV, such an assumption is itself not enough to suppress flavour violation below the experimental bounds when the flavour symmetry is broken at low scales. Namely, in this case the associated flavour gauge bosons can mediate dangerous FCNCs and their masses generally must be well above the TeV scale. In the minimal flavour breaking scenarios, where the masses of the gauge bosons are proportional to the SM Yukawa couplings (as the only sources of flavour breaking), they generate tree-level four-fermion operators proportional to inverse powers of the SM Yukawa couplings, enhancing FCNC among the first generations, and resulting in severe constraints from FCNC and $CP$ violation observables in the kaon sector.

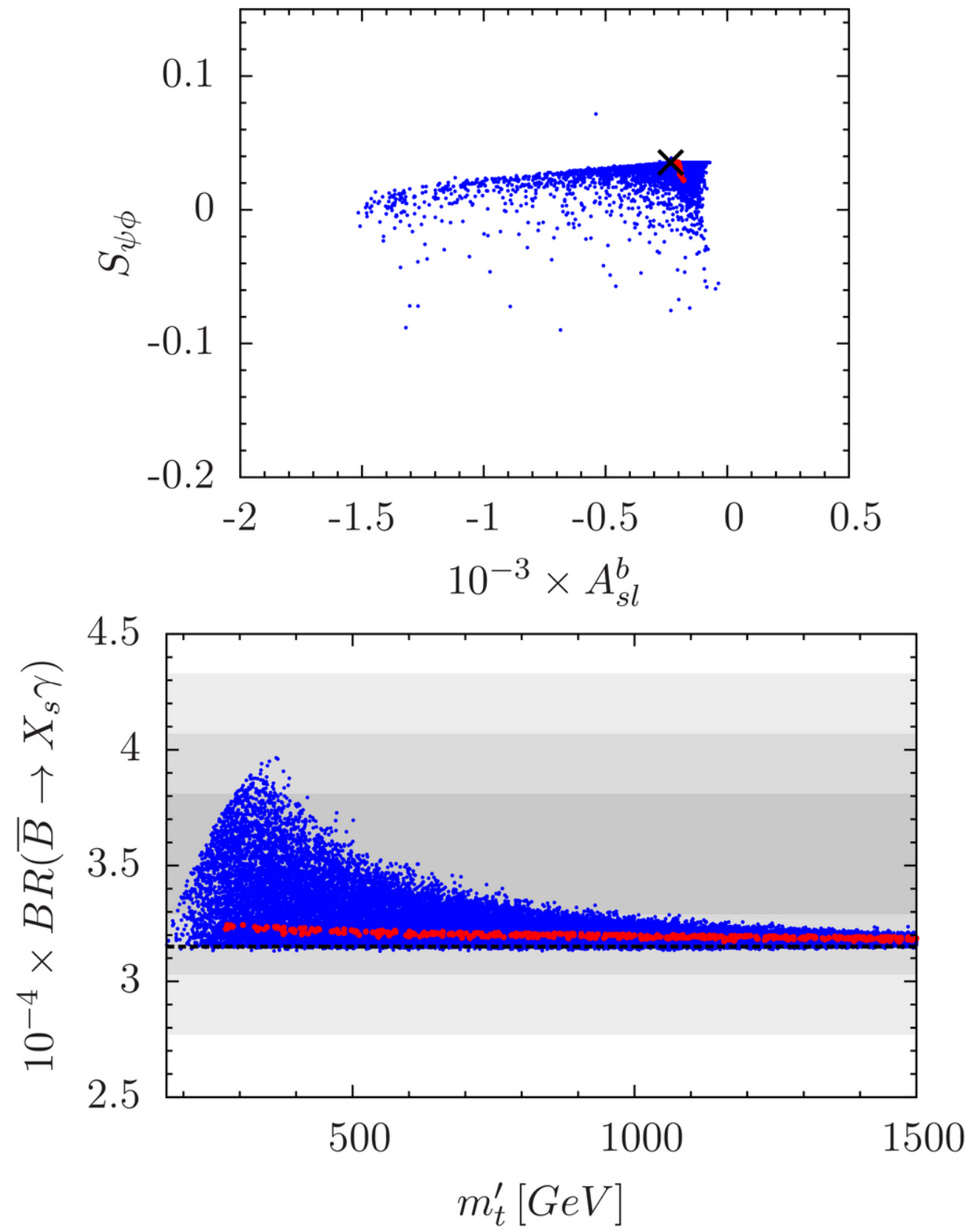

Fig. 218: Correlation plot of $S_{\psi\phi}$ and the $b$ semileptonic $CP$ asymmetry $A_{sl}^b$ on the left and $BR(\bar{B} \to X_s\gamma)$ and the mass of the lightest new fermionic resonance ($m_t'$) on the right in the minimal gauged $SU(3)^3$ flavour model (taken from Ref. [1975]). Grey regions refer to the present experimental error ranges. The big cross mark refers to the SM values reported in Ref. [1975]. In red the points for which the current experimental tension between $\Delta m_{B_{d,s}}$, $S_{\psi K_S^0}$ and $\epsilon_K$ is resolved, in blue all others.

A way out, exploited in Ref. [1973], is for the fields breaking the flavour symmetry to be instead proportional to the inverse of the SM Yukawa couplings. Then, the effective operators generated by integrating out the flavour gauge bosons will be roughly proportional to positive powers of the Yukawa couplings, suppressing flavour violating effects for the light generations, much like in MFV models. The spectrum of the extra flavour states will thus present an inverted hierarchy, with states associated to the third generation much lighter than those associated to the first two.

Another particularity of gauged flavour symmetry models is that in general extra flavourful fermions have to be added to cancel flavour gauge anomalies. Such fermions are also welcome as they can make the dynamical SM Yukawa terms arise from a renormalisable Lagrangian. In the quark sector, the smallest set of fermions cancelling all anomalies in the $SU(3)^3$ case was found in Ref. [1973] and leads automatically to the inverted hierarchy structure mentioned above (see also Ref. [1974] for examples of $SU(2)^3$ gauged models). The SM fermion masses arise via a see-saw like mechanism, after integrating out the extra fermions.

The strongest constraints on this kind of models do not necessarily come from flavour violating observables but also from electroweak precision tests (EWPT) and direct searches for new particles, opening the possibility for direct discoveries of flavour dynamics at the LHC [1973]. The lightest new states are the top partners in the quark sector and a few flavour gauge bosons that behave as flavour non-universal (but diagonal) $Z'$ (see also Sec. 17.6.1). Depending on the flavour gauge group a few flavour gauge bosons could lie in the TeV range. Most of the spectrum however is much heavier than a TeV and can only be probed through precision flavour observables.

An extensive analysis of $\Delta F = 2$ observables and of $B \to X_s\gamma$ in the minimal gauged quark $SU(3)^3$ flavour model was studied in detail in Ref. [1975]. The model allows in principle for significant deviations from the SM predictions for $\epsilon_K$ , $\Delta m_{B_{d,s}}$ , mixing induced $CP$ -asymmetries $S_{\psi K^0_S}$ and $S_{\psi\phi}$ and $B \to X_s\gamma$ decay. Some predicted correlations among Belle II relevant observables are shown in Fig. 218.

The gauging of the lepton flavour group was considered in Ref. [1976]. In contrast to the quark case, the unknown nature of neutrino masses allows for several possibilities for constructing a consistent model. In particular, the maximal lepton flavour symmetry group is $U(3)^3$ for the case of purely Dirac neutrinos, and $U(3)^2 \times O(3)$ for the Majorana case. In this later case, which results in a type-I inverse see-saw scenario, $\mu \to eee$ is at present generically the most sensitive flavour non-conserving channel. However, the model also predicts potentially observable effects in LFV tau decays (see Fig. 219).

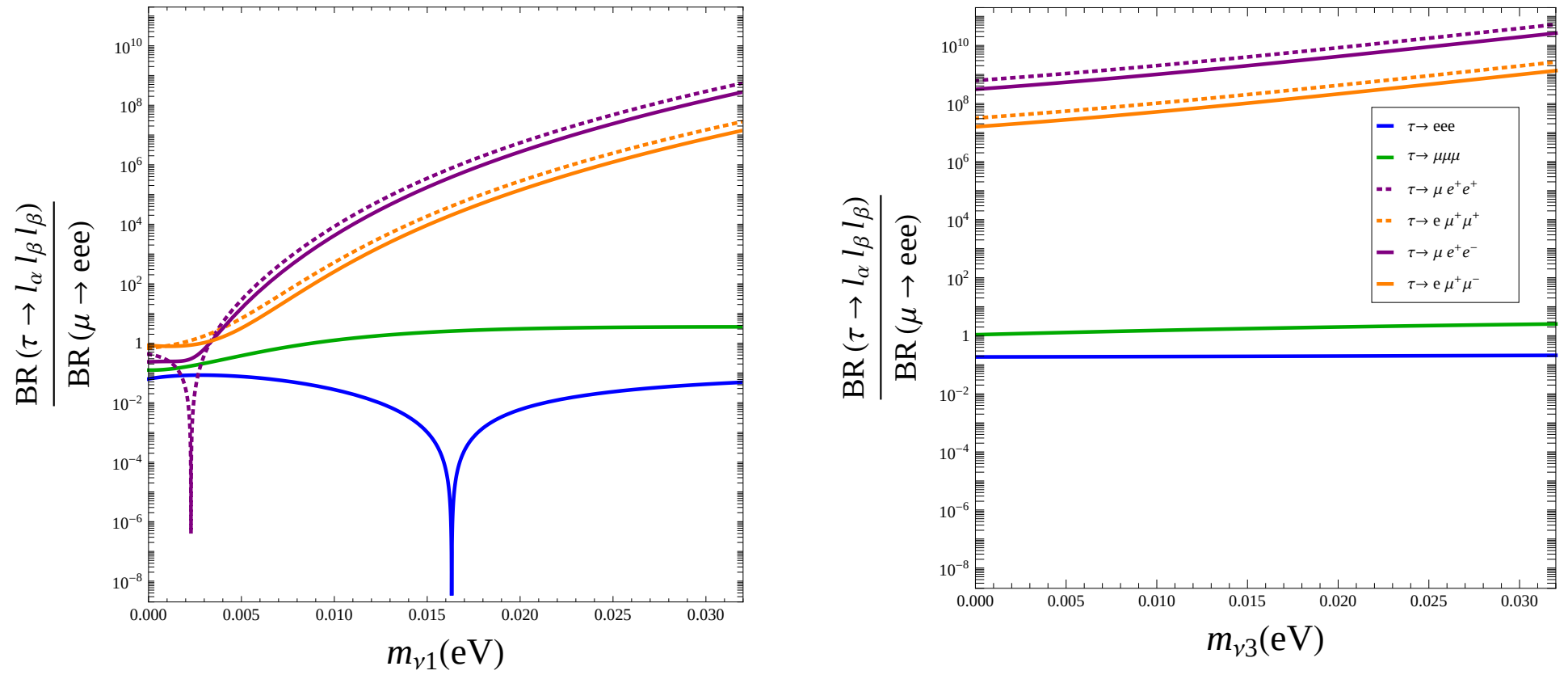

Fig. 219: Predictions of the gauged-flavour type-I see-saw scenario in a $CP$-even case (taken from Ref. [1976], to which the reader is referred for details): branching ratios for the different lepton rare decays over that for $\mu \to eee$ as functions of the lightest neutrino mass, for neutrino normal ordering ($m_{\nu_1}$, left) and inverted ordering ($m_{\nu_3}$, right).

### *17.6.3. 3-3-1 model.* *(Contributing authors: Andrzej Buras and Fulvia De Fazio)*

The name 331 generically indicates a set of models based on the gauge group $SU(3)_C \times SU(3)_L \times U(1)_X$ [1977, 1978]; this group is at first spontaneously broken to the SM group $SU(3)_C \times SU(2)_L \times U(1)_Y$ and then to $SU(3)_C \times U(1)_Q$. The enlargement of the gauge group with respect to SM has two interesting consequences. The first one is that the number of generations must necessarily be equal to the number of colours, if one requires anomaly cancellation as well as asymptotic freedom of QCD; this might be viewed as an explanation for the existence of three generations. Moreover, quark generations should have different transformation properties under the action of $SU(3)_L$. In particular, two quark generations should transform as triplets, one as an antitriplet. Identifying the latter with the third generation, this difference could be at the origin of the large top quark mass. This choice imposes that the leptons should transform as antitriplets. However, one could also choose to exchange the role of triplets and antitriplets, provided that the number of triplets equals that of antitriplets, in order to fulfil the anomaly cancellation requirement. As a consequence, different variants of the model are obtained corresponding to the way one fixes the fermion representations.

The following relation holds among some of the generators of the group:

$$Q = T_3 + \beta T_8 + X, \tag{599}$$

where $Q$ indicates the electric charge, $T_3$ and $T_8$ are the two diagonal generators of the $SU(3)_L$ and $X$ is the generator of $U(1)_X$. $\beta$ is a parameter that, together with the choice for the fermion representations, defines a specific version of the model.

Several new particles are predicted to exist in 331 models. Known SM fermions fill the two upper components of the (anti)triplets; the third one is in general a new heavy fermion with electric charges depending on $\beta$. (An exception is the model having $\beta = \sqrt{3}$, called minimal 331 model, where only new heavy quarks are present but no new leptons.) The Higgs system is also enlarged.

Five new gauge bosons exist due to the extension of the SM gauge group $SU(2)_L$ to $SU(3)_L$. A new neutral boson ($Z'$) is always present, together with other four that might be charged depending on the selected variant of the model. An important difference with respect to the SM is the existence of tree level flavour changing neutral currents (FCNC) mediated by $Z'$. These arise only in the quark sector, due to the universality of the coupling of the $Z'$ to leptons that guarantees that no FCNC show up in this case. Moreover, new FCNC are purely left-handed since universality is also realised in the $Z'$ couplings to right-handed quarks. Such tree level transitions turn out to be no larger than the corresponding loop induced SM contribution, due to the smallness of the relevant couplings. This is a very appealing feature of this model, since $Z'$ could be responsible for the anomalies recently emerged in the flavour sector.

As in the SM, quark mass eigenstates are obtained upon rotation of flavour eigenstates through two unitary matrices $V_L$ (for down-type quarks) and $U_L$ (for up-type quarks). In analogy with the SM case, the relation

$$V_{\rm CKM} = U_L^\dagger V_L \tag{600}$$

holds. However, while in the SM the two rotation matrices never appear individually and $V_{\rm CKM}$ enters only in charged current interactions, this is not the case in 331 models and both $U_L$ and $V_L$ enter in tree-level FCNCs mediated by $Z'$ in the up-quark and down-quark sector, respectively. Due to the relation (600), one can choose to write $U_L$ in terms of $V_L$ and $V_{CKM}$.

A suitable parametrisation for $V_L$ is

$$V_L = \begin{pmatrix} \tilde{c}_{12}\tilde{c}_{13} & \tilde{s}_{12}\tilde{c}_{23}e^{i\delta_3} - \tilde{c}_{12}\tilde{s}_{13}\tilde{s}_{23}e^{i(\delta_1-\delta_2)} & \tilde{c}_{12}\tilde{c}_{23}\tilde{s}_{13}e^{i\delta_1} + \tilde{s}_{12}\tilde{s}_{23}e^{i(\delta_2+\delta_3)} \\ -\tilde{c}_{13}\tilde{s}_{12}e^{-i\delta_3} & \tilde{c}_{12}\tilde{c}_{23} + \tilde{s}_{12}\tilde{s}_{13}\tilde{s}_{23}e^{i(\delta_1-\delta_2-\delta_3)} & -\tilde{s}_{12}\tilde{s}_{13}\tilde{c}_{23}e^{i(\delta_1-\delta_3)} - \tilde{c}_{12}\tilde{s}_{23}e^{i\delta_2} \\ -\tilde{s}_{13}e^{-i\delta_1} & -\tilde{c}_{13}\tilde{s}_{23}e^{-i\delta_2} & \tilde{c}_{13}\tilde{c}_{23} \end{pmatrix}. \tag{601}$$

The interaction Lagrangian describing $Z'$ couplings to down-quarks in 331 models can be written as follows:

$$i\mathcal{L}_L(Z') = i\left[\Delta_L^{sd}(Z')(\bar{s}\gamma^\mu P_L d) + \Delta_L^{bd}(Z')(\bar{b}\gamma^\mu P_L d) + \Delta_L^{bs}(Z')(\bar{b}\gamma^\mu P_L s)\right] Z'_\mu \tag{602}$$

with the first upper index denoting outgoing quark and the second incoming one. As a consequence

$$\Delta_L^{ji}(Z') = (\Delta_L^{ij}(Z'))^*. \tag{603}$$

With this parametrisation, the $Z'$ couplings to quarks, for the three meson systems, $K$, $B_d$ and $B_s$

$$\Delta_L^{sd}(Z'), \qquad \Delta_L^{bd}(Z'), \qquad \Delta_L^{bs}(Z') \tag{604}$$

being proportional to $v_{32}^* v_{31}$, $v_{33}^* v_{31}$ and $v_{33}^* v_{32}$, respectively, depend only on four new parameters (explicit formulae can be found in Ref. [1979]):

$$\tilde{s}_{13}, \quad \tilde{s}_{23}, \quad \delta_1, \quad \delta_2\,. \tag{605}$$

Here $\tilde{s}_{13}$ and $\tilde{s}_{23}$ are positive definite and $\delta_i$ vary in the range $[0, 2\pi]$. Therefore for fixed $M_{Z'}$ and $\beta$, the $Z'$ contributions to all processes discussed in the following depend only on these parameters and on their size, implying very strong correlations between NP contributions to various observables. Indeed, as seen in (601) the $B_d$ system involves only the parameters $\tilde{s}_{13}$ and $\delta_1$ while the $B_s$ system depends on $\tilde{s}_{23}$ and $\delta_2$. Furthermore, stringent correlations between observables in $B_{d,s}$ sectors and in the kaon sector are found since kaon physics depends on $\tilde{s}_{13}$, $\tilde{s}_{23}$ and $\delta_2 - \delta_1$. Additional non-negligible contributions come from tree-level $Z$ exchanges generated by the the $Z - Z'$ mixing that depends on an additional parameter $\tan\bar{\beta}$ [1980]. The fact that in 331 models deviations from SM predictions are mainly related to $Z'$ exchanges implies that NP effects in these models are likely to come from scales beyond the reach of the LHC. On the other hand they can be suitably investigated at Belle II, where the effects of a virtual $Z'$ can be detected even if its mass is too high to be detected at the LHC.

Extensive recent flavour analyses in these models can be found in Refs. [1979–1983]. References to earlier analysis of flavour physics in 331 models can be found there and in Refs. [1984, 1985]. In particular in Ref. [1980] 24 different models corresponding to four values of $\beta$, three values of $\tan\bar{\beta}$ and two fermion representations have been considered. With the help of electroweak precision data it was possible to reduce the number of these models to seven.

The most recent updated analyses in [1982, 1983] concentrated on the ratio $\varepsilon'/\varepsilon$ and its correlation with $\varepsilon_K$ and B-physics observables such as $\Delta M_{s,d}$, $B_s \to \mu^+\mu^-$ and the Wilson coefficient $C_9$. They were motivated by the anomalies in $\varepsilon'/\varepsilon$ [178–180, 1986], tension between $\varepsilon_K$ and $\Delta M_{s,d}$ within the SM [1928] implied by the recent lattice data [151] and in the case of $C_9$ by the LHCb anomalies in $B \to K^*(K)\mu^+\mu^-$ summarised in [524, 612]. We briefly recall the main results of these two papers putting the emphasis on the last analysis in [1983] which could take into account new lattice QCD results from Fermilab Lattice and MILC Collaborations [151] on $B^0_{s,d} - \bar{B}^0_{s,d}$ hadronic matrix elements.

The new analyses in [1982, 1983] show that the requirement of an enhancement of $\varepsilon'/\varepsilon$ has a significant impact on other flavour observables. Moreover, in [1983] it has also been shown that the results are rather sensitive to the value of $|V_{cb}|$, as has been illustrated by choosing two values: $|V_{cb}| = 0.040$ and $|V_{cb}| = 0.042$. There is also some sensitivity to $V_{ub}$ which is less precisely known than $V_{cb}$. In this context an improved determination of $V_{cb}$ and $V_{ub}$ at Belle II will allow higher precision with which predictions in 331 models can be made and to choose between various scenarios for these two CKM elements discussed below.

The main findings of [1982, 1983] for $M_{Z'} = 3$ TeV are as follows:

- Among seven 331 models selected in [1980] on the basis of the electroweak precision study only three (M8, M9, M16) can provide for both choices of $|V_{cb}|$, significant shift of $\varepsilon'/\varepsilon$, even though not larger than $6 \times 10^{-4}$.
- The tensions between $\Delta M_{s,d}$ and $\varepsilon_K$ pointed out in [1928] can be removed in these models (M8, M9, M16) for both values of $|V_{cb}|$.
- Two of them (M8 and M9) can simultaneously suppress $B_s \to \mu^+\mu^-$ by at most 10% and 20% for $|V_{cb}| = 0.042$ and $|V_{cb}| = 0.040$, respectively. This can still bring the theory within $1\sigma$ range of the combined result from CMS and LHCb and for $|V_{cb}| = 0.040$ one can even reach the present central experimental value of this rate. On the other hand the maximal deviations from SM in the Wilson coefficient $C_9$ are $C_9^{\rm NP} = -0.1$ and $C_9^{\rm NP} = -0.2$ for these two $|V_{cb}|$ values, respectively. Due to this moderate shift, these models do not really help in the case of $B_d \to K^*\mu^+\mu^-$ anomalies that require deviations as high as $C_9^{\rm NP} = -1.0$ [524, 612].
- In M16 the situation is reversed. It is possible to reduce the rate for $B_s \to \mu^+\mu^-$ for $M_{Z'} = 3$ TeV for the two $|V_{cb}|$ values by at most 3% and 10%, respectively but with the corresponding values $C_9^{\rm NP} = -0.3$ and $-0.5$ the anomaly in $B_d \to K^*\mu^+\mu^-$ can be significantly reduced.
- The maximal shifts in $\varepsilon'/\varepsilon$ decrease fast with increasing $M_{Z'}$ in the case of $|V_{cb}| = 0.042$ but are practically unchanged for $M_{Z'} = 10$ TeV when $|V_{cb}| = 0.040$ is used.
- On the other hand for larger values of $M_{Z'}$ the effects in $B_s \to \mu^+\mu^-$ and $B_d \to K^*\mu^+\mu^-$ are much smaller. NP effects in rare $K$ decays and $B \to K(K^*)\nu\bar{\nu}$ remain small in all 331 models even for $M_{Z'}$ of a few TeV. This could be challenged by NA62, KOTO and Belle II experiments in this decade.

We show these correlations for $M_{Z'} = 3$ TeV and $|V_{cb}| = 0.040$ in Fig. 220. In this figure we plot the shift $\Delta(\varepsilon'/\varepsilon)$ versus $\bar{B}(B_s \to \mu^+\mu^-)$ (left plots) or versus the NP contribution $Re\left[C_9^{\rm NP}\right]$ (right plots). Since recent data would favour a suppression of the former as well as a large negative shift in the latter, on the basis of this figure we conclude that the first requirement would select M8 (upper plots) while the second would favour M16 (lower ones). Therefore, a substantial improvement in the experimental measurements of the angular observables in $B \to K^*\mu^+\mu^-$, expected from Belle II, could select which one among the 331 models has a chance to survive. This in turn means that the size of the deviations in $B_s \to \mu^+\mu^-$ and in $\varepsilon'/\varepsilon$ can be assessed and contrasted with data.

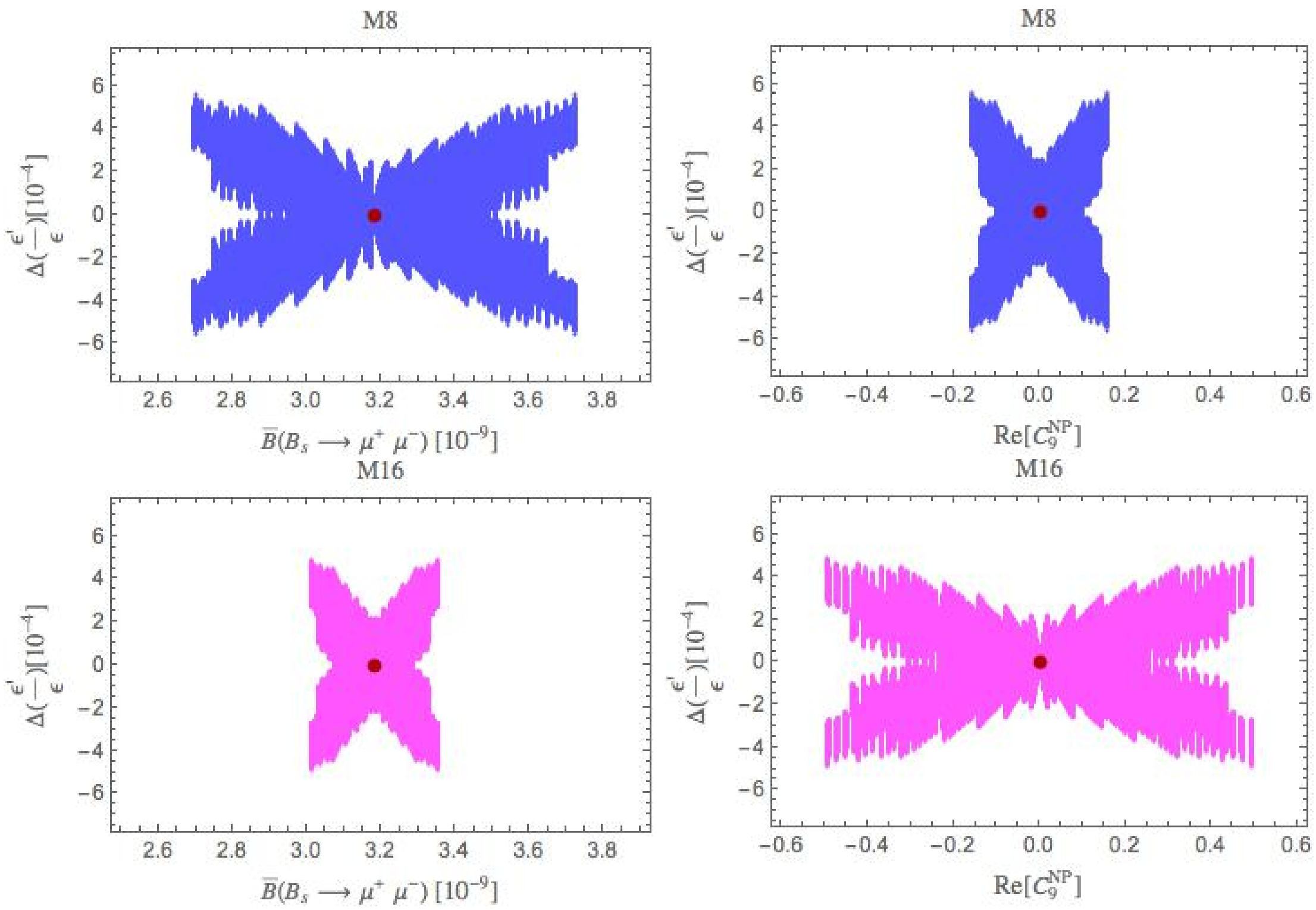

Fig. 220: *Correlations of $\Delta(\varepsilon'/\varepsilon)$ with $B_s \to \mu^+\mu^-$ (left panels) and with $C_9^{\rm NP}$ (right panels) for M8 and M16. Red dots represent central SM values. $M_{Z'} = 3$ TeV and $|V_{cb}| = 0.040$.*

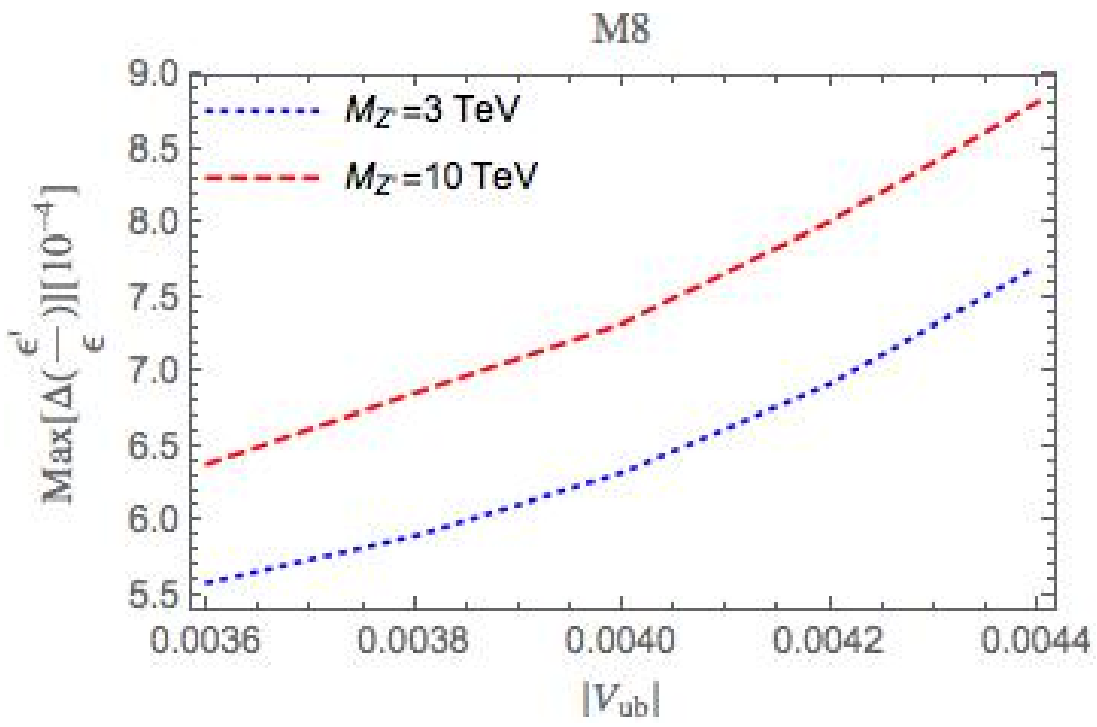

Fig. 221: *Maximal values of $\Delta(\varepsilon'/\varepsilon)$ for $|V_{cb}| = 0.040$ as function of $|V_{ub}|$ for $M_{Z'} = 3$ TeV and $M_{Z'} = 10$ TeV.*

All these statements are valid for $|V_{ub}| = 0.0036$ as favoured by the exclusive decays. For larger values of $|V_{ub}|$ the maximal shifts in $\varepsilon'/\varepsilon$ are larger. We illustrate this in Fig. 221 where we show these shifts as functions of $|V_{ub}|$ for $|V_{cb}| = 0.040$ and two values of $M_{Z'}$.

The main message from [1982, 1983] is that NP contributions in 331 models can simultaneously solve $\Delta F = 2$ tensions, enhance $\varepsilon'/\varepsilon$ and suppress either the rate for $B_s \to \mu^+\mu^-$ or the $C_9$ Wilson coefficient without any significant NP effects on $K^+ \to \pi^+\nu\bar{\nu}$ and $K_L \to \pi^0\nu\bar{\nu}$ and $b \to s\nu\bar{\nu}$ transitions. While sizeable NP effects in $\Delta F = 2$ observables and $\varepsilon'/\varepsilon$ can persist for $M_{Z'}$ outside the reach of the LHC, such effects in $B_s \to \mu^+\mu^-$ will only be detectable provided $Z'$ will be discovered soon.

Let us finally mention that, even though we have stressed that deviations from SM predictions in 331 models are mainly expected due to the existence of FCNC mediated by $Z'$, it is possible that also the other new gauge bosons present in these models can lead to interesting NP effects. Indeed, the gauge bosons denoted by $V^{\pm Q_V}$ and $Y^{\pm Q_Y}$ have electric charge depending on $\beta$ and lepton number $L = \mp 2$, but carry no lepton generation number so that the lepton generation number can be violated due to such new gauge boson mediation. Recent studies of the lepton sector in this scenario can be found in Refs. [1987, 1988].

*17.6.4. Left-right symmetry models.* *(Contributing author: Monika Blanke)*

In left-right symmetric models [1989–1992] the Standard Model gauge group is extended by an additional $SU(2)_R$ factor under which the right-handed quarks and leptons transform as doublets. Parity can hence be restored at high scales, and the model contains right-handed neutrinos, thereby giving rise to non-vanishing neutrino masses.

Despite extensive searches, the $W'$ and $Z'$ gauge bosons associated with $SU(2)_R$ have not yet been observed and the current direct limits from the LHC reach up to about 2.5 TeV [1993–1995]. A significant improvement is expected from the growing 13 TeV data set. Indirect constraints on the right-handed scale arise from electroweak precision constraints [1996] and from flavour violating decays [1997–2000].

In analogy to the SM, the $W'$ gauge boson mediates right-handed charged currents. The coupling strength of right-handed quarks is given by the unitary mixing matrix $V_R$. Unlike for the CKM matrix, all six complex phases of this matrix are physical.

The presence of right-handed flavour changing charged current interactions can be tested in semileptonic decays [2001–2004]. Processes which are used to determine the elements of the CKM matrix from tree level processes are sensitive to right-handed contributions. The crucial ingredient to identify right-handed contributions is their non-universal effect – different decays sensitive to the same CKM element are affected by right-handed charged currents in a different manner, manifesting itself in discrepancies between the various determinations of CKM elements. The current tensions between inclusive and exclusive determinations of $|V_{ub}|$ and $|V_{cb}|$ can however not be explained by the presence of right-handed currents, as the pattern of effects is inconsistent with the theoretical prediction [2005]. In addition, the size of effect necessary to remove the tension in $|V_{ub}|$ determinations is in tension with the constraints from meson mixing observables and only achievable with large fine-tuning [2000]. With improved theoretical description of semileptonic decays and their precise measurement at Belle II, it will be possible to put much tighter constraints on the presence of right-handed charged currents – or to unravel their presence. Measurements must directly examine propagator chirality through decay helicities.

In addition to the enhanced gauge symmetry, left-right symmetric models are often equipped with a discrete symmetry, usually parity, charge conjugation symmetry or $CP$, thus making the restoration of the corresponding symmetry at high scales manifest [2006–2010]. These scenarios imply specific structures for the right-handed mixing matrix, with a hierarchy close to the CKM one. These models have in common that a large amount of fine-tuning is required to satisfy the constraints from neutral kaon mixing [1999, 2011–2015].

In the absence of a discrete symmetry the structure of the $V_R$ matrix is not restricted theoretically and has to be determined from data. The most stringent constraints are obtained from meson mixing observables, allowing only for very specific structures of $V_R$. If in addition small fine-tuning is required, then all off-diagonal elements of $V_R$ are found to be close to zero [2000, 2016]. The most stringent constraints again stem from neutral kaon mixing, but also $B_{d,s}$ mixing obserbables play a prominent role. A further decrease of uncertainties will therefore be crucial for constraining the structure of $V_R$ [2017].

An important contribution to meson mixing observables in left-right models is generated by tree level flavour changing heavy Higgs exchanges [1997–2000]. These heavy Higgs particles are generally present as a remnant of the $SU(2)_R$ symmetry breaking. In minimal left-right models their contributions to $\Delta F = 2$ observables require to push their masses well above 10 TeV. Keeping simultaneously the $W'$ and $Z'$ gauge bosons at the few TeV scale then forces the Higgs potential to be close to the non-perturbative regime. While it can be rigorously proven that it is not possible to avoid tree-level flavour changing Higgs couplings in left-right models [2018], the model can be augmented by discrete symmetries that suppress these couplings to a safe level even for heavy Higgs bosons around the TeV scale [2019].

Another interesting probe of left-right symmetric models is given by $b \to s\gamma$ and $b \to d\gamma$ transitions [2000, 2020–2023]. While in a significant portion of the parameter space of these models the effects are rather modest, it is possible to generate a sizeable contribution to the chirality-flipped magnetic penguin operator $O_7'$. Its presence could be most easily detected in observables sensitive to the photon polarisation [504, 505].

Last but not least left-right symmetric models can also generate visible effects in lepton flavour violating $\tau$ decays [2024, 2025]. In models with Higgs triplet representations, which are theoretically appealing due to the natural realisation of TeV-scale right-handed Majorana neutrinos, the doubly-charged Higgs bosons mediate the decays $\tau \to \ell_i \ell_j \ell_k$ with branching ratios in the reach of Belle II.

### *17.6.5. E6-inspired models.* *(Contributing authors: Thomas Deppisch and Jakob Schwichtenberg)*

*$E_6$ Unification.* The exceptional rank 6 group $E_6$ is one of the most popular unification groups [2026–2029] due to, for example, the automatic absence of anomalies and the fact that all SM fermions of one generation live in the fundamental representation. Moreover, $E_6$ is in a unique position among the suitable groups, because it is not a member of an infinite family and the only exceptional group with complex representations (In contrast, $SU(5)$ is part of the infinite $SU(N)$ family and $SO(10)$ of the infinite $SO(N)$ family.) In addition, $E_6$ models are a rich source of inspiration for phenomenological studies [2030–2033].

*Vector-like Quarks.* A generic prediction of $E_6$ Models is the existence of a vector-like quark, $D$, in each generation, which lives in the $(3, 1, -\frac{1}{3})$ representation of $G_{SM} = SU(3)_C \times SU(2)_L \times U(1)_Y$. In contrast to chiral quarks in a sequential fourth generation, which are ruled out by precision electroweak measurements [2034] and the Higgs discovery [2035], such vector-like quarks are still a viable extension of the SM.

Mass terms for vector-like quarks are $G_{SM}$ invariant and therefore one usually expects that they are "superheavy" [2036]. Formulated differently, one expects that only fermions which can not get a $G_{SM}$ invariant mass term are light. This is known as the "Survival Hypothesis" [2037, 2038], which is an explanation for the lightness of the chiral SM fermions. However, the term "superheavy" is not very precise. The exotic quarks acquire their mass through the Higgs mechanism and therefore their mass depends on the breaking chain. For example, in the breaking chain

$$\begin{aligned} E_6 &\to SO(10) \\ &\to SU(4)_C \times SU(2)_L \times U(1)_R \\ &\to SU(3)_C \times SU(2)_L \times U(1)_R \times U(1)_X \\ &\to SU(3)_C \times SU(2)_L \times U(1)_Y \end{aligned} \tag{606}$$

the vector-like quarks can be the lightest exotic fermions and acquire their mass through the vacuum expectation value which breaks the $SU(4)_C \times SU(2)_L \times U(1)_R$ intermediate symmetry. The masses of the vector-like quarks are therefore proportional to the scale at which this symmetry breaks. Symmetry breaking scales in GUTs can be calculated by solving the renormalisation group equations for the gauge couplings. For the breaking chain in Eq. (606) this was recently done in [2039] with the result[69]: $M_{3221} = 10^3 - 10^{10.5}$ GeV, $M_{421} = 10^{10.5}$ GeV, $M_{SO10} = 10^{14.7}$ GeV. The $E_6$ scale can not be computed since the gauge couplings are already unified at the $SO(10)$ scale and therefore there is no boundary condition left. Although $M_{SO10} = 10^{14.7}$ GeV corresponds to a proton lifetime well below the present bound from Super-Kamiokande $\tau_p(p \to e^+\pi^0) > 10^{34}$ yrs [77], this does not necessarily excludes this scenario, because it is well known that threshold corrections can alter these results dramatically [2041].

To estimate the masses of the vector-like quarks, we observe that in $E_6$ models the Yukawa couplings of the SM fermions and the exotic fermions have a common origin. The general Yukawa sector above the $E_6$ scale reads [2042]

$$\mathcal{L}_{\mathrm{Y}} = \Psi^T i\sigma_2 \Psi (Y_{27}\varphi + Y_{351'}\phi + Y_{351}\xi) + h.c.\,, \tag{607}$$

where $\Psi$ denotes the fermionic 27, $Y_i$ Yukawa couplings and $\varphi$, $\phi$ and $\xi$ the scalar representations 27, $351'$, 351, respectively. Therefore the Yukawa couplings of the lightest generation of the vector-like quarks could be as small as the Yukawa couplings of the first SM generation. Together with $M_{421}$ from above this means that the masses for the lightest vector-like quarks could be in the region $10 - 100$ TeV, which will be probed in the near future through precision measurements of flavour observables [2043].

The pattern of flavour violation in models with vector-like quarks was recently discussed extensively in [1953]. Among the main findings is, for example, that tree-level $Z$ contributions can increase $\epsilon'/\epsilon$ sufficiently to remove the current tension between the SM prediction [178–181] and the latest data [2044]. The patterns of flavour violation through vector-like $E_6$ quarks are summarised in Tables 5,6 and 10 in [1953]. Significant effects are possible in $B \to K^*\nu\nu$ processes and could therefore be observed by Belle II.

[69] The authors of [2039] "find a sharp disagreement" with the result of an earlier study [2040] and "this difference brings [this breaking chain] back among the potentially realistic ones", because $M_{3221}$ can be sufficiently high to yield realistic light neutrino masses.

*Leptoquarks in $E_6$ inspired SUSY models.* In supersymmetric $E_6$ models [2029, 2045, 2046] the situation is quite different. Because of $\mathcal{R}$-parity conservation the vector-like quarks are not allowed to mix with the SM down-type quarks, but there can be interesting flavour signatures from other sectors. The flavour physics in an $E_6$ inspired SUSY model was recently discussed in Ref. [2047]. Of special phenomenological interest in this model are the leptoquark couplings in the superpotential

$$\mathcal{W}_{\text{leptoquark}} = \lambda_3 \, (L \cdot Q) \, D^c + \lambda_4 \, u^c e^c D. \tag{608}$$

With these fields in the multi-TeV range, the dominant contributions to flavour physics come from neutral current operators induced by the scalar leptoquarks. The effective Hamiltonian with dimension six operators then becomes

$$\begin{aligned} \mathcal{H}_{\text{eff}} = \frac{1}{2}\lambda_3^{ijm}(m_D^2)^{-1}_{mn}\lambda_4^{kln} \; \mathcal{O}_b \quad + \frac{1}{2}\lambda_3^{ijm}(m_D^2)^{-1}_{mn}\lambda_3^{*,kln} \; \mathcal{O}_c \\ + \frac{1}{2}\lambda_3^{ijm}(m_D^2)^{-1}_{mn}\lambda_3^{*,kln} \; \mathcal{O}_e \quad + \frac{1}{2}\lambda_4^{ijm}(m_D^2)^{-1}_{mn}\lambda_4^{*,kln} \; \mathcal{O}_f, \end{aligned} \tag{609}$$

with

$$\mathcal{O}_b = (\bar{u}_l P_L u_j)(\bar{e}_k P_L e_i) - \tfrac{1}{4}(\bar{u}_l \, \sigma^{\mu\nu} P_L u_j)(\bar{e}_k \, \sigma_{\mu\nu} P_L e_i), \tag{610}$$

$$\mathcal{O}_c = (\bar{\nu}_l \gamma^\mu P_L \nu_i)(\bar{d}_k \gamma_\mu P_L d_j), \tag{611}$$

$$\mathcal{O}_e = (\bar{e}_l \gamma^\mu P_L e_i)(\bar{u}_k \gamma_\mu P_L u_j), \tag{612}$$

$$\mathcal{O}_f = (\bar{e}_l \gamma^\mu P_R e_i)(\bar{u}_k \gamma_\mu P_R u_j). \tag{613}$$

Here $i, j, k, l, m, n$ are flavour indices and $(m_D^2)^{-1}$ is the inverse of the squared leptoquark mass matrix.

It is an interesting fact that these $E_6$ leptoquarks couple down-type quarks to neutrinos and up-type quarks to charged leptons. Further, these operators are lepton flavour non-universal. Therefore they open new (semi-)leptonic decay channels not present in the SM. Summing over all neutrino final states can also enhance decay rates in comparison to SM processes. Hence, improved measurements of the following rare decays may be sensitive to this kind of new physics:

- Leptonic decays of the $D^0$ meson are highly suppressed in the SM. The operators $\mathcal{O}_b$, $\mathcal{O}_e$ and $\mathcal{O}_f$ can significantly enhance these processes inducing also LFV decays.
- Operator $\mathcal{O}_c$ contributes to the decay modes $B \to X_s \nu\bar{\nu}$ and $K \to \pi\nu\bar{\nu}$. In these cases, future measurements can reach the sensitivity of the SM predictions. This may allow to either find or exclude $E_6$ leptoquarks for a large region of the parameter space. Figure 222 shows the allowed regions using current data [77]. The mean values $x$ and $y$ are defined by

$$\frac{\lambda_3^{\ell 21} \lambda_3^{*,3\ell' 1}}{2m_D^2} \equiv \frac{x + \mathrm{i}y}{2m_D^2}, \qquad \ell, \ell' = 1, 2, 3, \tag{614}$$

  and $m_D$ is the mass of the lightest leptoquark.

Apart from the neutral current operators, there are also charged current operators which have to compete with the contributions from three families of Higgs doublets and also with $W$ boson exchange. Searches for SUSY Higgs couplings, e.g. in $b \to s\gamma$, also put relevant constrains on this model.

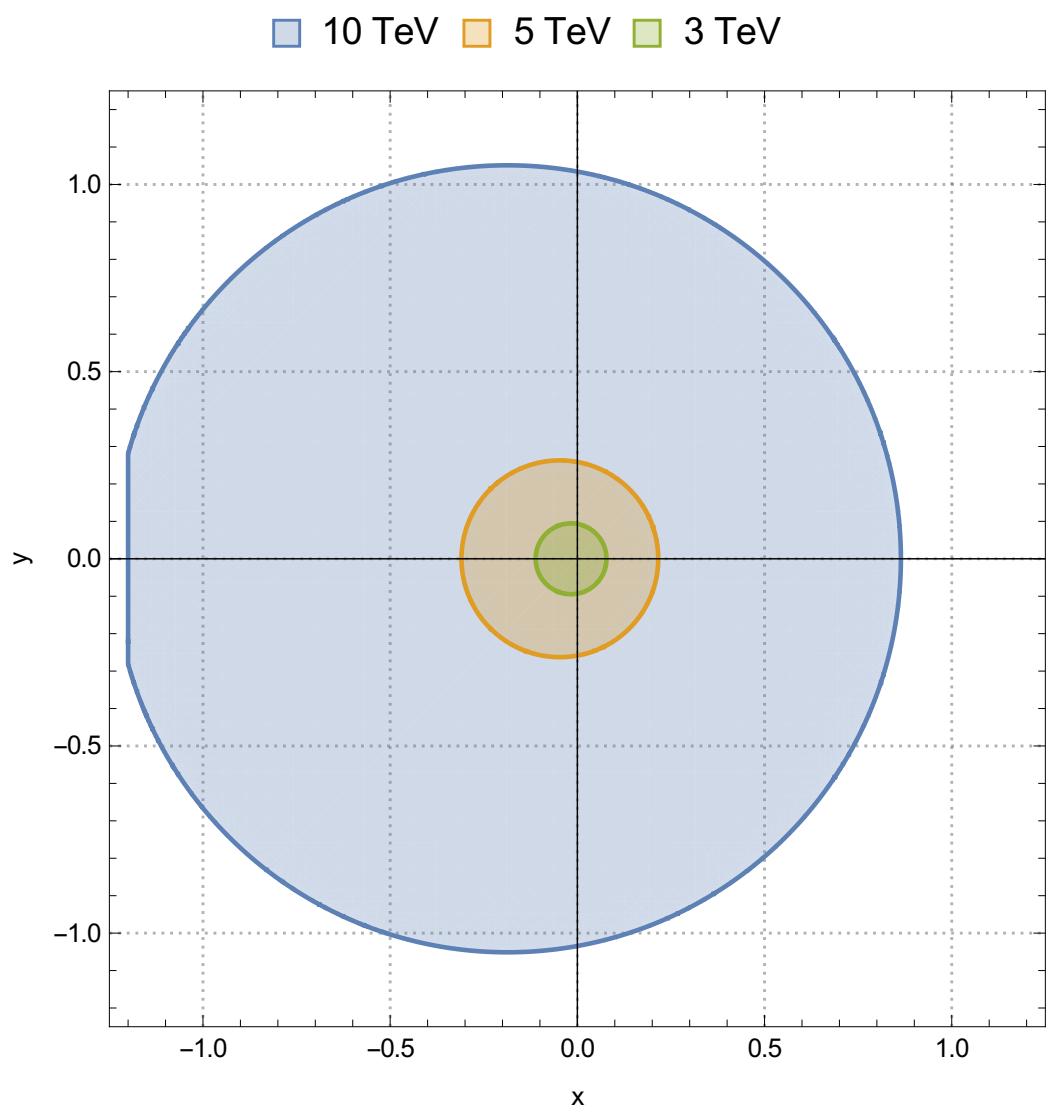

Fig. 222: **Constraints from current data of $B \to X_s \nu\bar{\nu}$ on $E_6$ leptoquarks**. Allowed regions for parameters $x$ and $y$ that parametrise the couplings are shown for several values of the leptoquark mass $m_D$.

## *17.7. Models of Compositeness*

*(Contributing author: David Straub)*

*Introduction.* Partially composite SM quarks or leptons are predicted by several new physics models. In composite Higgs models that are motivated by solving the SM's Higgs naturalness problem, they are a consequence of the linear coupling of the elementary fermion fields to composite operators that are required to generate fermion masses while avoiding the flavour problems of extended technicolor theories [2048]. In extra-dimensional models, e.g. of Randall-Sundrum type, partial compositeness in the four-dimensional dual picture arises from bulk fermions coupled to a bulk or brane Higgs [2049–2052]. If quarks were partially composite, the hierarchies in their masses and mixing angles could be a consequence of different degrees of compositeness (sticking to the 4D language for definiteness), the first generation corresponding to the "mostly elementary" and the third generation to the "significantly composite" quarks.

Most effects in flavour physics in these models arise from the presence of heavy spin-1 states (composite resonances or KK modes) with the quantum numbers of the SM gauge bosons and are thus closely related to $Z'$ models (see e.g. [2053–2056]). While the collider phenomenology is often distinct from renormalisable models with extended gauge sectors since the resonances tend to be broad, these differences are less relevant for flavour physics. The characteristic flavour effects arise from the flavour structure of the composite-elementary mixings.

Since some of the spin-1 resonances can mix with the elementary electroweak gauge bosons, also flavour-changing $Z$ couplings can be generated. This leads to an interplay between flavour and electroweak precision tests. While the strong constraints on modifications of the electroweak $T$ parameter suggest the presence of a global custodial symmetry $SU(2)_L \times$ $SU(2)_R$ in the strong sector, the constraints on modifications of the $Z \to b\bar{b}$ partial width from LEP require a "custodial protection" of the $Z\bar{b}b$ couplings within this class of models, which can be achieved by an appropriate choice of the representation of the composite fermions under this global symmetry [2057]. This in turn leads to forbidden or suppressed flavour-changing $Z$ couplings for one chirality of down-type quarks (see e.g. [2058, 2059]).

*Flavour structure.* An interesting possibility for the flavour structure is *flavour anarchy*, implying that the couplings in the strongly-coupled composite sector are structureless (*i.e.* do not exhibit pronounced hierarchies), while the CKM and quark mass hierarchies arise purely from the composite-elementary mixings. Writing the rotation between the composite-elementary basis and the mass basis as

$$\begin{pmatrix} \psi^{\text{light}} \\ \Psi^{\text{heavy}} \end{pmatrix} = \begin{pmatrix} \cos\phi & -\sin\phi \\ \sin\phi & \cos\phi \end{pmatrix} \begin{pmatrix} \psi^{\text{ele}} \\ \Psi^{\text{comp}} \end{pmatrix} \tag{615}$$

where $\psi^{\text{light}}$ is a SM fermion, $\Psi^{\text{heavy}}$ a new heavy fermion, and only the field $\Psi^{\text{comp}}$ couples to the Higgs field, $s_\psi \equiv \sin\phi$ is the degree of compositeness of $\psi^{\text{light}}$. In the case of flavour anarchy, one finds the approximate relations

$$m_{u^i} \sim v\, g_\rho s_{q_L^i} s_{u_R^i} \tag{616}$$

for the up-type quark masses where $g_\rho$ is a generic strong-sector coupling, and analogously for the down-type quark masses, while the off-diagonal elements of the CKM matrix are roughly given by

$$V_{ij} \sim s_{q_L^i}/s_{q_L^j}\,, \qquad i > j. \tag{617}$$

This dependence leads to an automatic parametric suppression of FCNCs. For instance, the $K^0$-$\bar{K}^0$ mixing operator $(\bar{s}_L \gamma_\mu d_L)^2$ is proportional to $s_{q_L^2}^2 s_{q_L^1}^2 \sim s_{q_L^3}^4 V_{td}^2 V_{ts}^2$, which is the same CKM-like suppression as in the SM. However, the model is not minimally flavour violating (MFV) as also FCNC operators with right-handed quarks are generated. Using the naive parametric counting, this leads to a constraint on the new physics scale of around 15 TeV from indirect $CP$ violation in $K^0$-$\bar{K}^0$ mixing [2054, 2055, 2060].

To ameliorate this problem, it has been suggested that the composite sector is exactly invariant under a large flavour symmetry which is only broken minimally (*i.e.* by the amount required to reproduce CKM mixing) by the composite-elementary mixings[70]. A maximal symmetry based on $U(3)$ rotations in three-generation space can be invoked to obtain MFV models [2070, 2071], but these tend to struggle with strong constraints from electroweak precision tests or quark compositeness searches as they tie the compositeness of light quarks to the top quark. An alternative is to restrict the flavour symmetry to the first two generations, *i.e.* use $U(2)$ rotations, which avoids these problems but still leads to CKM-like flavour violation [1934, 2056].

*Signals in flavour physics.* The most likely observables to be affected by models with partial quark compositeness strongly depend on the flavour structure and the implementation of custodial protection. We will discuss the most relevant cases in turn.

---

[70] For alternative mechanisms, see e.g. [2061–2069].

- *Anarchic flavour structure.* The most sensitive observables are expected to be $\epsilon_K$, hadronic electric dipole moments (EDMs), $\epsilon'/\epsilon$, and the mass differences and $CP$ violation in $B_d$ and $B_s$ mixing. While the mass differences have already been measured to a very good precision, their constraining power is currently limited by the limited knowledge of CKM elements from tree-level processes, e.g. $V_{cb}$ and $V_{ub}$ from semi-leptonic decays. Improved measurements of these decays by Belle II will thus play an important role in scrutinising these models.
- *MFV with left-handed compositeness.* In these models the flavour symmetry is maximal and only broken by the composite-elementary mixings of *right-handed* quarks. In this case, no FCNCs arise at tree-level. Constraints from loop level processes are expected to be weak given the strong bounds from electroweak precision tests. However there can be sizeable contributions to hadronic EDMs.
- *MFV with right-handed compositeness.* In this case the flavour symmetry is only broken by the composite-elementary mixings of *left-handed* quarks. Again, there can be sizeable contributions to hadronic EDMs, but also FCNCs are generated at tree-level. The most sensitive observables are expected to be $\epsilon_K$ as well as $B_d$ and $B_s$ mixing.
- *$U(2)$ models.* In these models, both $\Delta F = 1$ and $\Delta F = 2$ processes with kaons or $B$ mesons are sensitive to tree-level new physics effects. Up to subleading effects in the (flavour symmetry breaking) spurion expansion, effects arise in operators with the same chirality as those present in the SM, which potentially allows to distinguish these models from the anarchic ones. In the $\Delta F = 1$ sector, signals at Belle II could arise in $b \to s\ell\ell$ and $b \to s\nu\bar{\nu}$ transitions (see e.g. [613, 2059]).

## *17.8. Conclusions*

Belle II has a unique potential to reveal physics beyond the Standard Model (BSM). In this chapter we have described 10 BSM model classes and elucidated their imprints on the observables to be studied at Belle II. Tables 145-149 list 80 interesting observables which can be measured by Belle II and summarises the sensitivity of these observables to effects form the various BSM model classes. The plethora of measurements (giving complementary information) will eventually help to pin down many features of the next theory superseding the SM. We may hope that Belle II will guide the high-$p_T$ experiments to the discovery of new particles.

## 18. Global analyses

*Editors: F. Bernlochner, R. Itoh, J. Kamenik, V. Lubicz, U. Nierste, Y. Sato, L. Silvestrini*

*Additional section writers: W. Altmannshofer, F. Beaujean, M. Bona, M. Ciuchini, J.A. Evans, S. Jahn, F. Mahmoudi, A. Paul, J. Rosiek D. Shih, D. Straub P. Urquijo, D. Van Dyk, R. Watanabe*

### *18.1. Introduction*

To interpret results of Belle II measurements requires global analyses when those observables are related theoretically to the other observables which have or haven't yet been measured. Especially, when a signal of new physics is discovered, a consistency check has to be done throughly including numbers of observables. Obviously, this is not an easy task since each observable has very complicated theoretical expressions and inputs as we have seen throughout this book. Thus, developing tools is of primary importance to perform global analyses. In this final chapter, we present some results of the global analysis and also the tools developed for this purpose.

A very successful example of global analysis in our field is the CKM unitary matrix fit, which contains four free parameters and must explain all observed flavour changing phenomena, both CP conserving and violating, in the SM. Two tools have been developed independently, the CKMfitter and the UTfit packages. In section 18.2, we demonstrate the new physics sensitivity of Belle II experiment, using these tools: taking into account the future achievable precisions for various observables obtained in the previous chapters, we assess the reach of the precision in case of SM and a discovery potential of new physics at Belle II through the CKM unitarity triangle analysis.

In the SM, the flavour phenomena are described by using the effective Hamiltonian (see Chapter 7). The Wilson coefficients in this Hamiltonian include the information of the weak interaction (e.g. dependence of the $W$ boson and top quark masses) as well as QCD radiative corrections. A new physics effect can appear as deviations between the experimentally determined values of the Wilson coefficients and their SM expectations. Thus, the Wilson coefficient fit is the most useful method for global analyses. In section 18.3, we extend the effective Hamiltonian in a generic manner (model independent method), yet limiting the extension by focusing on the two kinds of flavour changing processes, where hints of new physics have been seen recently. We discuss what kind of deviations from SM appear in those Wilson coefficients and its expected sensitivity at the Belle II experiment.

There are more and more global analyses tools available for flavour physics and in the B2TiP framework, we organised dedicated workshops to discuss and compare them inviting their developers. The tools we have discussed are the ones which fit the Wilson coefficients or the new physics parameters (Flavio, HEPfit, EOS), the ones which compute the flavour physics observables including contributions from the supersymmetric models (SuperIso, SUSY-Flavour, FormFlavor) and the one which is a statistical tool (pypmc). In the section 18.4, we describe the characteristics of these tools which we have studied.

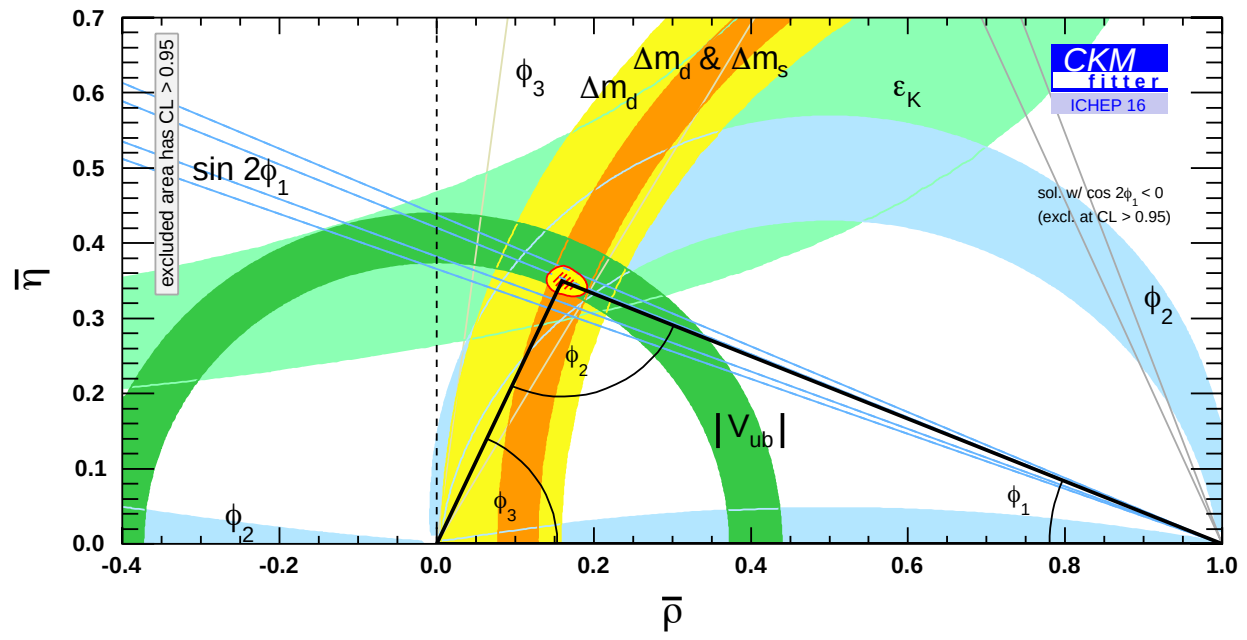

Fig. 223: UT fit today as determined with CKMFitter.

### 18.2. *CKM Unitarity Triangle global fits*

In the SM, the weak charged-current mixes different quark generations. The strength of such transitions are encoded in the unitary CKM matrix. In the case of three generations of quarks, the physical content of this matrix reduces to four real parameters, among which there is one phase, the only source of $CP$ violation in the quark sector (see Sec. 7). These four real parameters are defined in a phase-convention independent way,

$$\lambda^2 = \frac{|V_{us}|^2}{|V_{ud}|^2 + |V_{us}|^2}, \quad A^2\lambda^4 = \frac{|V_{cb}|^2}{|V_{ud}|^2 + |V_{us}|^2}, \quad \bar{\rho} + i\bar{\eta} = -\frac{V_{ud}V_{ub}^*}{V_{cd}V_{cb}^*}. \tag{618}$$

At Belle II the attention on the combined analysis of CKM Unitarity Triangle constraints will shift from the pure metrology of the SM to the investigation of deviations in flavour physics and manifestations of New Physics (NP).

In order to perform global analyses of the CKM Unitarity Triangle, two tools have been developed, CKMfitter and UTfit, which use different statistical methods. In this section, using these two packages, we present case studies to demonstrate the potential of discovering new physics through the CKM Unitarity Triangle analysis at the Belle II experiment.

#### 18.2.1. *CKMfitter.* *(Contributing author: Phillip Urquijo)*

Here we discuss the future prospects of Unitarity Triangle global analyses using the CKMfitter package, which uses a frequentist approach, based on the Rfit model to describe systematic uncertainties (see Ref. [91] for details). A table of key inputs used in the SM global fit is presented in Table 153. Low-energy strong interactions constitute a central issue in flavour physics, which explains the need for accurate inputs for hadronic quantities such as decay constants, form factors, and bag parameters. CKMfitter mostly relies on Lattice QCD simulations, with a specific averaging procedure to combine the results from different collaborations. A similar approach is followed in order to combine the inclusive and exclusive determinations of $|V_{ub}|$ and $|V_{cb}|$, which are not in excellent agreement. The current constraints on the Unitarity Triangle parameters are depicted in the $\bar{\rho} - \bar{\eta}$ plane in Fig. 223.

We consider two key scenarios in the Belle II era, defined as follows.

(1) World average scenario: the central values of the CKM matrix parameters stay at the world average (ca. 2016) central values while for the uncertainties we consider Belle II + LHCb + LQCD future projections using the world average values as of 2017.

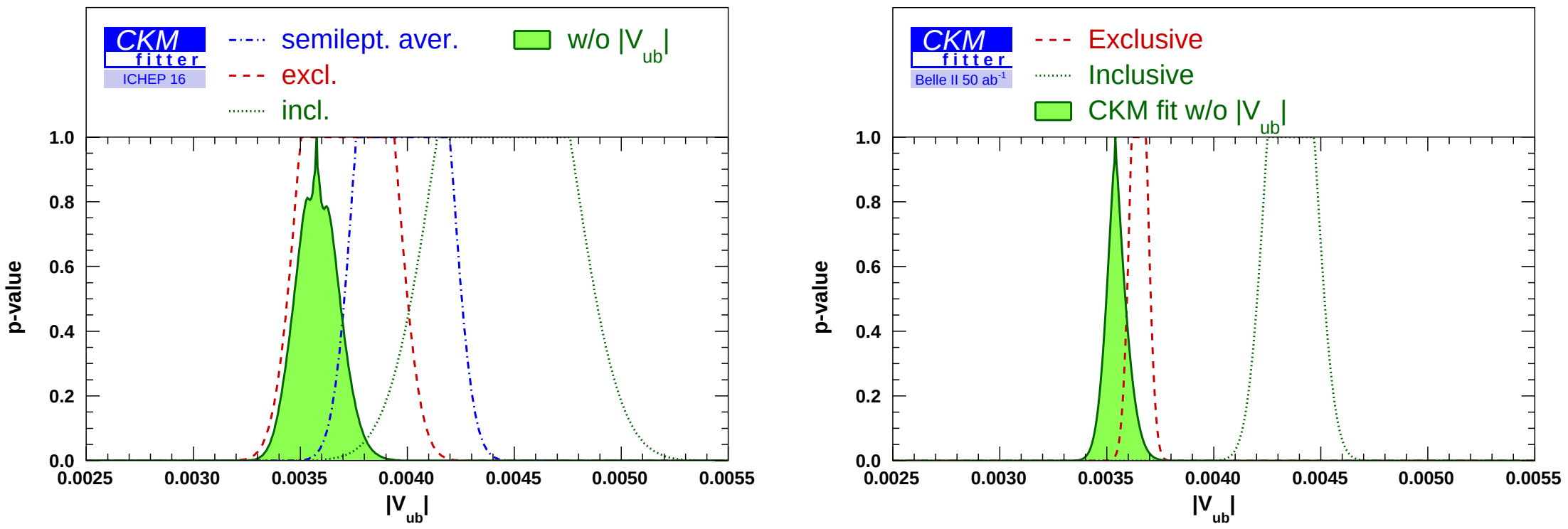

Fig. 224: $|V_{ub}|$ today (left) and extrapolated to the 50 ab$^{-1}$ scenario (right).

For reference we show the precision of Belle II only scenarios (rather than Belle II + LHCb).

(2) SM-like scenario: the central values are chosen such that they satisfy the SM (*i.e.* closed Unitarity Triangle) while the uncertainties are projected to the future Belle II + LHCb + LQCD data. For reference we show the precision of the Belle II only cases as well.

The input to the fits are shown in Table 153. While the projections include input from Belle II and LHCb, it is expected that Belle II will provide the most precise measurements of many key observables used in the determination of these parameters. The exceptions are $\phi_3$, which will be of similar precision at LHCb, and $B_s$ and $B$ mixing, which will be measured with greater precision at LHCb.

One of the most important inputs from Belle II will be the measurement of $|V_{ub}|$ from exclusive and inclusive semileptonic $B$ decays. Figure 224 shows the current and projected precision of exclusive and inclusive world averages, their combination performed by CKM-Fitter, and the expected value based on CKM unitarity. An interesting test in the Belle II era will be the comparison between $Br(B \to \tau\nu)$ and $\sin 2\phi_1$. We depict the projected precision for these inputs compared to the constraints from the global fit in Fig. 225.

The fit results of scenarios (1) and (2) are shown in Fig. 226, and summarised in Table 154. For scenario (1) we show the associated p-values for the fits. For scenario (2) we show the numerical precision of the CKM Unitarity Triangle (UT) parameters.

The current and projected (Belle II combined with LHCb) fits for the world average scenario with various data subsets are shown in Fig. 227. The plots show constraints from loop, tree, $CP$-conserving, and $CP$-violating scenarios respectively.

The CKMFitter group has performed analyses of new physics in mixing, in particular $\Delta B = 2$ operators, assuming that tree decays are not affected by NP effects. Within this framework, NP contributions to the $B_{d,s}$ mixing amplitudes can be parameterised as

$$M_{12}^{d,s} = (M_{12}^{d,s})_{\rm SM} \times \left(1 + h_{d,s} e^{2i\sigma_{d,s}}\right) \qquad (619)$$

The results of fits with current constraints, and with the full Belle II data set are shown in Fig. 228.

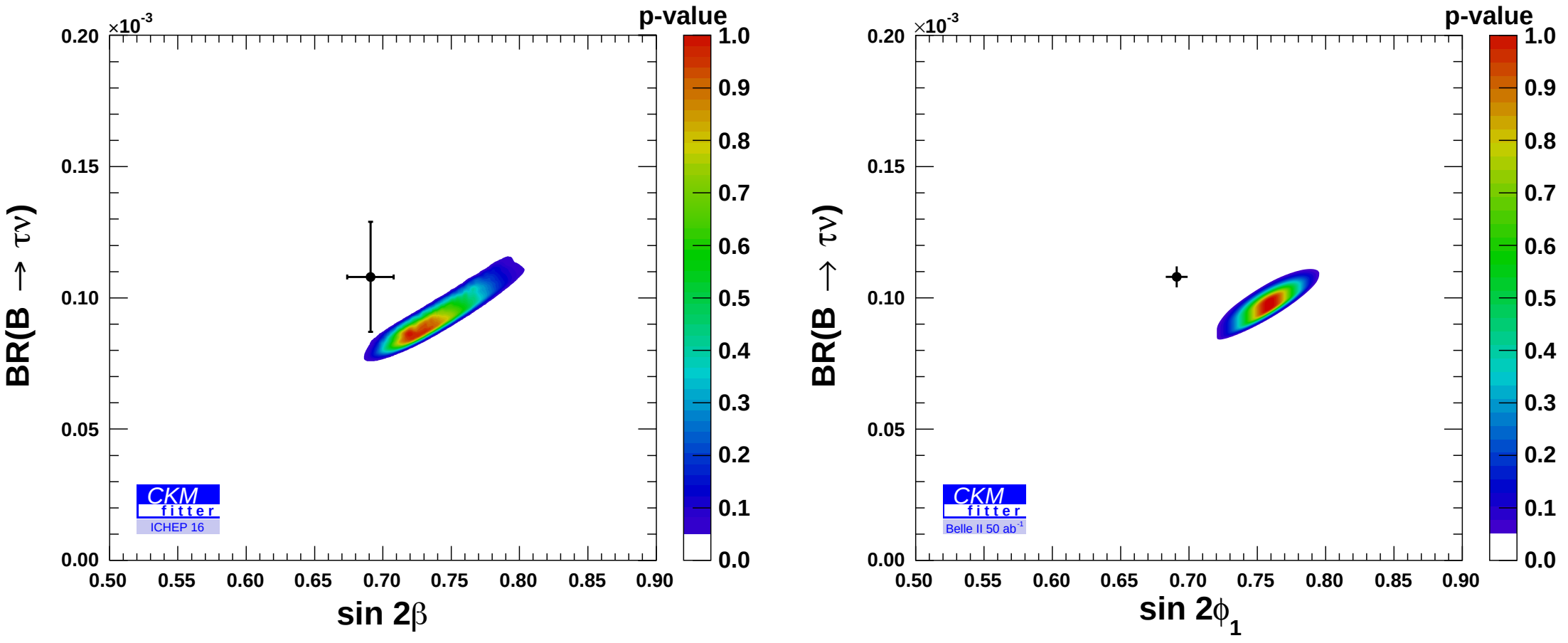

Fig. 225: $\sin 2\phi_1$ versus $Br(B \to \tau\nu)$ derived from the global fit (contour) and direct measurements (data points) for current world average values (left) and Belle II projections (right).

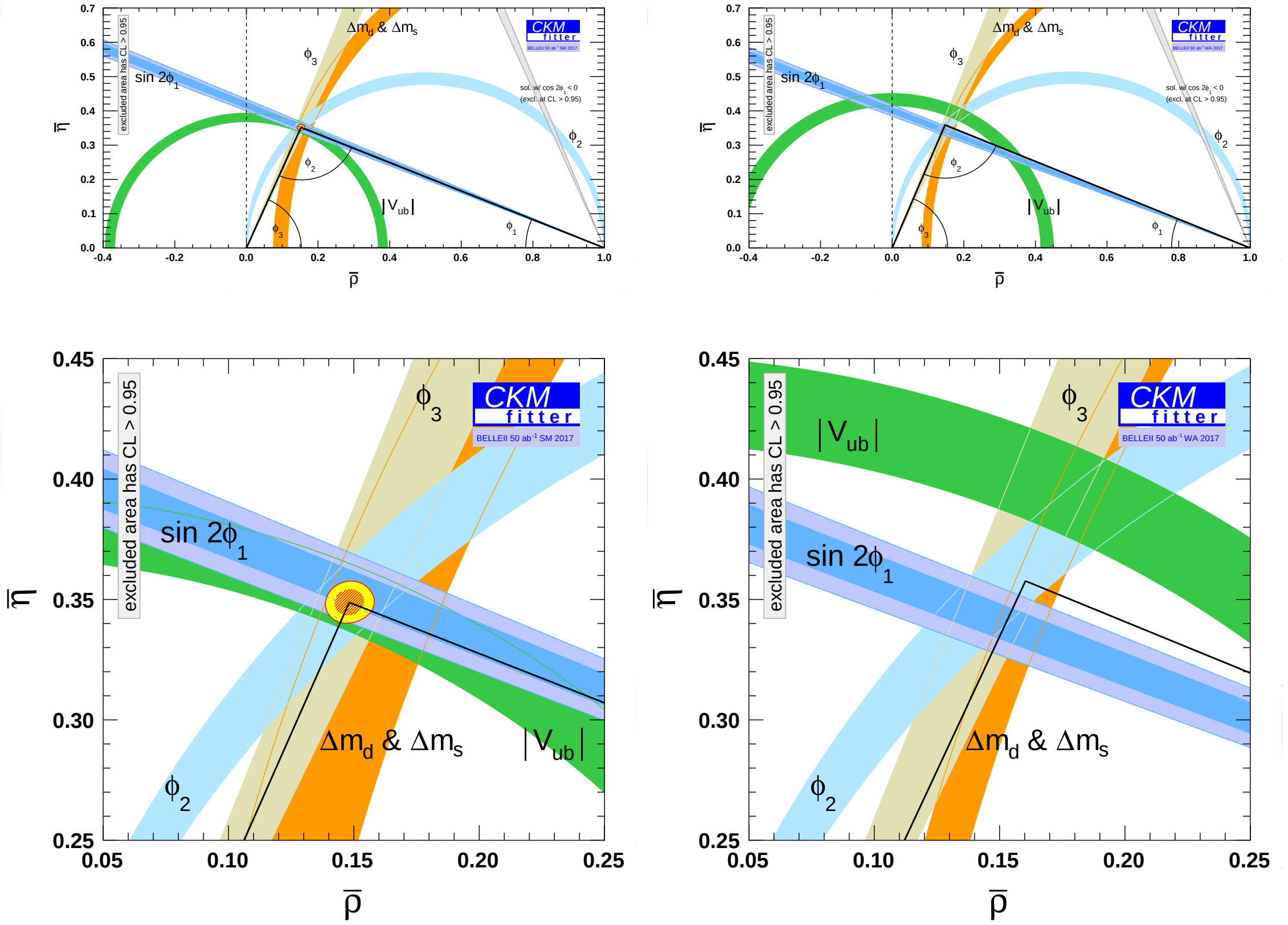

Fig. 226: UT fit today is extrapolated to the 50 $ab^{-1}$ scenario for an SM-like scenario (left) and world average values (right).

Table 153: The CKMFitter input parameter values for the current situation (as of 2016) and for each scenario, year 2025 for the world average scenario (scenario (1)), and year 2025 for the SM-like scenario (scenario (2)). The values in brackets are where Belle II and LHCb upgrade projections are combined.

| Input | Current 2016 | Scenario (1) Belle II (+LHCb) 2025 | Scenario (2) Belle II (+LHCb) 2025 |
|---|---|---|---|
| $\lvert V_{ub}\rvert$(semileptonic)$[10^{-3}]$ | $4.01 \pm 0.08 \pm 0.22$ | $\pm 0.10$ | $3.71 \pm 0.09$ |
| $\lvert V_{cb}\rvert$(semileptonic)$[10^{-3}]$ | $41.00 \pm 0.33 \pm 0.74$ | $\pm 0.57$ | $41.80 \pm 0.60$ |
| $\mathcal{B}(B \to \tau\nu)$ | $1.08 \pm 0.21$ | $\pm 0.04$ | $0.817 \pm 0.03$ |
| $\sin 2\phi_1$ | $0.691 \pm 0.017$ | $\pm 0.008$ | $0.710 \pm 0.008$ |
| $\phi_3[^\circ]$ | $73.2^{+6.3}_{-7.0}$ | $\pm 1.5$ ($\pm 1.0$) | $67 \pm 1.5$ ($\pm 1.0$) |
| $\phi_2[^\circ]$ | $87.6^{+3.5}_{-3.3}$ | $\pm 1.0$ | $90.4 \pm 1.0$ |
| $\Delta m_d$ | $0.510 \pm 0.003$ | - | - |
| $\Delta m_s$ | $17.757 \pm 0.021$ | - | - |
| $\mathcal{B}(B_s \to \mu\mu)$ | $2.8^{+0.7}_{-0.6}$ | ($\pm 0.5$) | $3.31^{+0.7}_{-0.6}$ ($\pm 0.5$) |
| $f_{B_s}$ | $0.224 \pm 0.001 \pm 0.002$ | 0.001 | - |
| $B_{B_s}$ | $1.320 \pm 0.016 \pm 0.030$ | 0.010 | - |
| $f_{B_s}/f_{B_d}$ | $1.205 \pm 0.003 \pm 0.006$ | 0.005 | - |
| $B_{B_s}/B_{B_d}$ | $1.023 \pm 0.013 \pm 0.014$ | 0.005 | - |
| $\lvert V_{cd}\rvert(\nu N)$ | $0.230 \pm 0.011$ | - | - |
| $\lvert V_{cs}\rvert(W \to c\bar{s})$ | $0.94^{+0.32}_{-0.26} \pm 0.13$ | - | - |
| $f_{D_s}/f_{D_d}$ | $1.175^{+0.001}_{-0.004}$ | - | - |
| $\mathcal{B}(D \to \mu\nu)$ | $0.374 \pm 0.017$ | $\pm 0.010$ | - |
| $\epsilon_K$ | $2.228 \pm 0.011$ | - | - |
| $\lvert V_{us}\rvert f_+^{K\to\pi}(0)$ | $0.2163 \pm 0.0005$ | - | $0.22449 \pm 0.0005$ |
| $\mathcal{B}(K \to e\nu)$ | $1.581 \pm 0.008$ | - | $1.5689 \pm 0.008$ |
| $\mathcal{B}(K \to \mu\nu)$ | $0.6355 \pm 0.0011$ | - | $0.6357 \pm 0.0011$ |
| $\mathcal{B}(\tau \to K\nu)$ | $0.6955 \pm 0.0096$ | - | $0.7170 \pm 0.0096$ |
| $\lvert V_{ud}\rvert$ | $0.97425 \pm 0.00022$ | - | - |

For a NP contribution to the mixing of a meson with $q_i\bar{q}_j$ flavour quantum numbers due to the operator

$$\frac{C_{ij}^2}{\Lambda^2}\left(\bar{q}_{i,L}\gamma^\mu q_{j,L}\right)^2, \tag{620}$$

one finds that

$$h \simeq 1.5\frac{|C_{ij}|^2}{|\lambda_{ij}^t|^2}\frac{(4\pi)^2}{G_F\Lambda^2} \simeq \frac{|C_{ij}|^2}{|\lambda_{ij}^t|^2}\left(\frac{4.5\text{ TeV}}{\Lambda}\right), \quad \sigma = \arg(C_{ij}\lambda_{ij}^{t*}), \tag{621}$$

where $\lambda_{ij}^t = V_{ti}^* V_{tj}$ and $V$ is the CKM matrix. The scales of the operators probed in $B_d$ mixing by the end of Belle II data taking will be 17 TeV and 1.4 TeV for CKM like couplings in tree and one-loop level NP interactions respectively. For scenarios with no hierarchy, *i.e.*

Table 154: CKMFitter results for the Wolfenstein parameters with current world averages, and with the SM-like scenario with Belle II precision and with Belle II combined with LHCb by the year 2025.

| Input | Current WA | SM value Belle II | SM value Belle II+LHCb |
|---|---|---|---|
| $A$ | $0.8227^{+0.0066}_{-0.0136}$ | $^{+0.0025}_{-0.0027}$ | $^{+0.0024}_{-0.0028}$ |
| $\lambda$ | $0.22543^{+0.00042}_{-0.00031}$ | $^{0.00036}_{-0.00030}$ | $^{0.00035}_{-0.00030}$ |
| $\bar{\rho}$ | $0.1504^{+0.0121}_{-0.0062}$ | $^{+0.0054}_{-0.0044}$ | $^{+0.0042}_{-0.0040}$ |
| $\bar{\eta}$ | $0.3540^{+0.00069}_{-0.0076}$ | $^{+0.0037}_{-0.00040}$ | $^{+0.0036}_{-0.00037}$ |

Table 155: Uncertainties on external input parameters in the 5 and 50 ab$^{-1}$ scenarios used in the UT Fit study. In the 5 ab$^{-1}$ study it is assumed that no improvement with respect to the present uncertainties is assumed.

| Parameter | Error (5ab$^{-1}$) | Error (50ab$^{-1}$) |
|---|---|---|
| $\alpha_s(M_Z)$ | $\pm 0.0012$ | $\pm 0.0004$ |
| $m_t$ (GeV) | $\pm 0.73$ | $\pm 0.6$ |
| $\lvert V_{us}\rvert$ | $\pm 0.0011$ | $\pm 0.0002$ |
| $B_K$ | $\pm 0.029$ | $\pm 0.002$ |
| $f_{B_s}$ (GeV) | $\pm 0.05$ | $\pm 0.001$ |
| $f_{B_s}/f_{B_d}$ | $\pm 0.013$ | $\pm 0.006$ |
| $B_{B_s}/B_{B_d}$ | $\pm 0.036$ | $\pm 0.007$ |
| $B_{B_s}$ | $\pm 0.053$ | $\pm 0.007$ |

$|C_{ij}| = 1$, corresponding scale of operators probed will be $2 \times 10^3$ TeV and $2 \times 10^2$ TeV in tree and one-loop level NP interactions respectively.

*18.2.2. UTfit.* *(Contributing authors: Marcella Bona, Marco Ciuchini)*

Here we discuss the impact of Belle II on the Unitarity Triangle analysis within and beyond the Standard Model in the Bayesian approach of the UT*fit* Collaboration [759, 2072–2075]. We consider the two scenarios from Table 1.3 of the Introduction. In particular, we present results using experimental uncertainties corresponding to 5 and 50 ab$^{-1}$ for $|V_{cb}|$, $|V_{ub}|$, $\sin 2\phi_1$, $\phi_3$, and $\phi_2$, while central values are tuned to the SM. For other input parameters, in the 50 ab$^{-1}$ scenario we use the uncertainties reported in Tab. 155, based on the extrapolation of Appendix B.2 of Ref. [2076].

The projected uncertainties of the SM fit for the CKM parameters, UT angles and $BR(B \to \tau\nu)$ (not used in the fit) are reported in Tab. 156 and Fig. 229. Generalising the analysis beyond the SM following the notation introduced in Ref. [2074]:

$$C_{B_d} e^{2i\phi_{B_d}} = \frac{\langle B_d^0 | H_{\rm eff}^{\rm full} | \bar{B}_d^0 \rangle}{\langle B_d^0 | H_{\rm eff}^{\rm SM} | \bar{B}_d^0 \rangle} \tag{622}$$

where $H_{\rm eff}^{\rm SM}$ includes only the SM box diagrams, while $H_{\rm eff}^{\rm full}$ includes also the NP contribution to the $B_d -^0 \bar{B}_d^0$ mixing, we obtain the uncertainties presented in Table. 157 and Fig. 230 for the CKM parameters and the parameters representing NP contributions to $B_d - \bar{B}_d$ mixing.

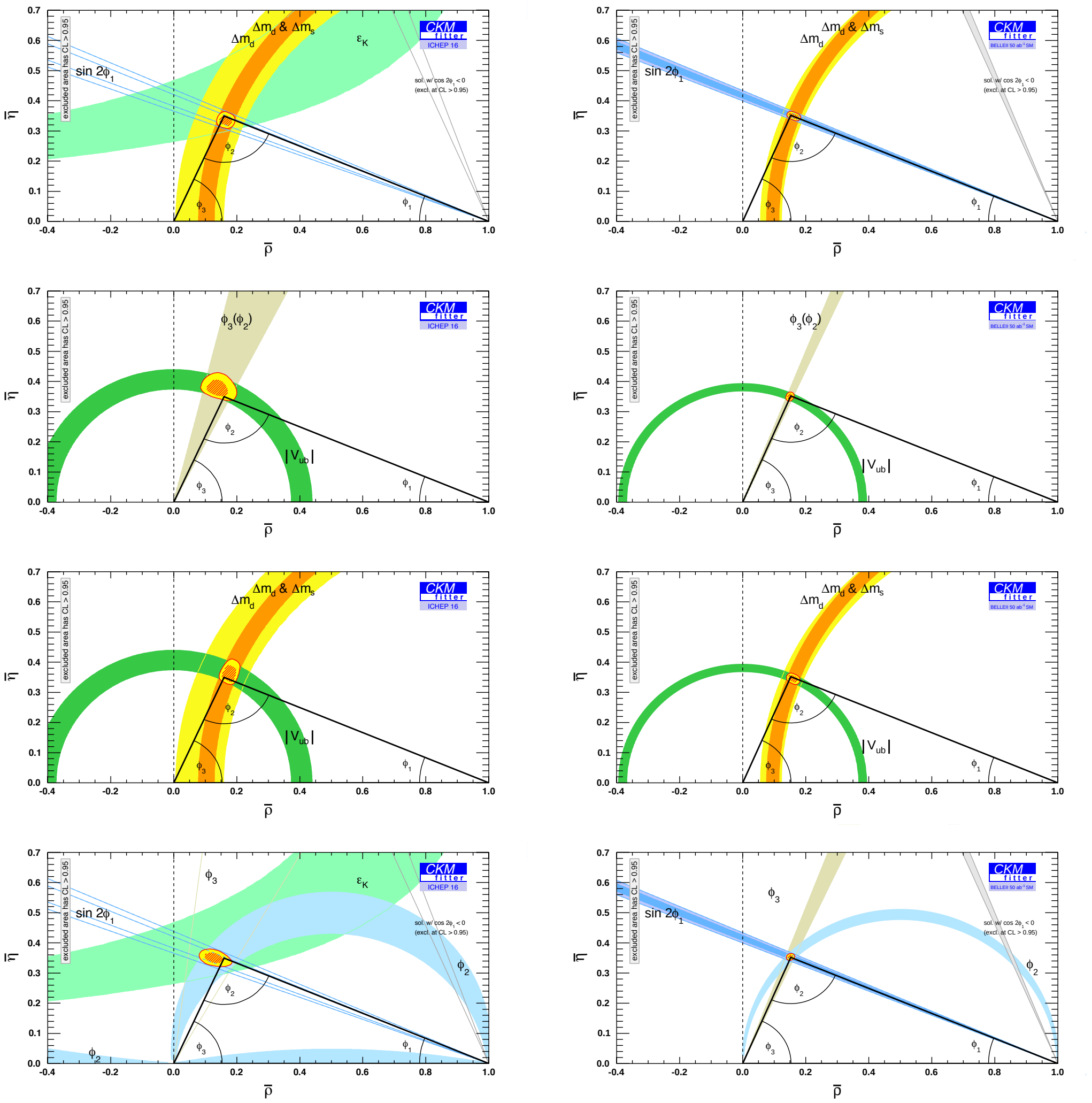

Fig. 227: UT fit today (left) and extrapolated to the 50 $ab^{-1}$ scenario for the SM-like scenario (right). Four sets of fits are shown using loop, tree, $CP$ conserving and $CP$ violating data subsets, respectively.

### *18.3. Model-independent analyses of new physics*

One can parametrise all possible types of new physics in terms of Wilson coefficients of the weak effective hamiltonian. In hadronic decays this approach involves too many coefficients to be feasible in practice. However, in some cases only a restricted set of Wilson coefficients contributes and such model-independent fits are possible. These cases are discussed in this section.

#### *18.3.1. Tree-level decays.* *(Contributing author: Ryoutaro Watanabe)*

(Semi-)leptonic $B$ meson decays are derived from the quark level process, $b \to q\ell\nu$ for $q = u$ and $c$. Belle II has sufficient sensitivity to precisely measure a variety of observables for $\bar{B} \to D^{(*)}\ell\bar{\nu}$, $\bar{B} \to \pi\ell\bar{\nu}$, and $\bar{B} \to \ell\bar{\nu}$ (for $\ell = \tau$, $\mu$, $e$). The observed $4\sigma$ discrepancy from the SM in $R_{D^{(*)}} \equiv \mathcal{B}(\bar{B} \to D^{(*)}\tau\bar{\nu})/\mathcal{B}(\bar{B} \to D^{(*)}\ell\bar{\nu})$ (for $\ell = \mu$ or $e$) must be characterised in terms of new physics scenarios.

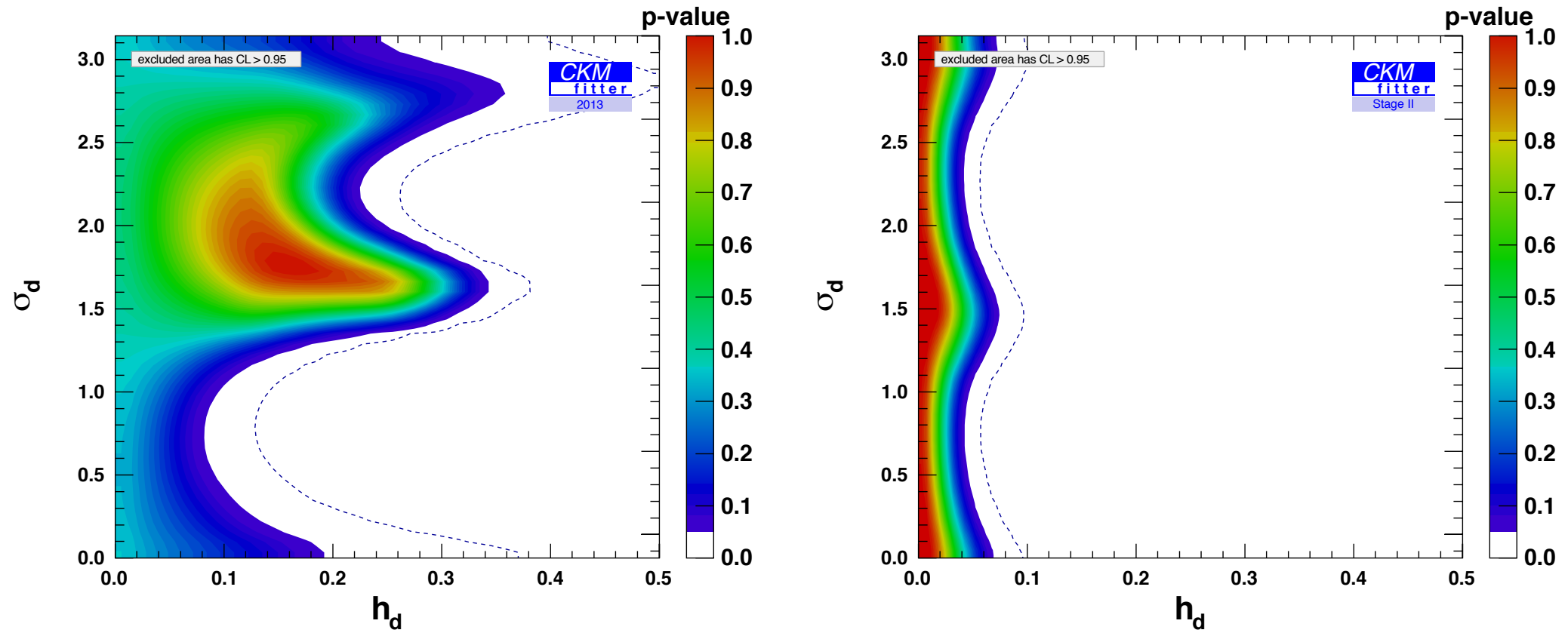

Fig. 228: Results of the fit to NP in mixing, for current constraints (left) and year 2025 constraints (right).

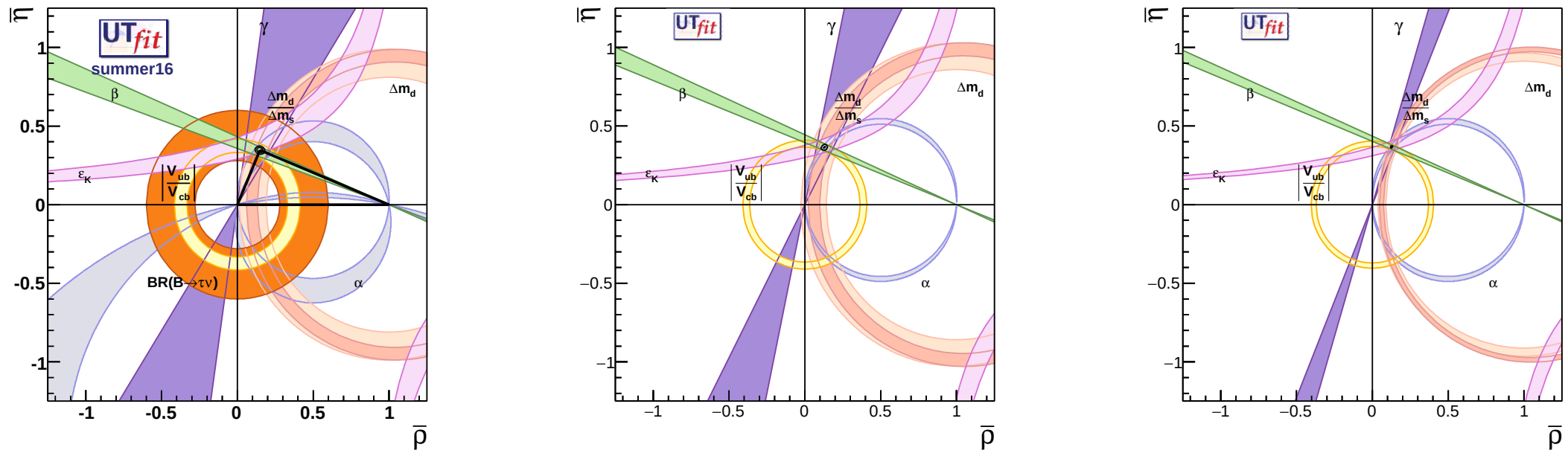

Fig. 229: SM Unitarity triangle fit today (left) and extrapolated to the 5 ab$^{-1}$ (centre) and 50 ab$^{-1}$ (right) gs.

In the presence of all possible NP contributions in the process $b \to q\tau\nu$, the effective Lagrangian can be described by

$$-\mathcal{L}_{\rm eff} = 2\sqrt{2}G_F V_{qb}\Big[(\delta_{\nu_\tau,\nu_\ell} + C_{V_1}^{(q,\nu_\ell)})\mathcal{O}_{V_1}^{(q,\nu_\ell)} + \sum_{X=}^{V_2,S_1,S_2,T} C_X^{(q,\nu_\ell)}\mathcal{O}_X^{(q,\nu_\ell)}\Big], \qquad (623)$$

Table 156: Extrapolated uncertainties of the fit in the 5 and 50 $ab^{-1}$ scenarios. For comparison, we also report the uncertainties of the current fit.

| Parameter | Error | | |
|---|---|---|---|
| | current | $5ab^{-1}$ | $50ab^{-1}$ |
| $\lambda$ | $\pm0.0007$ | $\pm0.0007$ | $\pm0.0002$ |
| $A$ | $\pm0.012$ | $\pm0.008$ | $\pm0.005$ |
| $\bar{\rho}$ | $\pm0.013$ | $\pm0.007$ | $\pm0.004$ |
| $\bar{\eta}$ | $\pm0.011$ | $\pm0.006$ | $\pm0.004$ |
| $R_b$ | $\pm0.013$ | $\pm0.007$ | $\pm0.005$ |
| $R_t$ | $\pm0.022$ | $\pm0.006$ | $\pm0.004$ |
| $\phi_2(^\circ)$ | $\pm2.0$ | $\pm0.9$ | $\pm0.6$ |
| $\phi_1(^\circ)$ | $\pm0.8$ | $\pm0.4$ | $\pm0.3$ |
| $\phi_3(^\circ)$ | $\pm1.9$ | $\pm1.0$ | $\pm0.6$ |
| $\beta_s(^\circ)$ | $\pm0.034$ | $\pm0.02$ | $\pm0.01$ |
| $J_{CP}$ | $\pm0.093$ | $\pm0.06$ | $\pm0.04$ |
| $BR(B\to\tau\nu)$ | $\pm0.06$ | $\pm0.05$ | $\pm0.02$ |

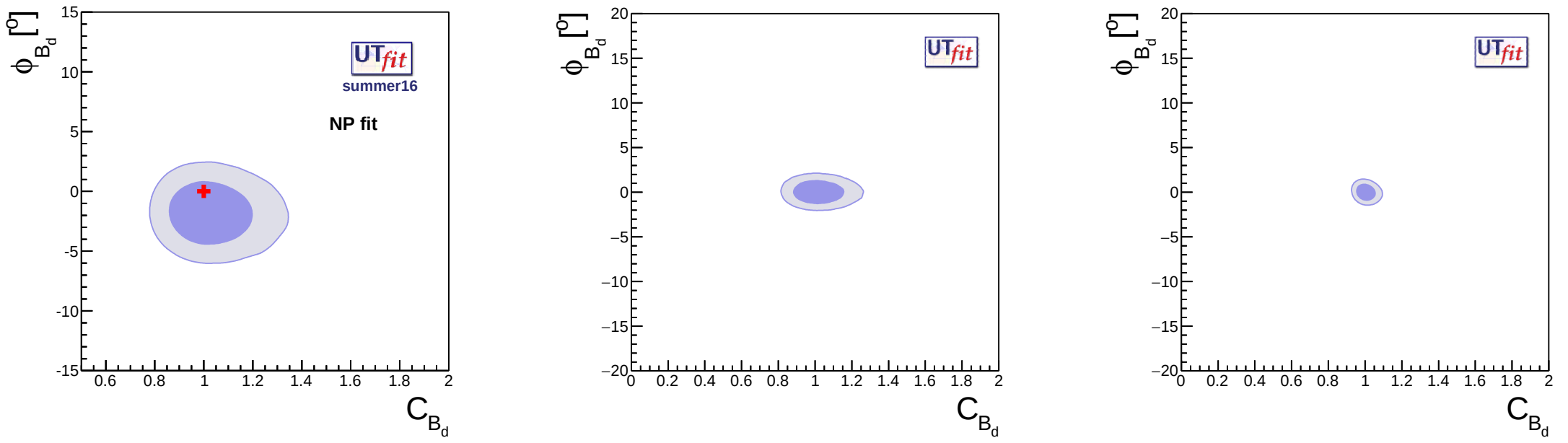

Fig. 230: Constraints on the NP parameters $C_{B_d}$ and $\phi_{B_d}$ today (left) and extrapolated to the 5 $ab^{-1}$ (centre) and 50 $ab^{-1}$ (right) scenarios.

Table 157: Extrapolated uncertainties on $\bar{\rho}$, $\bar{\eta}$ and the NP parameters $C_{B_d}$ and $\phi_{B_d}$ in the 5 and 50 $ab^{-1}$ scenarios. For comparison, we also report the uncertainties of the current fit.

| Parameter | Error | | |
|---|---|---|---|
| | current | $5ab^{-1}$ | $50ab^{-1}$ |
| $\bar{\rho}$ | $\pm0.027$ | $\pm0.008$ | $\pm0.006$ |
| $\bar{\eta}$ | $\pm0.025$ | $\pm0.009$ | $\pm0.007$ |
| $C_{B_d}$ | $\pm0.11$ | $\pm0.09$ | $\pm0.03$ |
| $\phi_{B_d}(^\circ)$ | $\pm1.7$ | $\pm0.8$ | $\pm0.6$ |

for $q = u$ and $c$, where the four-Fermi operators $\mathcal{O}_X$ are written as

$$\mathcal{O}_{V_1}^{(q,\nu_\ell)} = (\bar{q}\gamma^\mu P_L b)(\bar{\tau}\gamma_\mu P_L \nu_\ell)\,, \tag{624}$$

$$\mathcal{O}_{V_2}^{(q,\nu_\ell)} = (\bar{q}\gamma^\mu P_R b)(\bar{\tau}\gamma_\mu P_L \nu_\ell)\,, \tag{625}$$

$$\mathcal{O}_{S_1}^{(q,\nu_\ell)} = (\bar{q} P_R b)(\bar{\tau} P_L \nu_\ell)\,, \tag{626}$$

$$\mathcal{O}_{S_2}^{(q,\nu_\ell)} = (\bar{q} P_L b)(\bar{\tau} P_L \nu_\ell)\,, \tag{627}$$

$$\mathcal{O}_{T}^{(q,\nu_\ell)} = (\bar{q}\sigma^{\mu\nu} P_L b)(\bar{\tau}\sigma_{\mu\nu} P_L \nu_\ell)\,, \tag{628}$$

and $C_X$ denotes the Wilson coefficient of $\mathcal{O}_X$ normalised by $2\sqrt{2}G_F V_{qb}$. The SM contribution is presented as $\delta_{\nu_\tau,\nu_\ell}$ in Eq. (623). The superscript $(q,\nu_\ell)$ specifies the flavours of the quark $q$ and the neutrino $\nu_\ell$ in $b \to q\tau\nu_\ell$; $\mathcal{O}_X^{(c,\nu_\ell)}$ contributes to $\bar{B} \to D^{(*)}\tau\bar{\nu}$, whereas $\mathcal{O}_X^{(u,\nu_\ell)}$ to $\bar{B} \to \pi\tau\bar{\nu}$ and $\bar{B} \to \tau\bar{\nu}$. Note that it is not necessary that the neutrino flavour is the same as $\nu_\tau$ for new physics since it is not identified by the experiment. Eq. (623) is the most general form without considering the right-handed neutrinos.

In addition to each $V_1$, $V_2$, $S_1$, $S_2$, and $T$ scenario, specific scenarios, $C_{\mathrm{LQ}_{1,2}} \equiv C_{S_2} = \pm 4 C_T$, are also considered here. These specific combinations of the Wilson coefficients are realised in some leptoquark (LQ) models[71], e.g. see the $E^6$-inspired model in Sec. 17.6.5.

In the following subsection, we report measurable observables that can probe new physics in the processes and their potentials expected at Belle II.

*Ratio to the light-leptonic modes:.* As for the semi-tauonic $B$ decays, it is useful to define the ratios

$$R_{D^{(*)}} = \frac{\mathcal{B}(\bar{B} \to D^{(*)}\tau\bar{\nu})}{\mathcal{B}(\bar{B} \to D^{(*)}\ell\bar{\nu})}\,, R_\pi = \frac{\mathcal{B}(\bar{B} \to \pi\tau\bar{\nu})}{\mathcal{B}(\bar{B} \to \pi\ell\bar{\nu})}\,, \tag{629}$$

where $\ell = \mu$ or $e$, since the uncertainty in $|V_{qb}|$, which is dominant in the SM, is canceled out. These ratios also cancel out many experimental uncertainties The current experimental analyses result in $R_D^{\mathrm{ex}} = 0.397 \pm 0.040 \pm 0.028$ [230], $R_{D^*}^{\mathrm{ex}} = 0.316 \pm 0.016 \pm 0.010$ [230], and $R_\pi^{\mathrm{ex}} = 1.05 \pm 0.51$ [297]. On the other hand, the SM predicts $R_D^{\mathrm{sm}} = 0.305 \pm 0.012$, $R_{D^*}^{\mathrm{sm}} = 0.252 \pm 0.004$, and $R_\pi^{\mathrm{sm}} = 0.641 \pm 0.016$. Large deviations are seen in $R_{D^{(*)}}^{\mathrm{ex/sm}}$.Since it is expected that these observables will be measured with high accuracy at Belle II, they will ultimately become very powerful NP tests.

The dominant sources of theoretical uncertainty in the purely tauonic decay, $\bar{B} \to \tau\bar{\nu}$, are $f_B$ and $|V_{ub}|$. Then we have potentially two observables to reduce such uncertainties:

$$R_{\mathrm{ps}} = \frac{\tau_{B^0}}{\tau_{B^-}} \frac{\mathcal{B}(B^- \to \tau^-\bar{\nu}_\tau)}{\mathcal{B}(\bar{B}^0 \to \pi^+\ell^-\bar{\nu}_\ell)}\,, R_{\mathrm{pl}} = \frac{\mathcal{B}(B^- \to \tau^-\bar{\nu}_\tau)}{\mathcal{B}(B^- \to \mu^-\bar{\nu}_\mu)}\,. \tag{630}$$

The former has $\sim 10\%$ uncertainty [243], *e.g.*, $R_{\mathrm{ps}}^{\mathrm{sm}} = 0.574 \pm 0.046$ from $f_B$ and a form factor in $\bar{B} \to \pi\ell\bar{\nu}$, whereas the latter has a very accurate SM prediction, *e.g.*, $R_{\mathrm{pl}}^{\mathrm{sm}} = 222.36$. The experimental status is obtained as $R_{\mathrm{ps}}^{\mathrm{ex}} = 0.73 \pm 0.14$ while $R_{\mathrm{pl}}^{\mathrm{ex}}$ is not measured yet. These observables will be also good tools to test new physics scenarios in $b \to q\tau\nu$ at Belle II.

[71] To be precise, the relations are given as $C_{S_2} \simeq \pm 7.8 C_T$ at the $B$ meson scale while $C_{S_2} = \pm 4 C_T$ is obtained at the scale where the LQ model is defined ($\sim O(\mathrm{TeV})$).

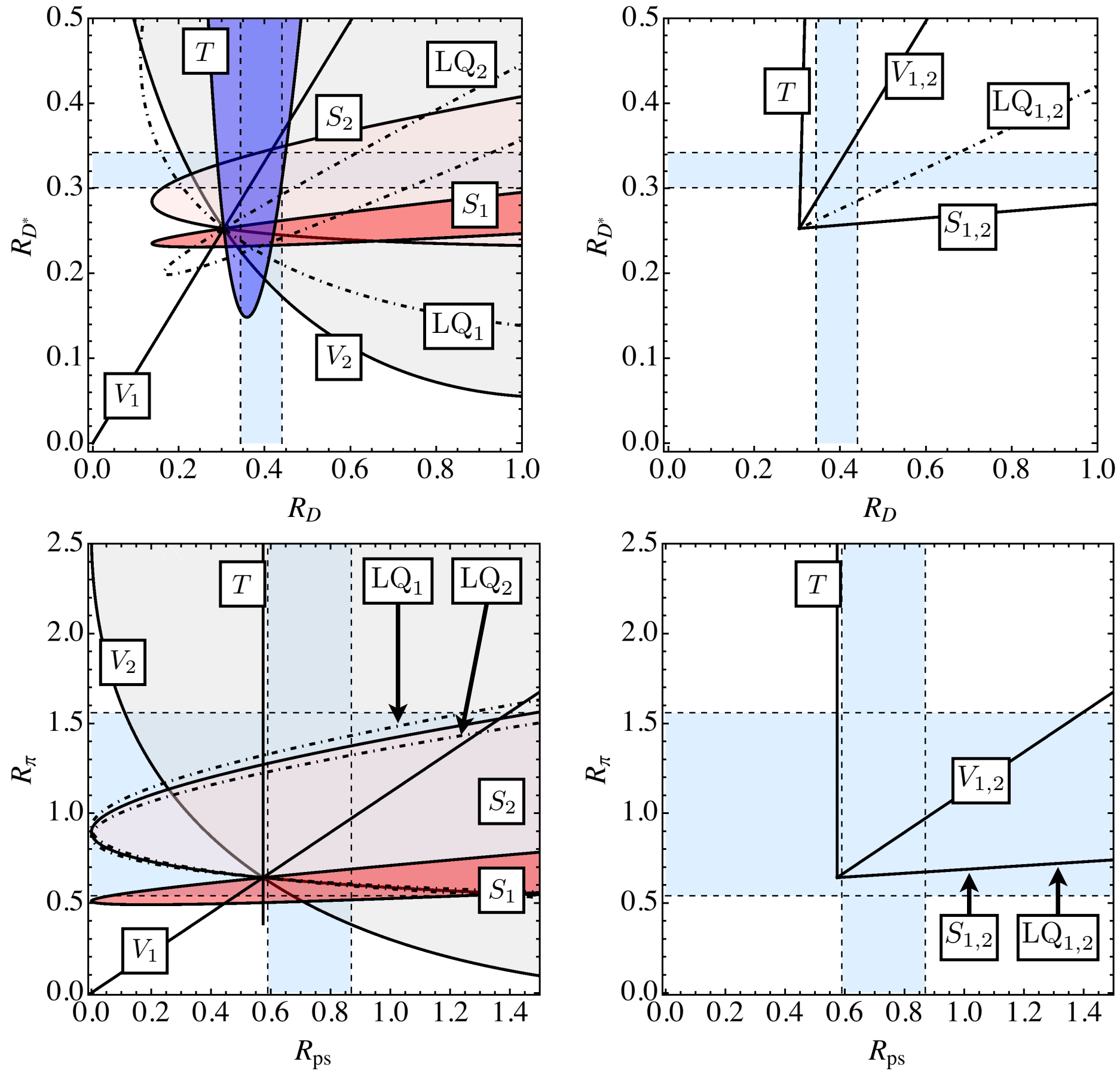

Fig. 231: (upper) Possible covered regions of $R_D$ and $R_{D^*}$ in the presence of one new physics operator of $\mathcal{O}_X^{(c,\nu_\tau)}$ and $\mathcal{O}_X^{(c,\nu_{\mu,e})}$ in the left and right panels, respectively. The boundaries for the LQ scenarios are also shown with dot-dashed lines. The light blue horizontal and vertical bands are the experimental values with $1\sigma$ ranges. (lower) Similar results for $R_{\rm ps}$ and $R_\pi$.

In the presence of one NP operator $\mathcal{O}_X^{(c,\nu_\ell)}$, $R_D$ and $R_{D^*}$ are correlated via the shared Wilson coefficient $C_X^{(c,\nu_\ell)}$. In Fig. 231, we show possible regions of $R_D$ and $R_{D^*}$. Each shaded region can be obtained by each NP operator (as indicated in the figure) with some value of $C_X^{(c,\nu_\ell)}$. We also show regions for the two LQ scenarios with dot-dashed boundaries. We can see that $R_D$ is sensitive to the scalars whereas $R_{D^*}$ to the tensor as it is reflecting spin properties of the charmed mesons. Thus precisely measuring these two ratios may provide us a hint of the type of existing NP operator if (one of) the measured values are deviated from the SM predictions.

A similar conclusion can be seen for the correlation between $R_\pi$ and $R_{\rm ps}$ in the presence of $\mathcal{O}_X^{(u,\nu_\ell)}$. For the tensor scenario, $R_{\rm ps}$ is fixed as the SM value since the tensor current does not contribute to $\bar{B} \to \tau\bar{\nu}$.

Next, we discuss a maximum reach of limits on NP contributions $C_X^{(q,\nu_\tau)}$ at Belle II. The reach can be evaluated by assuming *reference central values of experimental data to be the same as the SM values* and taking theoretical and expected uncertainties into account. For the evaluations, we have taken the NP contributions to be real and omitted regions which do not include the SM points ($C_X^{(q,\nu_\tau)} = 0$).

Table 158: 95% CL expected limits of the new physics contributions $C_X$ obtained by measuring $R_{D^{(*)}}$ at Belle II ($5\,\text{ab}^{-1}$, $50\,\text{ab}^{-1}$). The reference values of experimental data are given in the main text. The NP contribution $C_X$ is assumed to be real and the ranges which include the SM point ($C_X = 0$) are shown.

| | $R_D$ | | $R_{D^*}$ | |
|---|---|---|---|---|
| NP scenario | Belle II ($5\,\text{ab}^{-1}$) | Belle II ($50\,\text{ab}^{-1}$) | Belle II ($5\,\text{ab}^{-1}$) | Belle II ($50\,\text{ab}^{-1}$) |
| $C_{V_1}$ | $[-0.08, 0.09]$ | $[-0.05, 0.06]$ | $[-0.04, 0.04]$ | $[-0.02, 0.02]$ |
| $C_{V_2}$ | $[-0.08, 0.09]$ | $[-0.05, 0.06]$ | $[-0.04, 0.04]$ | $[-0.02, 0.02]$ |
| $C_{S_1}$ | $[-0.12, 0.12]$ | $[-0.07, 0.08]$ | $[-0.82, 0.52]$ | $[-0.26, 0.26]$ |
| $C_{S_2}$ | $[-0.12, 0.12]$ | $[-0.07, 0.08]$ | $[-0.52, 0.82]$ | $[-0.26, 0.26]$ |
| $C_T$ | $[-0.21, 0.17]$ | $[-0.12, 0.11]$ | $[-0.01, 0.02]$ | $[-0.007, 0.007]$ |
| $C_{\text{LQ}_1}$ | $[-0.11, 0.11]$ | $[-0.06, 0.07]$ | $[-0.10, 0.10]$ | $[-0.04, 0.04]$ |
| $C_{\text{LQ}_2}$ | $[-0.13, 0.13]$ | $[-0.08, 0.08]$ | $[-0.18, 0.14]$ | $[-0.07, 0.06]$ |

Table 159: 95% CL expected limits of the new physics contributions $C_X$ obtained by measuring $R_\pi$ and $R_{\text{ps}}$ at Belle II ($5\,\text{ab}^{-1}$, $50\,\text{ab}^{-1}$). The reference values of experimental data are given in the main text. The NP contribution $C_X$ is assumed to be real and the ranges which include the SM point ($C_X = 0$) are shown.

| | $R_\pi$ | | $R_{\text{ps}}$ | |
|---|---|---|---|---|
| NP scenario | Belle II ($5\,\text{ab}^{-1}$) | Belle II ($50\,\text{ab}^{-1}$) | Belle II ($5\,\text{ab}^{-1}$) | Belle II ($50\,\text{ab}^{-1}$) |
| $C_{V_1}$ | $[-0.45, 0.30]$ | $[-0.12, 0.11]$ | $[-0.13, 0.14]$ | $[-0.08, 0.10]$ |
| $C_{V_2}$ | $[-0.45, 0.30]$ | $[-0.12, 0.11]$ | $[-0.14, 0.13]$ | $[-0.10, 0.08]$ |
| $C_{S_1}$ | $[-1.26, 0.42]$ | $[-0.32, 0.17]$ | $[-0.03, 0.04]$ | $[-0.02, 0.03]$ |
| $C_{S_2}$ | $[-1.26, 0.42]$ | $[-0.32, 0.17]$ | $[-0.04, 0.03]$ | $[-0.03, 0.02]$ |
| $C_T$ | $[-1.30, 0.26]$ | $[-0.13, 0.10]$ | - | - |
| $C_{\text{LQ}_1}$ | $[-1.34, 0.40]$ | $[-0.23, 0.16]$ | $[-0.04, 0.03]$ | $[-0.03, 0.02]$ |
| $C_{\text{LQ}_2}$ | $[-1.18, 0.45]$ | $[-0.93, 0.19]$ | $[-0.04, 0.03]$ | $[-0.03, 0.02]$ |

In Table 158, we show the expected limits from $R_{D^{(*)}}$ obtained at the early and final stages of Belle II ($5\,\text{ab}^{-1}$, $50\,\text{ab}^{-1}$). We can make sure that the scalar scenarios are sensitive to $R_D$ whereas the tensor scenario to $R_{D^*}$ as is already mentioned above. At the early stage we see that the $V_{1,2}$ scenario with larger than 3% of the SM contribution, $|C_{V_{1,2}}| \gtrsim 0.03$, can be tested. Similarly, the NP scenarios with $|C_{S_{1,2}, \text{LQ}_{1,2}}| \gtrsim 0.07$–$0.08$ and $|C_T| \gtrsim 0.01$ can be examined. The limits will be further improved at the final stage. The expected ranges, however, are reduced only by half at most as the theoretical uncertainties in $R_{D^{(*)}}$ become dominant. Thus further precise evaluations of the form factors are necessary to exploit the maximum potential of Belle II for new physics in $R_{D^{(*)}}$.

In Table 159, we show the expected limits from $R_\pi$ and $R_{\text{ps}}$ obtained at Belle II with $5\,\text{ab}^{-1}$ and $50\,\text{ab}^{-1}$. At the early stage of Belle II, $R_\pi$ provides loose constraints on the NP contributions while the exclusion limits from $R_{\text{ps}}$ can be $|C_{V_{1,2}}| \gtrsim 0.14$ for the $V_{1,2}$ scenarios and $|C_{S_{1,2}}| \gtrsim 0.04$ for the $S_{1,2}$ scenarios. On the other hand, the $T$ scenario is constrained only from $R_\pi$ since the tensor operator does not contribute to $R_{\rm ps}$. Thus the $\mathrm{LQ}_{1,2}$ scenarios, the combination of $S_2$ and $T$, have the same contribution with $S_2$ for $R_{\rm ps}$. At the stage of Belle II with $50\,\mathrm{ab}^{-1}$, $|C_{V_{1,2}}| \gtrsim 0.1$ can be obtained from both $R_\pi$ and $R_{\rm ps}$. As for the $S_{1,2}$ and $\mathrm{LQ}_{1,2}$ scenarios, the limit is slightly improved as $|C_{S_{1,2}}| \gtrsim 0.03$ for $R_{\rm ps}$. Finally the $T$ scenario with $|C_T| \gtrsim 0.1$ can be tested by measuring $R_\pi$.

Table 160: Maximum $p$ values for the NP scenarios obtained from the fit to the BaBar $q^2$ distribution.

| Model | $\bar{B} \to D\tau\bar{\nu}$ | $\bar{B} \to D^*\tau\bar{\nu}$ | $\bar{B} \to (D + D^*)\tau\bar{\nu}$ |
|---|---|---|---|
| SM | 54% | 65% | 67% |
| $V_1$ | 54% | 65% | 67% |
| $V_2$ | 54% | 65% | 67% |
| $S_2$ | 0.02% | 37% | 0.1% |
| $T$ | 58% | 0.1% | 1.0% |
| $\mathrm{LQ}_1$ | 13% | 58% | 25% |
| $\mathrm{LQ}_2$ | 21% | 72% | 42% |

The ratio of purely leptonic decays, $R_{\rm pl}$, also provides exclusion limits of $C_X$. It is compared with $R_{\rm ps}$ and then it turns out that the sensitivity of $R_{\rm pl}$ is factor 2 weaker than that of $R_{\rm ps}$ [243]. Nevertheless, $R_{\rm pl}$ is good observable in the sense that it has the very accurate theoretical prediction and could be used as a consistency check.

To conclude, measuring the ratios, $R_D$, $R_{D^*}$, $R_\pi$, $R_{\rm ps}$, and $R_{\rm pl}$ at Belle II can probe new physics with the contributions up to $O(1\text{–}10\%)$ of the SM values.

*Distributions.* Besides integrated quantities such as the ratios shown above, several distributions and asymmetries are measurable in the semi-tauonic decays. As for $\bar{B} \to D^{(*)}\tau\bar{\nu}$, a variety of such observables has been proposed to test NP scenarios in the literature [225, 254–257, 264, 265, 276, 284, 296, 883, 884, 2077–2083]. Among them, the distribution of $q^2 = (p_B - p_{D^{(*)}})^2$ has been already analysed [251, 270] and thus is expected to be measured at a relatively early stage of Belle II, compared with the other observables. Below we illustrate potential of the $q^2$ distribution in $\bar{B} \to D^{(*)}\tau\bar{\nu}$ at Belle II for discriminating the NP scenarios.

In Ref. [270], BaBar measured background-subtracted and efficiency-corrected $q^2$ distributions for signal events of $\bar{B} \to D^{(*)}\tau\bar{\nu}$. Comparing it with those for the NP scenarios, we obtain the $p$ values as shown in Table 160. One finds that the $S_2$ and $T$ scenarios are disfavoured by the observed $q^2$ data while the others (including the SM) have larger (but not significant) $p$ values. This is totally different from what is obtained from the integrated quantities $R_{D^{(*)}}$. However, we should note that the given $q^2$ data from BaBar does not include systematic errors and the normalisations of the data are left as free parameter of the fit. Therefore, the results in the table are not conclusive although we can see that the $q^2$ distributions are useful to test the NP scenarios.

Table 161: Luminosity required to discriminate various simulated "data" and tested model at 99.9% C.L. using $R_{D^{(*)}}(q^2)$ (or $R_{D^{(*)}}$ in parentheses). (–) indicates that it is impossible to discriminate data and model.

| $\mathcal{L}$ [fb$^{-1}$] | | Model | | | | | | |
|---|---|---|---|---|---|---|---|---|
| | | SM | $V_1$ | $V_2$ | $S_2$ | $T$ | LQ$_1$ | LQ$_2$ |
| Data | $V_1$ | 1170 | | $10^6$ | 500 | 900 | 4140 | 2860 |
| | | (270) | | (–) | (–) | (–) | (–) | (1390) |
| | $V_2$ | 1140 | $10^6$ | | 510 | 910 | 4210 | 3370 |
| | | (270) | (–) | | (–) | (–) | (–) | (1960) |
| | $S_2$ | 560 | 560 | 540 | | 380 | 1310 | 730 |
| | | (290) | (13750) | (36450) | | (–) | (35720) | (4720) |
| | $T$ | 600 | 680 | 700 | 320 | | 620 | 550 |
| | | (270) | (–) | (–) | (–) | | (–) | (1980) |
| | LQ$_1$ | 1010 | 4820 | 4650 | 1510 | 800 | | 5920 |
| | | (270) | (–) | (–) | (–) | (–) | | (1940) |
| | LQ$_2$ | 1020 | 3420 | 3990 | 1040 | 650 | 5930 | |
| | | (250) | (1320) | (1820) | (20560) | (4110) | (1860) | |

The above analysis will be improved as data is accumulated at Belle II. A discriminative potential of $q^2$ distributions is discussed with the use of the following quantities [296]:

$$R_{D^{(*)}}(q^2) \equiv \frac{d\mathcal{B}(\bar{B} \to D^{(*)}\tau\bar{\nu})/dq^2}{d\mathcal{B}(\bar{B} \to D^{(*)}\ell\bar{\nu})/dq^2} N_{D^{(*)}}(q^2)\,, \tag{631}$$

where the normalisation $N_{D^{(*)}}(q^2)$ (to avoid rapid suppression of the phase spaces at $q^2 = m_\tau^2$) is defined in Ref. [296]. Here, we consider to see whether we can distinguish the NP scenarios by measuring $R_{D^{(*)}}(q^2)$ in the case that the present status of the anomalies on the integrated quantities, $R_{D^{(*)}}$, remains in future. In order to see this, we simulate "experimental data" for $R_{D^{(*)}}(q^2)$ assuming *one of the NP scenarios that can explain the present values of* $R_{D^{(*)}}$ and compare them with other NP scenarios. The $q^2$ distributions are binned as given in the BaBar hadronic tag analysis. Statistical uncertainties in each bin of $R_{D^{(*)}}(q_i^2)$ are approximately described by

$$\delta^{\text{stat}} R_{D^{(*)}}(q_i^2) \sim \frac{1}{\sqrt{N_{B\bar{B}}\epsilon_i^\tau}} \frac{\sqrt{\mathcal{B}_i^\tau}}{\mathcal{B}_i^\ell} N_{D^{(*)}}(q_i^2)\,, \tag{632}$$

where $N_{B\bar{B}} = \mathcal{L} \times \sigma(e^+e^- \to B\bar{B})$ is the number of produced $B\bar{B}$ pairs for an integrated luminosity, $\mathcal{B}_i^{\tau(\ell)}$ are the partial branching ratios of $\bar{B} \to D^{(*)}\tau\bar{\nu}$ ($\bar{B} \to D^{(*)}\ell\bar{\nu}$) for the $i$-th bin, and $\epsilon_i^\tau$ denotes the efficiency for the signal process $\bar{B} \to D^{(*)}\tau\bar{\nu}$. In the following test, the total experimental uncertainty (including systematic one) is assumed as $\delta^{\text{exp}} R_{D^{(*)}}(q_i^2) = 2\delta^{\text{stat}} R_{D^{(*)}}(q_i^2)$ and then the efficiency is taken universally as $\epsilon_i^\tau = 10^{-4}$. Theoretical uncertainties are correlated between the $q^2$ bins and then taken as appropriate.

Given the above setup, we evaluate required luminosities so that we can discriminate simulated data and the NP scenarios by measuring $R_{D^{(*)}}(q^2)$ at 99.9% CL. The result is shown in Table 161. As a comparison, we also show results obtained by measuring $R_{D^{(*)}}$ in the parentheses. One can see that some cases of "data"-model, such as "$S_2$"-$T$ or "$S_2$"-$V_{1,2}$, require only $\sim 500\,\mathrm{fb}^{-1}$ and thus can be already tested by $R_{D^{(*)}}(q^2)$ using the present data while it is not the case for $R_{D^{(*)}}$. One also finds that we need $1-6\,\mathrm{ab}^{-1}$ to test the LQ scenarios, which will be achieved at the early stage of Belle II as is pointed out. Interestingly, discriminative potential of $R_{D^{(*)}}(q^2)$ is different from and complementary to that of $R_{D^{(*)}}$ as is shown in the table.

*Connection to collider physics.* Here, we illustrate how the $R_{D^{(*)}}$ anomalies can be examined by the high-energy experiment at the 14 TeV LHC. A comprehensive study of such a test should be done for every specific model and is beyond the scope of this report. Instead that, we show a LHC study for a scalar leptoquark model, which leads contributions with the form of $C_{\mathrm{LQ}_2}$, as an example. The minimum requirement to accommodate the $R_{D^{(*)}}$ anomalies for this model is given in the Lagrangian as

$$\mathcal{L}_{S_1} = \left( g_{1L}^{33} \bar{Q}_L^{c,3} (i\sigma_2) L_L^3 + g_{1R}^{23} \bar{u}^{c,2} \ell_R^3 \right) S_1 \,, \tag{633}$$

where $S_1$ is a $SU(2)_L$ singlet scalar leptoquark, $Q^{c,3} = (t^c\ b^c)^T$, $L^3 = (\nu_\tau\ \tau)^T$, $u^{c,2} = c^c$, and $\ell^3 = \tau$. The contribution of $C_{\mathrm{LQ}_2}$ is presented by

$$2\sqrt{2} G_F V_{cb}\, C_{\mathrm{LQ}_2} = -\frac{g_{1L}^{33}\, g_{1R}^{23*}}{M_{S_1}^2} \,, \tag{634}$$

and then the central values of the present anomalies can be explained with $C_{\mathrm{LQ}_2} \simeq 0.26$. In this setup, one can see that the scalar LQ boson decays only into $S_1 \to c\tau$, $b\nu$, and $t\tau$. Since leptoquark bosons are dominantly pair-produced due to QCD interaction at the LHC, there are six possible final states. All of them are worth analysing at ATLAS and CMS to probe an evidence of the $R_{D^{(*)}}$ anomalies measured at Belle, BaBar, and LHCb.

In Ref. [2084], two final states, $(b\nu)(\bar{b}\bar{\nu})$ and $(c\tau)(\bar{c}\bar{\tau})$, have been studied in great detail by doing numerical simulations with adopting optimised cut analyses. To see a connection between LHC direct searches and the $R_{D^{(*)}}$ anomalies, it is assumed that $g_{1L}^{33}$ and $g_{1R}^{23}$ are related keeping $C_{\mathrm{LQ}_2} = 0.26$.

As a result, the 95% CL current and expected excluded regions in the $(g_{1L}^{33}, M_{S_1})$ plane are shown in Fig. 232. The shaded regions in blue-cyan and red are excluded at 95% CL by the current 8 TeV data of $(b\nu)(\bar{b}\bar{\nu})$ and $(c\tau)(\bar{c}\bar{\tau})$, respectively[72]. The blue and red curves show the 95% CL exclusion limits from $(b\nu)(\bar{b}\bar{\nu})$ and $(c\tau)(\bar{c}\bar{\tau})$ obtained at $\mathcal{L} = 300\,\mathrm{fb}^{-1}$ of the 14 TeV LHC. Note that for the $(c\tau)(\bar{c}\bar{\tau})$ analysis, the realistic values of tagging/mistagging efficiencies of $c$-jet are taken into account [2089]. Through the condition $C_{\mathrm{LQ}_2} = 0.26$ required by the $R_{D^{(*)}}$ anomalies, both the searches of $(b\nu)(\bar{b}\bar{\nu})$ and $(c\tau)(\bar{c}\bar{\tau})$ can constrain the model parameters ($g_{1L}^{33}$, $g_{1R}^{23}$, and $M_{S_1}$) and then one can see from the figure that we can probe the $S_1$ leptoquark up to at least $M_{S_1} = 0.8$ TeV at 14 TeV LHC.

Similar correlations between the collider study and the flavour anomalies should be functional in other specific models. It will enable us to improve searching for new physics that explain the $R_{D^{(*)}}$ anomalies and that may exist in $R_\pi$ and $R_{\mathrm{ps}}$.

[72] As for the blue and cyan regions, both translated bound from the $(b\tilde{\chi}_1^0)(\bar{b}\tilde{\chi}_1^0)$ searches [2085, 2086] and direct bound from the $(b\nu)(\bar{b}\bar{\nu})$ searches [2086, 2087] are taken into account. As for the red region, the CMS search for $(b\tau)(\bar{b}\bar{\tau})$ [2088] is recast as a bound for $(c\tau)(\bar{c}\bar{\tau})$ by evaluating mis-identification of $c$-jet as $b$-jet.

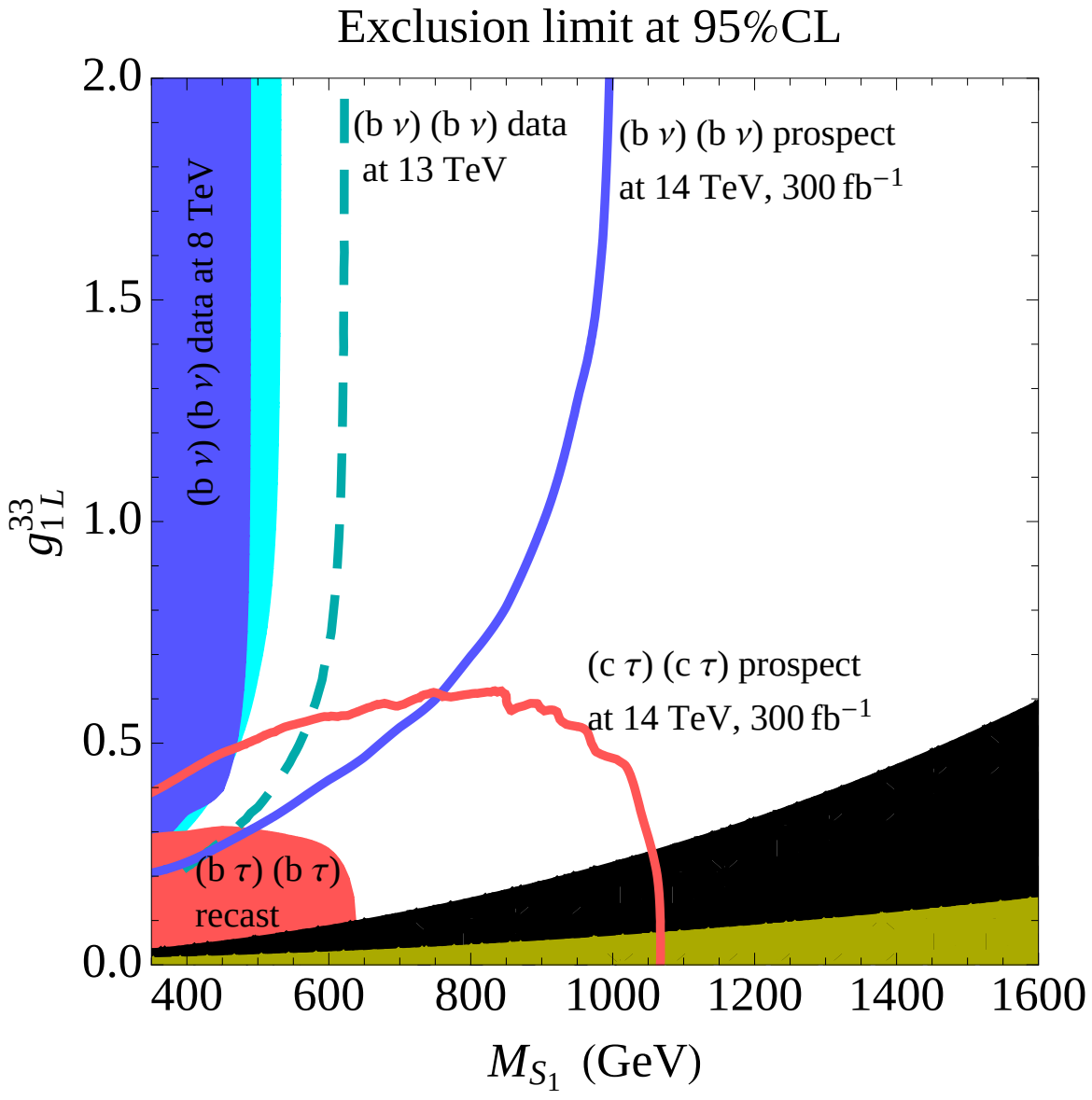

Fig. 232: Current excluded regions obtained from 8-13 TeV data and exclusion limits expected at $\mathcal{L} = 300\,\text{fb}^{-1}$ of the 14 TeV LHC in the $(g_{1L}^{33}, M_{S_1})$ plane at 95% CL. The black region indicates the area for $\Gamma_{S_1}/M_{S_1} > 0.2$ in which a narrow width approximation becomes invalid. The dark-yellow region is theoretically unacceptable due to $g_{1R}^{23} > 4\pi$.

#### *18.3.2. Loop-level decays.* *(Contributing author: Wolfgang Altmannshofer)*

*Theoretical Framework.* We extend the discussion of the effective Hamiltonian in section 9.1.1 to the most generic new physics scenario including the violation of lepton universality and/or lepton flavour. The effective Hamiltonian that enables the model independent studies of the leptonic decays $B^0 \to \ell^+\ell^-$ and $B_s \to \ell^+\ell^-$ as well as semileptonic transitions of the type $b \to d\ell\ell$, $b \to s\ell\ell$, $b \to d\nu\nu$ and $b \to s\nu\nu$ can be written as

$$\mathcal{H}_{\text{eff}} = \mathcal{H}_{\text{eff}}^{\text{SM}} - \frac{4G_F}{\sqrt{2}} V_{tq}^* V_{tb} \sum_i C_i^{\text{NP}} \mathcal{O}_i \,, \tag{635}$$

where $\mathcal{H}_{\text{eff}}^{\text{SM}}$ is the effective Hamiltonian of the SM, $C_i^{\text{NP}}$ are the Wilson coefficients encoding the effect of new physics and $\mathcal{O}_i$ are dimension 6 operators built from light SM particles.[73] Following the notation of [2090], the most relevant operators are dipole operators

$$(\mathcal{O}_7^{(\prime)})_q = \frac{e}{16\pi^2} m_b (\bar{q}\sigma^{\mu\nu} P_{R(L)} b) F_{\mu\nu} \,, \tag{636}$$

[73] Note that it is far from established if the SM particles are the only dynamical degrees of freedom below the electro-weak scale. If new light particles interact sufficiently weak with the SM, they can evade direct detection. Examples are axions, light Higgs particles, light dark matter, sterile neutrinos and dark photons. If such new degrees of freedom are lighter than $B$ mesons, novel exotic decay modes of $B$ mesons can open up that are not described by the effective Hamiltonian formalism but require the explicit addition of light new particles to the SM [616]. Exotic signatures include the decays of $B$ mesons into invisible particles or resonances in the di-lepton invariant mass spectra [1797].

$$(\mathcal{O}_8^{(\prime)})_q = \frac{g_s}{16\pi^2} m_b (\bar{q}\sigma^{\mu\nu} T^a P_{R(L)} b) G^a_{\mu\nu} \,, \tag{637}$$

as well as various semi-leptonic 4 fermion contact interactions

$$(\mathcal{O}_9^{(\prime)})_q^\ell = \frac{e^2}{16\pi^2} (\bar{q}\gamma_\mu P_{L(R)} b)(\bar{\ell}\gamma^\mu \ell) \,, \tag{638}$$

$$(\mathcal{O}_{10}^{(\prime)})_q^\ell = \frac{e^2}{16\pi^2} (\bar{q}\gamma_\mu P_{L(R)} b)(\bar{\ell}\gamma^\mu \gamma_5 \ell) \,, \tag{639}$$

$$(\mathcal{O}_\nu^{(\prime)})_q^\ell = \frac{e^2}{8\pi^2} (\bar{q}\gamma_\mu P_{L(R)} b)(\bar{\nu}_\ell \gamma^\mu P_L \nu_\ell) \,, \tag{640}$$

$$(\mathcal{O}_S^{(\prime)})_q^\ell = \frac{e^2}{16\pi^2} (\bar{q} P_{R(L)} b)(\bar{\ell}\ell) \,, \tag{641}$$

$$(\mathcal{O}_P^{(\prime)})_q^\ell = \frac{e^2}{16\pi^2} (\bar{q} P_{R(L)} b)(\bar{\ell}\gamma_5 \ell) \,, \tag{642}$$

$$(\mathcal{O}_T^{(\prime)})_q^\ell = \frac{e^2}{16\pi^2} (\bar{q}\sigma_{\mu\nu} P_{L(R)} b)(\bar{\ell}\sigma^{\mu\nu} P_{L(R)} \ell) \,, \tag{643}$$

where $\ell = e, \mu, \tau$. We explicitly allow for lepton universality violation. In the presence of lepton flavour violation, the semi-leptonic contact interactions can also contain leptons of different flavour, e.g.

$$(\mathcal{O}_9^{(\prime)})_q^{\ell\ell'} = \frac{e^2}{16\pi^2} (\bar{q}\gamma_\mu P_{L(R)} b)(\bar{\ell}\gamma^\mu \ell') \,, \tag{644}$$

and analogously for the other 4 fermion interactions.

Using the experimental result on leptonic and semi-leptonic $B$ decays, allowed ranges for the new physics Wilson coefficients can be determined in a model independent fashion [493, 497, 498, 524, 592, 612, 647, 1963, 2091–2104]. Reparameterising the effective Hamiltonian in the following way

$$\mathcal{H}_{\text{eff}} = \mathcal{H}_{\text{eff}}^{\text{SM}} - \sum_i \frac{1}{\Lambda_i^2} \mathcal{O}_i \,, \tag{645}$$

allows to translate information on the new physics Wilson coefficients $C_i^{\text{NP}}$ defined in Eq. (635) into constraints on the new physics scales $\Lambda_i$ that suppress the dimension 6 operators $\mathcal{O}_i$ [74].

In many classes of new physics, relations exists between the Wilson coefficients of the above operators. For example, in models with Minimal Flavour Violation (MFV) [621, 1867] and in models with a minimally broken $U(2)^3$ flavour symmetry [1934], the dominant Wilson coefficients are universal for down and strange quarks $(C_i)_d = (C_i)_s$. In models that do not contain any sources of lepton flavour universality violation beyond the SM Yukawa couplings, one has for instance $(C_i^{(\prime)})_q^e = (C_i^{(\prime)})_q^\mu = (C_i^{(\prime)})_q^\tau$ for $i = 9, 10, \nu$ while the scalar, pseudoscalar and tensor operators scale with the mass of the involved leptons, $(C_i^{(\prime)})_q^e / m_e = (C_i^{(\prime)})_q^\mu / m_\mu = (C_i^{(\prime)})_q^\tau / m_\tau$ for $i = S, P, T$. If new physics is heavy compared to the electro-weak scale and it can be described by the SM effective field theory (SMEFT) with a linearly realised Higgs boson [620] one finds relations among the scalar and pseudoscalar operators $(C_S)_q^\ell = -(C_P)_q^\ell$, $(C'_S)_q^\ell = (C'_P)_q^\ell$ and vanishing tensor operators $(C_T^{(\prime)})_q^\ell = 0$ [622][75]. Moreover, due to $SU(2)_L$ invariance, the operators involving neutrinos and left-handed charged leptons are related. At the level of $SU(2)_L$ invariant dimension 6 operators one has $(C_\nu^{(\prime)})_q = ((C_9^{(\prime)})_q - (C_{10}^{(\prime)})_q)/2$.

[74] See [2105] for a complementary discussion in the context of "simplified models"

[75] These relations can be violated in the presence of non-standard dynamics triggering electroweak symmetry breaking [2106].

*Leptonic Decays.* The leptonic decays $B^0 \to \ell^+\ell^-$ and $B_s \to \ell^+\ell^-$, with $\ell = e, \mu, \tau$ are very well known to be highly sensitive probes of new physics.

The existing measurements from LHCb, CMS and ATLAS of the branching ratios of the muonic decays $B^0 \to \mu^+\mu^-$ and $B_s \to \mu^+\mu^-$ can be interpreted in a model independent way as constraints on the Wilson coefficients of the operators $(\mathcal{O}_{10,S,P}^{(\prime)})_s^\mu$ [2097]. Under the assumption that one operator dominates, the corresponding constraints on the new physics scale $\Lambda$ are at the level of $\Lambda \gtrsim 30$ TeV for $(\mathcal{O}_{10}^{(\prime)})_s^\mu$ and $\Lambda \gtrsim 100$ TeV in the case the scalar and pseudoscalar operators $(\mathcal{O}_S^{(\prime)})_s^\mu$, $(\mathcal{O}_P^{(\prime)})_s^\mu$. The constraints on the operators involving the down quark instead of the strange quark are slightly stronger. These bounds demonstrate the exquisite sensitivity of these rare $B$ meson decays to new physics.

Combining the $B_s \to \mu^+\mu^-$ results with measurements of the decay rate and angular distribution of the semi-leptonic decay $B \to K\mu^+\mu^-$, allows to lift flat directions in new physics parameter space that can appear when more than one operators is considered simultaneously [2107].

In the absence of BSM sources of lepton flavour universality violation, the branching ratios of the leptonic decays scale with the lepton masses squared. Given the current and foreseeable experimental sensitivities, the di-muon decays are therefore the most sensitive probes of new physics in such scenarios. Searches for the di-electron and di-tau decays $B \to e^+e^-$ and $B \to \tau^+\tau^-$ are sensitive probes of new sources of lepton flavour universality violation and well motivated [537].

*Exclusive Semileptonic Decays.* Exclusive semileptonic decays like $B \to K^{(*)}\ell^+\ell^-$ and $B_s \to \phi\ell^+\ell^-$ give access to a plethora of observables. Beyond the corresponding decay rates, these decays offer for example angular distributions, $CP$ asymmetries and lepton flavour universality tests that all can be used to probe the SM and its extensions.

The differential decay distribution of the $B \to K\ell^+\ell^-$ decays as function of the angle $\theta_\ell$, defined as the angle between the direction of the $\ell^+$ and the direction of the $B$ in the dilepton rest-frame, reads [590]

$$\frac{d\Gamma}{dz} \propto \frac{3}{4}(1 - F_H)(1 - z^2) + \frac{1}{2}F_H + A_{\rm FB}z \ , \tag{646}$$

with $z = \cos\theta_\ell$. The forward-backward asymmetry $A_{\rm FB}$ and the flat-term $F_H$ are functions of the di-lepton invariant mass $q^2$. They are powerful probes of the scalar and tensor interactions $(\mathcal{O}_{S,P,T}^{(\prime)})_s^\ell$. Combined with the measurement of the branching ratio of $B_s \to \mu^+\mu^-$, existing data allows to constrain the corresponding complex-valued Wilson coefficients involving muons in one fit [2102].

The differential decay distribution of the $B \to K^*\ell^+\ell^-$ decays is more complex and involves three angles (see e.g. [588])

$$\frac{d^3\Gamma}{d\cos\theta_\ell d\cos\theta_K d\phi} = \frac{9}{32\pi}\sum_i I_i f_i(\theta_\ell, \theta_K, \phi) \ , \tag{647}$$

with angular coefficients $I_i$ that depend on the di-lepton invariant mass $q^2$. Analogous distributions describe the decays $B_s \to \phi\ell^+\ell^-$ and $B^0 \to \rho\ell^+\ell^-$. Many useful observables can be constructed out of the angular coefficients both at low $q^2$ and at high $q^2$, e.g. the $CP$ averaged angular coefficients $S_i$ [588], $CP$ asymmetries $A_i$ [2108] and the $P_i'$ observables [593]. Many of these observables are sensitive probes of new physics in the dipole-operators $(\mathcal{O}_7^{(\prime)})_q$ and the four fermion contact interactions $(\mathcal{O}_{9,10}^{(\prime)})_q^\ell$.

The $B \to K^* e^+ e^-$ decay provides also theoretically clean observables in the very low $q^2$ region $4m_e^2 < q^2 < 1$ GeV$^2$ [490, 507] that allow for interesting tests of new physics in the dipole operators $(\mathcal{O}_7^{(\prime)})_s$.

Among the various $CP$ violating observables there are the direct $CP$ asymmetries in $B \to K^{(*)}\ell^+\ell^-$ decays as well as $CP$ asymmetries of the angular coefficients in $B \to K^*\ell^+\ell^-$. Particularly interesting are the three T-odd angular observables $A_7$, $A_8$ and $A_9$ as they are not suppressed by small strong phases and therefore could be of $O(1)$ in the presence of $CP$ violating new physics [2108]. These $CP$ violating observables nearly vanish in the SM with very small uncertainties even in the presence of non-factorizable long distance effects. Observation of non-zero $CP$ asymmetries in $b \to s\ell^+\ell^-$ decays would be a clear signature of new physics. In the absence of a non-zero signal, precise measurements of the $CP$ asymmetries $A_{7,8,9}$ can provide important bounds on BSM sources of $CP$ violation in the form of imaginary parts of the $(C_9^{(\prime)})_q^\ell$ and $(C_{10}^{(\prime)})_q^\ell$ Wilson coefficients, that are still only weakly constrained at the moment. Interesting $CP$ observables can also be extracted from time-integrated and time-dependent analyses of decays into $CP$ eigenstates like $B^0 \to K^*(\to K_S^0\pi^0)\ell^+\ell^-$ [2109].

Already existing measurements of decay rates and angular observables in the exclusive semileptonic decays based on the $b \to s\mu\mu$ transition show an intriguing pattern of deviations from SM predictions that consistently point to non-standard effects in a single operator $(\mathcal{O}_9)_s^\mu$ [497, 647]. Assuming that hadronic effects are estimated in a sufficiently conservative way, the latest global fits [498, 524, 612, 2103] find preference for a new physics contribution $(C_9)_s^\mu \simeq -1$ at the level of $\sim 4\sigma$. Translated into a new physics scale in the effective Hamiltonian (645), this corresponds to $\Lambda \simeq 35$ TeV, a scale that is not far above the direct reach of future high energy colliders. However, unexpectedly large long distance effects can not be excluded as an explanation for the apparent discrepancies at this time [491, 2110].

Very interesting in this context are lepton flavour universality tests where hadronic effects cancel to a very high precision and the SM predictions are robust with accuracies at the 1% level or better [597]. Lepton flavour universality tests include ratios of branching ratios involving muons and electrons in the final state [598, 2111]

$$R_{K^{(*)}} = \frac{\mathrm{BR}(B \to K^{(*)}\mu\mu)}{\mathrm{BR}(B \to K^{(*)}ee)}, R_\phi = \frac{\mathrm{BR}(B_s \to \phi\mu\mu)}{\mathrm{BR}(B_s \to \phi ee)} . \tag{648}$$

Interestingly enough, LHCb has measured $R_K = 0.745^{+0.090}_{-0.074} \pm 0.036$ [391] for 1 GeV$^2 < q^2 <$ 6 GeV$^2$, which differs from the SM prediction $R_K^{\mathrm{SM}} \simeq 1$ by approximately $2.5\sigma$. This result is in striking agreement with NP explanations of the anomalies in the $b \to s\mu\mu$ transitions discussed above, assuming that the new physics affects the di-muon decays but not the di-electron decays, *i.e.* $(C_9)_s^\mu \simeq -1$ and $(C_9)_s^e \simeq 0$. Various explicit new physics models have been constructed that realise such a scenario (see e.g. [287, 292, 599, 623, 648, 2112]). Future measurements of the LFU ratios showing significant deviations from 1 would establish clean and robust evidence for new physics.

Other tests of LFU are given by ratios or differences of angular observables in decays to final states with di-electrons vs. di-muons [491, 524, 599, 600]. Examples are the difference of the forward backward asymmetries $A_{\rm FB}$ or the angular observables $S_5$ [599]

$$D_{A_{\rm FB}} = A_{\rm FB}(B \to K^* \mu\mu) - A_{\rm FB}(B \to K^* ee \ , \tag{649}$$

$$D_{S_5} = S_5(B \to K^* \mu\mu) - S_5(B \to K^* ee) \ . \tag{650}$$

Measurements of these LFU differences provide additional means to probe lepton flavour non-universal new physics in rare $B$ decays. For $(C_9)_s^\mu \simeq -1$ and $(C_9)_s^e \simeq 0$, non-standard effects at the level of $O(10\%)$ are generically predicted in these observables. The LFU differences might also serve as discriminants between new physics models if precision measurements are feasible [599].

Furthermore, measurements of double ratios of branching ratios like $R_{K^*}/R_K$ and $R_\phi/R_K$ provide a clean probe of flavour non-universal physics coupling to right-handed quarks [598, 2113].

New physics models that contain new sources of LFU violation typically also lead to distinct non-standard effects in decays involving taus in the final state like $B \to K^{(*)}\tau\tau$. If the new physics couples dominantly to the third generation of quarks and leptons, the tauonic decays could be enhanced by an order of magnitude or more compared to the SM predictions [287, 623]. Another class of new physics models that feature LFU violation are based on gauging $L_\mu - L_\tau$, the difference of muon-number and tau-number [648]. Such setups predict modifications in decays with muon and taus in the final state that are comparable in size but opposite in sign.

Some new physics scenarios that contain new sources of LFU violation also lead to lepton flavour violating (LFV) decays of $B$ mesons like $B \to K^{(*)}\mu e$, $B \to K^{(*)}\tau e$, and $B \to K^{(*)}\tau\mu$ [287, 623]. The rates for such decays could be just below current limits from BaBar [2114, 2115], that are at the level of $10^{-7}$ in the case of $\mu e$ and at the level of few $\times 10^{-5}$ in the case of final states containing taus. $B \to K^{(*)}\tau\mu$ branching ratios at the level of few $\times 10^{-7}$ are predicted in models with extended Higgs sectors that propose a new source of 1st and 2nd generation fermion masses [2116]. Interesting complementarity exists between LFV $B$ decays, flavour violating charged lepton decays like $\ell \to \ell'\gamma$, $\ell \to 3\ell'$ and $\mu \to e$ conversion in nuclei. Any observation of a LFV process would constitute indisputable evidence for new physics.

*Inclusive Semi-Leptonic Decays.* The inclusive decays $B \to X_s \ell^+\ell^-$ are expected to be theoretically cleaner compared to the exclusive decays that are limited by the knowledge of hadronic form-factors and non-factorisable long distance effects. SM predictions for decay rates and angular observables at low $q^2$ have reached an accuracy of $5\% - 10\%$ both for the muon and electron modes [567]. Due to the theoretically clean nature of the inclusive decays, Belle II measurements have the unique opportunity to establish lepton universal new physics in $b \to s\ell\ell$ transitions.

The double differential decay width of the $B \to X_s \ell^+\ell^-$ decay provides three independent observables, $H_T$, $H_L$, and $H_A$ [554]

$$\frac{d^2\Gamma}{dq^2 dz} = \frac{3}{8}\left[(1+z^2)H_T(q^2) + 2zH_A(q^2) + 2(1-z^2)H_L(q^2)\right] \ , \tag{651}$$

where, $z = \cos\theta$, with $\theta$ the angle between the $\ell^+$ and the $B$ meson momenta in the di-lepton rest frame. The observable $H_A$ is equivalent to the forward backward asymmetry and the differential decay rate is given by $H_T + H_L$. Precise measurements of these observables allows clean determinations of the Wilson coefficients $(C_7^{(\prime)})_s$, $(C_9^{(\prime)})_s^\ell$, and $(C_{10}^{(\prime)})_s^\ell$, and therefore crucial cross checks of the discrepancies in the recent LHCb data on the related exclusive modes. Global fits of the LHCb data (if interpreted as sign of new physics) predict a $\sim 25\%$ suppression of the $B \to X_s\mu^+\mu^-$ decay rate compared to the SM prediction, both at low and high di-muon invariant mass.

Similar to the case of the exclusive decays, LFU tests in $B \to X_s\ell^+\ell^-$ offer very clean probes of new physics. If the value of $R_K \simeq 0.75$ measured by LHCb is due to new physics, LVU violating effects of similar size can generically be expected in the inclusive decays. Measurements of the ratio of $B \to X_s\mu^+\mu^-$ and $B \to X_s e^+e^-$ branching ratios, $R_{X_s}$ [524, 598], as well as measurements of ratios or differences of lepton flavour specific angular coefficients $H_i$ [599] will help to distinguish the chirality structure of the underlying new physics interactions.

Interesting is also the tauonic $B \to X_s\tau^+\tau^-$ decay that could be enhanced by orders of magnitude by new physics, and the lepton flavour violating decay modes $B \to X_s\mu e$, $B \to X_s\tau e$, and $B \to X_s\tau\mu$. The LFV modes are absent in the SM and any observation of them would be an unambiguous sign of new physics.

*Decays with Neutrinos in the Final State.* New physics in decays with neutrinos in the final state is described by the operators $(\mathcal{O}_\nu^{(\prime)})_q^\ell$. Because the final state neutrinos cannot be detected in the experiment, there are only three observables that are accessible in the $B \to K^{(*)}\nu\bar\nu$ decays as function of the di-neutrino invariant mass (or equivalently as function of the missing energy). These are the two branching ratios $\mathrm{BR}(B \to K\nu\bar\nu)$ and $\mathrm{BR}(B \to K^*\nu\bar\nu)$ as well as $F_L$, the $K^*$ longitudinal polarisation fraction in the $B \to K^*\nu\bar\nu$ decay [614].

Assuming lepton flavour universality, new physics effects in all observables in $b \to s\nu\bar\nu$ transitions depend on two combinations of the complex Wilson coefficients $C_\nu$ and $C_\nu'$ [614, 2117]

$$\epsilon^2 \;=\; \frac{|C_\nu|^2 + |C_\nu'|^2}{|C_\nu^{\mathrm{SM}}|^2}\;, \eta = -\frac{\mathrm{Re}(C_\nu' C_\nu^*)}{|C_\nu|^2 + |C_\nu'|^2}\;. \tag{652}$$

Measurements of $B \to K^{(*)}\nu\bar\nu$ observables can be interpreted as constraints in the $\epsilon - \eta$ plane. Significant deviation from the SM point $(\epsilon, \eta) = (1, 0)$ signals the presence of new physics; a non-zero value of $\eta$ signals the presence of right-handed currents. Equivalently, one can look at the correlation of new physics effects in $\mathrm{BR}(B \to K^*\nu\bar\nu)$ and $\mathrm{BR}(B \to K\nu\bar\nu)$ to identify the presence of right-handed currents [613]. Current bounds on the $B \to K^{(*)}\nu\bar\nu$ branching ratios [624, 625] give the limit $0.5 \lesssim \epsilon \lesssim 3$, while $\eta$ is currently largely unconstrained. This corresponds to bounds on the new physics scale in (645) at the order of $\Lambda \sim 10$ TeV.

The $SU(2)_L$ gauge symmetry relates neutrinos to left-handed charged leptons and therefore new physics effects in $b \to s\nu\nu$ and $b \to s\ell\ell$ and transitions can be related as long as the new physics respects $SU(2)_L$ symmetry. Assuming lepton flavour universality, any new physics effect in $B \to K^{(*)}\nu\bar\nu$ decays necessarily implies new physics effects of the same order in $B \to K^{(*)}\mu^+\mu^-$. On the other hand, new physics in $B \to K^{(*)}\mu^+\mu^-$ does *not* necessarily imply new physics in $B \to K^{(*)}\nu\bar{\nu}$, if the new physics is specific to right-handed leptons. The decays involving neutrinos and charged leptons therefore give complementary information.

The complementarity between $b \to s\nu\nu$ and $b \to s\ell\ell$ transitions is even more pronounced if we consider the possibility of LFU violation [613]. For example, new physics models based on the gauged $L_\mu - L_\tau$ symmetry, predict effects in $b \to s\mu^+\mu^-$ and $b \to s\tau^+\tau^-$ transitions, while $B \to K^{(*)}\nu\bar{\nu}$ decays remain approximately SM-like, as the individual effects in $B \to K^{(*)}\nu_\mu\bar{\nu}_\mu$ and $B \to K^{(*)}\nu_\tau\bar{\nu}_\tau$ cancel in the sum over neutrino flavours [648]. Other new physics scenarios that are described by operators involving left-handed taus are best probed by the $B \to K^{(*)}\nu\bar{\nu}$ decays that are sensitive to the modified $B \to K^{(*)}\nu_\tau\bar{\nu}_\tau$ rates. New physics operators involving right-handed taus on the other hand can only be probed by searching for the $b \to s\tau^+\tau^-$ transitions.

The simultaneous study of the decays $B \to K^{(*)}\nu\bar{\nu}$, $B \to K^{(*)}\ell^+\ell^-$, $B \to X_s\ell^+\ell^-$ and $B \to \ell^+\ell^-$ will teach us a lot about possible new dynamics at scales in the reach of the LHC and beyond. Future Belle II results on the $B \to K^{(*)}\nu\bar{\nu}$ decays will give valuable complementary information and will be key to disentangle the $SU(2)_L$ structure of new physics in rare decays.

*Future Sensitivities with Belle II and LHCb.* *(Contributing author: F. U. Bernlochner)*

In this section, prospects for new physics searches in $b \to s$ transitions are studied under the SM hypothesis as well as in several NP scenarios in a global fit setting. Special attention is given to present anomalies and for the fits the `flavio` [618] framework further described in Section 18.4.2 is used. The future uncertainties for LHCb and all Figures and tables have been taken from Ref. [2118].

Most measurements included will be dominated by the statistical uncertainty for the studied luminosity milestones, with only a few exceptions as e.g. for the differential branching fractions $\mathrm{d}\mathcal{B}/\mathrm{d}q^2$ of $B \to K^{(*)}\mu\mu$, where the dominant systematic uncertainties arise from the branching ratio of the respective normalisation channels, the form factor models and data-simulation differences. Correlations between the systematic uncertainties are assumed to be negligible. The development of theoretical uncertainties is much harder to predict and an overall improvement of all form factor uncertainties by a factor of two is asumed by the end of the Belle II data taking. For the remaining uncertainties, in particular systematic uncertainties due to non-factorizable hadronic contributions, it is assumed that they will stay the same as at present.

In what follows three milestones are considered: Milestones I and II correspond to integrated luminosities of 5 $\mathrm{ab}^{-1}$ and 50 $\mathrm{ab}^{-1}$ or 8 $\mathrm{fb}^{-1}$ and 22 $\mathrm{fb}^{-1}$ for Belle II and LHCb, respectively. In addition an additonal milestone III is assessed that assumes a luminosity of 50 $\mathrm{fb}^{-1}$ for LHCb.

In the following considerations, the effective Wilson coefficient $C_7^{\mathrm{eff}}$ (see e.g. [2090]) is used instead of $C_7$ as this effective coefficient is independent of the regularisation scheme, where we define

$$C_7^{\mathrm{eff}} = C_7^{\mathrm{eff\,SM}} + C_7^{\mathrm{NP}}, \tag{653}$$

$$C_7'^{\mathrm{eff}} = C_7'^{\mathrm{eff\,SM}} + C_7'^{\mathrm{NP}}. \tag{654}$$

Table 162: New physics scenarios for LHCb, Belle II exclusive and Belle II inclusive Wilson coefficient scans. Contributions to the Wilson coefficients arising from new physics are given for each scan. The values are from Ref. [2118].

| | $(C_9^{\mathrm{NP}\mu\mu}, C_{10}^{\mathrm{NP}\mu\mu})$ | $(C_9'^{\mu\mu}, C_{10}'^{\mu\mu})$ | $(C_9^{\mathrm{NP}\mu\mu}, C_9^{\mathrm{NP}ee})$ | $(\mathcal{R}e\,(C_7'^{\mathrm{NP}})\,, \mathcal{I}m\,(C_7'^{\mathrm{NP}}))$ | $(\mathcal{R}e\,(C_7^{\mathrm{NP}})\,, \mathcal{I}m\,(C_7^{\mathrm{NP}}))$ |
|---|---|---|---|---|---|
| LHCb | $(-1.0, 0.0)$ | $(-0.2, -0.2)$ | $(-1.0, 0.0)$ | $(0.00, 0.04)$ | $(-0.075, 0.000)$ |
| Belle II exclusive | $(-1.4, 0.4)$ | $(0.4, 0.2)$ | $(-1.4, -0.7)$ | $(0.08, 0.00)$ | $(-0.050, 0.050)$ |
| Belle II inclusive | $(-0.8, 0.6)$ | $(0.8, 0.2)$ | $(-0.8, 0.4)$ | $(0.02, -0.06)$ | $(-0.050, -0.075)$ |

The impact of future measurements is studied by performing scans of the new physics contribution to the Wilson coefficients at a scale of $\mu = 4.8\,\mathrm{GeV}$, under the SM hypothesis and several different new physics scenarios, listed in Table 162. The measurements are separated depending on whether they are inclusive or exclusive. This allows for a proper comparison given their respective uncertainties have different origins. Various NP scenarios are chosen for each class of measurement and each scan parameter on the basis of existing global fits [493, 2103, 2119]. Scans to $C_S$ and $C_P$ (see e.g. [2090]) are omitted as these are dominated by contributions from purely leptonic $B \to \ell^+ \ell^-$ decays.

The scans of the electromagnetic dipole coefficients $C_7^{(\prime)}$ derive their sensitivity from measurements of the branching fractions of $B_s \to \phi\gamma$, $B^+ \to K^{*+}\gamma$, $B^0 \to K^{*0}\gamma$, $B \to X_s\gamma$, on $\mathcal{A}^{\Delta\Gamma}(B_s \to \phi\gamma)$ and $S_{K^*\gamma}$ as well as from $A_T^{(2)}$ and $A_T^{\mathrm{Im}}$ extracted from $B^0 \to K^{*0}e^+e^-$ decays at very low $q^2$. In addition, the angular observables $A_{7,8,9}$ in $B^0 \to K^{*0}\mu^+\mu^-$ constrain the imaginary part of $C_7^{(\prime)}$.

The measurements entering the scans of the semi-leptonic coefficients $C_{9,10}^{(\prime)}$ comprise the inclusive $\mathcal{B}(B \to X_s\mu^+\mu^-)$ at low and high $q^2$; the low $q^2$ range is split equally for extrapolations. The forward-backward asymmetry $A_{\mathrm{FB}}(B \to X_s\ell^+\ell^-)$ has been measured at low and high $q^2$, and extrapolations to future sensitivities are available in several low and high $q^2$ ranges. The differential branching fractions $d\mathcal{B}/dq^2$ of $B^+ \to K^+\mu^+\mu^-$, $B^0 \to K^{*0}\mu^+\mu^-$ and $B_s \to \phi\mu^+\mu^-$ decays in both low and high $q^2$ regions is included in the scans, as well as the angular observables $S_{3,4,5}, F_L, A_{\mathrm{FB}}$ in several bins of $q^2$ from LHCb. The angular observables available for Belle (II) are $P'_{4,5}(B^0 \to K^{*0}\mu^+\mu^-)$ in similar ranges. Scans of $C_{10}^{(\prime)}$ further include the branching fraction of the decay $B_s \to \mu^+\mu^-$.

In the scan of $C_9^{\mathrm{NP}\mu\mu}$ vs. $C_9^{\mathrm{NP}ee}$, the observables $P'_{4,5}$ extracted from $B^0 \to K^{*0}e^+e^-$ decays are included in addition to the muonic final state as Belle II will have good sensitivity to determine these. Information on electrons is further obtained from the ratios of branching fraction between muon and electron final states for $R(X_s)$, $R(K)$, $R(K^*)$ and $R(\phi)$. The results of the Belle collaboration on $R(K)$ and $R(K^*)$ in the region $0.0 < q^2 < 22.0\,\mathrm{GeV}^2$ were not considered as input in this scan as the charmonium region is included [602]. The inclusive measurement of $R(X_s)$ will become accessible at Belle II.

The result of scans of the unprimed and primed semi-leptonic and electromagnetic dipole Wilson coefficients of the five scenarios summarised in Table 162 are shown in Figures 233 and 234, respectively. Belle II and LHCb will be able to probe new physics in semi-leptonic decays with unprecedented precision. If new physics is present in $C_9^{\mathrm{NP}\mu\mu}$ and the current anomalies in $b \to s\ell^+\ell^-$ persist at a comparable strength, both experiments will be able to rule out the SM with great significance.

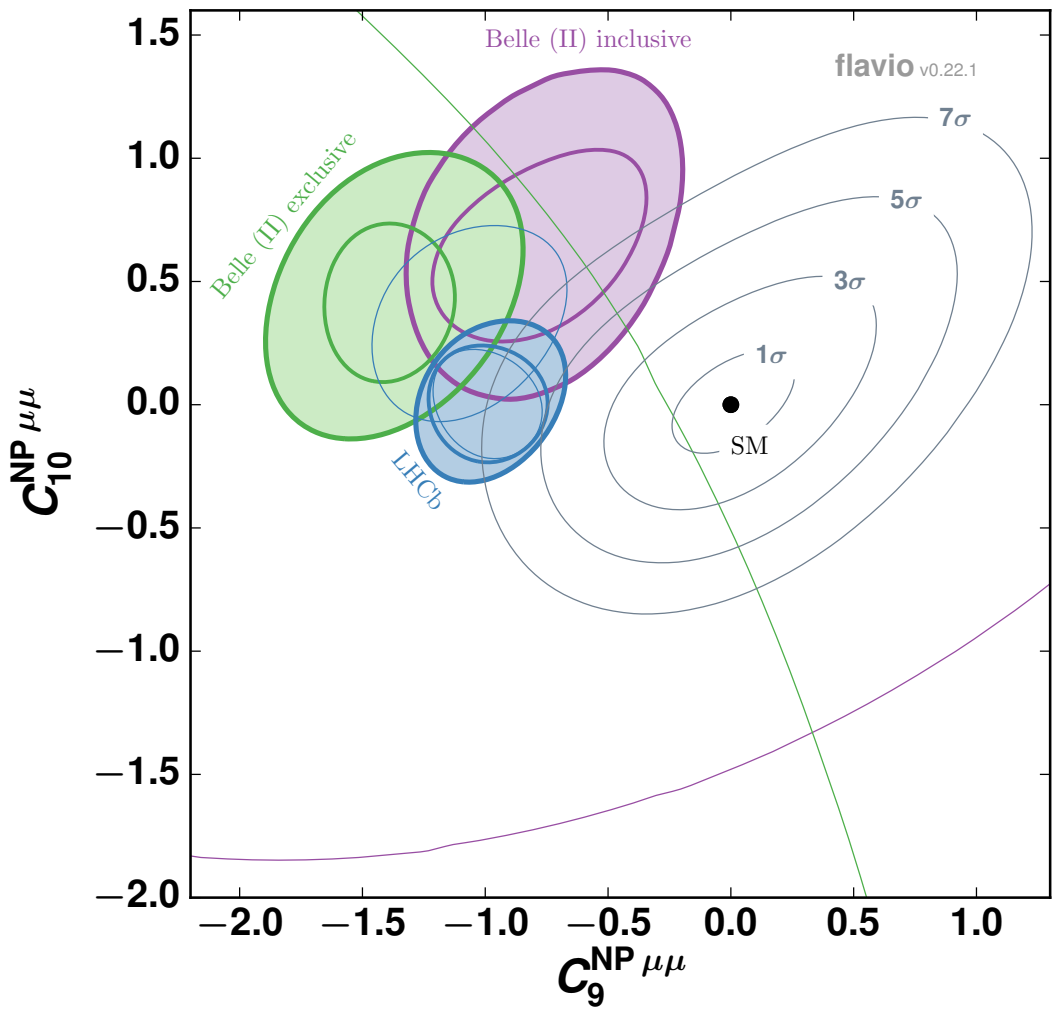

(a) $C_9^{\mathrm{NP}\,\mu\mu}$ versus $C_{10}^{\mathrm{NP}\,\mu\mu}$ .

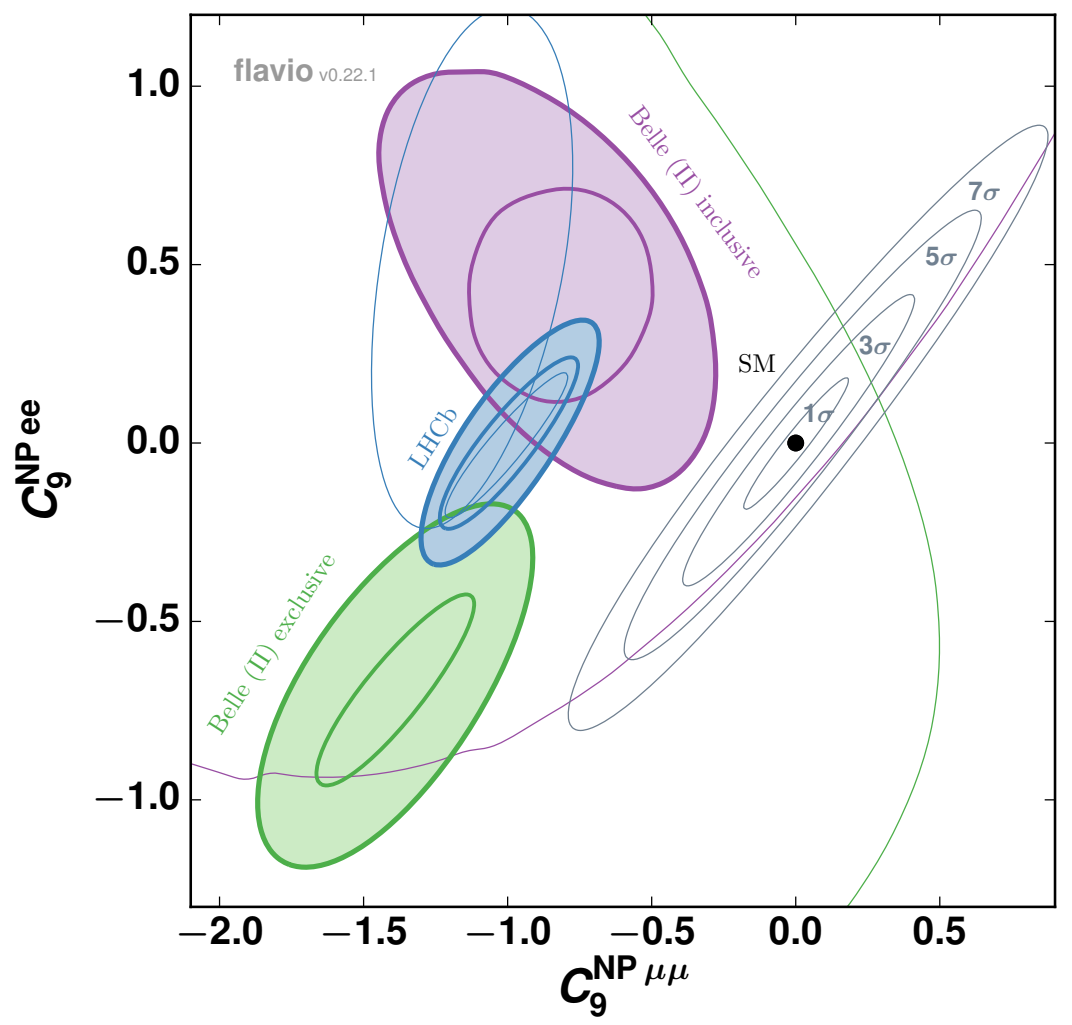

(b) $C_9^{\mathrm{NP}\,\mu\mu}$ versus $C_9^{\mathrm{NP}\,ee}$ .

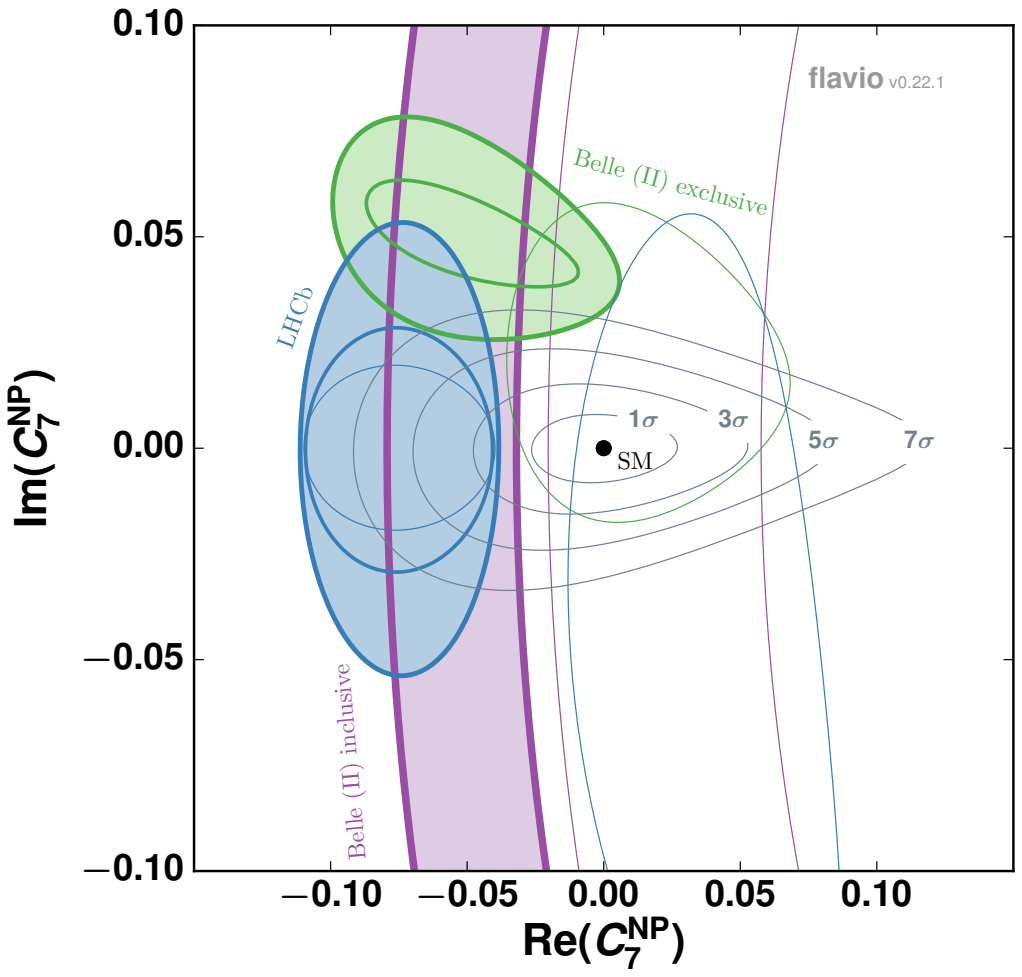

(c) $\mathcal{R}e\,(C_7^{\mathrm{NP}})$ versus $\mathcal{I}m\,(C_7^{\mathrm{NP}})$.

Fig. 233: In the two-dimensional scans of pairs of Wilson coefficients, the current average (not filled) as well as the extrapolations to future sensitivities (filled) of LHCb at milestones I, II and III (exclusive) and Belle II at milestones I and II (inclusive and exclusive) are given and are progressively overlaid. The central values of the extrapolations have been evaluated in the NP scenarios listed in Table 162. The future projections at milestones I, II and III are given by the filled contours The contours correspond to $1\sigma$ uncertainty bands. The Standard Model point (black dot) with the $1\sigma, 3\sigma$, $5\sigma$ and $7\sigma$ exclusion contours with a combined sensitivity of Belle II's $50\,\mathrm{ab}^{-1}$ and LHCb's $50\,\mathrm{fb}^{-1}$ datasets is indicated in light grey. The primed operators show no tensions with respect to the SM; hence no SM exclusions are provided. The Figures are taken from Ref. [2118].

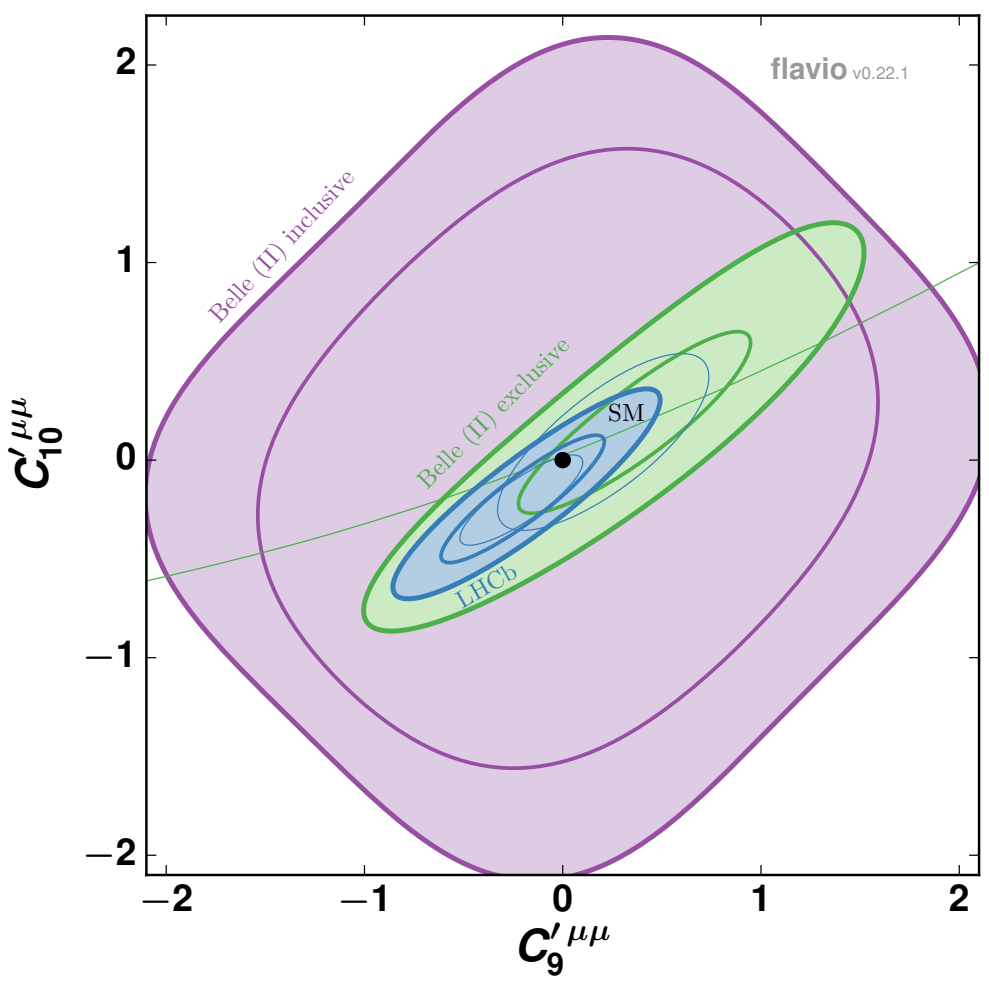

(a) $C'^{\mu\mu}_{9}$ versus $C'^{\mu\mu}_{10}$ .

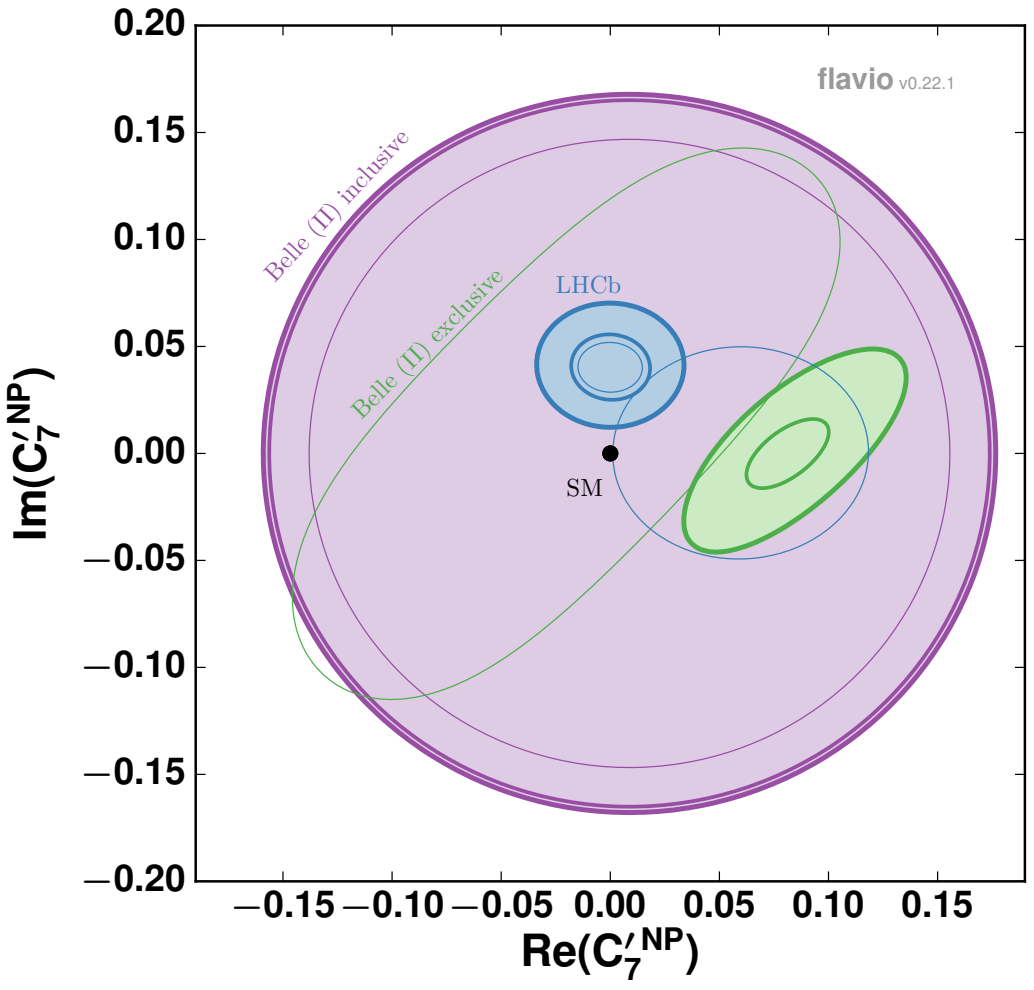

(b) $\mathcal{R}e\left(C'^{\rm NP}_7\right)$ versus $\mathcal{I}m\left(C'^{\rm NP}_7\right)$.

Fig. 234: A description of the plots can be found in the caption of Figure 233. The Figures are taken from Ref. [2118].

## *18.4. Global analysis tools*

### *18.4.1. SuperIso.* *(Contributing author: Farvah Mahmoudi)*

SuperIso [1877, 2120, 2121] is a public code written in C, dedicated to the calculation of flavour physics observables and the muon anomalous magnetic moment. In addition to the full calculations in the SM and generic implementation based on additional new physics contributions to the Wilson coefficients, SuperIso is able to perform the calculations in specific new physics models such as general 2HDM, general MSSM and NMSSM. An extension towards automatic calculations in a given new physics model is ongoing. The code incorporates the state of the art publicly available calculations, with a particular attention to avoid approximations and use the most accurate calculations available.

The SuperIso package can be downloaded from: `http://superiso.in2p3.fr`

A broad set of flavour physics observables sensitive to new physics contributions is implemented in SuperIso. This includes rare decays such as branching ratios of $B_{d,s} \to \ell^+\ell^-$ (with $\ell = e, \mu, \tau$), branching ratios of $B \to X_{d,s}\gamma$, isospin asymmetry of $B \to K^*\gamma$, inclusive and exclusive semileptonic $b \to s$ transitions namely branching ratios and angular observables of $B_s \to X_s\ell^+\ell^-$, $B \to K^{(*)}\ell^+\ell^-$, $B_s \to \phi\ell^+\ell^-$, in addition to leptonic and semileptonic decays such as the branching ratio of $B_u \to \tau\nu_\tau$, branching ratios of $B \to D^{(*)}\tau\nu_\tau$, branching ratio of $K \to \mu\nu_\mu$, branching ratio of $D \to \mu\nu_\mu$, and the branching ratios of $D_s \to \tau\nu_\tau$ and $D_s \to \mu\nu_\mu$, as well as meson mixings. The code is modular and other observables can be added easily.

The calculation of the Wilson coefficients is done in two steps. First they are calculated at the matching scale, $\mathcal{O}(M_W)$. They are subsequently evolved using the Renormalisation Group Equations (RGE) to a lower scale, $\mathcal{O}(m_b)$ relevant for the $B$ physics observables. The Wilson coefficients are calculated at the $\mu_W$ scale at NNLO in the SM and 2HDM and NLO in the MSSM (including some partial NNLO calculations). The RGEs are implemented at NNLO.

A particular care has been taken to avoid having hard coded values in the code so that the input parameters can be safely chosen by the users.

SuperIso can use a SUSY Les Houches Accord (SLHA) file [2122, 2123] as input, which is generated automatically by the program via a call to a spectrum generator or provided by the user. The direct calls are available for 2HDM (types I, II, III and IV), different supersymmetric scenarios, such as the Constrained MSSM (CMSSM), the Non-Universal Higgs Mass model (NUHM), the Anomaly Mediated Supersymmetry Breaking scenario (AMSB), the Hypercharge Anomaly Mediated Supersymmetry Breaking scenario (HCAMSB), the Mixed Modulus Anomaly Mediated Supersymmetry Breaking scenario (MMAMSB), the Gauge Mediated Supersymmetry Breaking scenario (GMSB) and the phenomenological MSSM (pMSSM), and for the NMSSM scenarios namely CNMSSM, NGMSB and NNUHM.

Several example main programs are given in the package providing the values of the observables in different models. SuperIso respects the Flavour Les Houches Accord (FLHA) [2124], which is a standard format for flavour related quantities. An output FLHA file can be subsequently generated.

In brief, the code first scans the SLHA file and transfers the needed parameters in a structure which is used by most of the internal routines. Alternatively, the structure can be filled directly in the main file. The observables which do not depend on the Wilson coefficients can then be computed directly. For the other observables, it is first necessary to use routines to compute the Wilson coefficients at the $\mu_W$ scale, then to use RGE routines to get Wilson coefficients at the $\mu_b$ scales. Alternatively, the Wilson coefficients can be directly given in the main program or through an FLHA interface. All the observables can then be computed with the adequate routines.

For $B \to K^* \ell^+ \ell^-$ and $B_s \to \phi \ell^+ \ell^-$ decays, both the full and soft form factor approaches are implemented, and several parametrisations for the non-factorisable power corrections are provided. In addition, the lepton flavour is generic, allowing for the computation of lepton flavour ratios such as $R_K^{(*)}$.

The correlations matrices between the observables dependent on the Wilson coefficient are also available. The theoretical correlations and errors have been computed in the SM by varying all the parameters in a Monte Carlo program, taking into account also the form factor correlations. The theoretical correlation matrices have been added to the latest experimental correlations matrices.

An automatic and parallel calculation of the errors and correlations is possible from the combination of elementary uncertainties. A function computing the $\chi^2$ is also available. The choice of observables included in the $\chi^2$ can be easily achieved by commenting/uncommenting the observables in the main file. Recent examples of model-independent fits and studies using SuperIso can be found in [2125–2127]. In addition, SuperIso is interfaced in the FlavBit module of GAMBIT [2128, 2129], which provides a thorough set of tools for performing fits.

The calculation duration for one point depends on the number of selected observables, a standard calculation with $\chi^2$ and about 100 observables takes less than one second on a laptop.

An extension of SuperIso including the dark matter relic density calculation as well as direct and indirect detection experiments, SuperIso Relic, is also available publicly [2130, 2131].

*18.4.2. Flavio.* *(Contributing author: David Straub)*

Flavio is an open source Python package to compute flavour physics observables in the SM and beyond. Rather than implementing specific new physics models, new physics contributions to all processes can be supplied as contributions to Wilson coefficients of local dimension six operators, while an interface to other codes (e.g. BSM Wilson coefficient calculators SARAH/FlavorKit [1944] or FormFlavor [2132]) is realized through the Wilson coefficient exchange format (WCxf) [2133]. In this way, flavio can serve as an interface between model building and precision flavour measurements.

The package not only includes numerical values and uncertainties of all relevant input parameters that allow to predict flavour observables including theoretical uncertainties, but also a library of experimental measurements of these observables that allows to construct likelihood functions. Statistical inference of Standard Model parameters or Wilson coefficients, using these likelihoods, is implemented both in a Bayesian and a frequentist framework. In the Bayesian case, interfaces to the Markov Chain Monte Carlo libraries `pypmc`[76] and `emcee` [2134] are implemented. In the frequentist case, flavio implements its own one- and two-dimensional likelihood profilers. Performing the same analysis with Bayesian or frequentist statistics within the same framework allows for powerful cross-checks of the dependence of the fit results on the statistical approach.

Being written in Python and thus not requiring compilation, a main feature of the code is that it can be run interactively and can be easily modified at run time, including parameter values but also parameterisations of quantities such as hadronic form factors. At the same time, parallelisation of computationally intensive routines makes it suitable for large-scale numerical analyses.

At present, flavio includes the following observables.

- Mass differences in $B^0$ and $B_s$ mixing and mixing-induced $CP$ asymmetries in $B^0 \to J/\psi K^0_S$ and $B_s \to J/\psi\phi$,
- $CP$ violation parameter $\epsilon_K$ in $K^0$ mixing,
- Binned and differential branching ratios, angular observables, and angular $CP$ asymmetries in rare $B_q \to M\ell^+\ell^-$ decays, where $M = K, K^*, \phi$,
- Binned and differential branching ratios and angular observables in $\Lambda_b \to \Lambda\ell^+\ell^-$ decays,
- Binned branching ratio and forward-backward asymmetry in $B_q \to X_s\ell^+\ell^-$ decays,
- Rare radiative decays $B \to X_s\gamma$, $B \to K^*\gamma$, $B_s \to \phi\gamma$,
- Rare leptonic $B$ decays $B_q \to \ell^+\ell^-$,
- Rare $B$ decays $B \to M\nu\bar{\nu}$ with $M = \pi, \rho, K, K^*$,
- Rare kaon decays $K^+ \to \pi^+\nu\bar{\nu}$ and $K^0_L \to \pi^0\nu\bar{\nu}$,
- Charged-current kaon and pion decays $K \to \pi\ell\nu$, $K \to \ell\nu$, $\pi \to e\nu$,
- Binned and differential branching ratios and angular observables in charged-current semi-leptonic $B$ decays $B \to M\ell\nu$ with $M = \pi, \rho, \omega, D, D^*$,
- Charged-current inclusive $B$ decay $B \to X_c\ell\nu$,
- Charged-current leptonic $B$ decays $B \to \ell\nu$ and $B_c \to \ell\nu$,
- Charged-current $D$ decays $D \to \ell\nu$ and $D_s \to \ell\nu$.

[76] https://pypi.org/project/pypmc/

In all processes with leptons in the final state, new physics contributions to the channels with different lepton flavour can be specified separately, allowing the analysis of models with violation of lepton flavour universality. In many cases, also charged lepton flavour violating final states, that are forbidden in the SM, are implemented.

Flavio has already been used in numerous publications[77]. Examples include fits of Wilson coefficients in radiative $B$ decays [493], model-independent analysis of new physics in $b \to s\ell\ell$ transitions [611, 2135] and an analysis of new physics in $b \to c\ell\nu$ transitions [262].

The installation instructions and online manual can be found on the flavio web site, `https://flav-io.github.io`. The public source code repository can be found at `https://github.com/flav-io/flavio`, where also code contributions can be submitted.

*18.4.3. HEPfit.* *(Contributing author: Ayan Paul)*

HEPfit is a computational tool for the combination of indirect and direct constraints on High Energy Physics models. The code is built in a modular structure so that one can select observables and models of interest. It can be used to build customised models and customised observables. It has a statistical framework based on Markov Chain Monte Carlo (MCMC) driven Bayesian analysis, however any statistical framework can be used as an alternative. HEPfit allows for the use of parametric and experimental correlations and can read likelihood distributions directly from ROOT histograms.

The goal of HEPfit is to implement electroweak, Higgs and flavour physics observables to the highest degree of precision with minimum theoretical assumptions built in. This has been done in the SM and in several models beyond SM, such as MSSM, THDM, L-R symmetric models, and several EFTs. Since the statistical treatment in HEPfit is based on MCMC, optimised computational time is of utmost importance. HEPfit is massively parallelised to run over a large number of CPUs using openMPI.

Here we focus on how HEPfit can be used for a B-Factory both by experimentalists for making predictions for observables and by theorists to fit model parameters to data. The list of observables implemented in HEPfit includes leptonic and semileptonic decays of the $B$ mesons, flavour violation in the lepton sector and oscillations and $CP$ violation in the $B$ and $K$ meson systems. Some of these have been implemented in models beyond the SM. We list all of these observables below.

HEPfit has a dedicated flavour program in which several $\Delta B = 2$, $\Delta B = 1$ [2110, 2136, 2137] and $\Delta S = 2$ observables have been implemented to state-of-the-art precision in the SM and models beyond the SM. HEPfit also includes observables with lepton flavour violation. In table 163 we list the processes and corresponding models that have either been fully implemented (✓) or are under development ($\circ$) currently. $H_{eff}$ refers to the implementation of a model with generalised Wilson coefficients at a given scale. Since HEPfit is continuously under development, the list of available observables keep increasing and a more complete list in the online resource.

Detailed documentation of the code along with instructions on how to install and run it can be found on the HEPfit website `http://hepfit.roma1.infn.it`. The MCMC core is implemented in BAT [2138].

[77] See `https://flav-io.github.io/papers.html` for a full list.

Table 163: The processes that have been implemented (✓) or are under development (○) in HEPfit for flavour physics.

| Processes | Standard Model | THDM | MSSM | $H_{eff}$ |
|---|---|---|---|---|
| $\Delta B = 2$ | ✓ | ✓ | ○ | ○ |
| $\Delta S = 2$ | ✓ | | ○ | ○ |
| $B \to \tau\nu$ | ✓ | ✓ | ○ | ○ |
| $B \to D^{(*)}\tau\nu$ | ○ | ✓ | | ○ |
| $B_{s/d} \to \mu\mu$ | ✓ | ○ | ○ | ○ |
| rare K decays | ○ | | | ○ |
| $B \to X_s\gamma$ | ✓ | ✓ | ○ | ○ |
| $B \to V\gamma$ | ✓ | | | ○ |
| $B \to P/V\ell^+\ell^-$ | ✓ | | | ○ |
| $B \to X_s\ell^+\ell^-$ | ○ | | | ○ |
| $B \to PP/PV$ | ○ | | | ○ |
| $\ell_i \to \ell_j\gamma$ | | | ✓ | |
| $\ell_i \to 3\ell_j$ | | | ✓ | |
| $(g-2)_\mu$ | | | ✓ | |

While the primary goal of HEPfit is to provide a fast multi-purpose MCMC based fitting framework with a host of models implemented, the code offers a few other options in case the user wants to implement their own statistical framework or use the code to generate a large set of values for the observables given certain ranges for the parameter set.

While HEPfit provides a set of models and observables of relevance it also allows the user to build their own standalone model or one that inherits from an existing model. The user can also define their own observables, with or without defining a new model.

Some example plots on Lepton Flavour Violating tau decays, which are produced by using the HEPfit, are shown in Fig. 235 and Fig. 236.

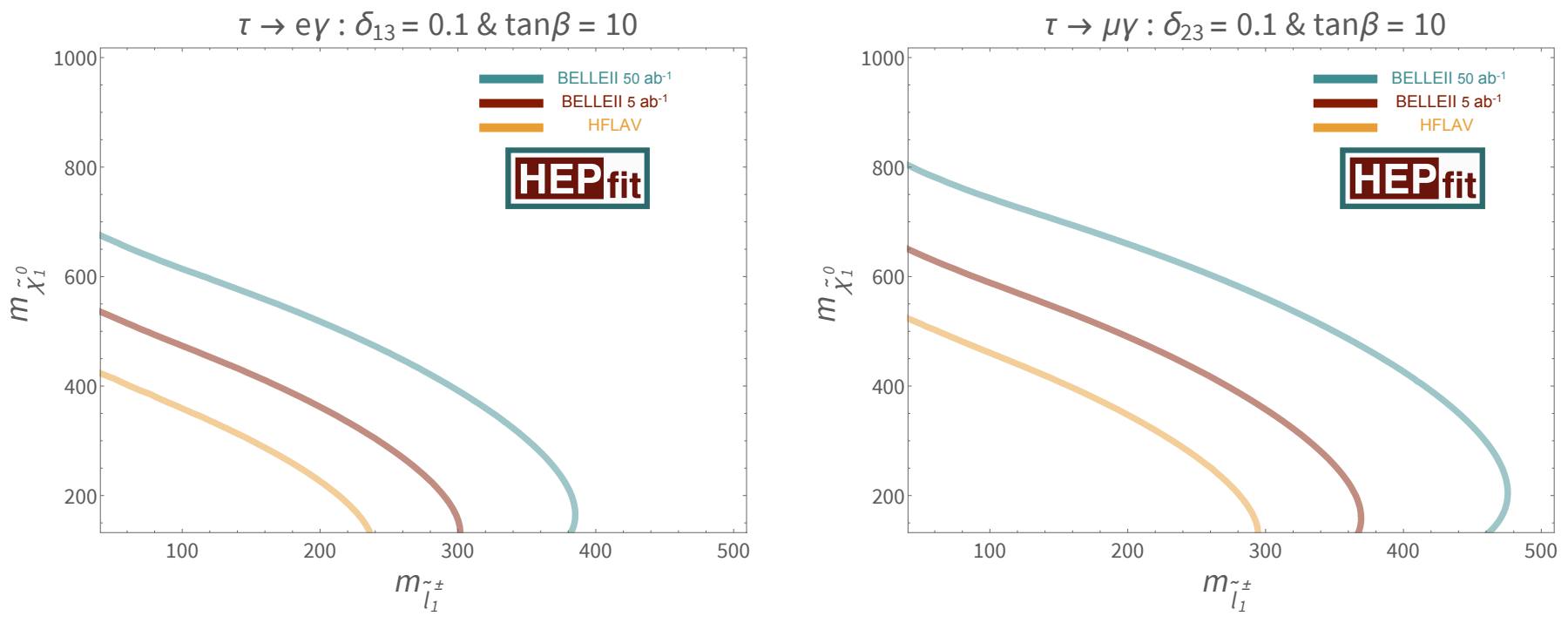

Fig. 235: Constraints on neutralino mass ($m_{\tilde{\chi}_1^0}$) vs. slepton mass ($m_{\tilde{l}_1^\pm}$) from $\tau \to e\gamma$ and $\tau \to \mu\gamma$ measurements at Belle II. The orange, red and green lines display the bounds at 90% CL from current HFLAV averages, 5 ab$^{-1}$ and 50 ab$^{-1}$ data respectively.

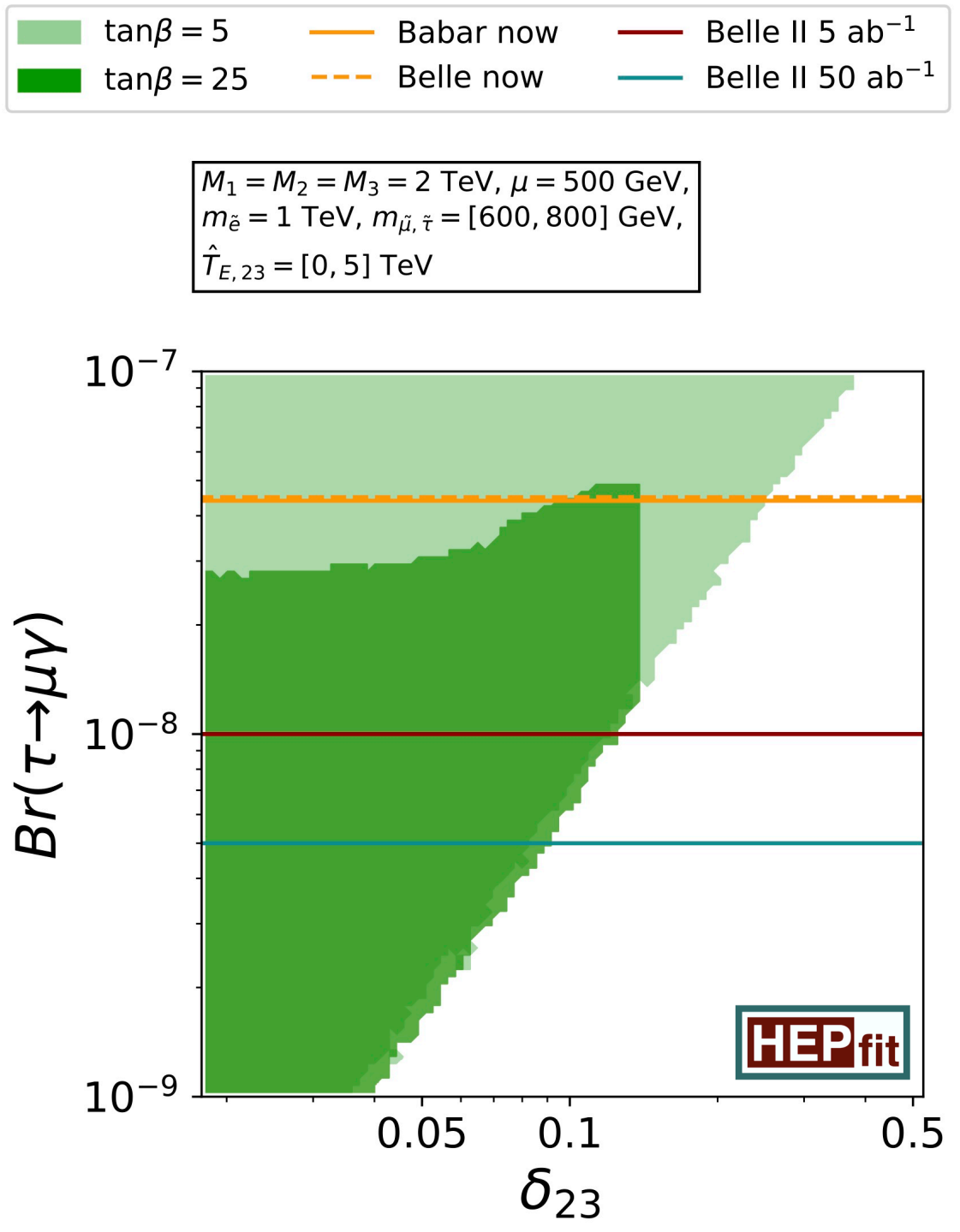

Fig. 236: Possible values for the branching ratio of the process $\tau \to \mu\gamma$ for the given MSSM inputs (see box), depending on $\delta_{23}$. The light green region features $\tan\beta = 5$, and for the dark green contour we set $\tan\beta = 25$. The current 90% C.L. upper limits by BaBar and Belle are given by the orange lines; the expected future limits by Belle II with 5 and 50 ab$^{-1}$ of data are marked by the red and green lines, respectively.

*18.4.4. SUSY_Flavour.* *(Contributing author: Janusz Rosiek)*

`SUSY_FLAVOR` [2139–2141] is a library of numerical routines designed to calculate over 30 low-energy observables related to flavour and CP violation within the most general $R$-parity conserving MSSM.

Due to its ability to calculate numerous observables simultaneously, the code is well equiped to carry out multi-process analyses. It can combine the input from high $p_T$ experiments, like FCNC decay constraints from the top or Higgs sector, with the range of measurements Belle II has particular sensitivty to ($b \to s, d$ or $c, s \to u$ transitions). This allows to test the SM and to connect direct and indirect constraints.

The main features of `SUSY_FLAVOR` are:

- Routines which calculate the observables summarised in Table 164.
- The code implements the fullz general structure of the MSSM, assuming only $R$-parity conservation. The implementation has no limitations on the size of complex phases or flavour violating entries of the soft term matrices. In addition, it takes into account the non-vanishing non-holomorphic trilinear soft terms (cf. [2149, 2150]):

$$L_{nh} = A'_l H_u^\dagger L E + A'_d H_u^\dagger Q D + A'_u H_d^\dagger Q U$$

- The program is able to perform resummation of the chirally enhanced corrections arising in the regime of large $\tan\beta$ and/or large trilinear SUSY breaking. The resummation is implemented to all order of the perturbation theory, including the case of the non-minimal flavour structure of soft breaking terms [2151].

| Observable | Reference |
|---|---|
| $\Delta F = 0$ | |
| $\frac{1}{2}(g-2)_\ell$, $\ell = \mu, \tau$ | |
| $EDM_\ell$, $\ell = e, \mu, \tau$ | [2142] |
| $EDM_{neutron}$ | [2142] |
| $\Delta F = 1$ | |
| $\mathrm{Br}(\mu \to e\gamma)$, $\mathrm{Br}(\tau \to e\gamma)$, $\mathrm{Br}(\tau \to \mu\gamma)$ | |
| $\mathrm{Br}(K_L \to \pi^0 \nu\nu)$, $\mathrm{Br}(K^+ \to \pi^+ \nu\nu)$ | [2143] |
| $\mathrm{Br}(B_d \to \ell\ell')$, $\ell, \ell' = e, \mu, \tau$ | [2144] |
| $\mathrm{Br}(B_s \to \ell\ell')$, $\ell, \ell' = e, \mu, \tau$ | [2144] |
| $\mathrm{Br}(B^+ \to \tau^+ \nu)$ | |
| $\mathrm{Br}(B \to D\tau\nu)/\mathrm{Br}(B \to Dl\nu)$ | |
| $\mathrm{Br}(B \to D^\star\tau\nu)/\mathrm{Br}(B \to D^\star l\nu)$ | |
| $\mathrm{Br}(B \to X_s\gamma)$ | [2145] |
| $\mathrm{Br}(t \to ch, uh)$ | [2146] |
| $\Delta F = 2$ | |
| $\lvert\epsilon_K\rvert$, $\Delta M_K$ | [2147, 2148] |
| $\Delta M_D$ | |
| $\Delta M_{B_d}$, $\Delta M_{B_s}$ | [2147, 2148] |

Table 164: Observables calculated by `SUSY_FLAVOR` v2.5 and references with details of calculations (if published).

- The output is written to SLHA2-like structured files, with custom defined blocks summarising the results for $\Delta F = 0, 1, 2$ transitions.

*18.4.5. EOS.* *(Contributing authors: Danny Van Dyk)* The open-source EOS [2152] software package fulfils multiple use cases. First, it provides estimates for flavour physics observables within the SM and in model-independent frameworks of Effective Field Theories (EFTs). Second, EOS can be used to infer parameters from flavour observables within a Bayesian statistics framework. Finally, EOS is capable of producing MC pseudo events for signal PDFs with a given theory uncertainty. While the complete set of observables that EOS features is too extensive to be listed here, we list the set of processes and a subset of the references to the corresponding theoretical predictions used in the package.

Semileptonic $b$ decays

- $B \to \ell\nu$, with $\ell = e, \mu, \tau$ [2153]
- $B_s \to K^*\ell\nu$, with $\ell = e, \mu$ [2153]
- $B \to \{\pi, D^{(*)}\}\,\ell\nu$, with $\ell = e, \mu, \tau$ [2153, 2154]
- $B \to \pi\pi\ell\nu$, with $\ell = e, \mu$ [909]

Rare $b$ decays

- $B_{(s)} \to \ell^+\ell^-$, with $\ell = e, \mu, \tau$ [2155]

- $B \to X_s\gamma$ [2156]
- $B \to K^*\gamma$ [484]
- $B \to X_s\ell^+\ell^-$, with $\ell = e, \mu$ [565]
- $B \to K^{(*)}\ell^+\ell^-$, with $\ell = e, \mu, \tau$ [484, 2157, 2158]
- $\Lambda_b \to \Lambda\ell^+\ell^-$, with $\ell = e, \mu$ [2159]

EOS features numerical implementations of QCD sum rules, which can be used to determine decay constants and form factors that arise in the computation of the above observables. Several parameterisations, and their default parameter values, of $B \to$ P(seudoscalar) and $B \to$ V(ector) form factors are implemented and selectable within EOS.

The remainder of this section gives an introduction to EOS and discusses three of its use cases.

For the first use case we consider the leptonic forward-backward asymmetry $A_{\mathrm{FB}}$ in the decay $B \to K^*\mu^+\mu^-$. Its prediction is provided by two different theory approaches, both implemented in EOS, which work in two different parts of the dilepton phase space. For small values of the dilepton mass square $q^2 \ll m_b^2$, or equivalently large recoil energy $E_{K^*} = \mathcal{O}(m_b)$ of the $K^*$ in the $B$ rest frame, the framework of QCD-improved factorisation is applicable [483, 484]. Conversely, for $q^2 \simeq m_b^2$, or equivalently low hadronic recoil of the $K^*$, an Operator Product Expansion exists [642, 2157].

Predictions can be obtained with either

- the central value and naive Gaussian uncertainty propagation, or
- a rigorous calculation of uncertainties, producing MC samples of either a prior- or posterior-predictive distribution for the observables of interest.

Here we use $A_{\mathrm{FB}}$ in the decay $B \to K^*\mu^+\mu^-$ as an example, which is binned in $s \equiv q^2$, the dilepton mass squared. The respective EOS method requires specification of the kinematic variables, for example $1\ \mathrm{GeV}^2 \leq q^2 \leq 4\ \mathrm{GeV}^2$. The prediction and uncertainty is then provided through one of the two above methods.

In order to facilitate the task of parameter inference, EOS provides a program to construct a (log)posterior distribution from univariate uniform, Gaussian and (asymmetric) Log Gamma distributions. The (log)likelihood can be constructed from either univariate uniform or Gaussian distributions, or from a data base of built-in multivariate constraints. The latter applies to both experimental measurements as well as theoretical constraints. The client program draws and stores random variates from the (log)posterior using a Markov-Chain-based Metropolis-Hastings algorithm or a mixture-density-based on a Population Monte Carlo algorithm. The application of the Population Monte Carlo algorithm allows for the calculation of the Bayesian *evidence* for the fit, which in turn can be used for model comparisons.

EOS has support to draw pseudo events from kinematic distributions. Sampling is carried out using the same Metropolis-Hastings implementation as above. Again we will consider the production of pseudo events for two decay processes, $B \to K^*\mu^+\mu^-$ and $B \to K^*e^+e^-$, for $1\ \mathrm{GeV}^2 \leq q^2 \leq 6\ \mathrm{GeV}^2$. Both are accessed using pre-defined signal PDFs for $B \to K^*\ell\ell$ at large recoil. Lepton flavour can be specified for each process. Wilson coefficients $\mathcal{C}_i^\ell$ for the $b \to s\ell^+\ell^-$ transitions can be configured as free parameters, which default to the SM values. Lepton-flavour non-universality in these transitions can be introduced by setting $\mathcal{C}_9^{\ell=\mu}$ to non-SM values.

In summary, EOS is a multi-purpose software framework (`https://github.com/eos/eos/`) with a library of numerical implementations for radiative and semileptonic $b$-hadron decays. It features methods for uncertainty propagation, parameter inference, and PDF sampling.

*18.4.6. pypmc.* *(Contributing authors: Frederik Beaujean, Stephan Jahn)*

In global fits of rare $B$ decays, one is interested in determining the most probable values of Wilson coefficients, as they are sensitive to new physics (NP) at short distances. A departure from the values predicted by the SM would provide clues as to the nature of NP. The overarching question in these global fits is whether the SM or some alternative NP model is favoured by the data. In the Bayesian approach to global fits, one seeks the marginal distributions of the Wilson coefficients to determine their probable ranges, and the Bayes factor, or ratio of evidences, between models. Typically only a handful of Wilson coefficients are studied in one fit but in order to fully account for theory uncertainties, it is necessary to include a number of nuisance parameters relating to, for example, quark masses, the CKM matrix, form factors, or non-perturbative corrections. Both the marginal posterior and the evidence have to be computed numerically, which requires integration of the posterior over some (marginal) or all (evidence) parameters in the respective model.

Evaluating the posterior can take up to seconds if many observables need to be predicted so an algorithm is preferred that produces samples from the posterior and estimates the evidence while exploiting a cluster of CPUs. Implementing a toolbox of such algorithms was the motivation to create `pypmc`, an open-source python package `https://github.com/fredRos/pypmc`.

Among others, `pypmc` implements the Metropolis-Hastings algorithm with an adaptive multivariate Gaussian function producing a random-walk Markov chain in parameter space. This algorithm is stable and efficient for unimodal problems with dozens of parameters but is known to fail if there are multiple modes or strong correlations among parameters.

*Adaptive importance sampling* Given the posterior $P$, the basic idea of importance sampling (IS) is to approximate the evidence as

$$\int \mathrm{d}x\, P(x) = \int \mathrm{d}x\, q(x) \frac{P(x)}{q(x)} \approx \frac{1}{N} \sum_{i=1}^{N} \frac{P(x_i)}{q(x_i)} \tag{655}$$

using a proposal function $q$. IS also produces independent weighted samples from $P$ as a by-product, whereas the Metropolis-Hastings algorithm only yields correlated samples from $P$ but not the evidence. Another advantage of IS is that it is trivially parallel, allowing to spread the costly evaluation of $P$ across a cluster of compute nodes. But for IS to be efficient, we need $q \approx P$. A generic and flexible proposal is a mixture density

$$q(x) = \sum_j \alpha_j q_j(x) \text{ with } \sum_j \alpha_j = 1, \tag{656}$$

where $q_j$ is either a Gaussian or Student's t distribution. The key challenge is to infer the parameters of $q_j$ and the weights $\alpha_j$ s.t. $q \approx P$. The core of `pypmc` revolves around updating a mixture based on samples using population Monte Carlo (PMC) or variational Bayes [2138]. Typically one would start to analyse $P$ with several Markov chains. These samples can be used to form an initial guess of the proposal, $q_0$, even if the chains don't mix or show strong autocorrelation. Then one can create additional samples through importance sampling with $q_0$ and update to $q_1$. In all cases of practical interest, only a handful of updates are needed to optimise $q$. Samples from all $q_t$ can be combined for the final inference step. This approach was used in [2102] to perform a global analysis to infer tensor and scalar Wilson coefficients by using a predecessor to the python interface to EOS, described elsewhere in this report. In total, the analysis had up to 60 parameters. Beyond that, the efficiency of IS degrades too much due to the curse of dimensionality.

*18.4.7. FormFlavor.* *(Contributing authors: Jared A. Evans, David Shih)*

`FormFlavor` is a powerful, modular, Mathematica-based tool for the evaluation of low-energy flavour and CP observables in BSM models of physics. `FormFlavor` can be viewed as three distinct, model-independent, modular components: `CalcAmps`, `FFWilson`, and `FFObservables`. A final component, the `FFModel`, contains elements pertinent to the particular BSM model being studied.

The `CalcAmps` package, built upon the machinery of `FeynArts` and `FormCalc` facilitates the automatic generation of one-loop Feynman diagrams. These loops are then converted into analytic expressions of the new physics contribution to the Wilson coefficients. The automated nature of `CalcAmps` greatly increases the reliability of the amplitudes, as an analog computation of the individual diagrams followed by a transcription into code, would be error-prone and need to be repeated for each and every model. `CalcAmps` only needs to be run once per model, as the output files are stored for later access. These output files contain analytic expressions for Wilson coefficients that can be manipulated with the `FormFlavor` machinery.

`FFWilson` allows one to compile the analytic output from `CalcAmps`. The compiled code can then be used to numerical evaluate the Wilson coefficients at particular parameter points. Within `FFWilson`, there are two separate compiling modes provided to allow for either a faster or more reliable determination of the Wilson coefficients. The compiling step should be performed once per session of using `FormFlavor`, while determining the numerical value of Wilson coefficients at a parameter point is performed once for each parameter point.

`FFObservables` takes as input numerical Wilson coefficients and converts these into flavour observables. Within `FFObservables`, there are additional routines that can be used to assess whether the parameter point is constrained by particular flavour observables. Adding new observables to the code is straightforward, and explicit instructions are provided in the manual [2132]. Translating the numerical Wilson coefficients to flavour observables via `FFObservables` is performed once for each parameter point. Importantly, `FFObservables` is its own standalone program, flavour observables can be evaluated just by passing numerical Wilson coefficients, without using a specific model or any of the infrastructure within `CalcAmps` and `FFWilson`.

The `FFModel` directories contains model specific elements that allow for the model independent routines of `CalcAmps` and `FFWilson` to interface with syntax of the `FeynArts`/`FormCalc` model, and the parameter input format. With the release of `FormFlavor`, one `FFModel`, the fully general MSSM, is included. The general MSSM model takes as input SLHA2 files or a Mathematica specific input format.

The most up-to-date version of the `FormFlavor` code and manual can be downloaded from http://formflavor.hepforge.org. The manual contains explicit instructions on how to get started with the code, add new observables, and utilize the more advanced machinery within the package. Additionally, the base package contains the Mathematica notebook `FormFlavor.nb` with a tutorial covering the basic usage of the code.

## *18.5. Conclusions*

The Belle II experiment has large varieties of observables sensitive to new physics. A new physics effect may manifest itself in a single observable or in several observables simultaneously. Either way, we need a consistent picture including new physics effects to interpret obtained experimental results, which requires a global analysis. Therefore, global analyses as well as developing tools for them will become more and more important in the future.

The successful example of the global analysis is the CKM matrix elements fit by using the CKMfitter or the UT fit packages. Although the CKM matrix elements are not predictable from the SM, measuring them independently from numbers of observables may exhibit an inconsistency. We demonstrated how much the future precision measurements by the Belle II experiment will improve the CKM matrix fit for testing the SM and, furthermore, have potential of discovering new physics.

The flavour changing observables are described by the Wilson coefficients of weak interaction in the SM. This description can be extended by including new physics operators with new Wilson coefficients. Global analyses are essential since the new physics effects would appear in many Wilson coefficients simultaneously with some patterns. Using the recent hints of new physics in the $b \to c\tau\nu_\tau$ and the $b \to s\ell\ell$ decays, we discussed how these effects appear in the Wilson coefficient fits and the future prospects at the Belle II experiments.

The ultimate goal for the tool development for the global analysis is to make the computation fully automatic but at the same time, flexible enough that users can choose different theoretical and experimental inputs. In order to achieve a higher reliability, independent groups developing the tools with the same functions is ideal. We introduced a few global analysis tools and their characteristics. We hope those tools, and the combinations of them, will be further developed in the future and will make it possible for wider users to perform new physics searches with Belle II data.

## 19. Summary

The Belle II detector will provide a major step forward in precision heavy flavour physics, quarkonium and exotic states, searches for dark sectors, and many other areas. The sensitivity to a large number of key observables can be improved by about an order of magnitude compared to the current measurements, and up to two orders in very clean search measurements. This increase in statistical precision arises not only due to the increased luminosity, but also from improved detector efficiency and precision for many channels. Many of the most interesting observables tend to have very small theoretical uncertainties that will therefore not limit the physics reach.

This book presented many new ideas for measurements, both to elucidate the nature of current anomalies seen in flavour, and to search for new phenomena in a plethora of observables that will become accessible with the Belle II dataset.

The simulation used for the studies in this book was state of the art at the time, though we are learning a lot more about the experiment during the commissioning period. The detector is in operation, and working spectacularly well.

## 20. Acknowledgements

This work was supported by the following funding sources: Australian Research Council and research grant Nos. DP180102629, DP170102389, DP170102204, DP150103061, FT130100303, FT130100018; Austrian Academy of Sciences, Austrian Science Fund No. P 31361-N36, and New Frontiers Program; Natural Sciences and Engineering Research Council of Canada and Compute Canada; Chinese Academy of Sciences and research grant Nos. QYZDB-SSW-SYS013 and QYZDJ-SSW-SLH011 and National Natural Science Foundation of China and research grant Nos. CRC 110 "Symmetries", PHY-1414345, PHY-1520966, PHY-1714253, PHY-1720252, 11521505, 11575017, 11675166, 11621131001 "the Emergence of Structure in QCD", 11761141009, 11475187, 11621131001, 11705209 and 11747601; the Ministry of Education, Youth and Sports of the Czech Republic under Contract No. LTT17020 and Charles University grants SVV 260448 and GAUK 404316; European Research Council, 7th Framework PIEF-GA-2013-622527, Horizon 2020 Marie Sklodowska-Curie grant agreement No. 700525 'NIOBE,' Horizon 2020 Marie Sklodowska-Curie RISE project JENNIFER grant agreement No. 644294, Horizon 2020 ERC-Advanced Grant No. 267104, and NewAve No. 638528 (European grants); L'Institut national de physique nucléaire et de physique des particules (IN2P3) du CNRS (France); BMBF, DFG, HGF, MPG and AvH Foundation (Germany); Department of Atomic Energy and Department of Science and Technology (India); Israel Science Foundation grant No. 2476/17 and United States-Israel Binational Science Foundation grant No. 2016113; Istituto Nazionale di Fisica Nucleare and the research grants BELLE2 and Iniziativa Specifica QFT-HEP; Japan Society for the Promotion of Science, Grant-in-Aid for Scientific Research grant Nos. 16H03968, 16H03993, 16K05323, 17H01133, 17H05405, 18H03710, 18K03621, 16H06492, 26400255, and 26220706, and the Ministry of Education, Culture, Sports, Science, and Technology (MEXT) of Japan; National Research Foundation (NRF) of Korea Grants Nos. 2015H1A2A1033649, 2016R1D1A1B01010135, 2016K1A3A7A09005603, 2016R1D1A1B02012900, 2018R1A2B-3003643, 2018R1A6A1A06024970, and 2018R1D1A1B07047294, Radiation Science Research Institute, Foreign Large-size Research Facility Application Supporting project, the Global Science Experimental Data Hub Center of the Korea Institute of Science and Technology Information, and KREONET/GLORIAD; Frontiers of Science Program contracts FOINS-296, CB-221329, CB-236394, CB-254409, CB-180023 and the Thematic Networks program (Mexico); Foundation for Fundamental Research of Matter No. 156, "Higgs as Probe and Portal" and National Organization for Scientific Research (Netherlands); National Science Center, Poland and research grant Nos. 2015/19/B/ST2/02848, 2017/25/B/ST2/00191, 2017/27/B/ST2/01391, and UMO-2015/18/M/ST2/00518, HARMONIA; Russian Ministry of Education and Science Funding No. 14.W03.31.0026; Slovenian Research Agency (research grant No. J1-8137 and research core funding No. P1-0135 No. P1-0035); Agencia Estatal de Investigacion, Spain grant Nos. FPA2014-55613-P and FPA2017-84445-P; CERN; Swiss National Science Foundation and research grant No. P300P2_167751; Ministry of Science and Technology and research grant No. MOST104-2628-M-002-014-MY4 and the Ministry of Education (Taiwan); Thailand Center of Excellence in Physics; TUBITAK ULAKBIM (Turkey); STFC No. ST/P000290/1 (United Kingdom); Ministry of Education and Science of Ukraine; the US National Science Foundation and research grant Nos. PHY-1614545, PHY-1714253, PHY-1414345, PHY-1720252, and PHY-1520966, and the US Department of Energy and research grant Nos. DE-SC0010118, DE-SC0012704, DE-SC0009973, DE-SC0012391, DE-SC0010504, DE-AC06-76RLO1830, DE-AC02-05CH11231, DE-AC05-06OR23177, DE-FG02-00ER41132, DE-SC0007983, DE-SC0011637, DE-SC0011726, and DE-SC0011784; International Centre of Physics (under the auspices of UNESCO), IOP, Hanoi, Viet Nam, and the National Foundation for Science and Technology Development (NAFOSTED) of Vietnam under contract No 103.01-2017.76. We thank the members of the advisory committee for useful input at the early stages of the document. We thank the KEK theory group and the EU-Rise Japan and Europe Network for Neutrino and Intensity Frontier Experimental Research for offering constant support for many theorists involved.

## References

[1] D. Boutigny et al., The BABAR physics book: Physics at an asymmetric $B$ factory, In *Workshop on Physics at an Asymmetric B Factory (BaBar Collaboration Meeting) Pasadena, California, September 22-24, 1997* (1998).

[2] A. J. Bevan et al., Belle, BaBar, Eur. Phys. J., **C74**, 3026 (2014), arXiv:1406.6311.

[3] J. Charles et al., Phys. Rev., **D89**(3), 033016 (2014), arXiv:1309.2293.

[4] Y. Ohnishi et al., PTEP, **2013**, 03A011 (2013).
[5] M. Bona et al., SuperB (2007), arXiv:0709.0451.
[6] T. Abe, Belle II (2010), arXiv:1011.0352.
[7] J. Kemmer et al., Nucl. Instrum. Meth., **A253**, 378–381 (1987).
[8] P. Fischer et al., Nucl. Instrum. Meth., **A582**, 843–848 (2007).
[9] M. Akatsu et al., Nucl. Instrum. Meth., **A440**, 124–135 (2000), arXiv:physics/9904009.
[10] M. Staric et al., Nucl. Instrum. Meth., **A595**, 252–255 (2008).
[11] M. Akatsu et al., Nucl. Instrum. Meth., **A528**, 763–775 (2004).
[12] L. L. Ruckman and G. S. Varner, Nucl. Instrum. Meth., **A602**, 438–445 (2009), arXiv:0805.2225.
[13] H. Kichimi et al., Belle, JINST, **5**, P03011 (2010), arXiv:1001.1194.
[14] T. Iijima et al., Nucl. Instrum. Meth., **A548**, 383–390 (2005), arXiv:physics/0504220.
[15] P. Krizan et al., Nucl. Instrum. Meth., **A565**, 457–462 (2006), arXiv:physics/0603022.
[16] S. Nishida et al., Nucl. Instrum. Meth., **A595**, 150–153 (2008).
[17] S. Nishida et al., Nucl. Instrum. Meth., **A766**, 28–31 (2014).
[18] S. Longo and J. M. Roney, JINST, **11**(08), P08017 (2016), arXiv:1608.07556.
[19] A. Kuzmin, Belle ECL, Nucl. Instrum. Meth., **A623**, 252–254 (2010).
[20] V. Balagura et al., Nucl. Instrum. Meth., **A564**, 590–596 (2006), arXiv:physics/0504194.
[21] D. J. Lange, Nucl. Instrum. Meth., **A462**, 152–155 (2001).
[22] T. Sjöstrand et al., Comput. Phys. Commun., **191**, 159–177 (2015), arXiv:1410.3012.
[23] S. Jadach et al., Comput.Phys.Commun., **130**, 260–325 (2000), arXiv:hep-ph/9912214.
[24] S. Jadach et al., Phys. Rev., **D63**, 113009 (2001), arXiv:hep-ph/0006359.
[25] N. Davidson et al., Comput. Phys. Commun., **183**, 821–843 (2012), arXiv:1002.0543.
[26] G. Balossini et al., Phys. Lett., **B663**, 209–213 (2008), arXiv:0801.3360.
[27] G. Balossini et al., Nucl. Phys., **B758**, 227–253 (2006), arXiv:hep-ph/0607181.
[28] C. M. Carloni Calame et al., Nucl. Phys. Proc. Suppl., **131**, 48–55, [,48(2003)] (2004), arXiv:hep-ph/0312014.
[29] Carlo M. Carloni C., Phys. Lett., **B520**, 16–24 (2001), arXiv:hep-ph/0103117.
[30] C. M. Carloni Calame et al., Nucl. Phys., **B584**, 459–479 (2000), arXiv:hep-ph/0003268.
[31] Frits A. Berends et al., Nucl. Phys., **B253**, 441 (1985).
[32] F.A. Berends et al., Computer Physics Communications, **40**(2), 285 – 307 (1986).
[33] Frits A. Berends et al., Nucl. Phys., **B253**, 421 (1985).
[34] N. Davidson et al., Comput. Phys. Commun., **199**, 86–101 (2016), arXiv:1011.0937.
[35] D. Bardin, Comput. Phys. Commun., **133**, 229–395 (2001), arXiv:hep-ph/9908433.
[36] A. B. Arbuzov and others., Comput. Phys. Commun., **174**, 728–758 (2006), arXiv:hep-ph/0507146.
[37] S. Banerjee et al., Phys. Rev., **D77**, 054012 (2008), arXiv:0706.3235.
[38] Strategic accelerator design (sad), `http://acc-physics.kek.jp/SAD` (), Accessed: 2016-05-06.
[39] S. Agostinelli et al., GEANT4, Nucl.Instrum.Meth., **A506**, 250–303 (2003).
[40] J. Allison et al., IEEE Trans. Nucl. Sci., **53**, 270 (2006).
[41] R. Brun et al., ROOT, an object oriented data analysis framework (2000).
[42] A. Dotti, Simulation of Showers with Geant4, In *Proceedings, International Conference on Calorimetry for the High Energy Frontier (CHEF 2013)*, pages 247–253 (2013).
[43] Geant4, *Physics reference manual Ver. 10.1* (Dec. 2014).
[44] D. H. Wright, Improving the Medium and Low Energy Physics Models in Geant4, In *Proceedings, International Conference on Calorimetry for the High Energy Frontier (CHEF 2013)*, pages 254–259 (2013).
[45] B. Andersson et al., Nucl. Phys., **B281**, 289–309 (1987).
[46] V. V. Uzhinsky, The Fritiof (FTF) Model in Geant4, In *Proceedings, International Conference on Calorimetry for the High Energy Frontier (CHEF 2013)*, pages 260–264 (2013).
[47] V. Ivanchenko et al., Recent Improvements in Geant4 Electromagnetic Physics Models and Interfaces, In *Progress in Nuclear Science and Technology*, volume 2, pages 898–903 (2011).
[48] Geant4, *User's Guide for Application Developers Ver. 10.2* (Dec. 2015).
[49] Y. Arimoto et al., Three Dimensional Field Analysis for Final Focus Magnet System at SuperKEKB, In *Proceedings, 5th International Particle Accelerator Conference (IPAC 2014)*, page WEPRI086 (2014).
[50] Cobham Technical Services, Vector Fields Software, Oxford, England, *Opera-3D User Guide Ver. 15R3* (Oct. 2012).
[51] R. Fruhwirth et al., Nucl. Instrum. Meth., **A732**, 95–98 (2013).
[52] T. Lueck et al., EPJ Web Conf., **150**, 00007 (2017).
[53] P.V.C. Hough, U.S. Patent 3069654 (1962).
[54] R. Fruhwirth and A. Strandlie, Comput. Phys. Commun., **120**(2-3), 197–214 (1999).
[55] R. Fruhwirth, Nucl. Instrum. Meth., **A262**, 444–450 (1987).

[56] C. Hoppner et al., Nucl. Instrum. Meth., **A620**, 518–525 (2010), arXiv:0911.1008.
[57] J. Rauch and T. Schlueter, J. Phys. Conf. Ser., **608**(1), 012042 (2015), arXiv:1410.3698.
[58] H. Ishino et al., Belle internal note BN715 (2005).
[59] W. Waltenberger et al., IEEE Transactions on Nuclear Science, **58**, 434 – 444 (2011).
[60] V. Blobel and C. Kleinwort (2002), arXiv:0208021.
[61] V. Blobel, Nucl. Instrum. Meth., **A566**, 5–13 (2006).
[62] C. Kleinwort, Nucl. Instrum. Meth., **A673**, 107–110 (2012).
[63] J. Rauch et al. (2014), arXiv:1410.3698.
[64] M. Staric, Nucl. Instrum. Meth., **A639**, 252–255 (2011).
[65] K. Nakamura et al., Particle Data Group, J. Phys. G, **37**, 075021 (2010).
[66] M. Staric, Nucl. Instrum. Meth., **A766**, 237–240 (2014).
[67] P. Baillon, Nucl. Instrum. Meth., **A238**, 341–346 (1985).
[68] R. Forty, LHCb, Nucl. Instrum. Meth., **A433**, 257–261 (1999).
[69] R. Fruhwirth, Nucl.Instrum.Meth., **262**, 444 (1987).
[70] T. Keck, CoRR, **arXiv:1609.06119** (2016).
[71] W. D. Hulsbergen, Nucl. Instrum. Meth., **A552**, 566–575 (2005), arXiv:physics/0503191.
[72] W. Waltenberger, CERN-CMS-NOTE-2008-033 (2008).
[73] D. Weyland, MSc thesis, Karlsruher Institut für Technologie (2017).
[74] D. M. Asner et al., CLEO, Phys. Rev., **D53**, 1039–1050 (1996), arXiv:hep-ex/9508004.
[75] G. C. Fox and S. Wolfram, Phys. Rev. Lett., **41**, 1581 (1978).
[76] F. Abudinén, *Development of a $B^0$ flavor tagger and performance study of a novel time-dependent CP analysis of the decay $B^0 \to \pi^0\pi^0$ at Belle II*, PhD thesis, Ludwig-Maximilians-Universität (2018), BELLE2-PTHESIS-2018-003.
[77] K. A. Olive et al., Particle Data Group, Chin. Phys., **C38**, 090001 (2014).
[78] S. Nissen, Implementation of a fast artificial neural network library (fann), Technical report, Department of Computer Science University of Copenhagen (DIKU) (2003), http://fann.sf.net.
[79] S. Pohl, *Track Reconstruction at the First Level Trigger of the Belle II Experiment*, PhD thesis, Ludwig-Maximilians-Universität (2017), BELLE2-PTHESIS-2018-001.
[80] I. Adachi et al., Phys. Rev. Lett., **108**, 171802 (2012), arXiv:1201.4643.
[81] H. Kakuno et al., Belle, Nucl. Instrum. Meth., **A533**, 516–531 (2004), arXiv:hep-ex/0403022.
[82] T. Keck et al., Fer (2018), arXiv:arXiv:1807.0868.
[83] M. Feindt et al., Nucl. Instrum. Meth., **A654**, 432–440 (201).
[84] A. Sibidanov et al., Belle, Phys. Rev., **D88**, 032005 (2013), arXiv:1306.2781.
[85] R. Glattauer et al., Belle, Phys. Rev., **D93**(3), 032006 (2016), arXiv:1510.03657.
[86] M. Kobayashi and T. Maskawa, Prog. Theor. Phys., **49**, 652–657 (1973).
[87] L-L. Chau and W-Y. Keung, Phys. Rev. Lett., **53**, 1802 (1984).
[88] C. Patrignani et al., Particle Data Group, Chin. Phys., **C40**(10), 100001 (2016).
[89] L. Wolfenstein, Phys. Rev. Lett., **51**, 1945 (1983).
[90] A. J. Buras et al., Phys. Rev., **D50**, 3433–3446 (1994), arXiv:hep-ph/9403384.
[91] J. Charles et al., CKMfitter Group, Eur. Phys. J., **C41**(1), 1–131 (2005), arXiv:hep-ph/0406184.
[92] I. I. Bigi (2015), arXiv:1509.03899.
[93] D. Boito et al., Phys. Rev., **D96**(11), 113003 (2017), arXiv:1709.09739.
[94] M. E. Peskin and D. V. Schroeder, *An Introduction to quantum field theory*, (Addison-Wesley, Reading, USA, 1995).
[95] M. Doring et al., Nucl. Phys., **A829**, 170–209 (2009), arXiv:0903.4337.
[96] V. Baru et al., Phys. Lett., **B586**, 53–61 (2004), arXiv:hep-ph/0308129.
[97] C. Hanhart et al., Phys. Rev., **D75**, 074015 (2007), arXiv:hep-ph/0701214.
[98] C Hanhart et al., Phys. Rev., **D76**, 034007 (2007), arXiv:0704.0605.
[99] A. Svarc et al., Phys. Rev., **C89**(4), 045205 (2014), arXiv:1401.1947.
[100] S. M. Flatte, Phys. Lett., **B63**, 224–227 (1976).
[101] S. Gardner and U-G. Meiß ner, Phys. Rev., **D65**, 094004 (2002), arXiv:hep-ph/0112281.
[102] D. Babusci et al., KLOE, Phys. Lett., **B718**, 910–914 (2013), arXiv:1209.4611.
[103] K. M. Watson, Phys. Rev., **95**, 228–236 (1954).
[104] F. Klingl et al., Z. Phys., **A356**, 193–206 (1996), arXiv:hep-ph/9607431.
[105] C. Hanhart, Phys. Lett., **B715**, 170–177 (2012), arXiv:1203.6839.
[106] F. Stollenwerk et al., Phys. Lett., **B707**, 184–190 (2012), arXiv:1108.2419.
[107] B. Kubis and J. Plenter, Eur. Phys. J., **C75**(6), 283 (2015), arXiv:1504.02588.
[108] J.-J. Wu et al., Phys. Rev. Lett., **108**, 081803 (2012), arXiv:1108.3772.
[109] A. P. Szczepaniak, Phys. Lett., **B747**, 410–416 (2015), arXiv:1501.01691.
[110] Q. Wang et al., Phys. Lett., **B725**(1-3), 106–110 (2013), arXiv:1305.1997.
[111] J. T. Daub et al., JHEP, **01**, 179 (2013), arXiv:1212.4408.

[112] M. Battaglieri et al., Acta Phys. Polon., **B46**, 257 (2015), arXiv:1412.6393.
[113] C. Hanhart et al., Phys. Rev. Lett., **115**(20), 202001 (2015), arXiv:1507.00382.
[114] F. K. Guo et al., Phys. Rev., **D93**(7), 074031 (2016), arXiv:1602.00940.
[115] G. Barton, *Introduction to dispersion techniques in field theory*, 1965).
[116] B. Ananthanarayan et al., Phys. Rept., **353**, 207–279 (2001), arXiv:hep-ph/0005297.
[117] R. Garcia-Martin et al., Phys. Rev., **D83**, 074004 (2011), arXiv:1102.2183.
[118] J. T. Daub et al., JHEP, **02**, 009 (2016), arXiv:1508.06841.
[119] R. Aaij et al., LHCb, Phys. Rev., **D90**(1), 012003 (2014), arXiv:1404.5673.
[120] R. Aaij et al., LHCb, Phys. Rev., **D89**(9), 092006 (2014), arXiv:1402.6248.
[121] J. F. Donoghue et al., Nucl. Phys., **B343**, 341–368 (1990).
[122] S. Eidelman and L. Lukaszuk, Phys. Lett., **B582**, 27–31 (2004), arXiv:hep-ph/0311366.
[123] A. Celis et al., Phys. Rev., **D89**, 013008 (2014), arXiv:1309.3564.
[124] A. Celis et al., Phys. Rev., **D89**(9), 095014 (2014), arXiv:1403.5781.
[125] N. N. Khuri and S. B. Treiman, Phys. Rev., **119**, 1115–1121 (1960).
[126] J. Kambor et al., Nucl. Phys., **B465**, 215–266 (1996), arXiv:hep-ph/9509374.
[127] A. V. Anisovich and H. Leutwyler, Phys. Lett., **B375**, 335–342 (1996), arXiv:hep-ph/9601237.
[128] S. P. Schneider et al., JHEP, **02**, 028 (2011), arXiv:1010.3946.
[129] S. P. Schneider et al., Phys. Rev., **D86**, 054013 (2012), arXiv:1206.3098.
[130] F. Niecknig et al., Eur. Phys. J., **C72**, 2014 (2012), arXiv:1203.2501.
[131] F. Niecknig and B. Kubis, JHEP, **10**, 142 (2015), arXiv:1509.03188.
[132] I. V. Danilkin et al., Phys. Rev., **D91**(9), 094029 (2015), arXiv:1409.7708.
[133] A. S. Kronfeld, Ann. Rev. Nucl. Part. Sci., **62**, 265–284 (2012), arXiv:1203.1204.
[134] Sz. Borsanyi et al., Science, **347**, 1452–1455 (2015), arXiv:1406.4088.
[135] R. Horsley et al., J. Phys., **G43**(10), 10LT02 (2016), arXiv:1508.06401.
[136] B. A. Thacker and G. P. Lepage, Phys. Rev., **D43**, 196–208 (1991).
[137] A. X. El-Khadra et al., Phys. Rev., **D55**, 3933–3957 (1997), arXiv:hep-lat/9604004.
[138] J. Heitger and R. Sommer, ALPHA, JHEP, **02**, 022 (2004), arXiv:hep-lat/0310035.
[139] B. Blossier et al., ETM, JHEP, **04**, 049 (2010), arXiv:0909.3187.
[140] S. Aoki et al., Flavour Lattice Averaging Group (FLAG), Eur. Phys. J., **C77**(2), 112 (2017), arXiv:1607.00299.
[141] J. M. Flynn et al., RBC/UKQCD, Phys. Rev., **D91**(7), 074510 (2015), arXiv:1501.05373.
[142] E. Dalgic et al., Phys. Rev., **D73**, 074502, [Erratum: Phys. Rev. D75, 119906 (2007)] (2006), arXiv:hep-lat/0601021.
[143] J. A. Bailey et al., Fermilab Lattice, MILC, Phys. Rev., **D92**(1), 014024 (2015), arXiv:1503.07839.
[144] J. A. Bailey et al., Fermilab Lattice, MILC, Phys. Rev., **D89**, 114504 (2014), arXiv:1403.0635.
[145] J. A. Bailey et al., Fermilab Lattice, MILC, Phys. Rev., **D92**(3), 034506 (2015), arXiv:1503.07237.
[146] H. Na et al., HPQCD, Phys. Rev., **D92**(5), 054510, [Erratum: Phys. Rev. D93, 119906 (2016)] (2015), arXiv:1505.03925.
[147] J. Harrison and Matthew others, HPQCD, Phys. Rev., **D97**(5), 054502 (2018), arXiv:1711.11013.
[148] E. Gamiz et al., HPQCD, Phys. Rev., **D80**, 014503 (2009), arXiv:0902.1815.
[149] N. Carrasco et al., ETM, JHEP, **03**, 016 (2014), arXiv:1308.1851.
[150] Y. Aoki et al., Phys. Rev., **D91**(11), 114505 (2015), arXiv:1406.6192.
[151] A. Bazavov et al., Fermilab Lattice, MILC (2016), arXiv:1602.03560.
[152] A. Patella, PoS, **LATTICE2016**, 020 (2017), arXiv:1702.03857.
[153] N. Carrasco et al., ETM, Phys. Rev., **D91**(7), 074506 (2015), arXiv:1502.00257.
[154] V. Lubicz et al., PoS, **LATTICE2016**, 290 (2016), arXiv:1610.09668.
[155] C. M. Bouchard et al., HPQCD, Phys. Rev., **D90**, 054506 (2014), arXiv:1406.2279.
[156] F. Bahr et al., Alpha, PoS, **LATTICE2016**, 295 (2016), arXiv:1701.03299.
[157] C. Bouchard et al., HPQCD, Phys. Rev., **D88**(5), 054509, [Erratum: Phys. Rev. D88, 079901 (2013)] (2013), arXiv:1306.2384.
[158] J. A. Bailey et al., Fermilab Lattice, MILC, Phys. Rev., **D93**(2), 025026 (2016), arXiv:1509.06235.
[159] J. A. Bailey et al., Fermilab Lattice, MILC, Phys. Rev. Lett., **115**(15), 152002 (2015), arXiv:1507.01618.
[160] W. Detmold et al., Phys. Rev., **D88**(1), 014512 (2013), arXiv:1306.0446.
[161] W. Detmold et al., Phys. Rev., **D92**(3), 034503 (2015), arXiv:1503.01421.
[162] W. Detmold and S. Meinel, Phys. Rev., **D93**(7), 074501 (2016), arXiv:1602.01399.
[163] M. Crisafulli et al., Nucl. Phys., **B457**, 594–618 (1995), arXiv:hep-ph/9506210.
[164] V. Gimenez et al., Phys. Lett., **B393**, 124–131 (1997), arXiv:hep-lat/9607018.
[165] V. Gimenez et al., Nucl. Phys., **B486**, 227–244 (1997), arXiv:hep-lat/9607055.
[166] A. Ali Khan et al., Phys. Rev., **D62**, 054505 (2000), arXiv:hep-lat/9912034.
[167] A. S. Kronfeld and J. N. Simone, Phys. Lett., **B490**, 228–235, [Erratum: Phys. Lett. B495, 441

(2000)] (2000), arXiv:hep-ph/0006345.
[168] S. Aoki et al., JLQCD, Phys. Rev., **D69**, 094512 (2004), arXiv:hep-lat/0305024.
[169] P. Gambino et al., Phys. Rev., **D96**(1), 014511 (2017), arXiv:1704.06105.
[170] S. Hashimoto (2017), arXiv:1703.01881.
[171] M. T. Hansen et al. (2017), arXiv:1704.08993.
[172] L. Maiani and M. Testa, Phys. Lett., **B245**, 585–590 (1990).
[173] D. J. Wilson et al., Phys. Rev., **D91**(5), 054008 (2015), arXiv:1411.2004.
[174] G. Moir et al., JHEP, **10**, 011 (2016), arXiv:1607.07093.
[175] M. Luscher, Commun. Math. Phys., **105**, 153–188 (1986).
[176] M. Luscher, Nucl. Phys., **B354**, 531–578 (1991).
[177] L. Lellouch and M. Luscher, Commun. Math. Phys., **219**, 31–44 (2001), arXiv:hep-lat/0003023.
[178] T. Blum et al., Phys. Rev., **D91**(7), 074502 (2015), arXiv:1502.00263.
[179] Z. Bai et al., RBC, UKQCD, Phys. Rev. Lett., **115**(21), 212001 (2015), arXiv:1505.07863.
[180] A. J. Buras et al., JHEP, **11**, 202 (2015), arXiv:1507.06345.
[181] T. Kitahara et al., JHEP, **12**, 078 (2016), arXiv:1607.06727.
[182] K. Rummukainen and S. A. Gottlieb, Nucl. Phys., **B450**, 397–436 (1995), arXiv:hep-lat/9503028.
[183] C. H. Kim et al., Nucl. Phys., **B727**, 218–243 (2005), arXiv:hep-lat/0507006.
[184] N. H. Christ et al., Phys. Rev., **D72**, 114506 (2005), arXiv:hep-lat/0507009.
[185] M. Lage et al., Phys. Lett., **B681**, 439–443 (2009), arXiv:0905.0069.
[186] S. He et al., JHEP, **07**, 011 (2005), arXiv:hep-lat/0504019.
[187] R. A. Briceno, Phys. Rev., **D89**(7), 074507 (2014), arXiv:1401.3312.
[188] K. Polejaeva and A. Rusetsky, Eur. Phys. J., **A48**, 67 (2012), arXiv:1203.1241.
[189] M. T. Hansen and S. R. Sharpe, Phys. Rev., **D86**, 016007 (2012), arXiv:1204.0826.
[190] R. A. Briceno and Z. Davoudi, Phys. Rev., **D88**(9), 094507 (2013), arXiv:1204.1110.
[191] Phys. Rev., **D95**(7), 074510 (2017), arXiv:1701.07465.
[192] N. Carrasco et al., Phys. Rev., **D90**(1), 014502 (2014), arXiv:1403.7302.
[193] N. Carrasco et al., ETM, Phys. Rev., **D92**(3), 034516 (2015), arXiv:1505.06639.
[194] N. H. Christ et al., RBC, UKQCD, Phys. Rev., **D93**(11), 114517 (2016), arXiv:1605.04442.
[195] Hadron Spectrum, JHEP, **12**, 089 (2016), arXiv:1610.01073.
[196] M. Lüscher, Nucl. Phys. B, **364**, 237 (1991).
[197] C. B. Lang et al., JHEP, **09**, 089 (2015), arXiv:1503.05363.
[198] S. Prelovsek and L. Leskovec, Phys. Rev. Lett., **111**, 192001 (2013), arXiv:1307.5172.
[199] S.-H. Lee et al., Fermilab Lattice, MILC (2014), arXiv:1411.1389.
[200] V. M. Abazov et al., D0, Phys. Rev. Lett., **117**(2), 022003 (2016), arXiv:1602.07588.
[201] C. B. Lang et al., Phys. Rev., **D94**, 074509 (2016), arXiv:1607.03185.
[202] R. Aaij et al., LHCb, Phys. Rev. Lett., **117**(15), 152003, [Addendum: Phys. Rev. Lett.118,no.10,109904(2017)] (2016), arXiv:1608.00435.
[203] R. A. Briceno et al., Phys. Rev., **D91**(3), 034501 (2015), arXiv:1406.5965.
[204] R. A. Briceno et al., Phys. Rev. Lett., **115**, 242001 (2015), arXiv:1507.06622.
[205] L. Leskovec et al., PoS, **LATTICE2016**, 159 (2016), arXiv:1611.00282.
[206] J. J. Dudek et al., Hadron Spectrum, Phys. Rev. Lett., **113**(18), 182001 (2014), arXiv:1406.4158.
[207] N. Ishii et al., HAL QCD, Phys. Lett., **B712**, 437–441 (2012), arXiv:1203.3642.
[208] Y. Ikeda et al. (2016), arXiv:1602.03465.
[209] K. J. Juge et al., Phys. Rev. Lett., **82**, 4400–4403 (1999), arXiv:hep-ph/9902336.
[210] A. Peters et al., J. Phys. Conf. Ser., **742**(1), 012006 (2016), arXiv:1602.07621.
[211] E. Braaten et al., Phys. Rev., **D90**(1), 014044 (2014), arXiv:1402.0438.
[212] A. Bussone et al., ETM, Phys. Rev., **D93**(11), 114505 (2016), arXiv:1603.04306.
[213] C. McNeile et al., HPQCD, Phys. Rev., **D85**, 031503 (2012), arXiv:1110.4510.
[214] H. Na et al., HPQCD, Phys. Rev., **D86**, 034506 (2012), arXiv:1202.4914.
[215] A. Bazavov et al., Fermilab Lattice, MILC, Phys. Rev., **D85**, 114506 (2012), arXiv:1112.3051.
[216] N. H. Christ et al., RBC/UKQCD, Phys. Rev., **D91**(5), 054502 (2015), arXiv:1404.4670.
[217] R. J. Dowdall et al., HPQCD, Phys. Rev. Lett., **110**(22), 222003 (2013), arXiv:1302.2644.
[218] C. Bourrely et al., Phys. Rev., **D79**, 013008], [Erratum: Phys. Rev. D82, 099902 (2010) (2009), arXiv:0807.2722.
[219] C. Bobeth et al., Phys. Rev. Lett., **112**, 101801 (2014), arXiv:1311.0903.
[220] G. Burdman et al., Phys. Rev., **D51**, 111–117 (1995), arXiv:hep-ph/9405425.
[221] B. A. Dobrescu and A. S. Kronfeld, Phys. Rev. Lett., **100**, 241802 (2008), arXiv:0803.0512.
[222] V. Cirigliano and I. Rosell, JHEP, **10**, 005 (2007), arXiv:0707.4464.
[223] M. Beneke and J. Rohrwild, Eur. Phys. J., **C71**, 1818 (2011), arXiv:1110.3228.
[224] J. G. Körner and G. A. Schuler, Z. Phys., **C46**, 93 (1990).
[225] Y. Sakaki et al., Phys. Rev., **D88**(9), 094012 (2013), arXiv:1309.0301.

[226] A. S. Kronfeld, PoS, **LATTICE2008**, 282 (2008), arXiv:0812.2030.
[227] Z. Ligeti et al., JHEP, **01**, 083 (2017), arXiv:1610.02045.
[228] D. Atwood and W. J. Marciano, Phys. Rev., **D41**, 1736 (1990).
[229] S. de Boer et al., Phys. Rev. Lett., **120**(26), 261804 (2018), arXiv:1803.05881.
[230] Y. Amhis et al., HFLAV, Eur. Phys. J., **C77**(12), 895 (2017), arXiv:1612.07233.
[231] A. Bazavov et al., Fermilab Lattice, MILC (2017), arXiv:1712.09262.
[232] K. Ikado et al., Belle, Phys. Rev. Lett., **97**, 251802 (2006), arXiv:hep-ex/0604018.
[233] K. Hara et al., Belle, Phys. Rev., **D82**, 071101 (2010), arXiv:1006.4201.
[234] I. Adachi et al., Belle, Phys. Rev. Lett., **110**(13), 131801 (2013), arXiv:1208.4678.
[235] B. Aubert et al., BaBar, Phys. Rev., **D81**, 051101 (2010), arXiv:0912.2453.
[236] J. P. Lees et al., BaBar, Phys. Rev., **D88**(3), 031102 (2013), arXiv:1207.0698.
[237] N. Satoyama et al., Belle, Phys. Lett., **B647**, 67–73 (2007), arXiv:hep-ex/0611045.
[238] B. Aubert et al., BaBar, Phys. Rev., **D79**, 091101 (2009), arXiv:0903.1220.
[239] A. Höcker et al., PoS, **ACAT**, 040 (2007), arXiv:physics/0703039.
[240] B. Kronenbitter et al., Belle, Phys. Rev., **D92**(5), 051102 (2015), arXiv:1503.05613.
[241] Y. Yook et al., Belle, Phys. Rev., **D91**(5), 052016 (2015), arXiv:1406.6356.
[242] A. Sibidanov et al., Belle (2017), arXiv:1712.04123.
[243] M. Tanaka and R. Watanabe, PTEP, **2017**(1), 013B05 (2017), arXiv:1608.05207.
[244] M. Beneke et al., Nucl. Phys., **B591**, 313–418 (2000), arXiv:hep-ph/0006124.
[245] T. Feldmann, Non-Leptonic Heavy Meson Decays - Theory Status, In *12th Conference on Flavor Physics and CP Violation (FPCP 2014) Marseille, France, May 26-30, 2014* (2014), arXiv:1408.0300.
[246] V. M. Braun and A. Khodjamirian, Phys. Lett., **B718**, 1014–1019 (2013), arXiv:1210.4453.
[247] A. Heller et al., Belle, Phys. Rev., **D91**(11), 112009 (2015), arXiv:1504.05831.
[248] M. Gelb et al., Belle, Phys. Rev., **D98**(11), 112016 (2018), arXiv:1810.12976.
[249] A. Matyja et al., Belle, Phys. Rev. Lett., **99**, 191807 (2007), arXiv:0706.4429.
[250] J. P. Lees et al., BaBar, Phys. Rev. Lett., **109**, 101802 (2012), arXiv:1205.5442.
[251] M. Huschle et al., Belle, Phys. Rev., **D92**(7), 072014 (2015), arXiv:1507.03233.
[252] R. Aaij et al., LHCb, Phys. Rev. Lett., **115**(11), 111803, [Addendum: Phys. Rev. Lett. 115, no.15, 159901 (2015)] (2015), arXiv:1506.08614.
[253] A. Vaquero Avilés-Casco et al., Fermilab Lattice, MILC, EPJ Web Conf., **175**, 13003 (2018), arXiv:1710.09817.
[254] U. Nierste et al., Phys. Rev., **D78**, 015006 (2008), arXiv:0801.4938.
[255] M. Tanaka and R. Watanabe, Phys. Rev., **D82**, 034027 (2010), arXiv:1005.4306.
[256] S. Fajfer et al., Phys. Rev., **D85**, 094025 (2012), arXiv:1203.2654.
[257] J. A. Bailey et al., Fermilab Lattice, MILC, Phys. Rev. Lett., **109**, 071802 (2012), arXiv:1206.4992.
[258] D. Bigi and P. Gambino, Phys. Rev., **D94**(9), 094008 (2016), arXiv:1606.08030.
[259] F. U. Bernlochner et al., Phys. Rev., **D95**, 115008, [Erratum: Phys. Rev. D 97, 059902(2018)] (2017), arXiv:1703.05330.
[260] D. Bigi et al., JHEP, **11**, 061 (2017), arXiv:1707.09509.
[261] S. Jaiswal et al., JHEP, **12**, 060 (2017), arXiv:1707.09977.
[262] M. Jung and D. M. Straub (2018), arXiv:1801.01112.
[263] R. Alonso et al., Phys. Rev., **D95**(9), 093006 (2017), arXiv:1702.02773.
[264] M. Tanaka and R. Watanabe, Phys. Rev., **D87**(3), 034028 (2013), arXiv:1212.1878.
[265] A. K. Alok et al. (2016), arXiv:1606.03164.
[266] A. Bozek et al., Belle, Phys. Rev., **D82**, 072005 (2010), arXiv:1005.2302.
[267] Y. Sato et al., Belle, Phys. Rev., **D94**(7), 072007 (2016), arXiv:1607.07923.
[268] S. Hirose et al., Belle, Phys. Rev. Lett., **118**, 211801 (2017), arXiv:1612.00529.
[269] A. Abdesselam et al., Belle (2019), arXiv:1904.08794.
[270] J. P. Lees et al., BaBar, Phys. Rev., **D88**(7), 072012 (2013), arXiv:1303.0571.
[271] R. Aaij et al., LHCb, Phys. Rev. Lett., **120**(17), 171802 (2018), arXiv:1708.08856.
[272] S. Hirose et al., Belle, Phys. Rev., **D97**(1), 012004 (2018), arXiv:1709.00129.
[273] I. Doršner et al., JHEP, **11**, 084 (2013), arXiv:1306.6493.
[274] P. Krawczyk and S. Pokorski, Phys. Rev. Lett., **60**, 182 (1988).
[275] W-S. Hou, Phys. Rev., **D48**, 2342–2344 (1993).
[276] M. Tanaka, Z. Phys., **C67**, 321–326 (1995), arXiv:hep-ph/9411405.
[277] T. Miki et al., Effects of charged Higgs boson and QCD corrections in $\bar{B} \to D\tau\bar{\nu}_\tau$, In *Higher luminosity B factories. Proceedings, 3rd Workshop, Shonan Village, Japan, August 6-7, 2002*, pages 116–124 (2002), arXiv:hep-ph/0210051.
[278] A. G. Akeroyd and S. Recksiegel, J. Phys., **G29**, 2311–2317 (2003), arXiv:hep-ph/0306037.
[279] A. Crivellin et al., Phys. Rev., **D87**(9), 094031 (2013), arXiv:1303.5877.
[280] M. Jung et al., JHEP, **11**, 003 (2010), arXiv:1006.0470.

[281] X.-D. Cheng et al., Eur. Phys. J., **C74**(10), 3081 (2014), arXiv:1401.6657.
[282] T. Enomoto and R. Watanabe, JHEP, **05**, 002 (2016), arXiv:1511.05066.
[283] A. Crivellin et al., Phys. Rev., **D86**, 054014 (2012), arXiv:1206.2634.
[284] A. Celis et al., JHEP, **01**, 054 (2013), arXiv:1210.8443.
[285] A. Crivellin et al., Phys. Rev. Lett., **116**(8), 081801 (2015), arXiv:1507.07567.
[286] B. Bhattacharya et al., Phys. Lett., **B742**, 370–374 (2015), arXiv:1412.7164.
[287] A. Greljo et al., JHEP, **07**, 142 (2015), arXiv:1506.01705.
[288] L. Calibbi et al., Phys. Rev. Lett., **115**, 181801 (2015), arXiv:1506.02661.
[289] S. M. Boucenna et al., Phys. Lett., **B760**, 214–219 (2016), arXiv:1604.03088.
[290] B. Bhattacharya et al., JHEP, **01**, 015 (2017), arXiv:1609.09078.
[291] S. Fajfer et al., Phys. Rev. Lett., **109**, 161801 (2012), arXiv:1206.1872.
[292] R. Alonso et al., JHEP, **10**, 184 (2015), arXiv:1505.05164.
[293] M. Freytsis et al., Phys. Rev., **D92**(5), 054018 (2015), arXiv:1506.08896.
[294] M. Bauer and M. Neubert, Phys. Rev. Lett., **116**(14), 141802 (2015), arXiv:1511.01900.
[295] S. Fajfer and N. Košnik, Phys. Lett., **B755**, 270–274 (2016), arXiv:1511.06024.
[296] Y. Sakaki et al., Phys. Rev., **D91**(11), 114028 (2015), arXiv:1412.3761.
[297] P. Hamer et al., Belle, Phys. Rev., **D93**(3), 032007 (2016), arXiv:1509.06521.
[298] P. del Amo Sanchez et al., BaBar, Phys. Rev., **D83**, 032007 (2011), arXiv:1005.3288.
[299] H. Ha et al., Belle, Phys. Rev., **D83**, 071101 (2011), arXiv:1012.0090.
[300] J. P. Lees et al., BaBar, Phys. Rev., **D86**, 092004 (2012), arXiv:1208.1253.
[301] Y. Grossman et al., Nucl. Phys., **B465**, 369–398, [Erratum: Nucl. Phys. B480, 753 (1996)] (1996), arXiv:hep-ph/9510378.
[302] Y. Grossman et al., Phys. Rev., **D55**, 2768–2773 (1997), arXiv:hep-ph/9607473.
[303] F. U. Bernlochner et al., Phys. Rev., **D85**, 094033 (2012), arXiv:1202.1834.
[304] Thomas Mannel, Aleksey V. Rusov, and Farnoush Shahriaran, Nucl. Phys., **B921**, 211–224 (2017), arXiv:1702.01089.
[305] Srimoy Bhattacharya, Soumitra Nandi, and Sunando Kumar Patra, Eur. Phys. J., **C79**(3), 268 (2019), arXiv:1805.08222.
[306] F. U. Bernlochner and Z. Ligeti, Phys. Rev., **D95**(1), 014022 (2016), arXiv:1606.09300.
[307] A. F. Falk et al., Phys. Lett., **B326**, 145–153 (1994), arXiv:hep-ph/9401226.
[308] S. Balk et al., Z. Phys., **C64**, 37–44 (1994), arXiv:hep-ph/9312220.
[309] L. Koyrakh, Phys. Rev., **D49**, 3379–3384 (1994), arXiv:hep-ph/9311215.
[310] Z. Ligeti and F. J. Tackmann, Phys. Rev., **D90**(3), 034021 (2014), arXiv:1406.7013.
[311] A. V. Manohar and M. B. Wise, Phys. Rev., **D49**, 1310–1329 (1994), arXiv:hep-ph/9308246.
[312] T. Mannel and F. J. Tackmann, Phys. Rev., **D71**, 034017 (2005), arXiv:hep-ph/0408273.
[313] F. J. Tackmann, Phys. Rev., **D72**, 034036 (2005), arXiv:hep-ph/0503095.
[314] Z. Ligeti et al., Phys. Rev., **D78**, 114014 (2008), arXiv:0807.1926.
[315] C. Bauer et al., Phys. Rev., **D70**, 094017 (2004), hep-ph/0408002.
[316] F. U. Bernlochner et al., SIMBA, PoS, **ICHEP2012**, 370 (2013), arXiv:1303.0958.
[317] I. Caprini et al., Nucl. Phys., **B530**, 153–181 (1998), arXiv:hep-ph/9712417.
[318] B. Aubert et al., BaBar, Phys. Rev., **D77**, 032002 (2008), arXiv:0705.4008.
[319] B. Aubert et al., BaBar, Phys. Rev., **D79**, 012002 (2009), arXiv:0809.0828.
[320] A. Abdesselam et al., Belle (2018), arXiv:1809.03290.
[321] B. Aubert et al., BaBar, Phys. Rev. Lett., **104**, 011802 (2010), arXiv:0904.4063.
[322] C. G. Boyd et al., Phys. Rev. Lett., **74**, 4603 (1995), arXiv:hep-ph/9412324.
[323] D. Bigi et al., Phys. Lett., **B769**, 441–445 (2017), arXiv:1703.06124.
[324] A. Abdesselam et al., Belle (2017), arXiv:1702.01521.
[325] W. Dungel et al., Belle, Phys. Rev., **D82**, 112007 (2010), arXiv:1010.5620.
[326] B. Aubert et al., BaBar, Phys. Rev., **D79**, 112004 (2009), arXiv:0901.1291.
[327] K. Abe et al., Belle, Phys. Rev., **D69**, 112002 (2004), hep-ex/0307021.
[328] V.M. Abazov et al., D0, Phys. Rev. Lett., **95**, 171803 (2005), hep-ex/0507046.
[329] J. Abdallah et al., DELPHI, Eur.Phys.J., **C45**, 35 (2006), hep-ex/0510024.
[330] B. Aubert et al., BaBar, Phys. Rev. Lett., **101**, 261802 (2008), arXiv:0808.0528.
[331] D. Liventsev et al., Belle, Phys. Rev., **D77**, 091503 (2008), arXiv:0711.3252.
[332] P. del Amo Sanchez et al., BaBar, Phys. Rev., **D82**, 111101 (2010), arXiv:1009.2076.
[333] R. Aaij et al., LHCb, JHEP, **1309**, 145 (2013), 1307.4556.
[334] R. Aaij et al., LHCb, Phys. Rev., **D91**(9), 092002, [Erratum: Phys. Rev. D93, 119901 (2016)] (2015), arXiv:1503.02995.
[335] R. Aaij et al., LHCb, Phys. Rev., **D92**(3), 032002 (2015), arXiv:1505.01710.
[336] R. Aaij et al., LHCb, Phys. Rev., **D94**, 072001 (2016), arXiv:1608.01289.
[337] A. Le Yaouanc et al., Phys. Lett., **B387**, 582 (1996), hep-ph/9607300.

[338] N. Uraltsev, Phys. Lett., **B501**, 86 (2001), hep-ph/0011124.
[339] V. Morenas et al., Phys. Lett., **B386**, 315 (1996), arXiv:hep-ph/9605206.
[340] V. Morenas et al., Phys. Rev., **D56**, 5668 (1997), hep-ph/9706265.
[341] D. Ebert et al., Phys. Lett., **B434**, 365 (1998), hep-ph/9805423.
[342] D. Ebert et al., Phys. Rev., **D61**, 014016 (2000), hep-ph/9906415.
[343] A. K. others Leibovich, Phys. Rev. Lett., **78**, 3995–3998 (1997), arXiv:hep-ph/9703213.
[344] I.I. Bigi et al., Eur.Phys.J., **C52**, 975 (2007), arXiv:0708.1621.
[345] J. Segovia et al., Phys. Rev., **D84**, 094029 (2011), arXiv:1107.4248.
[346] R. Klein et al., Phys. Rev., **D91**(9), 094034 (2015), arXiv:1503.00569.
[347] B. Aubert et al., BaBar, Phys. Rev. Lett., **100**, 151802 (2008), arXiv:0712.3503.
[348] T. Lueck, PoS, **EPS-HEP2015**, 562 (2015).
[349] M. Atoui, Lattice computation of $B \to D^*, D^{**} l\nu$ form factors at finite heavy masses, In *Proceedings, 48th Rencontres de Moriond on QCD and High Energy Interactions: La Thuile, Italy, March 9-16, 2013*, page 143 (2013), arXiv:1305.0462.
[350] M. Atoui et al., Eur. Phys. J., **C75**(8), 376 (2015), arXiv:1312.2914.
[351] B. Blossier, $B \to D^{**}$ – puzzle 1/2 vs 3/2, In *8th International Workshop on the CKM Unitarity Triangle (CKM 2014), Sept 8-12, Vienna, Austria, 2014* (2014), arXiv:1411.3563.
[352] P. Gambino and C. Schwanda, Phys. Rev., **D89**, 014022 (2014), arXiv:1307.4551.
[353] A. Alberti et al., Phys. Rev. Lett., **114**(6), 061802 (2015), arXiv:1411.6560.
[354] P. Gambino et al., Phys. Lett., **B763**, 60–65 (2016), arXiv:1606.06174.
[355] T. Mannel et al., JHEP, **11**, 109 (2010), arXiv:1009.4622.
[356] J. Heinonen and T. Mannel, Nucl. Phys., **B889**, 46 (2014), arXiv:1407.4384.
[357] J.P. Lees et al., BaBar, Phys. Rev., **D85**, 011101 (2012), arXiv:1110.5600.
[358] C. Oswald et al., Belle, Phys. Rev., **D87**(7), 072008 (2013), arXiv:1212.6400.
[359] M. Gronau and J. L. Rosner, Phys. Rev., **D83**, 034025 (2011), arXiv:1012.5098.
[360] I.I. Bigi et al., JHEP, **1109**, 012 (2011), arXiv:1105.4574.
[361] V.M. Abazov et al., D0, Phys. Rev. Lett., **102**, 051801 (2009), arXiv:0712.3789.
[362] R. Aaij et al., LHCb, Phys. Lett., **B698**, 14–20 (2011), arXiv:1102.0348.
[363] C. Oswald et al., Belle, Phys. Rev., **D92**(7), 072013 (2015), arXiv:1504.02004.
[364] M. Neubert, Phys. Rev., **D49**, 4623–4633 (1994), arXiv:hep-ph/9312311.
[365] I. I. Y. Bigi et al., Int. J. Mod. Phys., **A9**, 2467–2504 (1994), arXiv:hep-ph/9312359.
[366] C. W. Bauer et al., Phys. Rev., **D68**, 094001 (2003), arXiv:hep-ph/0102089.
[367] K. S. M. Lee and I. W. Stewart, Nucl. Phys., **B721**, 325–406 (2005), arXiv:hep-ph/0409045.
[368] P. Gambino et al., JHEP, **0710**, 058 (2007), arXiv:0707.2493.
[369] B. O. Lange et al., Phys. Rev., **D72**, 073006 (2005), arXiv:hep-ph/0504071.
[370] S.W. Bosch et al., Nucl. Phys., **B699**, 335–386 (2004), hep-ph/0402094.
[371] S. W. Bosch et al., JHEP, **11**, 073 (2004), arXiv:hep-ph/0409115.
[372] P. Gambino, K. J. Healey, and C. Mondino, Phys. Rev., **D94**(1), 014031 (2016), arXiv:1604.07598.
[373] J. R. Andersen and E. Gardi, JHEP, **0601**, 097 (2006), hep-ph/0509360.
[374] U. Aglietti and G. Ricciardi, Phys. Rev., **D70**, 114008 (2004), hep-ph/0407225.
[375] U. Aglietti et al., Nucl. Phys., **B768**, 85–115 (2007), hep-ph/0608047.
[376] U. Aglietti et al., Eur.Phys.J., **C59**, 831–840 (2009), 0711.0860.
[377] P. Urquijo et al., Belle, Phys. Rev. Lett., **104**, 021801 (2010), arXiv:0907.0379.
[378] J.P. Lees et al., BaBar, Phys. Rev., **D86**, 032004 (2012), arXiv:1112.0702.
[379] A. Bornheim et al., CLEO, Phys. Rev. Lett., **88**, 231803 (2002), hep-ex/0202019.
[380] A. Limosani et al., Belle, Phys. Lett., **B621**, 28–40 (2005), arXiv:hep-ex/0504046.
[381] J. P. Lees et al., BaBar, Phys. Rev., **D95**(7), 072001 (2017), arXiv:1611.05624.
[382] C. W. Bauer et al., Phys. Rev., **D64**, 113004 (2001), arXiv:hep-ph/0107074.
[383] I. Bigi et al., JHEP, **1004**, 073 (2010), arXiv:0911.3322.
[384] P. Gambino and J. F. Kamenik, Nucl. Phys., **B840**, 424 (2010), arXiv:1004.0114.
[385] Z. Ligeti et al., Phys. Rev., **D82**, 033003 (2010), arXiv:1003.1351.
[386] N. Brambilla et al., TUMQCD, Phys. Rev., **D97**(3), 034503 (2018), arXiv:1712.04983.
[387] A. Bazavov et al., Fermilab Lattice, MILC, TUMQCD (2018), arXiv:1802.04248.
[388] R. Aaij et al., LHCb, JHEP, **08**, 131 (2013), 1304.6325.
[389] R. Aaij et al., LHCb, JHEP, **02**, 104 (2016), arXiv:1512.04442.
[390] S. Wehle et al., Belle, Phys. Rev. Lett., **118**(11), 111801 (2017), arXiv:1612.05014.
[391] R. Aaij et al., LHCb, Phys. Rev. Lett., **113**, 151601 (2014), arXiv:1406.6482.
[392] R. Aaij et al., LHCb (2017), arXiv:1705.05802.
[393] V. Khachatryan et al., CMS, Phys. Lett., **B753**, 424–448 (2016), arXiv:1507.08126.
[394] G. Aad et al., Angular analysis of $B_d^0 \to K^* \mu^+ \mu^-$ decays in $pp$ collisions at $\sqrt{s} = 8$ TeV with the ATLAS detector, Technical Report ATLAS-CONF-2017-023, CERN, Geneva (Apr 2017).

[395] A. M Sirunyan et al., CMS (2017), arXiv:1710.02846.
[396] G. Buchalla et al., Rev. Mod. Phys., **68**, 1125–1144 (1996), arXiv:hep-ph/9512380.
[397] K. G. Chetyrkin et al., Phys. Lett., **B400**, 206–219, [Erratum: Phys. Lett. B425, 414 (1998)] (1997), arXiv:hep-ph/9612313.
[398] M. Czakon et al., JHEP, **03**, 008 (2007), arXiv:hep-ph/0612329.
[399] C. Bobeth et al., Nucl. Phys., **B574**, 291–330 (2000), arXiv:hep-ph/9910220.
[400] P. Gambino et al., Nucl. Phys., **B673**, 238–262 (2003), arXiv:hep-ph/0306079.
[401] M. Gorbahn and U. Haisch, Nucl. Phys., **B713**, 291–332 (2005), arXiv:hep-ph/0411071.
[402] M. Misiak and J. Urban, Phys. Lett., **B451**, 161–169 (1999), arXiv:hep-ph/9901278.
[403] G. Buchalla and A. J. Buras, Nucl. Phys., **B548**, 309–327 (1999), arXiv:hep-ph/9901288.
[404] T. Hermann et al., JHEP, **12**, 097 (2013), arXiv:1311.1347.
[405] J. Brod et al., Phys. Rev., **D83**, 034030 (2011), arXiv:1009.0947.
[406] C. Bobeth et al., Phys. Rev., **D89**(3), 034023 (2014), arXiv:1311.1348.
[407] B. Blok et al., Phys. Rev., **D49**, 3356, [Erratum: Phys. Rev. D50, 3572 (1994)] (1994), arXiv:hep-ph/9307247.
[408] M. Neubert, Phys. Rev., **D49**, 3392–3398 (1994), arXiv:hep-ph/9311325.
[409] S. J. Lee et al., Phys. Rev., **D75**, 114005 (2007), arXiv:hep-ph/0609224.
[410] M. Benzke et al., JHEP, **08**, 099 (2010), arXiv:1003.5012.
[411] S. Aoki et al., Flavour Lattice Averaging Group (FLAG), Eur. Phys. J., **C74**, 2890 (2014), arXiv:1310.8555.
[412] C. Bouchard et al., HPQCD, Phys. Rev. Lett., **111**(16), 162002, [Erratum: Phys. Rev. Lett. 112, 149902 (2014)] (2013), arXiv:1306.0434.
[413] R. R. Horgan et al., Phys. Rev., **D89**(9), 094501 (2014), arXiv:1310.3722.
[414] J. Flynn et al., Hadronic form factors for rare semileptonic $B$ decays, In *Proceedings, 33rd International Symposium on Lattice Field Theory (Lattice 2015)* (2015), arXiv:1511.06622.
[415] M. Beneke et al., Phys. Rev. Lett., **83**, 1914–1917 (1999), arXiv:hep-ph/9905312.
[416] M. Beneke et al., Nucl. Phys., **B606**, 245–321 (2001), arXiv:hep-ph/0104110.
[417] A. Khodjamirian and R. Rückl, Adv. Ser. Direct. High Energy Phys., **15**, 345–401 (1998), arXiv:hep-ph/9801443.
[418] P. Ball and R. Zwicky, Phys. Rev., **D71**, 014029 (2005), arXiv:hep-ph/0412079.
[419] A. Bharucha et al. (2015), arXiv:1503.05534.
[420] B. Aubert et al., BaBar, Phys. Rev., **D77**, 051103 (2008), arXiv:0711.4889.
[421] P. del Amo Sanchez et al., BaBar, Phys. Rev., **D82**, 051101 (2010), arXiv:1005.4087.
[422] J. P. Lees et al., BaBar, Phys. Rev., **D86**, 052012 (2012), arXiv:1207.2520.
[423] J. P. Lees et al., BaBar, Phys. Rev. Lett., **109**, 191801 (2012), arXiv:1207.2690.
[424] T. Saito et al., Belle, Phys. Rev., **D91**(5), 052004 (2015), arXiv:1411.7198.
[425] A. Abdesselam et al., Measurement of the inclusive $B \to X_{s+d}\gamma$ branching fraction, photon energy spectrum and HQE parameters, In *38th International Conference on High Energy Physics (ICHEP 2016) Chicago, IL, USA, August 03-10, 2016* (2016), arXiv:1608.02344.
[426] M. Czakon et al., JHEP, **04**, 168 (2015), arXiv:1503.01791.
[427] M. Misiak et al., Phys. Rev. Lett., **114**(22), 221801 (2015), arXiv:1503.01789.
[428] O. Buchmüller and H. Flächer, Phys. Rev., **D73**, 073008 (2006), arXiv:hep-ph/0507253.
[429] G. Paz, Theory of Inclusive Radiative B Decays, In *CKM unitarity triangle. Proceedings, 6th International Workshop, CKM 2010, Warwick, UK, September 6-10, 2010* (2010), arXiv:1011.4953.
[430] F. U. Bernlochner et al., Status of SIMBA, In *CKM unitarity triangle. Proceedings, 6th International Workshop, CKM 2010, Warwick, UK, September 6-10, 2010* (2011), arXiv:1101.3310.
[431] A. Crivellin and L. Mercolli, Phys. Rev., **D84**, 114005 (2011), arXiv:1106.5499.
[432] H. M. Asatrian and C. Greub, Phys. Rev., **D88**(7), 074014 (2013), arXiv:1305.6464.
[433] P. Gambino and M. Misiak, Nucl. Phys., **B611**, 338–366 (2001), arXiv:hep-ph/0104034.
[434] G. Buchalla et al., Nucl. Phys., **B511**, 594–610 (1998), arXiv:hep-ph/9705253.
[435] M. Kaminski et al., Phys. Rev., **D86**, 094004 (2012), arXiv:1209.0965.
[436] T. Huber et al., JHEP, **01**, 115 (2015), arXiv:1411.7677.
[437] I. R. Blokland et al., Phys. Rev., **D72**, 033014 (2005), arXiv:hep-ph/0506055.
[438] K. Melnikov and A. Mitov, Phys. Lett., **B620**, 69–79 (2005), arXiv:hep-ph/0505097.
[439] H. M. Asatrian et al., Phys. Lett., **B647**, 173–178 (2007), arXiv:hep-ph/0611123.
[440] T. Ewerth, Phys. Lett., **B669**, 167–172 (2008), arXiv:0805.3911.
[441] H. M. Asatrian et al., Phys. Rev., **D82**, 074006 (2010), arXiv:1005.5587.
[442] Z. Ligeti et al., Phys. Rev., **D60**, 034019 (1999), arXiv:hep-ph/9903305.
[443] A. Ferroglia and U. Haisch, Phys. Rev., **D82**, 094012 (2010), arXiv:1009.2144.
[444] M. Misiak and M. Poradzinski, Phys. Rev., **D83**, 014024 (2011), arXiv:1009.5685.
[445] S. J. Brodsky et al., Phys. Rev., **D28**, 228 (1983).

[446] K. Bieri et al., Phys. Rev., **D67**, 114019 (2003), arXiv:hep-ph/0302051.
[447] R. Boughezal et al., JHEP, **09**, 072 (2007), arXiv:0707.3090.
[448] M. Misiak and M. Steinhauser, Nucl. Phys., **B764**, 62–82 (2007), arXiv:hep-ph/0609241.
[449] M. Misiak and M. Steinhauser, Nucl. Phys., **B840**, 271–283 (2010), arXiv:1005.1173.
[450] Mikolaj Misiak, Abdur Rehman, and Matthias Steinhauser, Phys. Lett., **B770**, 431–439 (2017), arXiv:1702.07674.
[451] J. Charles et al., Phys. Rev., **D91**(7), 073007 (2015), arXiv:1501.05013.
[452] T. Ewerth et al., Nucl. Phys., **B830**, 278–290 (2010), arXiv:0911.2175.
[453] A. Alberti et al., JHEP, **01**, 147 (2014), arXiv:1311.7381.
[454] I. I. Y. Bigi et al., Phys. Lett., **B293**, 430–436, [Erratum: Phys. Lett. B297, 477 (1992)] (1992), arXiv:hep-ph/9207214.
[455] A. F. Falk et al., Phys. Rev., **D49**, 3367–3378 (1994), arXiv:hep-ph/9308288.
[456] C. W. Bauer, Phys. Rev., **D57**, 5611–5619, [Erratum: Phys. Rev. D60, 099907 (1999)] (1998), arXiv:hep-ph/9710513.
[457] T. Becher and M. Neubert, Phys. Lett., **B637**, 251–259 (2006), arXiv:hep-ph/0603140.
[458] M. Beneke et al., JHEP, **06**, 071 (2005), arXiv:hep-ph/0411395.
[459] G. Paz, JHEP, **06**, 083 (2009), arXiv:0903.3377.
[460] B. O. Lange et al., JHEP, **0510**, 084 (2005), hep-ph/0508178.
[461] B. O. Lange, JHEP, **01**, 104 (2006), arXiv:hep-ph/0511098.
[462] A. Ali and C. Greub, Phys. Lett., **B361**, 146–154 (1995), arXiv:hep-ph/9506374.
[463] A. Kapustin et al., Phys. Lett., **B357**, 653–658 (1995), arXiv:hep-ph/9507248.
[464] J. F. Donoghue and A. A. Petrov, Phys. Rev., **D53**, 3664–3671 (1996), arXiv:hep-ph/9510227.
[465] M. B. Voloshin, Phys. Lett., **B397**, 275–278 (1997), arXiv:hep-ph/9612483.
[466] Z. Ligeti et al., Phys. Lett., **B402**, 178–182 (1997), arXiv:hep-ph/9702322.
[467] A. K. Grant et al., Phys. Rev., **D56**, 3151–3154 (1997), arXiv:hep-ph/9702380.
[468] M. Benzke et al., Phys. Rev. Lett., **106**, 141801 (2011), arXiv:1012.3167.
[469] J. M. Soares, Nucl. Phys., **B367**, 575–590 (1991).
[470] A. Ali et al., Phys. Lett., **B429**, 87–98 (1998), arXiv:hep-ph/9803314.
[471] A. L. Kagan and M. Neubert, Phys. Rev., **D58**, 094012 (1998), arXiv:hep-ph/9803368.
[472] T. Hurth et al., Nucl. Phys., **B704**, 56–74 (2005), arXiv:hep-ph/0312260.
[473] M. Misiak, Acta Phys. Polon., **B40**, 2987–2996 (2009), arXiv:0911.1651.
[474] J. P. Lees et al., BaBar, Phys. Rev., **D86**, 112008 (2012), arXiv:1207.5772.
[475] A. Limosani et al., Belle, Phys. Rev. Lett., **103**, 241801 (2009), arXiv:0907.1384.
[476] S. Chen et al., CLEO, Phys. Rev. Lett., **87**, 251807 (2001), arXiv:hep-ex/0108032.
[477] B. Aubert et al., BaBar, Phys. Rev., **D72**, 052004 (2005), arXiv:hep-ex/0508004.
[478] S. Watanuki et al., Belle (2018), arXiv:1807.04236.
[479] B. Aubert et al., BaBar, Phys. Rev. Lett., **95**, 042001 (2005), arXiv:hep-ex/0504001.
[480] L. Pesantez et al., Belle, Phys. Rev. Lett., **114**(15), 151601 (2015), arXiv:1501.01702.
[481] J. P. Lees et al., BaBar, Phys. Rev., **D90**(9), 092001 (2014), arXiv:1406.0534.
[482] S. W. Bosch and G. Buchalla, Nucl. Phys., **B621**, 459–478 (2002), arXiv:hep-ph/0106081.
[483] M. Beneke et al., Nucl. Phys., **B612**, 25–58 (2001), arXiv:hep-ph/0106067.
[484] M. Beneke et al., Eur. Phys. J., **C41**, 173–188 (2005), arXiv:hep-ph/0412400.
[485] M. Dimou et al., Phys. Rev., **D87**(7), 074008 (2013), arXiv:1212.2242.
[486] A. Ali and V. M. Braun, Phys. Lett., **B359**, 223–235 (1995), arXiv:hep-ph/9506248.
[487] A. Khodjamirian et al., Phys. Lett., **B358**, 129–138 (1995), arXiv:hep-ph/9506242.
[488] J. Lyon and R. Zwicky, Phys. Rev., **D88**(9), 094004 (2013), arXiv:1305.4797.
[489] R. Aaij et al., LHCb, Phys. Rev. Lett., **112**(16), 161801 (2014), arXiv:1402.6852.
[490] D. Becirevic et al., JHEP, **08**, 090 (2012), arXiv:1206.1502.
[491] S. Jäger and J. Martin Camalich, Phys. Rev., **D93**(1), 014028 (2016), arXiv:1412.3183.
[492] M. Borsato, *Study of the* $B^0 \to K^{*0} e^+ e^-$ *decay with the LHCb detector and development of a novel concept of PID detector: the Focusing DIRC*, PhD thesis, Santiago de Compostela U. (2015).
[493] A. Paul and D. M. Straub, JHEP, **04**, 027 (2017), arXiv:1608.02556.
[494] P. Ball, G. W. Jones, and R. Zwicky, Phys. Rev., **D75**, 054004 (2007), arXiv:hep-ph/0612081.
[495] R. Aaij et al., LHCb, Phys. Rev., **D85**, 112013 (2012), arXiv:1202.6267.
[496] R. Aaij et al., LHCb, Nucl. Phys., **B867**, 1–18 (2013), arXiv:1209.0313.
[497] S. Descotes-Genon et al., Phys. Rev., **D88**, 074002 (2013), arXiv:1307.5683.
[498] W. Altmannshofer et al., Implications of $b \to s$ measurements, In *Proceedings, 50th Rencontres de Moriond Electroweak interactions and unified theories*, pages 333–338 (2015), arXiv:1503.06199.
[499] R. Aaij et al., LHCb, JHEP, **09**, 179 (2015), arXiv:1506.08777.
[500] A. Ali and A. Y. Parkhomenko, Eur. Phys. J., **C23**, 89–112 (2002), arXiv:hep-ph/0105302.
[501] A. L. Kagan and M. Neubert, Phys. Lett., **B539**, 227–234 (2002), arXiv:hep-ph/0110078.

[502] D. Atwood et al., Phys. Rev. Lett., **79**, 185–188 (1997), arXiv:hep-ph/9704272.
[503] F. Muheim et al., Phys. Lett., **B664**, 174–179 (2008), arXiv:0802.0876.
[504] E. Kou et al., JHEP, **12**, 102 (2013), arXiv:1305.3173.
[505] N. Haba et al., JHEP, **03**, 160 (2015), arXiv:1501.00668.
[506] B. Grinstein et al., Phys. Rev., **D71**, 011504 (2005), arXiv:hep-ph/0412019.
[507] S. Jäger and J. M. Camalich, JHEP, **05**, 043 (2013), arXiv:1212.2263.
[508] P. Ball and R. Zwicky, JHEP, **04**, 046 (2006), arXiv:hep-ph/0603232.
[509] A. Khodjamirian et al., JHEP, **09**, 089 (2010), arXiv:1006.4945.
[510] J. Gratrex and R. Zwicky (2018), arXiv:1804.09006.
[511] P. Ball and R. Zwicky, Phys. Lett., **B642**, 478–486 (2006), arXiv:hep-ph/0609037.
[512] M. Gronau et al., Phys. Rev. Lett., **88**, 051802 (2002), arXiv:hep-ph/0107254.
[513] M. Gronau and D. Pirjol, Phys. Rev., **D66**, 054008 (2002), arXiv:hep-ph/0205065.
[514] E. Kou et al., Phys. Rev., **D83**, 094007 (2011), arXiv:1011.6593.
[515] R. Aaij et al., LHCb, JHEP, **04**, 064 (2015), arXiv:1501.03038.
[516] F. Bishara and D. J. Robinson, JHEP, **09**, 013 (2015), arXiv:1505.00376.
[517] D. Atwood et al., Phys. Rev., **D71**, 076003 (2005), arXiv:hep-ph/0410036.
[518] R. Ammar et al., CLEO, Phys. Rev. Lett., **71**, 674–678 (1993).
[519] Y. Ushiroda et al., Belle, Phys. Rev., **D74**, 111104 (2006), arXiv:hep-ex/0608017.
[520] H. Nakano et al., Belle, Phys. Rev., **D97**(9), 092003 (2018), arXiv:1803.07774.
[521] J. Li et al., Belle, Phys. Rev. Lett., **101**, 251601 (2008), arXiv:0806.1980.
[522] H. Sahoo et al., Belle, Phys. Rev., **D84**, 071101 (2011), arXiv:1104.5590.
[523] T. Horiguchi et al., Belle, Phys. Rev. Lett., **119**(19), 191802 (2017), arXiv:1707.00394.
[524] W. Altmannshofer and D. M. Straub, Eur. Phys. J., **C75**(8), 382 (2015), arXiv:1411.3161.
[525] K. Abe et al., Belle, Phys. Rev. Lett., **96**, 221601 (2006), arXiv:hep-ex/0506079.
[526] Y. Ushiroda et al., Belle, Phys. Rev. Lett., **100**, 021602 (2008), arXiv:0709.2769.
[527] N. Taniguchi et al., Belle, Phys. Rev. Lett., **101**, 111801, [Erratum: Phys. Rev. Lett. 101, 129904 (2008)] (2008), arXiv:0804.4770.
[528] B. Aubert et al., BaBar, Phys. Rev., **D78**, 112001 (2008), arXiv:0808.1379.
[529] D. Dutta et al., Belle, Phys. Rev., **D91**(1), 011101 (2015), arXiv:1411.7771.
[530] A. Drutskoy et al., Belle, Phys. Rev. Lett., **98**, 052001 (2007), arXiv:hep-ex/0608015.
[531] P. del Amo Sanchez et al., BaBar, Phys. Rev., **D83**, 032006 (2011), arXiv:1010.2229.
[532] S. Villa et al., Belle, Phys. Rev., **D73**, 051107 (2006), arXiv:hep-ex/0507036.
[533] S. W. Bosch and G. Buchalla, JHEP, **08**, 054 (2002), arXiv:hep-ph/0208202.
[534] K. De Bruyn et al., Phys. Rev., **D86**, 014027 (2012), arXiv:1204.1735.
[535] S. W. Bosch, *Exclusive radiative decays of B mesons in QCD factorization*, PhD thesis, Munich, Max Planck Inst. (2002), arXiv:hep-ph/0208203.
[536] J. L. Hewett et al., editors, *The Discovery potential of a Super B Factory. Proceedings, SLAC Workshops, Stanford, USA, 2003* (2004), arXiv:hep-ph/0503261.
[537] C. Bobeth and U. Haisch, Acta Phys. Polon., **B44**, 127–176 (2013), arXiv:1109.1826.
[538] T. M. Aliev and G. Turan, Phys. Rev., **D48**, 1176–1184 (1993).
[539] T. M. Aliev et al., Nucl. Phys., **B515**, 321–341 (1998), arXiv:hep-ph/9708382.
[540] S. Bertolini and J. Matias, Phys. Rev., **D57**, 4197–4204 (1998), arXiv:hep-ph/9709330.
[541] I. I. Bigi et al., GESJ Phys., **2006N1**, 57–79 (2006), arXiv:hep-ph/0603160.
[542] A. Gemintern et al., Phys. Rev., **D70**, 035008 (2004), arXiv:hep-ph/0404152.
[543] H. M. Asatrian et al., Phys. Rev., **D85**, 014020 (2012), arXiv:1110.1251.
[544] H. M Asatrian and C. Greub, Phys. Rev., **D89**(9), 094028 (2014), arXiv:1403.4502.
[545] H. Simma and D. Wyler, Nucl. Phys., **B344**, 283–316 (1990).
[546] L. Reina et al., Phys. Lett., **B396**, 231–237 (1997), arXiv:hep-ph/9612387.
[547] L. Reina et al., Phys. Rev., **D56**, 5805–5815 (1997), arXiv:hep-ph/9706253.
[548] J. Cao et al., Phys. Rev., **D64**, 014012 (2001), arXiv:hep-ph/0103154.
[549] H. M. Asatrian et al., Phys. Rev., **D93**(1), 014037 (2016), arXiv:1511.00153.
[550] H. M. Asatrian et al., Phys. Rev., **D95**(5), 053006 (2017), arXiv:1611.08449.
[551] C-H. V. Chang et al., Phys. Lett., **B415**, 395–401 (1997), arXiv:hep-ph/9705345.
[552] I. Adachi et al., Belle, Phys. Lett., **B662**, 323–329 (2008), arXiv:hep-ex/0608037.
[553] R. Aaij et al., LHCb, JHEP, **1307**, 084 (2013), arXiv:1305.2168.
[554] K. S. M. Lee et al., Phys. Rev., **D75**, 034016 (2007), arXiv:hep-ph/0612156.
[555] A. Ali et al., Phys. Lett., **B273**, 505–512 (1991).
[556] H. H. Asatryan et al., Phys. Rev., **D65**, 074004 (2002), arXiv:hep-ph/0109140.
[557] H. H. Asatrian et al., Phys. Lett., **B507**, 162–172 (2001), arXiv:hep-ph/0103087.
[558] H. H. Asatryan et al., Phys. Rev., **D66**, 034009 (2002), arXiv:hep-ph/0204341.
[559] A. Ghinculov et al., Nucl. Phys., **B648**, 254–276 (2003), arXiv:hep-ph/0208088.

[560] H. M. Asatrian et al., Phys. Rev., **D66**, 094013 (2002), arXiv:hep-ph/0209006.
[561] H. M. Asatrian et al., Mod. Phys. Lett., **A19**, 603–614 (2004), arXiv:hep-ph/0311187.
[562] A. Ghinculov et al., Nucl. Phys., **B685**, 351–392 (2004), arXiv:hep-ph/0312128.
[563] C. Greub et al., JHEP, **12**, 040 (2008), arXiv:0810.4077.
[564] C. Bobeth et al., JHEP, **04**, 071 (2004), arXiv:hep-ph/0312090.
[565] T. Huber et al., Nucl. Phys., **B740**, 105–137 (2006), arXiv:hep-ph/0512066.
[566] T. Huber et al., Nucl. Phys., **B802**, 40–62 (2008), arXiv:0712.3009.
[567] T. Huber et al., JHEP, **06**, 176 (2015), arXiv:1503.04849.
[568] A. Ali et al., Phys. Rev., **D55**, 4105–4128 (1997), arXiv:hep-ph/9609449.
[569] J.-W. Chen et al., Phys. Lett., **B410**, 285–289 (1997), arXiv:hep-ph/9705219.
[570] G. Buchalla and G. Isidori, Nucl. Phys., **B525**, 333–349 (1998), arXiv:hep-ph/9801456.
[571] C. W. Bauer and C. N. Burrell, Phys. Rev., **D62**, 114028 (2000), arXiv:hep-ph/9911404.
[572] Z. Ligeti and F. J. Tackmann, Phys. Lett., **B653**, 404–410 (2007), arXiv:0707.1694.
[573] M. Benzke et al., JHEP, **10**, 031 (2017), arXiv:1705.10366.
[574] B. Aubert et al., BaBar, Phys. Rev. Lett., **93**, 081802 (2004), arXiv:hep-ex/0404006.
[575] J.P. Lees et al., BaBar (2013), arXiv:1312.5364.
[576] J. Kaneko et al., Belle, Phys. Rev. Lett., **90**, 021801 (2003), arXiv:hep-ex/0208029.
[577] M. Iwasaki et al., Belle, Phys. Rev., **D72**, 092005 (2005), arXiv:hep-ex/0503044.
[578] Y. Sato et al., Belle, Phys. Rev., **D93**(3), 032008, [Addendum: Phys. Rev.D93,no.5,059901(2016)] (2016), arXiv:1402.7134.
[579] K. S. M. Lee and I. W. Stewart, Phys. Rev., **D74**, 014005 (2006), arXiv:hep-ph/0511334.
[580] K. S. M. Lee et al., Phys. Rev., **D74**, 011501 (2006), arXiv:hep-ph/0512191.
[581] K. S. M. Lee and F. J. Tackmann, Phys. Rev., **D79**, 114021 (2009), arXiv:0812.0001.
[582] G. Bell et al., Nucl. Phys., **B843**, 143–176 (2011), arXiv:1007.3758.
[583] F. Krüger and L. M. Sehgal, Phys. Lett., **B380**, 199–204 (1996), arXiv:hep-ph/9603237.
[584] H. M. Asatrian et al., Phys. Rev., **D69**, 074007 (2004), arXiv:hep-ph/0312063.
[585] D. Seidel, Phys. Rev., **D70**, 094038 (2004), arXiv:hep-ph/0403185.
[586] G. Buchalla et al., Phys. Rev., **D63**, 014015 (2000), arXiv:hep-ph/0006136.
[587] F. Krüger et al., Phys. Rev., **D61**, 114028, [Erratum: Phys. Rev. D63, 019901 (2001)] (2000), arXiv:hep-ph/9907386.
[588] W. Altmannshofer et al., JHEP, **01**, 019 (2009), arXiv:0811.1214.
[589] J. Gratrex et al., Phys. Rev., **D93**(5), 054008 (2016), arXiv:1506.03970.
[590] C. Bobeth et al., JHEP, **12**, 040 (2007), arXiv:0709.4174.
[591] J. Matias et al., JHEP, **04**, 104 (2012), arXiv:1202.4266.
[592] S. Descotes-Genon et al., JHEP, **01**, 048 (2013), arXiv:1207.2753.
[593] S. Descotes-Genon et al., JHEP, **05**, 137 (2013), arXiv:1303.5794.
[594] F. Kruger and J. Matias, Phys. Rev., **D71**, 094009 (2005), arXiv:hep-ph/0502060.
[595] D. Becirevic and E. Schneider, Nucl. Phys., **B854**, 321–339 (2012), arXiv:1106.3283.
[596] Y. Grossman and D. Pirjol, JHEP, **06**, 029 (2000), arXiv:hep-ph/0005069.
[597] M. Bordone et al., Eur. Phys. J., **C76**(8), 440 (2016), arXiv:1605.07633.
[598] G. Hiller and M. Schmaltz, JHEP, **02**, 055 (2015), arXiv:1411.4773.
[599] W. Altmannshofer and I. Yavin, Phys. Rev., **D92**(7), 075022 (2015), arXiv:1508.07009.
[600] B. Capdevila et al., JHEP, **10**, 075 (2016), arXiv:1605.03156.
[601] J. P. Lees et al., BaBar, Phys. Rev., **D86**, 032012 (2012), arXiv:1204.3933.
[602] J.-T. Wei et al., Belle, Phys. Rev. Lett., **103**, 171801 (2009), arXiv:0904.0770.
[603] K. Abe et al., Belle, Phys. Rev. Lett., **88**, 021801 (2002), arXiv:hep-ex/0109026.
[604] A. Ishikawa et al., Belle, Phys. Rev. Lett., **91**, 261601 (2003), arXiv:hep-ex/0308044.
[605] A. Ishikawa et al., Belle, Phys. Rev. Lett., **96**, 251801 (2006), arXiv:hep-ex/0603018.
[606] T. Aaltonen et al., CDF, Phys. Rev. Lett., **108**, 081807 (2012), arXiv:1108.0695.
[607] J. P. Lees et al., BaBar, Phys. Rev., **D93**(5), 052015 (2016), arXiv:1508.07960.
[608] R. Aaij et al., LHCb, JHEP, **11**, 047 (2016), arXiv:1606.04731.
[609] V. Khachatryan et al., Measurement of the $P_1$ and $P_5'$ angular parameters of the decay $\mathrm{B}^0 \to \mathrm{K}^{*0}\mu^+\mu^-$ in proton-proton collisions at $\sqrt{s} = 8$ TeV, Technical Report CMS-PAS-BPH-15-008, CERN, Geneva (2017).
[610] R. Aaij et al., LHCb, Phys. Rev. Lett., **111**, 191801 (2013), arXiv:1308.1707.
[611] W. Altmannshofer et al., Eur. Phys. J., **C77**(6), 377 (2017), arXiv:1703.09189.
[612] S. Descotes-Genon et al., JHEP, **06**, 092 (2016), arXiv:1510.04239.
[613] A. J. Buras et al., JHEP, **02**, 184 (2015), arXiv:1409.4557.
[614] W. Altmannshofer et al., JHEP, **04**, 022 (2009), arXiv:0902.0160.
[615] J. F. Kamenik and C. Smith, Phys. Lett., **B680**, 471–475 (2009), arXiv:0908.1174.
[616] J. F. Kamenik and C. Smith, JHEP, **03**, 090 (2012), arXiv:1111.6402.

[617] R. R. Horgan et al., PoS, **LATTICE2014**, 372 (2015), arXiv:1501.00367.
[618] David Straub (2019), doi.org/10.5281/zenodo.375591.
[619] W. Buchmuller and D. Wyler, Nucl. Phys., **B268**, 621–653 (1986).
[620] B. Grzadkowski et al., JHEP, **10**, 085 (2010), arXiv:1008.4884.
[621] G. D'Ambrosio et al., Nucl. Phys., **B645**, 155–187 (2002), arXiv:hep-ph/0207036.
[622] R. Alonso et al., Phys. Rev. Lett., **113**, 241802 (2014), arXiv:1407.7044.
[623] S. L. Glashow et al., Phys. Rev. Lett., **114**, 091801 (2015), arXiv:1411.0565.
[624] J. P. Lees et al., BaBar, Phys. Rev., **D87**(11), 112005 (2013), arXiv:1303.7465.
[625] O. Lutz et al., Belle, Phys. Rev., **D87**, 111103 (2013), arXiv:1303.3719.
[626] P. del Amo Sanchez et al., BaBar, Phys. Rev., **D82**, 112002 (2010), arXiv:1009.1529.
[627] J. Grygier et al., Belle, Phys. Rev., **D96**(9), 091101 (2017), arXiv:1702.03224.
[628] J. P. Lees et al., BaBar, Phys. Rev., **D86**, 051105 (2012), arXiv:1206.2543.
[629] C. L. Hsu et al., Belle, Phys. Rev., **D86**, 032002 (2012), arXiv:1206.5948.
[630] Bhubanjyoti Bhattacharya, Cody M. Grant, and Alexey A. Petrov (2018), arXiv:1809.04606.
[631] M. Feindt et al., Nucl. Instrum. Meth., **A654**, 432–440 (2011), arXiv:1102.3876.
[632] R. D. Peccei and H. R. Quinn, Phys. Rev. Lett., **38**, 1440–1443 (1977).
[633] R. D. Peccei and H. R. Quinn, Phys. Rev., **D16**, 1791–1797 (1977).
[634] S. Weinberg, Phys. Rev. Lett., **40**, 223–226 (1978).
[635] F. Wilczek, Phys. Rev. Lett., **40**, 279–282 (1978).
[636] J. Jaeckel and A. Ringwald, Ann. Rev. Nucl. Part. Sci., **60**, 405–437 (2010), arXiv:1002.0329.
[637] S. Weinberg, Phys. Rev. Lett., **43**, 1566–1570 (1979).
[638] J. F. Kamenik and C. Smith, Phys. Rev., **D85**, 093017 (2012), arXiv:1201.4814.
[639] K. De Bruyn, Search for the rare decays $B^0_{(s)} \to \tau^+\tau^-$, Technical Report LHCb-CONF-2016-011. CERN-LHCb-CONF-2016-011, CERN, Geneva (Sep 2016).
[640] J. P. Lees et al., BaBar, Phys. Rev. Lett., **118**(3), 031802 (2017), arXiv:1605.09637.
[641] A. Morda, On the possibility of measuring br($b_s \to \tau\tau$) at lhcb, presentation at flavor of new physics in $b \to s$ transitions (2014), LHCb-TALK-2014-313.
[642] M. Beylich et al., Eur. Phys. J., **C71**, 1635 (2011), arXiv:1101.5118.
[643] J. Lyon and R. Zwicky (2014), arXiv:1406.0566.
[644] R. Aaij et al., LHCb, Phys. Rev. Lett., **111**, 112003 (2013), arXiv:1307.7595.
[645] D. Du et al., Fermilab Lattice, Phys. Rev., **D93**(3), 034005 (2016), arXiv:1510.02349.
[646] U. Haisch, (No) New physics in $B_s$ mixing and decay, In *2012 Electroweak Interactions and Unified Theories*, pages 219–226 (2012), arXiv:1206.1230.
[647] W. Altmannshofer and D. M. Straub, Eur. Phys. J., **C73**, 2646 (2013), arXiv:1308.1501.
[648] W. Altmannshofer et al., Phys. Rev., **D89**, 095033 (2014), arXiv:1403.1269.
[649] J. F. Kamenik and L. Vale others, Eur. Phys. J., **C77**(10), 701 (2017), arXiv:1705.11106.
[650] J. Charles et al., CKMfitter (2017), arXiv:1705.02981.
[651] D. London et al., Phys. Rev., **D60**, 074020 (1999), arXiv:hep-ph/9905404.
[652] F. J. Botella and J. P. Silva, Phys. Rev., **D71**, 094008 (2005), arXiv:hep-ph/0503136.
[653] T. Feldmann et al., JHEP, **08**, 066 (2008), arXiv:0803.3729.
[654] Y. Grossman, A. L. Kagan, and Z. Ligeti, Phys. Lett., **B538**, 327–334 (2002), arXiv:hep-ph/0204212.
[655] B. Aubert et al., BaBar, Phys. Rev., **D79**, 072009 (2009), arXiv:0902.1708.
[656] R. Aaij et al., LHCb, Phys. Rev. Lett., **115**(3), 031601 (2015), arXiv:1503.07089.
[657] A. Höcker et al., Eur. Phys. J., **C21**, 225–259 (2001), arXiv:hep-ph/0104062.
[658] M. Gronau, Phys. Rev. Lett., **63**, 1451 (1989).
[659] H. Boos, T. Mannel, and J. Reuter, Phys. Rev., **D70**, 036006 (2004), arXiv:hep-ph/0403085.
[660] H.-N. Li and S. Mishima, JHEP, **03**, 009 (2007), arXiv:hep-ph/0610120.
[661] M. Gronau and J. L. Rosner, Phys. Lett., **B672**, 349–353 (2009), arXiv:0812.4796.
[662] P. Frings et al., Phys. Rev. Lett., **115**, 061802 (2015), arXiv:1503.00859.
[663] R. Fleischer, Eur. Phys. J., **C10**, 299–306 (1999), arXiv:hep-ph/9903455.
[664] R. Fleischer, Nucl.Instrum.Meth., **A446**, 1–17 (2000), arXiv:hep-ph/9908340.
[665] K. De Bruyn and R. Fleischer, JHEP, **1503**, 145 (2015), arXiv:1412.6834.
[666] M. Ciuchini et al., Phys. Rev. Lett., **95**, 221804 (2005), arXiv:hep-ph/0507290.
[667] S. Faller et al., Phys. Rev., **D79**, 014030 (2009), arXiv:0809.0842.
[668] M. Ciuchini et al. (2011), arXiv:1102.0392.
[669] M. Jung, Phys. Rev., **D86**, 053008 (2012), arXiv:1206.2050.
[670] Z. Ligeti and D. J. Robinson, Phys. Rev. Lett., **115**(25), 251801 (2015), arXiv:1507.06671.
[671] M. Jung, Phys. Lett., **B753**, 187–190 (2016), arXiv:1510.03423.
[672] O. Long et al., Phys. Rev., **D68**, 034010 (2003), arXiv:hep-ex/0303030.
[673] B. Aubert et al., BaBar, Phys. Rev. Lett., **101**, 021801 (2008), arXiv:0804.0896.

[674] S. E. Lee et al., Belle, Phys. Rev., **D77**, 071101 (2008), arXiv:0708.0304.
[675] I. Adachi *et al.* (Belle), Phys. Rev. Lett., **108**, 171802 (2012).
[676] S. Faller et al., Phys. Rev., **D79**, 014005 (2009), arXiv:0810.4248.
[677] M. Jung and S. Schacht, Phys. Rev., **D91**(3), 034027 (2015), arXiv:1410.8396.
[678] L. Bel et al., JHEP, **07**, 108 (2015), arXiv:1505.01361.
[679] S. Stone and L. Zhang, Phys. Rev., **D79**, 074024 (2009), arXiv:0812.2832.
[680] R. Fleischer et al., Eur. Phys. J., **C71**, 1832 (2011), arXiv:1109.1112.
[681] M. Beneke, Phys. Lett., **B620**, 143–150 (2005), arXiv:hep-ph/0505075.
[682] M. Beneke et al., Eur. Phys. J., **C61**, 439–449 (2009), arXiv:0902.4446.
[683] M. Gronau et al., Phys. Lett., **B579**, 331–339 (2004), arXiv:hep-ph/0310020.
[684] Y. Grossman et al., Phys. Rev., **D68**, 015004 (2003), arXiv:hep-ph/0303171.
[685] M. Gronau et al., Phys. Lett., **B596**, 107–115 (2004), arXiv:hep-ph/0403287.
[686] M. Gronau and J. L. Rosner, Phys. Rev., **D71**, 074019 (2005), arXiv:hep-ph/0503131.
[687] C-W. Chiang et al., Phys. Rev., **D70**, 034020 (2004), arXiv:hep-ph/0404073.
[688] Y. Grossman et al., JHEP, **01**, 066 (2014), arXiv:1308.4143.
[689] S. Khalil and E. Kou, Phys. Rev. Lett., **91**, 241602 (2003), arXiv:hep-ph/0303214.
[690] D. Chang et al., Phys. Rev., **D67**, 075013 (2003), arXiv:hep-ph/0205111.
[691] J. Girrbach et al., JHEP, **06**, 044, [Erratum: JHEP 07, 001 (2011)] (2011), arXiv:1101.6047.
[692] R. Fleischer et al., Phys. Rev., **D78**, 111501 (2008), arXiv:0806.2900.
[693] M. Gronau and J. L. Rosner, Phys. Lett., **B666**, 467–471 (2008), arXiv:0807.3080.
[694] M. Neubert and J. L. Rosner, Phys. Rev. Lett., **81**, 5076–5079 (1998), arXiv:hep-ph/9809311.
[695] A. J. Buras and R. Fleischer, Eur. Phys. J., **C11**, 93–109 (1999), arXiv:hep-ph/9810260.
[696] M. Beneke and M. Neubert, Nucl. Phys., **B675**, 333–415 (2003), arXiv:hep-ph/0308039.
[697] A. R. Williamson and J. Zupan, Phys. Rev., **D74**, 014003, [Erratum: Phys. Rev. D74, 03901 (2006)] (2006), arXiv:hep-ph/0601214.
[698] R. Aaij et al., LHCb, Eur. Phys. J., **C73**(4), 2373 (2013), arXiv:1208.3355.
[699] J. P. Lees et al., BaBar, Phys. Rev., **D85**, 112010 (2012), arXiv:1201.5897.
[700] Y. Nakahama et al., Belle, Phys. Rev., **D82**, 073011 (2010), arXiv:1007.3848.
[701] T. E. Browder et al., CLEO, Phys. Rev. Lett., **81**, 1786–1790 (1998), arXiv:hep-ex/9804018.
[702] B. Aubert et al., BaBar, Phys. Rev., **D79**, 052003 (2009), arXiv:0809.1174.
[703] L. Santelj et al., Belle, JHEP, **10**, 165 (2014), arXiv:1408.5991.
[704] M. Gronau and D. London, Phys. Rev. Lett., **65**, 3381–3384 (1990).
[705] A. E. Snyder and H. R. Quinn, Phys. Rev., **D48**, 2139–2144 (1993).
[706] M. Gronau and J. L. Rosner (2016), arXiv:1608.06224.
[707] H. R. Quinn and J. P. Silva, Phys. Rev., **D62**, 054002 (2000), arXiv:hep-ph/0001290.
[708] M. Gronau and J. Zupan, Phys. Rev., **D70**, 074031 (2004), arXiv:hep-ph/0407002.
[709] M. Gronau and J. Zupan, Phys. Rev., **D71**, 074017 (2005), arXiv:hep-ph/0502139.
[710] P. Vanhoefer, PoS, **EPS-HEP2015**, 558 (2015), arXiv:1509.06548.
[711] J. P. Lees et al., BaBar, Phys. Rev., **D87**(5), 052009 (2013), arXiv:1206.3525.
[712] T. Julius et al., Belle, Phys. Rev., **D96**(3), 032007 (2017), arXiv:1705.02083.
[713] H. Ishino et al. (2007), arXiv:hep-ex/0703039.
[714] R. Aaij et al., LHCb, Phys. Lett., **B747**, 468–478 (2015), arXiv:1503.07770.
[715] B. Aubert et al., BaBar, Phys. Rev., **D78**, 071104 (2008), arXiv:0807.4977.
[716] J. Zhang et al., Belle, Phys. Rev. Lett., **91**, 221801 (2003), arXiv:hep-ex/0306007.
[717] J. P. Lees et al., BaBar, Phys. Rev., **D88**(1), 012003 (2013), arXiv:1304.3503.
[718] A. Kusaka et al., Belle, Phys. Rev. Lett., **98**, 221602 (2007), arXiv:hep-ex/0701015.
[719] H. J. Lipkin et al., Phys. Rev., **D44**, 1454–1460 (1991).
[720] M. Neubert and J. L. Rosner, Phys. Lett., **B441**, 403–409 (1998), arXiv:hep-ph/9808493.
[721] M. Gronau et al., Phys. Rev., **D60**, 034021, [Erratum: Phys. Rev. D69, 119901 (2004)] (1999), arXiv:hep-ph/9810482.
[722] S. Gardner, Phys. Rev., **D59**, 077502 (1999), arXiv:hep-ph/9806423.
[723] P. Kroll, Int. J. Mod. Phys., **A20**, 331–340 (2005), arXiv:hep-ph/0409141.
[724] M. Gronau et al., Phys. Rev., **D50**, 4529–4543 (1994), arXiv:hep-ph/9404283.
[725] P. Vanhoefer et al., Belle, Phys. Rev., **D93**(3), 032010, [Addendum: Phys. Rev.D94,no.9,099903(2016)] (2016), arXiv:1510.01245.
[726] A. F. Falk et al., Phys. Rev., **D69**, 011502 (2004), arXiv:hep-ph/0310242.
[727] Y. T. Duh et al., Belle, Phys. Rev., **D87**(3), 031103 (2013), arXiv:1210.1348.
[728] I. Adachi et al., Belle, Phys. Rev., **D88**(9), 092003 (2013), arXiv:1302.0551.
[729] I. Adachi et al., Belle, [Addendum: Phys. Rev.D89,no.11,119903(2014)] (2012), arXiv:1212.4015.
[730] M. Pivk et al., Eur. Phys. J., **C39**, 397–409 (2005), arXiv:hep-ph/0406263.
[731] B. Grinstein and D. Pirjol, Phys. Rev., **D73**, 014013 (2006), arXiv:hep-ph/0510104.

[732] A. Khodjamirian et al., Phys. Lett., **B402**, 167–177 (1997), arXiv:hep-ph/9702318.
[733] B. Aubert et al., BaBar, Phys. Rev., **D78**, 071102 (2008), arXiv:0807.3103.
[734] M. Gronau and D. London, Phys. Lett., **B253**, 483–488 (1991).
[735] M. Gronau and D. Wyler, Phys. Lett., **B265**, 172–176 (1991).
[736] D. Atwood et al., Phys. Rev. Lett., **78**, 3257–3260 (1997), arXiv:hep-ph/9612433.
[737] D. Atwood et al., Phys. Rev., **D63**, 036005 (2001), arXiv:hep-ph/0008090.
[738] A. Giri et al., Phys. Rev., **D68**, 054018 (2003), arXiv:hep-ph/0303187.
[739] Y. Grossman et al., Phys. Rev., **D67**, 071301 (2003), arXiv:hep-ph/0210433.
[740] J. P. Silva and A. Soffer, Phys. Rev., **D61**, 112001 (2000), arXiv:hep-ph/9912242.
[741] Y. Grossman et al., Phys. Rev., **D72**, 031501 (2005), arXiv:hep-ph/0505270.
[742] M. Gronau et al., Phys. Lett., **B649**, 61–66 (2007), arXiv:hep-ph/0702011.
[743] A. Bondar et al., Phys. Rev., **D82**, 034033 (2010), arXiv:1004.2350.
[744] W. Wang, Phys. Rev. Lett., **110**(6), 061802 (2013), arXiv:1211.4539.
[745] M. Martone and J. Zupan, Phys. Rev., **D87**(3), 034005 (2013), arXiv:1212.0165.
[746] B. Bhattacharya et al., Phys. Rev., **D87**(7), 074002 (2013), arXiv:1301.5631.
[747] A. Bondar et al., Eur. Phys. J., **C73**(6), 2476 (2013), arXiv:1303.6305.
[748] M. Rama, Phys. Rev., **D89**(1), 014021 (2014), arXiv:1307.4384.
[749] Y. Grossman and M. Savastio, JHEP, **03**, 008 (2014), arXiv:1311.3575.
[750] J. Brod and J. Zupan, JHEP, **01**, 051 (2014), arXiv:1308.5663.
[751] J. Brod, Phys. Lett., **B743**, 56–60 (2015), arXiv:1412.3173.
[752] C. Bobeth et al., JHEP, **06**, 040 (2014), arXiv:1404.2531.
[753] J. Brod et al., Phys. Rev., **D92**(3), 033002 (2015), arXiv:1412.1446.
[754] T. Jubb et al., Nucl. Phys., **B915**, 431–453 (2017), arXiv:1603.07770.
[755] A. Bondar, Proceedings of BINP special analysis meeting on Dalitz analysis, unpublished (2002).
[756] A. Poluektov et al., Belle, Phys. Rev., **D81**, 112002 (2010), arXiv:1003.3360.
[757] H. Aihara et al., Belle, Phys. Rev., **D85**, 112014 (2012), arXiv:1204.6561.
[758] R. Aaij et al., LHCb, JHEP, **10**, 097 (2014), arXiv:1408.2748.
[759] M. Bona et al., UTfit, JHEP, **10**, 081 (2006), arXiv:hep-ph/0606167.
[760] J. Libby, Direct CP violation in hadronic B decays, In *8th International Workshop on the CKM Unitarity Triangle (CKM2014) Vienna, Austria, September 8-12, 2014* (2014), arXiv:1412.4269.
[761] Y. Horii et al., Belle, Phys. Rev. Lett., **106**, 231803 (2011), arXiv:1103.5951.
[762] K. Trabelsi, Study of direct $CP$ in charmed $B$ decays and measurement of the CKM angle gamma at Belle, In *7th Workshop on the CKM Unitarity Triangle (CKM 2012) Cincinnati, Ohio, USA, September 28-October 2, 2012* (2013), arXiv:1301.2033.
[763] M. Nayak et al., Belle, Phys. Rev., **D88**(9), 091104 (2013), arXiv:1310.1741.
[764] M. Gersabeck, HFAG-charm averages, In *8th International Workshop on the CKM Unitarity Triangle (CKM 2014) Vienna, Austria, September 8-12, 2014* (2014), arXiv:1411.4595.
[765] G. Bonvicini et al., [CLEO Collaboration], Phys. Rev., **D89**(7), 072002, [Erratum: Phys. Rev.D91, 019903 (2015)] (2014), arXiv:1312.6775.
[766] E. White et al., Belle, Phys. Rev., **D88**(5), 051101 (2013), arXiv:1307.5935.
[767] J. Libby et al., Phys. Lett., **B731**, 197–203 (2014), arXiv:1401.1904.
[768] R. A. Briere (2014), arXiv:1411.7327.
[769] N. Lowrey et al., [CLEO Collaboration], Phys. Rev., **D80**, 031105 (2009), arXiv:0903.4853.
[770] J. Libby et al., CLEO, Phys. Rev., **D82**, 112006 (2010), arXiv:1010.2817.
[771] D. Ambrose, Measurement of the relative strong-phase difference between $D^0$ and $\overline{D}^0 \to K_S^0\pi^+\pi^-$, Presented on behalf of the BESIII Collaboration at the APS April Meeting, April 5th to 8th, 2014, Savannah, Georgia (2014).
[772] M. Nayak et al., Phys. Lett., **B740**, 1–7 (2015), arXiv:1410.3964.
[773] Resmi P. K. et al., PoS, **CKM2016**, 112 (2017), arXiv:1703.10317.
[774] R. Aaij et al., LHCb, JHEP, **12**, 087 (2016), arXiv:1611.03076.
[775] S. Malde, Synergy of BESIII and LHCb physics programmes, Technical Report LHCb-PUB-2016-025. CERN-LHCb-PUB-2016-025, CERN, Geneva (Oct 2016).
[776] T. Gershon and A. Poluektov, Phys. Rev., **D81**, 014025 (2010), arXiv:0910.5437.
[777] D. Zeppenfeld, Z. Phys., **C8**, 77 (1981).
[778] L. L. Chau and H. Y. Cheng, Phys. Rev. Lett., **56**, 1655–1658 (1986).
[779] M. J. Savage and M. B. Wise, Phys. Rev., **D39**, 3346, [Erratum: Phys. Rev. D40, 3127 (1989)] (1989).
[780] L.-L. Chau et al., Phys. Rev., **D43**, 2176–2192, [Erratum: Phys. Rev. D58, 019902 (1998)] (1991).
[781] H.-Y. Cheng, C.-W. Chiang, and A.-L. Kuo, Phys. Rev., **D91**, 014011 (2015), arXiv:1409.5026.
[782] T. Feldmann et al., Phys. Rev., **D58**, 114006 (1998), arXiv:hep-ph/9802409.

[783] C. Michael et al., ETM, Phys. Rev. Lett., **111**(18), 181602 (2013), arXiv:1310.1207.
[784] M. Beneke and S. Jager, Nucl. Phys., **B751**, 160–185 (2006), arXiv:hep-ph/0512351.
[785] M. Beneke et al., Nucl. Phys., **B832**, 109–151 (2010), arXiv:0911.3655.
[786] M. Beneke and S. Jager, Nucl. Phys., **B768**, 51–84 (2007), arXiv:hep-ph/0610322.
[787] H. J. Lipkin, Phys. Lett., **B433**, 117–124 (1998).
[788] M. Beneke and M. Neubert, Nucl. Phys., **B651**, 225–248 (2003), arXiv:hep-ph/0210085.
[789] H. J. Lipkin, Phys. Lett., **B254**, 247–252 (1991).
[790] L. Hofer et al., JHEP, **02**, 080 (2011), arXiv:1011.6319.
[791] F. Fichter, MSc thesis, Karlsruher Institut für Technologie (2016).
[792] R. Aaij et al., LHCb (2016), arXiv:1610.05187.
[793] S-H. Zhou et al. (2016), arXiv:1608.02819.
[794] M. Bauer et al., Z.Phys., **C34**, 103 (1987).
[795] A. L. Kagan, Phys. Lett., **B601**, 151–163 (2004), arXiv:hep-ph/0405134.
[796] M. Beneke et al., Nucl. Phys., **B774**, 64–101 (2007), arXiv:hep-ph/0612290.
[797] H.-Y. Cheng et al., Phys. Rev., **D77**, 014034 (2008), arXiv:0705.3079.
[798] H.-Y. Cheng and K.-C. Yang, Phys. Rev., **D76**, 114020 (2007), arXiv:0709.0137.
[799] H.-Y. Cheng and K.-C. Yang, Phys. Rev., **D78**, 094001, [Erratum: Phys. Rev. D79, 039903 (2009)] (2008), arXiv:0805.0329.
[800] H.-Y. Cheng and K.-C. Yang, Phys. Rev., **D83**, 034001 (2011), arXiv:1010.3309.
[801] Y. Y. Keum et al., Phys. Rev., **D63**, 054008 (2001), arXiv:hep-ph/0004173.
[802] C.-D. Lu et al., Phys. Rev., **D63**, 074009 (2001), arXiv:hep-ph/0004213.
[803] H.-Nan Li and S. Mishima, Phys. Rev., **D83**, 034023 (2011), arXiv:0901.1272.
[804] H.-N. Li and S. Mishima, Phys. Rev., **D90**(7), 074018 (2014), arXiv:1407.7647.
[805] .-Y. Cheng and S. Oh, JHEP, **09**, 024 (2011), arXiv:1104.4144.
[806] S. Descotes-Genon and C. T. Sachrajda, Nucl. Phys., **B650**, 356–390 (2003), arXiv:hep-ph/0209216.
[807] E. Lunghi et al., Nucl. Phys., **B649**, 349–364 (2003), arXiv:hep-ph/0210091.
[808] S. W. Bosch et al., Phys. Rev., **D67**, 094014 (2003), arXiv:hep-ph/0301123.
[809] A. G. Grozin and M. Neubert, Phys. Rev., **D55**, 272–290 (1997), arXiv:hep-ph/9607366.
[810] B. O. Lange and M. Neubert, Phys. Rev. Lett., **91**, 102001 (2003), arXiv:hep-ph/0303082.
[811] G. Bell et al., JHEP, **11**, 191 (2013), arXiv:1308.6114.
[812] V. M. Braun and A. N. Manashov, Phys. Lett., **B731**, 316–319 (2014), arXiv:1402.5822.
[813] H. Kawamura et al., Phys. Lett., **B523**, 111, [Erratum: Phys. Lett. B536, 344 (2002)] (2001), arXiv:hep-ph/0109181.
[814] V. M. Braun et al., Phys. Rev., **D69**, 034014 (2004), arXiv:hep-ph/0309330.
[815] S. J. Lee and M. Neubert, Phys. Rev., **D72**, 094028 (2005), arXiv:hep-ph/0509350.
[816] H. Kawamura and K. Tanaka, Phys. Lett., **B673**, 201–207 (2009), arXiv:0810.5628.
[817] T. Feldmann et al., Phys. Rev., **D89**(11), 114001 (2014), arXiv:1404.1343.
[818] A. Khodjamirian et al., Phys. Rev., **D72**, 094012 (2005), arXiv:hep-ph/0509049.
[819] Y.-Y. Keum et al., Phys. Lett., **B504**, 6–14 (2001), arXiv:hep-ph/0004004.
[820] M. Bartsch et al. (2008), arXiv:0810.0249.
[821] M. Ciuchini et al., Phys. Lett., **B674**, 197–203 (2009), arXiv:0811.0341.
[822] Q. Chang et al., Phys. Lett., **B740**, 56–60 (2015), arXiv:1409.2995.
[823] C. Bobeth et al., Eur. Phys. J., **C75**(7), 340 (2015), arXiv:1409.3252.
[824] J. Sun et al., Phys. Lett., **B743**, 444–450 (2015), arXiv:1412.2334.
[825] Q. Chang et al., Phys. Rev., **D91**, 074026 (2015), arXiv:1504.04907.
[826] Q. Chang et al., J. Phys., **G43**(10), 105004 (2016).
[827] K. Wang and G. Zhu, Phys. Rev., **D88**, 014043 (2013), arXiv:1304.7438.
[828] N. Kivel, JHEP, **05**, 019 (2007), arXiv:hep-ph/0608291.
[829] V. Pilipp, Nucl. Phys., **B794**, 154–188 (2008), arXiv:0709.3214.
[830] G. Bell, Nucl. Phys., **B795**, 1–26 (2008), arXiv:0705.3127.
[831] G. Bell, Nucl. Phys., **B822**, 172–200 (2009), arXiv:0902.1915.
[832] G. Bell and V. Pilipp, Phys. Rev., **D80**, 054024 (2009), arXiv:0907.1016.
[833] G. Bell et al., Phys. Lett., **B750**, 348–355 (2015), arXiv:1507.03700.
[834] C. S. Kim and Y. W. Yoon, JHEP, **11**, 003 (2011), arXiv:1107.1601.
[835] G. Bell and T. Huber, JHEP, **12**, 129 (2014), arXiv:1410.2804.
[836] M. Gronau, Phys. Lett., **B627**, 82–88 (2005), arXiv:hep-ph/0508047.
[837] M. Fujikawa et al., Belle, Phys. Rev., **D81**, 011101 (2010), arXiv:0809.4366.
[838] M. Kenzie, GammaCombo - A statistical analysis framework for combining measurements, fitting datasets and producing confidence intervals, https://gammacombo.github.io (2016).
[839] B. Pal et al., Belle, Phys. Rev. Lett., **116**, 161801 (2016), arXiv:1512.02145.
[840] C-H. Chen, Phys. Lett., **B520**, 33 (2001), arXiv:hep-ph/0107189.

[841] A. Ali et al., Phys. Rev., **D76**, 074018 (2007), arXiv:hep-ph/0703162.
[842] C.-K. Chua, Phys. Rev., **D78**, 076002 (2008), arXiv:0712.4187.
[843] J.-J. Wang et al., Phys. Rev., **D89**, 074046 (2014), arXiv:1402.6912.
[844] S. Baek et al., JHEP, **12**, 019 (2006), arXiv:hep-ph/0610109.
[845] A. Hayakawa et al., PTEP, **2014**, 023B04 (2014), arXiv:1311.5974.
[846] I. Dunietz, Phys. Rev., **D52**, 3048 (1995), arXiv:hep-ph/9501287.
[847] Y. Grossman, Phys. Lett., **B380**, 99 (1996), arXiv:hep-ph/9603244.
[848] J. G. Korner and G. R. Goldstein, Phys. Lett., **89B**, 105–110 (1979).
[849] A. Ali et al., Phys. Rev., **D58**, 094009 (1998), arXiv:hep-ph/9804363.
[850] H.-Y. Cheng and K.-C. Yang, Phys. Lett., **B511**, 40–48 (2001), arXiv:hep-ph/0104090.
[851] X.-Q. Li et al., Phys. Rev., **D68**, 114015, [Erratum: Phys. Rev. D71, 019902 (2005)] (2003), arXiv:hep-ph/0309136.
[852] K. F. Chen et al., Belle, Phys. Rev. Lett., **91**, 201801 (2003), arXiv:hep-ex/0307014.
[853] Bernard Aubert et al., BaBar, Phys. Rev., **D78**, 092008 (2008), arXiv:0808.3586.
[854] M. Beneke et al., Phys. Rev. Lett., **96**, 141801 (2006), arXiv:hep-ph/0512258.
[855] H.-Y. Cheng and C.-K. Chua, Phys. Rev., **D80**, 114008 (2009), arXiv:0909.5229.
[856] H.-Y. Cheng and C.-K. Chua, Phys. Rev., **D80**, 114026 (2009), arXiv:0910.5237.
[857] J. Zhu et al., Phys. Rev., **D72**, 054015 (2005), arXiv:hep-ph/0504187.
[858] J. Zhu et al., J. Phys., **G32**, 101–110 (2006), arXiv:hep-ph/0506316.
[859] Ying Li and Cai-Dian Lu, Phys. Rev., **D73**, 014024 (2006), arXiv:hep-ph/0508032.
[860] H.-W. Huang et al., Phys. Rev., **D73**, 014011 (2006), arXiv:hep-ph/0508080.
[861] Z.-T. Zou et al., Phys. Rev., **D91**, 054033 (2015), arXiv:1501.00784.
[862] P. Colangelo et al., Phys. Lett., **B597**, 291–298 (2004), arXiv:hep-ph/0406162.
[863] H.-Y. Cheng et al., Phys. Rev., **D71**, 014030 (2005), arXiv:hep-ph/0409317.
[864] H.-N. Li, Phys. Lett., **B622**, 63–68 (2005), arXiv:hep-ph/0411305.
[865] S. Baek et al., Phys. Rev., **D72**, 094008 (2005), arXiv:hep-ph/0508149.
[866] Q. Chang et al., JHEP, **06**, 038 (2007), arXiv:hep-ph/0610280.
[867] C. C. Chiang et al., Belle, Phys. Rev., **D78**, 111102 (2008), arXiv:0808.2576.
[868] M. Beneke et al., Eur. Phys. J., **C61**, 429–438 (2009), arXiv:0901.4841.
[869] G. Valencia, Phys. Rev., **D39**, 3339 (1989).
[870] A. Datta and D. London, Int. J. Mod. Phys., **A19**, 2505–2544 (2004), arXiv:hep-ph/0303159.
[871] A. Datta et al., Phys. Lett., **B701**, 357–362 (2011), arXiv:1103.2442.
[872] Y. Xie et al., JHEP, **09**, 074 (2009), arXiv:0908.3627.
[873] B. Bhattacharya et al., Phys. Rev., **D88**(1), 016007 (2013), arXiv:1306.1911.
[874] A. Datta et al., Phys. Rev., **D86**, 076011 (2012), arXiv:1207.4495.
[875] B. Aubert et al., BaBar, Phys. Rev., **D76**, 031102 (2007), arXiv:0704.0522.
[876] R. Itoh et al., Belle, Phys. Rev. Lett., **95**, 091601 (2005), arXiv:hep-ex/0504030.
[877] T. Aaltonen et al., CDF, Phys. Rev. Lett., **107**, 261802 (2011), arXiv:1107.4999.
[878] R. Aaij et al., LHCb, Phys. Lett., **B713**, 369–377 (2012), arXiv:1204.2813.
[879] W. Bensalem et al., Phys. Lett., **B538**, 309–320 (2002), arXiv:hep-ph/0205009.
[880] W. Bensalem et al., Phys. Rev., **D66**, 094004 (2002), arXiv:hep-ph/0208054.
[881] M. Gronau and J. L. Rosner, Phys. Lett., **B749**, 104–107 (2015), arXiv:1506.01346.
[882] Ashutosh K. Alok et al., JHEP, **11**, 122 (2011), arXiv:1103.5344.
[883] M. Duraisamy and A. Datta, JHEP, **09**, 059 (2013), arXiv:1302.7031.
[884] M. Duraisamy et al., Phys. Rev., **D90**(7), 074013 (2014), arXiv:1405.3719.
[885] R. Aaij et al., LHCb, Phys. Rev. Lett., **110**(24), 241802 (2013), arXiv:1303.7125.
[886] R. Aaij et al., LHCb, Phys. Rev. Lett., **108**, 101803 (2012), arXiv:1112.3183.
[887] R. Aaij et al., LHCb, Phys. Rev., **D90**(5), 052011 (2014), arXiv:1407.2222.
[888] R. Aaij et al., LHCb, JHEP, **07**, 166 (2015), arXiv:1503.05362.
[889] R. Aaij et al., LHCb, Submitted to: JHEP (2017), arXiv:1712.08683.
[890] J. Zhang et al., Belle (2005), arXiv:hep-ex/0505039.
[891] B. Aubert et al., BaBar, Phys. Rev. Lett., **97**, 201801 (2006), arXiv:hep-ex/0607057.
[892] R. Aaij et al., LHCb, Phys. Rev. Lett., **111**, 101801 (2013), arXiv:1306.1246.
[893] R. Aaij et al., LHCb, Phys. Rev. Lett., **112**(1), 011801 (2014), arXiv:1310.4740.
[894] R. Aaij et al., LHCb, Phys. Rev., **D90**(11), 112004 (2014), arXiv:1408.5373.
[895] J. Virto, PoS, **FPCP2016**, 007 (2017), arXiv:1609.07430.
[896] P. Guo et al., Eur. Phys. J., **A51**(10), 135 (2015), arXiv:1409.8652.
[897] I. Bediaga and P. C. Magalhaes (2015), arXiv:1512.09284.
[898] J. H. Alvarenga Nogueira et al., Few Body Syst., **58**(2), 98 (2017), arXiv:1609.01568.
[899] L. Wolfenstein, Phys. Rev., **D43**, 151–156 (1991).
[900] I. Bediaga et al., Phys. Rev., **D89**(9), 094013 (2014), arXiv:1307.8164.

[901] J. H. Alvarenga Nogueira et al., Phys. Rev., **D92**(5), 054010 (2015), arXiv:1506.08332.
[902] J. P. Dedonder et al., Acta Phys. Polon., **B42**, 2013 (2011), arXiv:1011.0960.
[903] H.-Y. Cheng et al., Phys. Rev., **D76**, 094006 (2007), arXiv:0704.1049.
[904] H.-Y. Cheng et al., Phys. Rev., **D94**(9), 094015 (2016), arXiv:1607.08313.
[905] M. Beneke, Quasi-Two-Body and Three-Body Decays in the Heavy Quark Expansion, Three-Body Charmless B Decays Workshop, Paris, France (2006).
[906] I. Stewart, SCET Methods for Quasi Two-Body and Three-Body Decays, Three-Body Charmless B Decays Workshop, Paris, France (2006).
[907] S. Krankl et al., Nucl. Phys., **B899**, 247–264 (2015), arXiv:1505.04111.
[908] M. V. Polyakov, Nucl. Phys., **B555**, 231 (1999), arXiv:hep-ph/9809483.
[909] S. Faller et al., Phys. Rev., **D89**(1), 014015 (2014), arXiv:1310.6660.
[910] C. Hambrock and A. Khodjamirian, Nucl. Phys., **B905**, 373–390 (2016), arXiv:1511.02509.
[911] S. Cheng et al., JHEP, **05**, 157 (2017), arXiv:1701.01633.
[912] A. Khodjamirian et al., Phys. Rev., **D75**, 054013 (2007), arXiv:hep-ph/0611193.
[913] P. Böer et al., JHEP, **02**, 133 (2017), arXiv:1608.07127.
[914] C. Patrignani et al., Particle Data Group, Chin. Phys., **C40**(10), 100001 (2016).
[915] H.-Y. Cheng and C.-K. Chua, Phys. Rev., **D88**, 114014 (2013), arXiv:1308.5139.
[916] C.-H. Chen and H.-N. Li, Phys. Lett., **B561**, 258–265 (2003), arXiv:hep-ph/0209043.
[917] W.-F. Wang et al., Phys. Rev., **D89**(7), 074031 (2014), arXiv:1402.5280.
[918] C. L. Y. Lee et al., Phys. Rev., **D46**, 5040–5048 (1992).
[919] A. Garmash et al., Belle, Phys. Rev., **D71**, 092003 (2005), arXiv:hep-ex/0412066.
[920] A. Garmash et al., Belle, Phys. Rev. Lett., **96**, 251803 (2006), arXiv:hep-ex/0512066.
[921] J. P. Lees et al., BaBar, Phys. Rev., **D96**(7), 072001 (2017), arXiv:1501.00705.
[922] H.-Y. Cheng et al., Phys. Rev., **D87**(11), 114001 (2013), arXiv:1303.4403.
[923] H.-Y. Cheng et al., Phys. Rev., **D73**, 014017 (2006), arXiv:hep-ph/0508104.
[924] B. Bhattacharya et al., Phys. Lett., **B726**, 337–343 (2013), arXiv:1306.2625.
[925] D. Xu et al., Int. J. Mod. Phys., **A29**, 1450011 (2014), arXiv:1307.7186.
[926] B. Aubert et al., BaBar, Phys. Rev., **D79**, 072006 (2009), arXiv:0902.2051.
[927] B. Aubert, BABAR, Phys. Rev. Lett., **100**, 051802 (Feb 2008).
[928] A. Bazavov et al., Fermilab Lattice, MILC, Phys. Rev., **D90**(7), 074509 (2014), arXiv:1407.3772.
[929] A. Vladikas, PoS, **FPCP2015**, 016 (2015), arXiv:1509.01155.
[930] J. Koponen et al. (2013), arXiv:1305.1462.
[931] S. Fajfer et al., Phys. Rev., **D91**(9), 094009 (2015), arXiv:1502.07488.
[932] G. C. Branco et al., Phys. Rept., **516**, 1–102 (2012), arXiv:1106.0034.
[933] A. Pich and P. Tuzon, Phys. Rev., **D80**, 091702 (2009), arXiv:0908.1554.
[934] A. Zupanc et al., Belle, JHEP, **09**, 139 (2013), arXiv:1307.6240.
[935] J. M. Link et al., FOCUS, Phys. Lett., **B535**, 43–51 (2002), arXiv:hep-ex/0203031.
[936] P. del Amo Sanchez et al., BaBar, Phys. Rev., **D83**, 072001 (2011), arXiv:1012.1810.
[937] M. Ablikim et al., BESIII, Phys. Rev., **D94**(3), 032001 (2016), arXiv:1512.08627.
[938] A. Zupanc, presentation at Belle II Theory Interface Platform Meeting, KEK, Oct 2014 (2014).
[939] Y. T. Lai et al., Belle, Phys. Rev., **D95**(1), 011102 (2017), arXiv:1611.09455.
[940] L. Widhalm et al., Belle, Phys. Rev. Lett., **97**, 061804 (2006), arXiv:hep-ex/0604049.
[941] B. Aubert et al., BABAR, Phys. Rev. Lett., **87**, 091801 (Aug 2001).
[942] L. et al Widhalm, Belle, Phys. Rev. Lett., **97**, 061804 (2006), arXiv:0604049.
[943] B. I. Eisenstein et al., CLEO, Phys. Rev., **D78**, 052003 (2008), arXiv:0806.2112.
[944] M. Ablikim et al., BESIII, Phys. Rev., **D89**(5), 051104 (2014), arXiv:1312.0374.
[945] M. Ablikim et al., BESIII, Phys. Rev., **D92**(7), 072012 (2015), arXiv:1508.07560.
[946] S. de Boer and G. Hiller, Phys. Rev., **D93**(7), 074001 (2016), arXiv:1510.00311.
[947] S. Fajfer et al., Phys. Rev., **D76**, 074010 (2007), arXiv:0706.1133.
[948] S. Fajfer and N. Košnik, Eur. Phys. J., **C75**(12), 567 (2015), arXiv:1510.00965.
[949] A. Paul et al., Phys. Rev., **D83**, 114006 (2011), arXiv:1101.6053.
[950] R. Aaij et al., LHCb, Phys. Lett., **B725**, 15–24 (2013), arXiv:1305.5059.
[951] R. Aaij et al., LHCb, Phys. Lett., **B754**, 167–175 (2016), arXiv:1512.00322.
[952] M. Petric et al., Belle, Phys. Rev., **D81**, 091102 (2010), arXiv:1003.2345.
[953] D. E. Hazard and A. A. Petrov, Phys. Rev., **D94**(7), 074023 (2016), arXiv:1607.00815.
[954] E. Golowich et al., Phys. Rev., **D79**, 114030 (2009), arXiv:0903.2830.
[955] A. Badin and A. A Petrov, Phys. Rev., **D82**, 034005 (2010), arXiv:1005.1277.
[956] M. Pospelov et al., Phys. Lett., **B662**, 53–61 (2008), arXiv:0711.4866.
[957] J. L. Feng and J. Kumar, Phys. Rev. Lett., **101**, 231301 (2008), arXiv:0803.4196.
[958] A. A. Petrov and A. E. Blechman, *Effective Field Theories*, (WSP, 2016).
[959] G. Isidori and J. F. Kamenik, Phys. Rev. Lett., **109**, 171801 (2012), arXiv:1205.3164.

[960] J. Lyon and R. Zwicky (2012), arXiv:1210.6546.

[961] G. Burdman et al., Phys. Rev., **D52**, 6383–6399 (1995), arXiv:hep-ph/9502329.

[962] S. Fajfer, Theoretical perspective on rare and radiative charm decays, In *7th International Workshop on Charm Physics (Charm 2015) Detroit, MI, USA, May 18-22, 2015* (2015), arXiv:1509.01997.

[963] Y. Grossman et al. (in preparation).

[964] A. Kagan, Talk at 7th International Workshop on Charm Physics (Charm 2015), Detroit, MI, May 18-22, 2015, `https://indico.fnal.gov/event/8909/session/17/contribution/69/material/slides/0.pdf` (2015).

[965] L. Silvestrini, Talk at Implications of LHCb measurements and future prospects, CERN, Nov 3-5, 2015, `https://indico.cern.ch/event/395704/contributions/1834102/attachments/1180736/1709272/silvestriniLHCb.pdf` (2015).

[966] G. C. Branco et al., Int. Ser. Monogr. Phys., **103**, 1–536 (1999).

[967] S. Bergmann and Y. Nir, JHEP, **09**, 031 (1999), arXiv:hep-ph/9909391.

[968] Y. Grossman et al., Phys. Rev., **D75**, 036008 (2007), arXiv:hep-ph/0609178.

[969] R. Aaij et al., LHCb, Phys. Rev. Lett., **108**, 111602 (2012), arXiv:1112.0938.

[970] Yosef Nir, Conf. Proc., **C9207131**, 81–136, [,81(1992)] (1992).

[971] M. Ciuchini et al., Phys. Lett., **B655**, 162–166 (2007), arXiv:hep-ph/0703204.

[972] Y. Grossman et al., Phys. Rev. Lett., **103**, 071602 (2009), arXiv:0904.0305.

[973] A. L. Kagan and M. D. Sokoloff, Phys. Rev., **D80**, 076008 (2009), arXiv:0907.3917.

[974] J. Brod et al., JHEP, **10**, 161 (2012), arXiv:1203.6659.

[975] J. Simone, PoS, **LATTICE2015**, 332 (2016).

[976] C. C. Chang, Hadronic matrix elements of neutral-meson mixing through lattice QCD (2015), arXiv:1509.07088.

[977] P. A. Boyle et al., RBC, UKQCD, Phys. Rev. Lett., **110**(15), 152001 (2013), arXiv:1212.1474.

[978] M. Peardon et al., Hadron Spectrum, Phys. Rev., **D80**, 054506 (2009), arXiv:0905.2160.

[979] C. Morningstar et al., Phys. Rev., **D83**, 114505 (2011), arXiv:1104.3870.

[980] J. J. Dudek, AIP Conf. Proc., **1735**, 020014 (2016).

[981] M. T. Hansen and S. R. Sharpe, Phys. Rev., **D90**(11), 116003 (2014), arXiv:1408.5933.

[982] M. T. Hansen and S. R. Sharpe, Phys. Rev., **D92**(11), 114509 (2015), arXiv:1504.04248.

[983] M. T. Hansen and S. R. Sharpe, Phys. Rev., **D93**(9), 096006 (2016), arXiv:1602.00324.

[984] Z. Bai et al., Phys. Rev. Lett., **113**, 112003 (2014), arXiv:1406.0916.

[985] N. H. Christ et al., Phys. Rev., **D91**(11), 114510 (2015), arXiv:1504.01170.

[986] M. Staric et al., BELLE, Phys. Rev. Lett., **98**, 211803, [,65(2007)] (2007), arXiv:hep-ex/0703036.

[987] B. Aubert et al., BaBar, Phys. Rev. Lett., **98**, 211802 (2007), arXiv:hep-ex/0703020.

[988] R Aaij et al., LHCb, Phys. Rev. Lett., **110**(10), 101802 (2013), arXiv:1211.1230.

[989] T. Aaltonen et al., CDF, Phys. Rev. Lett., **111**(23), 231802 (2013), arXiv:1309.4078.

[990] L. M. Zhang et al., Belle, Phys. Rev. Lett., **96**, 151801 (2006), arXiv:hep-ex/0601029.

[991] X. C. Tian et al., Belle, Phys. Rev. Lett., **95**, 231801 (2005), arXiv:hep-ex/0507071.

[992] B. Aubert et al., BaBar, Phys. Rev. Lett., **103**, 211801 (2009), arXiv:0807.4544.

[993] W.B. Yan L.K. Li and Z.P. Zhang, Belle Note 1381 (2016).

[994] C. Thomas and G. Wilkinson, JHEP, **10**, 185 (2012), arXiv:1209.0172.

[995] J. Insler et al., [CLEO Collaboration], Phys. Rev., **D85**, 092016 (2012), arXiv:1203.3804.

[996] R. Aaij et al., LHCb, Phys. Rev., **D93**(5), 052018 (2016), arXiv:1509.06628.

[997] S. Malde and G. Wilkinson, Phys. Lett., **B701**, 353–356 (2011), arXiv:1104.2731.

[998] F. Buccella et al., Phys. Lett., **B302**, 319–325 (1993), arXiv:hep-ph/9212253.

[999] Y. Grossman et al., Phys. Rev., **D85**, 114036 (2012), arXiv:1204.3557.

[1000] G. Isidori et al., Phys. Lett., **B711**, 46–51 (2012), arXiv:1111.4987.

[1001] J. Brod et al., Phys. Rev., **D86**, 014023 (2012), arXiv:1111.5000.

[1002] D. Pirtskhalava and P. Uttayarat, Phys. Lett., **B712**, 81–86 (2012), arXiv:1112.5451.

[1003] H.-Y. Cheng and C.-W. Chiang, Phys. Rev., **D85**, 034036, [Erratum: Phys. Rev. D85, 079903 (2012)] (2012), arXiv:1201.0785.

[1004] B. Bhattacharya et al., Phys. Rev., **D85**, 054014 (2012), arXiv:1201.2351.

[1005] W. Altmannshofer et al., JHEP, **04**, 049 (2012), arXiv:1202.2866.

[1006] T. Feldmann et al., JHEP, **06**, 007 (2012), arXiv:1202.3795.

[1007] H.-N. Li et al., Phys. Rev., **D86**, 036012 (2012), arXiv:1203.3120.

[1008] E. Franco et al., JHEP, **05**, 140 (2012), arXiv:1203.3131.

[1009] B. Bhattacharya et al., Nonleptonic Charm Decays and CP Violation, In *Proceedings, 5th International Workshop on Charm Physics (Charm 2012): Honolulu, Hawaii, USA, May 14-17, 2012* (2012), arXiv:1207.6390.

[1010] B. Bhattacharya et al. (2012), arXiv:1207.0761.

[1011] D. Atwood and A. Soni, PTEP, **2013**(9), 093B05 (2013), arXiv:1211.1026.

[1012] Y. Grossman and D. J. Robinson, JHEP, **04**, 067 (2013), arXiv:1211.3361.
[1013] G. Hiller et al., Phys. Rev., **D85**, 116008 (2012), arXiv:1204.1046.
[1014] F. Buccella et al., Phys. Rev., **D88**(7), 074011 (2013), arXiv:1305.7343.
[1015] M. Gronau, Phys. Lett., **B730**, 221–225, [Addendum: Phys. Lett.B735,282(2014)] (2014), arXiv:1311.1434.
[1016] A. Lenz, What did we learn in theory from the $\Delta A_{CP}$-saga?, In *Proceedings, 6th International Workshop on Charm Physics (Charm 2013): Manchester, UK, August 31-September 4, 2013* (2013), arXiv:1311.6447.
[1017] M. Gronau, Phys. Rev., **D91**(7), 076007 (2015), arXiv:1501.03272.
[1018] G. Hiller et al., Phys. Rev., **D87**(1), 014024 (2013), arXiv:1211.3734.
[1019] G. Hiller et al., PoS, **EPS-HEP2013**, 371 (2013), arXiv:1311.3883.
[1020] S. Müller et al., Phys. Rev., **D92**(1), 014004 (2015), arXiv:1503.06759.
[1021] S. Müller et al., Phys. Rev. Lett., **115**(25), 251802 (2015), arXiv:1506.04121.
[1022] U. Nierste and S. Schacht, Phys. Rev., **D92**(5), 054036 (2015), arXiv:1508.00074.
[1023] R. Aaij et al., LHCb, Phys. Lett., **B767**, 177–187 (2017), arXiv:1610.09476.
[1024] N. Dash et al. (2017), arXiv:1705.05966.
[1025] N. K. Nisar et al., Belle, Phys. Rev. Lett., **112**, 211601 (2014), arXiv:1404.1266.
[1026] R. Aaij et al., LHCb, JHEP, **10**, 25 (2014), arXiv:1406.2624.
[1027] B. R. Ko et al., Belle, Phys. Rev. Lett., **109**, 021601, [Erratum: Phys. Rev. Lett.109,119903(2012)] (2012), arXiv:1203.6409.
[1028] L.-L. C. Wang, Flavor Mixing and Charm Decay, In *Meson Spectra*, page 1218 (1980).
[1029] L.-L. Chau, Phys. Rept., **95**, 1–94 (1983).
[1030] J. J. de Swart, Rev. Mod. Phys., **35**, 916–939, [Erratum: Rev. Mod. Phys. 37, 326 (1965)] (1963).
[1031] T. A. Kaeding, Atom. Data Nucl. Data Tabl., **61**, 233 (1995), arXiv:nucl-th/9502037.
[1032] T. A. Kaeding and H. T. Williams, Comput. Phys. Commun., **98**, 398–414 (1996), arXiv:nucl-th/9511025.
[1033] M. Gronau et al., Phys. Rev., **D52**, 6356–6373 (1995), arXiv:hep-ph/9504326.
[1034] A. J. Buras et al., Nucl. Phys., **B268**, 16–48 (1986).
[1035] A. J. Buras and L. Silvestrini, Nucl. Phys., **B569**, 3–52 (2000), arXiv:hep-ph/9812392.
[1036] P. U. E. Onyisi et al., CLEO, Phys. Rev., **D88**(3), 032009 (2013), arXiv:1306.5363.
[1037] G. Bonvicini et al., CLEO, Phys. Rev., **D63**, 071101 (2001), arXiv:hep-ex/0012054.
[1038] H. Mendez et al., CLEO, Phys. Rev., **D81**, 052013 (2010), arXiv:0906.3198.
[1039] V. Babu et al., Belle, Phys. Rev., **D97**(1), 011101 (2018), arXiv:1712.00619.
[1040] R. Aaij et al., LHCb, JHEP, **10**, 055 (2015), arXiv:1508.06087.
[1041] B. Bhattacharya and J. L. Rosner, Phys. Rev., **D81**, 014026 (2010), arXiv:0911.2812.
[1042] H.-Y. Cheng and C.-W. Chiang, Phys. Rev., **D81**, 074021 (2010), arXiv:1001.0987.
[1043] D.-N. Gao, Phys. Rev., **D91**(1), 014019 (2015), arXiv:1411.0768.
[1044] Y. Hochberg and Y. Nir, Phys. Rev. Lett., **108**, 261601 (2012), arXiv:1112.5268.
[1045] A. J. Bevan et al., Phys. Rev., **D84**(11), 114009, [Erratum: Phys. Rev. D87, 039905 (2013)] (2011), arXiv:1106.5075.
[1046] M. Staric et al., Belle, Phys. Lett., **B670**, 190–195 (2008), arXiv:0807.0148.
[1047] P. del Amo Sanchez et al., BaBar, Phys. Rev., **D83**, 071103 (2011), arXiv:1011.5477.
[1048] B. R. Ko et al., Belle, JHEP, **02**, 098 (2013), arXiv:1212.6112.
[1049] J. P. Lees et al., BaBar, Phys. Rev., **D87**(1), 012004 (2013), arXiv:1209.3896.
[1050] S.K. Choi et al., Belle, Phys. Rev. Lett., **91**, 262001 (2003), arXiv:hep-ex/0309032.
[1051] N. Brambilla et al., Eur.Phys.J., **C74**(10), 2981 (2014), arXiv:1404.3723.
[1052] N. Brambilla et al., Eur.Phys.J., **C71**, 1534 (2011), arXiv:1010.5827.
[1053] N. Brambilla et al., Quarkonium Working Group (2004), arXiv:hep-ph/0412158.
[1054] N. Brambilla et al., Rev. Mod. Phys., **77**, 1423 (2005), hep-ph/0410047.
[1055] R. Aaij et al., LHCb, Phys. Rev. Lett., **115**(7), 072001 (2015), arXiv:1507.03414.
[1056] N. Brambilla et al., Phys. Lett., **B576**, 314–327 (2003), arXiv:hep-ph/0306107.
[1057] W. E. Caswell and G. P. Lepage, Phys. Lett., **B167**, 437 (1986).
[1058] G. T. Bodwin et al., Phys. Rev., **D51**, 1125–1171, [Erratum: Phys. Rev. D55, 5853 (1997)] (1995), arXiv:hep-ph/9407339.
[1059] N. Brambilla et al., Phys. Rev., **D79**, 074002, [Erratum: Phys. Rev. D83, 079904 (2011)] (2009), arXiv:0810.2259.
[1060] N. Brambilla et al., JHEP, **08**, 039, [Erratum: JHEP 04, 058 (2011)] (2006), arXiv:hep-ph/0604190.
[1061] G. T. Bodwin et al., Phys. Rev., **D65**, 054504 (2002), arXiv:hep-lat/0107011.
[1062] G. T. Bodwin, Theory of Charmonium Production, In *Proceedings, 5th International Workshop on Charm Physics (Charm 2012): Honolulu, Hawaii, USA, May 14-17, 2012* (2012), arXiv:1208.5506.
[1063] N. Brambilla et al., Nucl. Phys., **B566**, 275 (2000), arXiv:hep-ph/9907240.

[1064] A. Pineda and J. Soto, Nucl. Phys. Proc. Suppl., **64**, 428–432 (1998), arXiv:hep-ph/9707481.
[1065] N. Brambilla et al., Phys. Rev., **D63**, 014023 (2001), arXiv:hep-ph/0002250.
[1066] A. Pineda and A. Vairo, Phys. Rev., **D63**, 054007, [Erratum: Phys. Rev. D64, 039902 (2001)] (2001), arXiv:hep-ph/0009145.
[1067] C. Gattringer and C. B. Lang, Lect. Notes Phys., **788**, 1–343 (2010).
[1068] T. DeGrand and C. E. Detar, *Lattice methods for quantum chromodynamics*, (World Scientific, New Jersey, USA, 2006).
[1069] C. Liu, Review on Hadron Spectroscopy (2017), arXiv:1612.00103.
[1070] S. M. Ryan, EPJ Web Conf., **130**, 01002 (2016).
[1071] S. Prelovsek, Lattice studies of charmonia and exotics, In *7th International Workshop on Charm Physics (Charm 2015) Detroit, MI, USA, May 18-22, 2015* (2015), arXiv:1508.07322.
[1072] N. Brambilla et al., Phys. Lett., **B470**, 215 (1999), arXiv:hep-ph/9910238.
[1073] B. A. Kniehl et al., Nucl. Phys., **B635**, 357–383 (2002), arXiv:hep-ph/0203166.
[1074] A. V. Smirnov et al., Phys. Rev. Lett., **104**, 112002 (2010), arXiv:0911.4742.
[1075] C. Anzai et al., Phys. Rev. Lett., **104**, 112003 (2010), arXiv:0911.4335.
[1076] N. Brambilla et al., Phys. Rev., **D60**, 091502 (1999), arXiv:hep-ph/9903355.
[1077] N. Brambilla et al., Phys. Rev. Lett., **105**, 212001, [Erratum: Phys. Rev. Lett. 108, 269903 (2012)] (2010), arXiv:1006.2066.
[1078] A. Bazavov et al., Phys. Rev., **D86**, 114031 (2012), arXiv:1205.6155.
[1079] A. Bazavov et al., Phys. Rev., **D90**(7), 074038 (2014), arXiv:1407.8437.
[1080] A. Pineda and F. J. Yndurain, Phys. Rev., **D58**, 094022 (1998), arXiv:hep-ph/9711287.
[1081] S. Titard and F. J. Yndurain, Phys. Rev., **D49**, 6007–6025 (1994), arXiv:hep-ph/9310236.
[1082] Y. Kiyo and Y. Sumino, Nucl. Phys., **B889**, 156–191 (2014), arXiv:1408.5590.
[1083] A. H. Hoang et al., Phys. Rev., **D59**, 114014 (1999), arXiv:hep-ph/9804227.
[1084] M. Beneke, Phys. Lett., **B434**, 115–125 (1998), arXiv:hep-ph/9804241.
[1085] N. Brambilla et al., Phys. Lett., **B513**, 381–390 (2001), arXiv:hep-ph/0101305.
[1086] Y. Kiyo and Y. Sumino, Phys. Lett., **B730**, 76–80 (2014), arXiv:1309.6571.
[1087] B. A. Kniehl et al., Phys. Rev. Lett., **92**, 242001, [Erratum: Phys. Rev. Lett. 104, 199901 (2010)] (2004), arXiv:hep-ph/0312086.
[1088] S. Recksiegel and Y. Sumino, Phys. Lett., **B578**, 369–375 (2004), arXiv:hep-ph/0305178.
[1089] A. A. Penin et al., Phys. Lett., **B593**, 124–134, [Erratum: Phys. Lett. B677, 343 (2009)] (2004), arXiv:hep-ph/0403080.
[1090] A. H. Hoang (2000), arXiv:hep-ph/0008102.
[1091] N. Brambilla et al., Phys. Rev., **D65**, 034001 (2002), arXiv:hep-ph/0108084.
[1092] Y. Kiyo et al., Phys. Lett., **B752**, 122–127 (2016), arXiv:1510.07072.
[1093] C. Ayala et al., JHEP, **09**, 045 (2014), arXiv:1407.2128.
[1094] M. Beneke et al., Phys. Rev. Lett., **112**(15), 151801 (2014), arXiv:1401.3005.
[1095] M. B. Voloshin, Sov. J. Nucl. Phys., **36**, 143, [Yad. Fiz.36,247(1982)] (1982).
[1096] A. Pineda, Nucl. Phys., **B494**, 213–236 (1997), arXiv:hep-ph/9611388.
[1097] A. A. Penin et al., Nucl. Phys., **B699**, 183–206, [Erratum: Nucl. Phys. B829, 398 (2010)] (2004), arXiv:hep-ph/0406175.
[1098] Y. Kiyo et al., Nucl. Phys., **B841**, 231–256 (2010), arXiv:1006.2685.
[1099] N. Brambilla et al., Phys. Rev., **D73**, 054005 (2006), arXiv:hep-ph/0512369.
[1100] A. Pineda and J. Segovia, Phys. Rev., **D87**(7), 074024 (2013), arXiv:1302.3528.
[1101] E. Eichten et al., Rev. Mod. Phys., **80**, 1161–1193 (2008), arXiv:hep-ph/0701208.
[1102] N. Brambilla et al., Phys. Rev., **D85**, 094005 (2012), arXiv:1203.3020.
[1103] S. Steinbeißer and J. Segovia, pNRQCD determination of E1 radiative transitions, In *12th Conference on Quark Confinement and the Hadron Spectrum (Confinement XII) Thessaloniki, Greece, August 28-September 4, 2016* (2017), arXiv:1701.02513.
[1104] N. Brambilla et al., AIP Conf.Proc., **1343**, 418–420 (2011), arXiv:1012.0773.
[1105] Y. Koma et al., Phys. Rev. Lett., **97**, 122003 (2006), arXiv:hep-lat/0607009.
[1106] Y. Koma et al., PoS, **LAT2007**, 111 (2007), arXiv:0711.2322.
[1107] N. Brambilla et al., Phys. Rev., **D90**(11), 114032 (2014), arXiv:1407.7761.
[1108] N. Brambilla and A. Vairo, Phys. Rev., **D55**, 3974–3986 (1997), arXiv:hep-ph/9606344.
[1109] M. Baker et al., Phys. Rev., **D54**, 2829–2844, [Erratum: Phys. Rev. D56, 2475 (1997)] (1996), arXiv:hep-ph/9602419.
[1110] Y. Koma and M. Koma, PoS, **LATTICE2012**, 140 (2012), arXiv:1211.6795.
[1111] U. Tamponi et al., Belle, Phys. Rev. Lett., **115**(14), 142001 (2015), arXiv:1506.08914.
[1112] R. Mizuk et al., Belle, Phys. Rev. Lett., **109**, 232002 (2012), arXiv:1205.6351.
[1113] N. Brambilla et al., Phys. Rev., **D67**, 034018 (2003), arXiv:hep-ph/0208019.
[1114] M. Wurtz et al., Phys. Rev., **D92**(5), 054504 (2015), arXiv:1505.04410.

[1115] Y.-B. Yang et al. (2014), arXiv:1410.3343.
[1116] B. A. Galloway et al., PoS, **LATTICE2014**, 092 (2014), arXiv:1411.1318.
[1117] P. Perez-Rubio et al., Phys. Rev., **D92**(3), 034504 (2015), arXiv:1503.08440.
[1118] D. Mohler et al., Fermilab Lattice, MILC, PoS, **LATTICE2014**, 085 (2015), arXiv:1412.1057.
[1119] K. Cichy et al. (2016), arXiv:1603.06467.
[1120] L. Liu et al., Hadron Spectrum, JHEP, **1207**, 126 (2012), arXiv:1204.5425.
[1121] F.-K. Guo and U.-G. Meißner, Phys. Rev. Lett., **109**, 062001 (2012), arXiv:1203.1116.
[1122] C. Hughes et al., Phys. Rev., **D92**, 094501 (2015), arXiv:1508.01694.
[1123] R. Lewis and R. M. Woloshyn, Phys. Rev., **D84**, 094501 (2011), arXiv:1108.1137.
[1124] R. Lewis and R. M. Woloshyn, Phys. Rev., **D86**, 057501 (2012), arXiv:1207.3825.
[1125] J. J. Dudek et al., Phys. Rev., **D79**, 094504 (2009), arXiv:0902.2241.
[1126] D. Becirevic and F. Sanfilippo, JHEP, **01**, 028 (2013), arXiv:1206.1445.
[1127] D. Becirevic et al., JHEP, **05**, 014 (2015), arXiv:1411.6426.
[1128] Y. Chen et al., Phys. Rev., **D84**, 034503 (2011), arXiv:1104.2655.
[1129] K. J. Juge et al., Phys. Rev. Lett., **90**, 161601 (2003), arXiv:hep-lat/0207004.
[1130] G. S. Bali and A. Pineda, Phys. Rev., **D69**, 094001 (2004), arXiv:hep-ph/0310130.
[1131] E. Braaten and M. Kusunoki, Phys. Rev., **D69**, 074005 (2004), arXiv:hep-ph/0311147.
[1132] S. Fleming et al., Phys. Rev., **D76**, 034006 (2007), arXiv:hep-ph/0703168.
[1133] V. Baru et al., Phys. Rev., **D84**, 074029 (2011), arXiv:1108.5644.
[1134] E. Epelbaum et al., Rev. Mod. Phys., **81**, 1773–1825 (2009), arXiv:0811.1338.
[1135] P. Bicudo et al. (2016), arXiv:1612.02758.
[1136] A. Peters et al., Lattice QCD study of heavy-heavy-light-light tetraquark candidates, In *Proceedings, 34th International Symposium on Lattice Field Theory (Lattice 2016): Southampton, UK, July 24-30, 2016* (2016), arXiv:1609.00181.
[1137] M. Berwein et al., Phys. Rev., **D92**(11), 114019 (2015), arXiv:1510.04299.
[1138] S. K. Choi et al., Phys. Rev., **D84**, 052004 (2011), arXiv:1107.0163.
[1139] B. Aubert et al., BaBar, Phys. Rev., **D77**, 111101 (2008), arXiv:0803.2838.
[1140] D. Acosta et al., CDF, Phys. Rev. Lett., **93**, 072001 (2004), arXiv:hep-ex/0312021.
[1141] A. Abulencia et al., CDF, Phys. Rev. Lett., **98**, 132002 (2007), arXiv:hep-ex/0612053.
[1142] T. Aaltonen et al., CDF, Phys. Rev. Lett., **103**, 152001 (2009), arXiv:0906.5218.
[1143] V. M. Abazov et al., D0, Phys. Rev. Lett., **93**, 162002 (2004), arXiv:hep-ex/0405004.
[1144] R. Aaij et al., LHCb, Eur. Phys. J., **C72**, 1972 (2012), arXiv:1112.5310.
[1145] R. Aaij et al., LHCb, Phys. Rev. Lett., **110**, 222001 (2013), arXiv:1302.6269.
[1146] R. Aaij et al., LHCb, Phys. Rev., **D92**(1), 011102 (2015), arXiv:1504.06339.
[1147] S. Chatrchyan et al., CMS, JHEP, **04**, 154 (2013), arXiv:1302.3968.
[1148] M. Ablikim et al., BESIII, Phys. Rev. Lett., **112**(9), 092001 (2014), arXiv:1310.4101.
[1149] K. Abe et al., Evidence for $X(3872) \to \gamma J/\psi$ and the sub-threshold decay $X(3872) \to \omega J/\psi$, In *Lepton and photon interactions at high energies. Proceedings, 22nd International Symposium, LP 2005, Uppsala, Sweden, June 30-July 5, 2005* (2005), arXiv:hep-ex/0505037.
[1150] P. del Amo Sanchez et al., BaBar, Phys. Rev., **D82**, 011101 (2010), arXiv:1005.5190.
[1151] V. Bhardwaj et al., Belle, Phys. Rev. Lett., **107**, 091803 (2011), arXiv:1105.0177.
[1152] B. Aubert et al., BaBar, Phys. Rev., **D74**, 071101 (2006), arXiv:hep-ex/0607050.
[1153] B. Aubert et al., BaBar, Phys. Rev. Lett., **102**, 132001 (2009), arXiv:0809.0042.
[1154] R. Aaij et al., LHCb, Nucl. Phys., **B886**, 665–680 (2014), arXiv:1404.0275.
[1155] G. Gokhroo et al., Belle, Phys. Rev. Lett., **97**, 162002 (2006), arXiv:hep-ex/0606055.
[1156] T. Aushev et al., Belle, Phys. Rev., **D81**, 031103 (2010), arXiv:0810.0358.
[1157] B. Aubert et al., BaBar, Phys. Rev., **D77**, 011102 (2008), arXiv:0708.1565.
[1158] M. Ablikim et al., BESIII, Phys. Rev. Lett., **110**, 252001 (2013), arXiv:1303.5949.
[1159] Z. Q. Liu et al., Belle, Phys. Rev. Lett., **110**, 252002 (2013), arXiv:1304.0121.
[1160] T. Xiao et al., Phys. Lett., **B727**, 366–370 (2013), arXiv:1304.3036.
[1161] M. Ablikim et al., BESIII, Phys. Rev. Lett., **115**(11), 112003 (2015), arXiv:1506.06018.
[1162] M. Ablikim et al., BESIII, Phys. Rev. Lett., **111**(24), 242001 (2013), arXiv:1309.1896.
[1163] M. Ablikim et al., BESIII, Phys. Rev. Lett., **112**(2), 022001 (2014), arXiv:1310.1163.
[1164] M. Ablikim et al., BESIII, Phys. Rev., **D92**(9), 092006 (2015), arXiv:1509.01398.
[1165] M. Ablikim et al., BESIII, Phys. Rev. Lett., **115**(22), 222002 (2015), arXiv:1509.05620.
[1166] M. Ablikim et al., BESIII, Phys. Rev. Lett., **113**(21), 212002 (2014), arXiv:1409.6577.
[1167] X. L. Wang et al., Belle, Phys. Rev., **D91**, 112007 (2015), arXiv:1410.7641.
[1168] M. Ablikim et al., BESIII (2017), arXiv:1703.08787.
[1169] M. Ablikim et al., BESIII, Phys. Rev. Lett., **112**(13), 132001 (2014), arXiv:1308.2760.
[1170] M. Ablikim et al., BESIII, Phys. Rev. Lett., **115**(18), 182002 (2015), arXiv:1507.02404.
[1171] I. Adachi, Observation of two charged bottomonium-like resonances, In *Flavor physics and CP*

*violation. Proceedings, 9th International Conference, FPCP 2011, Maale HaChamisha, Israel, May 23-27, 2011* (2011), arXiv:1105.4583.
[1172] A. Bondar et al., Belle, Phys. Rev. Lett., **108**, 122001 (2012), arXiv:1110.2251.
[1173] A. Garmash et al., Belle, Phys. Rev., **D91**(7), 072003 (2015), arXiv:1403.0992.
[1174] P. Krokovny et al., Belle, Phys. Rev., **D88**(5), 052016 (2013), arXiv:1308.2646.
[1175] I. Adachi et al., Study of Three-Body Y(10860) Decays (2012), arXiv:1209.6450.
[1176] A. Garmash et al., Belle, Phys. Rev. Lett., **116**(21), 212001 (2016), arXiv:1512.07419.
[1177] A. Abdesselam et al., Belle, Phys. Rev. Lett., **117**(14), 142001 (2016), arXiv:1508.06562.
[1178] K. Abe et al., Belle, Phys. Rev. Lett., **93**, 051803 (2004), arXiv:hep-ex/0307061.
[1179] J. Brodzicka et al., Belle, Phys. Rev. Lett., **100**, 092001 (2008), arXiv:0707.3491.
[1180] J. Z. Bai et al., BES, Phys. Lett., **B605**, 63–71 (2005), arXiv:hep-ex/0307028.
[1181] N. E. Adam et al., CLEO, Phys. Rev. Lett., **96**, 082004 (2006), arXiv:hep-ex/0508023.
[1182] G. S. Adams et al., CLEO, Phys. Rev., **D73**, 012002 (2006), arXiv:hep-ex/0509011.
[1183] V. Bhardwaj et al., Belle, Phys. Rev. Lett., **111**(3), 032001 (2013), arXiv:1304.3975.
[1184] M. Ablikim et al., BESIII, Phys. Rev. Lett., **115**(1), 011803 (2015), arXiv:1503.08203.
[1185] K. Chilikin et al., Belle (2017), arXiv:1704.01872.
[1186] K. Abe et al., Belle, Phys. Rev. Lett., **94**, 182002 (2005), arXiv:hep-ex/0408126.
[1187] B. Aubert et al., BaBar, Phys. Rev. Lett., **101**, 082001 (2008), arXiv:0711.2047.
[1188] S. Uehara et al., Belle, Phys. Rev. Lett., **104**, 092001 (2010), arXiv:0912.4451.
[1189] J. P. Lees et al., BaBar, Phys. Rev., **D86**, 072002 (2012), arXiv:1207.2651.
[1190] S. Uehara et al., Belle, Phys. Rev. Lett., **96**, 082003 (2006), arXiv:hep-ex/0512035.
[1191] B. Aubert et al., BaBar, Phys. Rev., **D81**, 092003 (2010), arXiv:1002.0281.
[1192] K. Abe et al., Belle, Phys. Rev. Lett., **98**, 082001 (2007), arXiv:hep-ex/0507019.
[1193] P. Pakhlov et al., Belle, Phys. Rev. Lett., **100**, 202001 (2008), arXiv:0708.3812.
[1194] M. Ablikim et al., BESIII, Phys. Rev., **D86**, 071101 (2012), arXiv:1208.1857.
[1195] X. L. Wang et al., Belle, Phys. Rev., **D87**(5), 051101 (2013), arXiv:1210.7550.
[1196] R. Mizuk et al., Belle, Phys. Rev., **D78**, 072004 (2008), arXiv:0806.4098.
[1197] J. P. Lees et al., BaBar, Phys. Rev., **D85**, 052003 (2012), arXiv:1111.5919.
[1198] T. Aaltonen et al., CDF, Phys. Rev. Lett., **102**, 242002 (2009), arXiv:0903.2229.
[1199] T. Aaltonen et al., CDF, Mod. Phys. Lett., **A32**(26), 1750139 (2017), arXiv:1101.6058.
[1200] J. Brodzicka, Heavy Flavour Spectroscopy, In *24th International Symposium on Lepton-Photon Interactions at High Energy (LP09)* (2009).
[1201] R. Aaij et al., LHCb, Phys. Rev., **D85**, 091103 (2012), arXiv:1202.5087.
[1202] S. Chatrchyan et al., CMS, Phys. Lett., **B734**, 261–281 (2014), arXiv:1309.6920.
[1203] V. M. Abazov et al., D0, Phys. Rev., **D89**(1), 012004 (2014), arXiv:1309.6580.
[1204] J. P. Lees et al., BaBar, Phys. Rev., **D91**(1), 012003 (2015), arXiv:1407.7244.
[1205] R. Aaij et al., LHCb (2016), arXiv:1606.07895.
[1206] R. Aaij et al., LHCb (2016), arXiv:1606.07898.
[1207] V. M. Abazov et al., D0, Phys. Rev. Lett., **115**(23), 232001 (2015), arXiv:1508.07846.
[1208] M. Ablikim et al., BESIII, Phys. Rev., **D91**(11), 112005 (2015), arXiv:1503.06644.
[1209] K. Chilikin et al., Belle, Phys. Rev., **D90**(11), 112009 (2014), arXiv:1408.6457.
[1210] B. Aubert et al., BaBar, Phys. Rev. Lett., **95**, 142001 (2005), arXiv:hep-ex/0506081.
[1211] J. P. Lees et al., BaBar, Phys. Rev., **D86**, 051102 (2012), arXiv:1204.2158.
[1212] T. E. Coan et al., CLEO, Phys. Rev. Lett., **96**, 162003 (2006), arXiv:hep-ex/0602034.
[1213] Q. He et al., CLEO, Phys. Rev., **D74**, 091104 (2006), arXiv:hep-ex/0611021.
[1214] C. Z. Yuan et al., Belle, Phys. Rev. Lett., **99**, 182004 (2007), arXiv:0707.2541.
[1215] M. Ablikim et al., BESIII (2016), arXiv:1611.01317.
[1216] M. Ablikim et al., BESIII (2016), arXiv:1610.07044.
[1217] M. Ablikim et al., BESIII, Phys. Rev. Lett., **114**(9), 092003 (2015), arXiv:1410.6538.
[1218] C. P. Shen et al., Belle, Phys. Rev. Lett., **104**, 112004 (2010), arXiv:0912.2383.
[1219] X. L. Wang et al., Belle, Phys. Rev. Lett., **99**, 142002 (2007), arXiv:0707.3699.
[1220] J. P. Lees et al., BaBar, Phys. Rev., **D89**(11), 111103 (2014), arXiv:1211.6271.
[1221] M. Ablikim et al., BESIII, Phys. Rev., **D93**(1), 011102 (2016), arXiv:1511.08564.
[1222] G. Pakhlova et al., Belle, Phys. Rev. Lett., **100**, 062001 (2008), arXiv:0708.3313.
[1223] S.K. Choi et al., BELLE, Phys. Rev. Lett., **100**, 142001 (2008), arXiv:0708.1790.
[1224] R. Mizuk et al., Belle, Phys. Rev., **D80**, 031104 (2009), arXiv:0905.2869.
[1225] K. Chilikin et al., Belle, Phys. Rev., **D88**(7), 074026 (2013), arXiv:1306.4894.
[1226] B. Aubert et al., BaBar, Phys. Rev., **D79**, 112001 (2009), arXiv:0811.0564.
[1227] R. Aaij et al., LHCb, Phys. Rev. Lett., **112**(22), 222002 (2014), arXiv:1404.1903.
[1228] R. Aaij et al., LHCb, Phys. Rev., **D92**(11), 112009, [Phys. Rev.D92,112009(2015)] (2015), arXiv:1510.01951.

[1229] G. Pakhlova et al., Belle, Phys. Rev. Lett., **101**, 172001 (2008), arXiv:0807.4458.
[1230] B. Aubert et al., BaBar, Phys. Rev. Lett., **96**, 232001 (2006), arXiv:hep-ex/0604031.
[1231] B. Aubert et al., BaBar, Phys. Rev., **D78**, 112002 (2008), arXiv:0807.2014.
[1232] A. Sokolov et al., Belle, Phys. Rev., **D75**, 071103 (2007), arXiv:hep-ex/0611026.
[1233] A. Sokolov et al., Belle, Phys. Rev., **D79**, 051103 (2009), arXiv:0901.1431.
[1234] K. F. Chen et al., Belle, Phys. Rev. Lett., **100**, 112001 (2008), arXiv:0710.2577.
[1235] X. H. He et al., Belle, Phys. Rev. Lett., **113**(14), 142001 (2014), arXiv:1408.0504.
[1236] A. Abdesselam et al. (2016), arXiv:1609.08749.
[1237] D. Santel et al., Belle, Phys. Rev., **D93**(1), 011101 (2016), arXiv:1501.01137.
[1238] L. Maiani et al., Phys. Rev., **D89**, 114010 (2014), arXiv:1405.1551.
[1239] A. E. Bondar et al., Phys. Rev., **D84**, 054010 (2011), arXiv:1105.4473.
[1240] M. B. Voloshin, Phys. Rev., **D71**, 114003 (2005), arXiv:hep-ph/0504197.
[1241] F.-K. Guo and U. Meißner, Phys. Rev., **D86**, 091501 (2012), arXiv:1208.1134.
[1242] S. Olsen, Phys. Rev., **D91**(5), 057501 (2015), arXiv:1410.6534.
[1243] Z.-Y. Zhou et al., Phys. Rev. Lett., **115**(2), 022001 (2015), arXiv:1501.00879.
[1244] V. Baru et al. (2017), arXiv:1703.01230.
[1245] G. Cotugno et al., Phys. Rev. Lett., **104**, 132005 (2010), arXiv:0911.2178.
[1246] F.-K. Guo et al., Phys. Rev., **D82**, 094008 (2010), arXiv:1005.2055.
[1247] S. Godfrey and K. Moats, Phys. Rev., **D92**(5), 054034 (2015), arXiv:1507.00024.
[1248] A. G. Drutskoy et al., Eur. Phys. J., **A49**, 7–32 (2013), arXiv:1210.6623.
[1249] T. Mehen and J. Powell, Phys. Rev., **D88**(3), 034017 (2013), arXiv:1306.5459.
[1250] L. Maiani et al., Phys. Lett., **B749**, 289–291 (2015), arXiv:1507.04980.
[1251] L. Roca et al., Phys. Rev., **D92**(9), 094003 (2015), arXiv:1507.04249.
[1252] U. Meißner and J. A. Oller, Phys. Lett., **B751**, 59–62 (2015), arXiv:1507.07478.
[1253] F.-K. Guo et al., Phys. Rev., **D92**(7), 071502 (2015), arXiv:1507.04950.
[1254] C. J. Morningstar and M. J. Peardon, Phys. Rev., **D60**, 034509 (1999), arXiv:hep-lat/9901004.
[1255] T. Barnes et al., Phys. Rev., **D52**, 5242–5256 (1995), arXiv:hep-ph/9501405.
[1256] D. V. Bugg, Phys. Lett., **B598**, 8–14 (2004), arXiv:hep-ph/0406293.
[1257] D. V. Bugg, Europhys. Lett., **96**, 11002 (2011), arXiv:1105.5492.
[1258] D.-Y. Chen and X. Liu, Phys. Rev., **D84**, 094003 (2011), arXiv:1106.3798.
[1259] D.-Y. Chen and X. Liu, Phys. Rev., **D84**, 034032 (2011), arXiv:1106.5290.
[1260] C.-Y. Chen et al., Phys. Rev., **D84**, 074032 (2011), arXiv:1108.4458.
[1261] D.-Y. Chen et al., Phys. Rev., **D88**(3), 036008 (2013), arXiv:1304.5845.
[1262] D.-Y. Chen et al., Phys. Rev. Lett., **110**(23), 232001 (2013), arXiv:1303.6842.
[1263] E. S. Swanson, Phys. Rev., **D91**(3), 034009 (2015), arXiv:1409.3291.
[1264] F.-K. Guo et al., Phys. Rev., **D91**(5), 051504 (2015), arXiv:1411.5584.
[1265] E. S. Swanson (2015), arXiv:1504.07952.
[1266] V. Baru et al., Eur. Phys. J., **A44**, 93–103 (2010), arXiv:1001.0369.
[1267] C. Hanhart et al., Eur. Phys. J., **A47**, 101–110 (2011), arXiv:1106.1185.
[1268] P. G. Ortega et al., Phys. Rev., **D81**, 054023 (2010), arXiv:1001.3948.
[1269] P. G. Ortega et al., Phys. Rev., **D81**, 0054023 (2010), arXiv:0907.3997.
[1270] Z.-Y. Zhou and Z. Xiao, Eur. Phys. J., **A50**(10), 165 (2014), arXiv:1309.1949.
[1271] E. Cincioglu et al. (2016), arXiv:1606.03239.
[1272] M. Gell-Mann, Phys. Lett., **8**, 214–215 (1964).
[1273] C. Alexandrou et al., Phys. Rev. Lett., **97**, 222002 (2006), arXiv:hep-lat/0609004.
[1274] R. L. Jaffe and F. Wilczek, Phys. Rev. Lett., **91**, 232003 (2003), arXiv:hep-ph/0307341.
[1275] L. Maiani et al., Phys. Rev. Lett., **93**, 212002 (2004), arXiv:hep-ph/0407017.
[1276] G. 't Hooft et al., Phys. Lett., **B662**, 424–430 (2008), arXiv:0801.2288.
[1277] R. L. Jaffe, Phys. Rept., **409**, 1–45 (2005), arXiv:hep-ph/0409065.
[1278] L. Maiani et al., Phys. Rev., **D71**, 014028 (2005), arXiv:hep-ph/0412098.
[1279] L. Maiani et al. (2017), arXiv:1712.05296.
[1280] A. Ali et al., Eur. Phys. J., **C78**(1), 29 (2018), arXiv:1708.04650.
[1281] S. J. Brodsky et al., Phys. Rev. Lett., **113**(11), 112001 (2014), arXiv:1406.7281.
[1282] A. Ali et al., Phys. Rev., **D91**(1), 017502 (2015), arXiv:1412.2049.
[1283] G. C. Rossi and G. Veneziano, Nucl. Phys., **B123**, 507–545 (1977).
[1284] G. C. Rossi and G. Veneziano, Phys. Lett., **B70**, 255–259 (1977).
[1285] L. Montanet et al., Phys. Rept., **63**, 149–222 (1980).
[1286] L. Maiani et al., Phys. Rev., **D94**(5), 054026 (2016), arXiv:1607.02405.
[1287] A. Ali et al., Phys. Rev., **D85**, 054011 (2012), arXiv:1110.1333.
[1288] A. Esposito et al., Phys. Lett., **B758**, 292–295 (2016), arXiv:1603.07667.
[1289] M. Papinutto et al., A Tentative Description of $Z_c, b$ States in Terms of Metastable Feshbach

Resonances, In *Proceedings, 6th International Workshop on Charm Physics (Charm 2013)* (2013), arXiv:1311.7374.
[1290] A. L. Guerrieri et al., Phys. Rev., **D90**(3), 034003 (2014), arXiv:1405.7929.
[1291] C. Bignamini et al., Phys. Rev. Lett., **103**, 162001 (2009), arXiv:0906.0882.
[1292] A. Esposito et al., Phys. Rev., **D92**(3), 034028 (2015), arXiv:1508.00295.
[1293] P. Artoisenet and E. Braaten, Phys. Rev., **D81**, 114018 (2010), arXiv:0911.2016.
[1294] C. Bignamini et al., Phys. Lett., **B684**, 228–230 (2010), arXiv:0912.5064.
[1295] F.-K. Guo et al., Eur. Phys. J., **C74**(9), 3063 (2014), arXiv:1402.6236.
[1296] M. Albaladejo et al., Chin. Phys., **C41**(12), 121001 (2017), arXiv:1709.09101.
[1297] M. S. Chanowitz and S. R. Sharpe, Nucl. Phys., **B222**, 211–244, [Erratum: Nucl. Phys. B228, 588 (1983)] (1983).
[1298] T. Barnes et al., Nucl. Phys., **B224**, 241 (1983).
[1299] S. R. Cotanch and F. J. Llanes-Estrada, Nucl. Phys., **A689**, 481–484 (2001).
[1300] I. J. General et al., Eur. Phys. J., **C51**, 347–358 (2007), arXiv:hep-ph/0609115.
[1301] F. J. Llanes-Estrada and S. R. Cotanch, Phys. Lett., **B504**, 15–20 (2001), arXiv:hep-ph/0008337.
[1302] E. Abreu and P. Bicudo, J. Phys., **G34**, 195207 (2007), arXiv:hep-ph/0508281.
[1303] D. Horn and J. Mandula, Phys. Rev., **D17**, 898 (1978).
[1304] A. Le Yaouanc et al., Z. Phys., **C28**, 309–315 (1985).
[1305] F. Iddir et al., Phys. Lett., **B433**, 125–138 (1998), arXiv:hep-ph/9803470.
[1306] Y. A. Simonov, QCD and topics in hadron physics, In *QCD: Perturbative or nonperturbative? Proceedings, 17th Autumn School, Lisbon, Portugal, September 29-October 4, 1999*, pages 60–96 (1999), arXiv:hep-ph/9911237.
[1307] Y. A. Simonov, Phys. Atom. Nucl., **64**, 1876–1886, [Yad. Fiz.64,1959(2001)] (2001), arXiv:hep-ph/0110033.
[1308] Y. S. Kalashnikova and Y. B. Yufryakov, Phys. Lett., **B359**, 175–180 (1995), arXiv:hep-ph/9506269.
[1309] Y. A. Simonov, Phys. Atom. Nucl., **68**, 1294–1302, [Yad. Fiz.68,1347(2005)] (2005), arXiv:hep-ph/0406290.
[1310] Y. S. Kalashnikova and D. S. Kuzmenko, Phys. Atom. Nucl., **66**, 955–967, [Yad. Fiz.66,988(2003)] (2003), arXiv:hep-ph/0203128.
[1311] F. Buisseret and C. Semay, Phys. Rev., **D74**, 114018 (2006), arXiv:hep-ph/0610132.
[1312] F. Buisseret et al., Eur. Phys. J., **A32**, 123–126 (2007), arXiv:hep-ph/0703020.
[1313] Y. S. Kalashnikova and A. V. Nefediev, Phys. Rev., **D77**, 054025 (2008), arXiv:0801.2036.
[1314] Y. Kalashnikova et al., Phys. Rev., **D94**(11), 114007 (2016), arXiv:1611.10066.
[1315] N. Isgur and J. E. Paton, Phys. Rev., **D31**, 2910 (1985).
[1316] F. Buisseret and V. Mathieu, Eur. Phys. J., **A29**, 343–351 (2006), arXiv:hep-ph/0607083.
[1317] J. Merlin and J. E. Paton, Phys. Rev., **D35**, 1668 (1987).
[1318] E. Braaten et al., Phys. Rev. Lett., **112**, 222001 (2014), arXiv:1401.7351.
[1319] Y. S. Kalashnikova, Z. Phys., **C62**, 323–327 (1994).
[1320] E. Kou and O. Pene, Phys. Lett., **B631**, 164–169 (2005), arXiv:hep-ph/0507119.
[1321] N. Isgur, R. Kokoski, and J. Paton, Phys. Rev. Lett., **54**, 869 (1985).
[1322] Y. S. Kalashnikova, Phys. Rev., **D72**, 034010 (2005), arXiv:hep-ph/0506270.
[1323] E. Eichten et al., Phys. Rev., **D17**, 3090, [Erratum: Phys. Rev. D21, 313 (1980)] (1978).
[1324] Xin Li and M. B. Voloshin, Phys. Rev., **D88**(3), 034012 (2013), arXiv:1307.1072.
[1325] J. M. Torres-Rincon and F. J. Llanes-Estrada, Phys. Rev. Lett., **105**, 022003 (2010), arXiv:1003.5989.
[1326] F. E. Close, Phys. Lett., **B342**, 369–374 (1995), arXiv:hep-ph/9409203.
[1327] I. Dunietz et al., Eur. Phys. J., **C1**, 211–219 (1998), arXiv:hep-ph/9612421.
[1328] F. E. Close et al., Phys. Rev., **D57**, 5653–5657 (1998), arXiv:hep-ph/9708265.
[1329] G. Chiladze et al., Phys. Rev., **D58**, 034013 (1998), arXiv:hep-ph/9804248.
[1330] F. E. Close and P. R. Page, Phys. Lett., **B628**, 215–222 (2005), arXiv:hep-ph/0507199.
[1331] G. Pakhlova et al., Belle, Phys. Rev., **D80**, 091101 (2009), arXiv:0908.0231.
[1332] M. B. Voloshin, Prog. Part. Nucl. Phys., **61**, 455–511 (2008), arXiv:0711.4556.
[1333] S. Dubynskiy and M. B. Voloshin, Phys. Lett., **B666**, 344–346 (2008), arXiv:0803.2224.
[1334] X. Li and M. B. Voloshin, Mod. Phys. Lett., **A29**(12), 1450060 (2014), arXiv:1309.1681.
[1335] M. Cleven et al., Phys. Rev., **D92**(1), 014005 (2015), arXiv:1505.01771.
[1336] F.-K. Guo et al., Rev. Mod. Phys., **90**(1), 015004 (2018), arXiv:1705.00141.
[1337] S. Weinberg, Phys. Rev., **130**, 776–783 (1963).
[1338] E. Braaten and H. W. Hammer, Phys. Rept., **428**, 259–390 (2006), arXiv:cond-mat/0410417.
[1339] T. Hyodo, Int. J. Mod. Phys., **A28**, 1330045 (2013), arXiv:1310.1176.
[1340] T. Sekihara et al., PTEP, **2015**, 063D04 (2015), arXiv:1411.2308.
[1341] F. Aceti et al., Int. J. Mod. Phys. Conf. Ser., **26**, 1460077 (2014), arXiv:1301.2554.
[1342] Z.-H. Guo and J. A. Oller, Phys. Rev., **D93**(9), 096001 (2016), arXiv:1508.06400.

[1343] M. Cleven et al., Phys. Rev., **D90**(7), 074039 (2014), arXiv:1310.2190.
[1344] W. Qin et al., Phys. Rev., **D94**(5), 054035 (2016), arXiv:1605.02407.
[1345] C. Hanhart, Theory Concepts for Heavy Exotic Mesons, In *21st International Conference on Particles and Nuclei (PANIC 17) Beijing, China, September 1-5, 2017* (2017), arXiv:1712.01136.
[1346] A. A. Filin et al., Phys. Rev. Lett., **105**, 019101 (2010), arXiv:1004.4789.
[1347] F.-K. Guo and U. Meißner, Phys. Rev., **D84**, 014013 (2011), arXiv:1102.3536.
[1348] N. A. Törnqvist, Z. Phys., **C61**, 525–537 (1994), arXiv:hep-ph/9310247.
[1349] J. Nieves and M. P. Valderrama, Phys. Rev., **D86**, 056004 (2012), arXiv:1204.2790.
[1350] F.-K. Guo et al., Phys. Rev. Lett., **102**, 242004 (2009), arXiv:0904.3338.
[1351] C. Hidalgo-Duque et al., Phys. Lett., **B727**, 432–437 (2013), arXiv:1305.4487.
[1352] M.B. Voloshin, Phys. Rev., **D84**, 031502 (2011), arXiv:1105.5829.
[1353] T. Mehen and J. W. Powell, Phys. Rev., **D84**, 114013 (2011), arXiv:1109.3479.
[1354] F.-K. Guo et al., Phys. Rev., **D88**, 054007 (2013), arXiv:1303.6608.
[1355] V. Baru et al., Phys. Lett., **B763**, 20–28 (2016), arXiv:1605.09649.
[1356] V. Baru et al., JHEP, **06**, 158 (2017), arXiv:1704.07332.
[1357] M. Albaladejo et al., Eur. Phys. J., **C75**(11), 547 (2015), arXiv:1504.00861.
[1358] Q. Wang, V. Baru, A. A. Filin, C. Hanhart, A. V. Nefediev, and J. L Wynen, Phys. Rev., **D98**(7), 074023 (2018), arXiv:1805.07453.
[1359] T. Barnes et al., Phys. Rev., **D68**, 054006 (2003), arXiv:hep-ph/0305025.
[1360] E. van Beveren and G. Rupp, Phys. Rev. Lett., **91**, 012003 (2003), arXiv:hep-ph/0305035.
[1361] E. E. Kolomeitsev and M. F. M. Lutz, Phys. Lett., **B582**, 39–48 (2004), arXiv:hep-ph/0307133.
[1362] F.-K. Guo et al., Phys. Lett., **B641**, 278–285 (2006), arXiv:hep-ph/0603072.
[1363] Y.-J. Zhang et al., Phys. Rev., **D74**, 014013 (2006), arXiv:hep-ph/0604271.
[1364] M. Cleven et al., Eur. Phys. J., **A50**, 149 (2014), arXiv:1405.2242.
[1365] S. Fajfer and A. Prapotnik Brdnik, Phys. Rev., **D92**, 074047 (2015), arXiv:1506.02716.
[1366] S. Fajfer and A. Prapotnik Brdnik, Eur. Phys. J., **C76**(10), 537 (2016), arXiv:1606.06943.
[1367] L. Liu et al., Phys. Rev., **D87**(1), 014508 (2013), arXiv:1208.4535.
[1368] M. Albaladejo, Phys. Lett., **B767**, 465–469 (2017), arXiv:1610.06727.
[1369] M-L. Du et al., Phys. Rev., **D98**(9), 094018 (2018), arXiv:1712.07957.
[1370] R. Aaij et al., LHCb, Phys. Rev., **D90**(7), 072003 (2014), arXiv:1407.7712.
[1371] N. A. Tornqvist, Phys. Lett., **B590**, 209–215 (2004), arXiv:hep-ph/0402237.
[1372] E. S. Swanson, Phys. Lett., **B588**, 189–195 (2004), arXiv:hep-ph/0311229.
[1373] D. Gamermann and E. Oset, Phys. Rev., **D80**, 014003 (2009), arXiv:0905.0402.
[1374] T. Mehen, Phys. Rev., **D92**(3), 034019 (2015), arXiv:1503.02719.
[1375] Q. Wang et al., Phys. Rev. Lett., **111**(13), 132003 (2013), arXiv:1303.6355.
[1376] M. Cleven et al., Eur. Phys. J., **A47**, 120 (2011), arXiv:1107.0254.
[1377] M. Albaladejo et al., Phys. Lett., **B755**, 337–342 (2016), arXiv:1512.03638.
[1378] C. Hanhart et al., Phys. Lett., **B739**, 375–382 (2014), arXiv:1407.7452.
[1379] A. Esposito et al., Phys. Lett., **B746**, 194–201 (2015), arXiv:1409.3551.
[1380] F-K. Guo et al., Phys. Lett., **B725**, 127–133 (2013), arXiv:1306.3096.
[1381] G.-J. Ding, Phys. Rev., **D79**, 014001 (2009), arXiv:0809.4818.
[1382] Q. Wang et al., Phys. Rev., **D89**(3), 034001 (2014), arXiv:1309.4303.
[1383] L. Ma et al., Phys. Rev., **D91**(3), 034032 (2015), arXiv:1406.6879.
[1384] A. E. Bondar and M. B. Voloshin, Phys. Rev., **D93**(9), 094008 (2016), arXiv:1603.08436.
[1385] F.-K. Guo et al., Phys. Rev. Lett., **103**, 082003, [Erratum: Phys. Rev. Lett. 104, 109901 (2010)] (2009), arXiv:0907.0521.
[1386] F.-K. Guo and U.-G. Meißner, Phys. Rev. Lett., **108**, 112002 (2012), arXiv:1111.1151.
[1387] F.-K. Guo et al., Phys. Lett., **B760**, 417–421 (2016), arXiv:1604.00770.
[1388] F.-K. Guo et al., Phys. Rev., **D83**, 034013 (2011), arXiv:1008.3632.
[1389] F.-K. Guo et al., Phys. Rev. Lett., **105**, 162001 (2010), arXiv:1007.4682.
[1390] D.M.J. Lovelock et al., Phys. Rev. Lett., **54**, 377–380 (1985).
[1391] D. Besson et al., CLEO, Phys. Rev. Lett., **54**, 381 (1985).
[1392] N. A. Tornqvist, Phys. Rev. Lett., **53**, 878 (1984).
[1393] C. Meng and K.-T. Chao, Phys. Rev., **D77**, 074003 (2008), arXiv:0712.3595.
[1394] Y. A. Simonov and A. I. Veselov, Phys. Lett., **B671**, 55–59 (2009), arXiv:0805.4499.
[1395] R. Kaiser et al., Phys. Rev. Lett., **90**, 142001 (2003), arXiv:hep-ph/0208194.
[1396] M. B. Voloshin, Phys. Rev., **D85**, 034024 (2012), arXiv:1201.1222.
[1397] A. E. Bondar et al., Mod. Phys. Lett., **A32**(04), 1750025 (2017), arXiv:1610.01102.
[1398] Y. Lu et al., Phys. Rev., **D95**(3), 034018 (2017), arXiv:1701.00692.
[1399] V. Bernard et al., JHEP, **01**, 019 (2011), arXiv:1010.6018.
[1400] M. Göckeler et al., Phys. Rev., **D86**, 094513 (2012), arXiv:1206.4141.

[1401] M. Doring et al., Eur. Phys. J., **A47**, 139 (2011), arXiv:1107.3988.
[1402] M. Döring et al., Eur. Phys. J., **A47**, 163 (2011), arXiv:1108.0676.
[1403] M. Doring et al., Eur. Phys. J., **A48**, 114 (2012), arXiv:1205.4838.
[1404] Z-H. Guo et al., Phys. Rev., **D95**(5), 054004 (2017), arXiv:1609.08096.
[1405] D. Agadjanov et al., JHEP, **06**, 043 (2016), arXiv:1603.07205.
[1406] Private communication with C. DeTar and S.-H. Lee. (2014).
[1407] G. S. Bali et al., Phys. Rev., **D84**, 094506 (2011), arXiv:1110.2381.
[1408] S. Sasaki and T. Yamazaki, Phys. Rev., **D74**, 114507 (2006), arXiv:hep-lat/0610081.
[1409] V. Baru et al., Phys. Lett., **B726**, 537–543 (2013), arXiv:1306.4108.
[1410] M. Jansen et al., Phys. Rev., **D89**(1), 014033 (2014), arXiv:1310.6937.
[1411] M. Padmanath et al., Phys. Rev., **D92**(3), 034501 (2015), arXiv:1503.03257.
[1412] E.J. Garzon et al., Phys. Rev., **D89**, 014504 (2014), arXiv:1310.0972.
[1413] S. Ozaki and S. Sasaki, Phys. Rev., **D87**, 014506 (2013), arXiv:1211.5512.
[1414] S. Aoki et al., Phys. Rev., **D87**(3), 034512 (2013), arXiv:1212.4896.
[1415] Y. Chen et al., Phys. Rev., **D89**(9), 094506 (2014), arXiv:1403.1318.
[1416] S. Prelovsek et al., Phys. Rev., **D91**(1), 014504 (2015), arXiv:1405.7623.
[1417] M. Albaladejo et al., Eur. Phys. J., **C76**(10), 573 (2016), arXiv:1606.03008.
[1418] G. Bali and M. Hetzenegger, QCDSF, PoS, **LATTICE2010**, 142 (2010), arXiv:1011.0571.
[1419] G. Bali and M. Hetzenegger, QCDSF, PoS, **LATTICE2011**, 123 (2011), arXiv:1111.2222.
[1420] G. S. Bali et al., SESAM, Phys. Rev., **D71**, 114513 (2005), arXiv:hep-lat/0505012.
[1421] Z. S. Brown and K. Orginos, Phys. Rev., **D86**, 114506 (2012), arXiv:1210.1953.
[1422] P. Bicudo et al., Phys. Rev., **D92**(1), 014507 (2015), arXiv:1505.00613.
[1423] P. Bicudo et al., Phys. Rev., **D93**(3), 034501 (2016), arXiv:1510.03441.
[1424] Y. Ikeda et al., Phys. Lett., **B729**, 85–90 (2014), arXiv:1311.6214.
[1425] A. L. Guerrieri et al., PoS, **LATTICE2014**, 106 (2015), arXiv:1411.2247.
[1426] A. Francis et al., Phys. Rev. Lett., **118**(14), 142001 (2017), arXiv:1607.05214.
[1427] D. Janc and M. Rosina, Few Body Syst., **35**, 175–196 (2004), arXiv:hep-ph/0405208.
[1428] A. Del Fabbro et al., Phys. Rev., **D71**, 014008 (2005), arXiv:hep-ph/0408258.
[1429] M. Alberti et al., Phys. Rev., **D95**(7), 074501 (2017), arXiv:1608.06537.
[1430] X. Liao and T. Manke, Phys. Rev., **D65**, 074508 (2002), arXiv:hep-lat/0111049.
[1431] C. Michael (2003), arXiv:hep-ph/0308293.
[1432] Y. Liu and X.-Q. Luo, Phys. Rev., **D73**, 054510 (2006), arXiv:hep-lat/0511015.
[1433] X.-Q. Luo and Y. Liu, Phys. Rev., **D74**, 034502, [Erratum: Phys. Rev. D74, 039902 (2006)] (2006), arXiv:hep-lat/0512044.
[1434] T. Burch and C. Ehmann, Nucl. Phys., **A797**, 33–49 (2007), arXiv:hep-lat/0701001.
[1435] J. J. Dudek et al., Phys. Rev., **D77**, 034501 (2008), arXiv:0707.4162.
[1436] J. J. Dudek and E. Rrapaj, Phys. Rev., **D78**, 094504 (2008), arXiv:0809.2582.
[1437] M. Foster and C. Michael, UKQCD, Phys. Rev., **D59**, 094509 (1999), arXiv:hep-lat/9811010.
[1438] Phys. Rev., **D91**(11), 114503 (2015), arXiv:1410.7069.
[1439] J. J. Dudek et al., Hadron Spectrum, Phys. Rev., **D93**(9), 094506 (2016), arXiv:1602.05122.
[1440] R. A. Briceno et al., Chin. Phys., **C40**(4), 042001 (2016), arXiv:1511.06779.
[1441] A. Esposito et al., Int. J. Mod. Phys., **A30**, 1530002 (2015), arXiv:1411.5997.
[1442] R. Oncala and J. Soto, Phys. Rev., **D96**(1), 014004 (2017), arXiv:1702.03900.
[1443] N. Brambilla et al., Phys. Rev., **D93**(5), 054002 (2016), arXiv:1510.05895.
[1444] S. K. Choi et al., Belle, Phys. Rev. Lett., **89**, 102001, [Erratum: Phys. Rev. Lett. 89, 129901 (2002)] (2002), arXiv:hep-ex/0206002.
[1445] E. J. Eichten et al., Phys. Rev. Lett., **89**, 162002 (2002), arXiv:hep-ph/0206018.
[1446] T. Barnes et al., Phys. Rev., **D72**, 054026 (2005), arXiv:hep-ph/0505002.
[1447] A. Vinokurova et al., Belle, JHEP, **06**, 132 (2015), arXiv:1501.06351.
[1448] T. Iwashita et al., Belle, PTEP, **2014**(4), 043C01 (2014), arXiv:1310.2704.
[1449] V. Bhardwaj et al., Belle, Phys. Rev., **D93**(5), 052016 (2016), arXiv:1512.02672.
[1450] D.-Y. Chen, Eur. Phys. J., **C76**(12), 671 (2016), arXiv:1611.00109.
[1451] A. Bala et al., Belle, Phys. Rev., **D91**(5), 051101 (2015), arXiv:1501.06867.
[1452] B. Aubert et al., BaBar, Phys. Rev. Lett., **96**, 052002 (2006), arXiv:hep-ex/0510070.
[1453] V. N. Baier and V. S. Fadin, Phys. Lett., **B27**, 223–225 (1968).
[1454] A. B. Arbuzov et al., JHEP, **12**, 009 (1998), arXiv:hep-ph/9804430.
[1455] S. Binner et al., Phys. Lett., **B459**, 279–287 (1999), arXiv:hep-ph/9902399.
[1456] M. Benayoun et al., Mod. Phys. Lett., **A14**, 2605–2614 (1999), arXiv:hep-ph/9910523.
[1457] F.-K. Guo et al., Phys. Lett., **B665**, 26–29 (2008), arXiv:0803.1392.
[1458] K. Abe et al., Study of the Y (4260) resonance in $e^+e^-$ collisions with initial state radiation at Belle, In *Proceedings of the 33rd International Conference on High Energy Physics (ICHEP '06)* (2006),

arXiv:hep-ex/0612006.
[1459] B. Aubert et al., BaBar, Phys. Rev. Lett., **98**, 212001 (2007), arXiv:hep-ex/0610057.
[1460] C. P. Shen et al., Belle, Phys. Rev., **D89**(7), 072015 (2014), arXiv:1402.6578.
[1461] C.-Z. Yuan, Chin. Phys., **C38**, 043001 (2014), arXiv:1312.6399.
[1462] Phys. Rev., **D90**(11), 114021 (2014), arXiv:1407.7995.
[1463] V. M. Budnev et al., Phys. Rept., **15**, 181–281 (1975).
[1464] C.-N. Yang, Phys. Rev., **77**, 242–245 (1950).
[1465] K. Abe et al., Belle, Phys. Rev. Lett., **89**, 142001 (2002), arXiv:hep-ex/0205104.
[1466] E. Braaten and J. Lee, Phys. Rev., **D67**, 054007, [Erratum: Phys. Rev. D72, 099901 (2005)] (2003), arXiv:hep-ph/0211085.
[1467] J. P. Ma and Z. G. Si, Phys. Rev., **D70**, 074007 (2004), arXiv:hep-ph/0405111.
[1468] A. E. Bondar and V. L. Chernyak, Phys. Lett., **B612**, 215–222 (2005), arXiv:hep-ph/0412335.
[1469] Y.-J. Zhang et al., Phys. Rev. Lett., **96**, 092001 (2006), arXiv:hep-ph/0506076.
[1470] H.-M. Choi and C.-R. Ji, Phys. Rev., **D76**, 094010 (2007), arXiv:0707.1173.
[1471] K. Abe et al., Belle, Phys. Rev., **D70**, 071102 (2004), arXiv:hep-ex/0407009.
[1472] B. Aubert et al., BaBar, Phys. Rev., **D72**, 031101 (2005), arXiv:hep-ex/0506062.
[1473] P. Pakhlov et al., Belle, Phys. Rev., **D79**, 071101 (2009), arXiv:0901.2775.
[1474] A. V. Berezhnoy and A. K. Likhoded, Phys. Atom. Nucl., **67**, 757–761, [Yad. Fiz.67,778(2004)] (2004), arXiv:hep-ph/0303145.
[1475] A. V. Berezhnoy and A. K. Likhoded, Phys. Atom. Nucl., **70**, 478–484 (2007), arXiv:hep-ph/0602041.
[1476] Y. Kato et al., Belle, Phys. Rev., **D89**(5), 052003 (2014), arXiv:1312.1026.
[1477] B. Aubert et al., BaBar, Phys. Rev. Lett., **101**, 071801 (2008), arXiv:0807.1086.
[1478] B. Aubert et al., BaBar, Phys. Rev. Lett., **103**, 161801 (2009), arXiv:0903.1124.
[1479] G. Bonvicini et al., CLEO, Phys. Rev., **D81**, 031104 (2010), arXiv:0909.5474.
[1480] P. Krokovny, talk at ichep 2014 ().
[1481] S. Sandilya et al., Belle, Phys. Rev. Lett., **111**(11), 112001 (2013), arXiv:1306.6212.
[1482] S. Dobbs et al., Phys. Rev. Lett., **109**, 082001 (2012), arXiv:1204.4205.
[1483] M. Artuso et al., CLEO, Phys. Rev. Lett., **94**, 032001 (2005), arXiv:hep-ex/0411068.
[1484] H. S. Chung et al., Phys. Lett., **B697**, 48–51 (2011), arXiv:1011.1554.
[1485] M. B. Voloshin, Mod. Phys. Lett., **A19**, 2895–2898 (2004), arXiv:hep-ph/0410368.
[1486] S. E. Csorna et al., First observation of Upsilon (1D) states, In *High energy physics. Proceedings, 31st International Conference, ICHEP 2002, Amsterdam, Netherlands, July 25-31, 2002* (2002), arXiv:hep-ex/0207060.
[1487] G. Bonvicini et al., CLEO, Phys. Rev., **D70**, 032001 (2004), arXiv:hep-ex/0404021.
[1488] P. del Amo Sanchez et al., BaBar, Phys. Rev., **D82**, 111102 (2010), arXiv:1004.0175.
[1489] E. Rice et al., Phys. Rev. Lett., **48**, 906–910 (1982).
[1490] P. Moxhay and J. L. Rosner, Phys. Rev., **D28**, 1132 (1983).
[1491] A.M. Badalian et al., Phys.Atom.Nucl., **73**, 138–149 (2010), arXiv:0903.3643.
[1492] J.P. Lees et al., BaBar, Phys. Rev., **D84**, 091101 (2011), arXiv:1102.4565.
[1493] J.P. Lees et al., BaBar, Phys. Rev., **D84**, 011104 (2011), arXiv:1105.4234.
[1494] M. B. Voloshin, Sov.J.Nucl. Phys., **43**, 1011 (1986).
[1495] Y.-P. Kuang and T.-M. Yan, Phys. Rev., **D41**, 155 (1990).
[1496] S.F. Tuan, Mod. Phys. Lett., **A7**, 3527–3540 (1992).
[1497] D. M. Asner et al., CLEO, Phys. Rev., **D78**, 091103 (2008), arXiv:0808.0933.
[1498] S. Dobbs et al., Phys. Rev., **D86**, 052003 (2012), arXiv:1205.5070.
[1499] C.P. Shen et al., Belle, Phys. Rev., **D88**(1), 011102 (2013), arXiv:1305.5887.
[1500] D. Cronin-Hennessy et al., CLEO, Phys. Rev., **D76**, 072001 (2007), arXiv:0706.2317.
[1501] C. Cawlfield et al., CLEO, Phys. Rev., **D73**, 012003 (2006), arXiv:hep-ex/0511019.
[1502] X.-H. Liu et al., Eur. Phys. J., **C73**(1), 2284 (2013), arXiv:1212.4066.
[1503] F.-K. Guo et al., Phys. Lett., **B658**, 27–32 (2007), arXiv:hep-ph/0601120.
[1504] Y-H. Chen et al., Phys. Rev., **D95**(3), 034022 (2017), arXiv:1611.00913.
[1505] J.P. Lees et al., BaBar, Phys. Rev., **D84**, 072002 (2011), arXiv:1104.5254.
[1506] J. P. Lees et al., BaBar, Phys. Rev., **D90**(11), 112010 (2014), arXiv:1410.3902.
[1507] U. Heintz et al., Phys. Rev., **D46**, 1928–1940 (1992).
[1508] M. Kornicer et al., CLEO, Phys. Rev., **D83**, 054003 (2011), arXiv:1012.0589.
[1509] Y.-J. Gao et al. (2007), arXiv:hep-ph/0701009.
[1510] B. Aubert et al., BaBar, Phys. Rev. Lett., **102**, 012001 (2009), arXiv:0809.4120.
[1511] Seiji Ono, A. I. Sanda, and N. A. Tornqvist, Phys. Rev., **D34**, 186 (1986).
[1512] I. Adachi et al., Belle, Phys. Rev. Lett., **108**, 032001 (2012), arXiv:1103.3419.
[1513] C. B. Lang et al., Phys. Lett., **B750**, 17–21 (2015), arXiv:1501.01646.
[1514] M. B. Voloshin, Phys. Rev., **D95**(5), 054017 (2017), arXiv:1701.03064.

[1515] C. B. Lang et al., Phys. Rev., **D90**(3), 034510 (2014), arXiv:1403.8103.
[1516] D. Mohler et al., Phys. Rev. Lett., **111**(22), 222001 (2013), arXiv:1308.3175.
[1517] T.-P. Cheng and L.-F. Li, Phys. Rev., **D16**, 1425 (1977).
[1518] B. W. Lee and R. E. Shrock, Phys. Rev., **D16**, 1444 (1977).
[1519] J. R. Ellis et al., Eur. Phys. J., **C14**, 319–334 (2000), arXiv:hep-ph/9911459.
[1520] A. Dedes et al., Phys. Lett., **B549**, 159–169 (2002), arXiv:hep-ph/0209207.
[1521] A. Brignole and A. Rossi, Phys. Lett., **B566**, 217–225 (2003), arXiv:hep-ph/0304081.
[1522] A. Masiero et al., Nucl. Phys., **B649**, 189–204 (2003), arXiv:hep-ph/0209303.
[1523] J. Hisano et al., JHEP, **12**, 030 (2009), arXiv:0904.2080.
[1524] S. R. Choudhury et al., Phys. Rev., **D75**, 055011 (2007), arXiv:hep-ph/0612327.
[1525] G. Cvetic et al., Phys. Rev., **D66**, 034008, [Erratum: Phys. Rev. D68, 059901 (2003)] (2002), arXiv:hep-ph/0202212.
[1526] S. Davidson et al., Z. Phys., **C61**, 613–644 (1994), arXiv:hep-ph/9309310.
[1527] C.-X. Yue et al., Phys. Lett., **B547**, 252–256 (2002), arXiv:hep-ph/0209291.
[1528] A. G. Akeroyd et al., Phys. Rev., **D79**, 113010 (2009), arXiv:0904.3640.
[1529] R. Harnik et al., JHEP, **03**, 026 (2013), arXiv:1209.1397.
[1530] A. Celis et al., Phys. Rev., **D89**, 013008 (2014), arXiv:1309.3564.
[1531] Y. Omura, JHEP, **05**, 028 (2015), arXiv:1502.07824.
[1532] A. Goudelis, O. Lebedev, and J.-H. Park, Phys. Lett., **B707**, 369–374 (2012), arXiv:1111.1715.
[1533] A. M. Baldini et al., MEG, Eur. Phys. J., **C76**(8), 434 (2016), arXiv:1605.05081.
[1534] V. Khachatryan et al., CMS, Phys. Lett., **B749**, 337–362 (2015), arXiv:1502.07400.
[1535] CMS, CMS (2016).
[1536] G. Aad et al., ATLAS, Eur. Phys. J., **C77**(2), 70 (2017), arXiv:1604.07730.
[1537] J. R. Ellis et al., Phys. Rev., **D66**, 115013 (2002), arXiv:hep-ph/0206110.
[1538] K. Hayasaka et al., Belle, Phys. Lett., **B666**, 16–22 (2008), arXiv:0705.0650.
[1539] K. Hayasaka et al., Phys. Lett., **B687**, 139–143 (2010), arXiv:1001.3221.
[1540] Y. Miyazaki et al., Belle, Phys. Lett., **B648**, 341–350 (2007), arXiv:hep-ex/0703009.
[1541] I. I. Bigi and A. I. Sanda, Phys. Lett., **B625**, 47–52 (2005), arXiv:hep-ph/0506037.
[1542] G. Calderon et al., Phys. Rev., **D75**, 076001 (2007), arXiv:hep-ph/0702282.
[1543] Y. Grossman and Y. Nir, JHEP, **04**, 002 (2012), arXiv:1110.3790.
[1544] J. P. Lees et al., BaBar, Phys. Rev., **D85**, 031102, [Erratum: Phys. Rev. D85, 099904 (2012)] (2012), arXiv:1109.1527.
[1545] H. Zeen Devi et al., Phys. Rev., **D90**(1), 013016 (2014), arXiv:1308.4383.
[1546] V. Cirigliano et al., Phys. Rev. Lett., **120**(14), 141803 (2018), arXiv:1712.06595.
[1547] F.-S. Yu et al., Phys. Rev. Lett., **119**(18), 181802 (2017), arXiv:1707.09297.
[1548] J. H. Kuhn and E. Mirkes, Z.Phys., **C56**, 661–672 (1992).
[1549] J. H. Kuhn and E. Mirkes, Phys. Lett., **B398**, 407–414 (1997), arXiv:hep-ph/9609502.
[1550] Y. S. Tsai, Phys. Rev., **D51**, 3172–3181 (1995), arXiv:hep-ph/9410265.
[1551] C. A. Nelson et al., Phys. Rev., **D50**, 4544–4557 (1994).
[1552] S. Y. Choi et al., Phys. Rev., **D52**, 1614–1626 (1995), arXiv:hep-ph/9412203.
[1553] U. Kilian et al., Z. Phys., **C62**, 413–419 (1994).
[1554] K. Kiers et al., Phys. Rev., **D78**, 113008 (2008), arXiv:0808.1707.
[1555] N. Mileo et al., Phys. Rev., **D91**(7), 073006 (2015), arXiv:1410.1909.
[1556] L. Beldjoudi and Tran N. Truong, Phys. Lett., **B351**, 357–368 (1995), arXiv:hep-ph/9411423.
[1557] D. Kimura et al., CP violation in tau decays, In *Heavy Quarks and Leptons 2008 (HQ&L08) Melbourne, Australia, June 5-9, 2008* (2009), arXiv:0905.1802.
[1558] D. Kimura et al., PTEP, **2013**, 053B03, [Erratum: PTEP2014,no.8,089202(2014)] (2013), arXiv:1201.1794.
[1559] M. Jamin et al., Phys. Lett., **B640**, 176–181 (2006), arXiv:hep-ph/0605096.
[1560] M. Antonelli et al., FlaviaNet Working Group on Kaon Decays, Eur. Phys. J., **C69**, 399–424 (2010), arXiv:1005.2323.
[1561] M. Gourdin, Phys. Rept., **11**, 29 (1974).
[1562] H. Leutwyler, Electromagnetic form-factor of the pion, In *Continuous advances in QCD. Proceedings, Conference, Minneapolis, USA, May 17-23, 2002*, pages 23–40 (2002), arXiv:hep-ph/0212324.
[1563] G. Colangelo, Nucl. Phys. Proc. Suppl., **162**, 256–259, [,256(2006)] (2006).
[1564] V. Bernard, JHEP, **06**, 082 (2014), arXiv:1311.2569.
[1565] H. Neufeld and H. Rupertsberger, Z. Phys., **C68**, 91–102 (1995).
[1566] S. Nussinov and A. Soffer, Phys. Rev., **D78**, 033006 (2008), arXiv:0806.3922.
[1567] S. Descotes-Genon and B. Moussallam, Eur. Phys. J., **C74**, 2946 (2014), arXiv:1404.0251.
[1568] P. Avery et al., CLEO, Phys. Rev., **D64**, 092005 (2001), arXiv:hep-ex/0104009.
[1569] D. Aston et al., Nucl. Phys., **B296**, 493–526 (1988).

[1570] R. F. Dashen and M. Weinstein, Phys. Rev. Lett., **22**, 1337–1340 (1969).
[1571] J. Gasser and H. Leutwyler, Nucl. Phys., **B250**, 517–538 (1985).
[1572] Matthias Jamin et al., Nucl. Phys., **B622**, 279–308 (2002), arXiv:hep-ph/0110193.
[1573] D. Epifanov et al., Belle, Phys. Lett., **B654**, 65–73 (2007), arXiv:0706.2231.
[1574] V. Baru et al., Eur. Phys. J., **A23**, 523–533 (2005), arXiv:nucl-th/0410099.
[1575] V. Bernard et al., Phys. Rev., **D44**, 3698–3701 (1991).
[1576] M. Bischofberger et al., Belle, Phys. Rev. Lett., **107**, 131801 (2011), arXiv:1101.0349.
[1577] W. Fetscher et al., Phys. Lett., **B173**, 102–106 (1986).
[1578] W. Fetscher and H. J. Gerber, Adv. Ser. Direct. High Energy Phys., **14**, 657–705 (1995).
[1579] A. B. Arbuzov and T. V. Kopylova, JHEP, **09**, 109 (2016), arXiv:1605.06612.
[1580] A. Flores-Tlalpa et al., JHEP, **04**, 185 (2016), arXiv:1508.01822.
[1581] Y.-S. Tsai, Phys. Rev., **D4**, 2821, [Erratum: Phys. Rev. D13, 771 (1976)] (1971).
[1582] W. Fetscher, Phys. Rev., **D42**, 1544–1567 (1990).
[1583] K. Tamai, Nucl. Phys., **B668**, 385–402 (2003).
[1584] S. Jadach and Z. Was, Acta Phys. Polon., **B15**, 1151, [Erratum: Acta Phys. Polon. B16, 483 (1985)] (1984).
[1585] A. B. Arbuzov et al., JHEP, **10**, 001 (1997), arXiv:hep-ph/9702262.
[1586] D. A. Epifanov, Belle, Nucl. Part. Phys. Proc., **287-288**, 7–10 (2017).
[1587] N. Shimizu et al., Belle, PTEP, **2018**(2), 023C01 (2018), arXiv:1709.08833.
[1588] S. Weinberg, Phys. Rev., **112**, 1375–1379 (1958).
[1589] N. Severijns et al., Rev. Mod. Phys., **78**, 991–1040 (2006), arXiv:nucl-ex/0605029.
[1590] M. Bourquin et al., Bristol-Geneva-Heidelberg-Orsay-Rutherford-Strasbourg, Z. Phys., **C12**, 307 (1982).
[1591] C. Leroy and J. Pestieau, Phys. Lett., **B72**, 398–399 (1978).
[1592] B. Aubert et al., BaBar, Phys. Rev. Lett., **103**, 041802 (2009), arXiv:0904.3080.
[1593] N. Paver and Riazuddin, Phys. Rev., **D86**, 037302 (2012), arXiv:1205.6636.
[1594] A. Pich, Phys. Lett., **B196**, 561–565 (1987).
[1595] R. Escribano et al., Phys. Rev., **D94**(3), 034008 (2016), arXiv:1601.03989.
[1596] M. Fujikawa et al., Belle, Phys. Rev., **D78**, 072006 (2008), arXiv:0805.3773.
[1597] Z.-H. Guo and J. A. Oller, Phys. Rev., **D84**, 034005 (2011), arXiv:1104.2849.
[1598] C. Adolph et al., COMPASS, Phys. Lett., **B740**, 303–311 (2015), arXiv:1408.4286.
[1599] S. Nussinov and A. Soffer, Phys. Rev., **D80**, 033010 (2009), arXiv:0907.3628.
[1600] P. del Amo Sanchez et al., BaBar, Phys. Rev., **D83**, 032002 (2011), arXiv:1011.3917.
[1601] K. Hayasaka, Belle, PoS, **EPS-HEP2009**, 374 (2009).
[1602] E. A. GarcÈs et al., JHEP, **12**, 027 (2017), arXiv:1708.07802.
[1603] K. Inami et al., Belle, Phys. Lett., **B672**, 209–218 (2009), arXiv:0811.0088.
[1604] J. E. Bartelt et al., CLEO, Phys. Rev. Lett., **76**, 4119–4123 (1996).
[1605] B. Aubert et al., BaBar, Phys. Rev., **D77**, 112002 (2008), arXiv:0803.0772.
[1606] T. Bergfeld et al., CLEO, Phys. Rev. Lett., **79**, 2406–2410 (1997), arXiv:hep-ex/9706020.
[1607] S. U. Chung et al., Phys. Rev., **165**, 1491–1532 (1968).
[1608] A. Pich, Prog. Part. Nucl. Phys., **75**, 41–85 (2014), arXiv:1310.7922.
[1609] M. Ablikim et al., BESIII, Phys. Rev., **D90**(1), 012001 (2014), arXiv:1405.1076.
[1610] H. Albrecht et al., ARGUS, Phys. Lett., **B292**, 221–228 (1992).
[1611] G. Abbiendi et al., OPAL, Phys. Lett., **B492**, 23–31 (2000), arXiv:hep-ex/0005009.
[1612] K. Abe et al., Belle, Phys. Rev. Lett., **99**, 011801 (2007), arXiv:hep-ex/0608046.
[1613] D. Atwood and A. Soni, Phys. Rev., **D45**, 2405–2413 (1992).
[1614] K. Inami et al., Belle, Phys. Lett., **B551**, 16–26 (2003), arXiv:hep-ex/0210066.
[1615] S. Schael et al., ALEPH, Phys. Rept., **421**, 191–284 (2005), arXiv:hep-ex/0506072.
[1616] M. Davier et al., Rev. Mod. Phys., **78**, 1043–1109 (2006), arXiv:hep-ph/0507078.
[1617] M. Davier et al., Eur. Phys. J., **C74**(3), 2803 (2014), arXiv:1312.1501.
[1618] R. Barate et al., ALEPH, Eur. Phys. J., **C11**, 599–618 (1999), arXiv:hep-ex/9903015.
[1619] K. Ackerstaff et al., OPAL, Eur. Phys. J., **C7**, 571–593 (1999), arXiv:hep-ex/9808019.
[1620] G. Abbiendi et al., OPAL, Eur. Phys. J., **C35**, 437–455 (2004), arXiv:hep-ex/0406007.
[1621] A. Pich, Talk at 3rd Belle II Theory Interface Platform (B2TiP) Workshop, Tokyo, 26 29 October 2015, `https://kds.kek.jp/indico/event/19103/session/15/contribution/19/material/slides/0.pdf` (2015).
[1622] J. Erler, Rev. Mex. Fis., **50**, 200–202 (2004), arXiv:hep-ph/0211345.
[1623] S. Narison and A. Pich, Phys. Lett., **B211**, 183–188 (1988).
[1624] E. Braaten, Phys. Rev., **D39**, 1458 (1989).
[1625] E. Braaten et al., Nucl. Phys., **B373**, 581–612 (1992).
[1626] Workshop on the determination of the fundamental parameters in qcd, In *Mainz Institute for*

*Theoretical Physics workshop, Johannes Gutenberg University*, Mainz, Germany (2016).
[1627] M. A. Shifman et al., Nucl. Phys., **B147**, 385–447 (1979).
[1628] P. A. Baikov et al., Phys. Rev. Lett., **101**, 012002 (2008), arXiv:0801.1821.
[1629] A. A. Pivovarov, Z. Phys., **C53**, 461–464, [Yad. Fiz.54,1114(1991)] (1992), arXiv:hep-ph/0302003.
[1630] F. Le Diberder and A. Pich, Phys. Lett., **B286**, 147–152 (1992).
[1631] M. Beneke and M. Jamin, JHEP, **09**, 044 (2008), arXiv:0806.3156.
[1632] S. Menke (2009), arXiv:0904.1796.
[1633] K. G. Chetyrkin and A. Kwiatkowski, Z. Phys., **C59**, 525–532 (1993), arXiv:hep-ph/9805232.
[1634] A. Pich and A. Rodriguez-Sanchez, Phys. Rev., **D94**(3), 034027 (2016), arXiv:1605.06830.
[1635] D. Boito et al., Phys. Rev., **D95**(3), 034024 (2017), arXiv:1611.03457.
[1636] G. P. Lepage et al. (2014), arXiv:1404.0319.
[1637] D. Boito et al., Phys. Rev., **D91**(3), 034003 (2015), arXiv:1410.3528.
[1638] M. N. Achasov et al., J. Exp. Theor. Phys., **109**, 379–392, [Zh. Eksp. Teor. Fiz.136,442(2009)] (2009).
[1639] J. P. Lees et al., BaBar, Phys. Rev., **D96**(9), 092009 (2017), arXiv:1709.01171.
[1640] G. Ciezarek et al., Nature, **546**, 227–233 (2017), arXiv:1703.01766.
[1641] E. Gamiz et al., JHEP, **01**, 060 (2003), arXiv:hep-ph/0212230.
[1642] E. Gamiz et al., Phys. Rev. Lett., **94**, 011803 (2005), arXiv:hep-ph/0408044.
[1643] J. C. Hardy and I. S. Towner, Phys. Rev., **C91**(2), 025501 (2015), arXiv:1411.5987.
[1644] A. Pich and J. Prades, JHEP, **06**, 013 (1998), arXiv:hep-ph/9804462.
[1645] A. Pich and J. Prades, JHEP, **10**, 004 (1999), arXiv:hep-ph/9909244.
[1646] P. A. Baikov et al., Phys. Rev. Lett., **95**, 012003 (2005), arXiv:hep-ph/0412350.
[1647] E. Gamiz et al., PoS, **KAON**, 008 (2008), arXiv:0709.0282.
[1648] S. Chen et al., Eur. Phys. J., **C22**, 31–38 (2001), arXiv:hep-ph/0105253.
[1649] K. G. Chetyrkin et al., Nucl. Phys., **B533**, 473–493 (1998), arXiv:hep-ph/9805335.
[1650] J. G. Korner et al., Eur. Phys. J., **C20**, 259–269 (2001), arXiv:hep-ph/0003165.
[1651] K. Maltman, Phys. Rev., **D58**, 093015 (1998), arXiv:hep-ph/9804298.
[1652] J. Kambor and K. Maltman, Phys. Rev., **D62**, 093023 (2000), arXiv:hep-ph/0005156.
[1653] K. Maltman and J. Kambor, Phys. Rev., **D64**, 093014 (2001), arXiv:hep-ph/0107187.
[1654] K. Maltman and C. E. Wolfe, Phys. Lett., **B639**, 283–289 (2006), arXiv:hep-ph/0603215.
[1655] M. Antonelli et al., JHEP, **10**, 070 (2013), arXiv:1304.8134.
[1656] K. Maltman et al., Nucl. Phys. Proc. Suppl., **189**, 175–180 (2009), arXiv:0906.1386.
[1657] K. Maltman, Nucl. Phys. Proc. Suppl., **218**, 146–151 (2011), arXiv:1011.6391.
[1658] K. Maltman et al., Mod. Phys. Lett., **A31**(29), 1630030 (2016).
[1659] R. Hudspith et al., Phys. Lett., **B781**, 206–212 (2018), arXiv:1702.01767.
[1660] I. M. Nugent, BaBar, Nucl. Phys. Proc. Suppl., **253-255**, 38–41 (2014), arXiv:1301.7105.
[1661] S. Jadach et al., Eur.Phys.J., **C22**, 423–430 (2001), arXiv:hep-ph/9905452.
[1662] S. Jadach et al., Phys. Rev., **D94**(7), 074006 (2016), arXiv:1608.01260.
[1663] Stanislaw J. et al., Comput. Phys. Commun., **64**, 275 (1990).
[1664] M. Jezabek et al., Comput. Phys. Commun., **70**, 69 (1992).
[1665] S. Jadach et al., Comput. Phys. Commun., **76**, 361 (1993).
[1666] P. Golonka et al., Comput. Phys. Commun., **174**, 818–835 (2006), hep-ph/0312240.
[1667] E. Barberio et al., Comput. Phys. Commun., **66**, 115 (1991).
[1668] E. Barberio and Z. W*c*as, Comput. Phys. Commun., **79**, 291–308 (1994).
[1669] P. Golonka and Z. W*c* as, Eur. Phys. J., **C45**, 97–107 (2006), hep-ph/0506026.
[1670] S. Jadach et al., Comput. Phys. Commun., **140**, 475–512 (2001), arXiv:hep-ph/0104049.
[1671] O. Shekhovtsova, EPJ Web Conf., **80**, 00054 (2014), arXiv:1410.2428.
[1672] P. Roig, Nucl. Part. Phys. Proc., **260**, 41–46 (2015), arXiv:1410.8559.
[1673] D. M. Asner et al., CLEO, Phys. Rev., **D61**, 012002 (2000), arXiv:hep-ex/9902022.
[1674] T. E. Browder et al., CLEO, Phys. Rev., **D61**, 052004 (2000), arXiv:hep-ex/9908030.
[1675] S. Actis et al., Eur. Phys. J., **C66**, 585–686 (2010), arXiv:0912.0749.
[1676] O. Shekhovtsova et al., Phys. Rev., **D86**, 113008 (2012), arXiv:1203.3955.
[1677] I.M. Nugent et al., Phys. Rev., **D88**(9), 093012 (2013), arXiv:1310.1053.
[1678] Z. Was and J. Zaremba, Eur. Phys. J., **C75**(11), 566, [Erratum: Eur. Phys. J. C76, 465 (2016)] (2015), arXiv:1508.06424.
[1679] M. Chrzaszcz et al. (2016), arXiv:1609.04617.
[1680] A. E. Bondar et al., Comput. Phys. Commun., **146**, 139–153 (2002), hep-ph/0201149.
[1681] J. H. Kuhn and Z. W*c* as, Acta Phys.Polon., **B39**, 147–158 (2008), arXiv:hep-ph/0602162.
[1682] M. Fael et al., Phys. Rev., **D88**(9), 093011 (2013), arXiv:1310.1081.
[1683] M. Fael et al., JHEP, **07**, 153 (2015), arXiv:1506.03416.
[1684] J. Sasaki, Belle, Nucl. Part. Phys. Proc., **287-288**, 212–214 (2017).
[1685] A.B. Arbuzov et al., Eur.Phys.J., **C73**, 2625 (2013), arXiv:1212.6783.

[1686] T. Doan et al., Phys. Lett., **B725**, 92–96 (2013), arXiv:1303.2220.
[1687] S. Jadach et al., Phys. Rev., **D88**(11), 114022 (2013), arXiv:1307.4037.
[1688] G. Nanava et al., Eur. Phys. J., **C70**, 673–688 (2010), arXiv:0906.4052.
[1689] F. Gray, Muon $g-2$ experiment at Fermilab, In *12th Conference on the Intersections of Particle and Nuclear Physics (CIPANP 2015) Vail, Colorado, USA, May 19-24, 2015* (2015), arXiv:1510.00346.
[1690] N. Saito, J-PARC g-2/EDM, AIP Conf. Proc., **1467**, 45–56 (2012).
[1691] G. W. Bennett et al., Muon $g-2$, Phys. Rev., **D73**, 072003 (2006), arXiv:hep-ex/0602035.
[1692] D. Nomura, Acta Phys. Polon., **B46**(11), 2251 (2015).
[1693] F. Jegerlehner, EPJ Web Conf., **118**, 01016 (2016), arXiv:1511.04473.
[1694] F. Jegerlehner, Acta Phys. Polon., **B49**, 1157 (2018), arXiv:1804.07409.
[1695] A. Keshavarzi et al., Phys. Rev., **D97**(11), 114025 (2018), arXiv:1802.02995.
[1696] J. P. Lees et al., BaBar, Phys. Rev., **D86**, 032013 (2012), arXiv:1205.2228.
[1697] M. Ablikim et al., BESIII, Phys. Lett., **B753**, 629–638 (2016), arXiv:1507.08188.
[1698] D. Babusci et al., KLOE, Phys. Lett., **B720**, 336–343 (2013), arXiv:1212.4524.
[1699] F. Ambrosino et al., KLOE, Phys. Lett., **B700**, 102–110 (2011), arXiv:1006.5313.
[1700] F. Ambrosino et al., KLOE, Phys. Lett., **B670**, 285–291 (2009), arXiv:0809.3950.
[1701] R. R. Akhmetshin et al., CMD-2, Phys. Lett., **B648**, 28–38 (2007), arXiv:hep-ex/0610021.
[1702] M. N. Achasov et al., J. Exp. Theor. Phys., **103**, 380–384, [Zh. Eksp. Teor. Fiz.130,437(2006)] (2006), arXiv:hep-ex/0605013.
[1703] K. Hagiwara et al., J. Phys., **G38**, 085003 (2011), arXiv:1105.3149.
[1704] S. J. Brodsky and E. De Rafael, Phys. Rev., **168**, 1620–1622 (1968).
[1705] H. Czyz and J. H. Kuhn, Eur. Phys. J., **C18**, 497–509 (2001), arXiv:hep-ph/0008262.
[1706] M. Davier, Talk at $g-2$ Theory Initiative FermiLab, 3-6 June 2017, `https://indico.fnal.gov/event/13795/session/10/contribution/47/material/slides/0.pdf` (2017).
[1707] H. Czyz et al., Eur. Phys. J., **C27**, 563–575 (2003), arXiv:hep-ph/0212225.
[1708] F. Campanario et al., JHEP, **1402**, 114 (2014), arXiv:1312.3610.
[1709] V. L. Chernyak and A. R. Zhitnitsky, JETP Lett., **25**, 510, [Pisma Zh. Eksp. Teor. Fiz.25,544(1977)] (1977).
[1710] A. V. Efremov and A. V. Radyushkin, Theor. Math. Phys., **42**, 97–110, [Teor. Mat. Fiz.42,147(1980)] (1980).
[1711] G. P. Lepage and S. J. Brodsky, Phys. Lett., **B87**, 359–365 (1979).
[1712] I. I. Balitsky et al., Nucl. Phys., **B312**, 509–550 (1989).
[1713] V. L. Chernyak and I. R. Zhitnitsky, Nucl. Phys., **B345**, 137–172 (1990).
[1714] A. Khodjamirian, Eur. Phys. J., **C6**, 477–484 (1999), arXiv:hep-ph/9712451.
[1715] S. S. Agaev et al., Phys. Rev., **D83**, 054020 (2011), arXiv:1012.4671.
[1716] S. S. Agaev et al., Phys. Rev., **D86**, 077504 (2012), arXiv:1206.3968.
[1717] A. P. Bakulev et al., Phys. Rev., **D86**, 031501 (2012), arXiv:1205.3770.
[1718] S. S. Agaev et al., Phys. Rev., **D90**(7), 074019 (2014), arXiv:1409.4311.
[1719] S. V. Mikhailov et al., Phys. Rev., **D93**(11), 114018 (2016), arXiv:1604.06391.
[1720] H.-N. Li and G. F. Sterman, Nucl. Phys., **B381**, 129–140 (1992).
[1721] H.-C. Hu and H.-N. Li, Phys. Lett., **B718**, 1351–1357 (2013), arXiv:1204.6708.
[1722] H.-N. Li et al., JHEP, **01**, 004 (2014), arXiv:1310.3672.
[1723] B. Aubert et al., BaBar, Phys. Rev., **D80**, 052002 (2009), arXiv:0905.4778.
[1724] S. Uehara et al., Belle, Phys. Rev., **D86**, 092007 (2012), arXiv:1205.3249.
[1725] J. Gronberg et al., CLEO, Phys. Rev., **D57**, 33–54 (1998), arXiv:hep-ex/9707031.
[1726] V. M. Braun et al., Phys. Rev., **D74**, 074501 (2006), arXiv:hep-lat/0606012.
[1727] R. Arthur et al., Phys. Rev., **D83**, 074505 (2011), arXiv:1011.5906.
[1728] Phys. Rev., **D92**(1), 014504 (2015), arXiv:1503.03656.
[1729] B. Melic et al., Phys. Rev., **D68**, 014013 (2003), arXiv:hep-ph/0212346.
[1730] V. M. Braun et al., JHEP, **03**, 142 (2016), arXiv:1601.05937.
[1731] A. E. Blechman et al., Phys. Lett., **B608**, 77–86 (2005), arXiv:hep-ph/0410312.
[1732] L. A. Harland-Lang et al., Eur. Phys. J., **C73**, 2429 (2013), arXiv:1302.2004.
[1733] P. Ball and G.W. Jones, JHEP, **0708**, 025 (2007), arXiv:0706.3628.
[1734] Phys. Rev., **D88**(3), 034023 (2013), arXiv:1307.2797.
[1735] Y.-Y. Charng et al., Phys. Rev., **D74**, 074024, [Erratum: Phys. Rev. D78, 059901 (2008)] (2006), arXiv:hep-ph/0609165.
[1736] X. Liu et al., Phys. Rev., **D86**, 011501 (2012), arXiv:1205.1214.
[1737] J.-F. Hsu et al., Phys. Rev., **D78**, 014020 (2008), arXiv:0711.4987.
[1738] P. del Amo Sanchez et al., BaBar, Phys. Rev., **D84**, 052001 (2011), arXiv:1101.1142.
[1739] B. Aubert et al., BaBar, Phys. Rev., **D74**, 012002 (2006), arXiv:hep-ex/0605018.
[1740] Y. Grossmann, JHEP, **04**, 101 (2015), arXiv:1501.06569.

[1741] S. Alte et al., JHEP, **02**, 162 (2016), arXiv:1512.09135.
[1742] M. Masuda et al., Belle, Phys. Rev., **D93**(3), 032003 (2016), arXiv:1508.06757.
[1743] V. M. Braun et al., JHEP, **06**, 039 (2016), arXiv:1603.09154.
[1744] G. P. Lepage and S. J. Brodsky, Phys. Rev., **D22**, 2157 (1980).
[1745] G. Colangelo et al., JHEP, **09**, 091 (2014), arXiv:1402.7081.
[1746] G. Colangelo et al., Phys. Lett., **B738**, 6–12 (2014), arXiv:1408.2517.
[1747] G. Colangelo et al., JHEP, **09**, 074 (2015), arXiv:1506.01386.
[1748] G. Colangelo et al., Phys. Rev. Lett., **118**(23), 232001 (2017), arXiv:1701.06554.
[1749] G. Colangelo et al., JHEP, **04**, 161 (2017), arXiv:1702.07347.
[1750] M. Hoferichter et al., Phys. Rev., **D86**, 116009 (2012), arXiv:1210.6793.
[1751] M. Hoferichter et al., Phys. Rev., **D96**(11), 114016 (2017), arXiv:1710.00824.
[1752] M. Hoferichter et al., Eur. Phys. J., **C74**, 3180 (2014), arXiv:1410.4691.
[1753] C. Hanhart et al., Eur. Phys. J., **C73**(12), 2668, [Erratum: Eur. Phys. J. C75, 242 (2015)] (2013), arXiv:1307.5654.
[1754] C. W. Xiao et al. (2015), arXiv:1509.02194.
[1755] M. Hoferichter et al., Phys. Rev. Lett., **121**(11), 112002 (2018), arXiv:1805.01471.
[1756] M. Hoferichter et al., JHEP, **10**, 141 (2018), arXiv:1808.04823.
[1757] R. García-Martín and B. Moussallam, Eur. Phys. J., **C70**, 155–175 (2010), arXiv:1006.5373.
[1758] M. Hoferichter et al., Eur. Phys. J., **C71**, 1743 (2011), arXiv:1106.4147.
[1759] L.-Y. Dai and M. R. Pennington, Phys. Rev., **D90**(3), 036004 (2014), arXiv:1404.7524.
[1760] B. Moussallam, Eur. Phys. J., **C73**, 2539 (2013), arXiv:1305.3143.
[1761] M. Hoferichter et al., Int. J. Mod. Phys. Conf. Ser., **35**, 1460400 (2014), arXiv:1309.6877.
[1762] V. Pauk and M. Vanderhaeghen, Eur. Phys. J., **C74**(8), 3008 (2014), arXiv:1401.0832.
[1763] V. Pascalutsa, V. Pauk, and M. Vanderhaeghen, Phys. Rev., **D85**, 116001 (2012), arXiv:1204.0740.
[1764] I. Danilkin and M. Vanderhaeghen, Phys. Rev., **D95**(1), 014019 (2017), arXiv:1611.04646.
[1765] R. Essig et al., Working Group Report: New Light Weakly Coupled Particles, In *Community Summer Study 2013: Snowmass on the Mississippi (CSS2013) Minneapolis, MN, USA, July 29-August 6, 2013* (2013), arXiv:1311.0029.
[1766] C. Boehm et al., J. Phys., **G30**, 279–286 (2004), arXiv:astro-ph/0208458.
[1767] C. Boehm and P. Fayet, Nucl. Phys., **B683**, 219–263 (2004), arXiv:hep-ph/0305261.
[1768] P. Fayet, Phys. Rev., **D74**, 054034 (2006), arXiv:hep-ph/0607318.
[1769] P. Fayet, Phys. Rev., **D75**, 115017 (2007), arXiv:hep-ph/0702176.
[1770] O. Adriani et al., PAMELA, Nature, **458**, 607–609 (2009), arXiv:0810.4995.
[1771] A. A. Abdo et al., Fermi-LAT, Phys. Rev. Lett., **102**, 181101 (2009), arXiv:0905.0025.
[1772] M. Aguilar et al., AMS, Phys. Rev. Lett., **110**, 141102 (2013).
[1773] N. Arkani-Hamed et al., Phys. Rev., **D79**, 015014 (2009), arXiv:0810.0713.
[1774] M. Pospelov and A. Ritz, Phys. Lett., **B671**, 391–397 (2009), arXiv:0810.1502.
[1775] Y. Nomura and J. Thaler, Phys. Rev., **D79**, 075008 (2009), arXiv:0810.5397.
[1776] P. J. Fox and E. Poppitz, Phys. Rev., **D79**, 083528 (2009), arXiv:0811.0399.
[1777] D. N. Spergel et al., Phys. Rev. Lett., **84**, 3760–3763 (2000), arXiv:astro-ph/9909386.
[1778] M. R. Buckley and P. J. Fox, Phys. Rev., **D81**, 083522 (2010), arXiv:0911.3898.
[1779] A. Loeb and N. Weiner, Phys. Rev. Lett., **106**, 171302 (2011), arXiv:1011.6374.
[1780] S. Tulin et al., Phys. Rev., **D87**(11), 115007 (2013), arXiv:1302.3898.
[1781] M. Vogelsberger et al., Mon. Not. Roy. Astron. Soc., **423**, 3740 (2012), arXiv:1201.5892.
[1782] M. Duerr et al., JHEP, **09**, 042 (2016), arXiv:1606.07609.
[1783] M. Pospelov and J. Pradler, Ann. Rev. Nucl. Part. Sci., **60**, 539–568 (2010), arXiv:1011.1054.
[1784] R. Essig et al., Phys. Rev. Lett., **109**, 021301 (2012), arXiv:1206.2644.
[1785] C. Kouvaris and J. Pradler, Phys. Rev. Lett., **118**(3), 031803 (2017), arXiv:1607.01789.
[1786] B. Aubert et al., Search for Invisible Decays of a Light Scalar in Radiative Transitions $\Upsilon_{3S} \to \gamma A^0$, In *Proceedings, 34th International Conference on High Energy Physics (ICHEP 2008)* (2008), arXiv:0808.0017.
[1787] P. del Amo Sanchez et al., BaBar, Phys. Rev. Lett., **107**, 021804 (2011), arXiv:1007.4646.
[1788] R. Essig et al., JHEP, **11**, 167 (2013), arXiv:1309.5084.
[1789] E. Izaguirre et al., Phys. Rev., **D88**, 114015 (2013), arXiv:1307.6554.
[1790] B. Batell et al., Phys. Rev., **D90**(11), 115014 (2014), arXiv:1405.7049.
[1791] P. Coloma et al., JHEP, **04**, 047 (2016), arXiv:1512.03852.
[1792] B. Batell et al., Phys. Rev. Lett., **113**(17), 171802 (2014), arXiv:1406.2698.
[1793] S. Alekhin et al., Rept. Prog. Phys., **79**(12), 124201 (2016), arXiv:1504.04855.
[1794] M. Battaglieri et al., BDX (2016), arXiv:1607.01390.
[1795] B. Batell et al., Phys. Rev., **D80**, 095024 (2009), arXiv:0906.5614.
[1796] B. Holdom, Phys. Lett., **B166**, 196–198 (1986).

[1797] M. Freytsis et al., Phys. Rev., **D81**, 034001 (2010), arXiv:0911.5355.
[1798] M. Cicoli et al., JHEP, **10**, 146 (2012), arXiv:1206.0819.
[1799] M. Millea et al., Phys. Rev., **D92**(2), 023010 (2015), arXiv:1501.04097.
[1800] W. J. Marciano et al., Phys. Rev., **D94**(11), 115033 (2016), arXiv:1607.01022.
[1801] J. Jaeckel and M. Spannowsky, Phys. Lett., **B753**, 482–487 (2016), arXiv:1509.00476.
[1802] I. Brivio et al., Eur. Phys. J., **C77**(8), 572 (2017), arXiv:1701.05379.
[1803] S. Knapen et al., Phys. Rev. Lett., **118**(17), 171801 (2017), arXiv:1607.06083.
[1804] B. Doebrich et al., JHEP, **02**, 018, [JHEP02,018(2016)] (2016), arXiv:1512.03069.
[1805] E. Izaguirre et al., Phys. Rev. Lett., **118**(11), 111802 (2017), arXiv:1611.09355.
[1806] M. Reece and L.-T. Wang, JHEP, **07**, 051 (2009), arXiv:0904.1743.
[1807] R. Essig et al., Phys. Rev., **D80**, 015003 (2009), arXiv:0903.3941.
[1808] B. Batell et al., Phys. Rev., **D79**, 115008 (2009), arXiv:0903.0363.
[1809] J. D. Clarke et al., JHEP, **02**, 123 (2014), arXiv:1310.8042.
[1810] M. J. Dolan et al., JHEP, **03**, 171, [Erratum: JHEP 07, 103 (2015)] (2015), arXiv:1412.5174.
[1811] B. Batell et al., Phys. Rev., **D95**(7), 075003 (2017), arXiv:1606.04943.
[1812] J. P. Lees et al., BaBar, Phys. Rev., **D87**(3), 031102, [Erratum: Phys. Rev. D87, 059903 (2013)] (2013), arXiv:1210.0287.
[1813] J. P. Lees et al., BaBar, Phys. Rev., **D88**(7), 071102 (2013), arXiv:1210.5669.
[1814] J. P. Lees et al., BaBar, Phys. Rev. Lett., **107**, 221803 (2011), arXiv:1108.3549.
[1815] J. P. Lees et al., BaBar, Phys. Rev., **D88**(3), 031701 (2013), arXiv:1307.5306.
[1816] B. Batell et al., Phys. Rev., **D83**, 054005 (2011), arXiv:0911.4938.
[1817] K. Schmidt-Hoberg et al., Phys. Lett., **B727**, 506–510 (2013), arXiv:1310.6752.
[1818] G. Krnjaic, Phys. Rev., **D94**(7), 073009 (2016), arXiv:1512.04119.
[1819] R. Aaij et al., LHCb, Phys. Rev. Lett., **115**(16), 161802 (2015), arXiv:1508.04094.
[1820] N. Fernandez et al., Phys. Rev., **D90**(1), 015029 (2014), arXiv:1404.6599.
[1821] G. K. Yeghiyan, Phys. Rev., **D80**, 115019 (2009), arXiv:0909.4919.
[1822] N. Fernandez et al., Phys. Rev., **D93**(5), 054023 (2016), arXiv:1511.03728.
[1823] S. Godfrey and H. E. Logan, Phys. Rev., **D93**(5), 055014 (2016), arXiv:1510.04659.
[1824] J. P. Lees et al., BaBar, Phys. Rev. Lett., **108**, 211801 (2012), arXiv:1202.1313.
[1825] I. Jaegle, Belle, Phys. Rev. Lett., **114**(21), 211801 (2015), arXiv:1502.00084.
[1826] J. Alexander et al. (2016), arXiv:1608.08632.
[1827] J. P. Lees et al., BaBar, Phys. Rev. Lett., **119**(13), 131804 (2017), arXiv:1702.03327.
[1828] Y. Hochberg et al., Phys. Rev., **D97**(5), 055030 (2018), arXiv:1706.05008.
[1829] C. Boehm et al., JCAP, **1405**, 009 (2014), arXiv:1401.6458.
[1830] M. Ackermann et al., Fermi-LAT, Phys. Rev. Lett., **115**(23), 231301 (2015), arXiv:1503.02641.
[1831] J. P. Lees et al., BaBar, Phys. Rev. Lett., **113**(20), 201801 (2014), arXiv:1406.2980.
[1832] J. Alwall et al., J. High Energy Phys., **07**(arXiv:1405.0301. CERN-PH-TH-2014-064. CP3-14-18. LPN14-066. MCNET-14-09. ZU-TH 14-14), 079. 158 p, Comments: 158 pages, 27 figures; a few references have been added (May 2014).
[1833] R. Essig et al., Phys. Rev., **D82**, 113008 (2010), arXiv:1008.0636.
[1834] S. Jadach et al., Phys. Lett., **B390**, 298–308 (1997), arXiv:hep-ph/9608412.
[1835] D. Karlen, Nucl. Phys., **B289**, 23–35 (1987).
[1836] H. Ikeda et al., Belle, Nucl. Instrum. Meth., **A441**, 401–426 (2000).
[1837] K. Mimasu and V. Sanz, JHEP, **06**, 173 (2015), arXiv:1409.4792.
[1838] M. J. Dolan et al., JHEP, **12**, 094 (2017), arXiv:1709.00009.
[1839] B. Aubert et al., BaBar, Phys. Rev., **D76**, 091102 (2007), arXiv:0707.2798.
[1840] T. Ferber, Acta Phys. Polon., **B46**(11), pp.2067–2414 (2015).
[1841] X. G. He et al., Phys. Rev. D, **43**, R22–R24 (Jan 1991).
[1842] X.-G. He et al., Phys. Rev. D, **44**, 2118–2132 (Oct 1991).
[1843] J. P. Lees et al., BaBar, Phys. Rev., **D94**(1), 011102 (2016), arXiv:1606.03501.
[1844] P. Fayet, Phys. Rev., **D81**, 054025 (2010), arXiv:0910.2587.
[1845] H. Albrecht et al., ARGUS, Phys. Lett., **B179**, 403 (1986).
[1846] D. Besson et al., CLEO, Phys. Rev., **D30**, 1433 (1984).
[1847] O. Tajima et al., Belle, Phys. Rev. Lett., **98**, 132001 (2007), arXiv:hep-ex/0611041.
[1848] B. Aubert et al., BaBar, Phys. Rev. Lett., **103**, 251801 (2009), arXiv:0908.2840.
[1849] F. Wilczek, Phys. Rev. Lett., **39**, 1304 (1977).
[1850] R. E. Shrock and M. Suzuki, Phys. Lett., **B110**, 250 (1982).
[1851] G. Bonneau and F. Martin, Nucl. Phys., **B27**, 381–397 (1971).
[1852] V. N. Baier et al., Nucl. Phys., **B65**, 381–396 (1973).
[1853] G. Rodrigo et al., Eur. Phys. J., **C22**, 81–88 (2001), arXiv:hep-ph/0106132.
[1854] B. McElrath, Phys. Rev., **D72**, 103508 (2005), arXiv:hep-ph/0506151.

[1855] A. David et al., LHC Higgs Cross Section Working Group (2012), arXiv:1209.0040.
[1856] A. Rashed et al., Phys. Rev., **D82**, 054031 (2010), arXiv:1004.5419.
[1857] J. F. Gunion and H. E. Haber, Phys. Rev., **D67**, 075019 (2003), arXiv:hep-ph/0207010.
[1858] P. Bechtle et al., Eur. Phys. J., **C75**(9), 421 (2015), arXiv:1507.06706.
[1859] P. Franzini et al., Phys. Rev., **D35**, 2883–2886 (1987).
[1860] B. Aubert et al., Search for Dimuon Decays of a Light Scalar in Radiative Transitions tau(3S) —¿ gamma A0, In *2009 Aspen Winter Conference on Astronomy: THIRTY YEARS OF MAGNETARS: NEW FRONTIERS Aspen, Colorado, February 1-7, 2009* (2009), arXiv:0902.2176.
[1861] B. Aubert et al., BaBar, Phys. Rev. Lett., **103**, 181801 (2009), arXiv:0906.2219.
[1862] M. A. Sanchis-Lozano, Mod. Phys. Lett., **A17**, 2265–2276 (2002), arXiv:hep-ph/0206156.
[1863] M.-A. Sanchis-Lozano, J. Phys. Soc. Jap., **76**, 044101–044200 (2007), arXiv:hep-ph/0610046.
[1864] J. P. Alexander et al., CLEO, Phys. Rev., **D58**, 052004 (1998), arXiv:hep-ex/9802024.
[1865] G. S. Adams et al., CLEO, Phys. Rev. Lett., **94**, 012001 (2005), arXiv:hep-ex/0409027.
[1866] D. Besson et al., CLEO, Phys. Rev. Lett., **98**, 052002 (2007), arXiv:hep-ex/0607019.
[1867] R. S. Chivukula and H. Georgi, Phys. Lett., **B188**, 99 (1987).
[1868] B. Grinstein and M. B. Wise, Phys. Lett., **B201**, 274–278 (1988).
[1869] M. Misiak and M. Steinhauser, Eur. Phys. J., **C77**(3), 201 (2017), arXiv:1702.04571.
[1870] S. L. Glashow and S. Weinberg, Phys. Rev., **D15**, 1958 (1977).
[1871] W.-S. Hou, Phys. Lett., **B296**, 179–184 (1992).
[1872] F. Mahmoudi and O. Stal, Phys. Rev., **D81**, 035016 (2010), arXiv:0907.1791.
[1873] G. Aad et al., ATLAS, CMS, JHEP, **08**, 045 (2016), arXiv:1606.02266.
[1874] T. P. Cheng and M. Sher, Phys. Rev., **D35**, 3484 (1987).
[1875] R. Harnik et al., JHEP, **03**, 026 (2013), arXiv:1209.1397.
[1876] D. Chang et al., Phys. Rev., **D48**, 217–224 (1993), arXiv:hep-ph/9302267.
[1877] F. Mahmoudi, Comput. Phys. Commun., **180**, 1579–1613 (2009), arXiv:0808.3144.
[1878] M. Aaboud et al., ATLAS, Phys. Lett., **B759**, 555–574 (2016), arXiv:1603.09203.
[1879] A. J. Buras, Weak Hamiltonian, CP violation and rare decays, In *Probing the standard model of particle interactions. Proceedings, Summer School in Theoretical Physics, NATO Advanced Study Institute, 68th session, Les Houches, France, July 28-September 5, 1997. Pt. 1, 2*, pages 281–539 (1998), arXiv:hep-ph/9806471.
[1880] L. Mercolli and C. Smith, Nucl. Phys., **B817**, 1–24 (2009), arXiv:0902.1949.
[1881] C. Smith, Acta Phys. Polon. Supp., **3**, 53–64 (2010), arXiv:0909.4444.
[1882] C. Smith, *Minimal flavor violation*, Habilitation à Diriger des Recherches, Université Grenoble Alpes (2015).
[1883] J. M. Gerard, Z. Phys., **C18**, 145 (1983).
[1884] G. Colangelo et al., Eur. Phys. J., **C59**, 75–98 (2009), arXiv:0807.0801.
[1885] P. Paradisi et al., Phys. Lett., **B668**, 202–209 (2008), arXiv:0805.3989.
[1886] J. Bernon and C. Smith, JHEP, **07**, 038 (2014), arXiv:1404.5496.
[1887] P. Mertens and C. Smith, JHEP, **08**, 069 (2011), arXiv:1103.5992.
[1888] L. J. Hall and L. Randall, Phys. Rev. Lett., **65**, 2939–2942 (1990).
[1889] R. Brommer et al., Eur. Phys. J., **C74**(9), 3059 (2014), arXiv:1402.4024.
[1890] G. Colangelo and G. Isidori, JHEP, **09**, 009 (1998), arXiv:hep-ph/9808487.
[1891] A. J. Buras et al., Nucl. Phys., **B566**, 3–32 (2000), arXiv:hep-ph/9908371.
[1892] G. Isidori et al., JHEP, **08**, 064 (2006), arXiv:hep-ph/0604074.
[1893] E. Nikolidakis and C. Smith, Phys. Rev., **D77**, 015021 (2008), arXiv:0710.3129.
[1894] C. Smith, Minimal Flavor Violation as an alternative to R-parity, In *Proceedings, 34th International Conference on High Energy Physics (ICHEP 2008): Philadelphia, Pennsylvania, July 30-August 5, 2008* (2008), arXiv:0809.3152.
[1895] C. Csaki et al., Phys. Rev., **D85**, 095009 (2012), arXiv:1111.1239.
[1896] C. Smith, Phys. Rev., **D85**, 036005 (2012), arXiv:1105.1723.
[1897] G. Durieux and C. Smith, JHEP, **10**, 068 (2013), arXiv:1307.1355.
[1898] V. Cirigliano et al., Nucl. Phys., **B728**, 121–134 (2005), arXiv:hep-ph/0507001.
[1899] J. A. Casas and A. Ibarra, Nucl. Phys., **B618**, 171–204 (2001), arXiv:hep-ph/0103065.
[1900] A. de Gouvea and P. Vogel, Prog. Part. Nucl. Phys., **71**, 75–92 (2013), arXiv:1303.4097.
[1901] R. Alonso et al., JHEP, **04**, 159 (2014), arXiv:1312.2014.
[1902] F. Feruglio et al., Phys. Rev. Lett., **118**(1), 011801 (2017), arXiv:1606.00524.
[1903] M. Carpentier and S. Davidson, Eur. Phys. J., **C70**, 1071–1090 (2010), arXiv:1008.0280.
[1904] A. Abada et al., Phys. Rev., **D91**(11), 113013 (2015), arXiv:1503.04159.
[1905] Y. Kuno and Y. Okada, Rev. Mod. Phys., **73**, 151–202 (2001), arXiv:hep-ph/9909265.
[1906] B. M. Dassinger et al., JHEP, **10**, 039 (2007), arXiv:0707.0988.
[1907] R. Brüser et al., JHEP, **10**, 082 (2015), arXiv:1506.07786.

[1908] A. A. Petrov and D. V. Zhuridov, Phys. Rev., **D89**(3), 033005 (2014), arXiv:1308.6561.
[1909] A. Brignole and A. Rossi, Nucl. Phys., **B701**, 3–53 (2004), arXiv:hep-ph/0404211.
[1910] R. Mohanta, Eur. Phys. J., **C71**, 1625 (2011), arXiv:1011.4184.
[1911] P. Langacker and M. Plumacher, Phys. Rev., **D62**, 013006 (2000), arXiv:hep-ph/0001204.
[1912] A. Crivellin et al., Phys. Rev., **D92**(5), 054013 (2015), arXiv:1504.07928.
[1913] W. Altmannshofer et al., Phys. Lett., **B762**, 389–398 (2016), arXiv:1607.06832.
[1914] I. Doršner et al., Phys. Rept., **641**, 1–68 (2016), arXiv:1603.04993.
[1915] E. Bečirević et al. (2016), arXiv:1608.07583.
[1916] I. Doršner et al., JHEP, **06**, 108 (2015), arXiv:1502.07784.
[1917] A. Abada et al., JHEP, **12**, 061 (2007), arXiv:0707.4058.
[1918] A. Abada et al., Phys. Rev., **D78**, 033007 (2008), arXiv:0803.0481.
[1919] J. F. Kamenik and M. Nemevsek, JHEP, **11**, 023, [Erratum: JHEP 03, 033 (2014)] (2009), arXiv:0908.3451.
[1920] J. Hisano et al., Phys. Rev., **D53**, 2442–2459 (1996), arXiv:hep-ph/9510309.
[1921] T. Feldmann et al., JHEP, **11**, 078 (2016), arXiv:1608.04124.
[1922] M. Beneke et al., Nucl. Phys., **B906**, 561–614 (2016), arXiv:1508.01705.
[1923] A. Atre et al., JHEP, **05**, 030 (2009), arXiv:0901.3589.
[1924] W. Altmannshofer et al., JHEP, **11**, 202 (2013), arXiv:1308.3653.
[1925] E. Lunghi and A. Soni, Phys. Lett., **B666**, 162–165 (2008), arXiv:0803.4340.
[1926] A. J. Buras and D. Guadagnoli, Phys. Rev., **D78**, 033005 (2008), arXiv:0805.3887.
[1927] W. Altmannshofer et al., Nucl. Phys., **B830**, 17–94 (2010), arXiv:0909.1333.
[1928] M. Blanke and A. J. Buras, Eur. Phys. J., **C76**(4), 197 (2016), arXiv:1602.04020.
[1929] R. Barbieri et al., Eur. Phys. J., **C71**, 1725 (2011), arXiv:1105.2296.
[1930] R. Barbieri et al., Phys. Lett., **B377**, 76–82 (1996), arXiv:hep-ph/9512388.
[1931] A. Pomarol and D. Tommasini, Nucl. Phys., **B466**, 3–24 (1996), arXiv:hep-ph/9507462.
[1932] LHCb, LHCb (2016).
[1933] R. Barbieri et al., Eur. Phys. J., **C71**, 1812 (2011), arXiv:1108.5125.
[1934] R. Barbieri et al., JHEP, **07**, 181 (2012), arXiv:1203.4218.
[1935] R. Barbieri et al., JHEP, **10**, 040 (2012), arXiv:1206.1327.
[1936] A. J. Buras and J. Girrbach, JHEP, **01**, 007 (2013), arXiv:1206.3878.
[1937] S. Dimopoulos and G. F. Giudice, Phys. Lett., **B357**, 573–578 (1995), arXiv:hep-ph/9507282.
[1938] A. G. Cohen et al., Phys. Lett., **B388**, 588–598 (1996), arXiv:hep-ph/9607394.
[1939] G. Blankenburg and J. Jones-Pérez, Eur. Phys. J., **C73**(2), 2289 (2013), arXiv:1210.4561.
[1940] R. Barbieri et al., JHEP, **05**, 105 (2014), arXiv:1402.6677.
[1941] W. Porod, Comput. Phys. Commun., **153**, 275–315 (2003), arXiv:hep-ph/0301101.
[1942] W. Porod and F. Staub, Comput. Phys. Commun., **183**, 2458–2469 (2012), arXiv:1104.1573.
[1943] F. Staub, Comput. Phys. Commun., **185**, 1773–1790 (2014), arXiv:1309.7223.
[1944] W. Porod et al., Eur. Phys. J., **C74**(8), 2992 (2014), arXiv:1405.1434.
[1945] F. Staub et al., Comput. Phys. Commun., **183**, 2165–2206 (2012), arXiv:1109.5147.
[1946] B. Blankenburg et al., Eur. Phys. J., **C72**, 2126 (2012), arXiv:1204.0688.
[1947] J. Jones-Pérez, Eur. Phys. J., **C74**(2), 2772, [Erratum: Eur. Phys. J. C74, 2886 (2014)] (2014), arXiv:1311.1837.
[1948] B. Aubert et al., BaBar, Phys. Rev. Lett., **104**, 021802 (2010), arXiv:0908.2381.
[1949] A. M. Baldini et al. (2013), arXiv:1301.7225.
[1950] A. Crivellin et al., Phys. Rev., **D91**(7), 075006 (2015), arXiv:1503.03477.
[1951] A. Celis et al., Phys. Rev., **D92**(1), 015007 (2015), arXiv:1505.03079.
[1952] A. Crivellin et al., Phys. Rev. Lett., **114**, 151801 (2015), arXiv:1501.00993.
[1953] C. Bobeth et al., JHEP, **04**, 079 (2017), arXiv:1609.04783.
[1954] W. Altmannshofer et al., Phys. Rev., **D94**(9), 095026 (2016), arXiv:1604.08221.
[1955] X. G. He et al., Phys. Rev., **D43**, 22–24 (1991).
[1956] RobR.ert Foot, Mod. Phys. Lett., **A6**, 527–530 (1991).
[1957] X.-G. He et al., Phys. Rev., **D44**, 2118–2132 (1991).
[1958] P. Binetruy et al., Nucl. Phys., **B496**, 3–23 (1997), arXiv:hep-ph/9610481.
[1959] N. F. Bell and R. R. Volkas, Phys. Rev., **D63**, 013006 (2001), arXiv:hep-ph/0008177.
[1960] S. Choubey and W. Rodejohann, Eur. Phys. J., **C40**, 259–268 (2005), arXiv:hep-ph/0411190.
[1961] A. J. Buras and J. Girrbach, JHEP, **12**, 009 (2013), arXiv:1309.2466.
[1962] A. J. Buras et al., JHEP, **02**, 116 (2013), arXiv:1211.1896.
[1963] T. Hurth et al., JHEP, **12**, 053 (2014), arXiv:1410.4545.
[1964] R. Aaij et al., LHCb, JHEP, **06**, 115 (2015), arXiv:1503.07138.
[1965] S. Meinel and D. van Dyk, Phys. Rev., **D94**(1), 013007 (2016), arXiv:1603.02974.
[1966] D. Becirevic et al., Eur. Phys. J., **C76**(3), 134 (2016), arXiv:1602.00881.

[1967] M. Gronau and J. L. Rosner, Phys. Rev., **D59**, 113002 (1999), arXiv:hep-ph/9809384.
[1968] A. J. Buras et al., Nucl. Phys., **B697**, 133–206 (2004), arXiv:hep-ph/0402112.
[1969] S. Baek and D. London, Phys. Lett., **B653**, 249–253 (2007), arXiv:hep-ph/0701181.
[1970] R. Fleischer et al., Eur. Phys. J., **C51**, 55–61 (2007), arXiv:hep-ph/0702275.
[1971] L. Hofer and L. Vernazza, Status of the $B \to \pi K$ puzzle and its relation to $B_s \to \phi\pi$ and $B_s \to \phi\rho$ decays, In *7th International Workshop on the CKM Unitarity Triangle (CKM 2012) Cincinnati, Ohio, USA, September 28-October 2, 2012* (2012), arXiv:1212.4785.
[1972] N. Arkani-Hamed et al., Phys. Rev., **D61**, 116003 (2000), arXiv:hep-ph/9909326.
[1973] B. Grinstein et al., JHEP, **11**, 067 (2010), arXiv:1009.2049.
[1974] R. T. D'Agnolo and D. M. Straub, JHEP, **05**, 034 (2012), arXiv:1202.4759.
[1975] A. J. Buras et al., JHEP, **03**, 088 (2012), arXiv:1112.4477.
[1976] R. Alonso et al., JHEP, **12**, 119 (2016), arXiv:1609.05902.
[1977] F. Pisano and V. Pleitez, Phys. Rev., **D46**, 410–417 (1992), arXiv:hep-ph/9206242.
[1978] P. H. Frampton, Phys. Rev. Lett., **69**, 2889–2891 (1992).
[1979] A. J. Buras et al., JHEP, **1302**, 023 (2013), arXiv:1211.1237.
[1980] A. J. Buras et al., JHEP, **1408**, 039 (2014), arXiv:1405.3850.
[1981] A. J. Buras et al., JHEP, **1402**, 112 (2014), arXiv:1311.6729.
[1982] A. J. Buras and F. De Fazio, JHEP, **03**, 010 (2016), arXiv:1512.02869.
[1983] A. J. Buras and F. De Fazio, JHEP, **08**, 115 (2016), arXiv:1604.02344.
[1984] R. A. Diaz et al., Phys. Rev., **D72**, 035018 (2005), arXiv:hep-ph/0411263.
[1985] A.E. Carcamo Hernandez et al., Phys. Rev., **D73**, 035007 (2006), arXiv:hep-ph/0510421.
[1986] A. J. Buras and J.-M. Gerard, JHEP, **12**, 008 (2015), arXiv:1507.06326.
[1987] J. M. Cabarcas et al., Int. J. Mod. Phys., **A29**, 1450015 (2014), arXiv:1310.1407.
[1988] S. M. Boucenna et al., Phys. Rev., **D92**(5), 053001 (2015), arXiv:1502.07546.
[1989] Jogesh C. Pati and A. Salam, Phys. Rev., **D10**, 275–289, [Erratum: Phys. Rev. D11, 703 (1975)] (1974).
[1990] R. N. Mohapatra and J. C. Pati, Phys. Rev., **D11**, 566–571 (1975).
[1991] R. N. Mohapatra and J. C. Pati, Phys. Rev., **D11**, 2558 (1975).
[1992] G. Senjanovic and R. N. Mohapatra, Phys. Rev., **D12**, 1502 (1975).
[1993] S. Patra et al., Phys. Lett., **B752**, 186–190 (2016), arXiv:1506.03456.
[1994] G. Aad et al., ATLAS, Phys. Lett., **B754**, 302–322 (2016), arXiv:1512.01530.
[1995] V. Khachatryan et al., CMS, Phys. Rev. Lett., **116**(7), 071801 (2016), arXiv:1512.01224.
[1996] K. Hsieh et al., Phys. Rev., **D82**, 035011 (2010), arXiv:1003.3482.
[1997] P. Langacker and S. U. Sankar, Phys. Rev., **D40**, 1569–1585 (1989).
[1998] G. Barenboim et al., Phys. Rev., **D55**, 4213–4221 (1997), arXiv:hep-ph/9611347.
[1999] K. Kiers et al., Phys. Rev., **D66**, 095002 (2002), arXiv:hep-ph/0205082.
[2000] M. Blanke et al., JHEP, **03**, 024 (2012), arXiv:1111.5014.
[2001] A. Crivellin, Phys. Rev., **D81**, 031301 (2010), arXiv:0907.2461.
[2002] C-H. Chen and S-H. Nam, Phys. Lett., **B666**, 462–466 (2008), arXiv:0807.0896.
[2003] R. Feger et al., Phys. Rev., **D82**, 073002 (2010), arXiv:1003.4022.
[2004] A. J. Buras et al., Nucl. Phys., **B843**, 107–142 (2011), arXiv:1007.1993.
[2005] A. Crivellin and S. Pokorski, Phys. Rev. Lett., **114**(1), 011802 (2015), arXiv:1407.1320.
[2006] G. Senjanovic, Nucl. Phys., **B153**, 334–364 (1979).
[2007] R. N. Mohapatra et al., Phys. Rev., **D17**, 2462 (1978).
[2008] D. Chang, Nucl. Phys., **B214**, 435–451 (1983).
[2009] G. C. Branco et al., Nucl. Phys., **B221**, 317–330 (1983).
[2010] H. Harari and M. Leurer, Nucl. Phys., **B233**, 221–231 (1984).
[2011] G. Beall et al., Phys. Rev. Lett., **48**, 848 (1982).
[2012] G. Ecker and W. Grimus, Phys. Lett., **B153**, 279–285, [,359(1985)] (1985).
[2013] J. M. Frere et al., Phys. Rev., **D46**, 337–353, [,329(1991)] (1992).
[2014] G. Barenboim et al., Nucl. Phys., **B478**, 527–543 (1996), arXiv:hep-ph/9608450.
[2015] W. Dekens and D. Boer, Nucl. Phys., **B889**, 727–756 (2014), arXiv:1409.4052.
[2016] T. Heidsieck, *Four Generations versus Left-Right Symmetry: A Comparative Numerical Analysis*, PhD thesis, Munich, Tech. U. (2012).
[2017] S. Bertolini et al., Phys. Rev., **D89**(9), 095028 (2014), arXiv:1403.7112.
[2018] W. Noll, *Alles oder Nichts – Flavorverletzung im Kontext der links-rechts symmetrischen Vereinigung*, PhD thesis, Karlsruher Institut fur Technologie (2016), in German.
[2019] D. Guadagnoli and R. N. Mohapatra, Phys. Lett., **B694**, 386–392 (2011), arXiv:1008.1074.
[2020] G. M. Asatrian and A. N. Ionnisian, Mod. Phys. Lett., **A5**, 1089–1096 (1990).
[2021] G. M. Asatryan and A. N. Ioannisyan, Sov. J. Nucl. Phys., **51**, 858–860, [Yad. Fiz.51,1350(1990)] (1990).

[2022] D. Cocolicchio et al., Phys. Rev., **D40**, 1477 (1989).
[2023] E. Guadagnoli et al., JHEP, **04**, 093 (2011), arXiv:1103.4170.
[2024] R. N. Mohapatra, Phys. Rev., **D46**, 2990–2995 (1992).
[2025] A. G. Akeroyd et al., Phys. Rev., **D76**, 013004 (2007), arXiv:hep-ph/0610344.
[2026] F. Gursey et al., Phys. Lett., **60B**, 177–180 (1976).
[2027] Q. Shafi, Phys. Lett., **79B**, 301–303 (1978).
[2028] Y. Achiman and B. Stech, Phys. Lett., **77B**, 389–393 (1978).
[2029] K. S. Babu et al., JHEP, **05**, 108 (2015), arXiv:1504.00904.
[2030] J. L. Rosner, Comments Nucl. Part. Phys., **15**(5), 195–221 (1986).
[2031] J. L. Hewett and T. G. Rizzo, Phys. Rept., **183**, 193 (1989).
[2032] P. Athron et al., JHEP, **01**, 153 (2015), arXiv:1410.6288.
[2033] J. L. Rosner, Phys. Rev., **D90**(3), 035005 (2014), arXiv:1404.5198.
[2034] O. Eberhardt et al., Phys. Rev. Lett., **109**, 241802 (2012), arXiv:1209.1101.
[2035] E. Kuflik et al., Phys. Rev. Lett., **110**(9), 091801 (2013), arXiv:1204.1975.
[2036] R. Barbieri and D. V. Nanopoulos, Phys. Lett., **91B**, 369–375 (1980).
[2037] H. Georgi, Nucl. Phys., **B156**, 126–134 (1979).
[2038] R. Barbieri et al., Phys. Lett., **90B**, 91–97 (1980).
[2039] S. Bertolini et al., Phys. Rev., **D80**, 015013 (2009), arXiv:0903.4049.
[2040] N. G. Deshpande et al., Phys. Rev., **D47**, 2892–2896 (1993), arXiv:hep-ph/9211232.
[2041] V. V. Dixit and M. Sher, Phys. Rev., **D40**, 3765 (1989).
[2042] B. Stech and Z. Tavartkiladze, Phys. Rev., **D70**, 035002 (2004), arXiv:hep-ph/0311161.
[2043] K. Ishiwata, Z. Ligeti, and M. B. Wise, JHEP, **10**, 027 (2015), arXiv:1506.03484.
[2044] E. Abouzaid et al., KTeV, Phys. Rev., **D83**, 092001 (2011), arXiv:1011.0127.
[2045] J. Kang et al., Phys. Rev., **D77**, 035003 (2008), arXiv:0708.2701.
[2046] F. F. Deppisch et al., Phys. Rev., **D94**(1), 013003 (2016), arXiv:1603.07672.
[2047] F. F. Deppisch et al., Phys. Lett., **B762**, 432–440 (2016), arXiv:1608.05334.
[2048] D. B. Kaplan, Nucl. Phys., **B365**, 259–278 (1991).
[2049] Y. Grossman and M. Neubert, Phys. Lett., **B474**, 361–371 (2000), arXiv:hep-ph/9912408.
[2050] T. Gherghetta and A. Pomarol, Nucl. Phys., **B586**, 141–162 (2000), arXiv:hep-ph/0003129.
[2051] S. J. Huber and Q. Shafi, Phys. Lett., **B498**, 256–262 (2001), arXiv:hep-ph/0010195.
[2052] C. Csaki et al., Phys. Rev., **D70**, 015012 (2004), arXiv:hep-ph/0310355.
[2053] K. Agashe et al., Phys. Rev., **D71**, 016002 (2005), arXiv:hep-ph/0408134.
[2054] M. Blanke et al., JHEP, **03**, 001 (2009), arXiv:0809.1073.
[2055] M. Bauer et al., JHEP, **09**, 017 (2010), arXiv:0912.1625.
[2056] R. Barbieri et al., JHEP, **05**, 069 (2013), arXiv:1211.5085.
[2057] K. Agashe et al., Phys. Lett., **B641**, 62–66 (2006), arXiv:hep-ph/0605341.
[2058] A. J. Buras et al., JHEP, **09**, 076 (2009), arXiv:0905.2318.
[2059] David M. Straub, JHEP, **08**, 108 (2013), arXiv:1302.4651.
[2060] C. Csaki et al., JHEP, **09**, 008 (2008), arXiv:0804.1954.
[2061] A. L. Fitzpatrick et al., Phys. Rev. Lett., **100**, 171604 (2008), arXiv:0710.1869.
[2062] C. Csaki et al., Phys. Rev., **D80**, 016001 (2009), arXiv:0806.3757.
[2063] J. Santiago, JHEP, **12**, 046 (2008), arXiv:0806.1230.
[2064] C. Csaki et al., Phys. Rev., **D81**, 075025 (2010), arXiv:0907.0474.
[2065] C. Delaunay et al., Phys. Rev., **D83**, 115003 (2011), arXiv:1007.0243.
[2066] M. Bauer et al., Phys. Rev. Lett., **108**, 081603 (2012), arXiv:1110.0471.
[2067] M. Redi, Eur. Phys. J., **C72**, 2030 (2012), arXiv:1203.4220.
[2068] O. Matsedonskyi, JHEP, **02**, 154 (2015), arXiv:1411.4638.
[2069] G. Cacciapaglia et al., JHEP, **06**, 085 (2015), arXiv:1501.03818.
[2070] G. Cacciapaglia et al., JHEP, **04**, 006 (2008), arXiv:0709.1714.
[2071] M. Redi and A. Weiler, JHEP, **11**, 108 (2011), arXiv:1106.6357.
[2072] M. Ciuchini et al., JHEP, **07**, 013 (2001), arXiv:hep-ph/0012308.
[2073] M. Bona et al., UTfit, JHEP, **07**, 028 (2005), arXiv:hep-ph/0501199.
[2074] M. Bona et al., UTfit, JHEP, **03**, 080 (2006), arXiv:hep-ph/0509219.
[2075] M. Bona et al., UTfit, JHEP, **03**, 049 (2008), arXiv:0707.0636.
[2076] A. Andreazza et al., Frascati Phys. Ser., **60**, 1–302 (2015).
[2077] Y. Sakaki and H. Tanaka, Phys. Rev., **D87**(5), 054002 (2013), arXiv:1205.4908.
[2078] A. Datta et al., Phys. Rev., **D86**, 034027 (2012), arXiv:1206.3760.
[2079] P. Biancofiore et al., Phys. Rev., **D87**(7), 074010 (2013), arXiv:1302.1042.
[2080] K. Hagiwara et al., Phys. Rev., **D89**(9), 094009 (2014), arXiv:1403.5892.
[2081] S. Bhattacharya et al., Phys. Rev., **D93**(3), 034011 (2016), arXiv:1509.07259.
[2082] D. Becirevic et al. (2016), arXiv:1602.03030.

[2083] R. Alonso et al., Phys. Rev., **D94**(9), 094021 (2016), arXiv:1602.07671.
[2084] B. Dumont et al., Phys. Rev., **D94**(3), 034001 (2016), arXiv:1603.05248.
[2085] G. Aad et al., ATLAS, JHEP, **10**, 189 (2013), arXiv:1308.2631.
[2086] V. Khachatryan et al., CMS, JHEP, **06**, 116 (2015), arXiv:1503.08037.
[2087] G. Aad et al., ATLAS, Eur. Phys. J., **C76**(1), 5 (2016), arXiv:1508.04735.
[2088] V. Khachatryan et al., CMS, Phys. Lett., **B739**, 229–249 (2014), arXiv:1408.0806.
[2089] G. Aad et al., ATLAS, Phys. Rev. Lett., **114**(16), 161801 (2015), arXiv:1501.01325.
[2090] T. Blake and others., Prog. Part. Nucl. Phys., **92**, 50–91 (2017), arXiv:1606.00916.
[2091] C. Bobeth et al., JHEP, **07**, 098 (2010), arXiv:1006.5013.
[2092] S. Descotes-Genon et al., JHEP, **06**, 099 (2011), arXiv:1104.3342.
[2093] C. Bobeth et al., JHEP, **07**, 067 (2011), arXiv:1105.0376.
[2094] W. Altmannshofer et al., JHEP, **04**, 008 (2012), arXiv:1111.1257.
[2095] C. Bobeth et al., JHEP, **01**, 107 (2012), arXiv:1111.2558.
[2096] F. Beaujean et al., JHEP, **08**, 030 (2012), arXiv:1205.1838.
[2097] W. Altmannshofer and D. M. Straub, JHEP, **08**, 121 (2012), arXiv:1206.0273.
[2098] F. Beaujean et al., Eur. Phys. J., **C74**, 2897, [Erratum: Eur. Phys. J. C74, 3179 (2014)] (2014), arXiv:1310.2478.
[2099] R. R. Horgan et al., Phys. Rev. Lett., **112**, 212003 (2014), arXiv:1310.3887.
[2100] T. Hurth and F. Mahmoudi, JHEP, **04**, 097 (2014), arXiv:1312.5267.
[2101] D. Ghosh et al., JHEP, **12**, 131 (2014), arXiv:1408.4097.
[2102] F. Beaujean et al., Eur. Phys. J., **C75**(9), 456 (2015), arXiv:1508.01526.
[2103] T. Hurth et al., Nucl. Phys., **B909**, 737–777 (2016), arXiv:1603.00865.
[2104] A. Karan et al., Phys. Rev., **D95**(11), 114006 (2017), arXiv:1603.04355.
[2105] A. J. Buras et al., JHEP, **11**, 121 (2014), arXiv:1408.0728.
[2106] O. Cata and M. Jung, Phys. Rev., **D92**(5), 055018 (2015), arXiv:1505.05804.
[2107] D. Becirevic et al., Phys. Rev., **D86**, 034034 (2012), arXiv:1205.5811.
[2108] C. Bobeth et al., JHEP, **07**, 106 (2008), arXiv:0805.2525.
[2109] S. Descotes-Genon and J. Virto, JHEP, **04**, 045, [Erratum: JHE P07, 049 (2015)] (2015), arXiv:1502.05509.
[2110] M. Ciuchini et al., JHEP, **06**, 116 (2016), arXiv:1512.07157.
[2111] G. Hiller and F. Kruger, Phys. Rev., **D69**, 074020 (2004), arXiv:hep-ph/0310219.
[2112] C. Niehoff et al., Phys. Lett., **B747**, 182–186 (2015), arXiv:1503.03865.
[2113] D. Becirevic et al., Phys. Rev., **D92**(1), 014016 (2015), arXiv:1503.09024.
[2114] B. Aubert et al., BaBar, Phys. Rev., **D73**, 092001 (2006), arXiv:hep-ex/0604007.
[2115] J. P. Lees et al., BaBar, Phys. Rev., **D86**, 012004 (2012), arXiv:1204.2852.
[2116] W. Altmannshofer et al., Phys. Rev., **D93**(3), 031301 (2016), arXiv:1507.07927.
[2117] D. Melikhov et al., Phys. Lett., **B428**, 171–178 (1998), arXiv:hep-ph/9803269.
[2118] J. Albrecht et al. (2017), arXiv:1709.10308.
[2119] B. Capdevila and Javier others, JHEP, **01**, 093 (2018), arXiv:1704.05340.
[2120] F. Mahmoudi, Comput. Phys. Commun., **178**, 745–754 (2008), arXiv:0710.2067.
[2121] F. Mahmoudi, Comput. Phys. Commun., **180**, 1718–1719 (2009).
[2122] P. Z. Skands et al., JHEP, **07**, 036 (2004), arXiv:hep-ph/0311123.
[2123] B. C. Allanach et al., Comput. Phys. Commun., **180**, 8–25 (2009), arXiv:0801.0045.
[2124] F. Mahmoudi et al., Comput. Phys. Commun., **183**, 285–298 (2012), arXiv:1008.0762.
[2125] V. G. Chobanova et al., JHEP, **07**, 025 (2017), arXiv:1702.02234.
[2126] T. Hurth et al., Phys. Rev., **D96**(9), 095034 (2017), arXiv:1705.06274.
[2127] T. Hurth et al., JHEP, **11**, 176 (2017), arXiv:1708.04474.
[2128] F. U. Bernlochner et al., The GAMBIT Flavour Workgroup, Eur. Phys. J., **C77**(11), 786 (2017), arXiv:1705.07933.
[2129] P. Athron et al., GAMBIT, Eur. Phys. J., **C77**(11), 784, [Addendum: Eur. Phys. J.C78,no.2,98(2018)] (2017), arXiv:1705.07908.
[2130] A. Arbey and F. Mahmoudi, Comput. Phys. Commun., **181**, 1277–1292 (2010), arXiv:0906.0369.
[2131] A. Arbey and F. Mahmoudi, Comput. Phys. Commun., **182**, 1582–1583 (2011).
[2132] J. A. Evans and D. Shih (2016), arXiv:1606.00003.
[2133] J. Aebischer et al., Comput. Phys. Commun., **232**, 71–83 (2018), arXiv:1712.05298.
[2134] D. Foreman-Mackey et al., Publ. Astron. Soc. Pac., **125**, 306–312 (2013), arXiv:1202.3665.
[2135] W. Altmannshofer et al., Phys. Rev., **D96**(5), 055008 (2017), arXiv:1704.05435.
[2136] M. Ciuchini et al., PoS, **ICHEP2016**, 584 (2016), arXiv:1611.04338.
[2137] V. Cacchio et al., JHEP, **11**, 026 (2016), arXiv:1609.01290.
[2138] F. Beaujean et al., J. Phys. Conf. Ser., **664**(7), 072003 (2015).
[2139] J. Rosiek et al., Comput. Phys. Commun., **181**, 2180–2205 (2010), arXiv:1003.4260.

[2140] A. Crivellin et al., Comput. Phys. Commun., **184**, 1004–1032 (2013), arXiv:1203.5023.
[2141] J. Rosiek, Comput. Phys. Commun., **188**, 208–210 (2015), arXiv:1410.0606.
[2142] S. Pokorski et al., Nucl. Phys., **B570**, 81–116 (2000), arXiv:hep-ph/9906206.
[2143] A. J. Buras et al., Nucl. Phys., **B714**, 103–136 (2005), arXiv:hep-ph/0408142.
[2144] A. Dedes et al., Phys. Rev., **D79**, 055006 (2009), arXiv:0812.4320.
[2145] M. Misiak et al., Adv. Ser. Direct. High Energy Phys., **15**, 795–828 (1998), arXiv:hep-ph/9703442.
[2146] A. Dedes et al., JHEP, **11**, 137 (2014), arXiv:1409.6546.
[2147] A. J. Buras et al., Nucl. Phys., **B619**, 434–466 (2001), arXiv:hep-ph/0107048.
[2148] A. J. Buras et al., Nucl. Phys., **B659**, 3 (2003), arXiv:hep-ph/0210145.
[2149] J. Rosiek, Phys. Rev., **D41**, 3464 (1990).
[2150] J. Rosiek (1995), arXiv:hep-ph/9511250.
[2151] A. Crivellin et al., JHEP, **07**, 017 (2011), arXiv:1103.4272.
[2152] D. van Dyk et al., Eos (“btoplxnu” release) zenodo (May 2016).
[2153] T. Feldmann et al., Phys. Rev., **D92**(3), 034013 (2015), arXiv:1503.09063.
[2154] M. Bordone et al., Eur. Phys. J., **C76**(7), 360 (2016), arXiv:1602.06143.
[2155] R. Fleischer, Nucl. Phys. Proc. Suppl., **241-242**, 135–140 (2013), arXiv:1208.2843.
[2156] A. J. Buras et al., Nucl. Phys., **B631**, 219–238 (2002), arXiv:hep-ph/0203135.
[2157] B. Grinstein and D. Pirjol, Phys. Rev., **D70**, 114005 (2004), hep-ph/0404250.
[2158] C. Bobeth et al., Phys. Rev., **D87**(3), 034016, [Phys. Rev.D87,034016(2013)] (2013), arXiv:1212.2321.
[2159] P. Boer et al., JHEP, **01**, 155 (2015), arXiv:1410.2115.